\documentclass[acmtog,nonacm]{acmart}

\usepackage{booktabs} %
\usepackage{graphicx}
\usepackage{amsmath}
\usepackage{multirow}
\usepackage{subfigure}
\usepackage{array}
\usepackage{multicol}
\usepackage{xcolor}
\usepackage{graphbox}
\usepackage{epstopdf}

\usepackage[export]{adjustbox}
\usepackage{tikz}
\usetikzlibrary{calc,fit,arrows.meta}
\newlength\aelength

\AppendGraphicsExtensions{.gif}

\usepackage[ruled]{algorithm2e} %

\SetAlFnt{\small}
\SetAlCapFnt{\small}
\SetAlCapNameFnt{\small}
\SetAlCapHSkip{0pt}

\newcommand{\kaichun}[1]{{\color{black}#1}}
\newcommand{\yj}[1]{{\color{black}#1}}

\newcommand{\yjr}[1]{{\color{black}#1}}

\newcommand{\yjrr}[1]{{\color{black}#1}}

\newcommand{\eg}{\textit{e.g. }}
\newcommand{\ie}{\textit{i.e. }}
\newcommand{\vs}{\textit{v.s. }}
\newcommand{\etal}{\textit{et al.}}
\newcommand{\tabincell}[2]{\begin{tabular}
{@{}#1@{}}#2\end{tabular}}

\makeatletter
\def\subsubsection{\@startsection{subsubsection}{3}%
  \z@{.5\linespacing\@plus.7\linespacing}{.1\linespacing}%
  {\normalfont\itshape}}
\makeatother
\acmJournal{TOG}
\acmYear{2022} 
\acmVolume{1} 
\acmNumber{1} 
\acmArticle{1} 
\acmMonth{8} 
\acmPrice{15.00}
\setcopyright{acmcopyright}

\acmDOI{10.1145/3526212}

\begin{document}
\title{%
DSG-Net: Learning Disentangled Structure and Geometry for 3D Shape Generation 
}

\author{Jie Yang}

\authornotemark[1]
\affiliation{%
 \institution{Institute of Computing Technology, CAS and University of Chinese Academy of Sciences}}

\email{yangjie01@ict.ac.cn}

\author{Kaichun Mo}
\authornotemark[1]
\affiliation{%
\institution{Stanford University}}
 
\email{kaichun@cs.stanford.edu}

\author{Yu-Kun Lai}

\affiliation{%
\institution{Cardiff University}}
\email{LaiY4@cardiff.ac.uk}

\author{Leonidas J. Guibas}
\affiliation{%
\institution{Stanford University}}
\email{guibas@cs.stanford.edu}

\author{Lin Gao}
\authornotemark[2]
\affiliation{%
\institution{Institute of Computing Technology, CAS and University of Chinese Academy of Sciences}}
\email{gaolin@ict.ac.cn}

\authorsaddresses{
$\ast$ Authors contributed equally.\\
$\dag$ Corresponding author.\\
Project webpage: \url{http://geometrylearning.com/dsg-net/}. This is the author's version of the work. It is posted here for your personal use. Not for redistribution. }

\begin{teaserfigure}
\centering
{
  \includegraphics[width=0.95\linewidth]{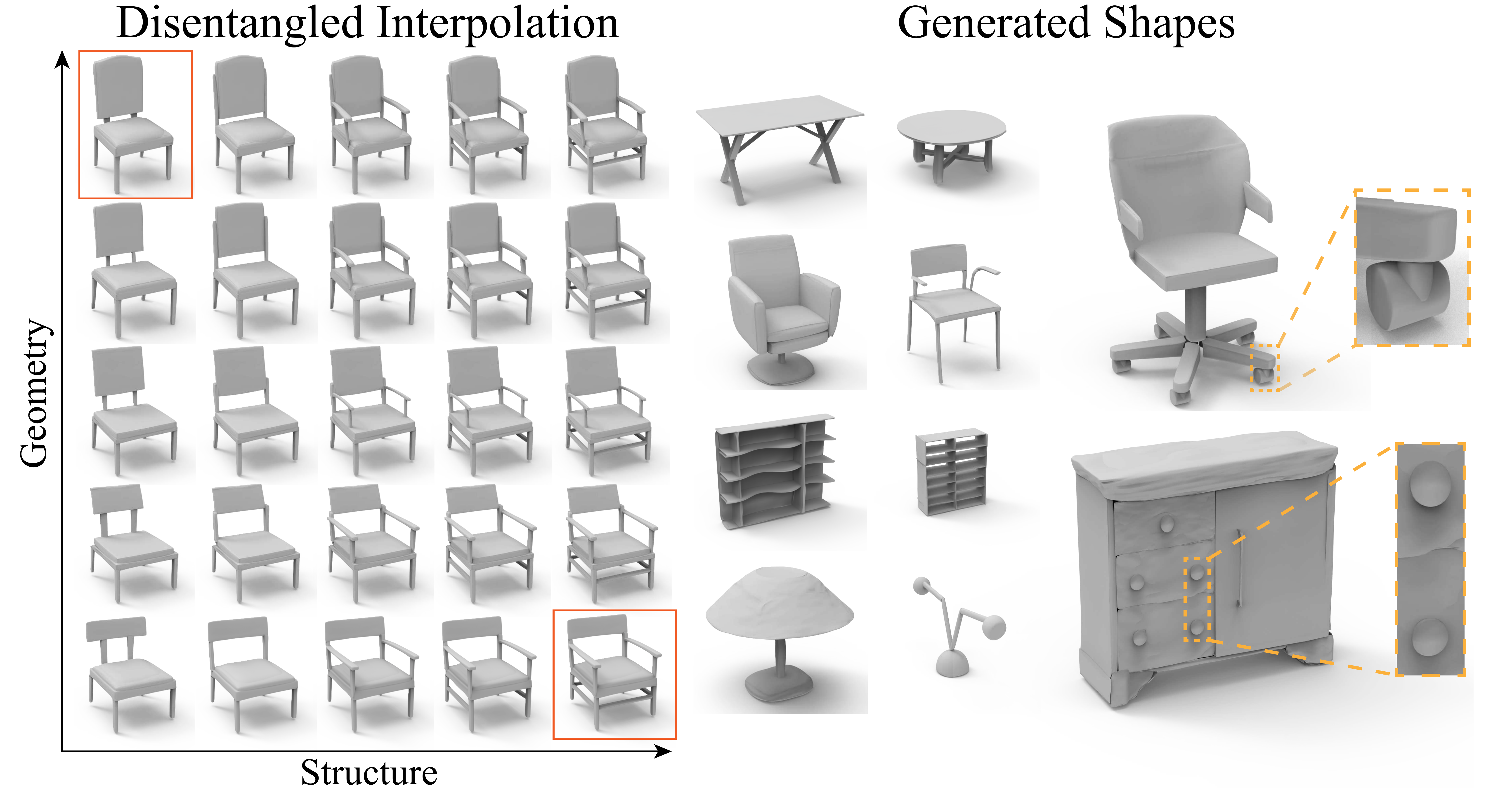}
}
\vspace{-3mm}
\caption{\yjr{Our deep generative network DSG-Net encodes 3D shapes with complex structure and fine geometry in a representation that leverages the synergy between geometry and structure, while disentangling these two aspects as much as possible. 
This enables novel modes of controllable generation for high-quality shapes. %
Left: results of disentangled interpolation.
Here, the top left and bottom right chairs (highlighted with red rectangles) are the input shapes.
The remaining chairs are generated automatically with our DSG-Net, where in each row, the \emph{structure} of the shapes is interpolated while keeping the geometry unchanged, whereas in each column, the \emph{geometry} is interpolated while retaining the structure.
Right: shape generation results with complex structure and fine geometry details by our DSG-Net. We show close-up views in dashed yellow rectangles to highlight local details.}}
\label{fig:teaser}
\Description{teaser figure of DSG-NET}
\end{teaserfigure}

\begin{abstract}
3D shape generation is a fundamental operation in computer graphics. While significant progress has been made, especially with recent deep generative models, it remains a challenge to synthesize high-quality shapes with rich geometric details and complex structures, in a controllable manner. To tackle this, \yj{we introduce DSG-Net, a deep neural network that learns a disentangled structured \& geometric mesh representation for 3D shapes}, where two key aspects of shapes, geometry and structure, are encoded in a synergistic manner to ensure plausibility of the generated shapes, while also being disentangled as much as possible. This supports a range of novel shape generation applications with disentangled control, such as interpolation of structure (geometry) while keeping geometry (structure) unchanged. To achieve this, we simultaneously learn structure and geometry through variational autoencoders (VAEs) in a hierarchical manner for both, with bijective mappings at each level. In this manner, we effectively encode geometry and structure in separate latent spaces, while ensuring their compatibility: the structure is used to guide the geometry and vice versa. At the leaf level, the part geometry is represented using a conditional part VAE, to encode high-quality geometric details, guided by the structure context as the condition. Our method not only supports controllable generation applications, but also produces high-quality synthesized shapes, outperforming state-of-the-art methods. 

\end{abstract}




\maketitle

\section{Introduction}

3D shapes are widely used in computer graphics and computer vision, with applications ranging from modeling, and recognition to rendering. Synthesizing high-quality shapes is therefore highly demanded for many downstream applications. Ideally, the synthesized shapes should be able to contain fine geometric details and complex structures, and the generation process needs to provide high-level control to ensure desired shapes are produced.

Shape generation has been extensively researched in recent years, benefiting especially from the capabilities of deep generative models. This has been true across a variety of 3D representations used to represent generated shapes, including point clouds, voxels, implicit fields, meshes, etc. However, existing methods still have  limitations in representing both complex shape structure as well as geometry details, which is what is required for many downstream applications. 

Moreover, for high-level control in shape generation, it is important to decompose shapes into multiple aspects that can be independently manipulated -- typically geometry and structure (\ie how different parts are related to form the overall shape). On the one hand, geometry and structure are synergistic: the structure of an object may restrict the specific geometric shapes that are plausible, and vice versa. On the other hand, to support high-level control, it is beneficial to derive a representation that disentangles these two aspects as much as possible. Such disentangled and synergistic representations offer significant benefits, including controllable generation of new shapes, \eg interpolating or transferring structure while keeping geometry unchanged, or manipulating geometry while retaining the structure.

Disentangled representations have been widely studied in image generation, allowing different aspects, such as different facial attributes (\eg expression, age, gender) to be manipulated separately, either in supervised~\cite{Xiao2018dnagan,Xiao2018elegant} or unsupervised~\cite{Chen2016infogan} manners. 
For disentanglement of 3D shapes, existing works either focus on specific data kinds such as human faces~\cite{DAbrevaya}, %
or are restricted to intrinsic/extrinsic decomposition, where shape geometry and poses are considered~\cite{Aumentado-Armstrong19}. 
None of these methods can handle more general shape geometry and structure disentanglement.

Most existing deep shape generation works produce synthesized shapes as a whole. This makes it particularly difficult to control the generation, either in a topology or geometry-aware manner. 
Recently, some pioneering works have addressed this shortcoming by considering shape generation using parts and their compositions, leading to improved geometric detail~\cite{gaosdmnet2019} and better handling of complex structure~\cite{mo2019structurenet}. However, neither is able to generate shapes with both complex structures and detailed geometry. Moreover, disentanglement of structure and geometry is not addressed in these works. 
\yj{A notable work SAG-Net~\cite{wu2019sagnet} pioneers the study of 3D shape geometry and structure disentangled/controllable generation where an attention-based Gated Recurrent Unit (GRU) network is proposed to jointly encode shape geometry as voxel maps and structure as fully connected graphs into a single latent space. However, it remains challenging to deal with more fine-grained parts and hierarchical shape structures, as well as fine geometric details. 
\yjr{And also, it achieves the controllable generation by the two-branch variational autoencoder (VAE) and optimization in a single latent space, rather than two independent latent spaces, which do not require optimizations for a disentangled generation.}
}

\begin{figure}[!t]
    \centering
    \includegraphics[width=0.95\linewidth]{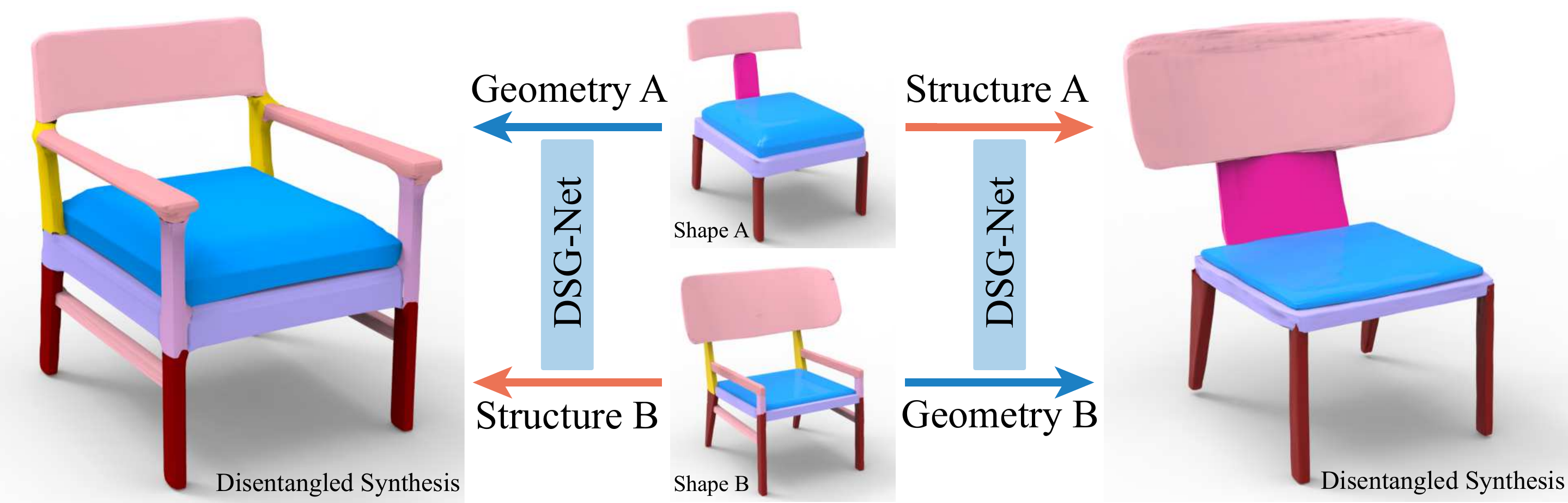}
    \vspace{-3mm}
    \caption{\yj{Our method learns a disentangled latent representation which maps a shape into two separate spaces (the geometric space and the structure space), enabling new disentangled shape manipulation applications, such as disentangled shape reconstruction, as shown in this figure, that combines the structure and geometry from different shapes while ensuring the high quality of shape generation.}}
    \label{fig:motivation}
    \vspace{-3mm}
\end{figure}

\yj{In this paper, we introduce Disentangled Structure \& Geometry Net
(DSG-Net), a novel deep generative model which learns to generate high-quality shape meshes while disentangling the shape geometry and structure generation of two independent spaces as much as possible, enabling many novel disentangled shape manipulation applications, as shown in Figure~\ref{fig:motivation}.
Our work uses the fine-grained shape part hierarchies in the PartNet~\cite{mo2019partnet} dataset.
In our disentangled shape representation, shape structure only includes the hierarchical part graphs with symbolic part semantics and relationships, whereas shape geometry contains the detailed part geometry.
For our network design, we encode both the structure and geometry hierarchies with an $n$-ary tree using separate variational autoencoders (VAEs) with recursive neural network architectures.
Both the geometry and structure information flows along the edges of hierarchical graphs and is aggregated into two latent spaces, allowing these two key aspects to be encoded \emph{separately} in a disentangled manner. 
\yjr{Furthermore, \yjrr{we design a \textit{Cycled Disentanglement}} mechanism to decompose the shape space into two separate spaces and further improve the performance of disentanglement. 
During the training, a new shape that combines the geometry and structure of any two input shapes will be synthesized, and our new design will encourage its geometry and structure to be the same as the two input shapes as much as possible.
Compared to the original framework, it only decomposes the shape with two separate encoders without additional supervision.
Our new designs are capable of improving the disentanglement performance efficiently.
}
Though disentangled, the two latent spaces need to be correlated to ensure the plausibility of the generated shapes. We simultaneously train both structure and geometry VAEs while ensuring necessary communications between them: the geometry follows the part semantics and the inter-part relationship edges in the structure, while the structure requires knowing part geometry for training. }

Our novel solution allows shapes with complex structure and delicate geometry to be represented and synthesized, outperforming state-of-the-art methods, \eg~\cite{mo2019structurenet,gaosdmnet2019}. The disentangled and synergistic formulation allows novel applications, such as shape generation and interpolation with separate control of structure and geometry. New shapes can also be synthesized by mixing structure and geometry from different examples.

Figure~\ref{fig:teaser} demonstrates the capability of our DSG-Net to interpolate shapes with rich geometry and complex structure in the geometry and structure spaces, separately where each row shows interpolation of a structure while keeping geometry unchanged, and each column presents interpolation of geometry while retaining the same structure. 
Through extensive evaluations and comparisons with the state-of-the-art deep neural generative models, our method shows significant advantages and superiority in various shape categories. Our method supports traditional applications such as shape generation, synthesis, and interpolation, but now with independent control on the shape structure and geometry detail, facilitating the design process.

In summary, we make the following key contributions:

\begin{itemize}

\item We propose a novel DSG-Net that learns to decompose shape space into two disentangled latent spaces, encoding the geometry and structure of shapes respectively \yj{(as illustrated in Figure~\ref{fig:motivation})}. \yjr{Furthermore, we design a \yjrr{\textit{Cycled Disentanglement (CycD)} mechanism} to further disentangle the structure and geometry of a shape in a self-supervised manner};

\item DSG-Net enables novel shape synthesis applications that exploit disentangled control of structure and geometry;

\item DSG-Net also allows high-quality shapes with complex structure and fine geometric details to be effectively represented and synthesized, outperforming state-of-the-art methods.

\end{itemize}

\section{Related Work}
\label{sec:related}

3D shape generation is a key research topic in 3D computer vision and graphics.
In this section, we give a brief review on recent advances in 3D shape representations, modeling 3D shape geometry and structure, as well as disentangled representation learning.

\subsection{3D Shape Representations}
In contrast to reaching a great consensus on representing 2D images as pixel grids,
researchers have been exploring a big variation of representations for 3D data.
\yj{
Recent works have developed deep learning frameworks for 
3D voxel grids~\cite{maturana2015voxnet,choy20163d}, 
multi-view 2D rendering of 3D data~\cite{su2015multi,kalogerakis20173d,su2018splatnet}, 
3D point clouds~\cite{qi2017pointnet,qi2017pointnet++,fan2017point,li2018point,achlioptas2018learning,huang2020hausdorff}, 
3D polygonal meshes~\cite{wang2018pixel2mesh,groueix2018atlasnet,chen2019bsp}, and 
3D implicit functions~\cite{park2019deepsdf,mescheder2019occupancy,chen2019learning}.
For more detailed discussion and comparison, we refer the readers to these surveys~\cite{ioannidou2017deep,bronstein2017geometric,ahmed2018deep,xiao2020survey,jin20203d,yuan2021revisit}.

There is a recent trend of studying part-based and structure-aware 3D shape representations, since 3D shapes naturally exhibit compositional part structures.
Part-based shape modeling decomposes complicated shapes into simpler parts for geometric modeling and organizes parts as part sequences or part hierarchies that encode shape part relationships and structures.
Many previous works investigated parsing 3D shapes into parts~\cite{huang2011joint,yi2017learning,tulsiani2017learning,zou20173d}, 
representing 3D shapes as part sequences or hierarchies~\cite{Kim13,van2013co,zhu2018scores,wu2019pq,wang2011symmetry,mo2019structurenet,niu2018im2struct,ganapathi2018parsing}, and 
generating 3D shapes with part structures~\cite{wu2019sagnet,mo2020pt2pc,gaosdmnet2019,li2017grass,kalogerakis2012probabilistic}.
We refer to survey papers~\cite{mitra2014structure,egstar2020_struct,xu2016data} for more comprehensive discussion.
}

\subsection{Modeling Shape Geometry}
There are several different approaches to generating detailed 3D shape geometry: direct methods, patch-based methods, deformation-based methods, and others.
Direct methods exploit decoder networks that output 3D contents in direct feed-forward procedures.
For instance, Choy~et~al.~\shortcite{choy20163d} and Tatarchenko~et~al.~\shortcite{tatarchenko2017octree} directly generate 3D voxel grids using 3D convolutional neural networks.
Fan~et~al.~\shortcite{fan2017point} and Achlioptas~et~al.~\shortcite{achlioptas2018learning} use Multi-layer Perceptrons (MLPs) to directly generate 3D point clouds.
Patch-based methods generate 3D shapes by assembling many local 3D surface patches.
AtlasNet~\cite{groueix2018atlasnet} and Deprelle~et~al.~\shortcite{Deprelle19} learn to reconstruct each 3D shape by a collection of local surface elements or point clouds.
Recent papers~\cite{Genova2019LearningST,jiang2020local} learn local implicit functions that are aggregated together to generate 3D shapes.
Deformation-based methods train neural networks to deform an initial shape template to the output shape.
For example, FoldingNet~\cite{yang2018foldingnet} and Pixel2Mesh~\cite{wang2018pixel2mesh} learn to deform 2D grid surfaces and 3D sphere manifolds to reconstruct 3D target outputs.

In our paper, we choose a deformation-based mesh representation for leaf-node parts, where
we deform a unified unit cube mesh with 5,402 vertices to describe leaf-node part geometry.
Representing 3D shapes as fine-grained part hierarchies~\cite{mo2019partnet,mo2019structurenet}, we find that it is effective and efficient for preserving geometry details for leaf-node parts, as previously shown in the recent works~\cite{gaosdmnet2019,gao2019sparse}. 
\yj{Different from SDM-Net~\cite{gaosdmnet2019}, we introduce a structure-conditioned part geometry VAE, that substantially improves data efficiency and reconstruction performance. Second, we build up bijective mappings between the structure and geometry nodes for synergistic joint learning, which enables disentangled representations for shape structure and geometry.
}
Compared to StructureNet~\cite{mo2019structurenet} that generates 3D point clouds for leaf-node parts, we find our method generates 3D part geometry with sharper edges and more details.

\vspace{-0.5mm}
\subsection{Modeling Shape Structure}
3D objects, especially \yj{man-made} ones, are highly compositional and structured.
Previous works attempt to infer the underlying shape grammars~\cite{kalogerakis2012probabilistic,Chaudhuri2011ProbabilisticRF,wu2016learning}, part-based templates~\cite{Ovsjanikov2011ExplorationOC,Kim13,ganapathi2018parsing}, and shape programs~\cite{Tian2019LearningTI,Sharma2018CSGNetNS}.
\yj{Then, Elena~\etal~\shortcite{balashova2018structure} proposed a structure-aware and voxel-based shape synthesis model that respects structure constraints (landmark points), which are predicted by a learned structure detector.}
There are also many papers investigating generating shapes in the part-by-part manner using consistent part semantics~\cite{wu2019sagnet,schor2019componet,Dubrovina2019CompositeSM,li2019learning} and sequential part instances~\cite{sung2017complementme,wu2019pq}.

Recently, researchers have been investigating representing shapes as part hierarchy, extending part granularity to more fine-grained scales.
\yj{The pioneering work to encode tree structure of object }
GRASS~\cite{li2017grass} uses binary part hierarchies and advocates to use recursive neural networks (RvNN) to hierarchically encode and decode parts along the tree structure.
A follow-up work StructureNet~\cite{mo2019structurenet} further extends the framework to handle $n$-ary part hierarchies with consistent part semantics for an object category~\cite{mo2019partnet}.
SDM-NET~\cite{gao2019sparse} learns to generate structured meshes with deformable parts by leveraging a part graph with rich support and symmetry relations.
Sun~et~al.~\shortcite{Sun2019LearningAH} and Paschalidou~et~al.~\shortcite{Paschalidou2020LearningUH} explore learning hierarchical part decompositions in unsupervised settings.

Our work adopts the hierarchical part representation introduced in StructureNet~\cite{mo2019structurenet} that can represent ShapeNet~\cite{chang2015shapenet} shapes with complicated structures and fine-grained leaf-node parts.
Different from StructureNet where shape geometry and structure are jointly modeled in one RvNN, we learn a pair of separate geometry RvNN and structure RvNN in a disentangled but synergistic fashion, which enables exploring geometric (structural) changes while keeping shape structure (geometry) unchanged.
We also find that by combining the state-of-the-art structure learning modules from StructureNet~\cite{mo2019structurenet} and the latest techniques in modeling detailed part geometry from SDM-Net~\cite{gaosdmnet2019} in an effective way, we achieve the best from both worlds that beats both StructureNet and SDM-Net in performance.

\subsection{Disentangled Analysis in Deep Learning}
\yj{

For 2D image generation, the architecture proposed in~\cite{karras2019style} enables intuitive, scale-specific control of high-level attributes for high-resolution image synthesis by automatic unsupervised separation.
HoloGAN~\cite{nguyen2019hologan} improves the visual quality of generation and allows manipulations by utilizing explicit 3D features to disentangle the shape and appearance in an end-to-end manner from unlabeled 2D images only. 

In 3D shape processing, generative modeling becomes a mainstream topic thanks to deep learning and tremendous public 3D datasets, some of which contain rich realistic textures.
Levinson~et~al.~\shortcite{levinson2019latent} propose a supervised generative model to achieve accurate disentanglement of pose and shape in a large-scale human mesh dataset, as well as successfully incorporating techniques such as pose and shape transfer.
Moreover, CFAN-VAE~\cite{tatro2020unsupervised} proposes to use the intrinsic conformal factor and extrinsic normal feature to achieve geometric disentanglement (pose and identity of human shapes) in an unsupervised way.
For general textured objects datasets, VON~\cite{zhu2018visual} presents a fully differentiable 3D-aware generative model with a disentangled 3D representation (shape, viewpoint, and texture) for image and shape synthesis. 

Compared to the above works, our work displays a rather novel capability - \emph{disentanglement of structure and geometry} of 3D shapes. In this work, we learn a disentangled structured mesh representation for 3D shapes, where the disentanglement is entirely between two explicitly defined factors, namely \emph{structure} and \emph{geometry}. 
\yjrr{Furthermore, the cycle consistency first enables the translation of images and shapes with unpaired examples in an unsupervised manner~\cite{CycleGAN2017,yi2017dualgan,gaovcgan2019}.
For the enhancement of disentangled learning, we adopt the cycle consistency into our framework to explicitly encourage latent-space disentanglement. 
}
Our network can not only be used to generate shapes with improved geometric details but also allows us to exploit independent control of structure and geometry with the disentangled latent spaces.
}

\section{Methodology}
\label{sec:method}

\begin{figure}[t]
    \centering
    \includegraphics[width=0.99\linewidth]{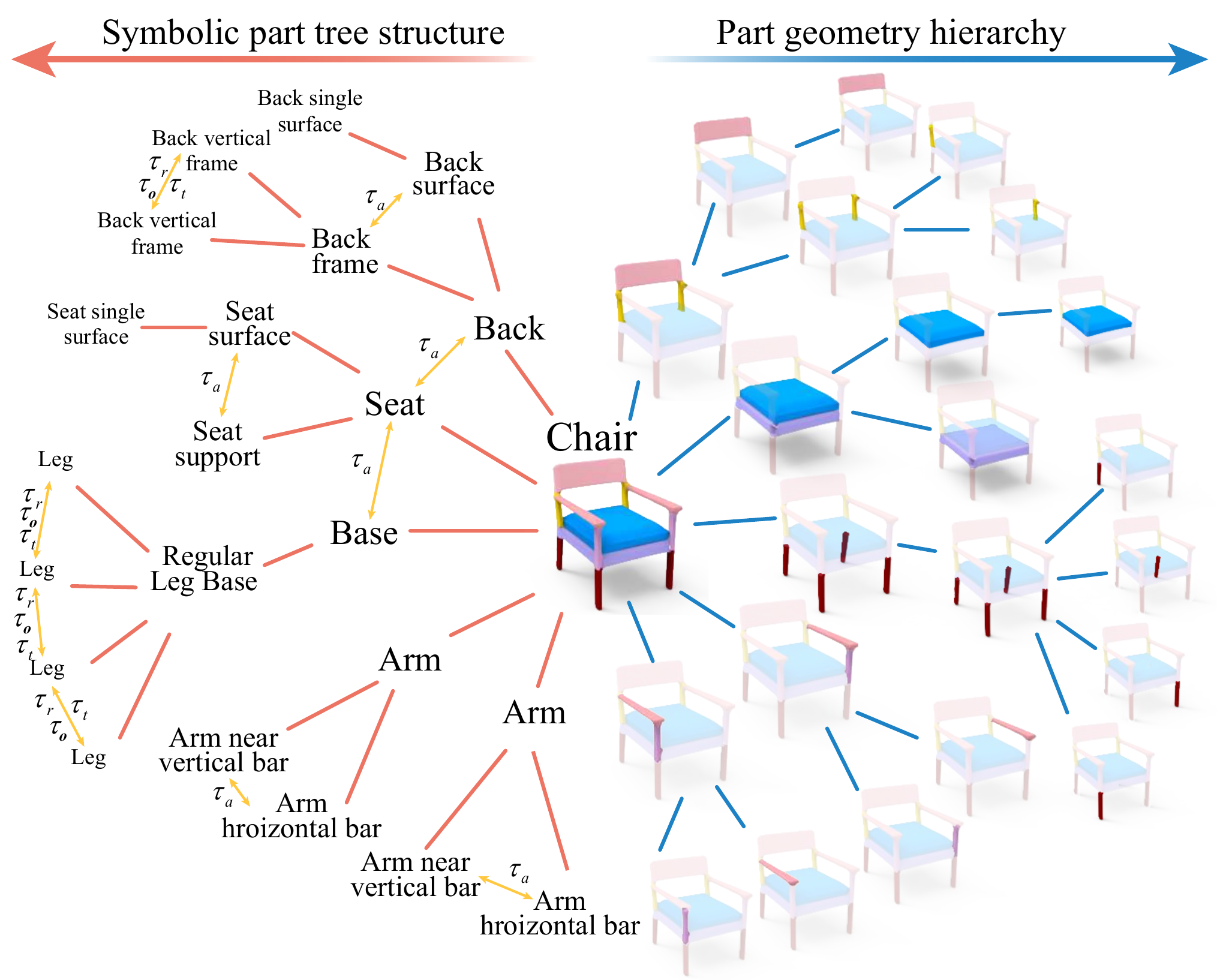}
    \vspace{-3mm}
    \caption{An example showing the proposed disentangled but highly synergistic representation of shape geometry and structure hierarchies. There is a bijective mapping between the tree nodes in the two hierarchies. In the structure hierarchy, we consider symbolic part semantics and a rich set of part relationships (orange arrows), such as adjacency $(\tau_a)$, \yj{translational} symmetry $(\tau_t)$, reflective symmetry $(\tau_r)$ and rotational symmetry $(\tau_o)$. In the part geometry hierarchy, the part geometry is represented by meshes.
    }
    \label{fig:representation}
    \vspace{-3mm}
\end{figure}

\yj{In this work, every 3D shape is decomposed into semantically consistent part instances that are organized by an $n$-ary part hierarchy covering parts at different granularities, ranging from coarse-grained parts (\eg chair back and base) to fine-grained ones (\eg back bars and legs).
We propose a \textit{disentangled} but \textit{synergistic} hierarchical representation (Figure~\ref{fig:representation}) and a learning framework \yjrr{with our Cycled Disentanglement mechanism} (Figure~\ref{fig:global-vae}), enabling disentangled control of shape geometry and structure in the shape generation procedure.}

In the following subsections, we first describe the detailed definitions for our disentangled shape representation of structure hierarchy and geometry hierarchy.
Then, we introduce a conditional part geometry VAE on encoding and decoding the fine-grained part geometry using a unified deformable mesh.
Next, we present our \yjr{disentangled} network architecture designs for the geometry and structure VAEs and discuss how to learn the disentangled shape geometry and structure latent spaces simultaneously where the geometry and structure VAEs guide the learning processes for each other.
Finally, a post-processing procedure is introduced for result refinement.

\subsection{Disentangled Shape Representation}

We adapt the hierarchical part segmentation in PartNet~\cite{mo2019partnet} for ShapeNet models~\cite{chang2015shapenet}, where each shape is decomposed into a set of parts~$\mathbf{P}$ and organized in a part hierarchy~$\mathbf{H}$ (\ie, the vertical parent-child part relationships) with rich part relationships~$\mathbf{R}$ (\ie, the horizontal among-sibling symmetry or adjacency part relationships).
Each part $P_i$ is associated with a semantic label $l_i$ (\eg chair back, chair leg) defined for a certain object class, as well as the detailed part geometry $G_i$.

We introduce a disentangled but synergistic shape representation for shape structure and geometry, where we represent each 3D shape as a pair of a structured hierarchy and a geometry hierarchy.
In our disentangled representation (see Figure~\ref{fig:representation}), a structure hierarchy abstracts away the part geometry and only describes a symbolic part hierarchy with part structures and relationships, namely $(\langle l_1, l_2, \cdots, l_N\rangle, \mathbf{H}, \mathbf{R})$, 
while a geometry hierarchy describes the part geometry $\langle G_1, G_2, \cdots, G_N\rangle$.
There is a bijective mapping between the tree nodes of the structure and geometry hierarchies where the part semantic label $l_i$ defined in the structure hierarchy corresponds to the part geometry $G_i$ included in the geometry hierarchy.
Also, the geometry hierarchy implicitly follows the same part hierarchy $\mathbf{H}$ and part relationships $\mathbf{R}$ as specified in the structure hierarchy.

\begin{figure}
    \centering
    \includegraphics[width=0.99\linewidth]{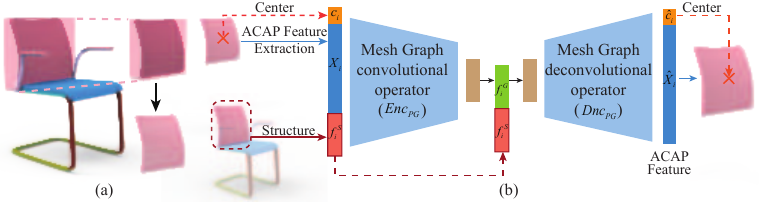}
    \vspace{-3mm}
    \caption{\yj{We present: (a) the non-rigid part mesh registration process, and (b) the architecture of our conditional part geometry variational autoencoder.
    In (a), we deform a box mesh to any given part geometry and then extract ACAP~\cite{gao2019sparse} feature based on the registration.
    In (b), for a single part mesh geometry, the encoder maps the part ACAP feature and its center position into a 128-dimensional geometric latent code, while the decoder reconstructs the part geometry by decoding the ACAP feature and the center vector. Both networks are conditioned on the part structure information along the structure hierarchy to generate specialized part geometry for different structure contexts.
    }}
    \label{fig:part-geo}
    \vspace{-3mm}
\end{figure}

\paragraph{Part Geometry Representation.}
For each part geometry $G_i$, we use a mesh representation to capture more geometric details, such as the decorative patterns and sharp boundary edges, than the point cloud representation used in StructureNet~\cite{mo2019structurenet}.
Given a closed box mesh manifold $G_{box}$ with $V=5402$ vertices, we first calculate the oriented bounding box (OBB) $B_i$ of each part $P_i$ and deform $G_{box}$, initialized with the shape $B_i$, to the target part geometry $G_i$ by adjusting the vertex positions through a non-rigid registration procedure.
Then, for each part, we use the ACAP (as-consistent-as-possible) feature~~\cite{gao2019sparse,gaosdmnet2019} $X_i$ as the representation of the deformed box mesh.
The ACAP feature $X_i\in\mathbb{R}^{V\times9}$ captures the local rotation and scale information in a one-ring neighbor patch of every vertex on the mesh and is capable of capturing large-scale local geometric deformations (\eg rotation greater than $180^{\circ}$). %
We show an example registration result in \yj{Figure~\ref{fig:part-geo} (a)}.
For the detailed calculation, please refer to the work~\cite{gao2019sparse}. 
Since the ACAP feature is invariant to spatial translation of the part, we incorporate an additional 3-dimensional vector to describe the part center $c_i$.
Overall, each part geometry is represented as a pair of an ACAP feature $X_i$ and a part center vector $c_i$, as shown in Figure~\ref{fig:part-geo} (b), \ie, $G_i=(X_i, c_i)$.

\paragraph{Geometry Hierarchy.} The geometry hierarchy for a 3D shape is a hierarchy of part geometries $\langle G_1, G_2, \cdots, G_N\rangle$ (Figure~\ref{fig:representation} right).
It decomposes a complicated shape geometry into a hierarchy of parts ranging from coarse-grained levels to fine-grained levels.
Each part geometry $G_i$ in the geometry hierarchy corresponds to a tree node in the structure hierarchy and gives a concrete geometric realization given the context of the entire shape structure to generate.
The geometry hierarchy implicitly follows the structural hierarchy and part relationships $\mathbf{H}$ and $\mathbf{R}$ defined in the structure hierarchy.

\paragraph{Structure Hierarchy.} We consider a symbolic structure hierarchy $(\langle l_1, l_2, \cdots, l_N\rangle, \mathbf{H}, \mathbf{R})$ as the structure representation for a shape, inspired by a recent work PT2PC~\cite{mo2020pt2pc}.
Figure~\ref{fig:representation} (left) presents an example for the symbolic structure hierarchy.
It only includes the semantic information of shape parts and the relationships between parts, while abstracting away the concrete part geometry.
PT2PC learns to generate 3D point cloud shapes conditioned on a given symbolic structure hierarchy as a fixed skeleton for shape generation.
In this work, we extend PT2PC to consider encoding and decoding the symbolic structure hierarchy and investigate its disentangled but synergistic relationship to the geometry hierarchy.

In the symbolic structure hierarchy, we represent each part with a semantic label $l_i$ (\eg chair back, chair leg) without having a concrete part geometry in the representation.
We include the rich sets of part relationships defined in the PartNet dataset in the symbolic structure hierarchy representation.
There are two kinds of part relationships: the vertical parent-child inclusion relationships (\eg a chair back and its sub-component chair back bars), as defined in $\mathbf{H}$, and the horizontal among-sibling part symmetry and adjacency relationships (\eg chair back bars have translational symmetry), as denoted in $\mathbf{R}$.
We use the part relationships $\mathbf{H}$ and $\mathbf{R}$ as provided in StructureNet~\cite{mo2019structurenet}.

\paragraph{Coupling Geometry and Structure Hierarchies.} 
Even though we are attempting a disentangled shape representation, the structure and geometry need to be compatible with each other for generating plausible and realistic shapes.
On the one hand, shape structure provides a high-level guidance for part geometry.
If four legs of a chair are specified to be symmetric to each other in the structure hierarchy, the four legs should have identical part geometry to satisfy the structural requirement.
On the other hand, given a certain type of part geometry, only certain kinds of shape structures are possible.
For example, it is nearly impossible to manufacture a swivel chair if no lift handle or gas cylinder parts are provided.

Concretely, in our disentangled shape representation, the geometry hierarchy $\langle G_1, G_2, \cdots, G_N\rangle$ and the structure hierarchy $(\langle l_1, l_2, \cdots, l_N\rangle, \mathbf{H}, \mathbf{R})$ of a shape are highly correlated and tightly coupled.
There is a bijective mapping between each part geometry node $G_i$ and the part structure symbolic node $l_i$.
We set up communication channels between the two hierarchies in the joint learning process.
The geometry hierarchy uses the part hierarchy $\mathbf{H}$ and relationship $\mathbf{R}$ in the encoding and decoding stages for passing messages and synchronizing geometry generation among related nodes.
To train the decoding stage of the structure hierarchy, we leverage the corresponding geometry nodes to help match the prediction to the ground-truth parts.
Thus, the synergy between the structure and geometry hierarchies is essential for simultaneously learning the embedding spaces.

\subsection{Conditional Part Geometry VAE}
\label{sec:condpartvae}
In the geometry hierarchy of a 3D shape, each part geometry $G_i$ is represented as a pair of ACAP feature $X_i\in\mathbb{R}^{V\times9}$ and the part center $c_i\in\mathbb{R}^3$.
We propose a part geometry conditional variational autoencoder (VAE) with a conditional part geometry encoder $Enc_{PG}$ that maps the part geometry $G_i=(X_i, c_i)$ into a 128-dimensional latent feature and a conditional part geometry decoder $Dec_{PG}$ which reconstructs $\hat{G}_i$ from the latent code.
Both the encoder and decoder are conditioned on the part semantics and its current structural context, in order to generate part geometry that is synergistic to the current structure tree nodes. \yj{We use the mesh graph convolutional operator to aggregate the local features around the vertex, which is also suitable for shape analysis~\cite{monti2017geometric,wang2020mgcn}. }

Figure~\ref{fig:part-geo} \yj{(b)} illustrates the proposed part geometry conditional VAE architecture.
The encoder network $Enc_{PG}$ performs two sequential mesh graph convolutional operations over the $X_i\in\mathbb{R}^{V\times9}$ feature map within local one-ring neighborhood around each vertex, extracts a global part geometry feature via a single fully-connected layer, which is then concatenated with the part center vector $c_i$, and finally predicts a 128-dim geometry feature $f^G_i$ for part $P_i$.
The decoder network $Dec_{PG}$ decodes the part ACAP feature $\hat{X}_i$ and the part center $\hat{c}_i$ through fully-connected and mesh-based convolutional layers.
Then, the decoded ACAP feature $\hat{X}_i$ is applied on every vertex of the closed box mesh $G_{box}$ to reconstruct the part mesh $\hat{G}_i$ and the reconstructed center $\hat{c}_i$ moves the part mesh to the correct position in the shape space.
\yj{Both the encoder and decoder are  conditioned on a structure code condition $f^S_i$ summarizing certain part semantics and its structural context.}

Different from SDM-NET where they train separate PartVAEs for different part semantics, we propose to use a single shared PartVAE to encode and decode shape part geometry that is conditional on the part structure information $f^S_i$.
The reason is three-fold:
firstly, PartNet gives far more part semantic labels than the SDM-NET data, where training separate networks for different part semantics is extremely costly and empirically hard to converge;
secondly, the data sample for some rare part categories is not sufficient to train a separate network;
lastly, our conditional PartVAE can be conditioned on structure codes summarizing the part semantics and sub-hierarchy information, allowing effective specialization for part geometry generation given different structure contexts.

To train the proposed conditional PartVAE, we define the loss as follows:
\begin{equation}
\mathcal{L}_\text{cond-PartVAE} = \lambda_1 \mathcal{L}^\text{recon}_\text{cond-PartVAE} + \mathcal{L}^\text{KL}_\text{cond-PartVAE}
\label{eq:losscondpartvae}
\end{equation}
where $\mathcal{L}^\text{recon}_\text{cond-PartVAE} = \lVert \hat{X}_i - X_i \rVert_2^2 + \lVert \hat{c}_i - c_i \rVert_2^2$ is the reconstruction loss and $\mathcal{L}^\text{KL}_\text{cond-PartVAE}$ is the standard KL divergence loss to encourage the learned embedding space to be close to a unit multivariate Gaussian distribution.

\begin{figure}
    \centering
    \includegraphics[width=0.99\linewidth]{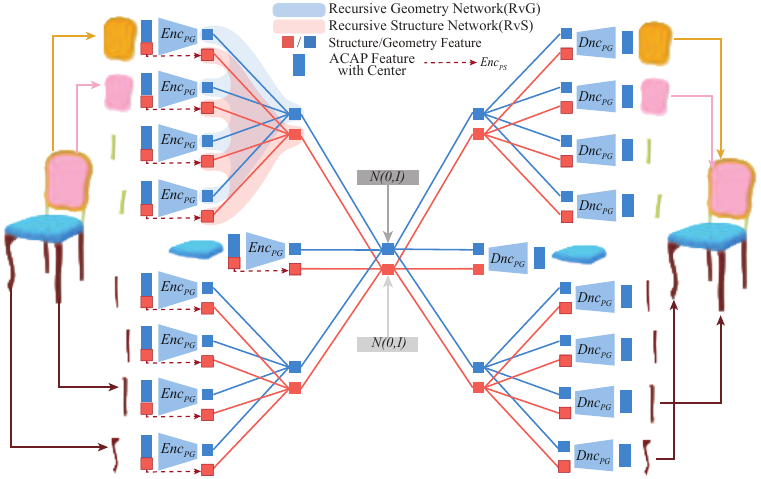}
    \vspace{-3mm}
    \caption{\yj{We train two \textit{disentangled} but \textit{synergistic} geometry and structure variational autoencoders (VAEs) with recursive encoders and decoders to learn disentangled latent spaces for shape geometry and structure. The figure illustrates the joint learning procedure of the structure VAE (red) and the geometry VAE (blue). In the encoding stage, the structure features summarize the symbolic part semantics and recursively compute sub-hierarchy structure contexts, while the geometry features encode the detailed part geometry for leaf nodes and propagate the geometry information along the same hierarchy. The decoding procedures of the VAEs are supervised to reconstruct the hierarchical structure and geometry information in an inverse manner.}}
    \label{fig:global-vae}
    \vspace{-3mm}
\end{figure}

\begin{figure*}
    \centering
    \includegraphics[width=0.99\linewidth]{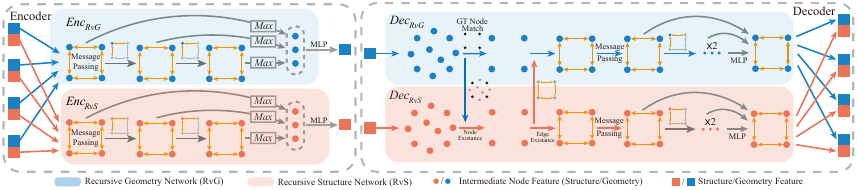}
    \vspace{-2mm}
    \caption{\yj{
    We illustrate the detailed architecture of recursive graph encoders ($Enc_{RvG}$, $Enc_{RvS}$) and recursive graph decoders ($Dec_{RvG}$, $Dec_{RvS}$).
    We show the geometry encoder and decoder in blue branches while presenting the structure ones in red.
    On the left, the geometry and structure encoders operate independently in summarizing features among children's part nodes to the parent nodes, while sharing the same structure information (part relationships) for graph message-passing iterations.
    On the right, the geometry and structure decoders are trained to reconstruct the geometry and structure of children nodes from the given parent nodes in a disentangled fashion, but with rich communications between the two that the decoded set of part geometry determines the structure while the graph convolution operations for the geometry branch are guided by the predicted structure.
    }
    }
    \label{fig:global-vae1}
    \vspace{-2mm}
\end{figure*}

\subsection{Disentangled Geometry and Structure VAEs}
To learn disentangled latent spaces for shape geometry and structure, we design two Variational Autoencoders (VAEs) with Recursive Neural Network (RvNN) encoders and decoders that are trained in a disentangled but synergistic manner.
\yjr{Besides, we also propose a novel Cycle Disentanglement for disentanglement learning of the shape space in a self-supervised fashion.
}
\yjr{Figure~\ref{fig:new-pipeline}} and Figure~\ref{fig:global-vae} provide an overview for the proposed disentangled \yjr{framework, including the disentangled VAEs}.
The geometry VAE (the blue part) and the structure VAE (the red part) learn two disentangled latent spaces for shape geometry and structure.
Though disentangled, the structure and geometry VAEs are jointly learned in a highly synergistic manner.

\subsubsection{Structure VAE}

\noindent Given a structure hierarchy $(\langle l_1, l_2, \cdots, l_N\rangle, \mathbf{H}, \mathbf{R})$ describing a symbolic tree with part semantics, hierarchy and relationships, 
the structure VAE is trained to learn a structure latent space.
For the encoding process, a part structure encoder $Enc_{PS}$ first summarizes the leaf-node part semantics and then a recursive graph structure encoder $Enc_{RvS}$ propagates features from the leaf nodes to the root in a bottom-up manner according to the part hierarchy $\mathbf{H}$ and relationships $\mathbf{R}$.
Inversely, the decoding process contains a recursive graph structure decoder $Dec_{RvS}$ that hierarchically predicts the structure features from the root to the leaf nodes in a top-down fashion and a part structure decoder $Dec_{PS}$ that decodes part semantic labels for the leaf nodes.

The structure VAE uses a similar recursive neural network architecture to StructureNet~\cite{mo2019structurenet}, 
but we are encoding and decoding symbolic structure hierarchies with no concrete part geometry.
It is thus difficult to train the decoding procedure given no part geometry since we are not able to perform node matching between a set of decoded children and the set of ground-truth parts.
To address this challenge, we borrow the corresponding part geometry decoded from the geometry VAE to perform the node matching for the training, where a communication channel between the structure and geometry VAEs is established.
Below, we discuss more details on the four network components for the structure VAE.

\paragraph{Encoders.}
To encode a symbolic structure hierarchy represented as $(\langle l_1, l_2, \cdots, l_N\rangle, \mathbf{H}, \mathbf{R})$, we need to introduce an additional part instance identifier for each part $d_i$, where $d_i=0,1,2,\cdots$, similar to PT2PC~\cite{mo2020pt2pc}.
Part instance identifiers help differentiate the part instances with the same part semantics for a parent node.
For example, if a chair base contains four chair legs, we mark them with part instance identifiers $0,1,2,3$.
The part instance identifiers are only necessary for the encoding stage and will be ignored in the decoding procedure.

For each leaf node part $P_i$, the part structure encoder $Enc_{PS}$ encodes the part semantics $l_i$ and its part instance identifier $d_i$ into a part structure latent code $f^S_i$.
\begin{equation}
    f^S_i=Enc_{PS}\left(\left[l_i;d_i\right]\right)
\end{equation}
where $Enc_{PS}$ is simply a fully-connected layer, $[;]$ denotes the vector concatenation, and we represent both $d_i$ and $l_i$ as one-hot vectors.

\yj{For the non-leaf part $P_i$, the recursive graph structure encoder $Enc_{RvS}$ gathers all children node features, performs graph message-passing along the part relationships defined in $\mathbf{R}$ among the children nodes, and finally computes $f^S_i$ by aggregating the children nodes' features, \yj{as illustrated in the red branch of Figure~\ref{fig:global-vae1} (left).}} Namely, 
\begin{equation}
    f^S_i=Enc_{RvS}\left(\left\{f_j^S\right\}_{(P_i, P_j)\in \mathbf{H}}, l_i, d_i\right)
    \label{eq:enc_rvs}
\end{equation}
where $(P_i, P_j)\in \mathbf{H}$ denotes that part $P_j$ is a child of $P_i$.
The module $Enc_{RvS}$ is composed of two iterations of graph message-passing similar to StructureNet~\cite{mo2019structurenet}, a max-pooling operation over the obtained node features and a fully-connected layer producing the part structure feature $f^S_i$ given the pooled feature and the part identifiers $[l_i;d_i]$ for the part.
Here, please note that the part instance identifiers are necessary, due to the max-pooling operation, to distinguish and count the different occurrences of part instances with the same part semantics, and such crucial information would otherwise be lost.

We repeatedly apply the part structure encoder $Enc_{RvS}$ until reaching the root node $P_\text{root}$.
The final root node structure feature $f^S_{root}$ is then mapped to the final structure embedding space through a fully-connected layer.
We use a KL divergence loss to encourage the learned structure latent space to be close to a unit multivariate Gaussian distribution.

\paragraph{Decoders.}
The decoding process of a structure VAE takes a structure latent code as input and recursively decodes a symbolic structure hierarchy $(\langle \hat{l}_1, \hat{l}_2, \cdots, \hat{l}_N\rangle, \mathbf{\hat{H}}, \mathbf{\hat{R}})$ as the output.
The part instance identifiers are not involved in the decoding procedure.

The recursive graph structure decoder $Dec_{RvS}$ consumes the parent structure feature $\hat{f}^S_i$ and infers a set of children node structure features $\{\tilde{f}^S_{i,1}, \tilde{f}^S_{i,2}, \cdots, \tilde{f}^S_{i,10}\}$, where we assume there is a maximum of 10 children parts per parent node.
Following StructureNet~\cite{mo2019structurenet}, we predict a semantic label and an existence probability for each part, by another fully-connected layer followed by classification output layers.
Besides the node prediction, by connecting all pairs of parts, we also predict a set of symmetric or adjacent edges $\hat{\textbf{R}}_i$ among the existing nodes.
Along the predicted edges, node features $\{\tilde{f}^S_{i,k}\}_{k}$ are updated via two graph message-passing operations and finally we decode a set of structure part nodes $\{\hat{f}^S_{j_1}, \hat{f}^S_{j_2}, \cdots, \hat{f}^S_{j_{K_i}}\}$, where $K_i$ denotes the number of existing nodes for part $P_i$. 
We refer the readers to StructureNet~\cite{mo2019structurenet} for more details. \yj{The red branch in Figure~\ref{fig:global-vae1} (right) illustrates this process.} More formally, 
\begin{equation}
  \left\{\hat{f}^S_{j_1}, \hat{f}^S_{j_2}, \cdots, \hat{f}^S_{j_{K_i}}, \hat{\textbf{R}}_i\right\}=Dec_{RvS}\left(\hat{f}^S_i\right)
\end{equation}

We repeat the recursive structure decoding procedure until reaching the leaf nodes.
For a leaf node part $\hat{P}_i$, the part structure decoder $Dec_{PS}$ simply decodes the part semantic label via a fully-connected layer followed by outputting a likelihood score for each part semantic label. Finally, we get
\begin{equation}
  \hat{l}_i=Dec_{PS}\left(\hat{f}^S_i\right)
\end{equation}

To train the hierarchical decoding process, StructureNet~\cite{mo2019structurenet} predicts part geometry for the intermediate nodes and establishes a correspondence between the predicted set of parts and the ground-truth set of parts.
However, it is difficult to directly adapt this training procedure to decode the symbolic structure hierarchy by matching the part semantic labels.
We resolve this challenge by building a communication channel between the structure hierarchy and the geometry one and borrowing the corresponding part geometry decoded in the geometry VAE for the matching procedure.

In our implementation, we utilize the conditional part geometry decoder $Dec_{PG}$ introduced in Sec.~\ref{sec:condpartvae} and predict an oriented bounding box (OBB) geometry $\hat{B}_j$ for each part $\hat{P}_j$ where $j=j_1,j_2,\cdots, j_{K_i}$.
We choose to use the OBB geometry for the matching process instead of the mesh geometry $\hat{G}_i$ since we observe a decreased accuracy for registering the box mesh $G_{box}$ to an intermediate part geometry, which is usually more complex than leaf-node parts.

To train the part existence scores, part edge predictions and the part semantic labels, we follow StructureNet~\cite{mo2019structurenet} and refer the readers to the paper for more details.

\subsubsection{Geometry VAE}

\noindent Given a geometry hierarchy $\langle G_1, G_2, \cdots, G_N\rangle$ encoding the part geometry of shape parts,
the geometry VAE learns to map the shape geometry to a geometry latent space, disentangled from the structure latent space.
The geometry latent space is also modeled to be a unit multivariate Gaussian distribution.

The geometry VAE shares a similar network architecture to the structure VAE. 
The encoding process starts from extracting part geometry features for all leaf-node parts via a part geometry encoder $Enc_{PG}$ and then recursively propagates the geometry features along the hierarchy to the root node, summarizing the geometry information for the entire shape through a recursive graph part geometry encoder $Enc_{RvG}$.
For the decoding process, we first use a recursive graph geometry decoder $Dec_{RvG}$ that hierarchically decodes the geometry features from the root to the leaf-node parts in an inversely recursive manner.
Then, we leverage a part geometry decoder $Dec_{PG}$ to reconstruct the part geometry for leaf-node parts.

There are two communication channels that allow the synergistic structure hierarchy to guide the geometry VAE encoding and decoding procedures.
Firstly, the part geometry encoder $Enc_{PG}$ and decoder $Dec_{PG}$ are conditioned on the structure context produced by the structure VAE, which allows for generating different kinds of part geometry according to different part semantics and shape structures.
Secondly, the graph message-passing procedures in the recursive graph geometry encoder $Enc_{RvG}$ and decoder $Dec_{RvG}$ borrow the part hierarchy and relationships defined in the structure hierarchy.

As follows, we describe the encoding and decoding stages for learning geometry VAE in more detail.

\paragraph{Encoders.}
We start from encoding each leaf node part geometry $G_i=(X_i, c_i)$ into a latent part geometry feature space.
We use the conditional part geometry encoder $Enc_{PG}$ introduced in Sec.~\ref{sec:condpartvae} that maps the part ACAP feature $X_i$ and the part center $c_i$ to a 128-dimensional feature $f^G_i$, namely,
\begin{equation}
    f^G_i=Enc_{PG}\left(\left[X_i;c_i\right], f^S_i\right)
\end{equation}
The network is conditioned on the structure code $f^S_i$ generated in the structure VAE, in order to gain some structural context on what the semantics for the current part is and what role the part plays in generating the final shape.

For each sub-hierarchy of the part geometry, we recursively produce the intermediate part geometry node feature $f^G_i$ by aggregating its children geometry node features $\{f^G_{j}\}_{j}$ through the recursive graph geometry encoder $Enc_{RvG}$, \yj{as illustrated in the blue branch of Figure~\ref{fig:global-vae1} (left)}.
Similar to the design of $Enc_{RvS}$ for structure VAE, it performs two iterations of graph message-passing operations among the children geometry node features based on the part relationships between sibling part nodes, and conduct a simple max-pooling operation to compute $f^G_i$, where we have
\begin{equation}
    f^G_i=Enc_{RvG}\left(\left\{f_j^S\right\}_{(P_i, P_j)\in \mathbf{H}}\right)
\end{equation}
Different from $Enc_{RvS}$ as shown in Eq.~\ref{eq:enc_rvs}, we do not encode the part geometry for the non-leaf node since the geometry is more complex and the registration to a box mesh is less accurate. 
The increased geometric complexity also makes it harder to effectively embed them in a low-dimensional latent space.
For the message-passing operations, we borrow the part relationships defined in the structure hierarchy.
This is achieved by maintaining a bijective mapping among the tree nodes in the structure and geometry hierarchies.
We repeatedly apply the recursive graph geometry encoder $Enc_{RvG}$ until reaching the root node $P_\text{root}$.
The final root node geometry feature $f^G_{root}$ is then mapped to the final geometry embedding space through a fully-connected layer.

\paragraph{Decoders.}
The decoding process of a geometry VAE takes a geometry latent code as input and recursively decodes a geometry hierarchy $\langle \hat{G}_1, \hat{G}_2, \cdots, \hat{G}_N\rangle$ for a shape.

\yj{As illustrated in the blue branch in Figure~\ref{fig:global-vae1} (right),} the recursive graph geometry decoder $Dec_{RvG}$ takes the parent geometry feature $\hat{f}^G_i$ as input and decodes a set of children node geometry features $\{\tilde{f}^G_{i,1}, \tilde{f}^G_{i,2}, \cdots, \tilde{f}^G_{i,10}\}$.
Then, based on the structural predictions on part existence scores, part semantic labels and part edge information from the synergistic structure VAE, 
we conduct two iterations of graph message-passing over the children node geometry features along the predicted pairwise part relationships $\hat{\textbf{R}}_i$. 
The decoder $Dec_{RvG}$ then produces a final set of children nodes with the predicted part geometry features.
\begin{equation}
  \left\{\hat{f}^G_{j_1}, \hat{f}^G_{j_2}, \cdots, \hat{f}^G_{j_{K_i}}\right\}=Dec_{RvG}\left(\hat{f}^S_i, \mathbf{\hat{R}}_i\right)
\end{equation}
where $Dec_{RvG}$ is conditioned on the decoded part relationships $\mathbf{\hat{R}}_i$ in the structure VAE and $K_i$ denotes the number of existing part nodes predicted by the recursive graph structure decoder $Dec_{RvS}$.

We repeat the recursive graph geometry decoding procedure until reaching the leaf nodes.
For a leaf node part $\hat{P}_i$, we use the conditional part geometry decoder $Dec_{PG}$ introduced in Sec.~\ref{sec:condpartvae} that reconstructs $\hat{G}_i=(\hat{X}_i, \hat{c}_i)$ from an input part geometry feature $\hat{f}^G_i$.
Formally, we have
\begin{equation}
  \hat{G}_i=Dec_{PG}\left(\hat{f}^G_i, \hat{f}^S_i\right)
\end{equation}
Notice that the network $Dec_{PG}$ is conditioned on the part structure code $\hat{f}^S_i$ predicted in the coupled structure VAE decoding procedure.

The geometry VAE is trained jointly with the structure VAE and the conditional part geometry VAE.
To supervise the reconstruction of the leaf-node part geometry in the decoding process, we simply adapt the loss terms defined in Eq.~\ref{eq:losscondpartvae} from Sec.~\ref{sec:condpartvae}.
We also add a KL divergence loss term to train the geometric latent space to get closer to the unit multivariate Gaussian distribution.

\begin{figure}
    \centering
    \includegraphics[width=0.99\linewidth]{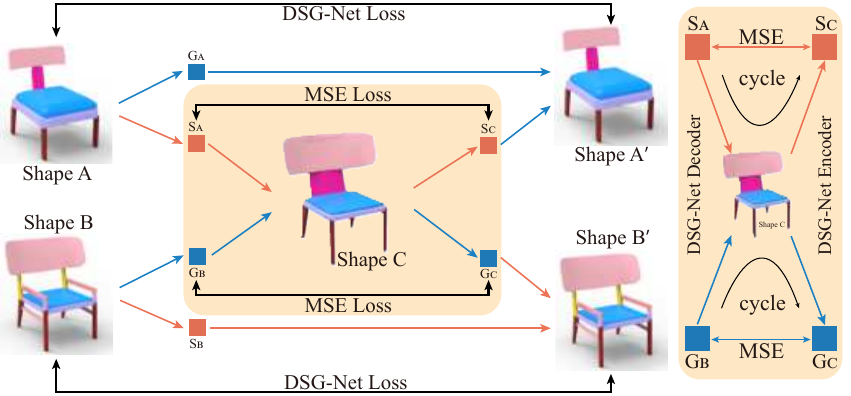}
    \vspace{-3mm}
    \caption{\yjr{We propose \yjrr{a new Cycled Disentanglement mechanism} to further
    disentangle the geometry and structure of shapes. Based on the pre-trained DSG-Net, we decouple two shapes ($A$ and $B$) into geometry ($G_{A}$, $G_{B}$) and structure ($S_{A}$, $S_{B}$) features respectively. And we can combine the two features ($S_{A}$, $G_{B}$) to synthesize new shape $C$, which fuses the structure of shape $A$ and geometry of shape $B$. 
    For the successful disentanglement, we extract the structure $S_{C}$ and geometry $G_{C}$ features of the newly synthesized shape C to encourage $S_{C}=S_{A}, G_{C}=G_{B}$ as much as possible by the MSE loss, where there are two cycles (as shown in the right part of the figure). Furthermore, we also constrain the reconstructed shapes $A^{\prime}, B^{\prime}$ are as close as possible to the input shapes by the original DSG-Net loss.}}
    \label{fig:new-pipeline}
    \vspace{-3mm}
\end{figure}

\yjr{
\subsection{Cycled Disentanglement Mechanism (CycD) for Disentangled Geometry \& Structure VAEs}
\noindent Furthermore, in order to ensure that the geometry and structure are effectively disentangled and improve the performance of disentanglement, 
we propose \yjrr{a new Cycled Disentanglement mechanism} to further disentangle the geometry and structure of shapes. 
Figure~\ref{fig:new-pipeline} illustrates our pipeline.

For any two input shapes ($A$, $B$), our new framework extracts their geometry and structure features ($\{G_{A}, S_{A}\},\{G_{B}, S_{B}\}$) by DSG-Net.
Then, we can synthesize a new shape $C$ by combining features from structure and geometry of the two shapes, such as $S_{A}$ and $G_{B}$.
So, the newly synthesized shape $C$ has the geometric features from shape $A$ and the structural features from shape $B$.
The disentangled performance of our framework can be ensured and improved via paired self-supervised losses.
During training, the loss terms $\mathcal{L}_{struct}, \mathcal{L}_{geo}$ encourage the geometry code $G_{C}$ and structure code $S_{C}$ of Shape $C$ to be the same as $G_{B}$ and $S_{A}$ under the MSE metric, i.e.,
\begin{equation}
\mathcal{L}_{struct} = ||S_{C} - S_{A}||_2^2, \mathcal{L}_{geo} = ||G_{C} - G_{B}||_2^2
\end{equation}
Furthermore, the geometry code $G_{C}$ and structure code $S_{C}$ of shape $C$ can be used to reconstruct the original shape $A$ and $B$ as $A^{\prime}, B^{\prime}$.
Hence, in addition to the constraints on the shape geometry and structure codes, we have added the supervisions on the shapes $A^{\prime}, B^{\prime}$, which aims to make the disentanglement more successful in a shape-aware manner.
}

\subsection{Post-Processing for Detached Parts}\label{sec:post}
\yj{
Though explicitly considering part relationships as soft constraints in the modeling already helps generate shapes whose parts are well structured and connected, we still observe some occasional failure cases of floating parts or part disconnections. 
We thus propose to use a post-processing module that directly optimizes the position of each part and further resolves the issue of detached parts.

Concretely, we fix the pre-trained parameters of the encoder and decoder, except for the center prediction MLP in the decoder network, so that we only optimize the center position of each part.
Then, taking as input an object with disconnected parts, we optimize for the final location for each part by adjusting the part center positions.
We employ two loss terms for the optimization: a structure loss, which enforces the adjacent parts to get closer to each other and the symmetric parts to satisfy the symmetric constraints, and 
an identity loss, which encourages the optimized shape to be similar to the input shape geometry, avoiding the degraded case that all parts are clustered together.
During the optimization process, we balance the two loss terms with weights 1 and 100.
}

\section{Experiments}
\label{sec:exps}

Learning disentangled latent spaces for shape structure and geometry allows us to generate high-quality 3D shape meshes with complex structure and detailed geometry in a controllable manner.
DSG-Net not only demonstrates the state-of-the-art performance for structured shape generative modeling, but also enables generating shape meshes with controllable structure and geometry factors.

In this section, we present extensive experiments on the tasks of shape reconstruction, generation and interpolation, where we show the superior performance of our proposed method on the PartNet dataset~\cite{mo2019partnet}, compared to several strong baselines, including StructureNet~\cite{mo2019structurenet}, SDM-Net~\cite{gaosdmnet2019}, IM-Net~\cite{chen2019learning} and BSP-Net~\cite{chen2019bsp}).
We also propose and formulate the tasks of disentangled shape reconstruction, generation and interpolation, where we manipulate one factor of shape structure and geometry while keeping the other unchanged.
We further benchmark our performance for disentangled shape reconstruction on a synthetic dataset.
\yj{In the end, ablation studies for some key designs of our network are presented.}
All experiments were carried out on a computer with an i9-9900K CPU, 64GB RAM, and a GTX 2080Ti GPU.

\subsection{Data Preparation}\label{sec:data-pare}
We primarily use the PartNet dataset~\cite{mo2019partnet} for the majority of our experiments.
PartNet provides fine-grained, multi-scale and hierarchical shape part segmentation for ShapeNet~\cite{chang2015shapenet} models.
We use the four biggest and most commonly used object categories for our experiments: chairs, tables, cabinets and lamps.
We follow the official training and test data splits.

All the PartNet shapes from the same object category share a canonical part template with consistent part semantics.
The vertical parent-child relationships are defined consistently according to the shared part semantics set, while the horizontal part symmetry and adjacency relationships are detected from the part annotations that provide different part structures for different shapes.
We directly follow the part semantics, hierarchy and relationships introduced in StructureNet~\cite{mo2019structurenet}, but we disentangle the unified part hierarchy into two disentangled but synergistic structure and geometry hierarchies (Figure~\ref{fig:representation}).
Following StructureNet~\cite{mo2019structurenet}, we only use the shapes where each parent part has a maximum of 10 children parts.

\begin{figure}[t]
    \centering
    \subfigure[Chair]{
    \begin{minipage}[b]{0.48\linewidth}
    {\includegraphics[width=0.48\linewidth]{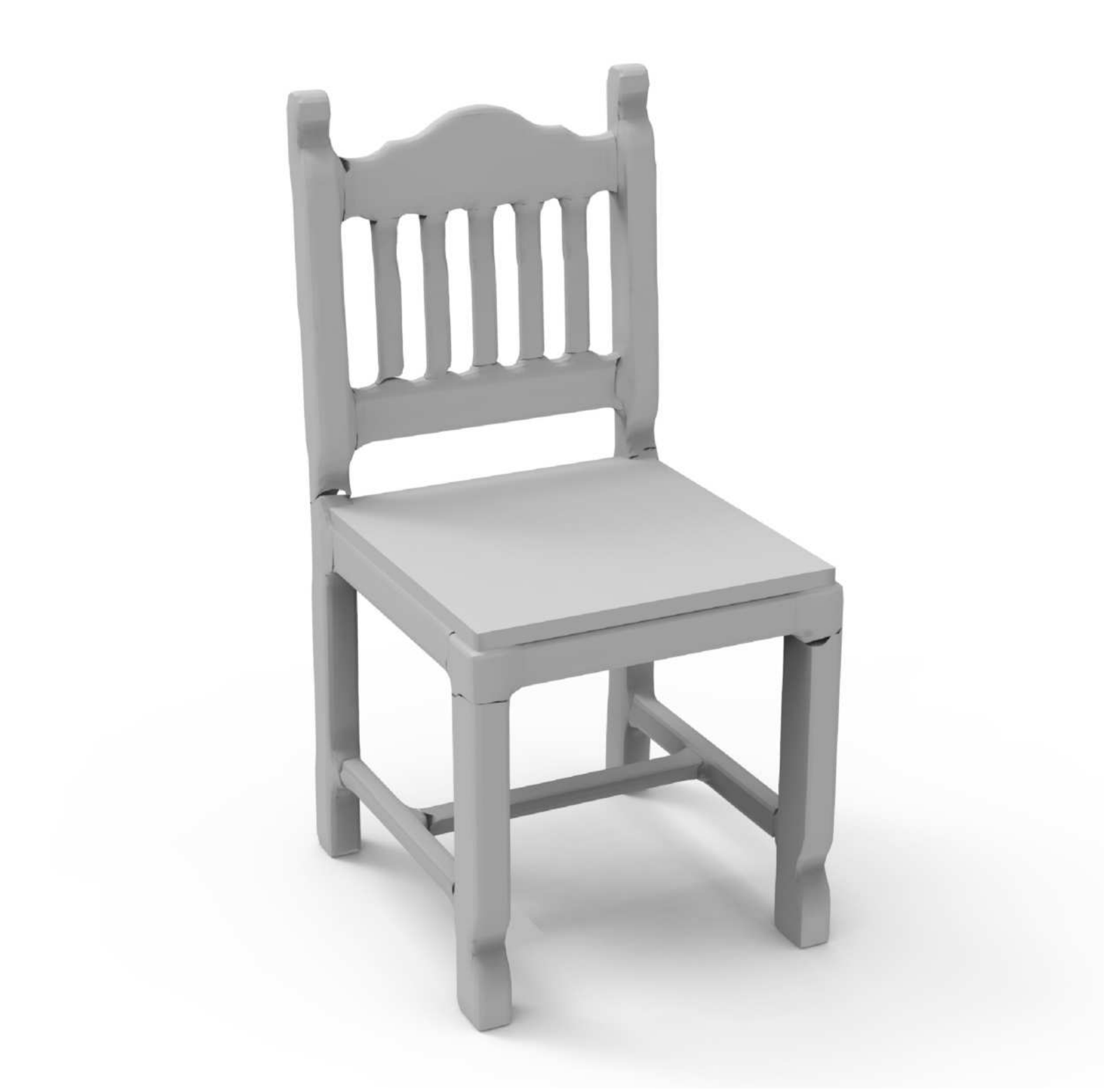}
    \includegraphics[width=0.48\linewidth]{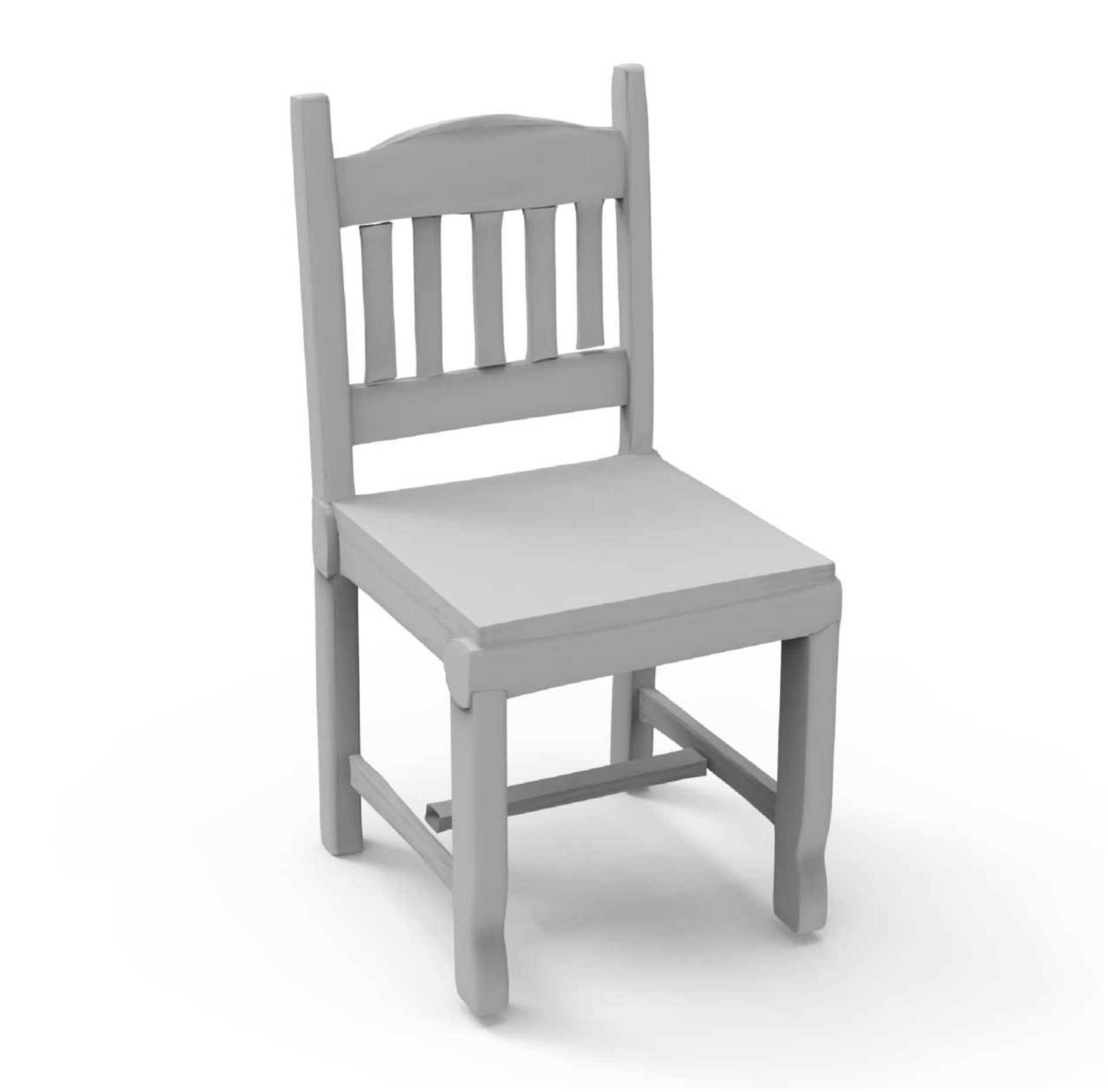}
    }
    \end{minipage}}
    \subfigure[Table]{
    \begin{minipage}[b]{0.48\linewidth}
    {
    \includegraphics[width=0.48\linewidth]{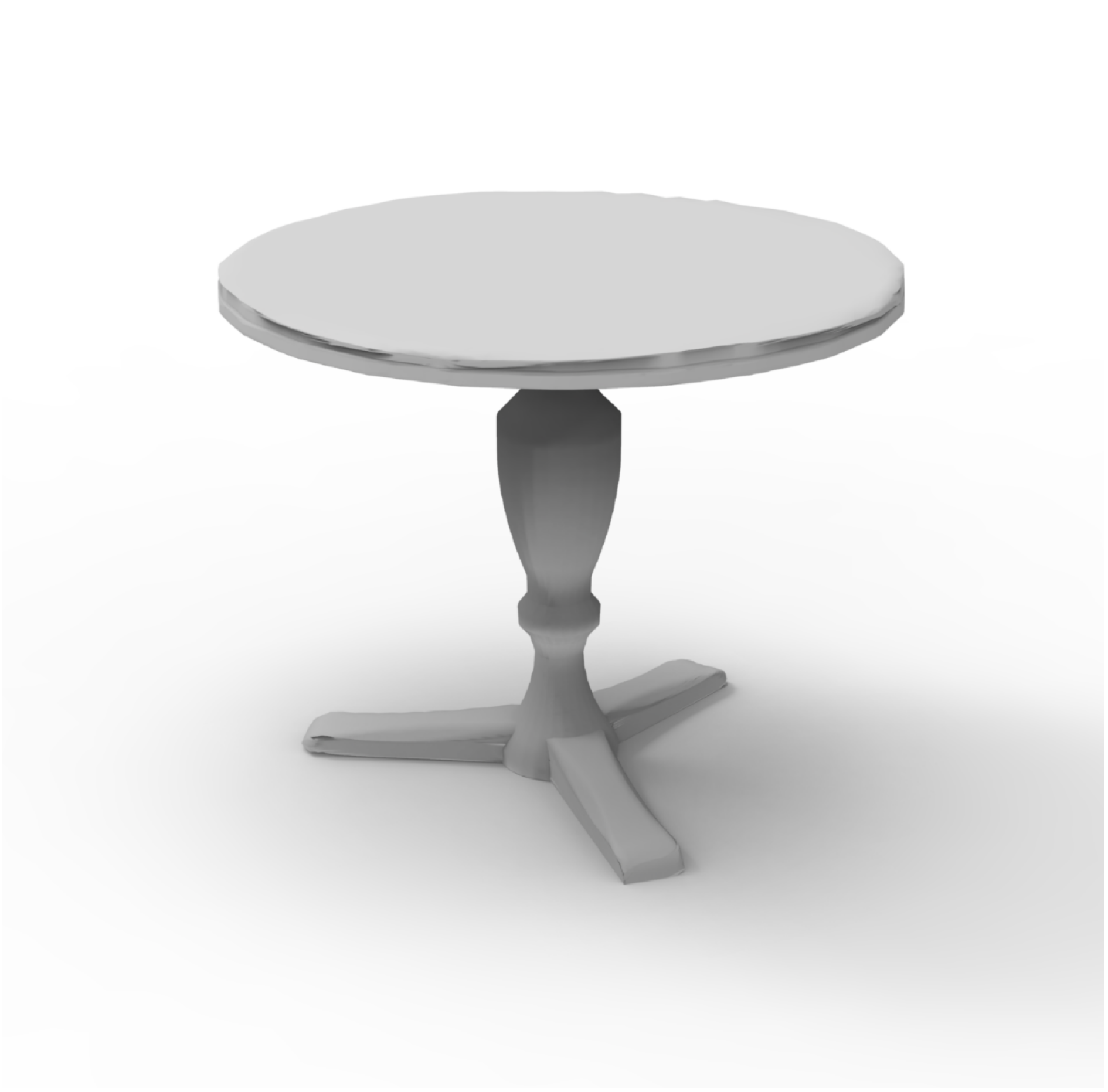}
    \includegraphics[width=0.48\linewidth]{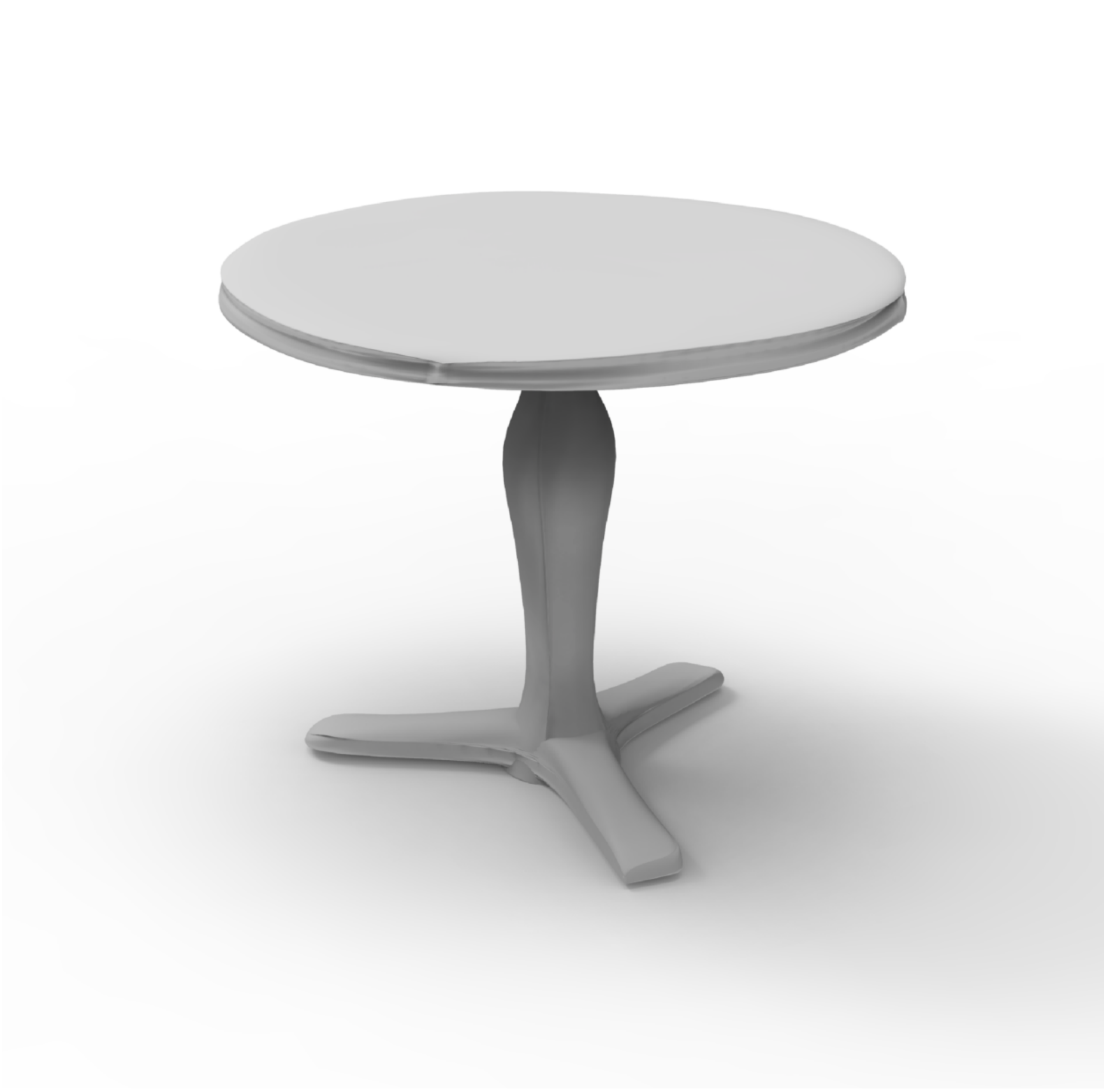}
    }
    \end{minipage}}
    \subfigure[Cabinet]{
    \begin{minipage}[b]{0.48\linewidth}
    {
    \includegraphics[width=0.48\linewidth]{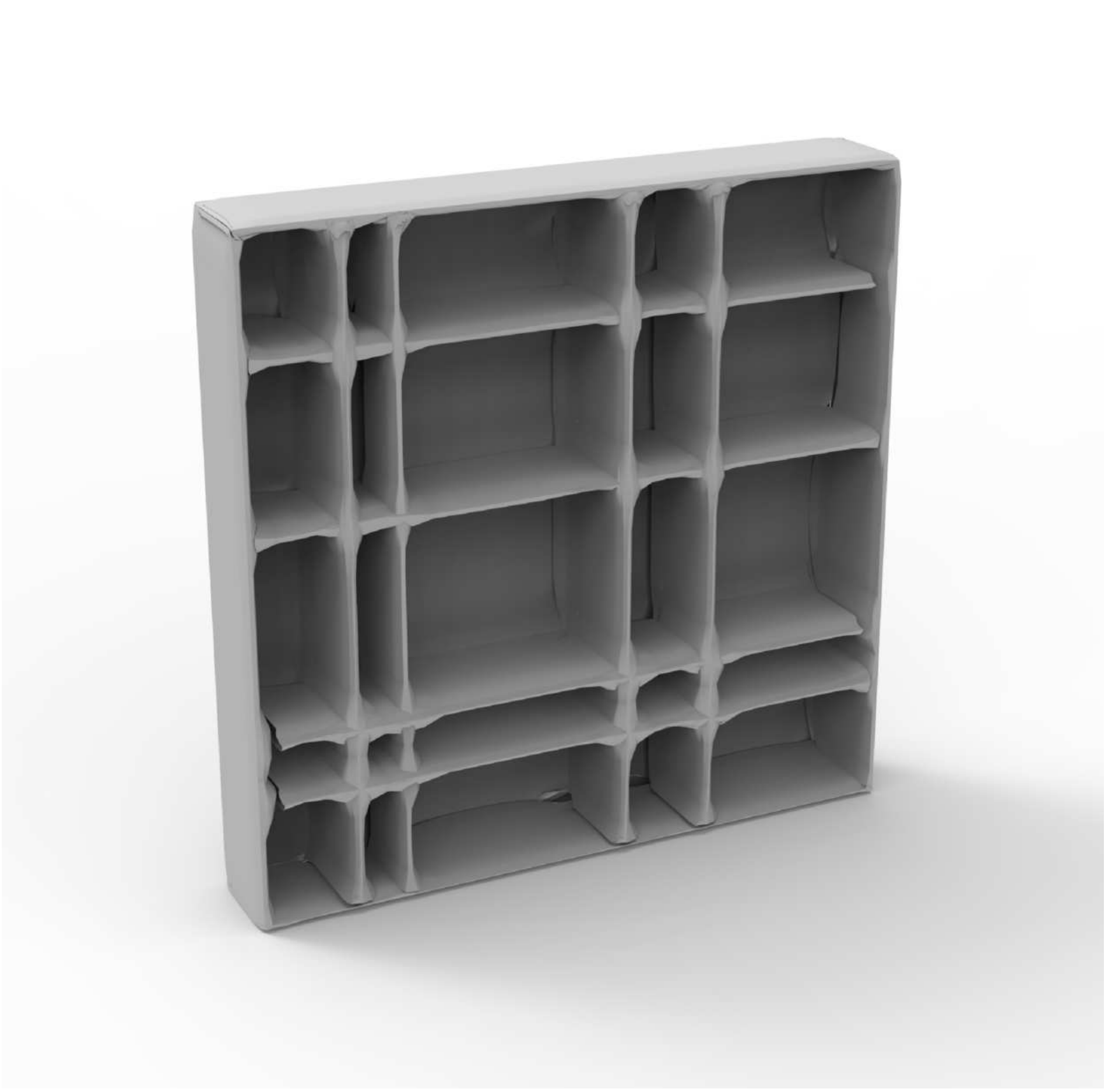}
    \includegraphics[width=0.48\linewidth]{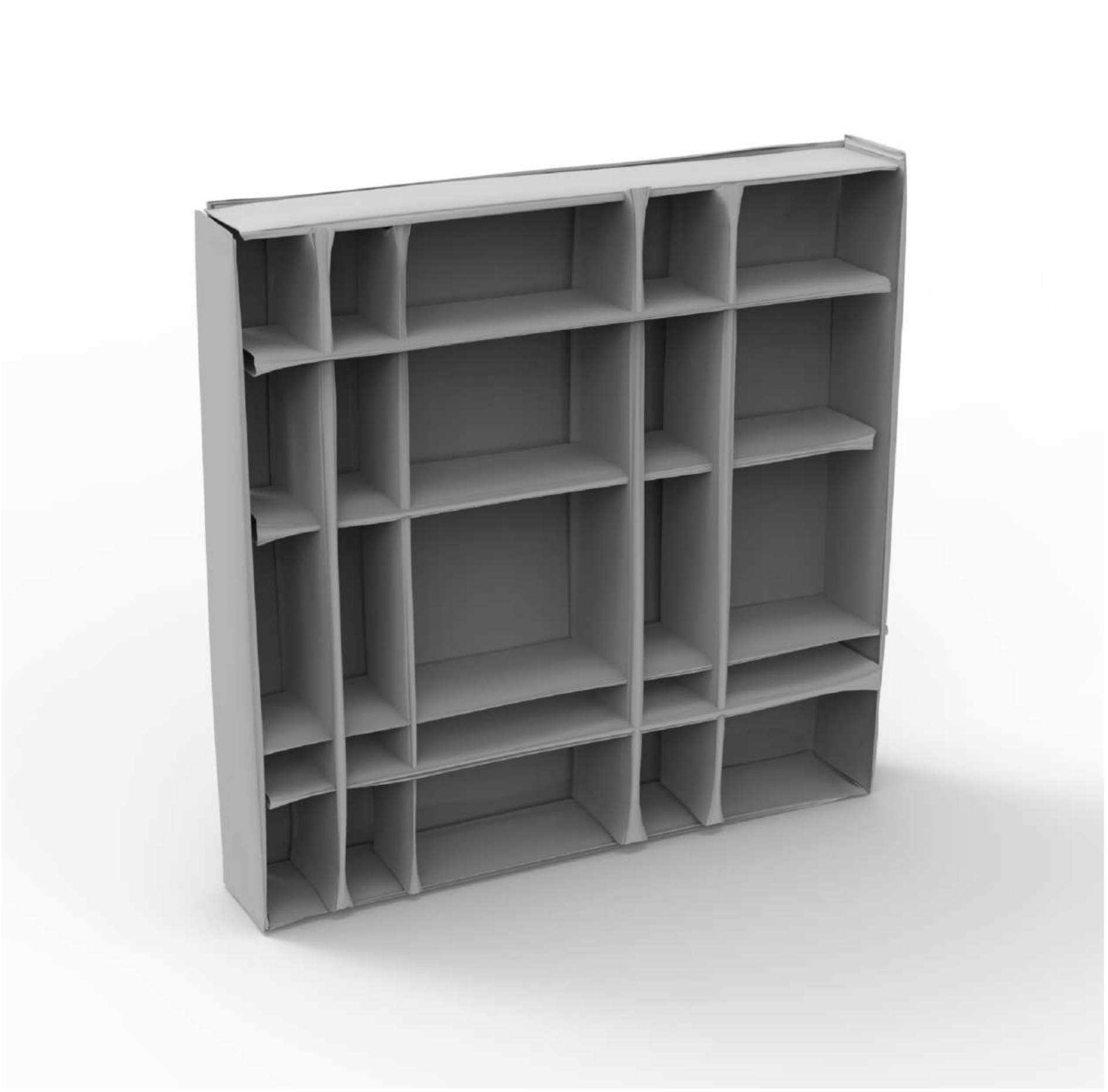}
    }
    \end{minipage}}
    \subfigure[Lamp]{
    \begin{minipage}[b]{0.48\linewidth}
    {
    \includegraphics[width=0.48\linewidth]{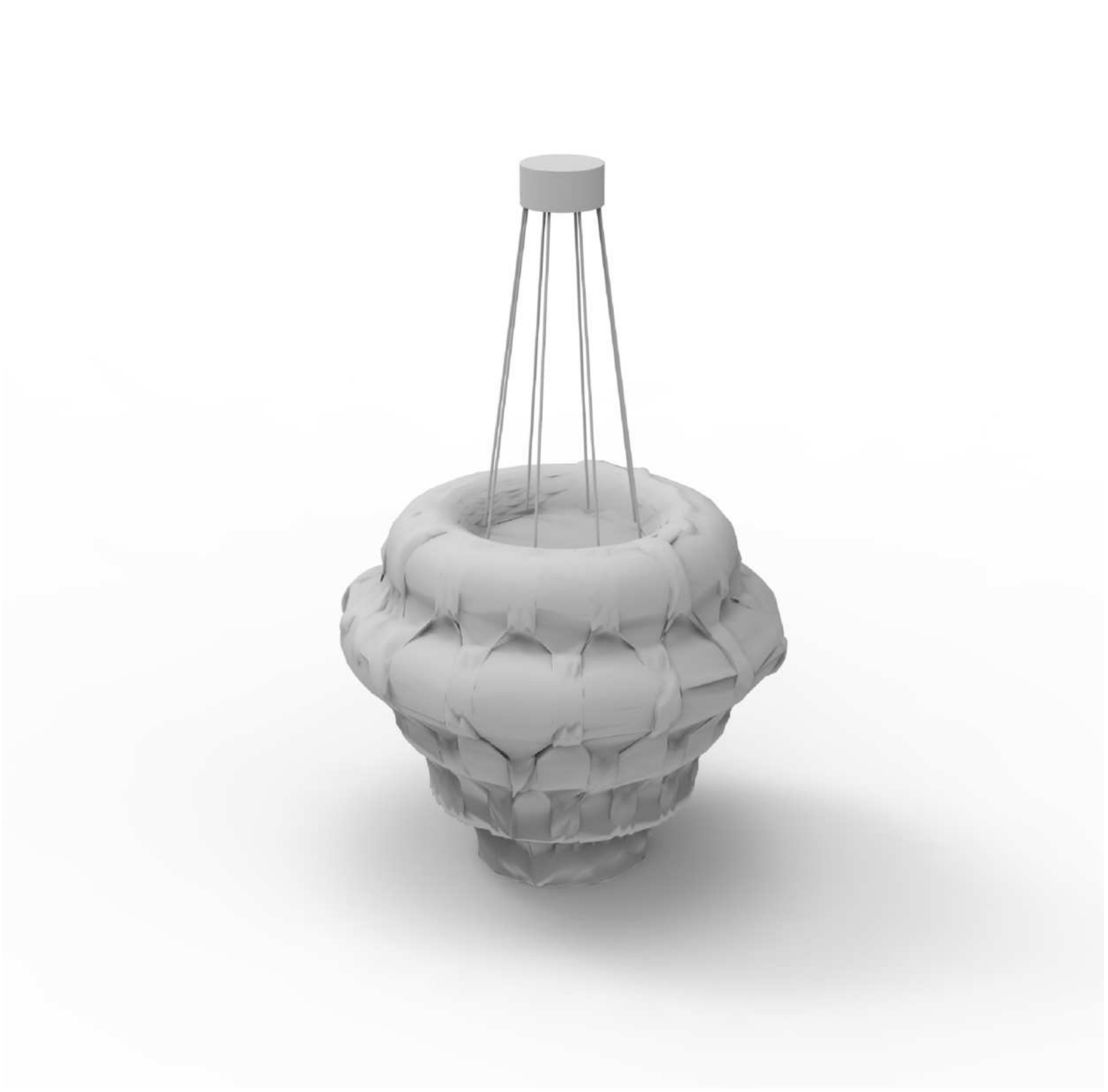}
    \includegraphics[width=0.48\linewidth]{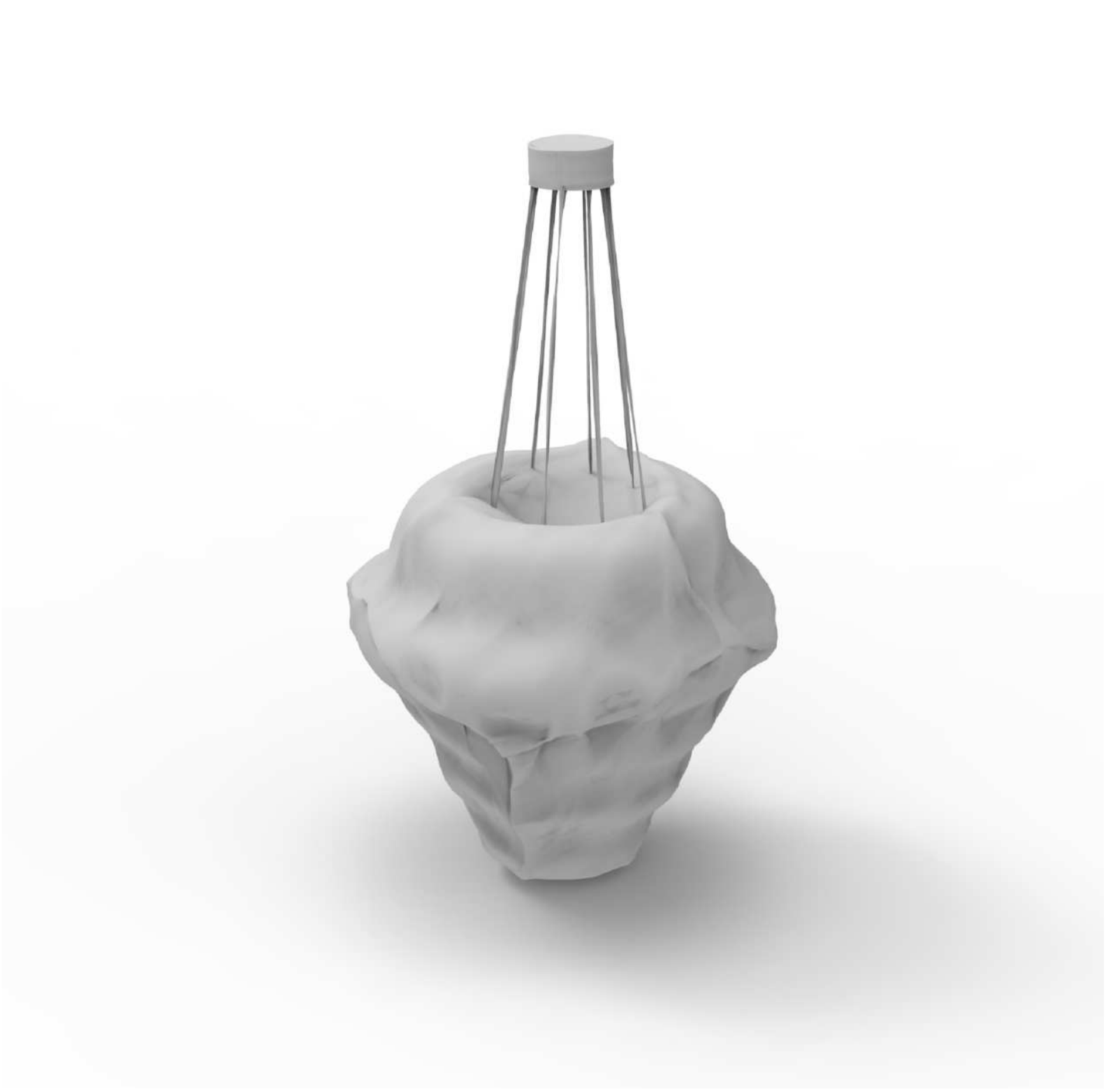}
    }
    \end{minipage}}
    \vspace{-4mm}
    \caption{\yjr{The gallery of shape reconstruction results on PartNet. For each set of results, the left column shows the ground-truth targets and the right column presents our reconstruction results. We observe that our method can capture both complex shape structures and detailed part geometry. }
    }
    \label{fig:reconstruction}
    \vspace{-3mm}
\end{figure}

\yj{Moreover, for quantitatively evaluating the task of disentangled shape reconstruction, 
we further introduce a synthetic dataset that contains 10,800 shapes
 with 54 kinds of shape structures and 200 geometric variations. 
Each shape is generated by picking one shape structure and one geometric variation, granting us access to the ground-truth shape synthesis outcome for every configuration pair.
The dataset is divided into the training and test sets with a ratio of 3:1.
\yj{For the detailed implementation, dataset description and training of our network, please refer to the supplementary material.}
We will release the code and data for facilitating future research.}

\subsection{Shape Reconstruction}
In this section, we present the shape reconstruction performance of our DSG-Net and provide quantitative and qualitative comparisons to the state-of-the-art 3D shape generative models.
Figure~\ref{fig:reconstruction} shows the shape reconstruction results for our DSG-Net on the four shape categories in PartNet. 
\yj{We present more reconstructed results in the supplementary material.}
We observe that our method successfully captures both the complex shape structure and the fine-grained geometry details.
Next, we present qualitative results on PartNet and provide quantitative evaluations on the synthetic dataset where we are provided with the ground-truth re-synthesized outputs.
\yjr{Please refer to the supplementary material for more results on shape reconstruction.}

\paragraph{Baselines.} 
We compare DSG-Net to four state-of-the-art methods for learning 3D shape representations -- IM-Net~\cite{chen2019learning}, BSP-Net~\cite{chen2019bsp}, StructureNet~\cite{mo2019structurenet}, and SDM-Net~\cite{gaosdmnet2019}, \yj{as well as an ablated version (SN + Mesh) of our method}.
IM-Net learns an implicit function representation for encoding 3D shapes, while BSP-Net puts attention on designing a compact mesh representation for 3D shapes.
They both represent a shape as a whole, without explicit modeling of shape parts and structures.
StructureNet and SDM-Net are more relevant baselines to our method since they both explicitly represent shapes as part hierarchies.
StructureNet uses point cloud representation for the part geometry, which we empirically find less effective on generating fine-grained shape geometry details.
SDM-Net represents shapes with shallower part hierarchies, which prevents it from generating shapes with complicated structures.
\yjr{The SN + Mesh is a naive combination of StructureNet~\cite{mo2019structurenet} backbone and ACAP mesh representation~\cite{gao2019sparse,gaosdmnet2019}, to validate that our proposed disentangled structure and geometry representation and the cycled disentanglement indeed help improve the performance for learning 3D shape generative models. \yjrr{For the performance of SN+Mesh, please refer to supplementary material.}}
\yj{All the methods are trained on the same data for the four object categories.}
\yj{We also compare to SAGNet~\cite{wu2019sagnet} in the supplementary material.}

\begin{table}[t]
\fontsize{8}{12}\selectfont
  \centering
  \caption{Shape reconstruction quantitative evaluations. \yj{We use two geometry metrics (CD and EMD) and one structure metric (HierInsSeg)}. 
  \yjr{DSG-Net achieves the best geometry and structure performance compared to all baseline methods. Thanks to the cycled disentanglement, DSG-Net achieves the best HierInsSeg scores, followed by StructureNet
  , 
  and DSG-Net outperforms them in terms of the geometric metrics by a large margin. 
  }
  }
    \begin{tabular}{ccccc}
    \toprule[1pt]
    \multirow{2}[4]{*}{DataSet} & \multirow{2}[4]{*}{Method} & \multicolumn{2}{c}{Geometry Metrics} & Structure Metrics \\
    \cmidrule{3-5}          &       & CD{\scriptsize$\times 10^{-3}$}$\downarrow$ & EMD{\scriptsize$\times 10^{-2}$}$\downarrow$ & HierInsSeg (HIS) $\downarrow$\\
    \midrule
    \midrule
    \multirow{6}[2]{*}{Chair} & StructureNet & 9.34 & 6.45 & 0.51 \\
          & IM-Net & 3.39 & 1.45 & 0.60 \\
          & BSP-Net & 8.27 & 2.07 & 0.78 \\
          & SDM-Net & 8.64 & 3.15 & 0.94 \\
          & {Ours}  & {\textbf{1.98}} & {\textbf{0.73}} & {\textbf{0.39}} \\
          & GT    &       &       & \underline{0.32} \\
    \midrule
    \midrule
    \multirow{6}[2]{*}{Table} & StructureNet & 14.63 & 5.68 & 0.97 \\
          & IM-Net & 5.04 & 2.08 & 1.13 \\
          & BSP-Net & 10.62 & 4.11 & 1.20 \\
          & SDM-Net & 9.73 & 4.63 & 1.38 \\
          & {Ours}  & {\textbf{3.42}} & {\textbf{0.75}}
          & {\textbf{0.85}} \\
          & GT    &       &       & \underline{0.65} \\
    \midrule
    \midrule
    \multirow{6}[2]{*}{Cabinet} & StructureNet & 16.34 & 5.74 & 0.57 \\
          & IM-Net & 4.73 & 3.82 & 0.72 \\
          & BSP-Net & 6.67 & 4.65 & 0.84 \\
          & SDM-Net & 18.02 & 7.9 & 1.38 \\
          & {Ours}  & {\textbf{2.96}} & {\textbf{0.97}} & {0.45} \\
          & GT    &       &       & \underline{0.35} \\
    \midrule
    \midrule
    \multirow{6}[2]{*}{Lamp} & StructureNet & 17.31 &  7.12 & 0.71 \\
          & IM-Net & 13.20 & 5.11 & 0.73 \\
          & BSP-Net & 17.17 & 7.56 & 0.98 \\
          & SDM-Net & 51.21 & 8.72 & 0.76 \\
          & {Ours}  & {\textbf{7.15}} & {\textbf{1.63}} & {\textbf{0.61}} \\
          & GT    &       &       & \underline{0.54} \\
    \bottomrule[1pt]
    \end{tabular}%
  \label{tab:reconeval}%
  \vspace{-2mm}
\end{table}%

\begin{figure}[h]
    \centering
    \includegraphics[width=0.155\linewidth]{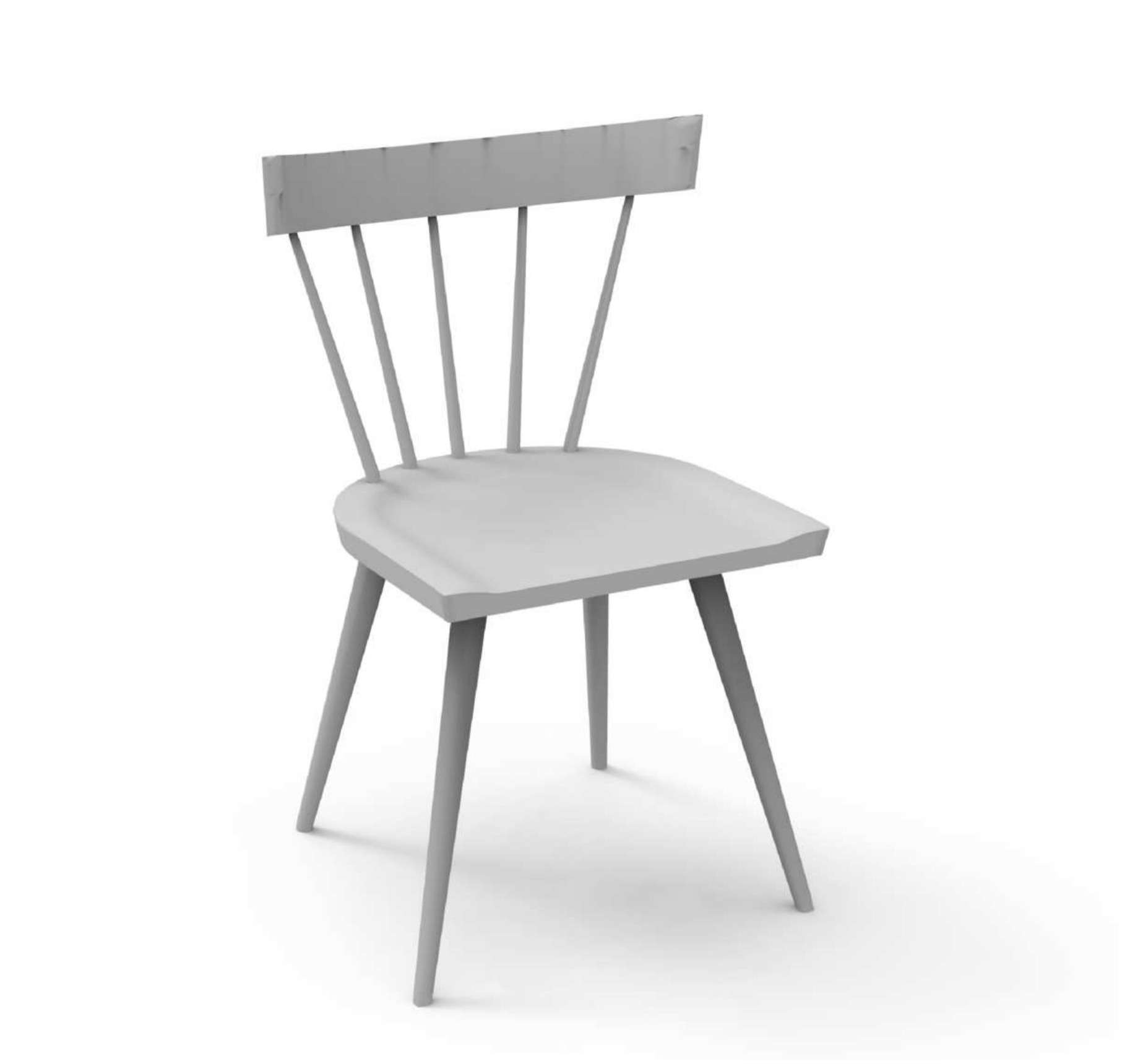}
    \includegraphics[width=0.155\linewidth]{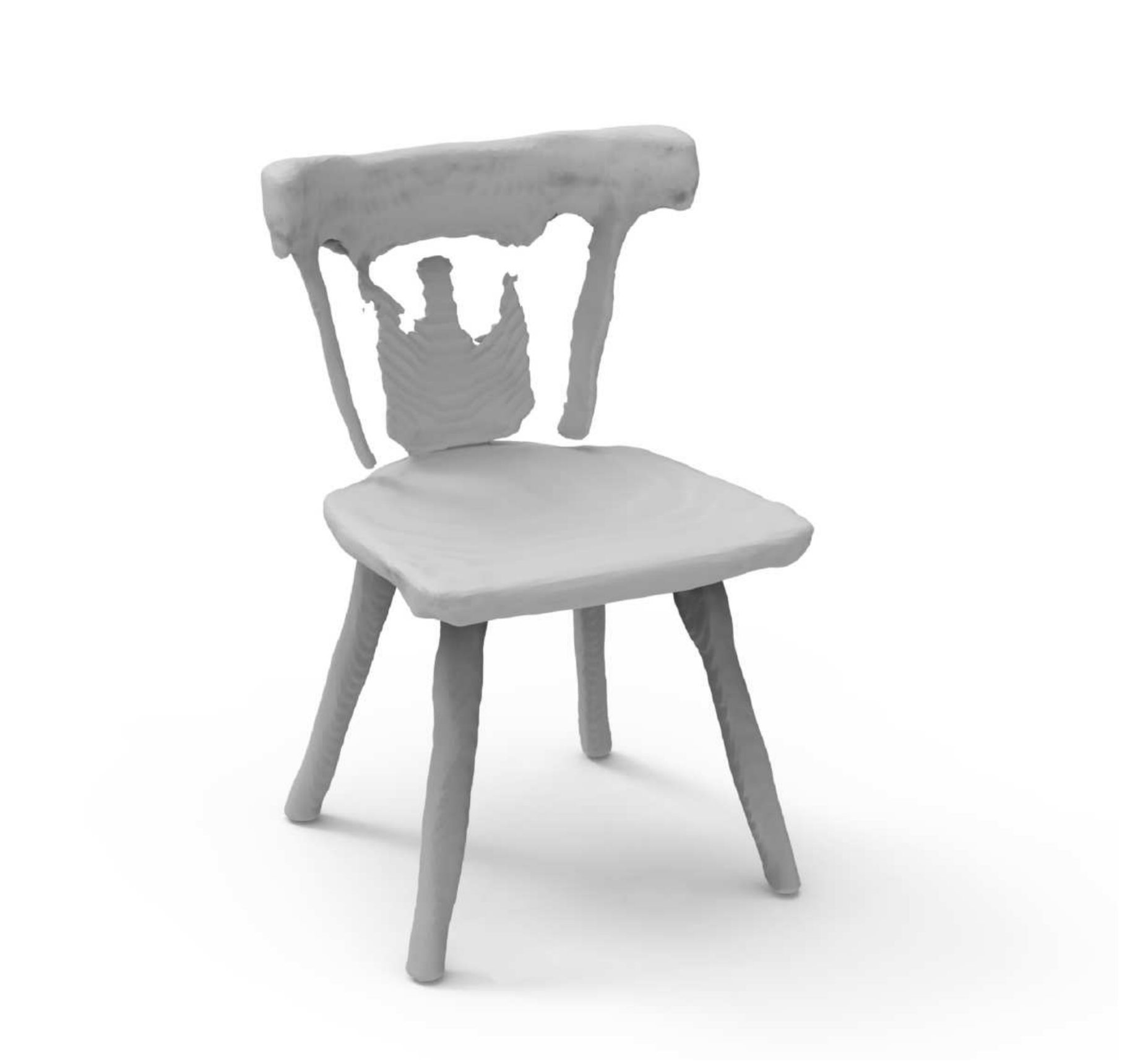}
    \includegraphics[width=0.155\linewidth]{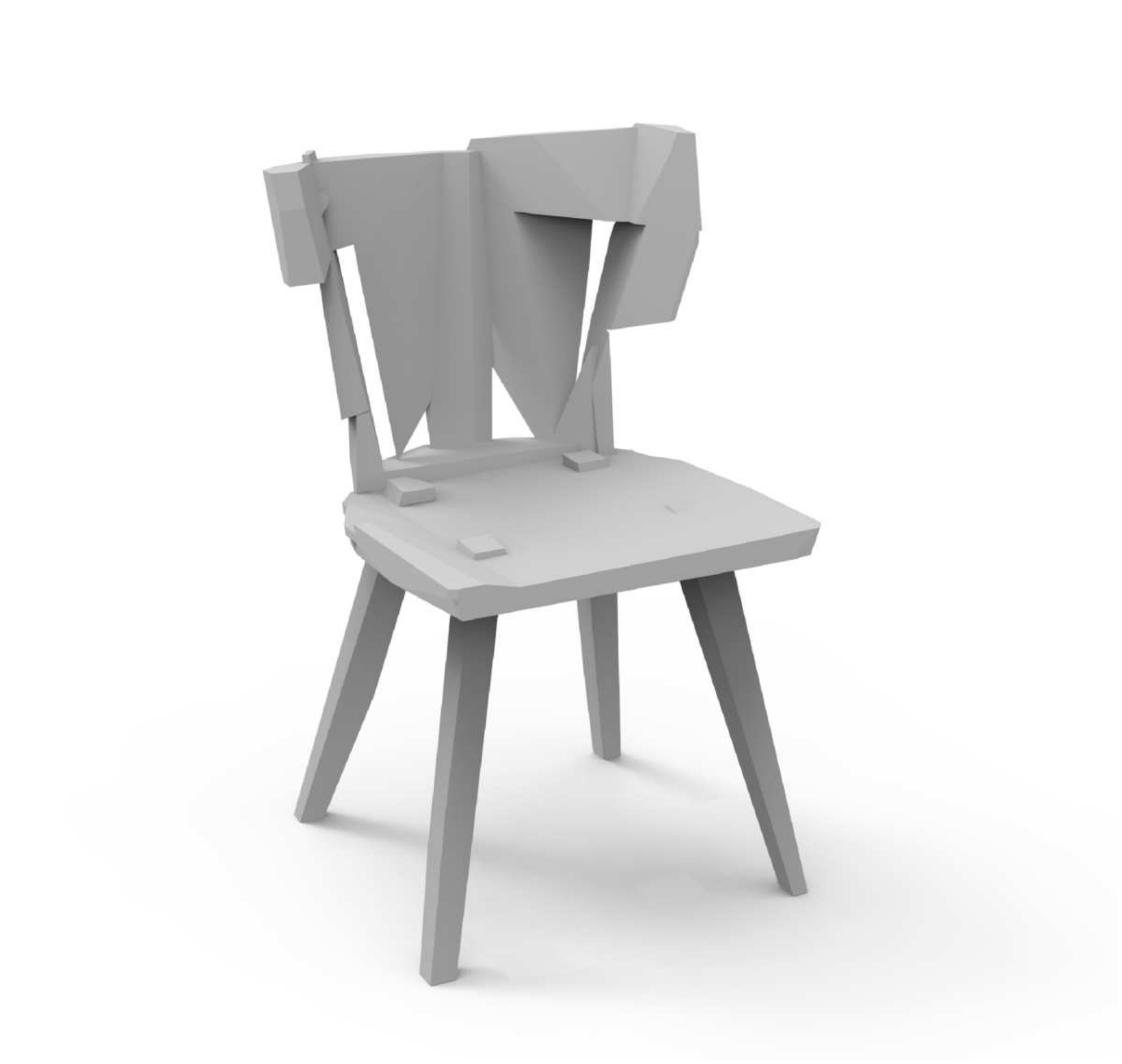}
    \includegraphics[width=0.155\linewidth]{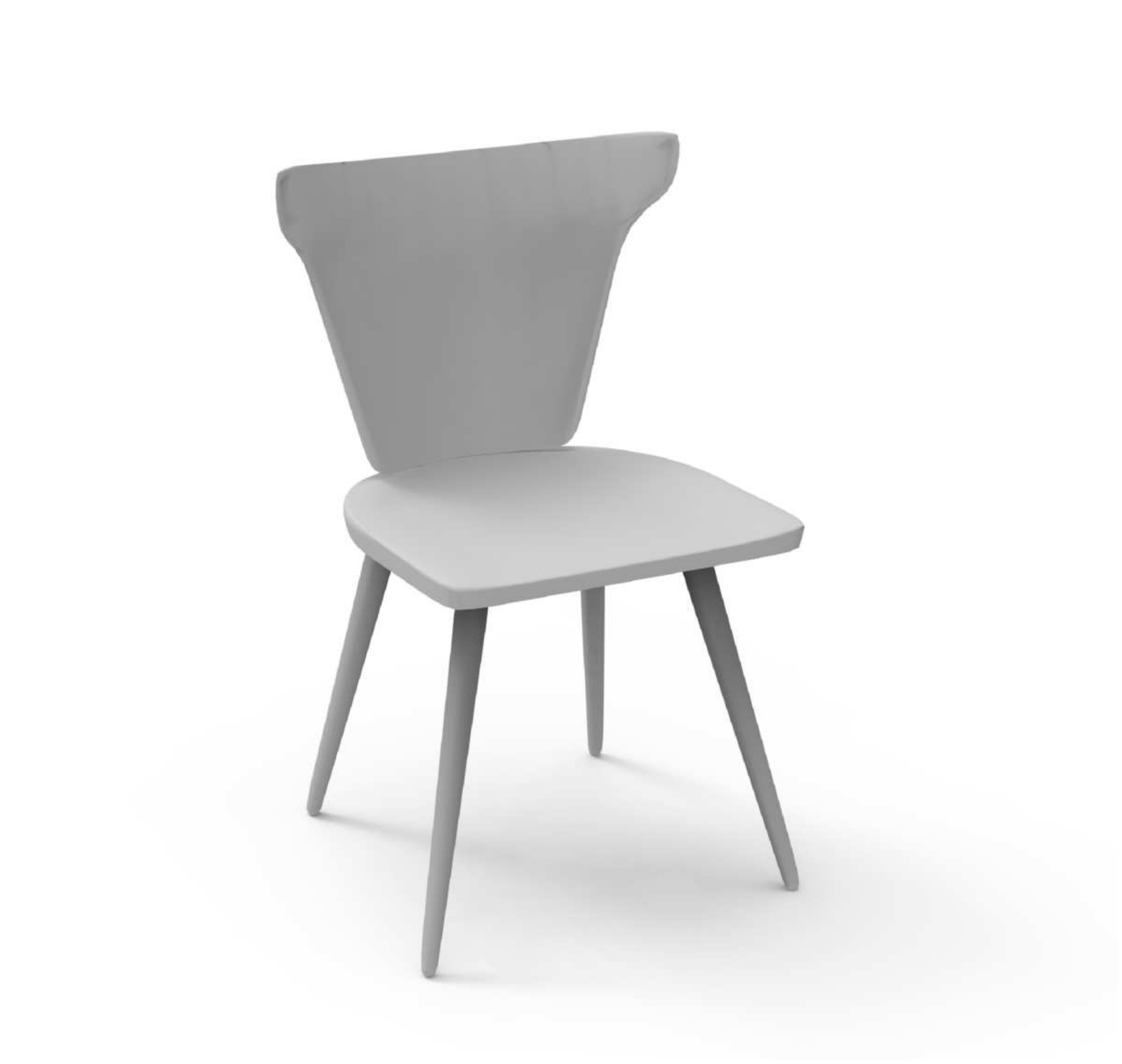}
    \includegraphics[width=0.155\linewidth]{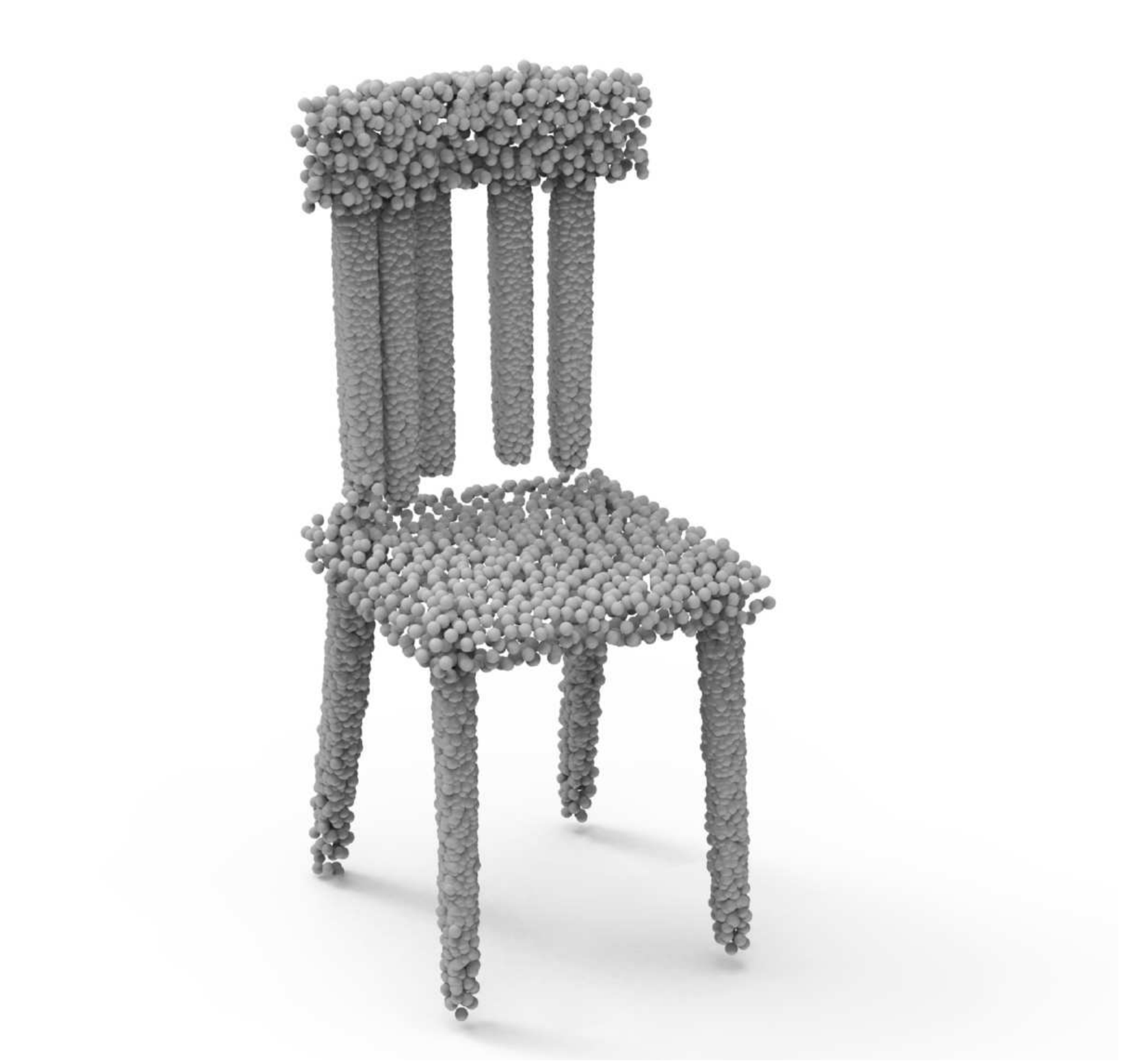}
    \includegraphics[width=0.155\linewidth]{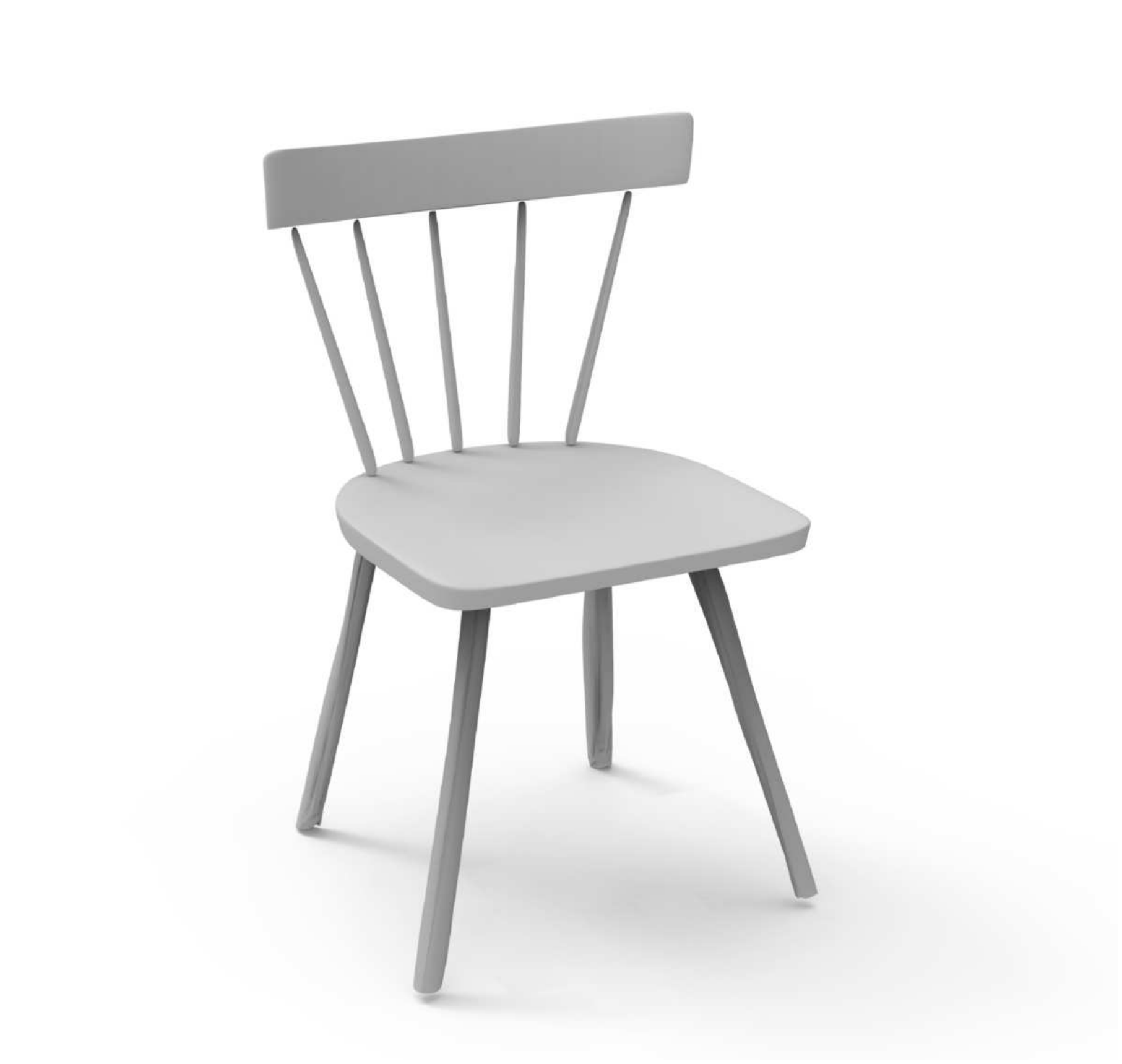}\\
    \hspace*{\fill}
    \includegraphics[width=0.145\linewidth]{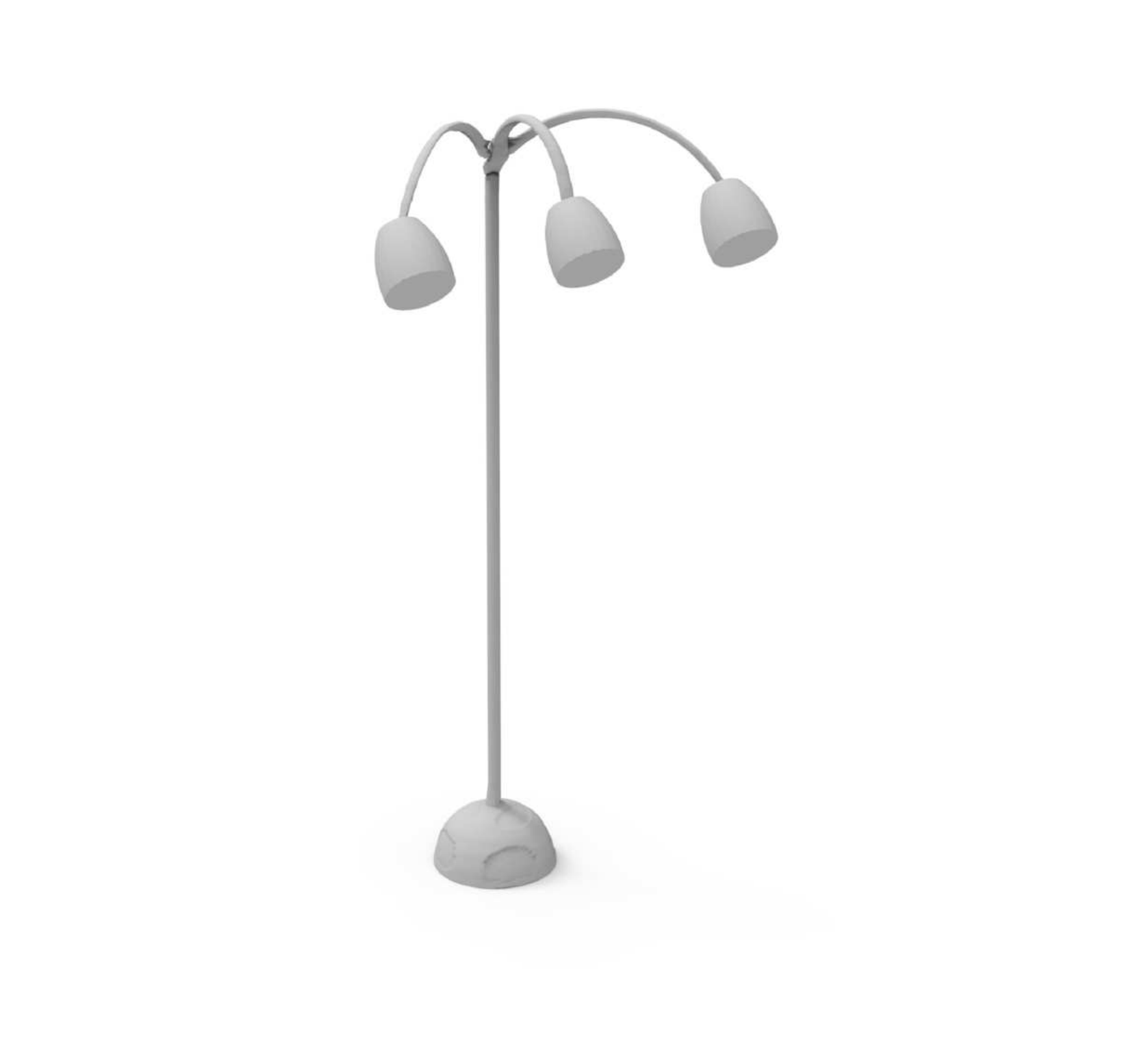}\hspace*{\fill}
    \includegraphics[width=0.145\linewidth]{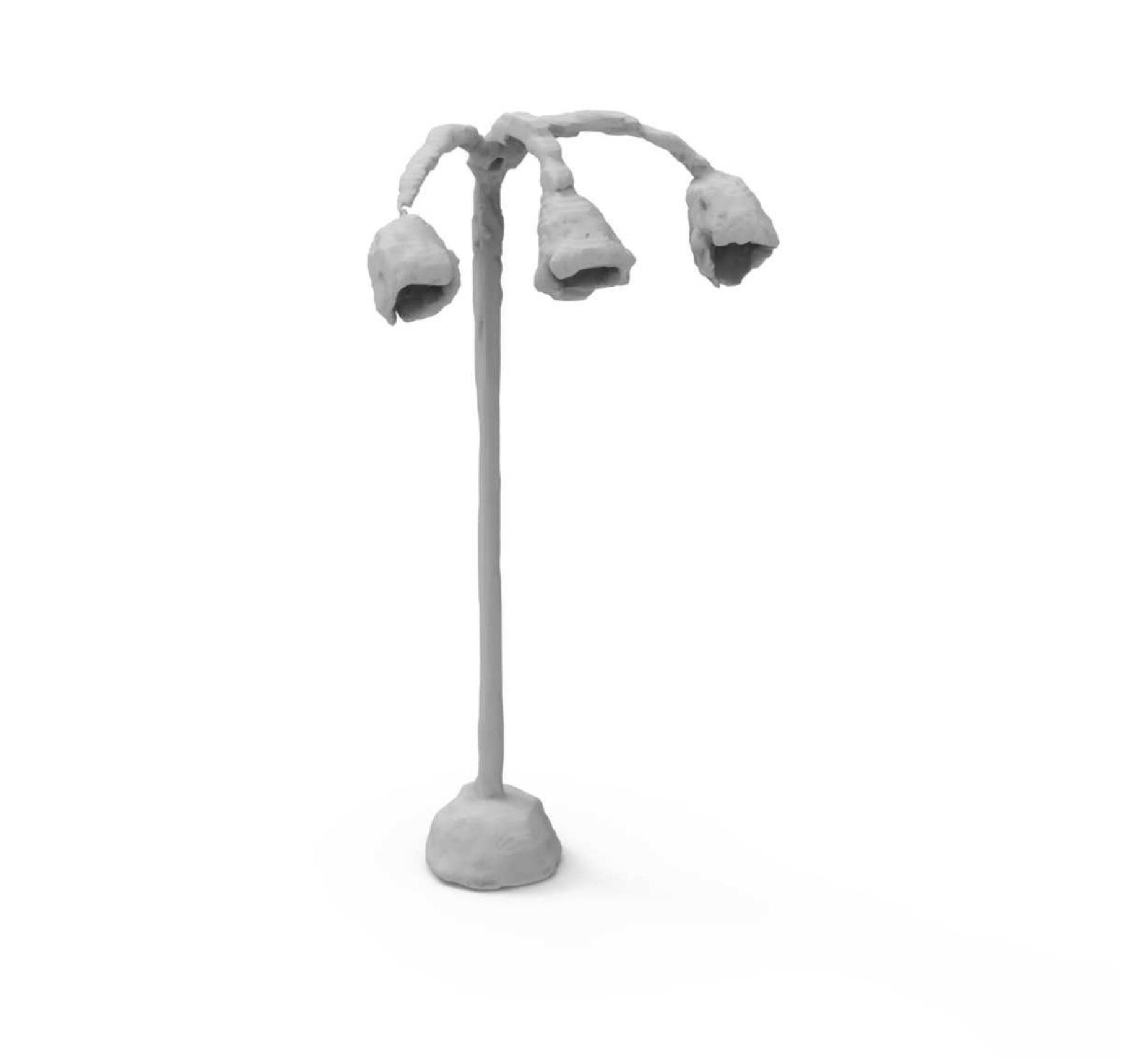}\hspace*{\fill}
    \includegraphics[width=0.145\linewidth]{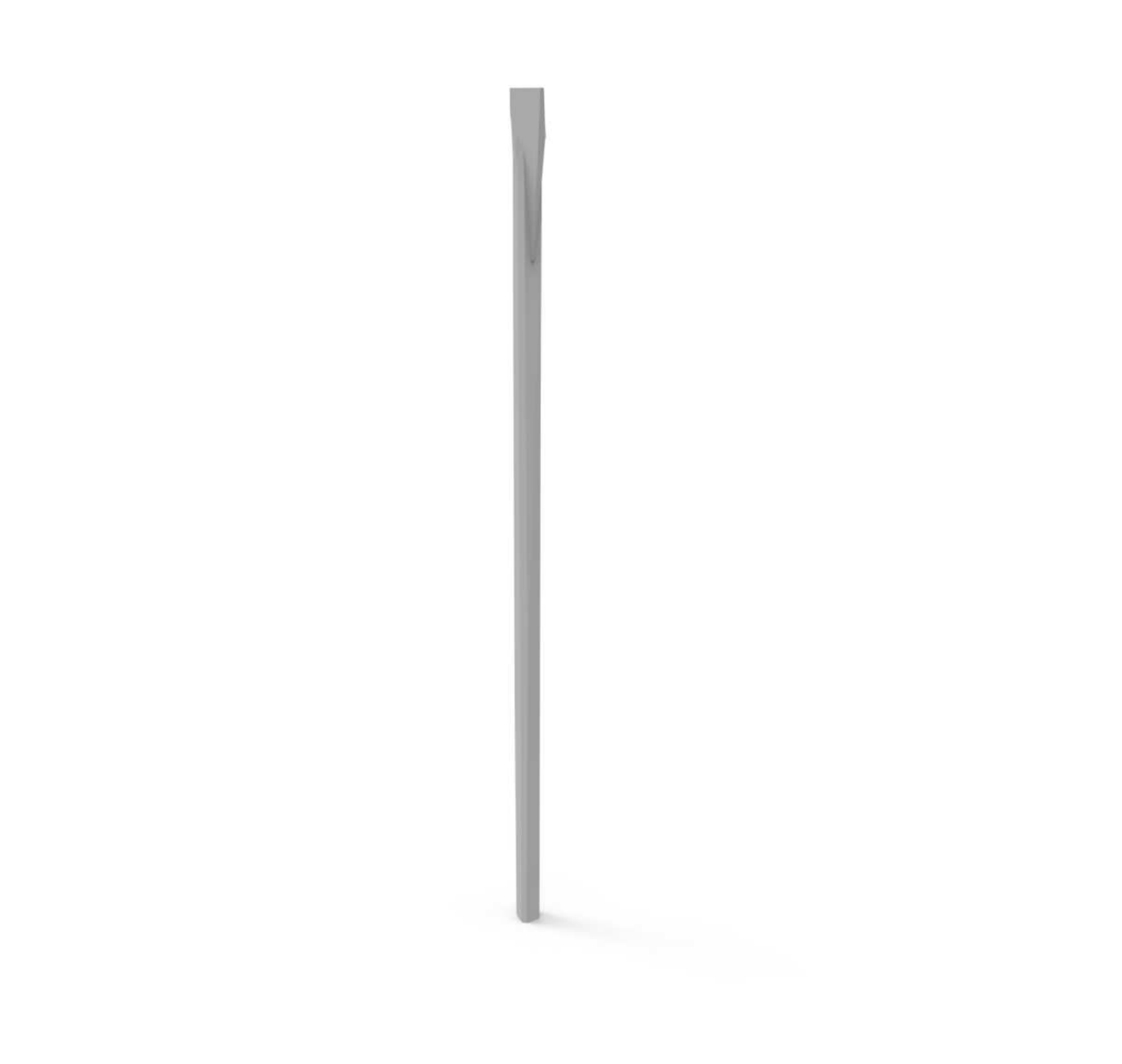}\hspace*{\fill}
    \includegraphics[width=0.145\linewidth]{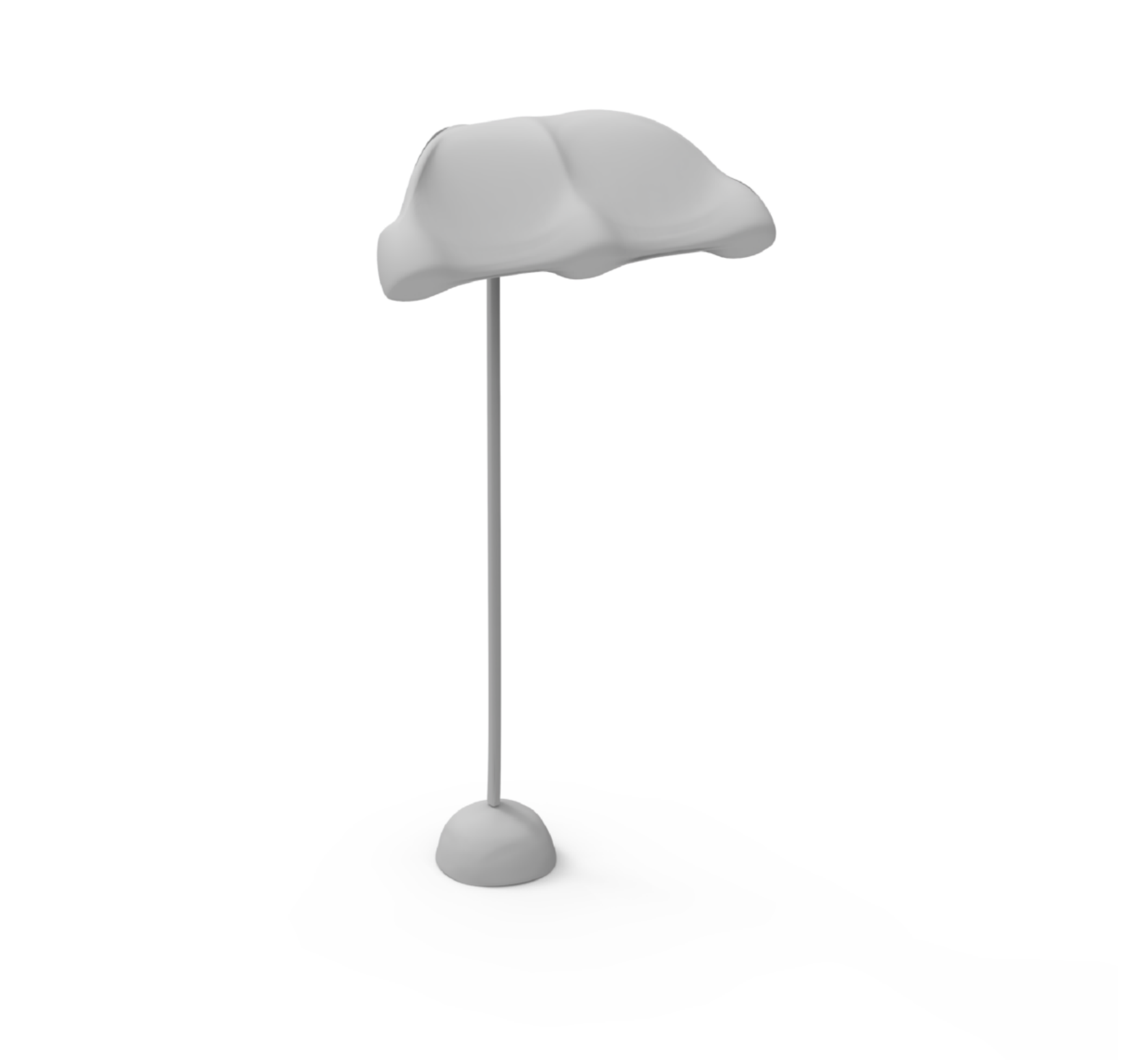}\hspace*{\fill}
    \includegraphics[width=0.145\linewidth]{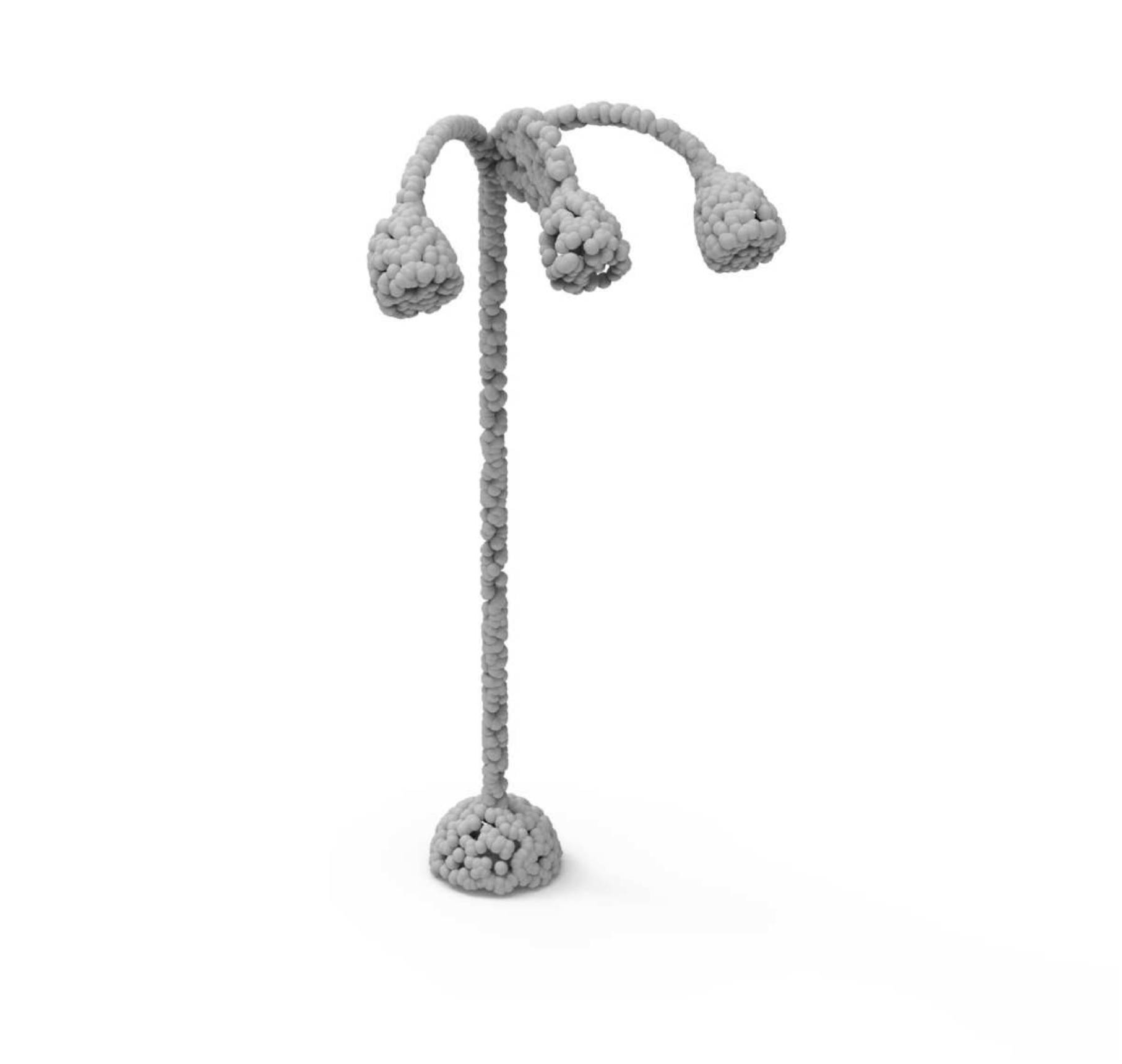}\hspace*{\fill}
    \includegraphics[width=0.145\linewidth]{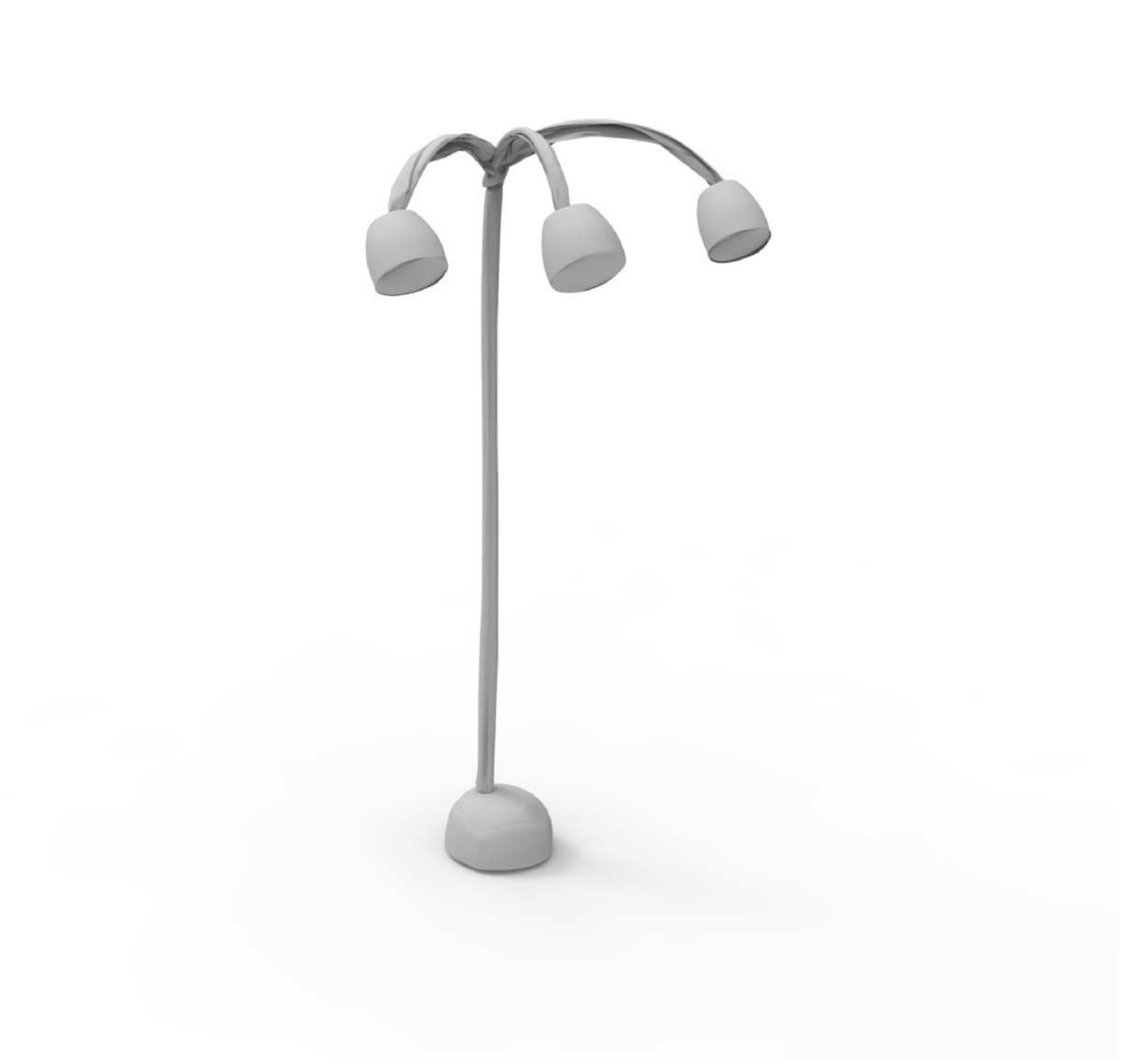}\hspace*{\fill}
    \\
    \vspace{-3mm}
    \includegraphics[width=0.145\linewidth]{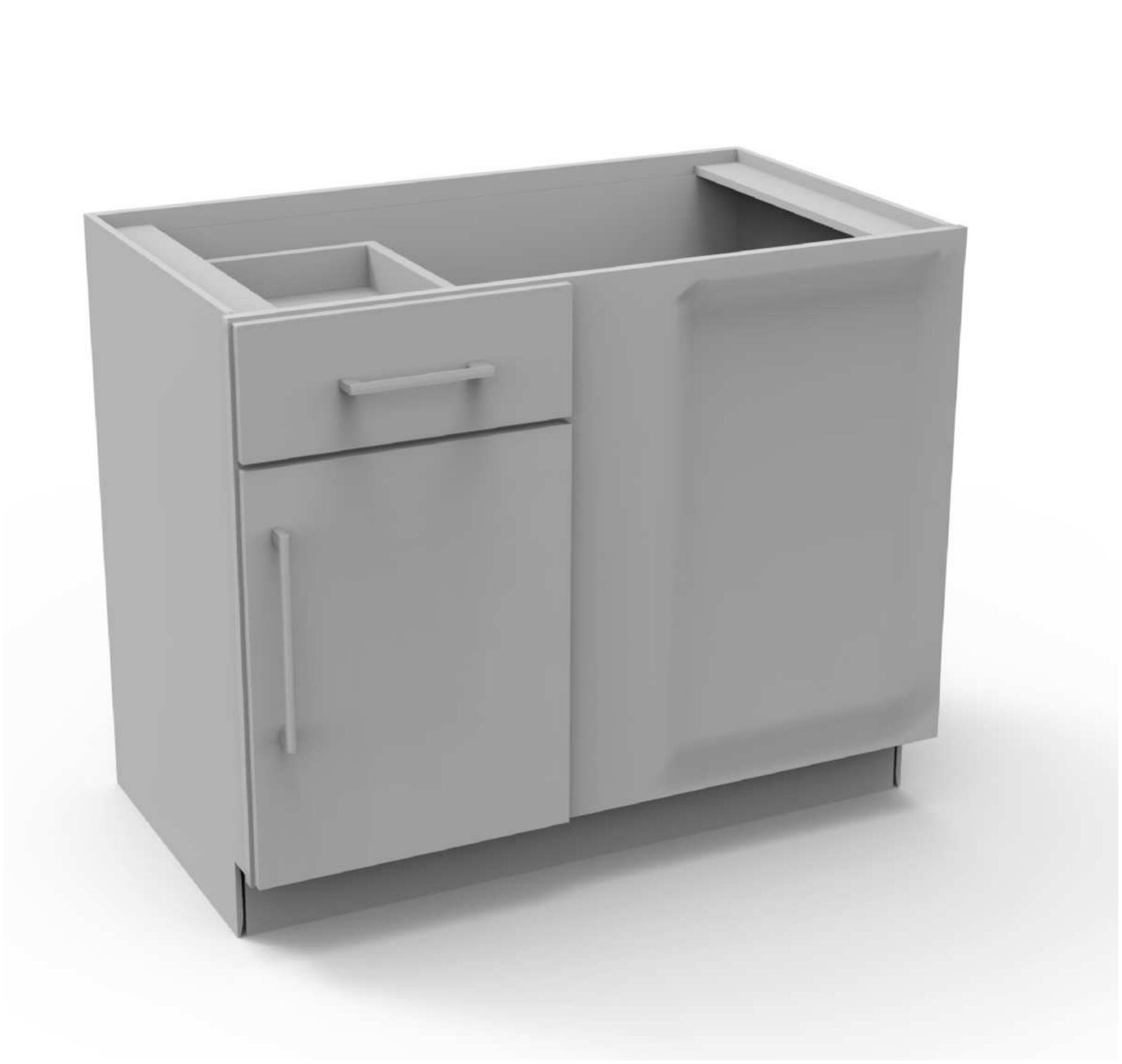}\hspace{1.3mm}
    \includegraphics[width=0.145\linewidth]{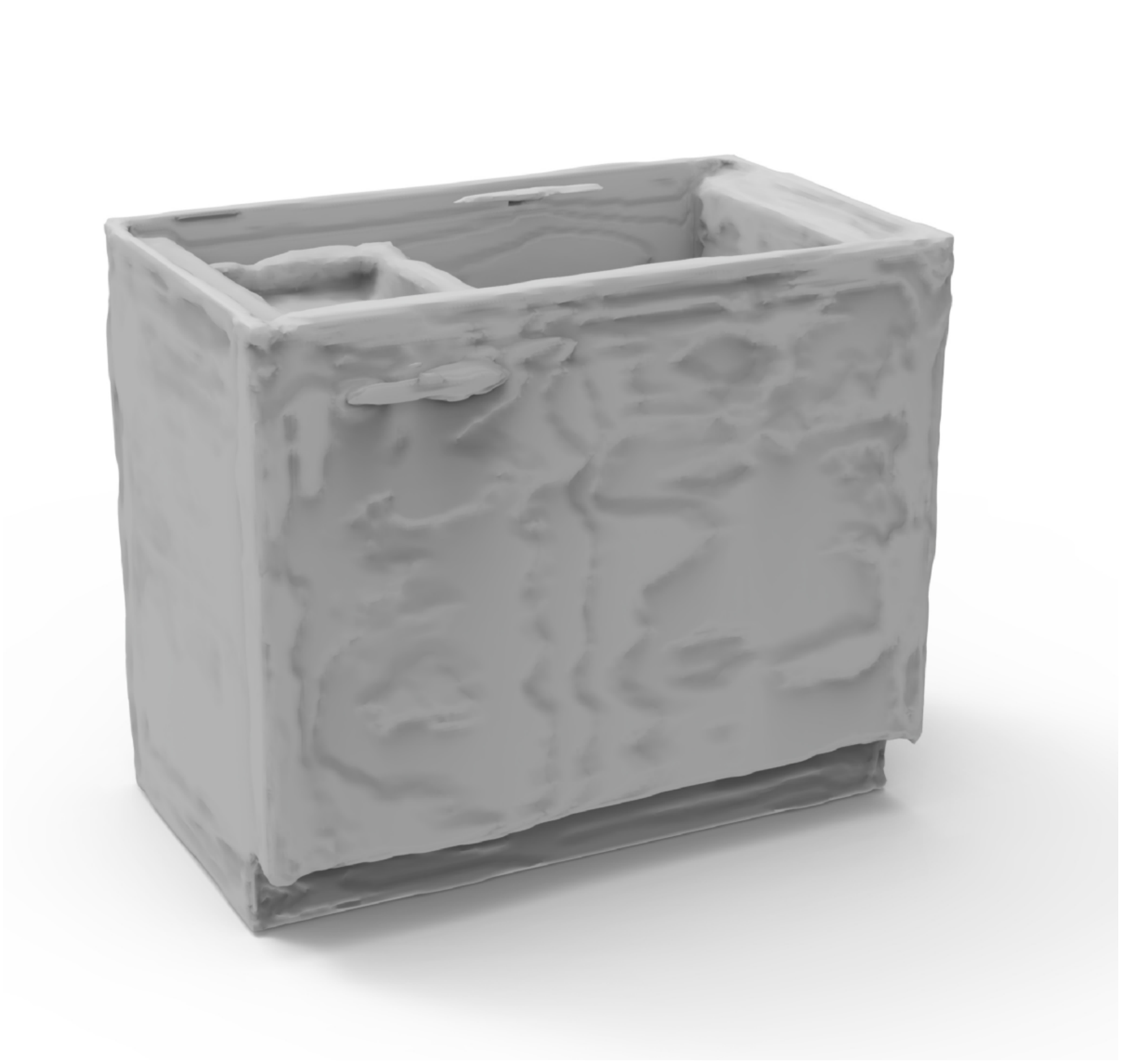}\hspace{1.3mm}
    \includegraphics[width=0.145\linewidth]{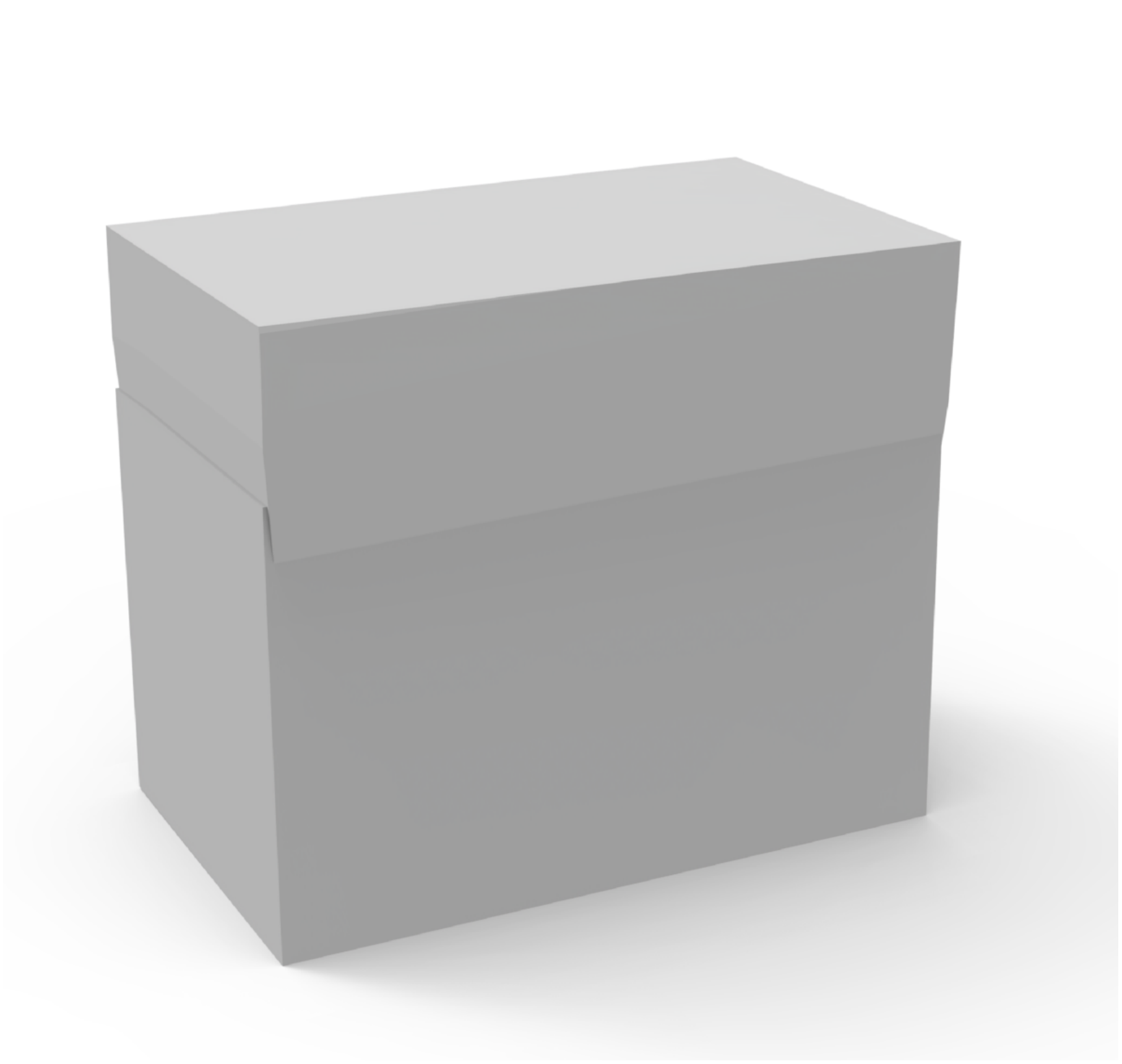}\hspace{1.3mm}
    \includegraphics[width=0.145\linewidth]{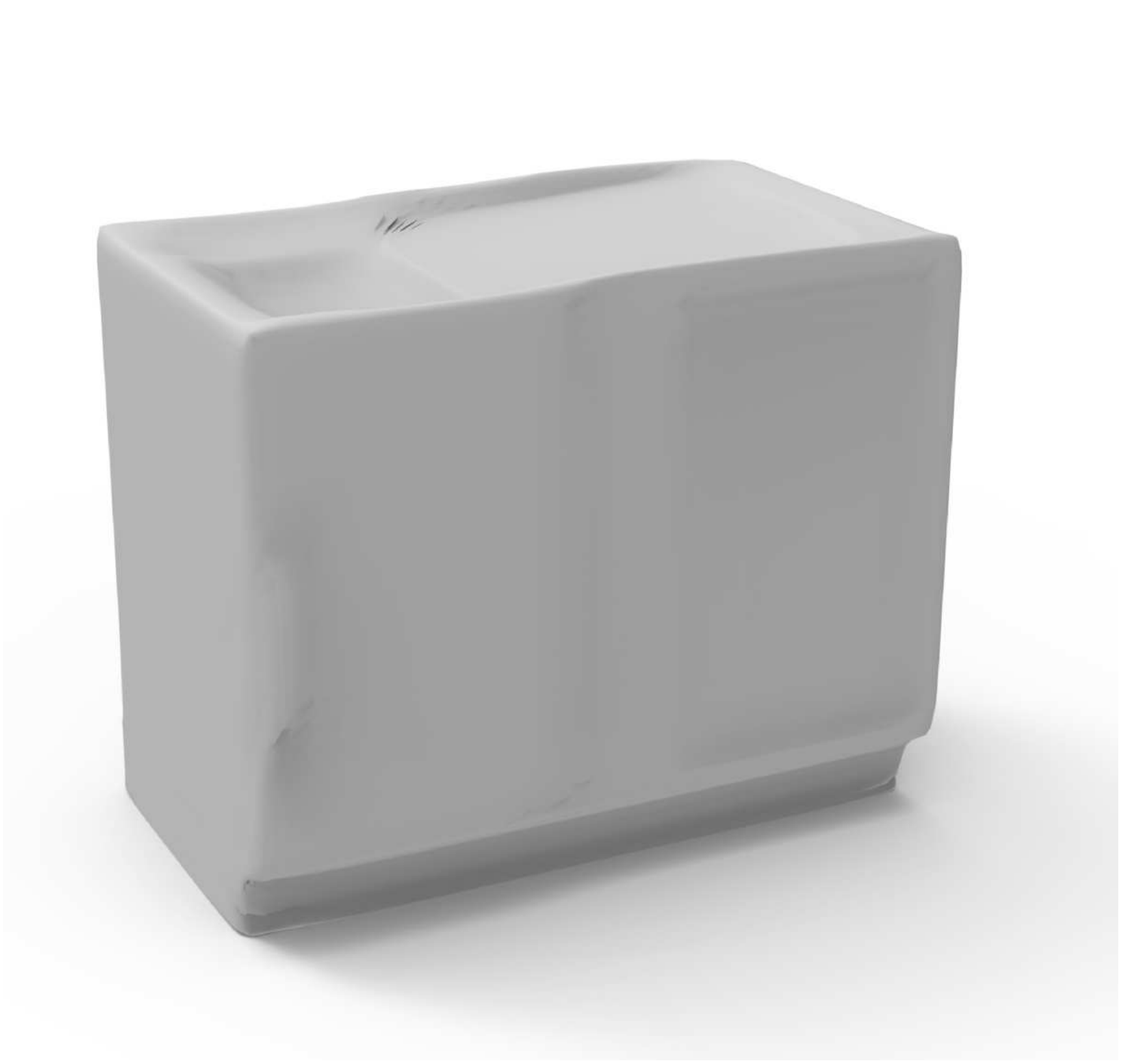}\hspace{1.3mm}
    \includegraphics[width=0.145\linewidth]{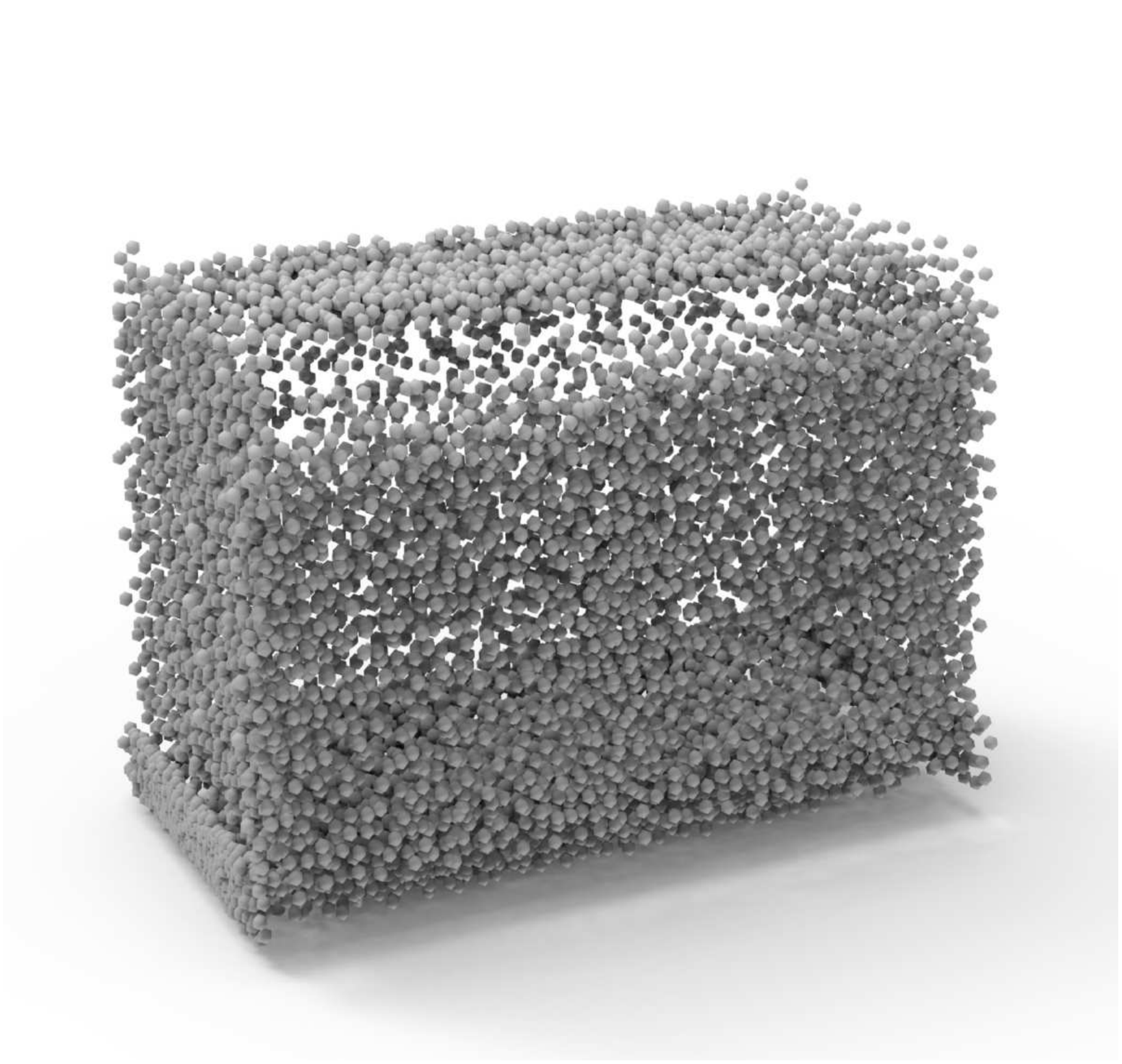}\hspace{1.3mm}
    \includegraphics[width=0.145\linewidth]{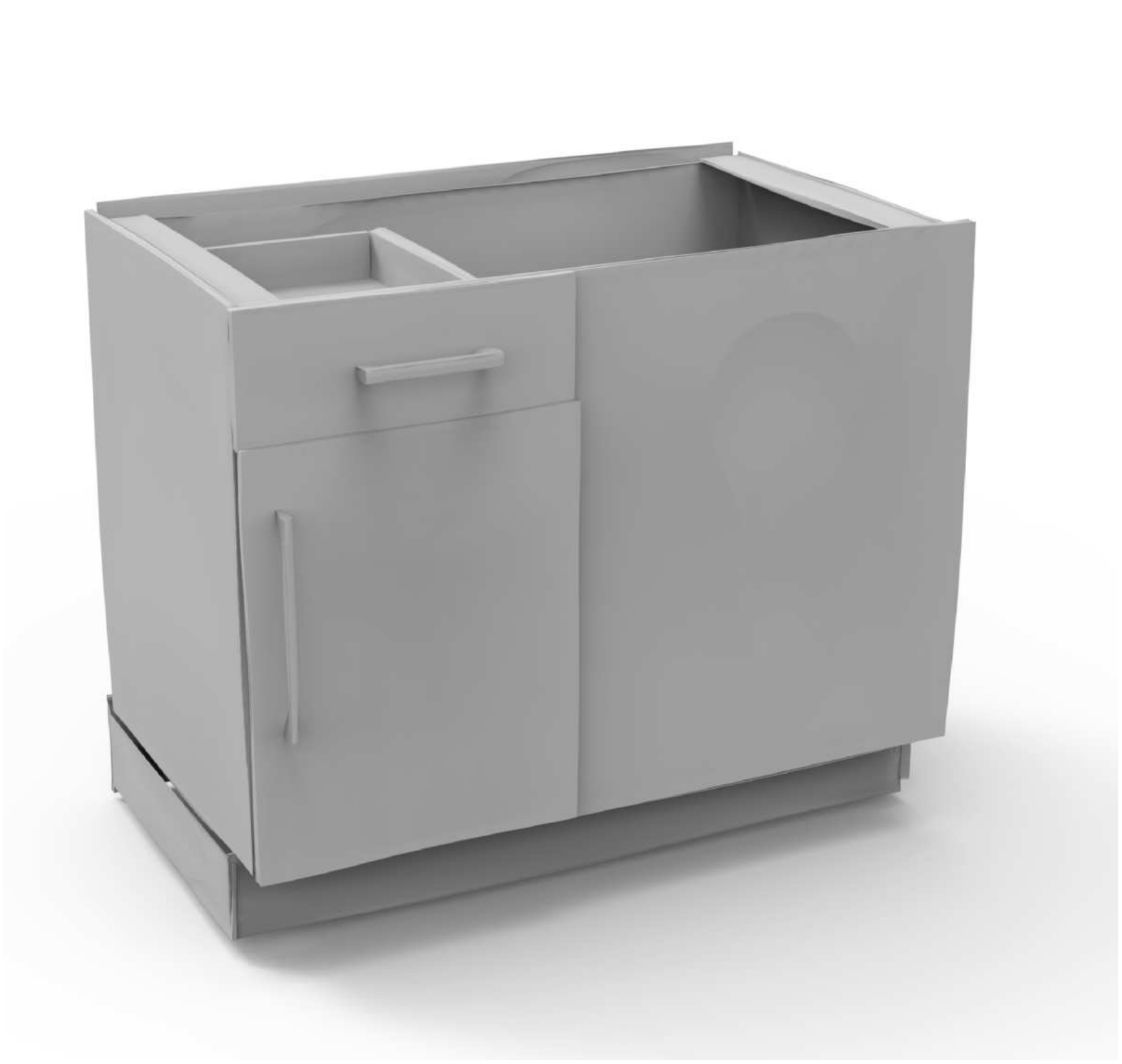}\\
    \vspace{-2mm}
    \subfigure[Input]{
    \includegraphics[width=0.15\linewidth]{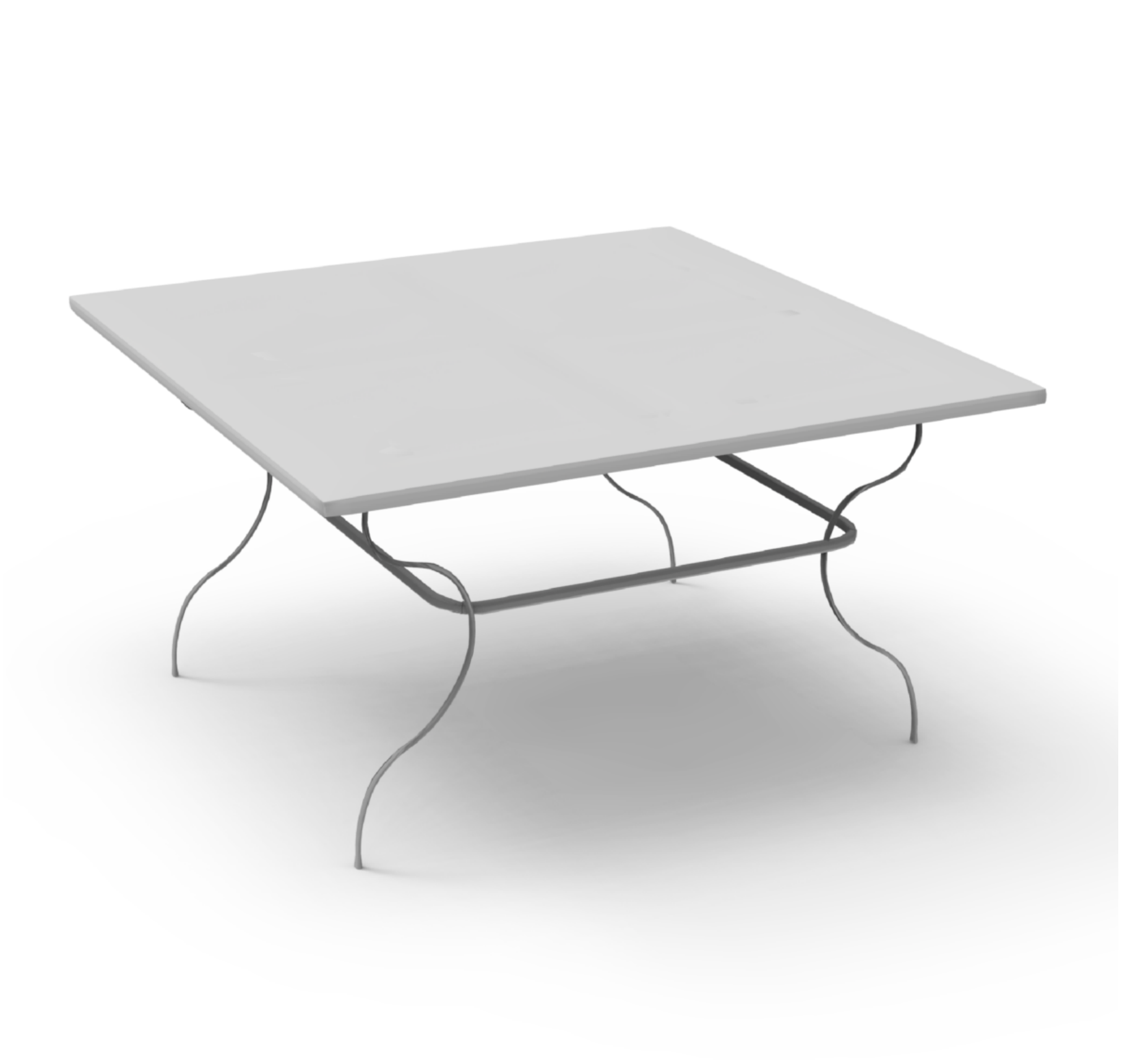}}\vspace{-1.3mm}
    \subfigure[IM-Net]{
    \includegraphics[width=0.15\linewidth]{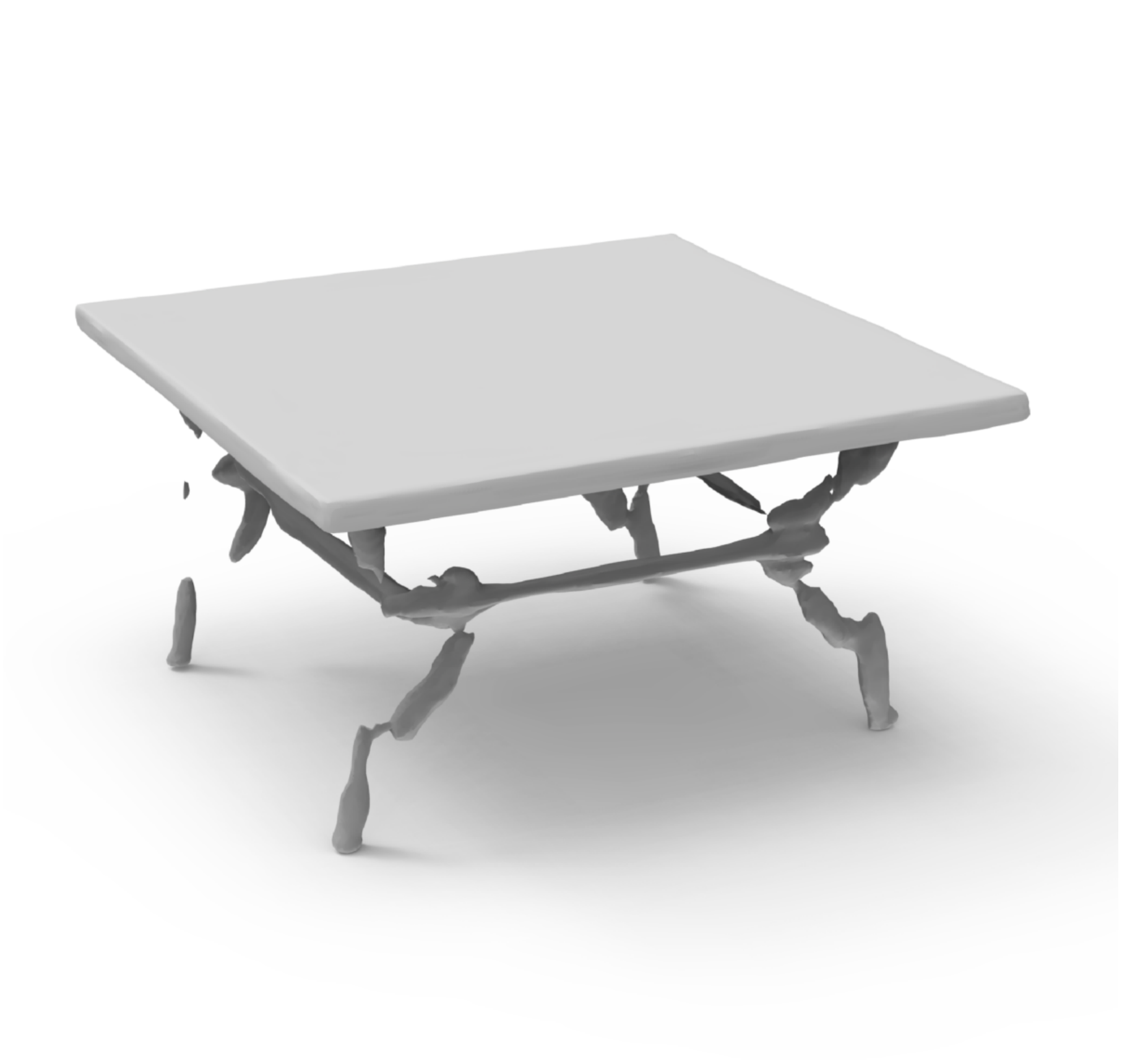}}\vspace{-1.3mm}
    \subfigure[BSP-Net]{
    \includegraphics[width=0.15\linewidth]{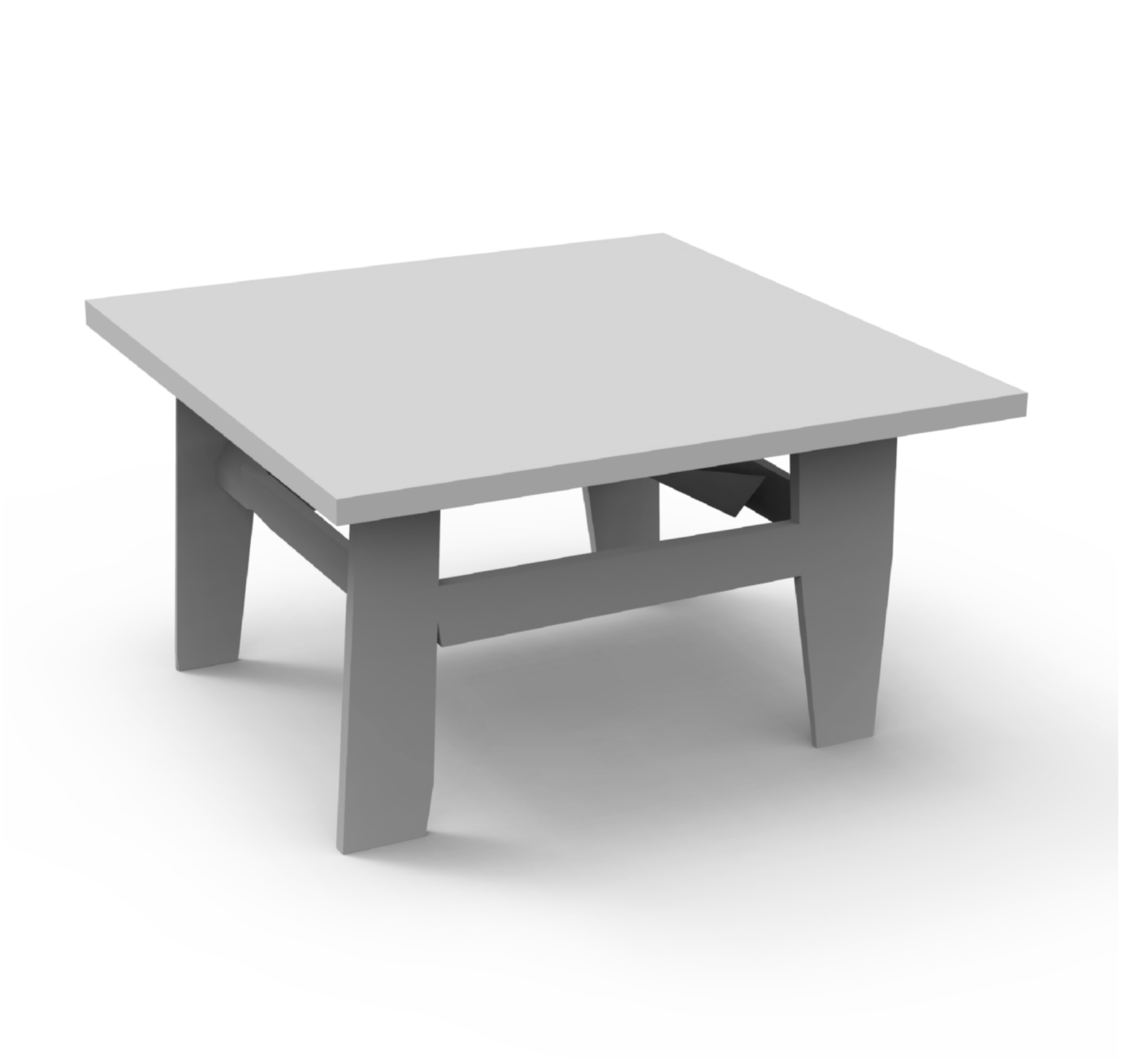}}\vspace{-1.3mm}
    \subfigure[SDM-Net]{
    \includegraphics[width=0.15\linewidth]{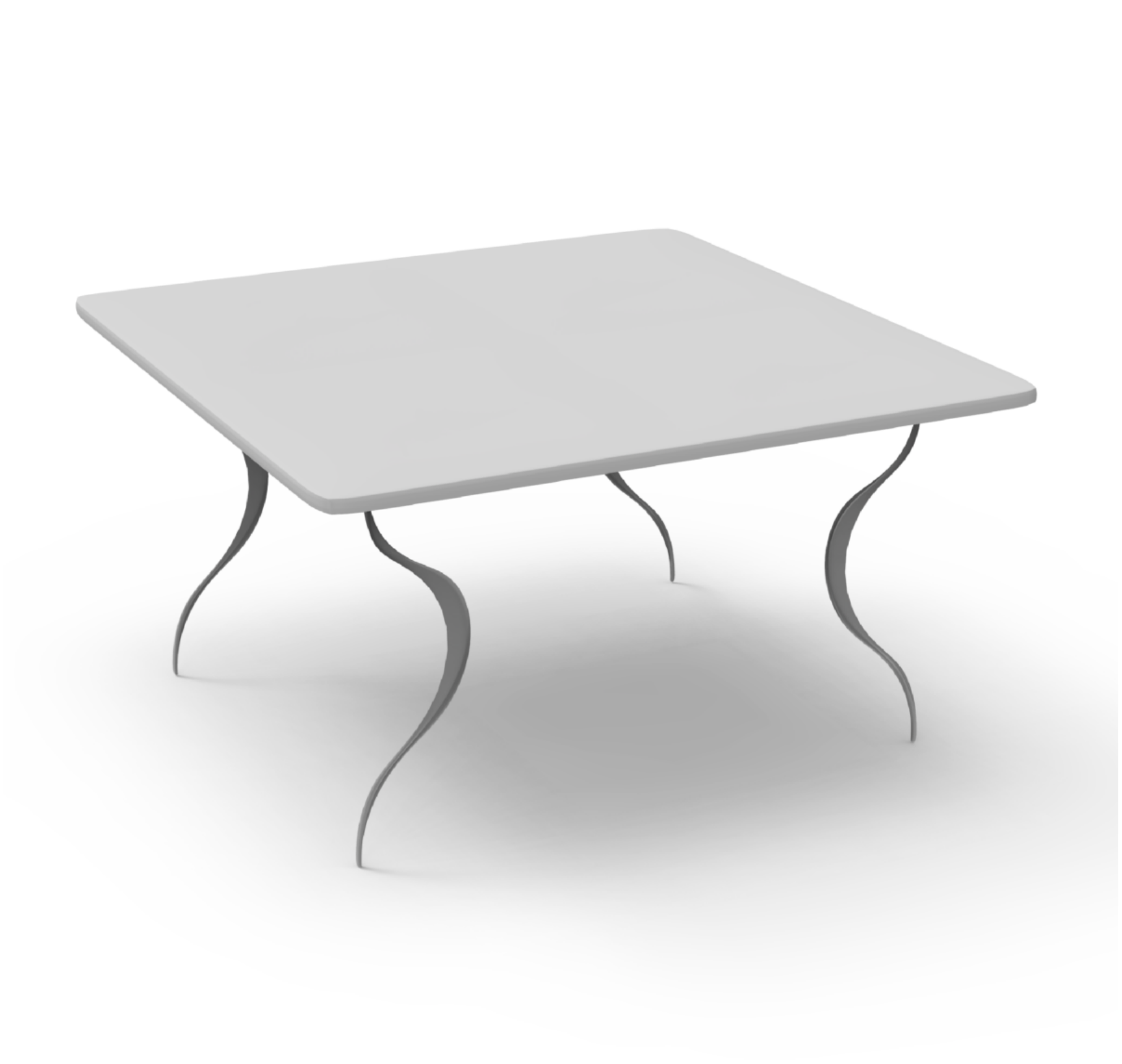}}\vspace{-1.3mm}
    \subfigure[SN]{
    \includegraphics[width=0.15\linewidth]{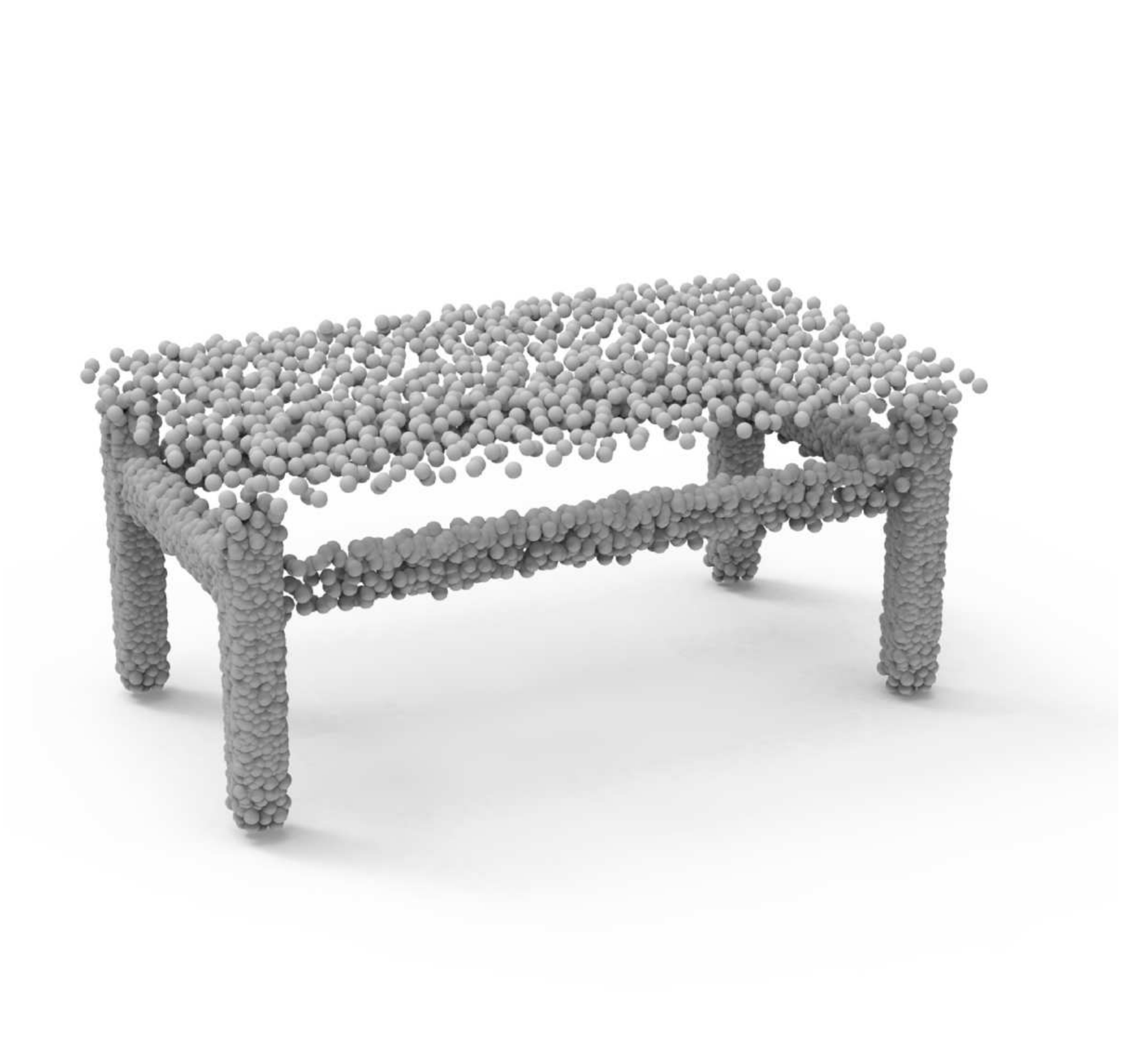}}\vspace{-1.3mm}
    \subfigure[Ours]{
    \includegraphics[width=0.15\linewidth]{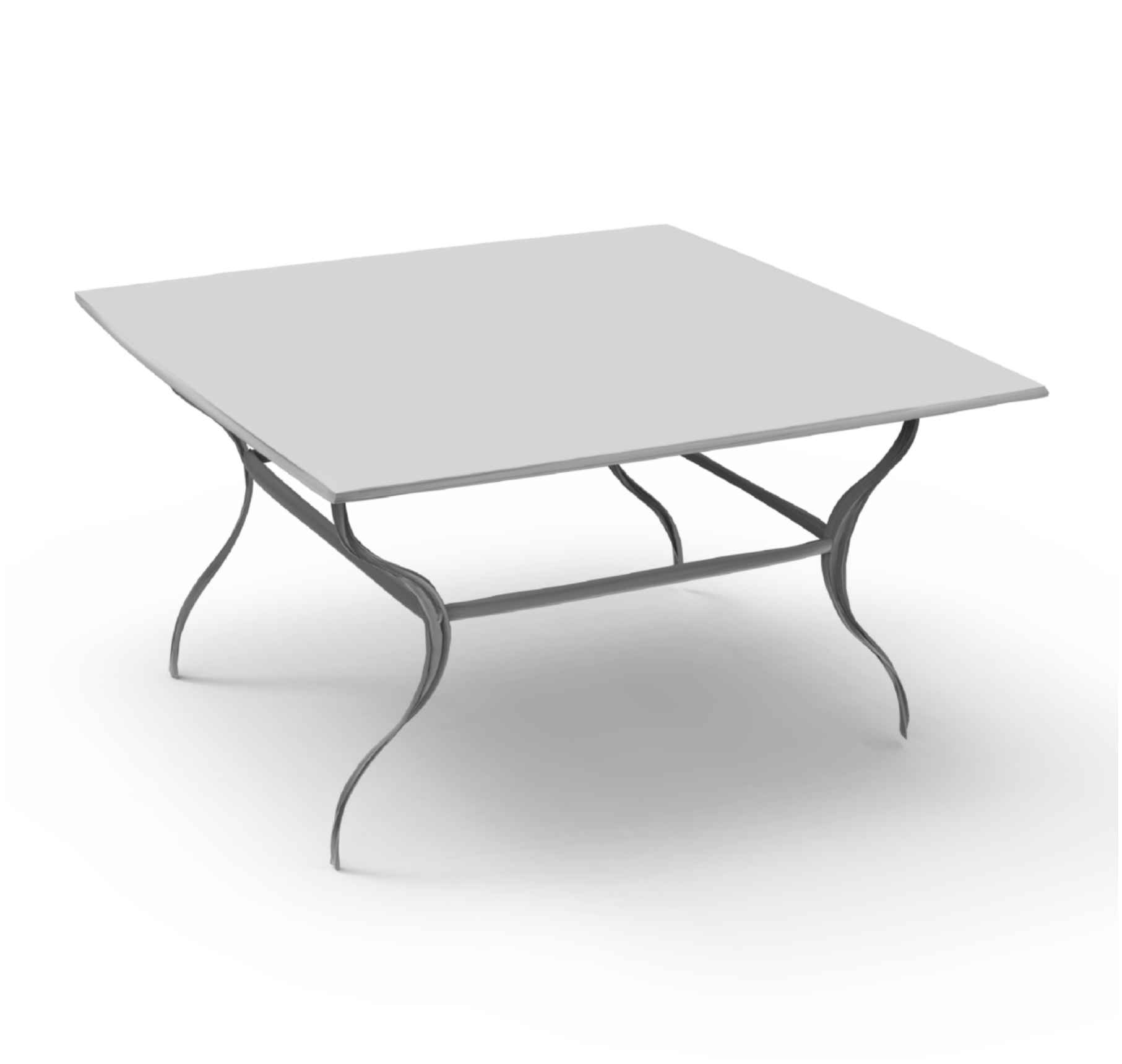}}
    \vspace{2mm}
    \caption{\yjr{Shape reconstruction comparison with the baseline methods. DSG-Net can reconstruct high-quality shape meshes with complex shape structures and detailed part geometry. IM-Net, BSP-Net, and SDM-Net fail to reconstruct the complicated shape structures (\eg chair back bars and table leg stretchers), while StructureNet (SN) generates point cloud shapes with less part geometry details and inaccurate part geometry. For instance, StructureNet fails to reconstruct the slanted bars for the chair in the first row and loses accuracy for the aspect ratio of the table top surface in the last row.}
    }
    \label{fig:comparison2}
    \vspace{-3mm}
\end{figure}

\paragraph{Metrics.} We adopt two kinds of metrics for quantitative comparisons to alternative methods: the geometry metrics and the structure metrics.
For the geometry metrics, we compare the reconstructed shapes against the input shapes without explicitly considering the shape parts and structures.
\yj{We follow the commonly used metrics in the literature: Chamfer Distance (CD)~\cite{barrow1977parametric} and \textit{Earth Mover's} Distance (EMD)~\cite{rubner2000earth}. 
The CD and EMD are two permutation-invariant metrics for evaluating the difference of two unordered point sets, which have been used in the literature~\cite{fan2017point}.
The CD measures the nearest distance for each point in one set to another point set. The EMD solves an optimization for bijective mapping between two point sets.}
For the structure metric, we use the HierInsSeg score proposed in PT2PC~\cite{mo2020pt2pc}.
To compute the HierInsSeg score, Mo~et~al.~\shortcite{mo2020pt2pc} first parse the reconstructed shape point cloud into the PartNet part hierarchy leveraging a pre-trained shape hierarchical instance segmentation network, and then compute the normalized tree-editing distance between the reconstructed and ground-truth part hierarchies.
We refer the readers to Fan~et~al.~\shortcite{fan2017point} and Mo~et~al.~\shortcite{mo2020pt2pc} for more details on the definitions of the metrics.

\paragraph{Results.}
Table~\ref{tab:reconeval} presents the quantitative comparisons between DSG-Net and the alternative methods.
Our method performs the best on the geometry metrics, indicating that DSG-Net captures better shape geometry.
\yjr{
Our DSG-Net also outperforms IM-Net, BSP-Net, and SDM-Net on the structure metric by significantly large margins, while achieving slightly better performance than StructureNet thanks to our cycled disentanglement. %
}
We observe that StructureNet achieves a comparable HierInsSeg score since it tends to generate parts that are more disconnected as shown in Figure~\ref{fig:comparison2} (e), which is beneficial to make the part structure clearer but is detrimental to the overall shape geometric appearance.
Figure~\ref{fig:comparison2} presents the qualitative comparison to the baseline methods.
It is clear to obverse that IM-Net, BSP-Net and SDM-Net fail to generate complicated shape structures, while our method can successfully capture these complex shape structures.
Compared with StructureNet, we reconstruct the shape geometry more accurately.

\begin{figure}[ht]
    \centering
    \includegraphics[width=0.15\linewidth]{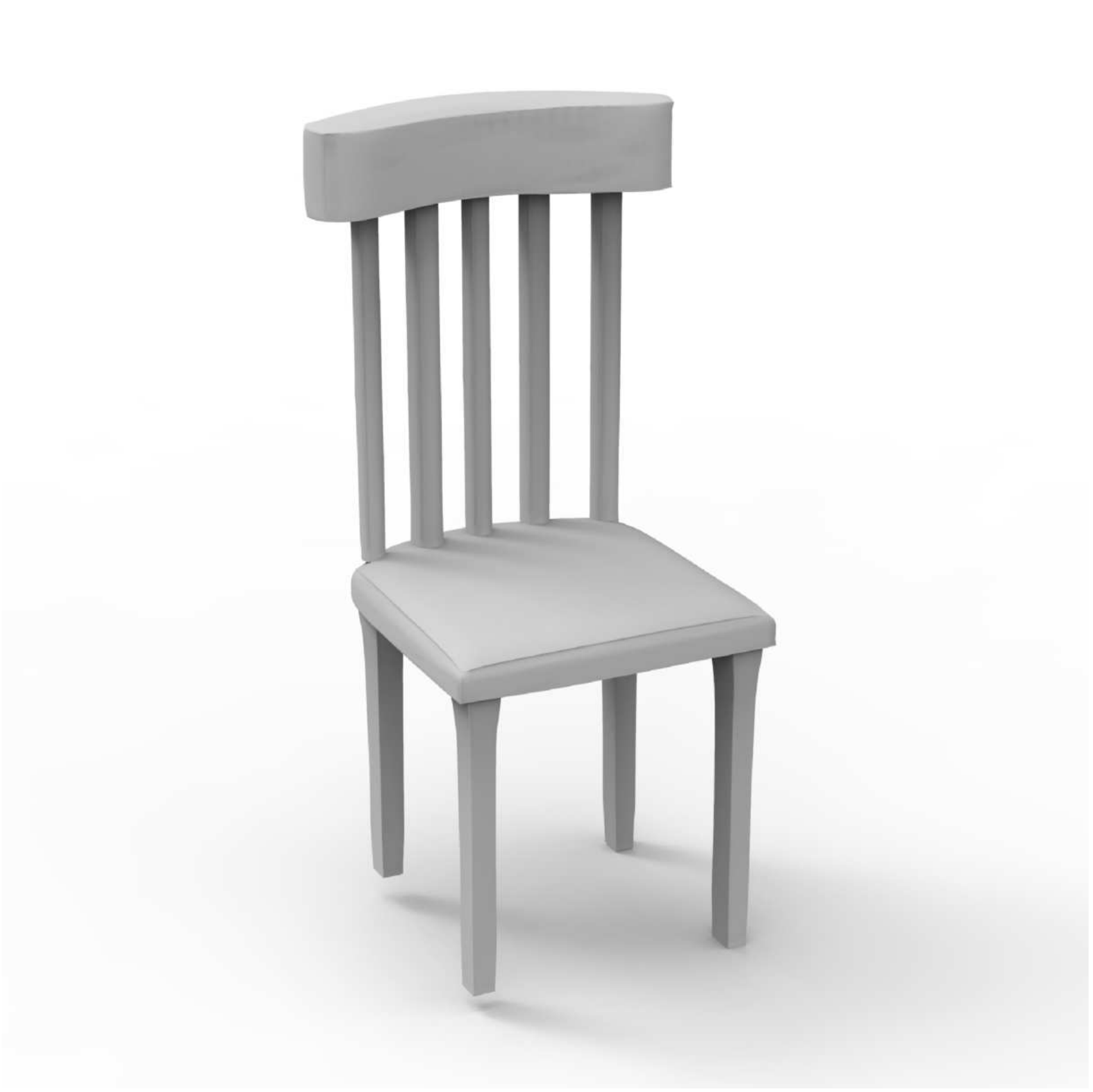}
    \includegraphics[width=0.15\linewidth]{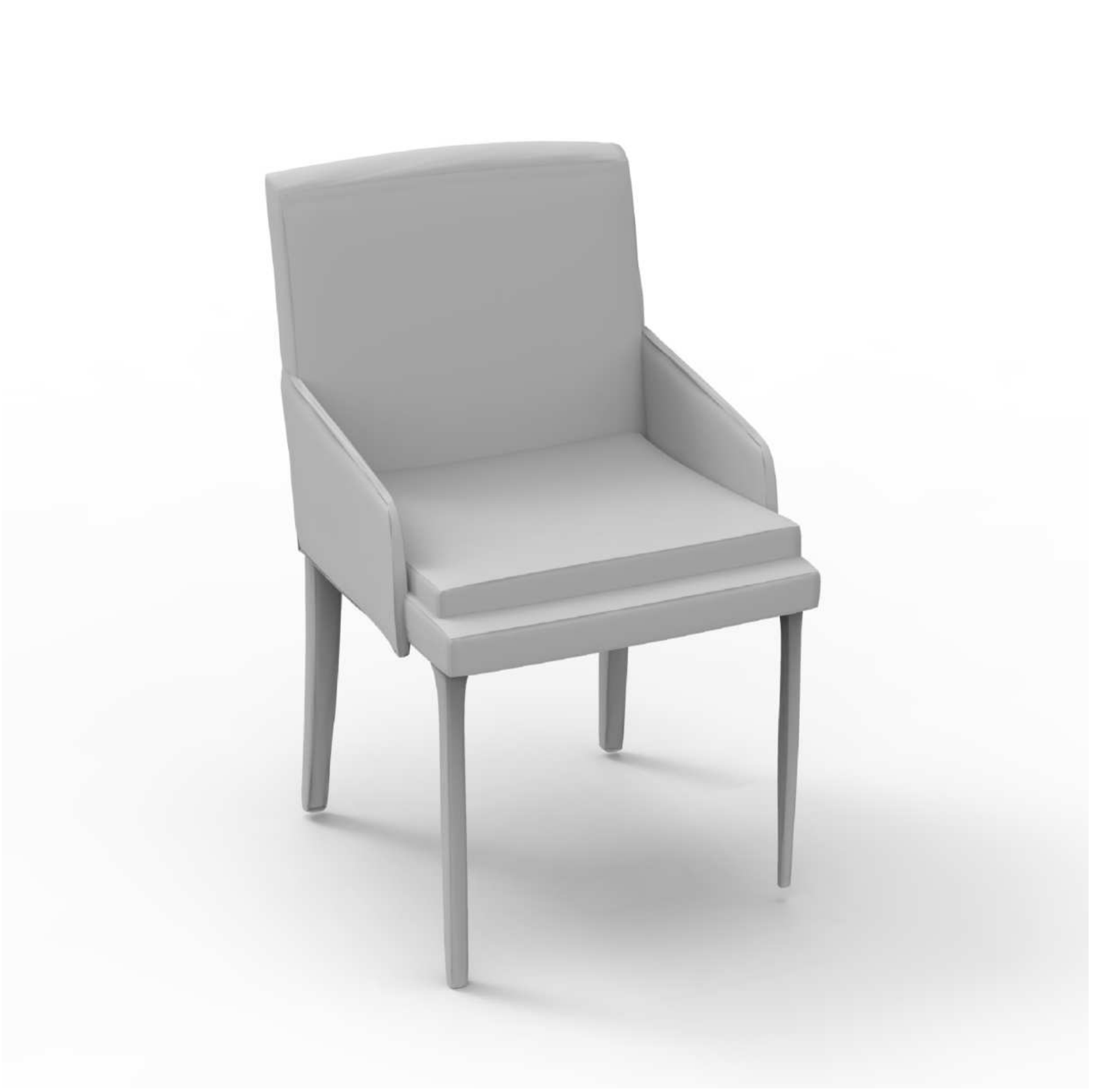}
    \includegraphics[width=0.15\linewidth]{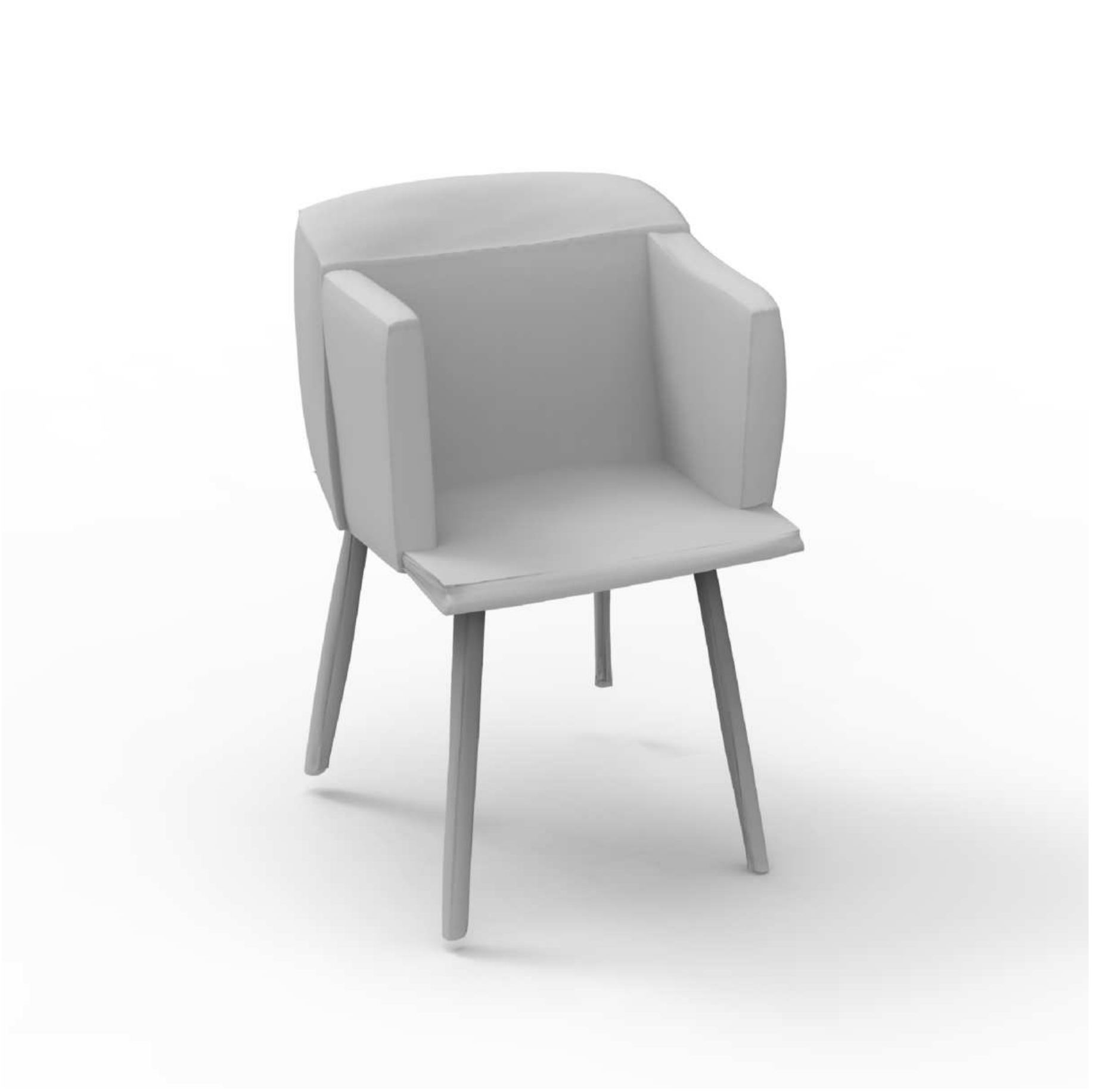}
    \includegraphics[width=0.15\linewidth]{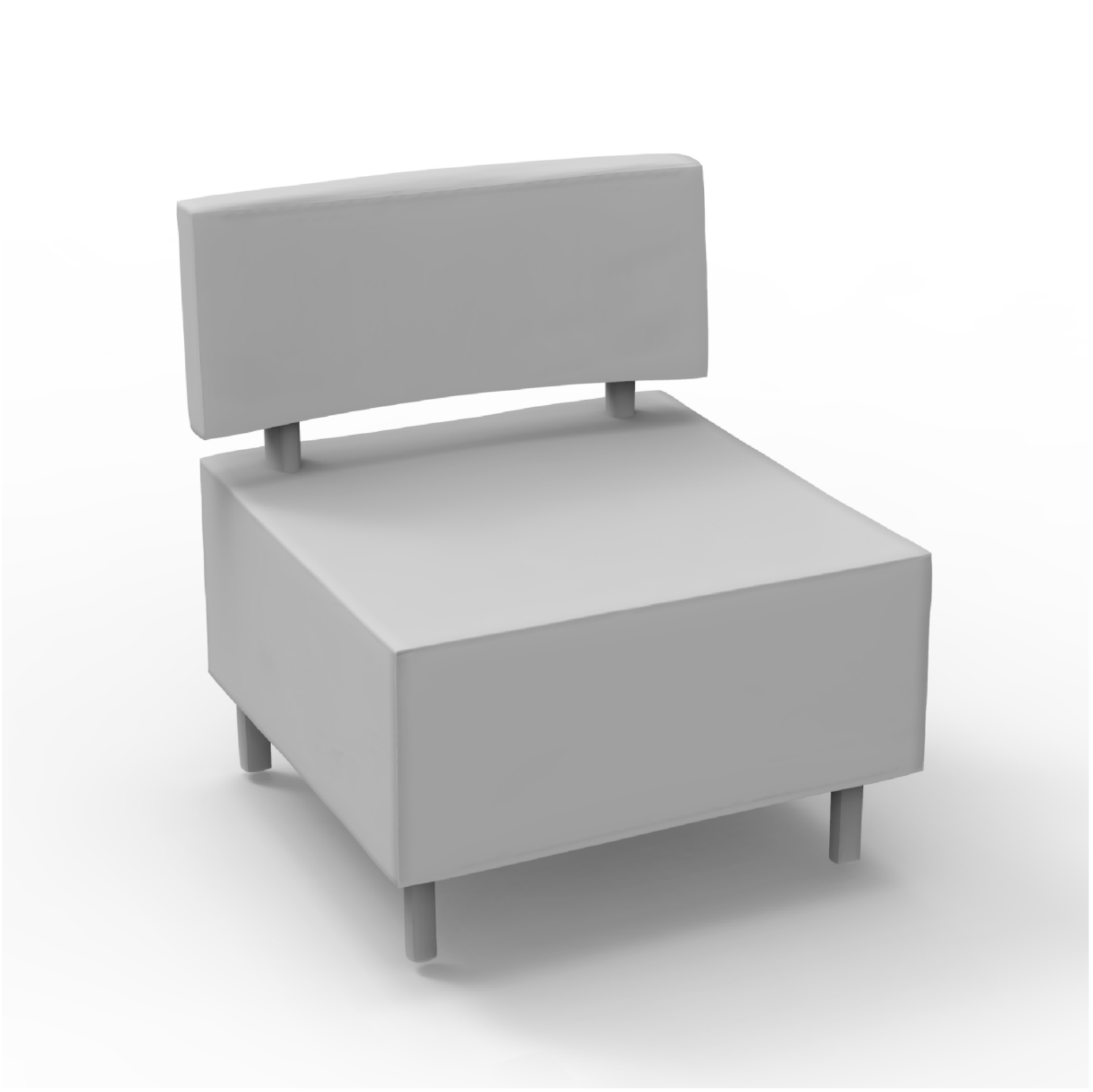}
    \includegraphics[width=0.15\linewidth]{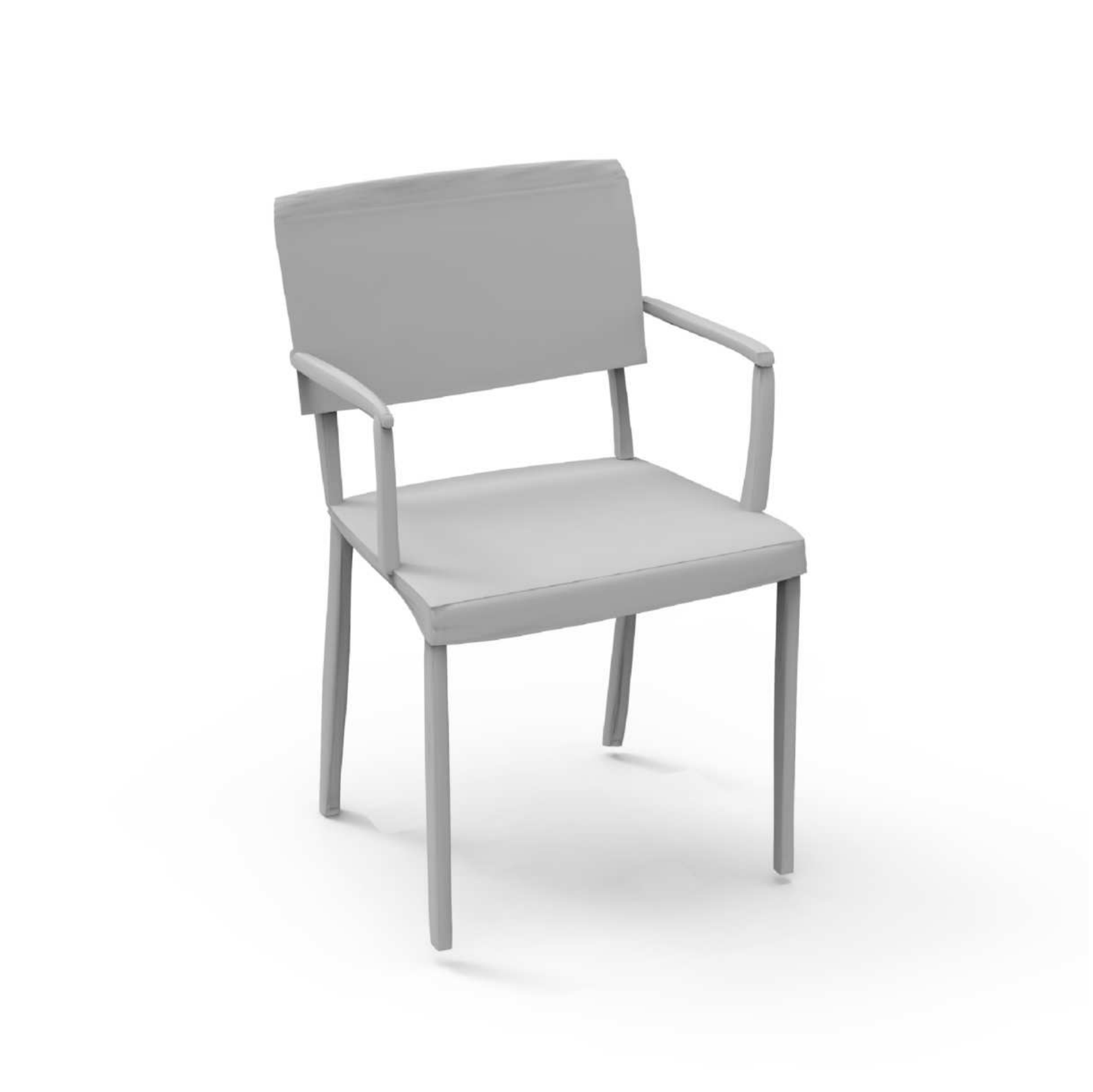}
    \includegraphics[width=0.15\linewidth]{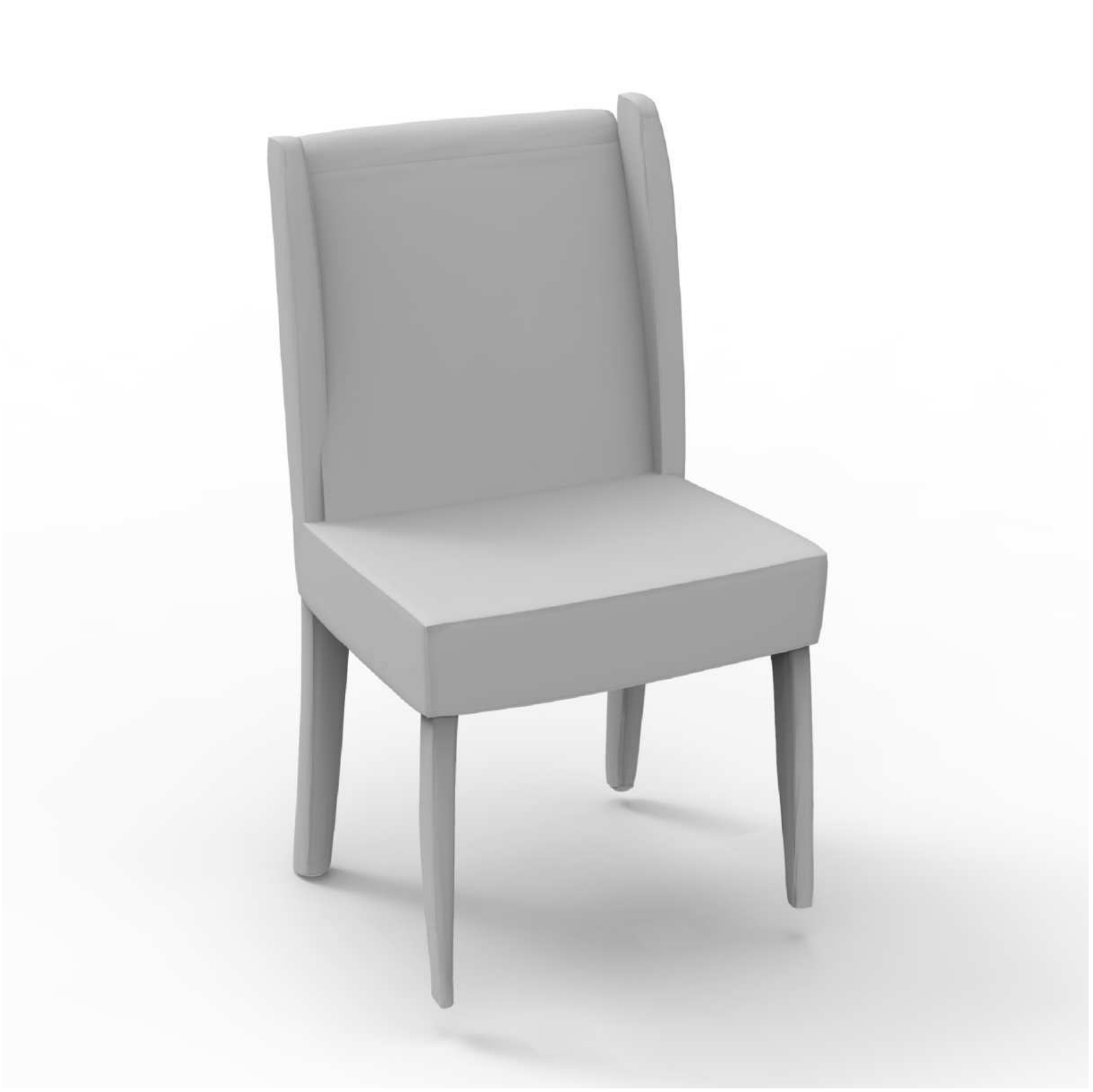}\\
    \includegraphics[width=0.15\linewidth]{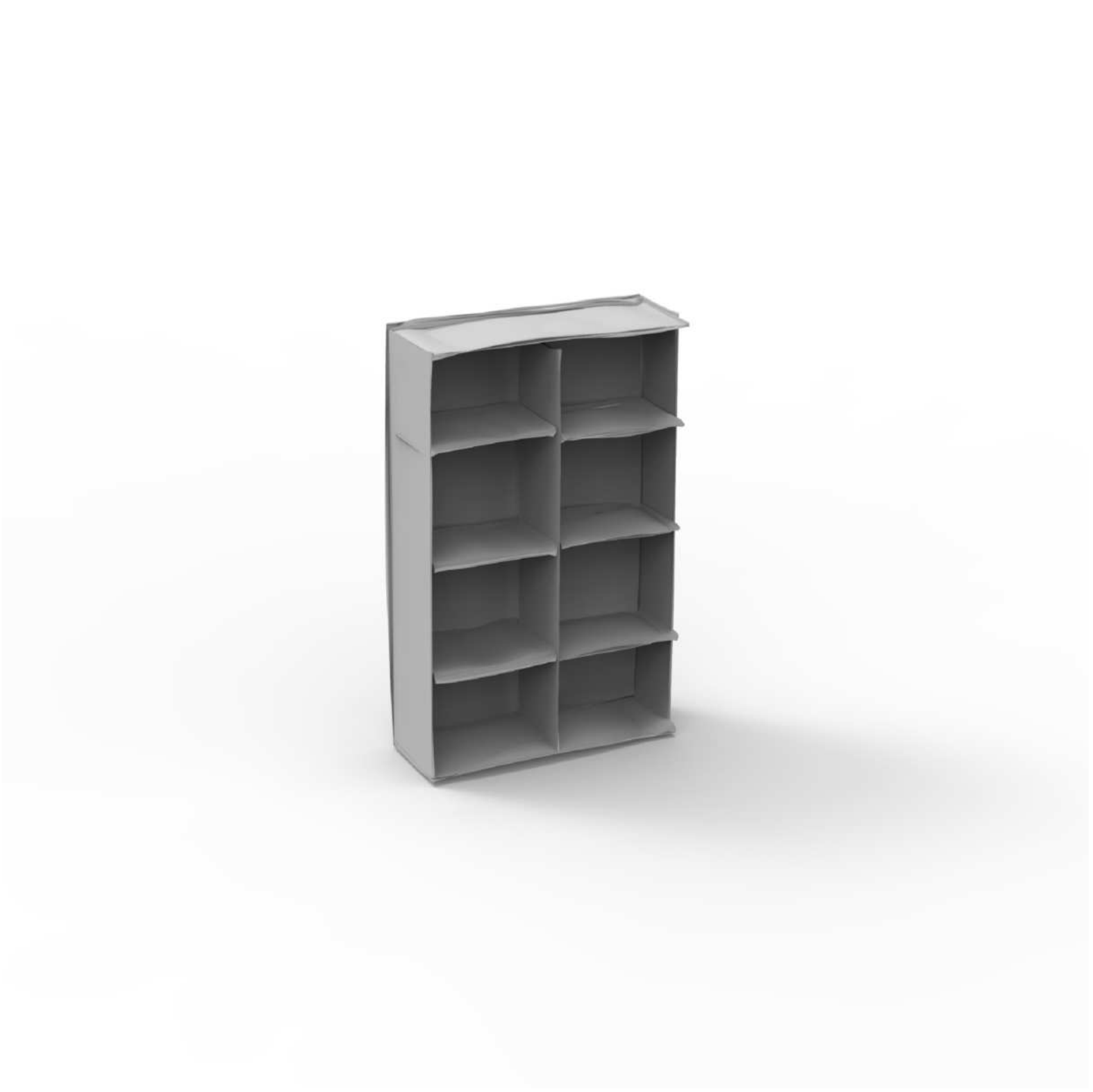}
    \includegraphics[width=0.15\linewidth]{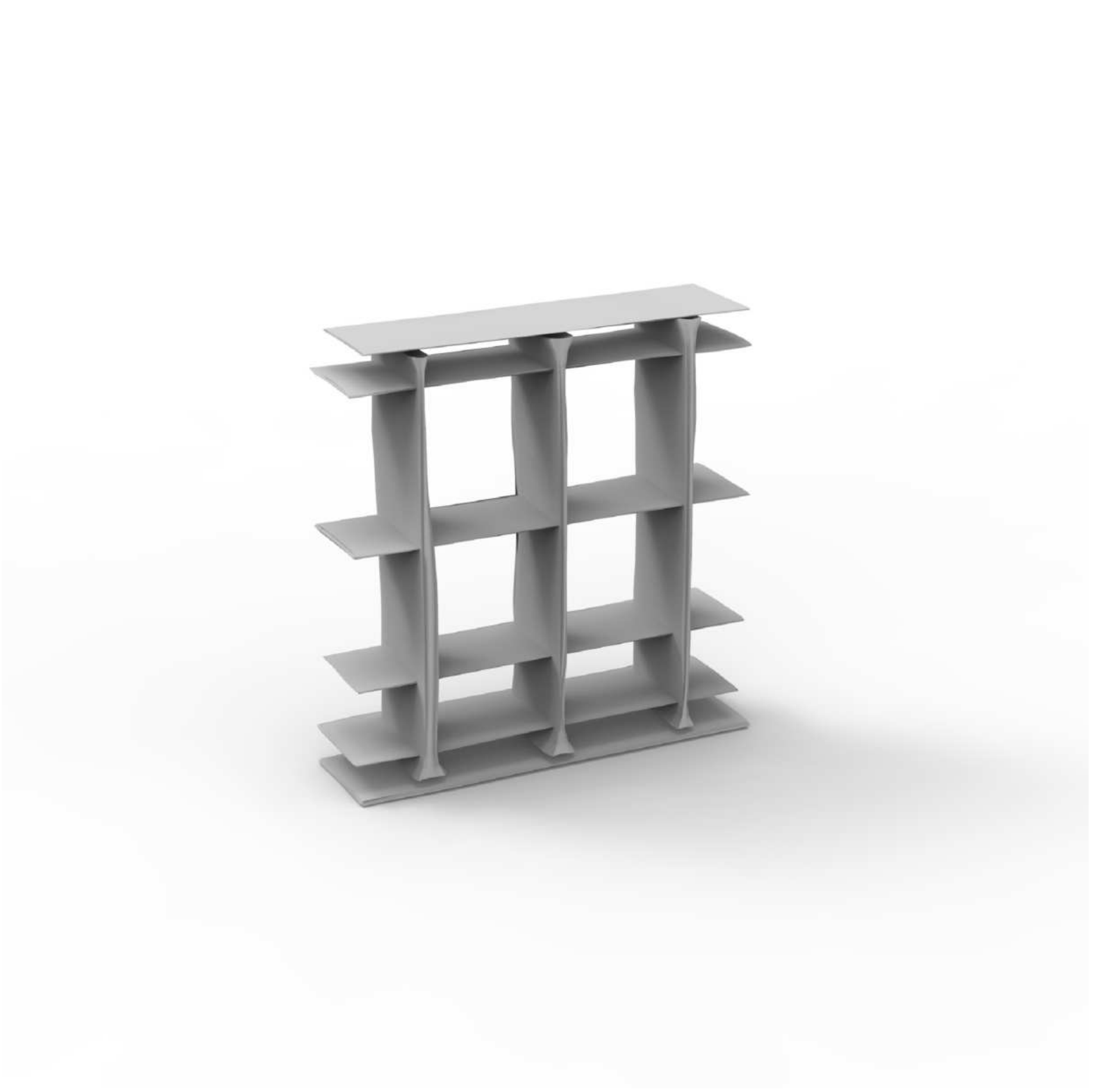}
    \includegraphics[width=0.15\linewidth]{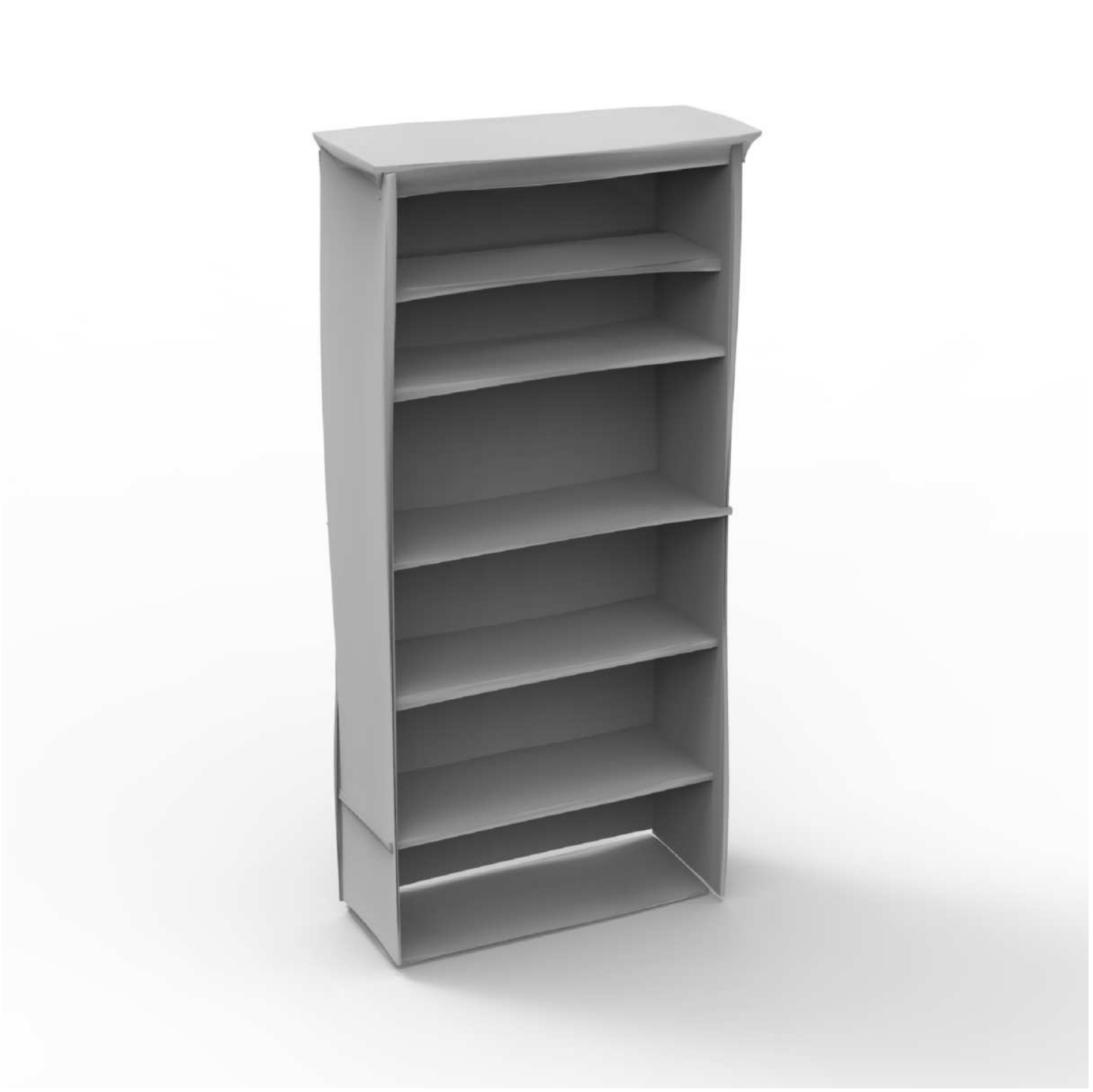}
    \includegraphics[width=0.15\linewidth]{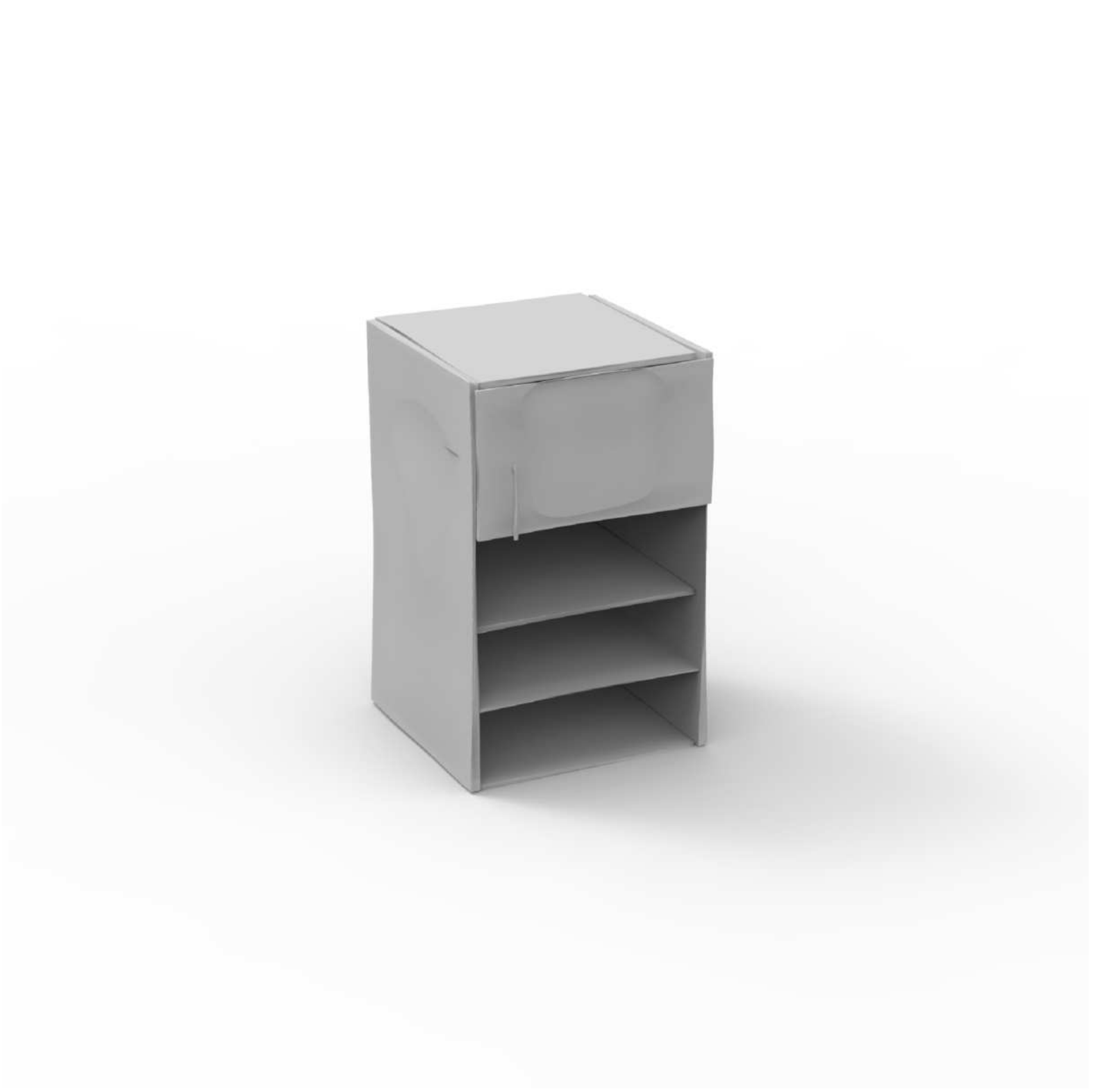}
    \includegraphics[width=0.15\linewidth]{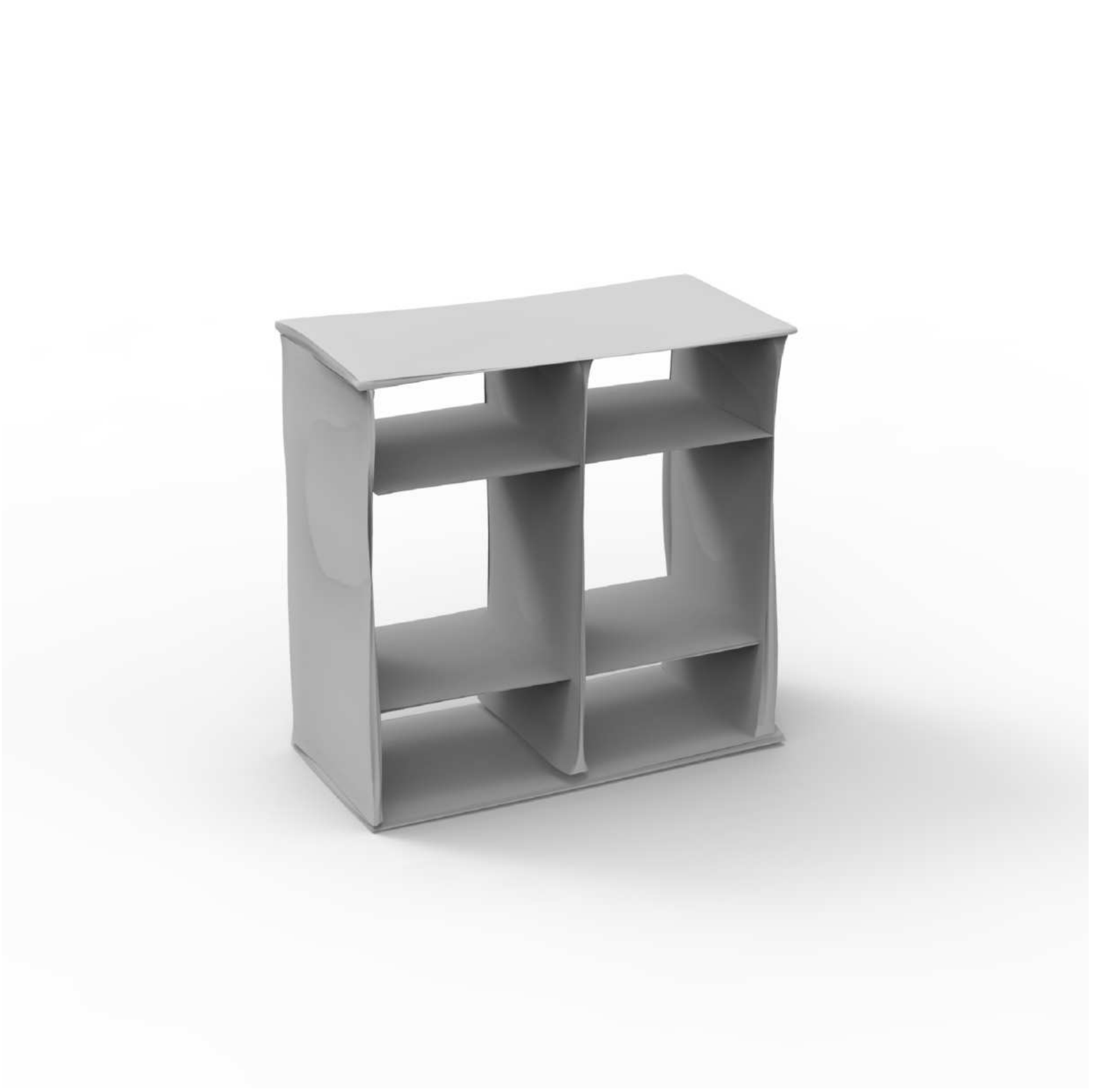}
    \includegraphics[width=0.15\linewidth]{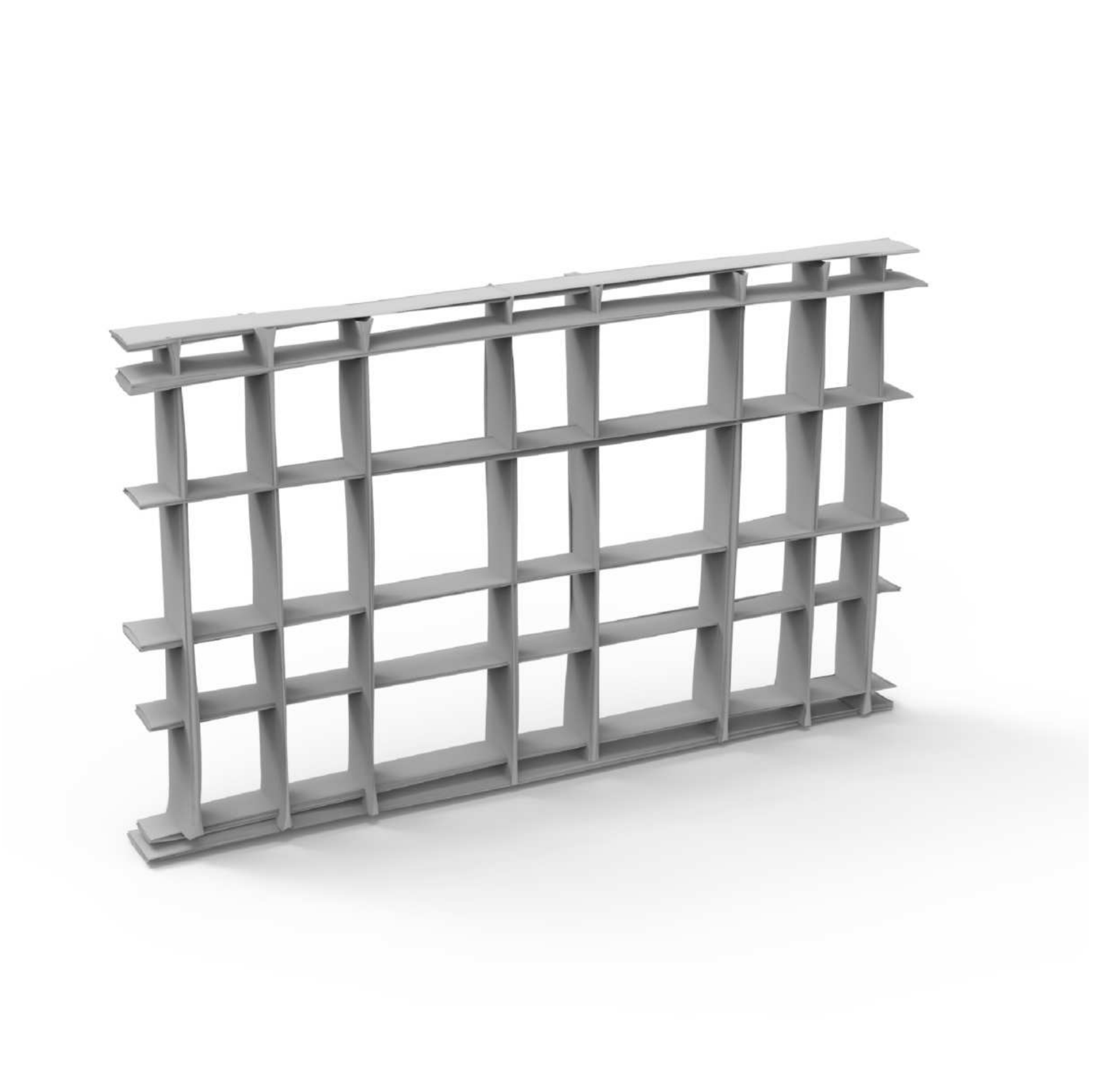}\\
    \includegraphics[width=0.15\linewidth]{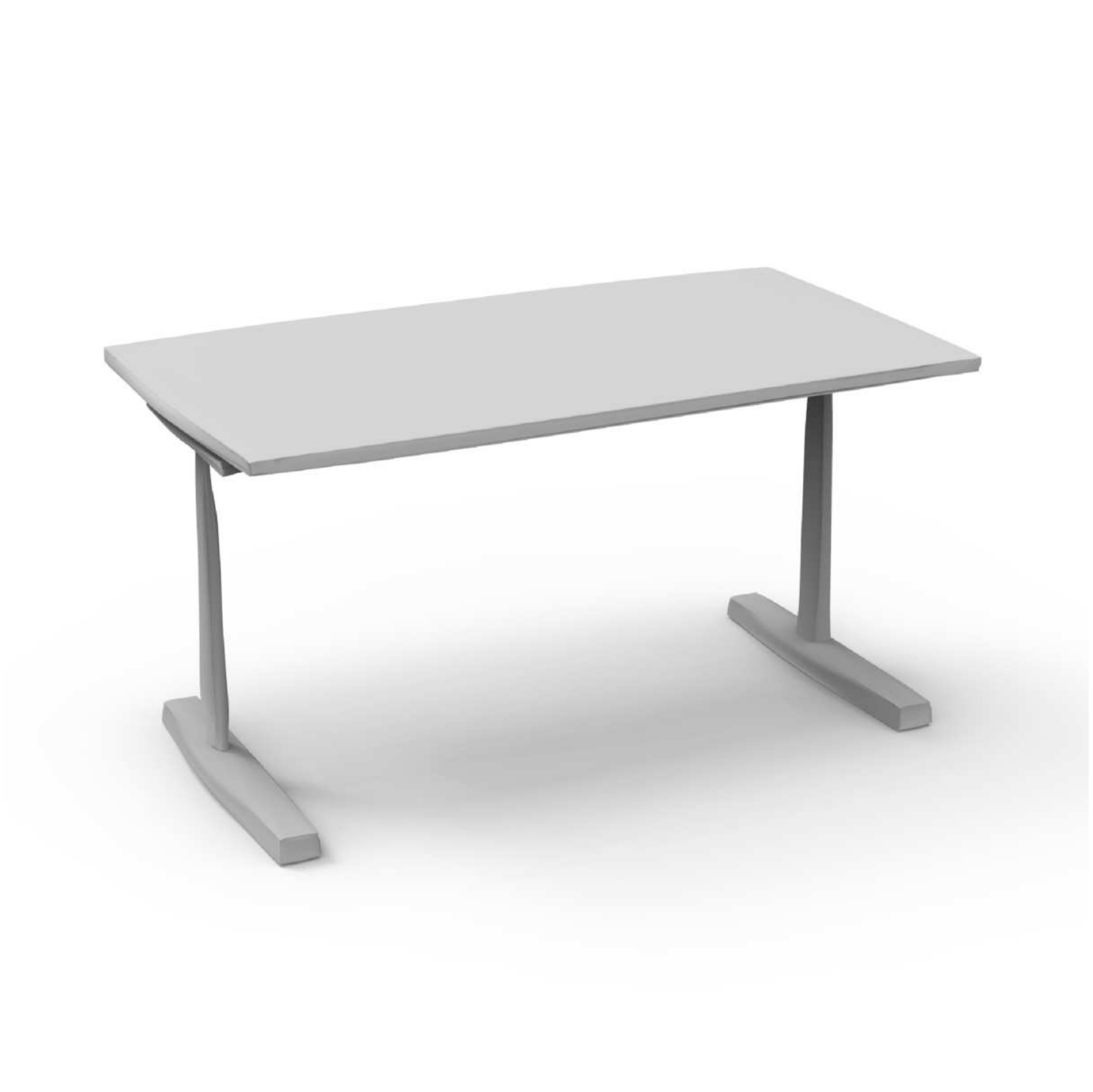}
    \includegraphics[width=0.15\linewidth]{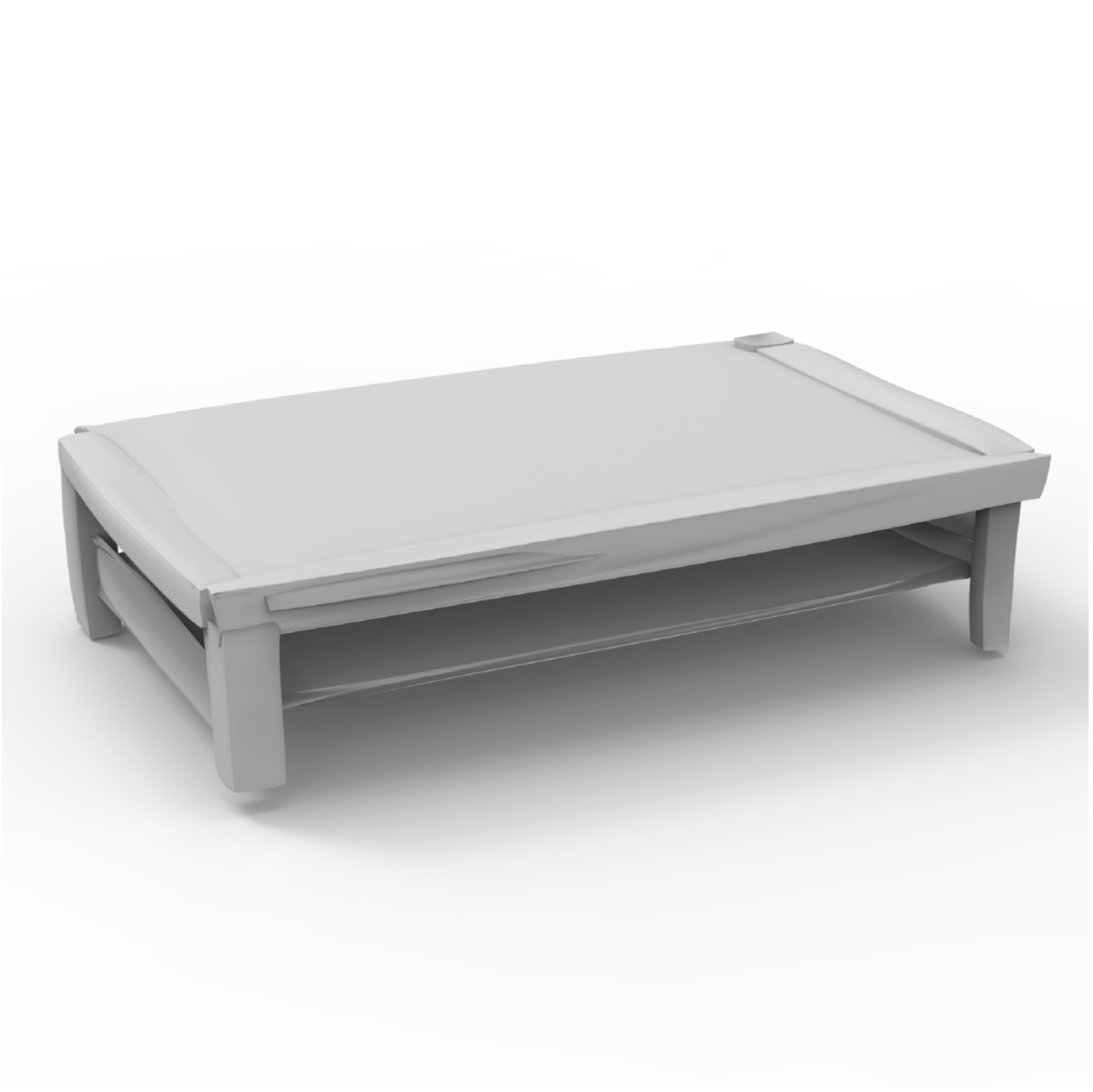}
    \includegraphics[width=0.15\linewidth]{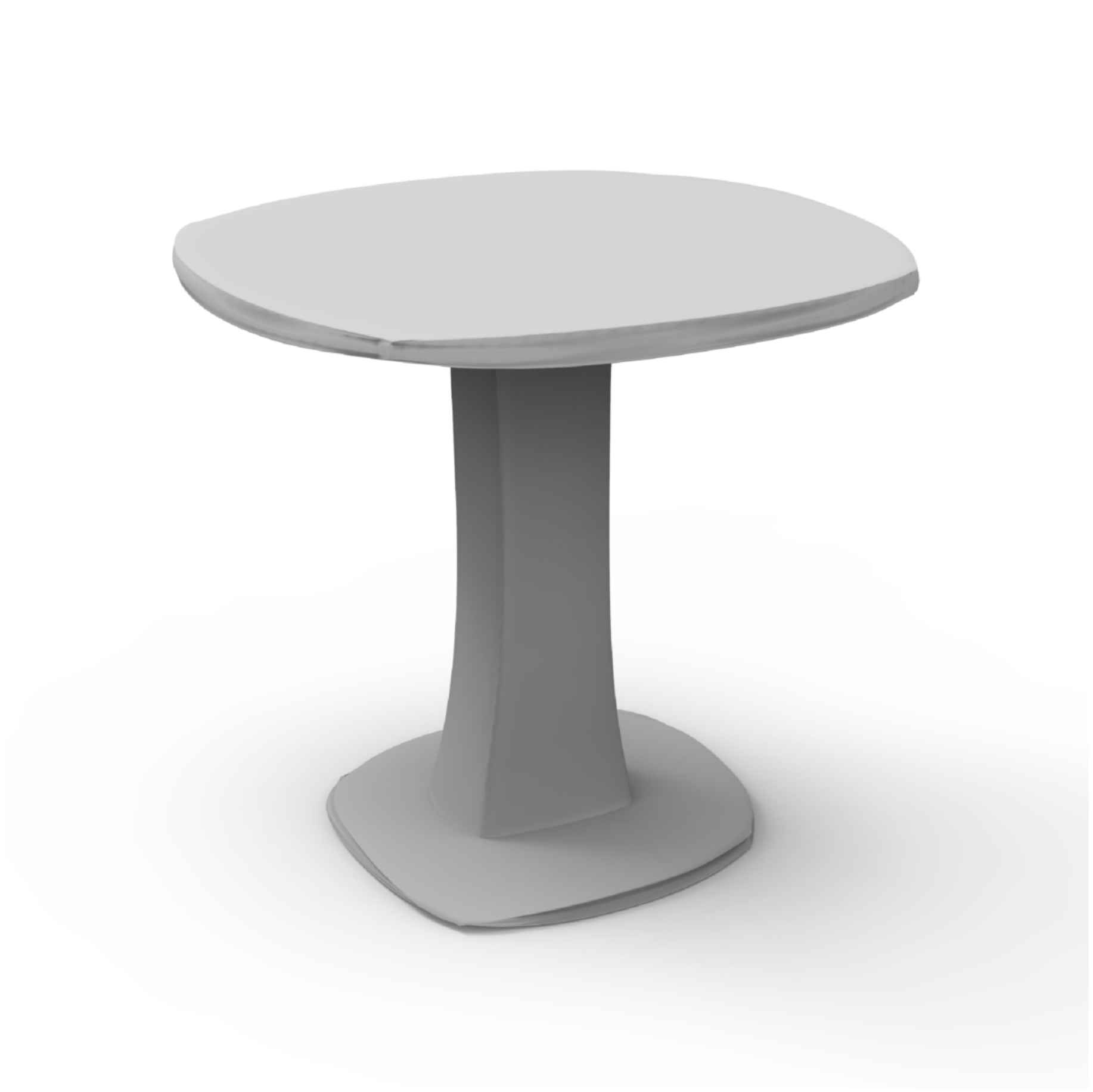}
    \includegraphics[width=0.15\linewidth]{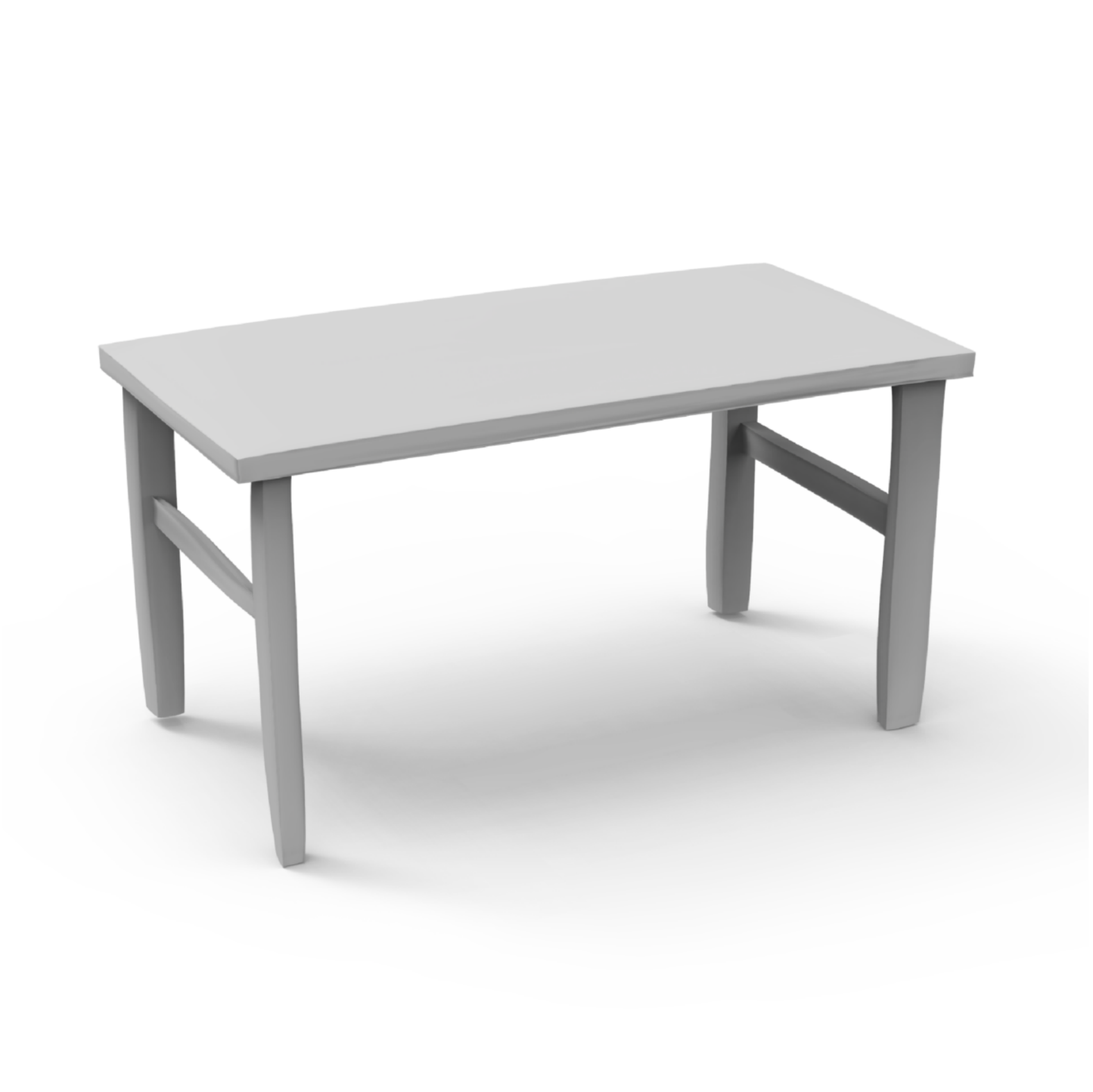}
    \includegraphics[width=0.15\linewidth]{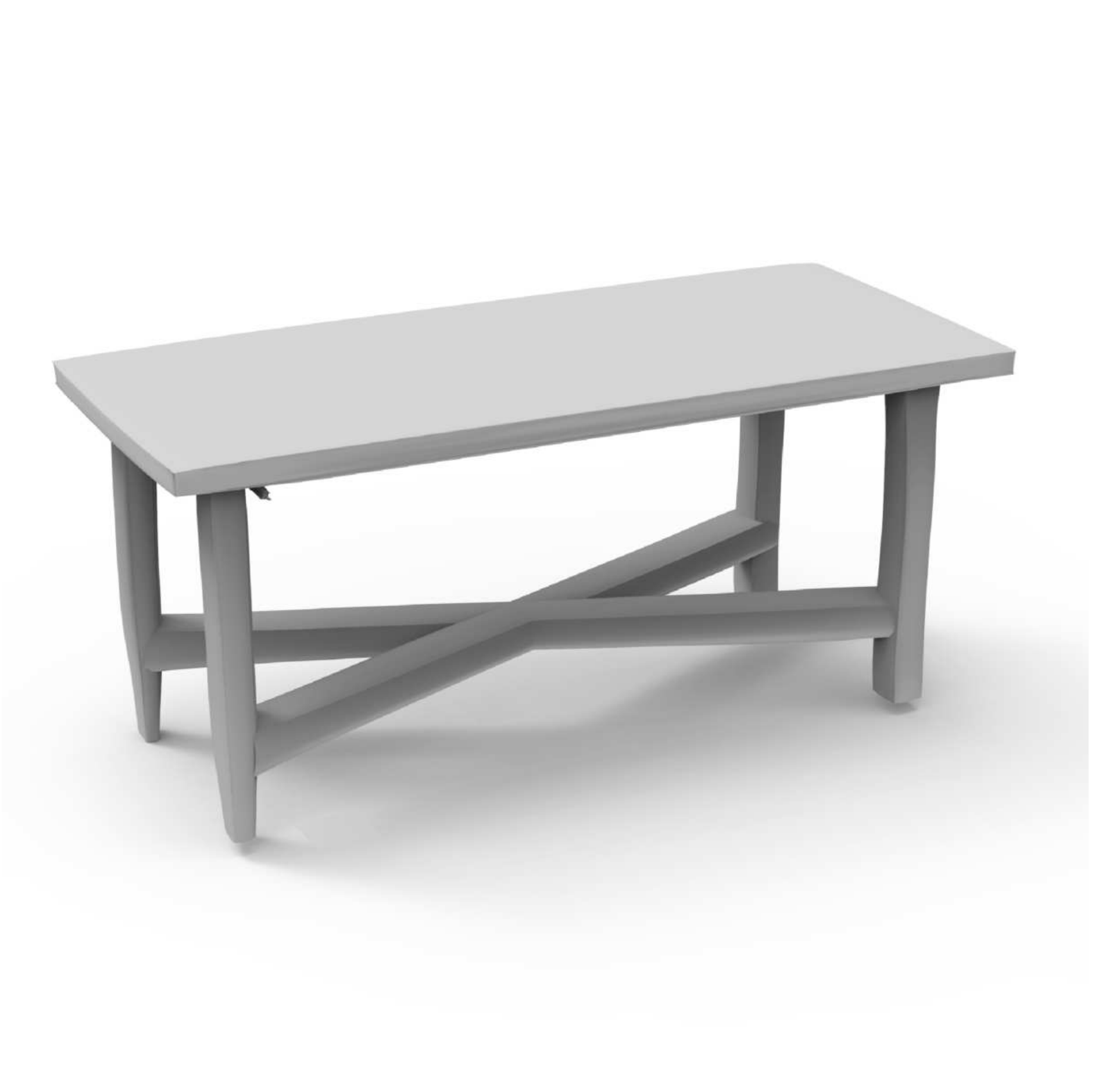}
    \includegraphics[width=0.15\linewidth]{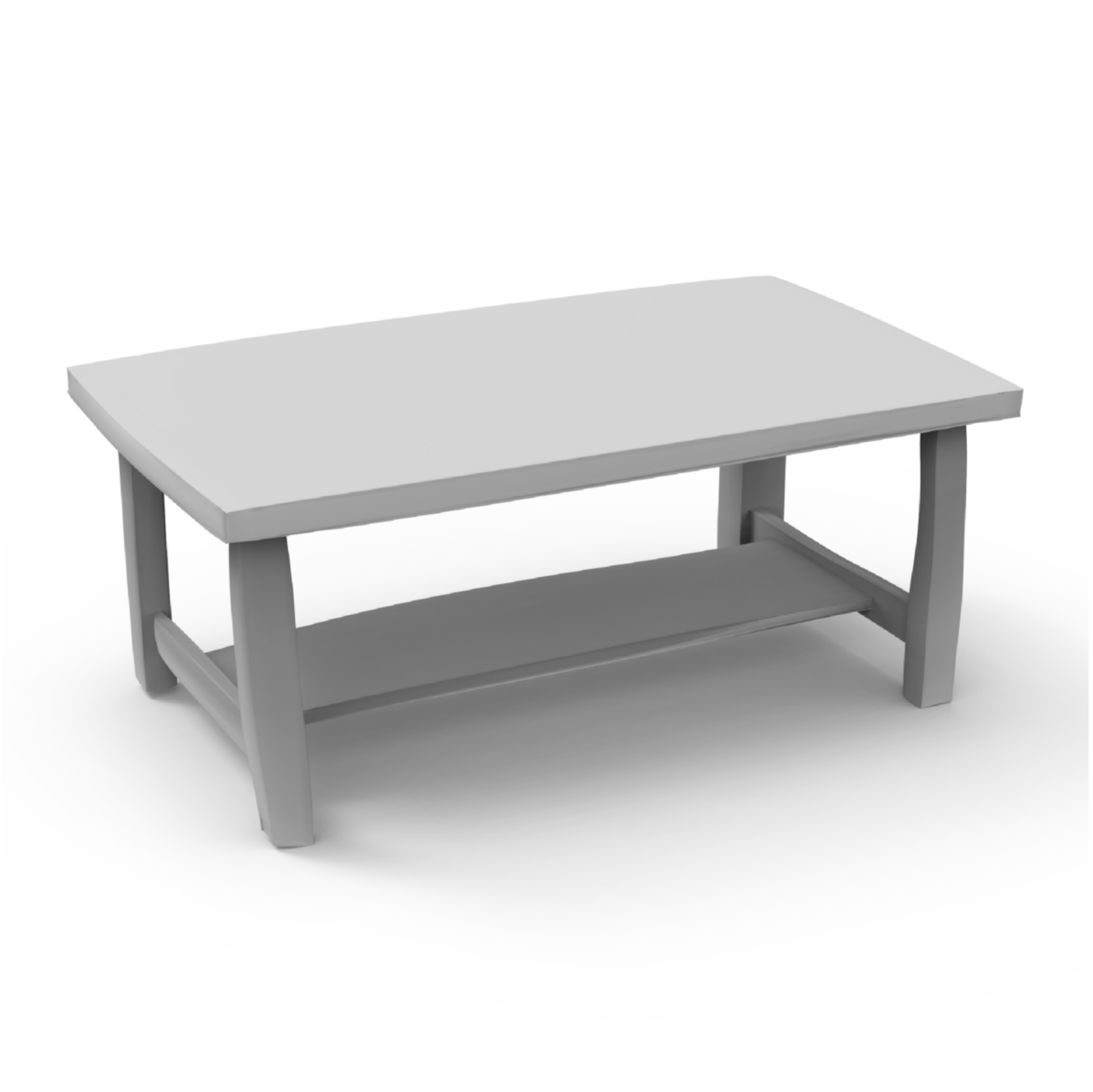}\\
    \includegraphics[width=0.15\linewidth]{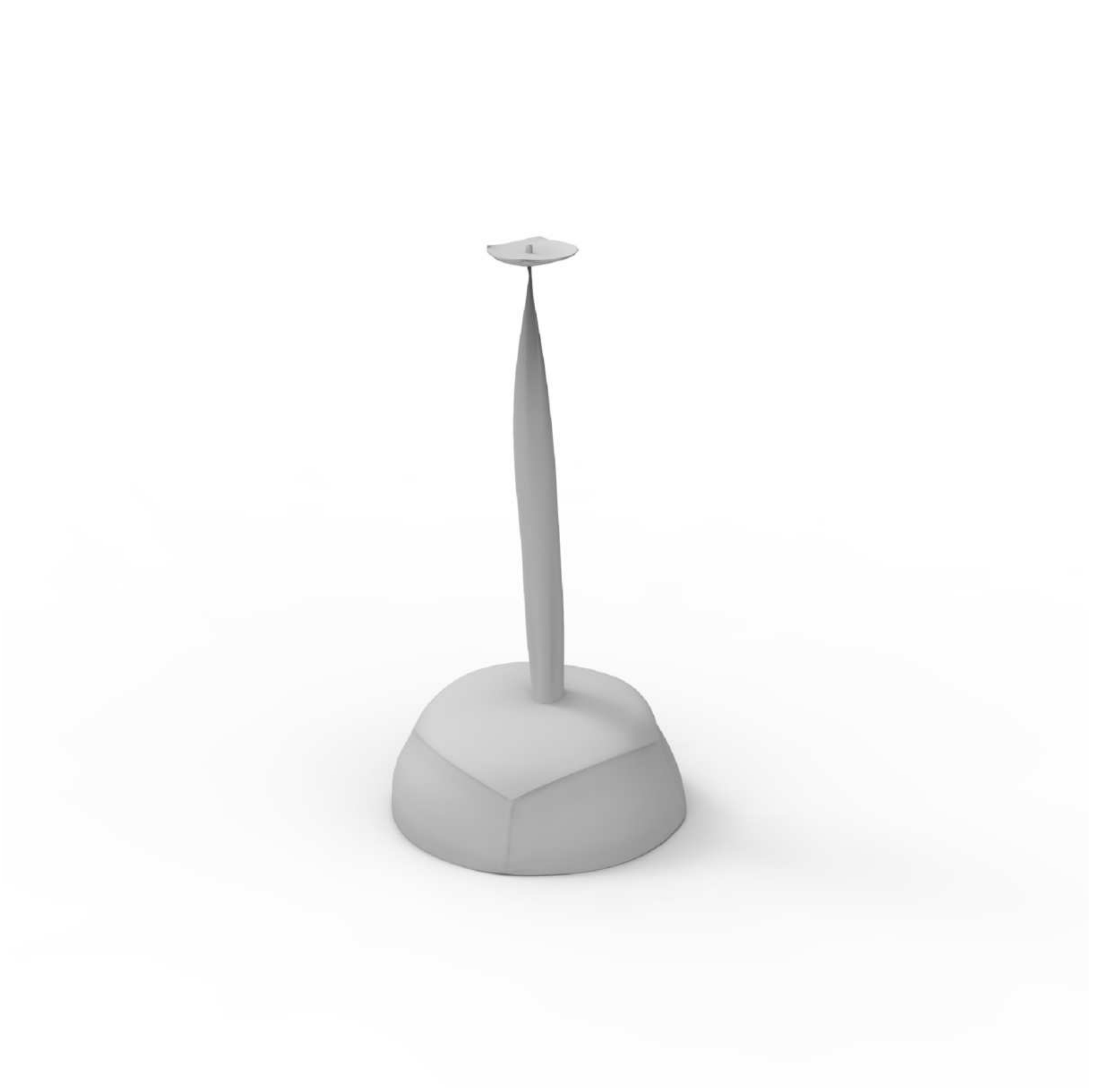}
    \includegraphics[width=0.15\linewidth]{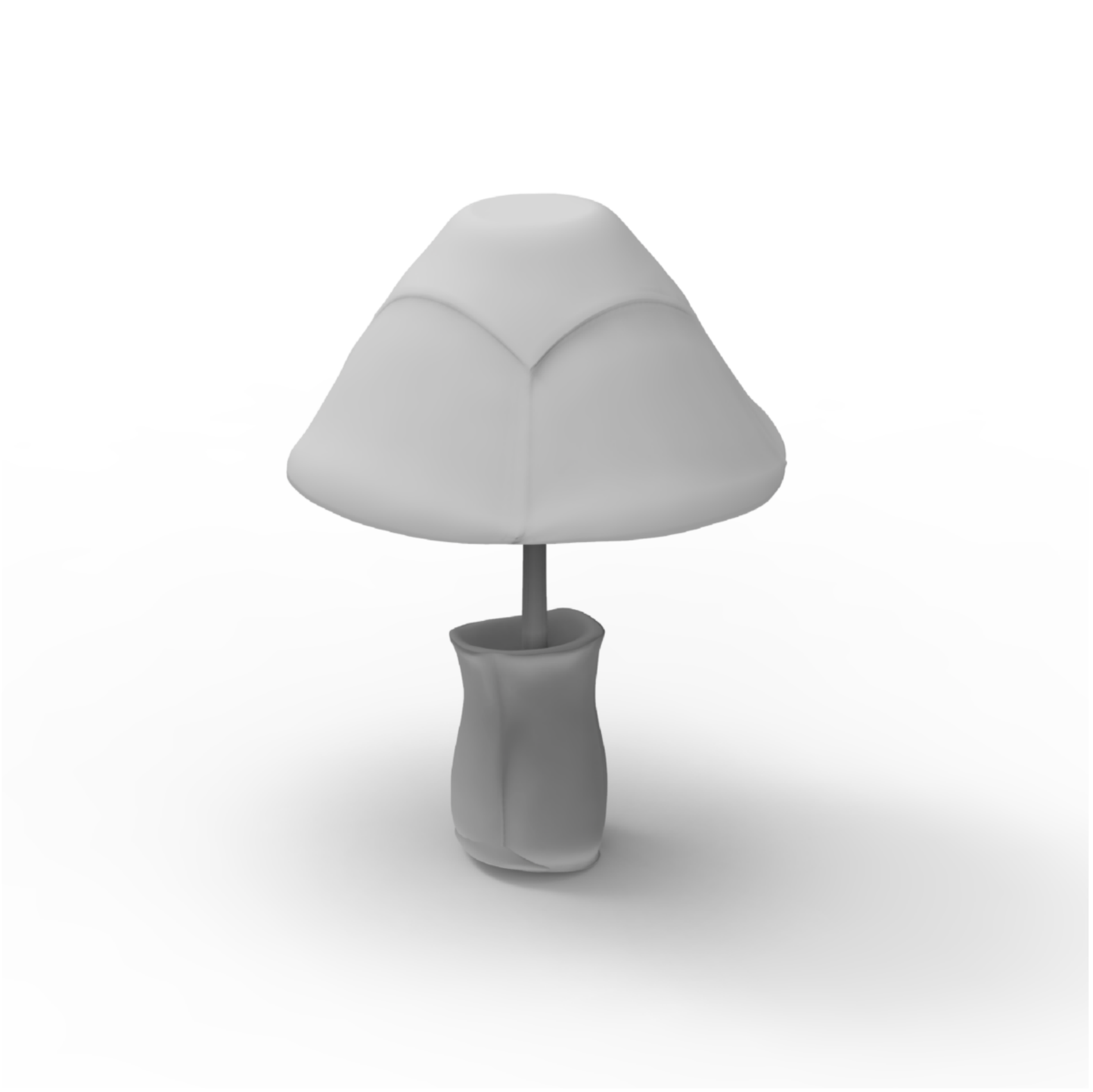}
    \includegraphics[width=0.15\linewidth]{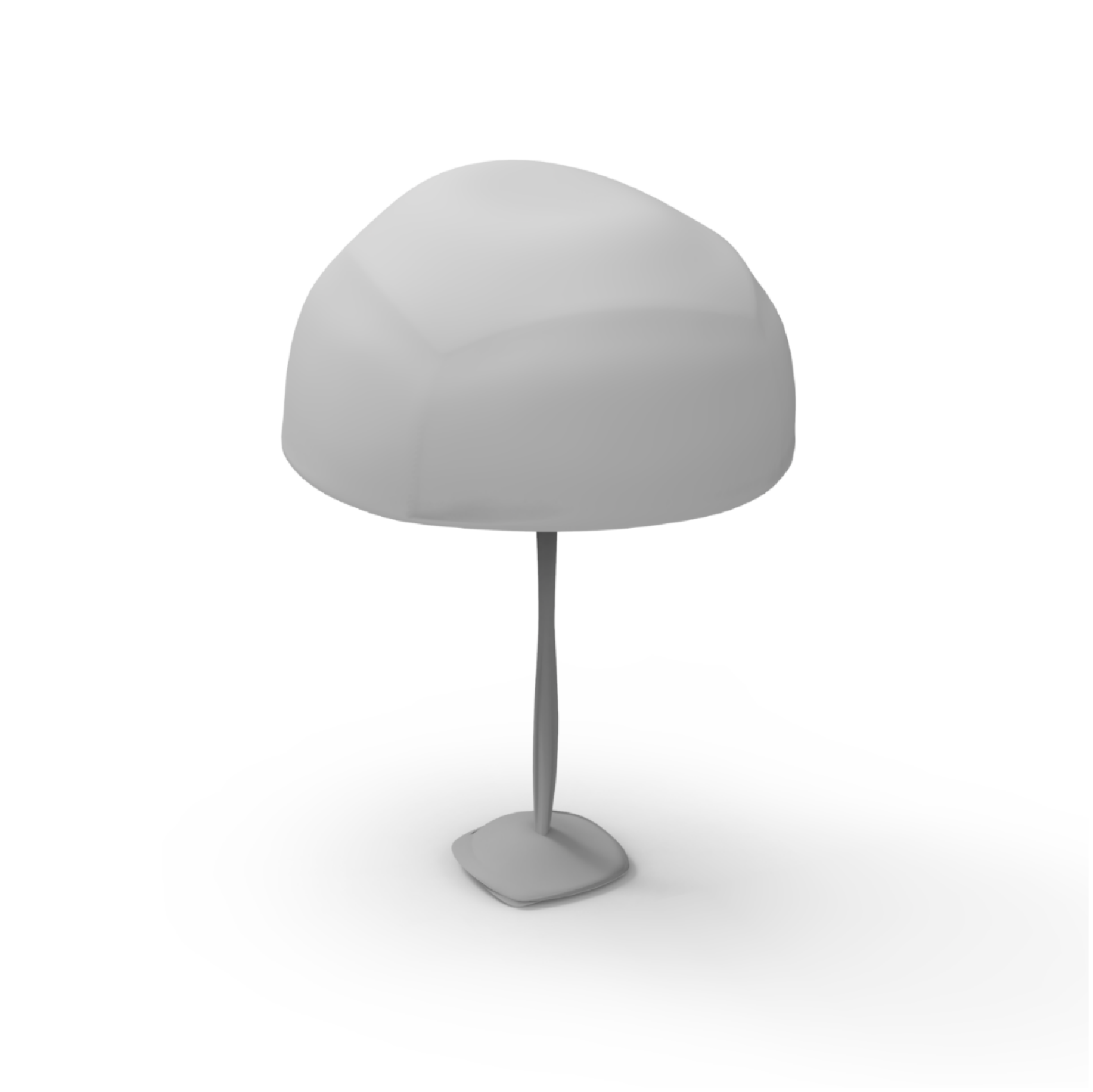}
    \includegraphics[width=0.15\linewidth]{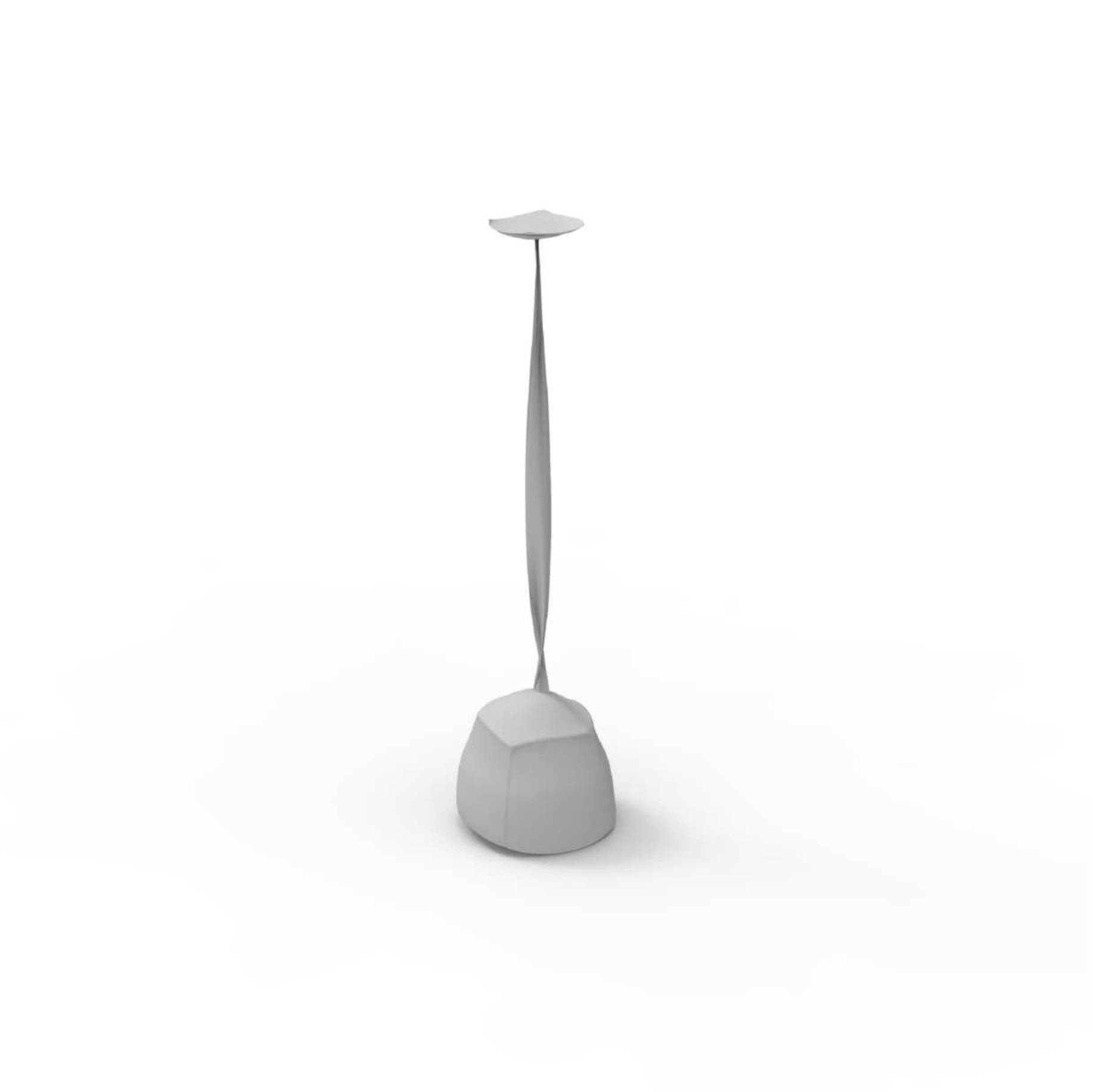}
    \includegraphics[width=0.15\linewidth]{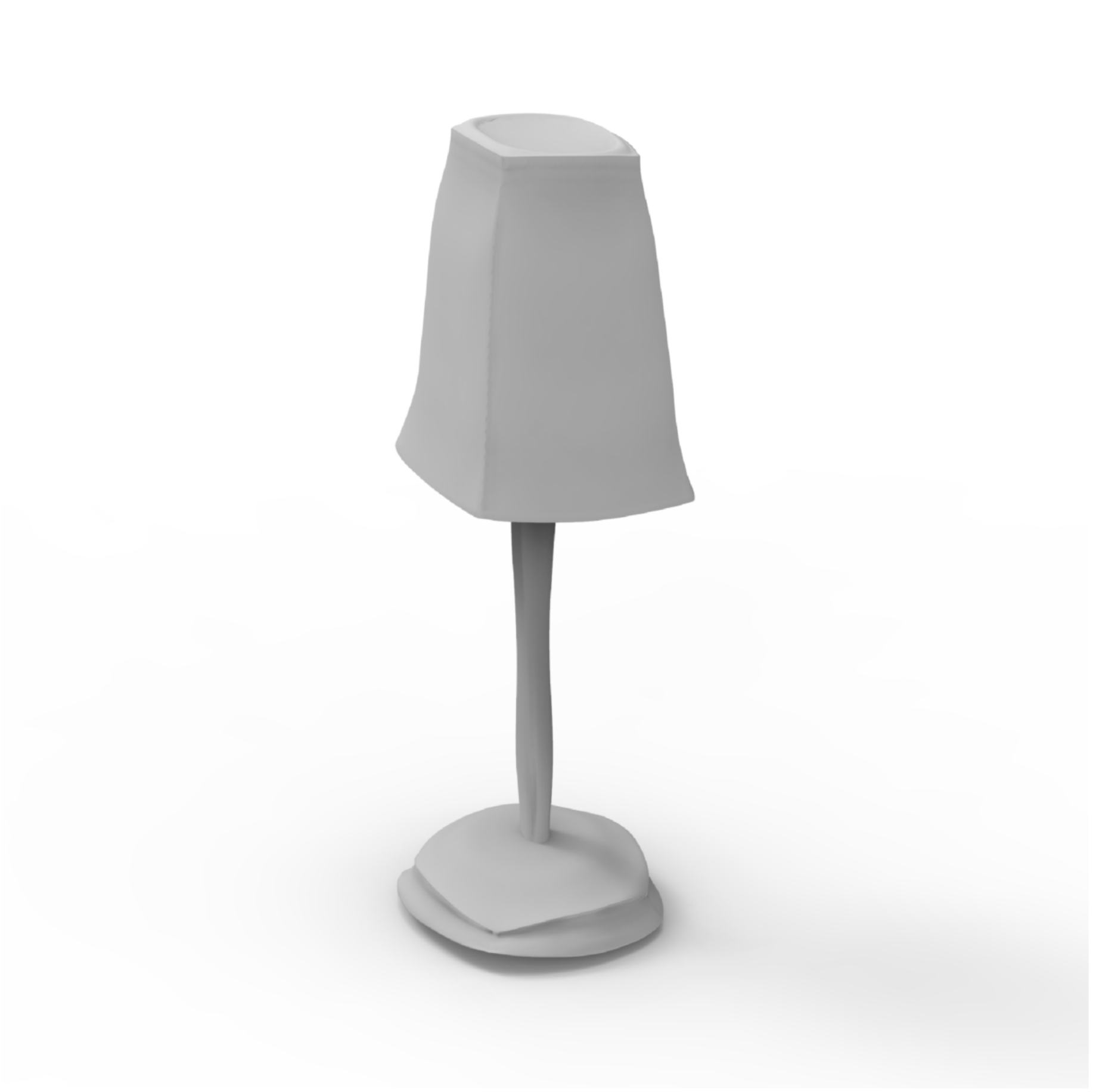}
    \includegraphics[width=0.15\linewidth]{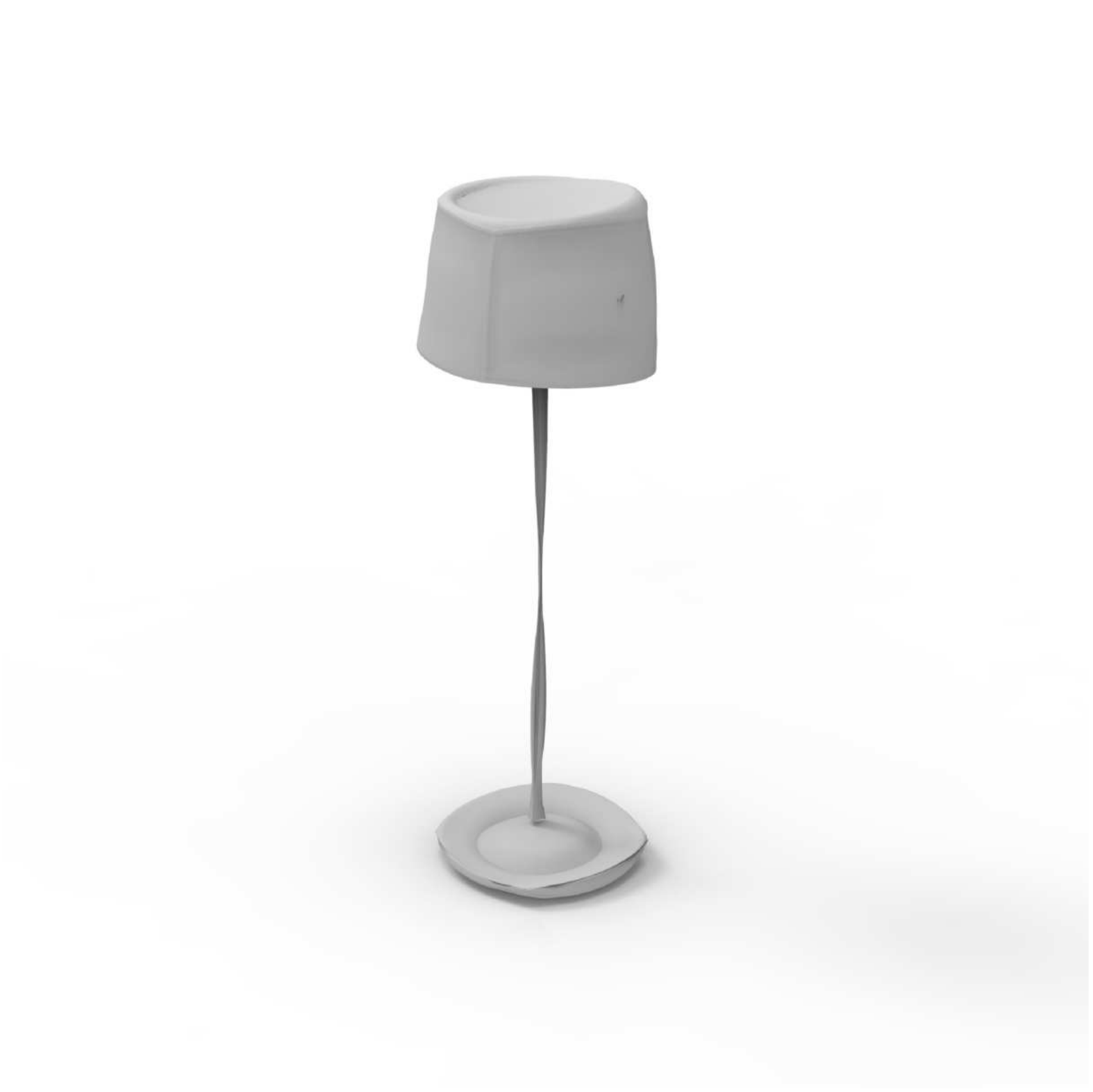}
    \vspace{-3mm}
    \caption{\yjr{Shape generation results. We sample random Gaussian noises in both latent spaces of shape structure and geometry and use DSG-Net to generate realistic shapes with complex structures and detailed geometry. We show six generation results for each of the four object categories in PartNet.}}
    \label{fig:generation}
    \vspace{-3mm}
\end{figure}

\subsection{Shape Generation}
The main goal of DSG-Net is to generate high-quality shapes with complex structures and fine-grained geometry.
Given a noise vector sampled from a unit Gaussian distribution, a 3D shape generative model maps it to a realistic 3D shape.
We evaluate the shape generation performance of DSG-Net and compare to several state-of-the-art baseline methods.
Quantitative evaluations and user-study results further validate our superior performance over baselines.
Equipped with two disentangled latent spaces for shape structure and geometry, DSG-Net also enables a novel task of generating shapes with a given structure or geometry pattern.
\yjr{Please refer to the supplementary material for more results on shape generation.}

\begin{figure}[!t]
    \centering
    \subfigure[Random Generation]{
    \begin{minipage}[b]{0.14\linewidth}
    \includegraphics[width=0.95\linewidth]{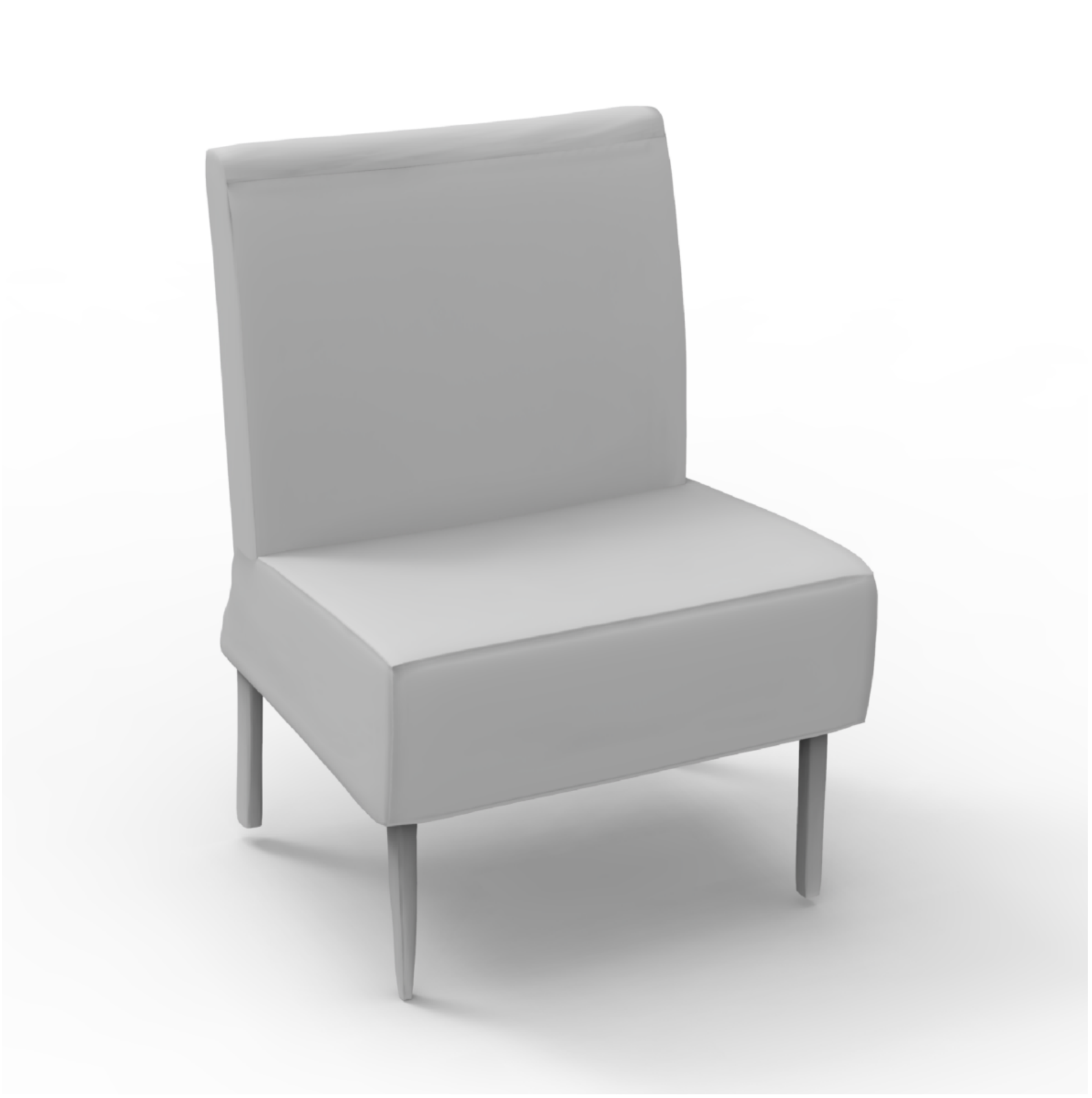}\\
    \includegraphics[width=0.95\linewidth]{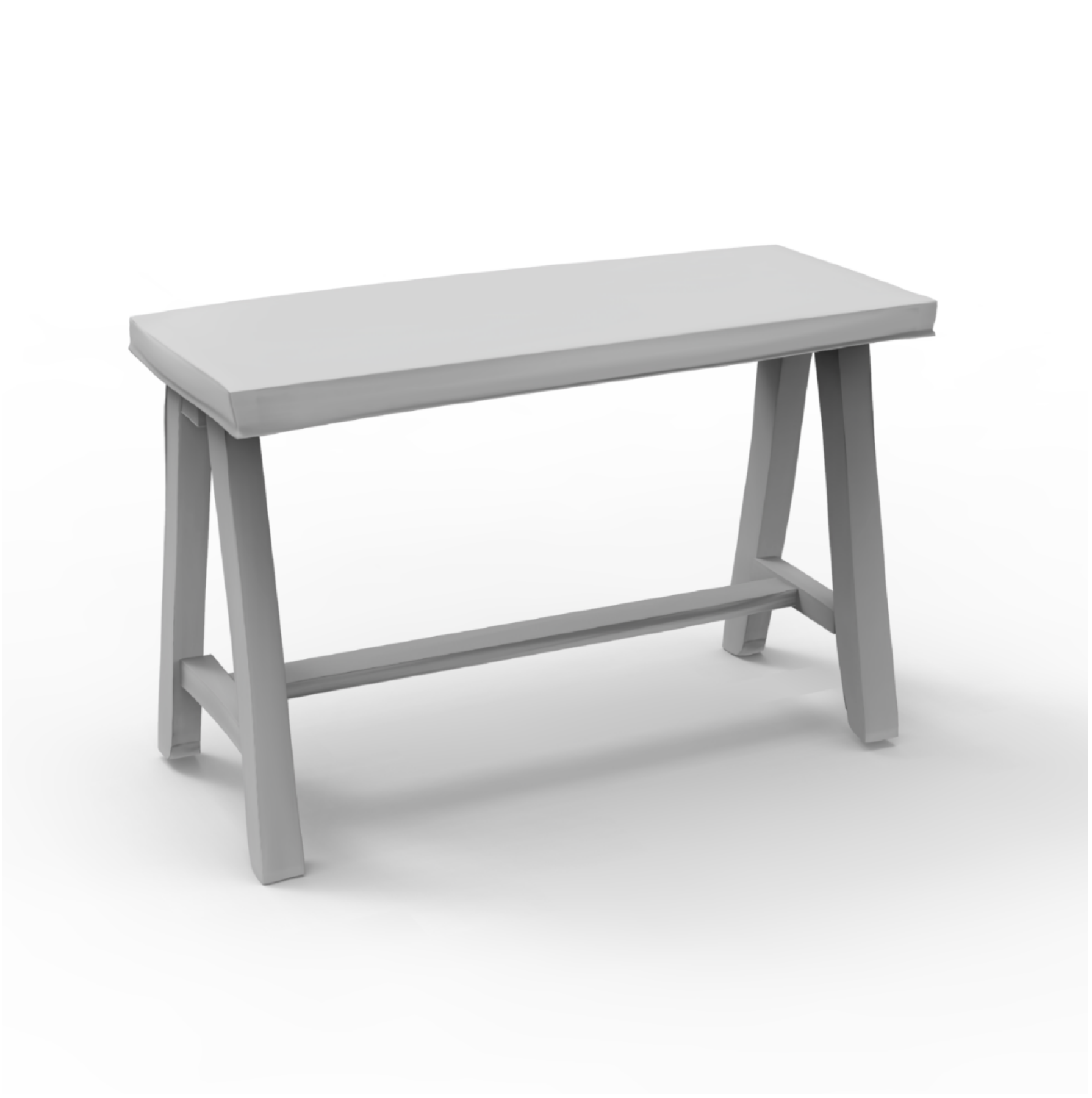}\vspace{1mm}\\
    \includegraphics[width=0.95\linewidth]{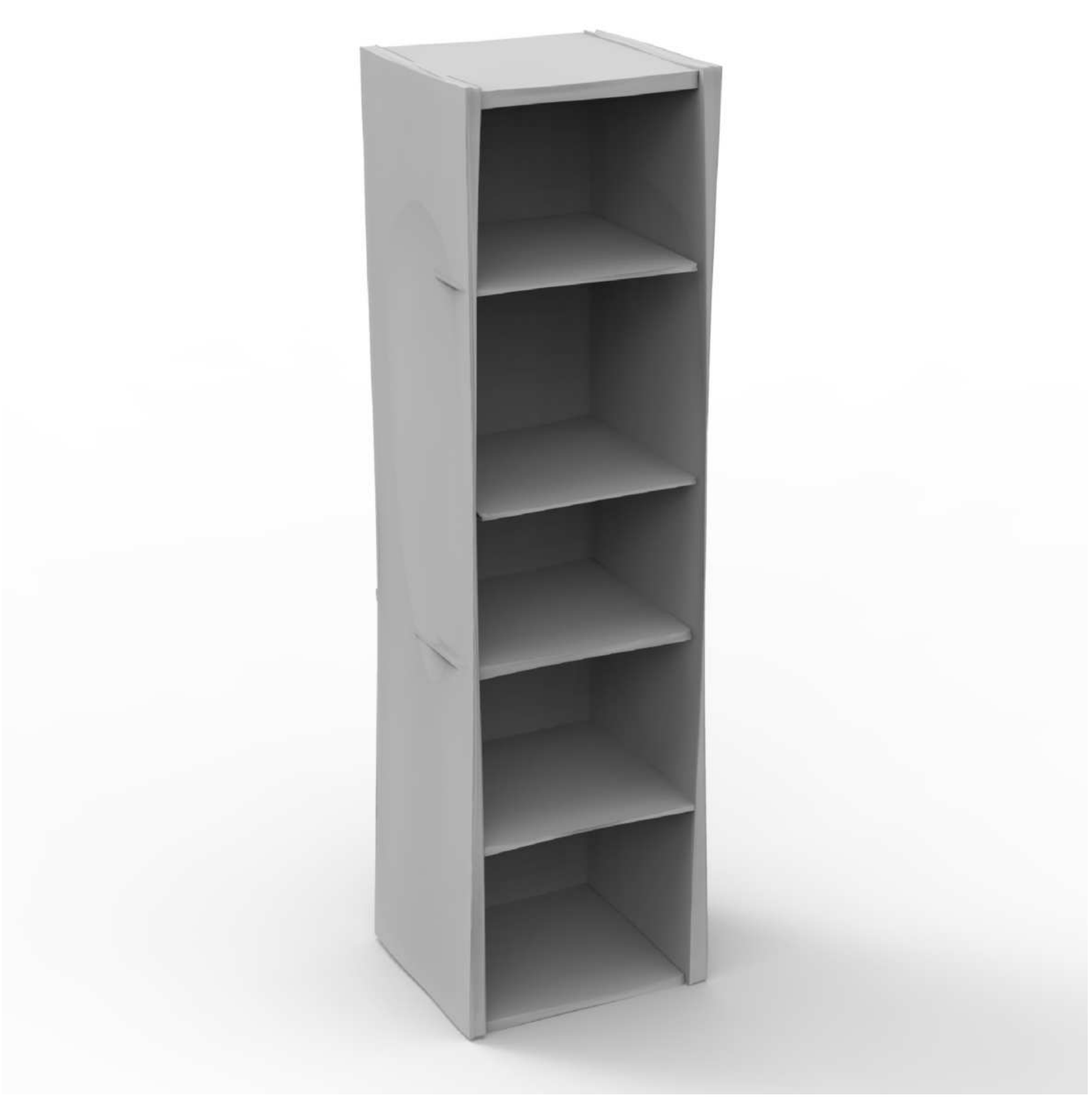}\\
    \includegraphics[width=0.95\linewidth]{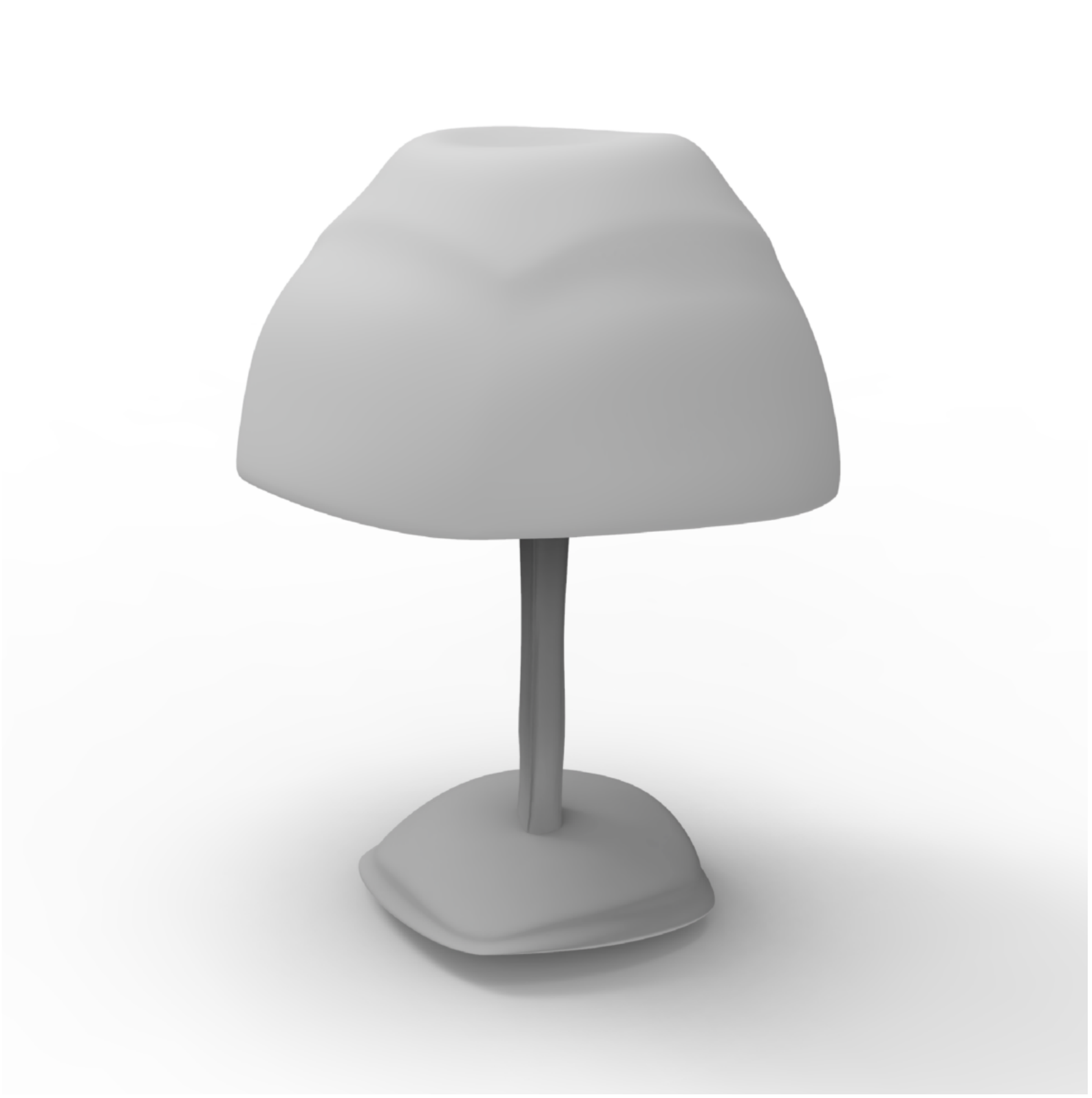}
    \end{minipage}\vline}
    \subfigure[Top-5 retrieved shapes in training sets]{
    \begin{minipage}[b]{0.73\linewidth}
    \includegraphics[width=0.19\linewidth]{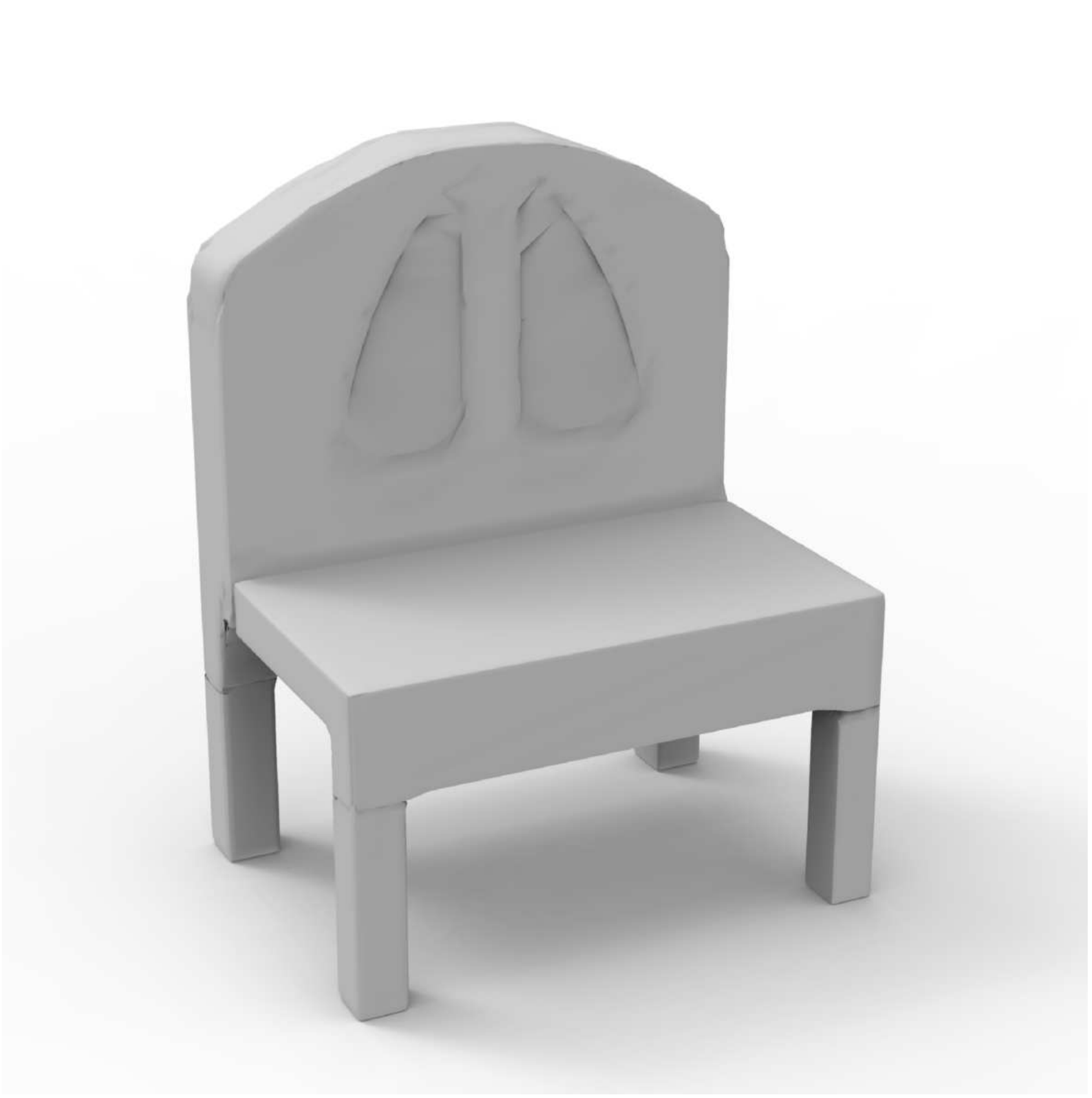}
    \includegraphics[width=0.19\linewidth]{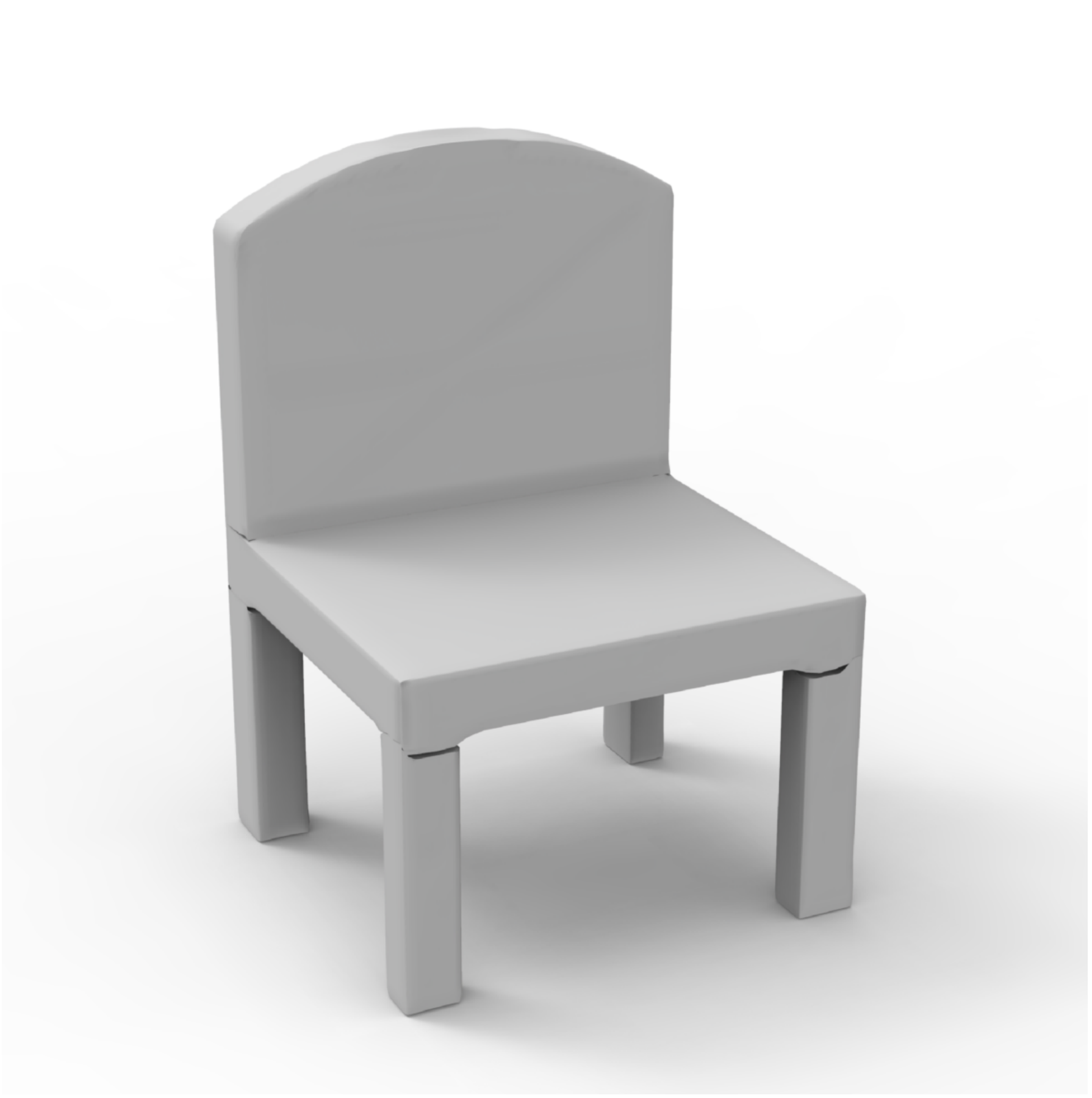}
    \includegraphics[width=0.19\linewidth]{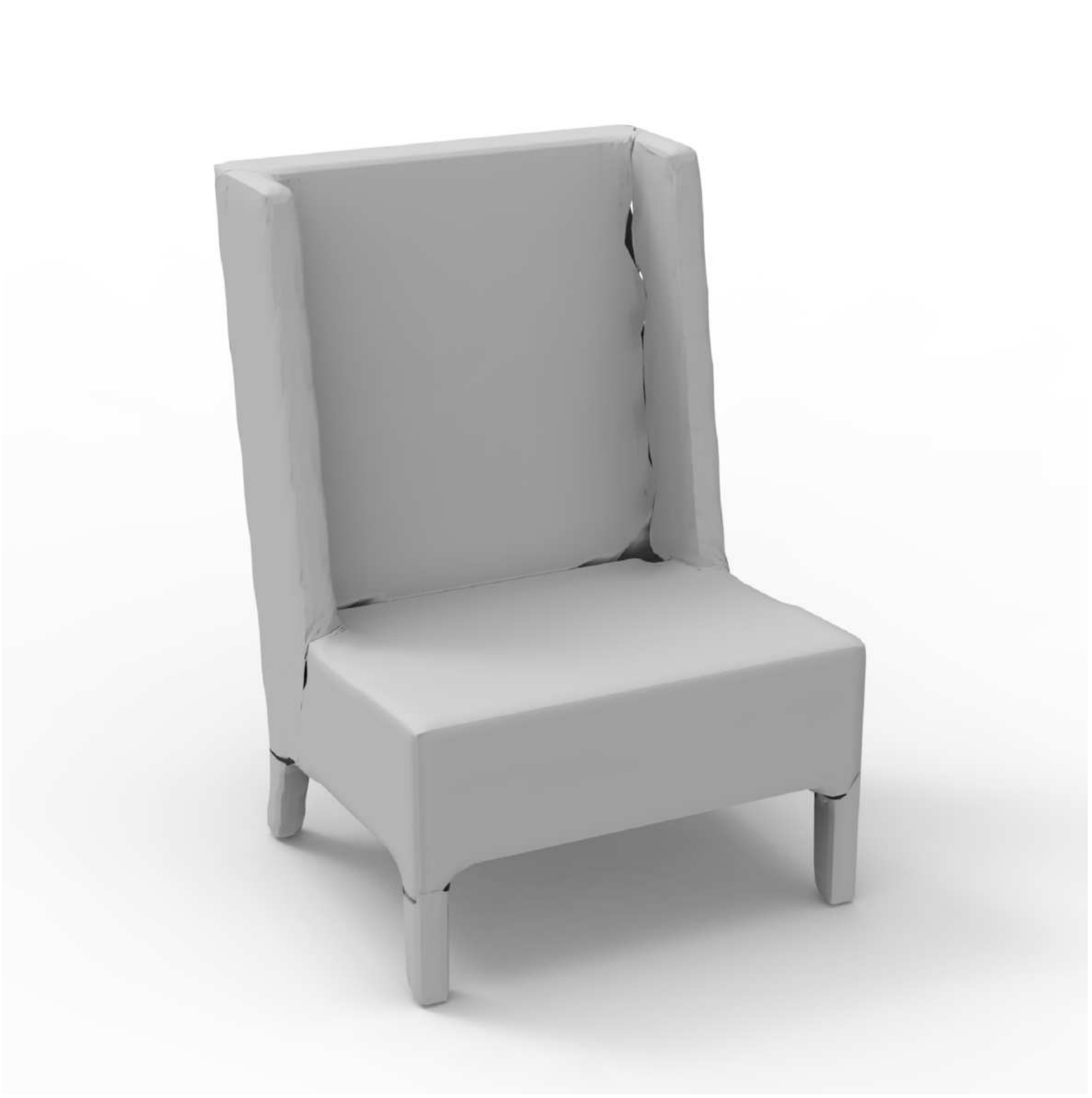}
    \includegraphics[width=0.19\linewidth]{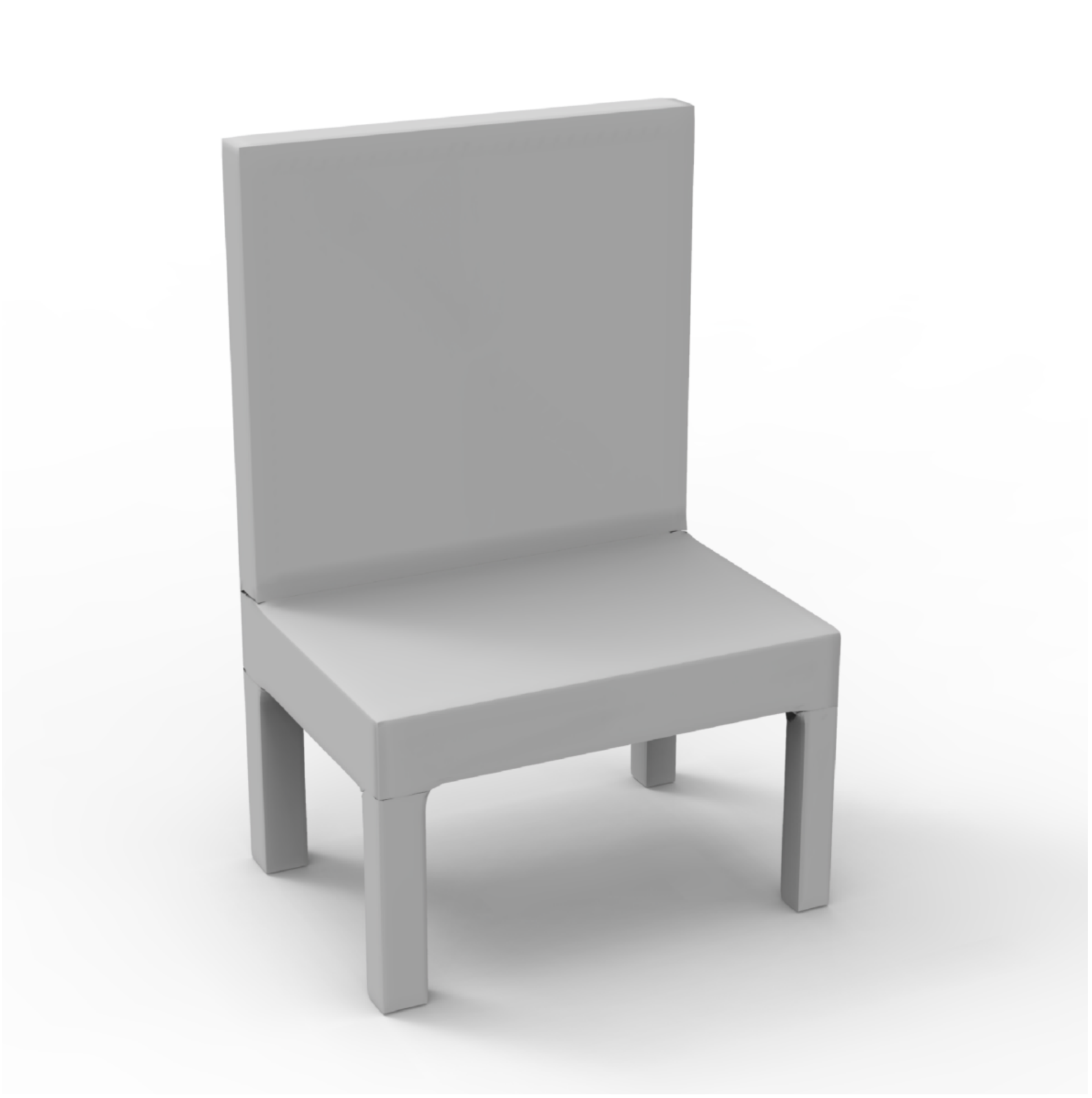}
    \includegraphics[width=0.19\linewidth]{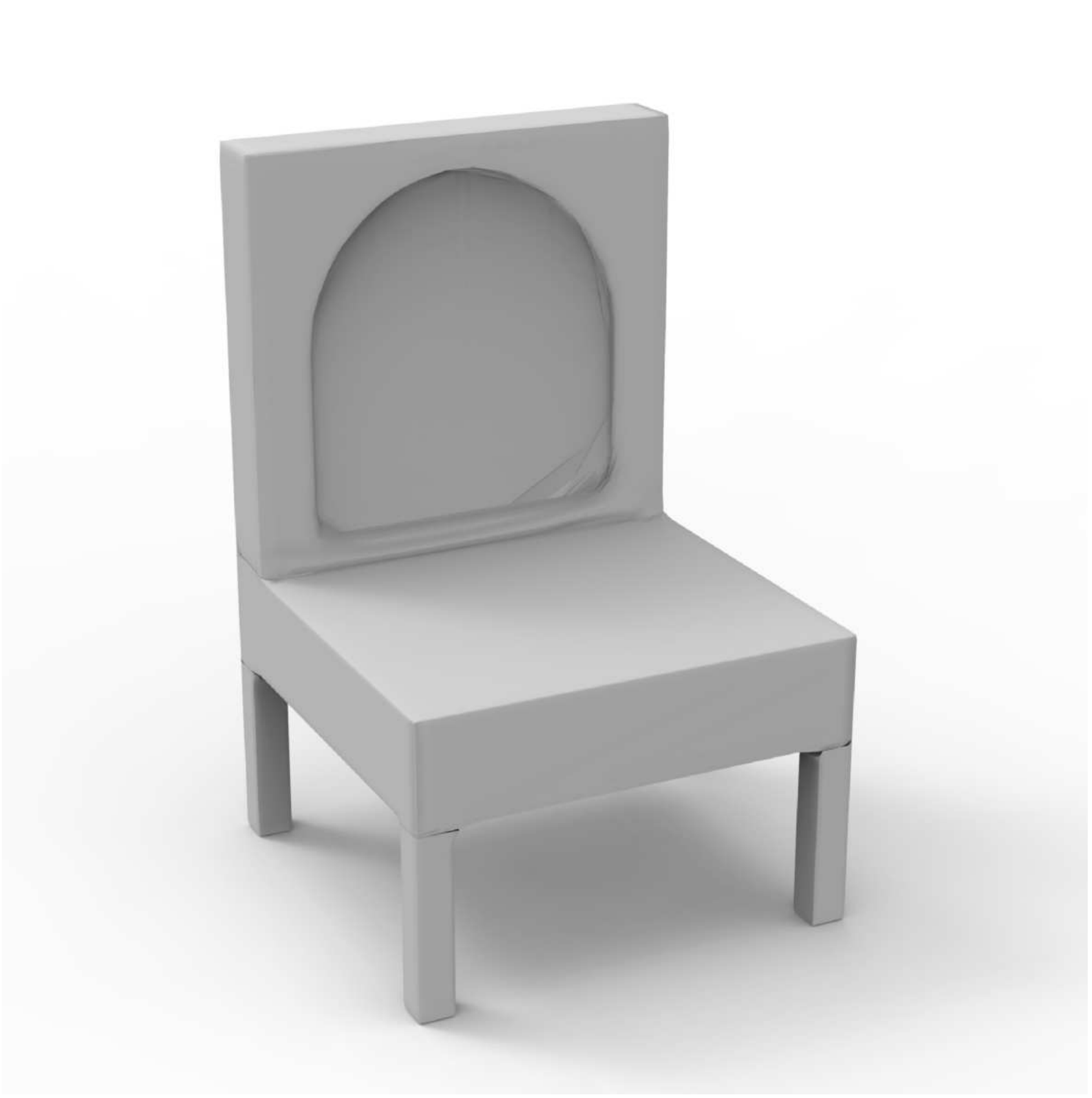}\\
    \includegraphics[width=0.19\linewidth]{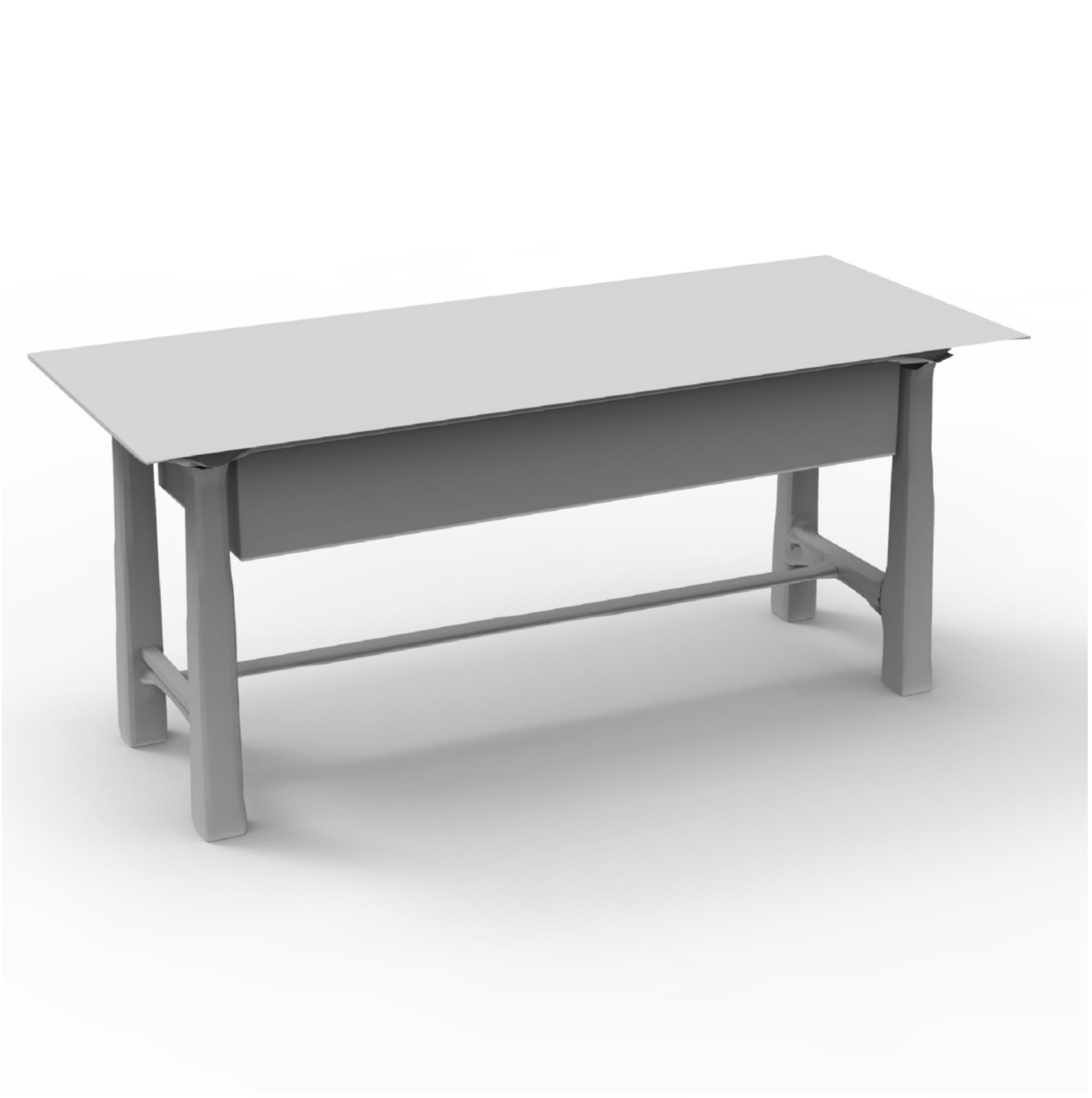}
    \includegraphics[width=0.19\linewidth]{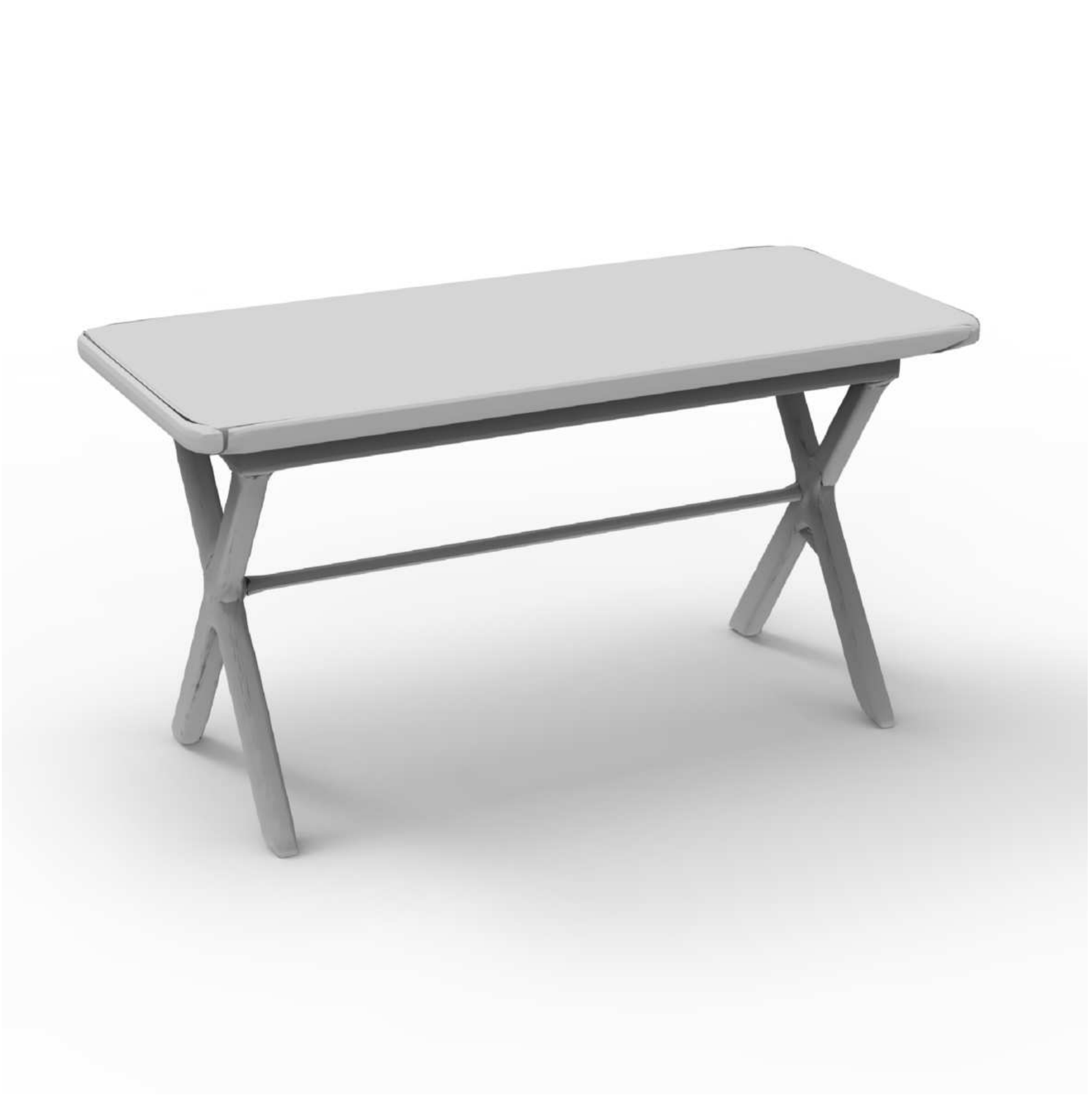}
    \includegraphics[width=0.19\linewidth]{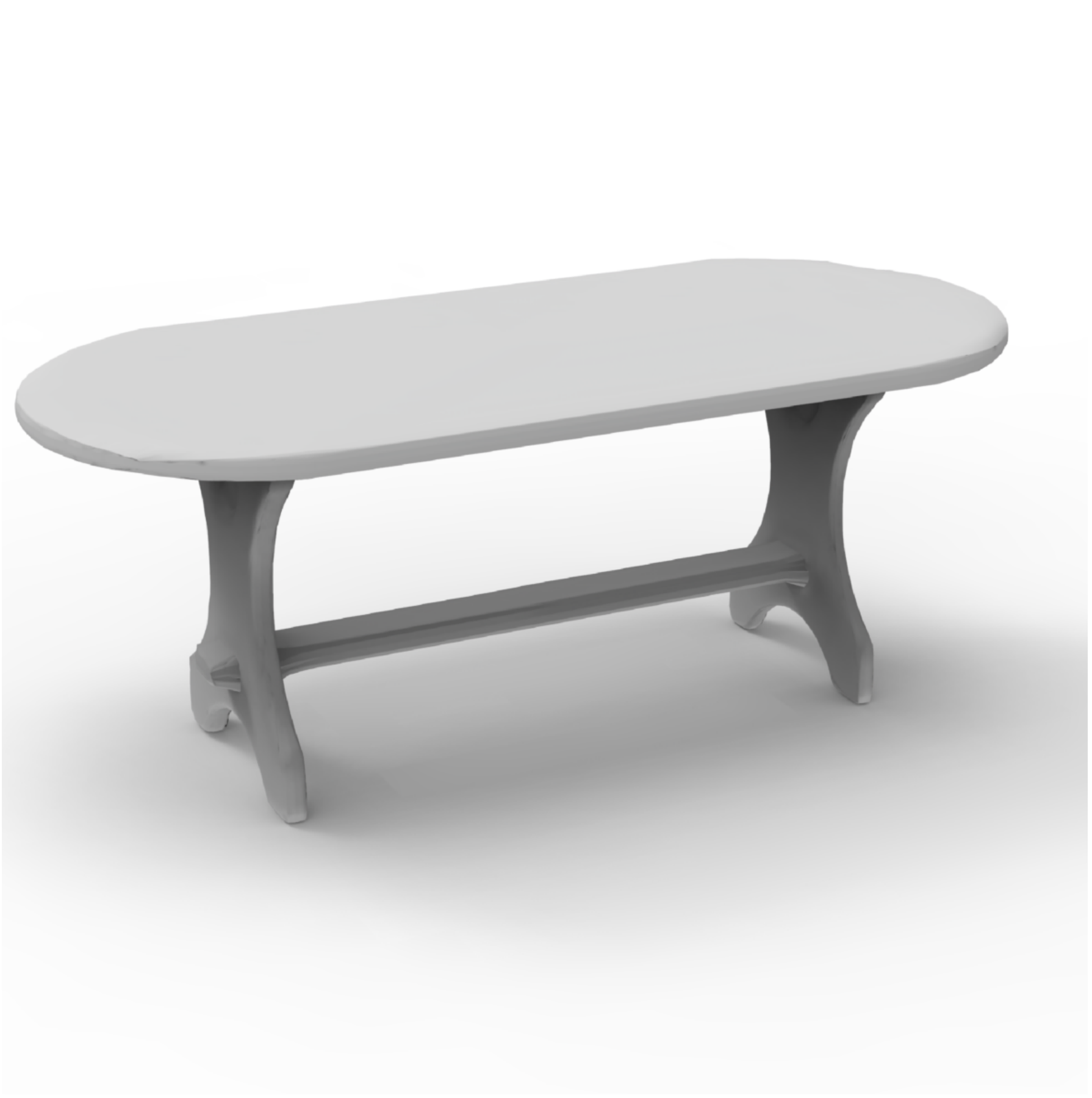}
    \includegraphics[width=0.19\linewidth]{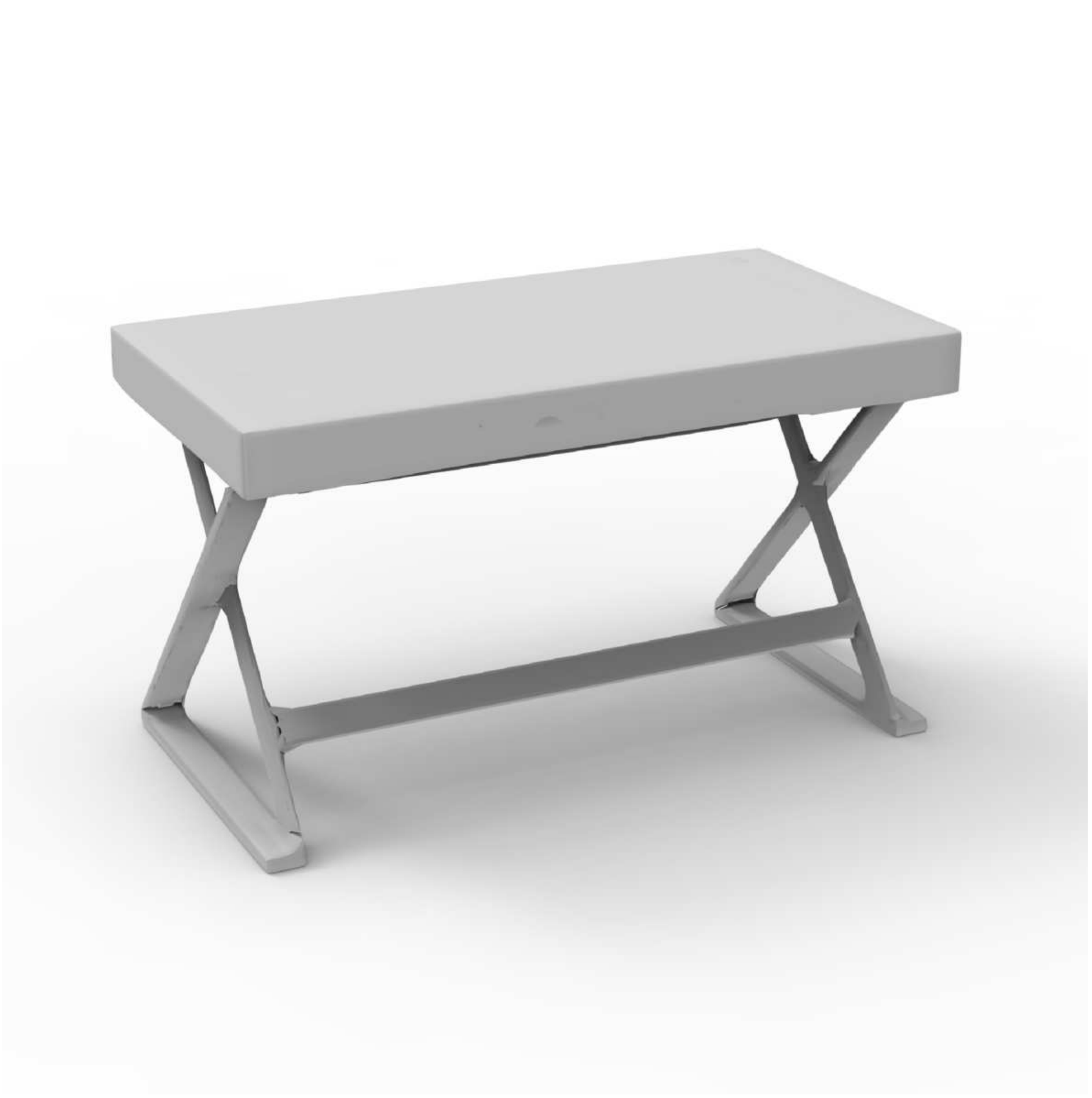}
    \includegraphics[width=0.19\linewidth]{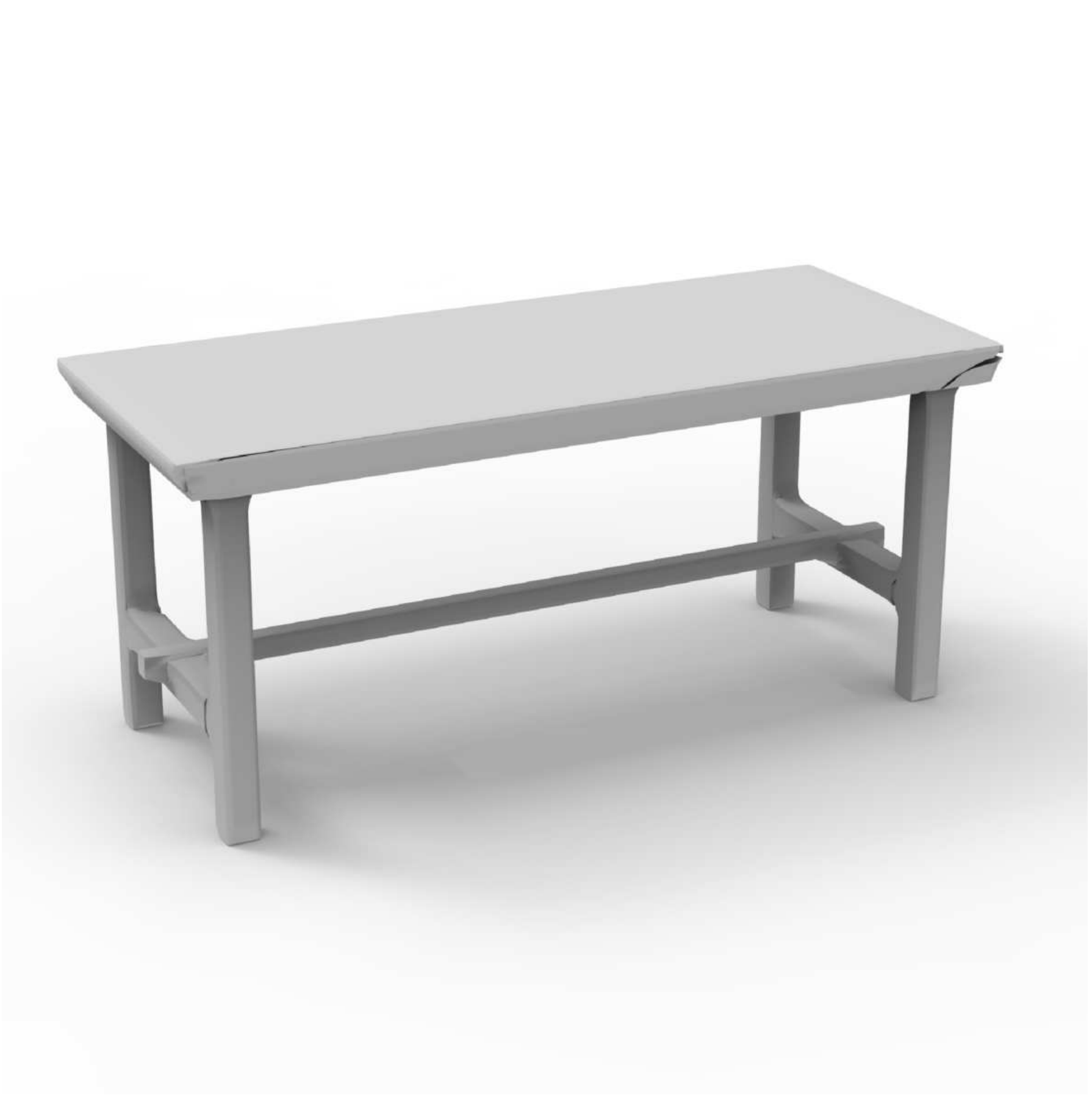}\\
    \includegraphics[width=0.19\linewidth]{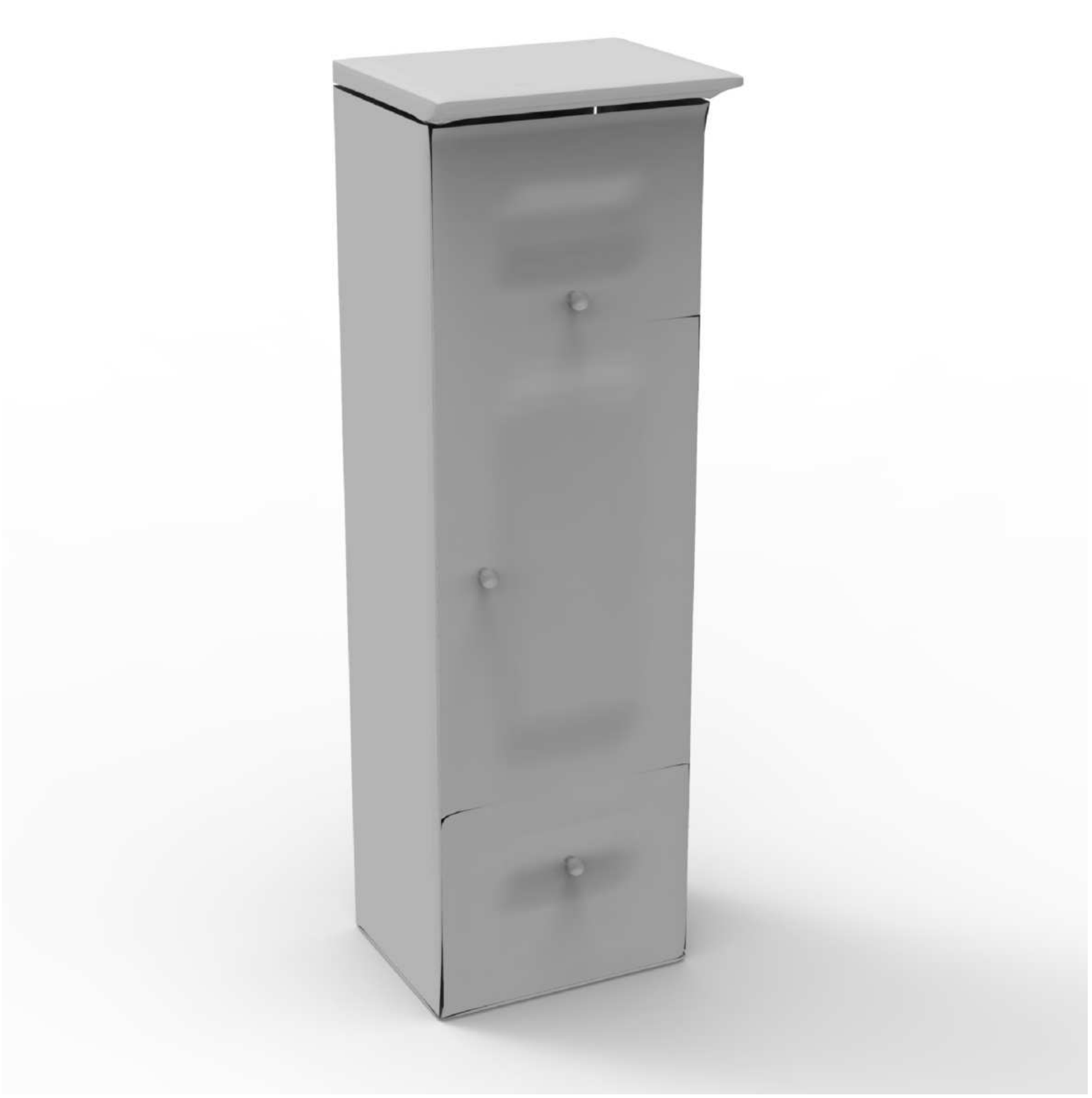}
    \includegraphics[width=0.19\linewidth]{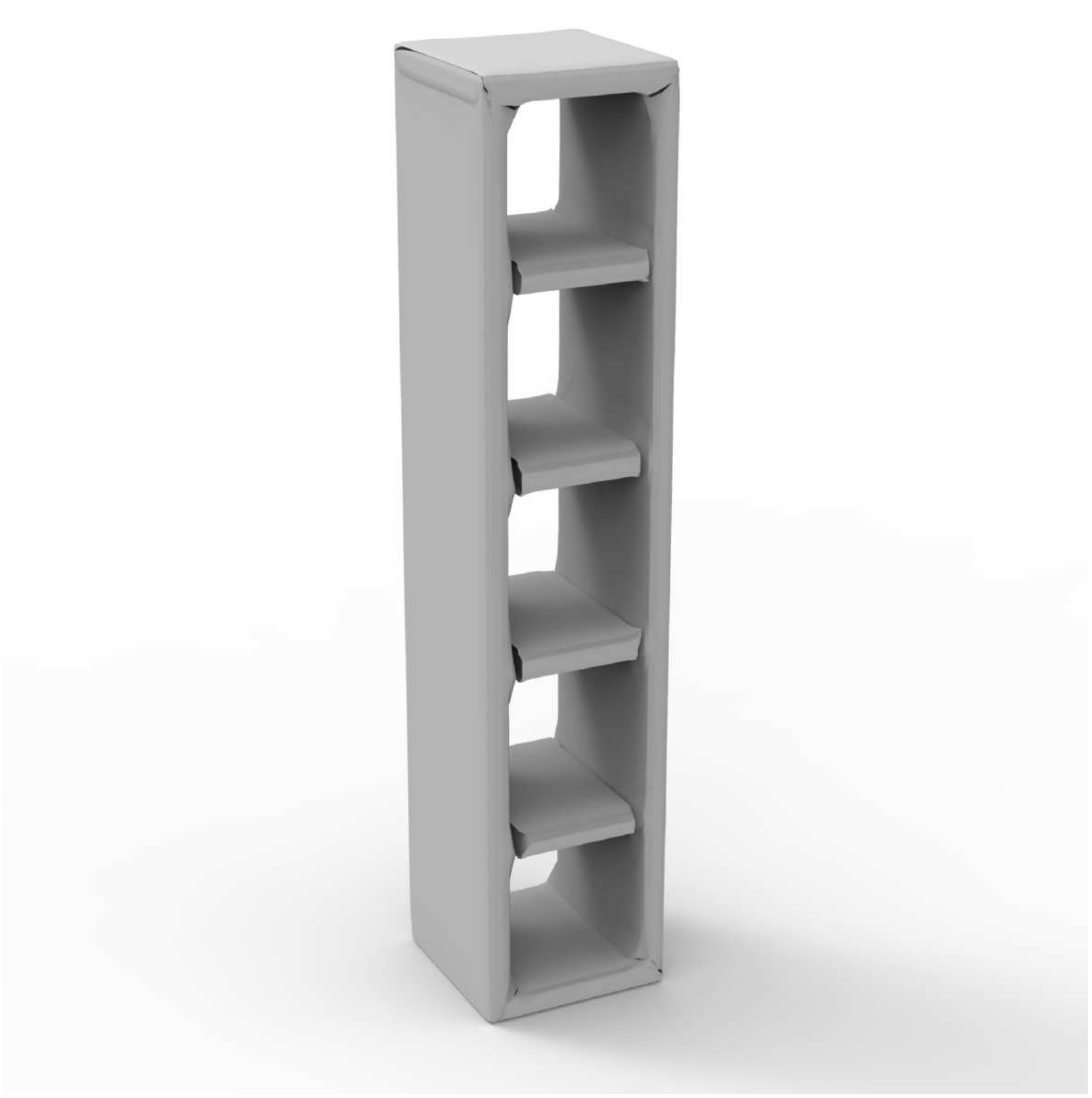}
    \includegraphics[width=0.19\linewidth]{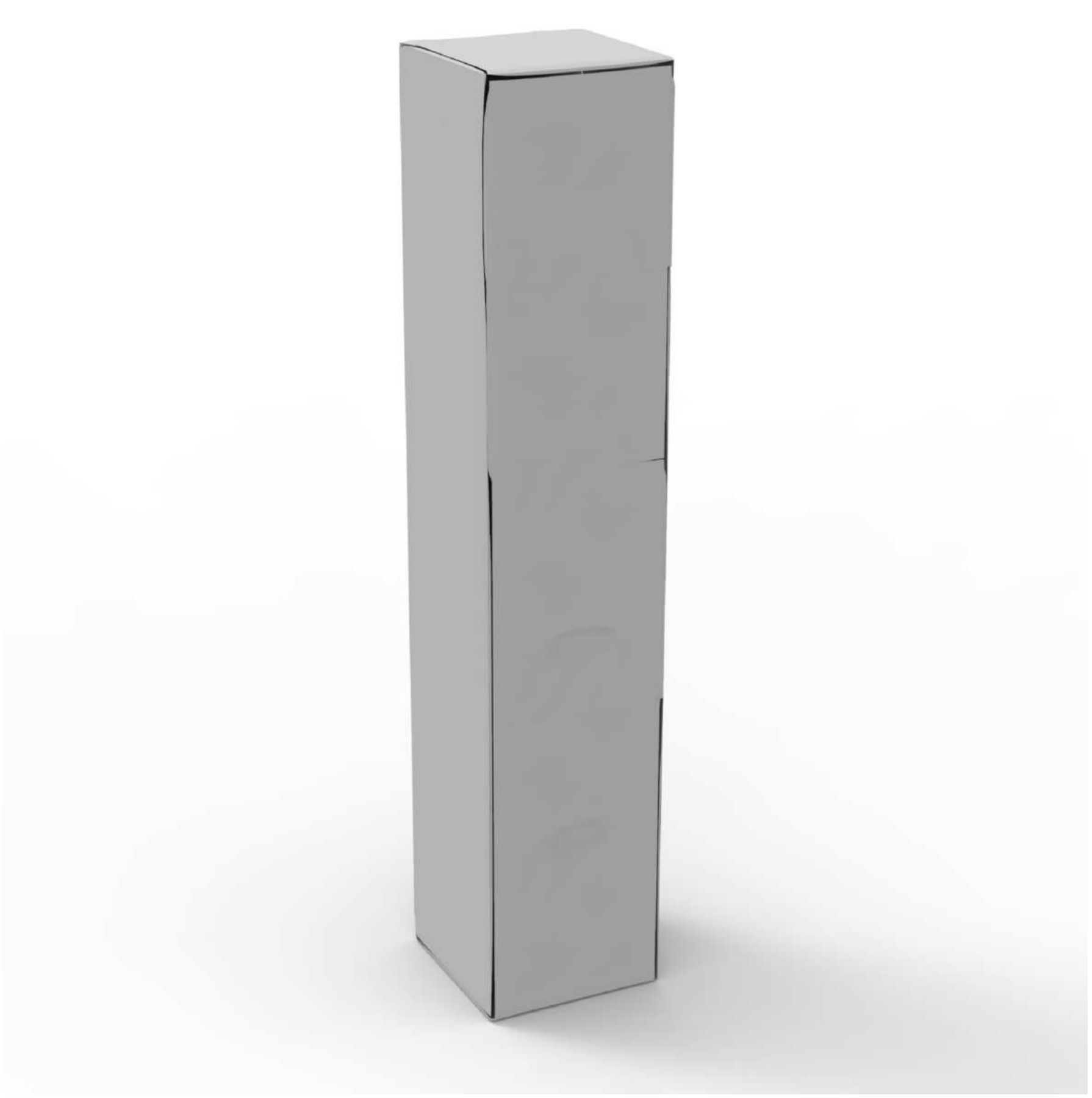}
    \includegraphics[width=0.19\linewidth]{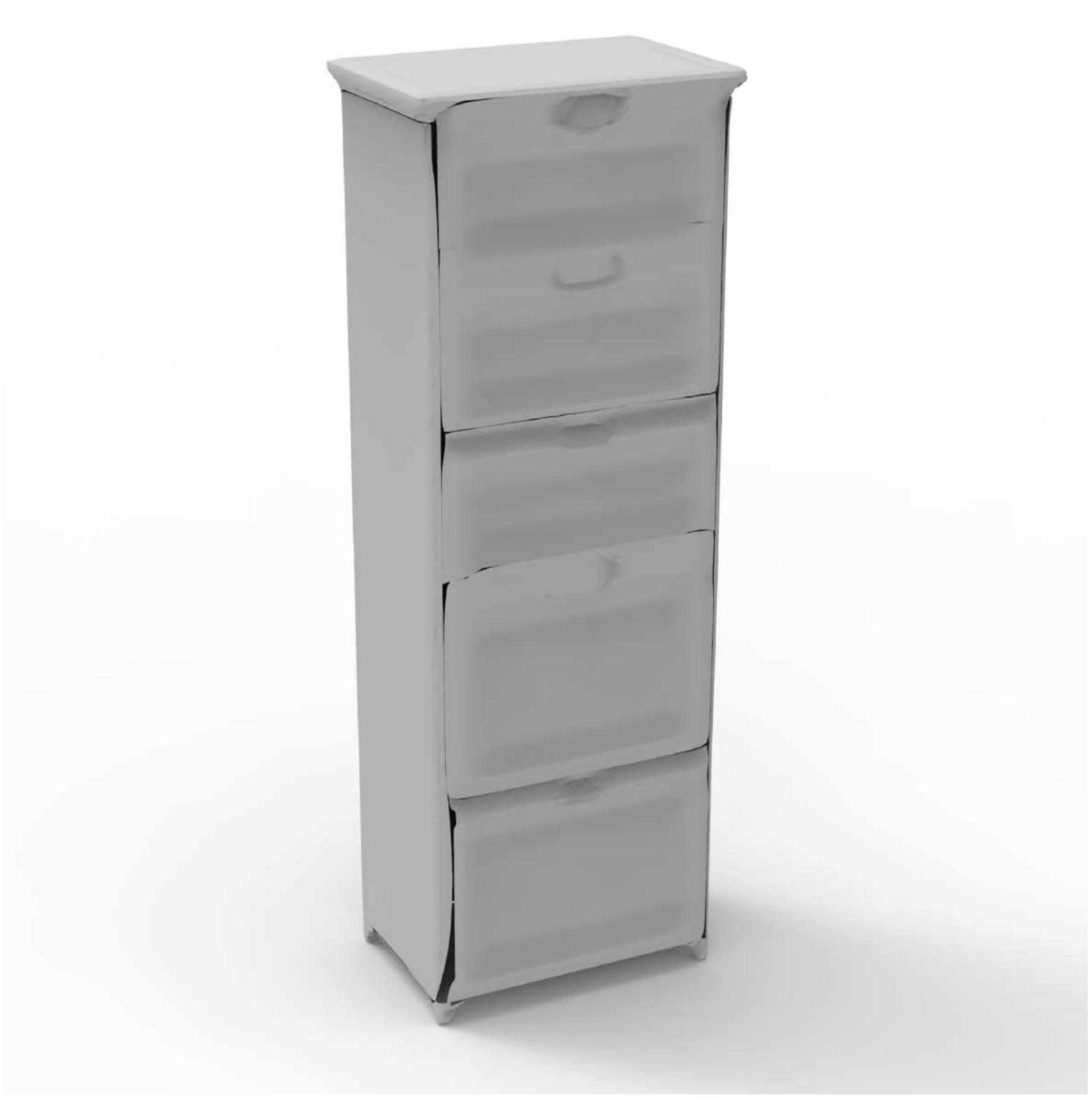}
    \includegraphics[width=0.19\linewidth]{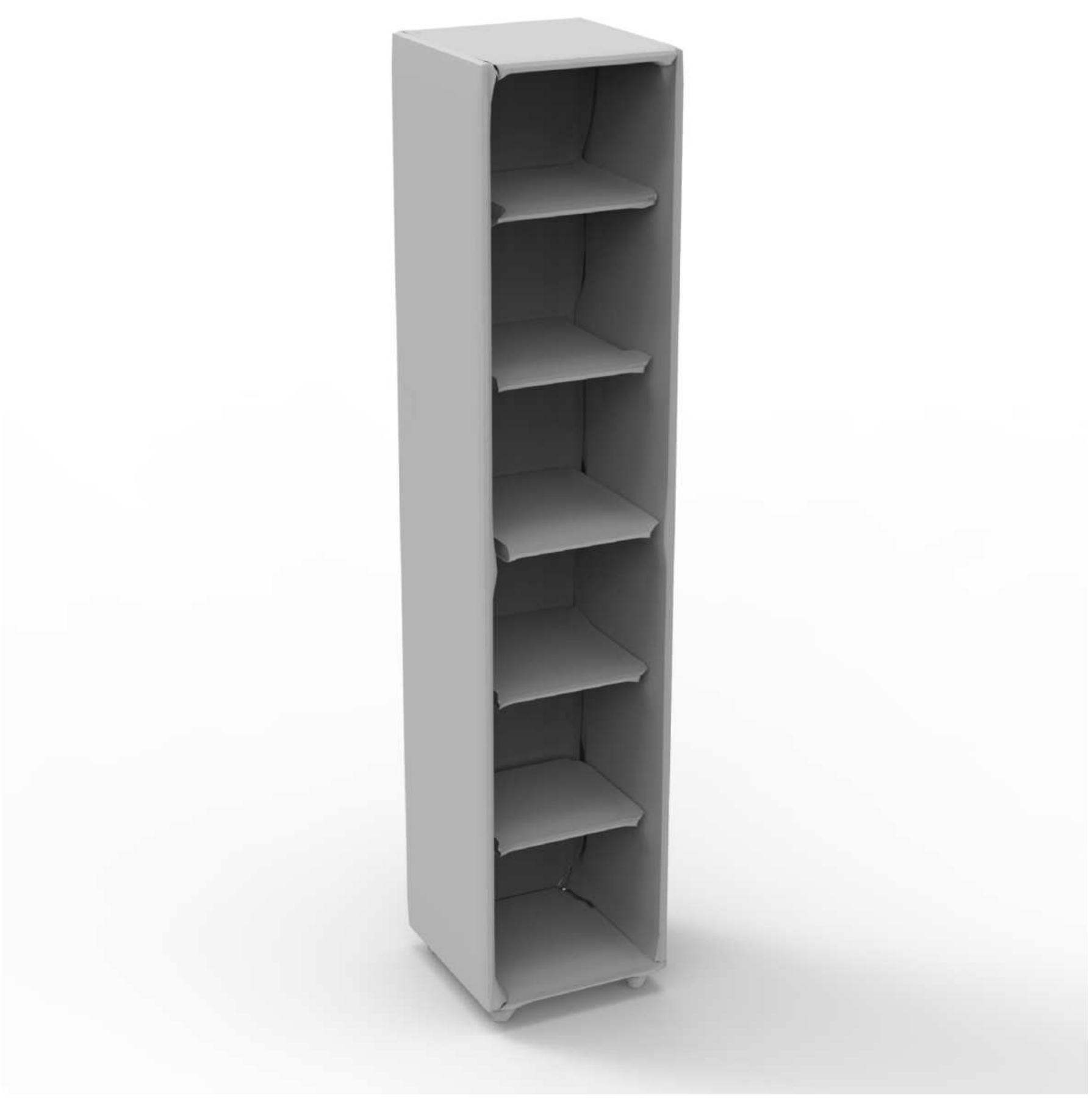}\\
    \includegraphics[width=0.19\linewidth]{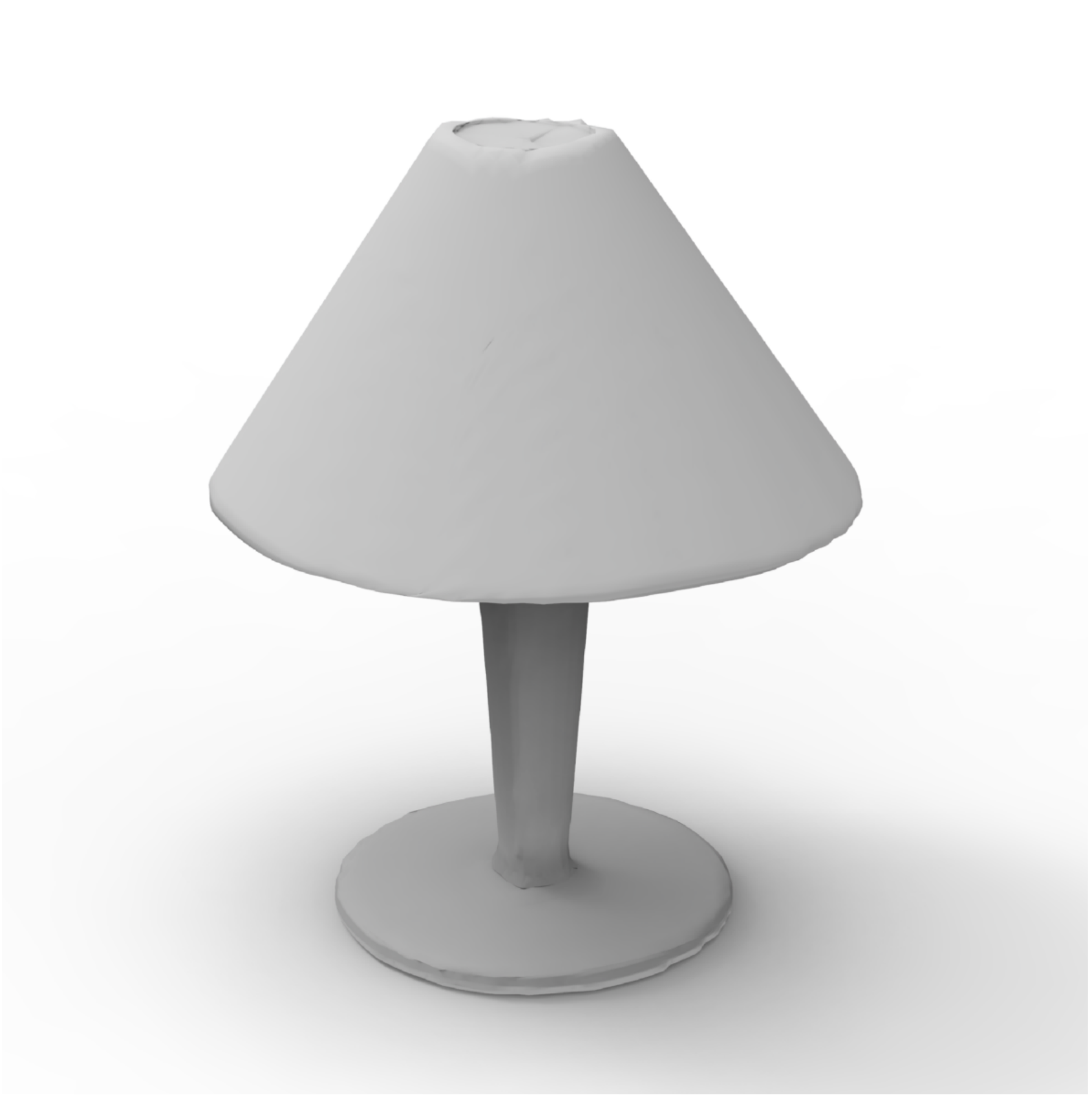}
    \includegraphics[width=0.19\linewidth]{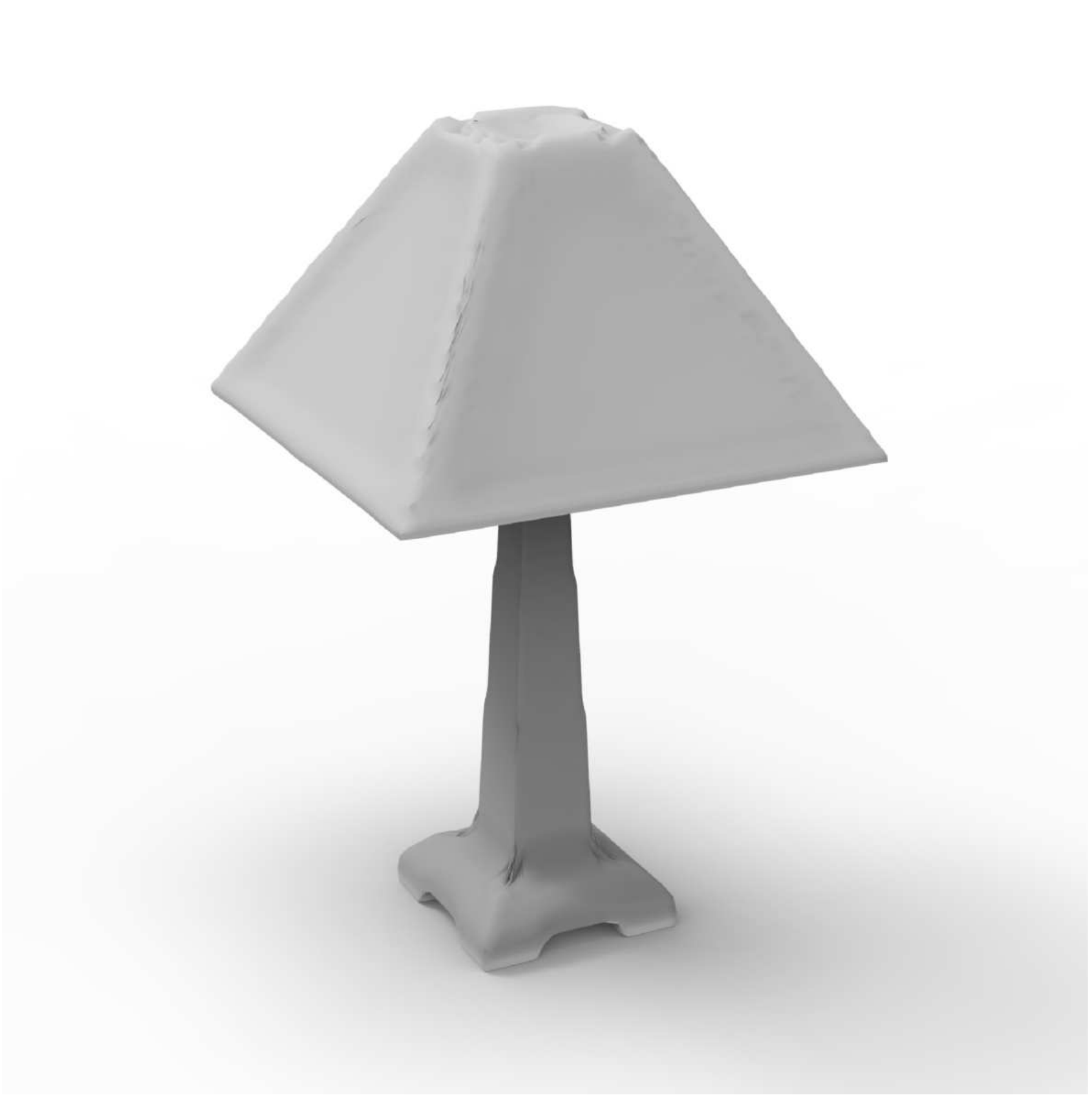}
    \includegraphics[width=0.19\linewidth]{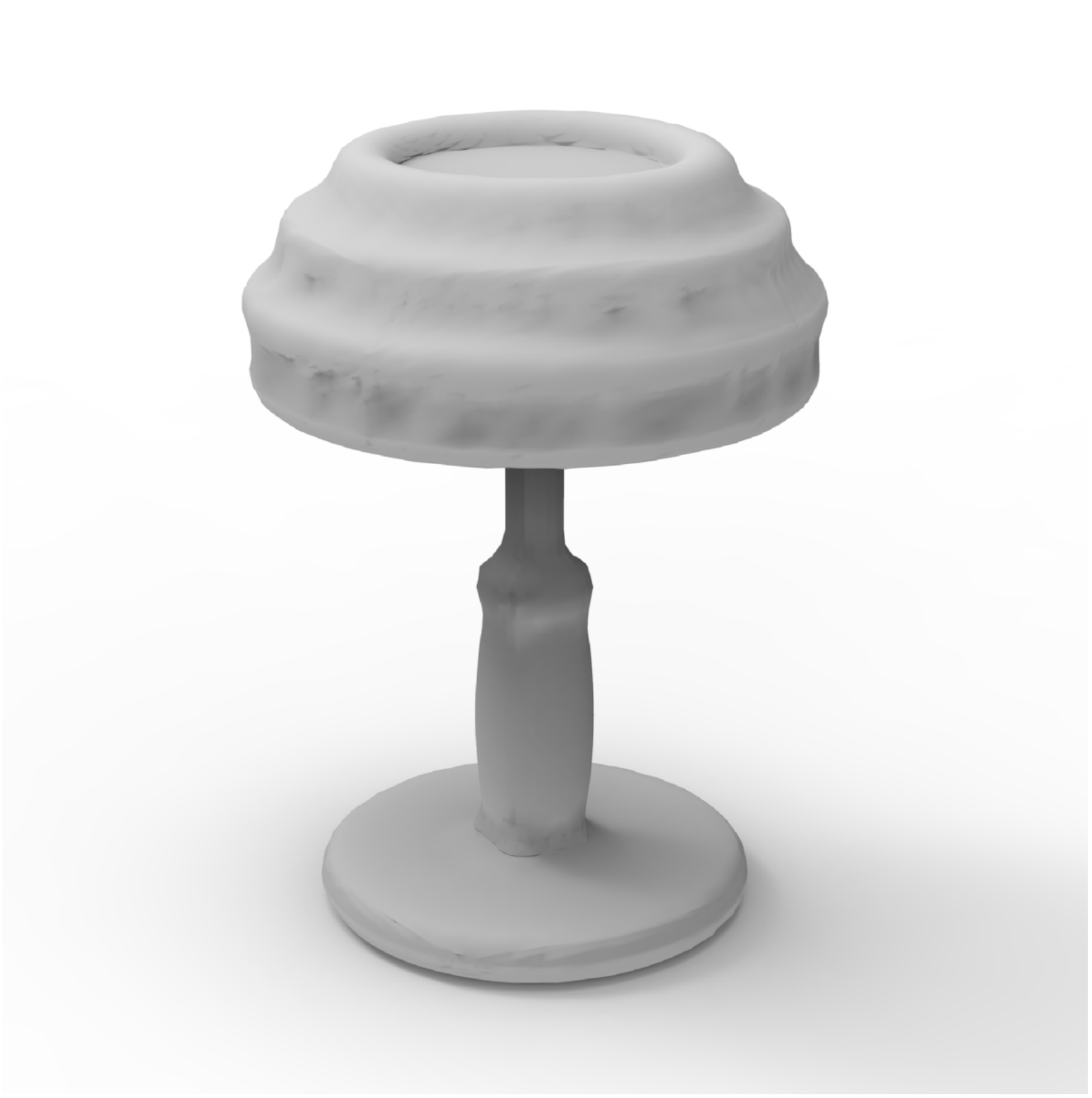}
    \includegraphics[width=0.19\linewidth]{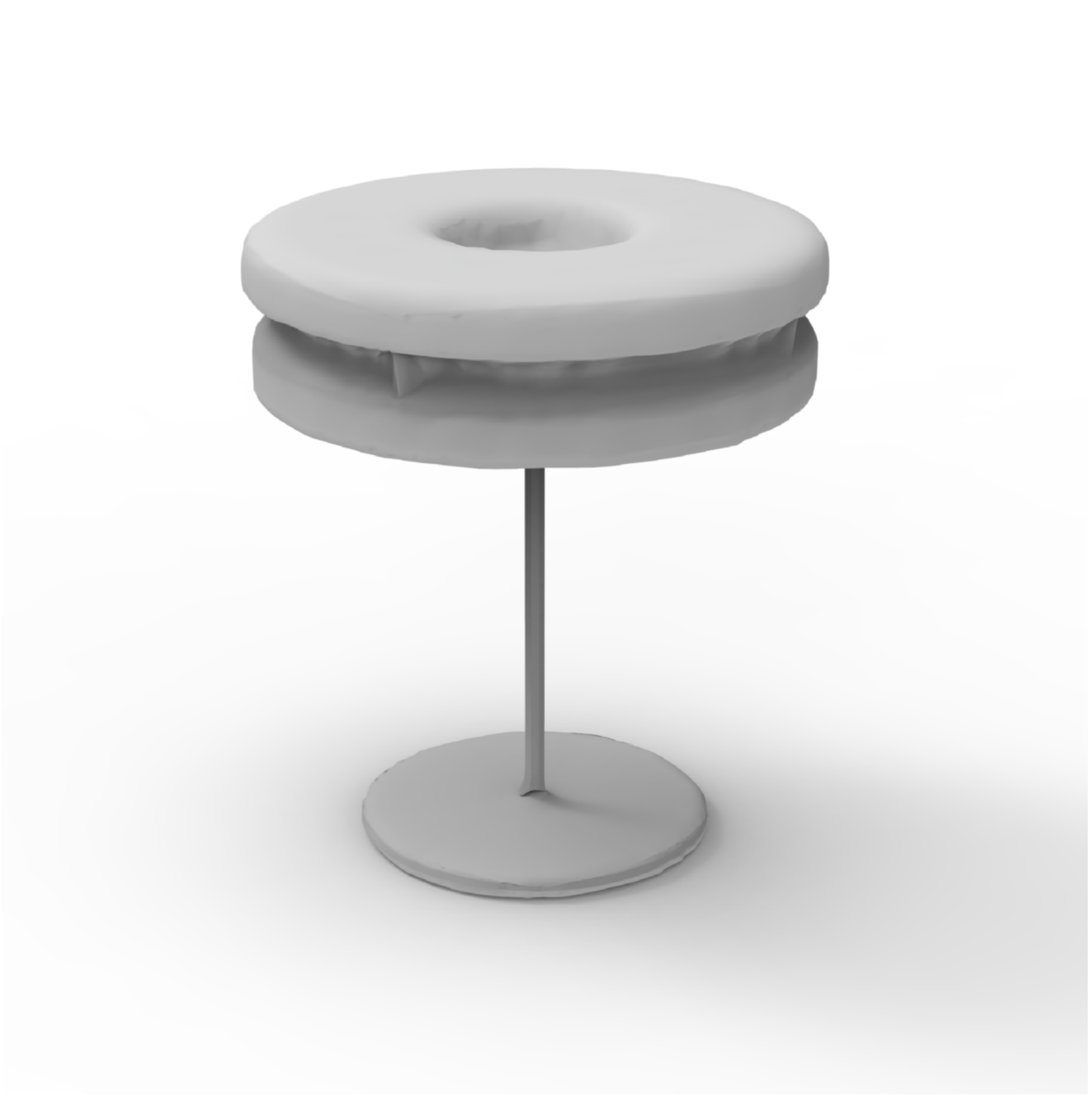}
    \includegraphics[width=0.19\linewidth]{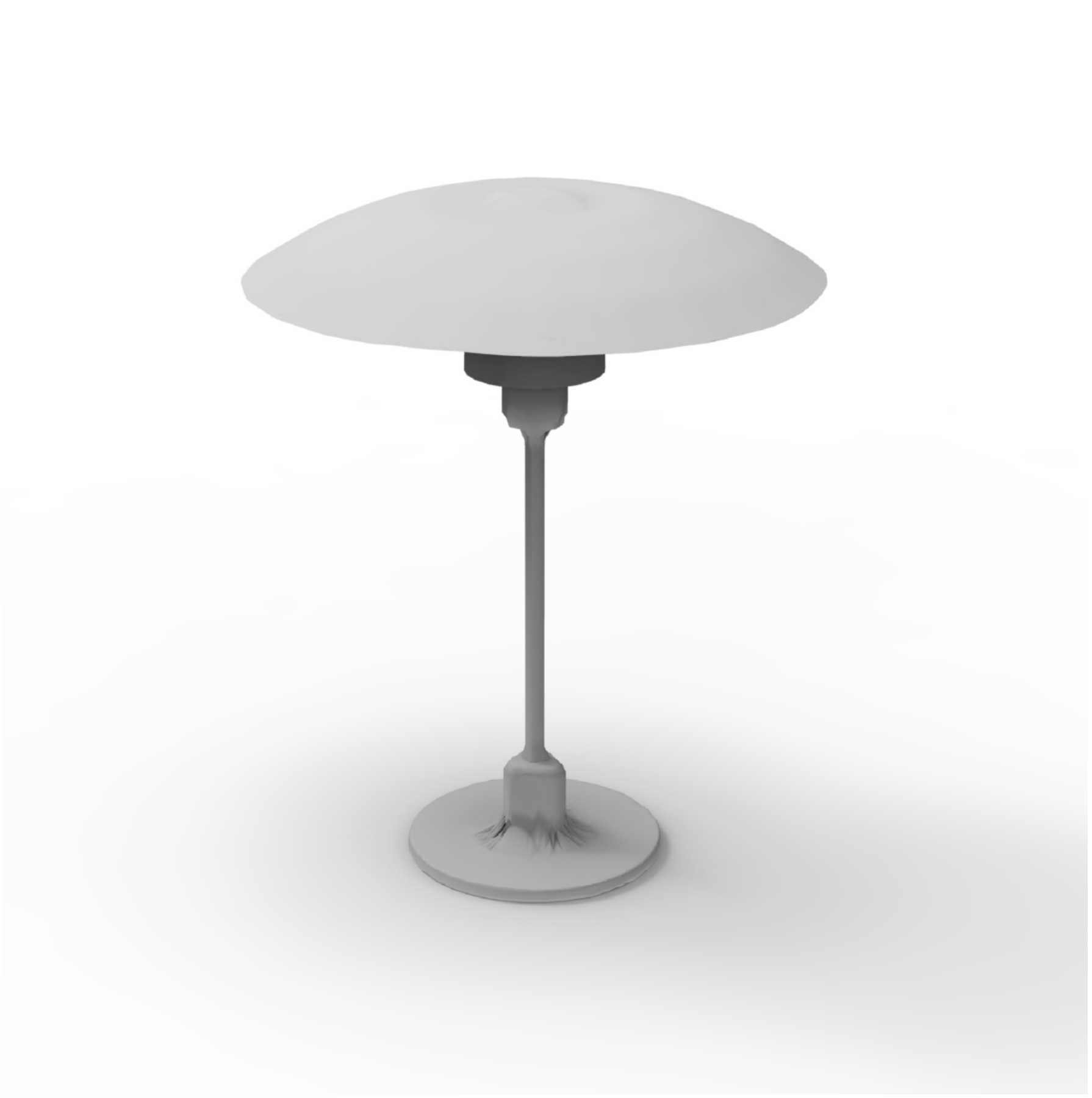}
    \end{minipage}}
    \vspace{-3mm}
    \caption{\yjr{Examples demonstrating generation of novel shapes. We present top-five retrieved shapes in the training sets, with CD as the retrieval metric, to our generated results shown in the left most column. Our generated shapes are different from the shapes in training sets, showing that DSG-Net does not simply overfit the training data.}}
    \label{fig:generation_nn}
    \vspace{-3mm}
\end{figure}

\begin{figure}[!t]
    \centering
    \subfigure[SDM-Net]{
    \includegraphics[width=0.11\linewidth]{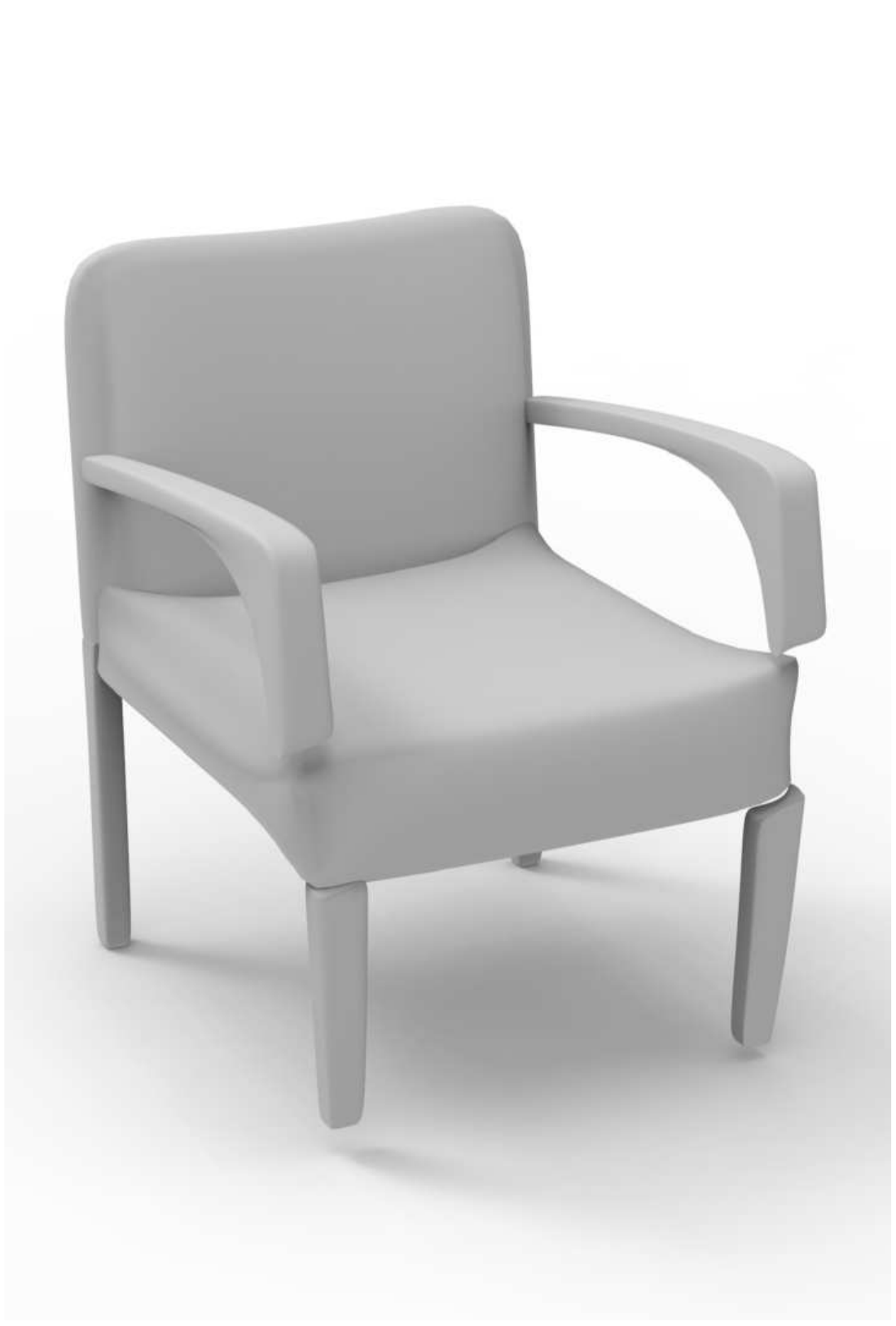}
    \includegraphics[width=0.11\linewidth]{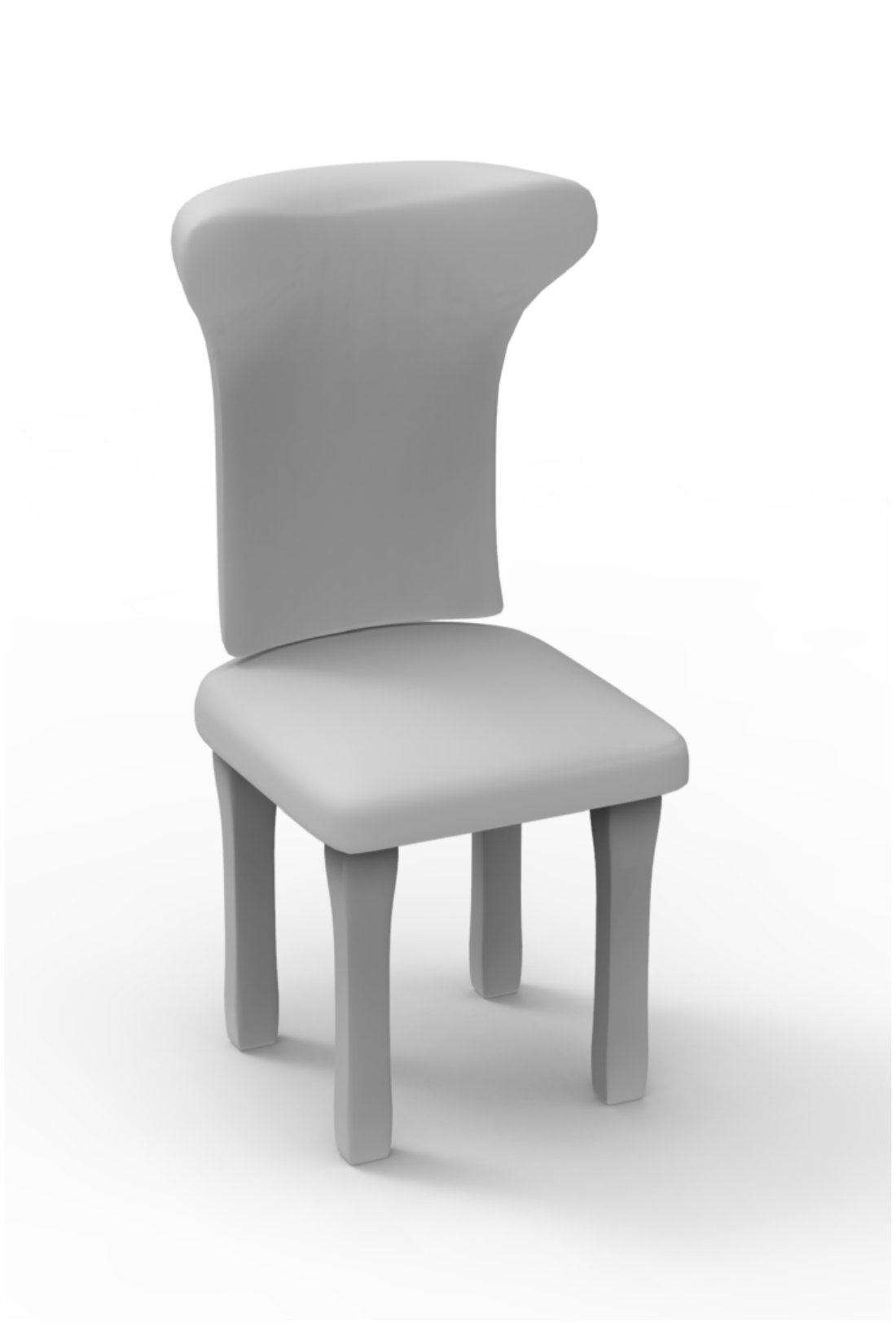}
    }
    \subfigure[StructureNet]{
    \includegraphics[width=0.11\linewidth]{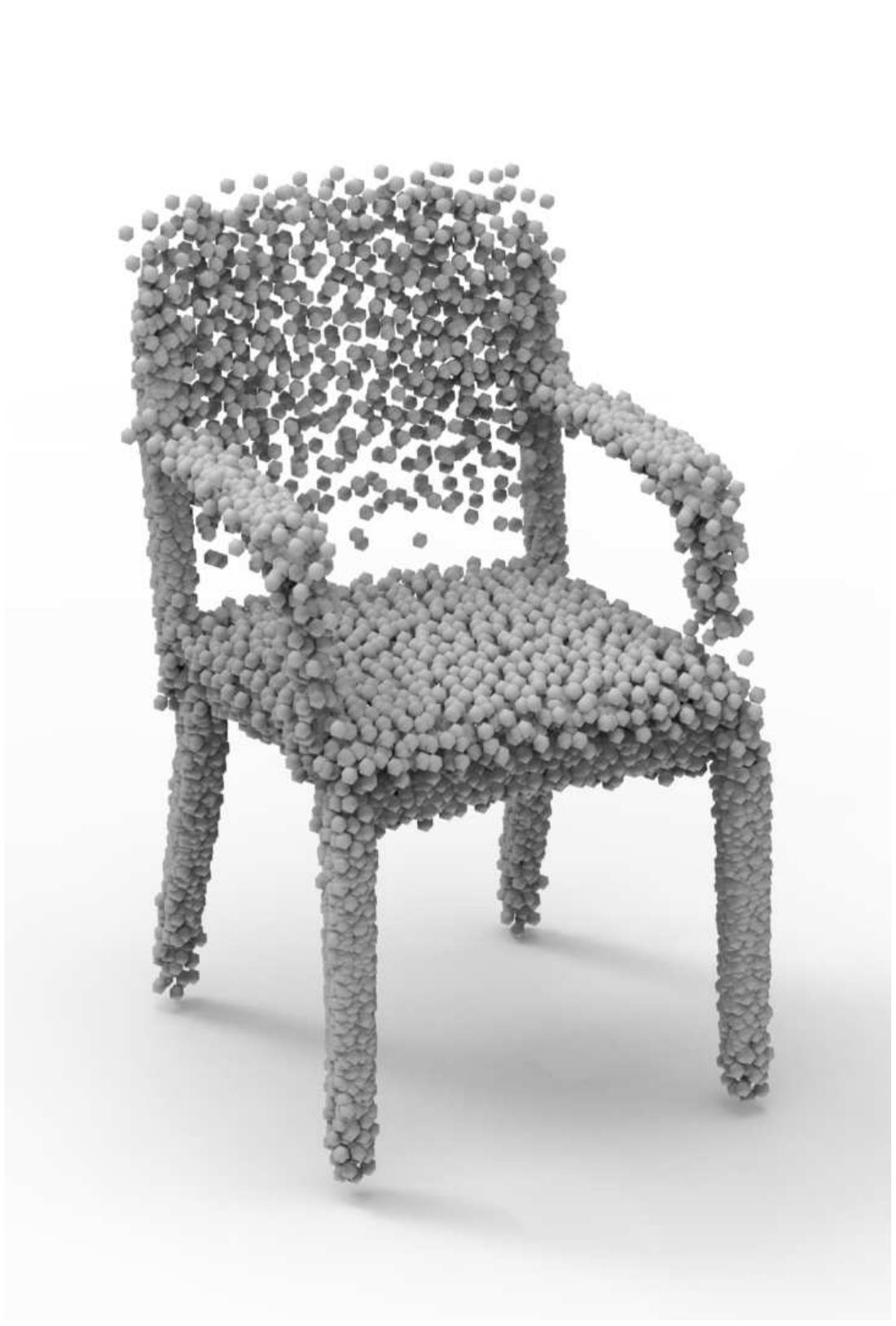}
    \includegraphics[width=0.11\linewidth]{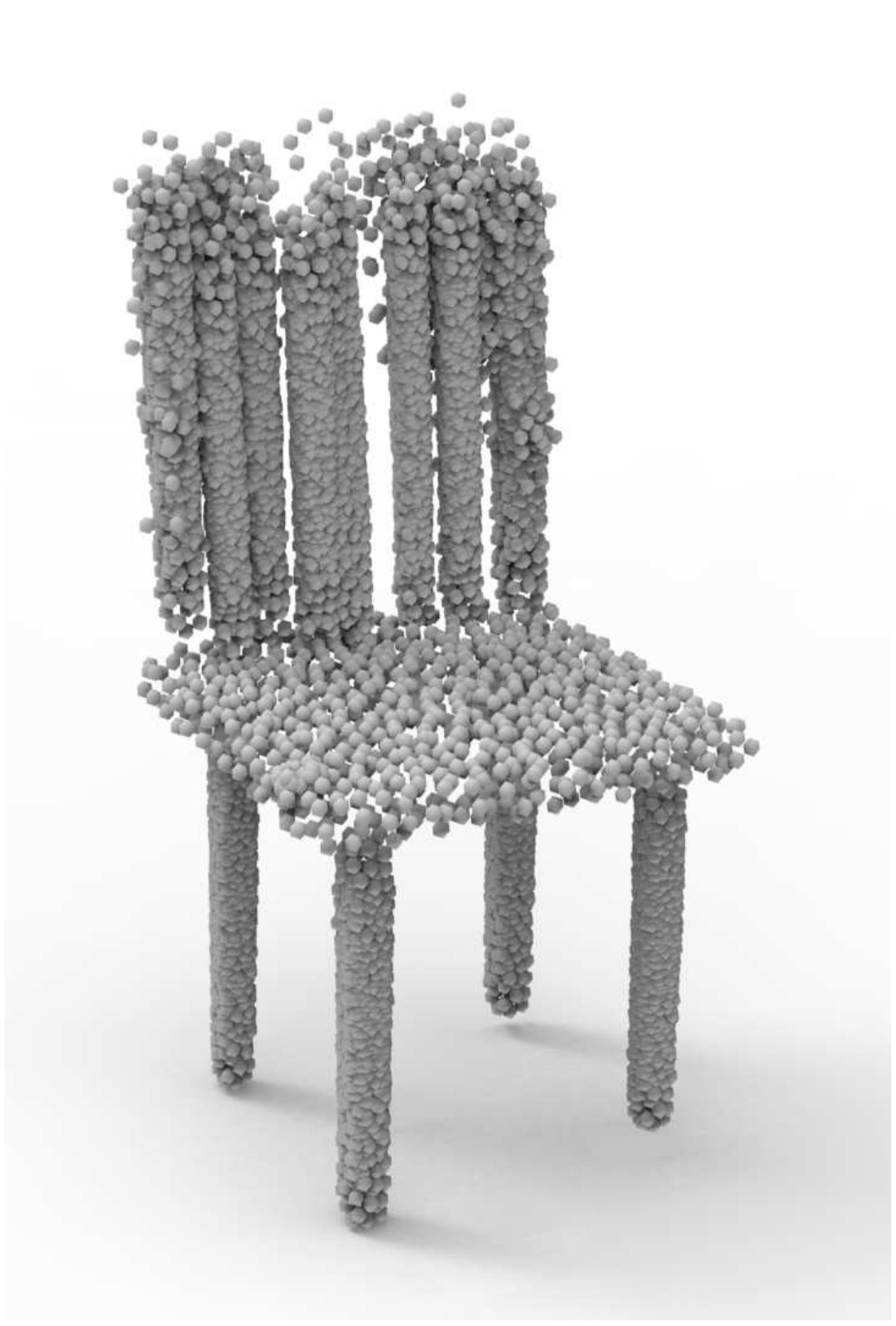}
    }
    \subfigure[SN+Mesh]{
    \includegraphics[width=0.11\linewidth]{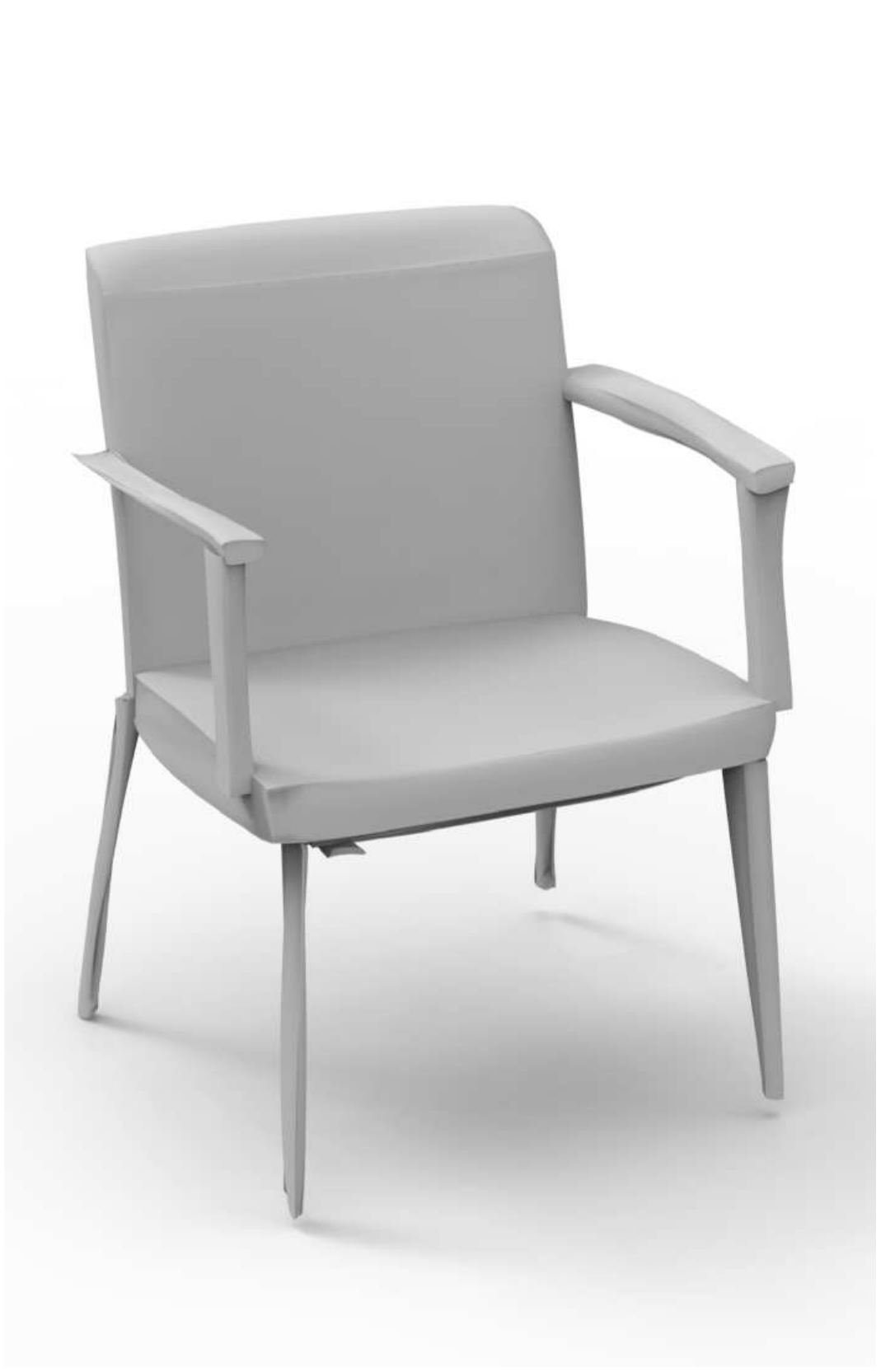}
    \includegraphics[width=0.10\linewidth]{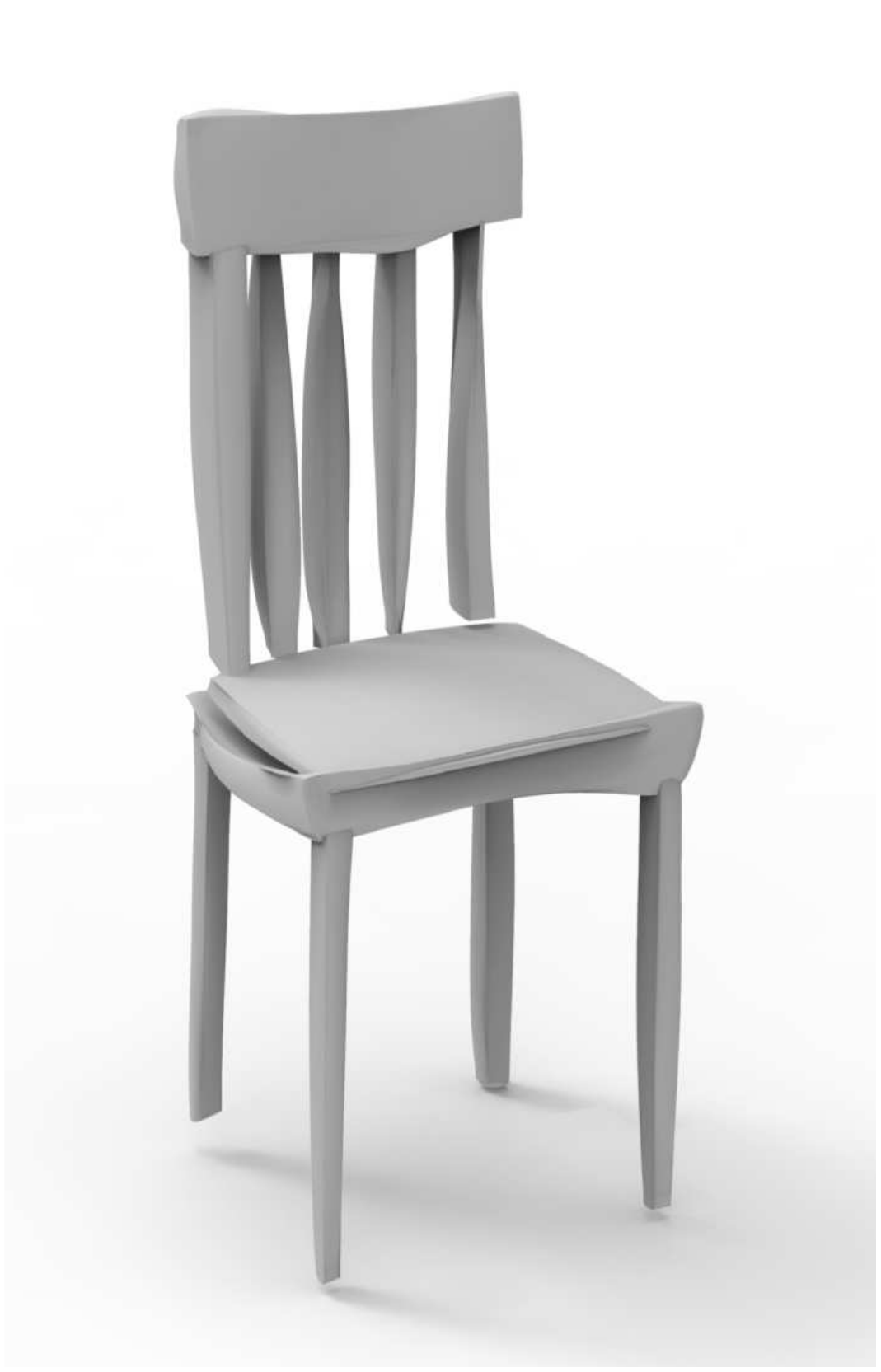}
    }
    \subfigure[Ours]{
    \includegraphics[width=0.11\linewidth]{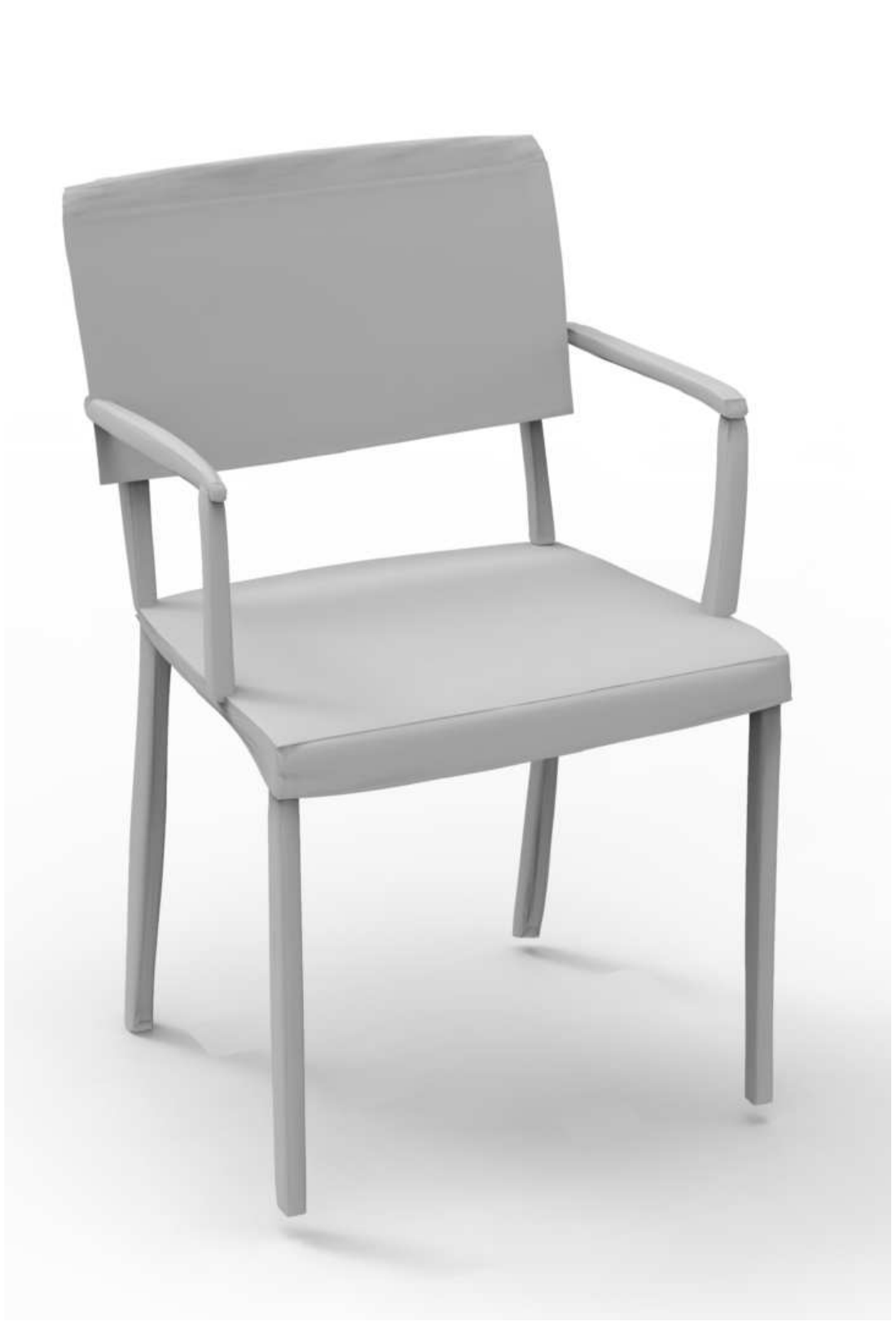}
    \includegraphics[width=0.10\linewidth]{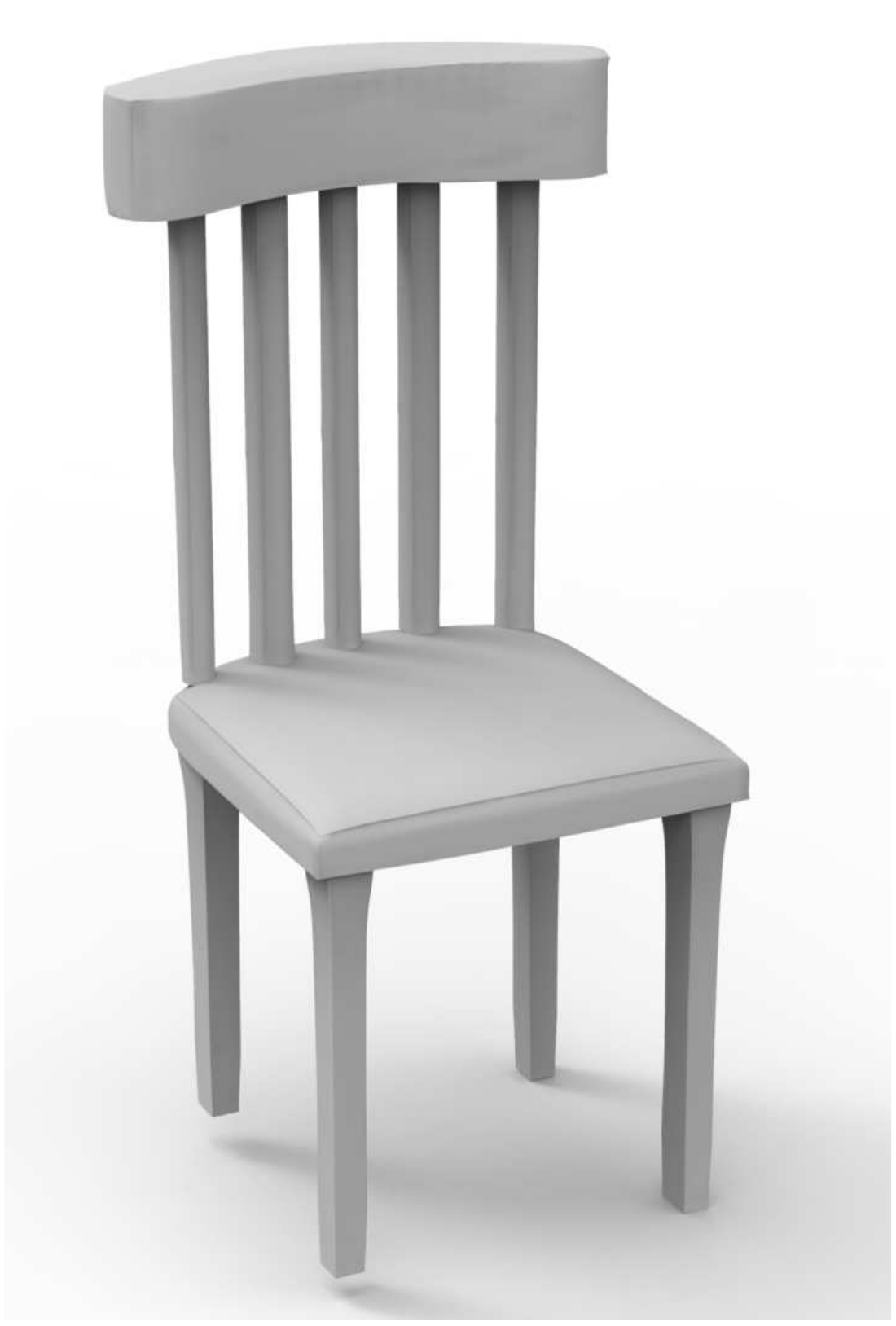}
    }
    \vspace{-3mm}
    \caption{\yjr{Qualitative comparisons on shape generation. We compare our generated shapes to the baseline methods and show that our method learns to generate shapes with complex structures and fine-grained geometry. %
    \yj{For each example, we show shapes generated by alternative methods which are nearest to our results under the CD metric. 
    }
    We can observe that, compared to our DSG-Net, StructureNet fails to generate high-quality shape geometry, SDM-Net cannot generate shapes with complex part structures, and the SN+Mesh is not able to generate shapes with compatible structure relationships.
    }
    }
    \label{fig:comparison_gen}
    \vspace{-3mm}
\end{figure}

\paragraph{Metrics.} The shape generation task aims to generate diverse and realistic shapes with complex structure and geometry.
Following StructureNet~\cite{mo2019structurenet}, we measure shape generation performance by the coverage and quality scores.
The coverage score computes the average distance from a real shape to the closest generated shape, while the quality score calculates the average distance from a generated shape to the closest real shape.
The coverage score reflects if the diversity of the generated results is large enough to cover all real samples, and the quality score measures if the generated results contain bad examples that are far from the real data distribution.
To compare with the baseline methods, we generate 1000 shapes and compute the coverage and quality scores regarding the geometry metric (CD) and the structure metric (HierInsSeg).

\paragraph{Results.} 
\yj{
Figure~\ref{fig:generation} shows our shape generation results, where we observe the complicated structure and fine mesh geometry are generated at the same time.
In Figure~\ref{fig:generation_nn}, we further validate the novelty of the generated shapes by comparing them to the top five retrieved training shapes.
}
In Figure~\ref{fig:comparison_gen}, we compare our method to SDM-Net, StructureNet and our ablated version (SN+Mesh), where we clearly see that DSG-Net generates better shape geometry than StructureNet and produces shapes with more complicated structures than SDM-Net.
We further show quantitative comparisons in Table~\ref{tab:gen_eval}, where we see that DSG-Net obtains clear improvements over the baselines.
In addition, we conduct a user study to further evaluate how realistic the generated shapes are for humans.
We render the shapes into images and ask the users to rank the three algorithms according to three different criteria (geometry, structure and overall). 
In Table~\ref{tab:userstudy}, we observe that our generated shapes perform the best to human users in all the three criteria. 
See supplementary for more results.

\begin{table}[t]
  \centering
  \caption{\yjr{Quantitative comparison on Shape generation. We report the coverage and quality scores relative to DSG-Net (\ie all the reported scores are divided by the corresponding DSG-Net scores for normalization) under the geometry metric (Chamfer-Distance) and the structure metric (HierInsSeg), compared to StructureNet, SDM-Net, and our ablated version (SN+Mesh). 
  \yj{Meanwhile, we also adopt Frech\'{e}t Point-cloud Distance (FPD)~\cite{shu20193d}  to evaluate the variety, coverage, and quality of generated shapes, which can be seen as an extension of \textit{Inception Score}~\cite{salimans2016improved} to point clouds.
  We follow PT2PC~\cite{mo2020pt2pc} to calculate the FPD on point clouds, which is the same as SAG-Net. The lower FPD score, the better.
  } 
  We observe that DSG-Net achieves the best performance across all metrics.}}
  \vspace{-3mm}
  \begin{adjustbox}{width={0.475\textwidth},keepaspectratio}
    \begin{tabular}{cccccc}
    \toprule[1pt]
    \multirow{2}[4]{*}{Method} & \multicolumn{2}{c}{Geometry} & \multicolumn{2}{c}{Structure} & \multicolumn{1}{c}{\multirow{2}[4]{*}{FPD $\downarrow$}\vspace{1mm}} \\
    \cmidrule{2-5}          & \multicolumn{1}{c}{Coverage $\uparrow$} & \multicolumn{1}{c}{Quality $\uparrow$} & \multicolumn{1}{c}{Coverage $\uparrow$} & \multicolumn{1}{c}{Quality $\uparrow$} & \\
    \midrule
    \midrule
    SDM-Net & 0.59 & 0.23 & 0.42 & 0.48 & 18.20 \\
    StructureNet & 0.70 & 0.77 & 0.76 & 0.98 & 12.57 \\
    SN+Mesh & 0.80 & 0.93 & 0.78 & 0.93 & 10.94 \\
    Ours    & \textbf{1.00} & \textbf{1.00} & \textbf{1.00} & \textbf{1.00} & \textbf{9.73} \\
    \bottomrule[1pt]
    \end{tabular}%
    \end{adjustbox}
  \label{tab:gen_eval}%
  \vspace{-3mm}
\end{table}%

\begin{table}[t]
  \centering
  \caption{\yjr{User study results on shape generation. We show the average ranking scores of the four methods: SDM-Net, StructureNet, SN+Mesh, and ours. The ranking ranges from 1 (the best) to 4 (the worst). The results are calculated based on 119 trials. We see that our method achieves the best on all metrics.}}
  \vspace{-3mm}
    \begin{tabular}{cccc}
    \toprule[1pt]
    Method & \multicolumn{1}{c}{Structure} & \multicolumn{1}{c}{Geometry} & \multicolumn{1}{c}{Overall} \\
    \midrule
    SDM-Net & 3.56 & 2.53 & 2.44 \\
    StructureNet & 2.64 & 3.73 & 3.65 \\
    SN+Mesh & 2.05 & 2.04 & 2.11 \\
    Ours  & \textbf{1.75} & \textbf{1.70} & \textbf{1.80} \\
    \bottomrule[1pt]
    \end{tabular}%
  \label{tab:userstudy}%
  \vspace{-3mm}
\end{table}%

\begin{figure}[t]
\centering

\begin{minipage}{0.21\linewidth}
\centering
\includegraphics[width=0.99\linewidth]{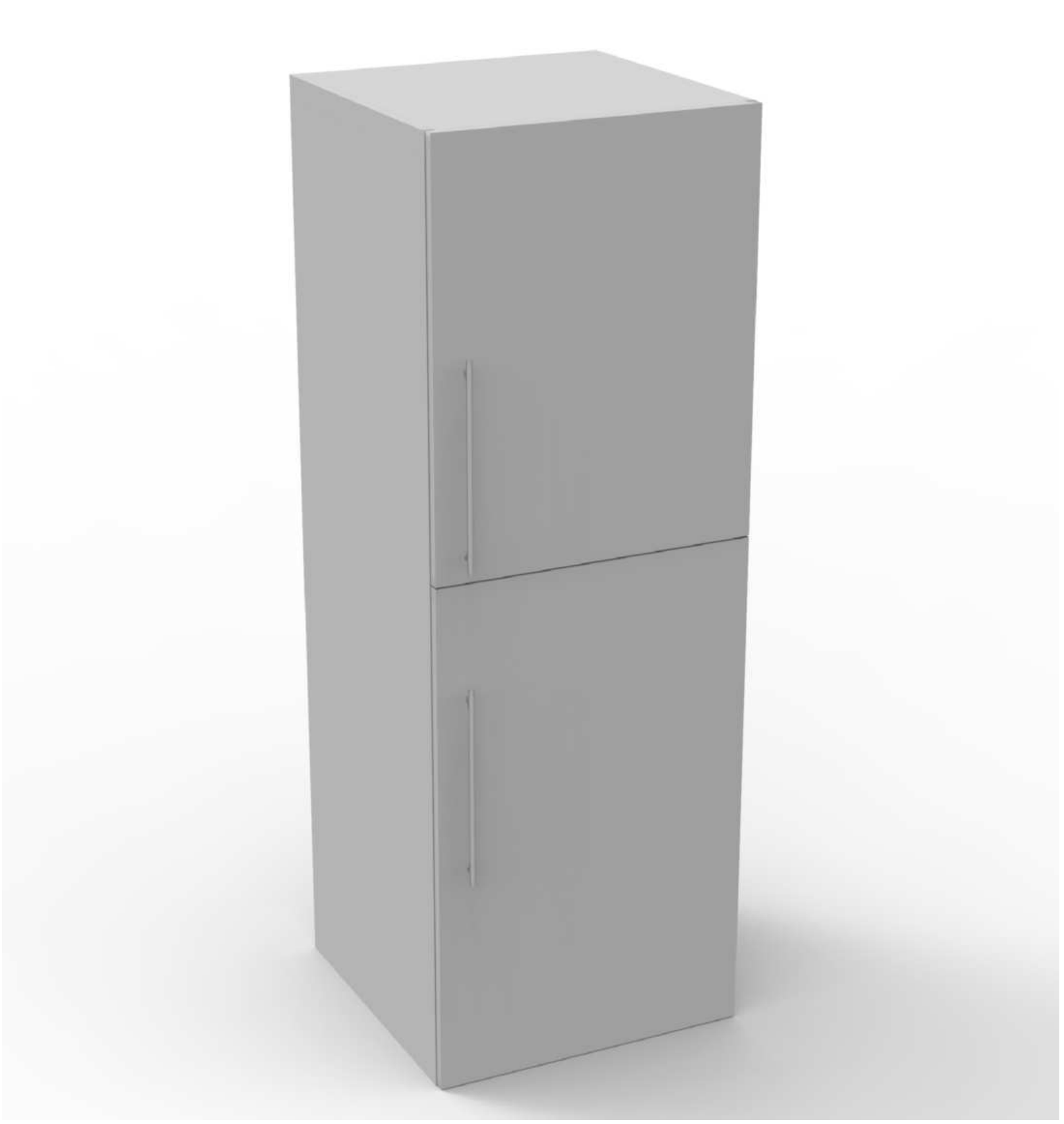}
\end{minipage}
\begin{minipage}{0.71\linewidth}
\centering
    \includegraphics[width=0.23\linewidth]{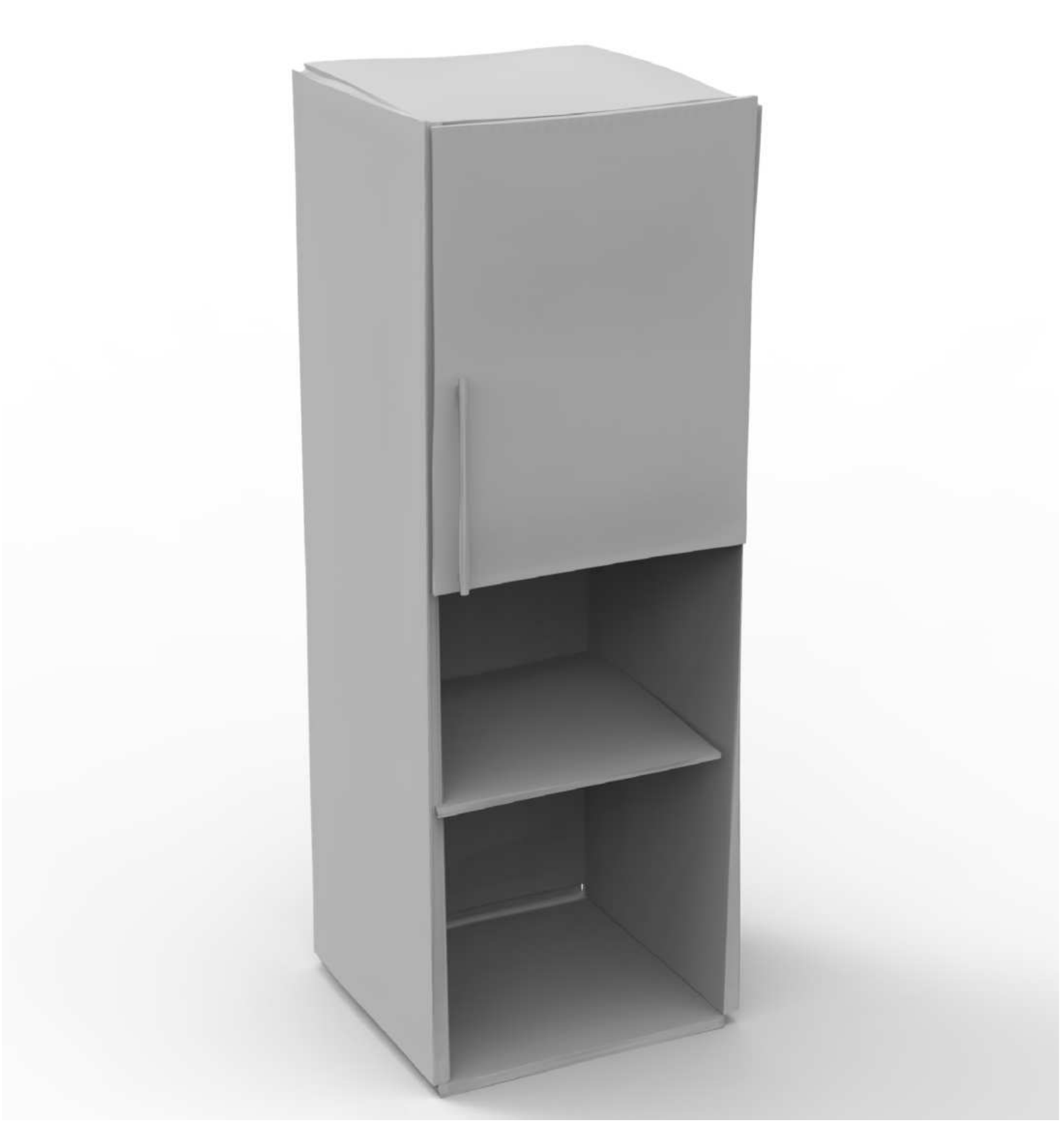}
    \includegraphics[width=0.23\linewidth]{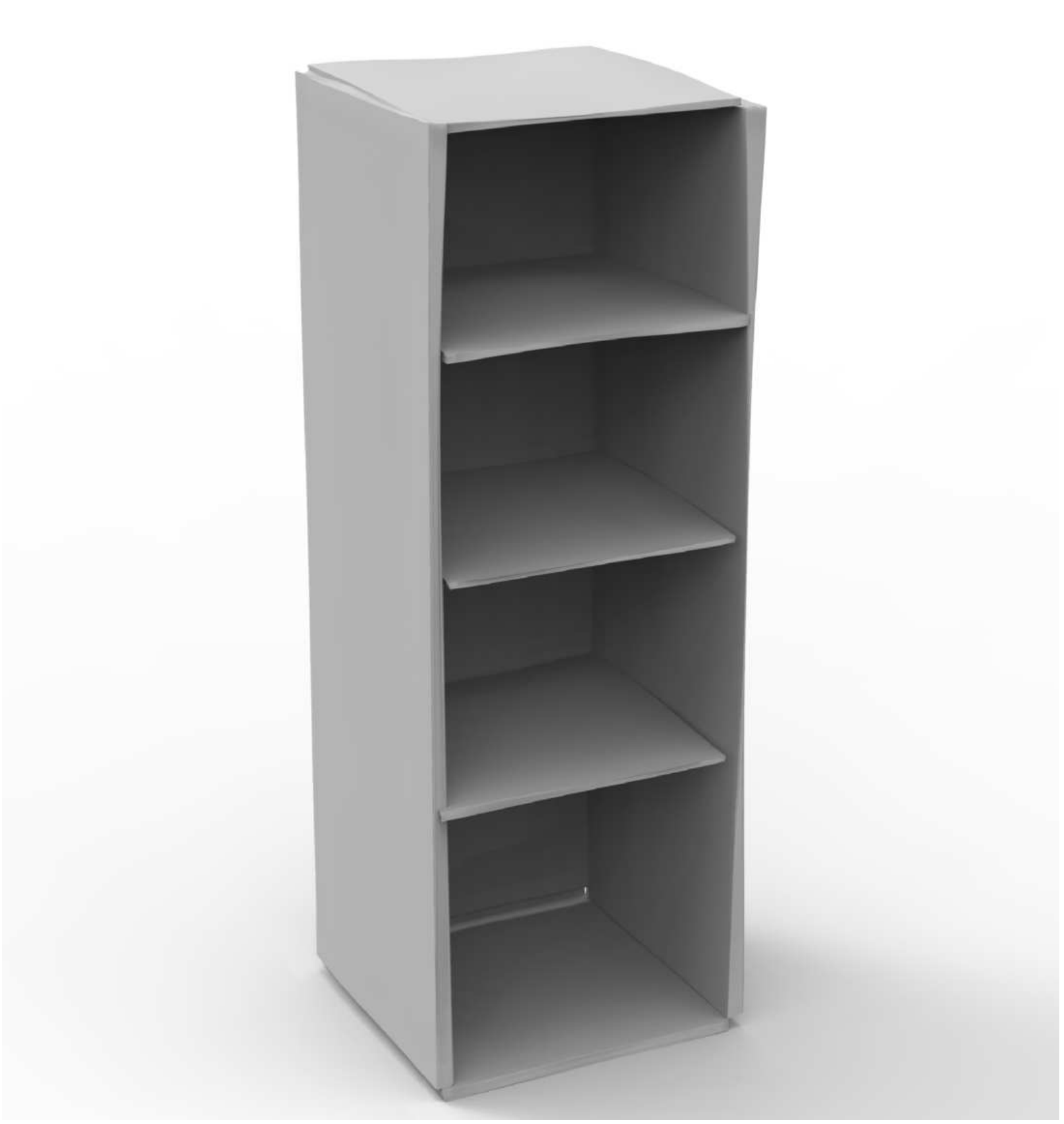}
    \includegraphics[width=0.23\linewidth]{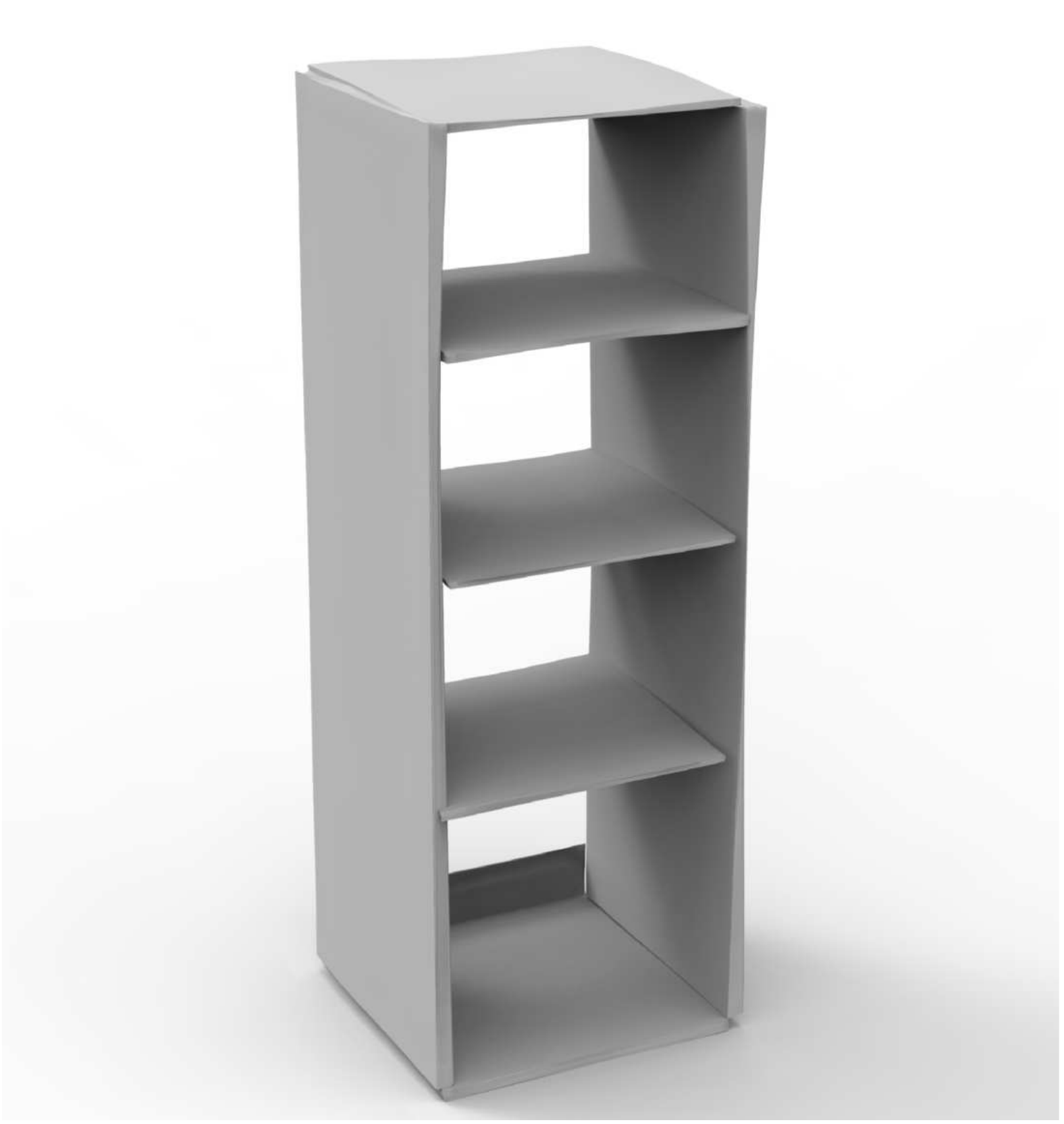}
    \includegraphics[width=0.23\linewidth]{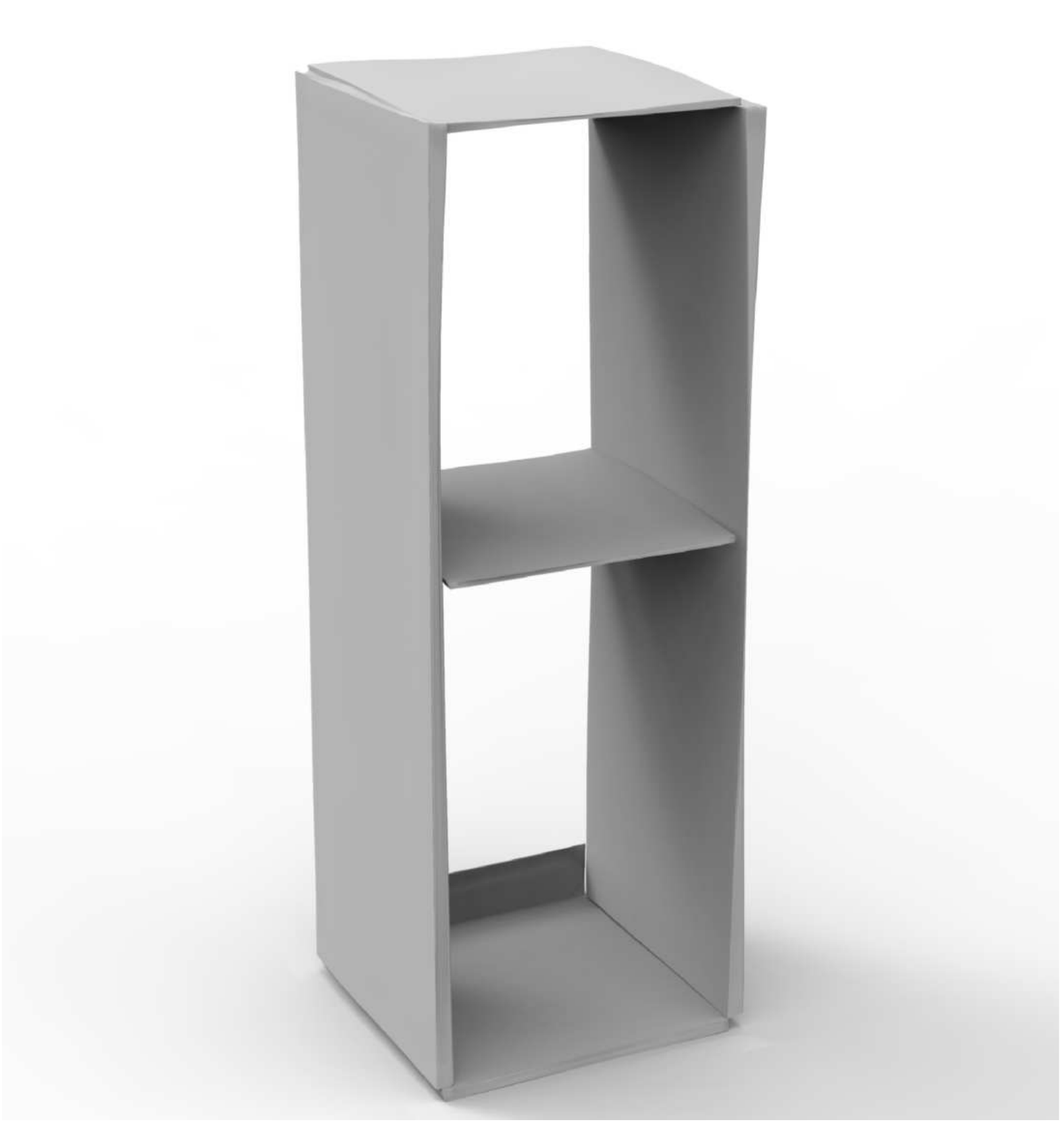}
    \\
    \includegraphics[width=0.23\linewidth]{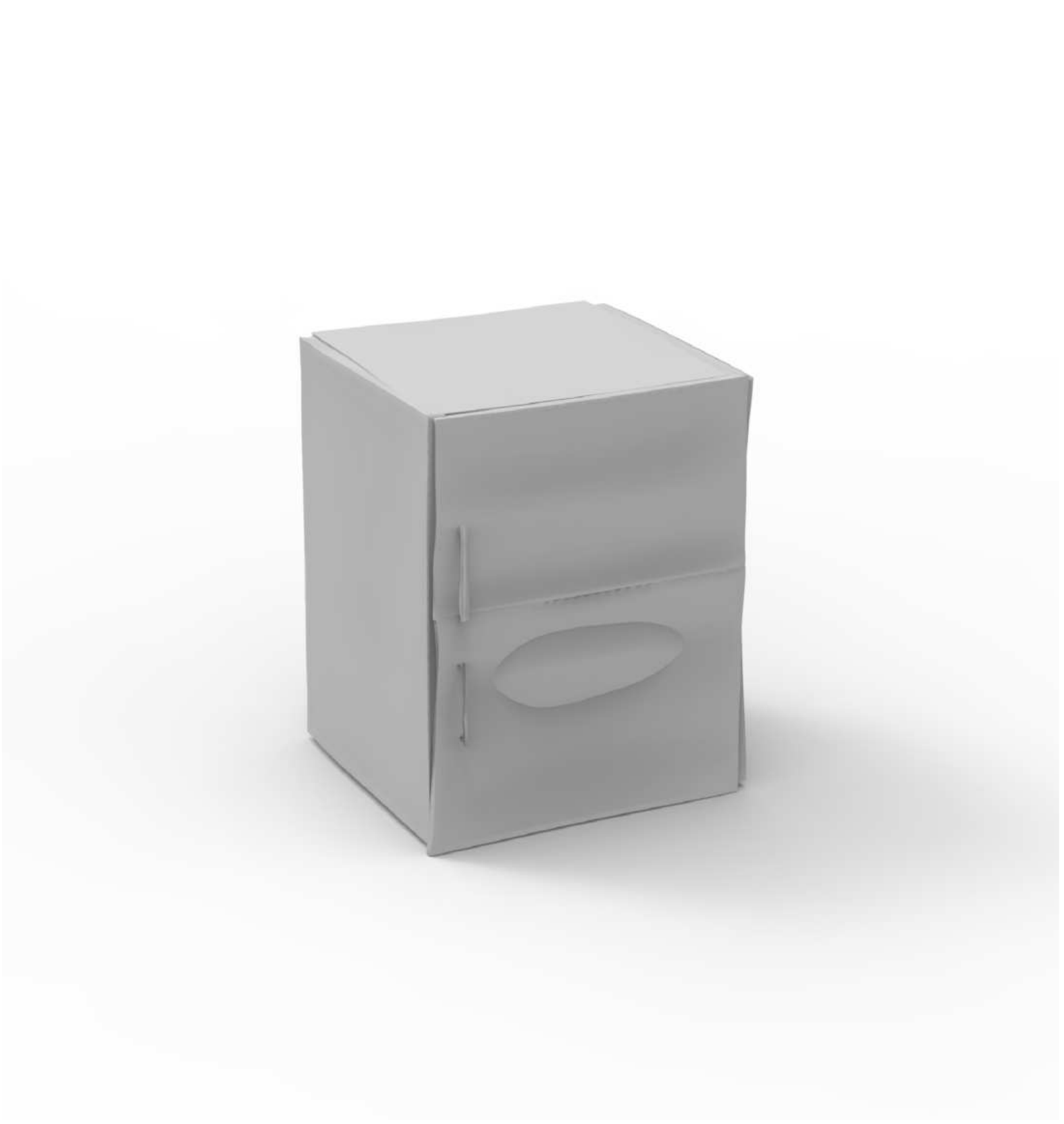}
    \includegraphics[width=0.23\linewidth]{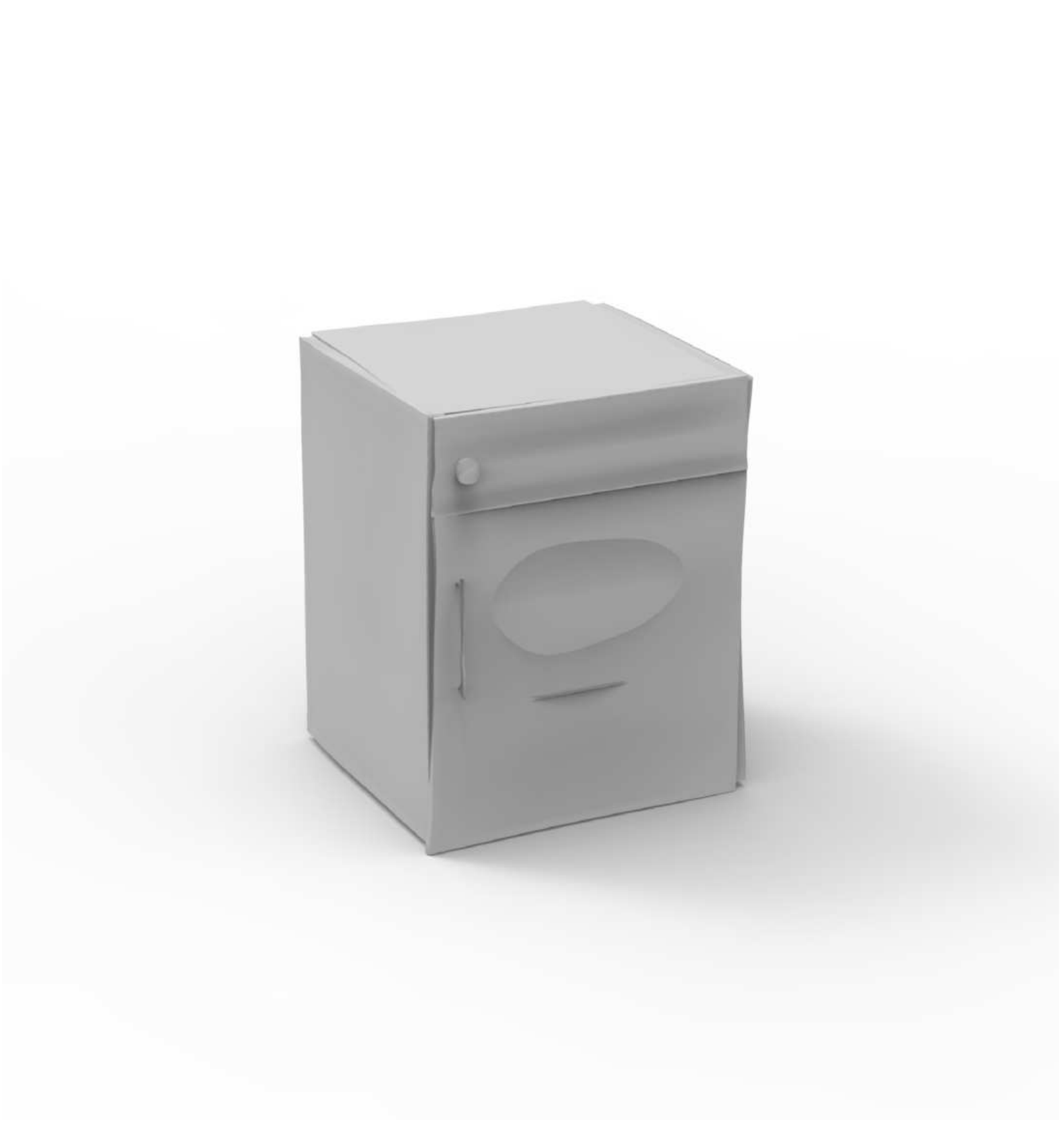}
    \includegraphics[width=0.23\linewidth]{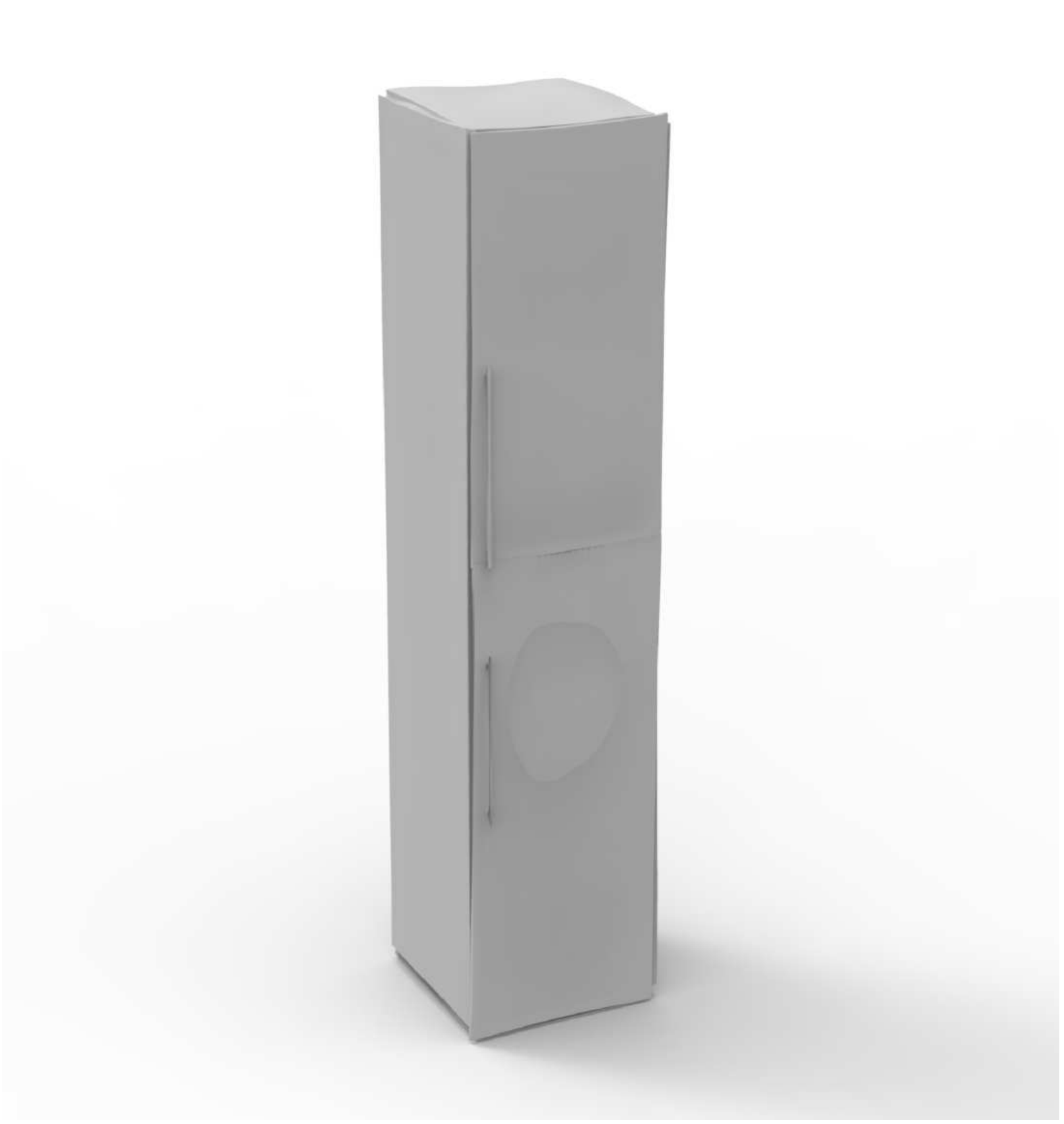}
    \includegraphics[width=0.23\linewidth]{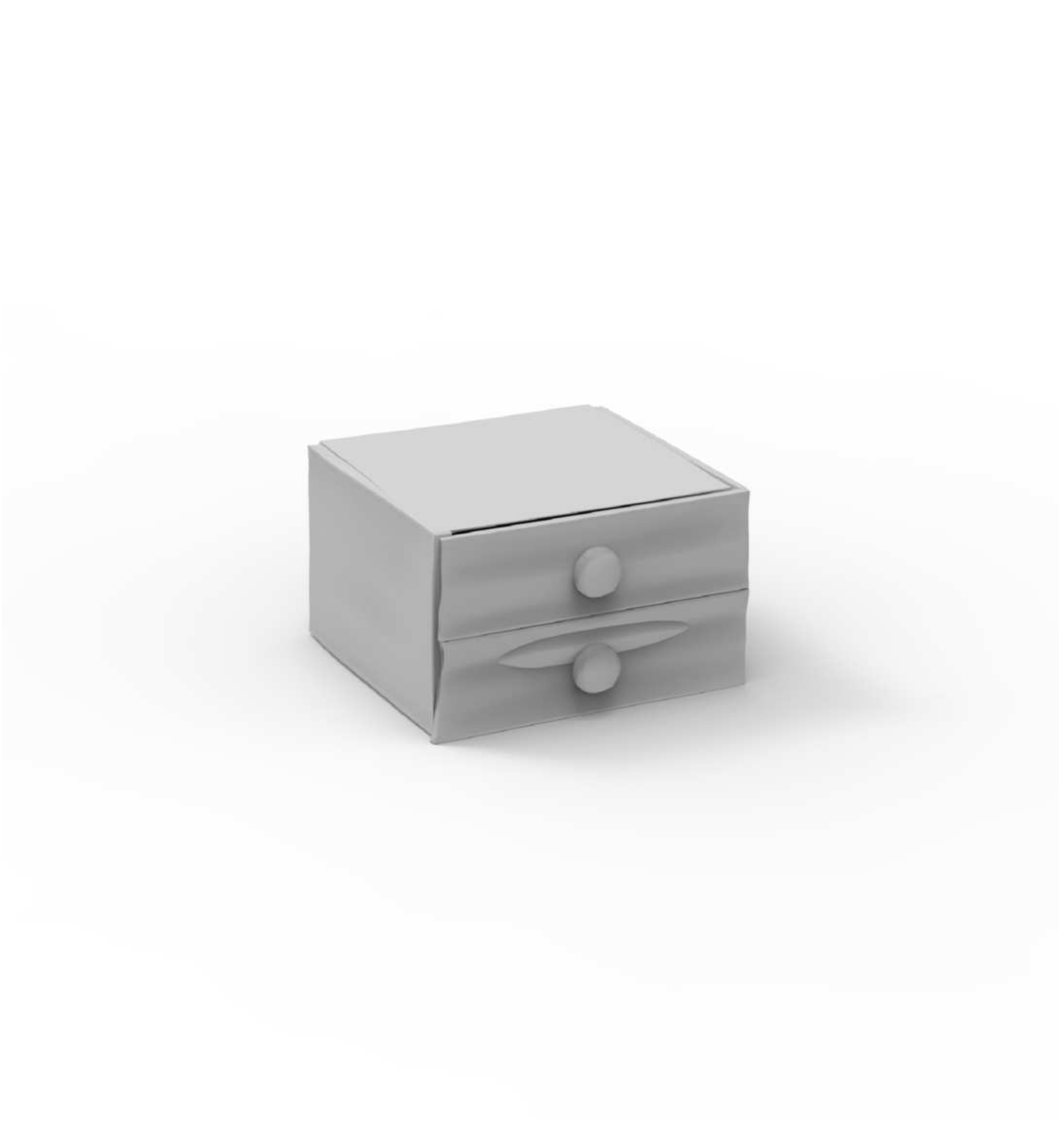}
\end{minipage}

\begin{minipage}{0.21\linewidth}
\centering
\includegraphics[width=0.99\linewidth]{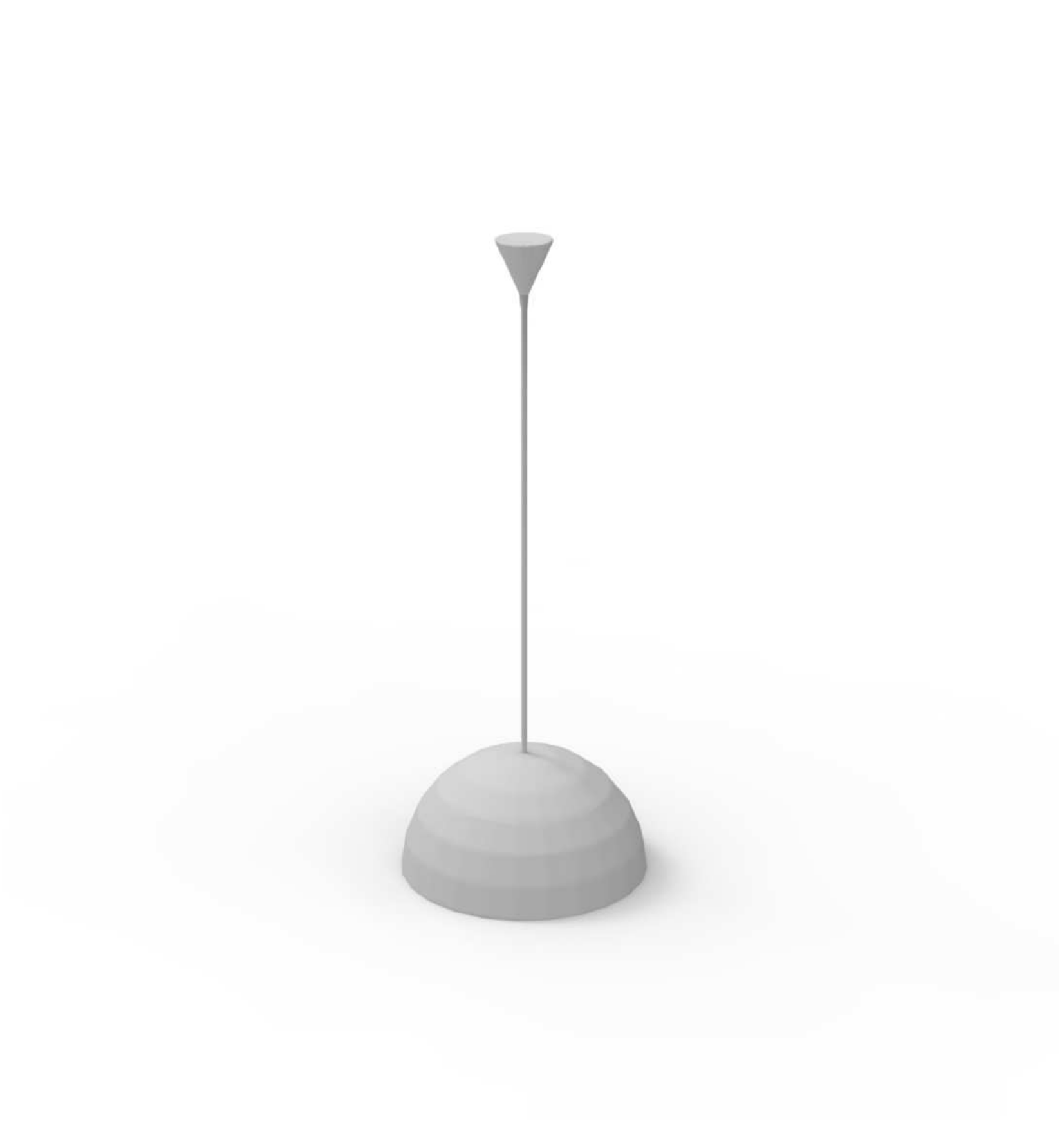}
\end{minipage}
\begin{minipage}{0.71\linewidth}
\centering
    \includegraphics[width=0.23\linewidth]{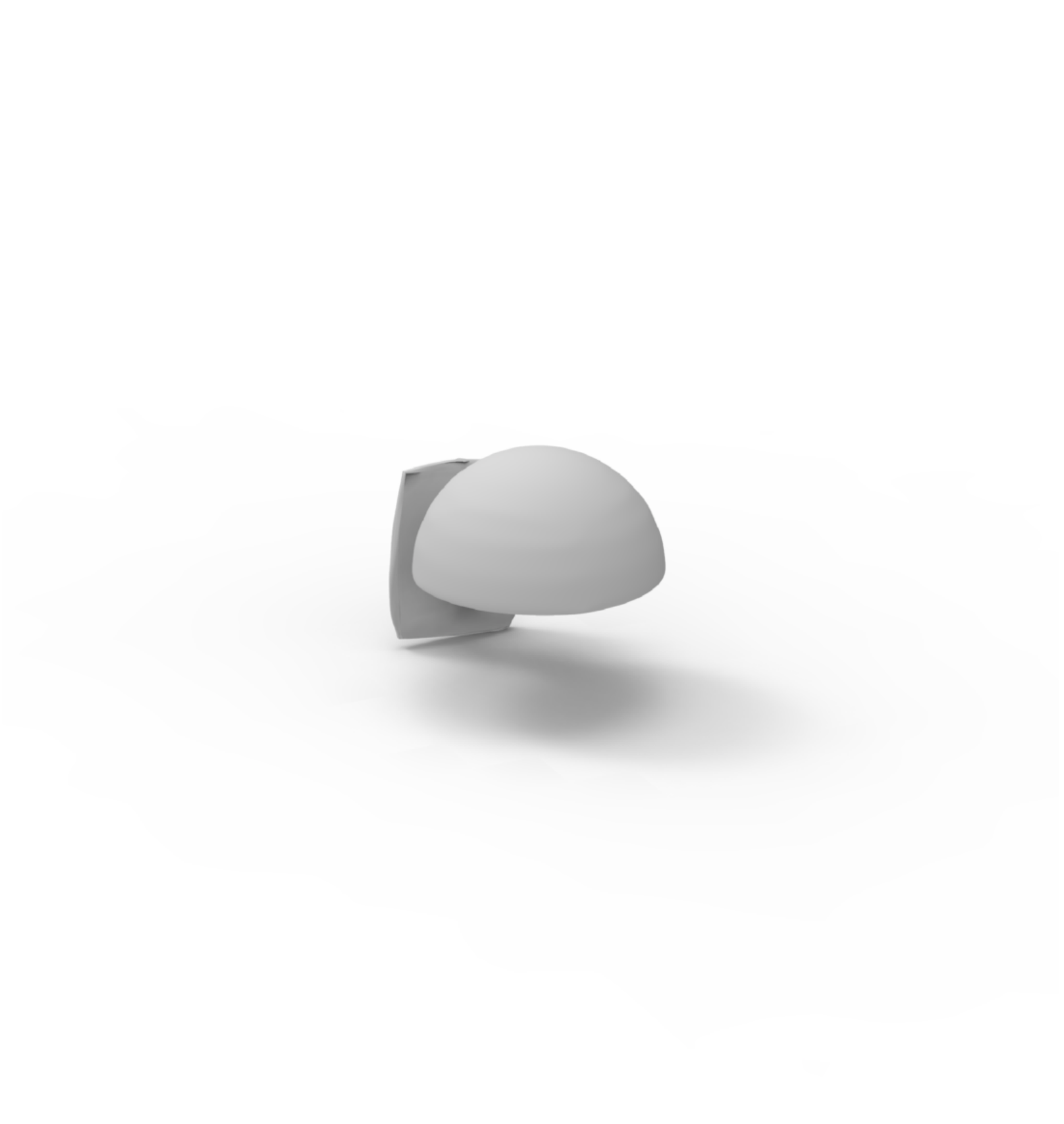}
    \includegraphics[width=0.23\linewidth]{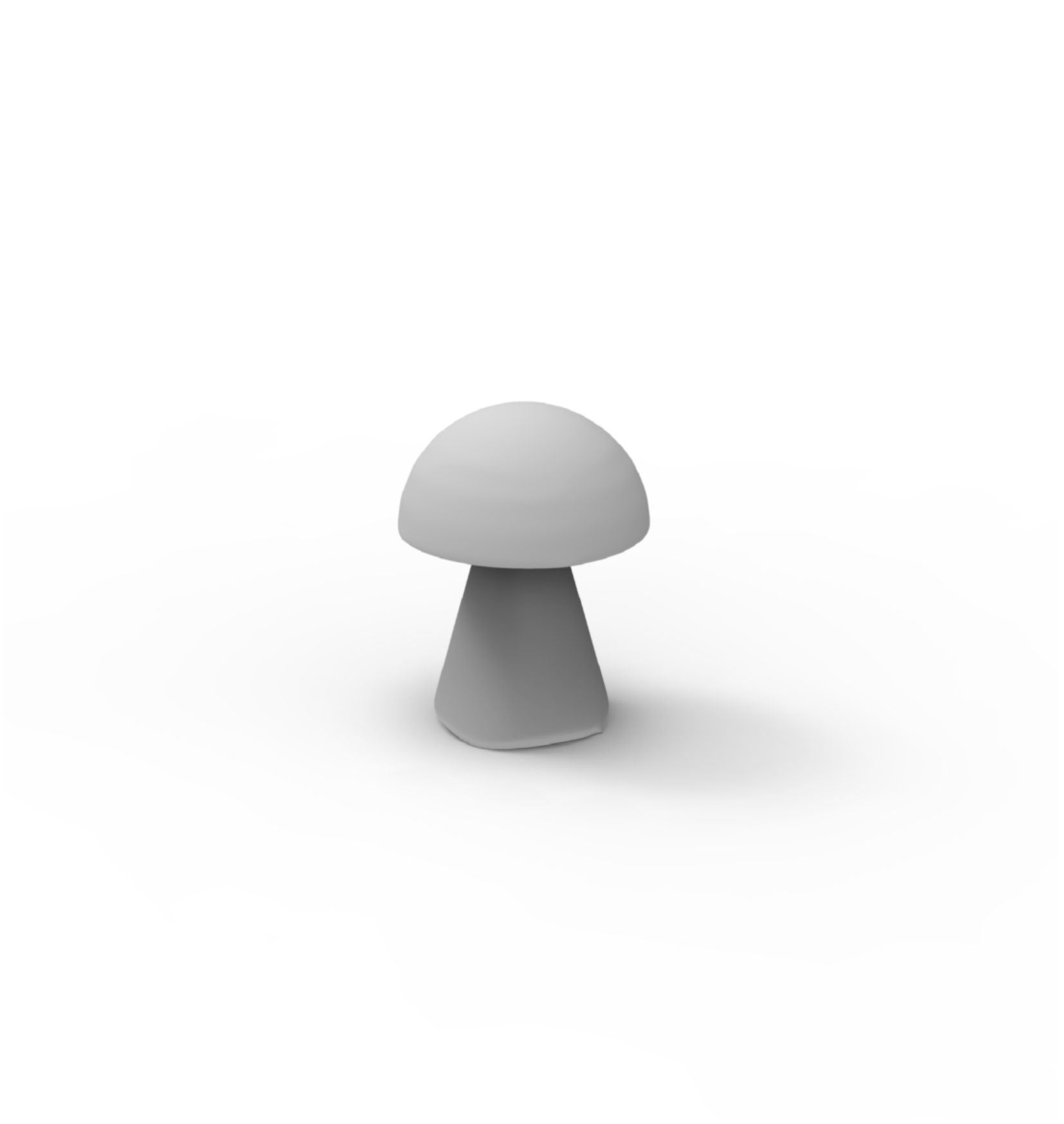}
    \includegraphics[width=0.23\linewidth]{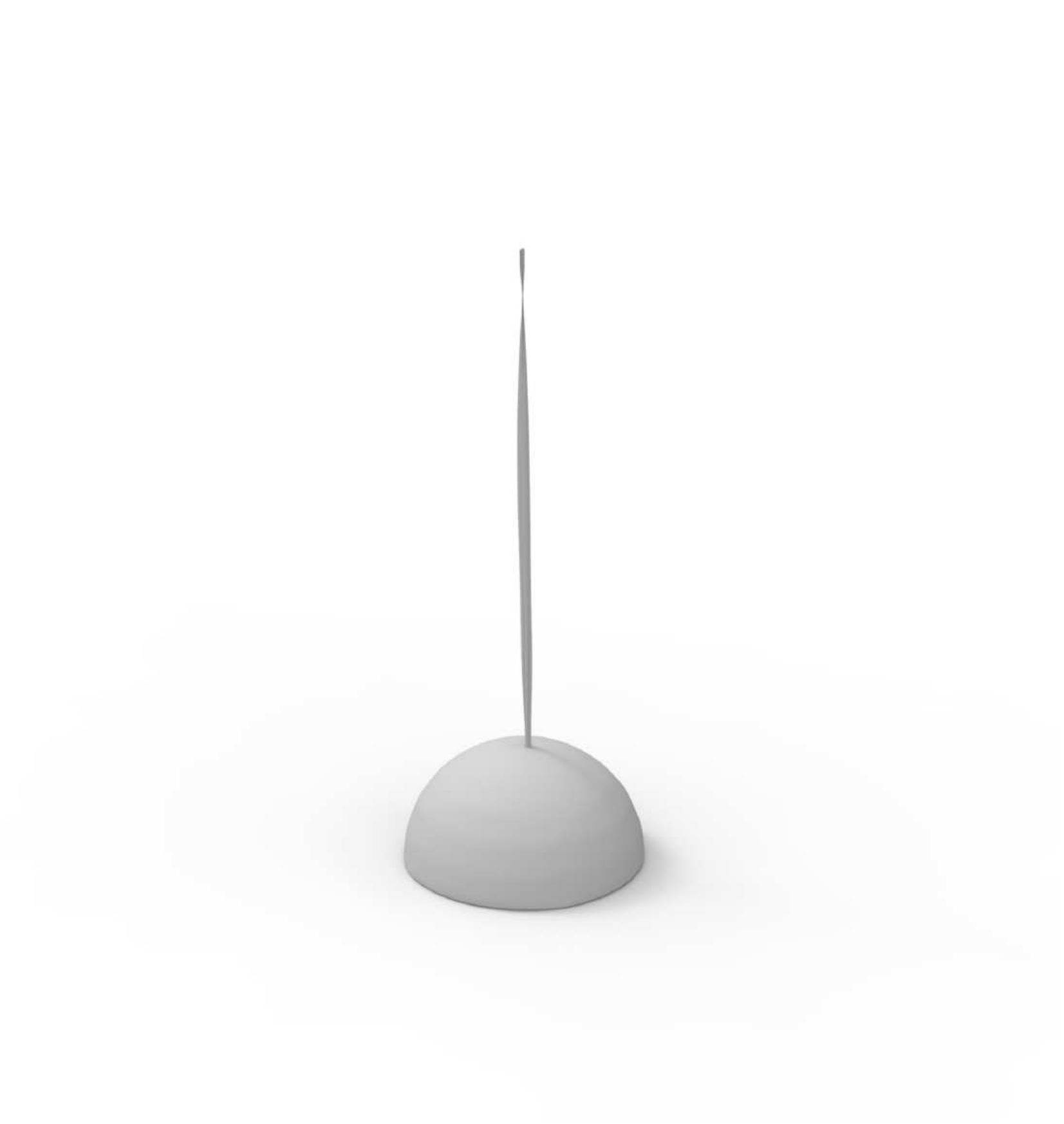}
    \includegraphics[width=0.23\linewidth]{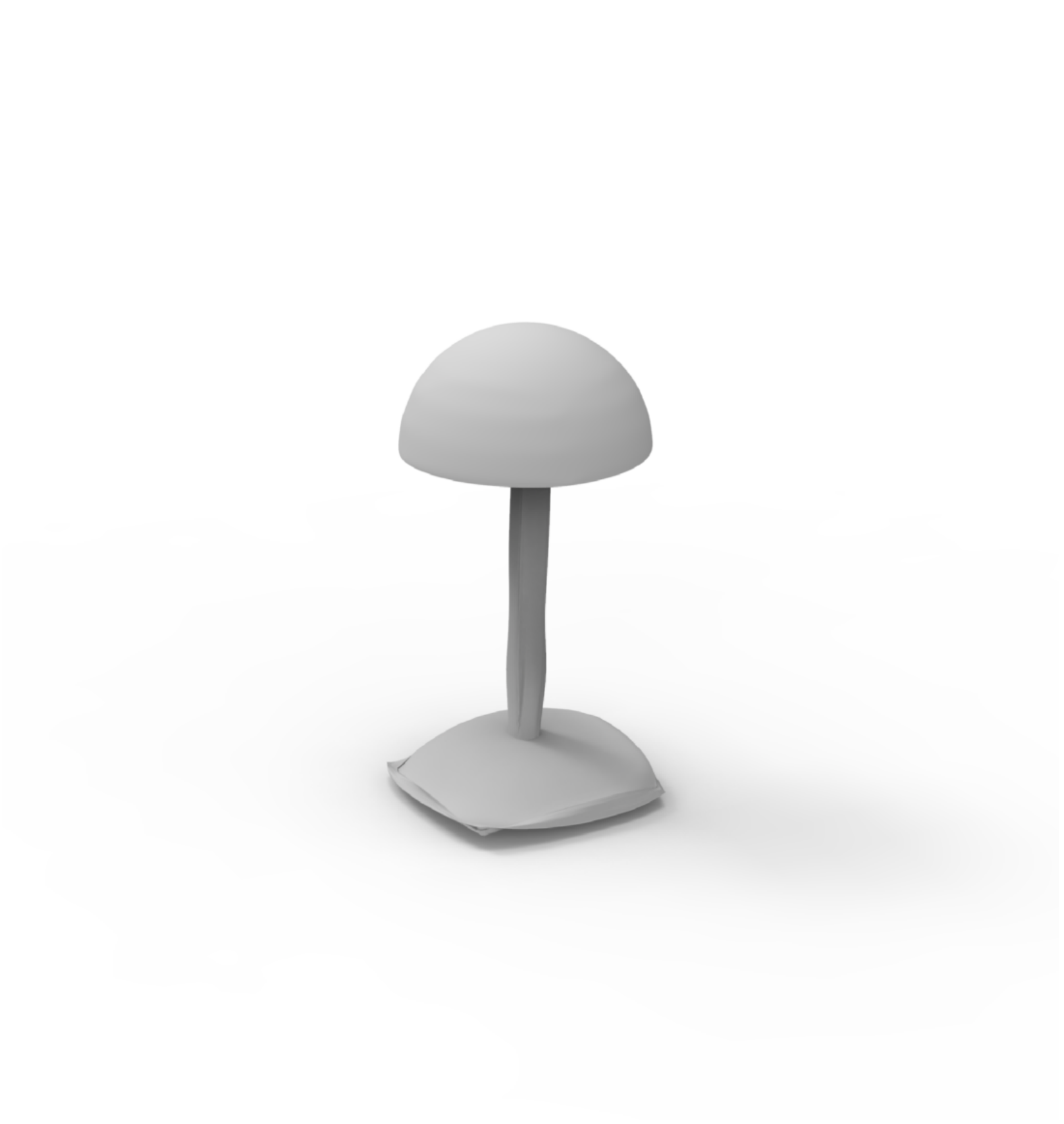}
    \\
    \includegraphics[width=0.23\linewidth]{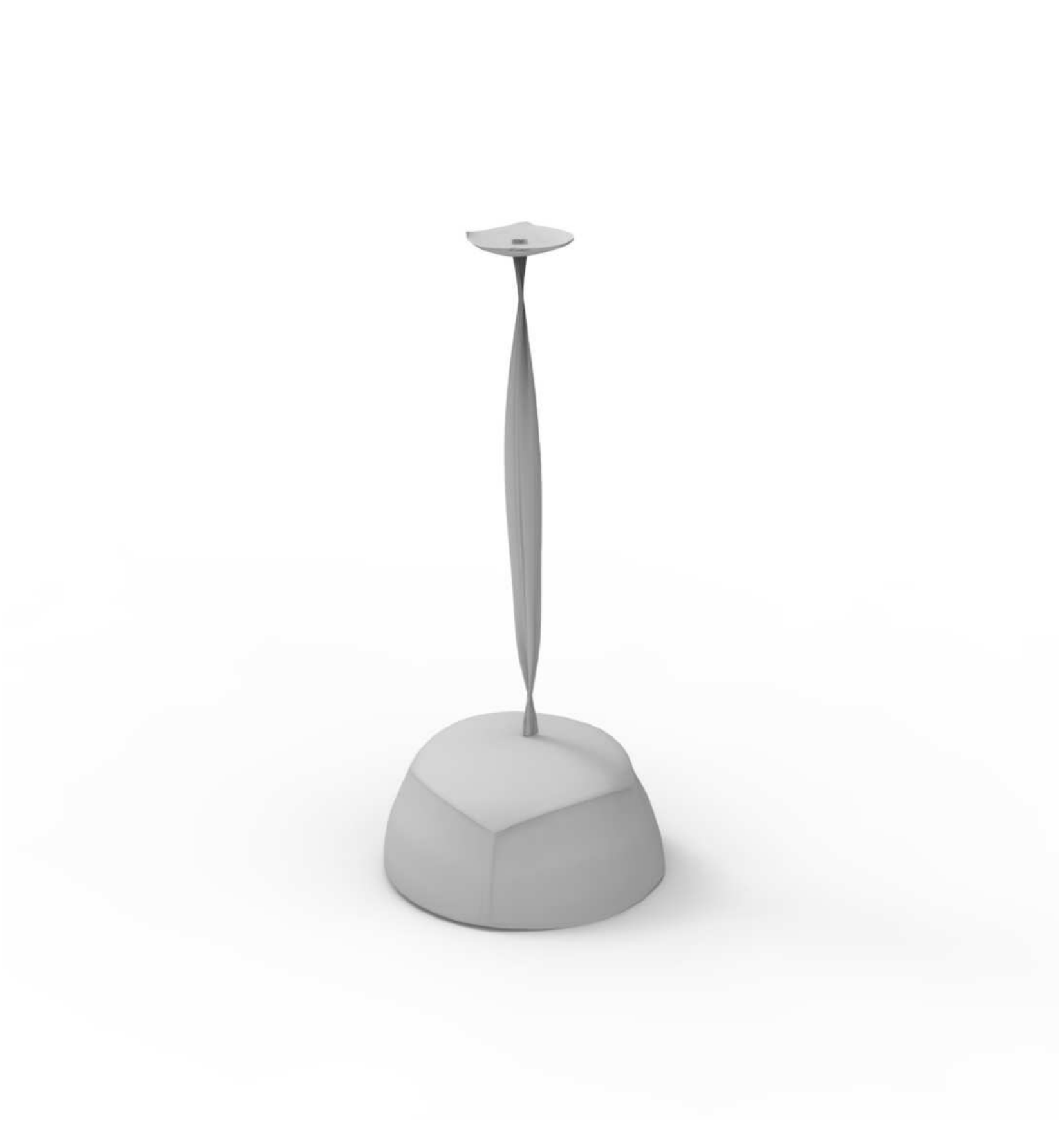}
    \includegraphics[width=0.23\linewidth]{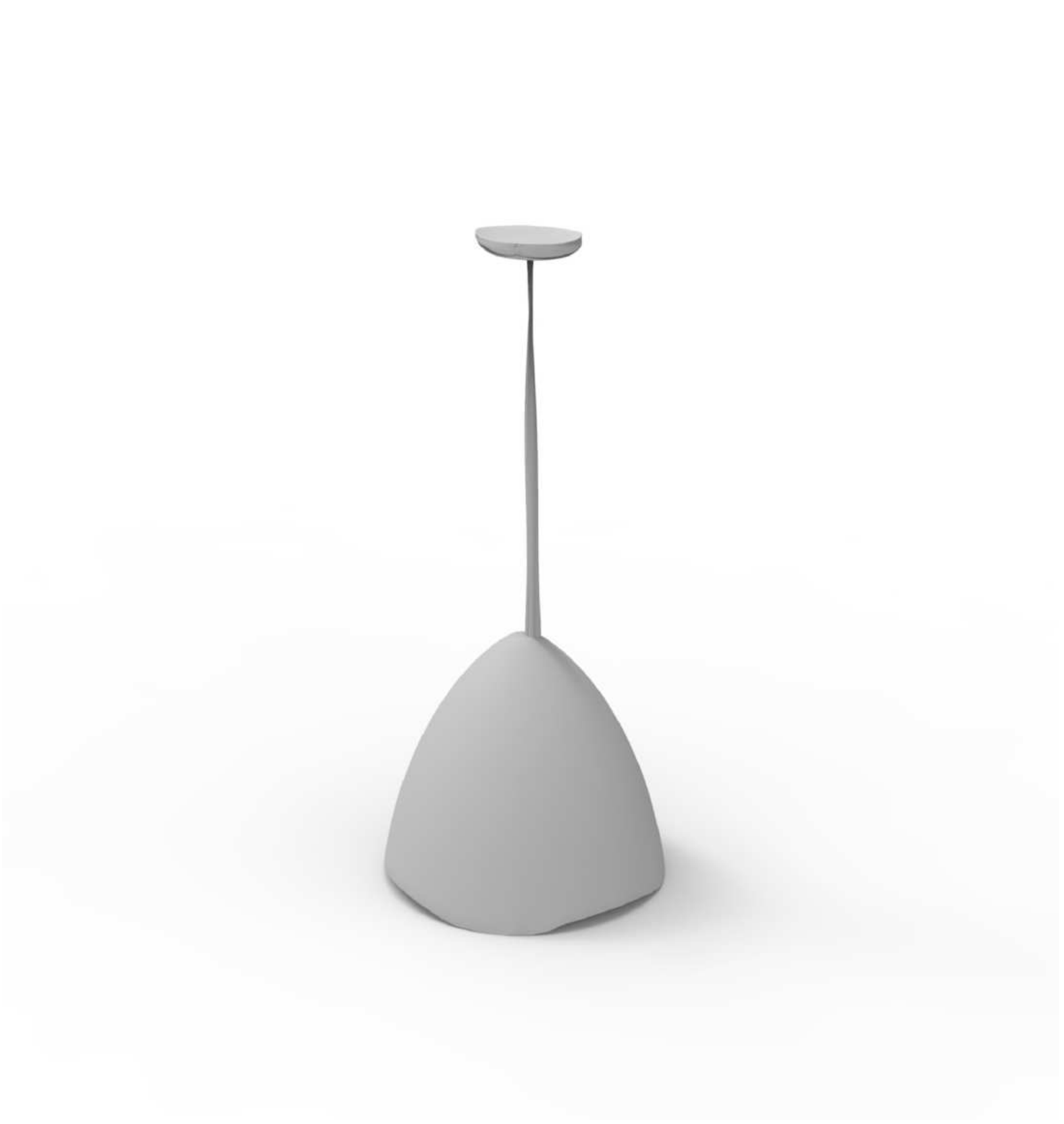}
    \includegraphics[width=0.23\linewidth]{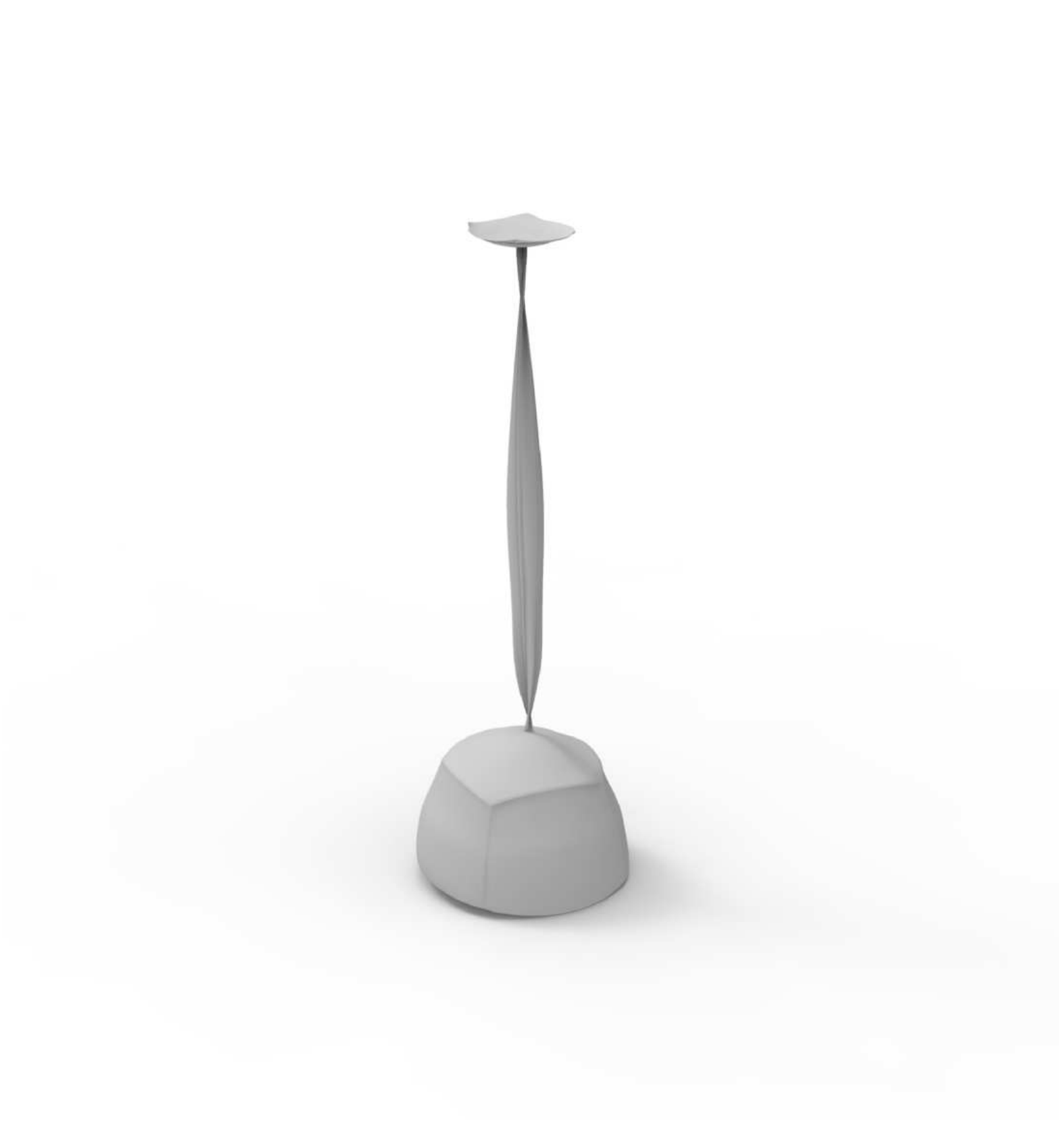}
    \includegraphics[width=0.23\linewidth]{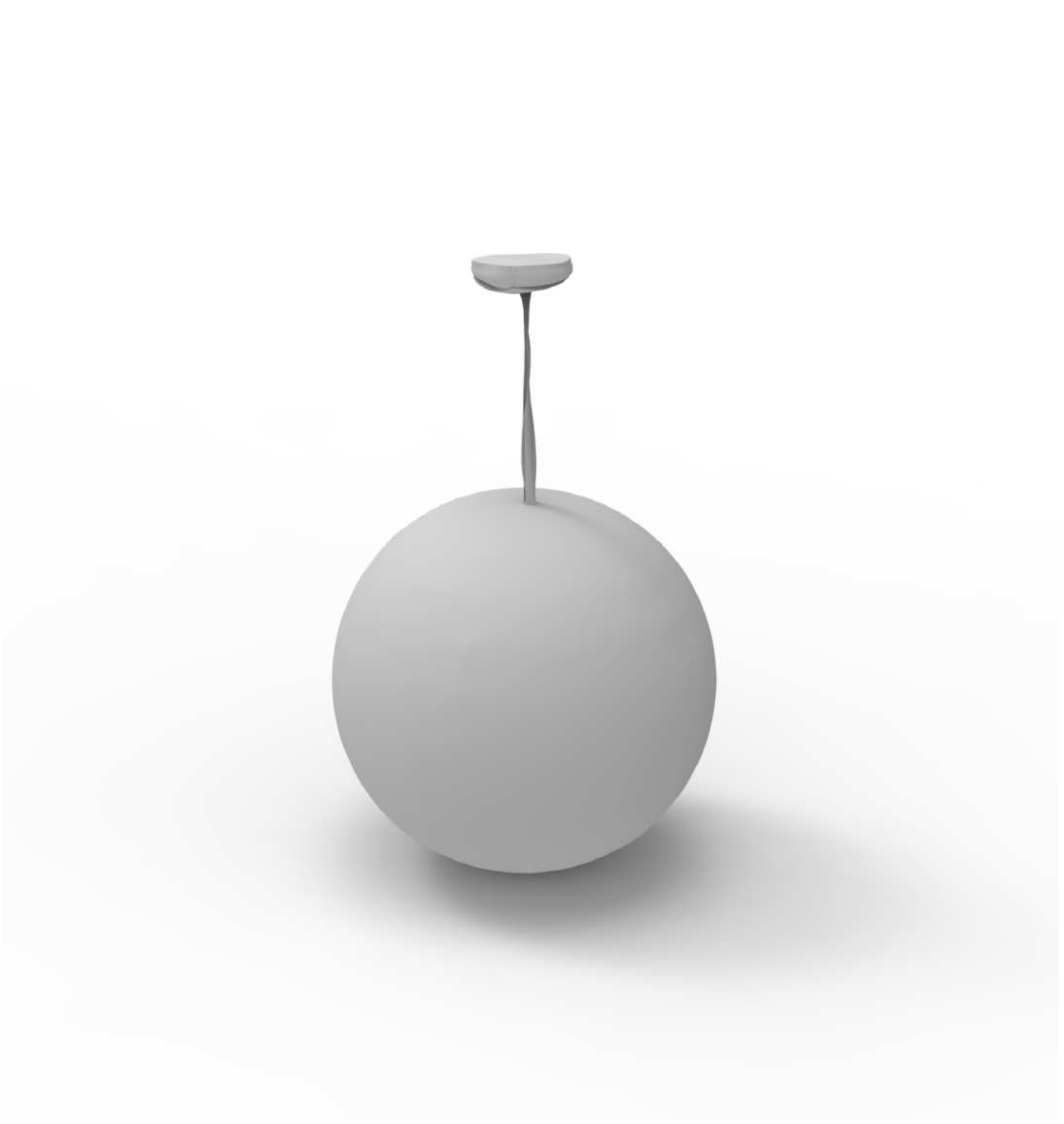}
\end{minipage}

\begin{minipage}{0.21\linewidth}
\centering
\includegraphics[width=0.99\linewidth]{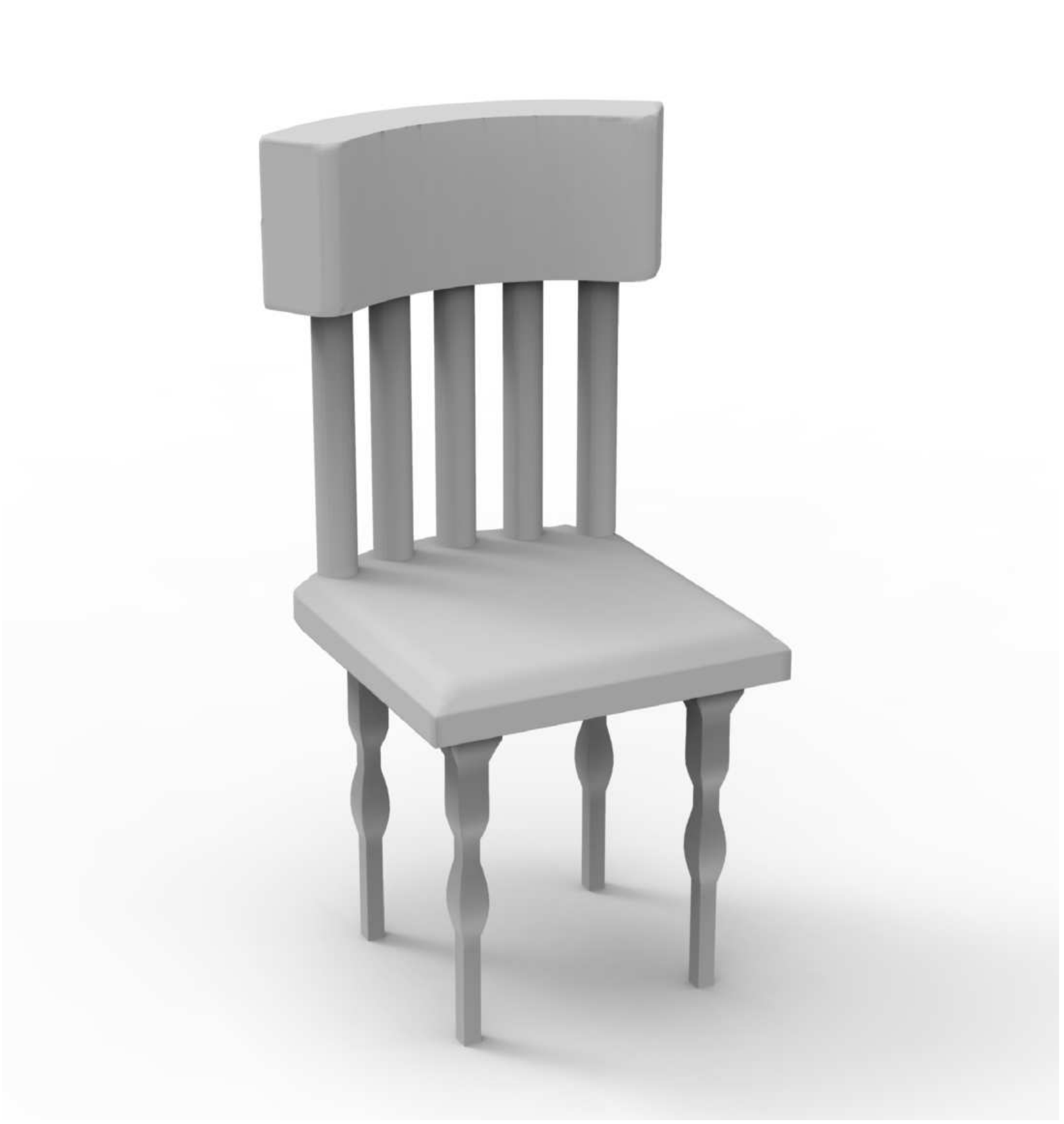}
\end{minipage}
\begin{minipage}{0.71\linewidth}
\centering
    \includegraphics[width=0.23\linewidth]{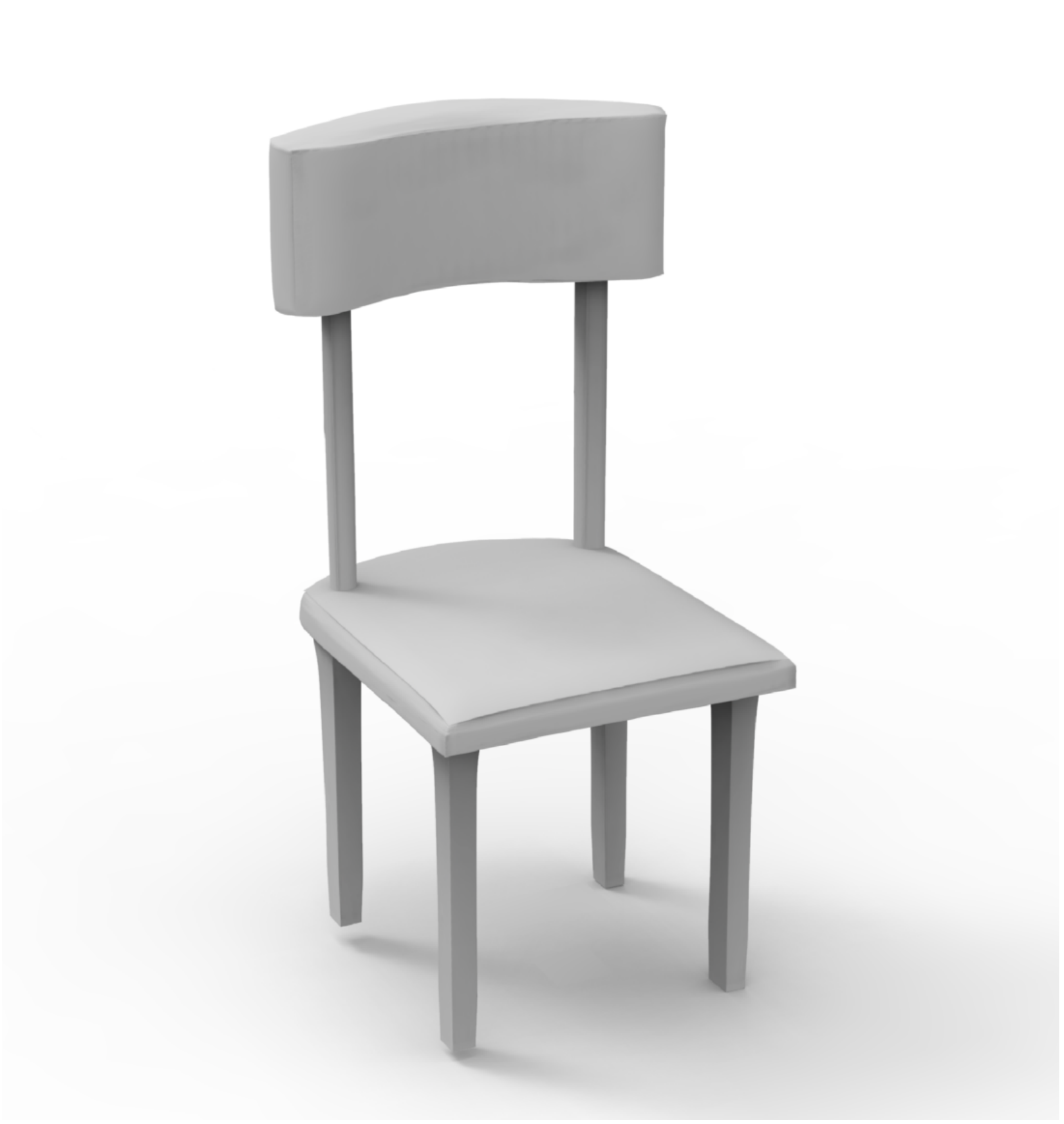}
    \includegraphics[width=0.23\linewidth]{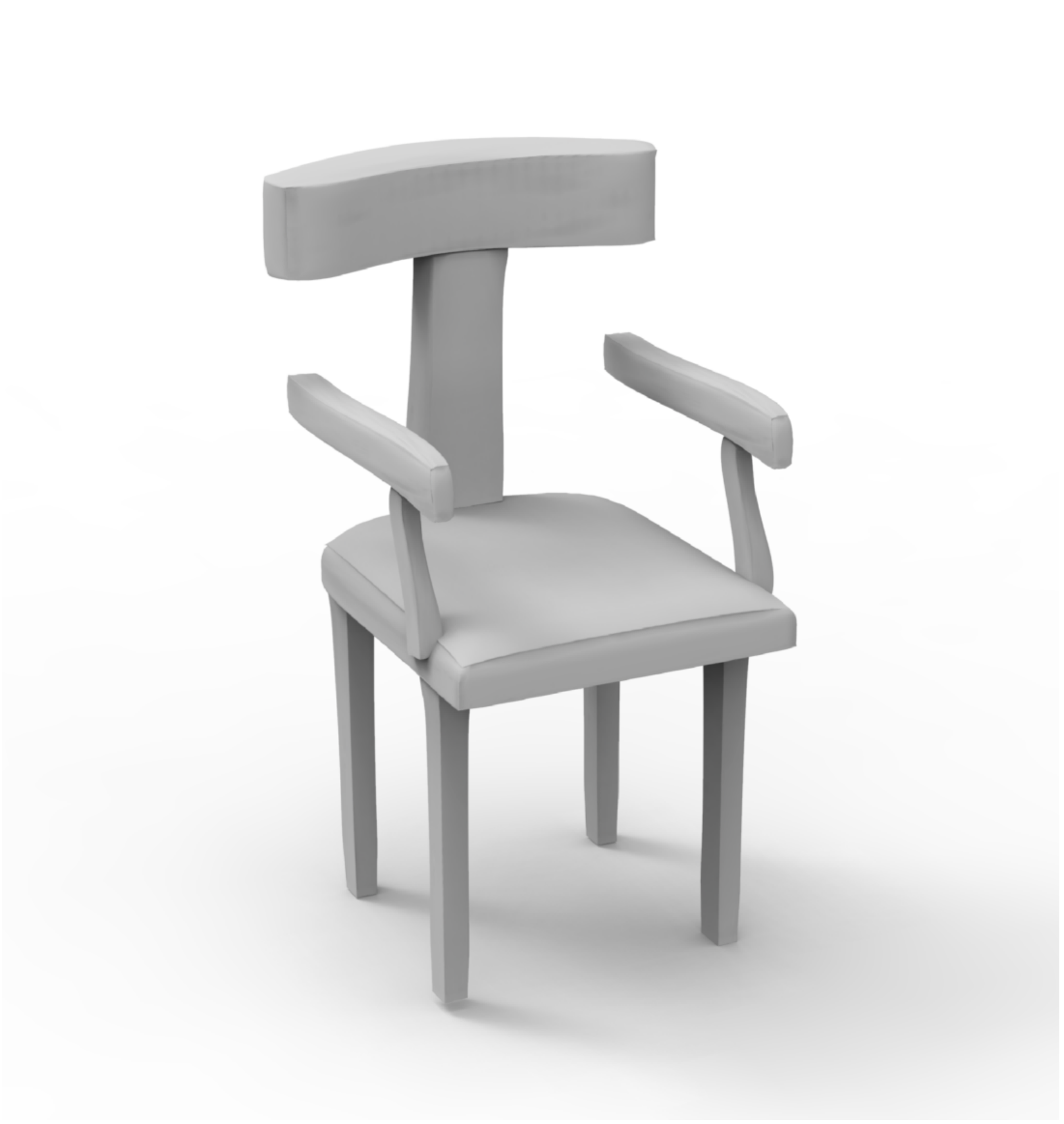}
    \includegraphics[width=0.23\linewidth]{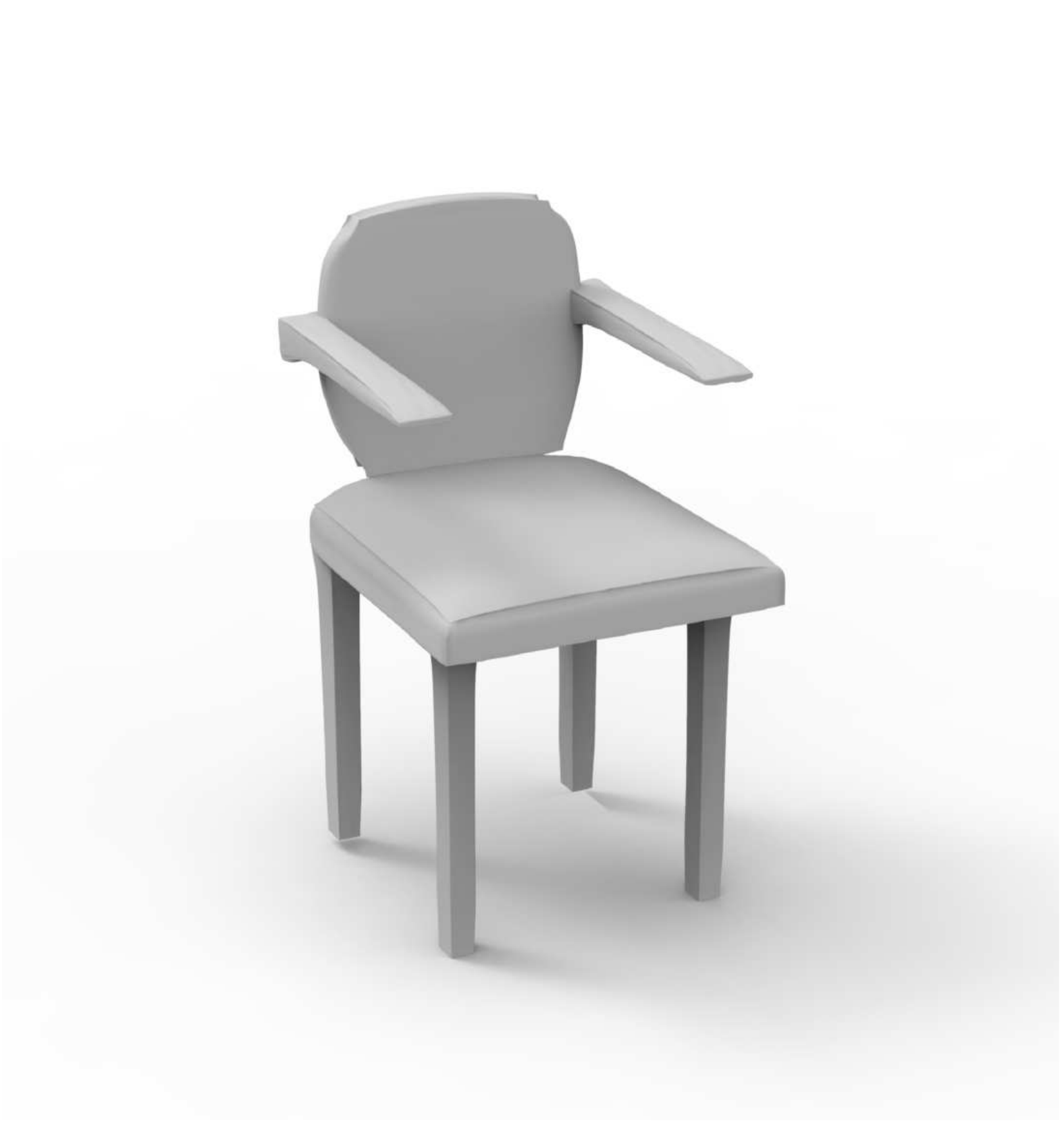}
    \includegraphics[width=0.23\linewidth]{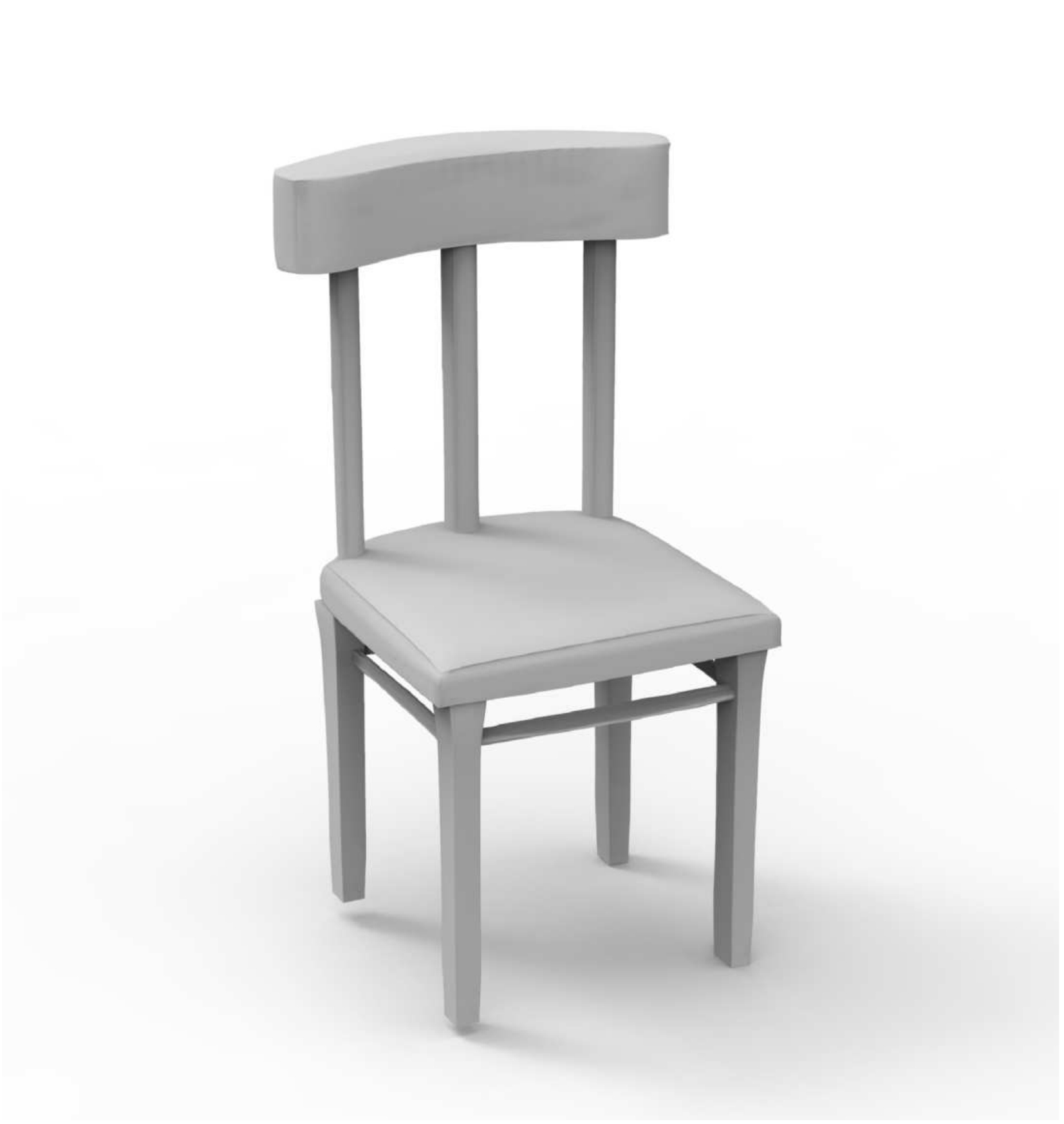}
    \\
    \includegraphics[width=0.23\linewidth]{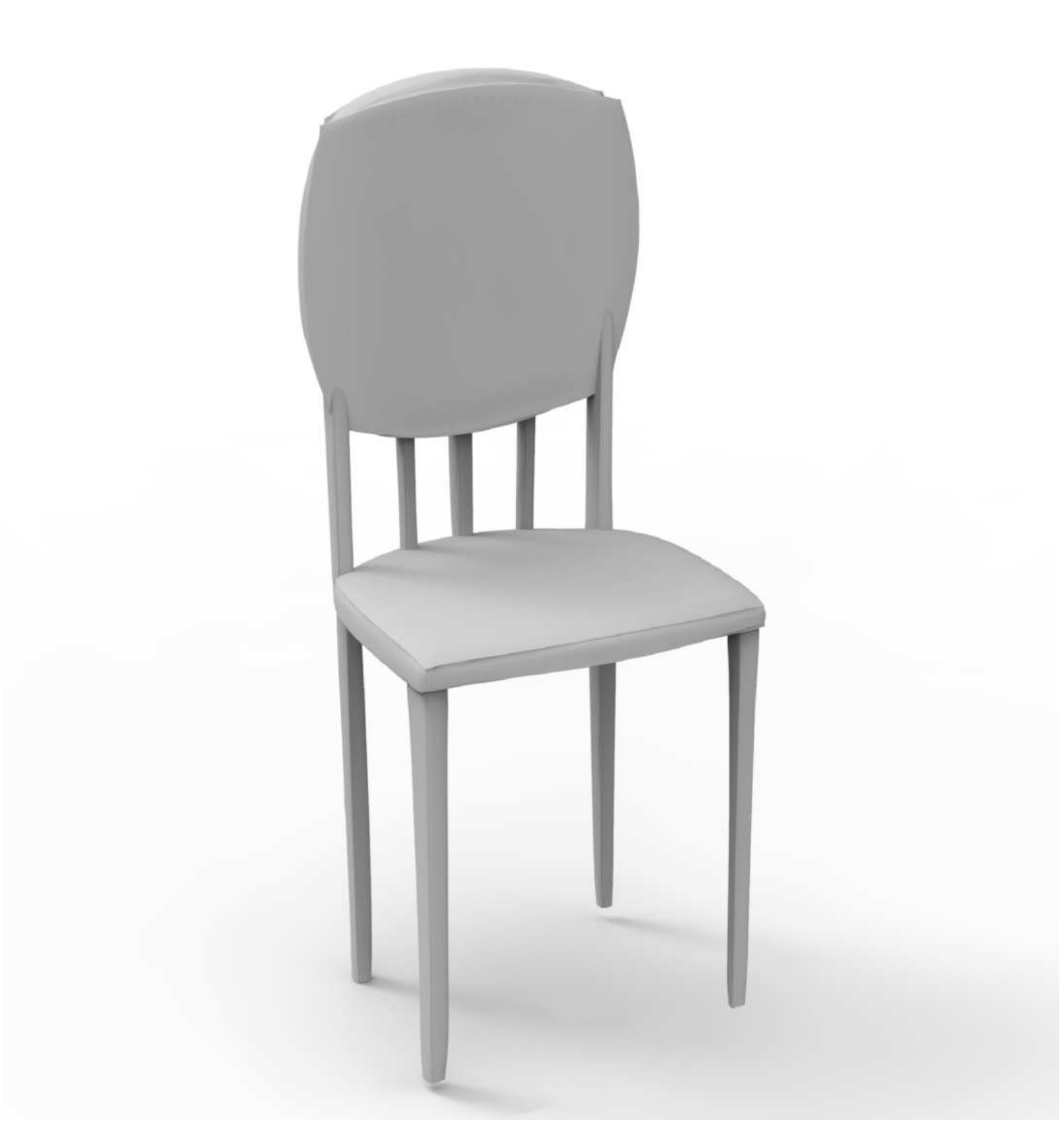}
    \includegraphics[width=0.23\linewidth]{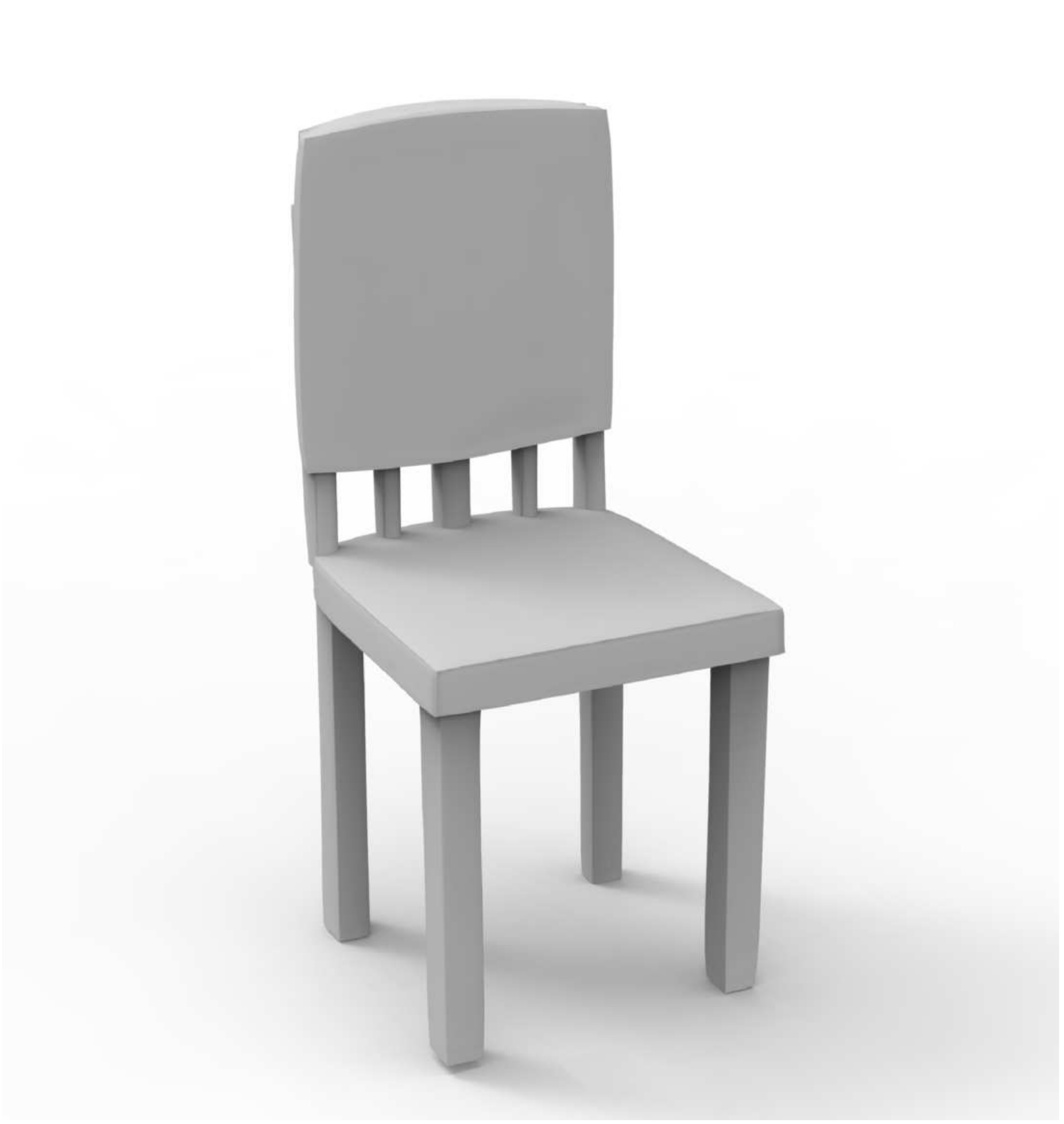}
    \includegraphics[width=0.23\linewidth]{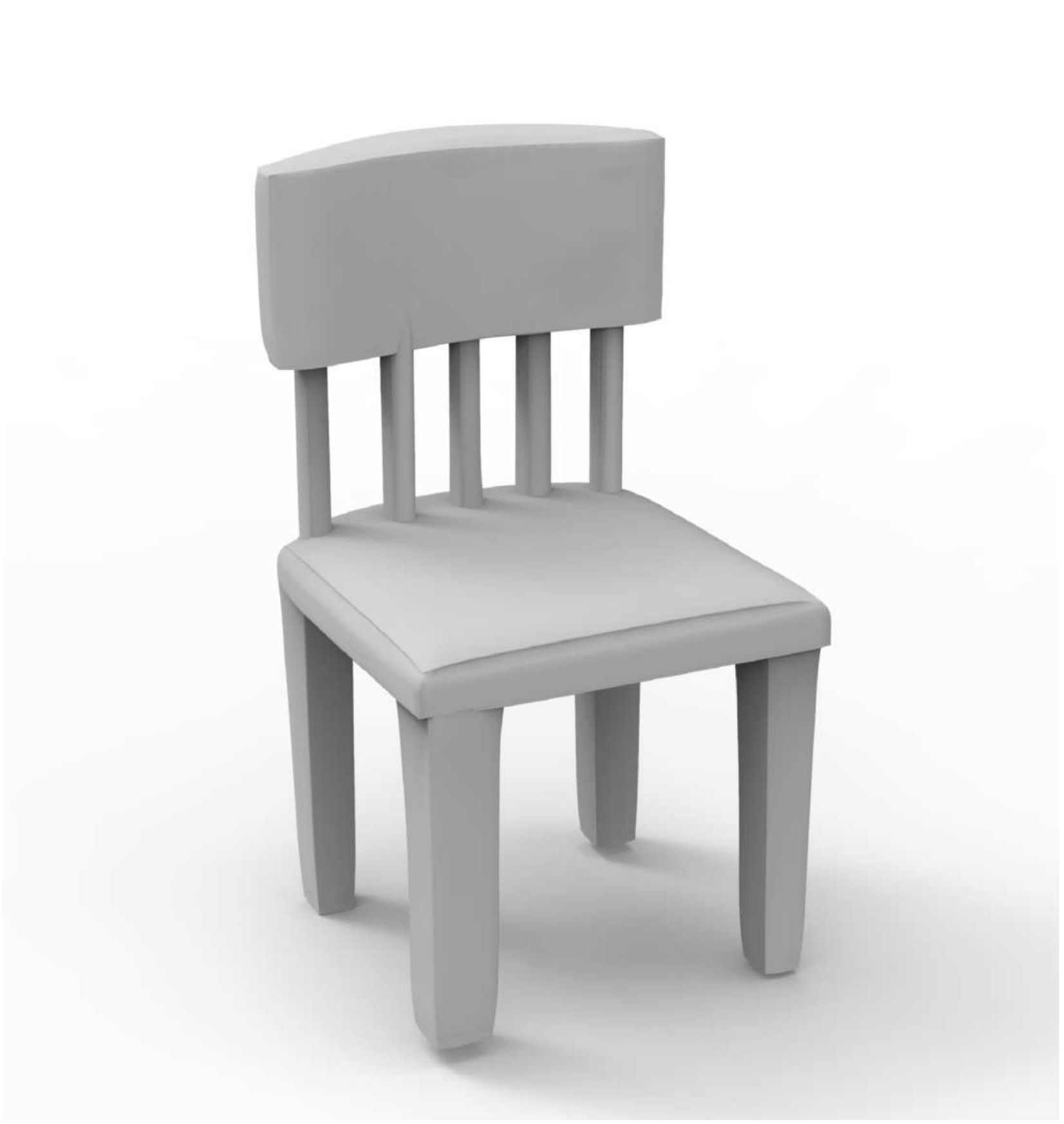}
    \includegraphics[width=0.23\linewidth]{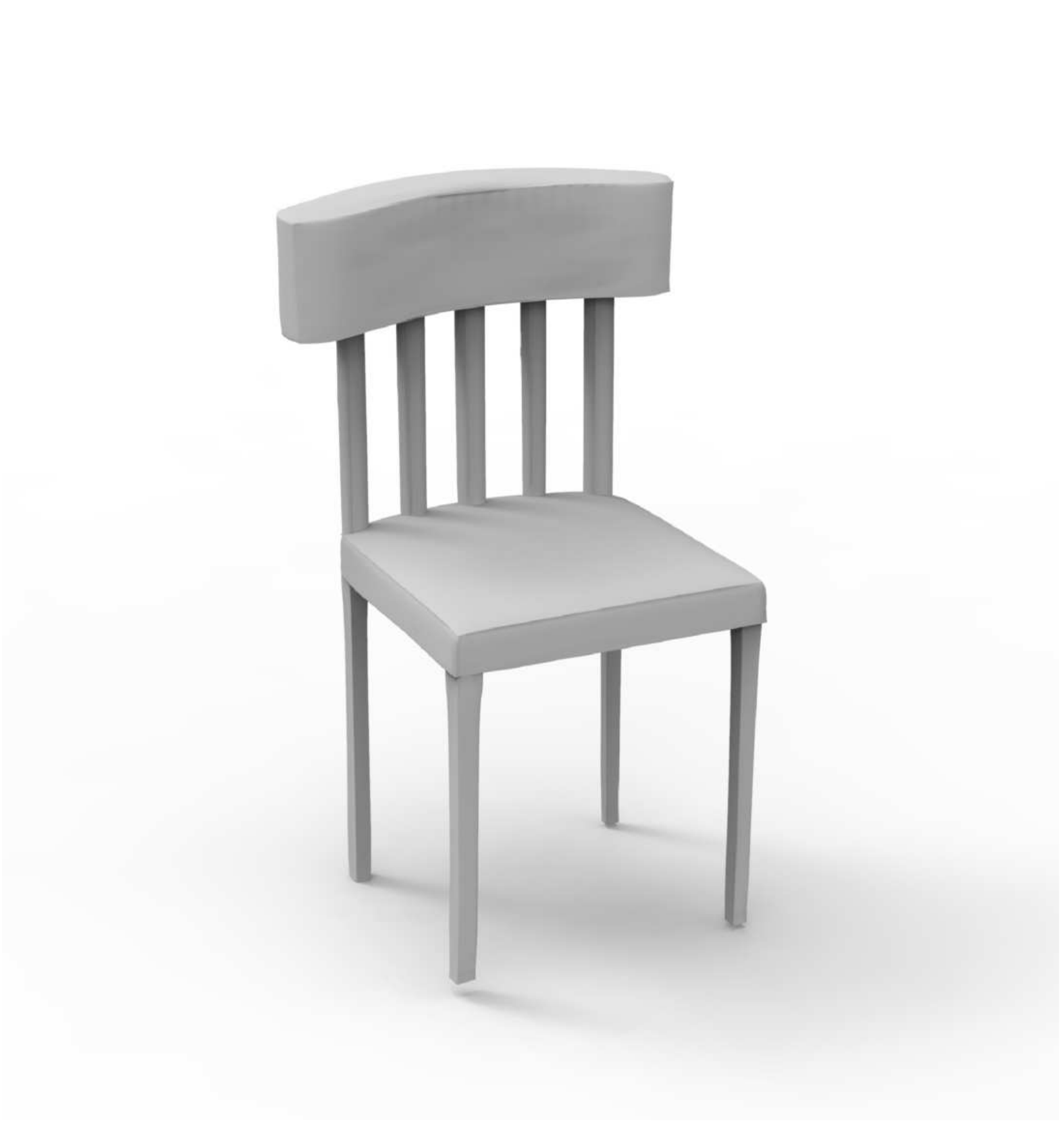}
\end{minipage}

\subfigure[Given Shape]{
\begin{minipage}{0.21\linewidth}
\centering
\includegraphics[width=0.99\linewidth]{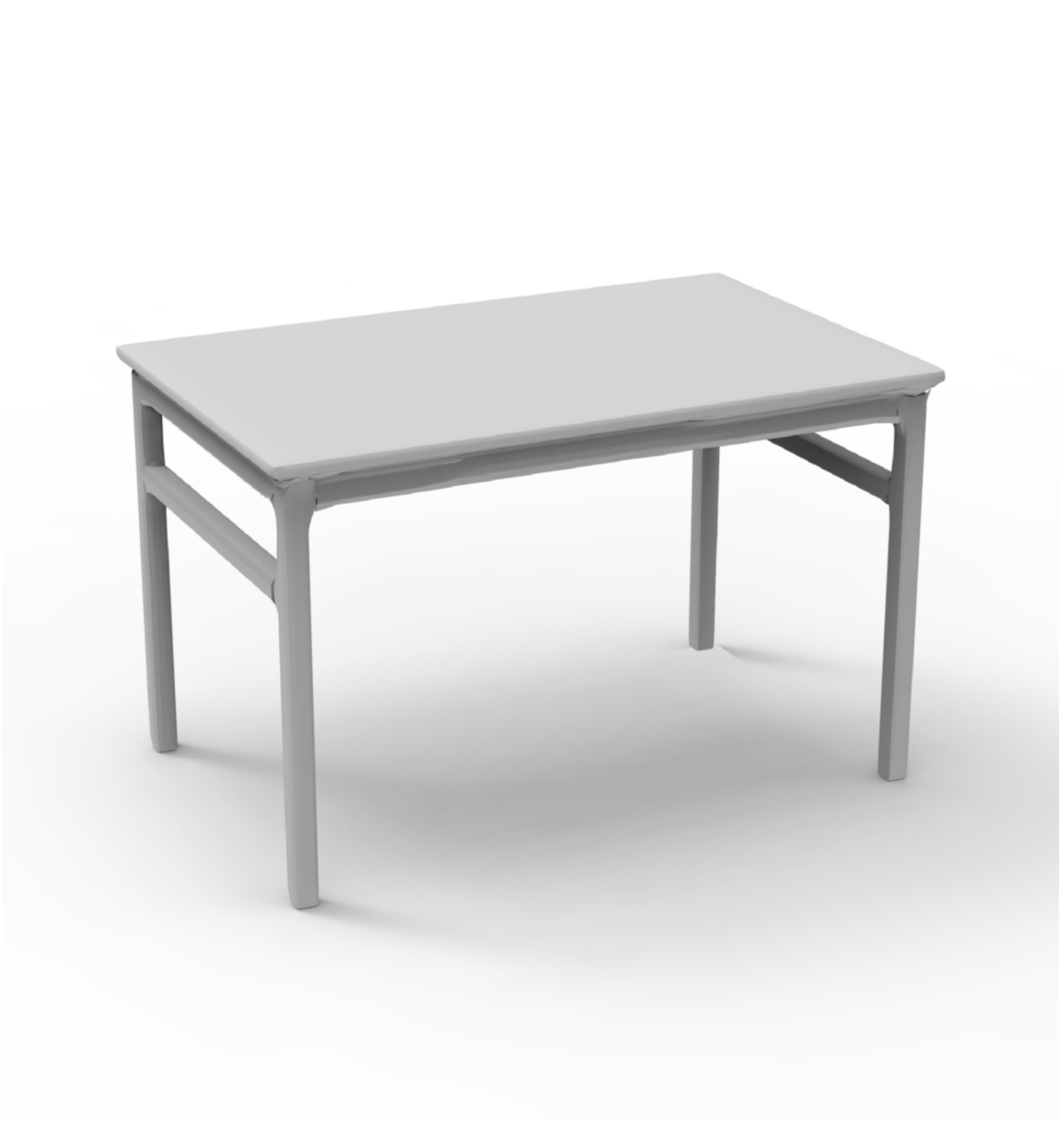}
\end{minipage}
}
\subfigure[Random generation]{
\begin{minipage}{0.71\linewidth}
\centering
    \includegraphics[width=0.23\linewidth]{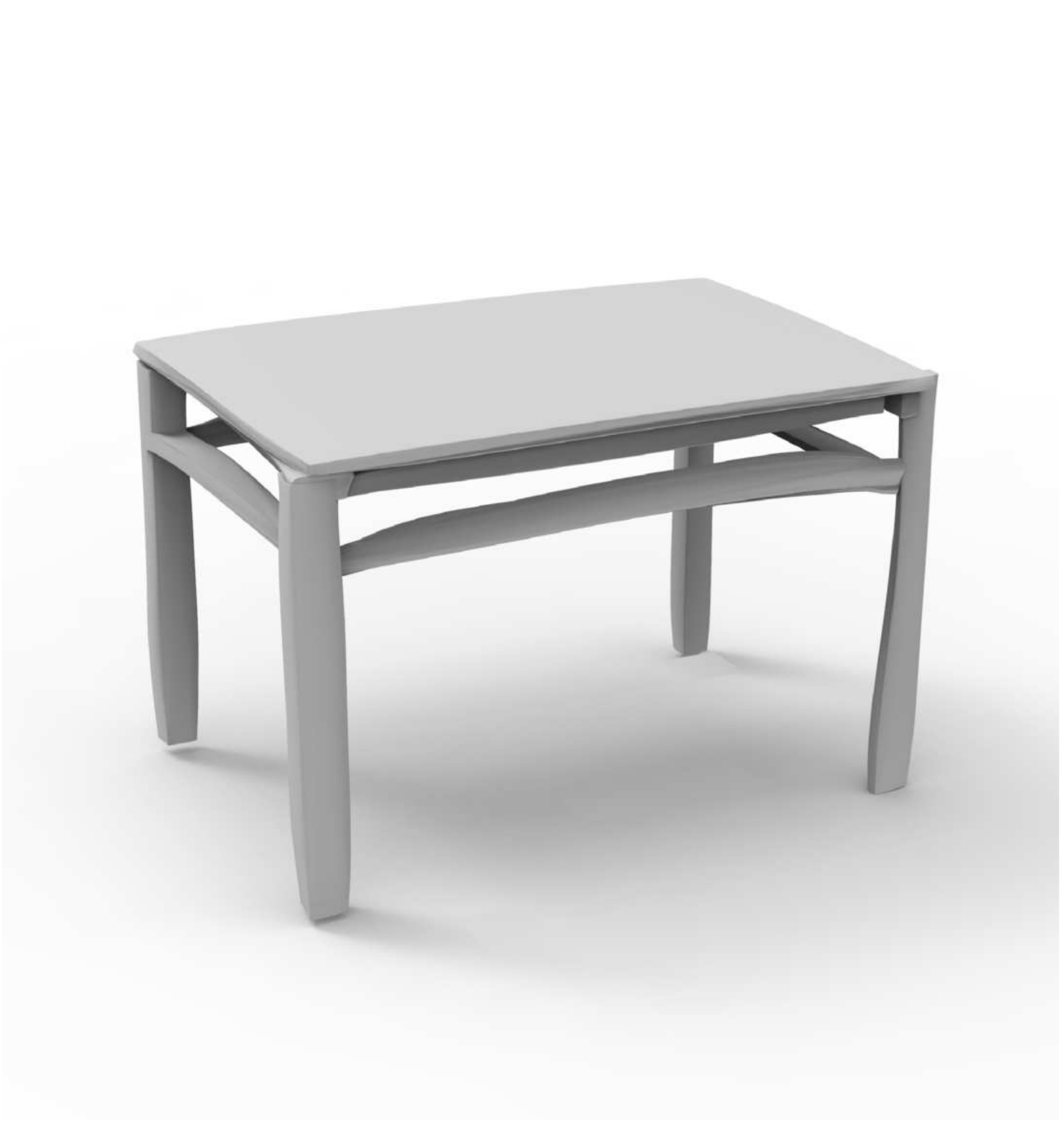}
    \includegraphics[width=0.23\linewidth]{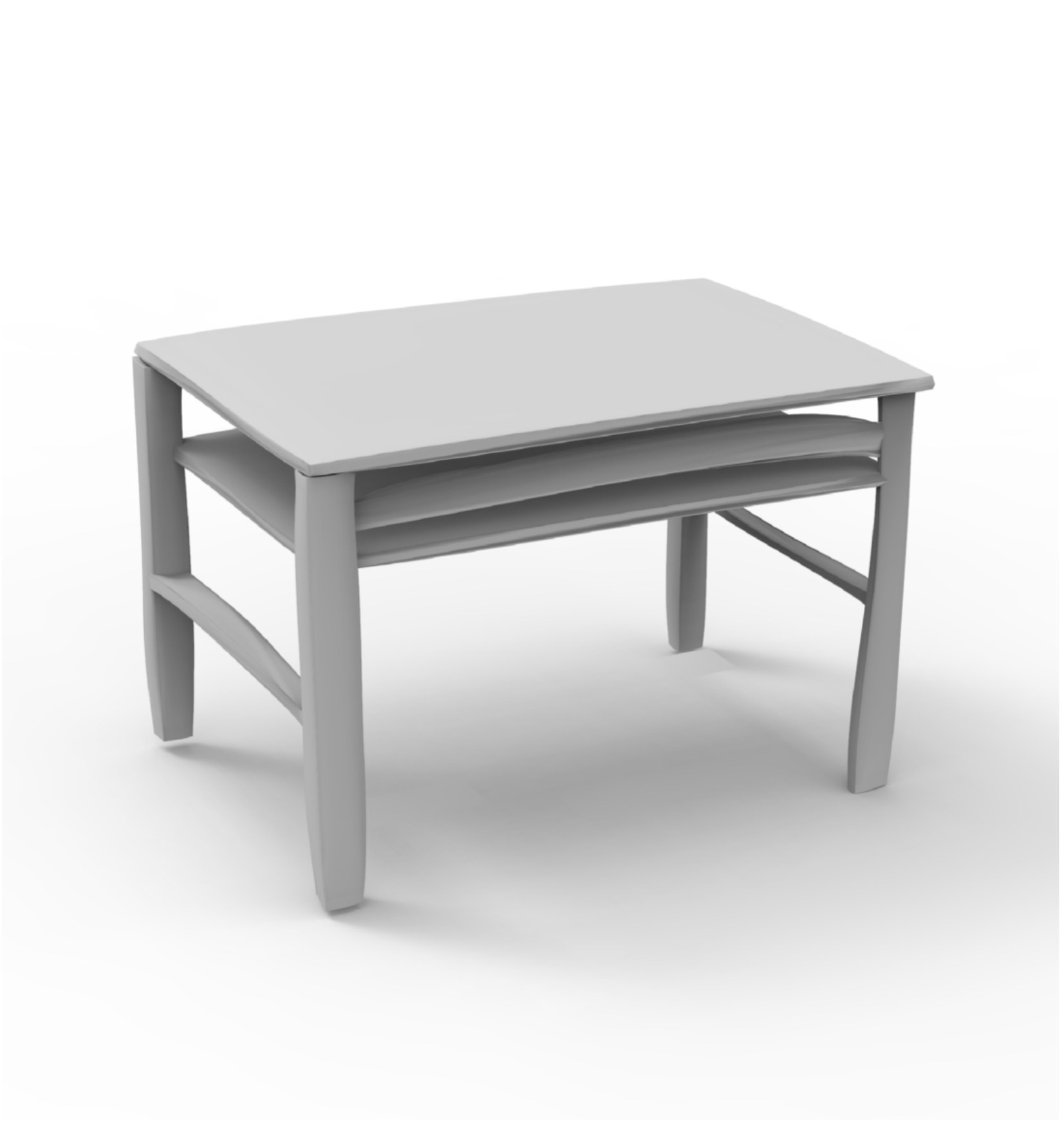}
    \includegraphics[width=0.23\linewidth]{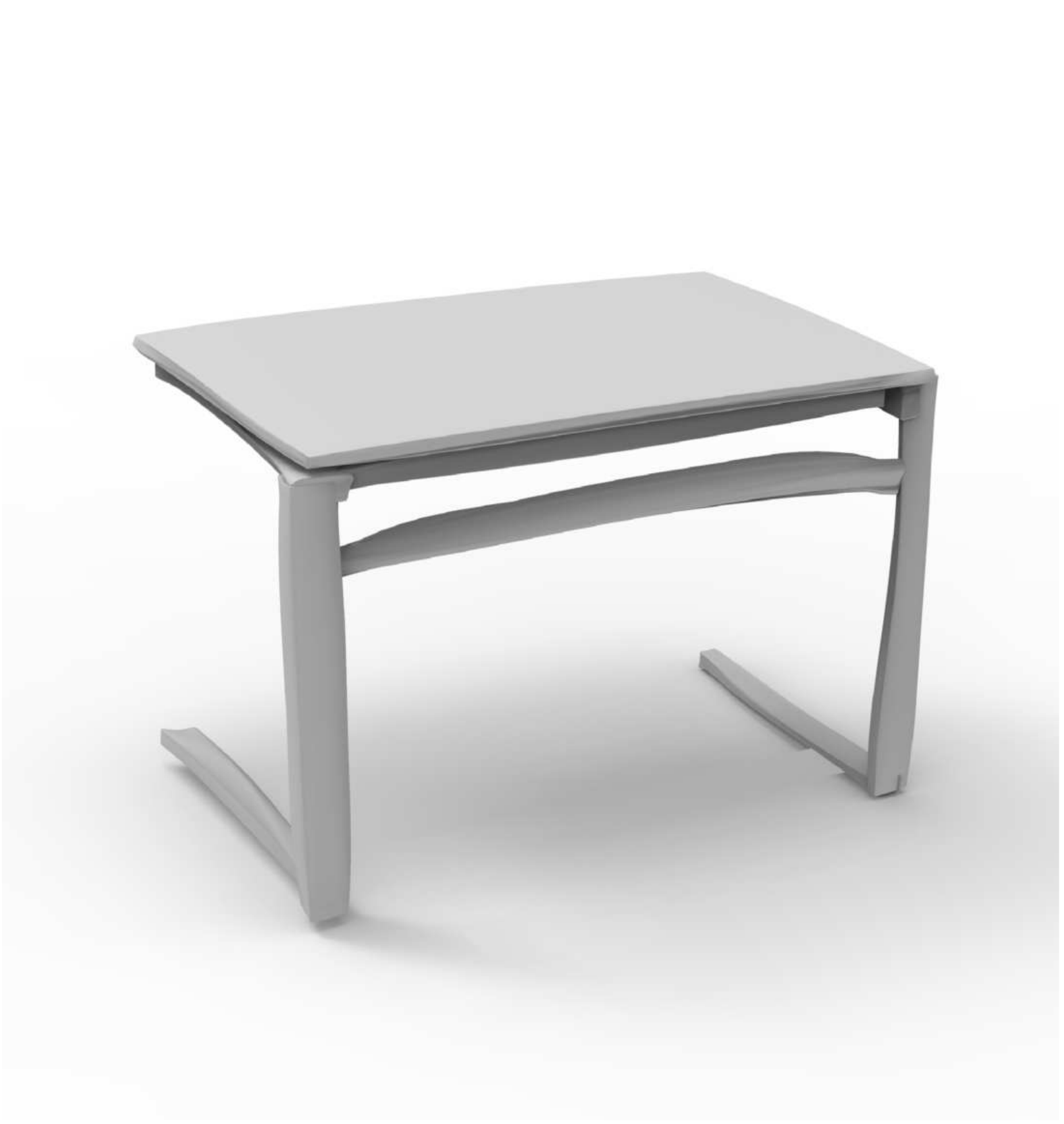}
    \includegraphics[width=0.23\linewidth]{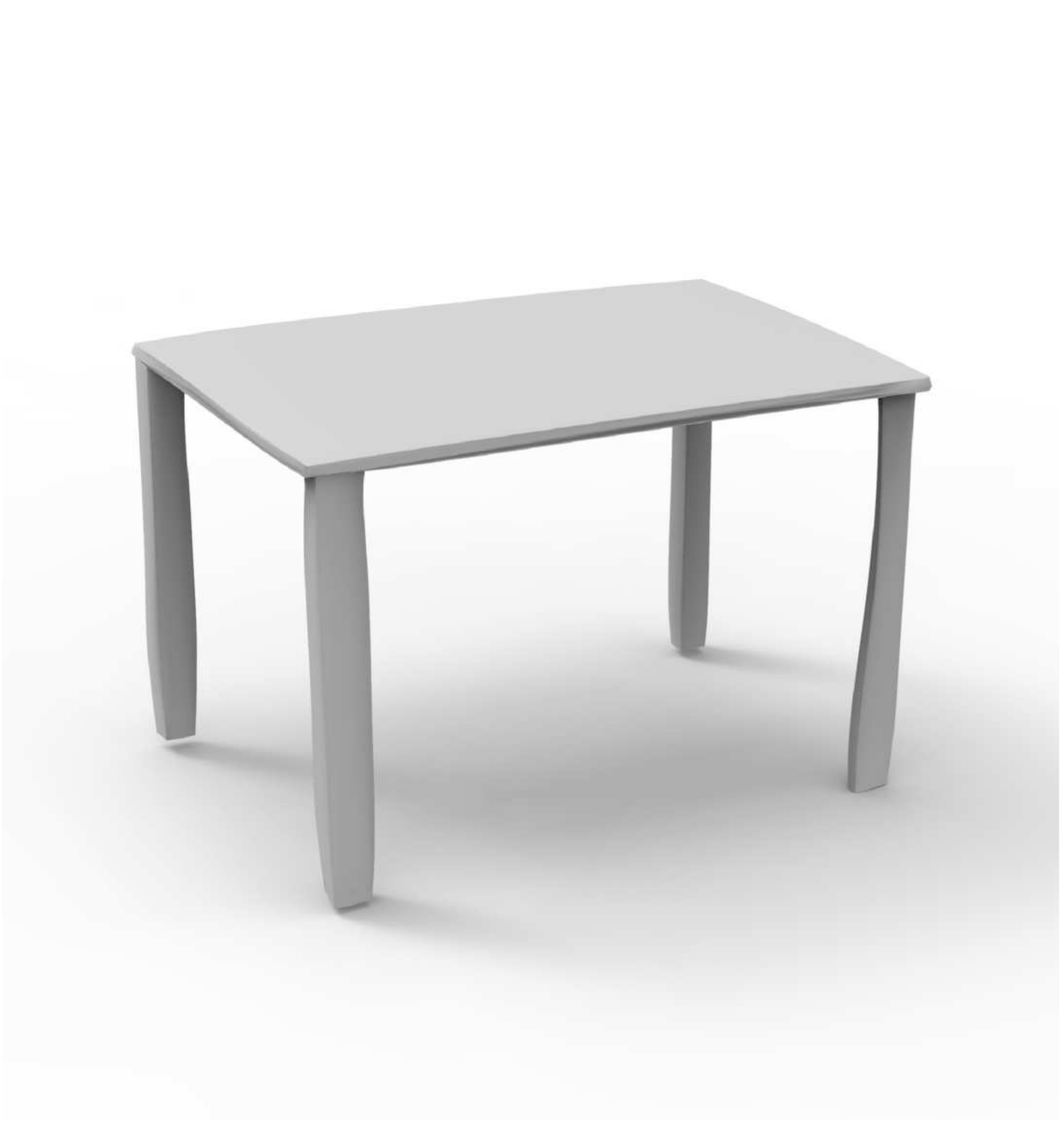}
    \\
    \includegraphics[width=0.23\linewidth]{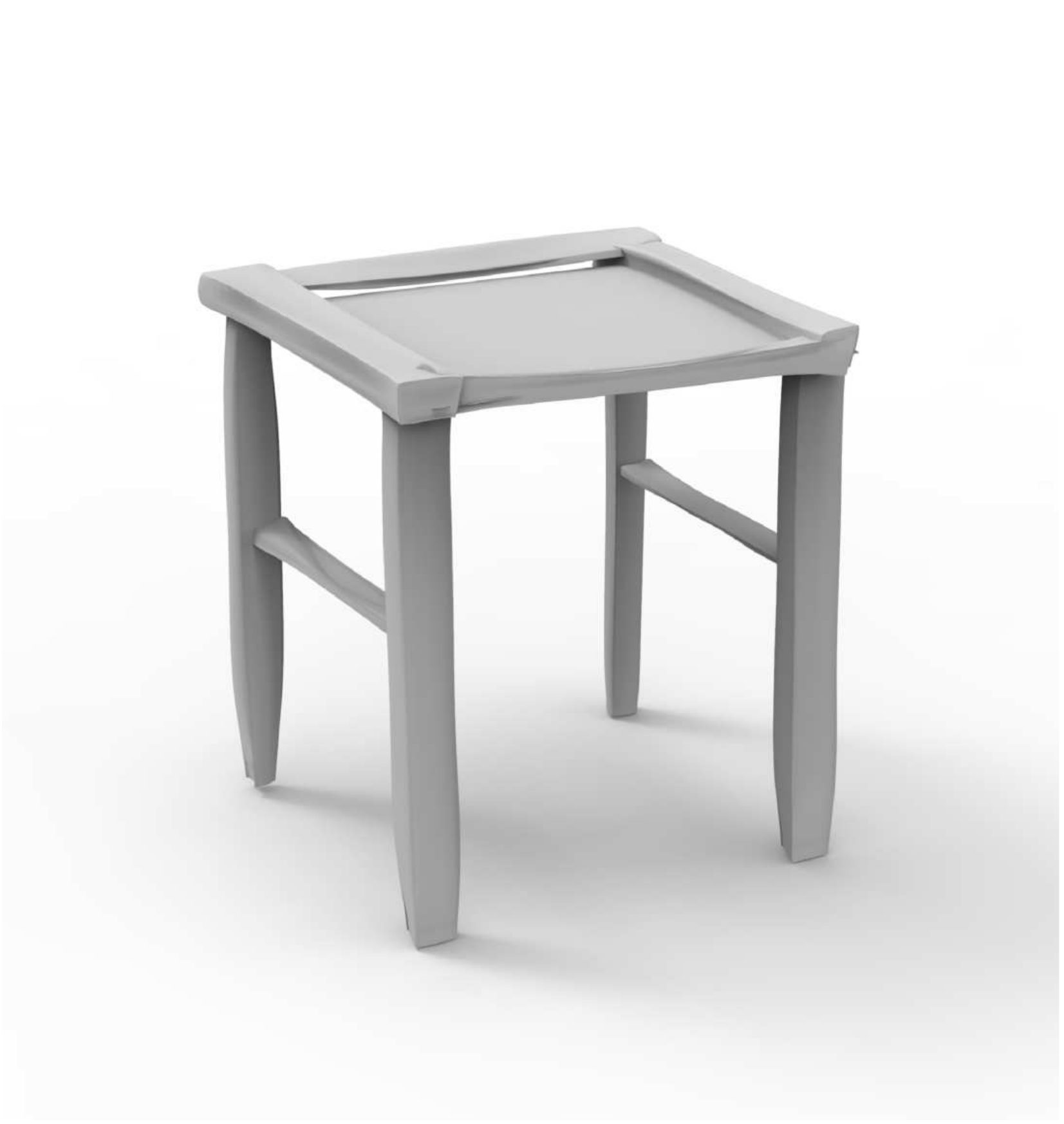}
    \includegraphics[width=0.23\linewidth]{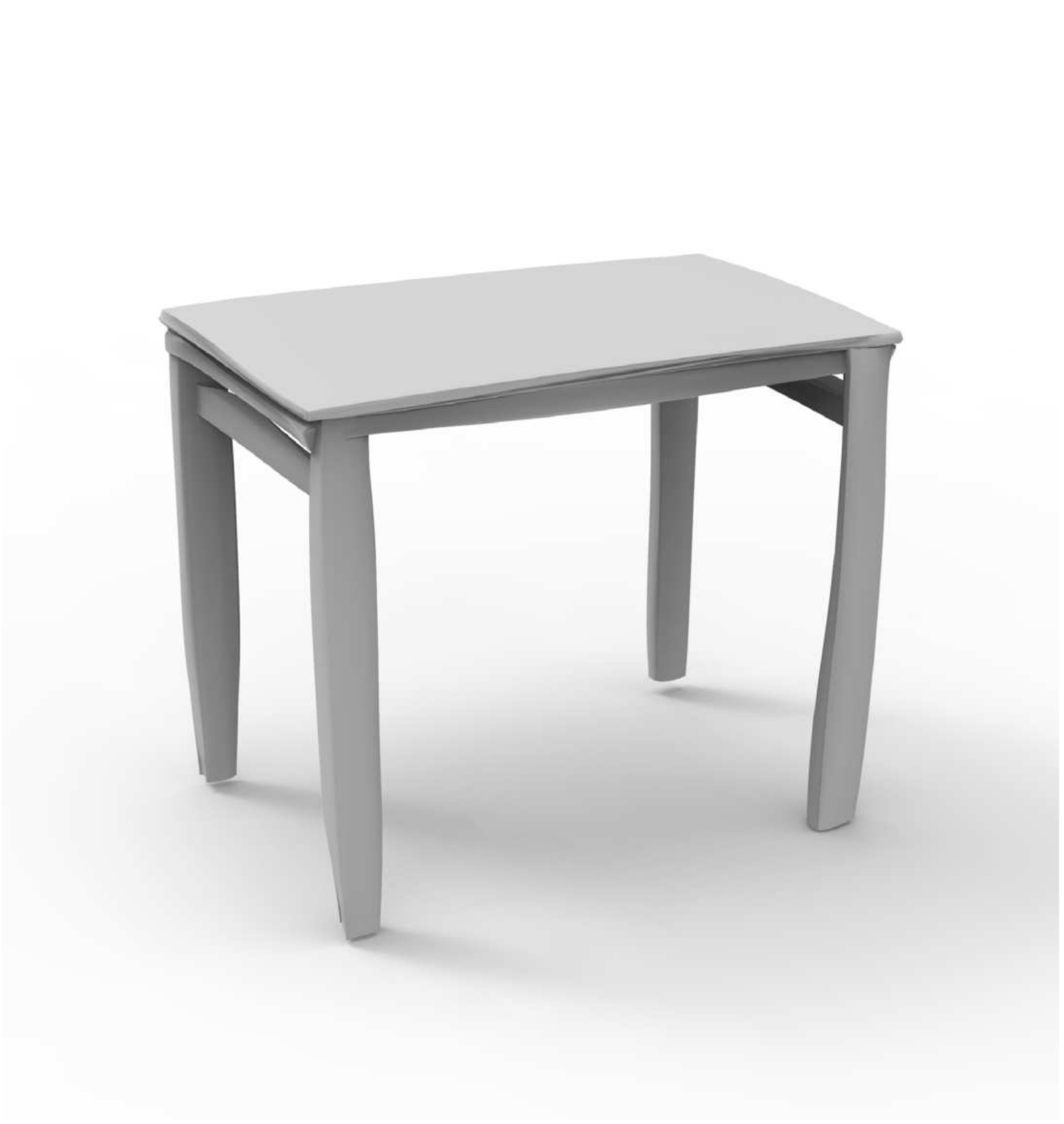}
    \includegraphics[width=0.23\linewidth]{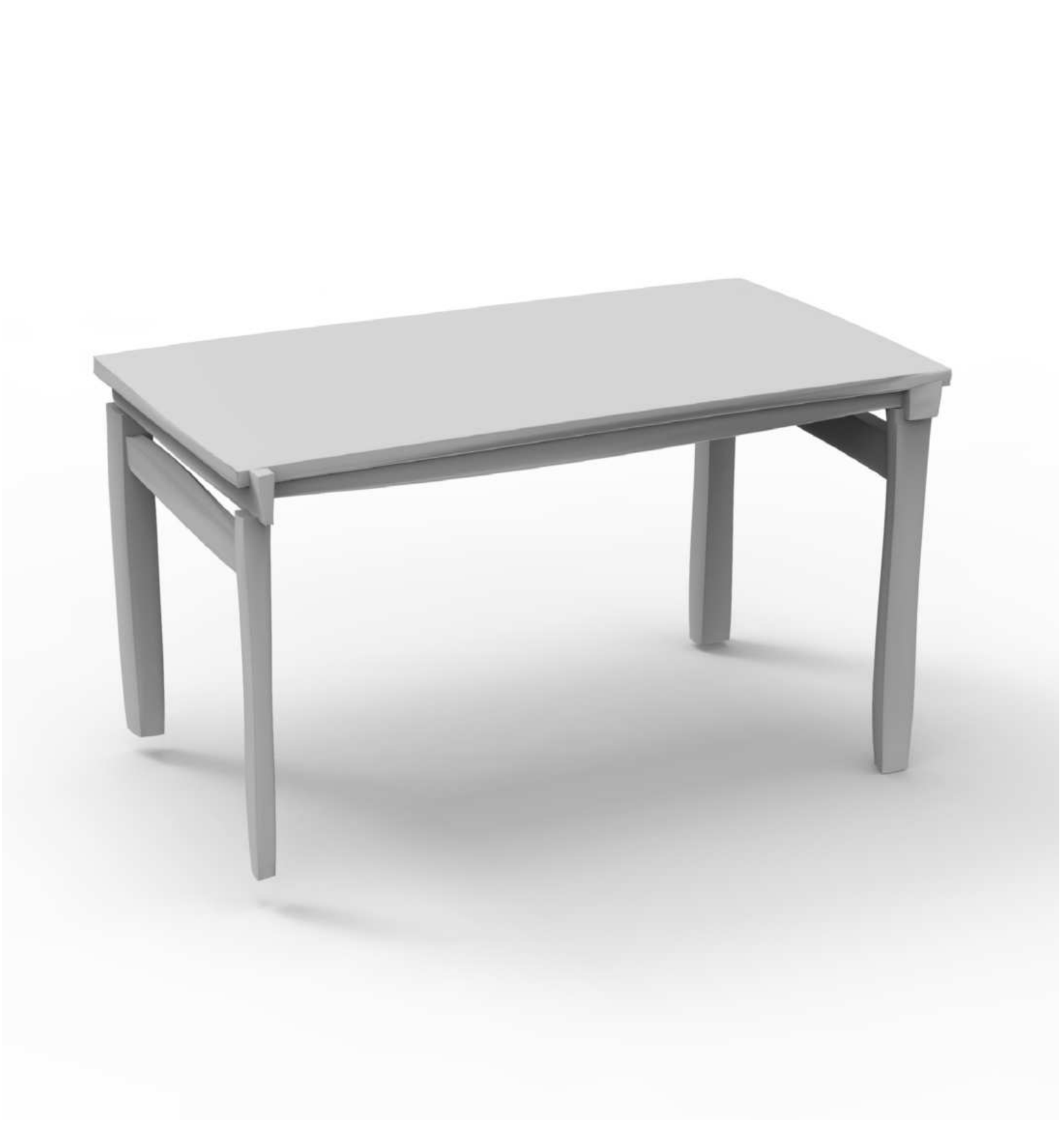}
    \includegraphics[width=0.23\linewidth]{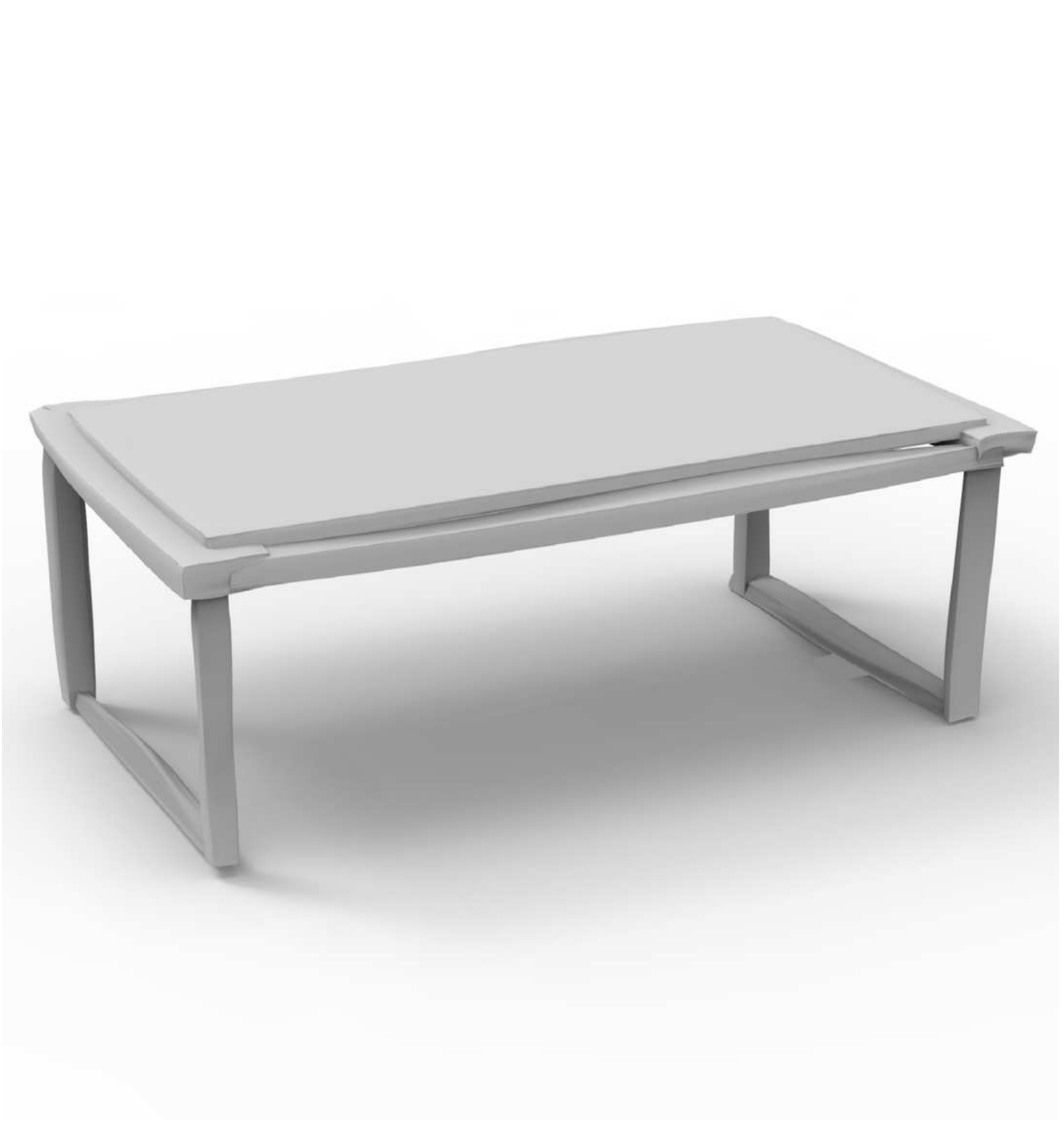}
\end{minipage}
}
\vspace{-3mm}
    \caption{\yjr{Qualitative results for disentangled shape generation on four categories. Given an input shape (a), we extract the geometry code and structure code. We fix one of them, and randomly sample code in the other latent space to generate new shapes (b). For the first row of (b), we keep the geometry code unchanged and randomly explore the structure latent space. And, for the second row, we keep the structure code unchanged and randomly sample over the geometry latent space.
    \yj{For the table example, we can see that the generated shapes share a similar size and length of table surface and legs in the first row but the other parts and shape structure are changing, while in the second row, the geometry of all parts is changing but the structure remains unchanged.
    We can also observe similar results for the cabinet and lamp examples.}}
    }
    \label{fig:generation_pt2pc2}
    \vspace{-3mm}
\end{figure}

\paragraph{Disentangled Shape Generation.} 
DSG-Net learns two disentangled latent spaces for modeling shape structure and geometry, which enables a novel task of generating shapes with a given shape structure or geometry pattern.
We demonstrate that given an input shape, DSG-Net can extract the structure code from the shape and pair it with a random geometry code, which allows us to explore shape geometry variations satisfying a certain shape structure.
It also works well to explore structure variations while keeping the geometry code unchanged.
\yj{
We show four controllable generation results on the four categories in Figure~\ref{fig:generation_pt2pc2}. In the experiments, given an input shape, the geometry code and structure code are extracted by running it through the encoding procedures. 
And then, we can keep one of them unchanged and randomly sample in the other latent space. %
We see that when we preserve the geometry code, the chair/table legs usually maintain similar width and length to the input shapes.
And, when we keep the structure code unchanged, we are generating shapes with geometric variations but satisfying the same symbolic structure hierarchy.
Please refer to supplementary material for more disentangled generation results.
}

\begin{figure}[t]
    \centering
    \includegraphics[width=0.19\linewidth]{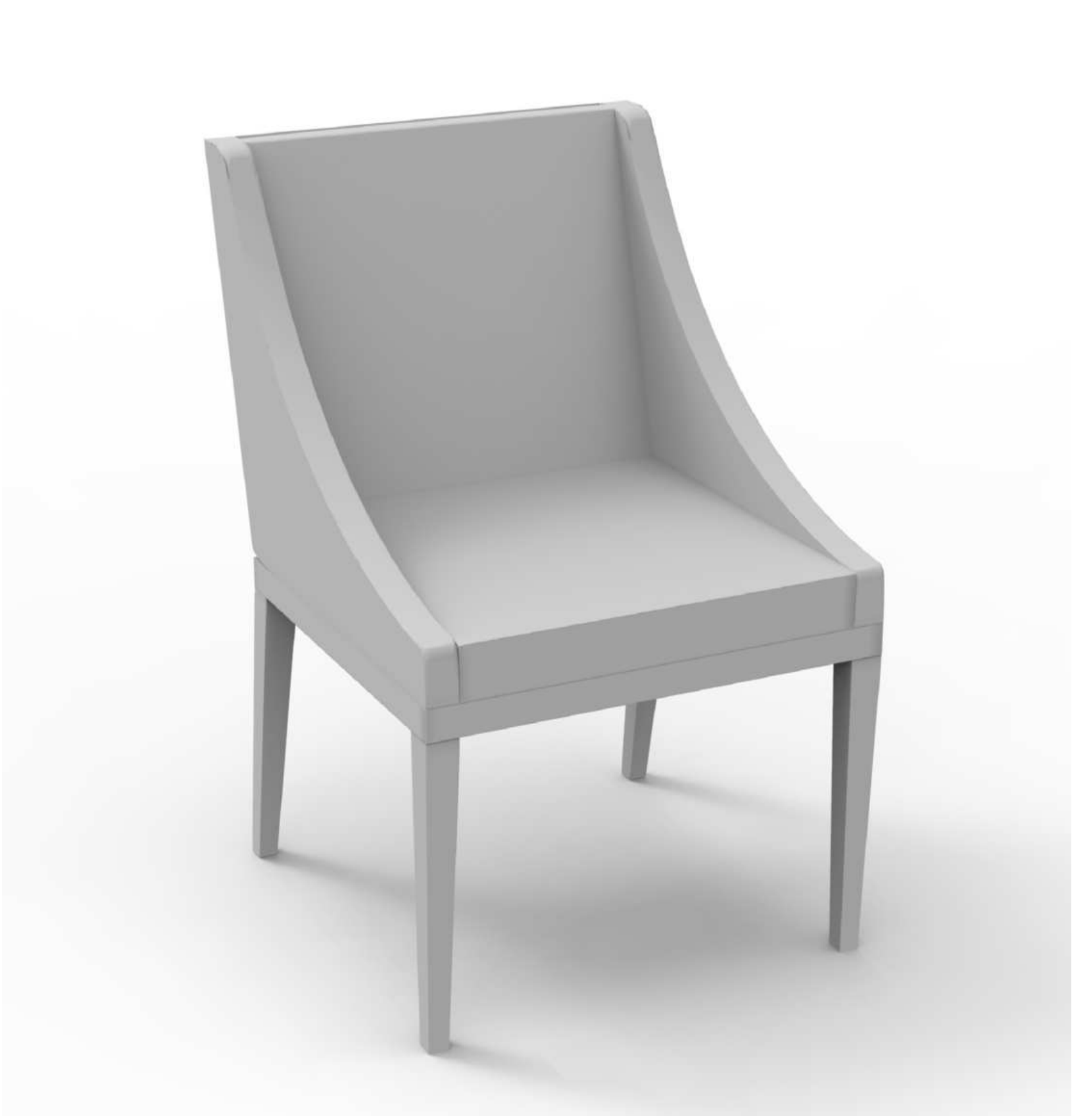}
    \includegraphics[width=0.19\linewidth]{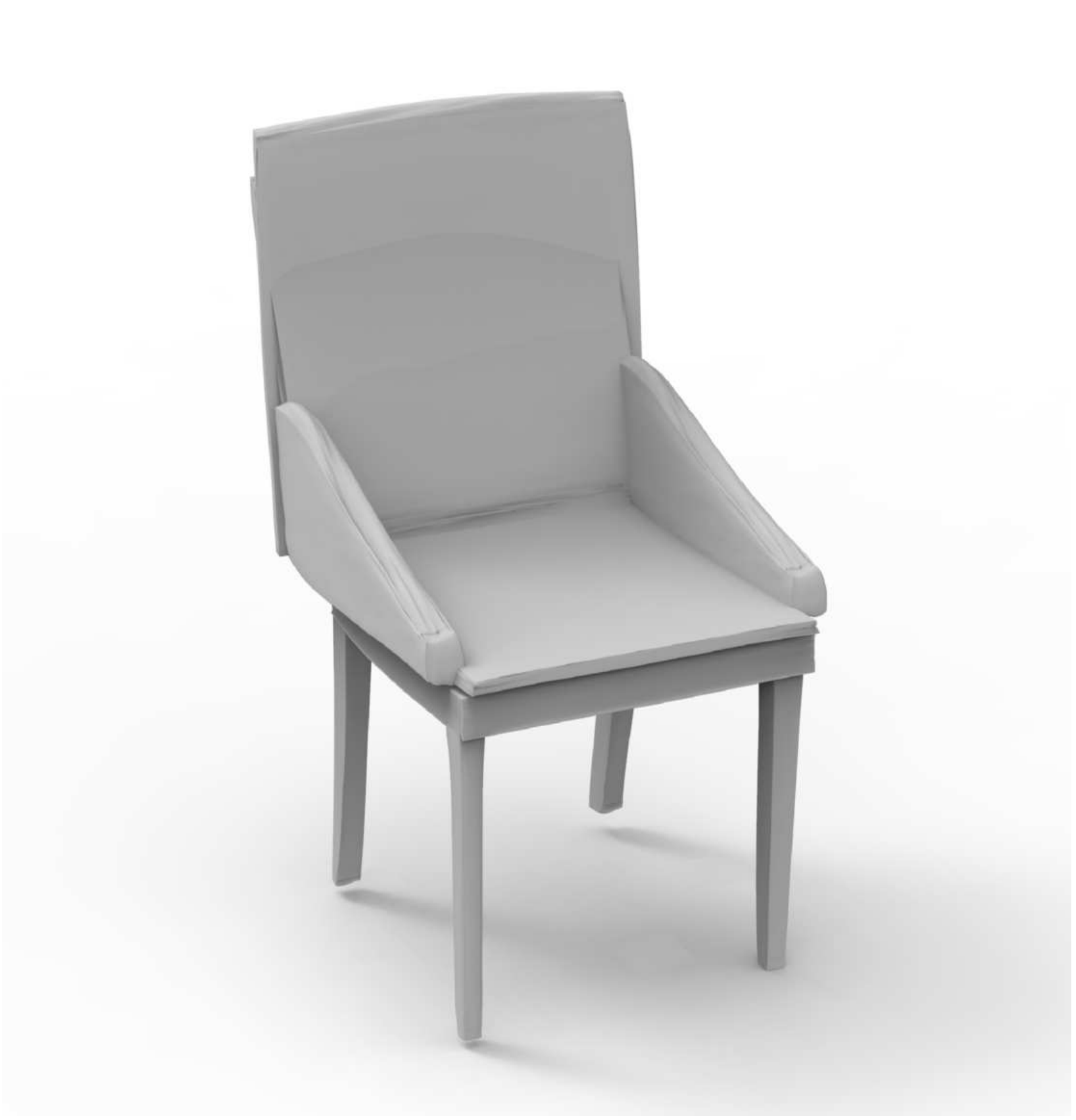}
    \includegraphics[width=0.19\linewidth]{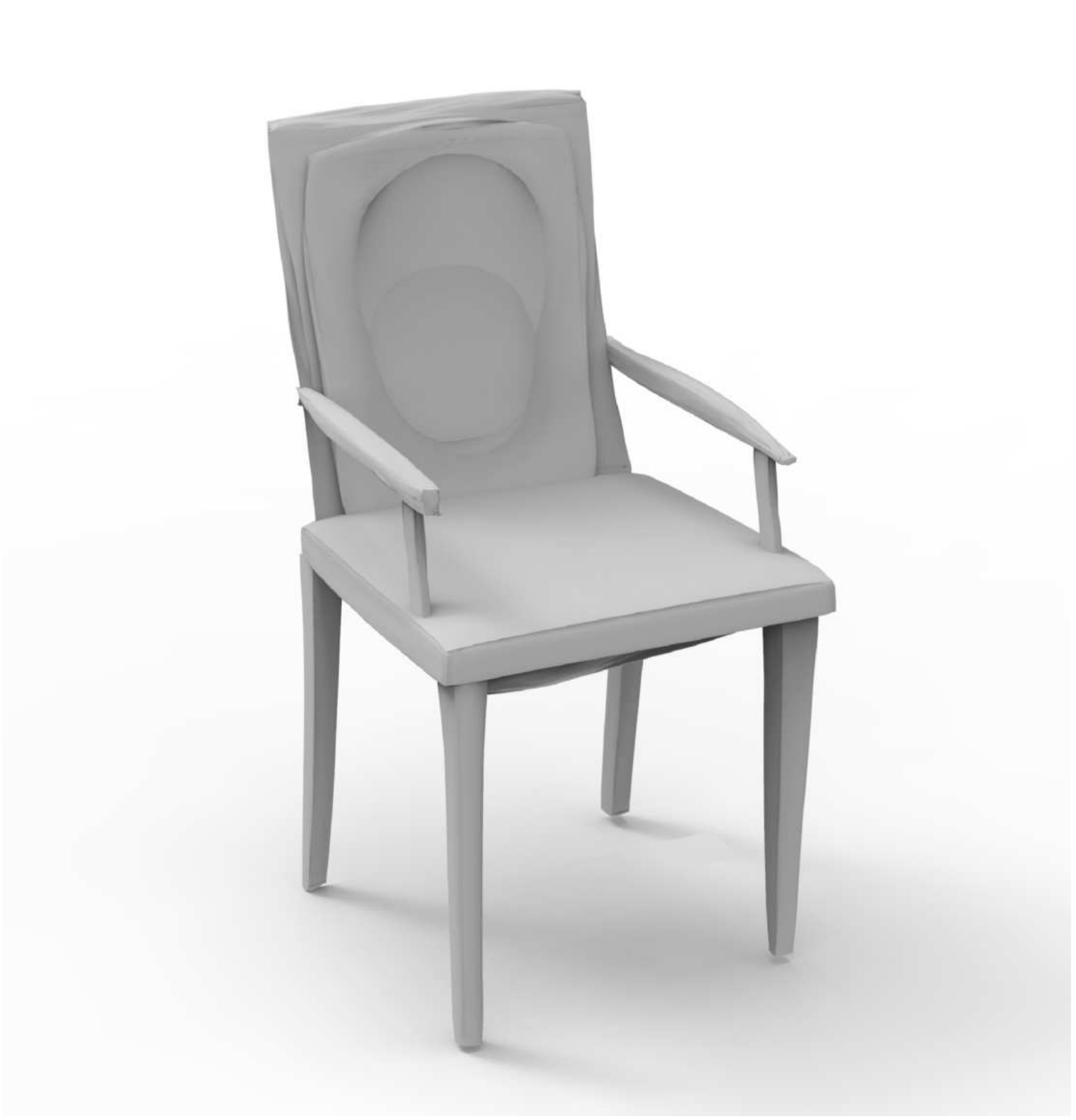}
    \includegraphics[width=0.19\linewidth]{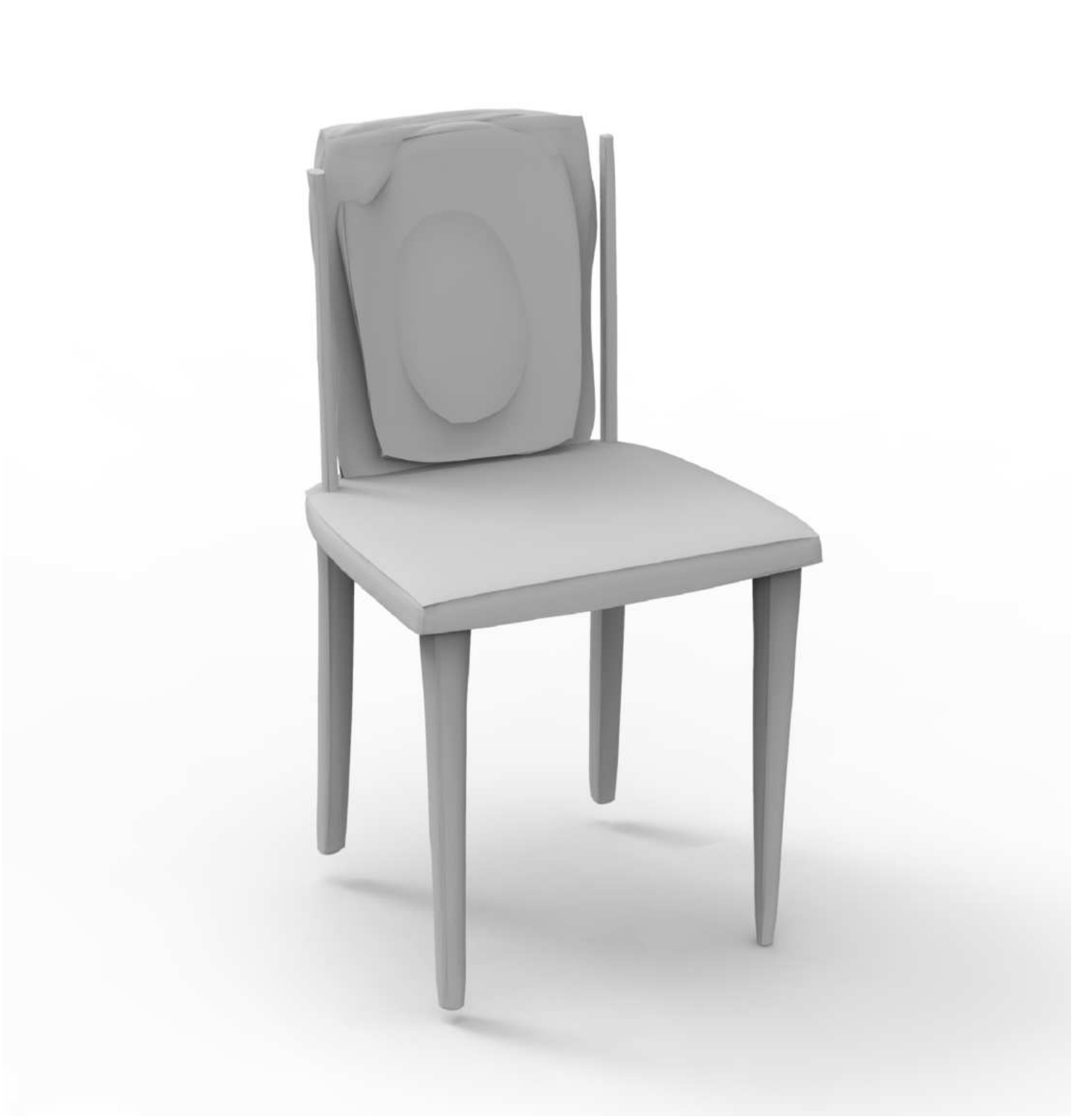}
    \includegraphics[width=0.19\linewidth]{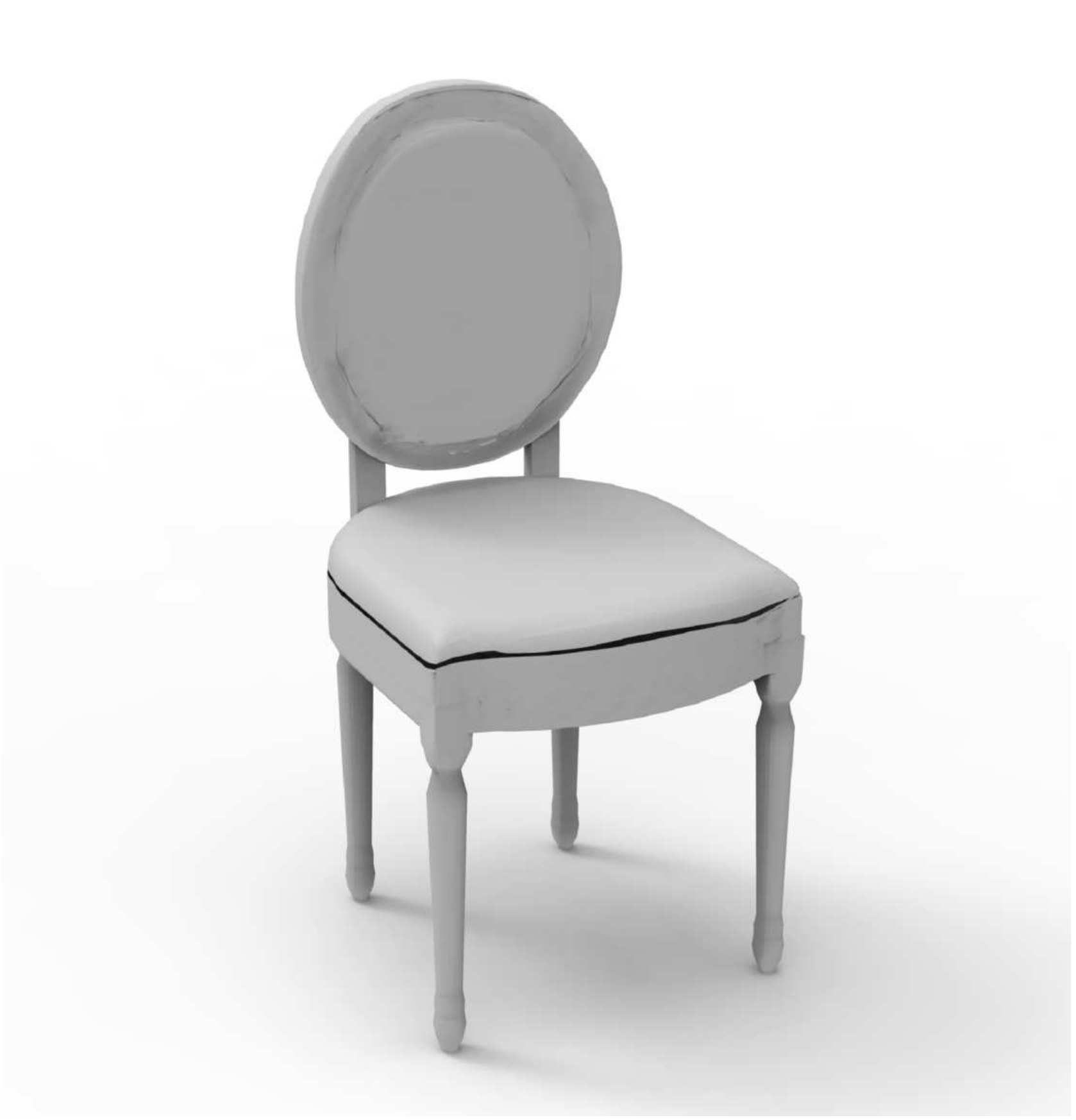}\\
    \includegraphics[width=0.19\linewidth]{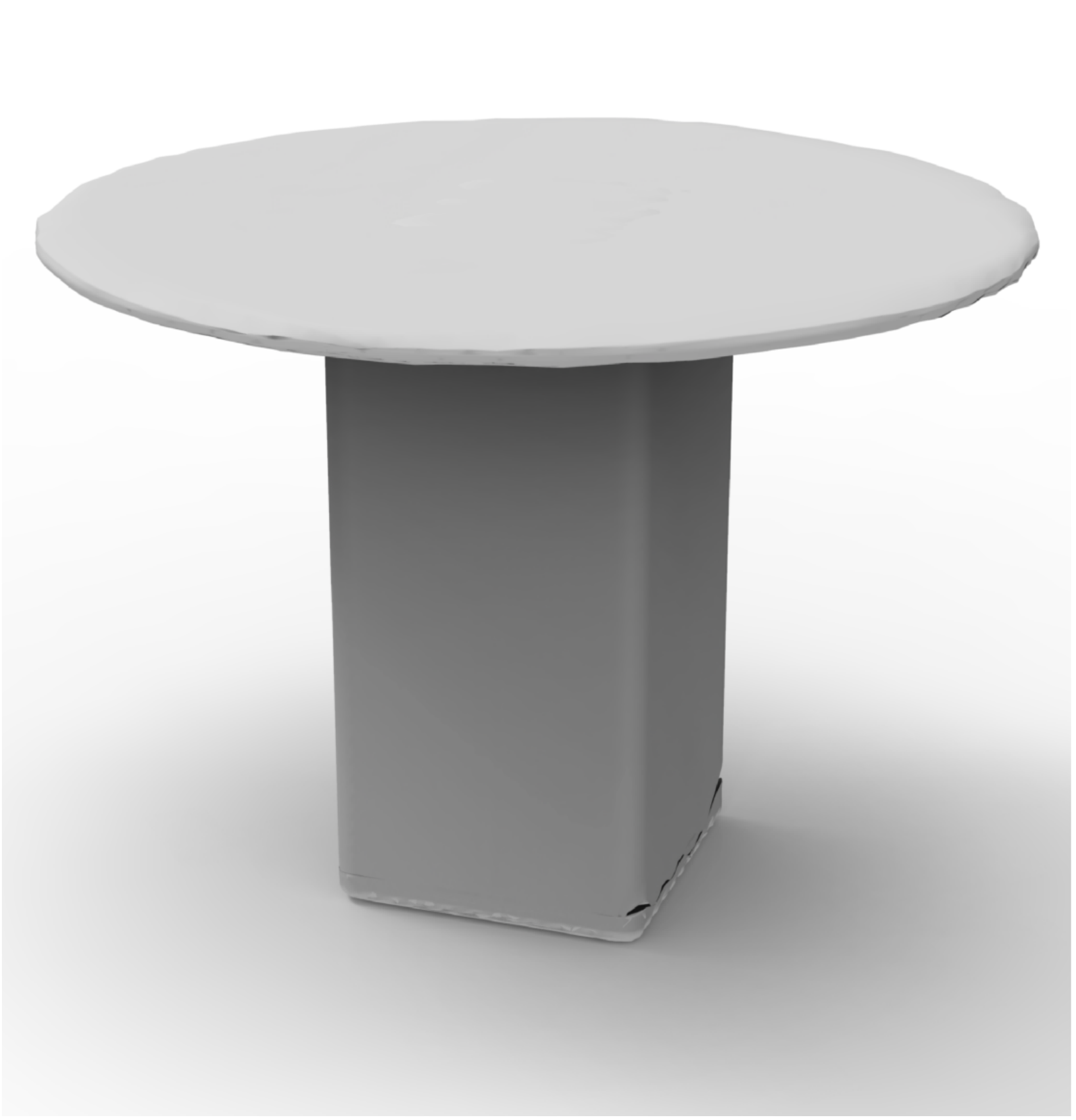}
    \includegraphics[width=0.19\linewidth]{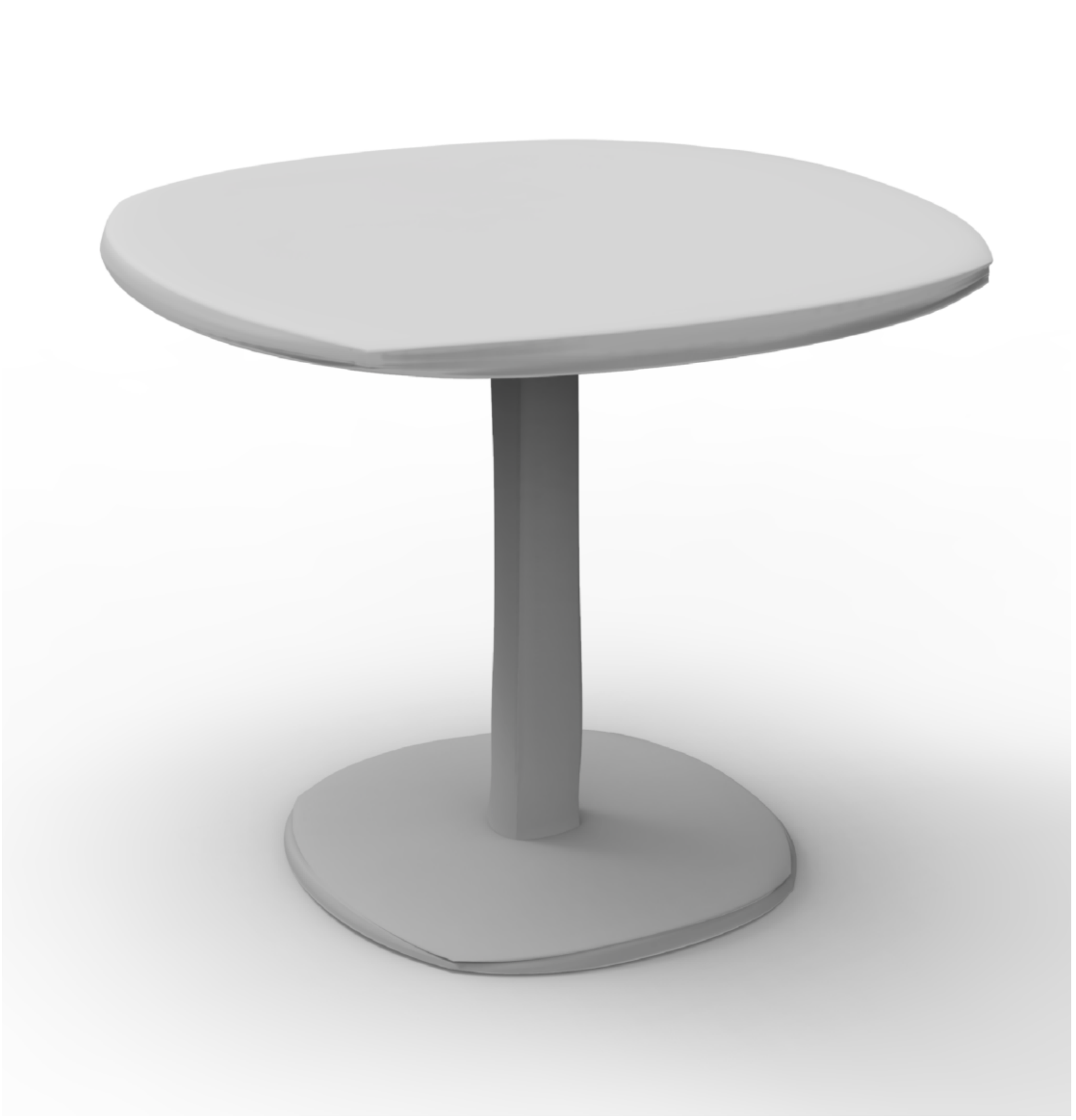}
    \includegraphics[width=0.19\linewidth]{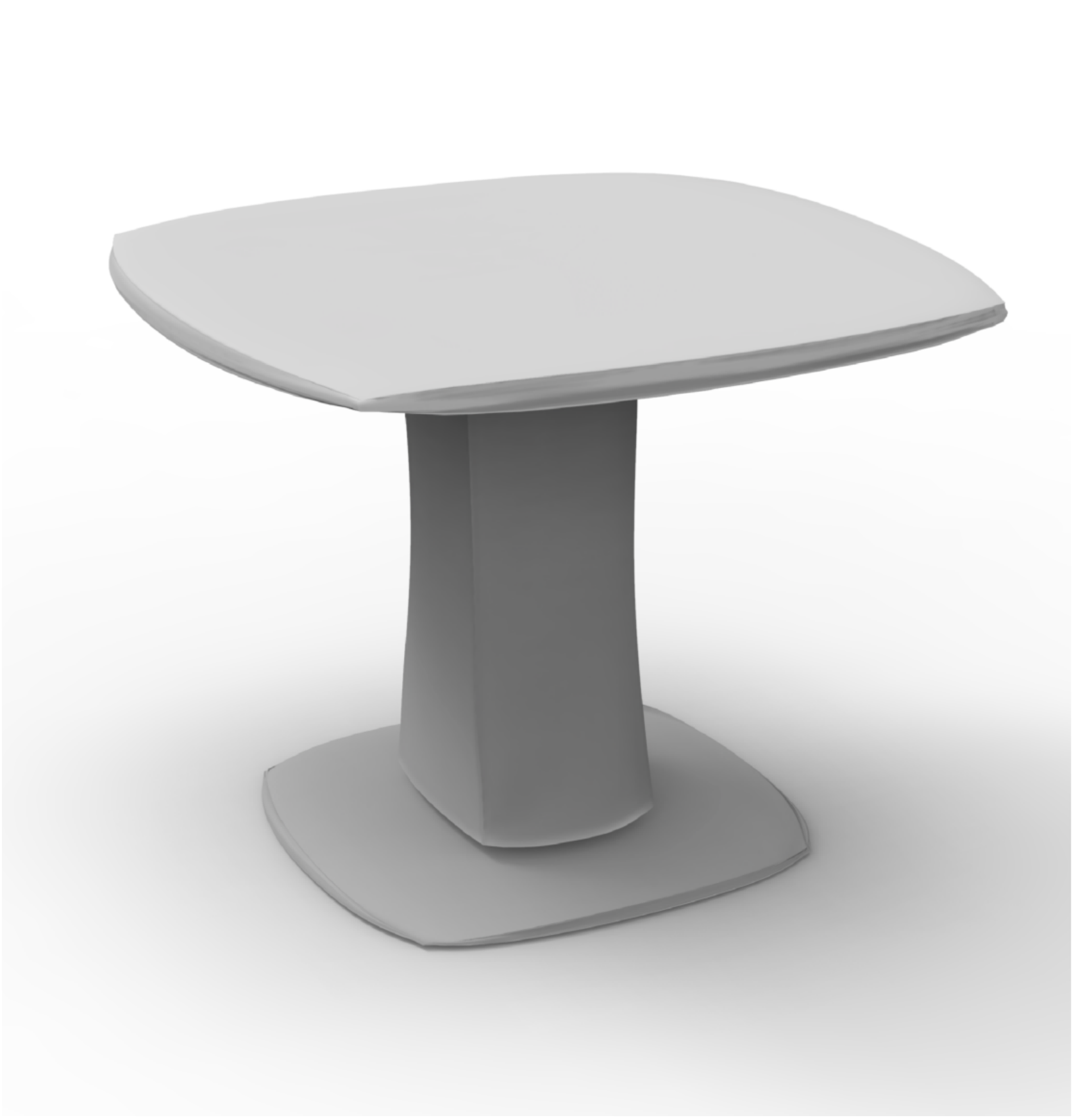}
    \includegraphics[width=0.19\linewidth]{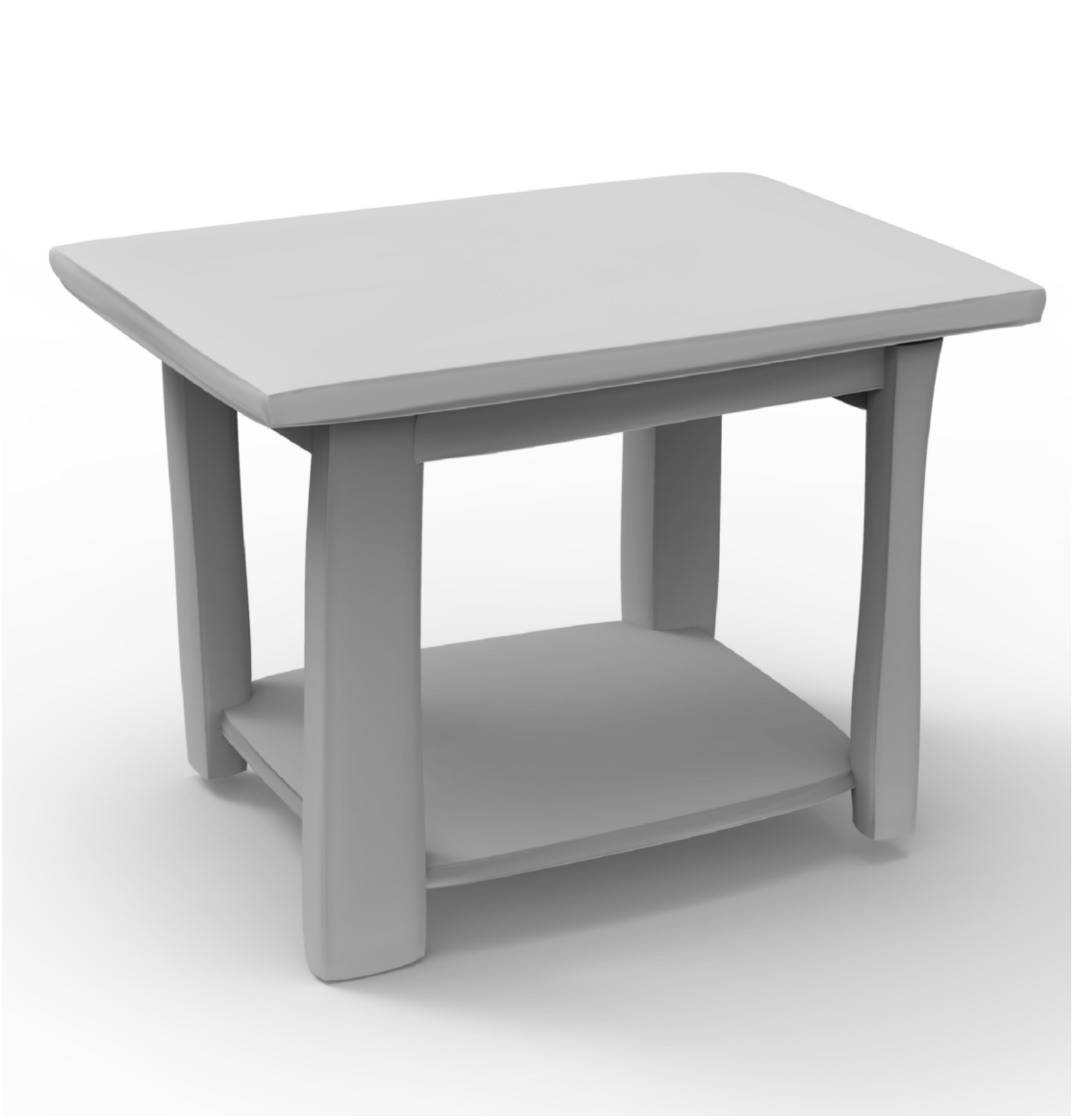}
    \includegraphics[width=0.19\linewidth]{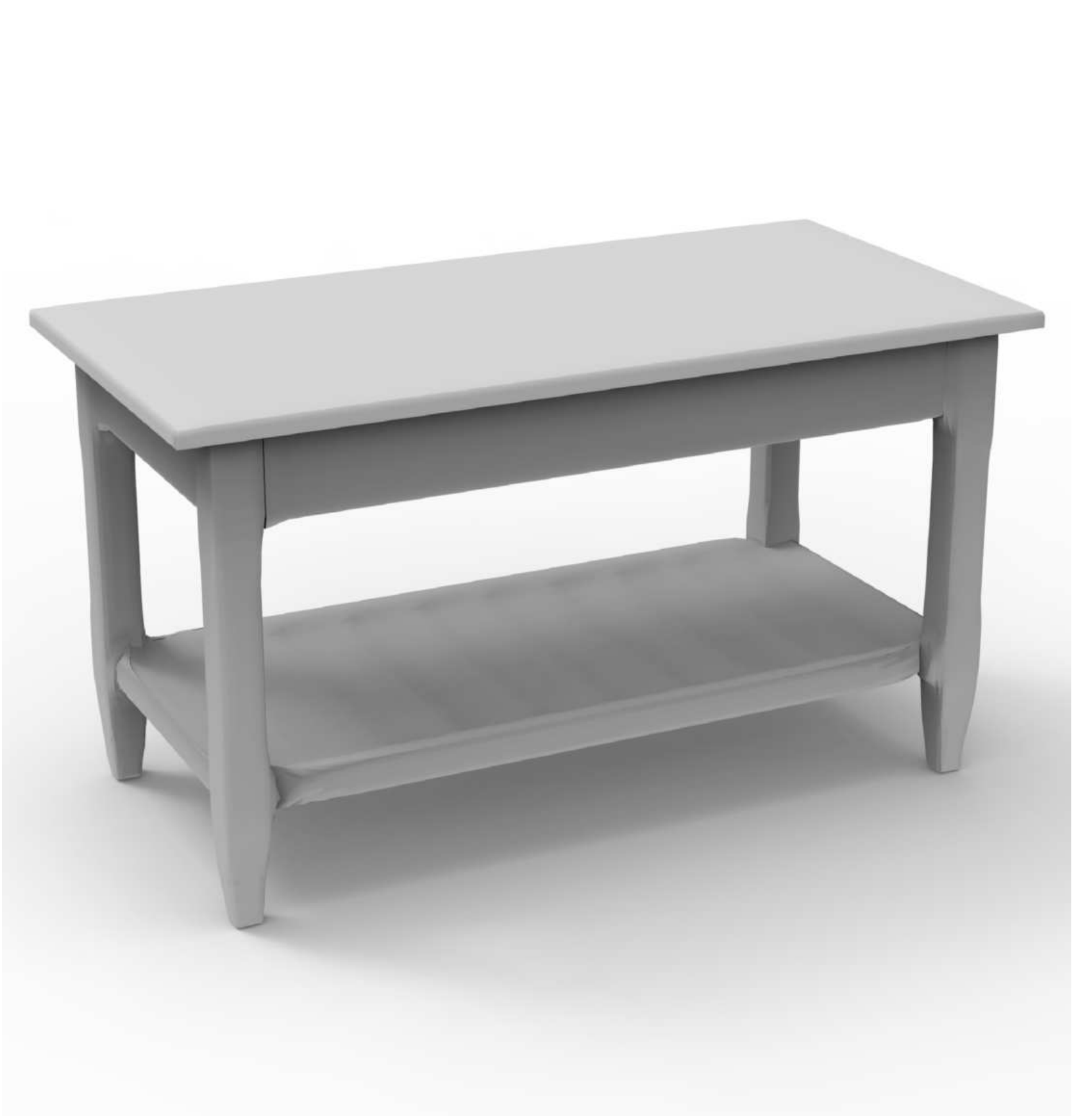}\\
    \includegraphics[width=0.19\linewidth]{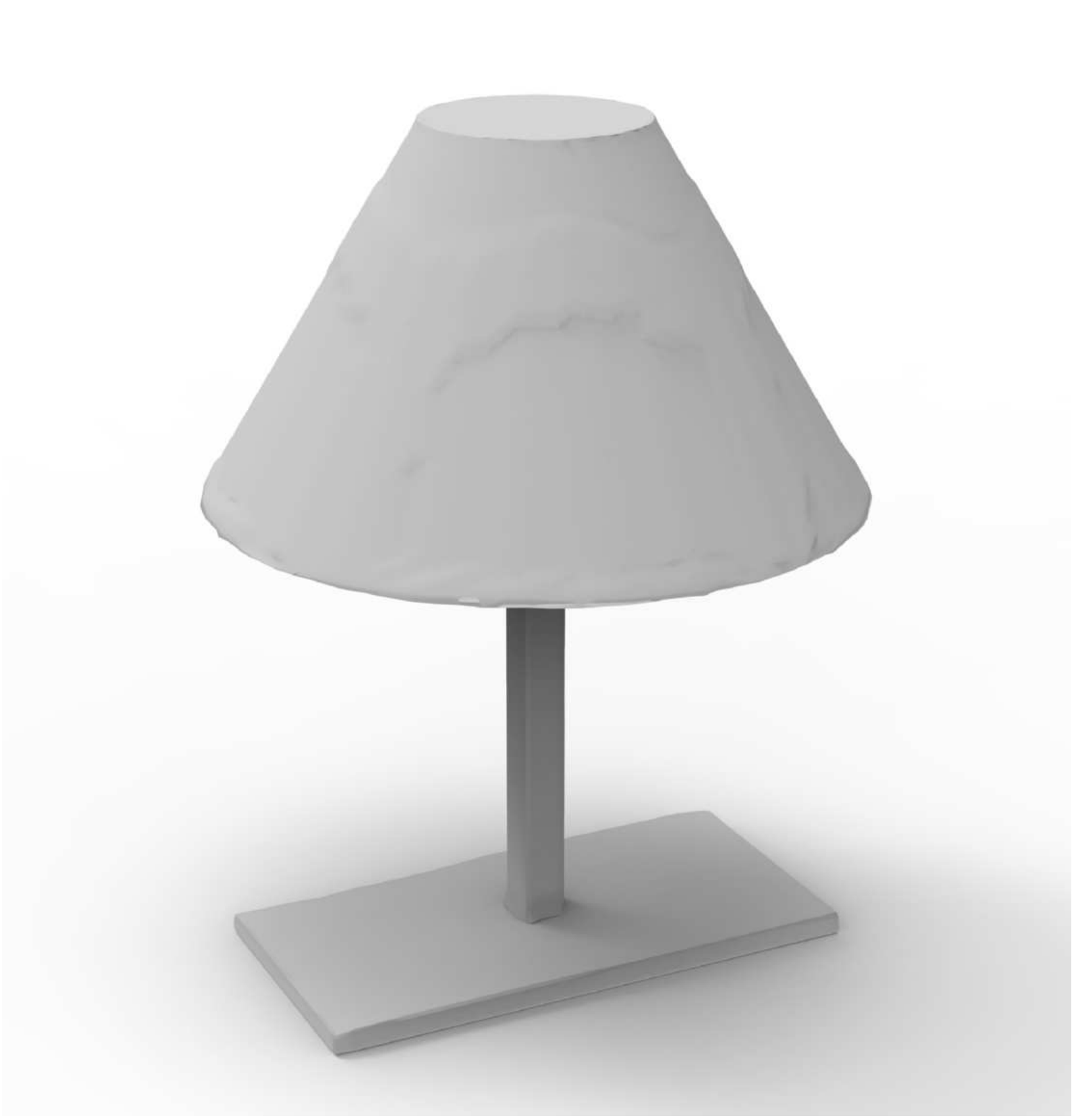}
    \includegraphics[width=0.19\linewidth]{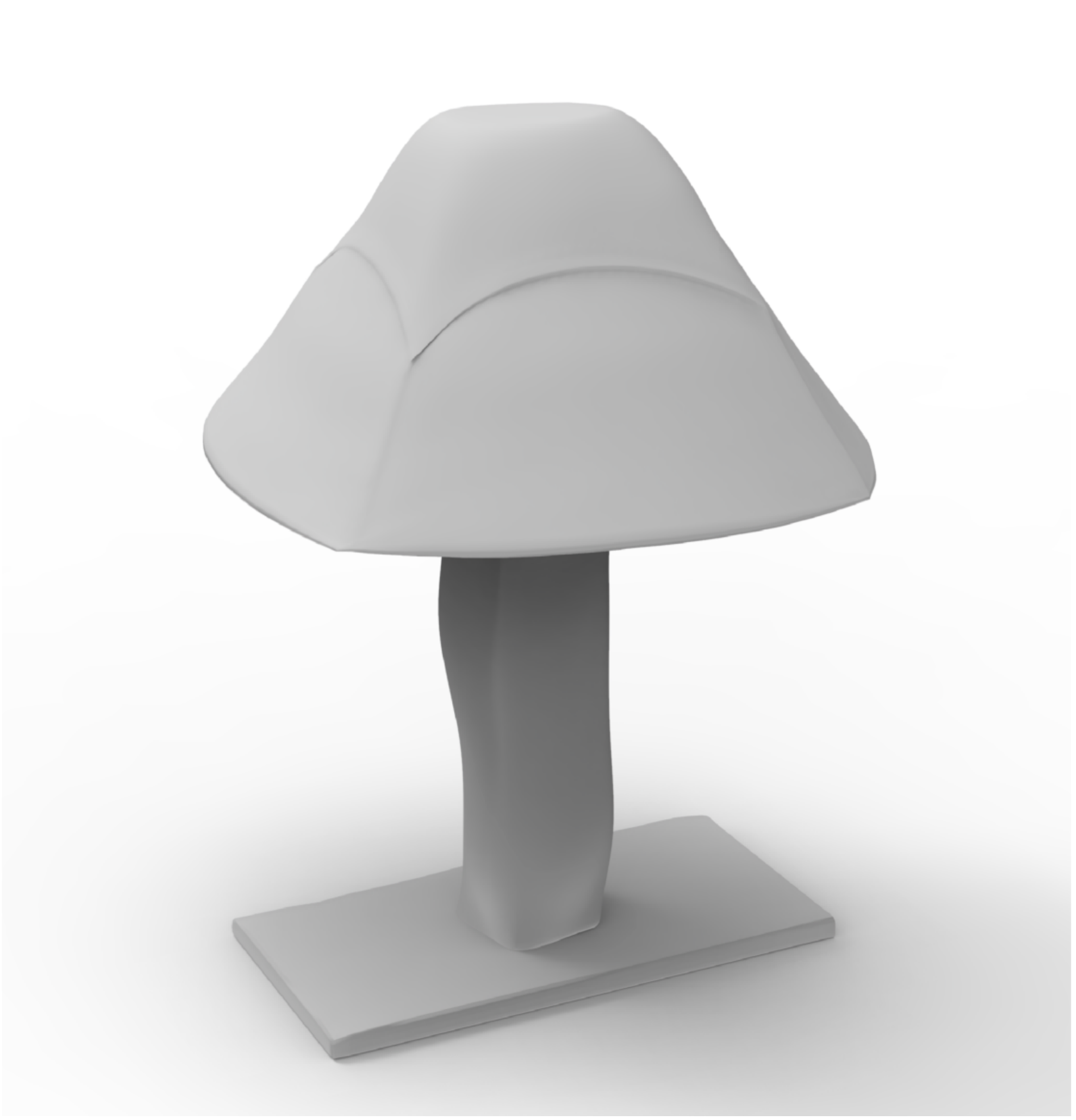}
    \includegraphics[width=0.19\linewidth]{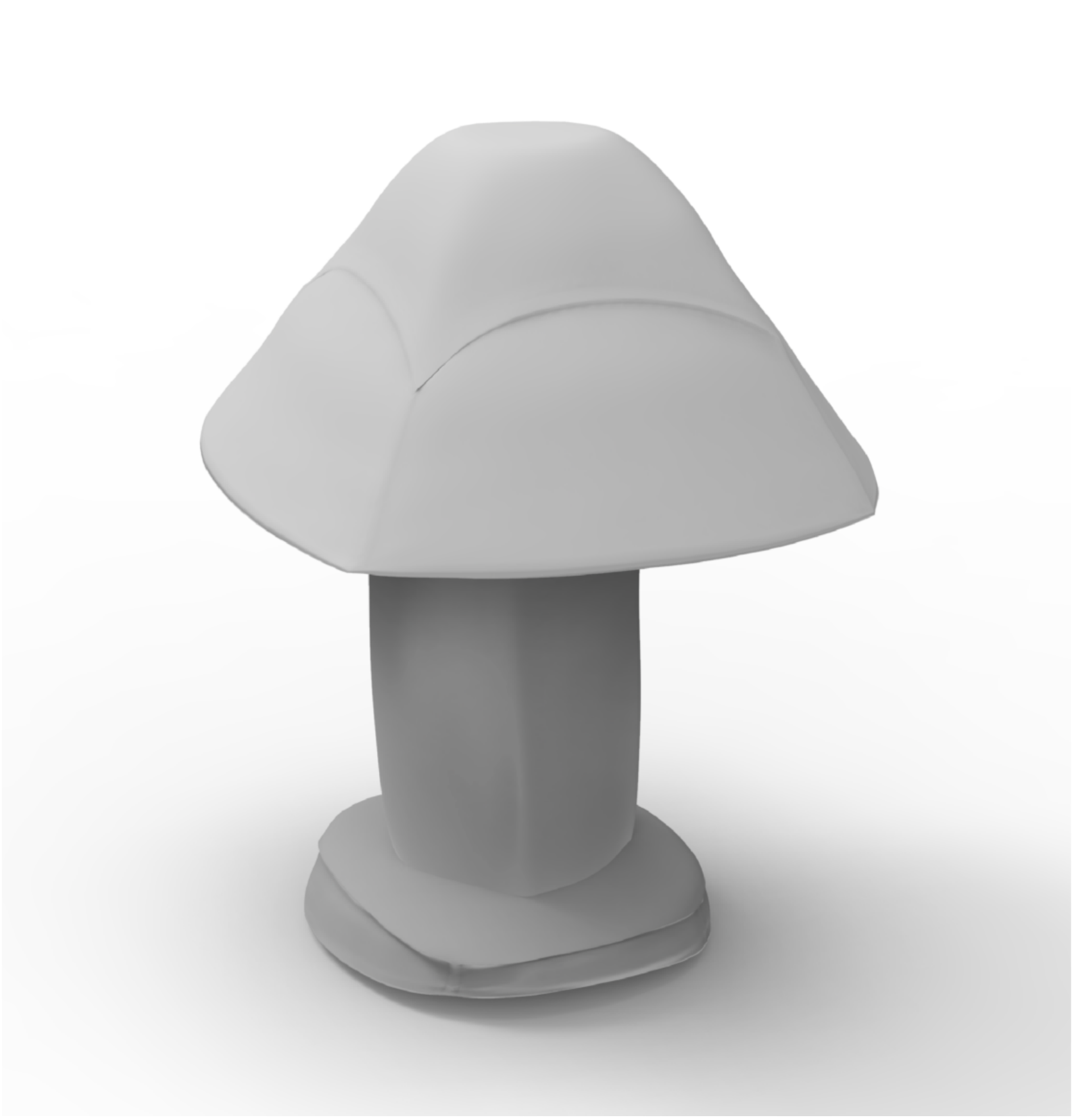}
    \includegraphics[width=0.19\linewidth]{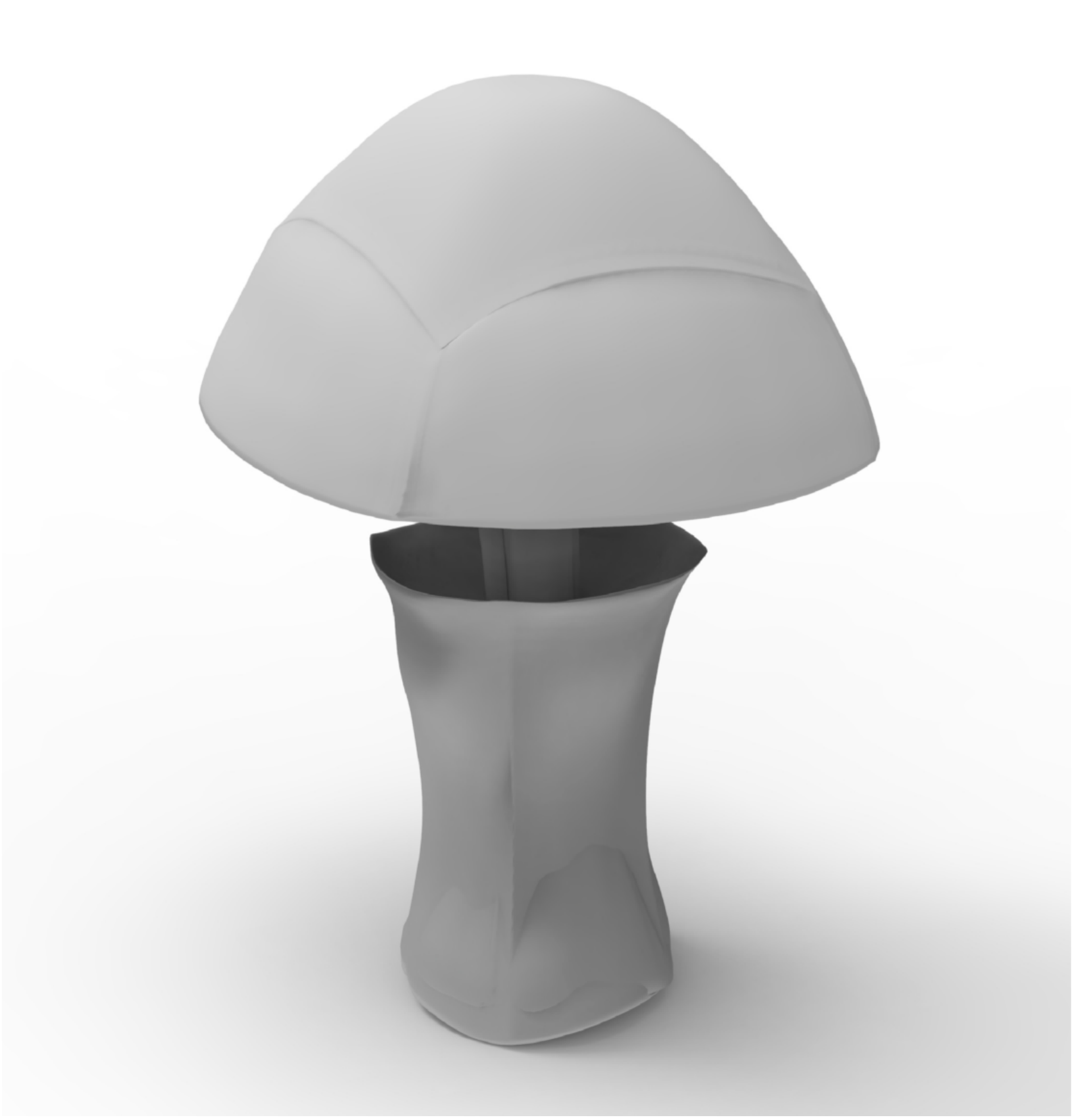}
    \includegraphics[width=0.19\linewidth]{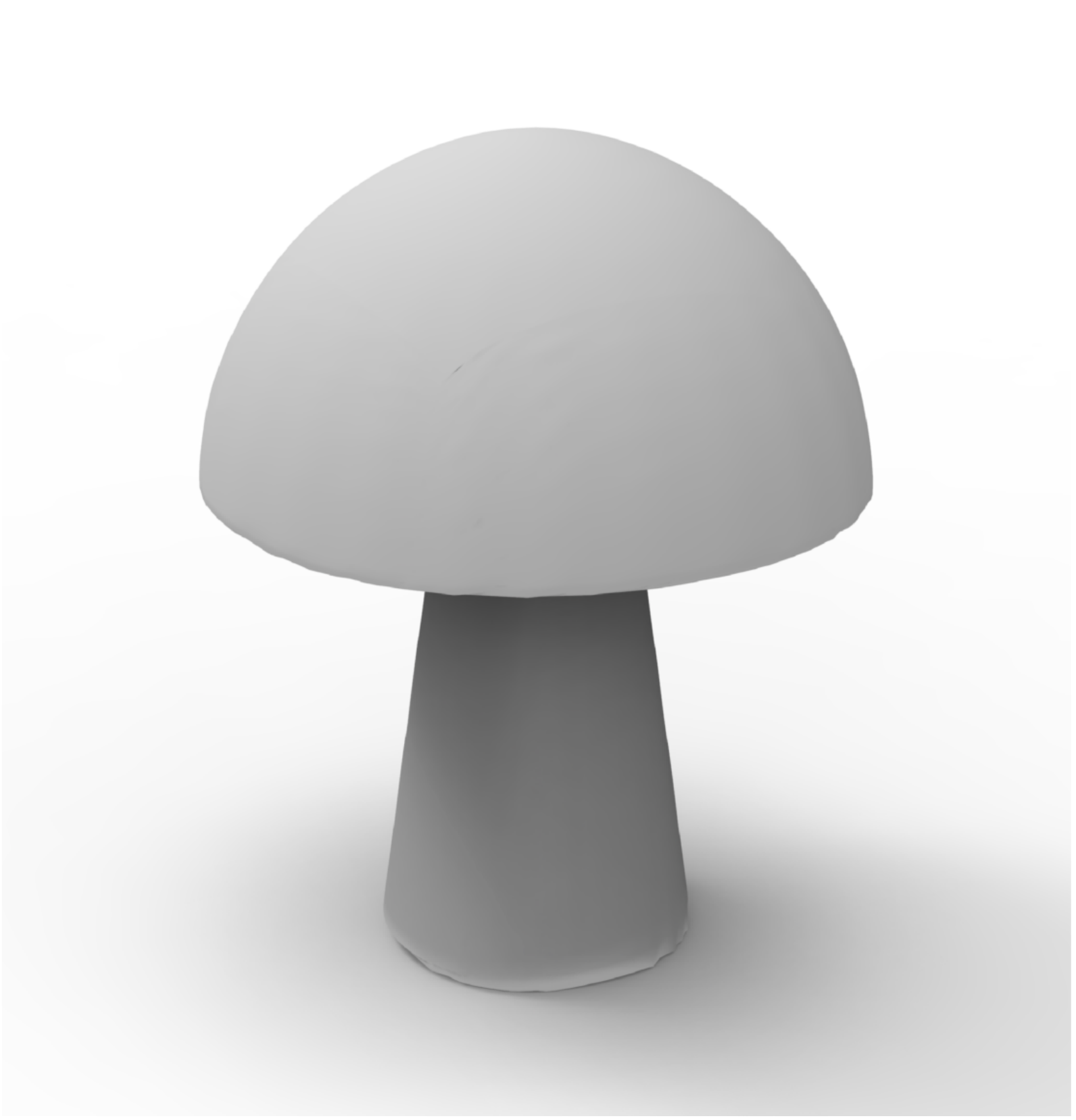}\\
    \includegraphics[width=0.19\linewidth]{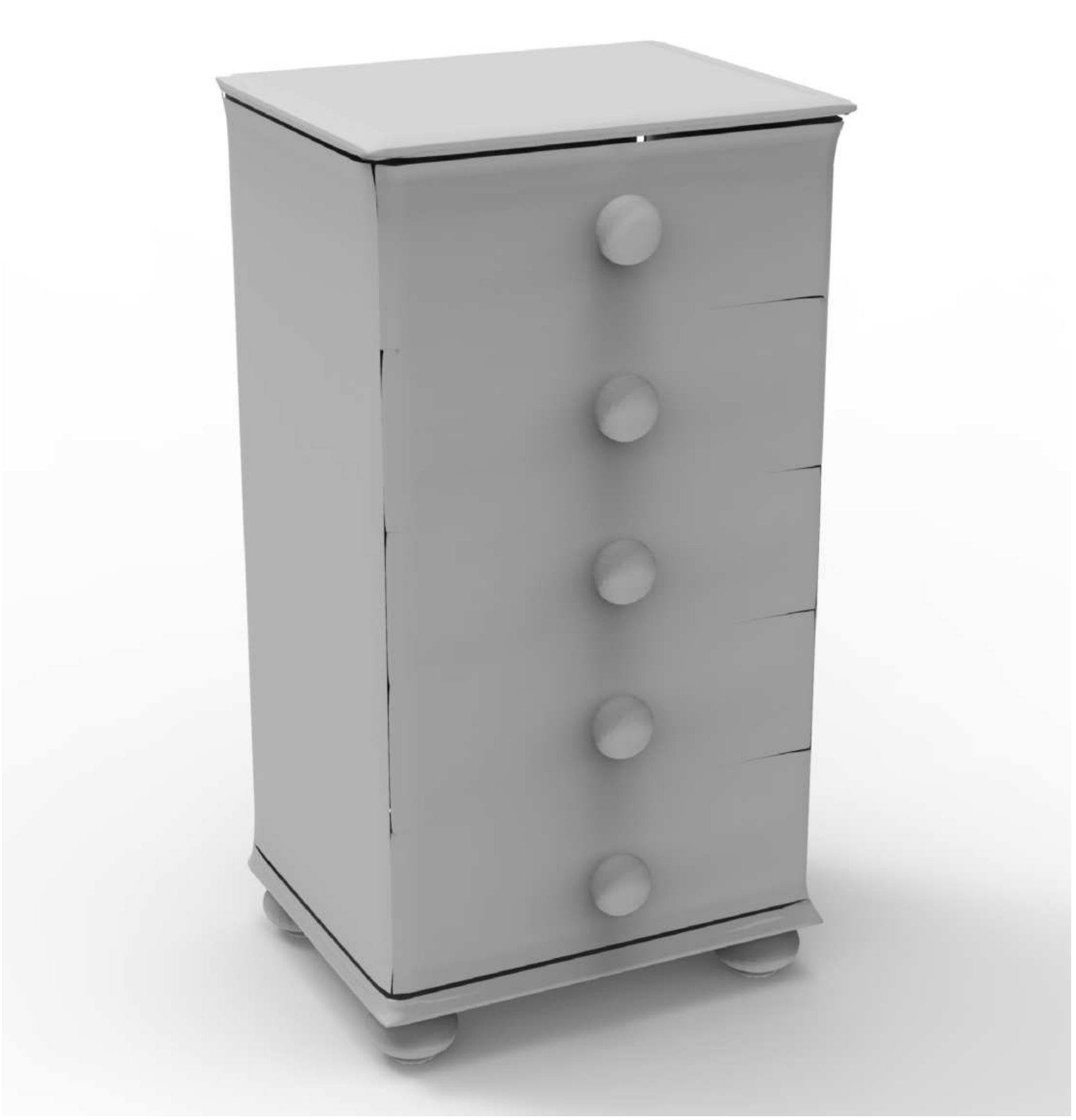}
    \includegraphics[width=0.19\linewidth]{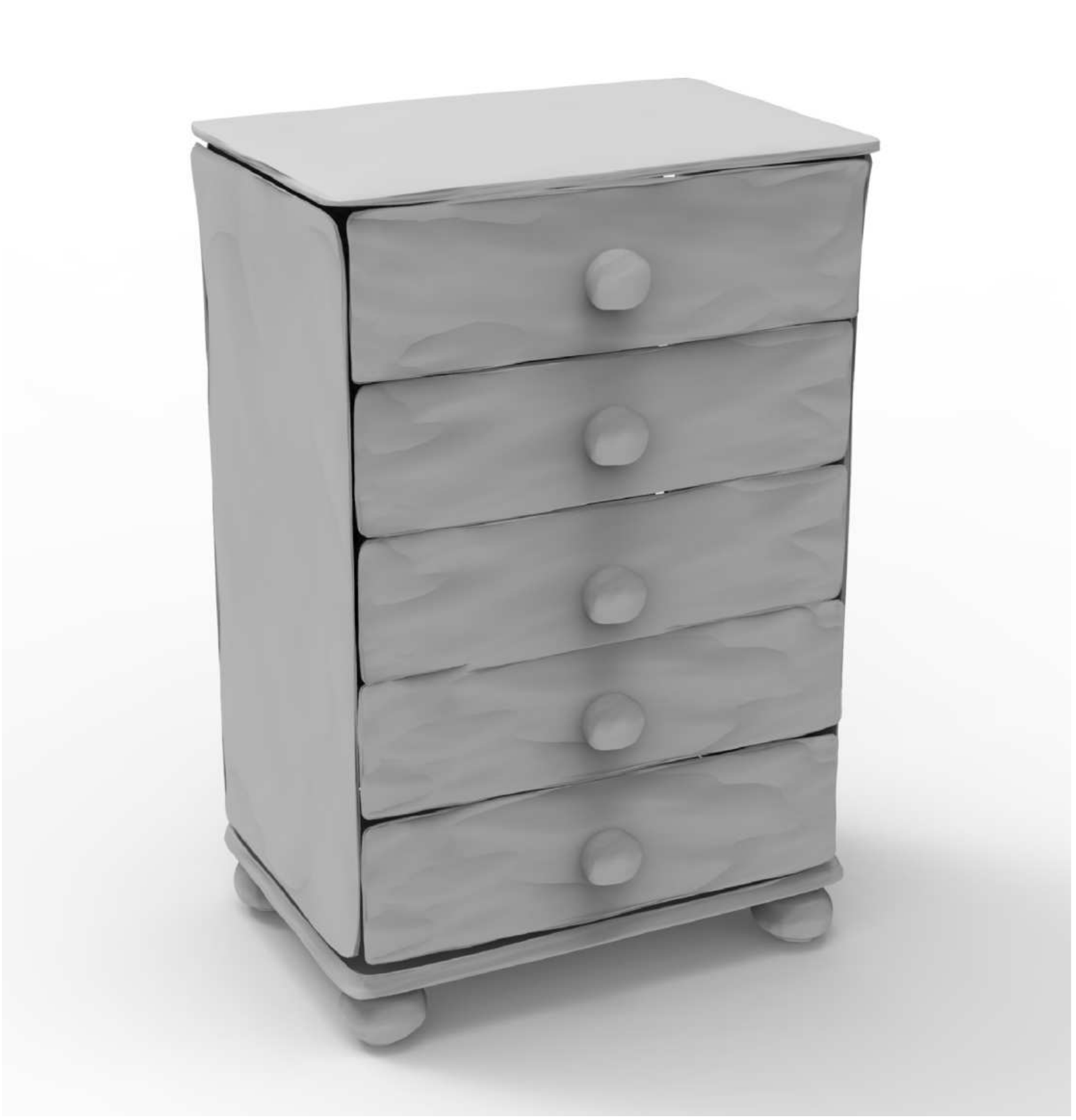}
    \includegraphics[width=0.19\linewidth]{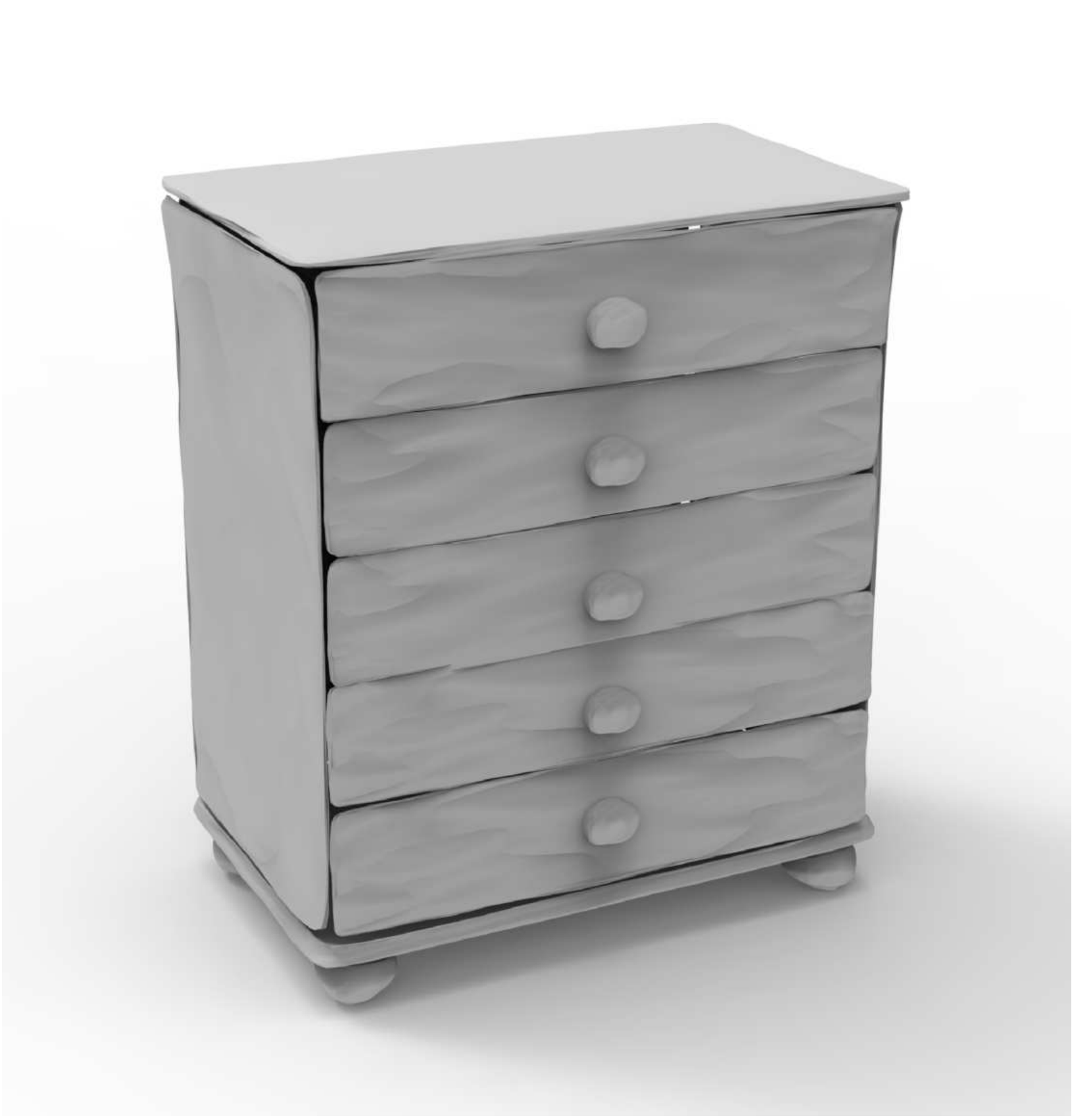}
    \includegraphics[width=0.19\linewidth]{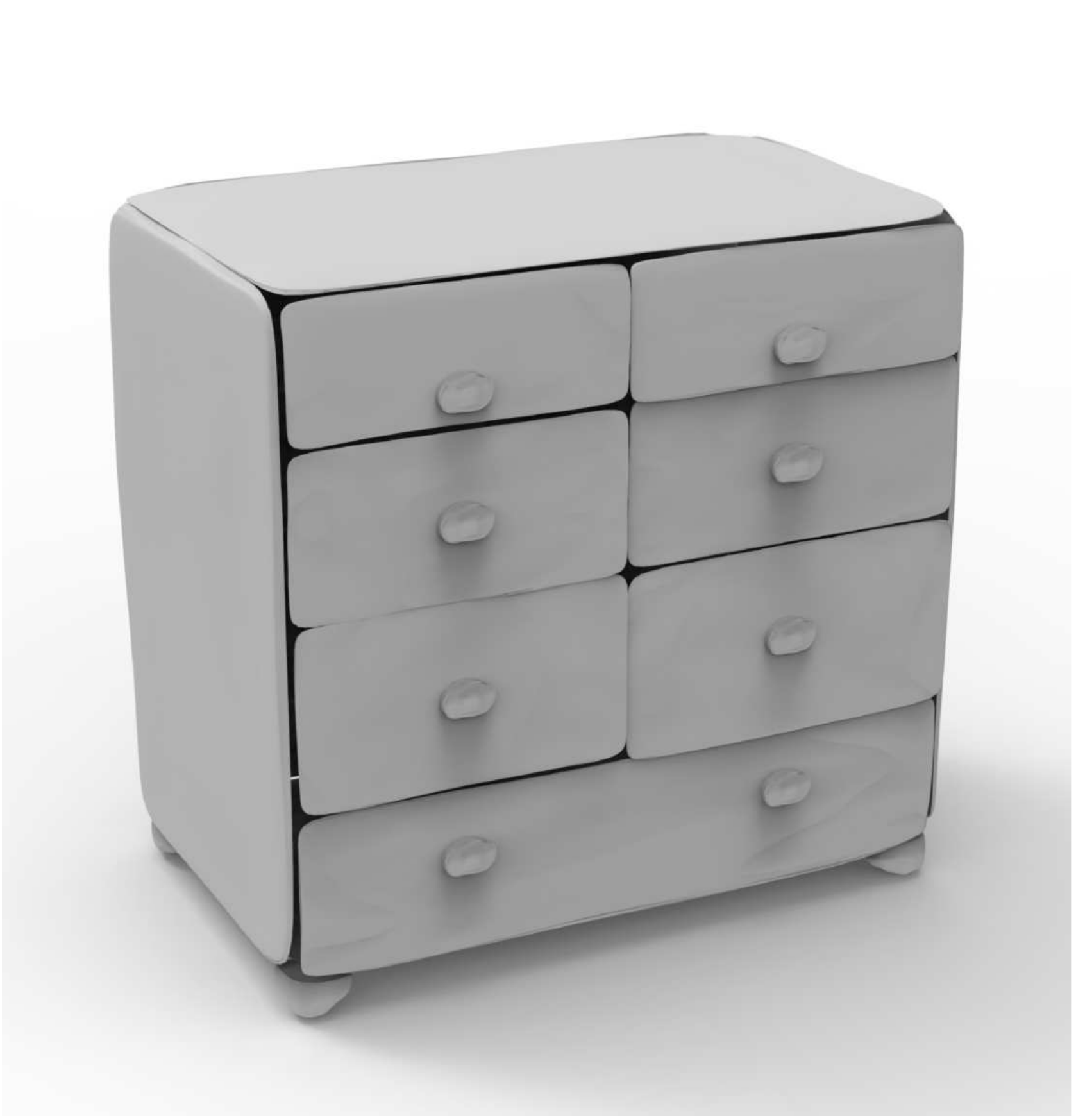}
    \includegraphics[width=0.19\linewidth]{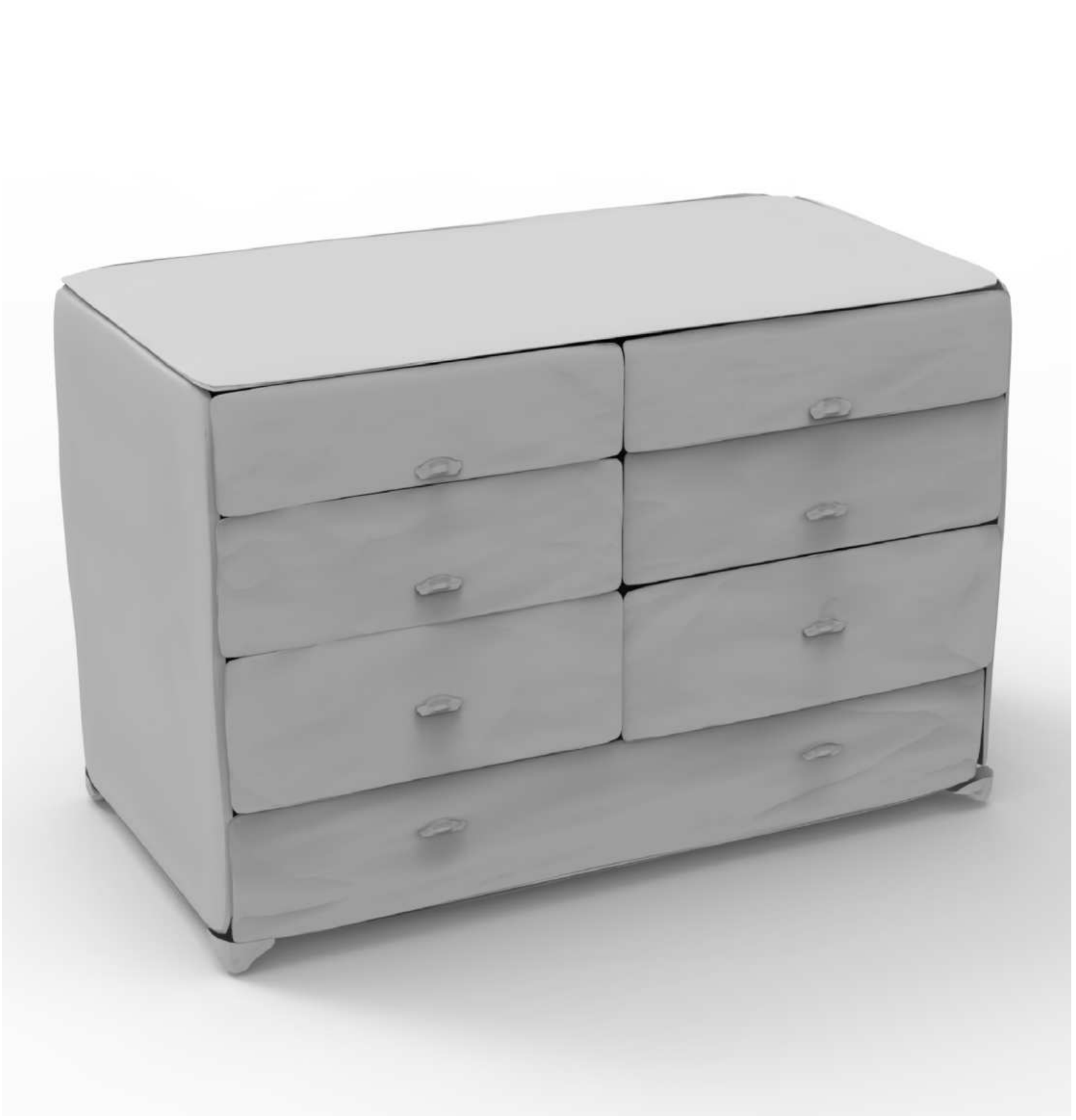}
    \vspace{-3mm}
    \caption{\yjr{Shape interpolation results. \yj{We linearly interpolate between input shape pairs (the left most and the right most shapes) jointly in the structure and geometry latent spaces. We see both continuous geometry variations and discrete structure changes. For the chair examples, in the first row, we see that the armrests become smaller and then disappear while the backrest changes from a square to round fashion in a more natural manner, while in the second row, the backrest gradually becomes square, while the supporter disappears form the first chair to the second chair.
    We observe similar behaviors for the table, lamp, and cabinet results.}}}
    \label{fig:interp_storage}
    \vspace{-3mm}
\end{figure}

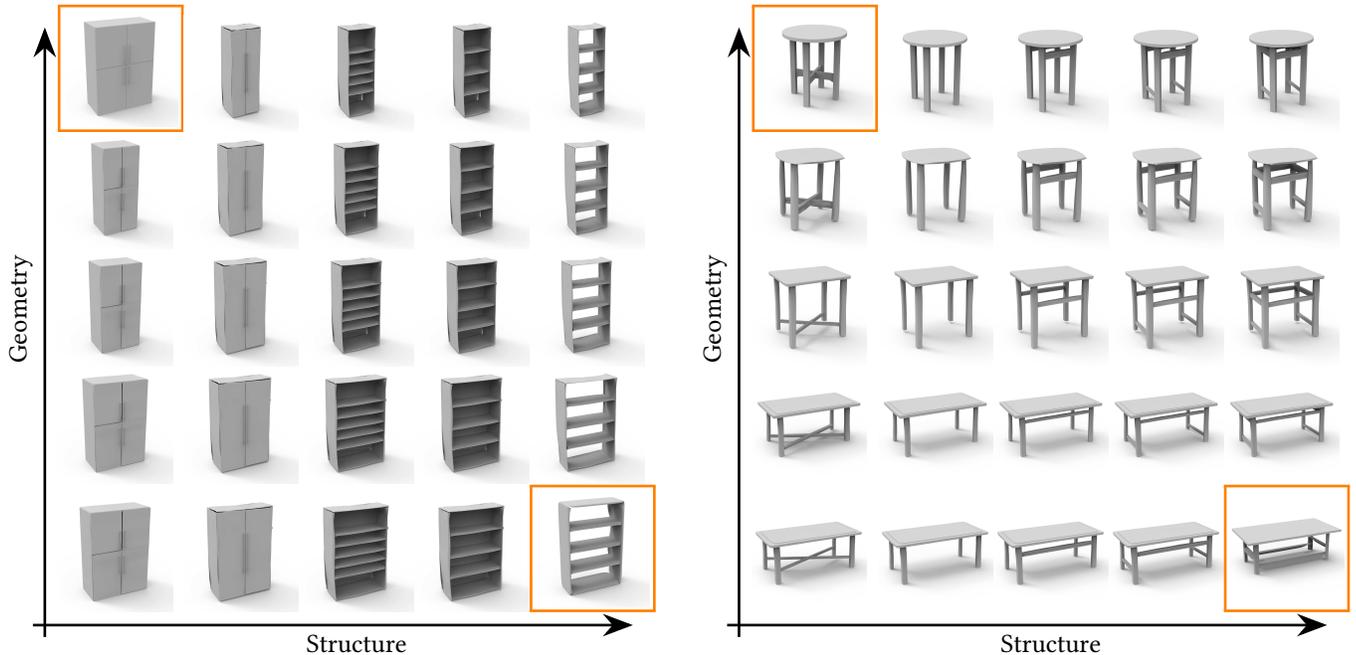
\begin{figure*}[t]
\begin{minipage}[b]{0.45\linewidth}
\centering
  \begin{tikzpicture}
  \matrix[nodes={anchor=south west,inner sep=0pt}]{ 
    \node (A1) {\includegraphics[width=0.19\linewidth, cfbox=orange 1pt 1pt]{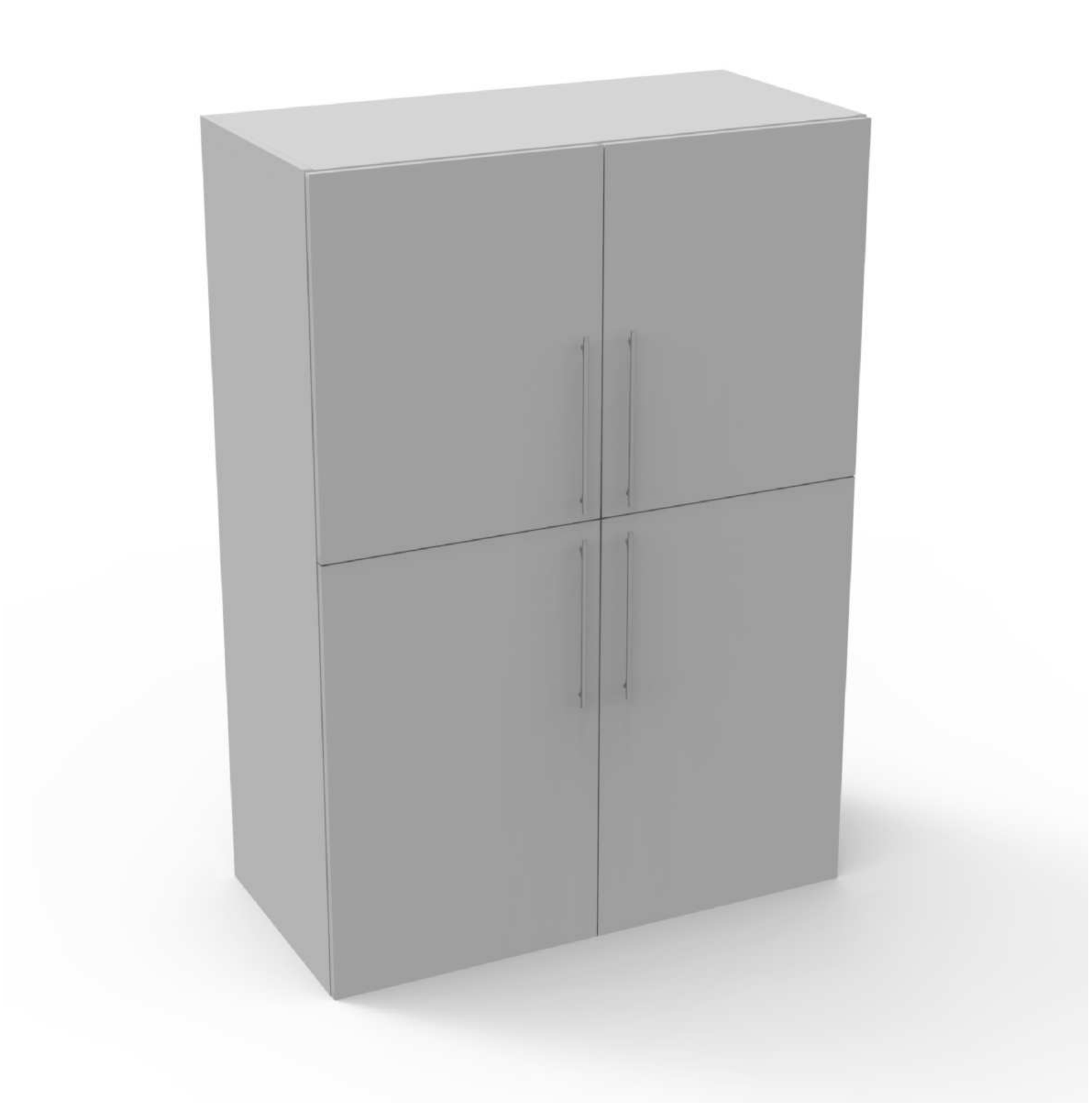}}; &
    \node (A2) {\includegraphics[width=0.19\linewidth]{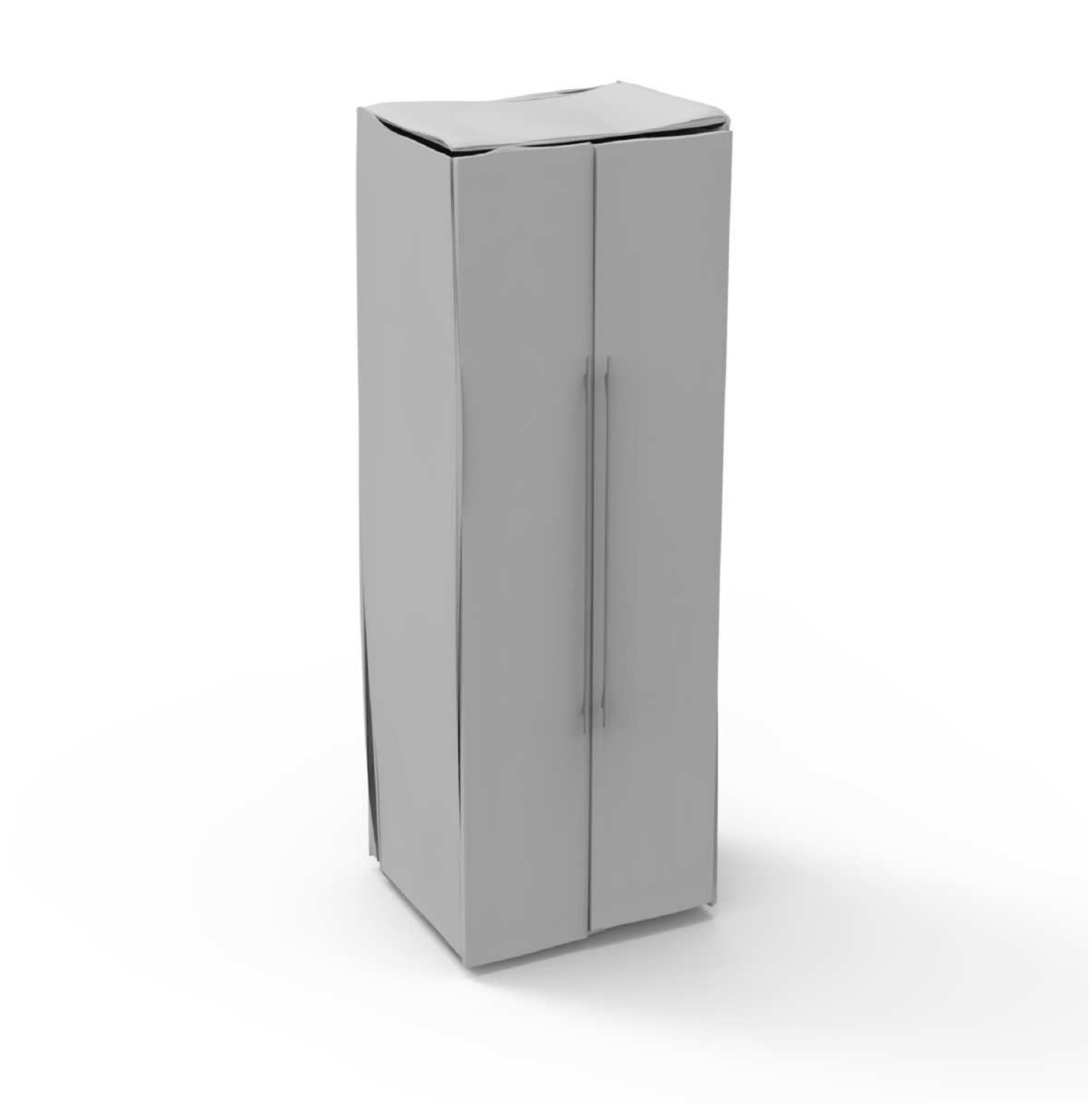}}; &
    \node (A3) {\includegraphics[width=0.19\linewidth]{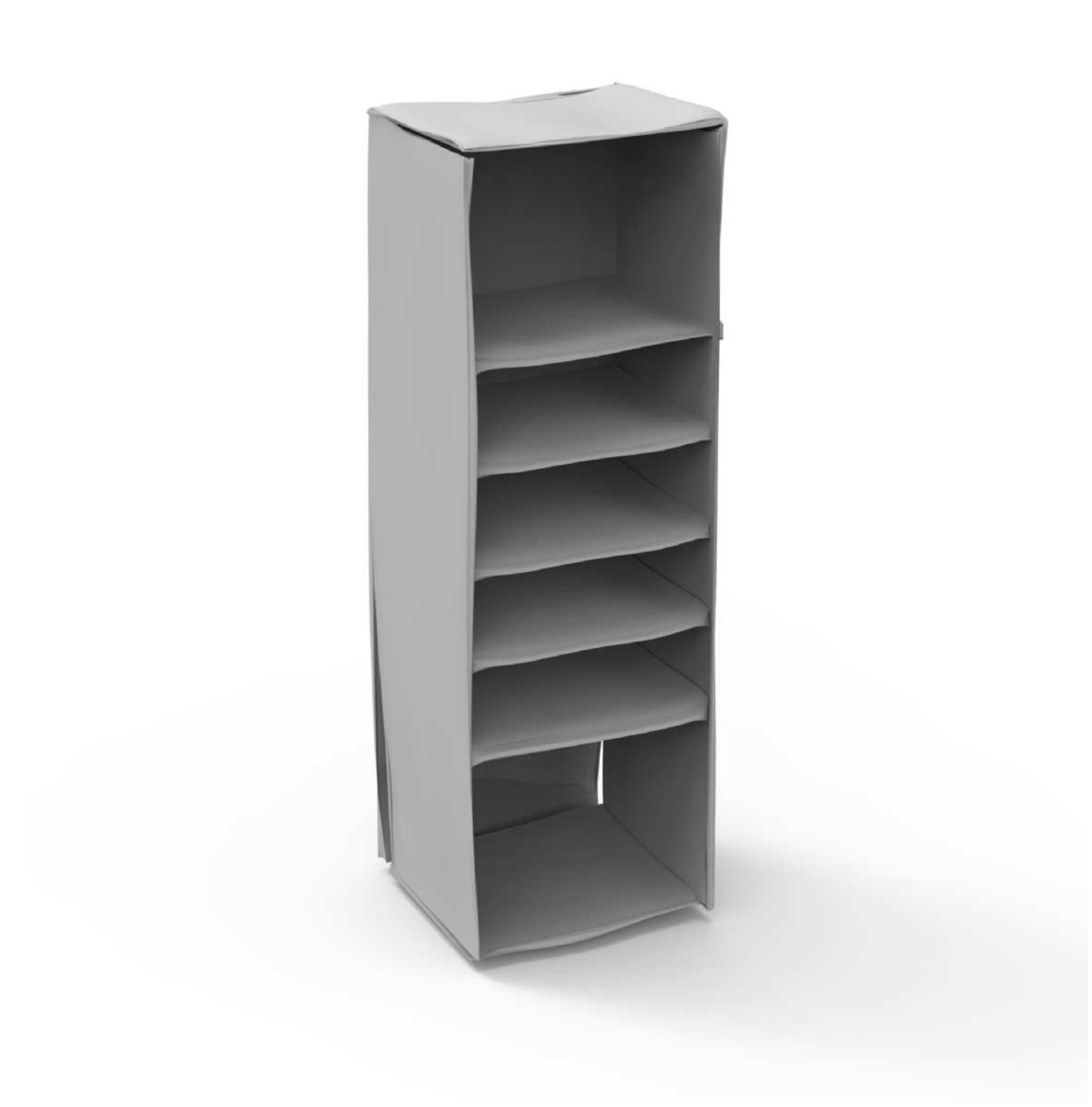}}; &
    \node (A4) {\includegraphics[width=0.19\linewidth]{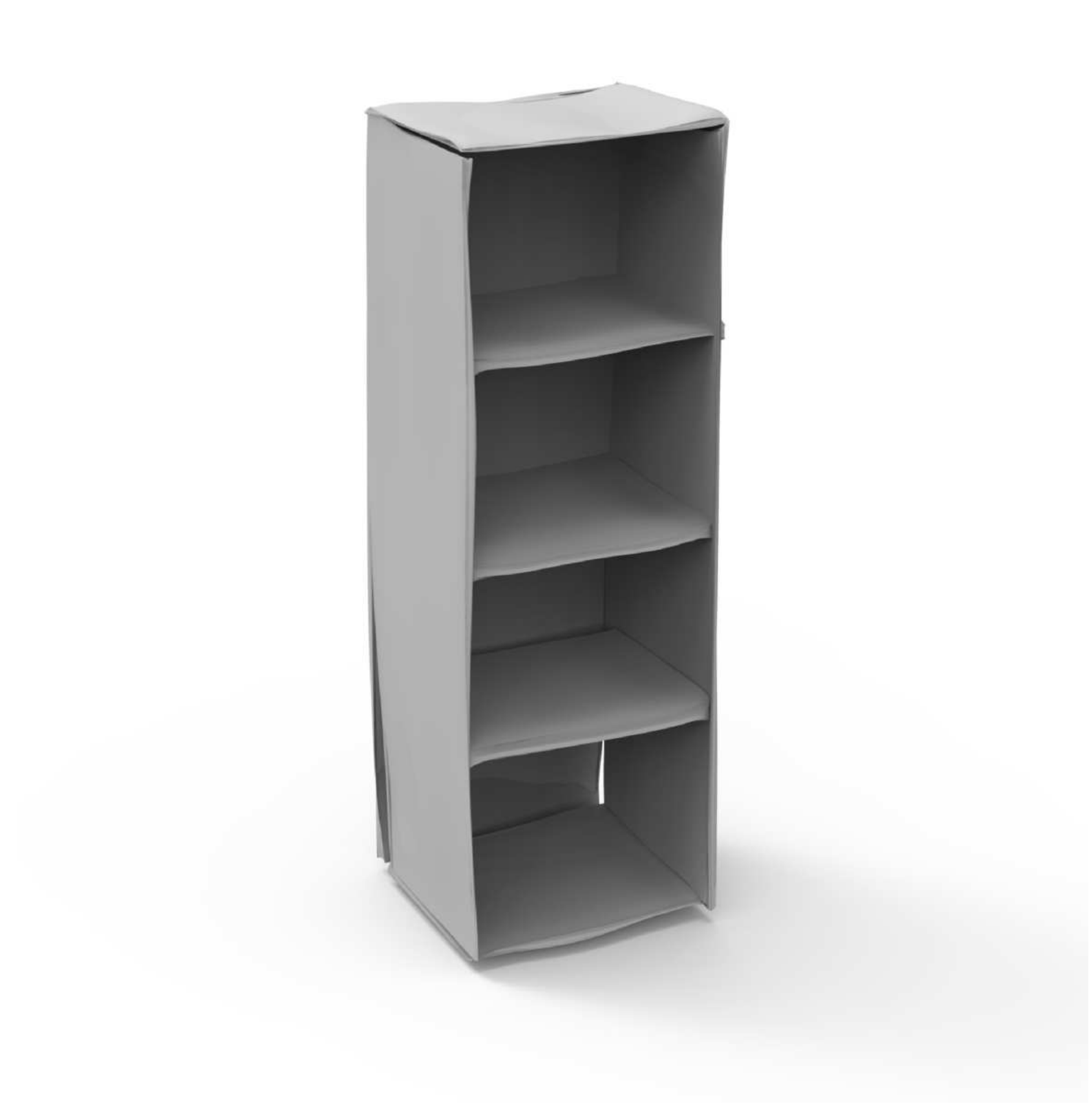}}; &
    \node (A5) {\includegraphics[width=0.19\linewidth]{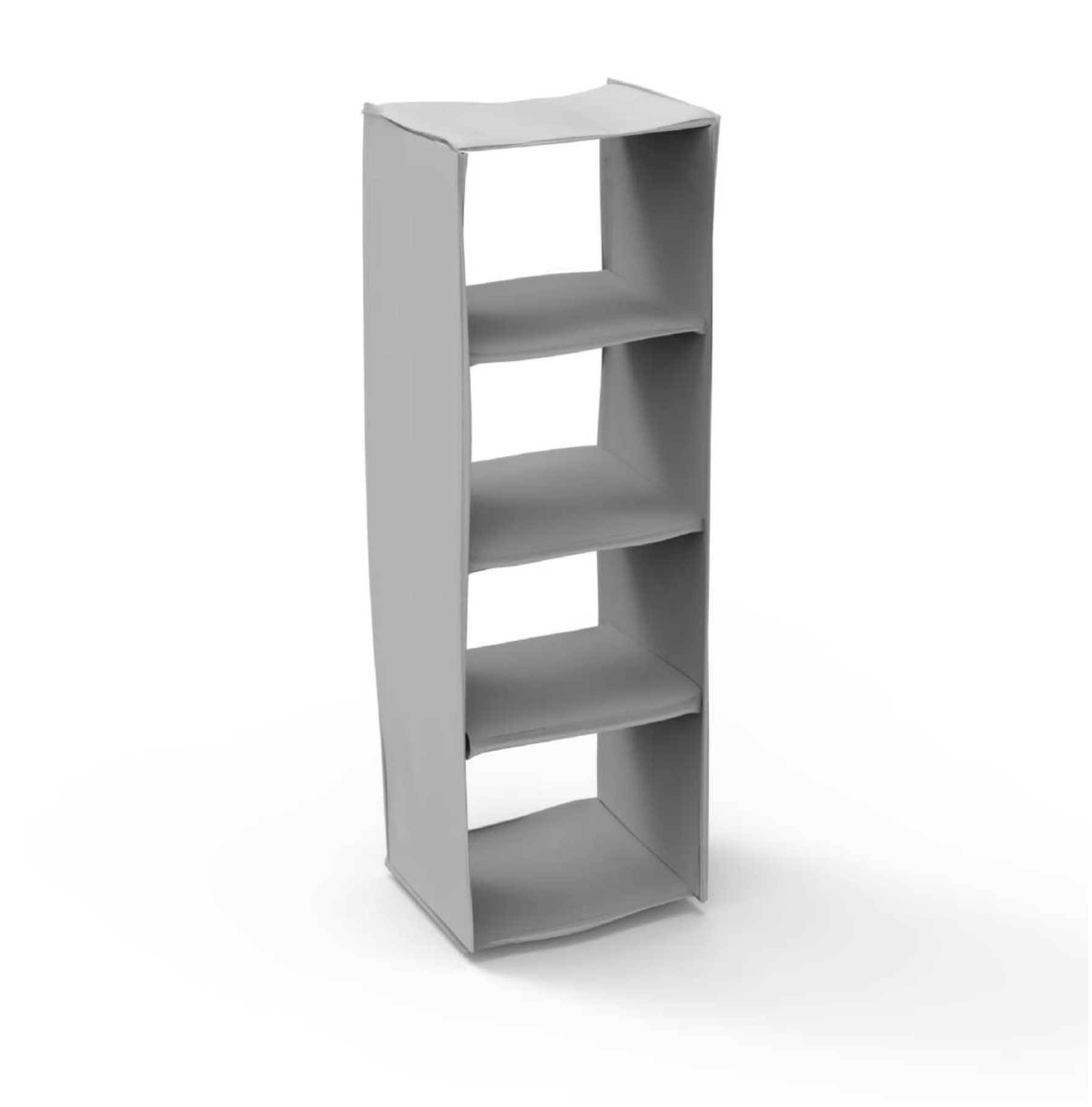}}; \\
    \node (B1) {\includegraphics[width=0.19\linewidth]{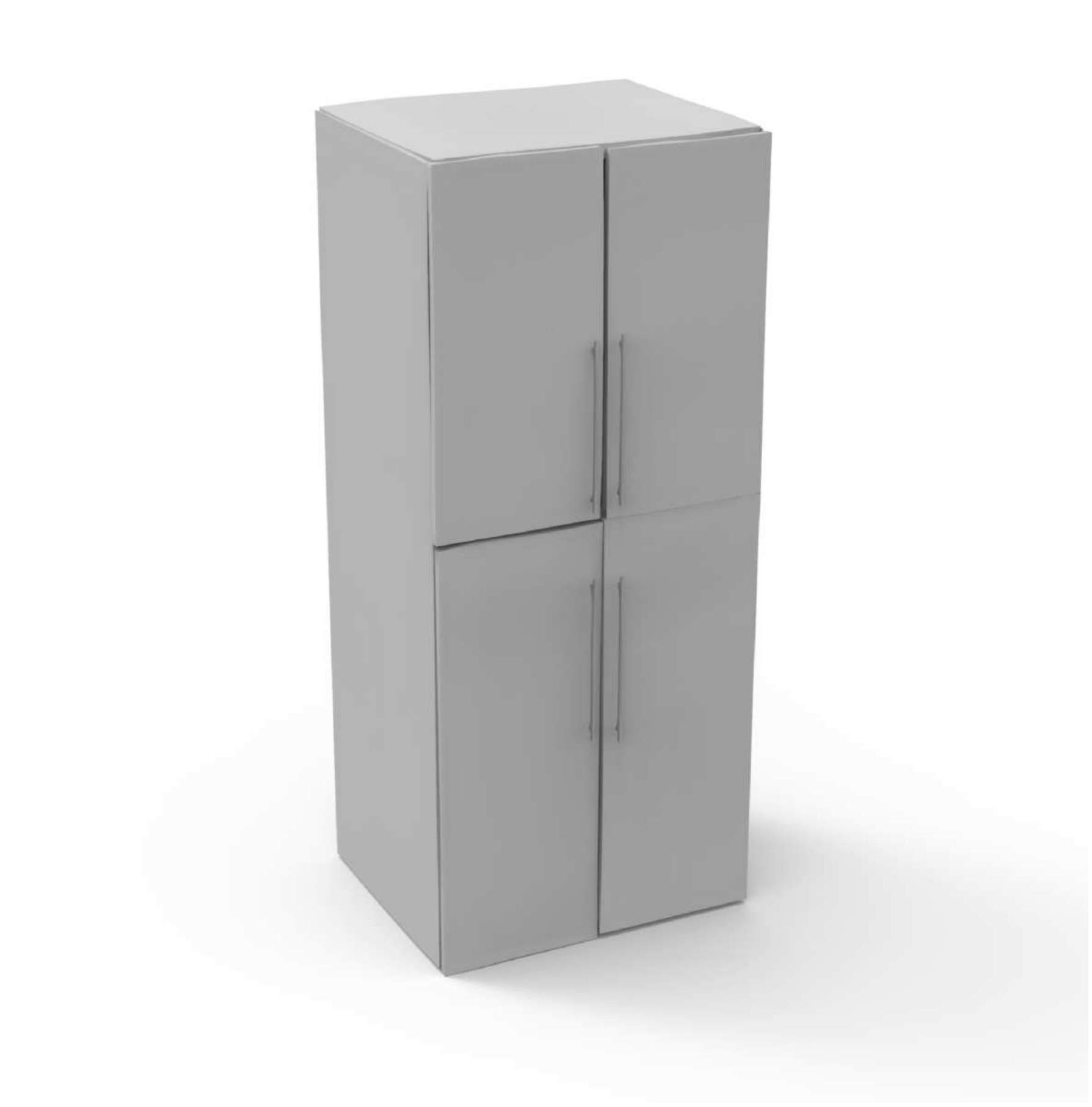}}; &
    \node (B2) {\includegraphics[width=0.19\linewidth]{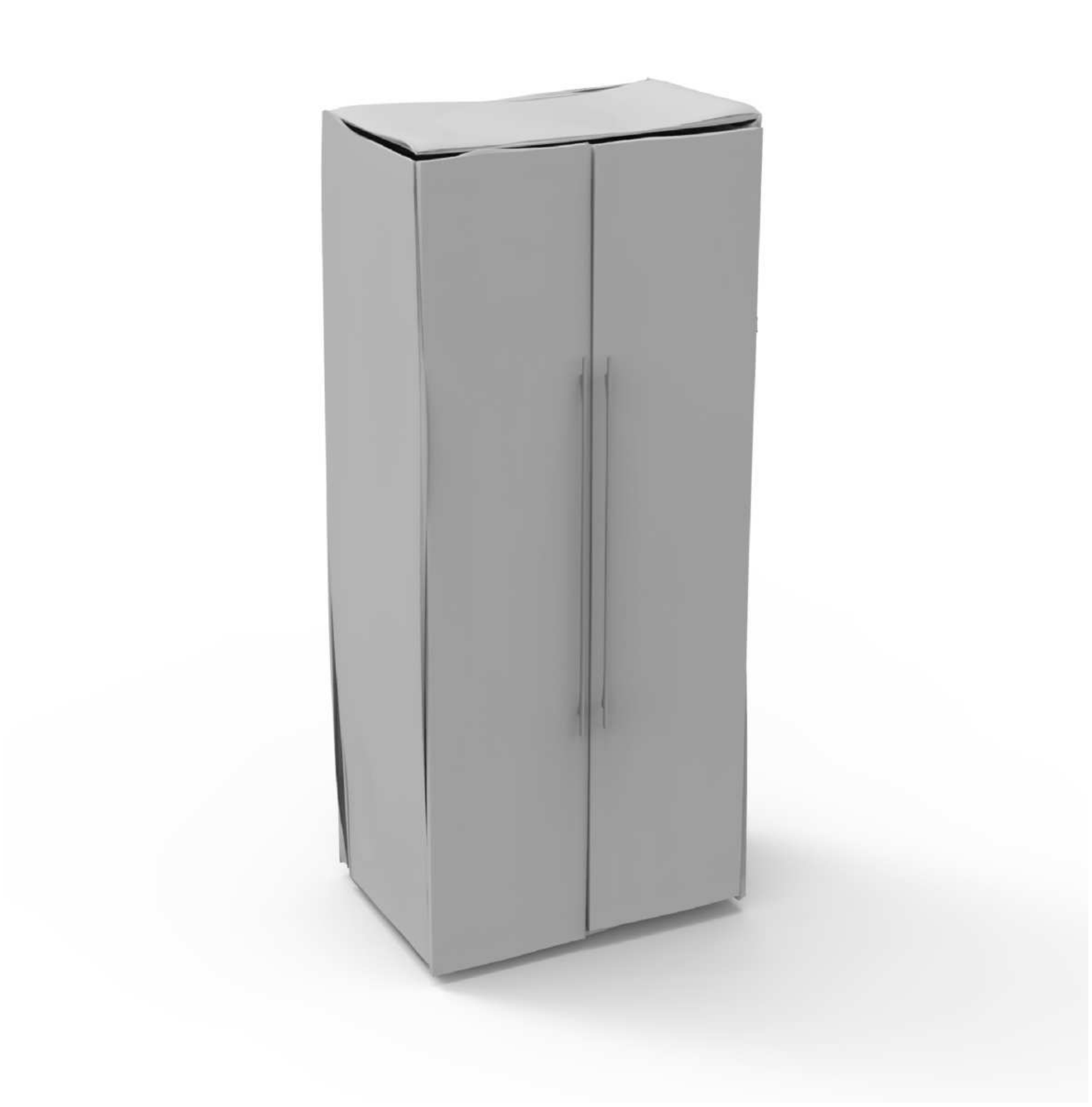}}; &
    \node (B3) {\includegraphics[width=0.19\linewidth]{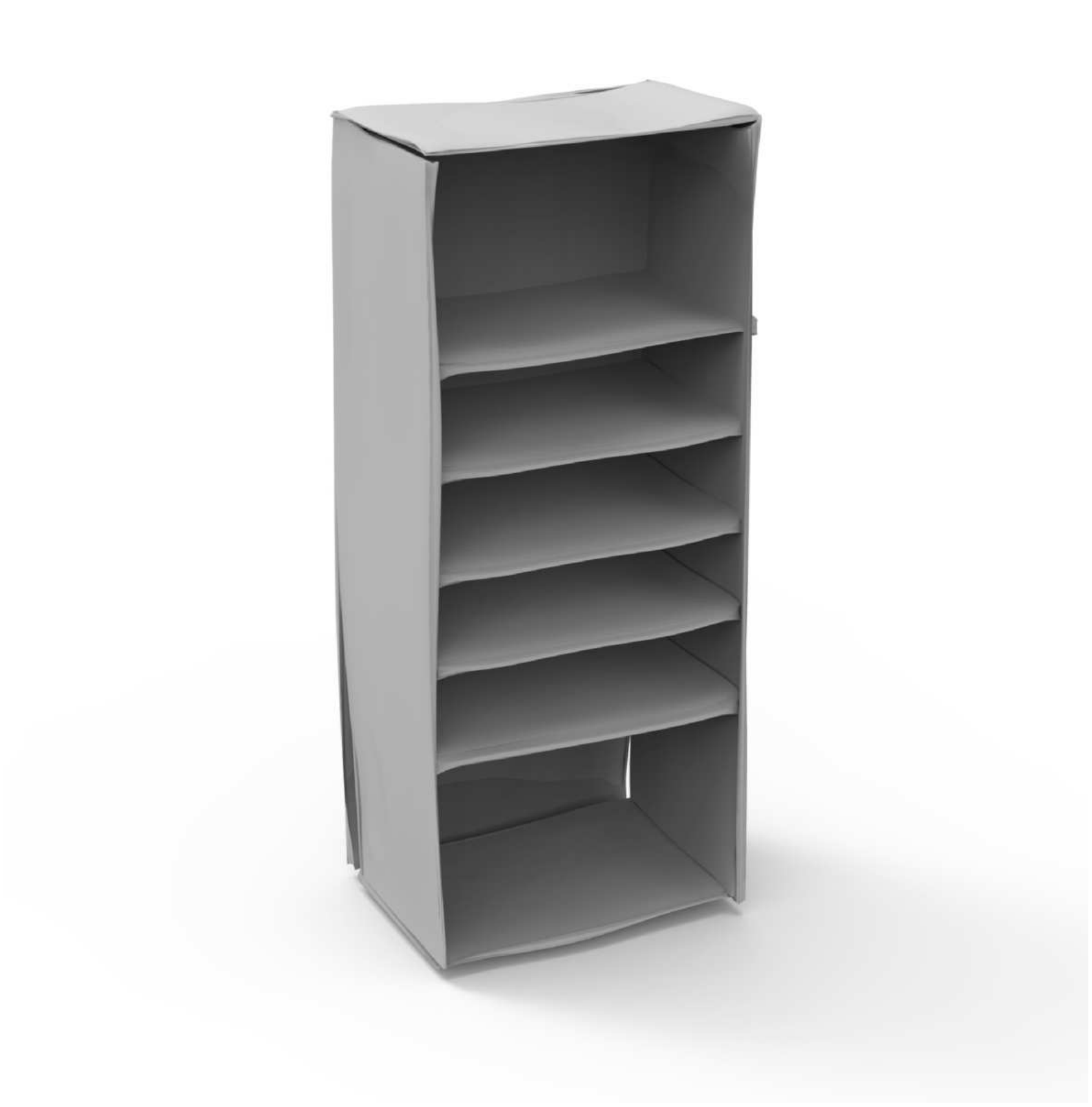}}; &
    \node (B4) {\includegraphics[width=0.19\linewidth]{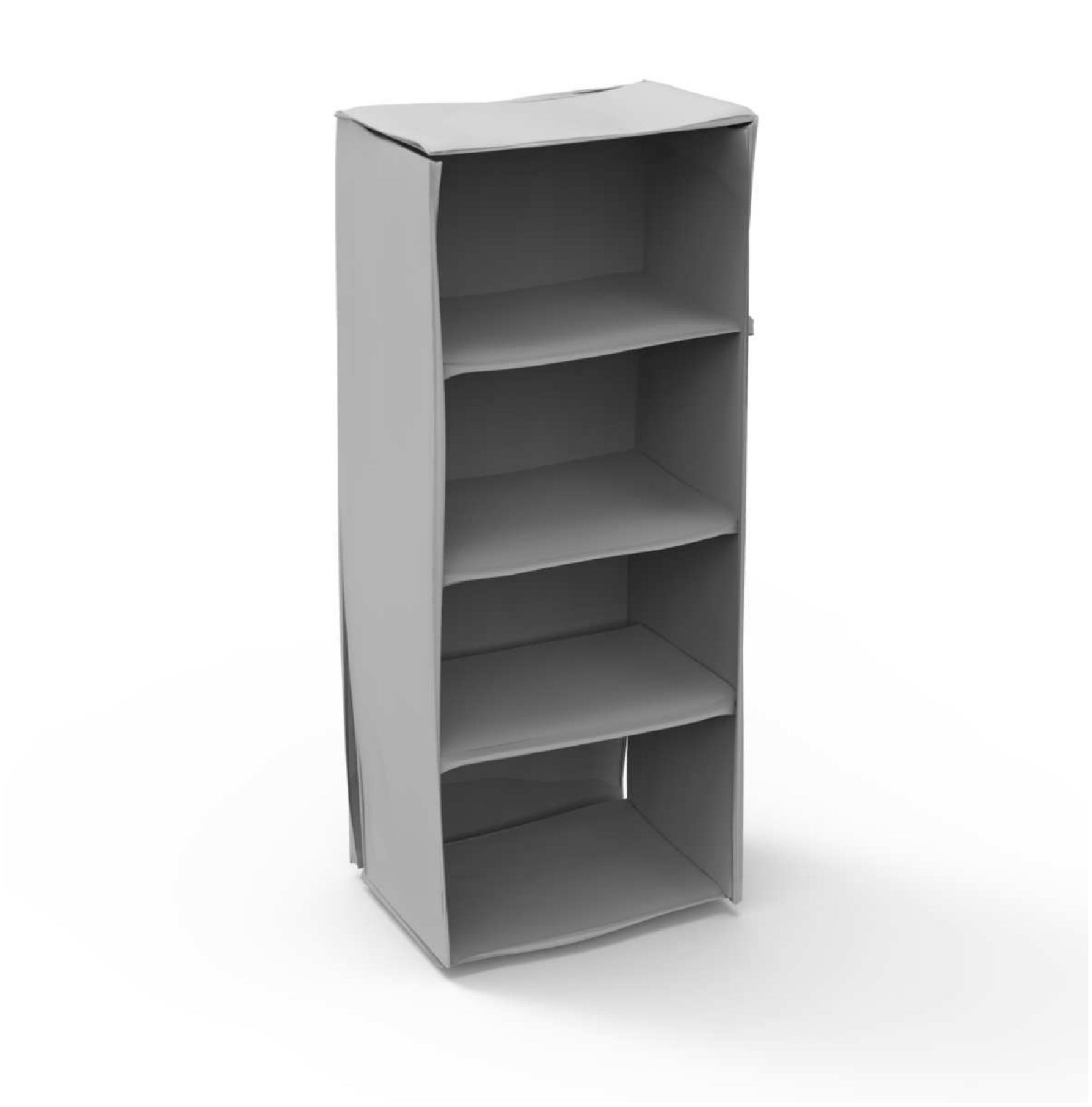}}; &
    \node (B5) {\includegraphics[width=0.19\linewidth]{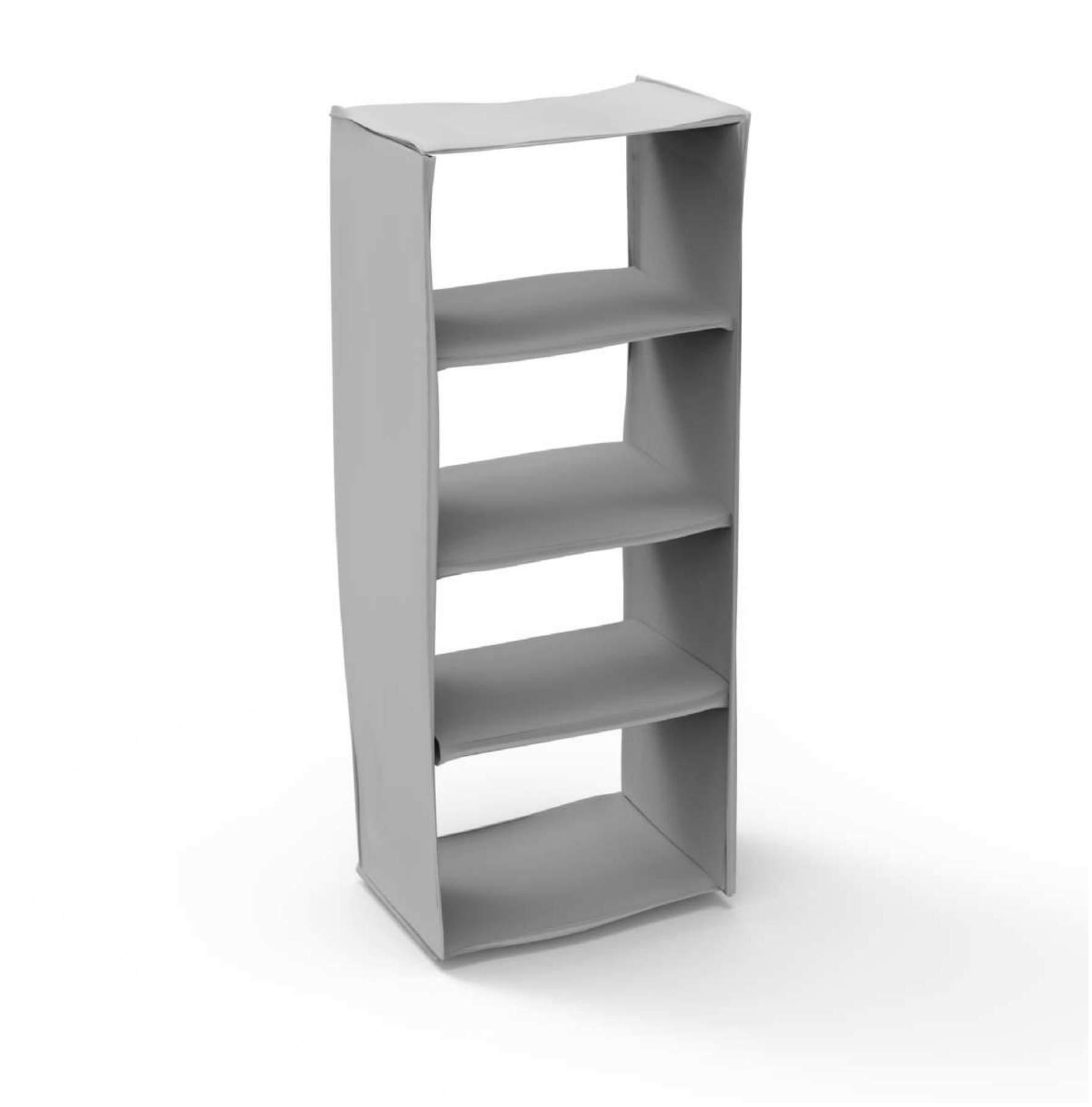}}; \\
    \node (C1) {\includegraphics[width=0.19\linewidth]{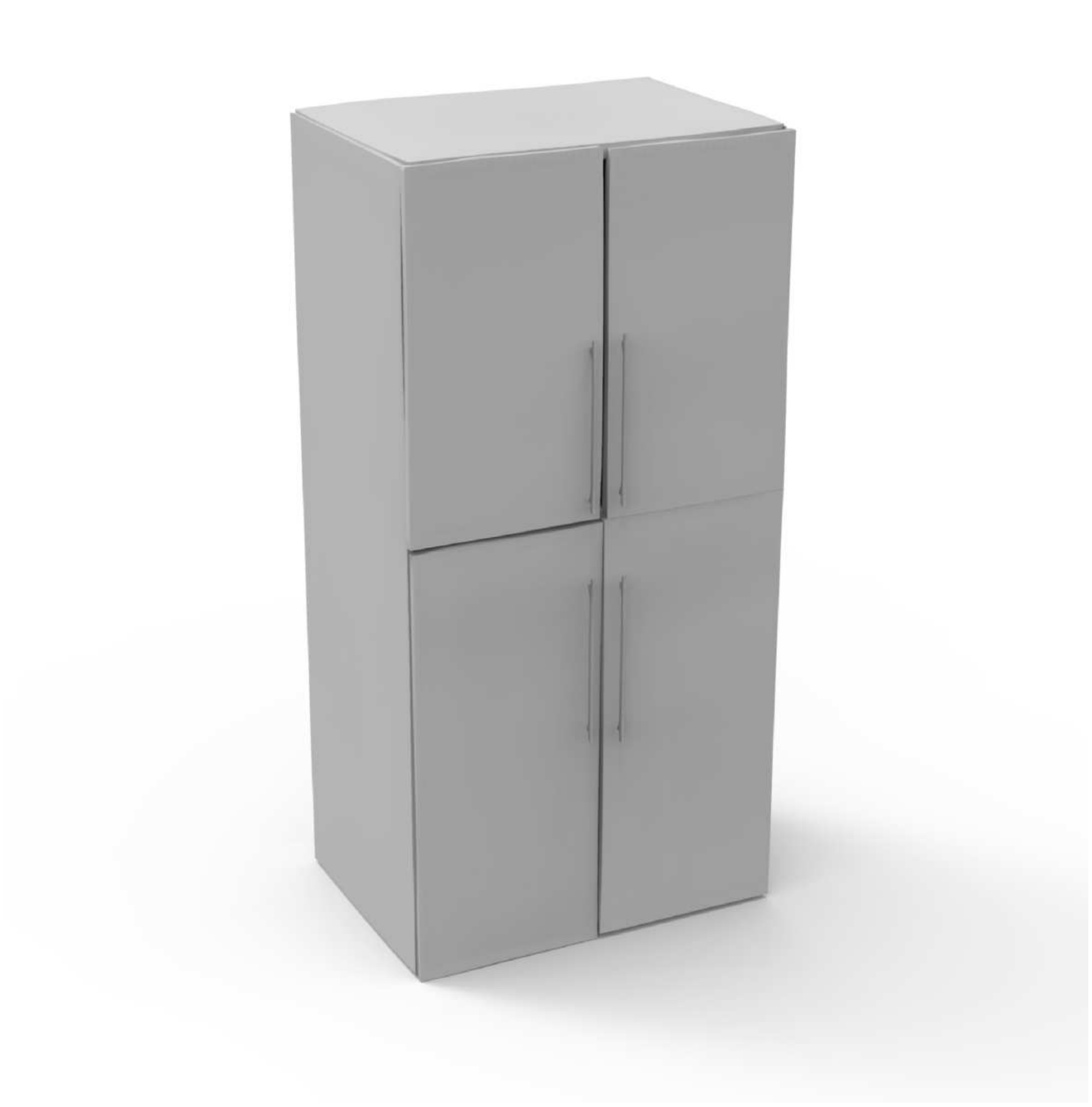}}; & 
    \node (C2) {\includegraphics[width=0.19\linewidth]{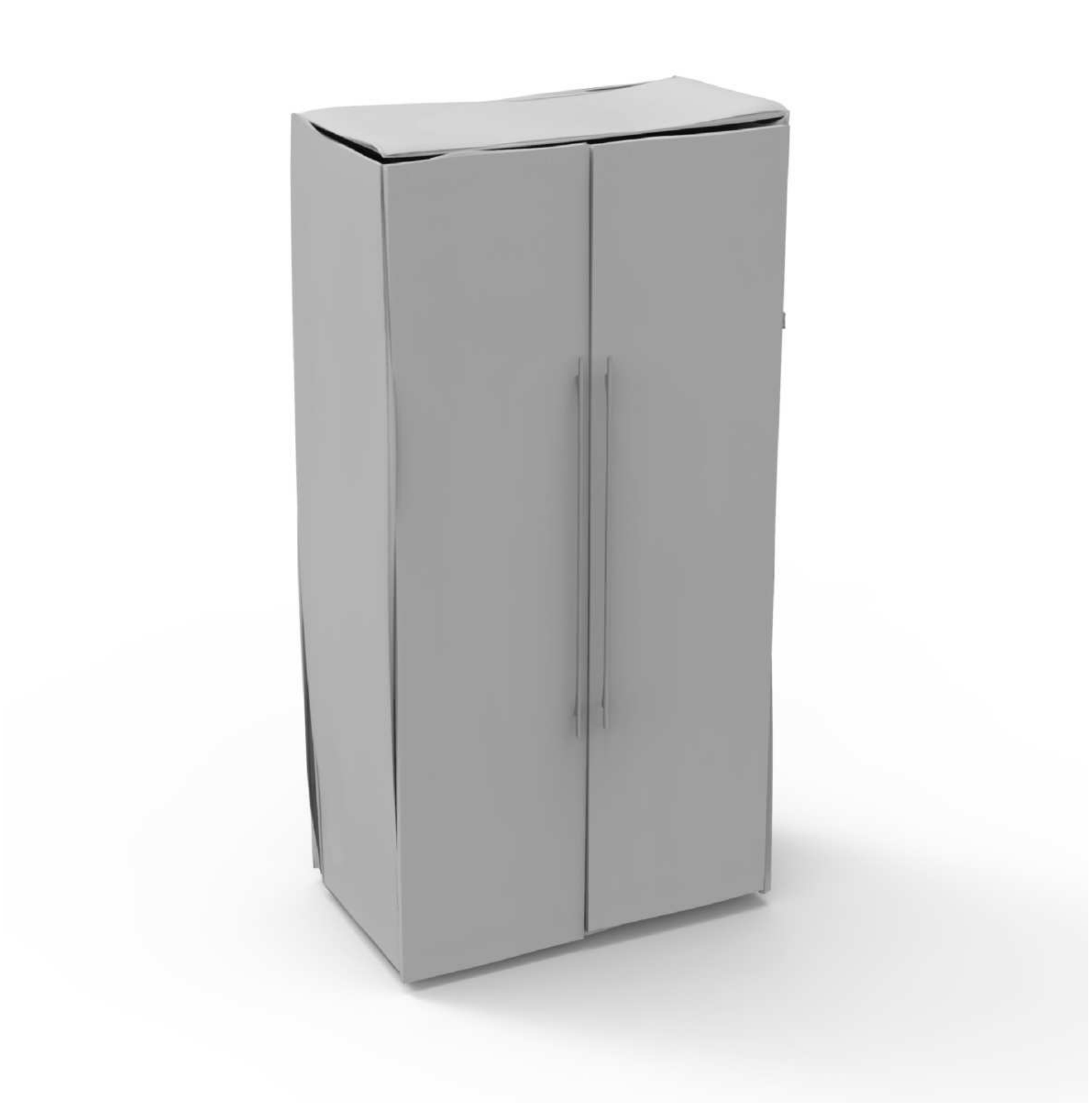}}; &
    \node (C3) {\includegraphics[width=0.19\linewidth]{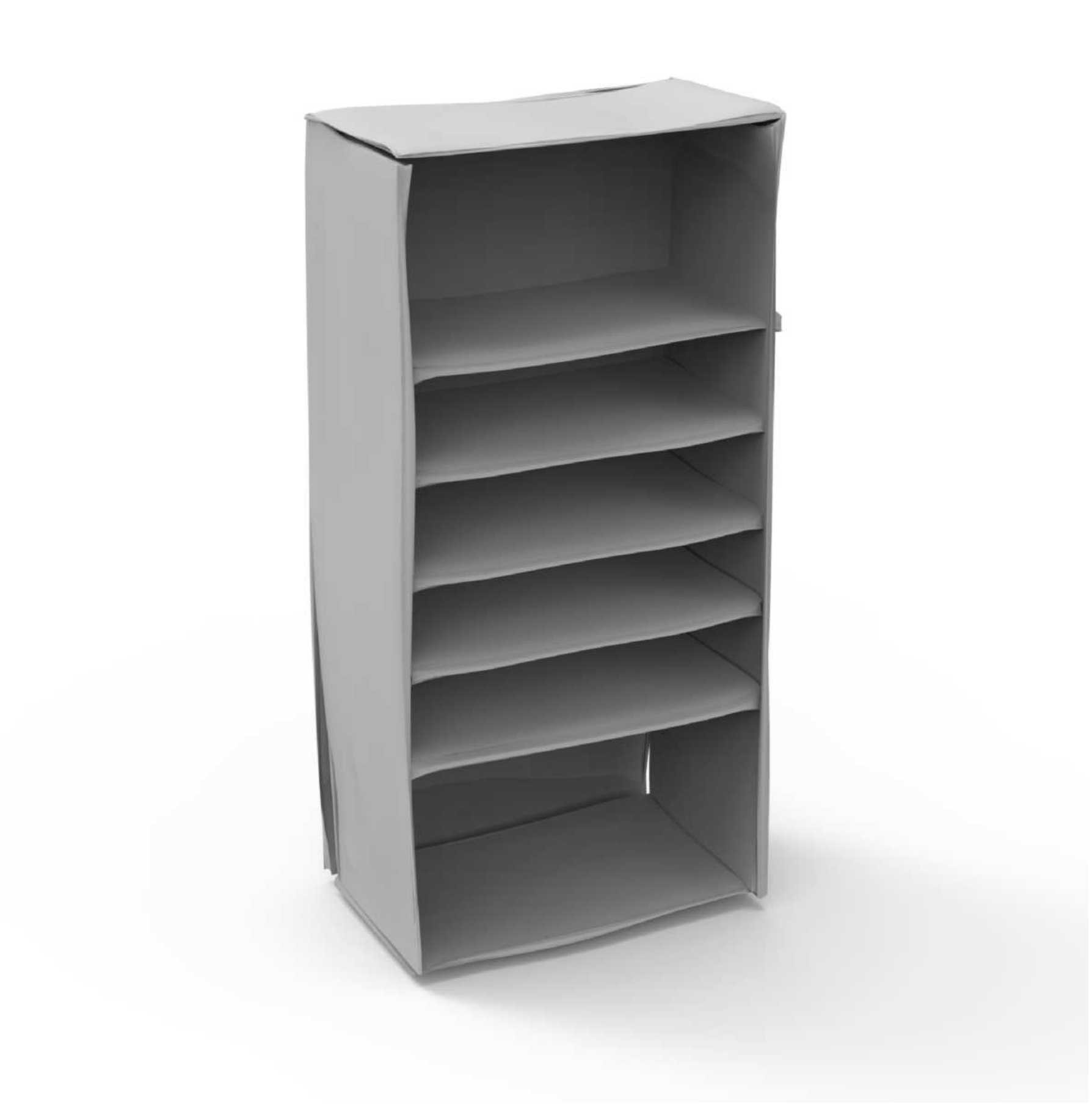}}; &
    \node (C4) {\includegraphics[width=0.19\linewidth]{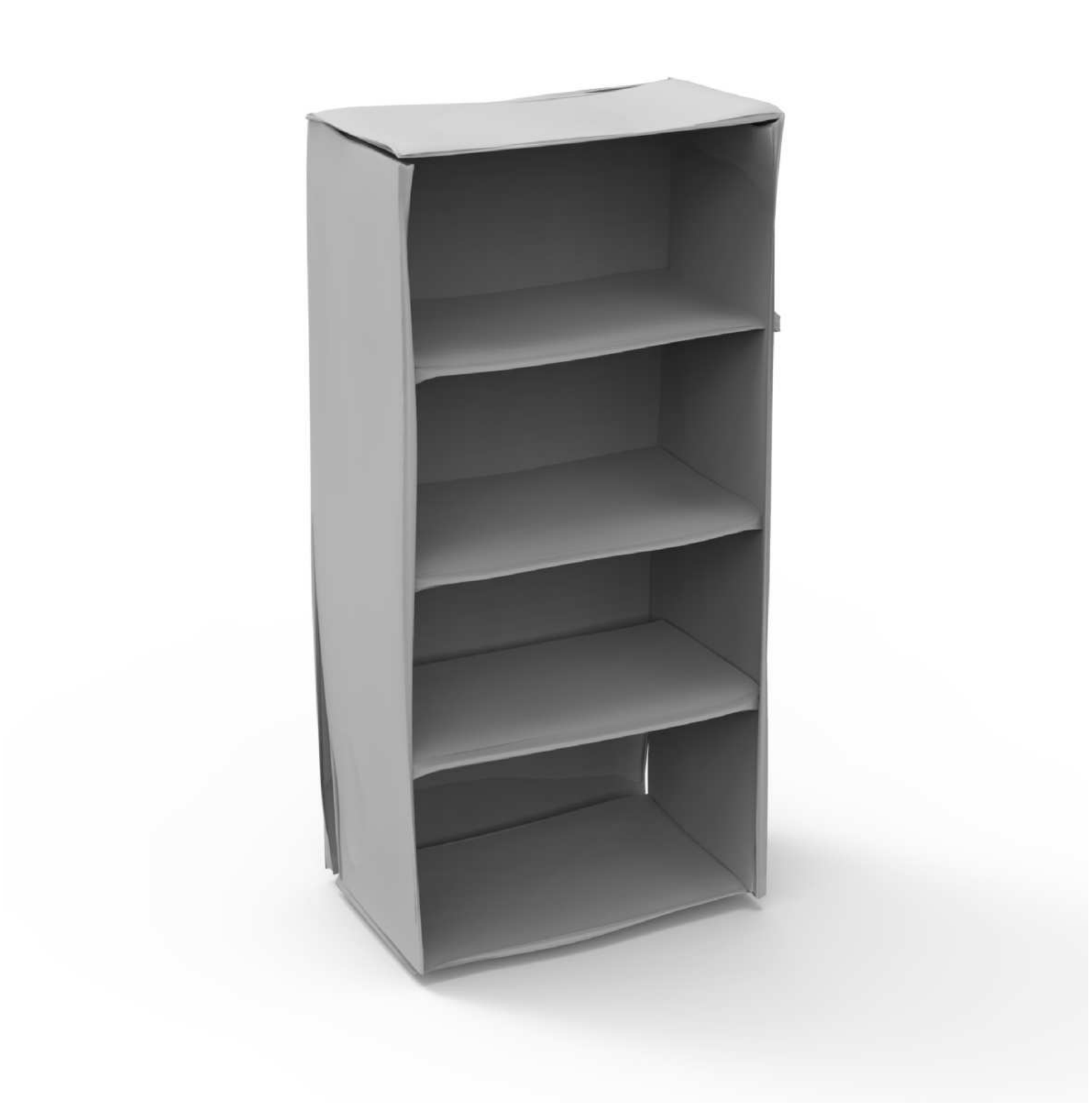}}; &
    \node (C5) {\includegraphics[width=0.19\linewidth]{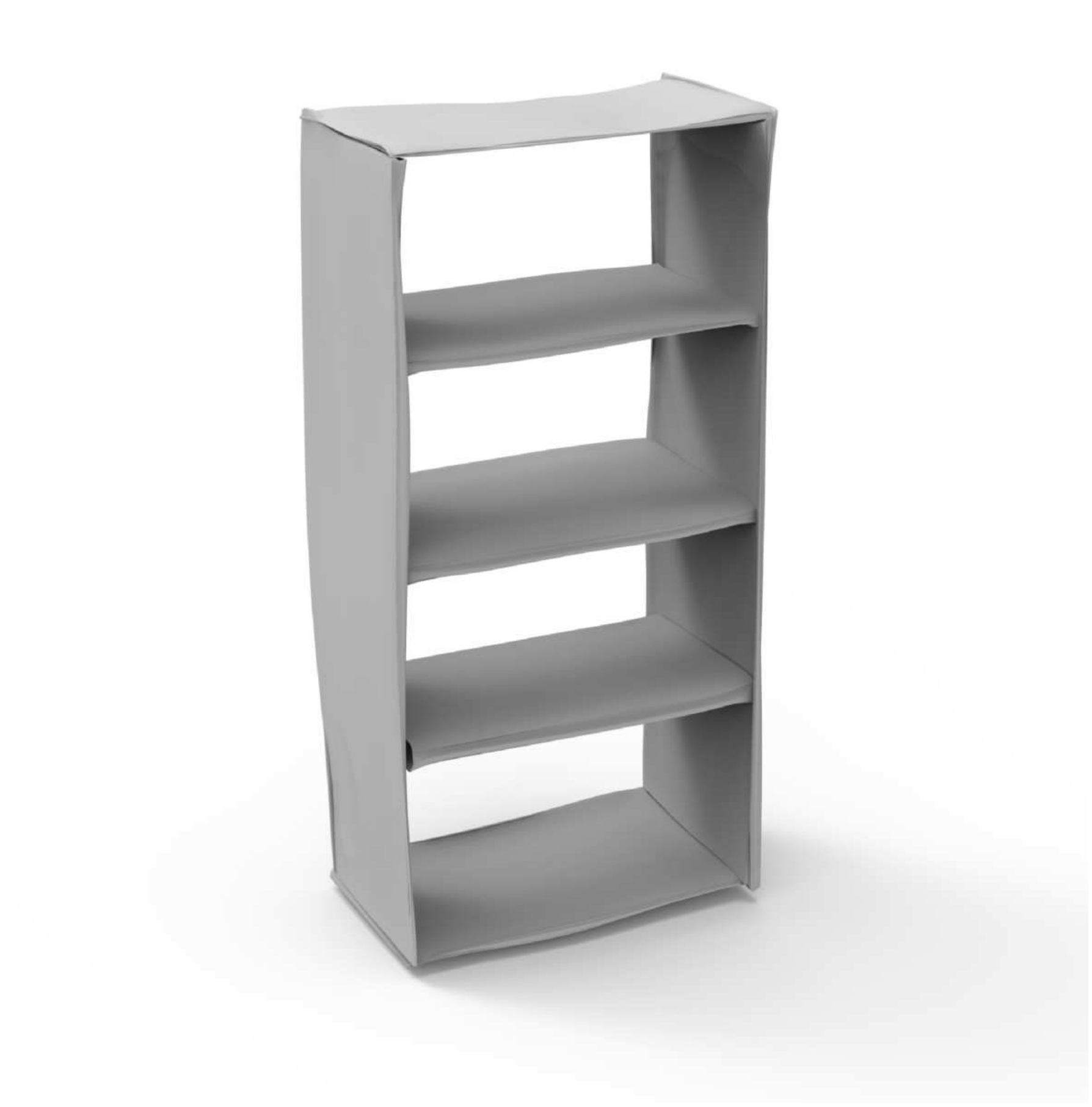}}; \\
    \node (D1) {\includegraphics[width=0.19\linewidth]{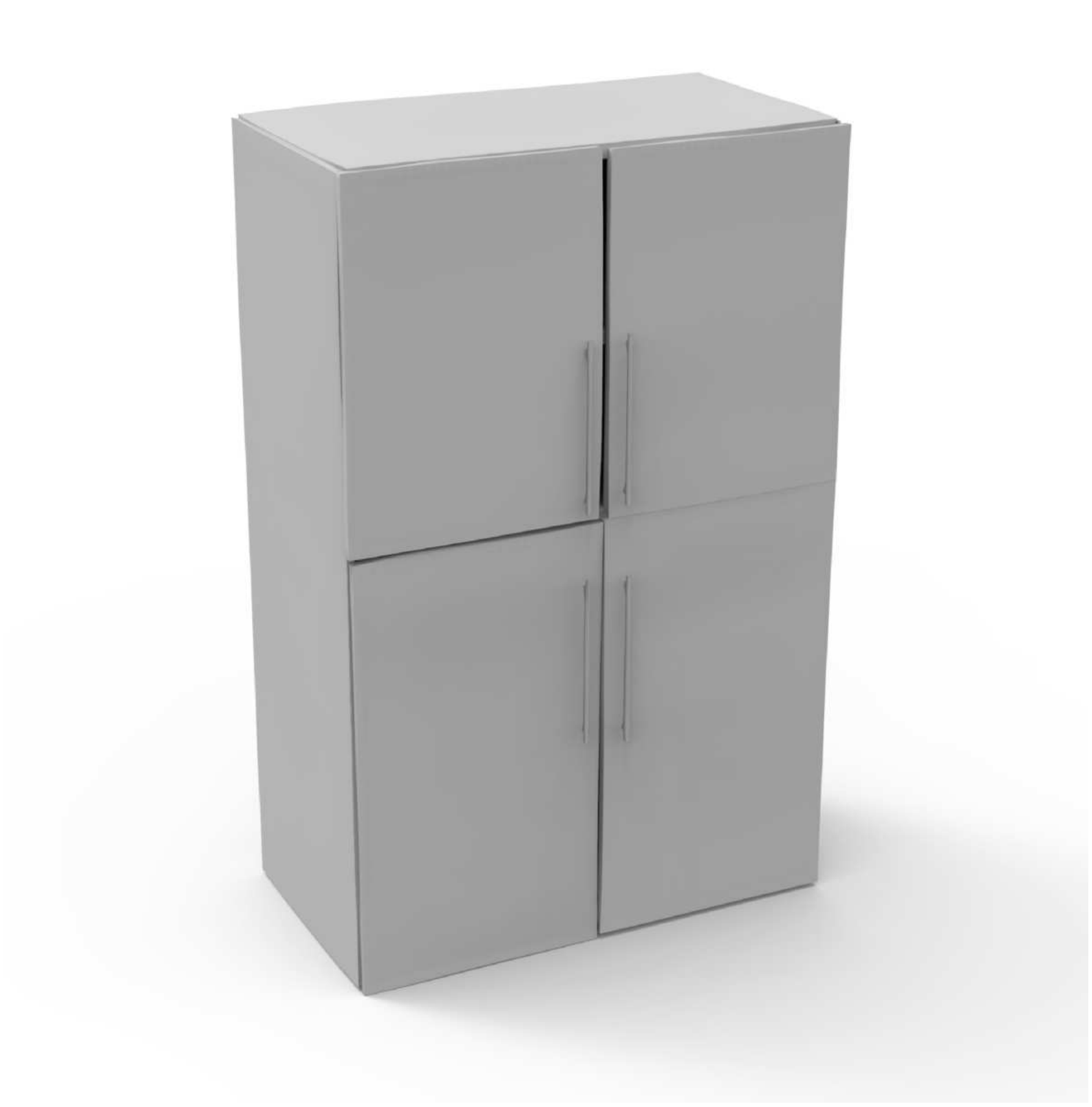}}; & 
    \node (D2) {\includegraphics[width=0.19\linewidth]{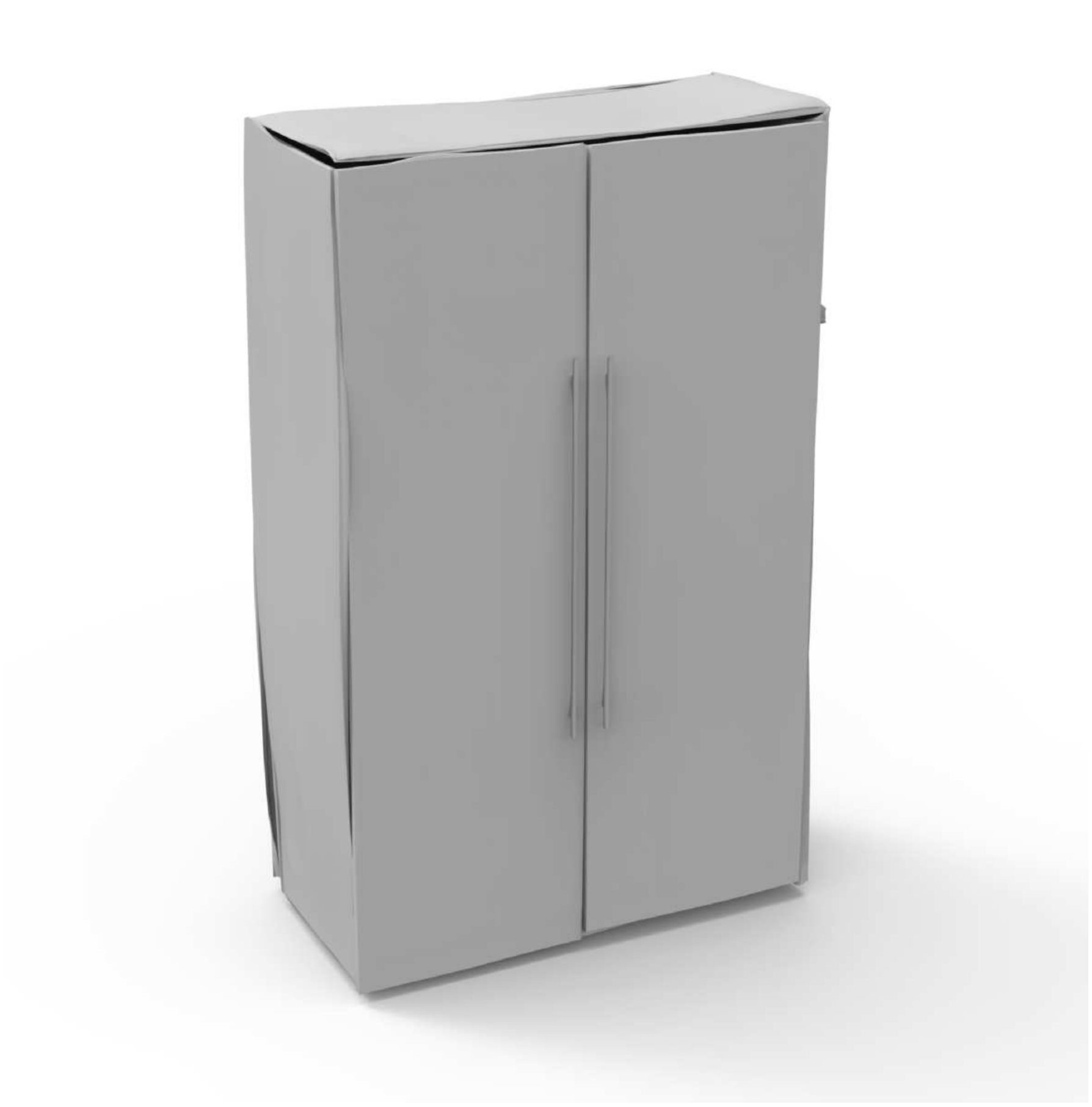}}; &
    \node (D3) {\includegraphics[width=0.19\linewidth]{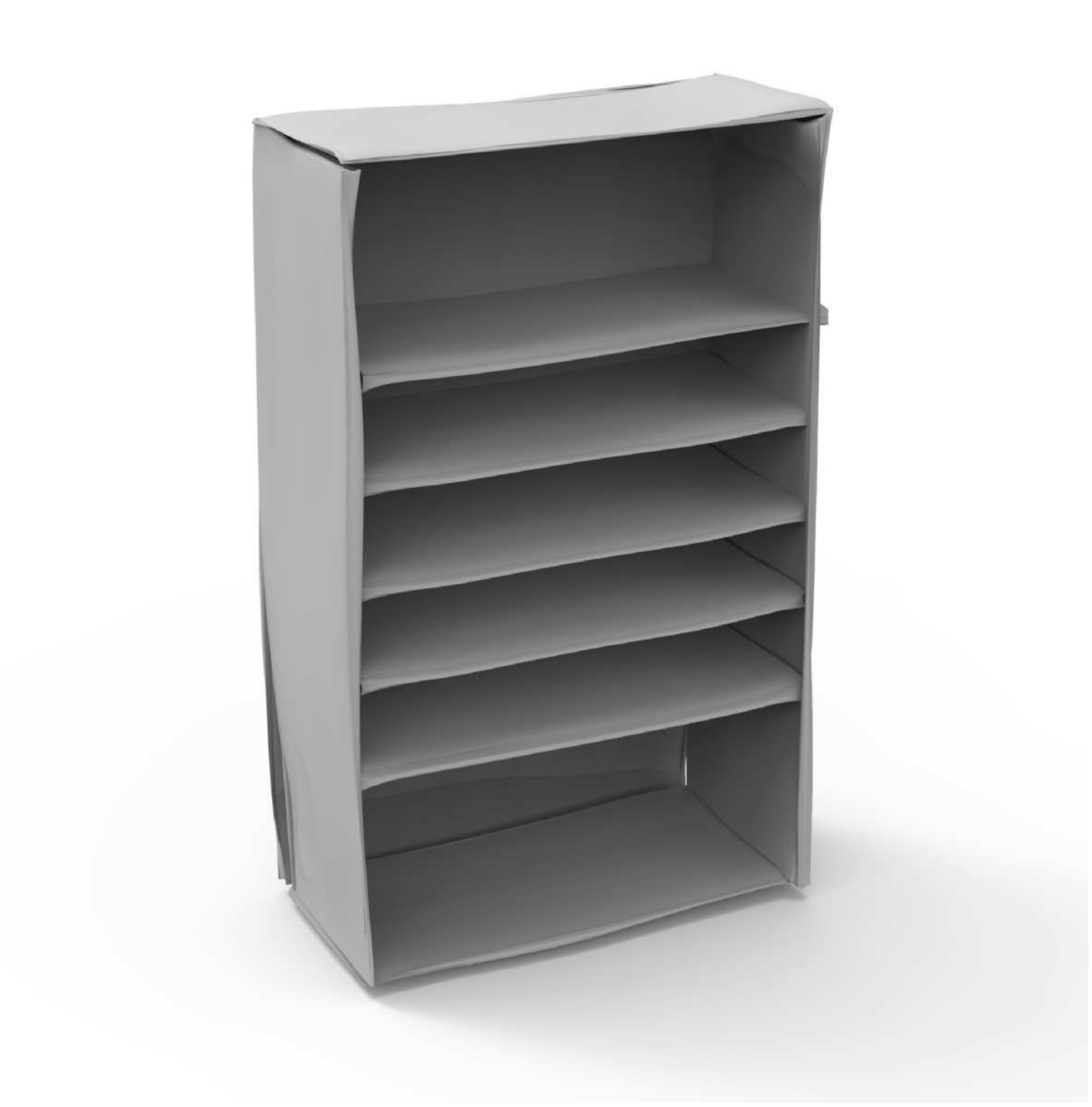}}; &
    \node (D4) {\includegraphics[width=0.19\linewidth]{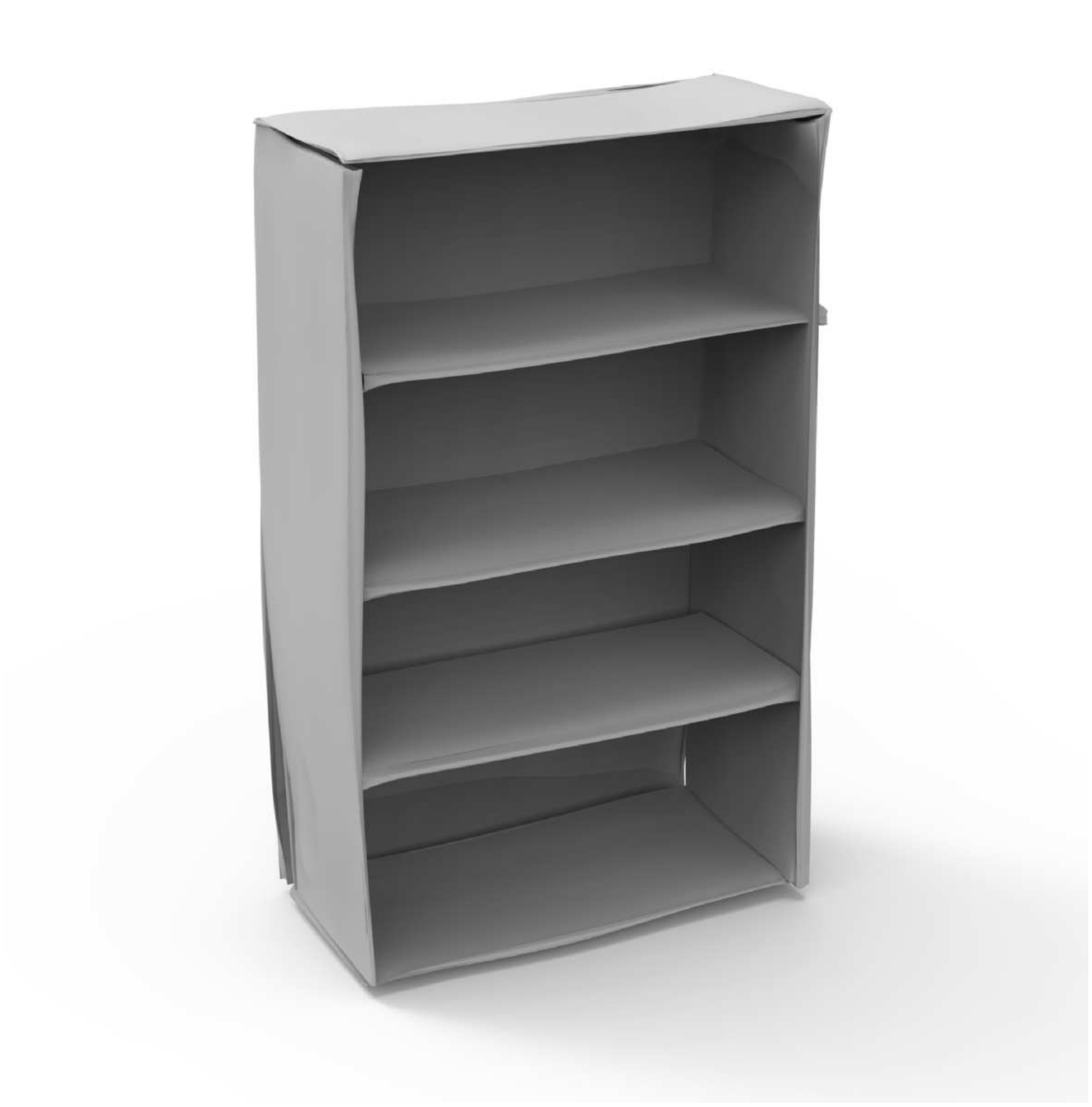}}; &
    \node (D5) {\includegraphics[width=0.19\linewidth]{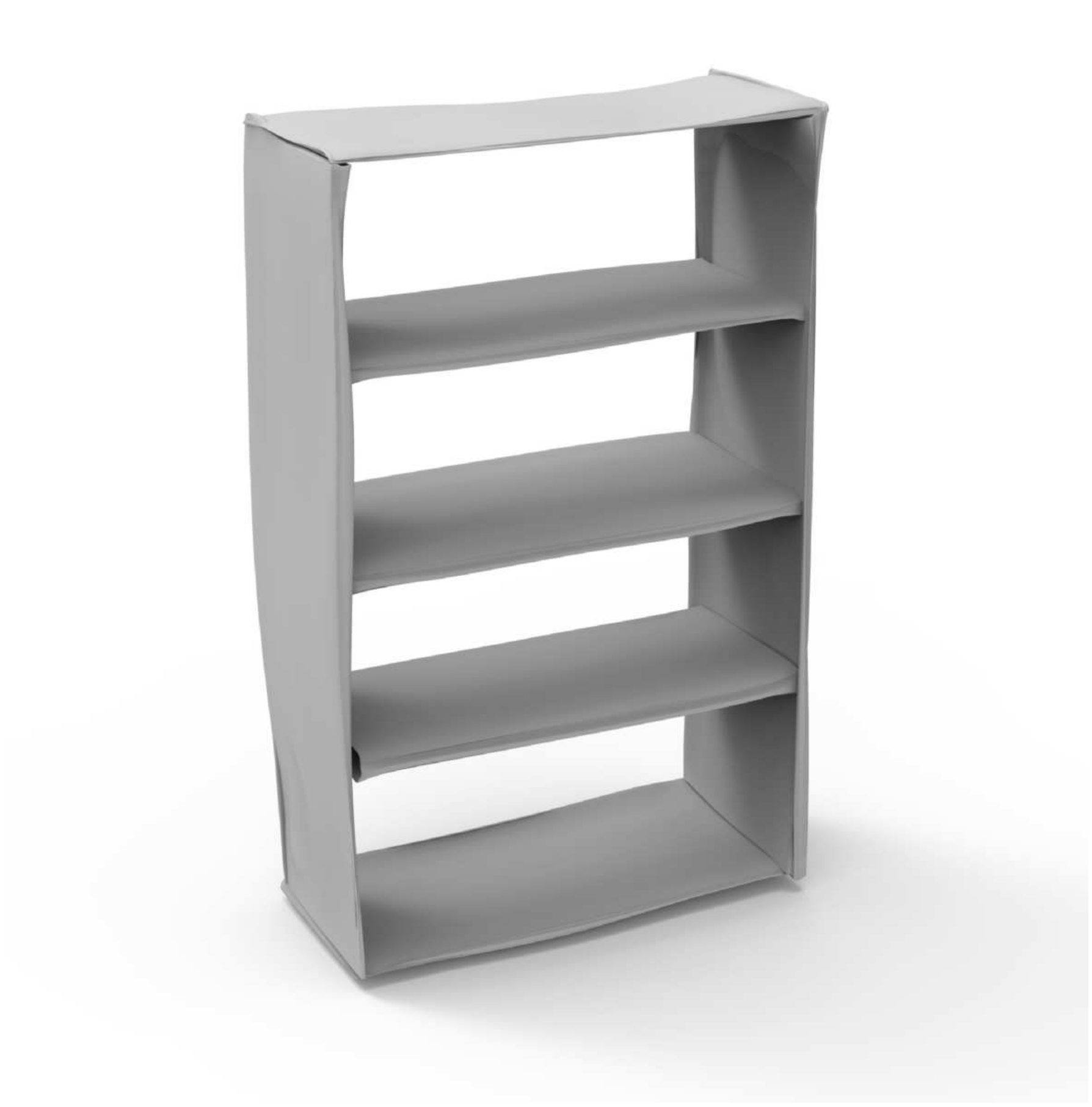}}; \\
    \node (F1) {\includegraphics[width=0.19\linewidth]{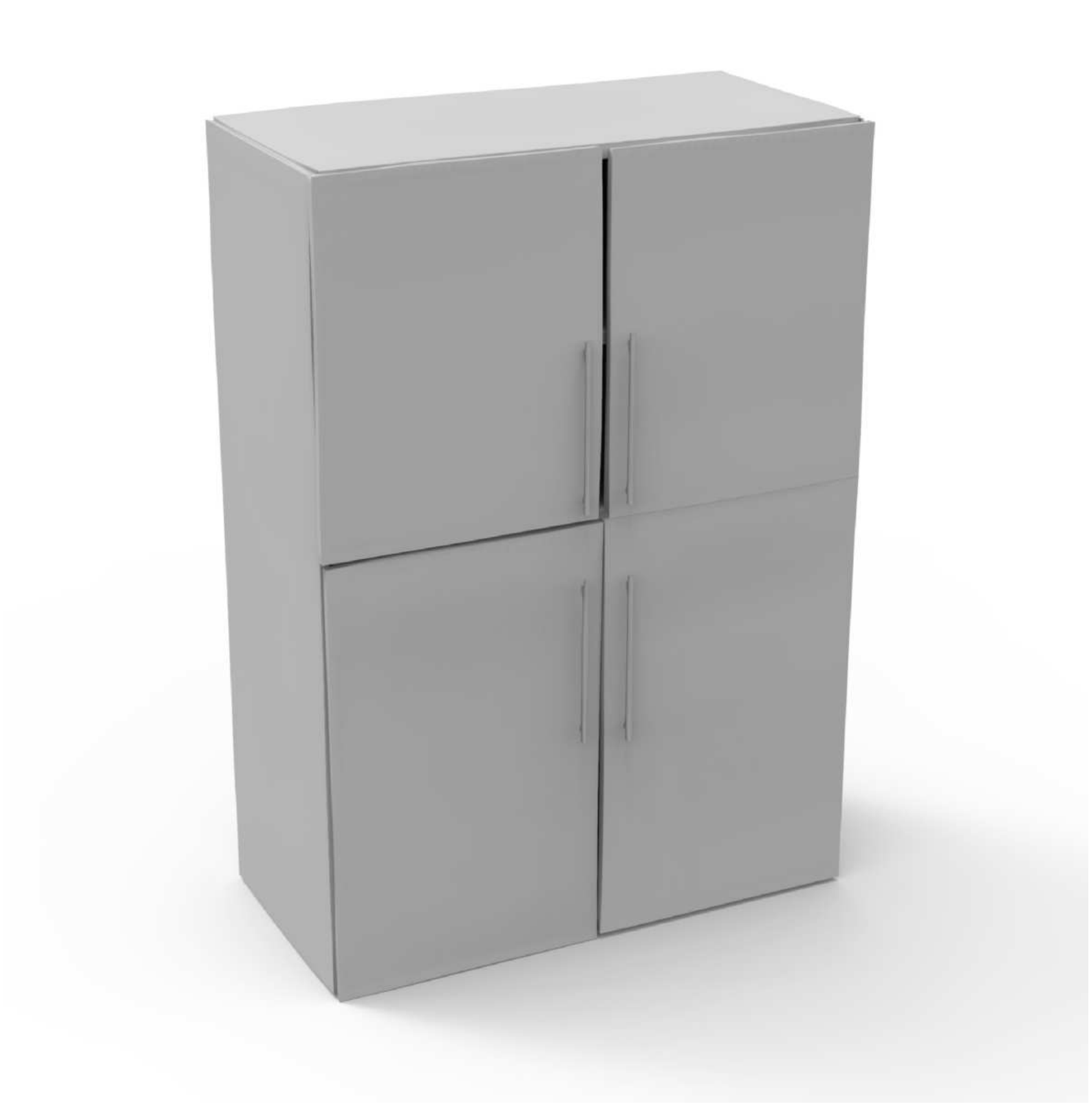}}; & 
    \node (F2) {\includegraphics[width=0.19\linewidth]{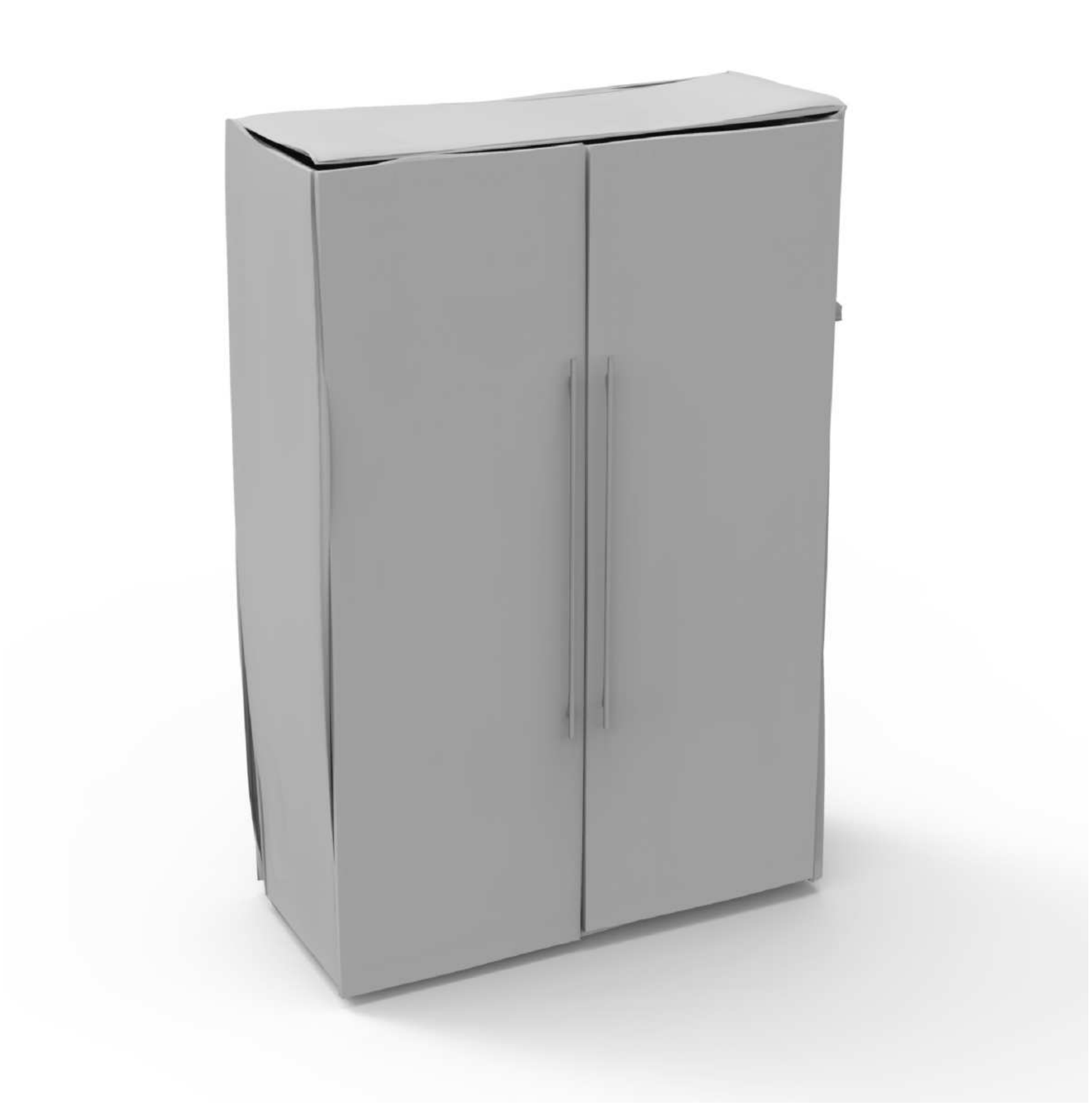}}; &
    \node (F3) {\includegraphics[width=0.19\linewidth]{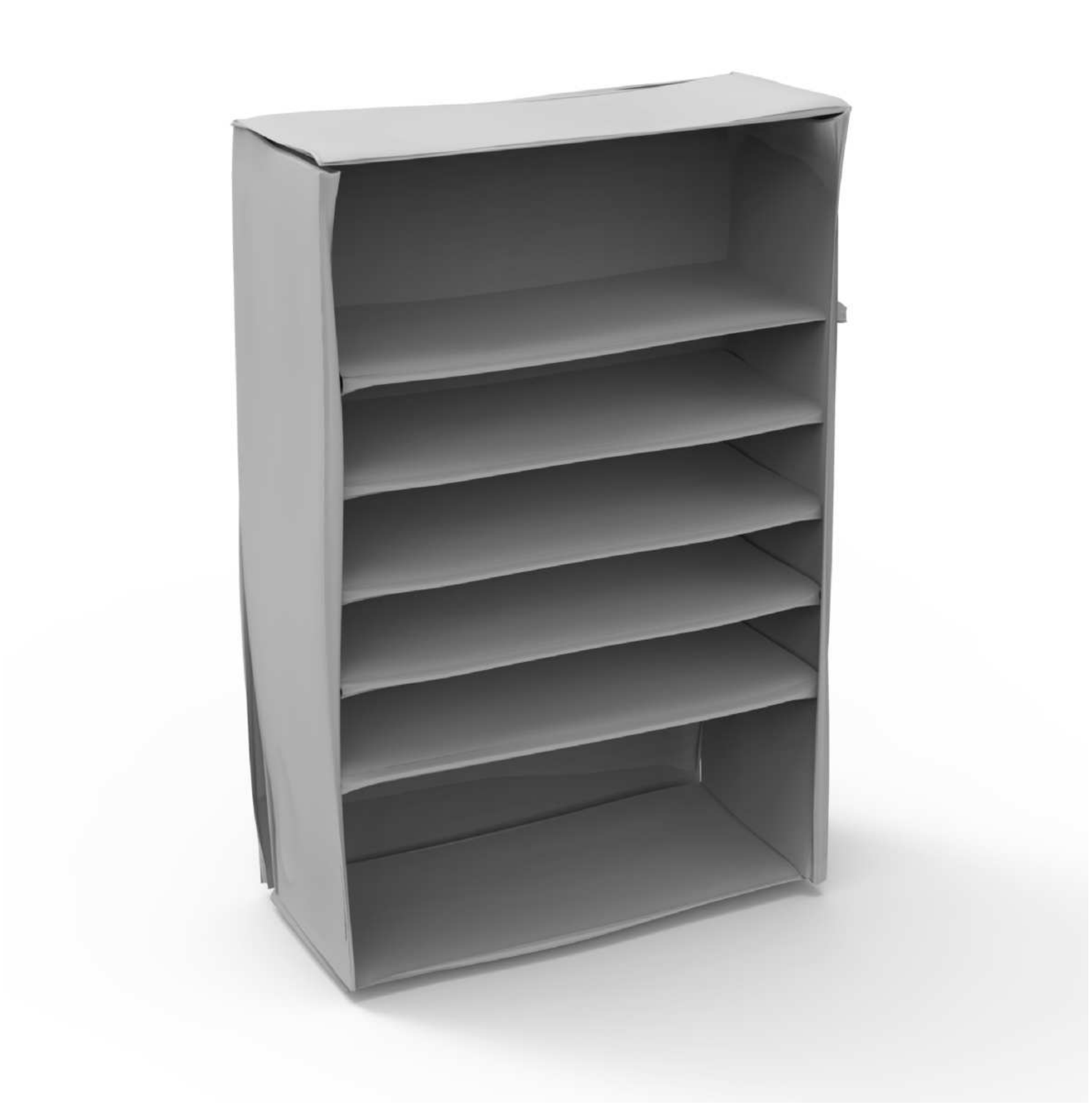}}; &
    \node (F4) {\includegraphics[width=0.19\linewidth]{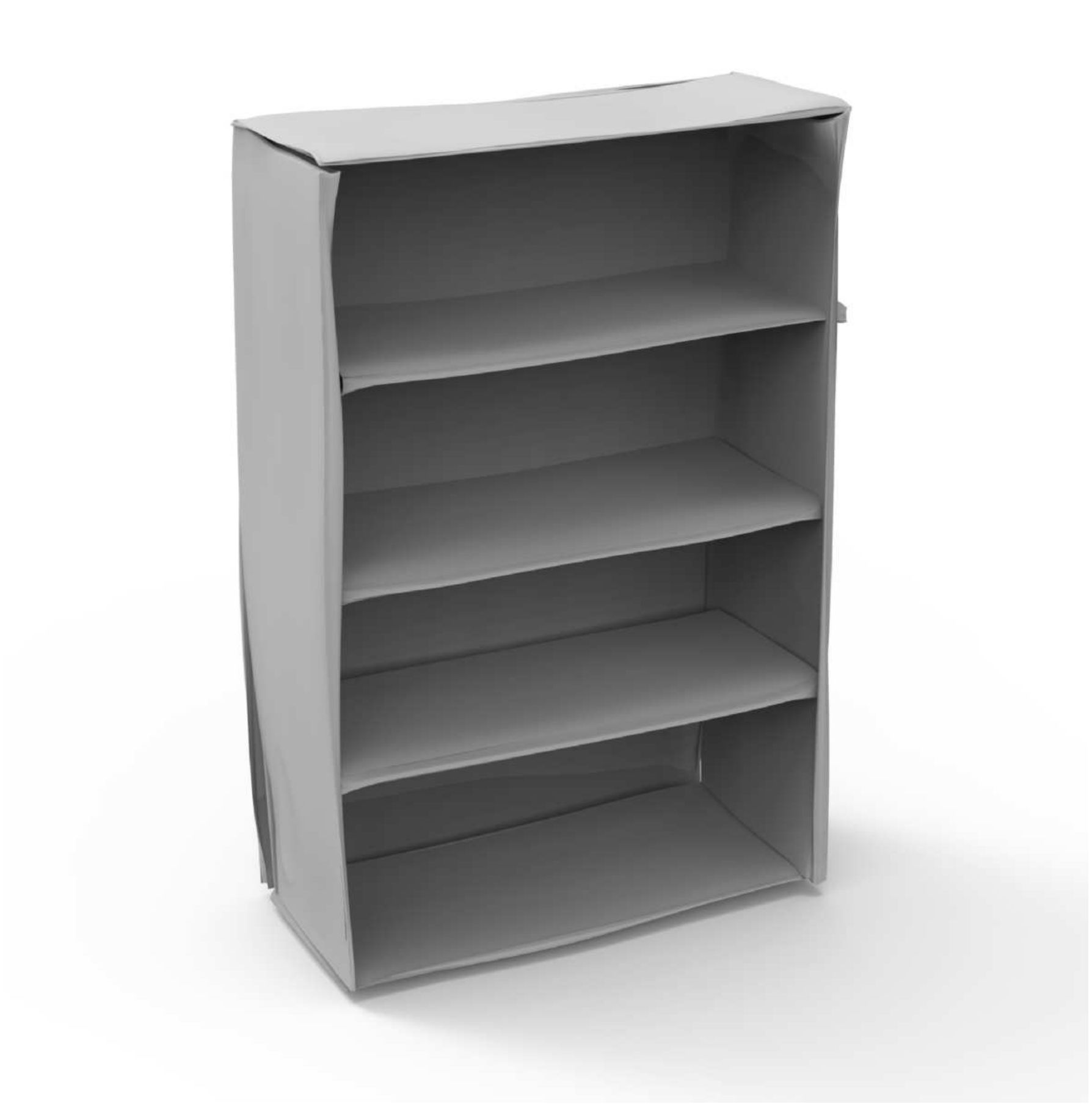}}; &
    \node (F5) {\includegraphics[width=0.19\linewidth, cfbox=orange 1pt 1pt]{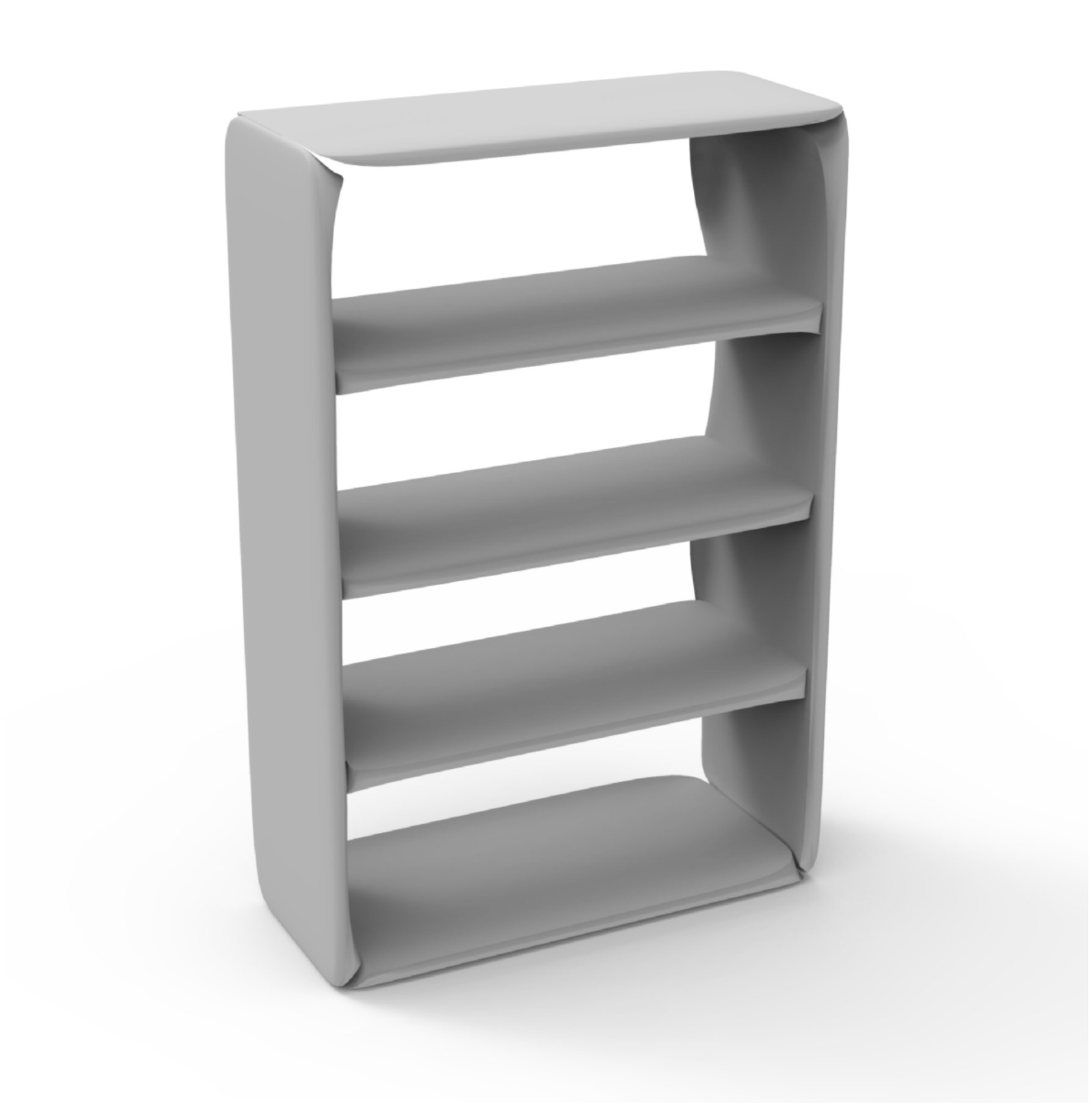}}; \\
  };
  \node[fit=(A1) (A2) (A3) (A4) (A5)
            (B1) (B2) (B3) (B4) (B5)
            (C1) (C2) (C3) (C4) (C5)
            (D1) (D2) (D3) (D4) (D5)
            (F1) (F2) (F3) (F4) (F5),
            inner sep=0pt,
            ] (PIC) {};

  \draw[line width=1pt,arrows={-Stealth[length=4mm]}] ([xshift=-1em,yshift=-0.5em]PIC.south west) -- ([xshift=-1em,yshift=-0.5em]PIC.south east);
  \draw[line width=1pt,arrows={-Stealth[length=4mm]}] ([yshift=-1em,xshift=-0.5em]PIC.south west) -- ([yshift=-1em,xshift=-0.5em]PIC.north west);

  \node[anchor=south] (label) [font=\fontsize{10}{10}\selectfont]at ([yshift=-2.em]A3|-PIC.south) {Structure};
  \node[anchor=center,rotate=90] (label) [font=\fontsize{10}{10}\selectfont] at ([xshift=-1.5em]C1-|PIC.west) {Geometry};

\end{tikzpicture}
    \end{minipage}
    \hspace{1.cm}
    \begin{minipage}[b]{0.45\linewidth}
    \centering
  \begin{tikzpicture}
  \matrix[nodes={anchor=south west,inner sep=0pt}]{ 
    \node (A1) {\includegraphics[width=0.19\linewidth, cfbox=orange 1pt 1pt]{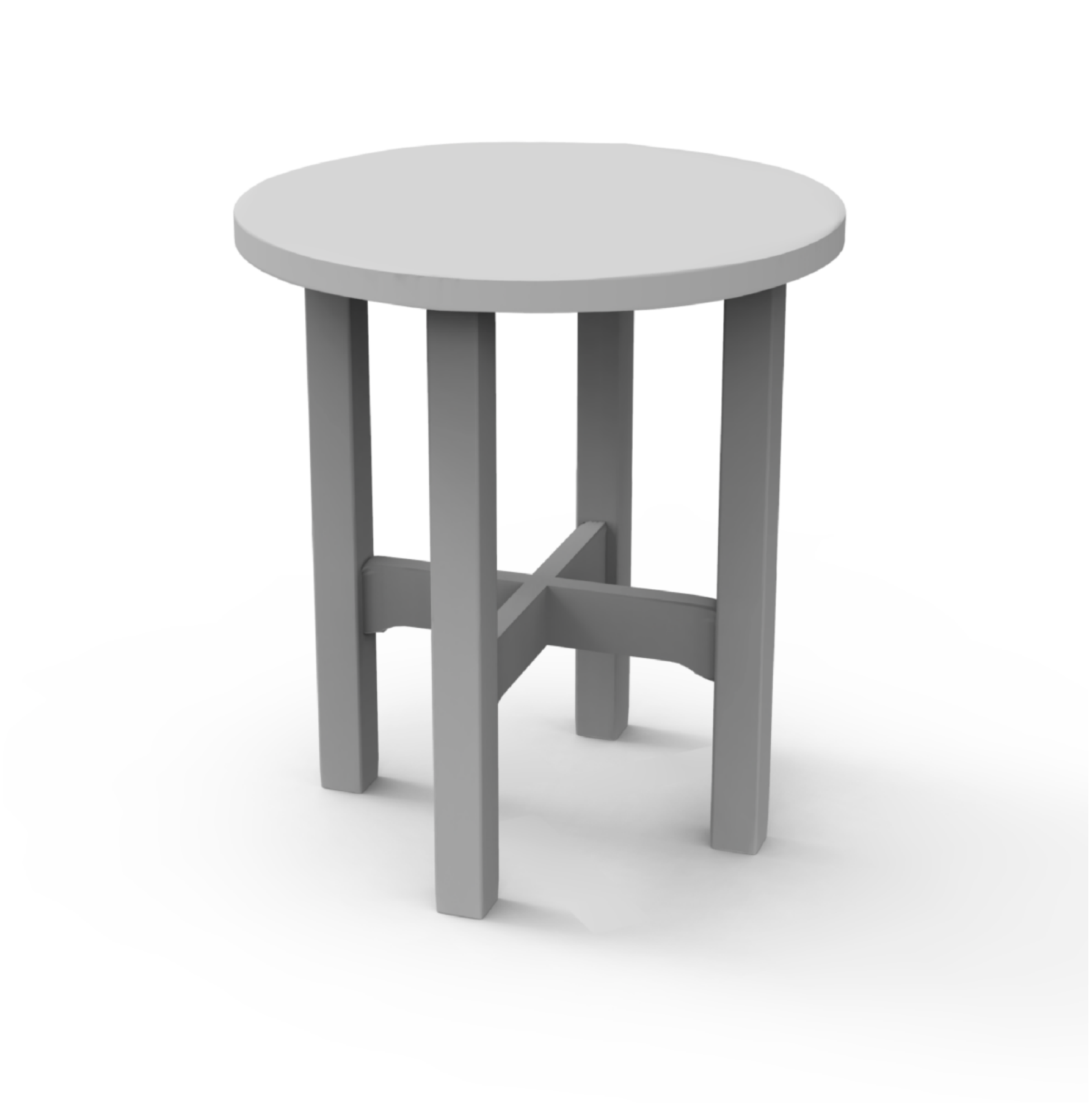}};  & 
    \node (A2) {\includegraphics[width=0.19\linewidth]{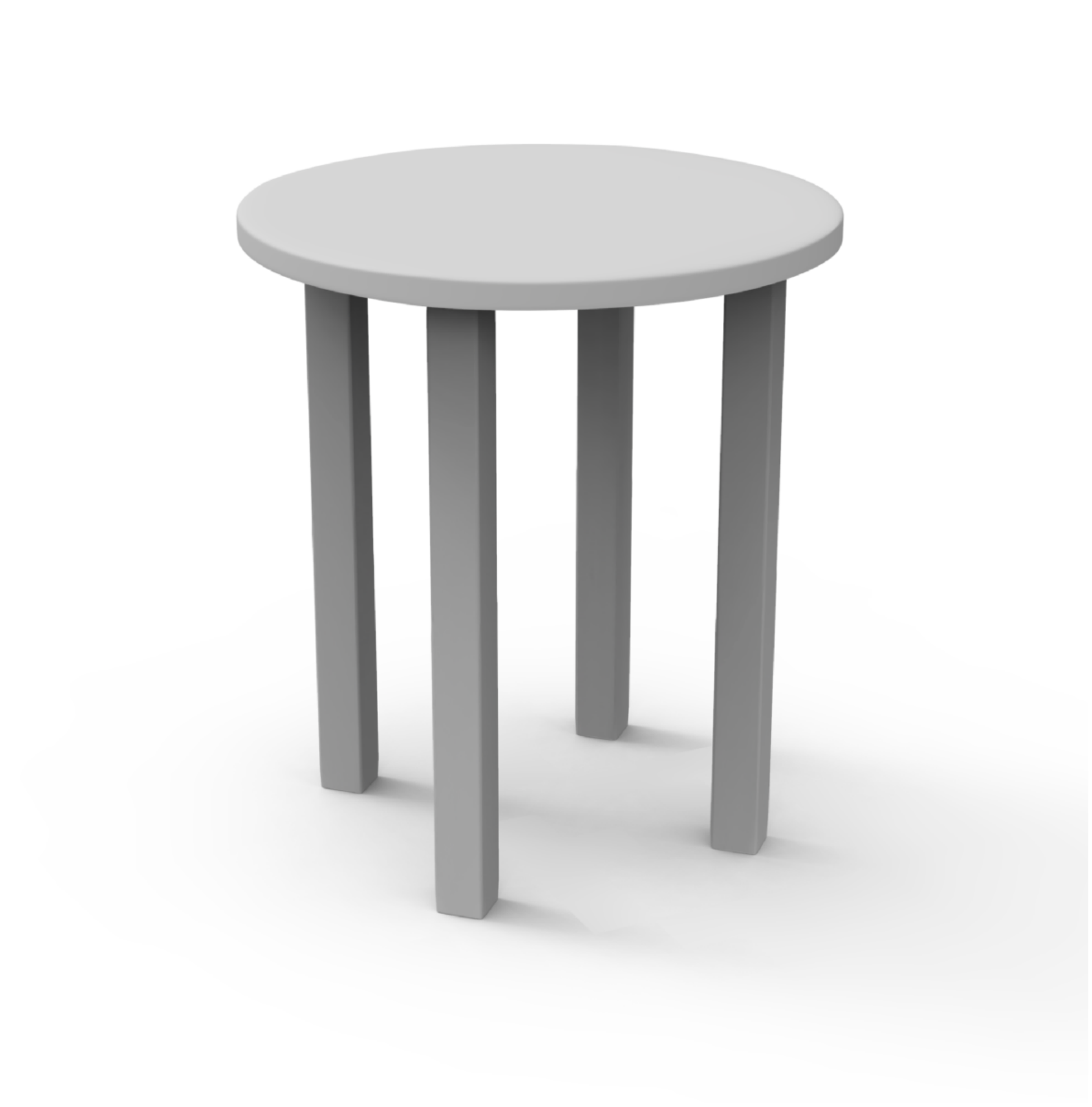}};  &
    \node (A3) {\includegraphics[width=0.19\linewidth]{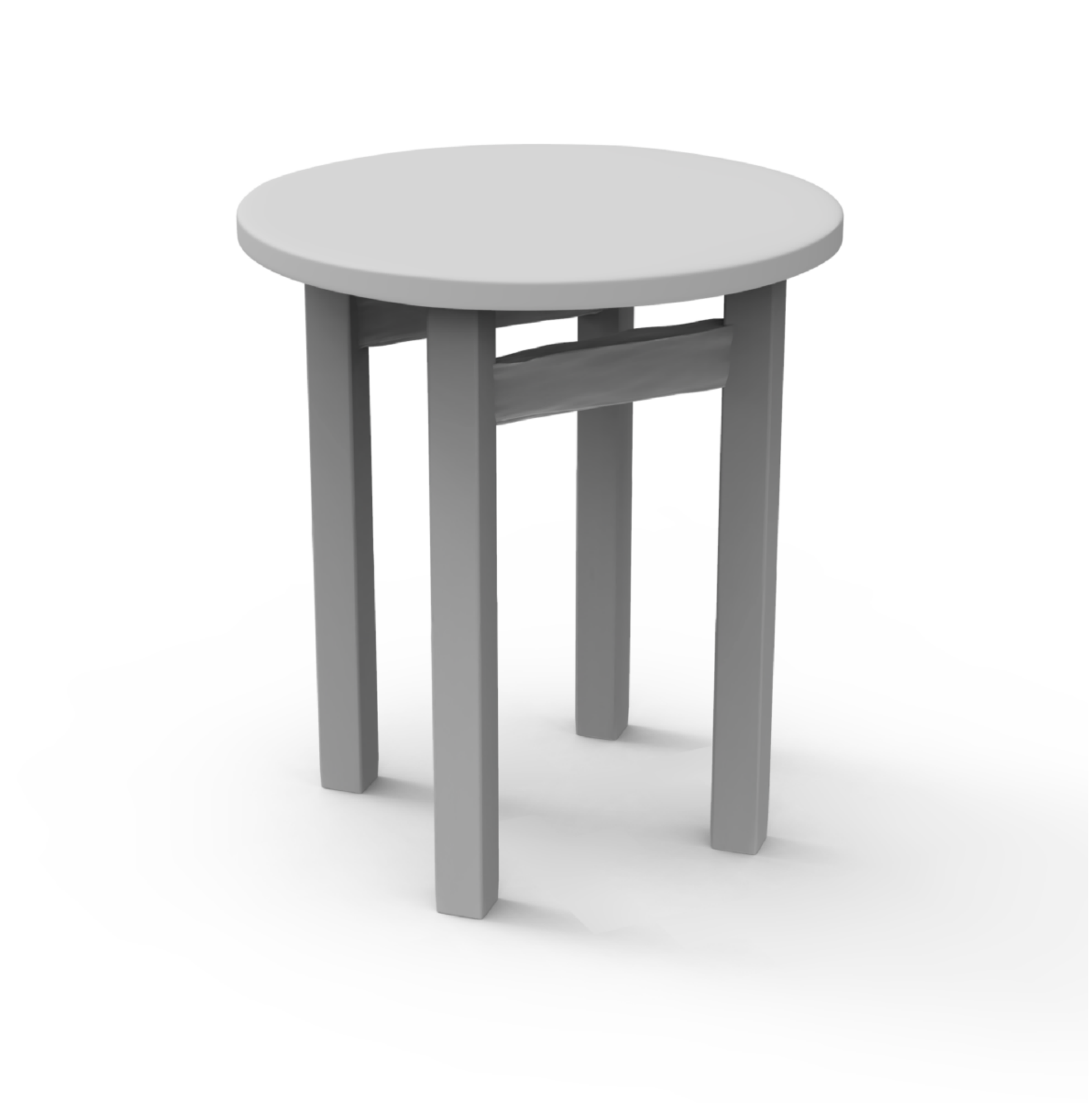}};  &
    \node (A4) {\includegraphics[width=0.19\linewidth]{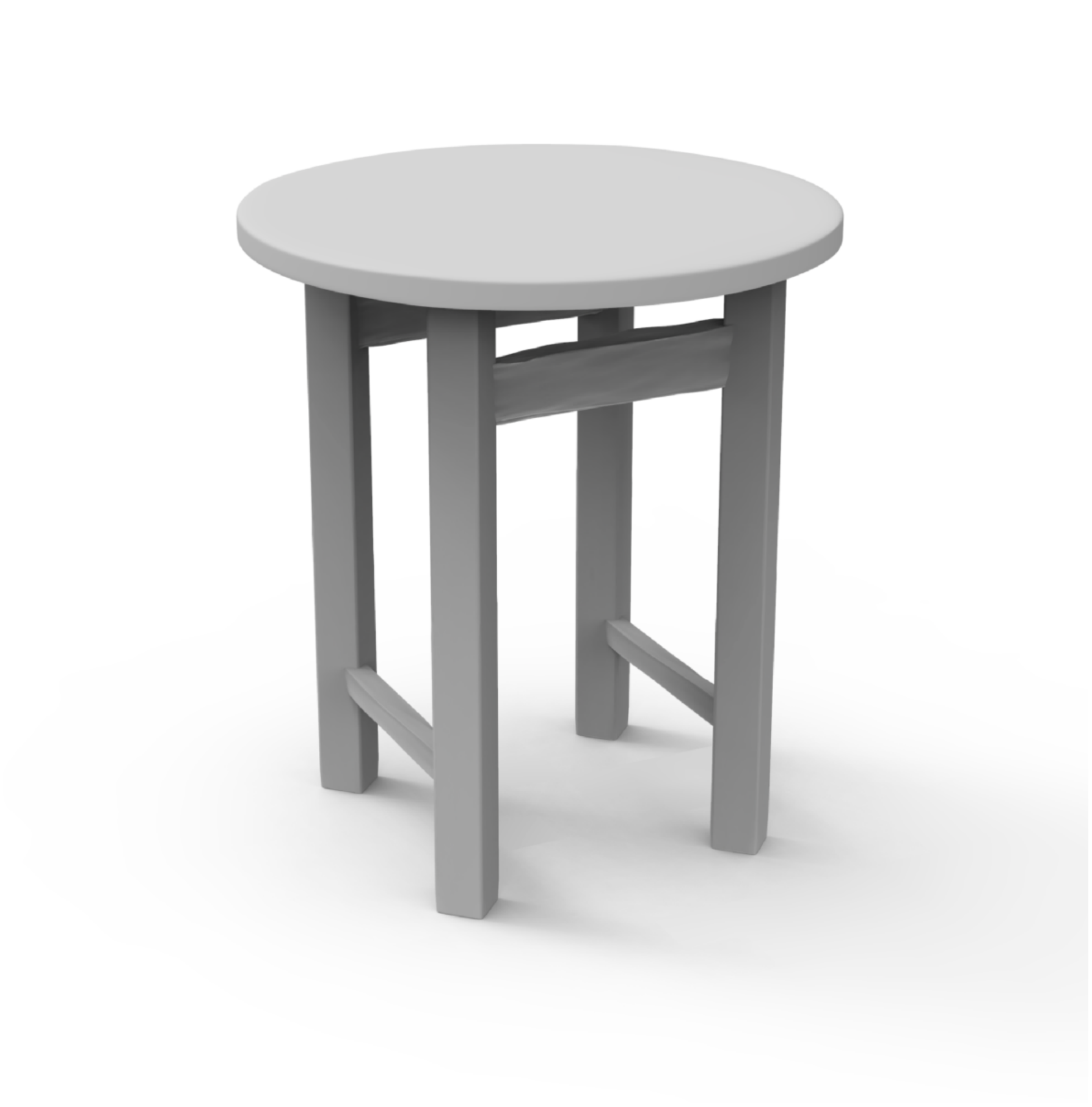}};  & 
    \node (A5) {\includegraphics[width=0.19\linewidth]{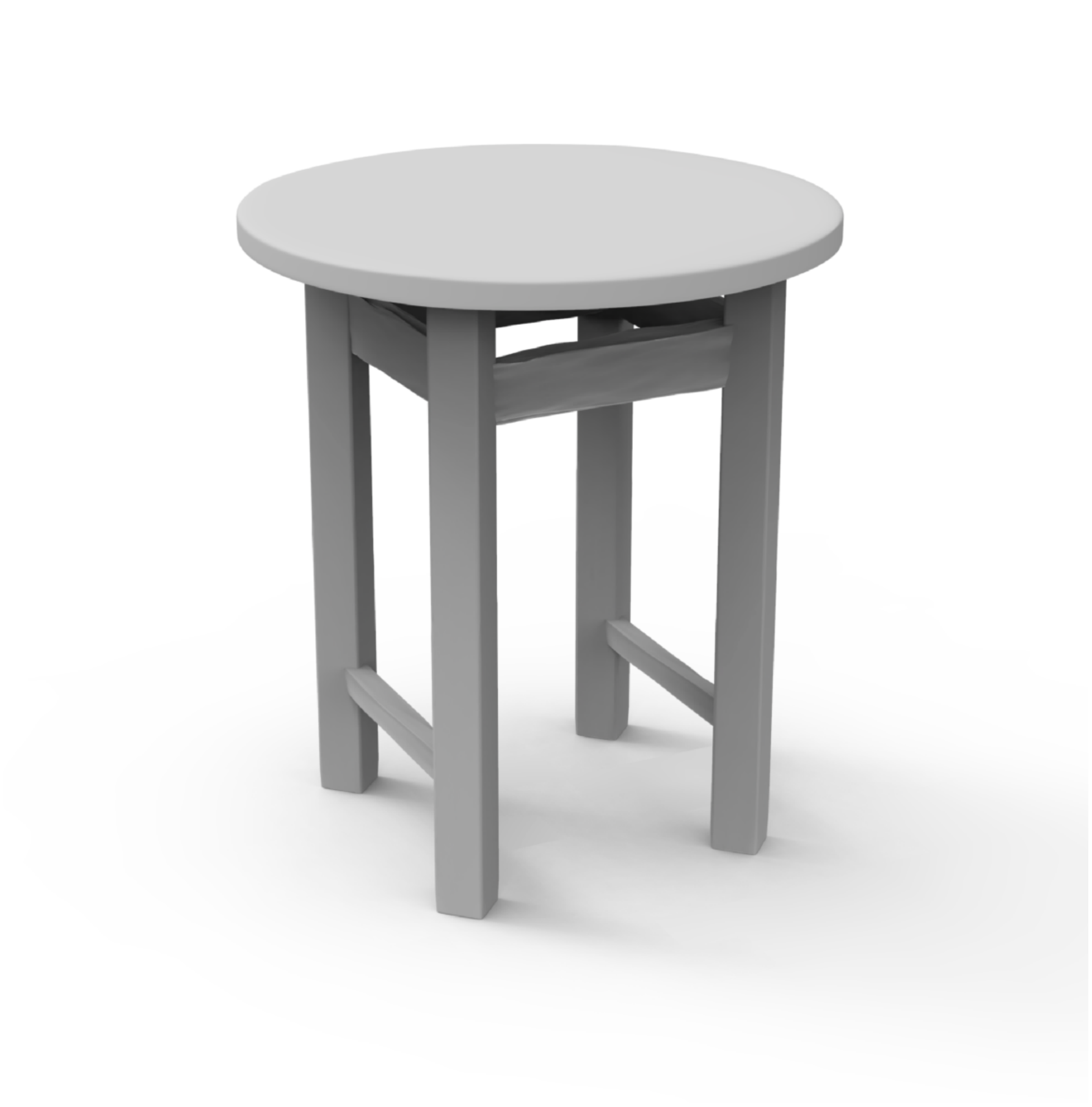}}; \\
    \node (B1) {\includegraphics[width=0.19\linewidth]{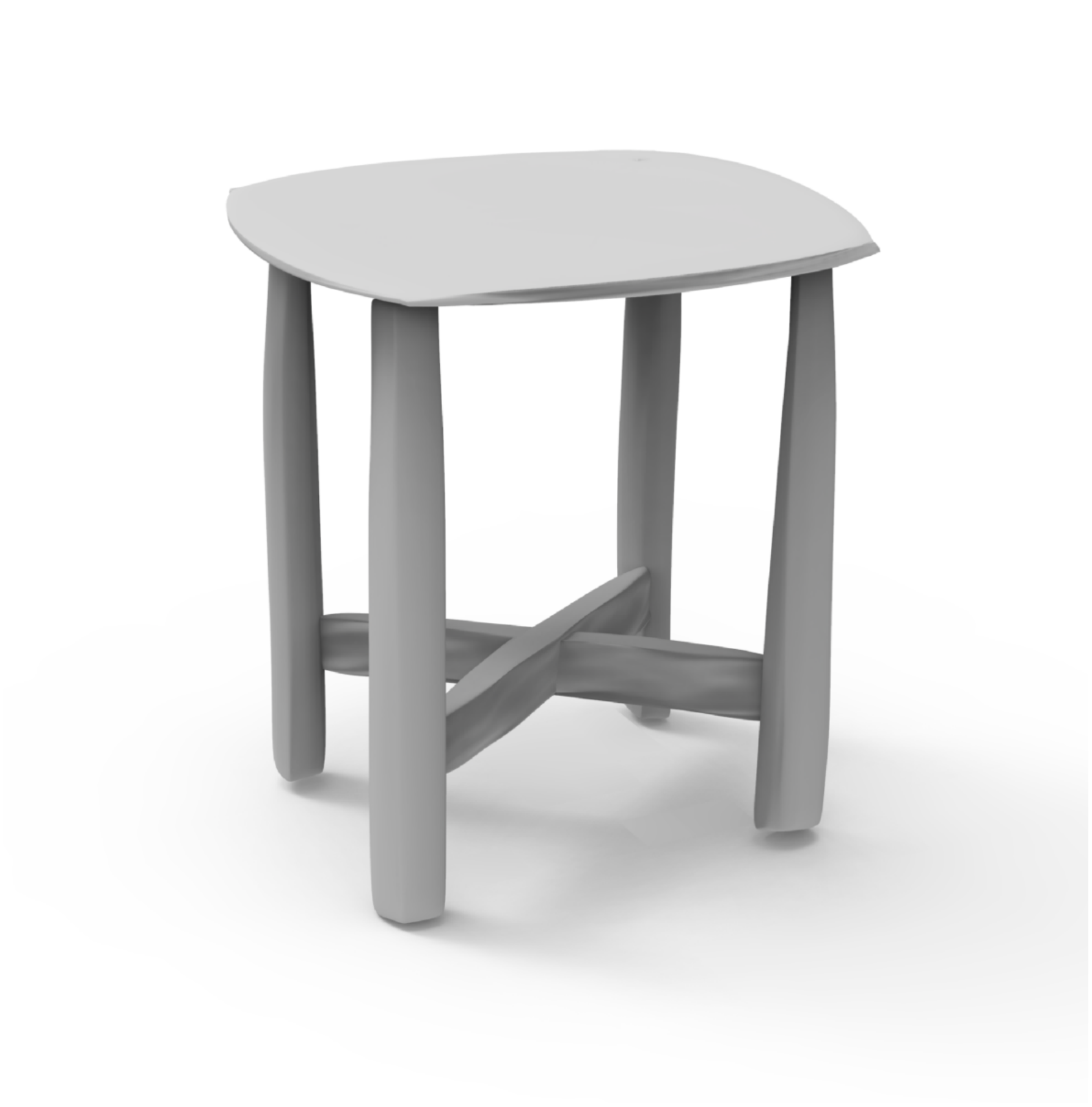}}; & 
    \node (B2) {\includegraphics[width=0.19\linewidth]{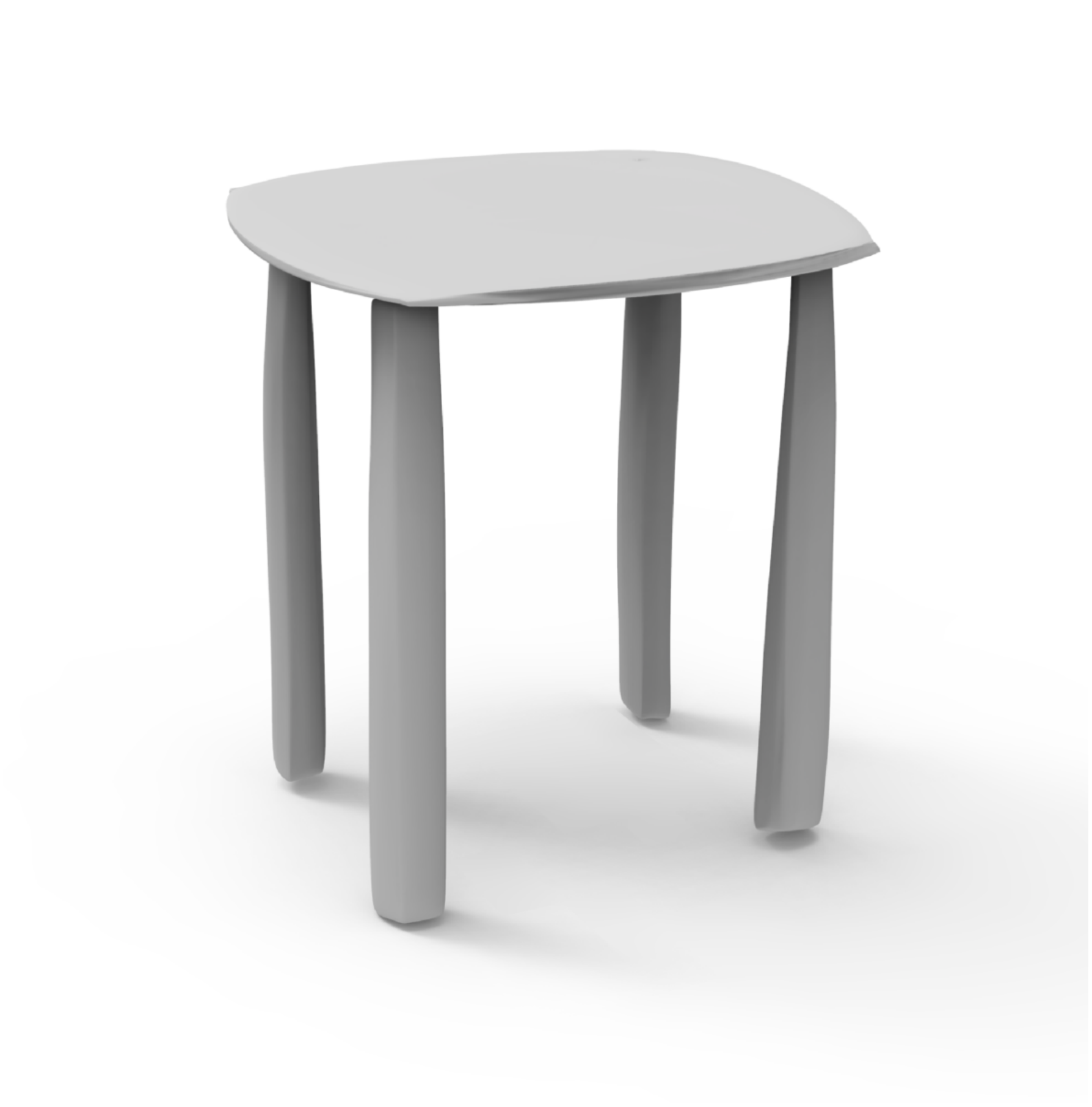}}; &
    \node (B3) {\includegraphics[width=0.19\linewidth]{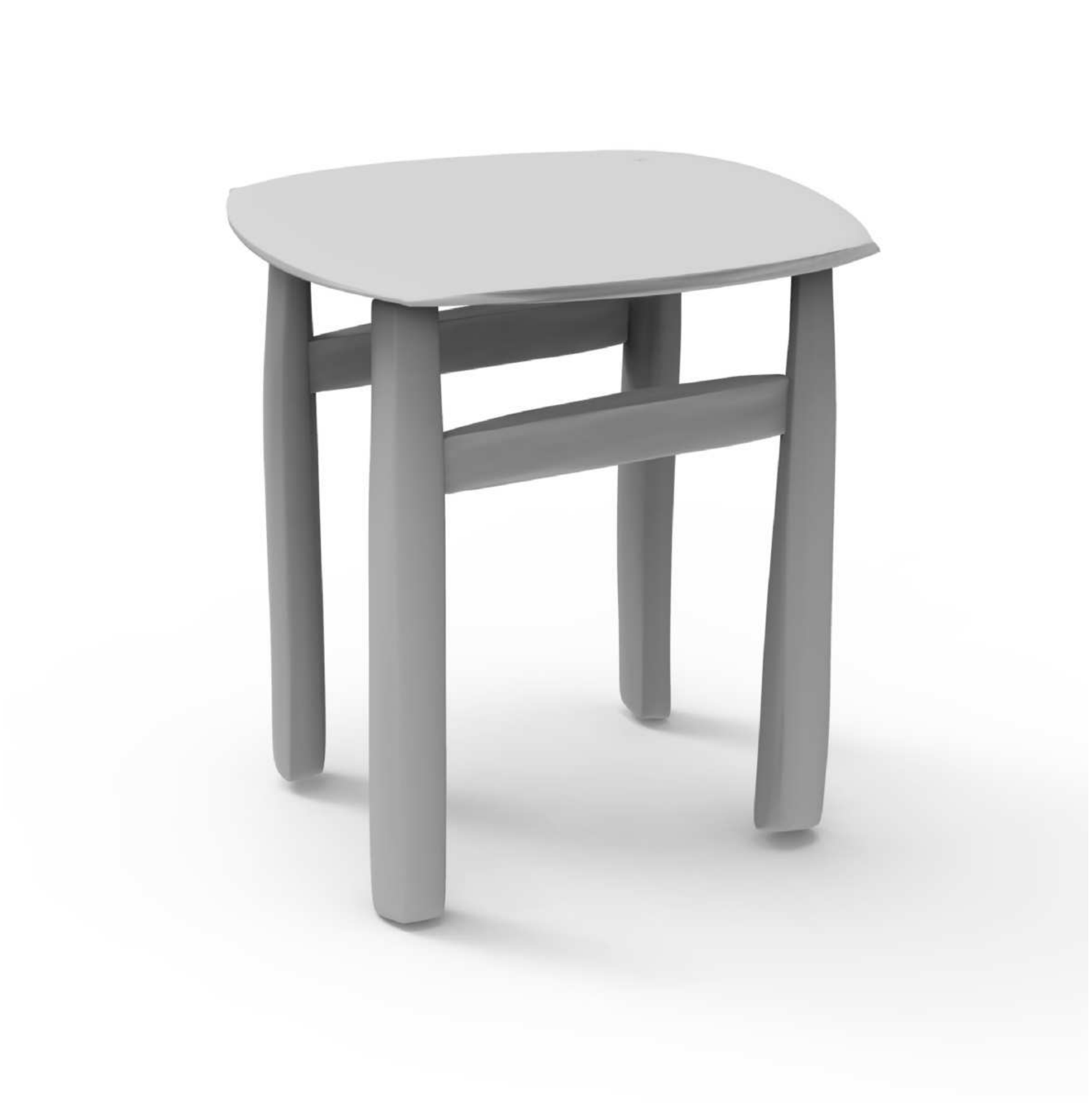}}; &
    \node (B4) {\includegraphics[width=0.19\linewidth]{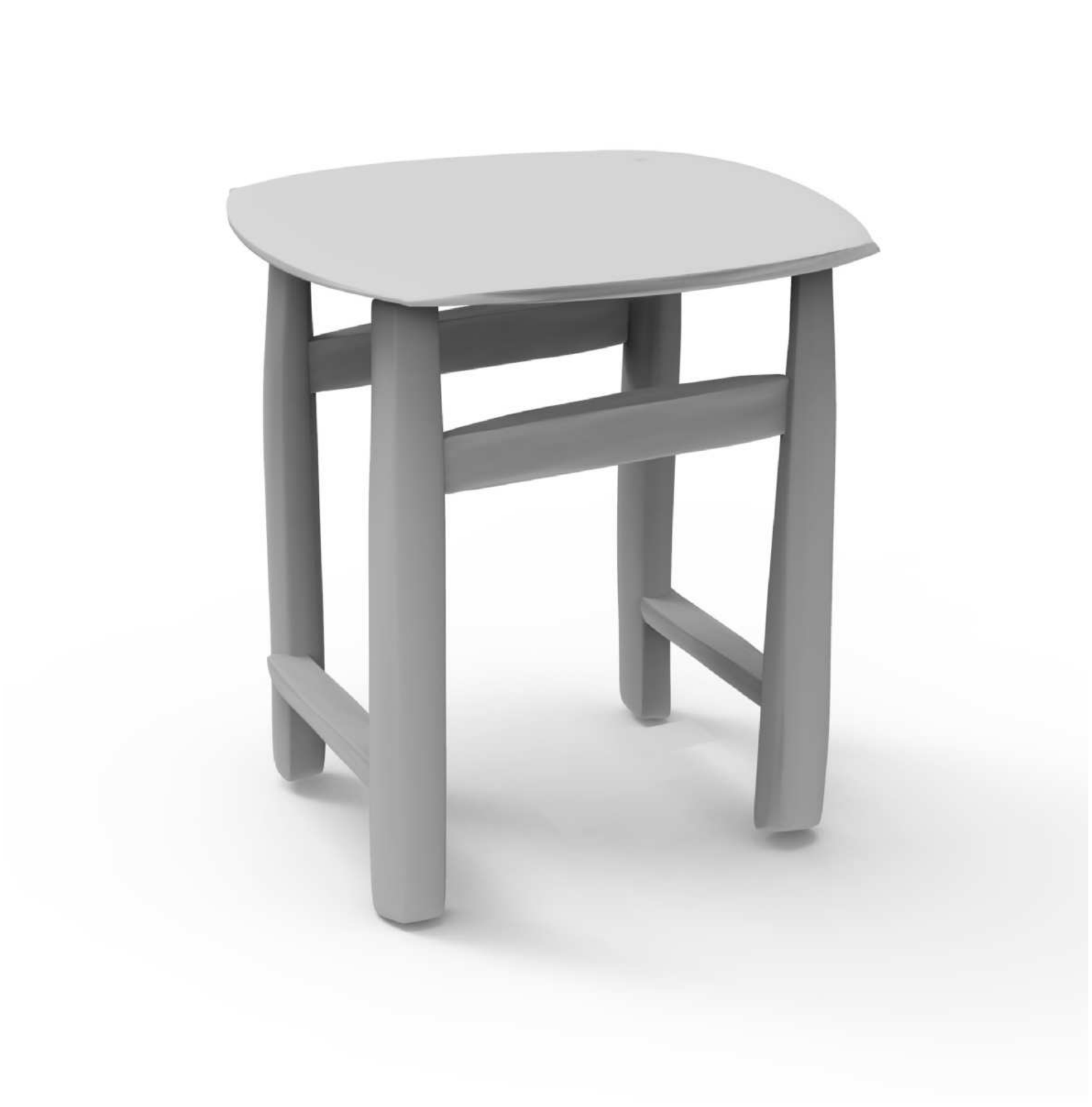}}; &
    \node (B5) {\includegraphics[width=0.19\linewidth]{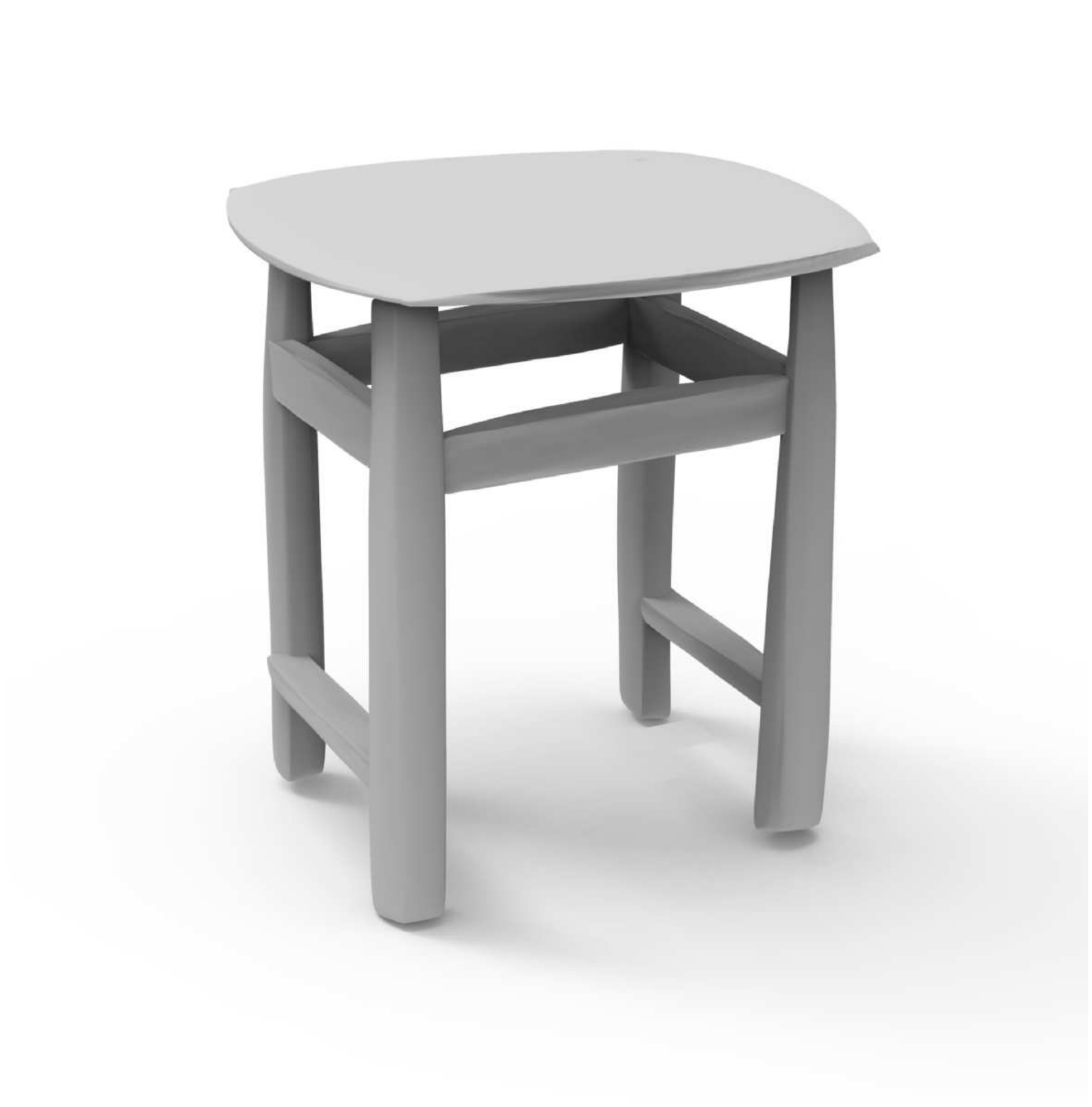}}; \\
    \node (C1) {\includegraphics[width=0.19\linewidth]{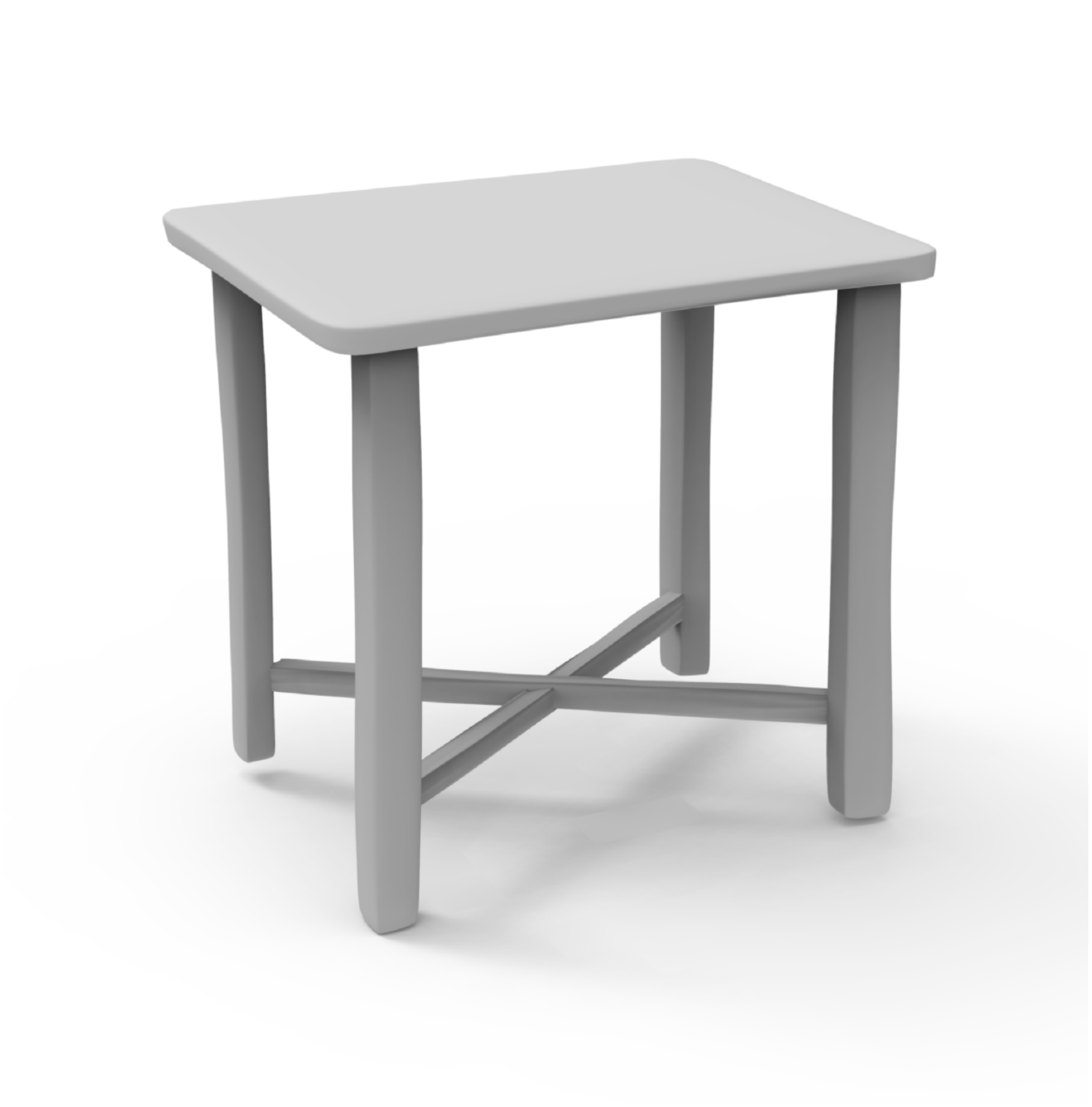}}; & 
    \node (C2) {\includegraphics[width=0.19\linewidth]{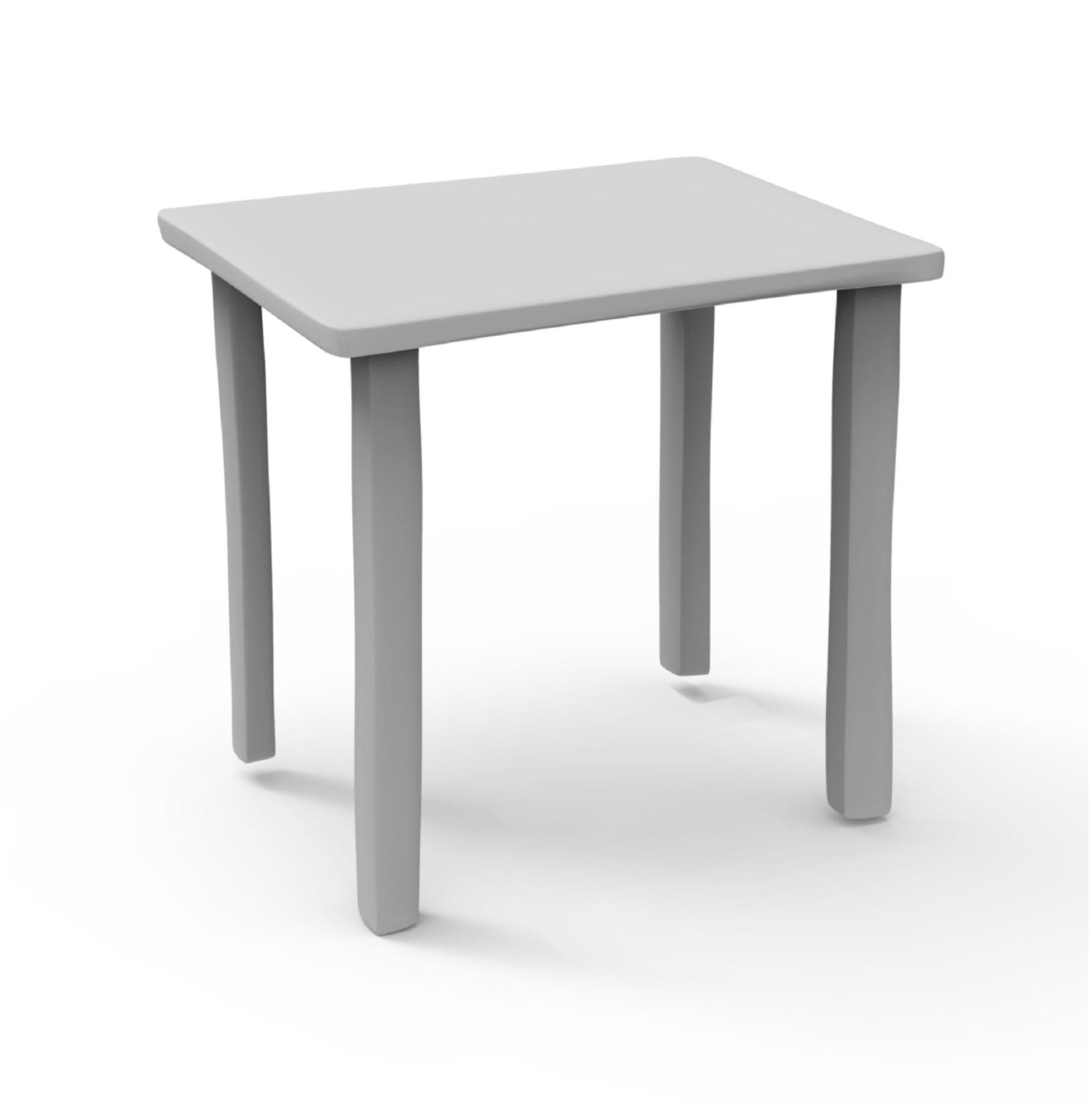}}; &
    \node (C3) {\includegraphics[width=0.19\linewidth]{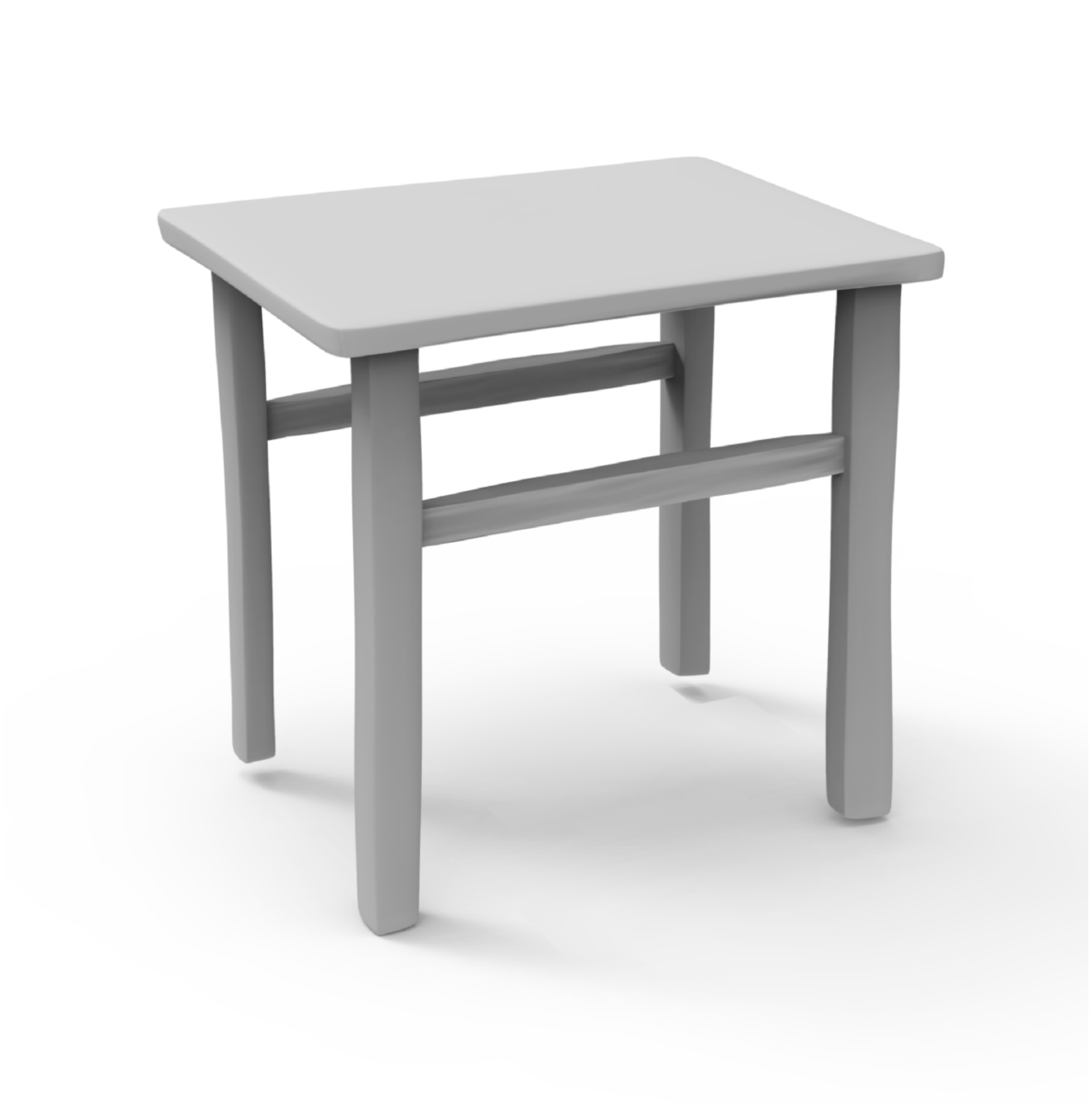}}; &
    \node (C4) {\includegraphics[width=0.19\linewidth]{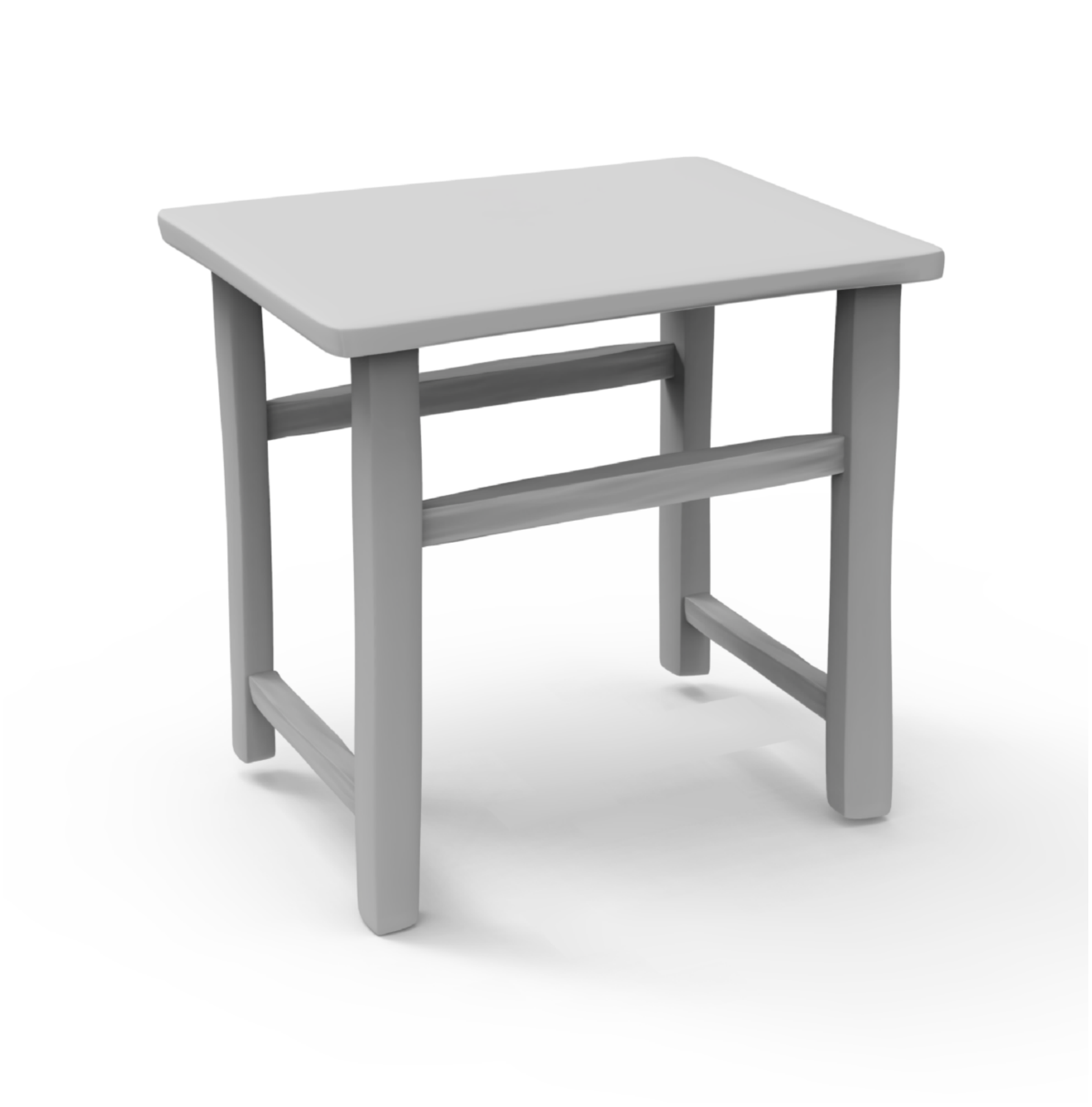}}; &
    \node (C5) {\includegraphics[width=0.19\linewidth]{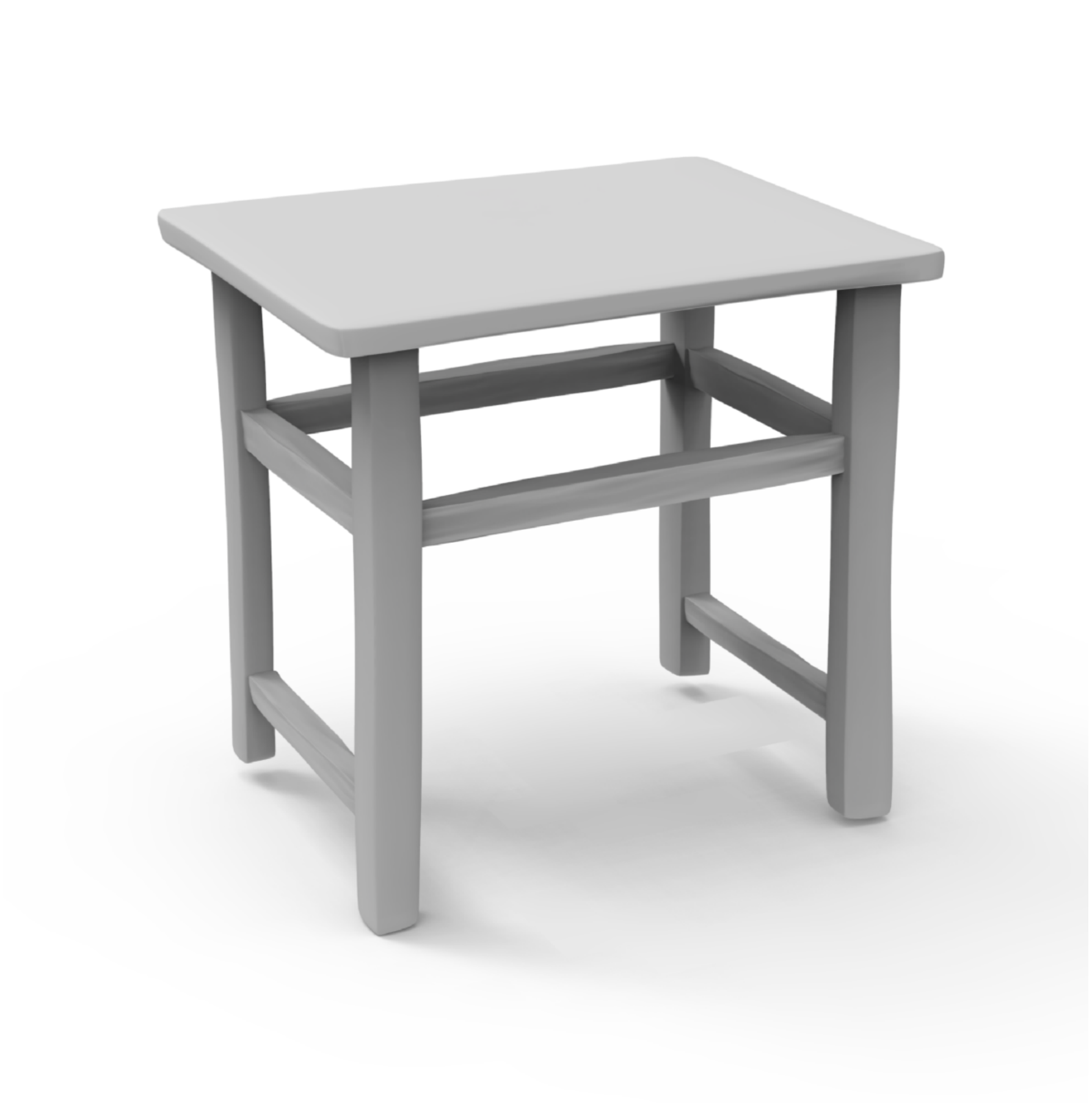}}; \\
    \node (D1) {\includegraphics[width=0.19\linewidth]{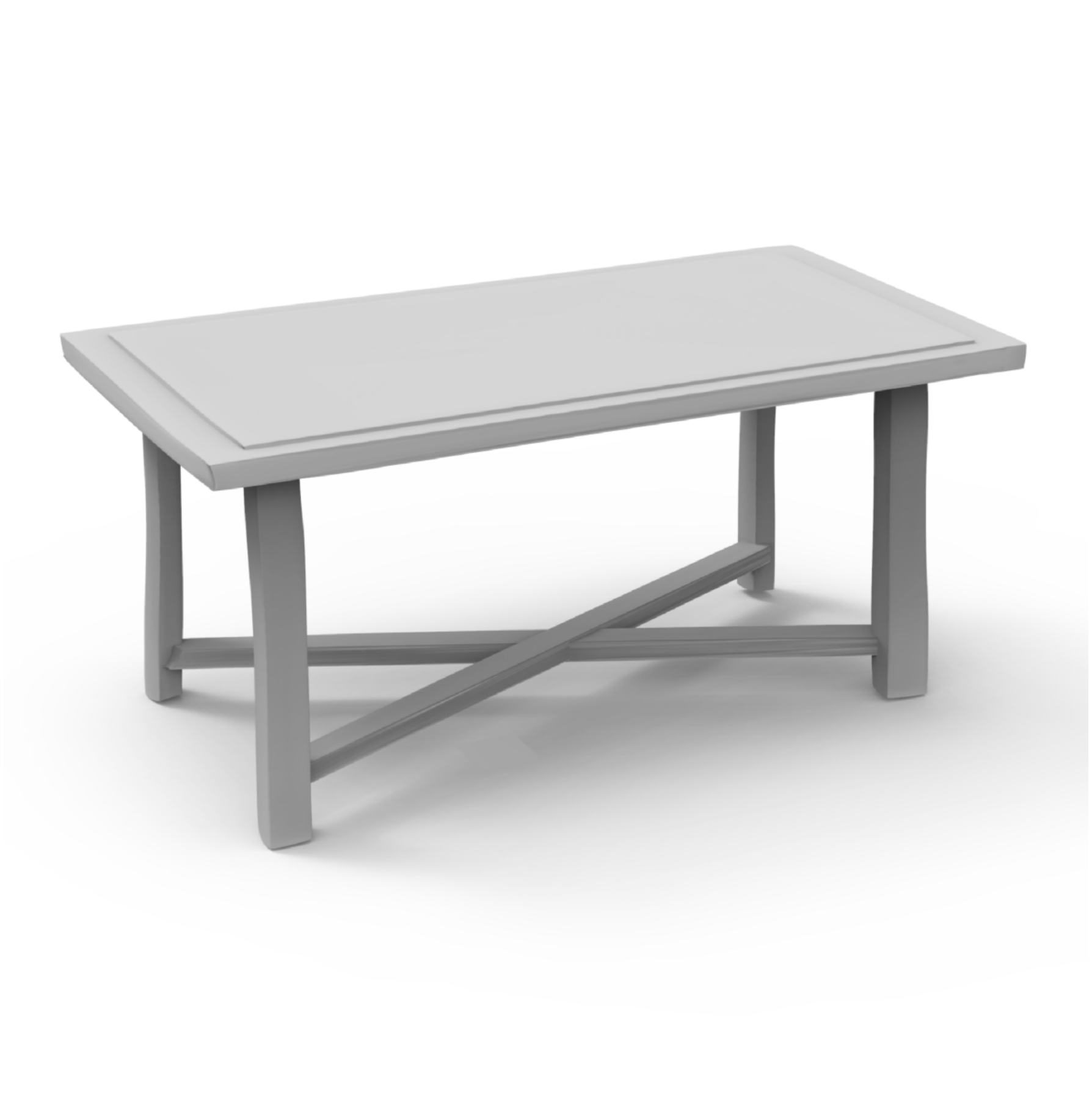}}; & 
    \node (D2) {\includegraphics[width=0.19\linewidth]{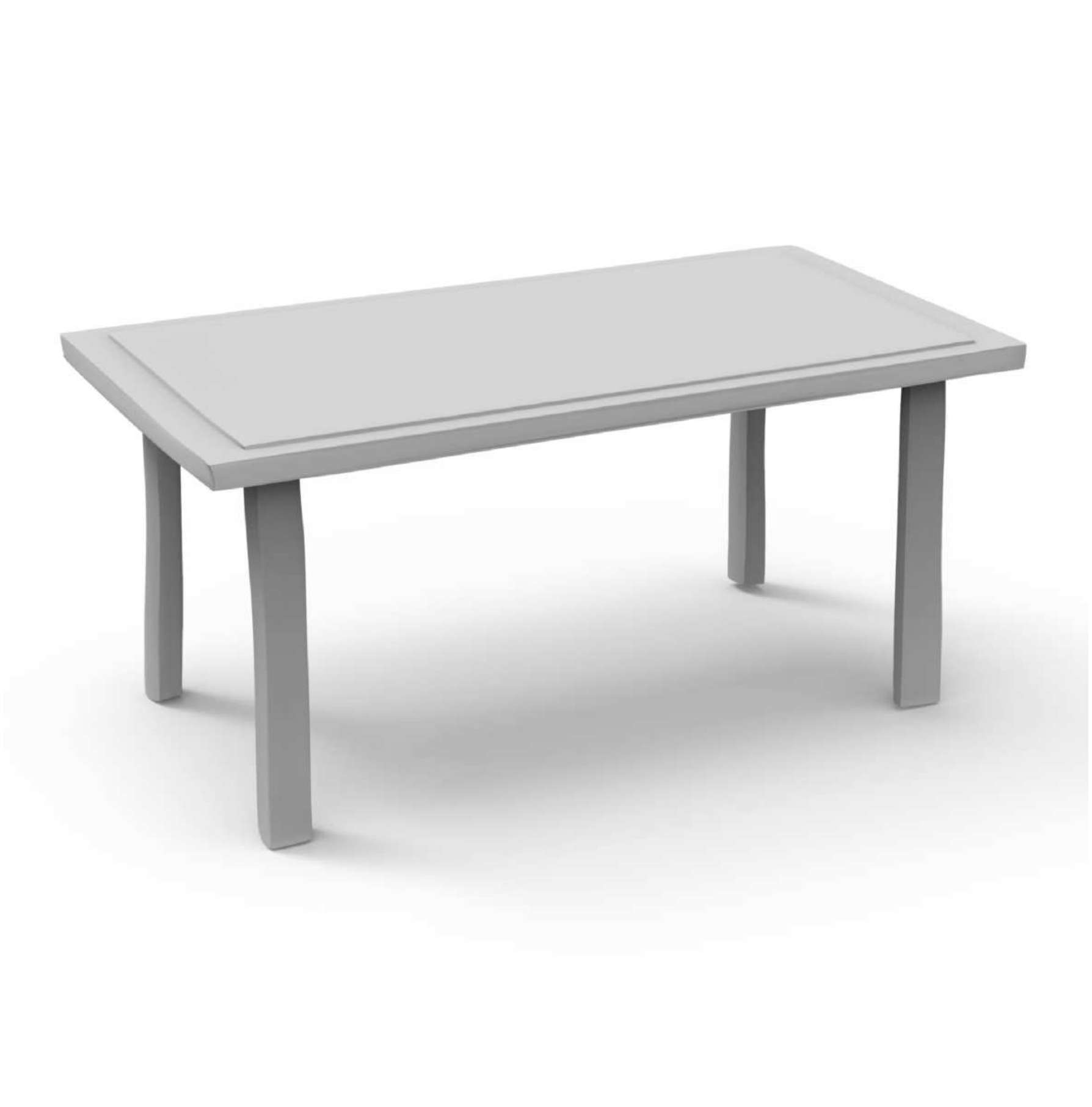}}; &
    \node (D3) {\includegraphics[width=0.19\linewidth]{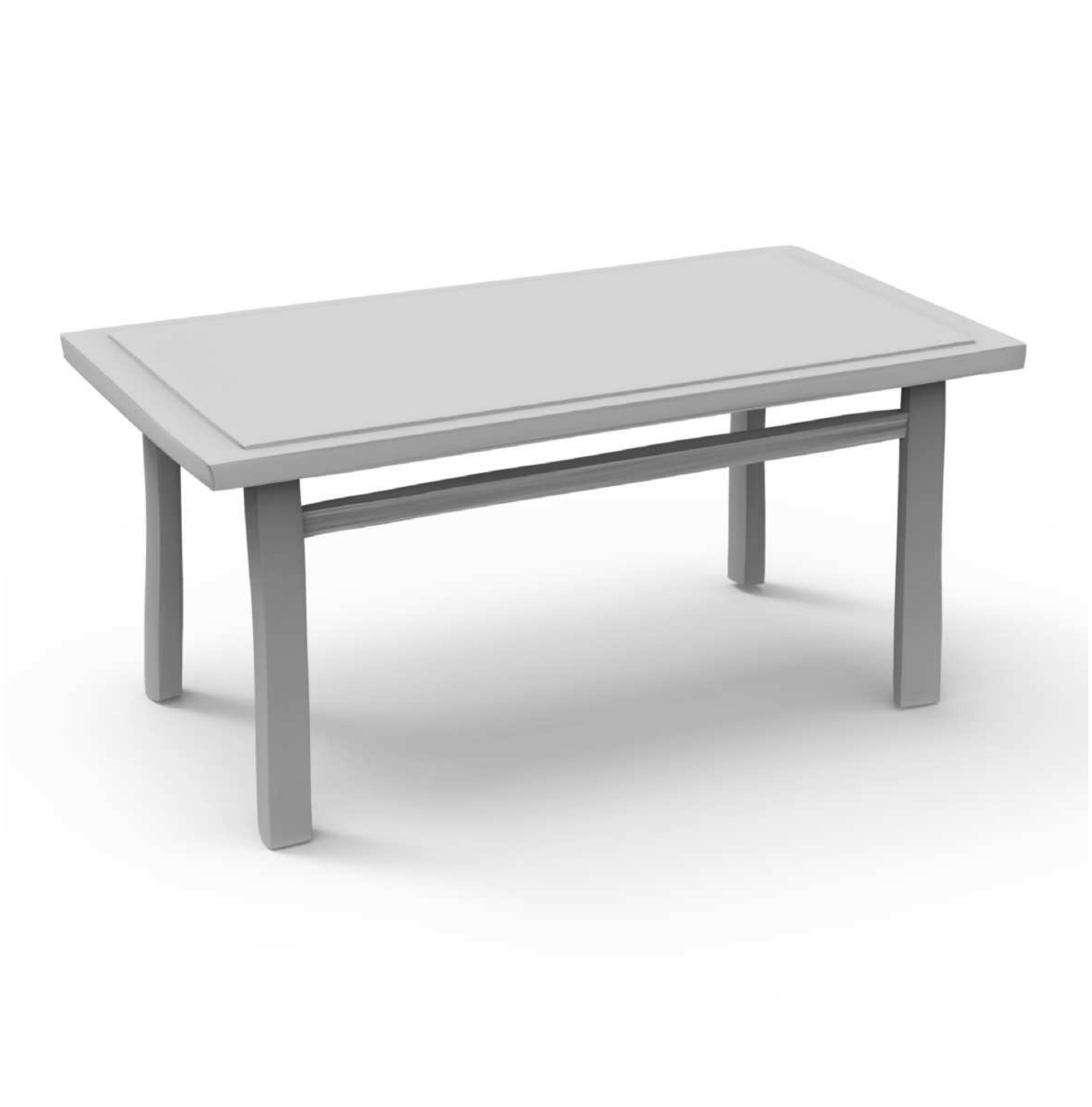}}; &
    \node (D4) {\includegraphics[width=0.19\linewidth]{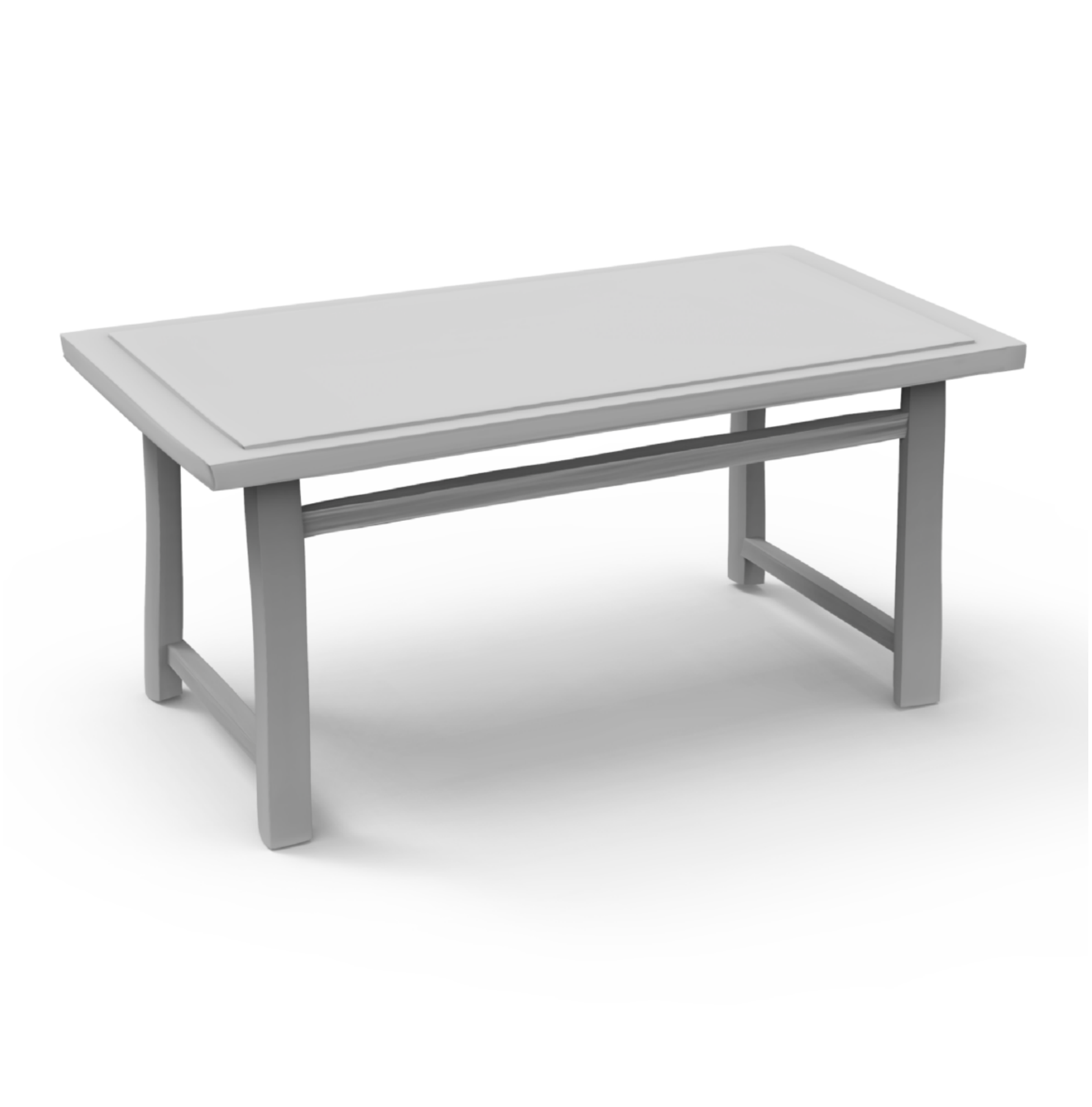}}; &
    \node (D5) {\includegraphics[width=0.19\linewidth]{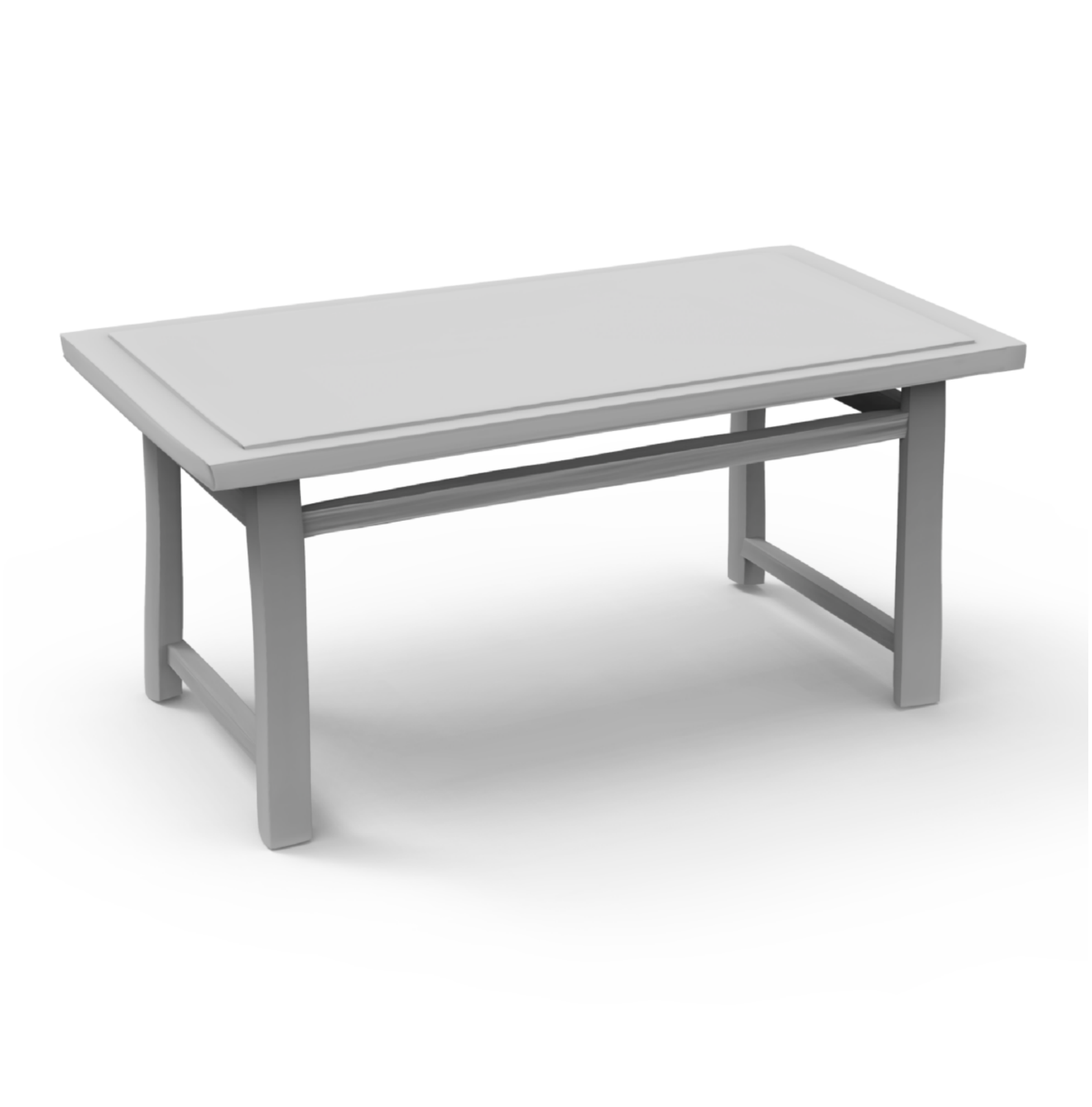}}; \\
    \node (F1) {\includegraphics[width=0.19\linewidth]{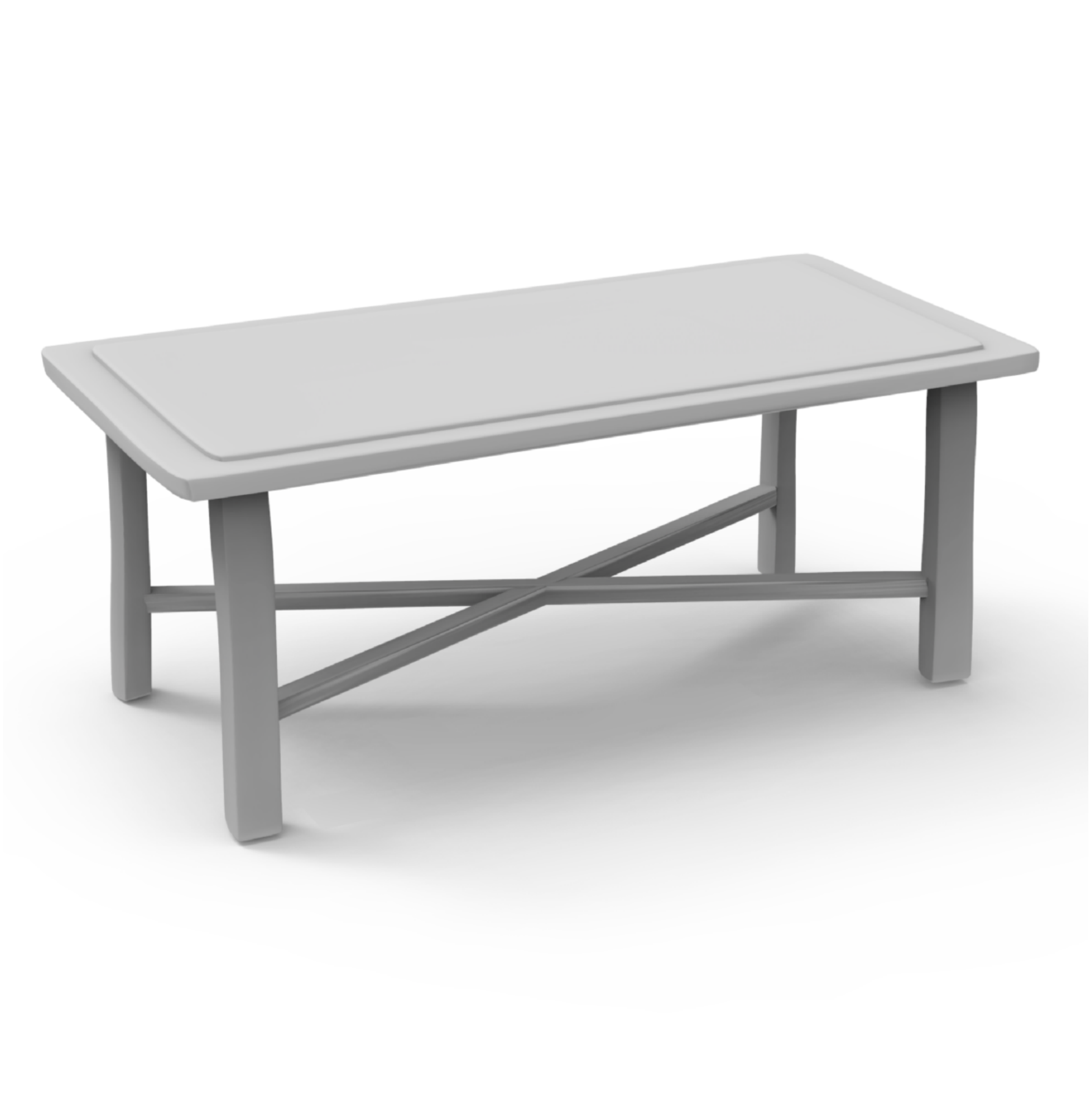}}; & 
    \node (F2) {\includegraphics[width=0.19\linewidth]{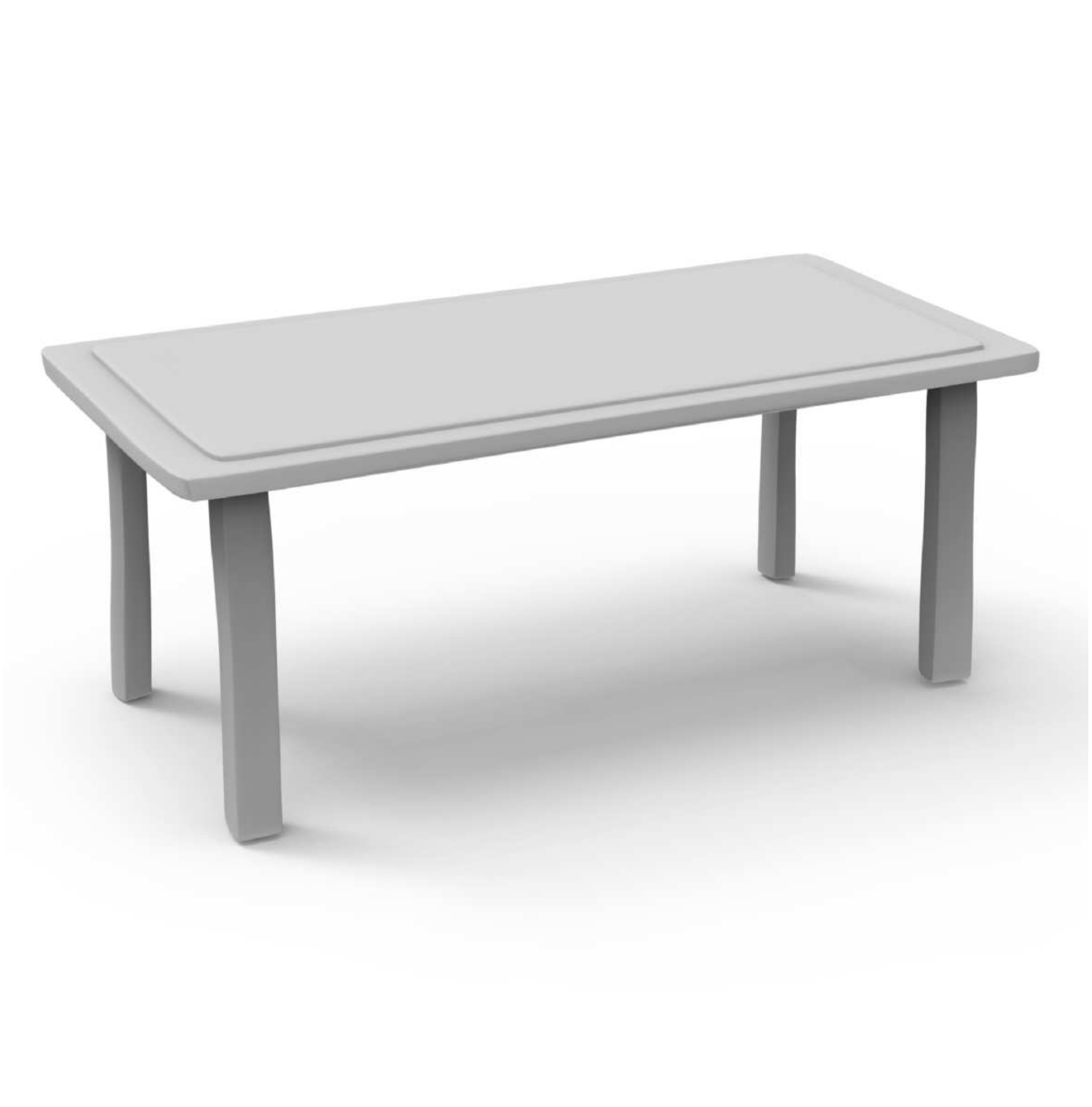}}; &
    \node (F3) {\includegraphics[width=0.19\linewidth]{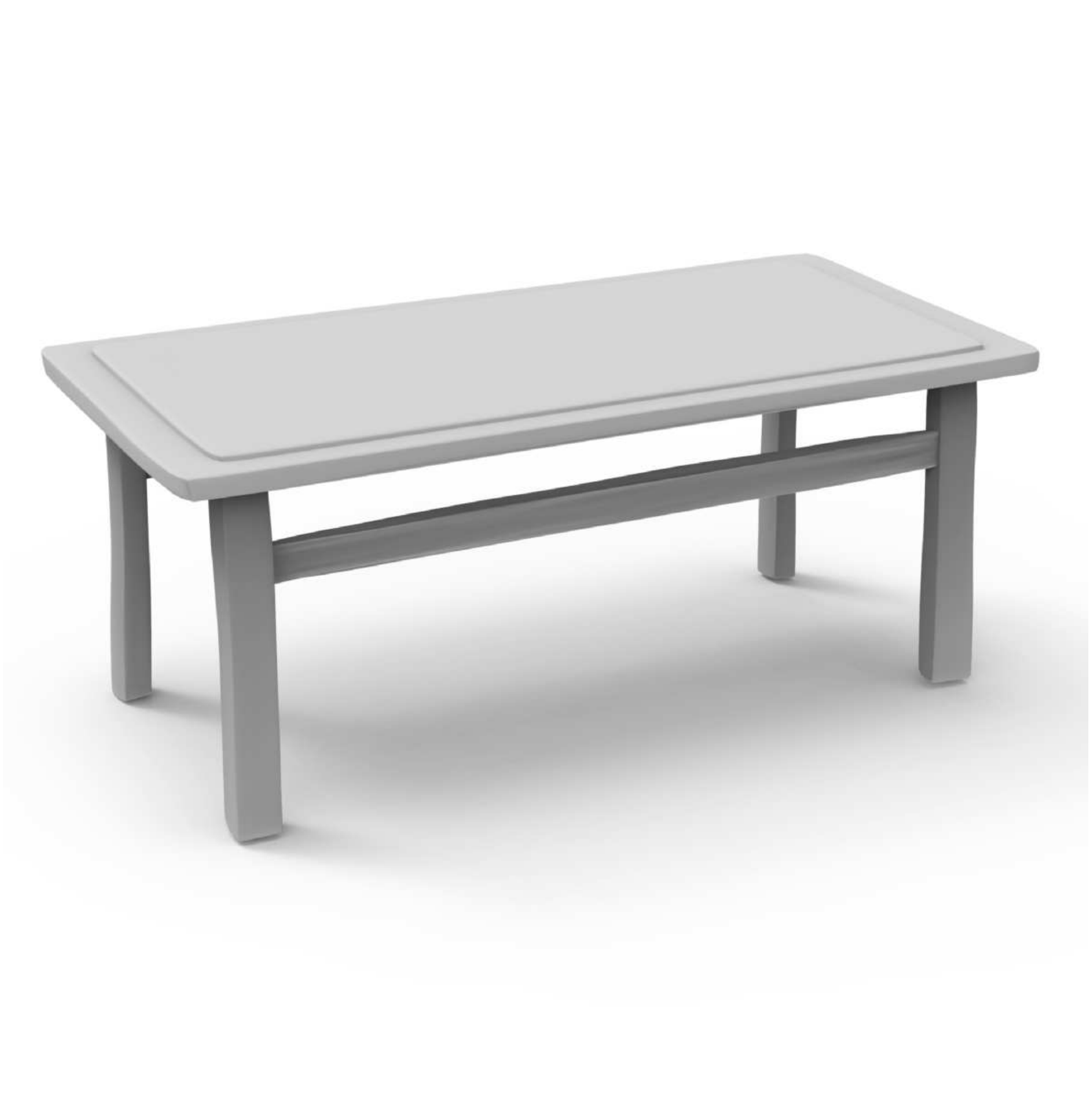}}; &
    \node (F4) {\includegraphics[width=0.19\linewidth]{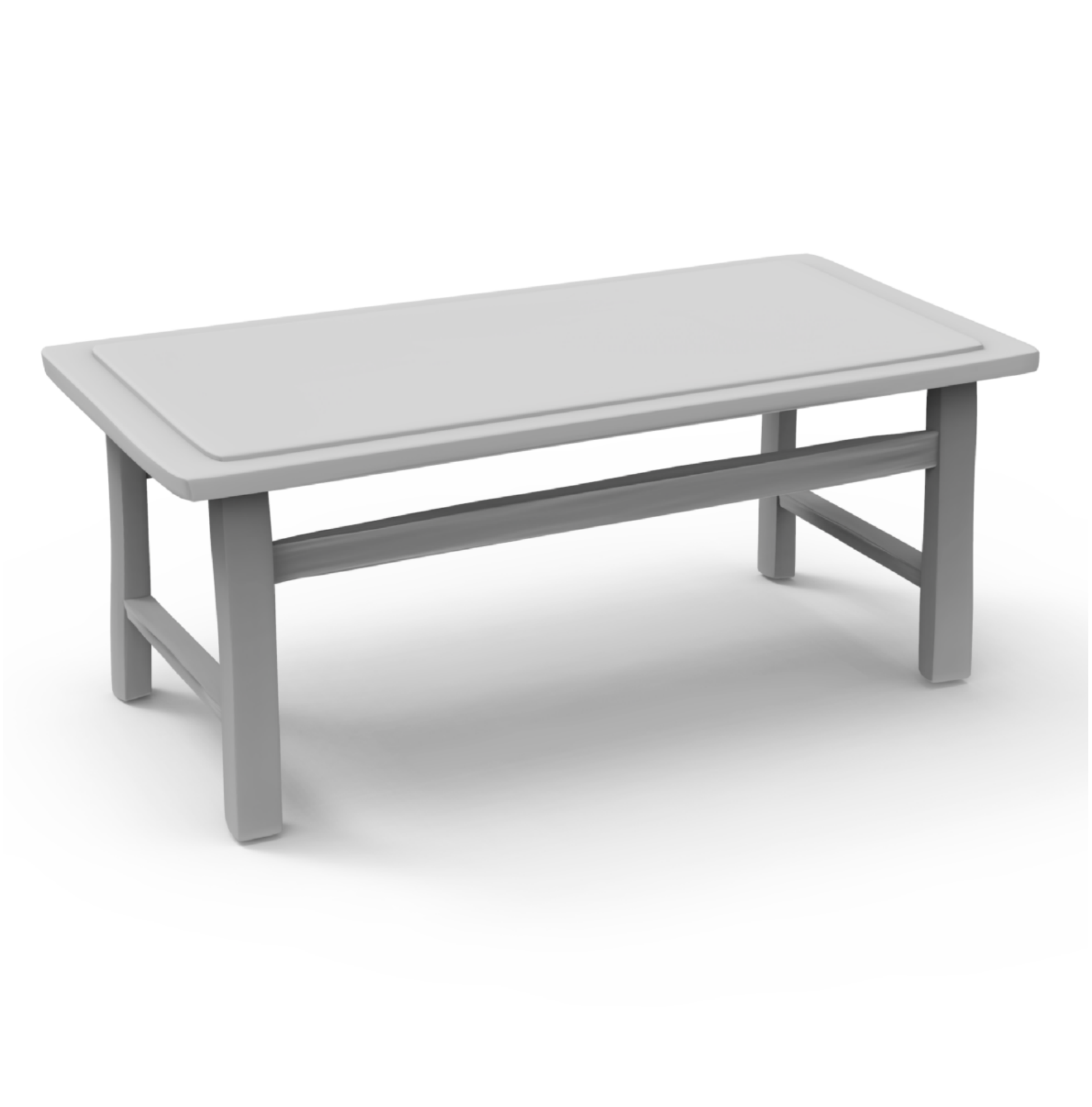}}; &
    \node (F5) {\includegraphics[width=0.19\linewidth, cfbox=orange 1pt 1pt]{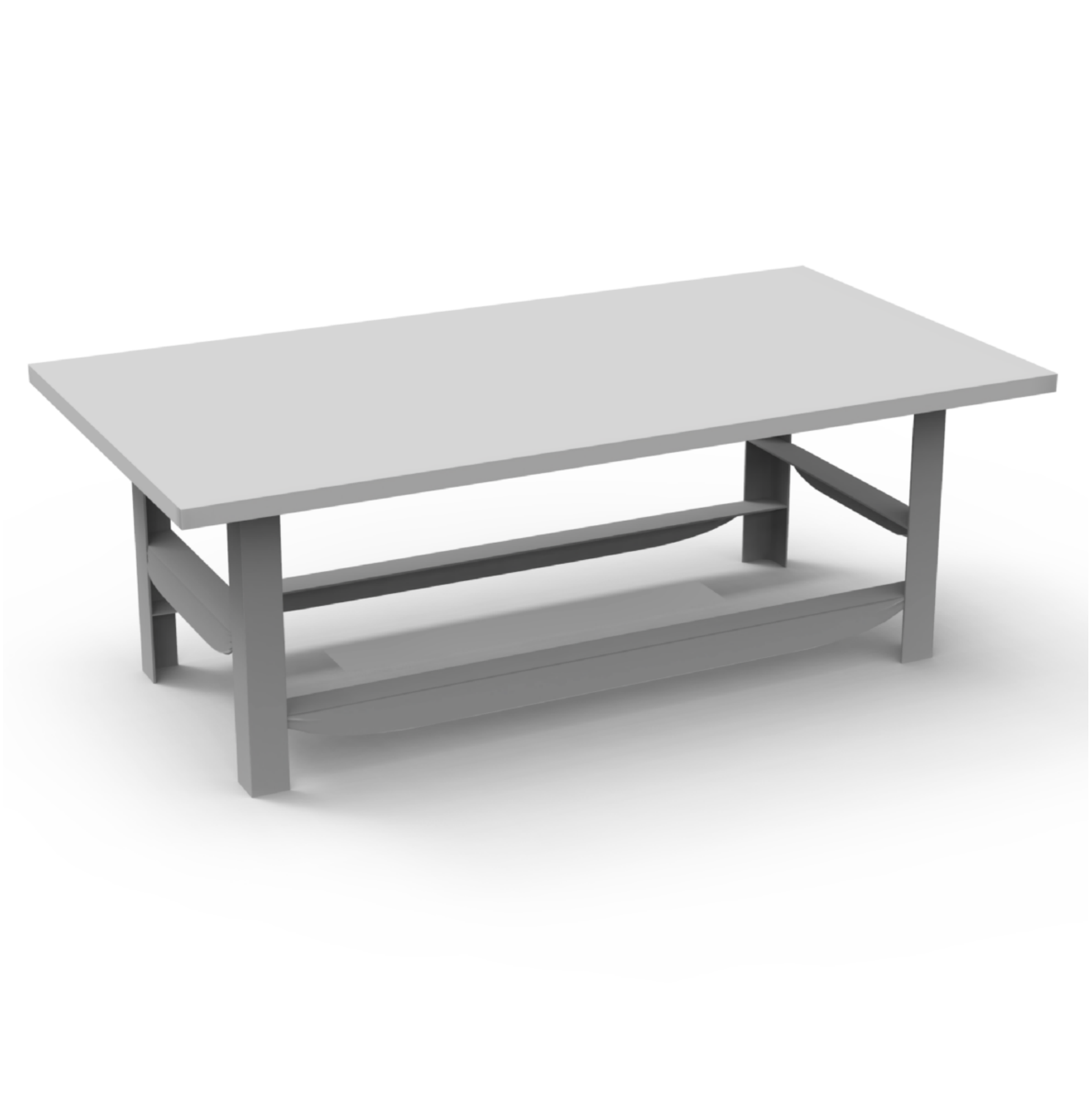}}; \\
  };
  \node[fit=(A1) (A2) (A3) (A4) (A5)
            (B1) (B2) (B3) (B4) (B5)
            (C1) (C2) (C3) (C4) (C5)
            (D1) (D2) (D3) (D4) (D5)
            (F1) (F2) (F3) (F4) (F5),
            inner sep=0pt,
            ] (PIC) {};

  \draw[line width=1pt,arrows={-Stealth[length=4mm]}] ([xshift=-1em,yshift=-0.5em]PIC.south west) -- ([xshift=-1em,yshift=-0.5em]PIC.south east);
  \draw[line width=1pt,arrows={-Stealth[length=4mm]}] ([yshift=-1em,xshift=-0.5em]PIC.south west) -- ([yshift=-1em,xshift=-0.5em]PIC.north west);

  \node[anchor=south] (label) [font=\fontsize{10}{10}\selectfont]at ([yshift=-2.em]A3|-PIC.south) {Structure};
  \node[anchor=center,rotate=90] (label) [font=\fontsize{10}{10}\selectfont] at ([xshift=-1.5em]C1-|PIC.west) {Geometry};

\end{tikzpicture}
    \end{minipage}
    \vspace{-3mm}
    \caption{
    \yjr{Disentangled shape reconstruction and interpolation results on PartNet Cabinet and Table. Here, the top left and bottom right shapes (highlighted with orange boxes) are the input shapes. The remaining shapes are generated automatically with our DSG-Net, where in each row, the \emph{structure} of the shapes is interpolated while keeping the geometry factor unchanged, whereas in each column, the \emph{geometry} is interpolated while retaining the structure. \yj{The vertical axis and horizontal axis represent the variation of structure and geometry respectively.}}
    }
    \label{fig:decouple_chair}
    \vspace{-3mm}
\end{figure*}

\subsection{Shape Interpolation}
We evaluate our DSG-Net on shape interpolation and demonstrate that our network learns smooth latent spaces.
\yjrr{Next, we propose a novel task of disentangled shape reconstruction that takes two shapes as inputs to re-synthesize a novel shape with ingredients of the structure of one shape and the geometry of the other shape.}
Moreover, with the help of \yjrr{the novel cycled} disentanglement and our learned disentangled latent spaces for shape structure and geometry, we can also achieve controllable interpolation between two shapes, varying shape structure while keeping geometry unchanged and vice versa.
\yjr{Please refer to the supplementary material for more results on shape interpolation.}

\paragraph{Results.} 
Figure~\ref{fig:interp_storage} shows some interpolated results that interpolate jointly in the structure and geometry latent spaces. 
We see both continuous geometry variations and discrete structure changes in the interpolation, which validates that DSG-Net learns a smooth manifold for shape generation.

\begin{figure}[t]
\centering

{\raisebox{-0.4\height}{\includegraphics[width=0.16\linewidth]{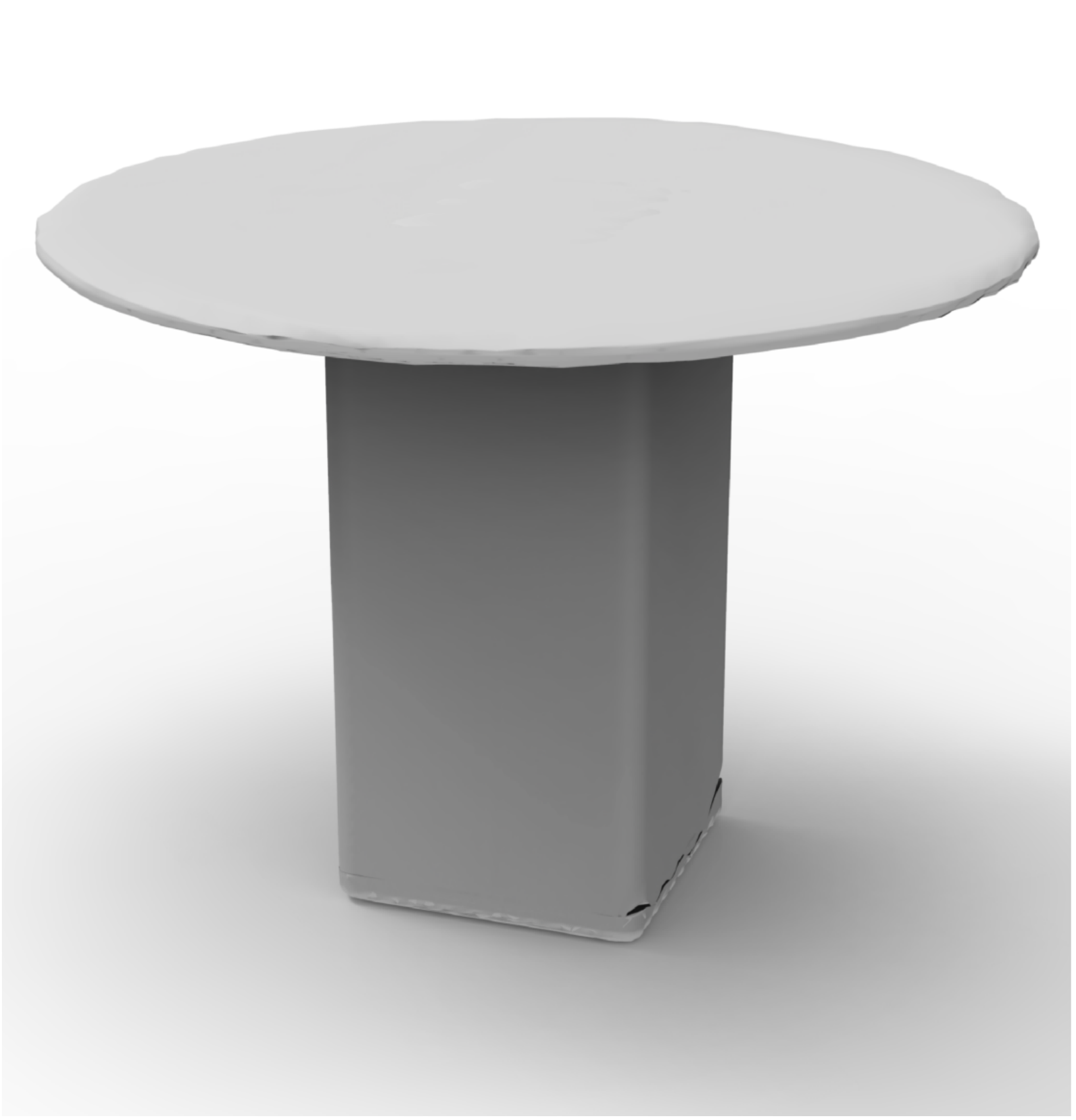}}}
\begin{minipage}{0.65\linewidth}
\centering
    \includegraphics[width=0.18\linewidth]{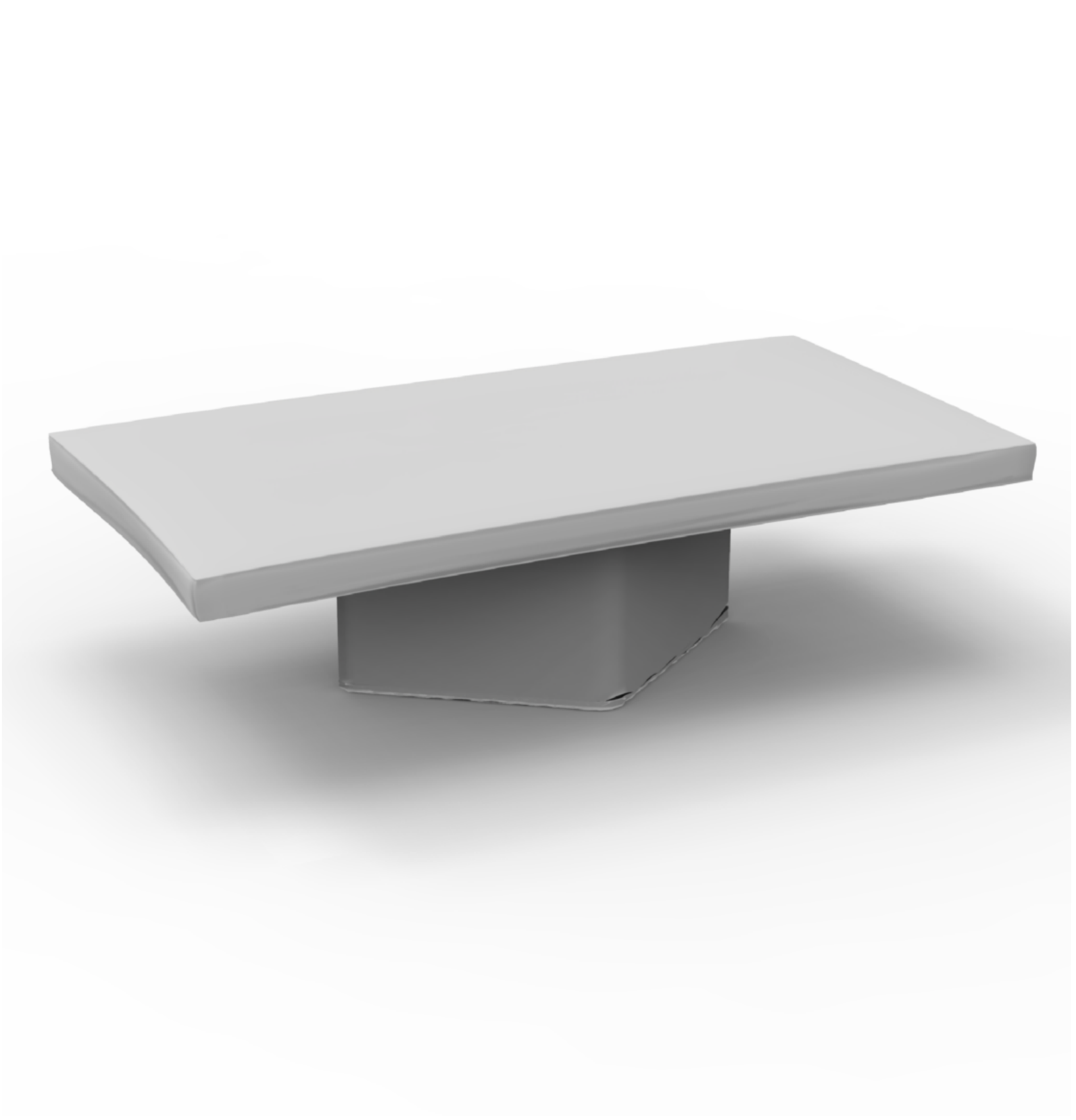}
    \includegraphics[width=0.18\linewidth]{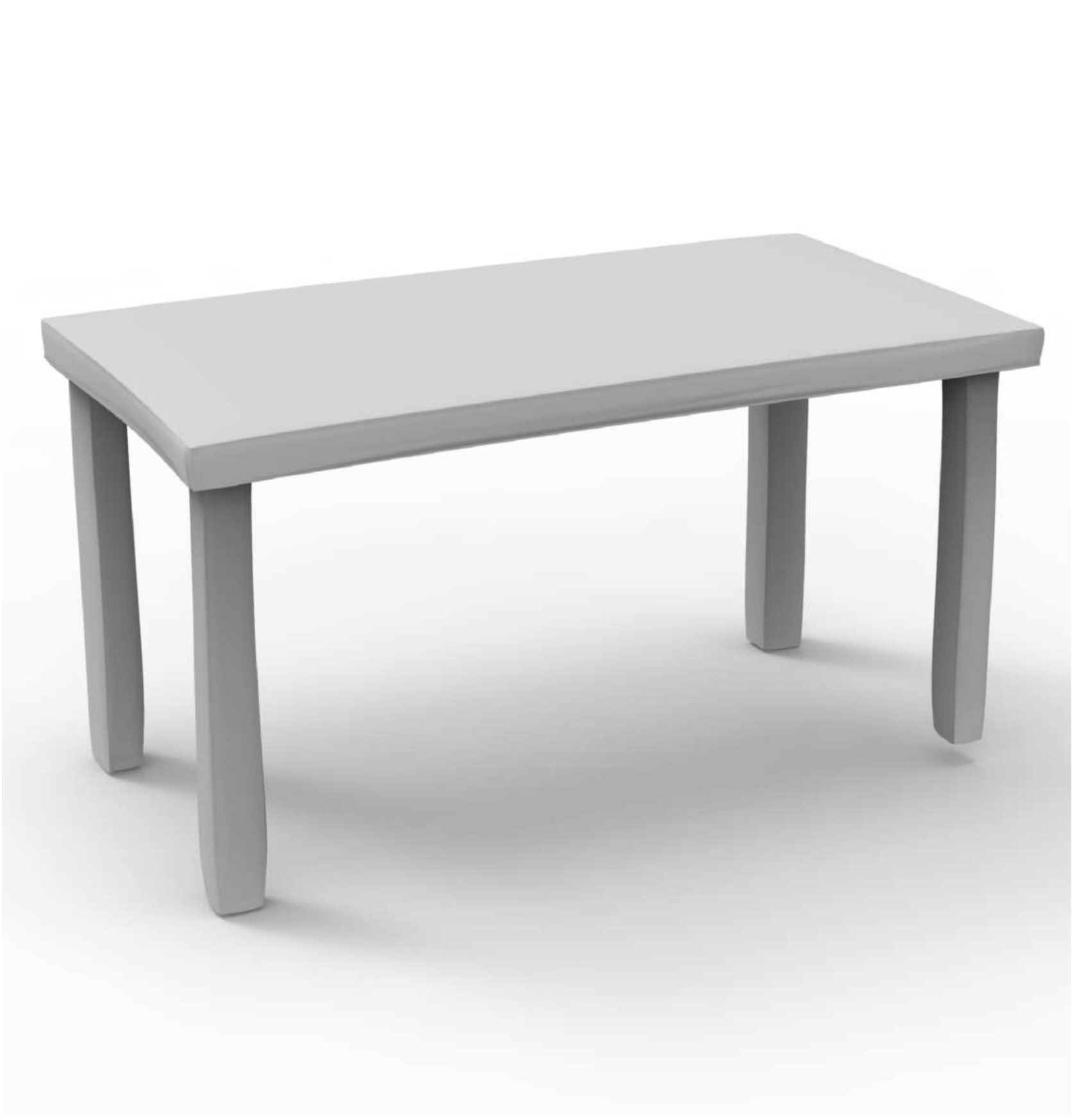}
    \includegraphics[width=0.18\linewidth]{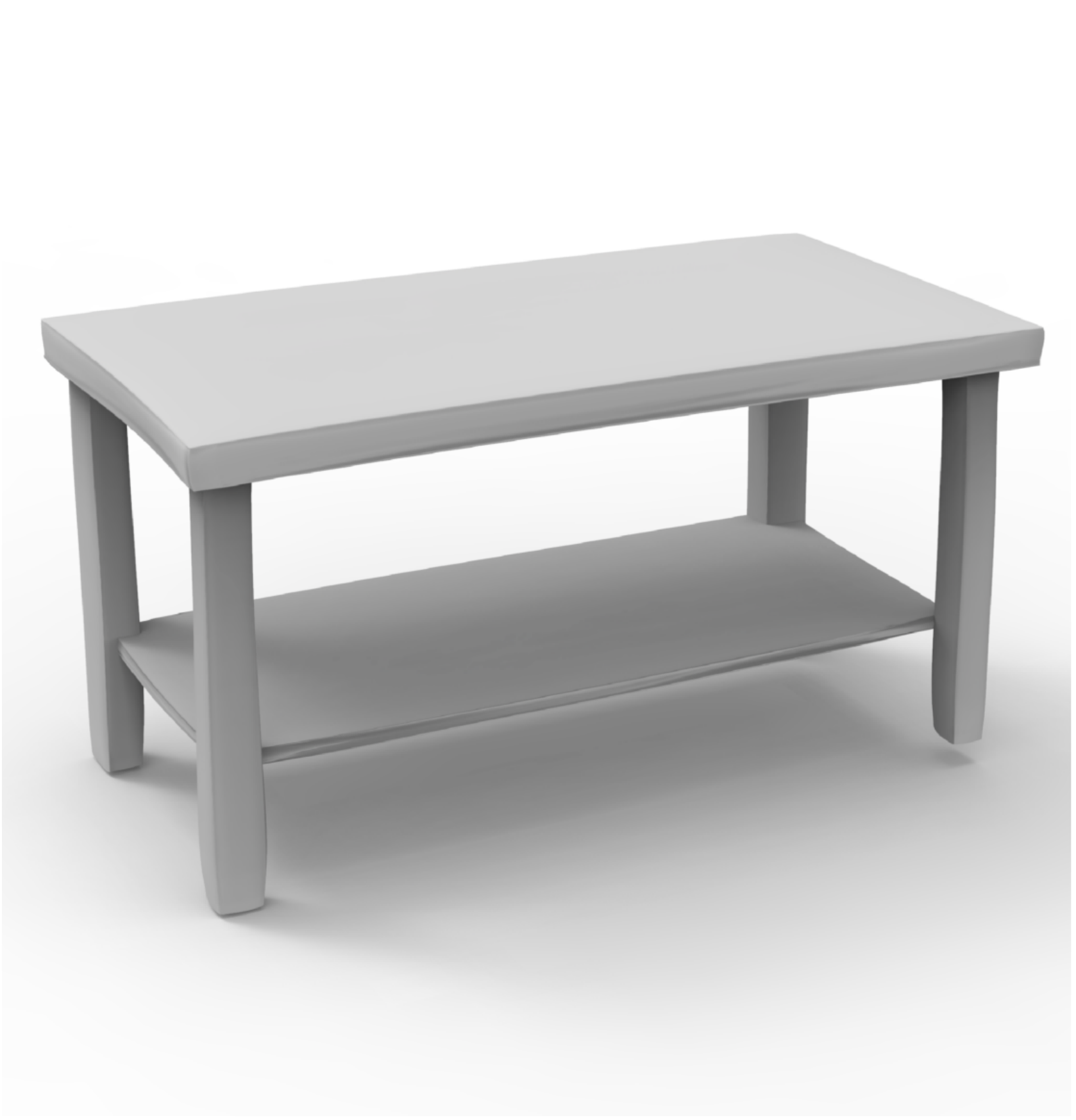}
    \includegraphics[width=0.18\linewidth]{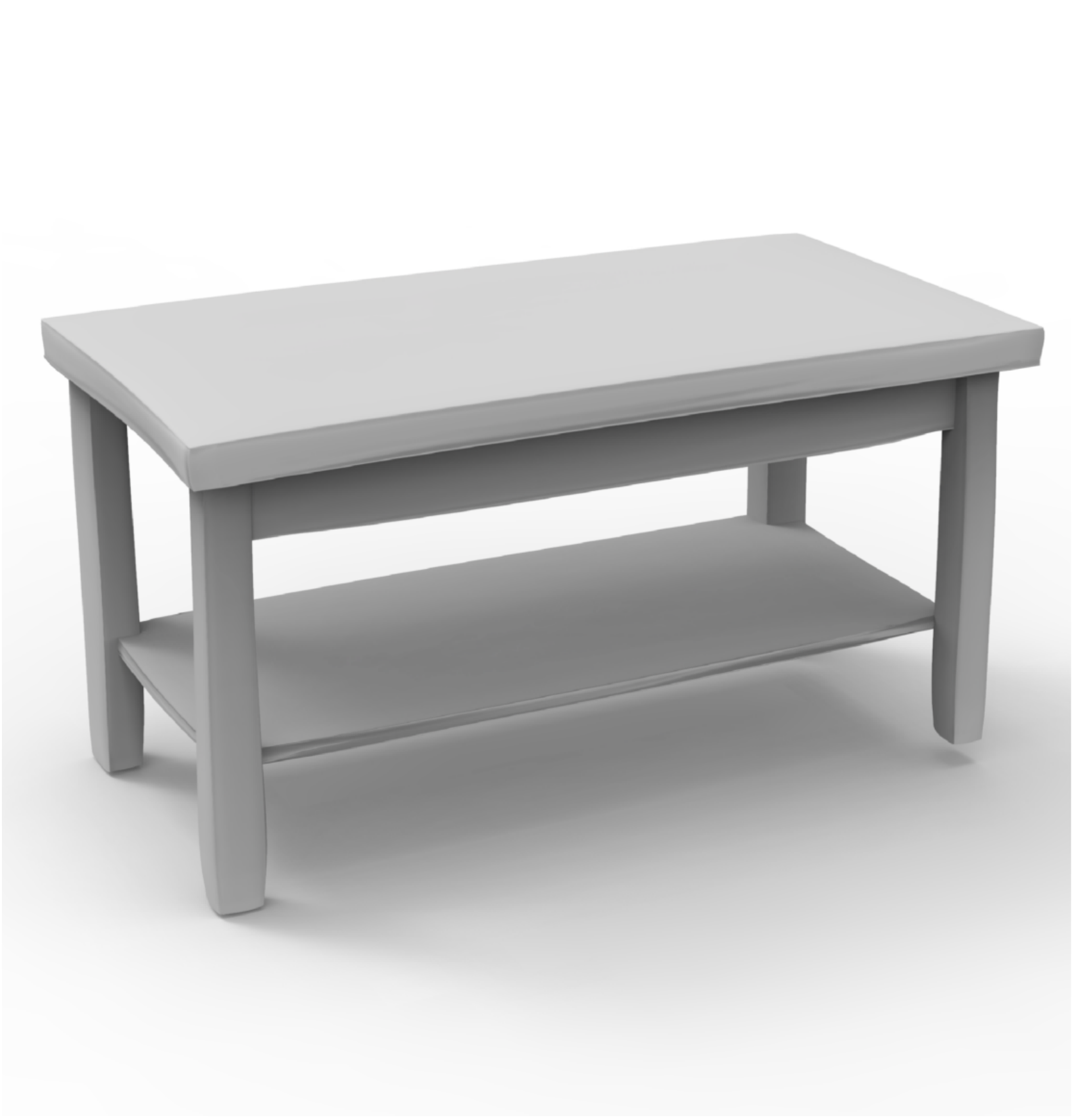}
    \includegraphics[width=0.18\linewidth]{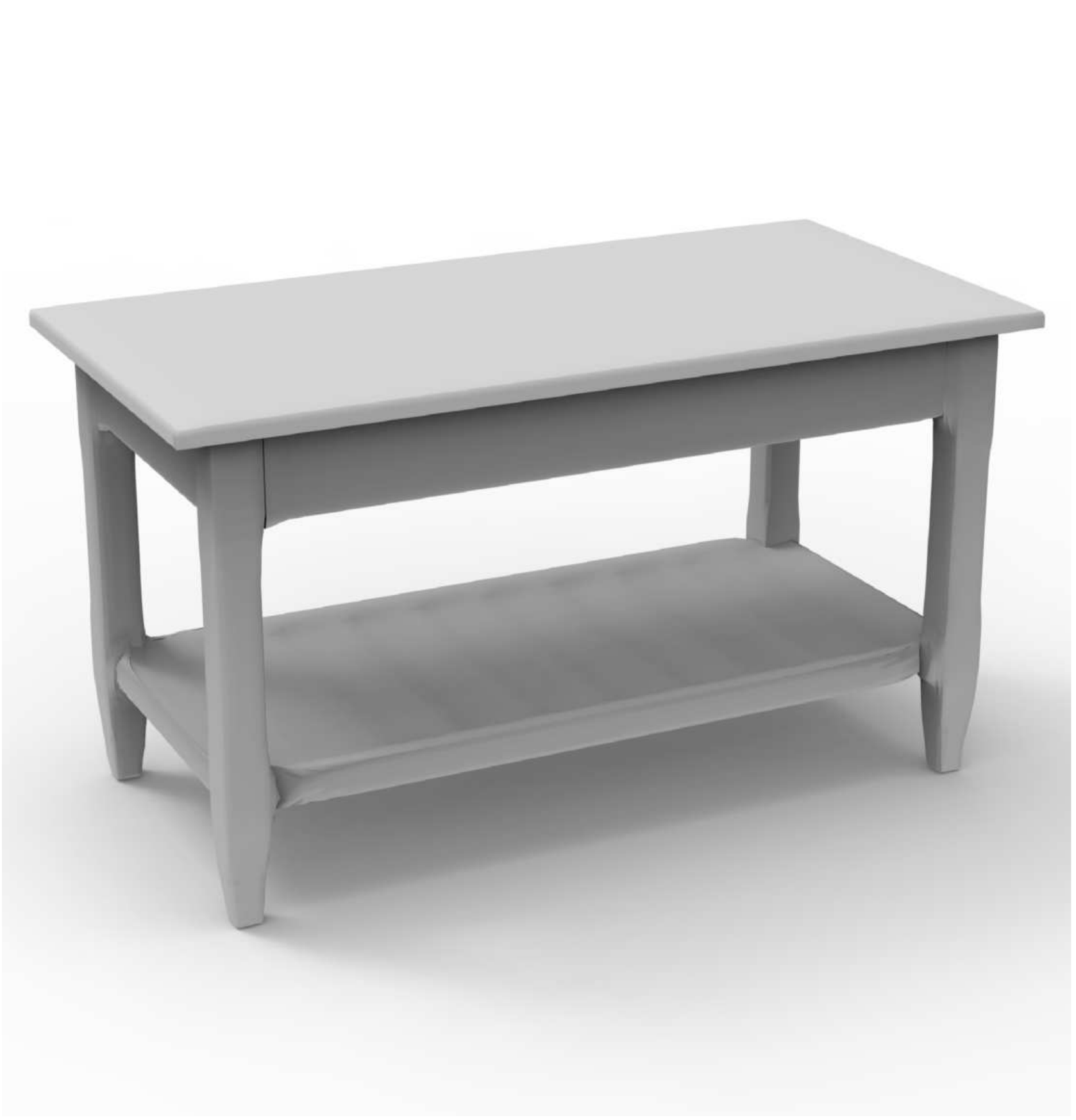}\\
    \includegraphics[width=0.18\linewidth]{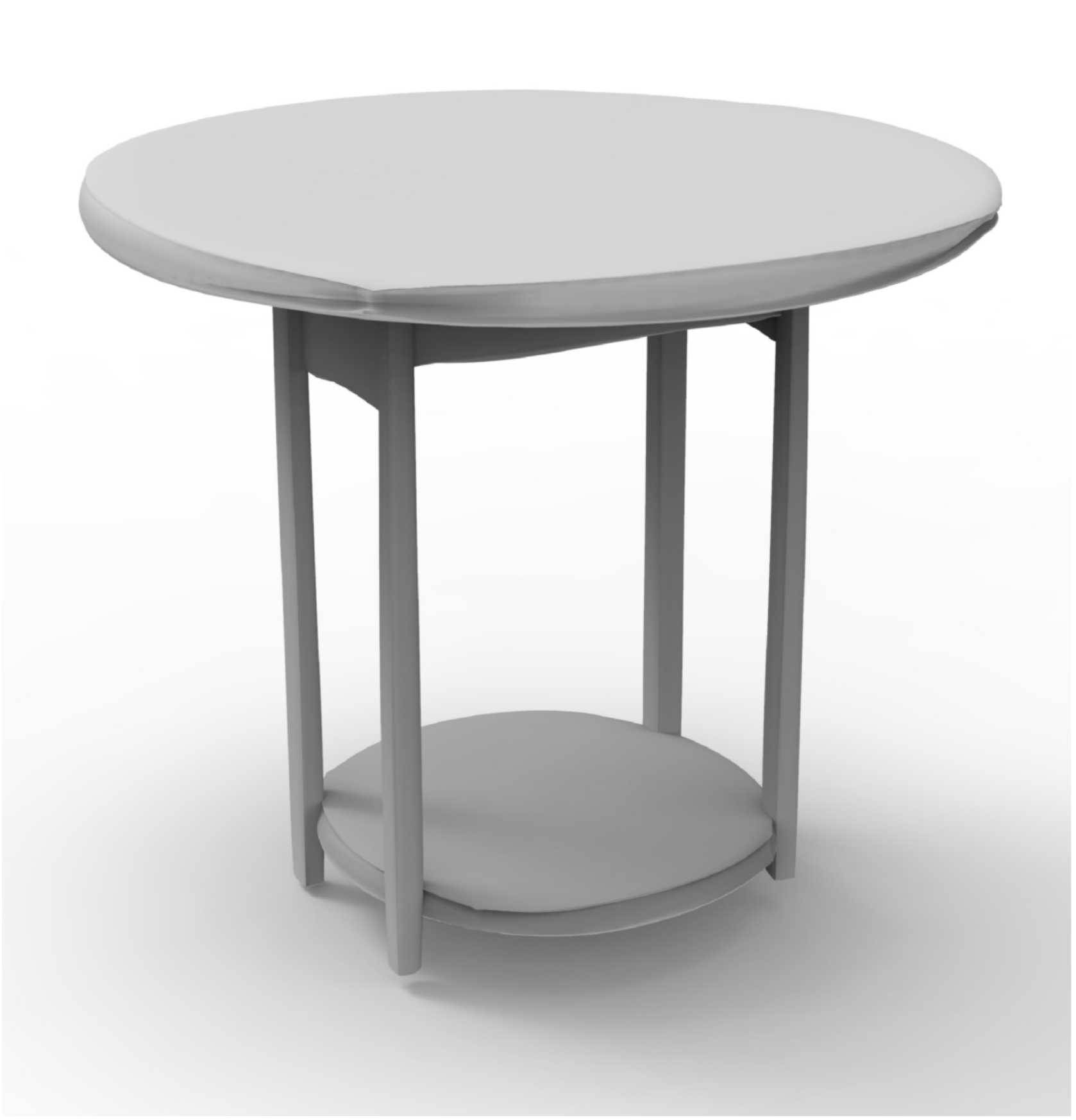}
    \includegraphics[width=0.18\linewidth]{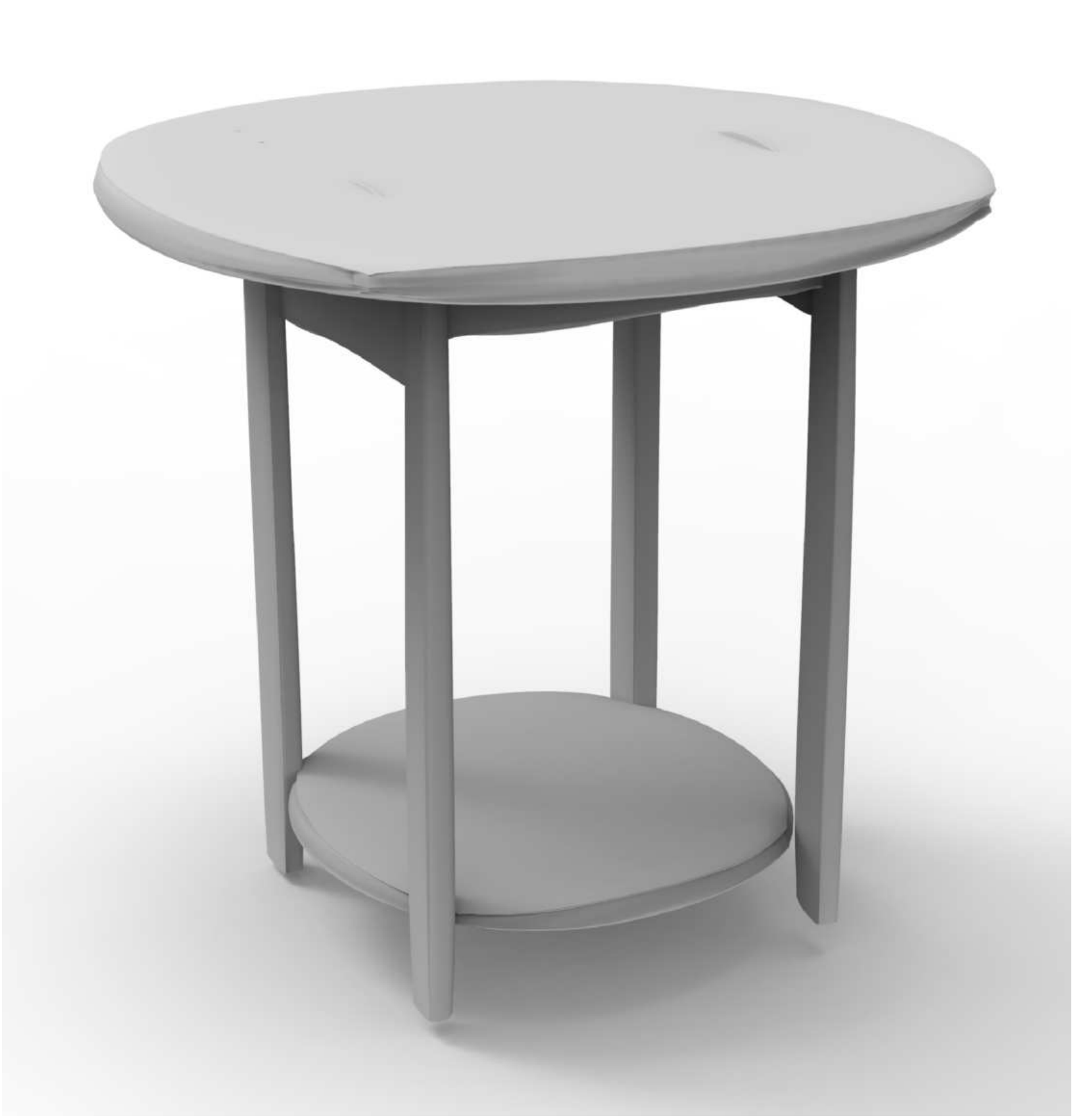}
    \includegraphics[width=0.18\linewidth]{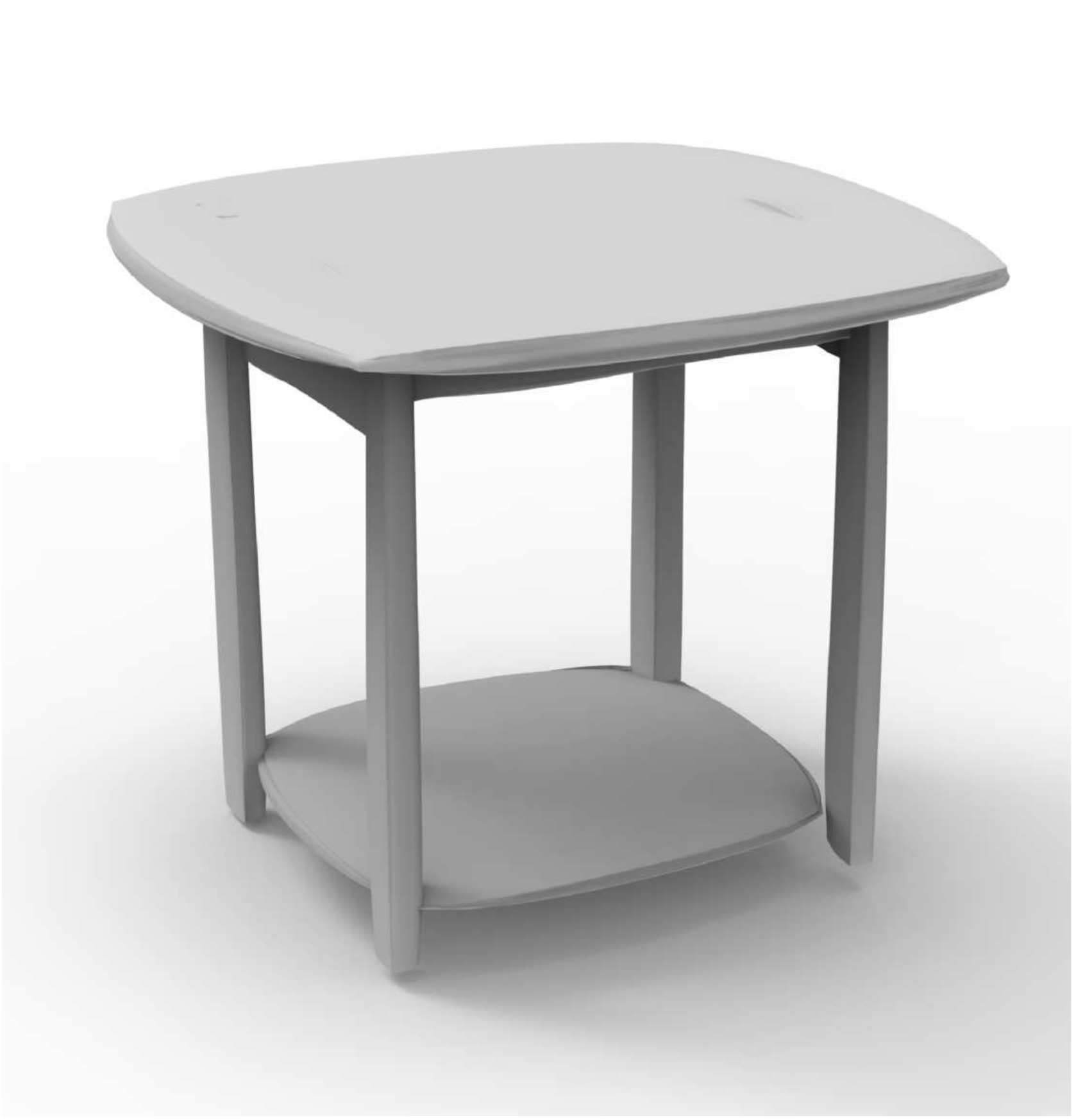}
    \includegraphics[width=0.18\linewidth]{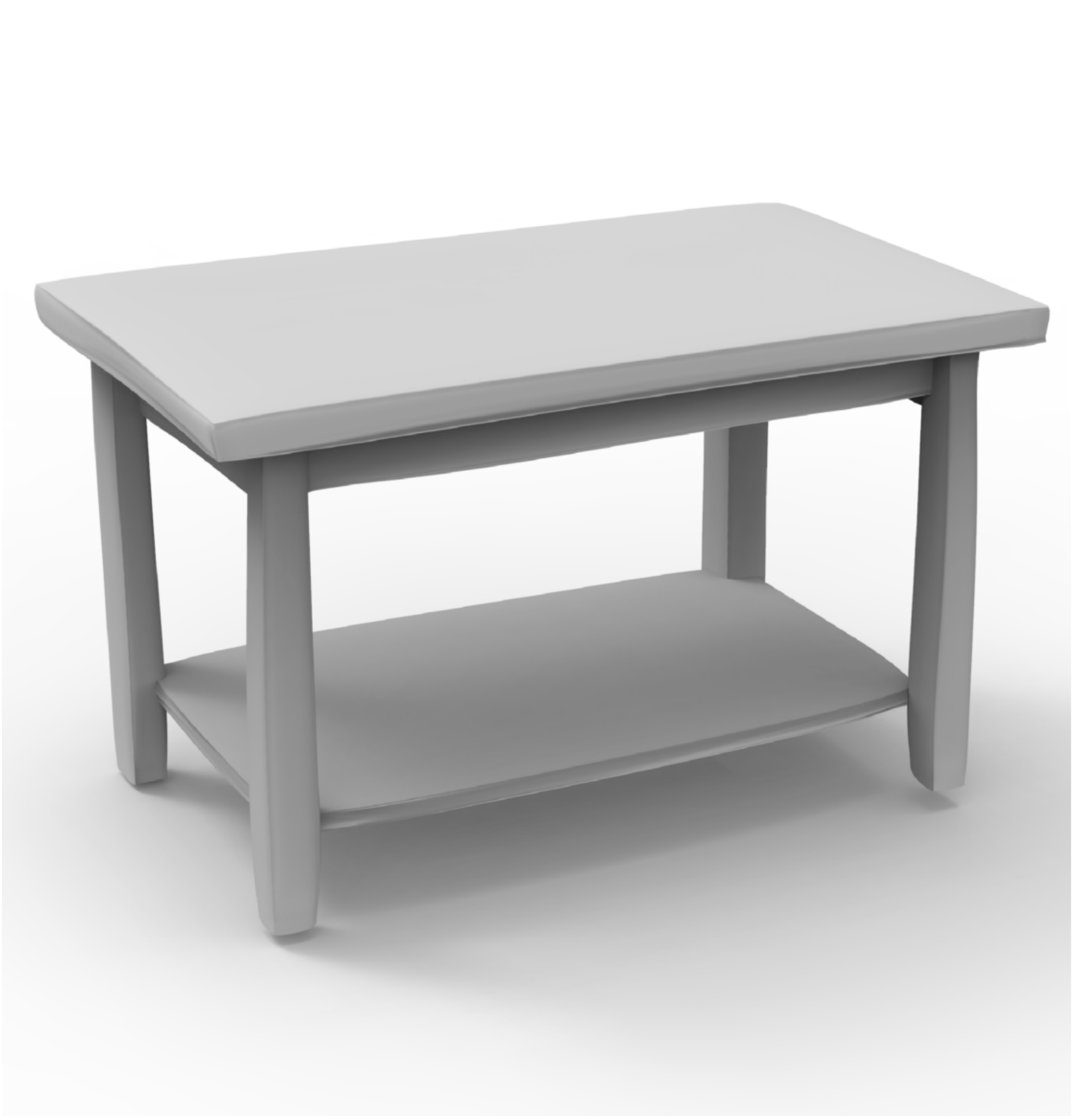}
    \includegraphics[width=0.18\linewidth]{imgs/interpolation/table/21490--28641/21490_0.pdf}
\end{minipage}
{\raisebox{-0.4\height}{\includegraphics[width=0.16\linewidth]{imgs/interpolation/table/21490--28641/21490_0.pdf}}}

{\raisebox{-0.4\height}{\includegraphics[width=0.16\linewidth]{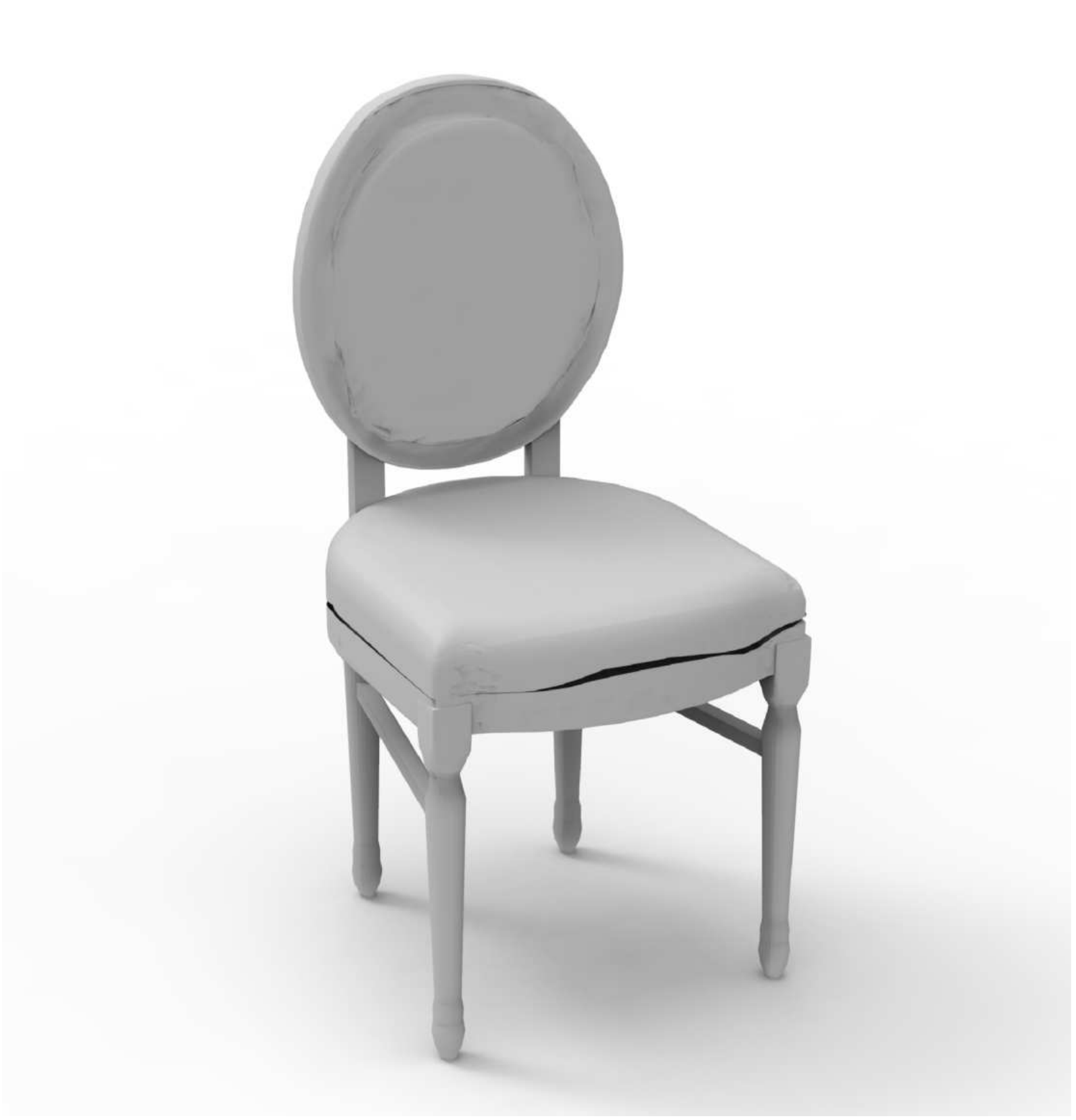}}}
\begin{minipage}{0.65\linewidth}
\centering
    \includegraphics[width=0.18\linewidth]{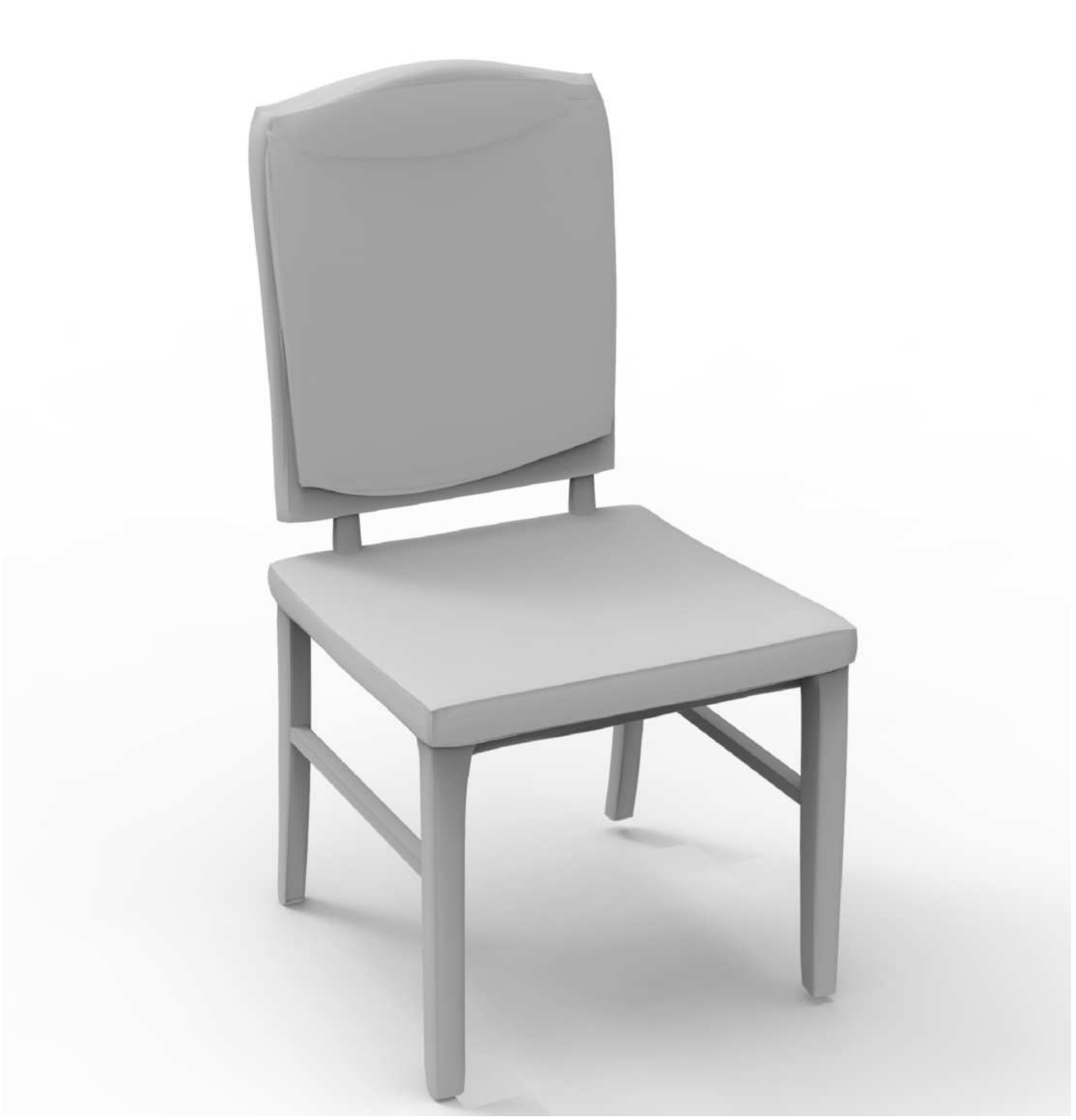}
    \includegraphics[width=0.18\linewidth]{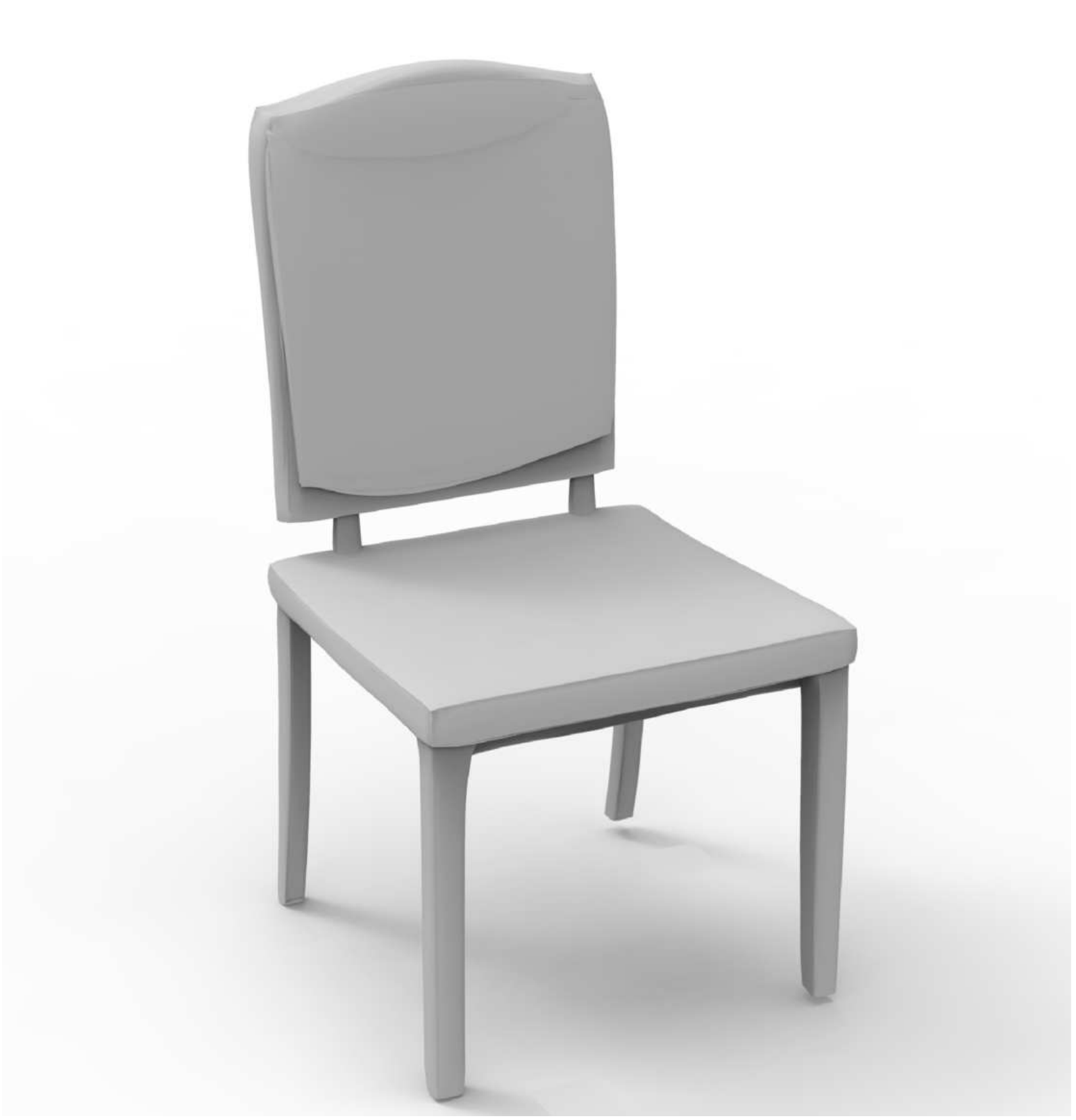}
    \includegraphics[width=0.18\linewidth]{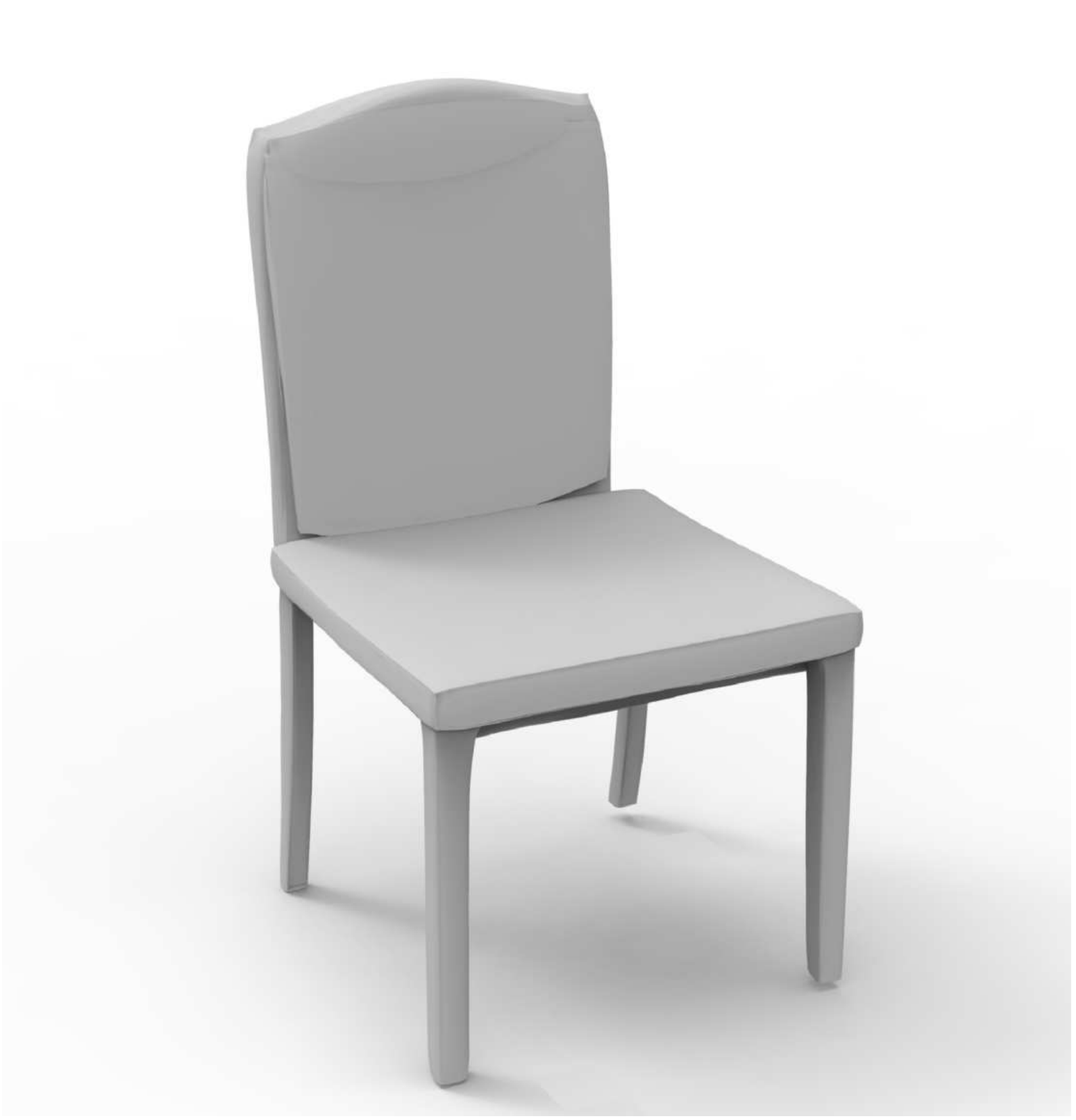}
    \includegraphics[width=0.18\linewidth]{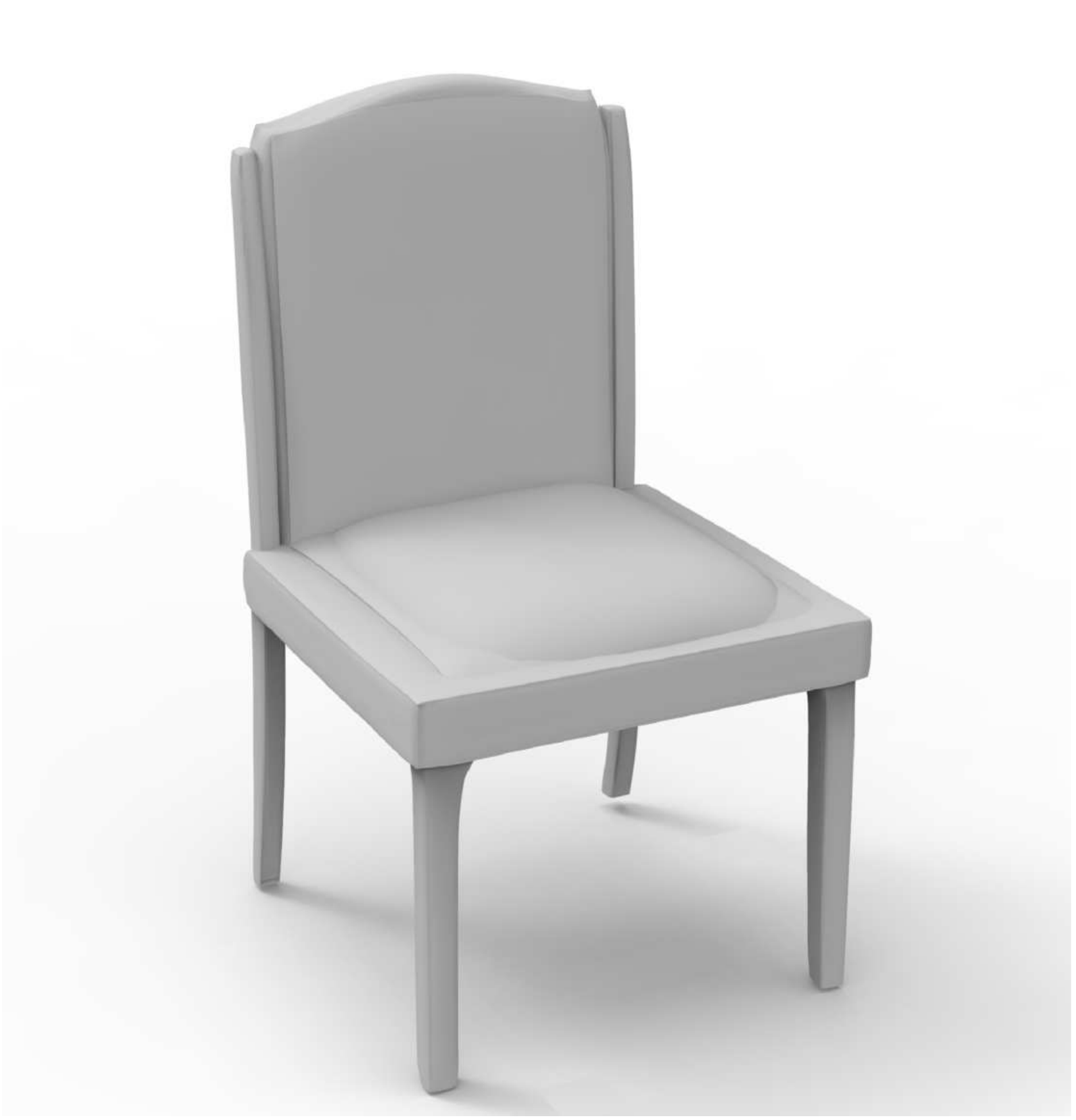}
    \includegraphics[width=0.18\linewidth]{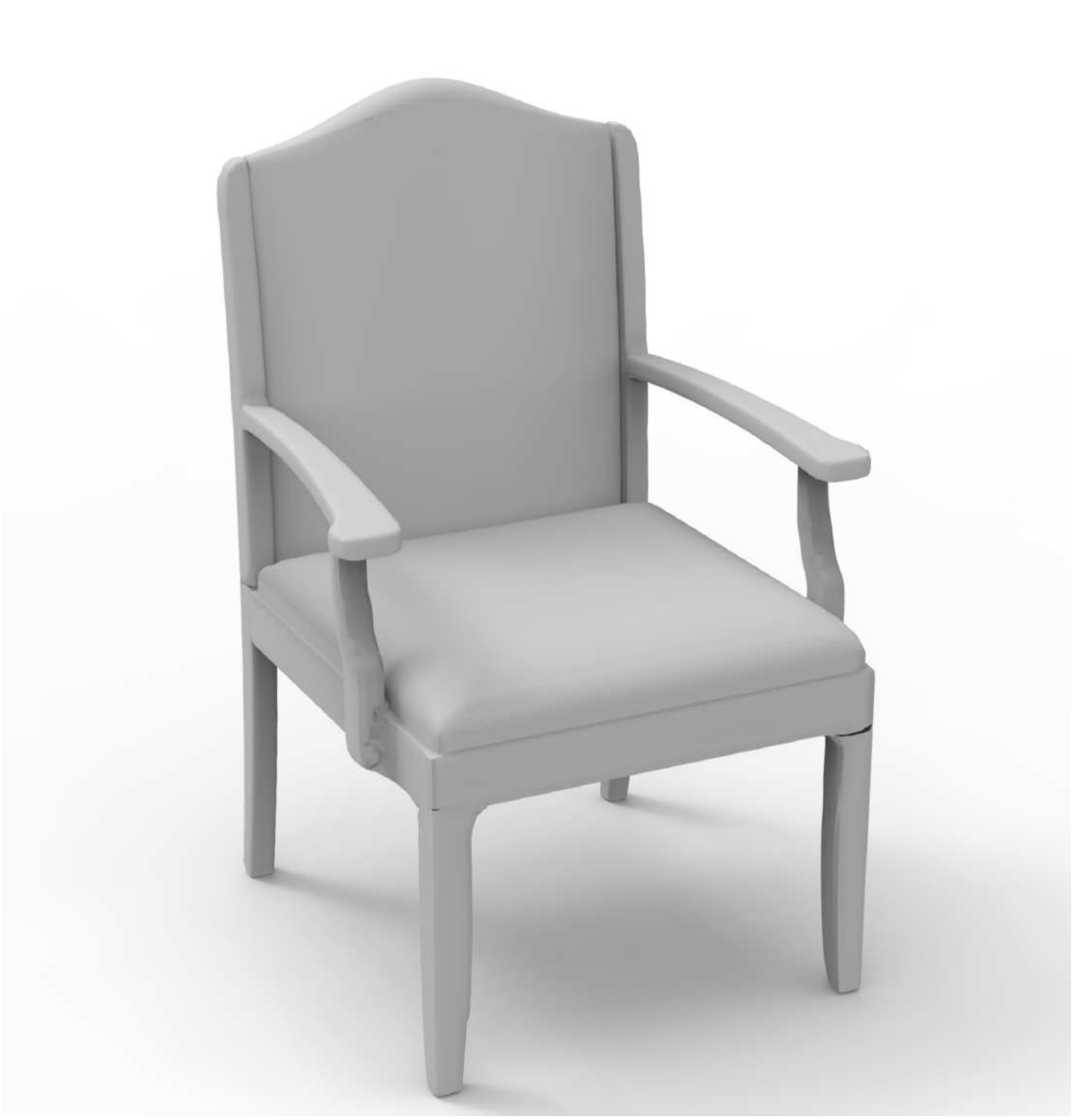}\\
    \includegraphics[width=0.18\linewidth]{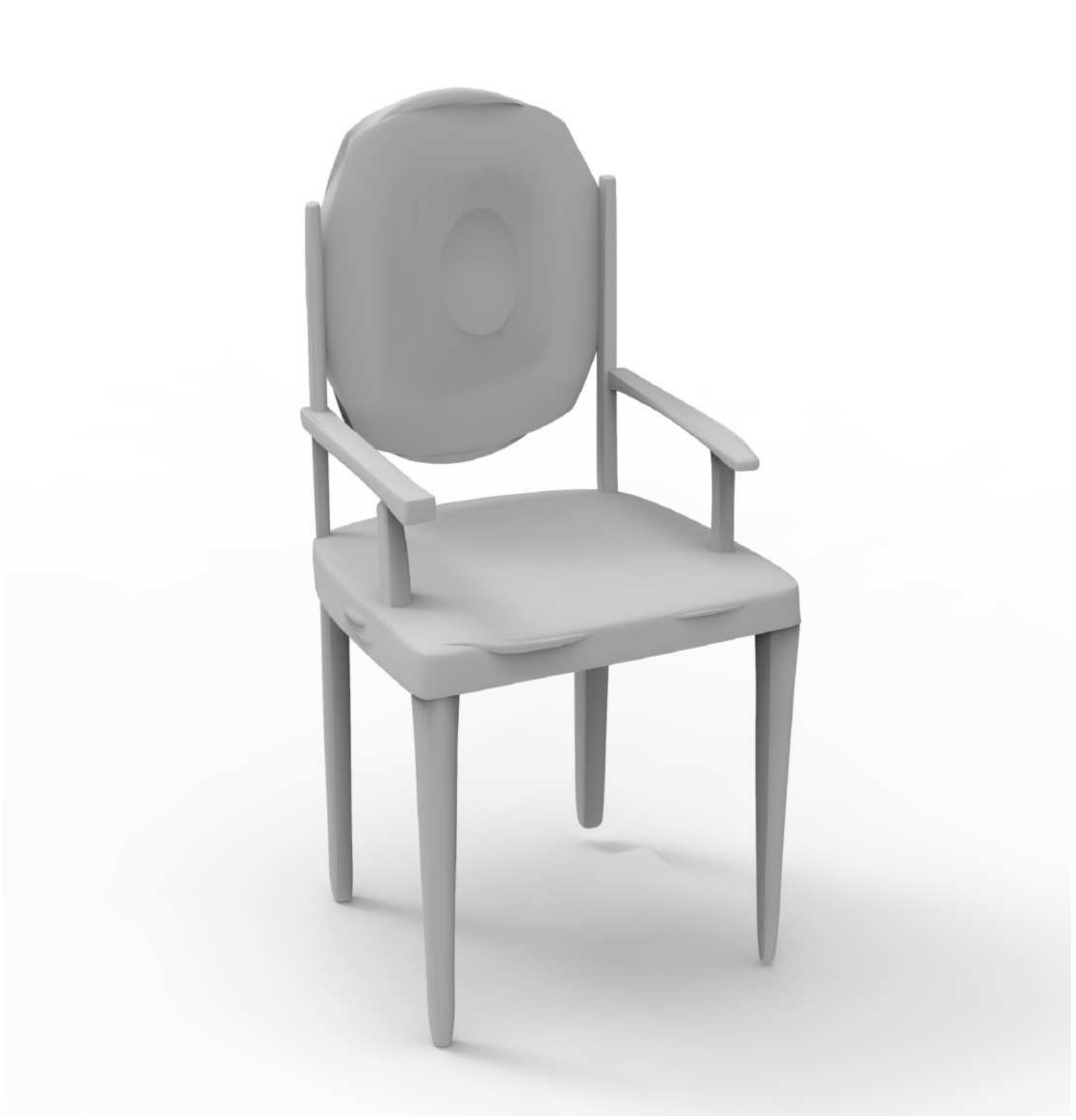}
    \includegraphics[width=0.18\linewidth]{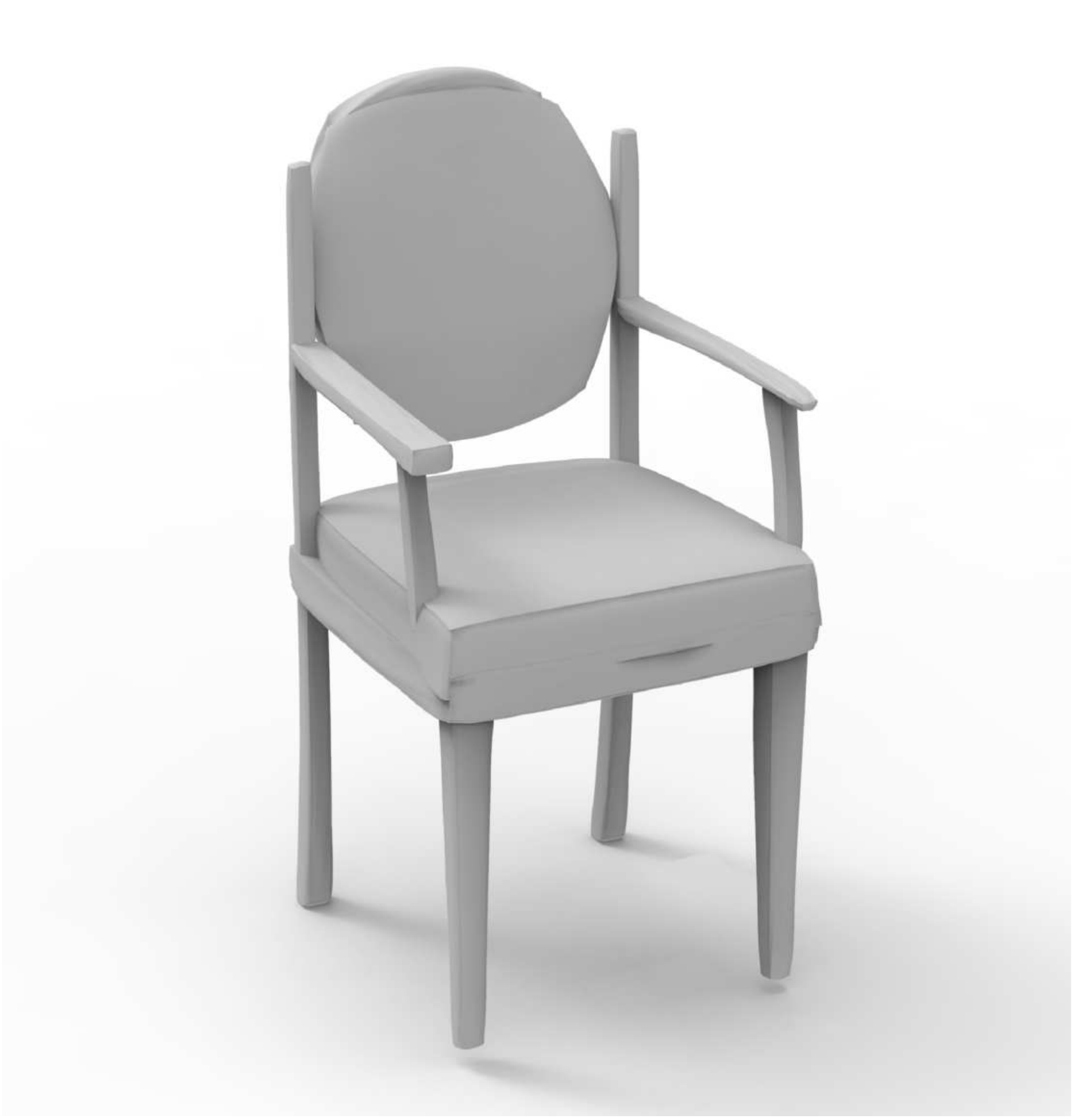}
    \includegraphics[width=0.18\linewidth]{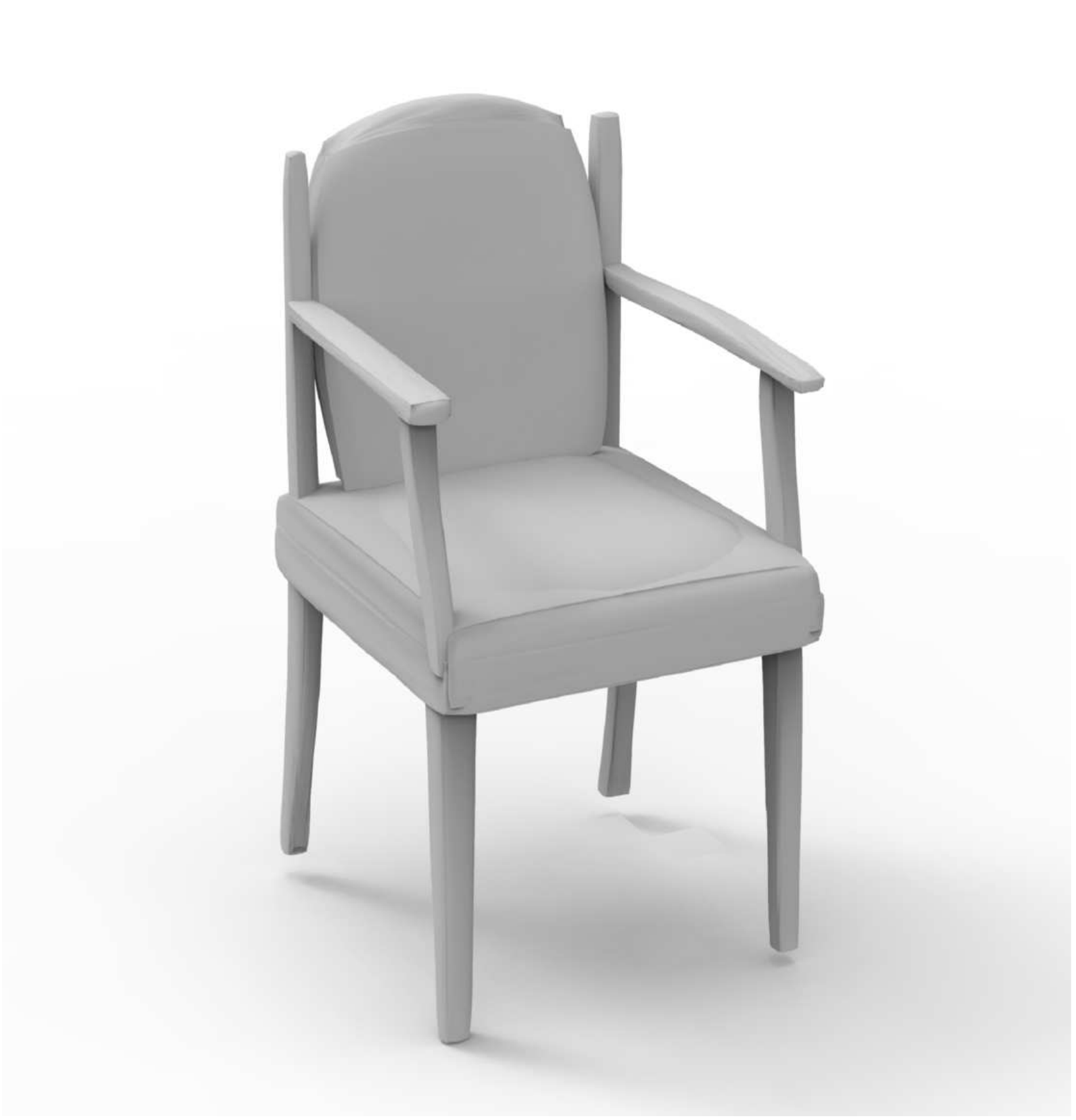}
    \includegraphics[width=0.18\linewidth]{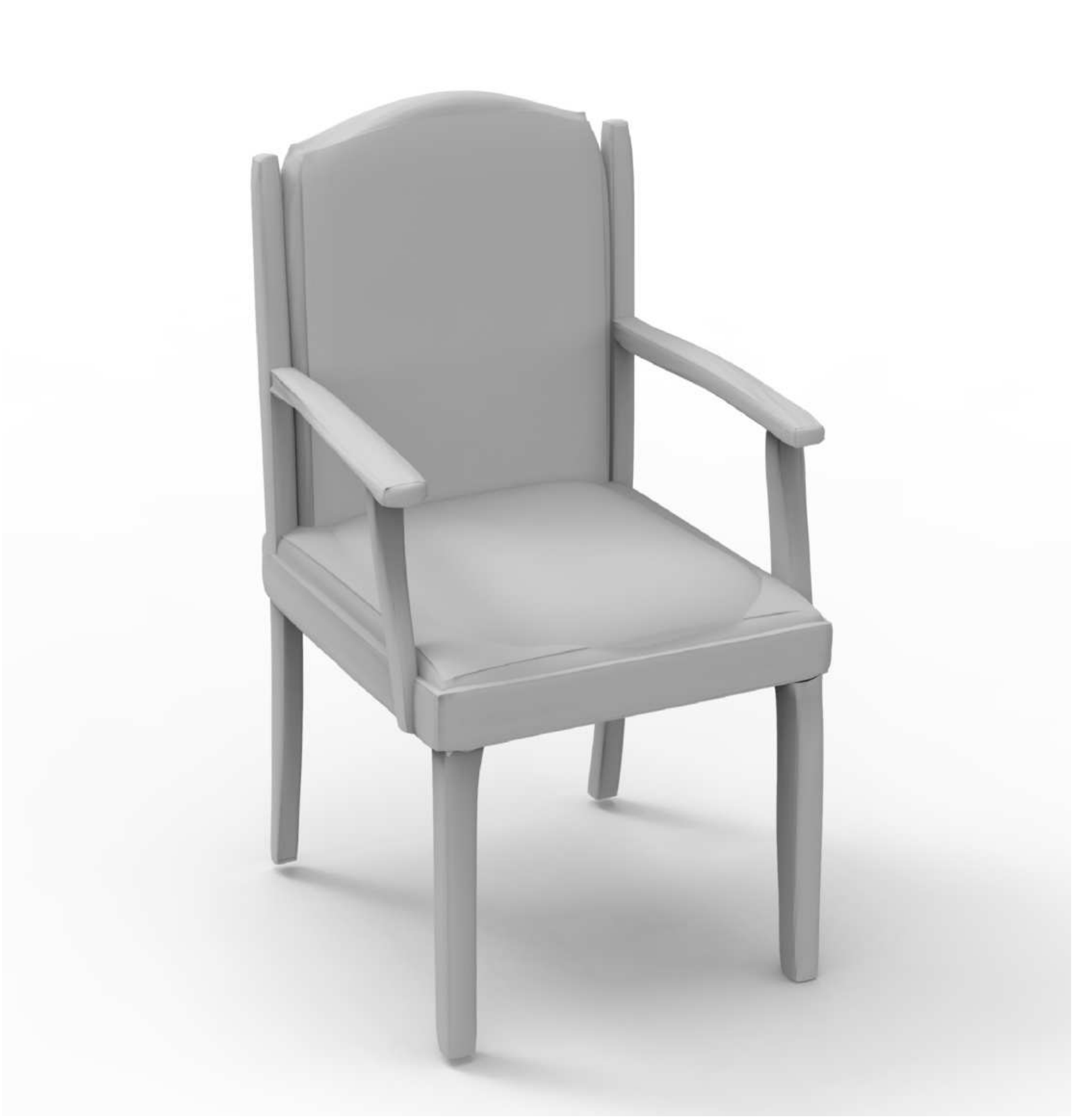}
    \includegraphics[width=0.18\linewidth]{imgs/interpolation/chair/44234--38791/44234_0.pdf}
\end{minipage}
{\raisebox{-0.4\height}{\includegraphics[width=0.16\linewidth]{imgs/interpolation/chair/44234--38791/44234_0.pdf}}}

\subfigure[Source]{\raisebox{-0.4\height}{\includegraphics[width=0.16\linewidth]{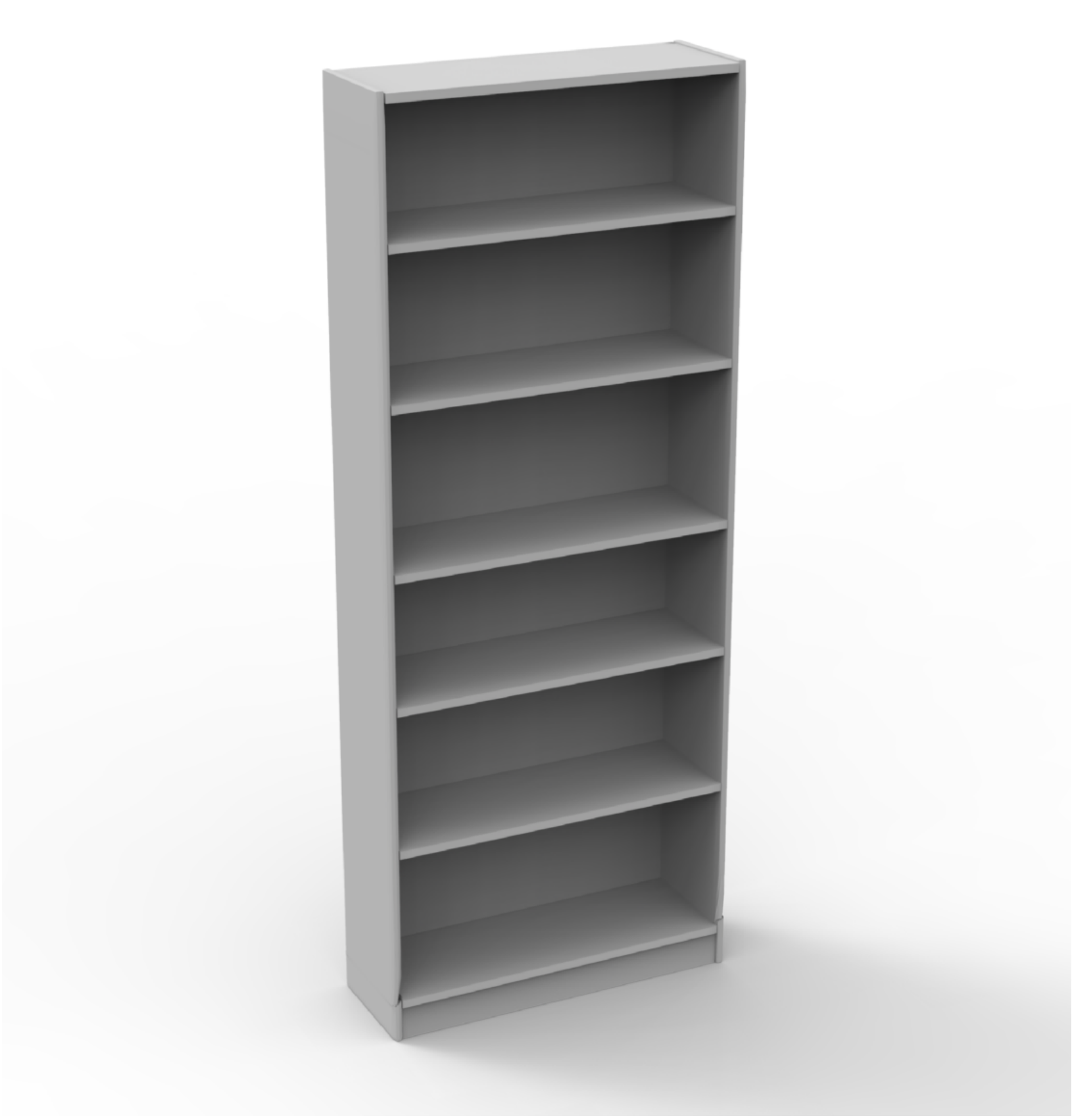}}}
\begin{minipage}{0.65\linewidth}
\centering
    \includegraphics[width=0.18\linewidth]{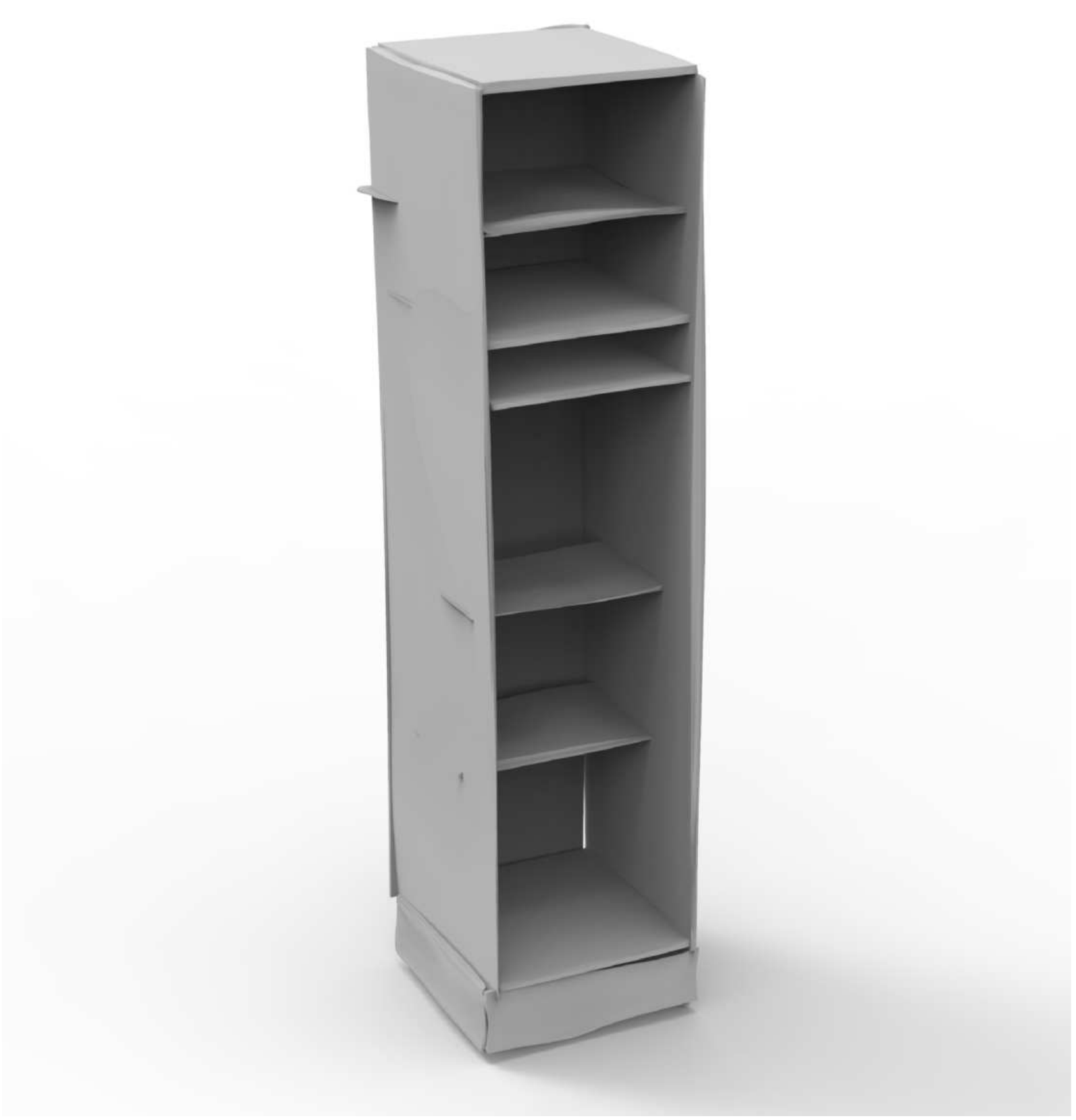}
    \includegraphics[width=0.18\linewidth]{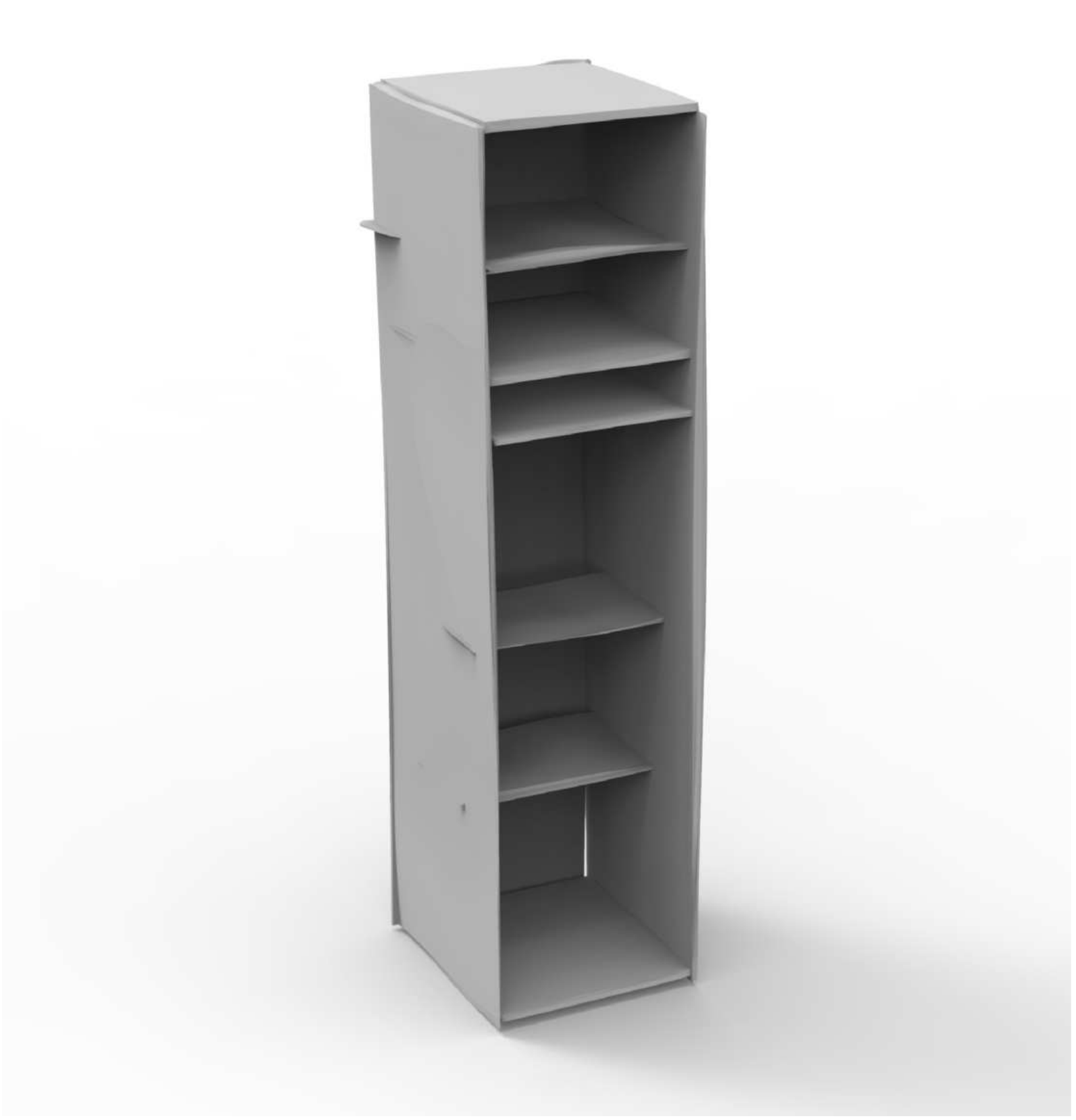}
    \includegraphics[width=0.18\linewidth]{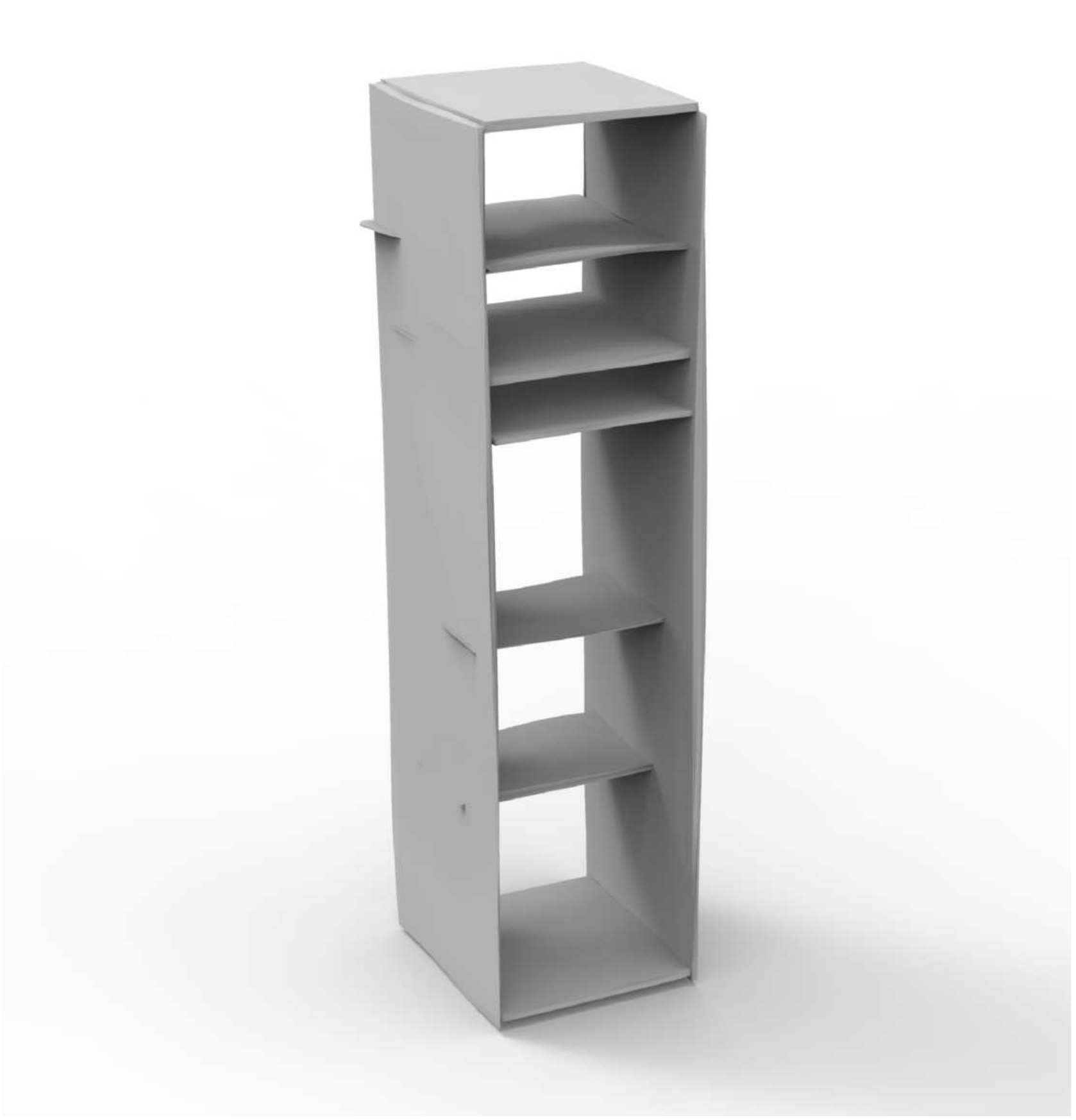}
    \includegraphics[width=0.18\linewidth]{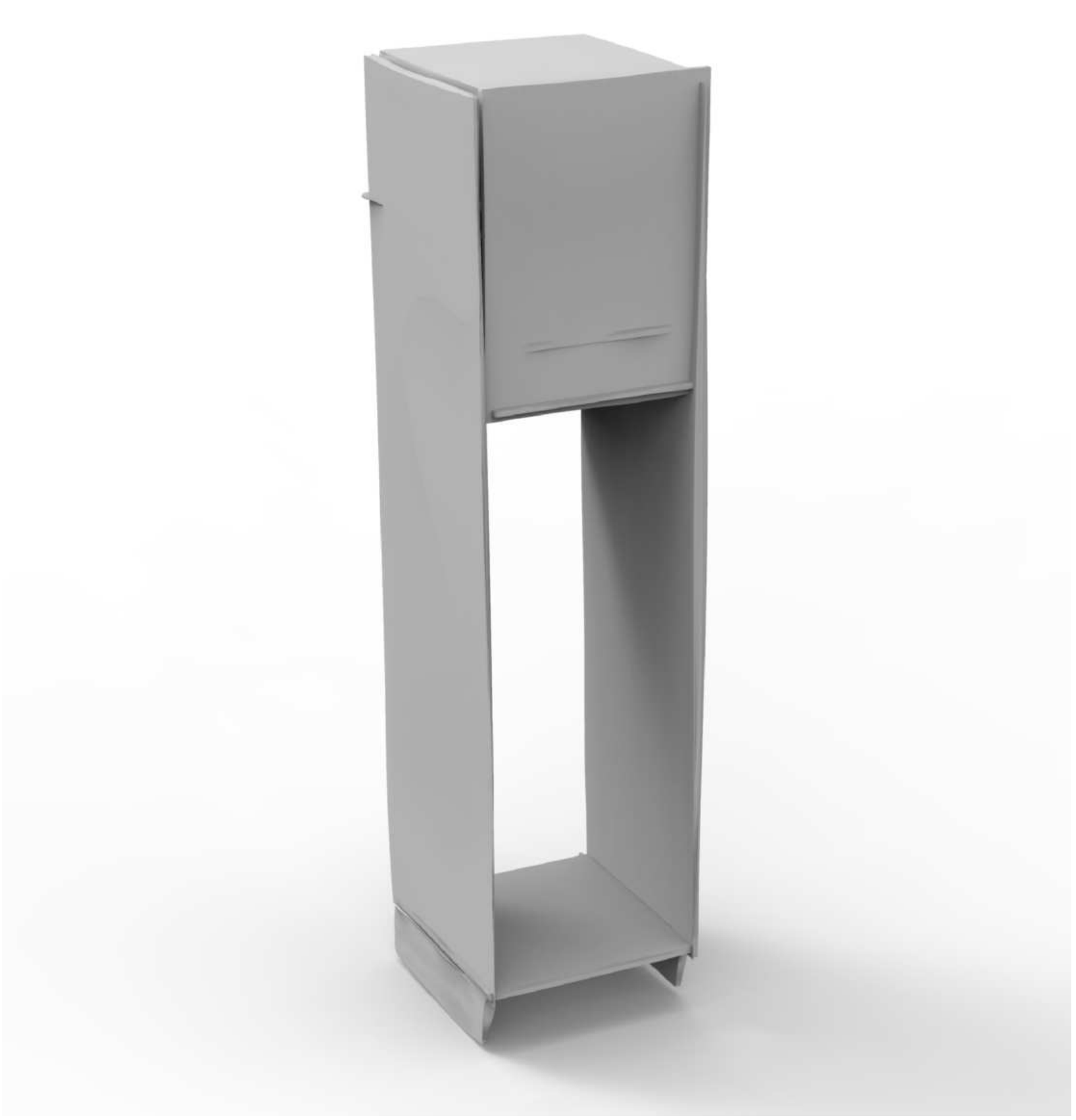}
    \includegraphics[width=0.18\linewidth]{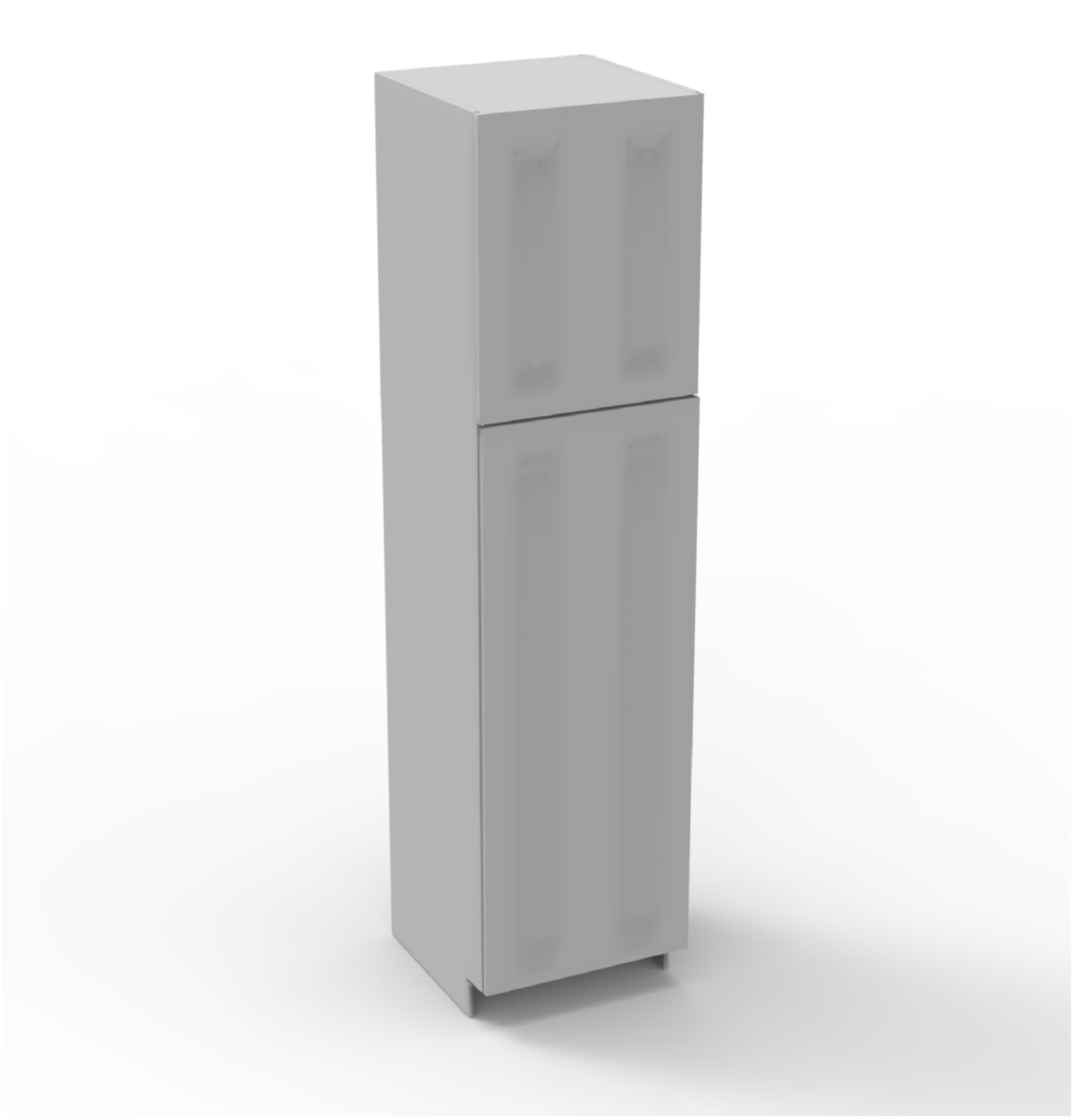}\\
    \includegraphics[width=0.18\linewidth]{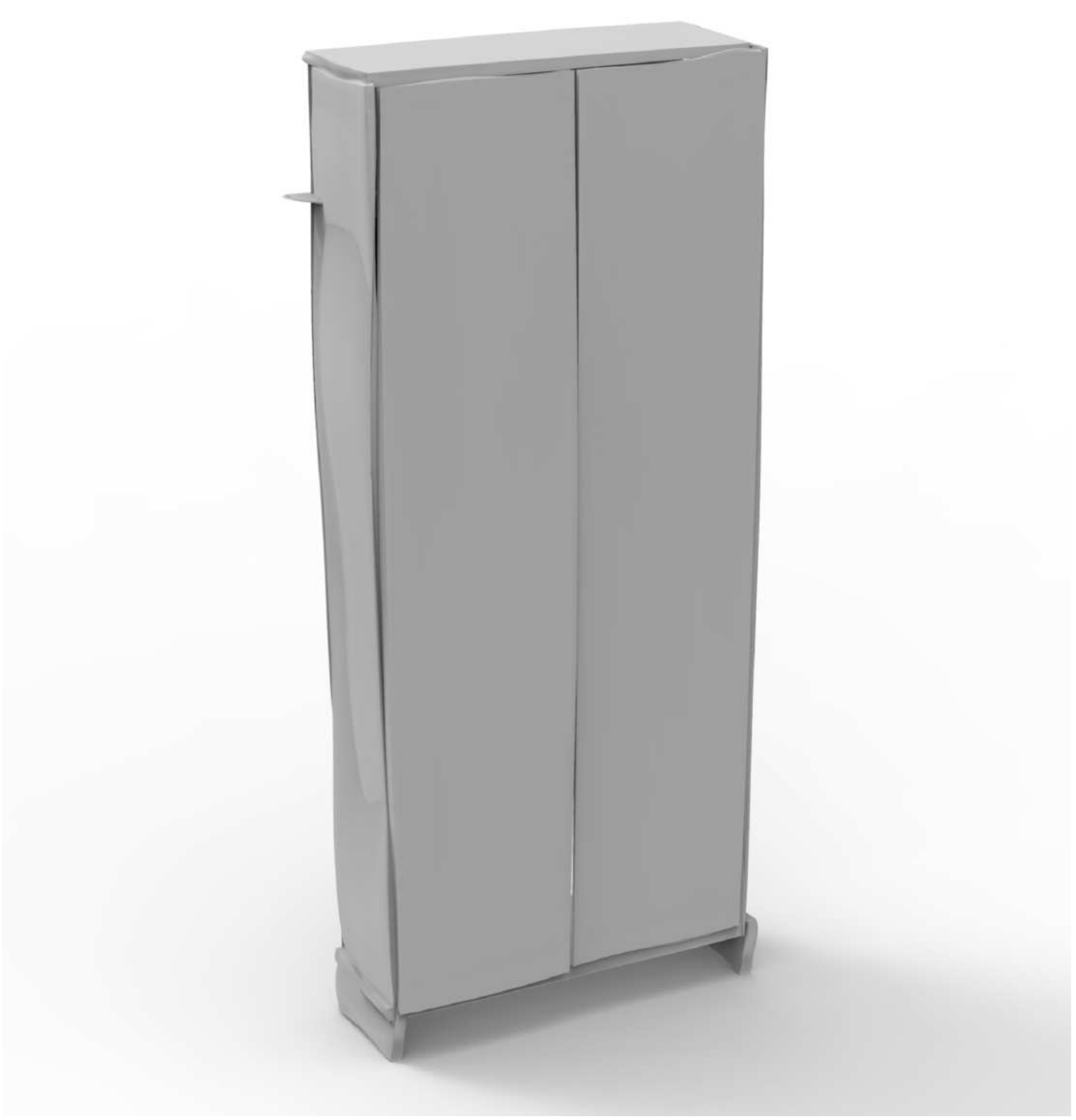}
    \includegraphics[width=0.18\linewidth]{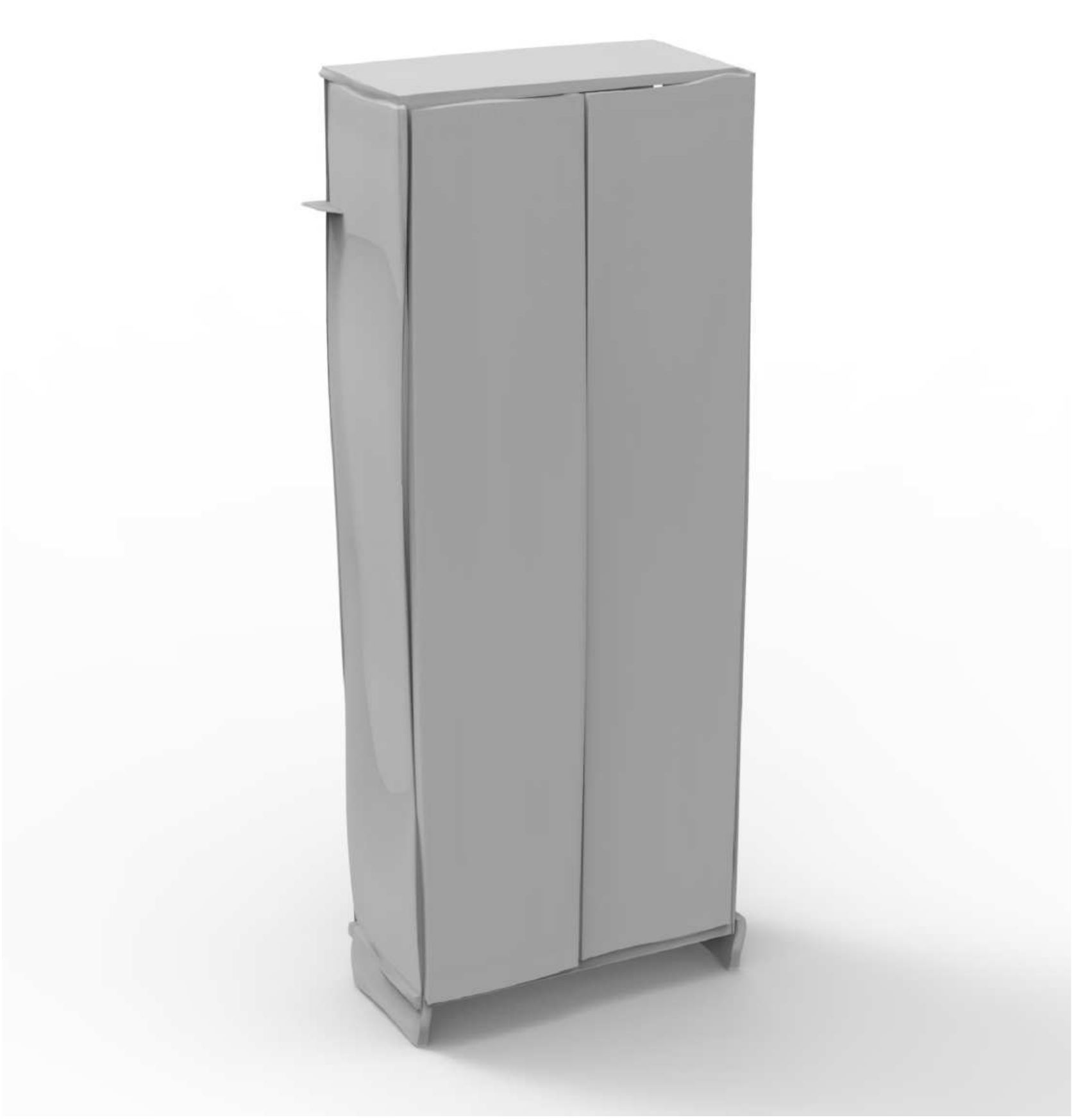}
    \includegraphics[width=0.18\linewidth]{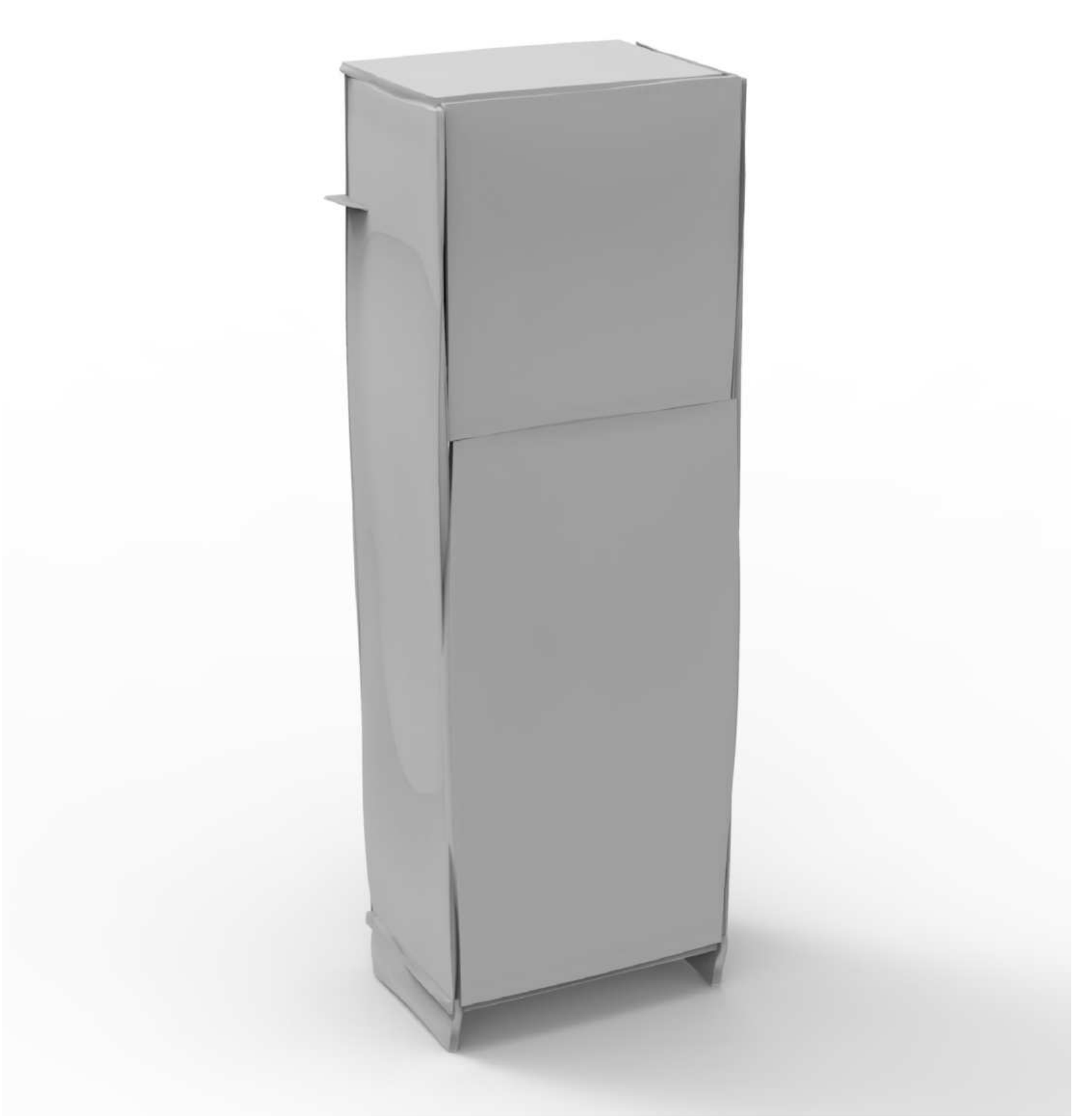}
    \includegraphics[width=0.18\linewidth]{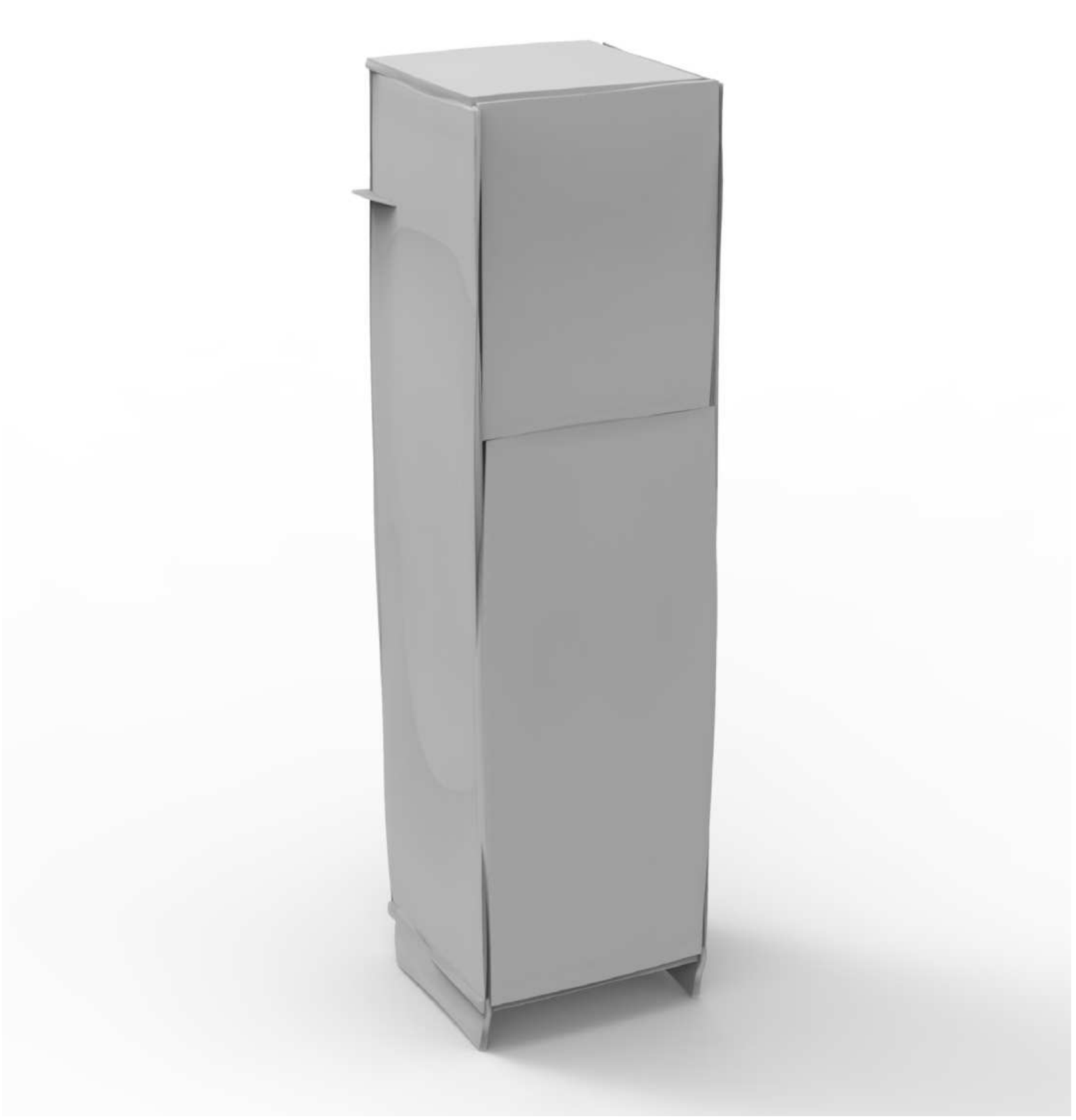}
    \includegraphics[width=0.18\linewidth]{imgs/interpolation/storage/46157--47273/46157_0.pdf}
\end{minipage}
\subfigure[Target]{\raisebox{-0.4\height}{\includegraphics[width=0.16\linewidth]{imgs/interpolation/storage/46157--47273/46157_0.pdf}}}
    \vspace{-3mm}
    \caption{\yjr{Qualitative results for disentangled shape interpolation. Columns (a) and (b) respectively show the source and target shapes. For the two result rows in the middle, the first row interpolates the structure between two shapes while using the geometry code of the target shape, while the second row interpolates the geometry code between two shapes while fixing the structure code of the target shape. We see a clear disentanglement of the shape structure and geometry in the interpolated results.}}
    \label{fig:interp_table}
    \vspace{-3mm}
\end{figure}

\paragraph{Disentangled Shape Interpolation.} 
Our disentangled representations for shape structure and geometry also allow us to achieve controllable interpolation between two shapes that one may keep the structure or geometry factor unchanged while interpolating the other factor.
Figure~\ref{fig:interp_table} shows some disentangled shape interpolation results.
From the results, we find that DSG-Net can achieve disentangled shape interpolation and every interpolated step produces a very realistic and reasonable shape.

\begin{table}[t]
  \centering
  \caption{\yjr{Quantitative evaluations of disentangled shape reconstruction on the synthetic data (for details please refer to Sec.~\ref{sec:data-pare}). We compare two ablated versions of our method, namely ours (w/o edge) and ours (w/o CycD) since there is no applicable external baseline method for this novel task. We observe that allowing edge communications between the structure and geometry hierarchies and novel cycled disentanglement are essential.}}
    \begin{tabular}{cccc}
    \toprule[1pt]
    \multirow{2}[3]{*}{Method} & \multicolumn{2}{c}{Geometry Metrics} & Structure Metrics \\
\cmidrule{2-4}          & CD{\scriptsize$\times 10^{-3}$}$\downarrow$ & EMD{\scriptsize$\times 10^{-2}$}$\downarrow$ & HierInsSeg(HIS) $\downarrow$\\
    \midrule
    \midrule
    Ours (w/o edge) & 1.50 & 1.39 & 1.92 \\
    Ours (w/o CycD) & 1.29& 0.61 & 1.87 \\
    Ours  & \textbf{1.02} & \textbf{0.58} & \textbf{1.43} \\
    GT    &       &       & \underline{1.19} \\
    \bottomrule[1pt]
    \end{tabular}%
  \label{tab:syndataeval}%
  \vspace{-2mm}
\end{table}%

\paragraph{Disentangled Shape Reconstruction} 
Our methods learn two disentangled latent manifolds (structure and geometry) for shape representations, which opens up new possibilities for controllable shape editing and re-synthesis tasks.
Given two input shapes, one can push the two shapes through our structure and geometry VAE encoders and obtains the structure and geometry features for both shapes.
Then, by re-combining the structure code of one shape and the geometry code of the other shape, DSG-Net is able to re-synthesize a novel shape that follows the structure of the first shape and the geometry of the second shape.

Figure~\ref{fig:decouple_chair} shows a set of qualitative results on the PartNet dataset (Cabinet and Table). 
The shapes in each row share the same geometry code while the shapes in every column have the same structural feature.
Here, the top left and bottom right shapes are the inputs.
The remaining shapes are generated with our DSG-Net, where in each row, the structure of the shapes is interpolated while keeping the geometry factor unchanged, whereas in each column, the geometry is interpolated while retaining the structure. 
The figure demonstrates that our method is able to re-synthesize novel shapes with pairs of geometry and structure configurations.

We further quantitatively benchmark the performance of DSG-Net for disentangled shape reconstruction on the synthetic dataset, where we have access to the ground-truth reconstruction results given a pair of structure and geometry configurations.
Table~\ref{tab:syndataeval} shows the quantitative results of the synthetic data. 
Since there is no applicable baseline method for this novel task, we compare two ablated versions of our method: 1. ours (w/o edge), where we ignore the part relationships from the part hierarchies and remove the graph message-passing procedures, which further reduces the communication between the structure and geometry;
2. ours (w/o CycD), where we remove the novel cycled disentanglement and losses ($\mathcal{L}_{struct}$ and $\mathcal{L}_{geo}$), which further reduces the capability of the disentanglement of geometry and structure.
We see that removing edge communications or cycled disentanglement provides us with worse performance, which proves the importance of maintaining the synergy between the disentangled structure and geometry hierarchies.
For more results on disentangled interpolation on Lamp and Chair categories and some qualitative results on the synthetic data, please refer to supplementary material.

\subsection{Ablation Studies}\label{sec:abla}

\kaichun{We perform four sets of ablation studies to demonstrate the necessity and effectiveness of the key components and training strategies for our method.}
\yj{First, the evaluation on the post-processing procedure shows its effectiveness for resolving the issue of detached parts.}
\kaichun{Then, we compare cascaded training and end-to-end training for the part geometry VAE and the disentangled backbone VAEs. We observe similar performances for the two strategies. We take the end-to-end training approach due to its simplicity.}
\yj{We also validate the design choice of learning a unified conditional part geometry VAE, instead of training separate VAEs for each part semantics as used in SDM-Net~\cite{gaosdmnet2019}.
\yjr{Finally, we demonstrate that explicitly considering part relationships and conducting graph message-passing operations along the edges are important. Removing the edge components from our network gives significantly worse results.
In terms of the cycled disentanglement, we also demonstrate its importance for our disentangled representation of shape structure and geometry on shape reconstruction, including the disentangled shape reconstruction on synthetic data.
\yjrr{Moreover, the ablated version (StructureNet + Mesh) is a naive combination of StructureNet~\cite{mo2019structurenet} backbone and ACAP mesh representation~\cite{gao2019sparse,gaosdmnet2019}. We also evaluate it to validate that our proposed disentangled structure and geometry representation and the cycled disentanglement indeed help improve the performance for learning 3D shape generative models.
}
}
We validate the post-processing stage in the main paper, while presenting the other four in the supplementary material.}

\begin{figure}[t]
    \centering
    \subfigure[Input Shape with Detached Parts]{\includegraphics[width=0.3\linewidth]{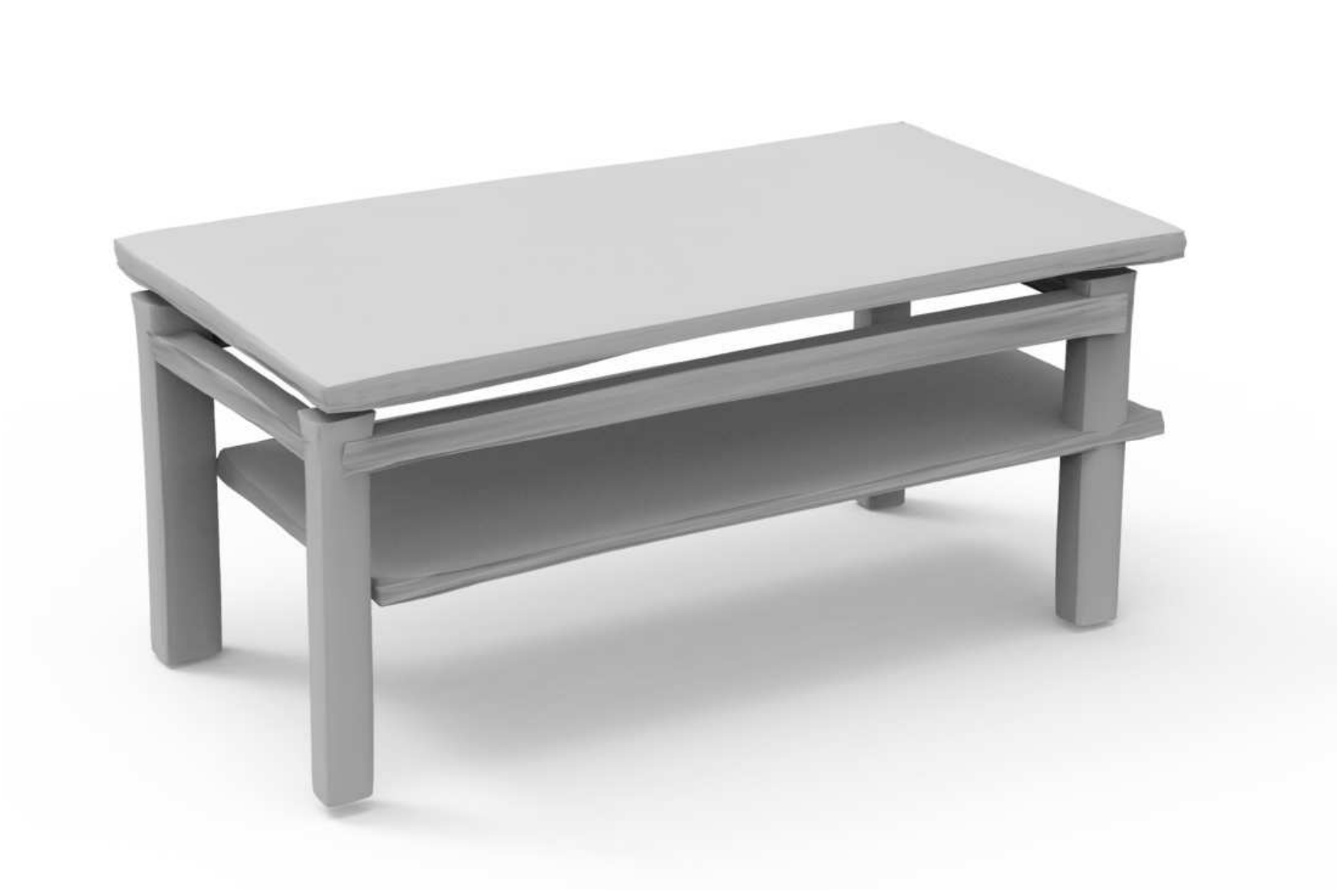}}\hspace{2mm}
    \subfigure[Input Shape with Detached Parts (manually move the detached parts)]{\includegraphics[width=0.3\linewidth]{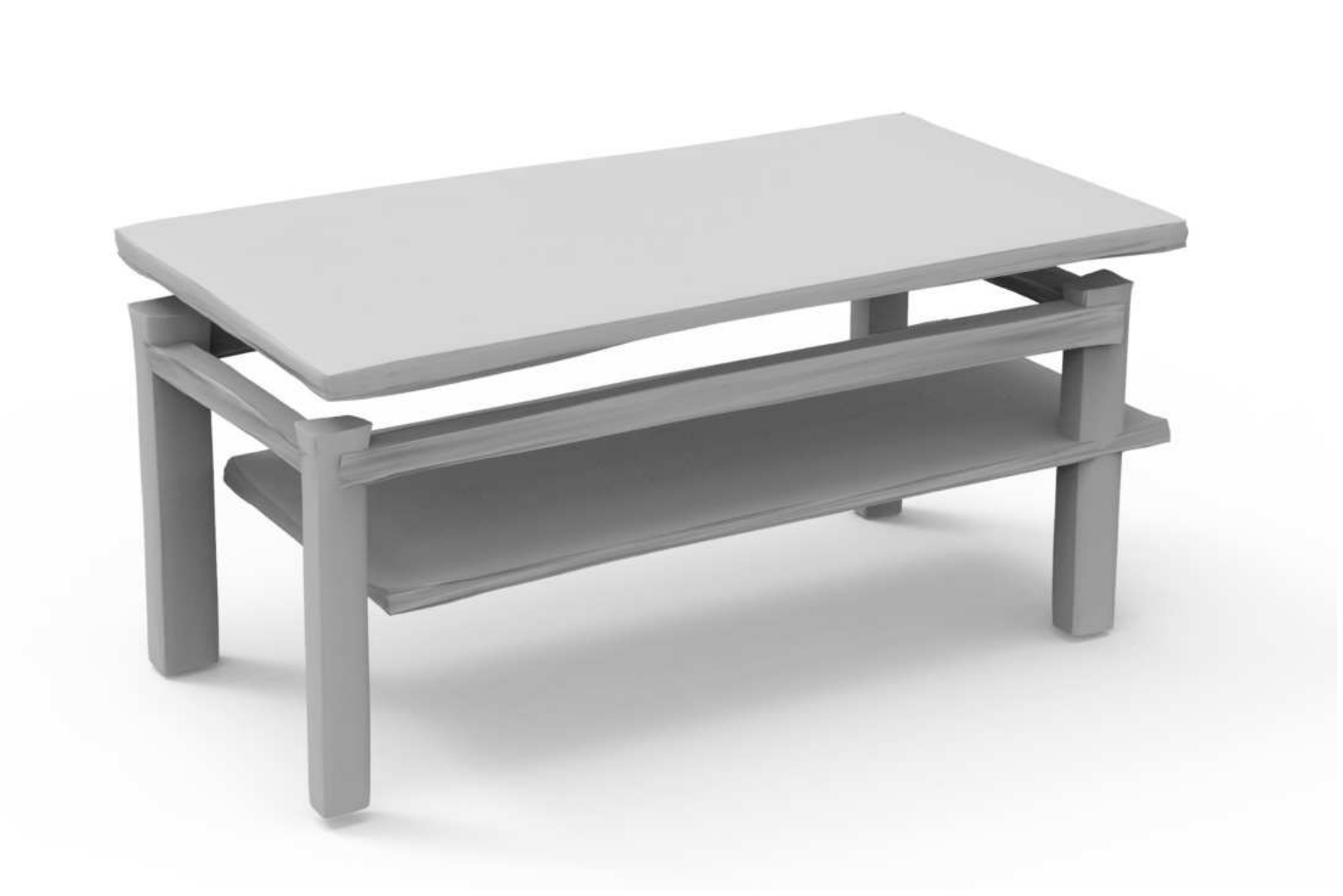}}\hspace{2mm}
    \subfigure[Optimized Results]{\includegraphics[width=0.3\linewidth]{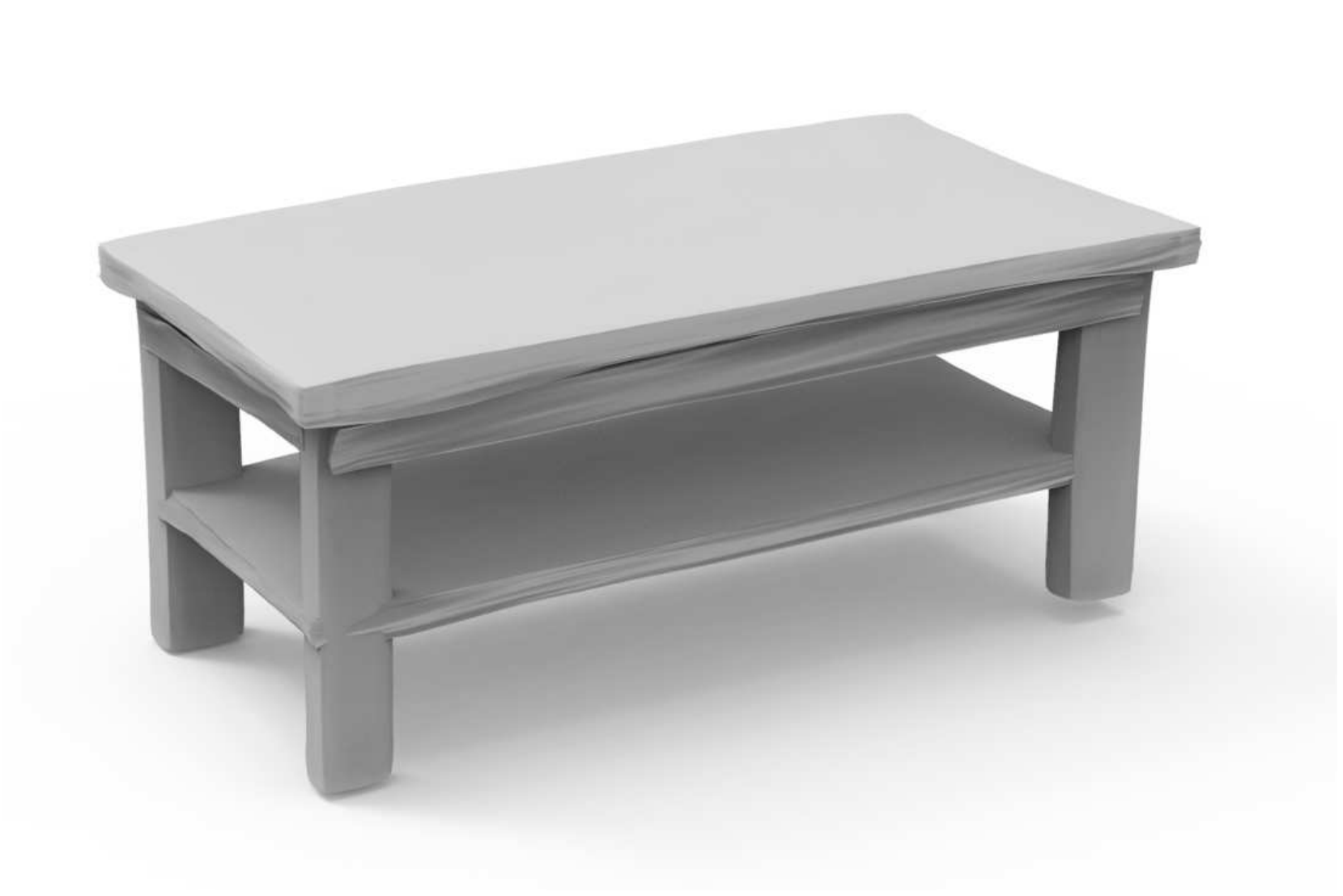}}
    \vspace{-3mm}
    \caption{Qualitative result of the post-processing optimization. We optimize the center positions of the disconnected parts for our generated results. (a) presents one output table shape produced by our network, and (b) shows a shape where we magnify the part detachment issue by manually adjusting the part positions for better illustrating the effectiveness of the post-processing optimization. We show the final optimized shape output using the post-processing optimization in (c), where we observe that the post-processing has the ability to fix the part connectivity issue.}
    \label{fig:abla_post}
    \vspace{-3mm}
\end{figure}

\paragraph{Effectiveness of post-processing on detached parts}
\yj{Although we explicitly consider the part relations for minimizing the gaps between the adjacent parts, there are still some failure cases of floating parts or part disconnections.
We thus proposed a post-processing module to address the part connectivity issue in Sec.~\ref{sec:post}.
From the result shown in Figure~\ref{fig:abla_post}, we see that the post-processing optimization procedure successfully removes the unwanted part gaps and fixes the part connectivity issue in most cases.
Furthermore, we notice that some other post-optimization methods (\eg as the one introduced in the SDM-Net paper) may also help address this issue. 
There is also a recent work COALESCE~\cite{yin2020coalesce} that aims to synthesize part connections and joints. 
We leave it as future work to incorporate these techniques in end-to-end learning pipelines with 3D shape generative models.
}

\section{Limitations and Future Works}
\label{sec:limit}

\yj{
\yj{Our method depends on heavily annotated shape structure hierarchies and fine-grained part geometric annotations for a large-scale collection of 3D shapes as input to our networks.} It is a non-trivial task to obtain such data from automatic algorithms. One may consider predicting such hierarchies by training hierarchical part instance segmentation networks (as shown in PartNet~\cite{mo2019partnet} Sec. 5.3 and StructureNet~\cite{mo2019structurenet} shape abstraction experiments). But, these methods all require a large-scale training dataset of fine-grained part and structure annotations. For unsupervised methods, although recent works, \eg Cuboid Abstraction~\cite{sun2019learning}, show promising results for learning such fine-grained shape parts and structures, it still remains a challenging topic in the research community. 
%\yj{In future work, we will implement our proposed approach in Jittor~\cite{hu2020jittor}, which is a fully just-in-time (JIT) compiled deep learning framework with higher performance.}

}

\begin{figure}[t]
    \centering
    \subfigure[]{\includegraphics[width=0.11\linewidth]{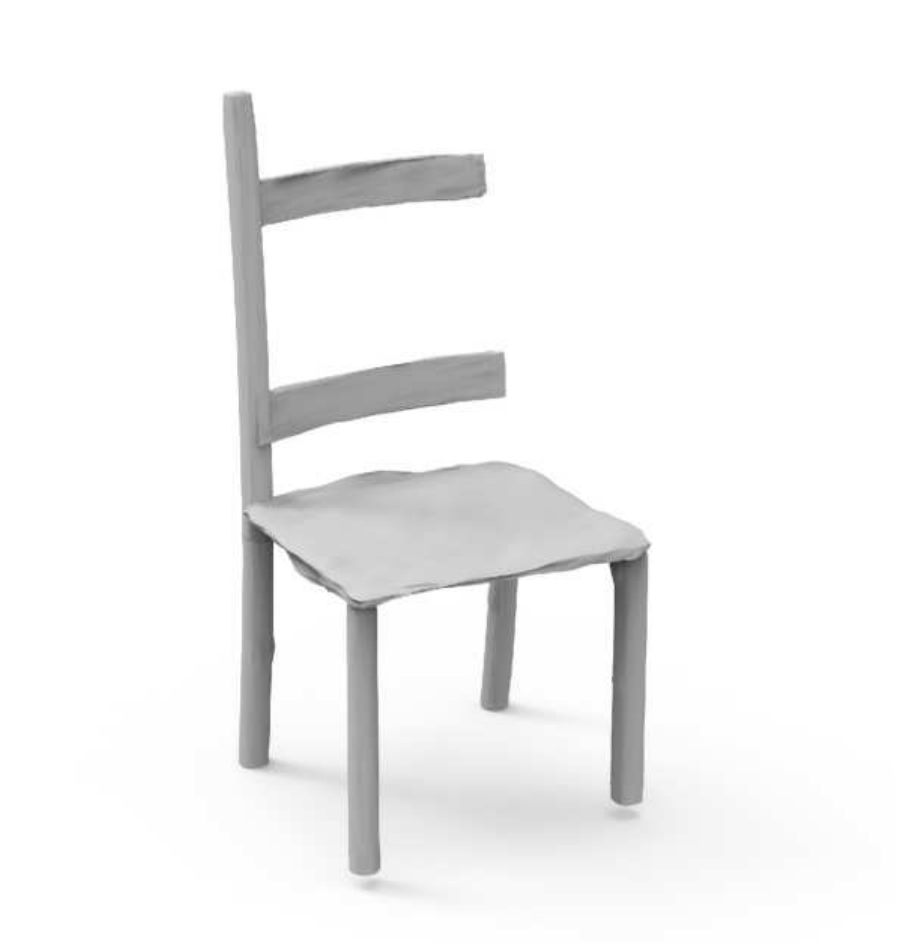}
    \includegraphics[width=0.11\linewidth]{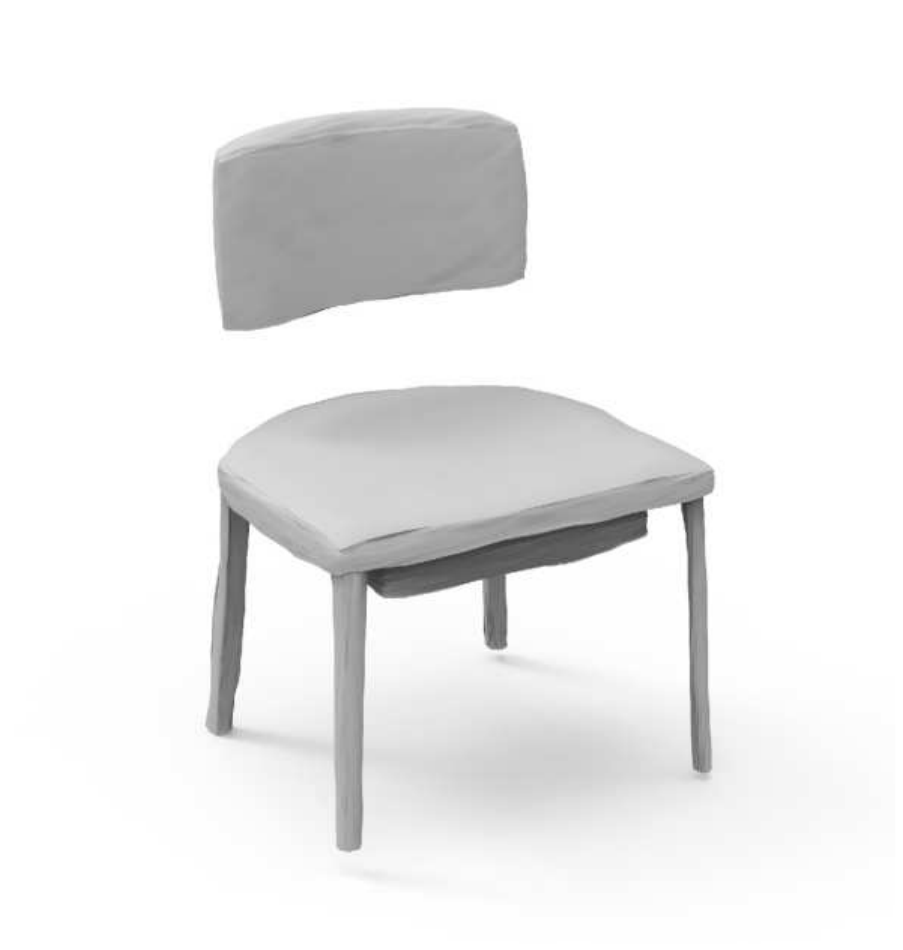}}
    \subfigure[]{\includegraphics[width=0.11\linewidth]{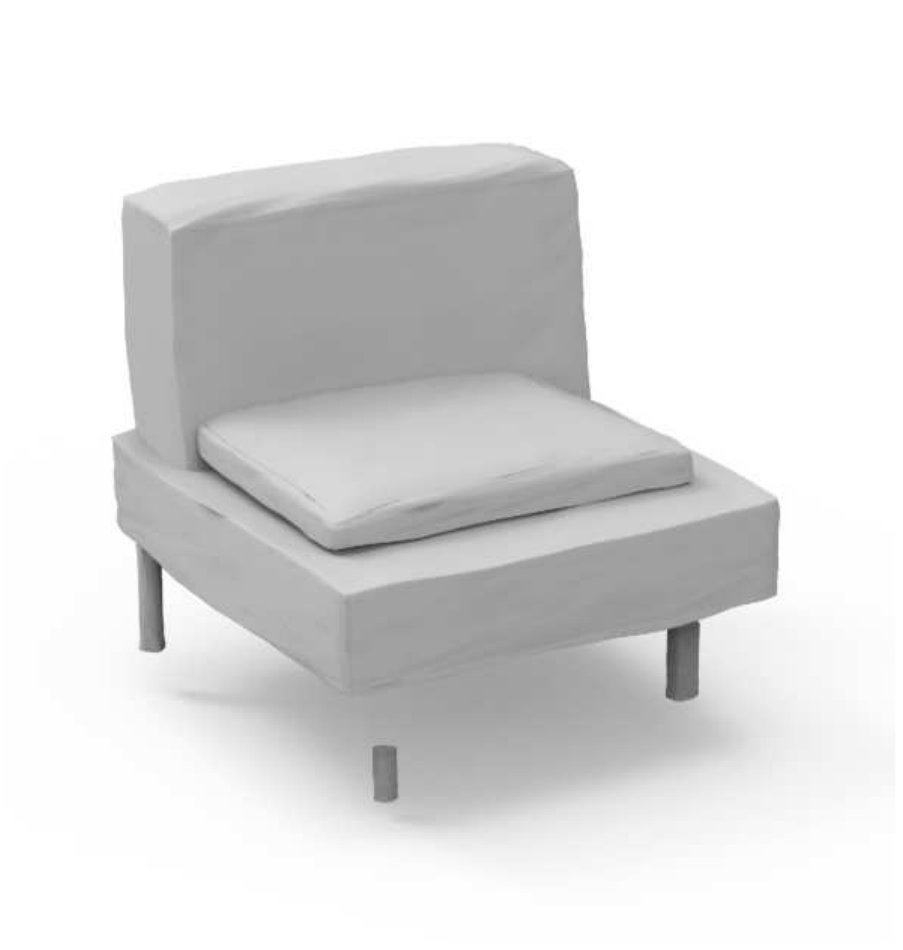}}
    \subfigure[]{\includegraphics[width=0.11\linewidth]{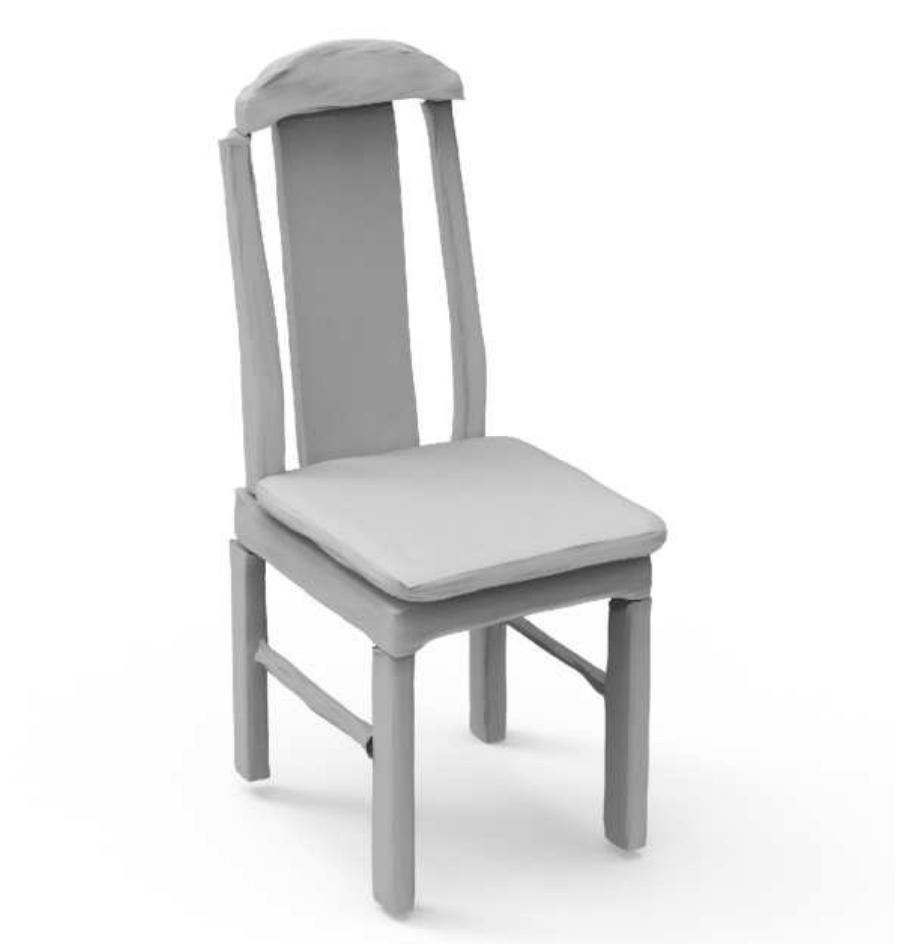}
    \includegraphics[width=0.11\linewidth]{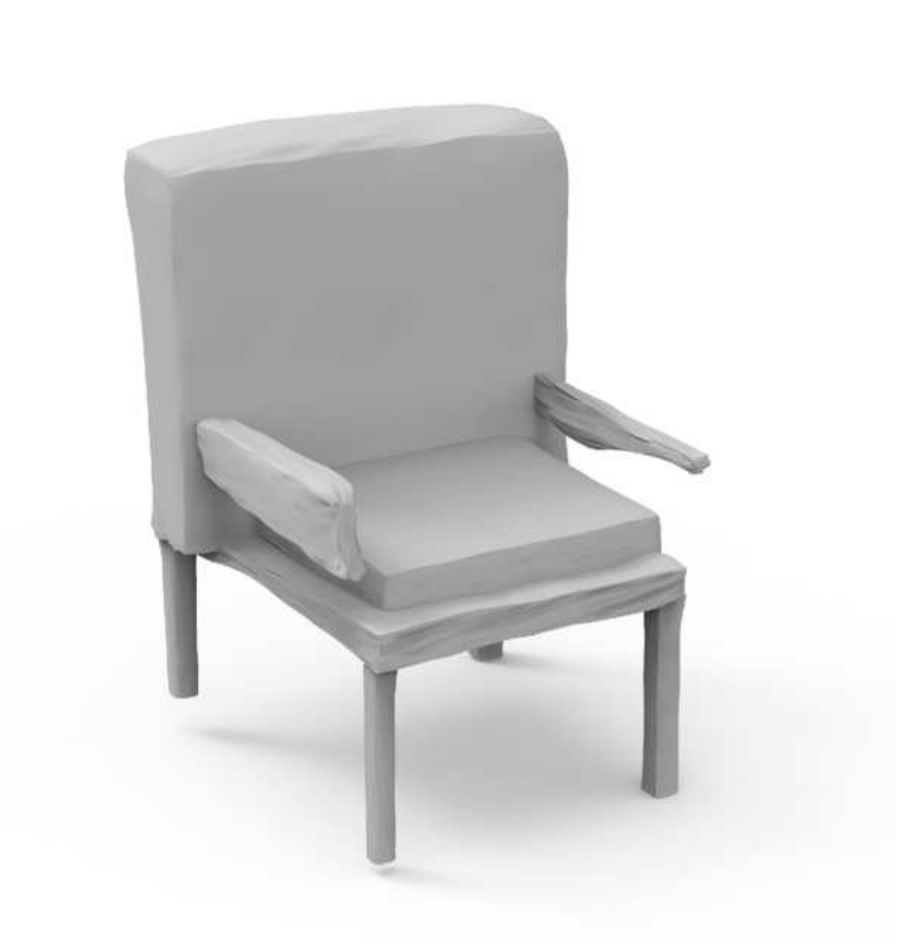}}
    \subfigure[]{\includegraphics[width=0.11\linewidth]{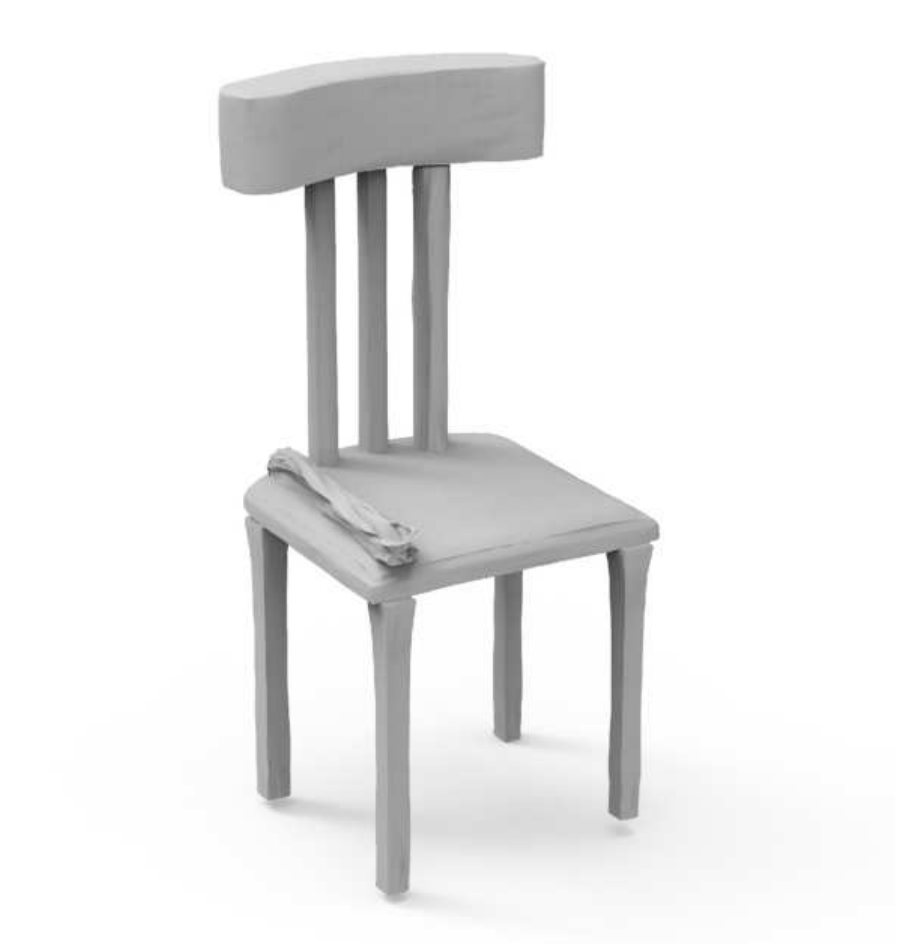}}
    \subfigure[]{\includegraphics[width=0.11\linewidth]{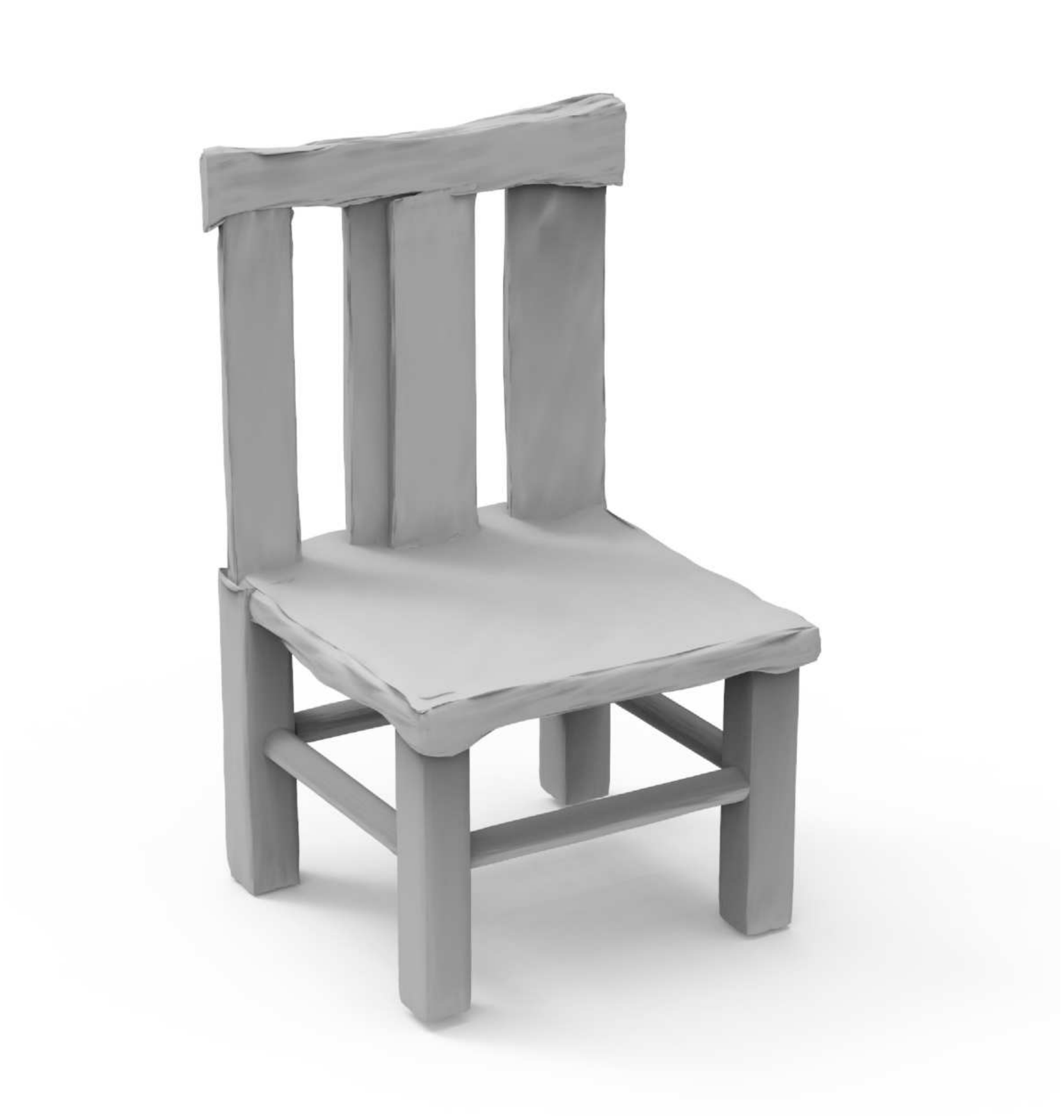}}
    \subfigure[]{\includegraphics[width=0.11\linewidth]{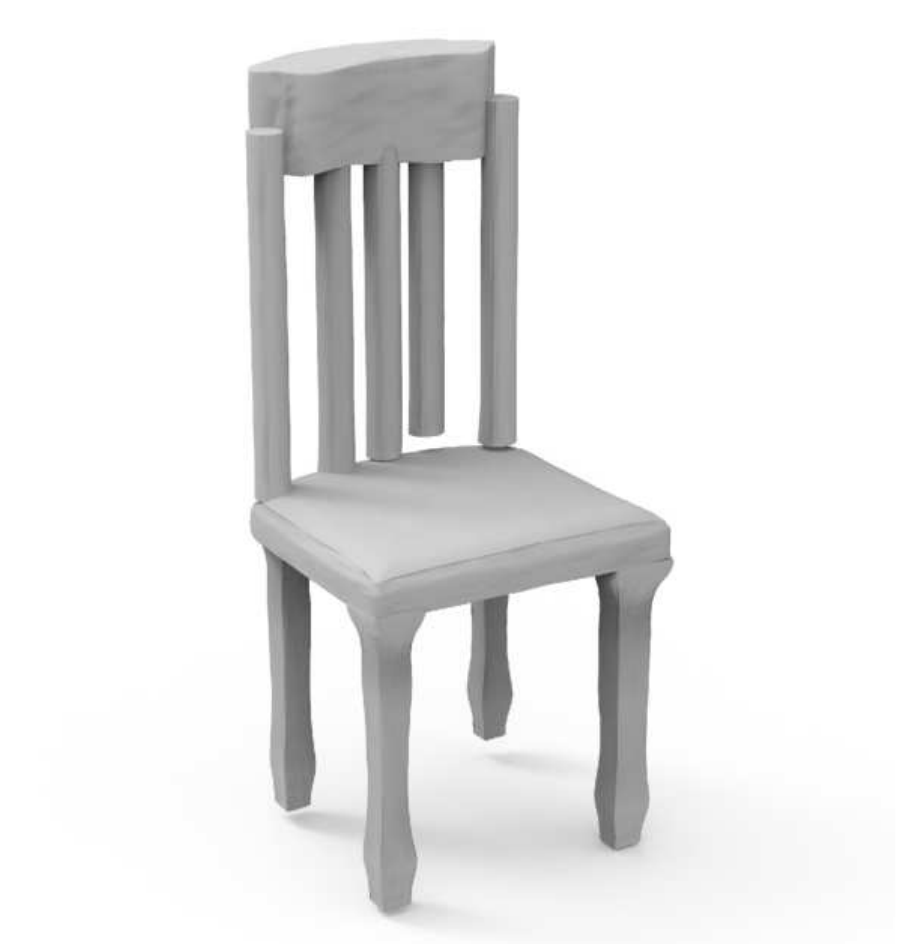}}
    \vspace{-3mm}
    \caption{\yjrr{Failure cases. We present some problematic generation results, such as missing parts (a), detached parts (b), asymmetric parts (c),
    extra parts (d), duplicate parts (e), and incompatible size of parts (f).}}
    \label{fig:failurecase}
    \vspace{-3mm}
\end{figure}

\yj{Finally, although our method achieves the state-of-the-art performance in generating shapes with fine geometry and complex structure, 
\yj{our method still has some failure cases as shown in Figure~\ref{fig:failurecase}: \yjrr{(1) 
some generated shapes may have detached parts and asymmetric parts (c), especially for rare shapes, as well as missing parts (a), detached parts (b), extra parts (d), duplicate parts (e), or incompatible size of parts (f), etc.; 
(2) our network restricts the maximum number of siblings to 10, which is the same as StructureNet. }
This is to improve the efficiency of memory consumption. But as a result, our network cannot handle shapes with more than 10 siblings for a part.
Future works can work on addressing these issues.
}
}

\section{Conclusion}
\label{sec:concl}

In this paper, we have presented DSG-Net, a novel deep generative model that learns to represent and generate 3D shapes in disentangled latent spaces of geometry and structure, while considering their synergy to ensure the plausibility of generated shapes. Through extensive evaluation, our method produces high-quality shapes with complex structures and fine geometric details, outperforming state-of-the-art methods. Our method also enables disentangled control of geometry and structure in shape generation, supporting novel applications such as interpolation of geometry (structure) while keeping structure (geometry) unchanged.

\begin{acks}

This work was supported by the Beijing Municipal Natural Science Foundation for Distinguished Young Scholars (No. JQ21013), the National Natural Science Foundation of China (No. 62061136007 and No. 61872440), Royal Society Newton Advanced Fellowship (No. NAF\verb|\|R2\verb|\|192151) and the Youth Innovation Promotion Association CAS. Kaichun Mo and Leonidas J. Guibas were supported by a grant from the Samsung GRO program, NSF grant CHS-1528025, ARL grant W911NF-21-2-0104, a Vannevar Bush Faculty fellowship, and gifts from the Autodesk and Snap corporations.

\end{acks}

\bibliographystyle{ACM-Reference-Format}
\bibliography{bibliography}

\clearpage
\appendix
\setcounter{section}{-1}

\section{\centering Supplementary Material}
This supplementary material accompanies the main paper, 
which presents 
quantitative and qualitative comparisons to SAG-Net~\cite{wu2019sagnet}, data preparation, more network training and implementation details, \yj{ablation studies for our key designs,} 
and more visualization results for shape reconstruction, interpolation and generation. 

All sections are listed as follows:

\begin{figure*}[]
    \centering
    \vspace{3mm}
    \includegraphics[width=0.135\linewidth]{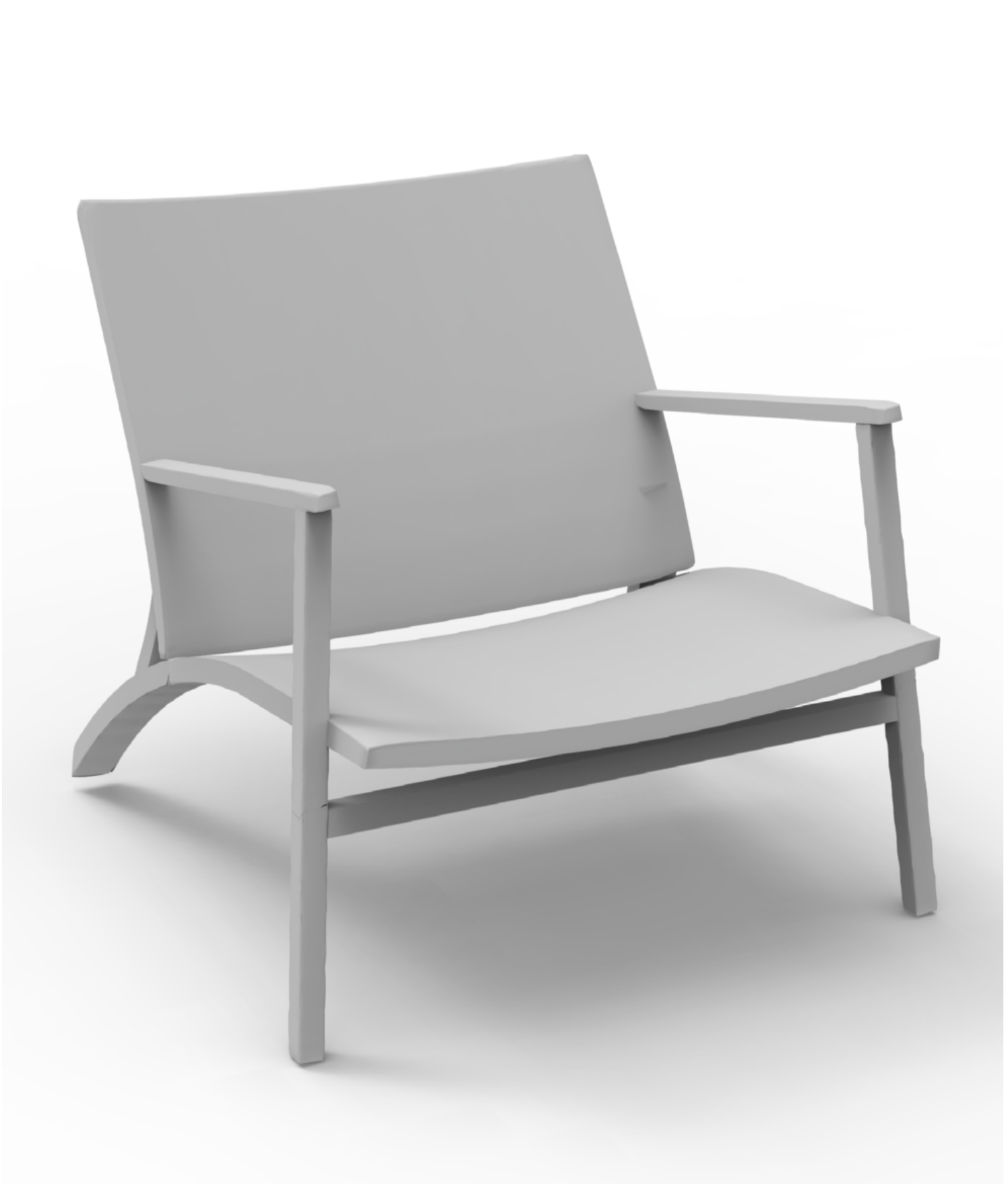}
    \includegraphics[width=0.135\linewidth]{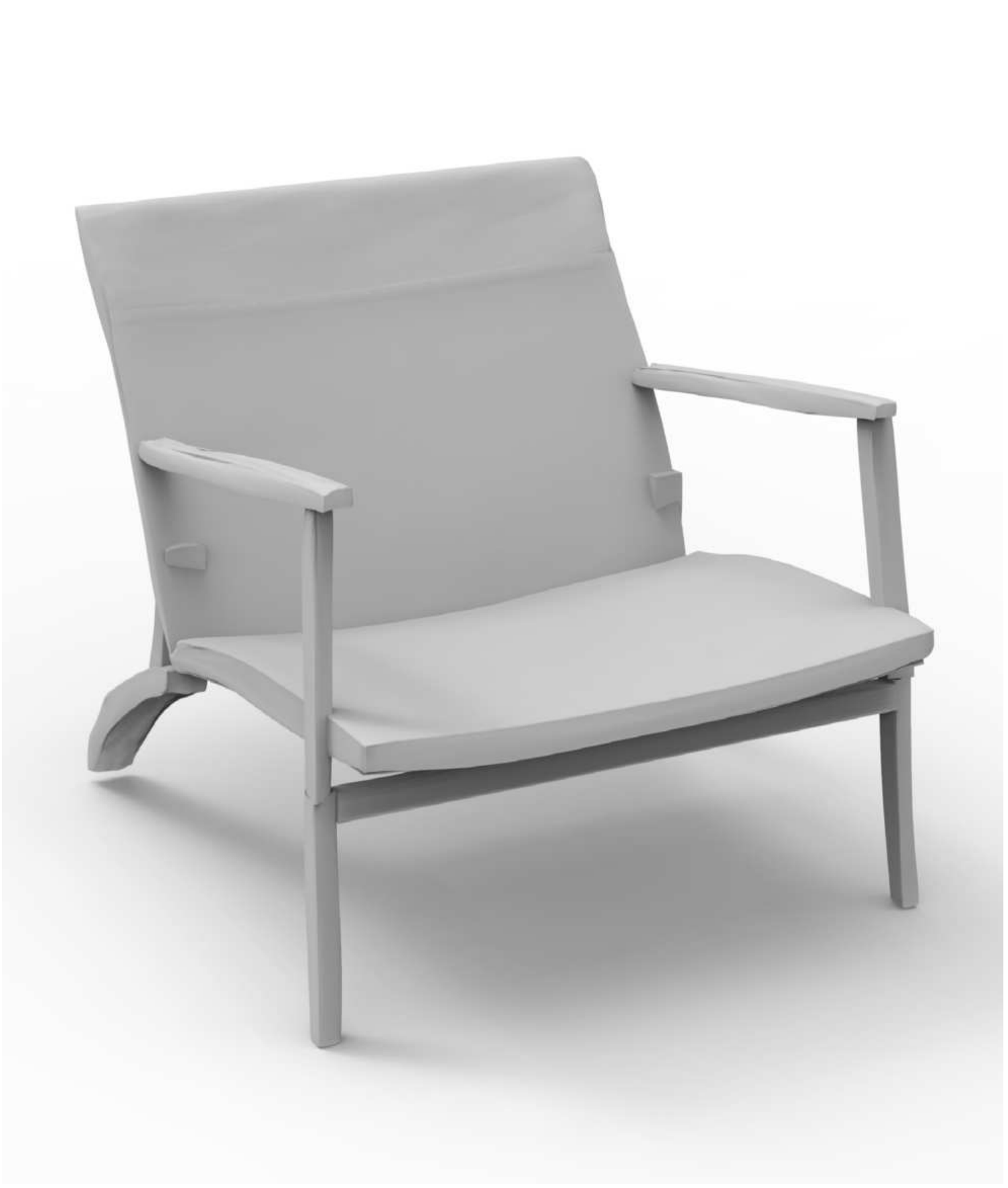}
    \includegraphics[width=0.135\linewidth]{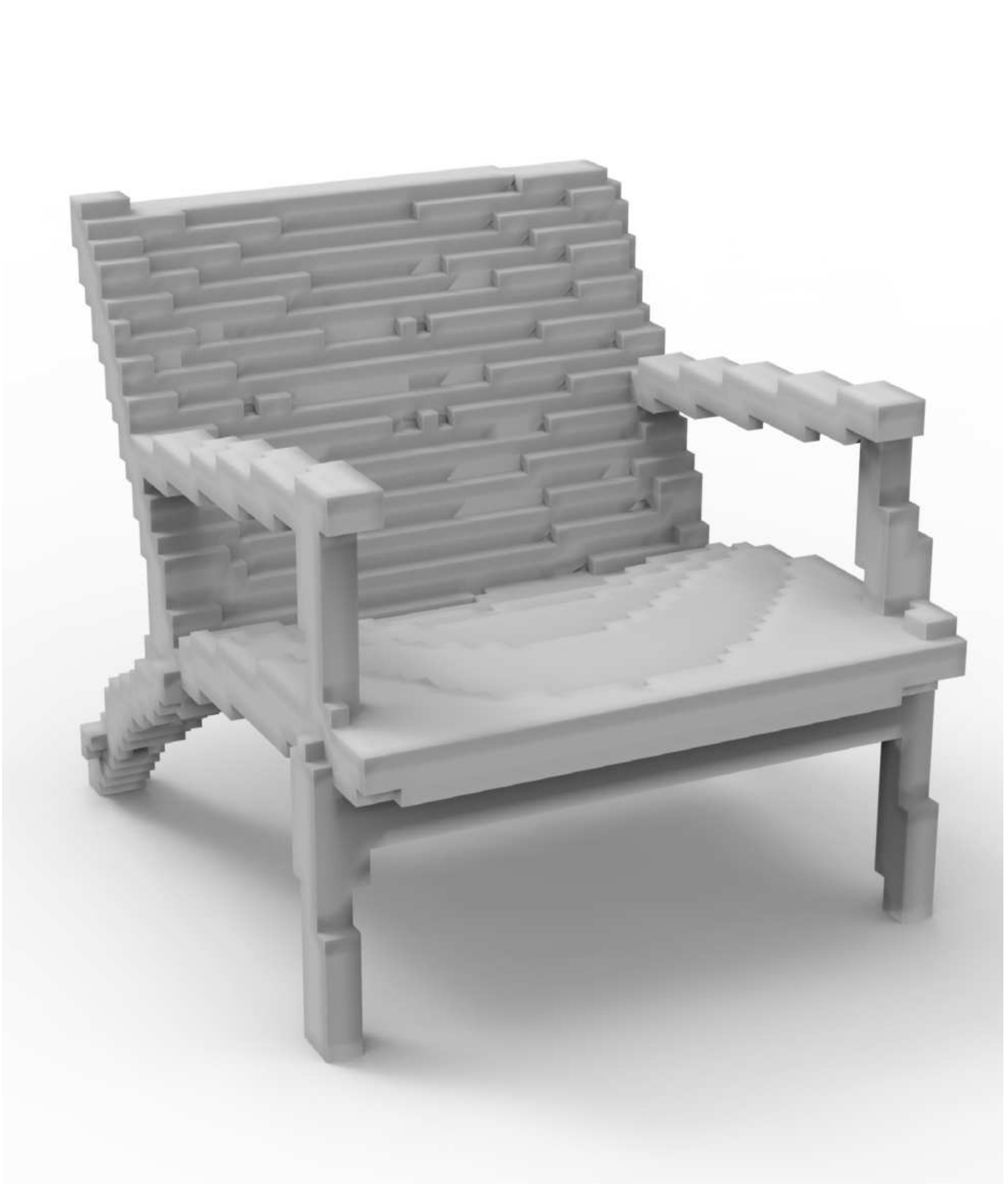}
    \includegraphics[width=0.135\linewidth]{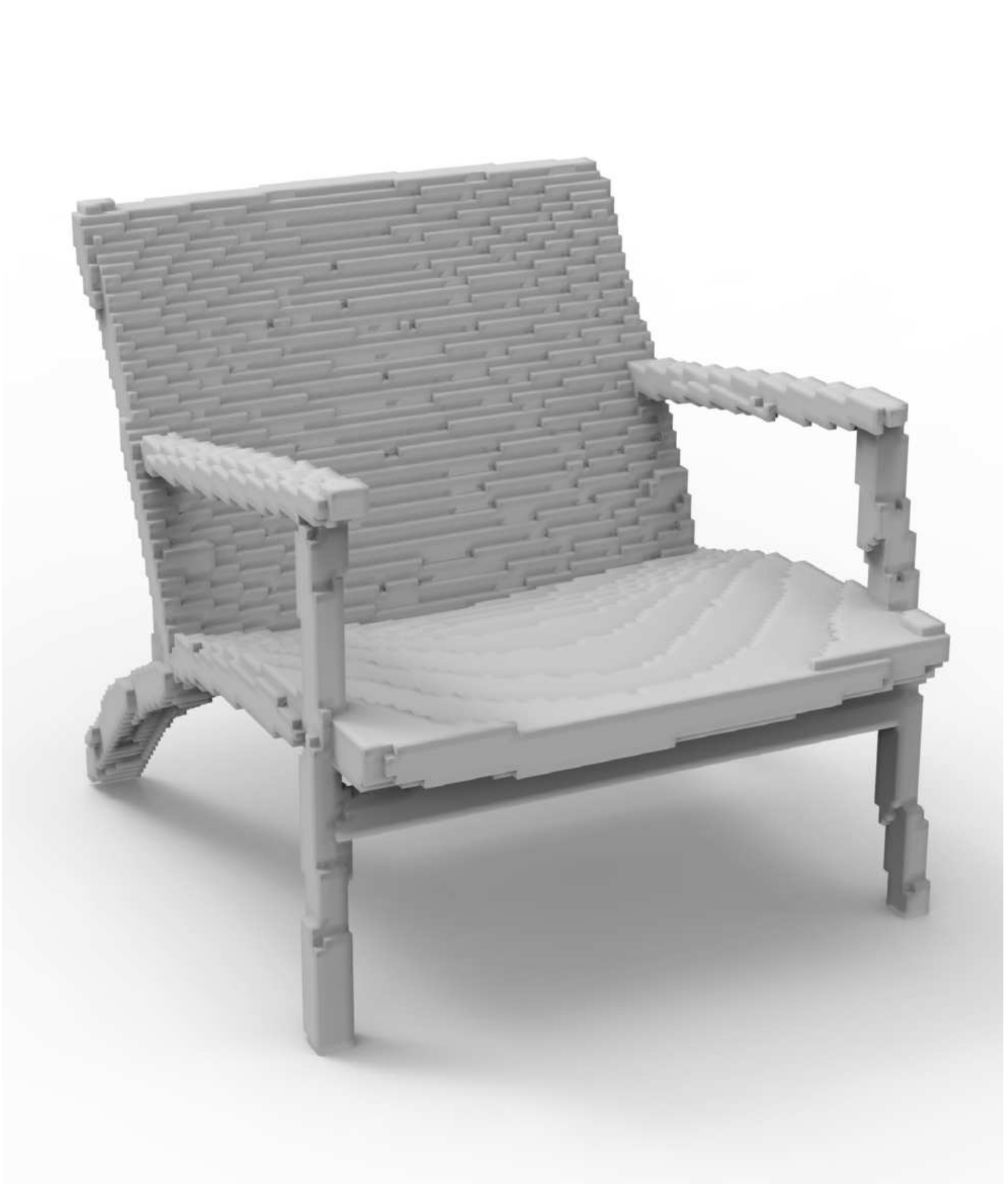}
    \includegraphics[width=0.135\linewidth]{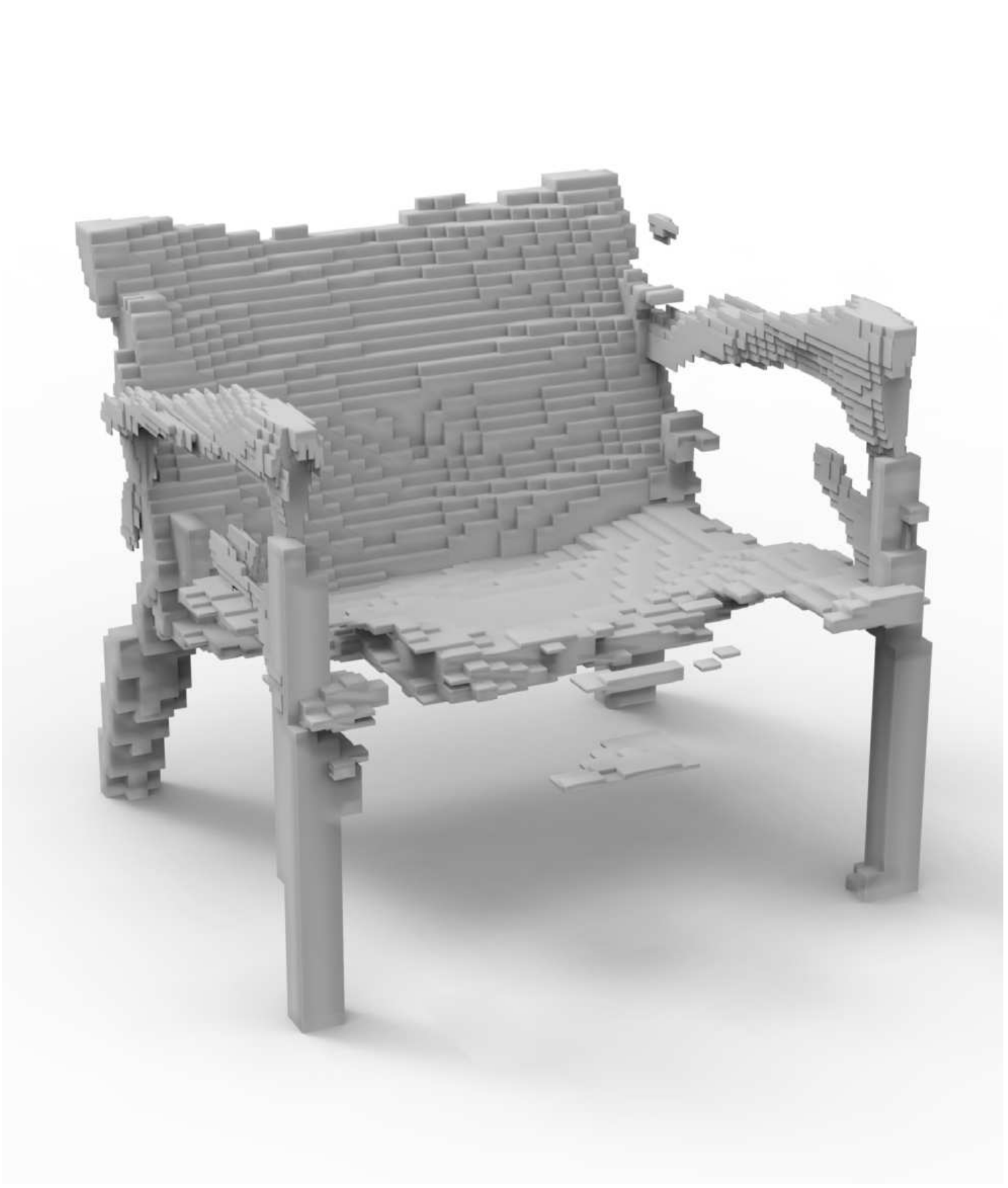}
    \includegraphics[width=0.135\linewidth]{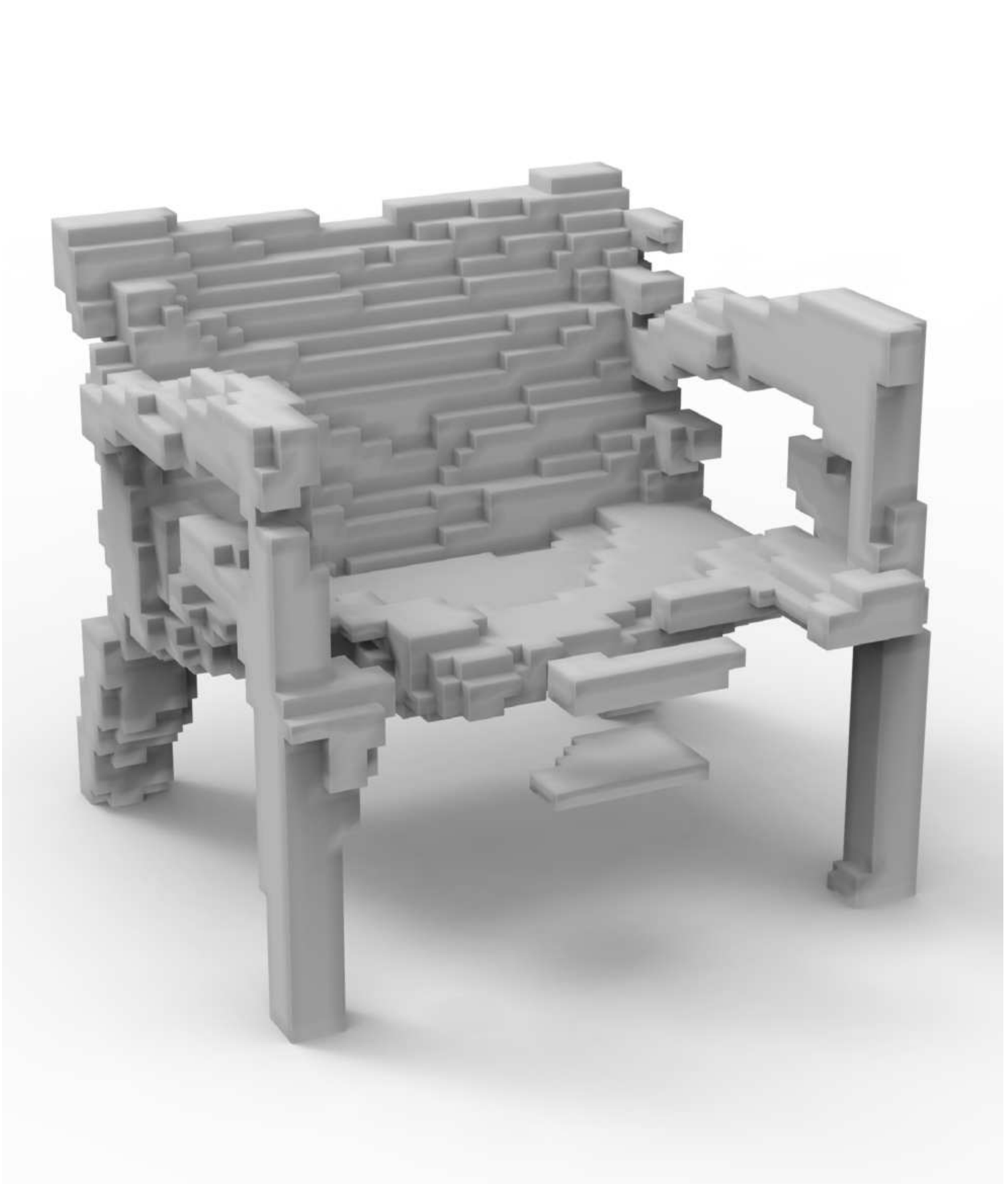}
    \includegraphics[width=0.135\linewidth]{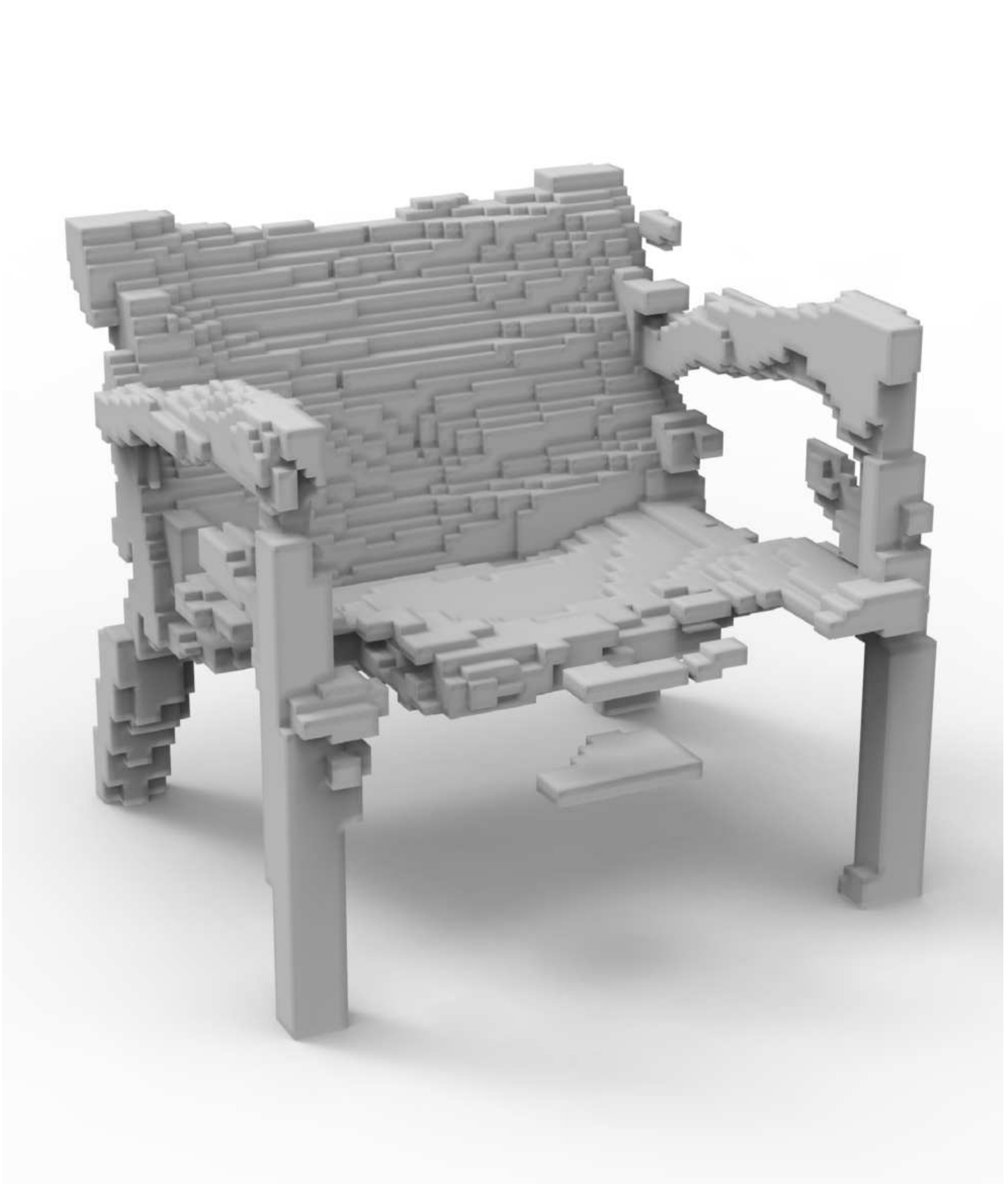}\\
    \vspace{2mm}
    \includegraphics[width=0.135\linewidth]{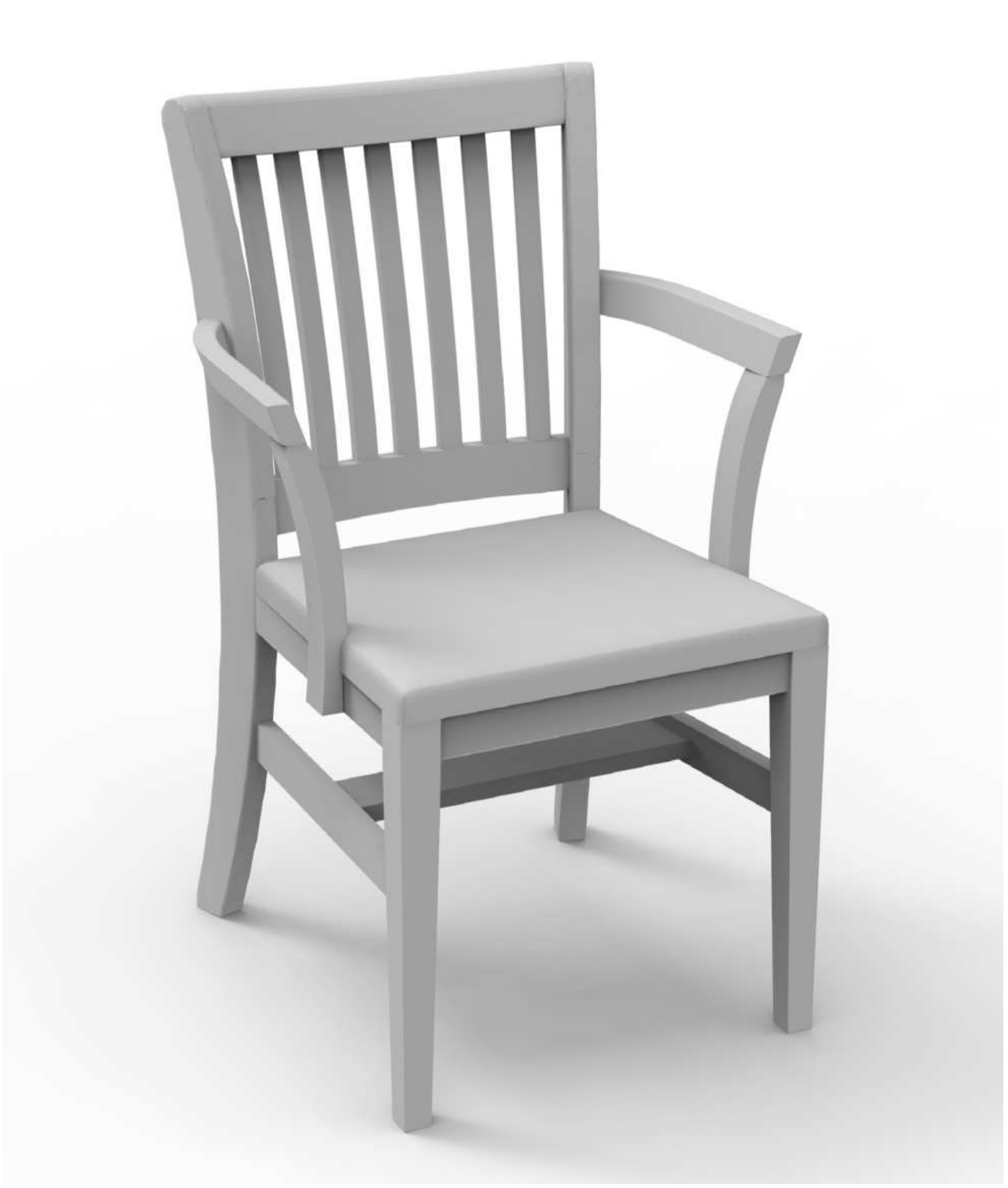}
    \includegraphics[width=0.135\linewidth]{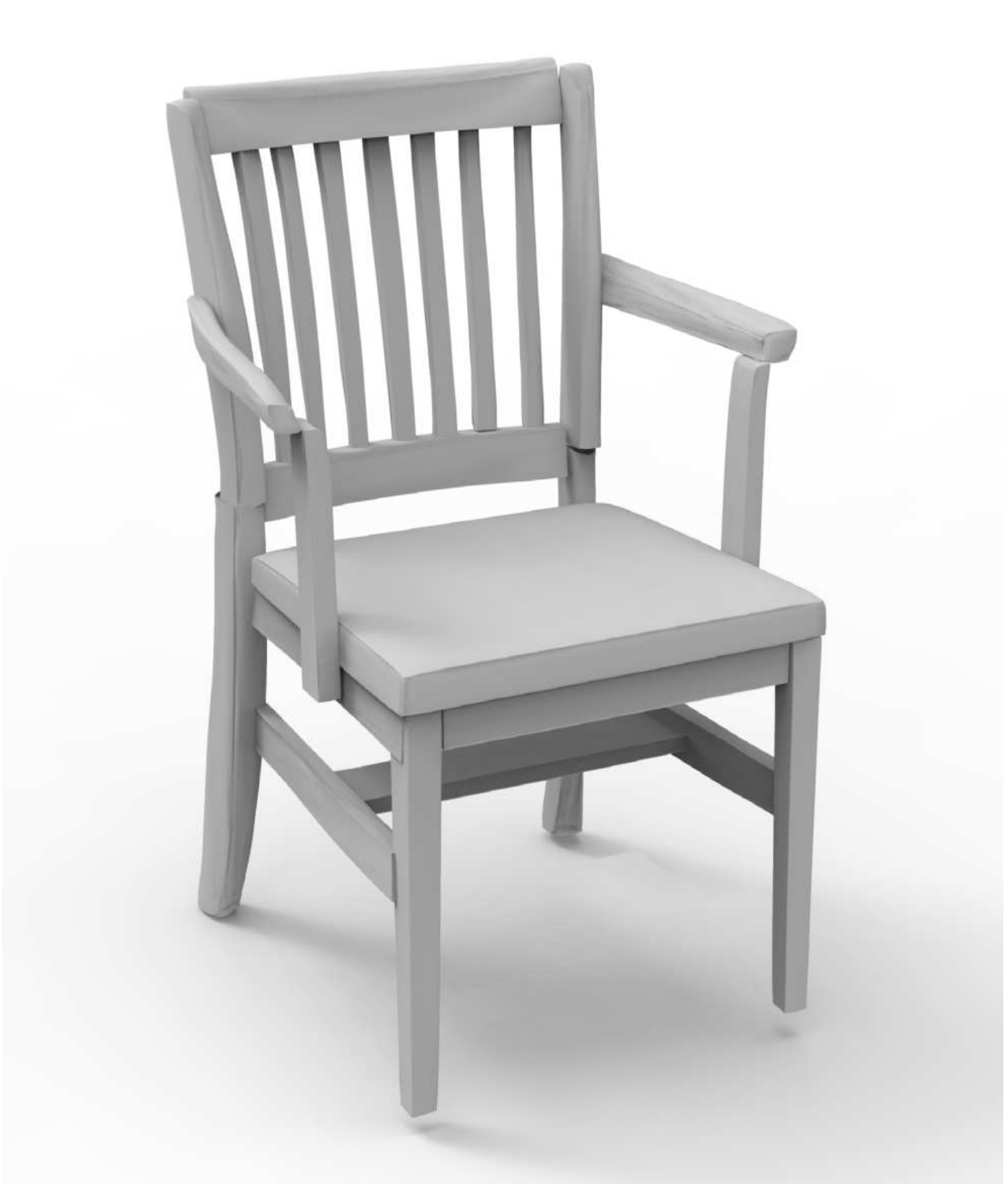}
    \includegraphics[width=0.135\linewidth]{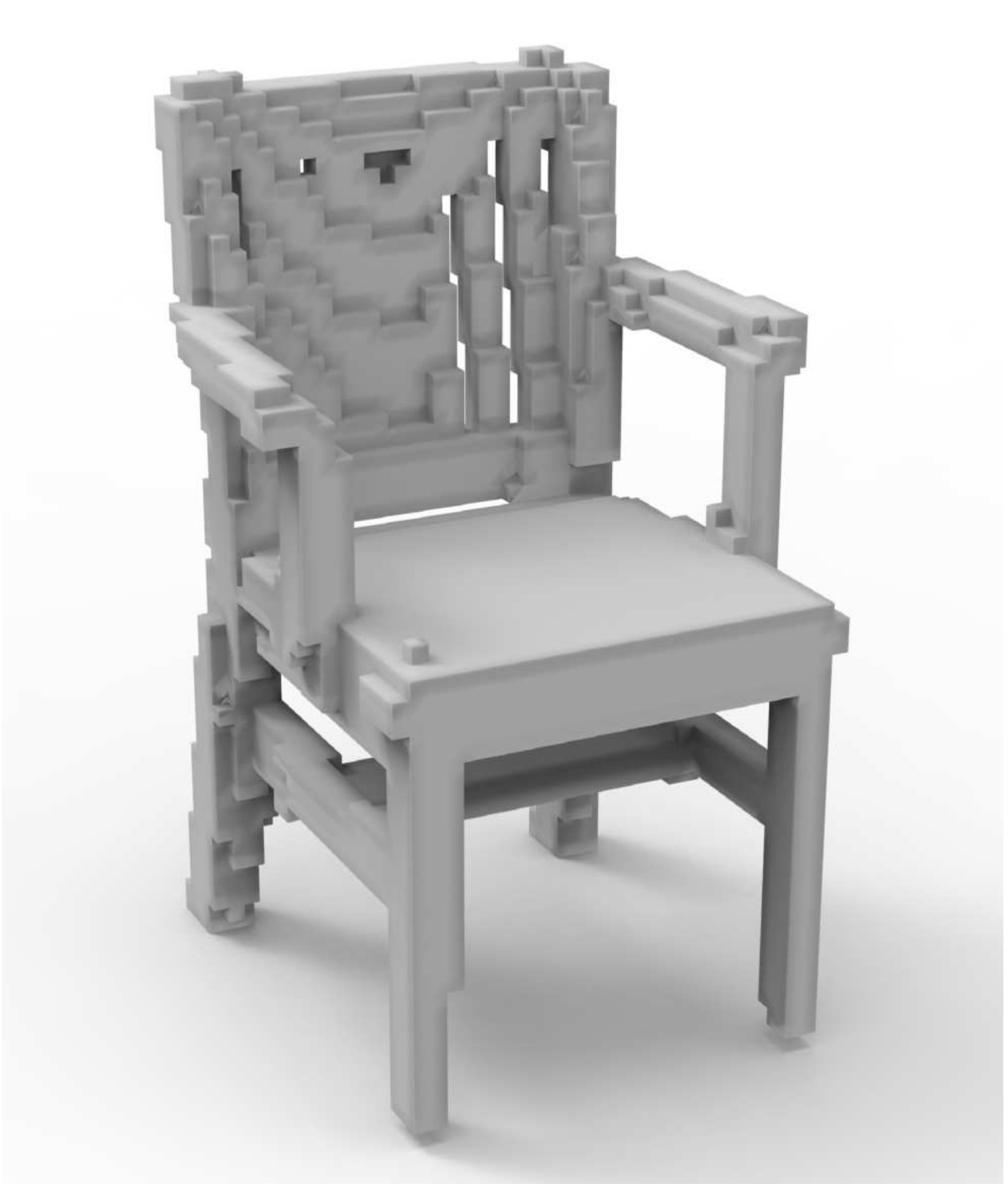}
    \includegraphics[width=0.135\linewidth]{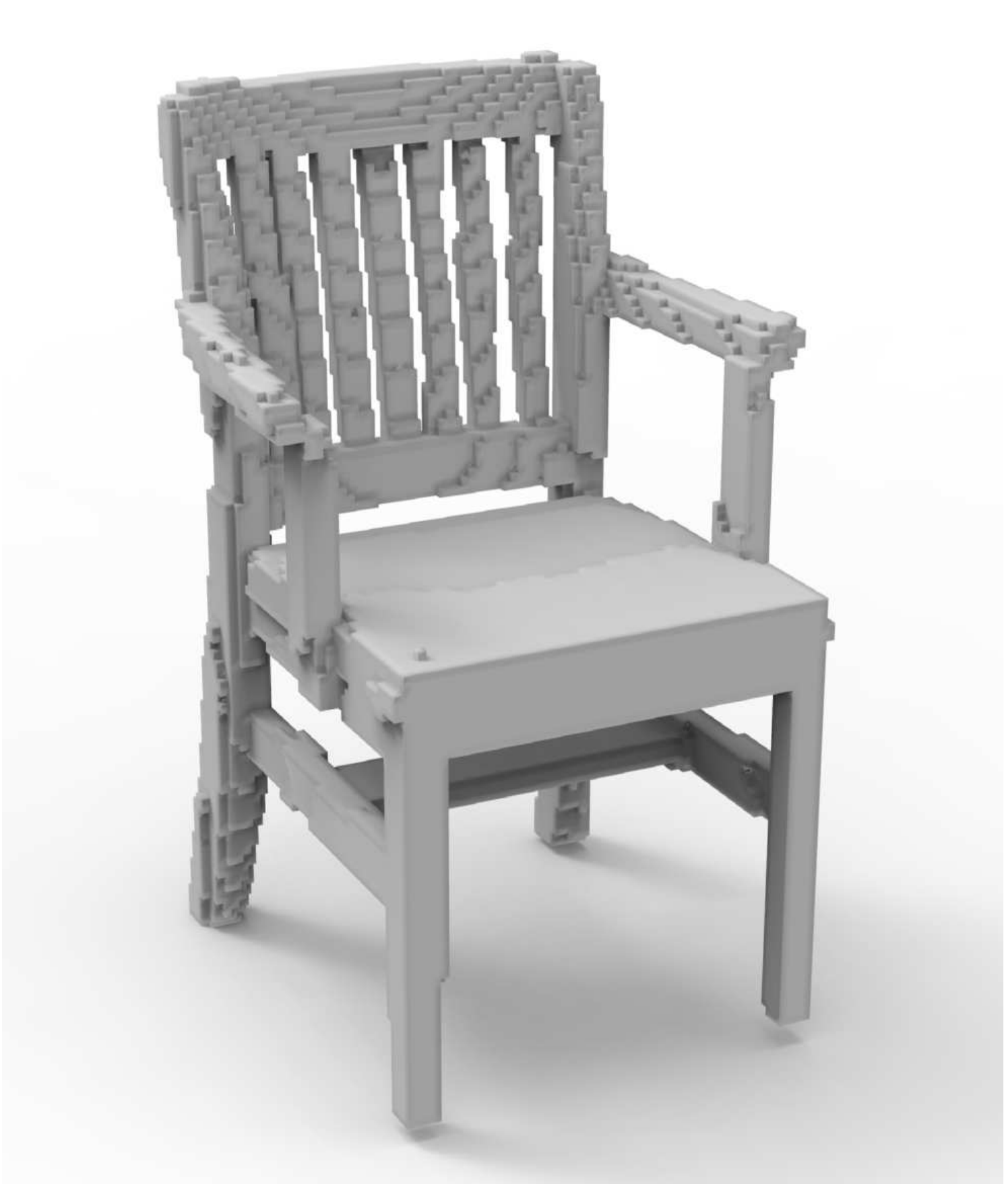}
    \includegraphics[width=0.135\linewidth]{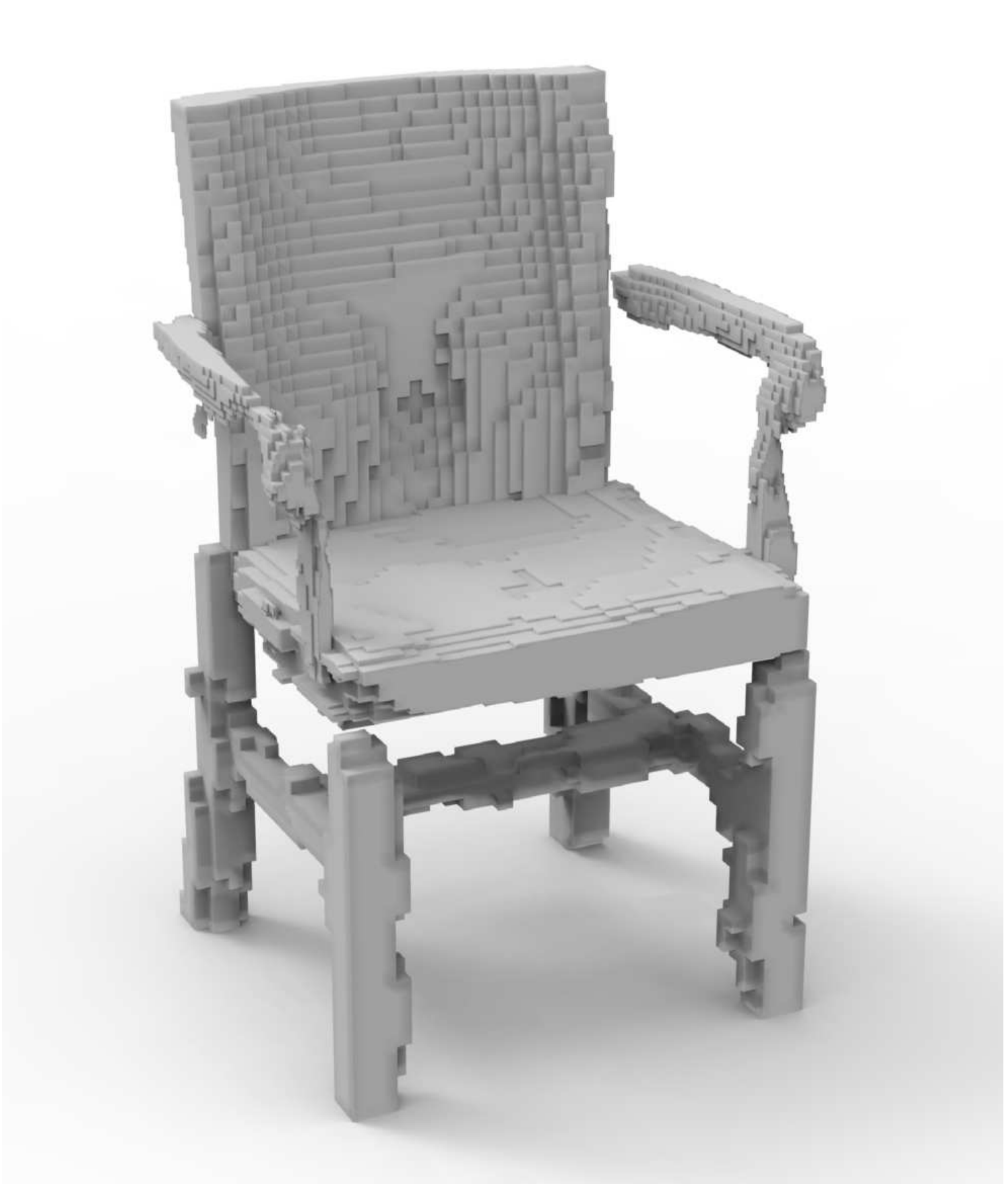}
    \includegraphics[width=0.135\linewidth]{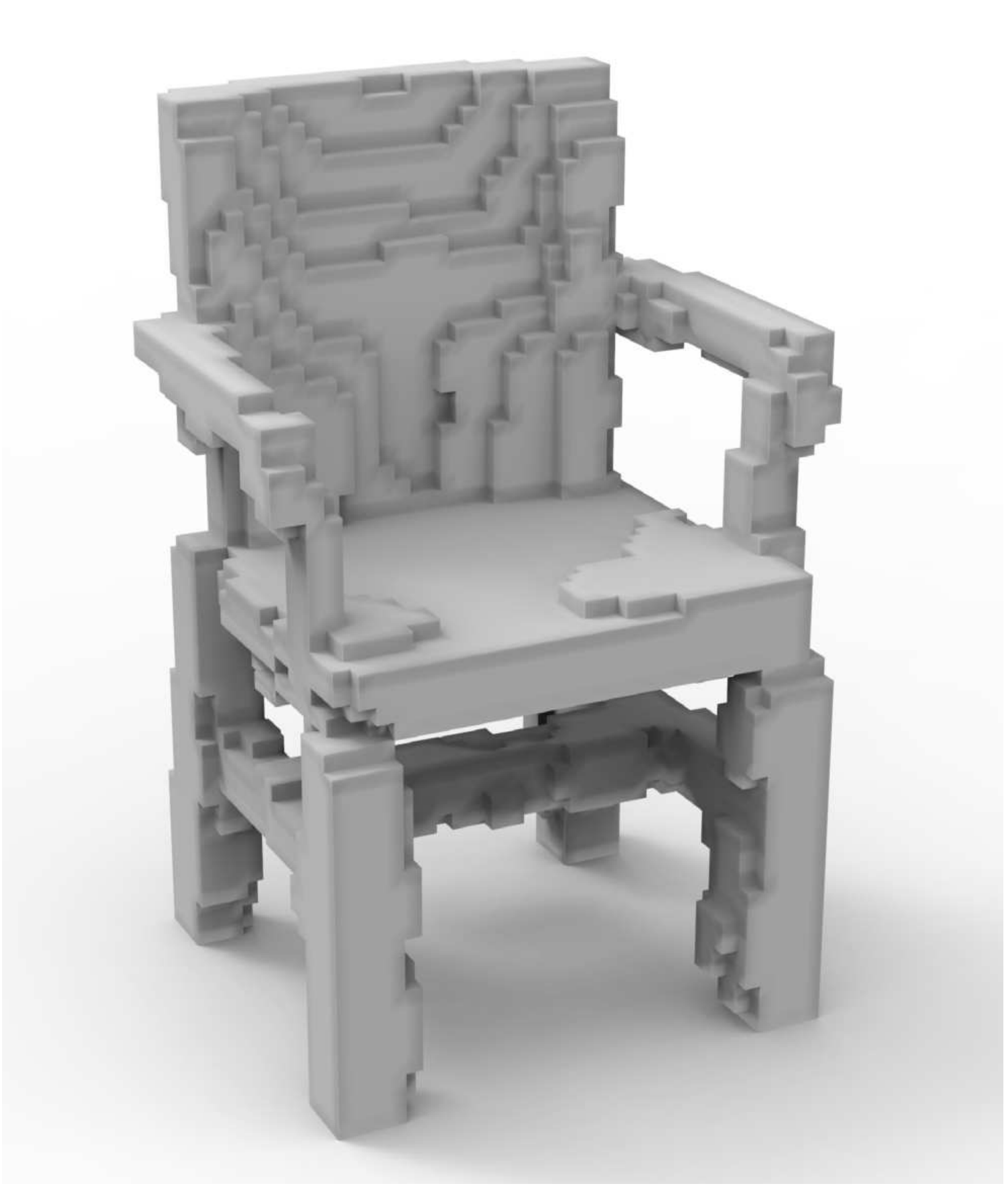}
    \includegraphics[width=0.135\linewidth]{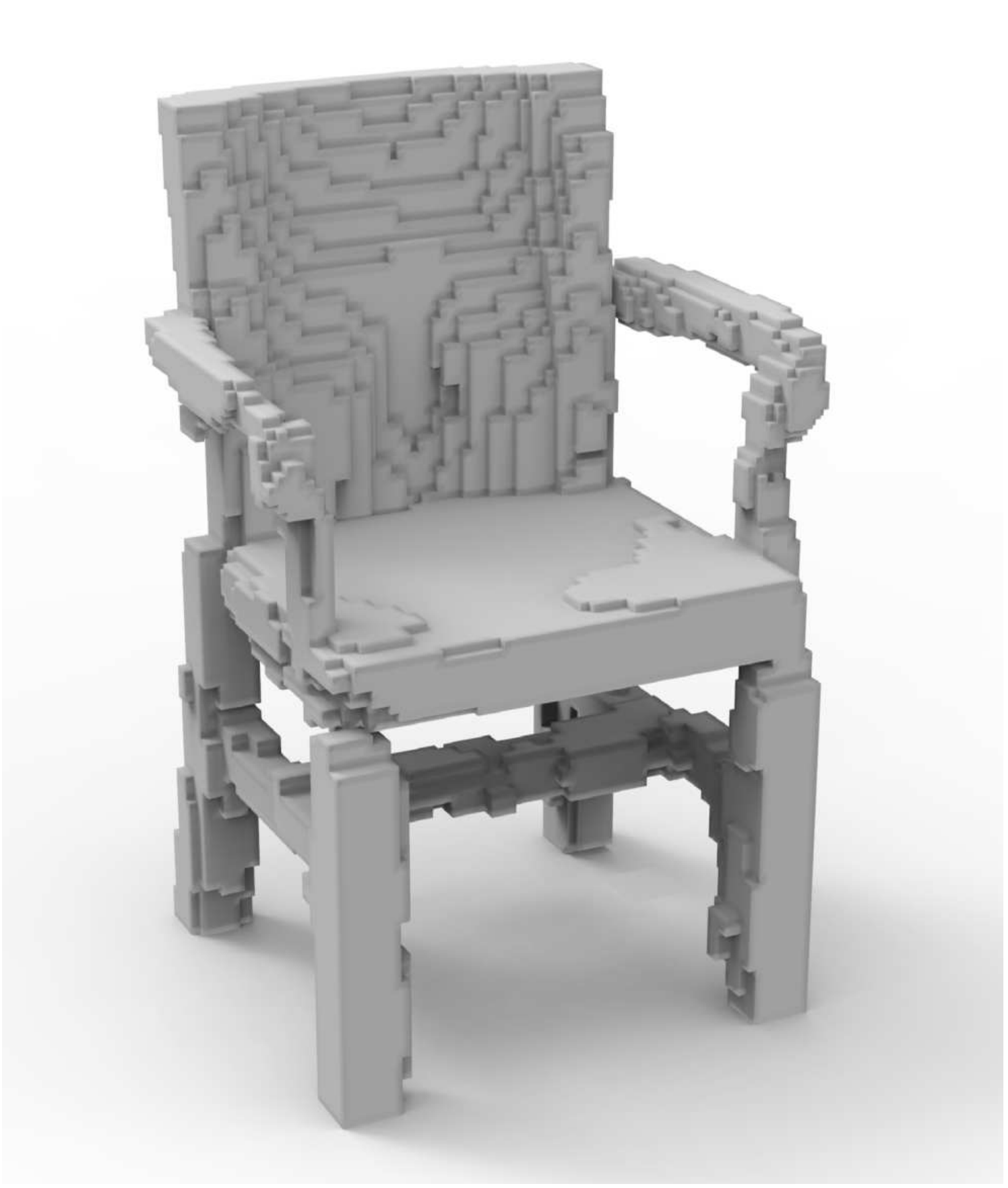}\\
    \vspace{2mm}
    \includegraphics[width=0.135\linewidth]{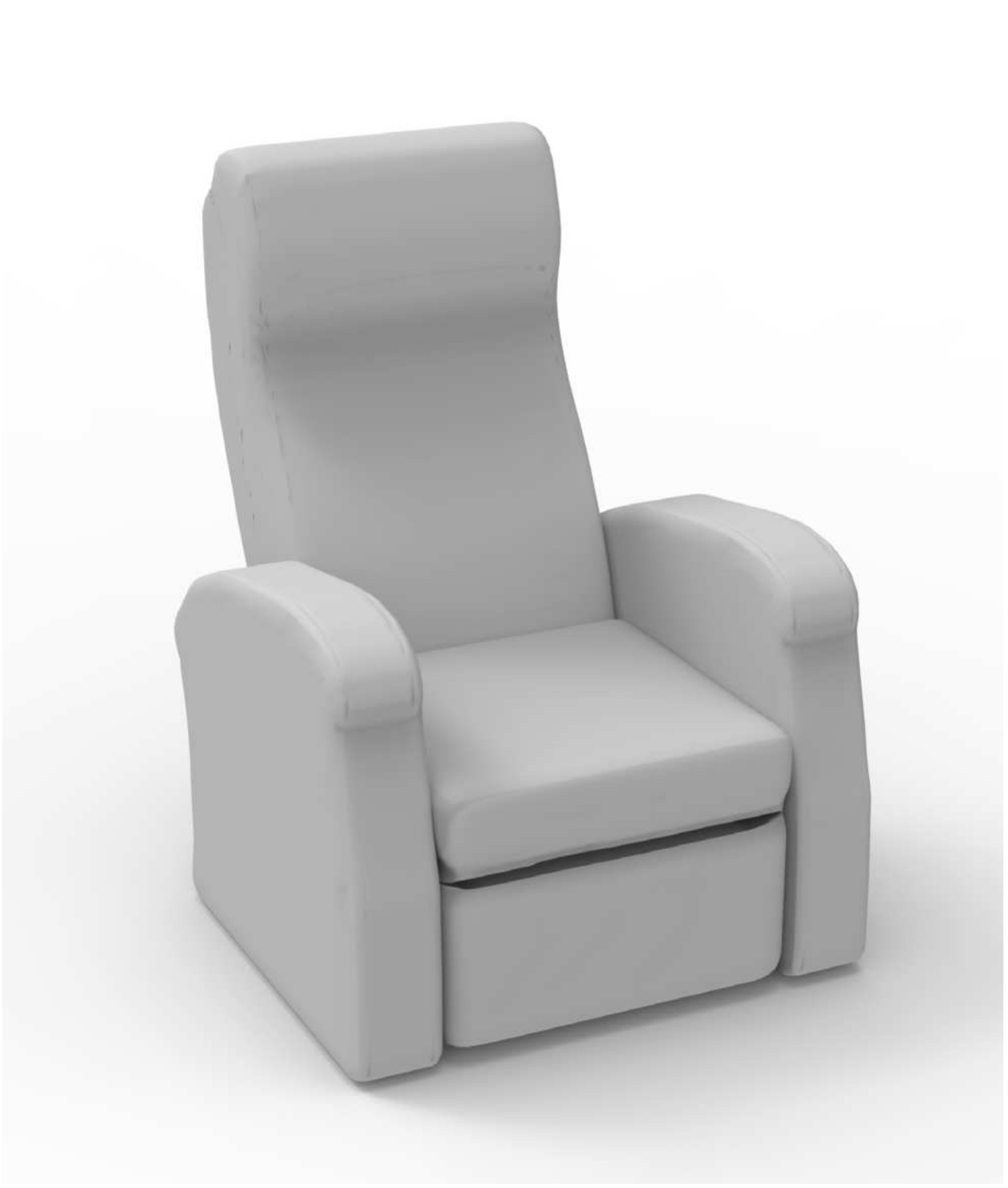}
    \includegraphics[width=0.135\linewidth]{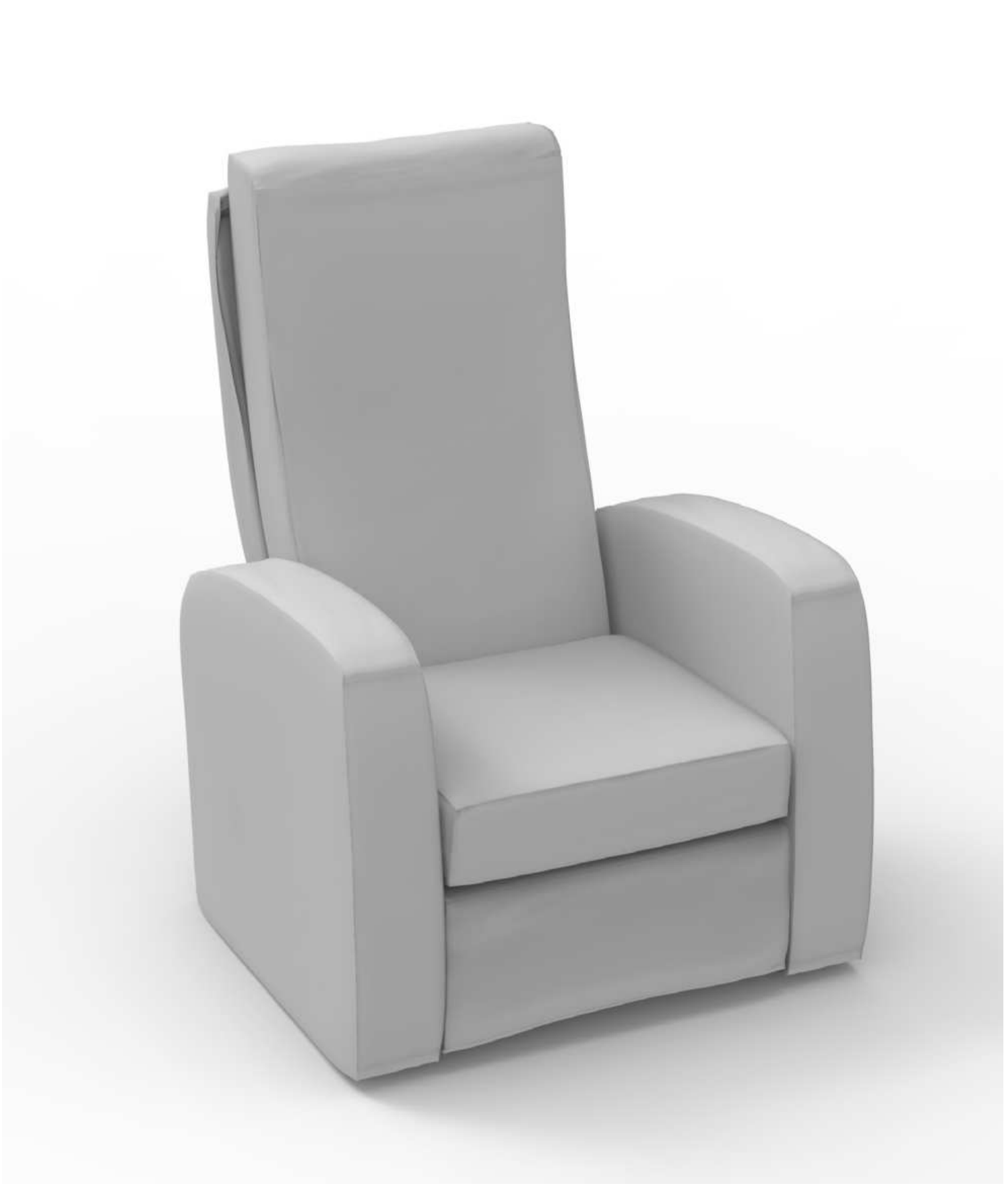}
    \includegraphics[width=0.135\linewidth]{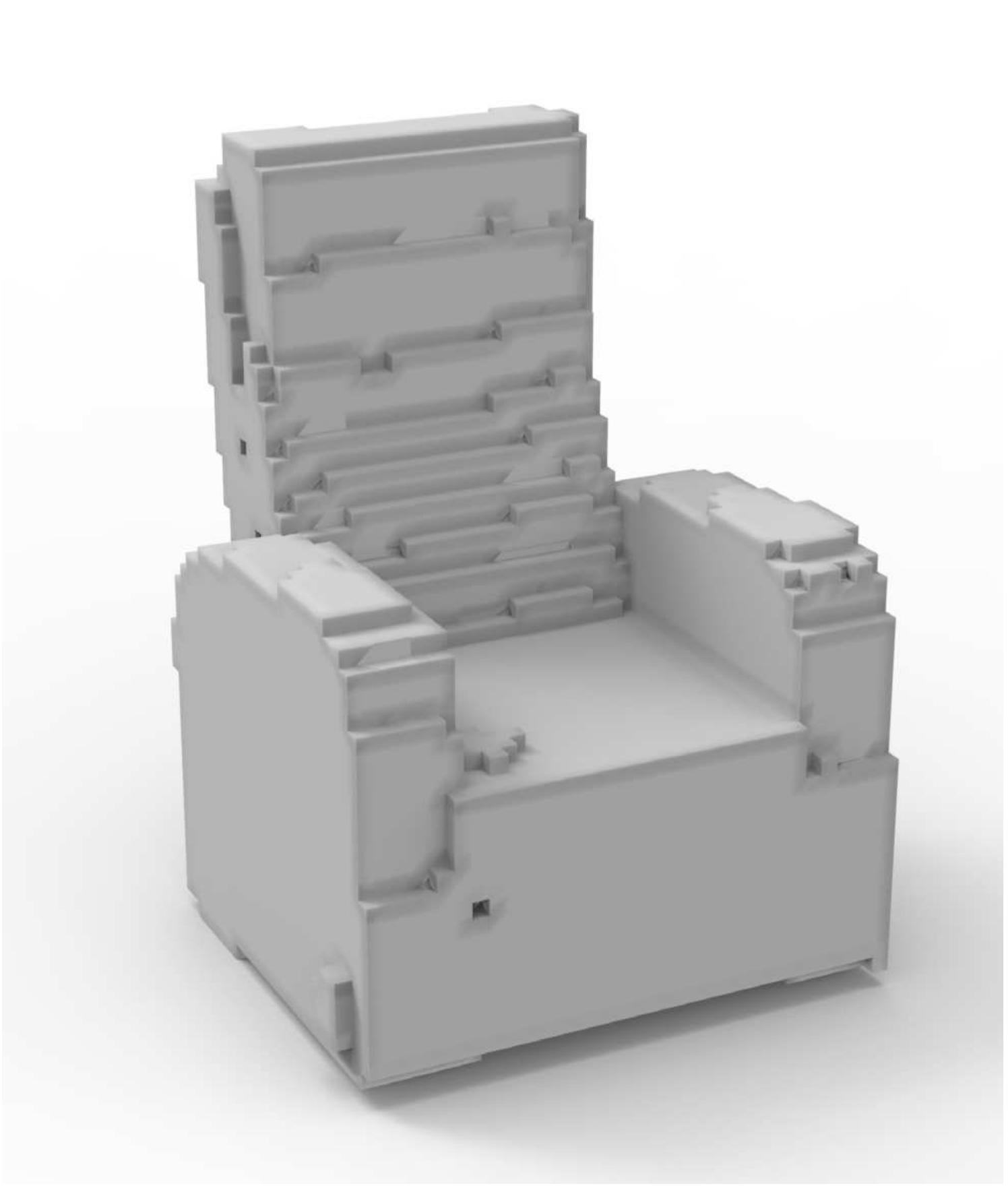}
    \includegraphics[width=0.135\linewidth]{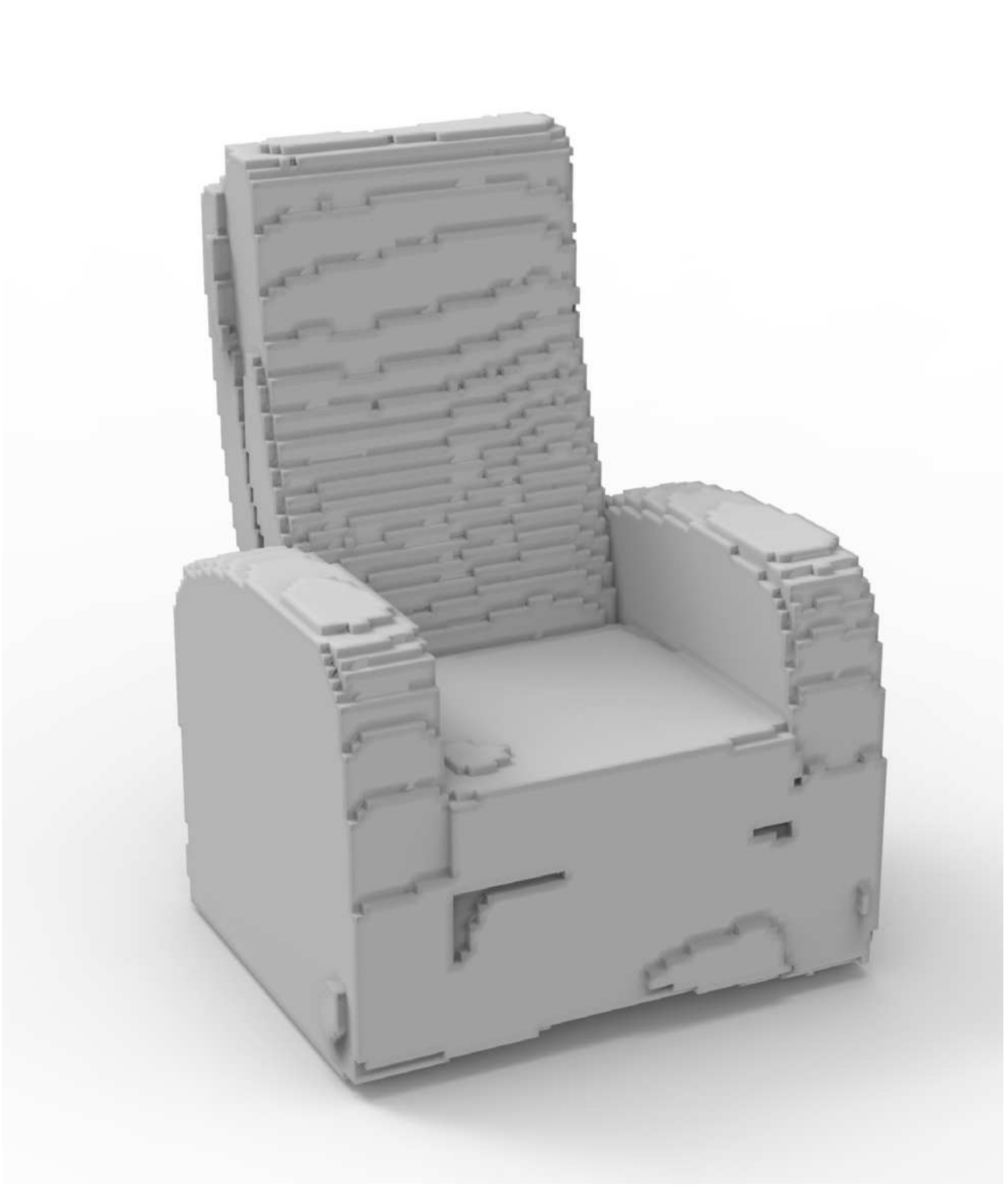}
    \includegraphics[width=0.135\linewidth]{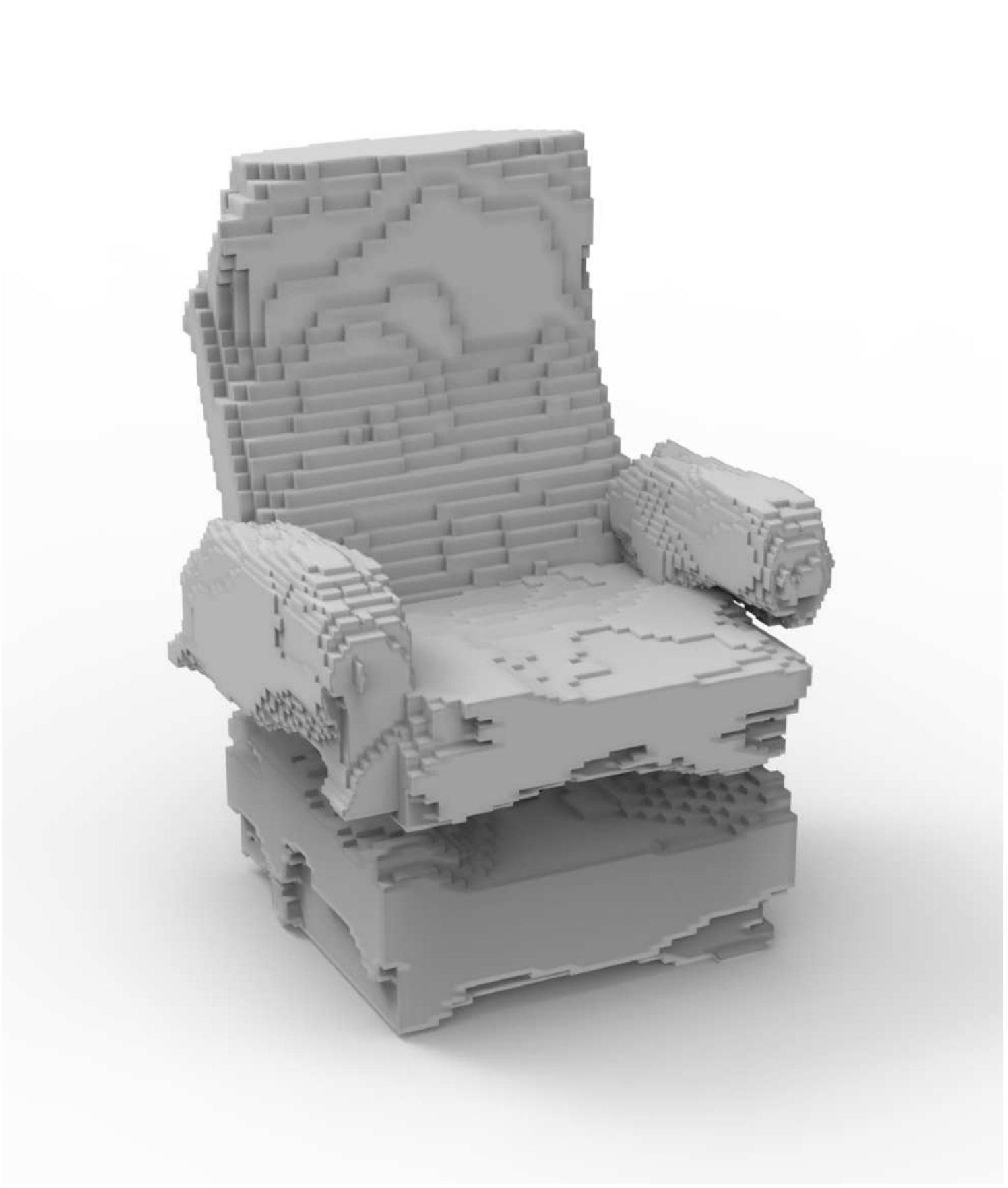}
    \includegraphics[width=0.135\linewidth]{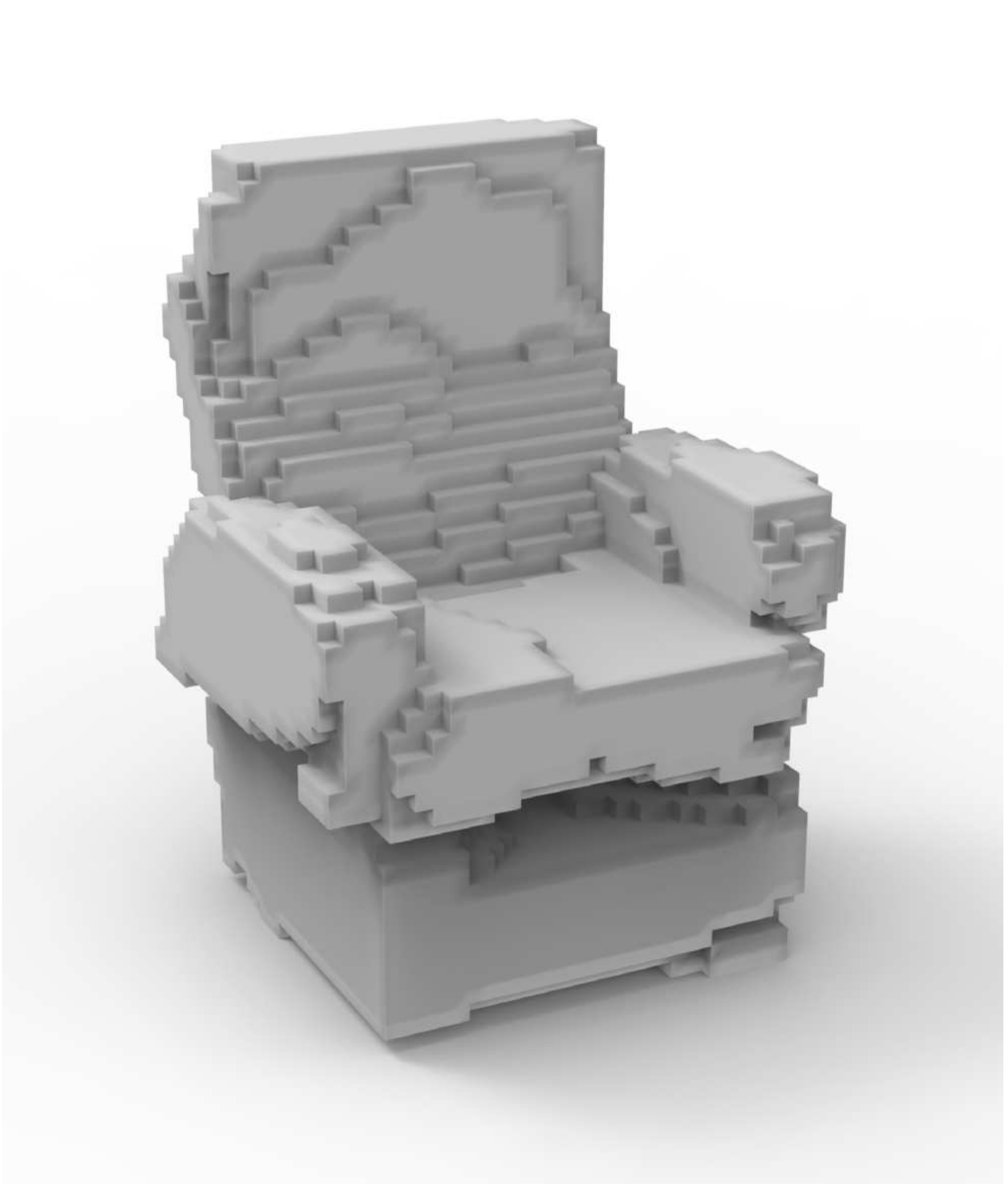}
    \includegraphics[width=0.135\linewidth]{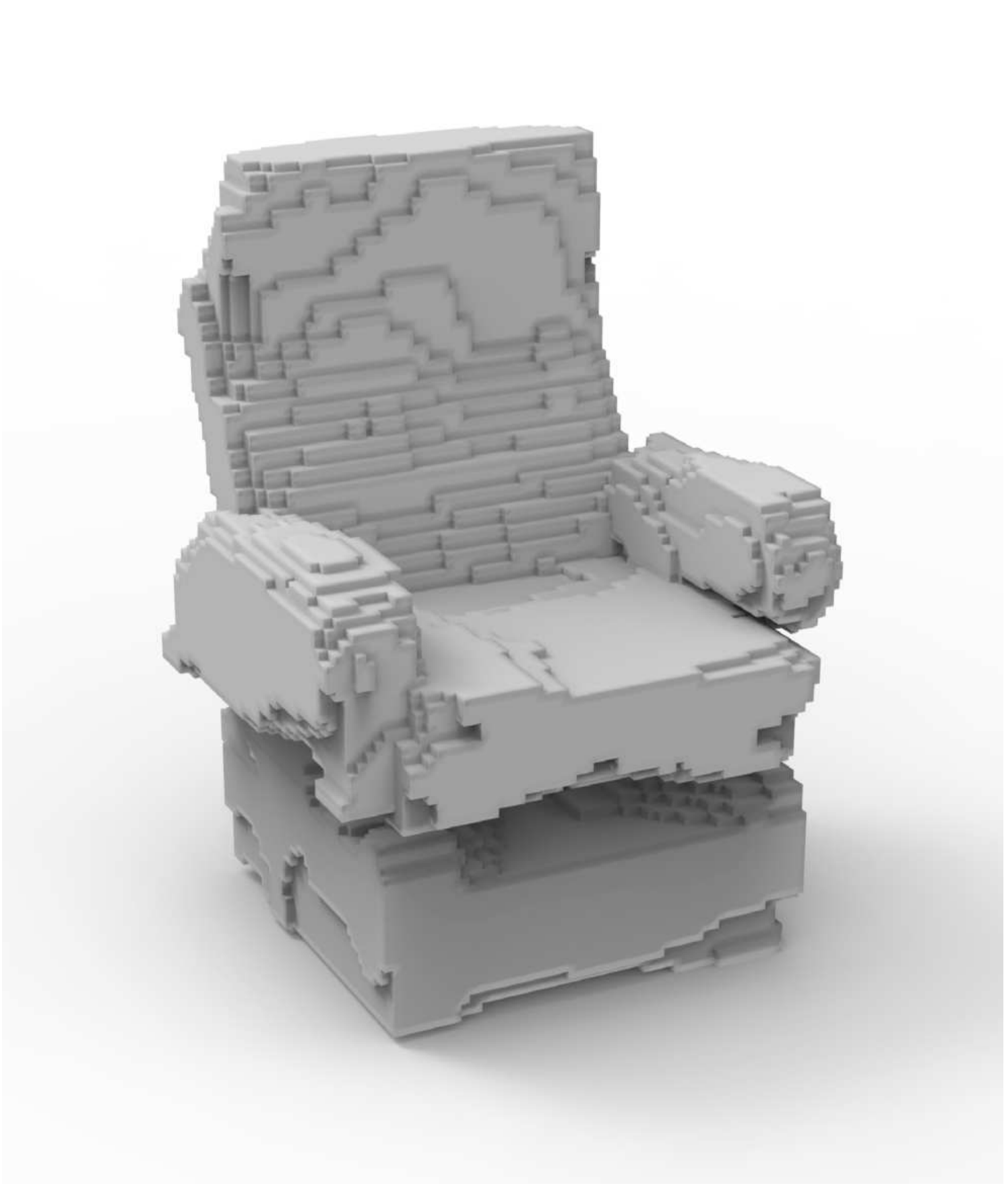}\\
    \vspace{2mm}
    \includegraphics[width=0.135\linewidth]{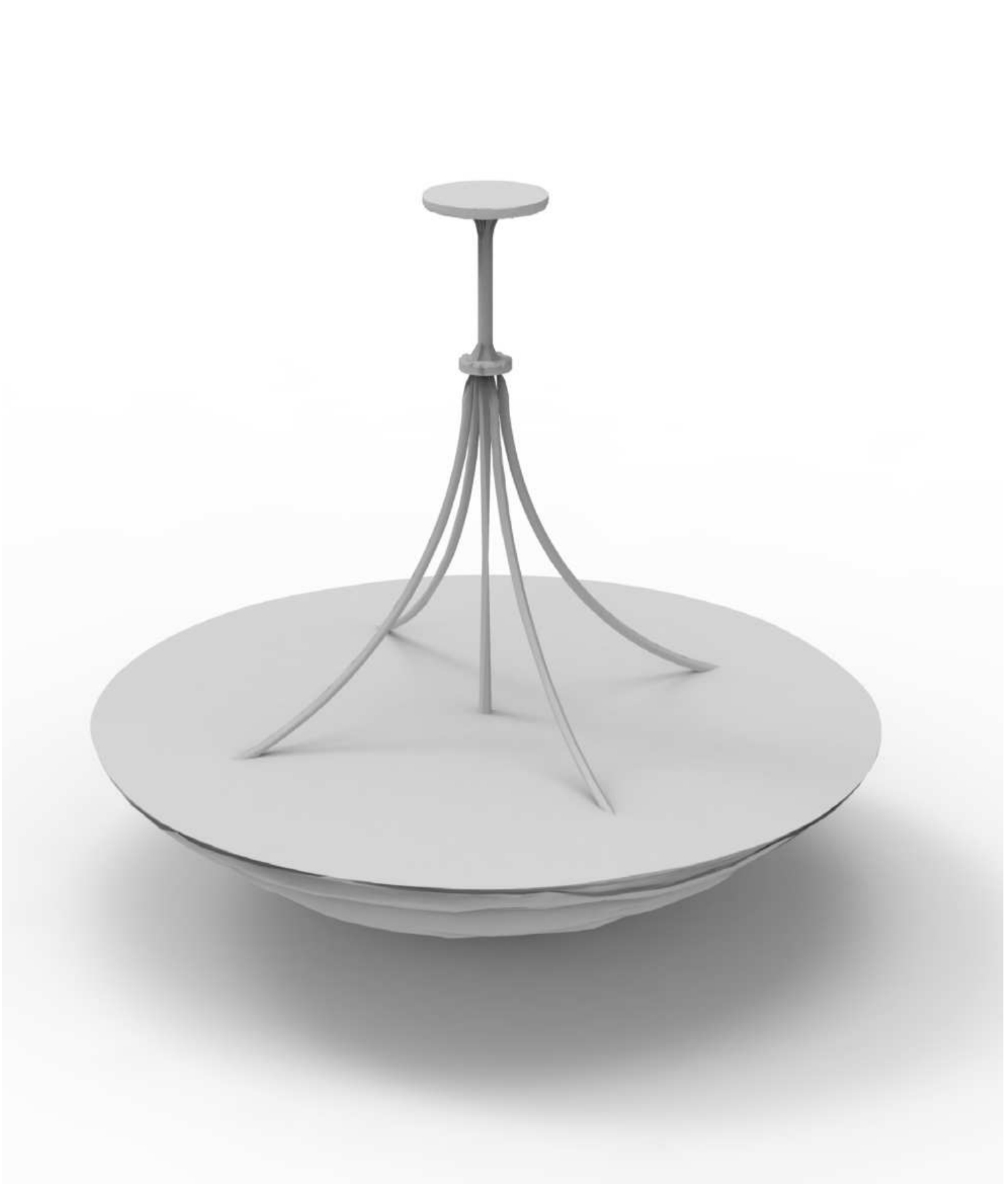}
    \includegraphics[width=0.135\linewidth]{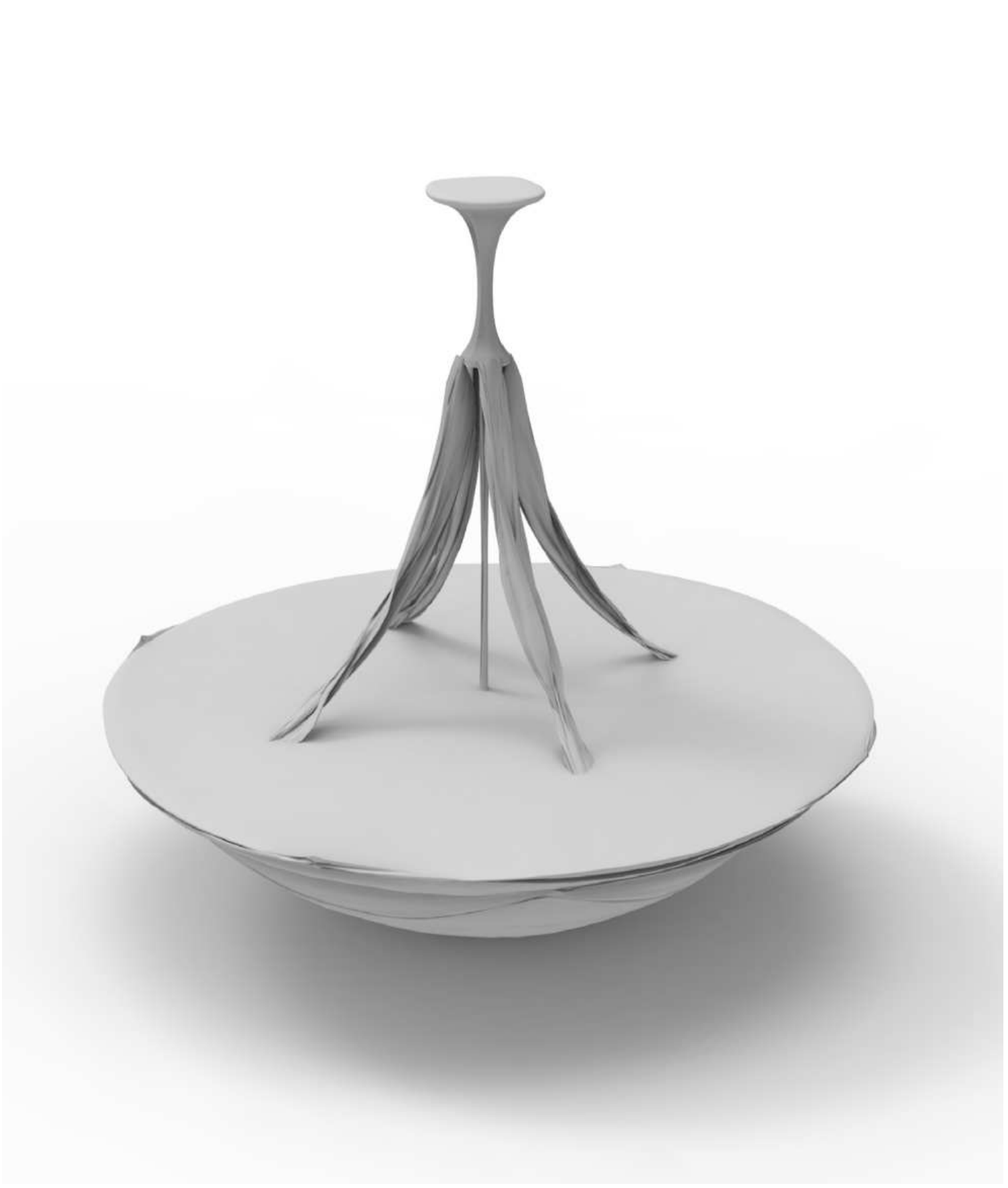}
    \includegraphics[width=0.135\linewidth]{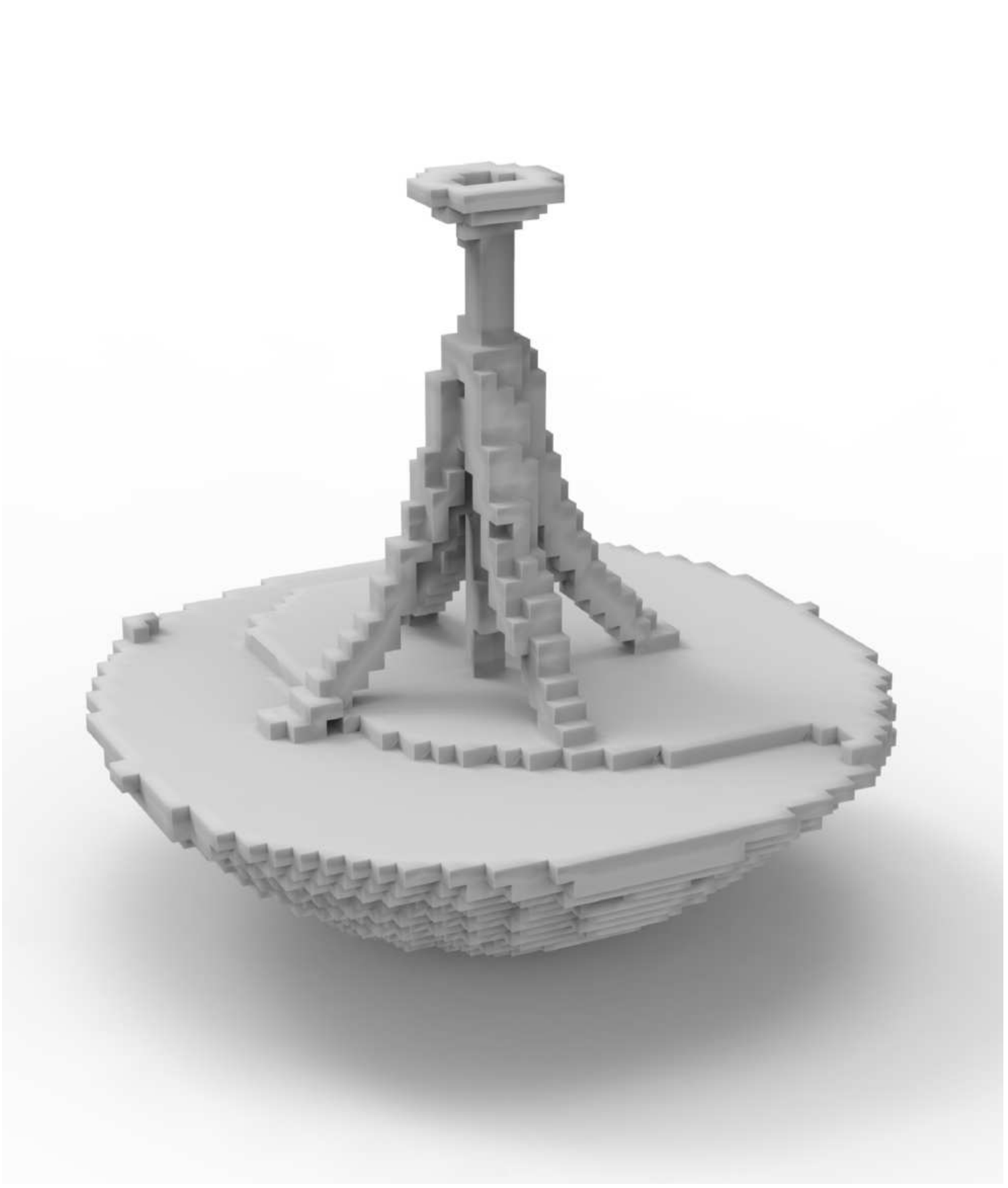}
    \includegraphics[width=0.135\linewidth]{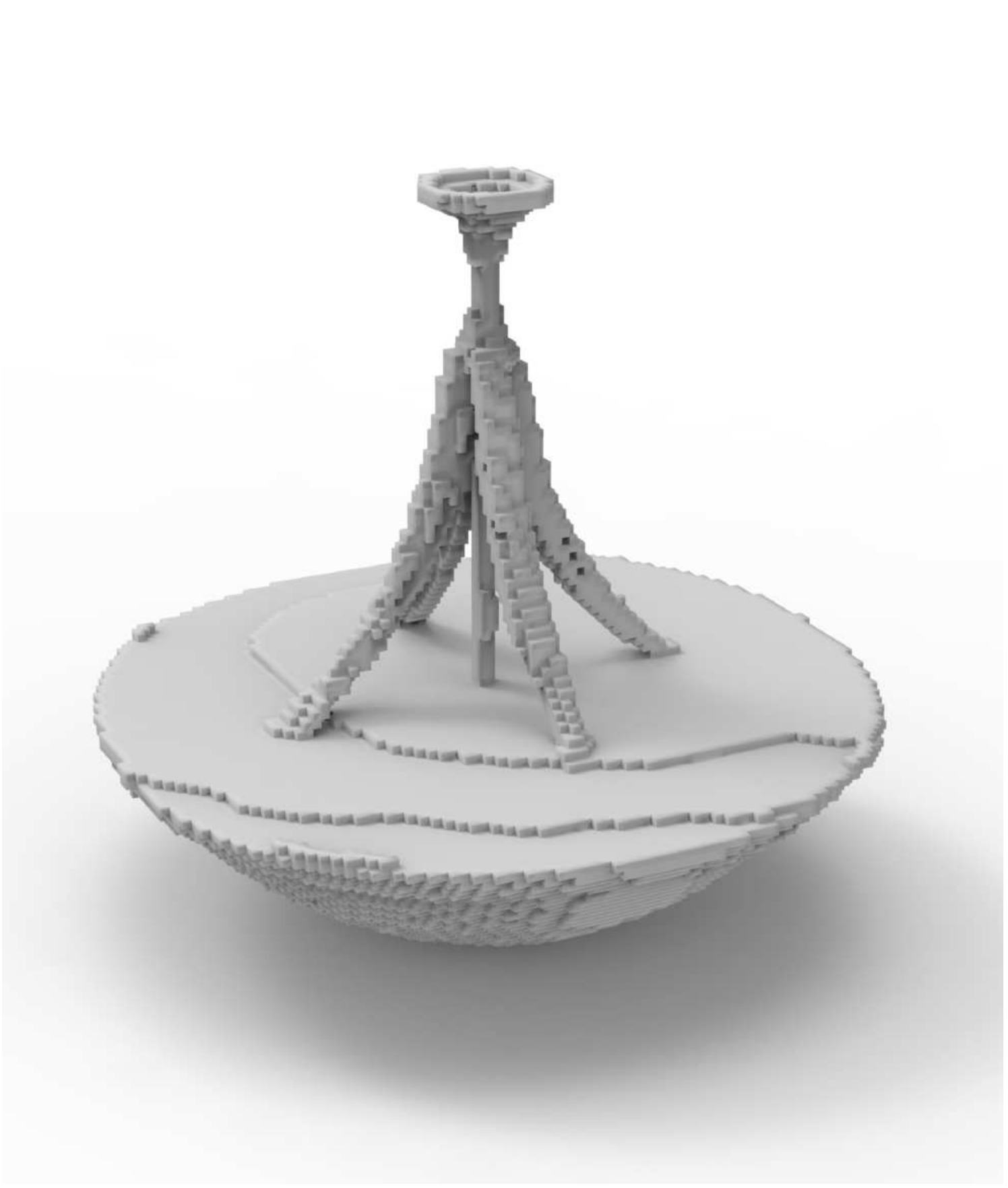}
    \includegraphics[width=0.135\linewidth]{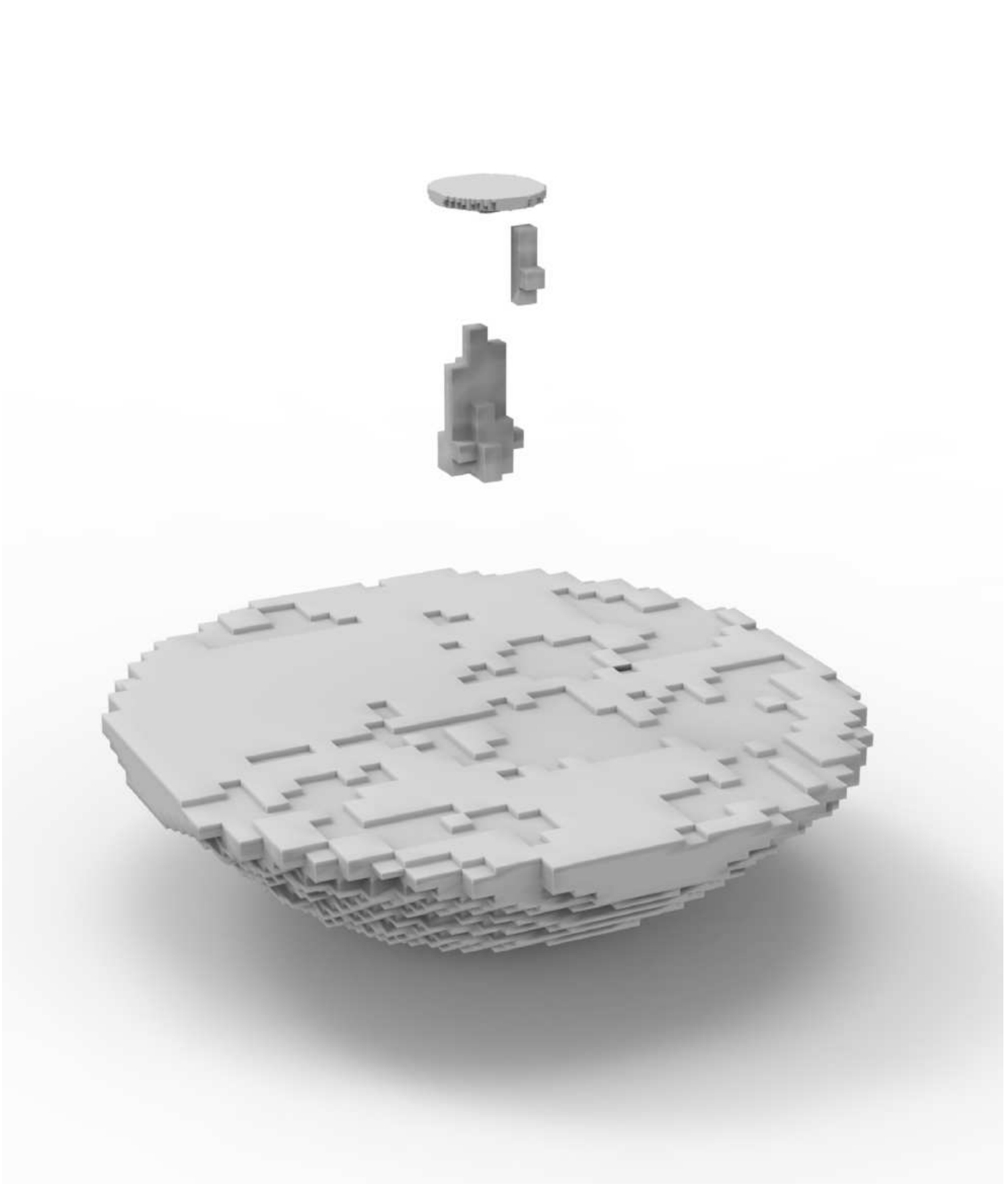}
    \includegraphics[width=0.135\linewidth]{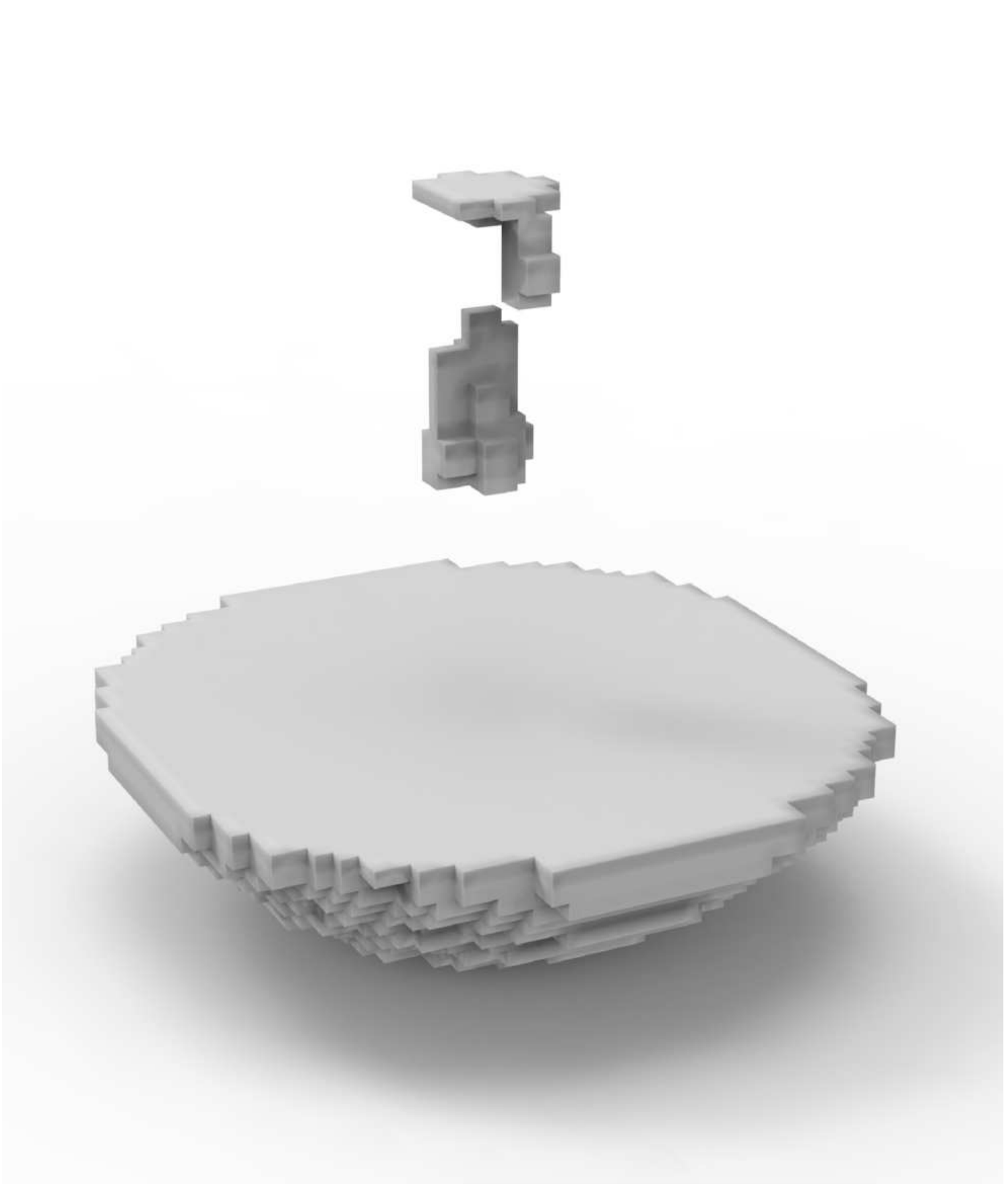}
    \includegraphics[width=0.135\linewidth]{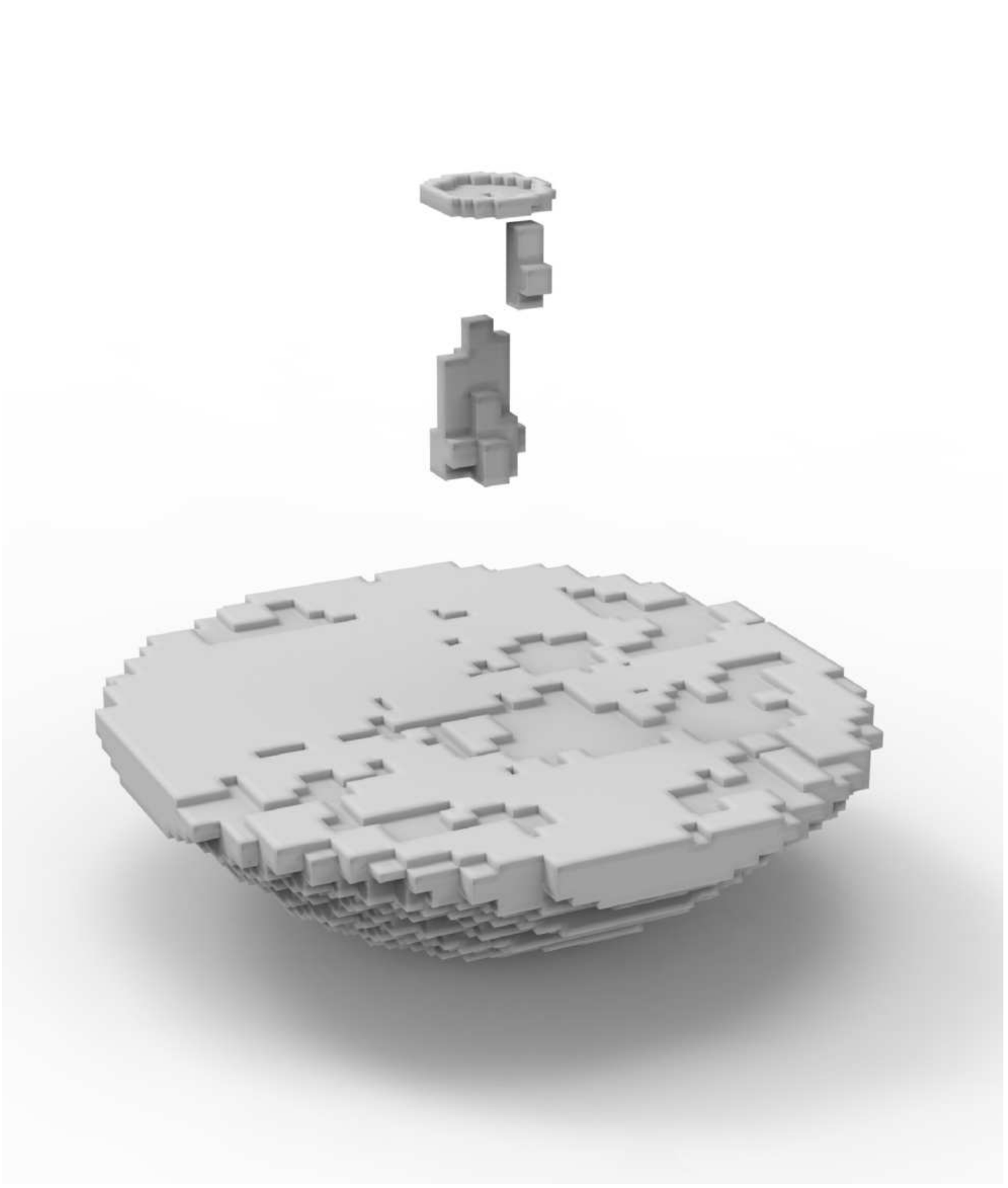}\\
    \vspace{2mm}
    \includegraphics[width=0.135\linewidth]{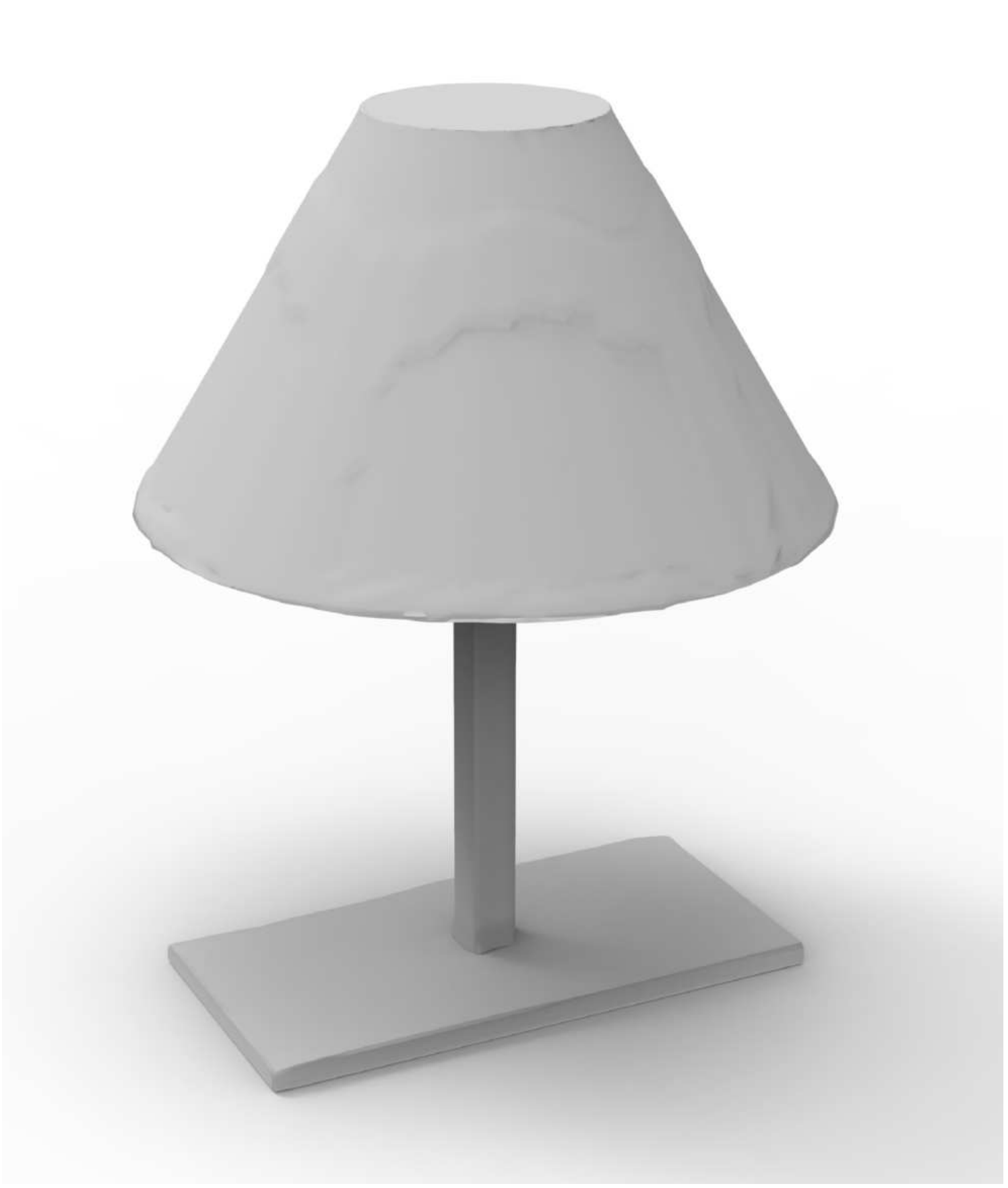}
    \includegraphics[width=0.135\linewidth]{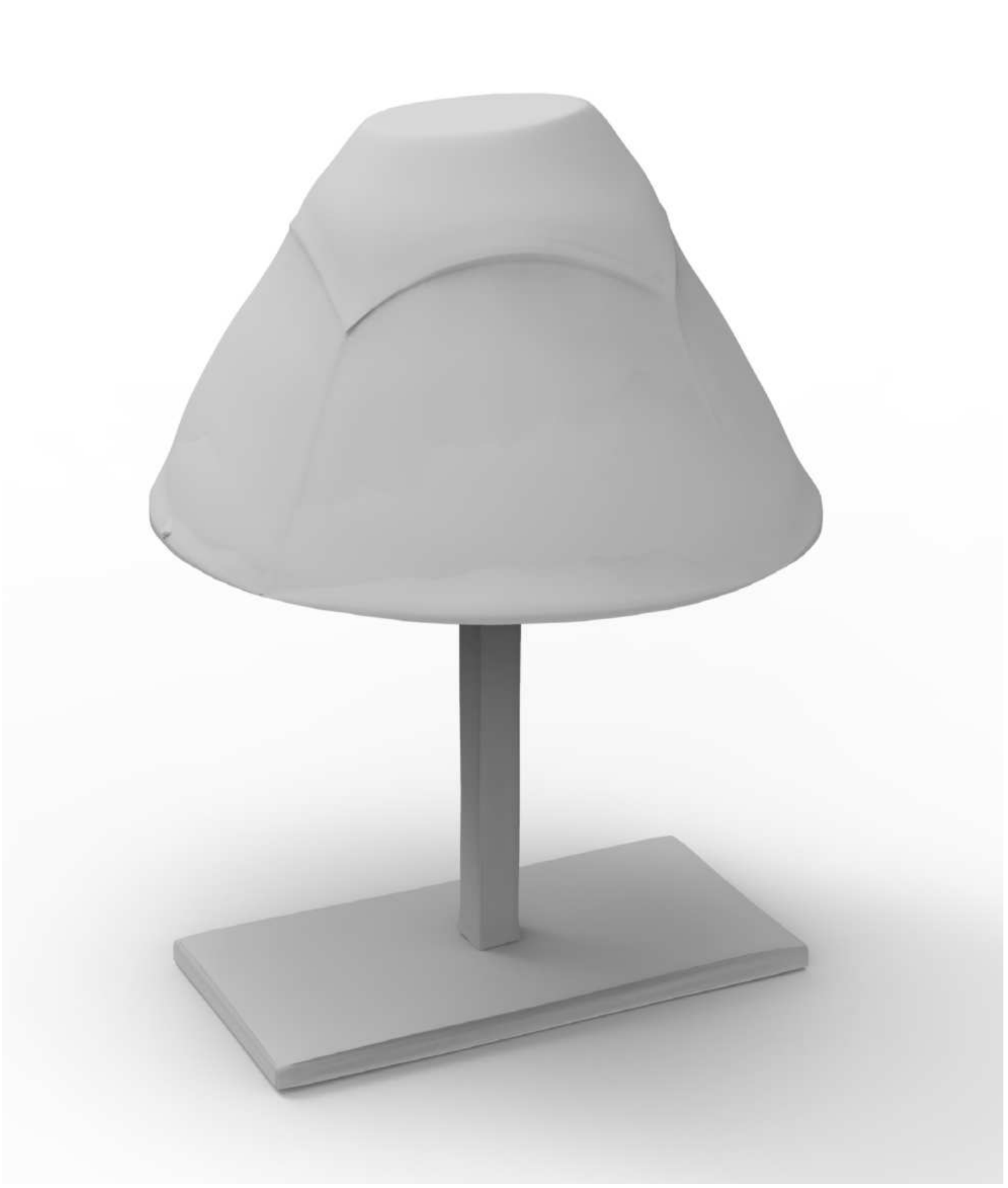}
    \includegraphics[width=0.135\linewidth]{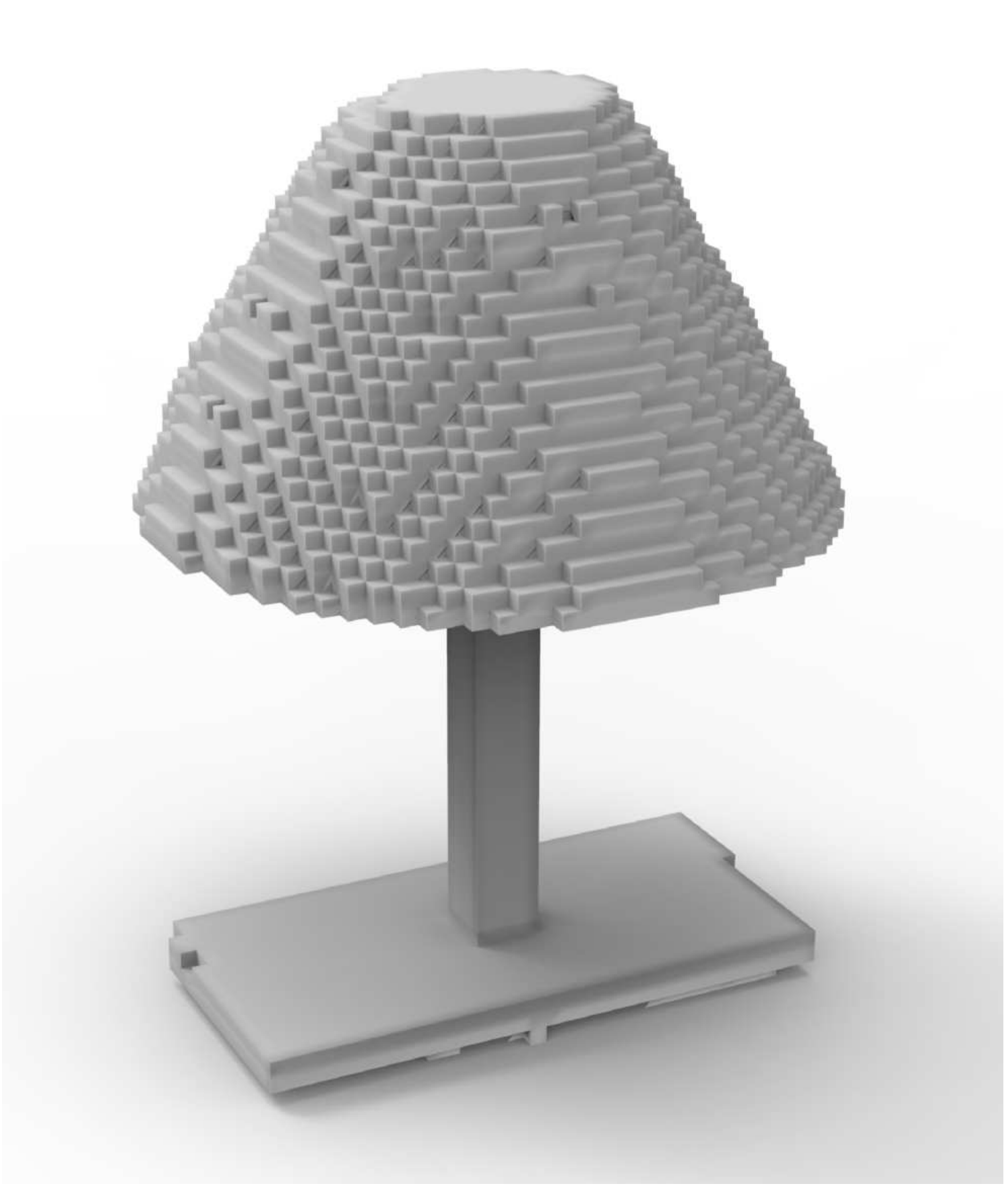}
    \includegraphics[width=0.135\linewidth]{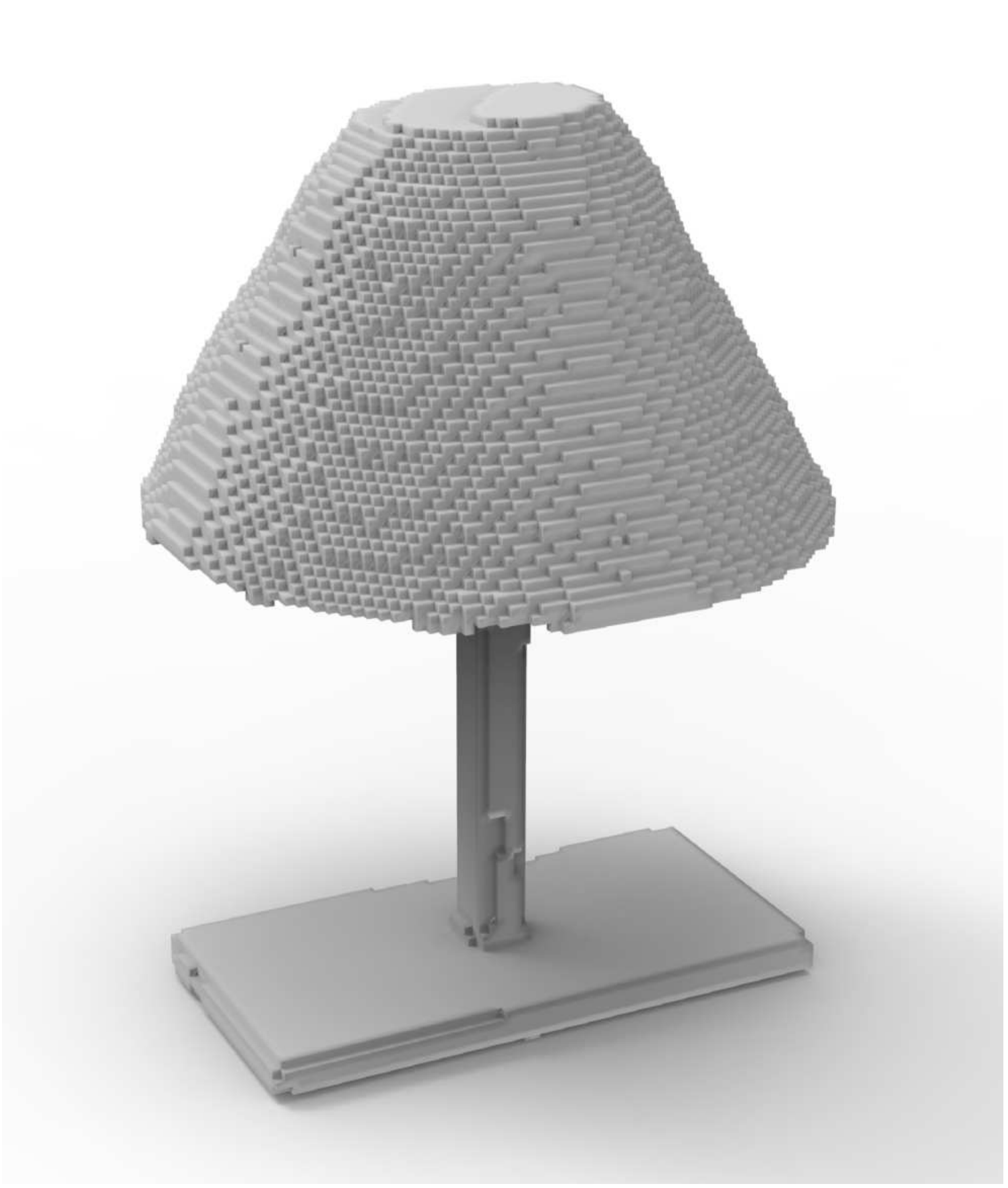}
    \includegraphics[width=0.135\linewidth]{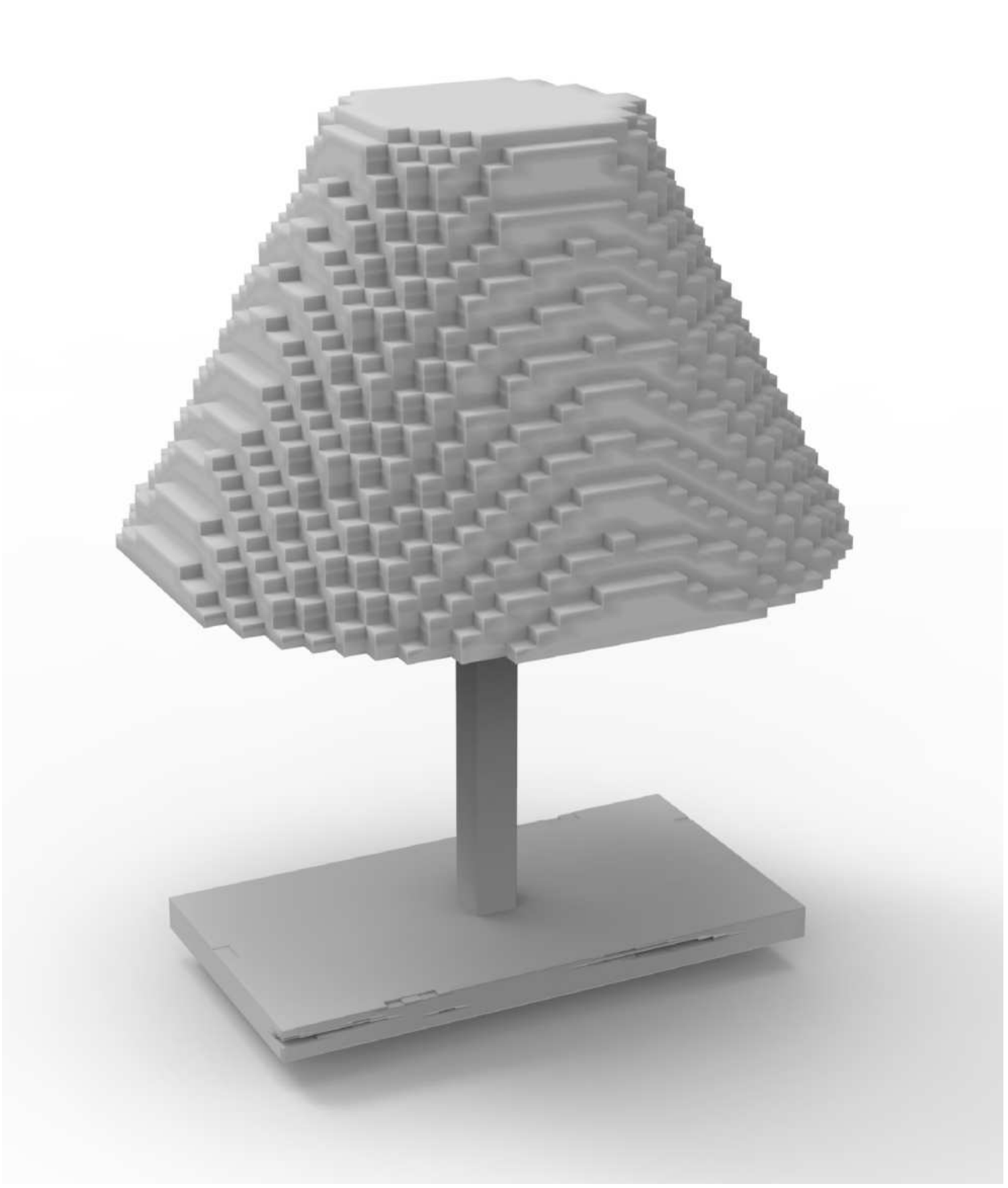}
    \includegraphics[width=0.135\linewidth]{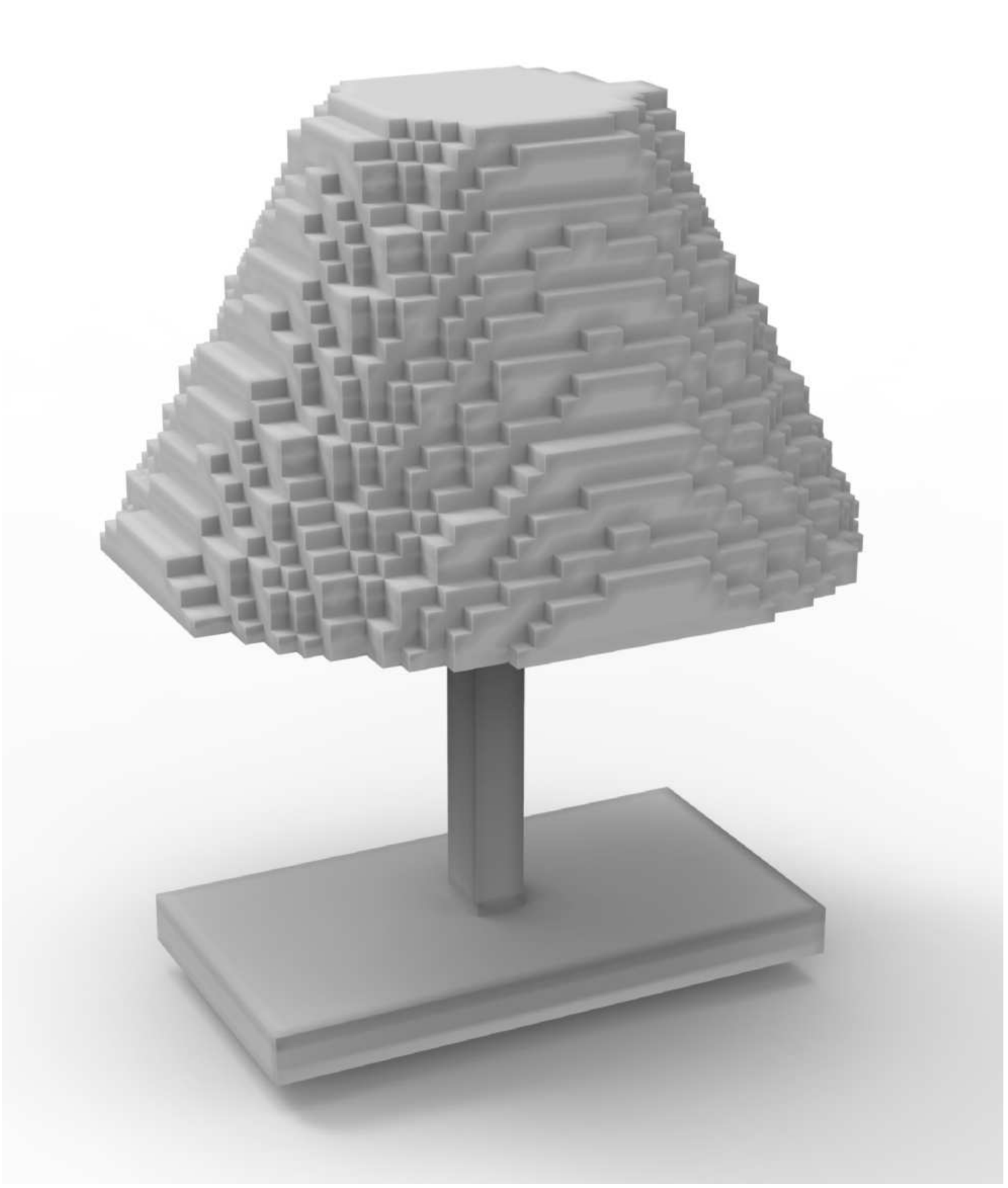}
    \includegraphics[width=0.135\linewidth]{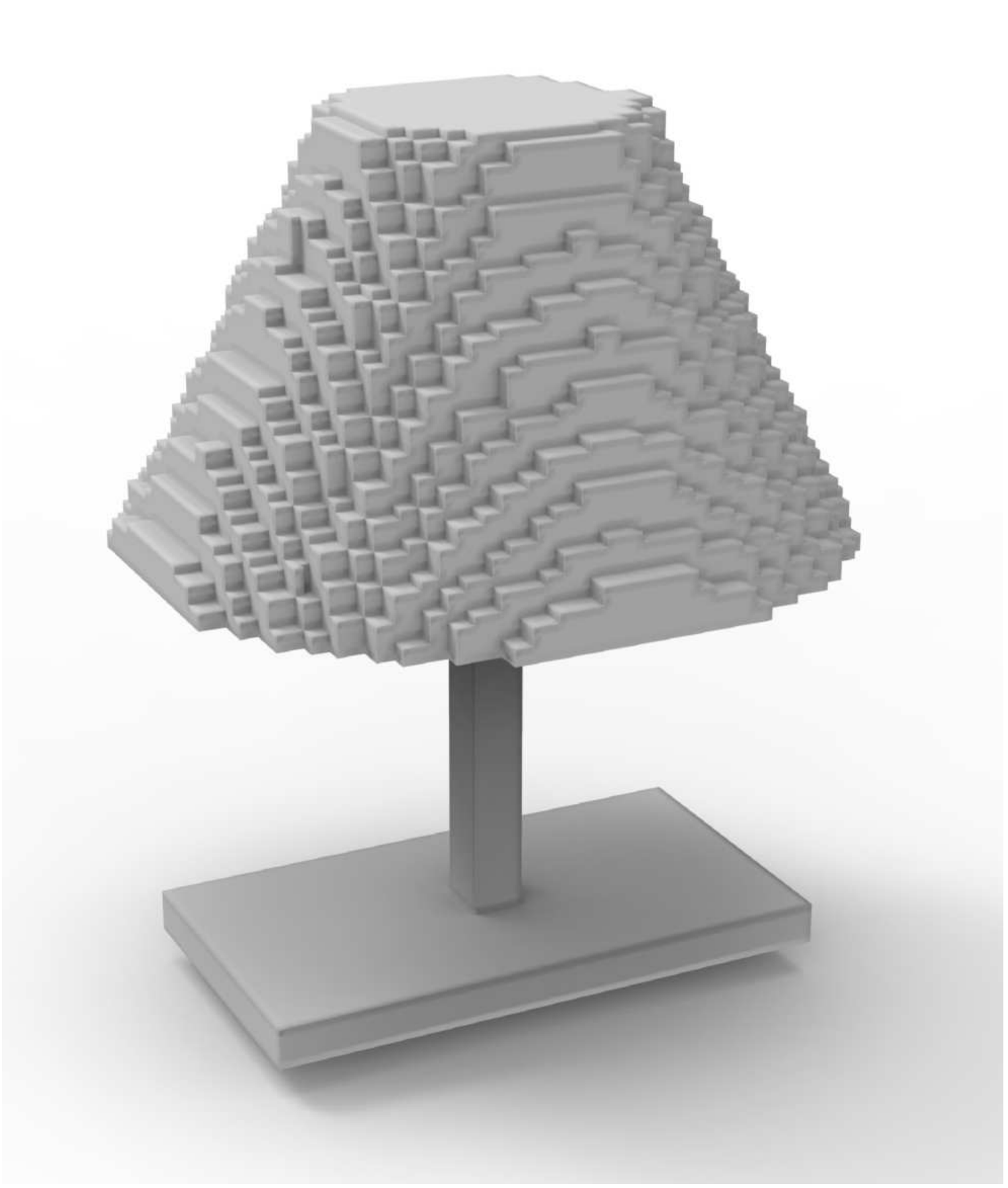}\\
    \vspace{2mm}
    \subfigure[Input Shape]{
    \includegraphics[width=0.135\linewidth]{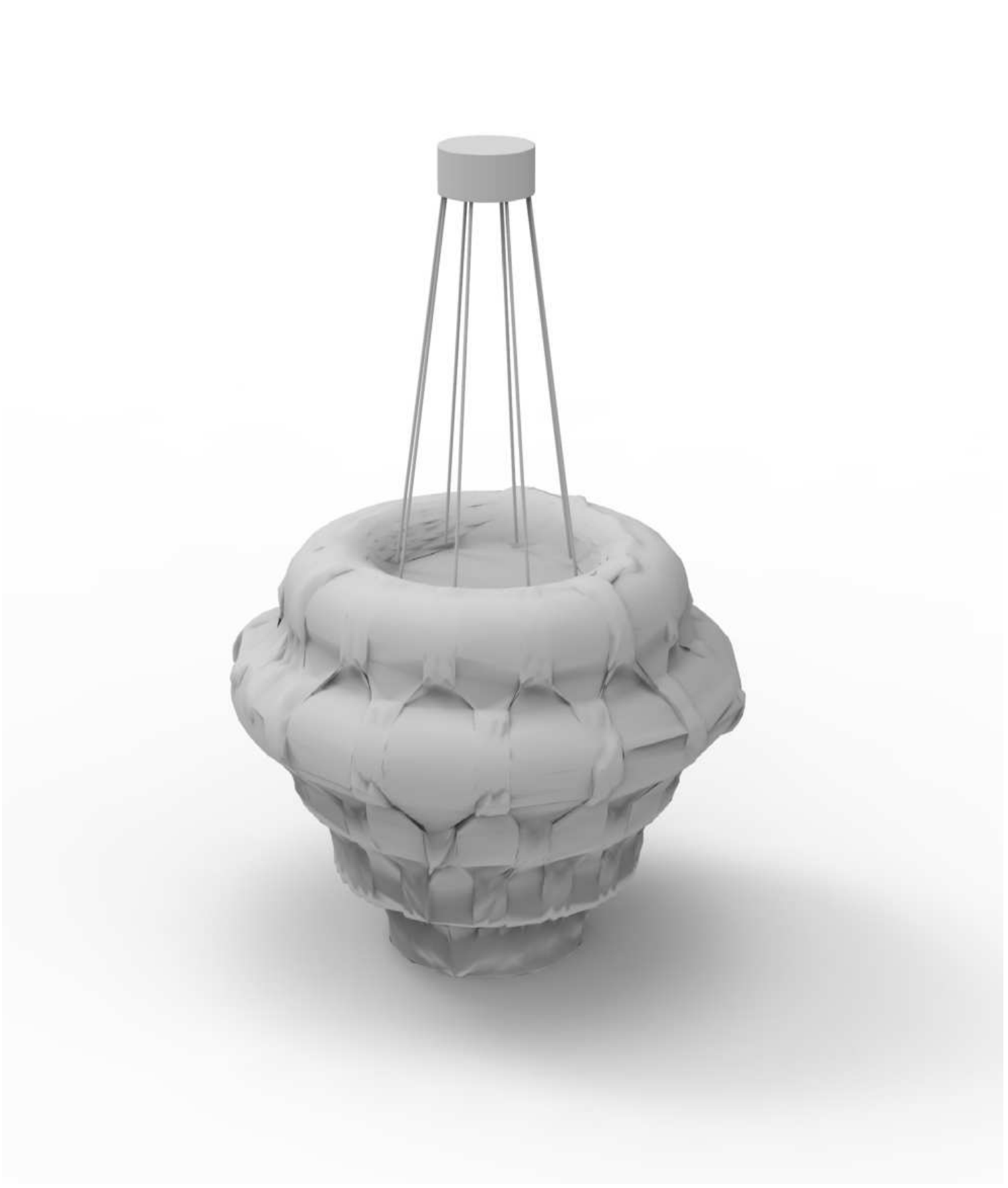}}\hspace{-1mm}
    \subfigure[Ours]{
    \includegraphics[width=0.135\linewidth]{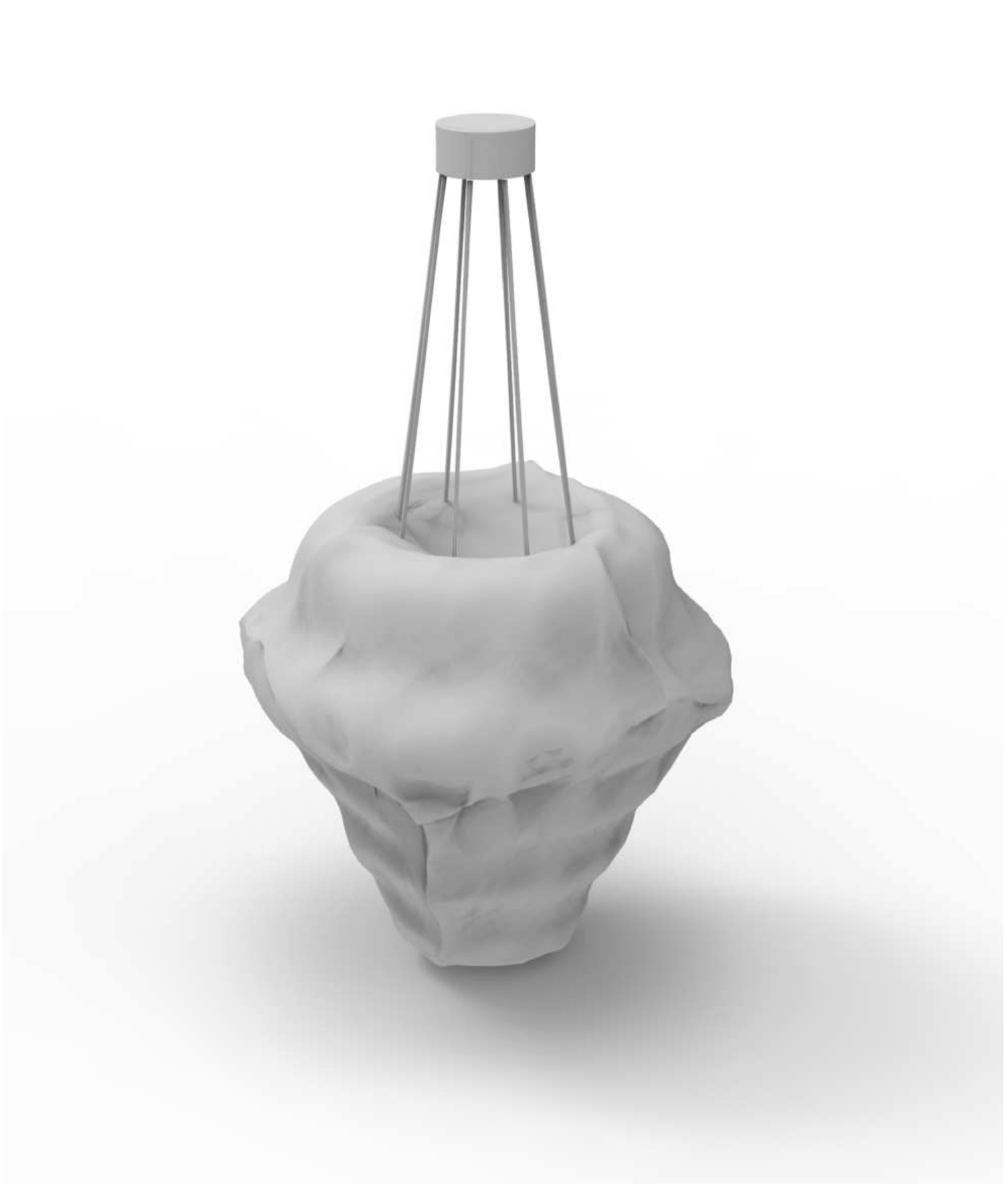}}\hspace{-1mm}
    \subfigure[Ours-$64^3$]{
    \includegraphics[width=0.135\linewidth]{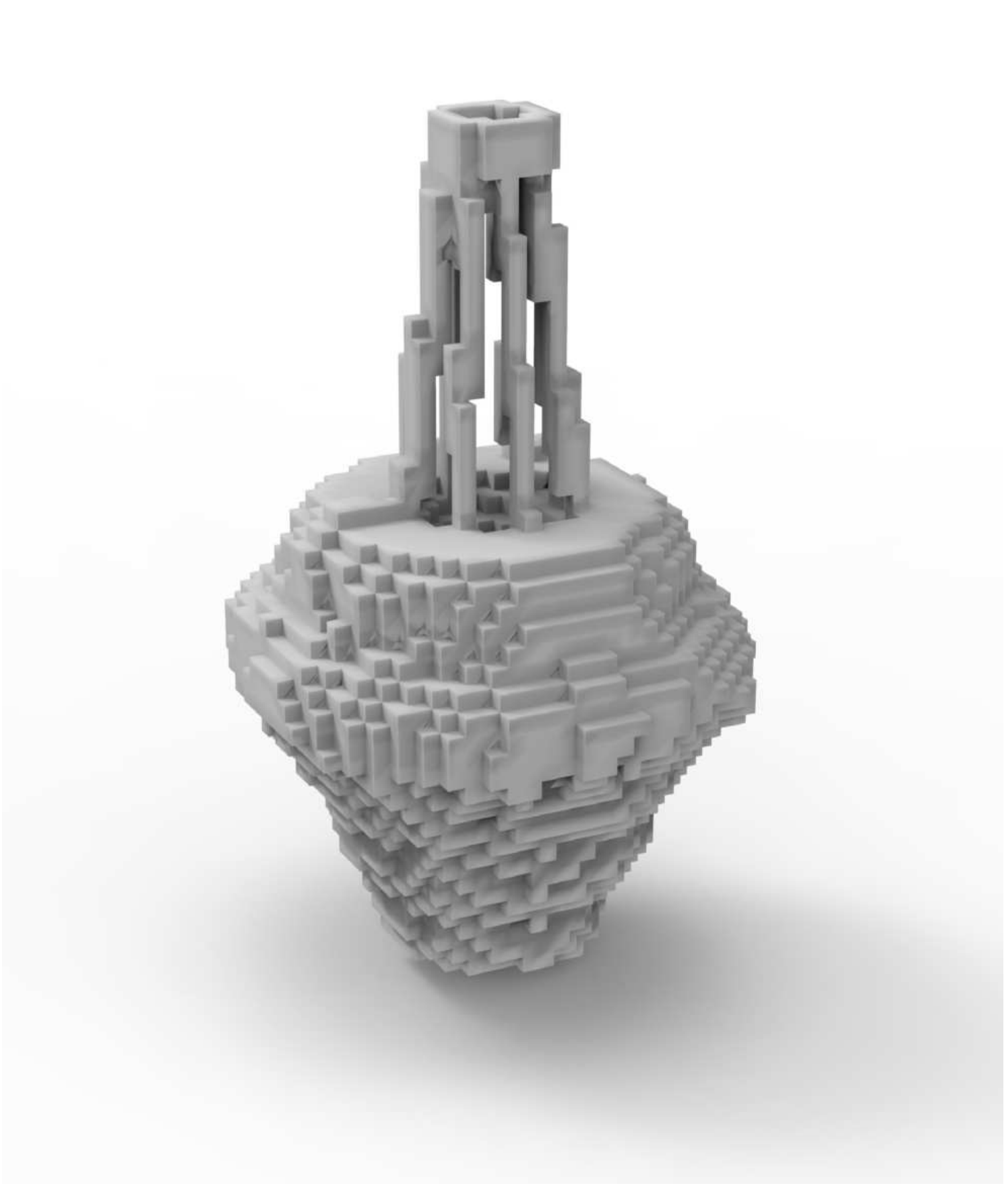}}\hspace{-1mm}
    \subfigure[Ours-$128^3$]{
    \includegraphics[width=0.135\linewidth]{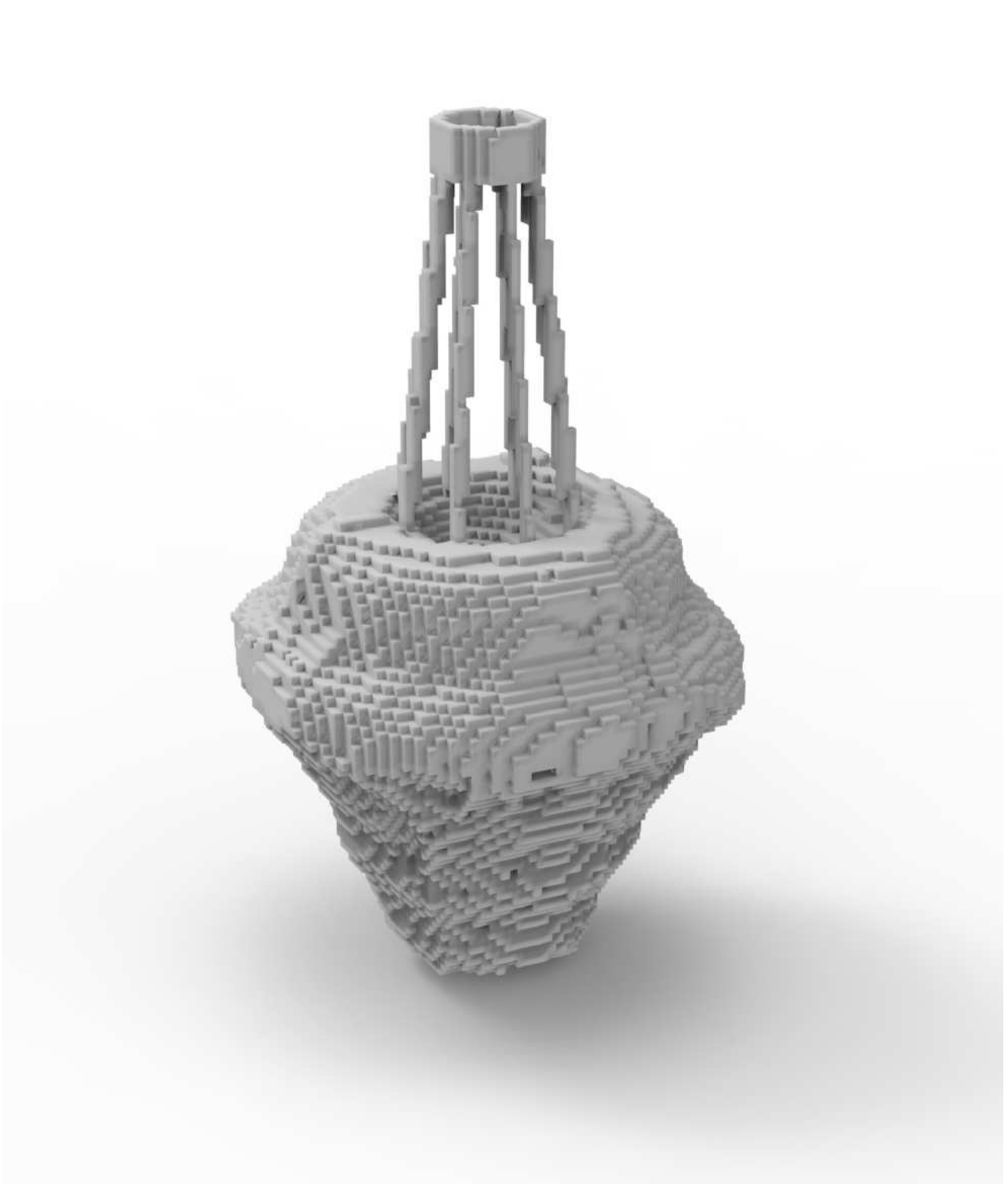}}\hspace{-1mm}
    \subfigure[SAGNet]{
    \includegraphics[width=0.135\linewidth]{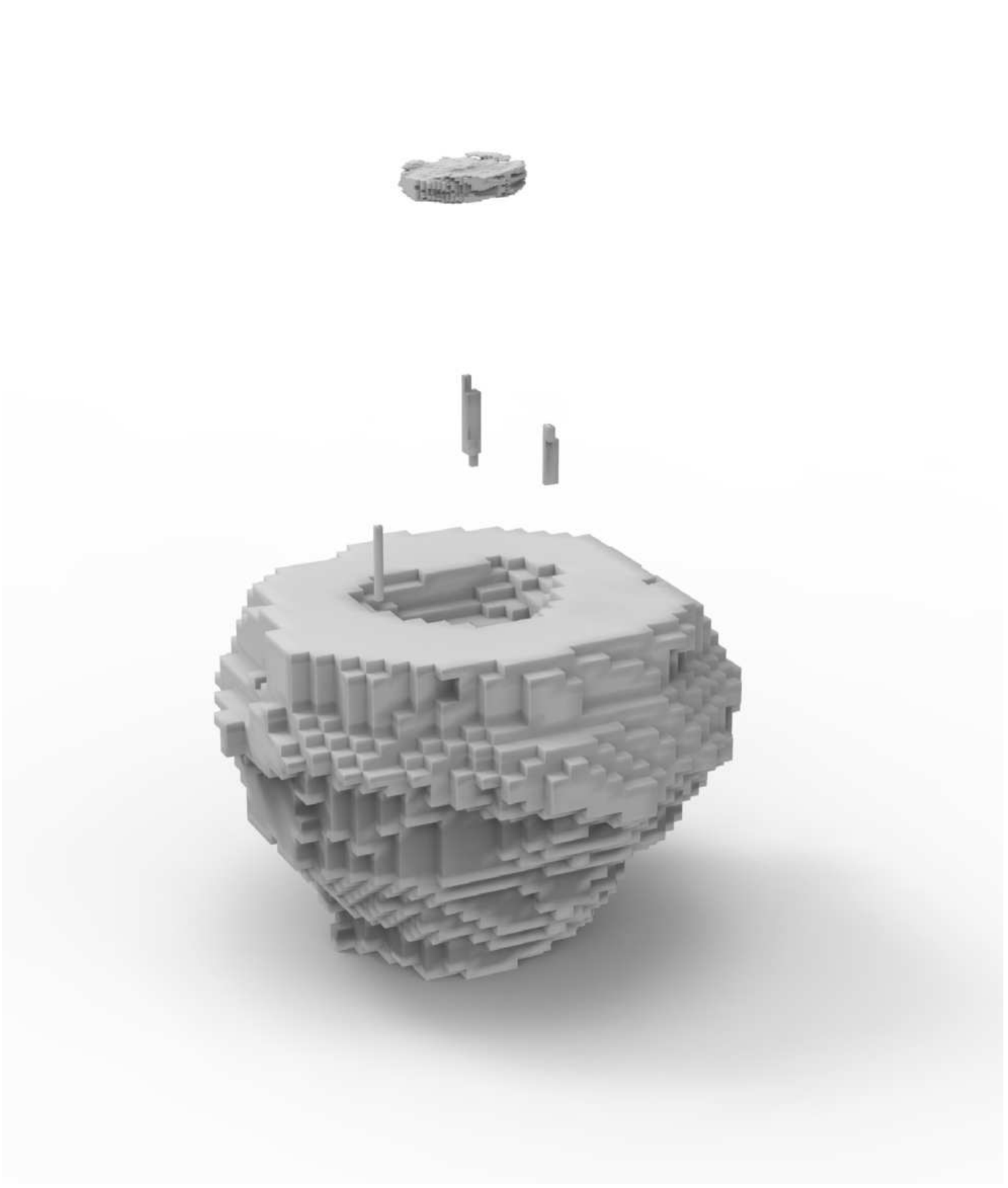}}\hspace{-1mm}
    \subfigure[SAG-$64^3$]{
    \includegraphics[width=0.135\linewidth]{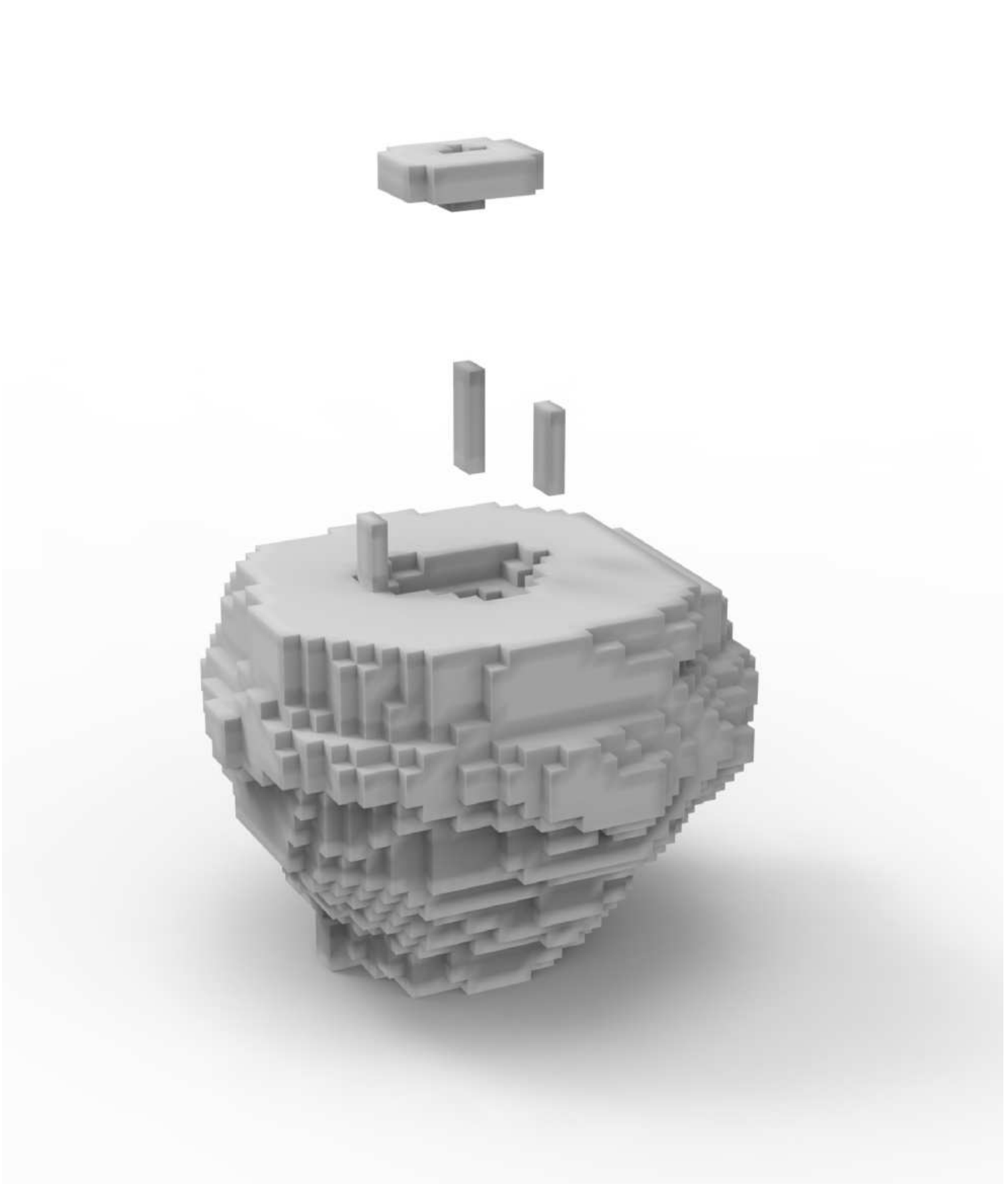}}\hspace{-1mm}
    \subfigure[SAG-$128^3$]{
    \includegraphics[width=0.135\linewidth]{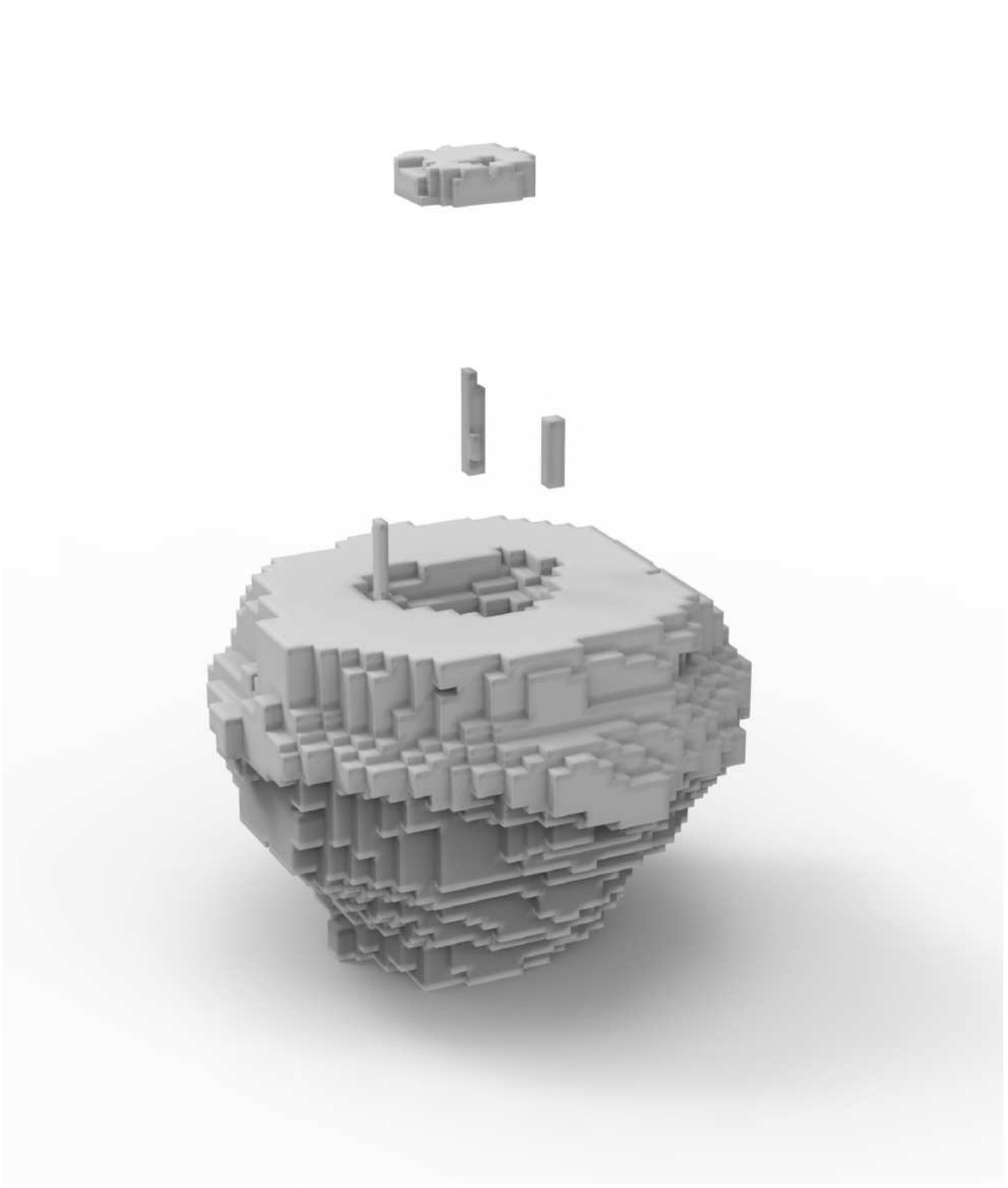}}
    \caption{\textbf{Comparison with SAG-Net.} \yjr{We evaluate the shape reconstruction on two overlapping shape categories (Lamp and Chair). 
    For a fair comparison, we convert the results of both methods into voxel maps with the same resolutions ($64^3$ and $128^3$). 
    For each category, we display three examples with different voxel resolutions. 
    (b) and (e) are the original outputs of the DSG-Net and SAGNet. (c), (d), (f) and (g) are the voxelized shapes with $64^3$ and $128^3$ resolutions. 
    We can clearly see that our results achieve better performance than SAGNet visually, such as those parts with thin structures (horizontal stretcher of chairs and chains of lamps).}}
    \label{fig:comp-sagnet}
\end{figure*}

\begin{itemize}
	
	\item \yj{Section~\ref{sec:comp_sag} provides quantitative and qualitative comparisons to SAG-Net~\cite{wu2019sagnet} for shape reconstruction and generation tasks;}

	\yjr{\item Section~\ref{sec:abla1} provides four ablation studies for some key components of our network architecture, including Removing Part Relationships and Edges, Training Separate Part Geometry VAEs, Training Strategy (Cascaded \vs End-to-end), \yjrr{StructureNet (SN)~\cite{mo2019structurenet}+Mesh \vs Ours,} and Removing Cycled Disentanglement and Losses ($\mathcal{L}_{struct}$ and $\mathcal{L}_{geo}$);}

	\item \yj{Section~\ref{sec:data} provides more details on the dataset creation;}
	
	\item \yj{Section~\ref{sec:impl} provides more training and implementation details for our network;}

	\item \yjr{Section~\ref{sec:more_recon},  Section~\ref{sec:more_gen},
	Section~\ref{sec:more_inter},
	Section~\ref{sec:more_disinter}, and Section~\ref{sec:more_disgen} provide more visualization results on shape reconstruction, generation and interpolation, especially for our newly proposed disentangled shape interpolation and generation tasks.}
\end{itemize}

\begin{table}[b]
  \centering
  \caption{\yjr{Shape reconstruction quantitative comparison to SAG-Net. We use two geometry metrics (CD and EMD) and one structure metric (HierInsSeg). DSG-Net achieves better geometry/structure reconstruction performance compared to SAG-Net.}}
    \begin{tabular}{ccccc}
    \toprule[1pt]
    \multirow{2}[4]{*}{DataSet} & \multirow{2}[4]{*}{Method} & \multicolumn{2}{c}{Geometry Metrics} & Structure Metrics \\
    \cmidrule{3-5}          &       & CD{\scriptsize$\times 10^{-3}$}$\downarrow$   & EMD{\scriptsize$\times 10^{-2}$}$\downarrow$  & HierInsSeg(HIS) $\downarrow$ \\
    \midrule
    \midrule
    \multirow{3}[2]{*}{Chair} & SAG-Net & 8.62 & 6.87 & 0.55 \\
          & {Ours}  & {\textbf{1.98}} & {\textbf{0.73}} & {\textbf{0.39}} \\
          & GT    &       &       & \underline{0.32} \\
    \midrule
    \midrule
    \multirow{3}[2]{*}{Lamp} & SAG-Net & 18.29 & 6.49 & 0.83 \\
          & {Ours}  & {\textbf{7.15}} & {\textbf{1.63}} & {\textbf{0.61}} \\
          & GT    &       &       & \underline{0.55} \\
    \bottomrule[1pt]
    \end{tabular}%
  \label{tab:comp_sag_recon}%
\end{table}%

\begin{table}[t]
  \centering
  \small
  \caption{\yjr{Shape generation quantitative comparison to SAG-Net. We report the coverage and quality scores relative to DSG-Net (\ie all the reported scores are divided by the corresponding DSG-Net scores for normalization) under the geometry metric (Chamfer-Distance) and the structure metric (HierInsSeg), compared to SAG-Net as a baseline. 
  \yj{Meanwhile, we also adopt Frech\'{e}t Point-cloud Distance~\cite{shu20193d}  to evaluate the variety, coverage, and quality of generated shapes, which can be seen as an extension of \textit{Inception Score}~\cite{salimans2016improved} to point clouds.
  We follow PT2PC~\cite{mo2020pt2pc} to calculate the Frech\'{e}t Point-cloud Distance (FPD)~\cite{shu20193d} on point clouds, which is the same as SAG-Net. The lower FPD score, the better.
  } 
  We observe that DSG-Net achieves better performance across all metrics.}}
    \begin{tabular}{cccccc}
    \toprule[1pt]
    \multirow{2}[4]{*}{Method} & \multicolumn{2}{c}{Geometry} & \multicolumn{2}{c}{Structure} & \multicolumn{1}{c}{\multirow{2}[4]{*}{FPD $\downarrow$}\vspace{1mm}} \\
    \cmidrule{2-5}          & \multicolumn{1}{c}{Coverage $\uparrow$} & \multicolumn{1}{c}{Quality $\uparrow$} & \multicolumn{1}{c}{Coverage $\uparrow$} & \multicolumn{1}{c}{Quality $\uparrow$} & \\
    \midrule
    \midrule
    SAG-Net & 0.60 & 0.30 & 0.39 & 0.45 & 18.34 \\
    Ours  & \textbf{1.00} & \textbf{1.00} & \textbf{1.00} & \textbf{1.00} & \textbf{9.73} \\
    \bottomrule[1pt]
    \end{tabular}%
  \label{tab:comp_sag_gen}%
\end{table}%

\section{Comparison to SAG-Net~\cite{wu2019sagnet}}\label{sec:comp_sag}
In the pioneering work SAG-Net~\cite{wu2019sagnet}, geometry and structure features are jointly encoded in a single latent code, which is different from our disentangled geometry and structure representations for 3D shapes.
Besides, instead of generating mesh geometry and exploiting hierarchical shape structure, SAG-Net geometry is represented with a voxel-based representation and the graph structure is represented by a fully connected graph. 

Our DSG-NET is fundamentally different in that geometry and structure are encoded in two \textit{disentangled} latent codes and we model shape parts in a \textit{hierarchical} manner. 
The hierarchy of encoded geometry guided by the structure is the key in our work that achieves disentanglement while ensuring structure/geometry compatibility, which does not appear in SAG-NET. 
This novel design enables DSG-NET to disentangle geometry and structure while keeping the two informed of each other, and thus synthesizes 3D mesh models with complex structure and compatible fine geometry, advancing the state-of-the-art in neural shape representations.

We compare to SAG-Net on the shape reconstruction and generation tasks.
Since SAG-Net requires different shape part segmentation from our data, we only compare to SAG-Net on two overlapping categories (\ie chairs and lamps). 
We compare our shape reconstruction and generation performance to SAG-Net on the chair and lamp categories in Tables~\ref{tab:comp_sag_recon} and \ref{tab:comp_sag_gen}. 
\yj{For shape reconstruction, we evaluate the performance on three metrics, including two geometric metrics (CD and EMD) and one structure metric (HIS). 
For shape generation, we also evaluate the performance using five scores, including coverage/quality of geometry, structure and Frech\'{e}t Point-cloud Distance (FPD)~\cite{shu20193d}. 
Note that, the FPD score measures the variety, coverage and quality of generated shapes. Following SAG-Net, we convert the voxel maps/meshes to point clouds. 
Then, we adopt the same calculation of Frech\'{e}t Point-cloud Distance~\cite{shu20193d} as used in PT2PC~\cite{mo2020pt2pc}, where a PointNet is trained for 3D shape classification on ModelNet40.
The FPD mainly computes distribution distance between the real point cloud shape features and fake (generated) point cloud shape features extracted by the PointNet classifier.
It is defined as:
\begin{equation}
\mathrm{FPD}=\left|\mu_{r}-\mu_{f}\right|^{2}+\operatorname{tr}\left(\Sigma_{r}+\Sigma_{f}-2\left(\Sigma_{r} \Sigma_{f}\right)^{1 / 2}\right)
\end{equation}
where $\mu$ and $\Sigma$ are the mean vector and the covariance matrix of PointNet features of real data distribution $r$ and fake (generated) data distribution $f$. $Tr(\cdot)$ is the matrix trace.
}
We observe that DSG-Net achieves better performance than SAG-Net under all metrics for the shape reconstruction and generation tasks.

For the visual comparison with SAG-Net, 
we voxelize the results of ours and SAG-Net with same resolutions ($64^3$ and $128^3$). 
We show the evaluation on shape reconstruction in Figure.~\ref{fig:comp-sagnet}. 
From the results, we can see that our results can achieve high performance and maintain the detailed geometry, such as some part with detailed or thin part structures (\eg horizontal stretcher of chairs and chains of lamps).

\section{Ablation Studies}\label{sec:abla1}

\begin{figure*}[t]
    \centering
    \subfigure[Input Mesh]{
    \begin{minipage}[b]{0.23\linewidth}
    {\includegraphics[width=0.48\linewidth]{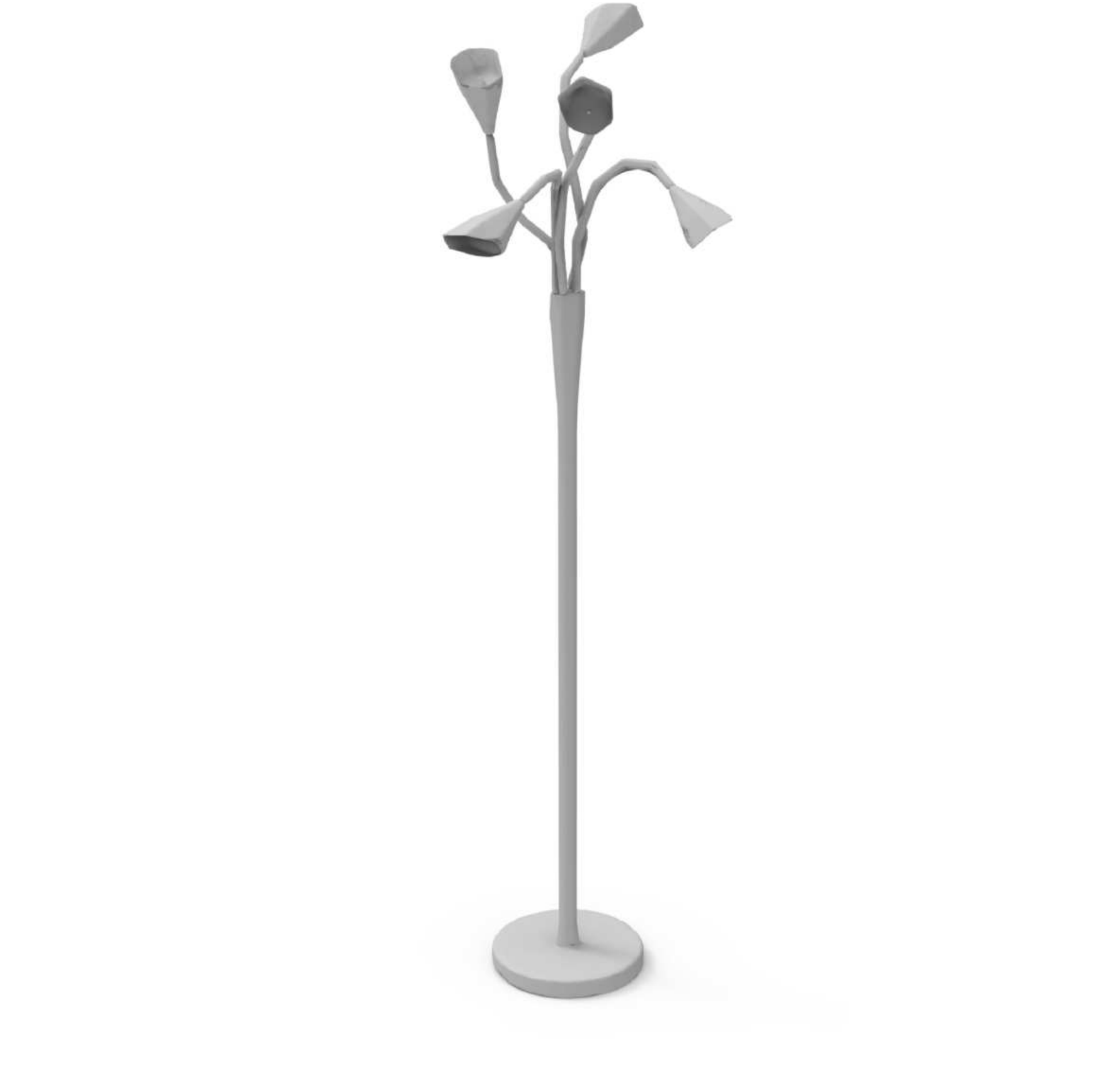}
    \includegraphics[width=0.48\linewidth]{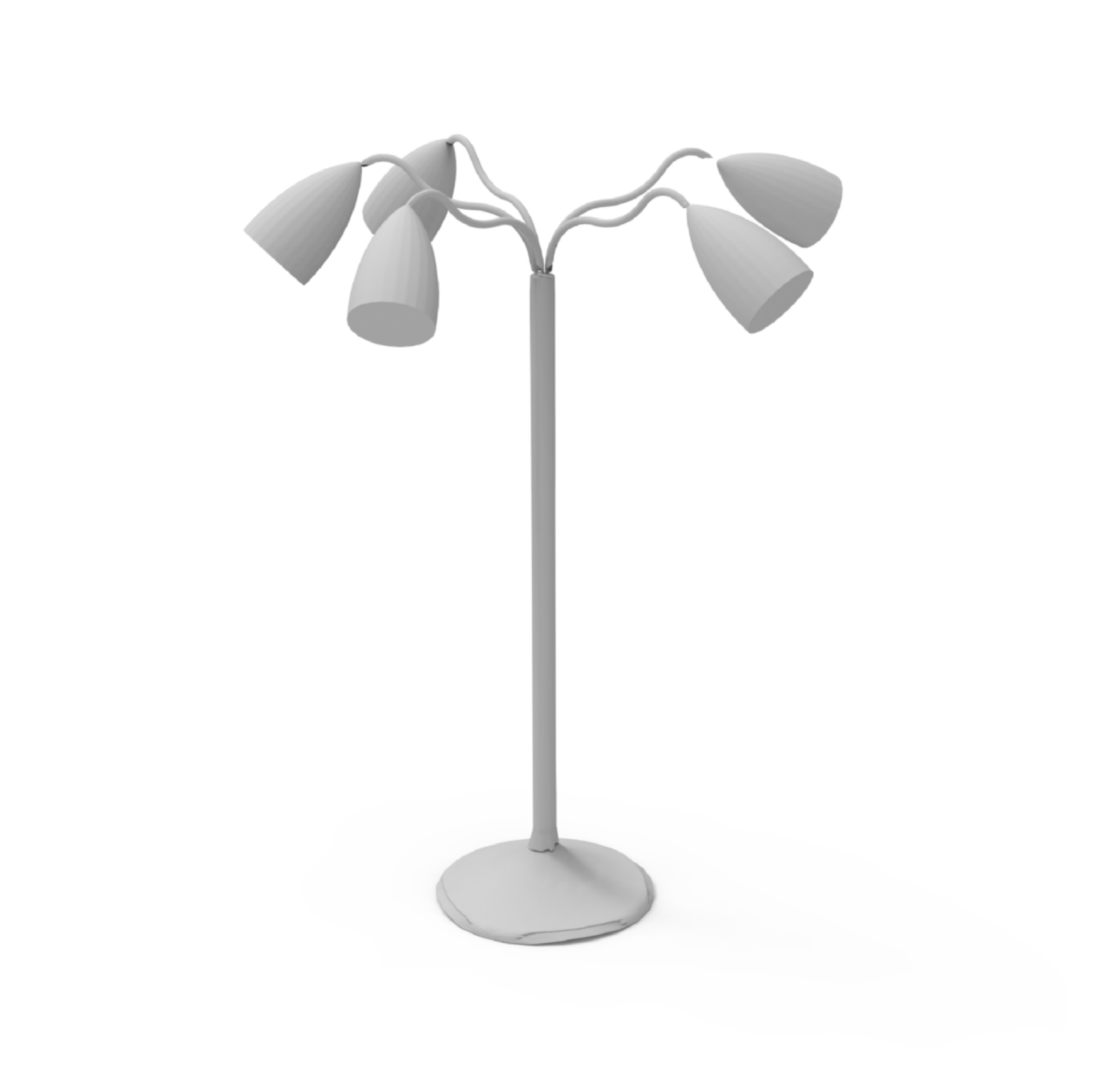}\\
    \includegraphics[width=0.48\linewidth]{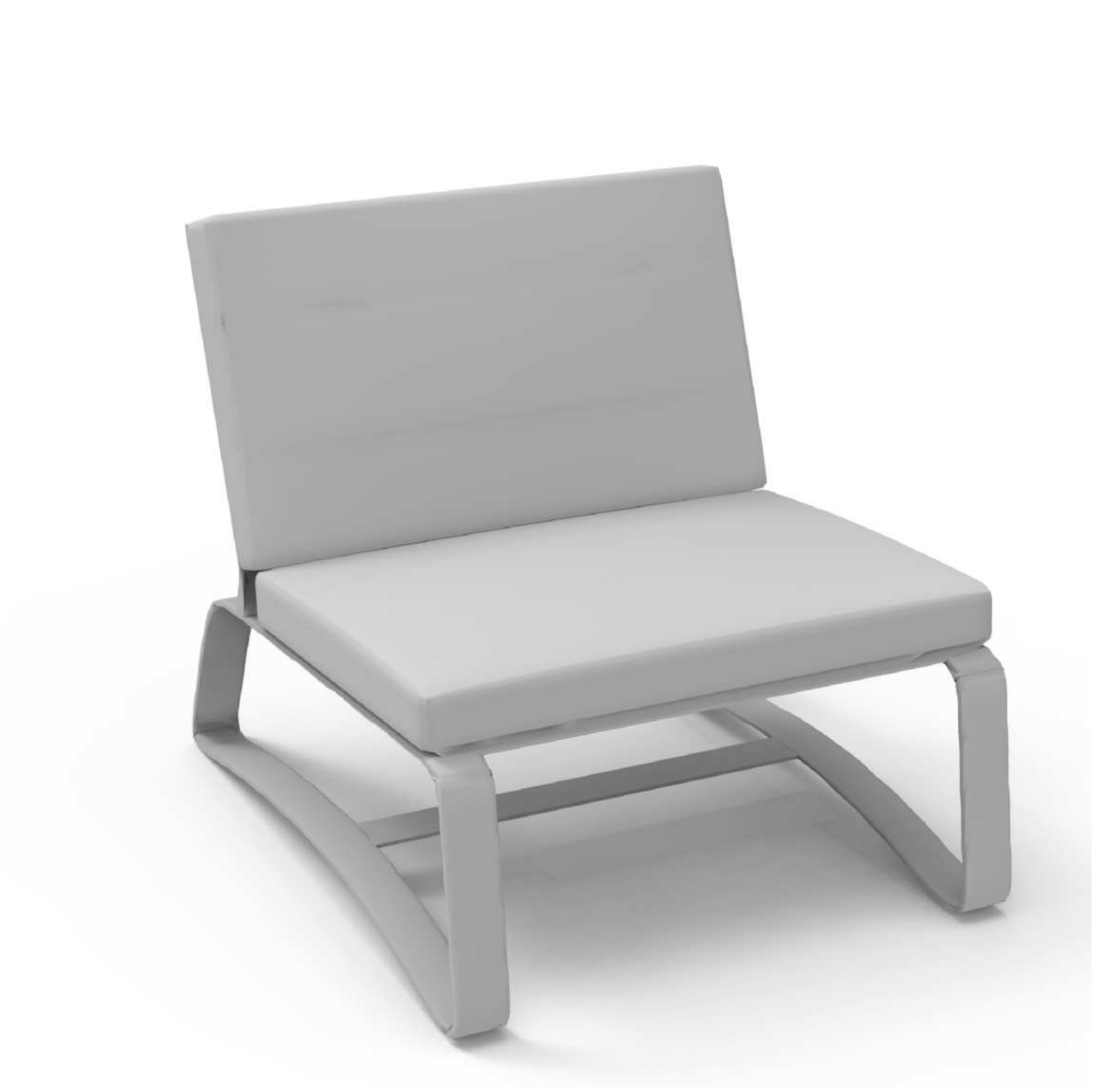}
    \includegraphics[width=0.48\linewidth]{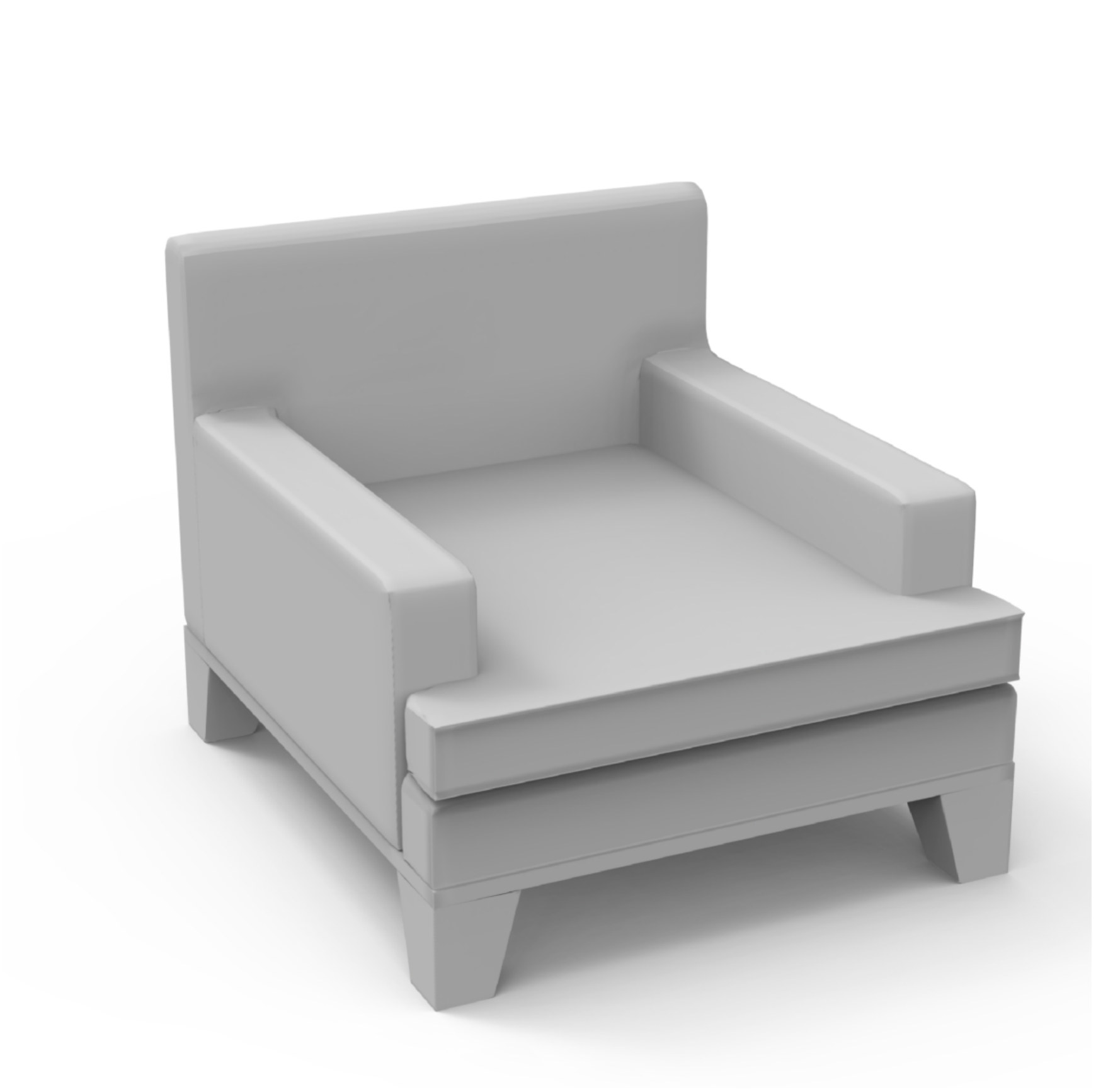}}
    \end{minipage}}
    \subfigure[SN+Mesh]{
    \begin{minipage}[b]{0.23\linewidth}
    {\includegraphics[width=0.48\linewidth]{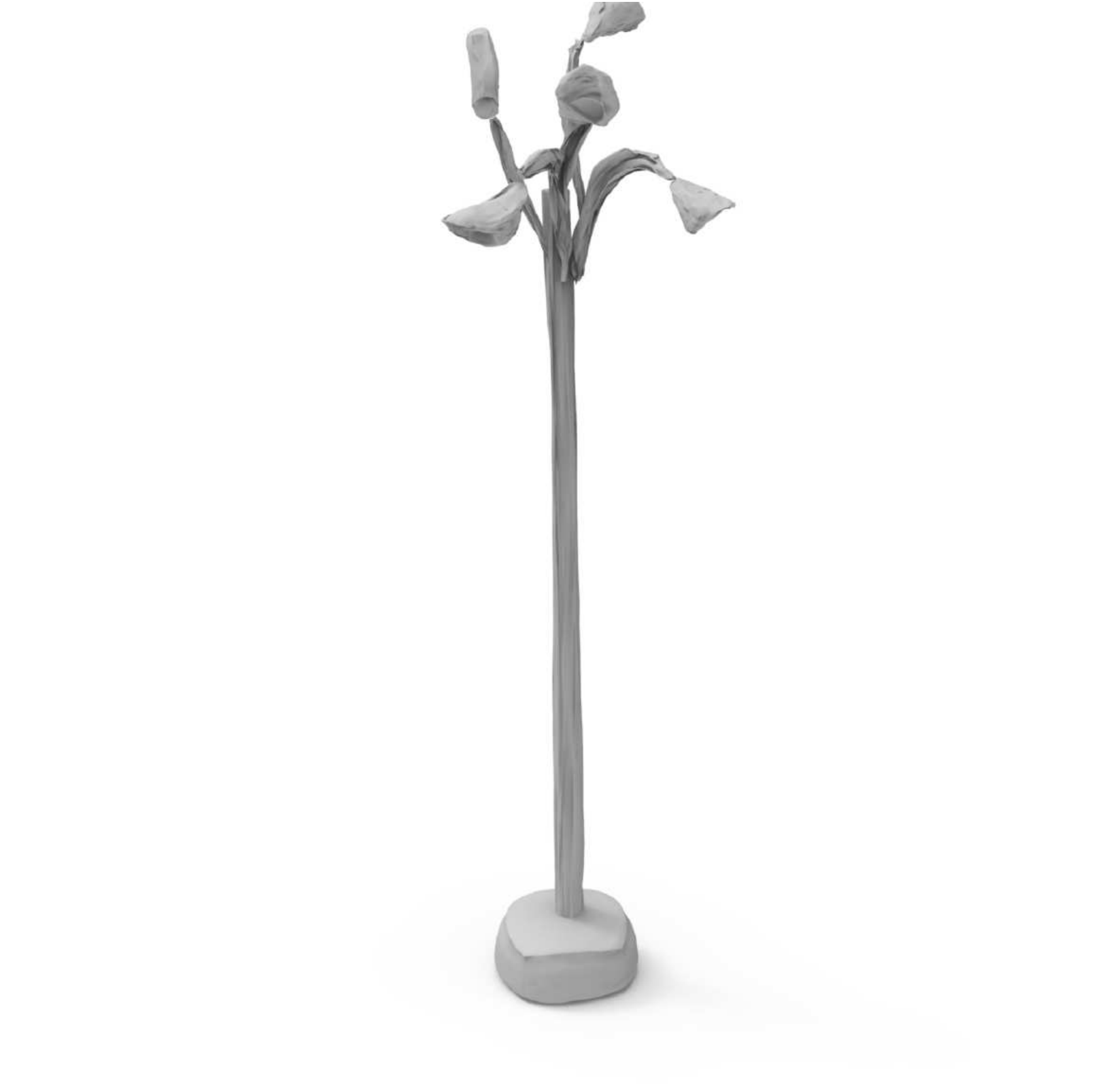}
    \includegraphics[width=0.48\linewidth]{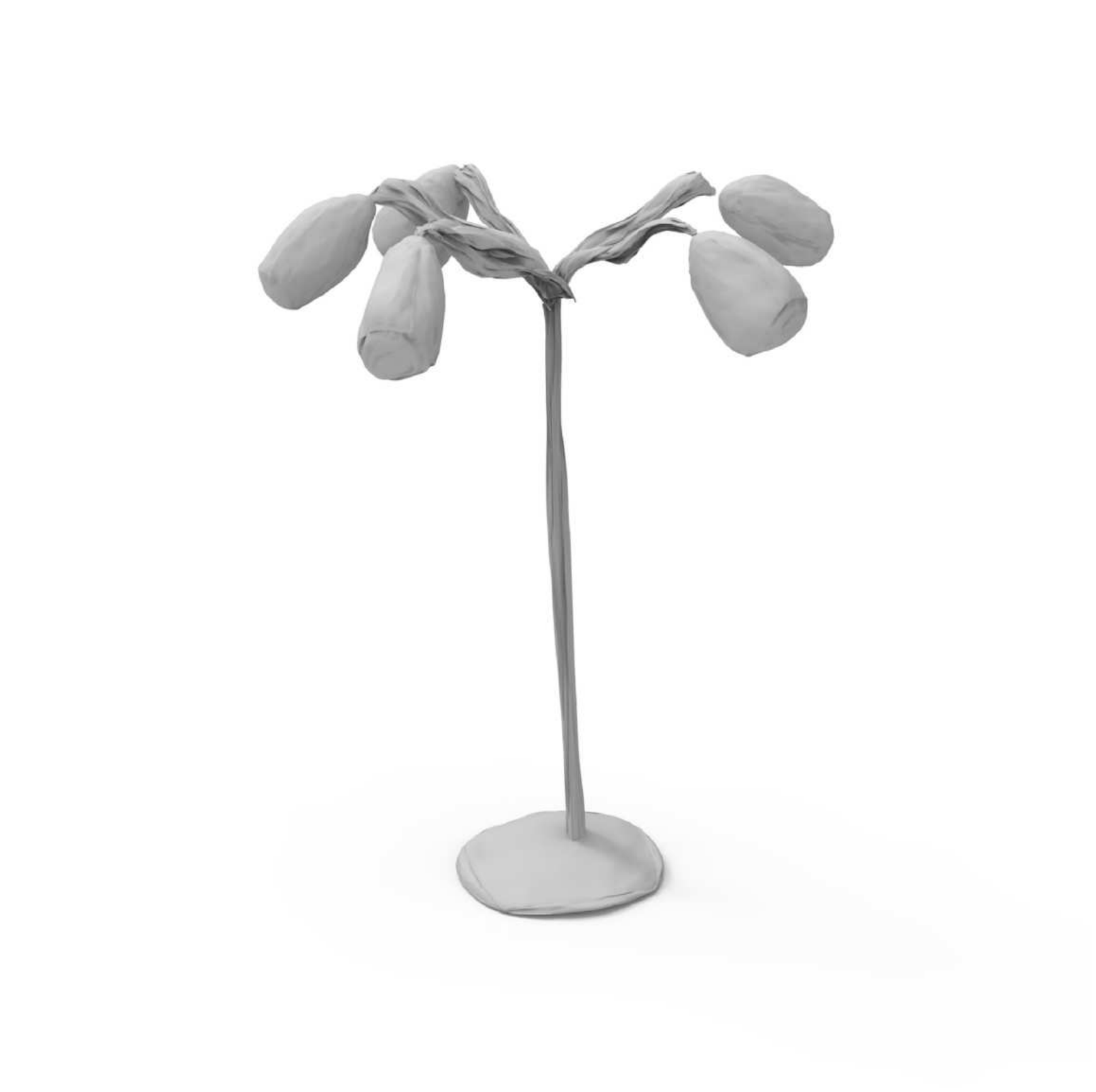}\\
    \includegraphics[width=0.48\linewidth]{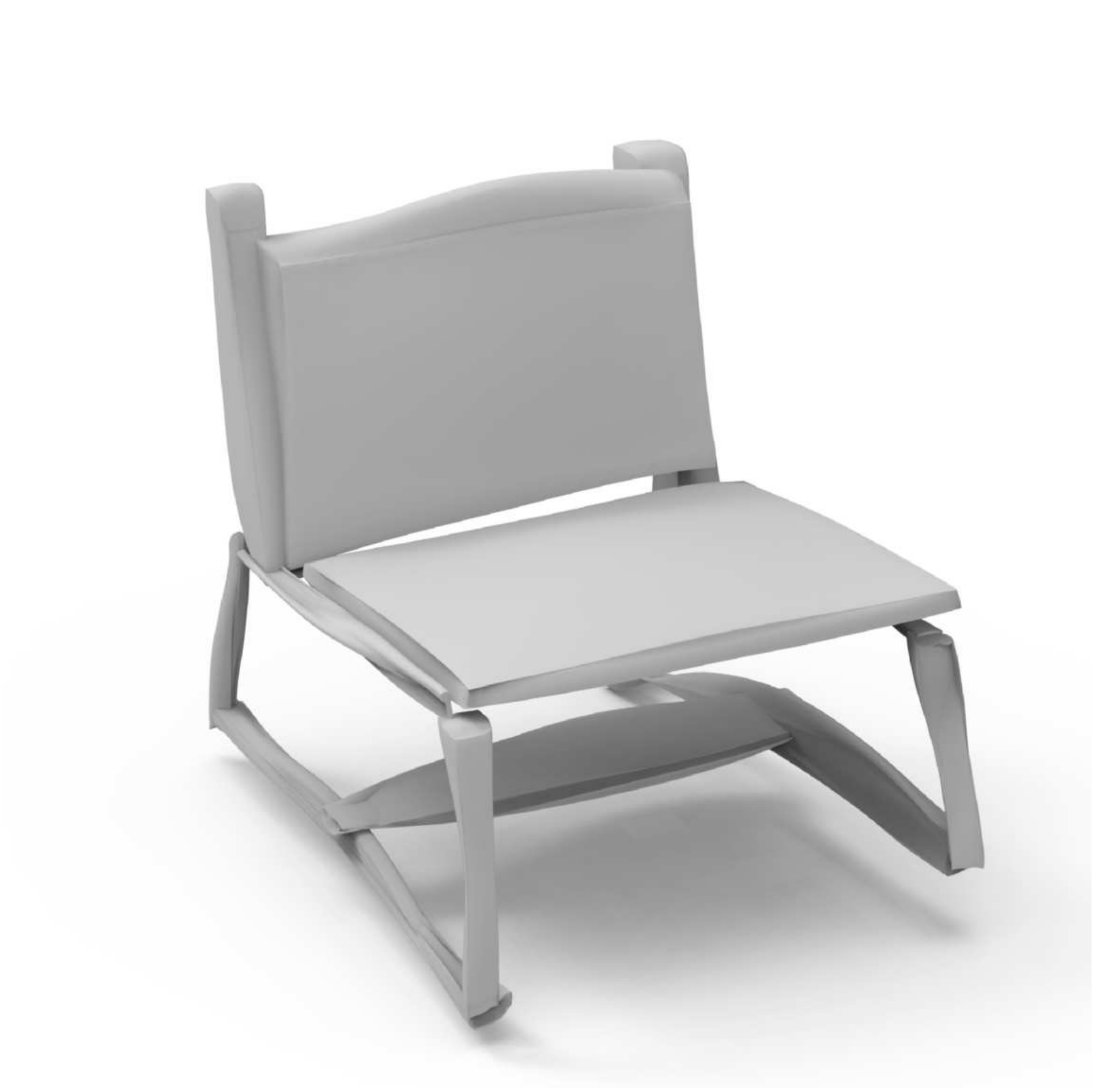}
    \includegraphics[width=0.48\linewidth]{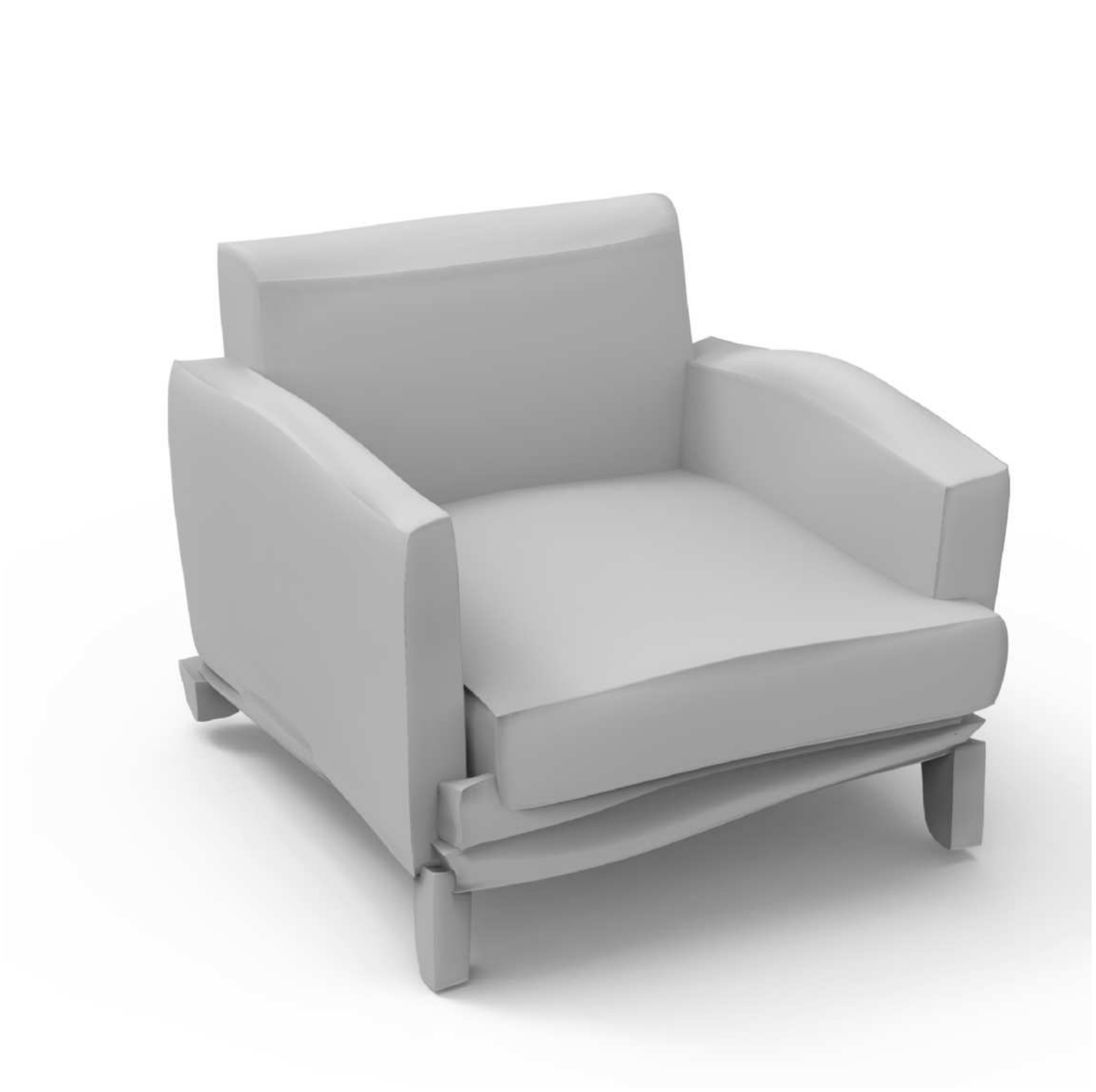}}
    \end{minipage}}
    \subfigure[Ours (w/o CycD)]{
    \begin{minipage}[b]{0.23\linewidth}
    {\includegraphics[width=0.48\linewidth]{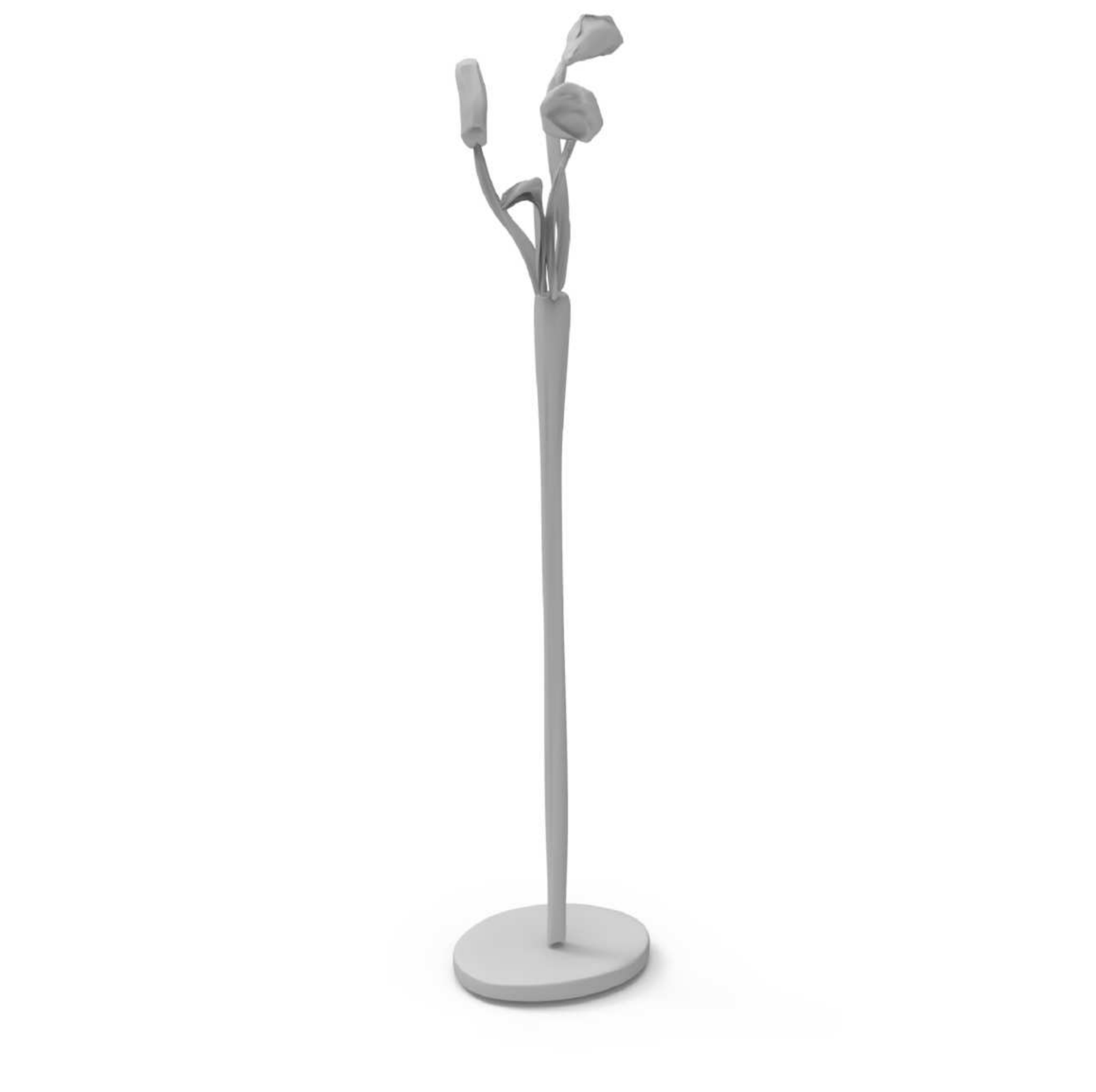}
    \includegraphics[width=0.48\linewidth]{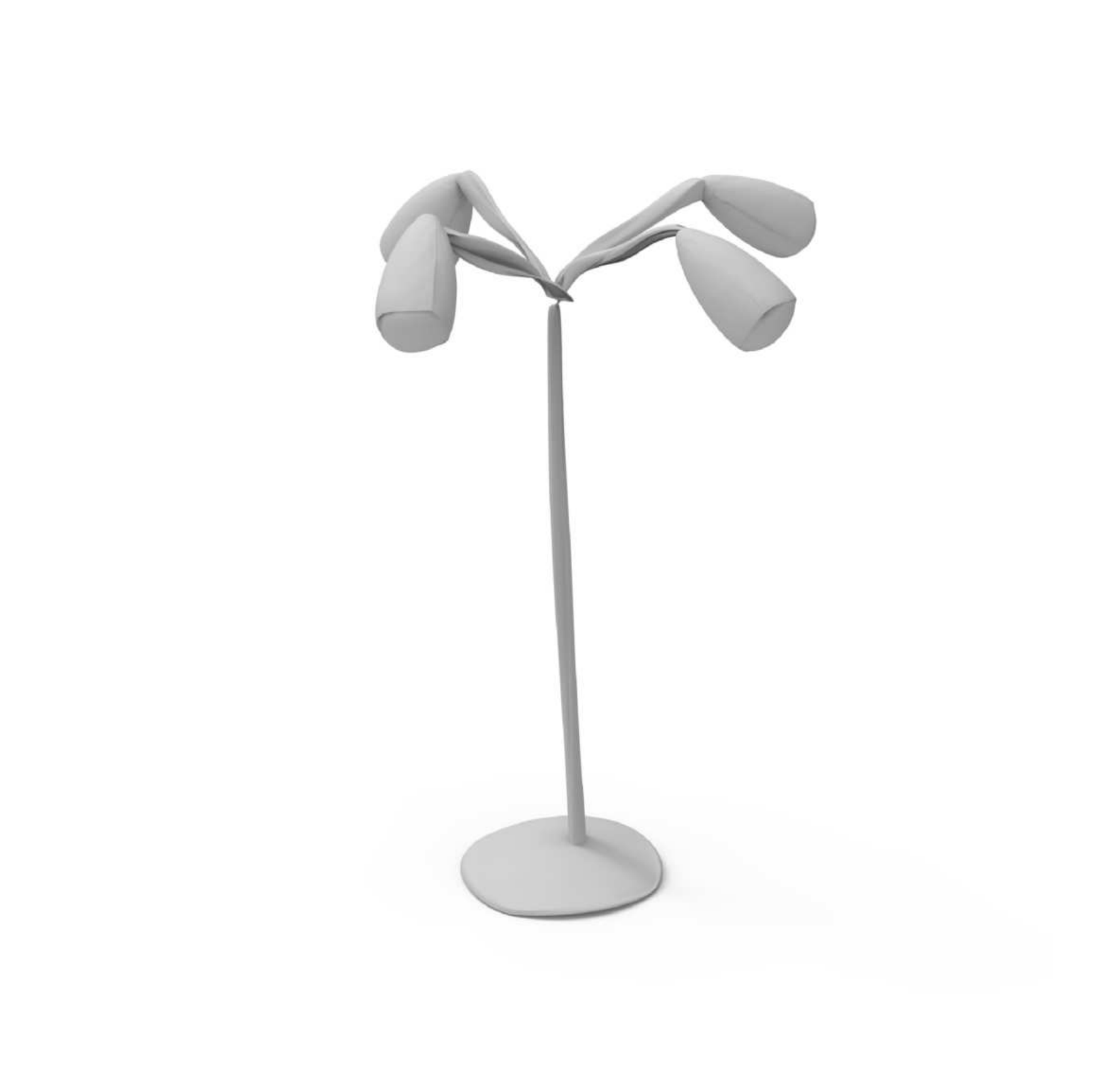}\\
    \includegraphics[width=0.48\linewidth]{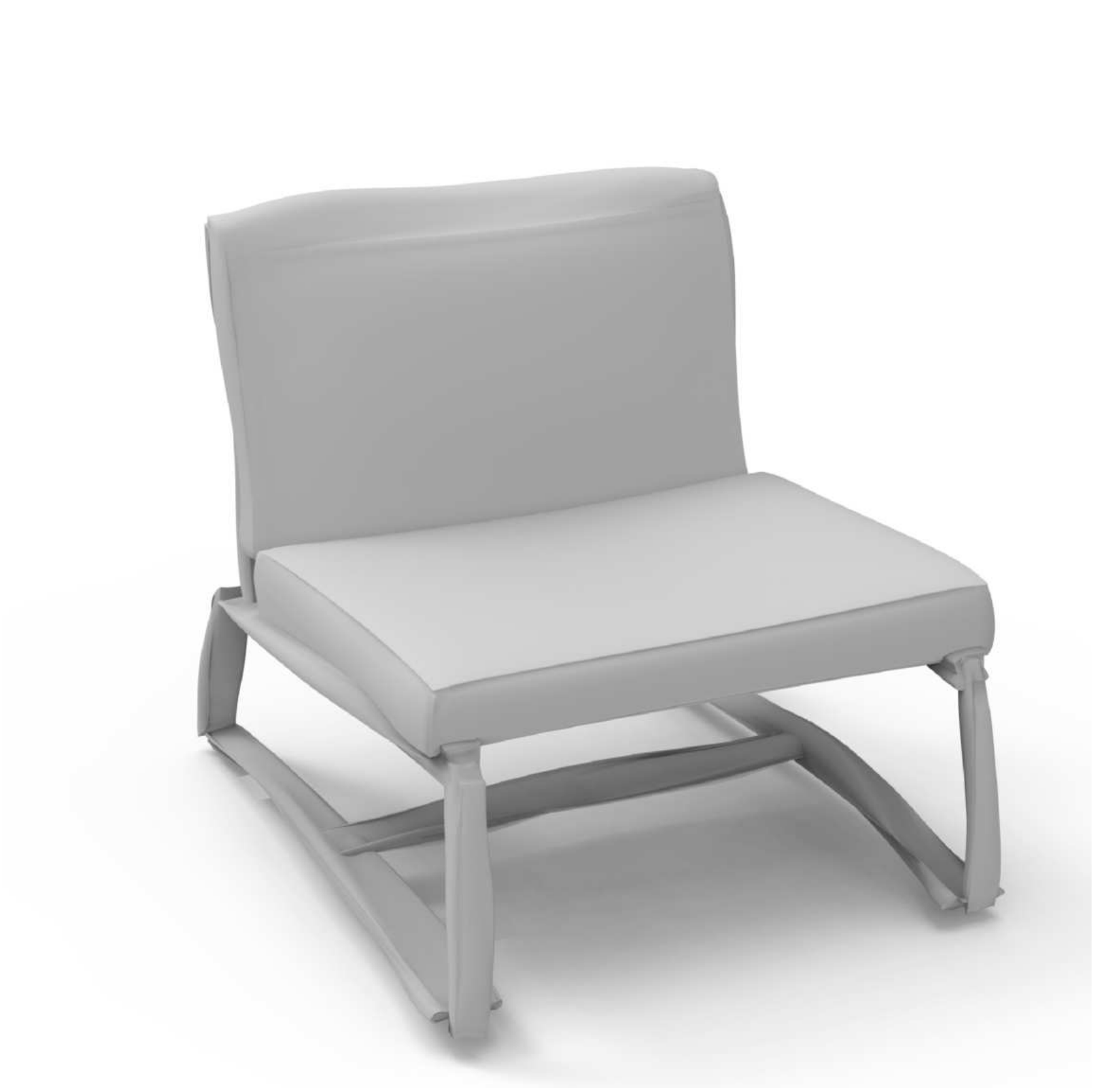}
    \includegraphics[width=0.48\linewidth]{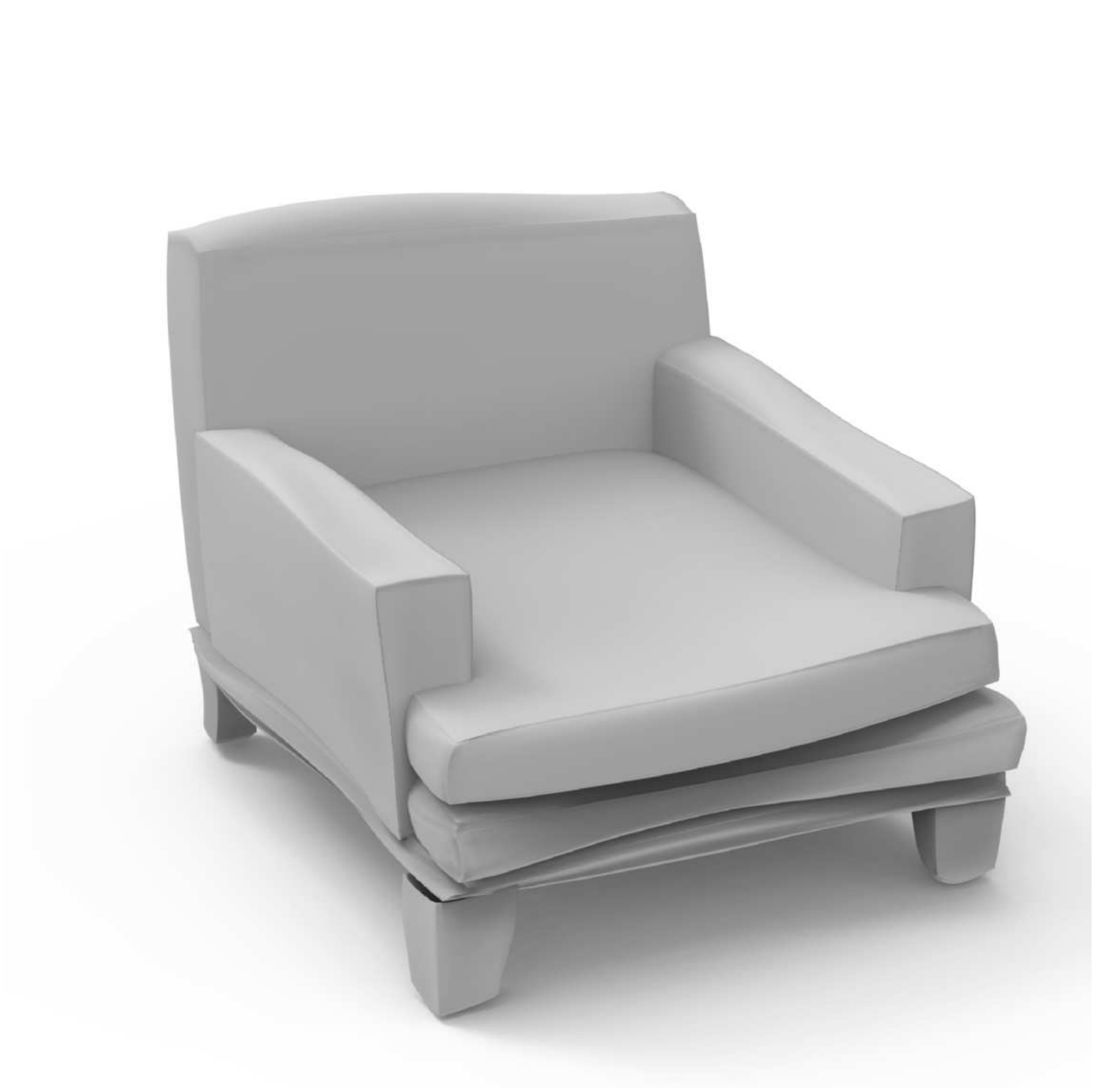}}
    \end{minipage}}
    \subfigure[Ours]{
    \begin{minipage}[b]{0.23\linewidth}
    {\includegraphics[width=0.48\linewidth]{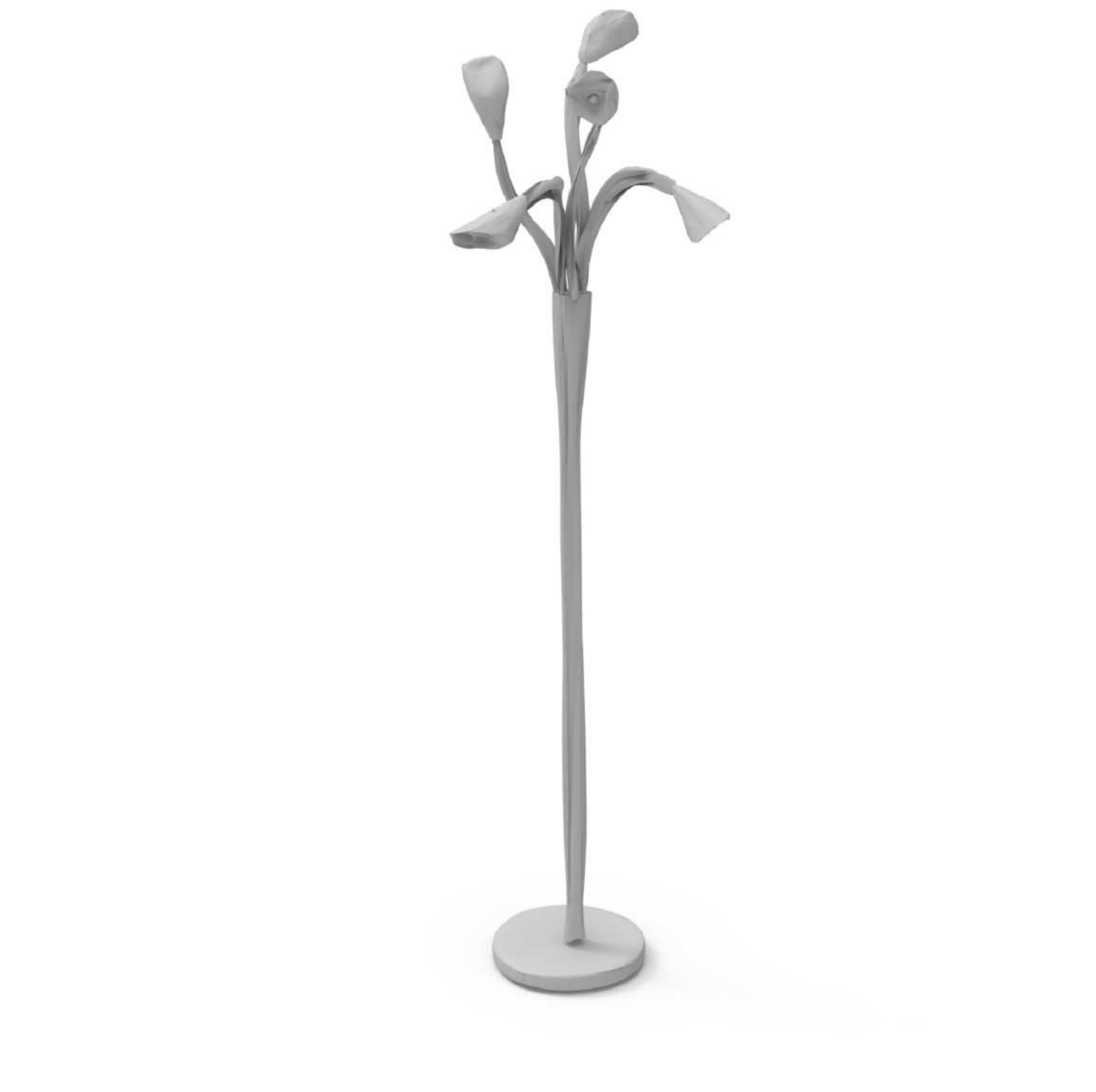}
    \includegraphics[width=0.48\linewidth]{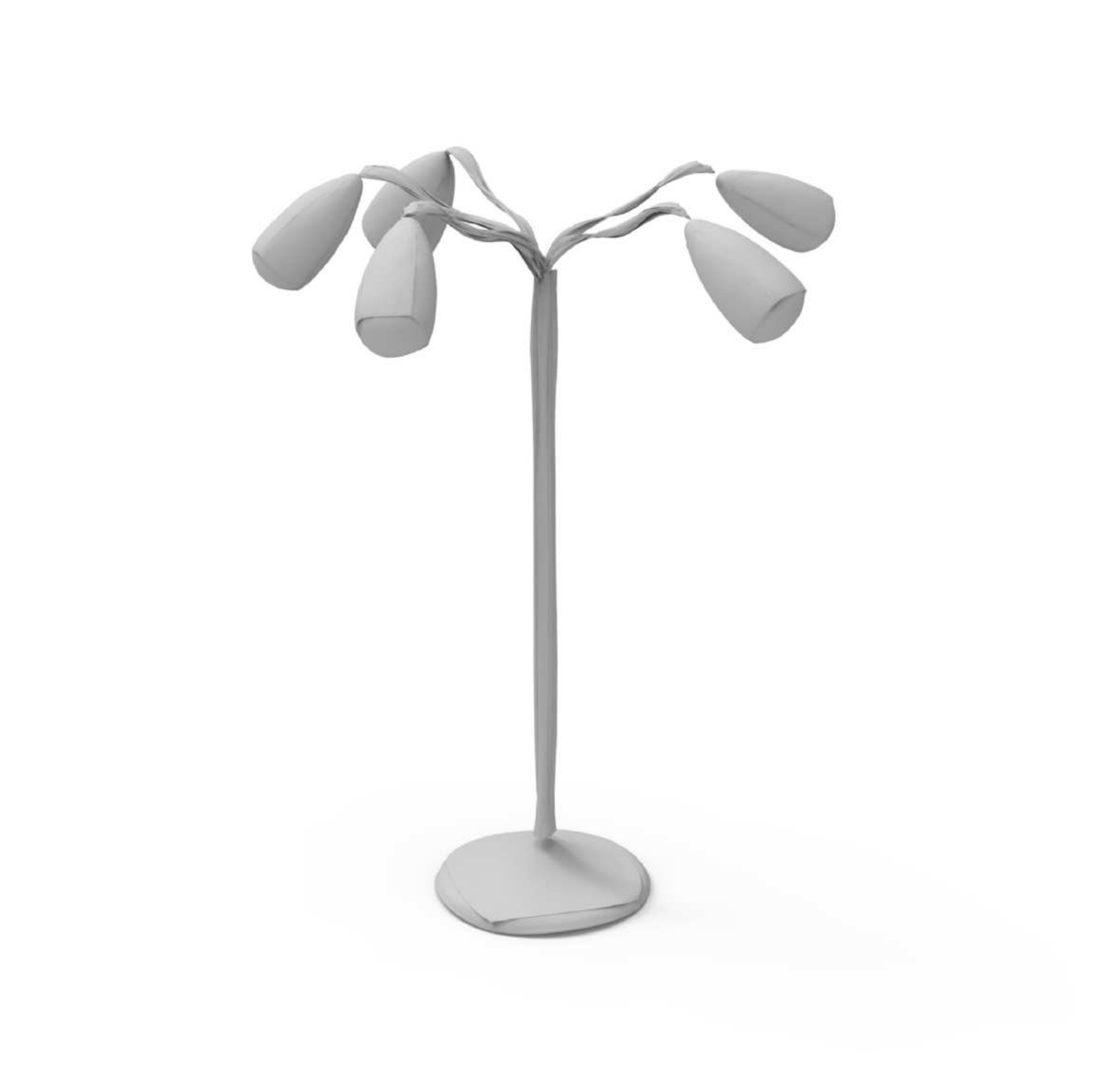}\\
    \includegraphics[width=0.48\linewidth]{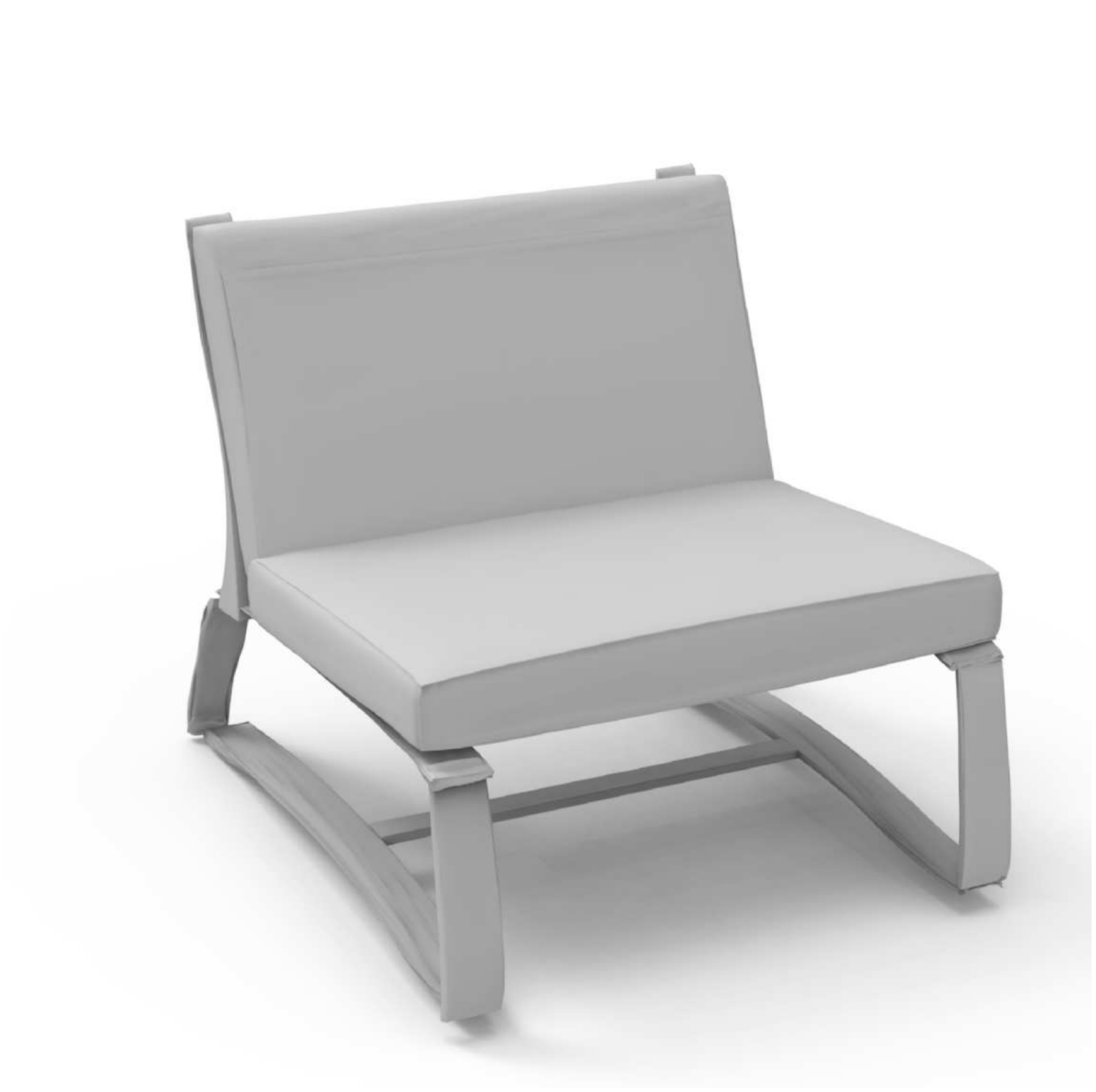}
    \includegraphics[width=0.48\linewidth]{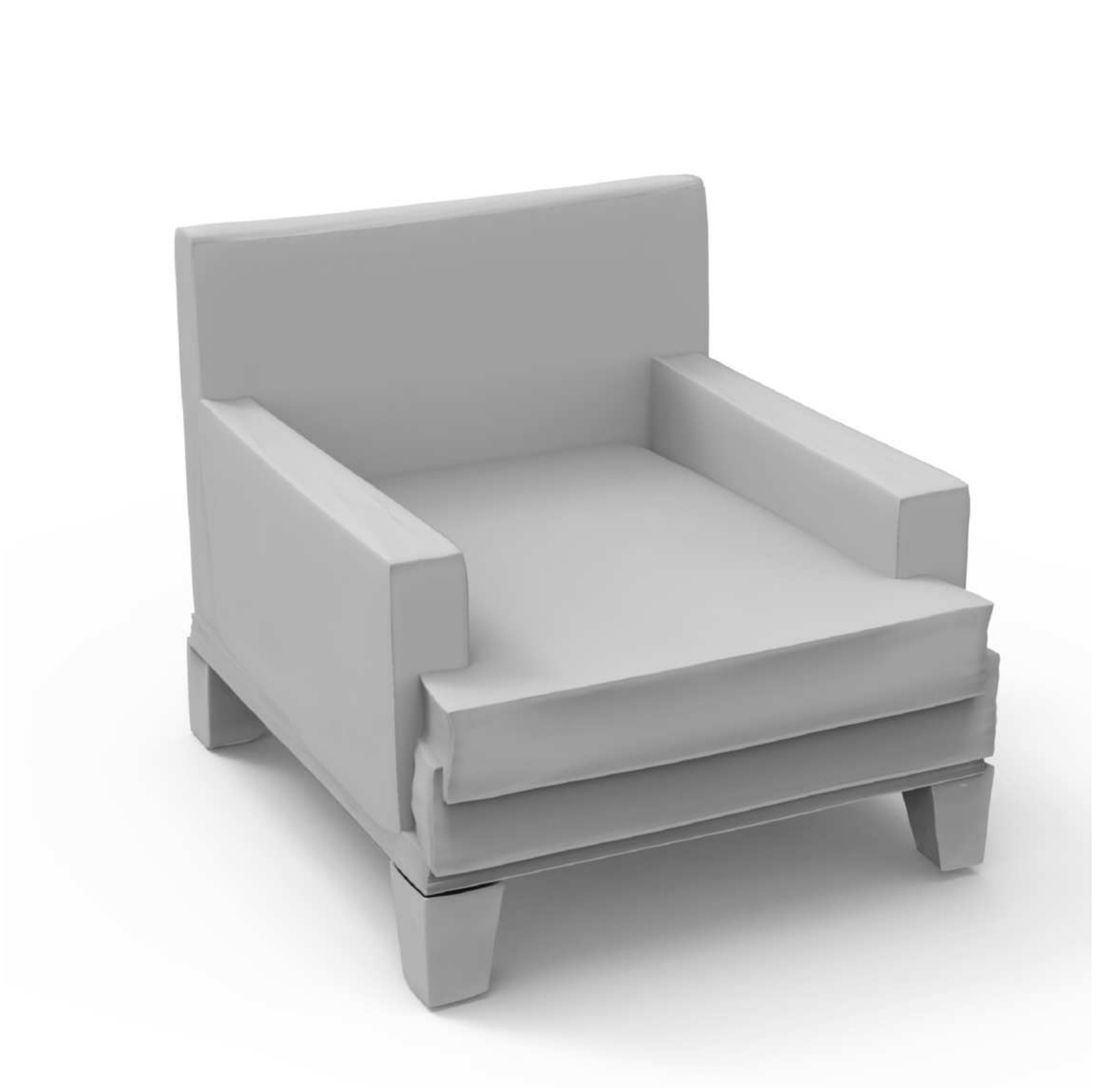}}
    \end{minipage}}
   \caption{\yjr{We show some qualitative comparisons between our final DSG-Net (the disentangled pipeline) with two ablated versions: one is the approach that naively combines StructureNet and ACAP mesh representation (without disentanglement design), namely SN+Mesh, and the other is DSG-Net without the cycled disentanglement, namely Ours (w/o CycD).
   While three methods demonstrate strong performance for shape reconstruction, we still observe that our final version achieves higher accurate reconstruction regarding some part geometric details, such as the supporting pole of the left lamp, the chain of the right lamp, the back of the left chair, and the arm of the right sofa.
   This explains the performance boost in terms of the geometric scores in Table 6.
   From these visual results, the cycled disentanglement and our disentangled shape representation benefit the performance of shape reconstruction (especially for the structure score) and they enable more interesting applications, such as disentangled shape generation/interpolation.}
   }
   \label{fig:abla_dis-mechanism}
\end{figure*}

\kaichun{We perform four sets of ablation studies to demonstrate the necessity and effectiveness of the key components and training strategies for our method.}
First, we demonstrate that explicitly considering part relationships and conducting graph message-passing operations along the edges are important. Removing the edge components from our network gives significantly worse results.
We also validate the design choice of learning a unified conditional part geometry VAE, instead of training separate VAEs for each part semantics as used in SDM-Net~\cite{gaosdmnet2019}.
\yjr{Our proposed cycled disentanglement has been demonstrated successful on the shape reconstruction and disentangled shape reconstruction quantitatively and qualitatively.}
\yjrr{We also evaluate and compare with the ablated version SN+Mesh, that is a naive combination of StructureNet~\cite{mo2019structurenet} backbone and ACAP mesh representation~\cite{gao2019sparse,gaosdmnet2019}, 
to validate that our proposed disentangled structure and geometry representation and the cycled disentanglement indeed help improve the performance for learning 3D shape generative models.}
\yj{Finally, we compare cascaded training and end-to-end training for the part geometry VAE and the disentangled backbone VAEs. We observe similar performances for the two strategies. We take the end-to-end training approach due to its simplicity.}

\begin{figure}[h]
    \centering
    \includegraphics[width=0.95\linewidth]{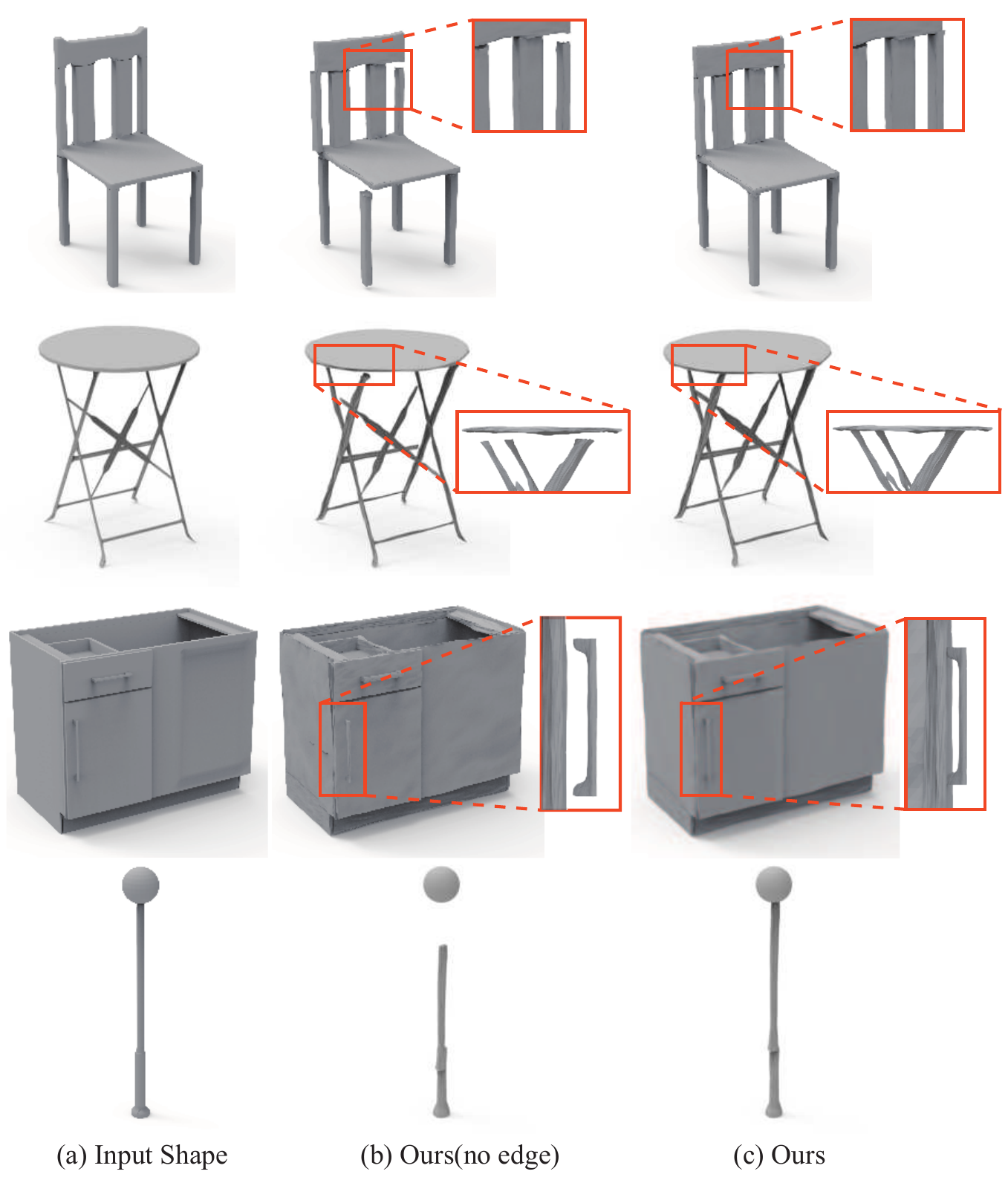}
    \caption{Qualitative comparison on shape reconstruction with the no-edge version of our method. We can see that removing edges introduces disconnected parts in the reconstructed shapes.}
    \label{fig:abla_edge}
\end{figure}

\begin{table}[h]
\fontsize{8}{11}\selectfont
  \centering
  \caption{\yjrr{Quantitative shape reconstruction performance comparing our full pipeline to three ablated versions: 1. Ours (w/o edge), that removes the edge components and the graph message-passing modules;
  2. Ours (w/o CycD), which removes the cycled disentanglement and losses ($\mathcal{L}_{struct}$ and $\mathcal{L}_{geo}$);
  3. SN+Mesh, that naively combines the StructureNet backbone and SDM-Net ACAP mesh representation.
  We observe worse performances in terms of the geometry metrics by large margins when we remove edges from the part hierarchies or disable the cycled disentanglement,
  or naively replace the point cloud representation with SDM-Net ACAP mesh representation
  while achieving comparable performance in terms of the HierInsSeg structure metric.}}
 \begin{adjustbox}{width={0.48\textwidth},keepaspectratio}
    \begin{tabular}{ccccc}
    \toprule[1pt]
    \multirow{2}[2]{*}{DataSet} & \multirow{2}[2]{*}{Method} & \multicolumn{2}{c}{Geometry Metrics} & \multicolumn{1}{c}{Structure Metrics} \\
\cmidrule{3-5}          &       & CD{\scriptsize$\times 10^{-3}$}$\downarrow$ & EMD{\scriptsize$\times 10^{-2}$}$\downarrow$ & HierInsSeg(HIS) $\downarrow$ \\
    \midrule
    \midrule
    \multirow{5}[0]{*}{Chair} & StructureNet & 9.73 & 6.46 & {0.51} \\
          & SN+Mesh & 3.32 & 0.89 & 0.51 \\
          & Ours (w/o edge) & 3.59 & 0.94 & 0.59 \\
          & Ours (w/o CycD) & 2.39 & 0.82 & 0.53 \\
          & {Ours}  & {\textbf{1.98}} & {\textbf{0.73}} & {\textbf{0.39}} \\
          & GT    &       &       & \underline{0.32} \\
    \midrule
    \midrule
    \multirow{5}[0]{*}{Table} & StructureNet & 14.63 & 5.68 & {0.97} \\
          & SN+Mesh & 4.79 & 0.96 & 0.96 \\
          & Ours (w/o edge) & 5.07 & 1.02 & 1.03 \\
          & Ours (w/o CycD) & 4.05 & 0.84 & 0.99 \\
          & {Ours}  & {\textbf{3.42}} & {\textbf{0.75}} &
          {\textbf{0.85}} \\
          & GT    &       &       & \underline{0.65} \\
    \midrule
    \midrule
    \multirow{5}[0]{*}{Cabinet} & StructureNet & 16.34 & 5.74 & 0.57 \\
          & SN+Mesh & 4.02 & 1.59 & 0.58 \\
          & Ours (w/o edge) & 4.59 & 1.83 & 0.66 \\
          & Ours (w/o CycD) & 3.49 & 1.10 & 0.58 \\
          & {Ours}  & {\textbf{2.96}} & {\textbf{0.97}} & \textbf{0.45} \\
          & GT    &       &       & \underline{0.35} \\
    \midrule
    \midrule
    \multirow{5}[0]{*}{Lamp} & StructureNet & 17.31 &  7.12 & 0.70 \\
          & SN+Mesh & 12.93 & 2.60 & 0.72 \\
          & Ours (w/o edge) & 12.40 & 3.49 & 0.73 \\
          & Ours (w/o CycD) & 10.43 & 1.97 & 0.69 \\
          & {Ours}  & {\textbf{7.15}} & {\textbf{1.63}} & {\textbf{0.61}} \\
          & GT    &       &       & \underline{0.54} \\
    \bottomrule[1pt]
    \end{tabular}%
    \end{adjustbox}
  \label{tab:abla_edge}%
\end{table}%

\paragraph{Removing Part Relationships and Edges.} 
Our network explicitly models the part relationships as horizontal edges among sibling nodes in the shape part hierarchy.
Graph message-passing operations are conducted along the edges in both encoding and decoding stages.
In this experiment, we compare to a no-edge version of our network where we remove the edge components and the message-passing modules.
In Table~\ref{tab:abla_edge}, we see that removing the edge components gives worse results than our full pipeline.
Figure~\ref{fig:abla_edge} illustrates three example reconstructed shapes for our method with and without edge components, where we see clearly that removing edge components creates more artifacts, such as disconnected parts.

\begin{figure}[t]
    \centering
    \includegraphics[width=0.49\linewidth]{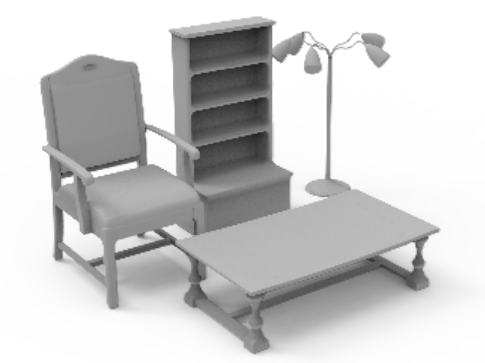}
    \includegraphics[width=0.49\linewidth]{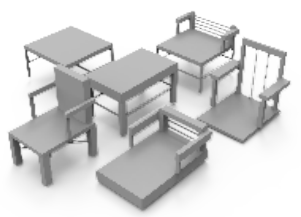}
    \caption{\yj{Example shapes in the PartNet dataset (left) and the synthetic dataset (right).}}
    \label{fig:data_viz}
\end{figure}

\yjr{
\paragraph{Removing Cycled Disentanglement \& Losses ($\mathcal{L}_{struct}$ and $\mathcal{L}_{geo}$)}
In our proposed \yjrr{novel disentanglement}, the goal of cycled disentanglement ensures the structure and geometry of any shape can be decoupled as much as possible.
In this ablation, we aim to demonstrate its importance quantitatively and qualitatively for our disentangled representation of shape structure and geometry on shape reconstruction, including the disentangled shape reconstruction on synthetic data. 
In Table~\ref{tab:abla_edge}, we validate the performance on shape reconstruction for 4 categories of PartNet, and the quantitative results demonstrate that the cycled disentanglement has a very large improvement, especially compared to our ablated version (SN+Mesh).
Figure~\ref{fig:abla_dis-mechanism} also displays some examples on Lamp and Chair datasets, where there is a clear visual difference in the geometric details.
The full version can learn more fine-grained part geometry details.

Besides, we also demonstrate the performance on the disentangle shape reconstruction for synthetic data, where we have ground truth (GT) data.
In Table 2 of our main paper, the performance without CycD is significantly worse than our final version (with CycD) for the structure score (HIS). Due to the simple geometric details in the synthetic dataset (made up of boxes of different scales, for details please refer to Sec.~\ref{sec:data} and Fig.~\ref{fig:data_viz}),
the improvement on geometric is smaller.
}

\begin{figure}[t]
    \centering
    \subfigure[Input shape]{\includegraphics[width=0.29\linewidth]{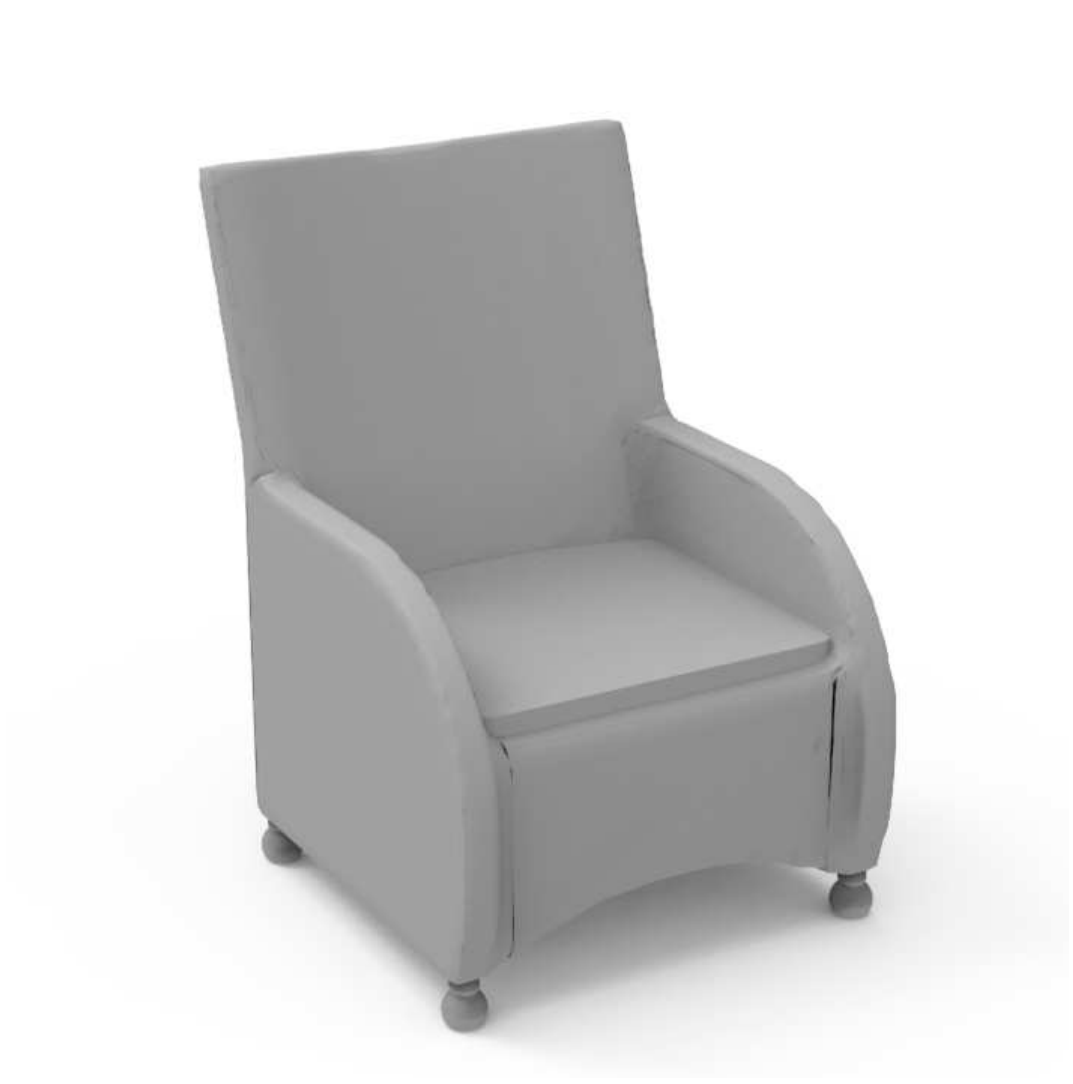}}
    \hspace{3mm}
    \subfigure[Reconstruction by using separate part geometry VAEs for different part semantics]{\includegraphics[width=0.29\linewidth]{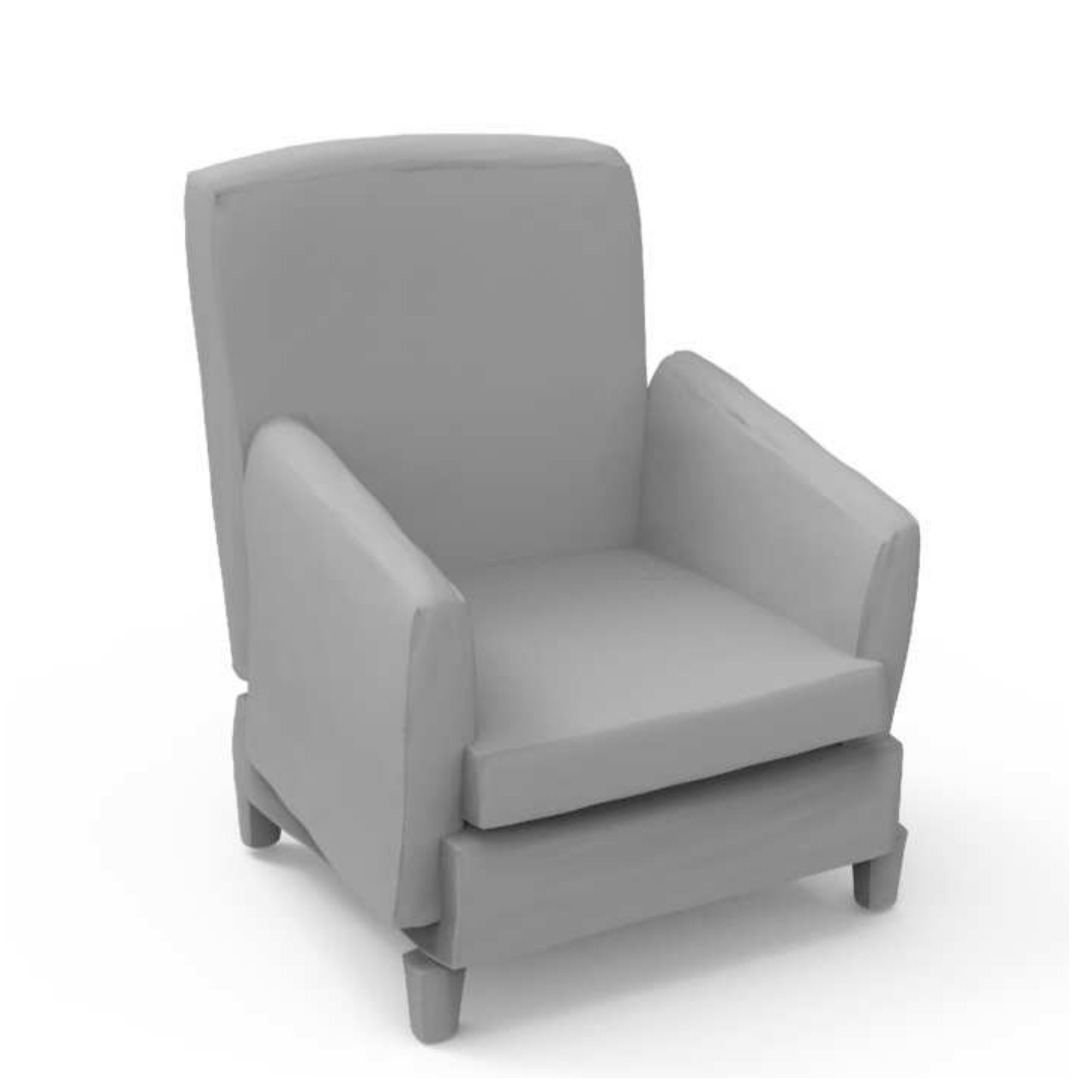}}
    \hspace{3mm}
    \subfigure[Reconstruction by using a single conditional part geometry VAE for all part semantics]{\includegraphics[width=0.29\linewidth]{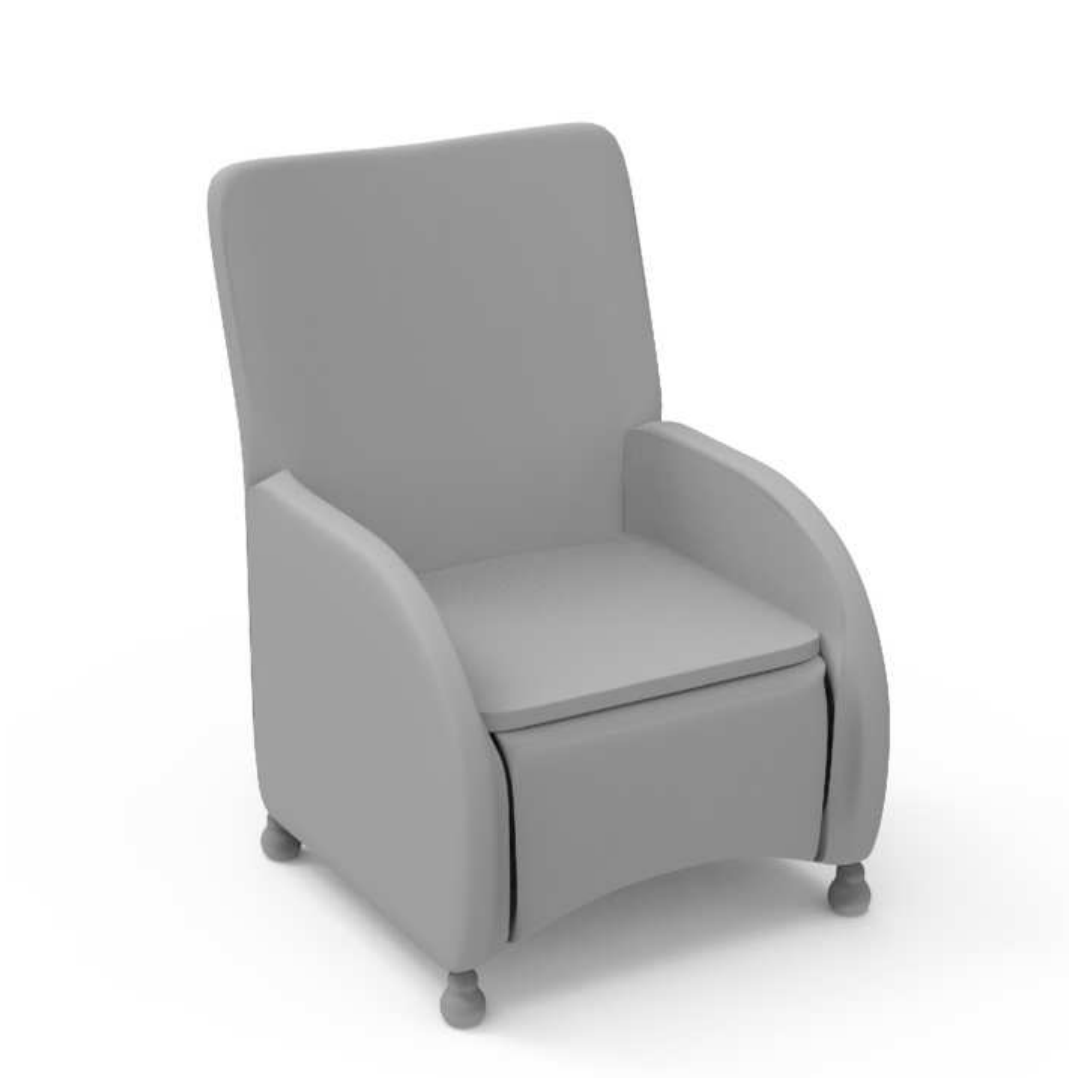}}
    
    \caption{We compare the reconstructed results for a chair using a unified conditional part geometry VAE or using separate VAEs for different semantic parts. We find that using a single VAE reconstructs part geometry with more fine-grained part geometry details, \eg the curvy armrests and the delicate sofa foots.}
    \label{fig:cpgvae_single}
\end{figure}

\paragraph{Training Separate Part Geometry VAEs.}
In our network design, for encoding and decoding leaf-node part geometry, we train a unified part geometry VAE that is conditional on the part semantic labels and structure contexts.
SDM-Net, however, proposes to use separate part geometry VAEs for different part semantics.
We argue that while it is preferable in the SDM-Net experiments, it is very costly and ineffective to train separate networks on the PartNet data, where we have more fine-grained part categories than the SDM-Net dataset.
In Table~\ref{tab:cpgvae_single}, we try the alternative method of training our network with 57 separate part geometry VAEs for PartNet chairs and show that the performance of shape reconstruction is significantly worse than training a unified conditional part geometry VAE.
Figure~\ref{fig:cpgvae_single} compares the reconstructed shapes of a chair and we see that a unified part geometry VAE learns to reconstruct part geometry with more fine-grained part geometry details, \eg the curvy armrests and the delicate sofa foots.

\begin{table}[htbp]
  \centering
  \caption{\yjr{Shape reconstruction performance by our DSG-Net using a single conditional part geometry VAE for all semantic parts or using separate VAEs for different semantic parts. We see that training one single conditional part geometry VAE is more data-efficient and thus works much better on the PartNet dataset.}}
  \begin{adjustbox}{width={0.48\textwidth},keepaspectratio}
    \begin{tabular}{cccc}
    \toprule[1pt]
    \multirow{2}[4]{*}{Method} & \multicolumn{2}{c}{Geometry Metrics} & \multicolumn{1}{c}{Structure Metrics} \\
\cmidrule{2-4}          & CD{\scriptsize$\times 10^{-3}$}$\downarrow$ & EMD{\scriptsize$\times 10^{-2}$}$\downarrow$ & HierInsSeg(HIS) $\downarrow$ \\
    \midrule
    \midrule
    \multirow{2}{*}{\tabincell{c}{Ours\\(not share one PG VAE)}} & \multirow{2}{*}{14.22} & \multirow{2}{*}{5.67} & \multirow{2}{*}{0.88} \\
    &       &       &     \\
    \midrule
    \multirow{2}{*}{\tabincell{c}{Ours\\(share one PG VAE)}} & \multirow{2}{*}{\textbf{1.98}} & \multirow{2}{*}{\textbf{0.73}} & \multirow{2}{*}{\textbf{0.39}} \\
    &       &       &     \\
    \midrule
    GT    &       &       & \underline{0.32} \\
    \bottomrule[1pt]
    \end{tabular}%
    \end{adjustbox}
  \label{tab:cpgvae_single}%
\end{table}%

\paragraph{StructureNet+Mesh (SN+Mesh) \vs Ours.}
\yjrr{
We also find that DSG-Net achieves higher geometric accuracy than our ablated version (SN + Mesh) while having comparable performance on the structure side.
While both methods demonstrate strong performance for shape reconstruction, we still observe that our final version achieves more accurate reconstruction regarding some part geometric details (see Figure~\ref{fig:abla_dis-mechanism} caption for more details), which explains the performance boost in terms of the geometric scores in Table~\ref{tab:abla_edge}.
Also, by learning the disentangled shape geometry and structure spaces, DSG-Net enables novel controllable shape manipulation applications that explicitly disentangle the geometry and structure control, 
while the SN+Mesh cannot accomplish these.
}

\paragraph{Cascaded \vs End-to-end Training.}
\yj{In our network, we have multiple VAEs for predicting the structure and geometry of shape and part geometric details. 
In our method, we train all network modules, including the part geometry VAE and the coupled hierarchical graph VAEs, in an end-to-end manner.
We compare to a cascaded training scheme where we first train the part geometry VAE and then train the rest of our model.
In Table~\ref{tab:abla_train}, we evaluate the influence of two training strategies on the chair category.
We observe similar performance for the two training schemes. So finally, we picked the end-to-end solution for simplicity. 
}

\begin{table}[h]
  \centering
  \caption{\yjr{Shape reconstruction quantitative evaluations on training strategy. We evaluate two training strategies on the chair category: 1. cascaded training: we pre-train the PG VAE firstly, and then train the coupled hierarchical VAEs based on the pre-trained PG VAE; 2. end-to-end training: train two networks simultaneously. The separate training has similar performance to our task on geometry. So, for simplicity of training network, we choose the end-to-end solution.}}
  \begin{adjustbox}{width={0.48\textwidth},keepaspectratio}
    \begin{tabular}{cccc}
    \toprule[1pt]
    \multirow{2}[4]{*}{Method} & \multicolumn{2}{c}{Geometry Metrics} & Structure Metrics \\
\cmidrule{2-4}          & CD{\scriptsize$\times 10^{-3}$}$\downarrow$ & EMD{\scriptsize$\times 10^{-2}$}$\downarrow$ & HierInsSeg(HIS) $\downarrow$\\
    \midrule
    \midrule
    Ours (Cascaded Training) & 2.10 & 0.77 & 0.51 \\
    Ours (End-to-end Training) & 1.98 & 0.73 & 0.39 \\
    GT    &       &       & \underline{0.32} \\
    \bottomrule[1pt]
    \end{tabular}%
  \end{adjustbox}
  \label{tab:abla_train}%
\end{table}%

\begin{figure*}[t]
    \centering
    \subfigure[Chair]{
    \begin{minipage}[b]{0.24\linewidth}
    {
    \includegraphics[width=0.49\linewidth]{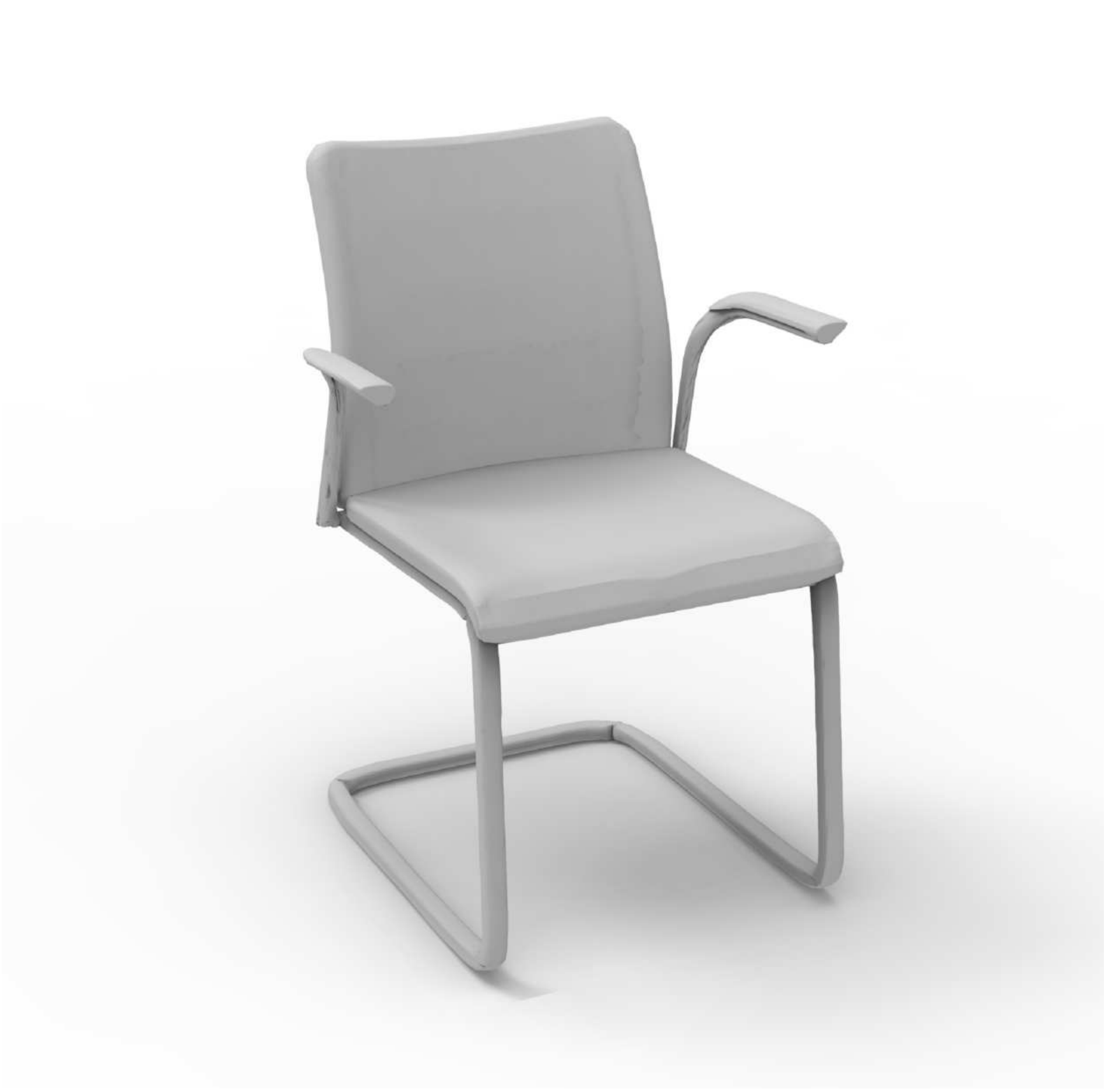}
    \includegraphics[width=0.49\linewidth]{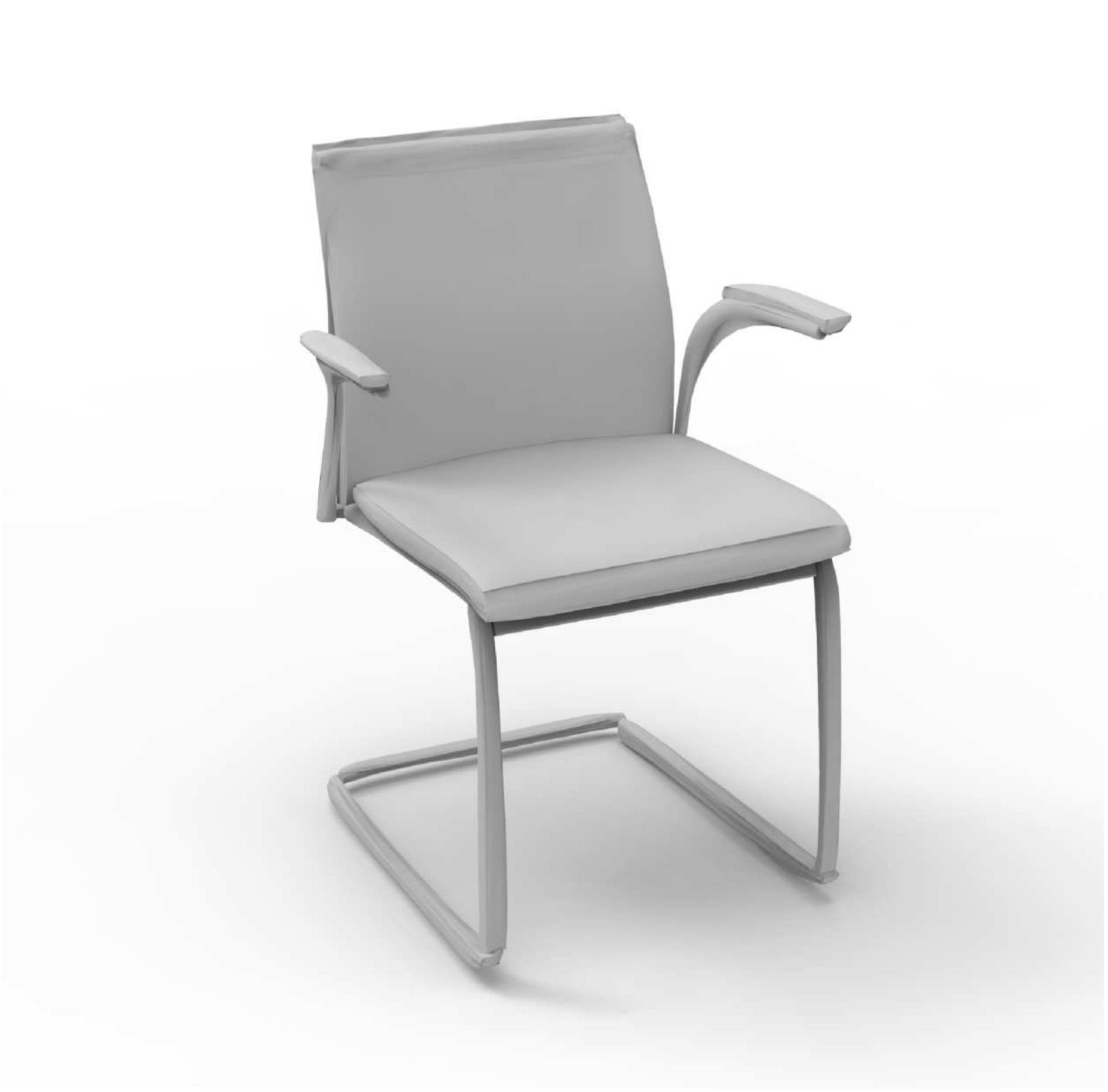}\\
    \includegraphics[width=0.49\linewidth]{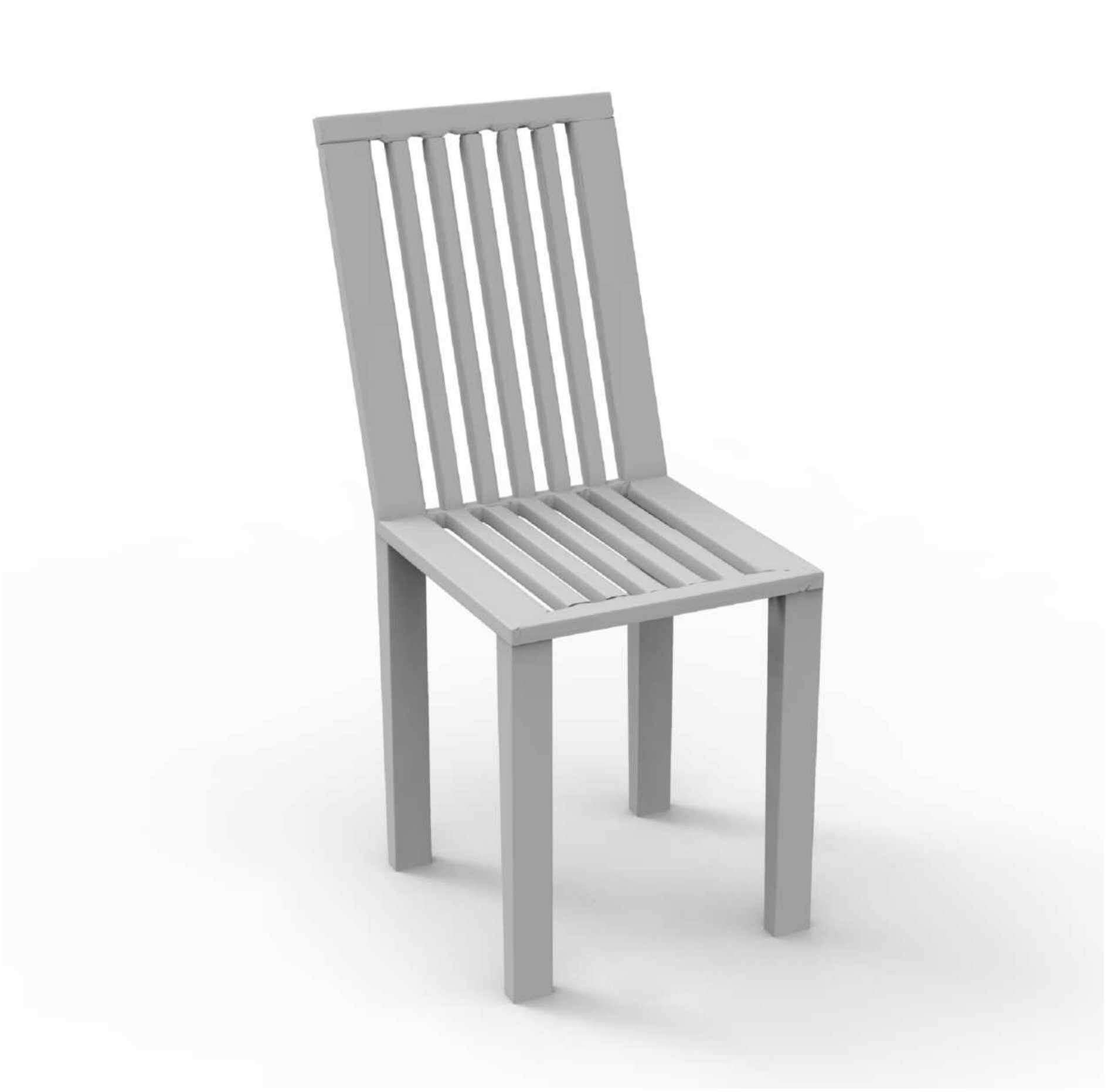}
    \includegraphics[width=0.49\linewidth]{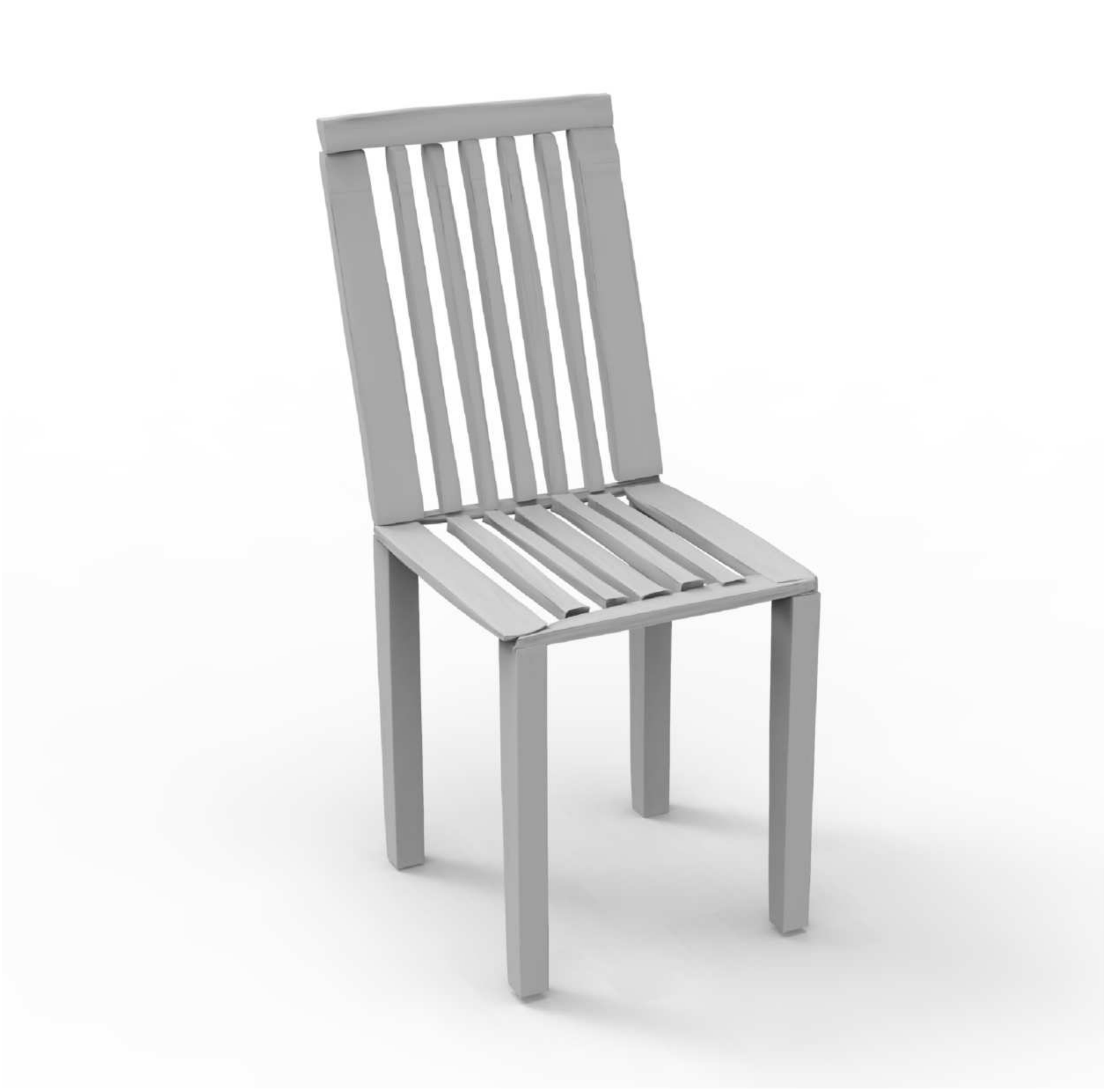}\\
    \includegraphics[width=0.48\linewidth]{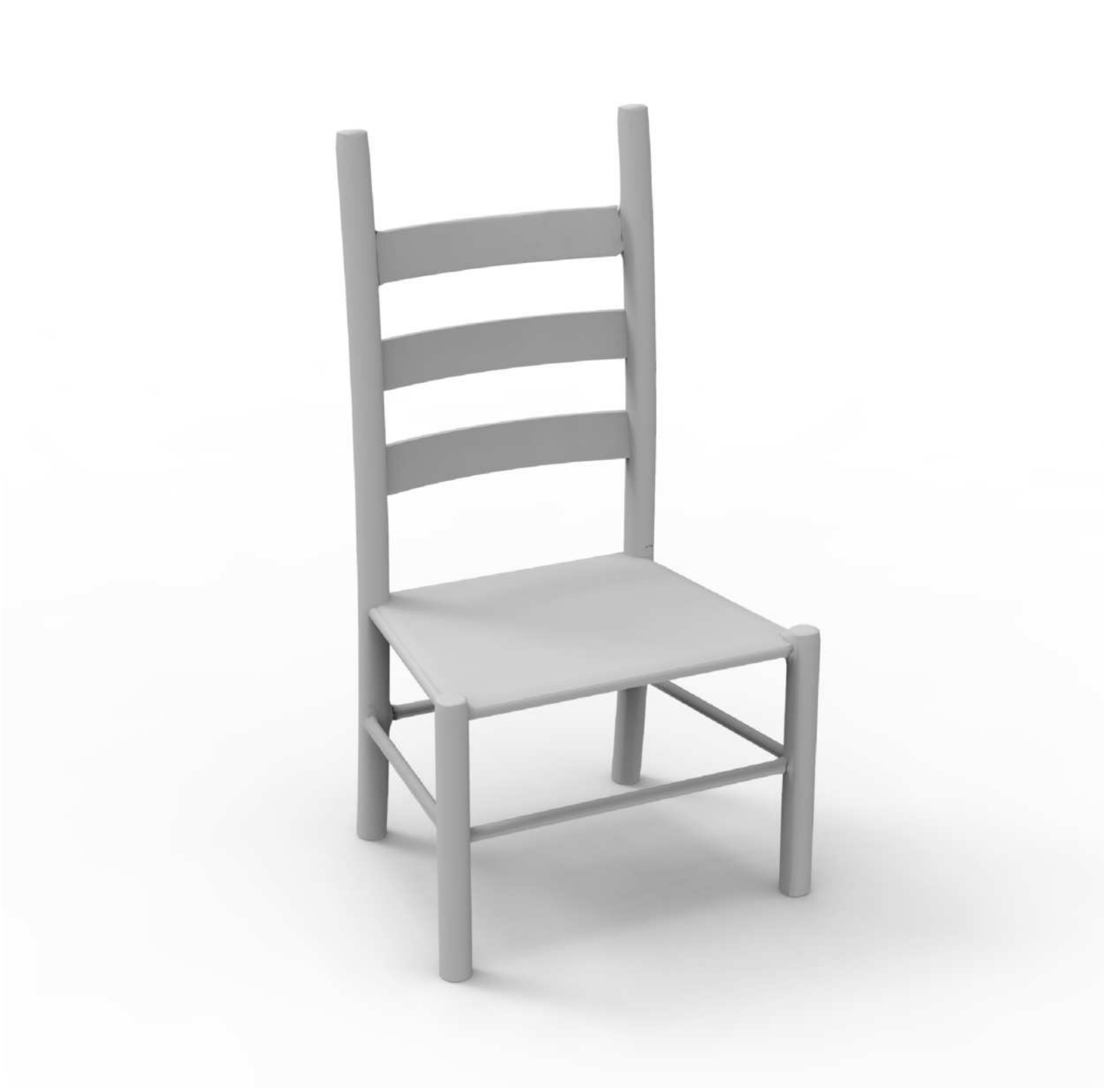}
    \includegraphics[width=0.48\linewidth]{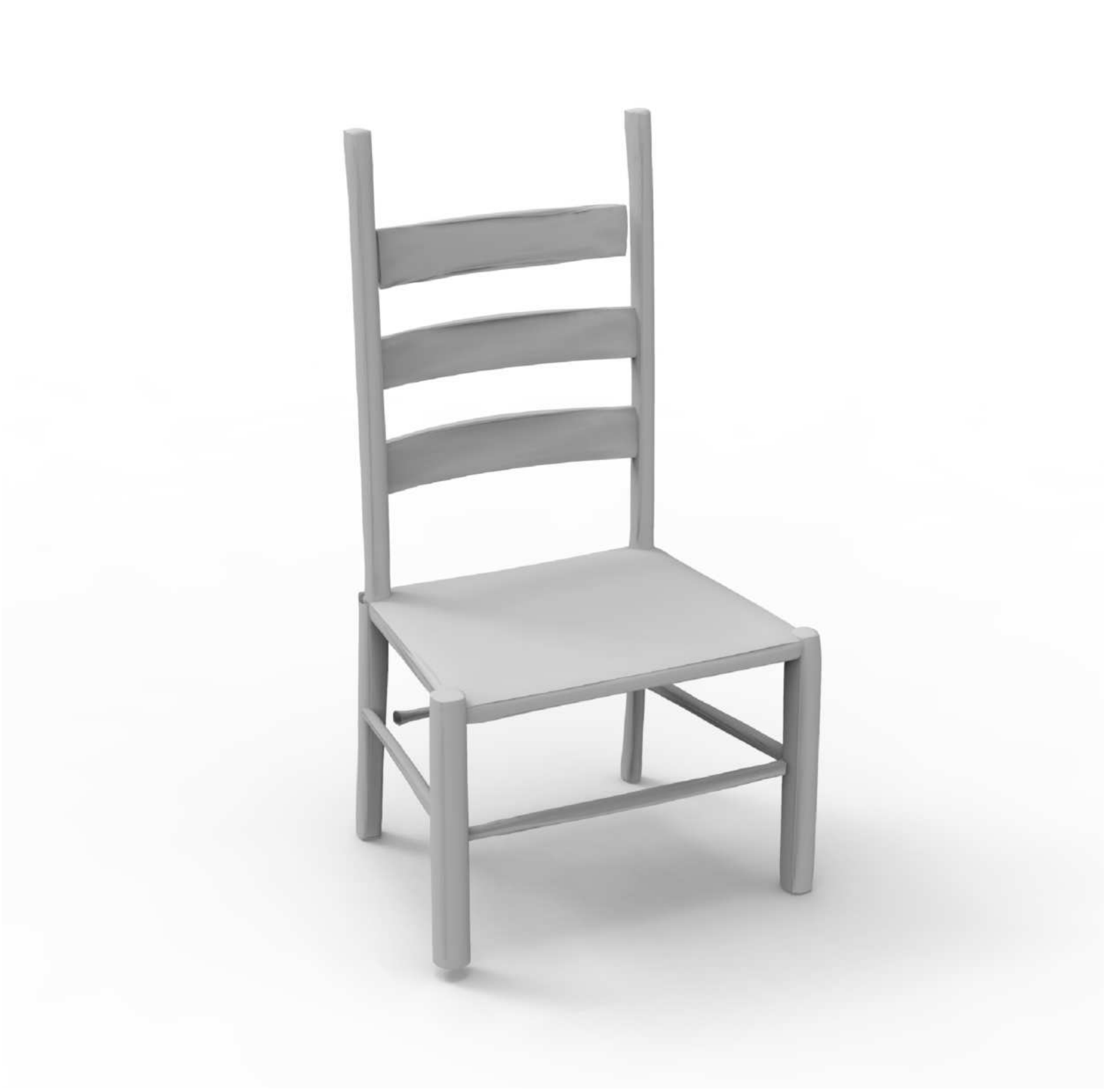}
    }
    \end{minipage}}
    \subfigure[Table]{
    \begin{minipage}[b]{0.24\linewidth}
    {
    \includegraphics[width=0.49\linewidth]{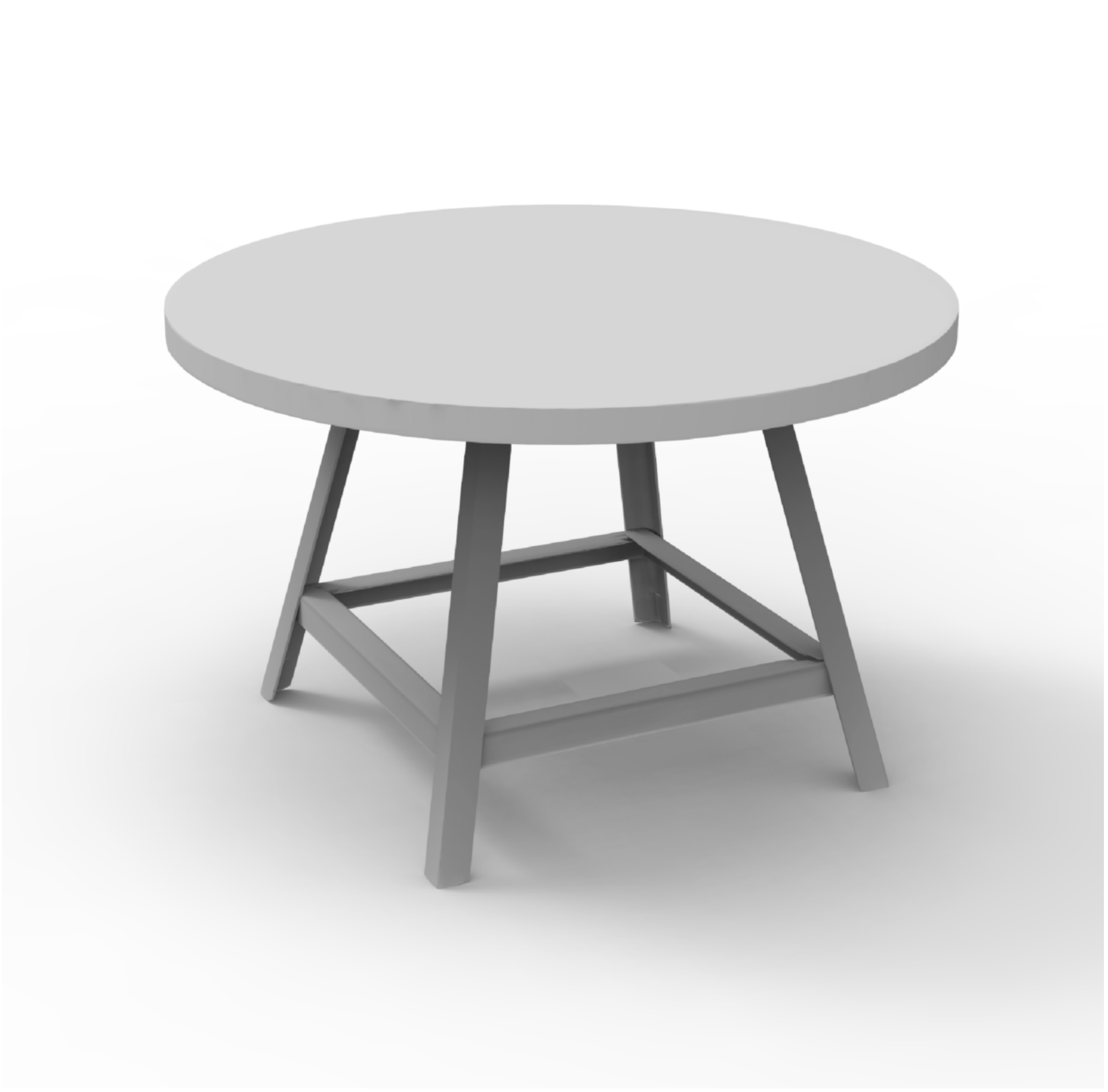}
    \includegraphics[width=0.49\linewidth]{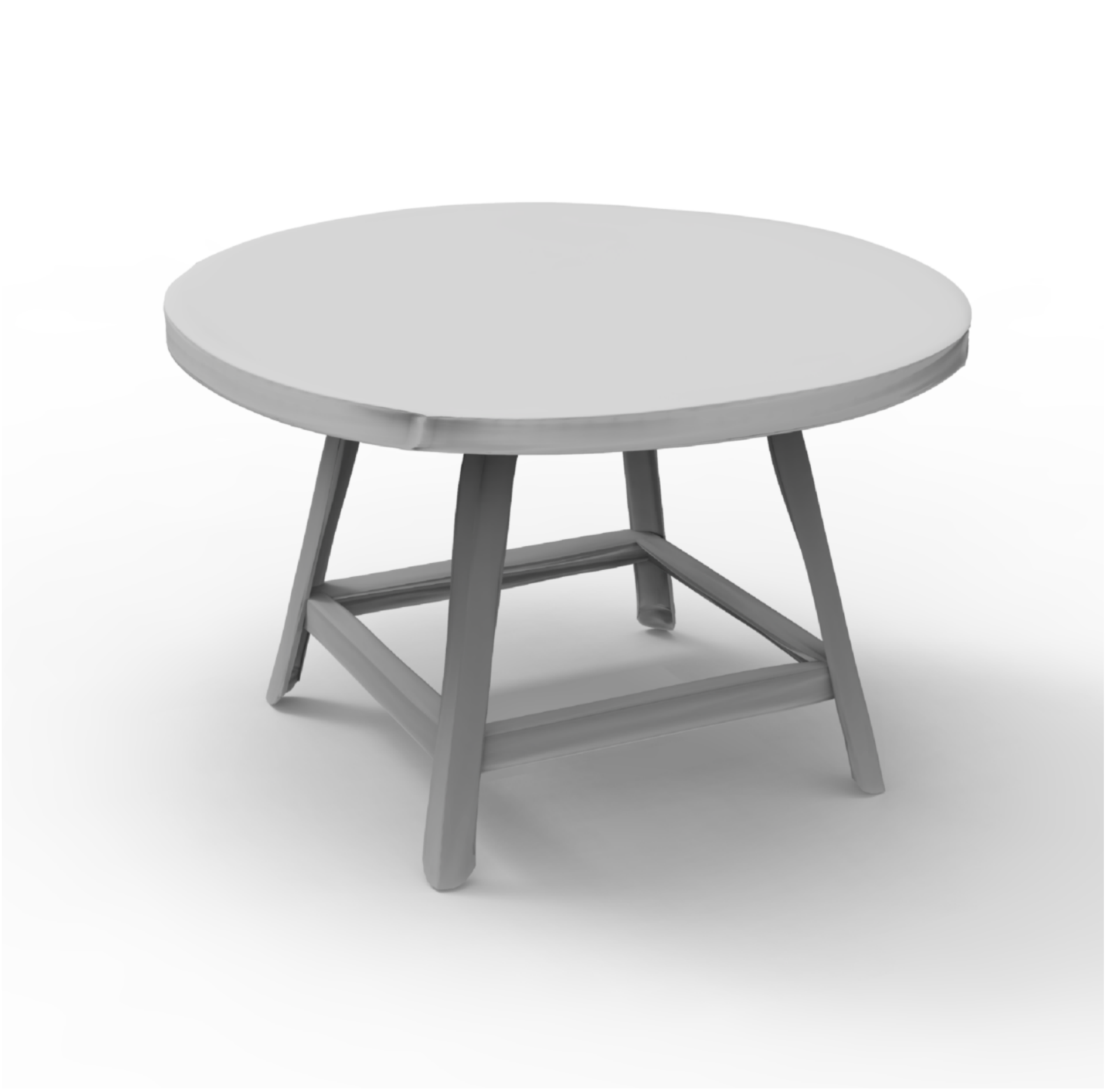}\\
    \includegraphics[width=0.48\linewidth]{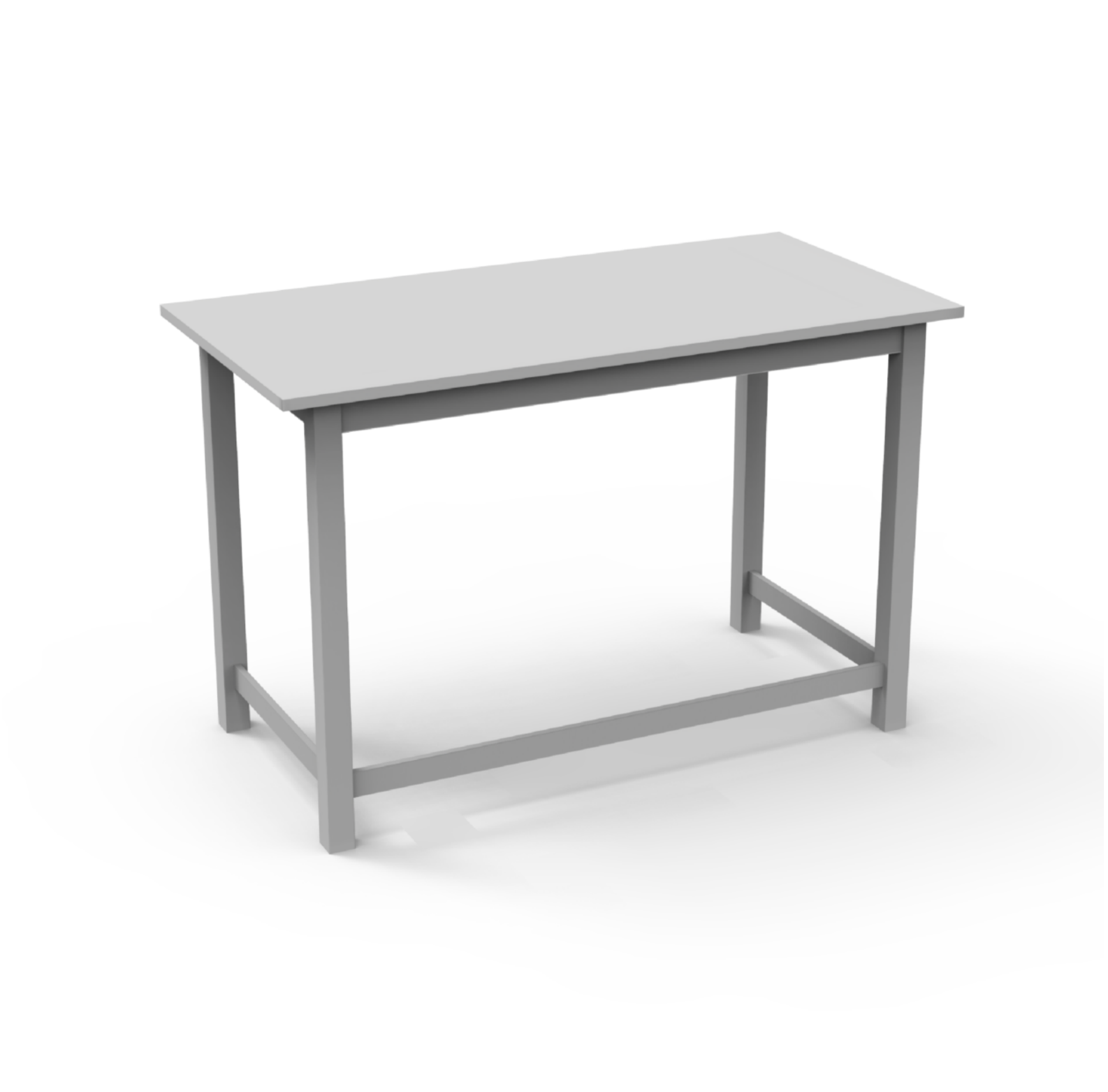}
    \includegraphics[width=0.48\linewidth]{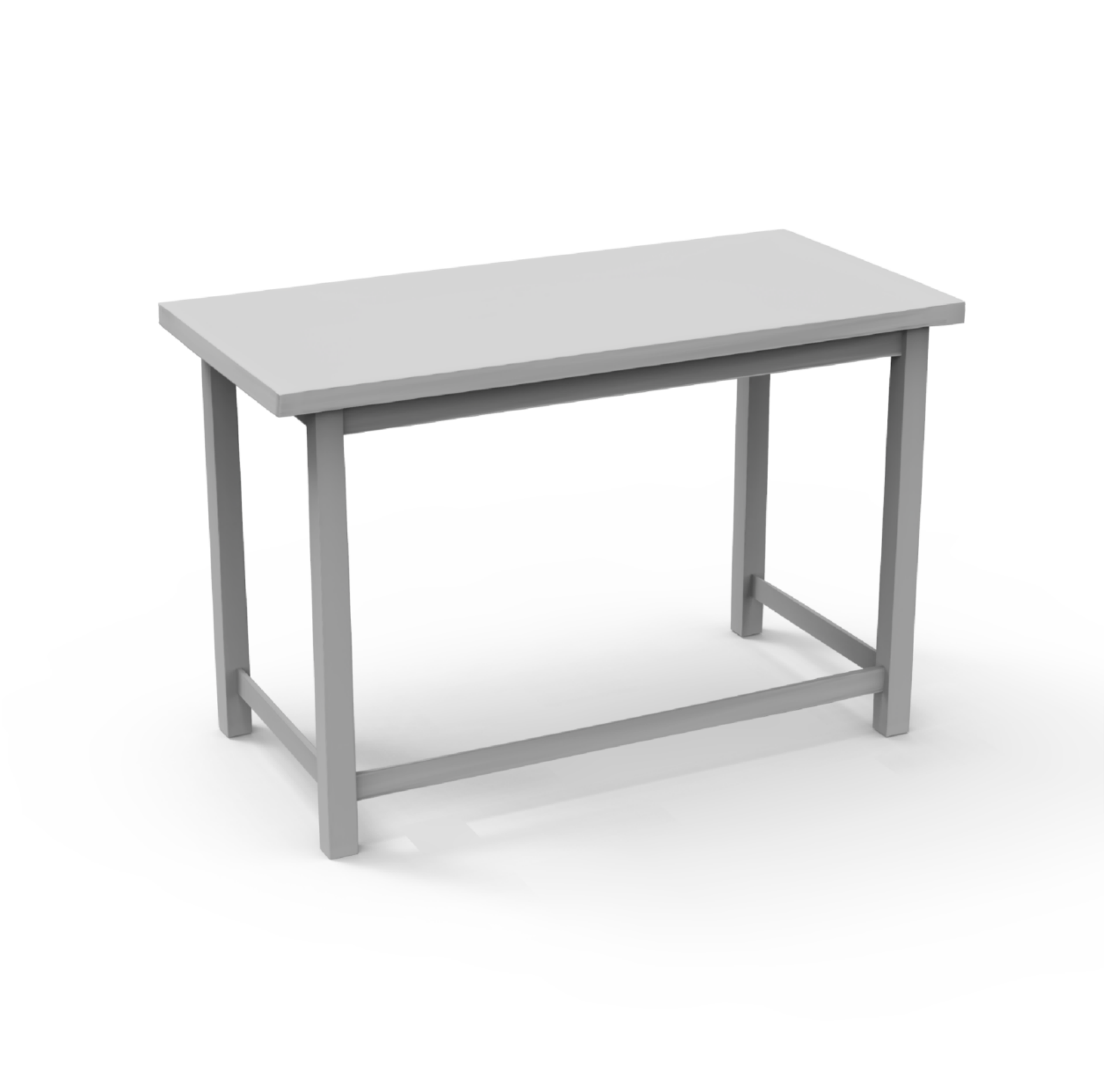}\\
    \includegraphics[width=0.48\linewidth]{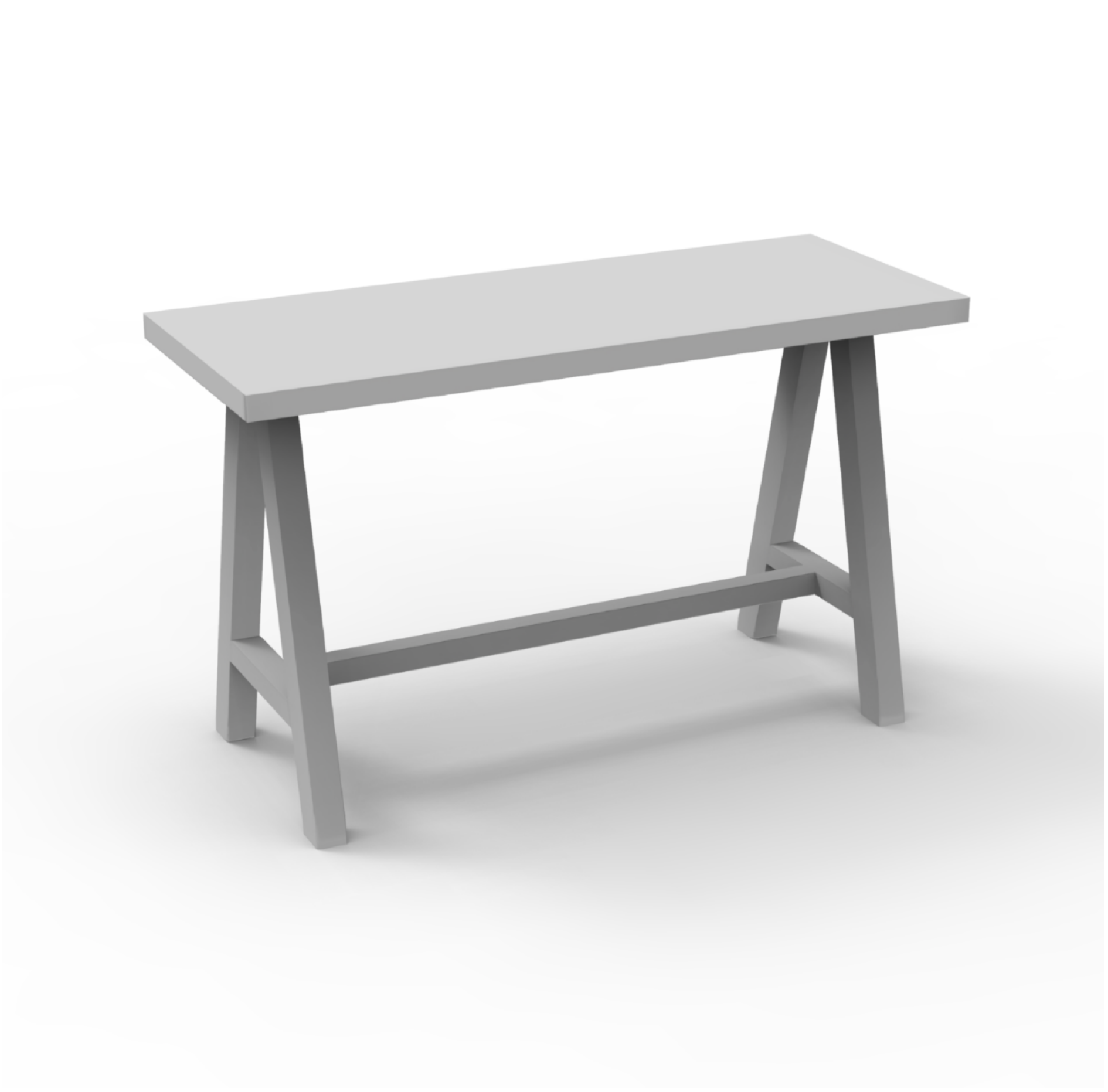}
    \includegraphics[width=0.48\linewidth]{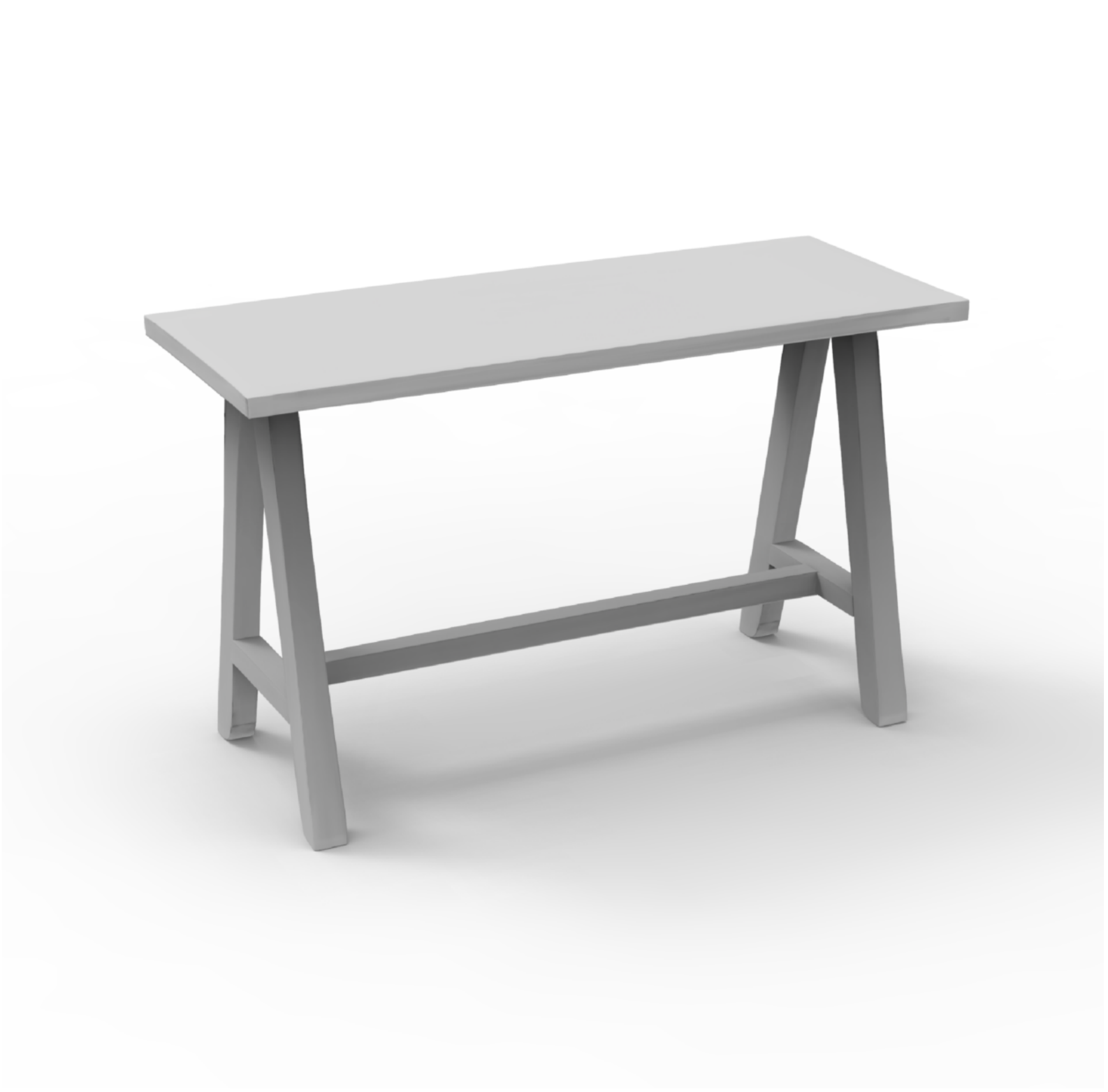}
    }
    \end{minipage}}
    \subfigure[Cabinet]{
    \begin{minipage}[b]{0.24\linewidth}
    {
    \includegraphics[width=0.49\linewidth]{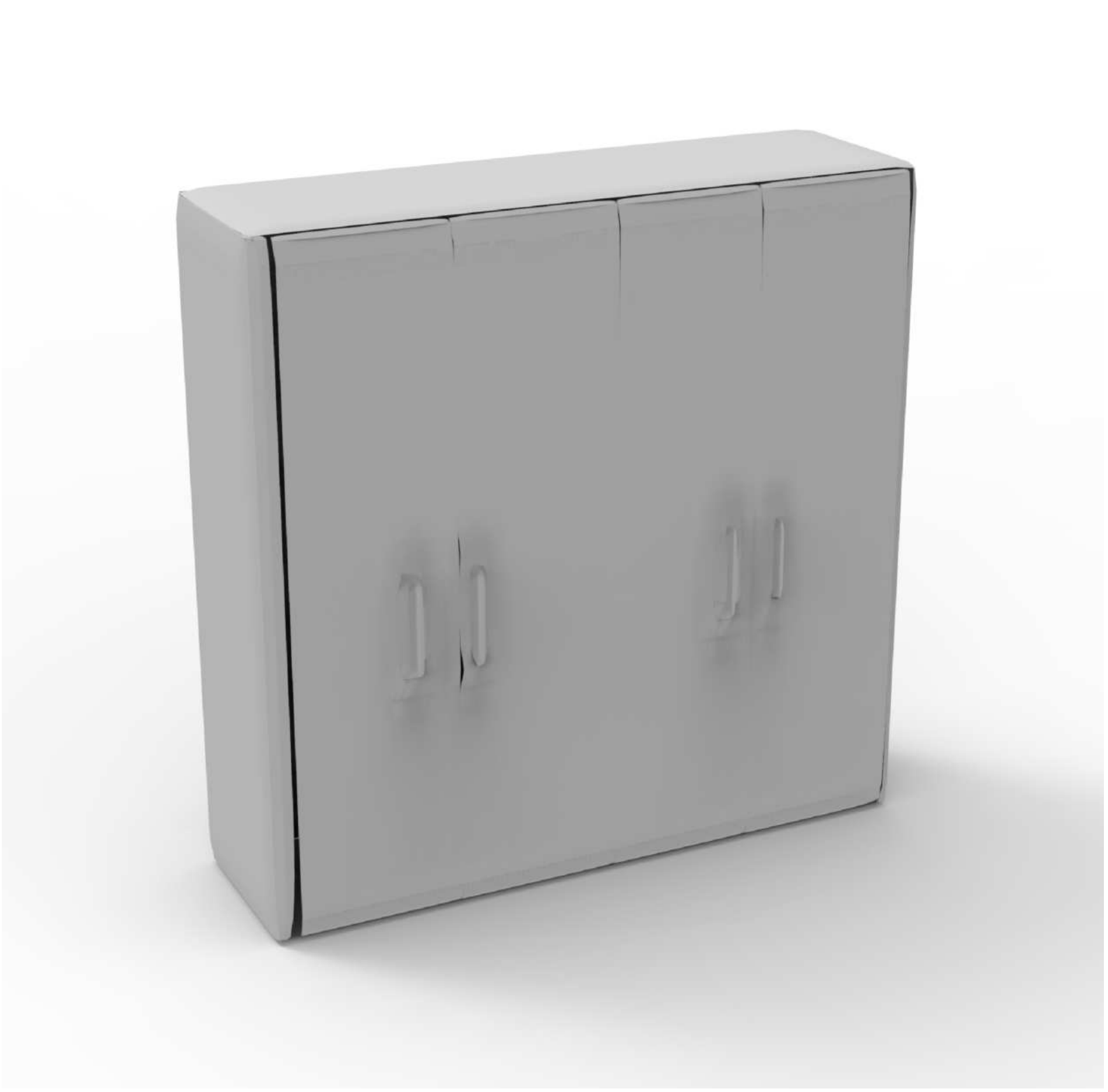}
    \includegraphics[width=0.49\linewidth]{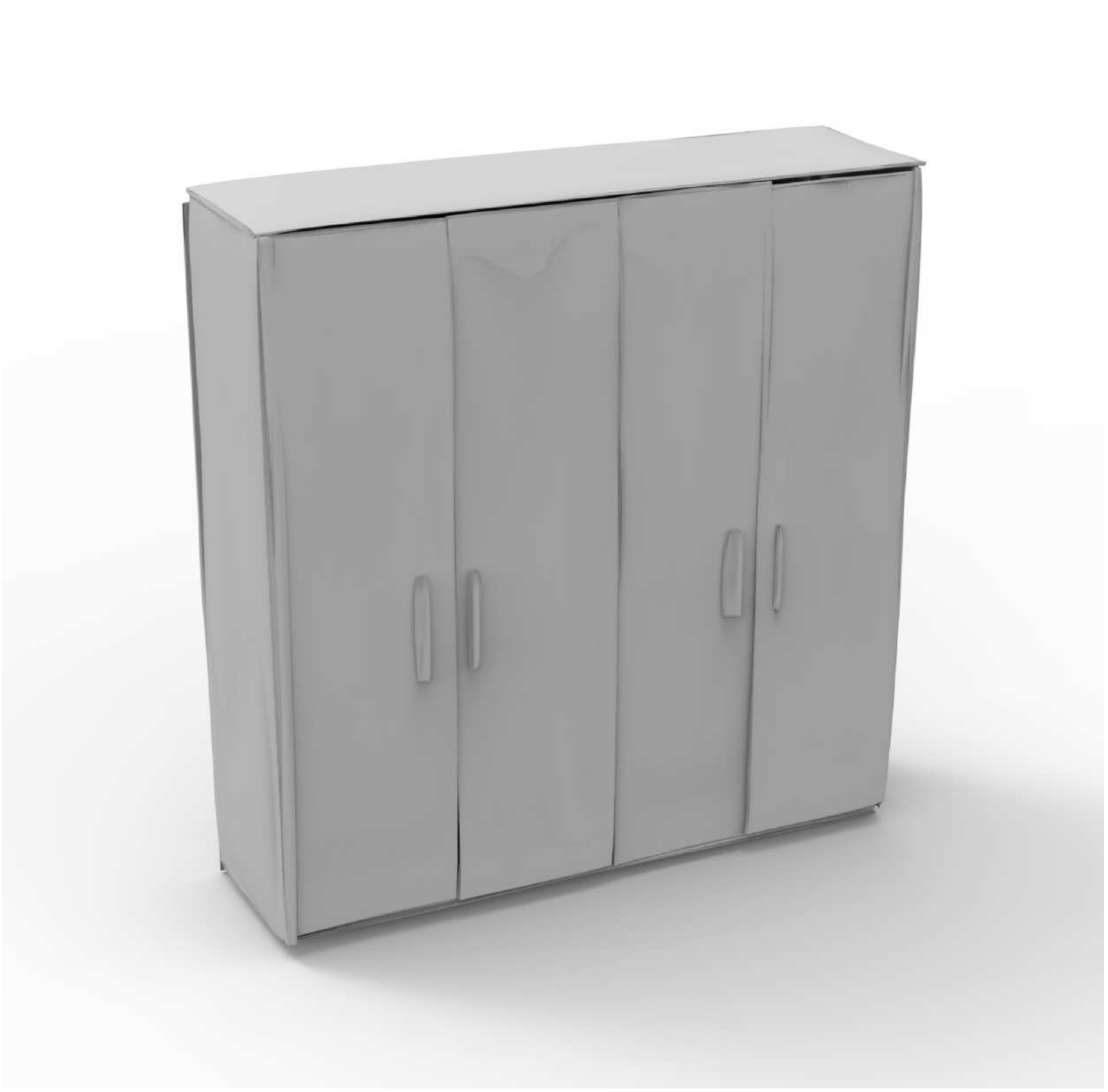}\\
    \includegraphics[width=0.49\linewidth]{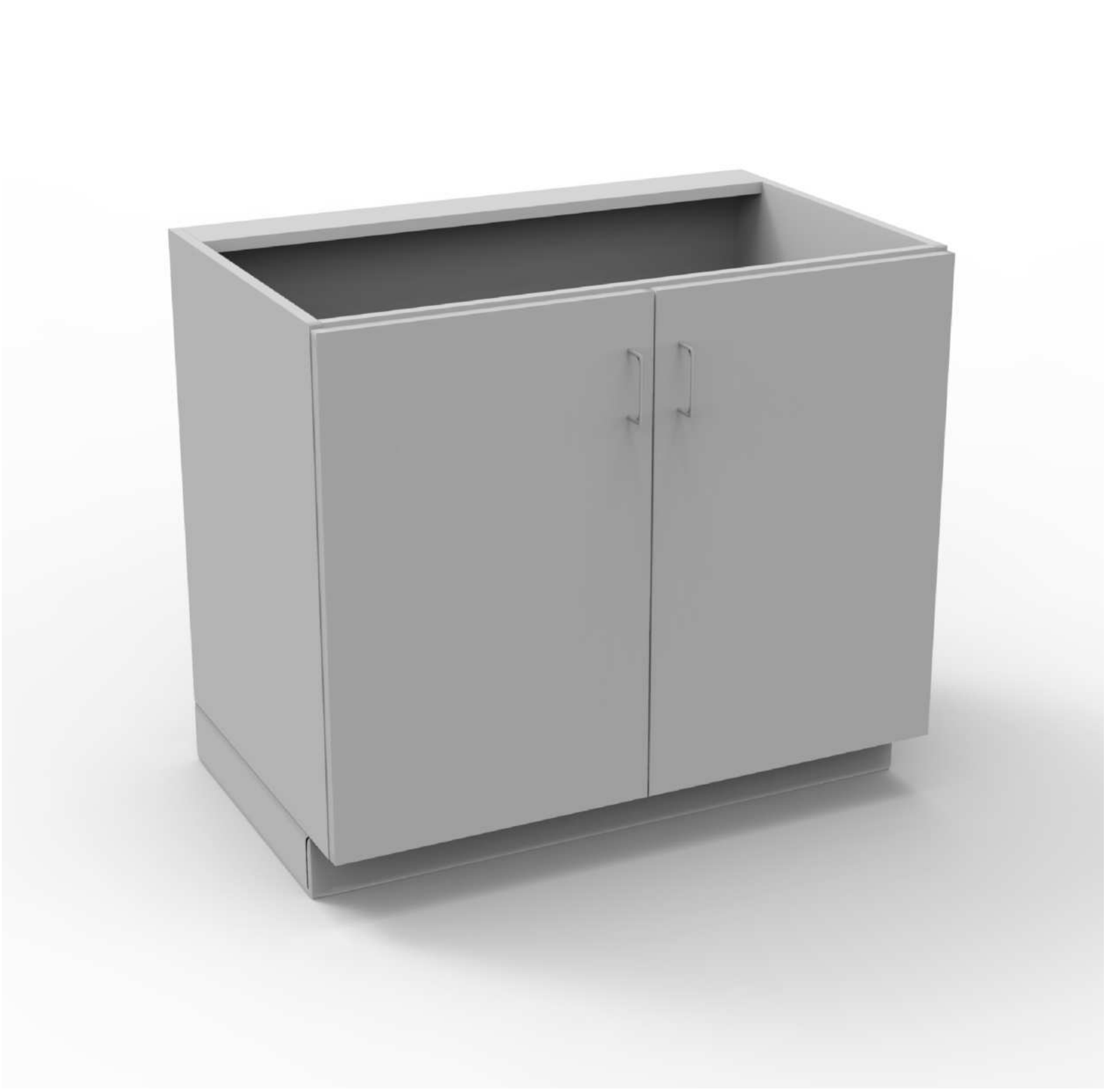}
    \includegraphics[width=0.49\linewidth]{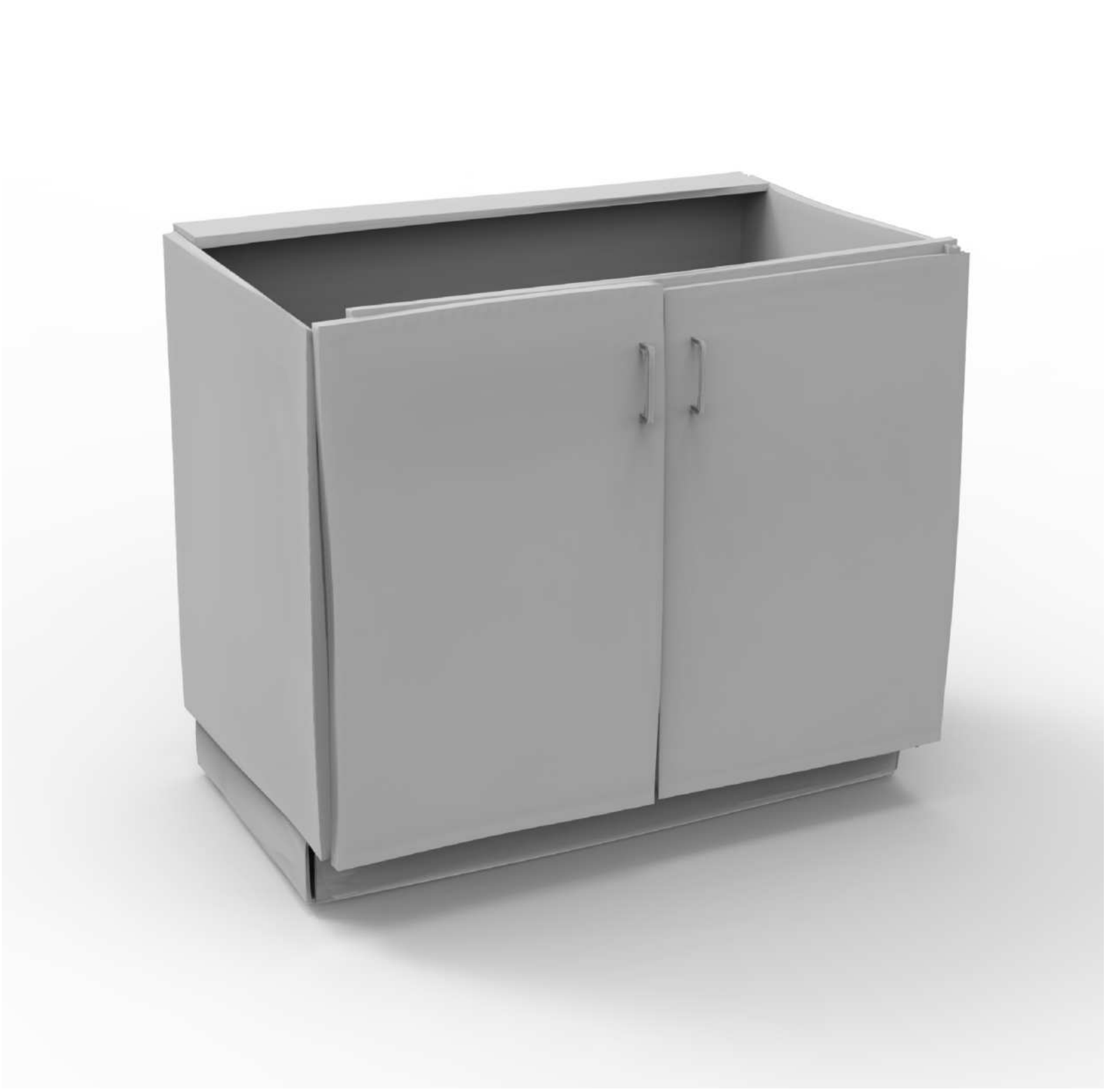}\\
    \includegraphics[width=0.48\linewidth]{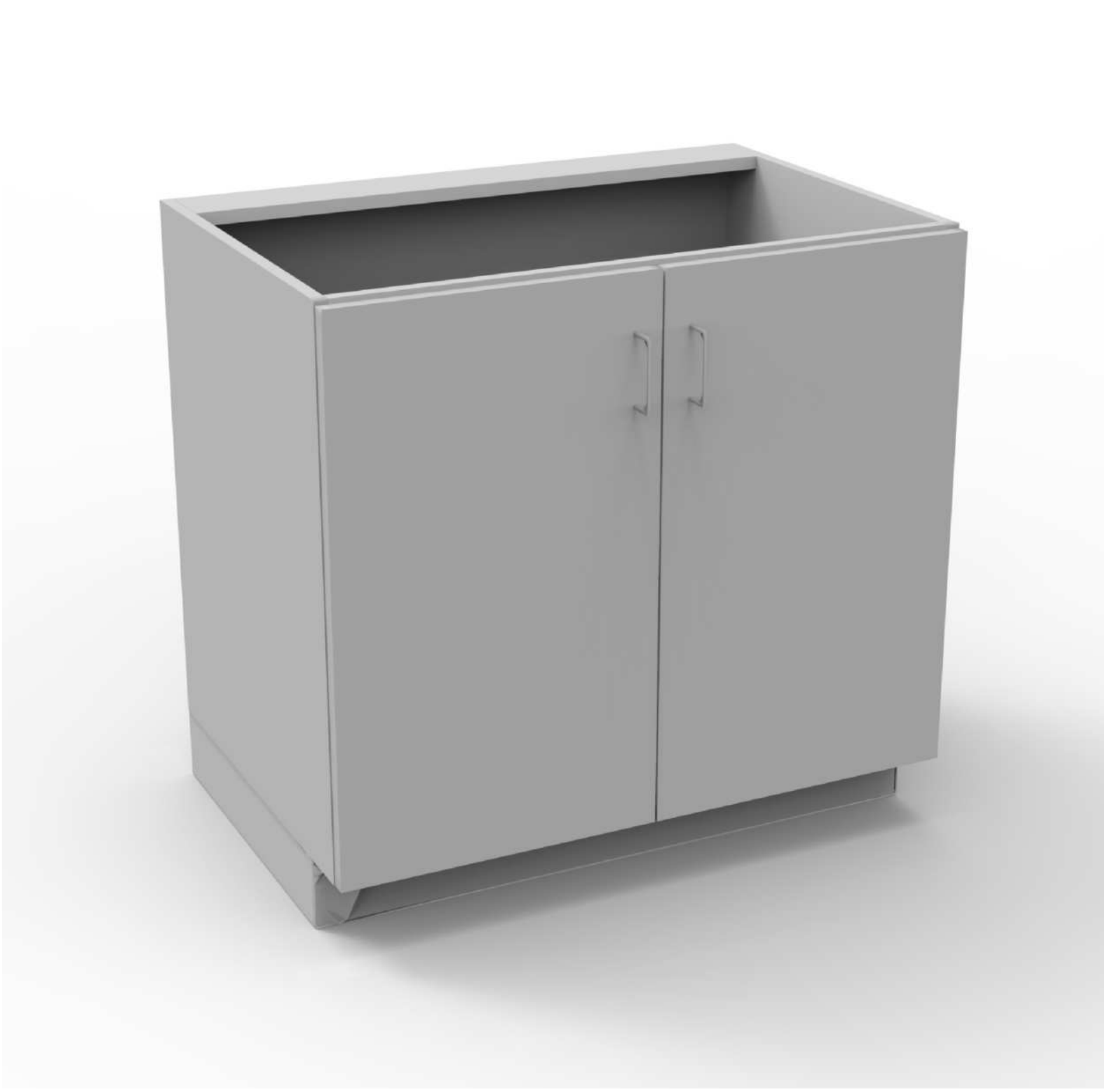}
    \includegraphics[width=0.48\linewidth]{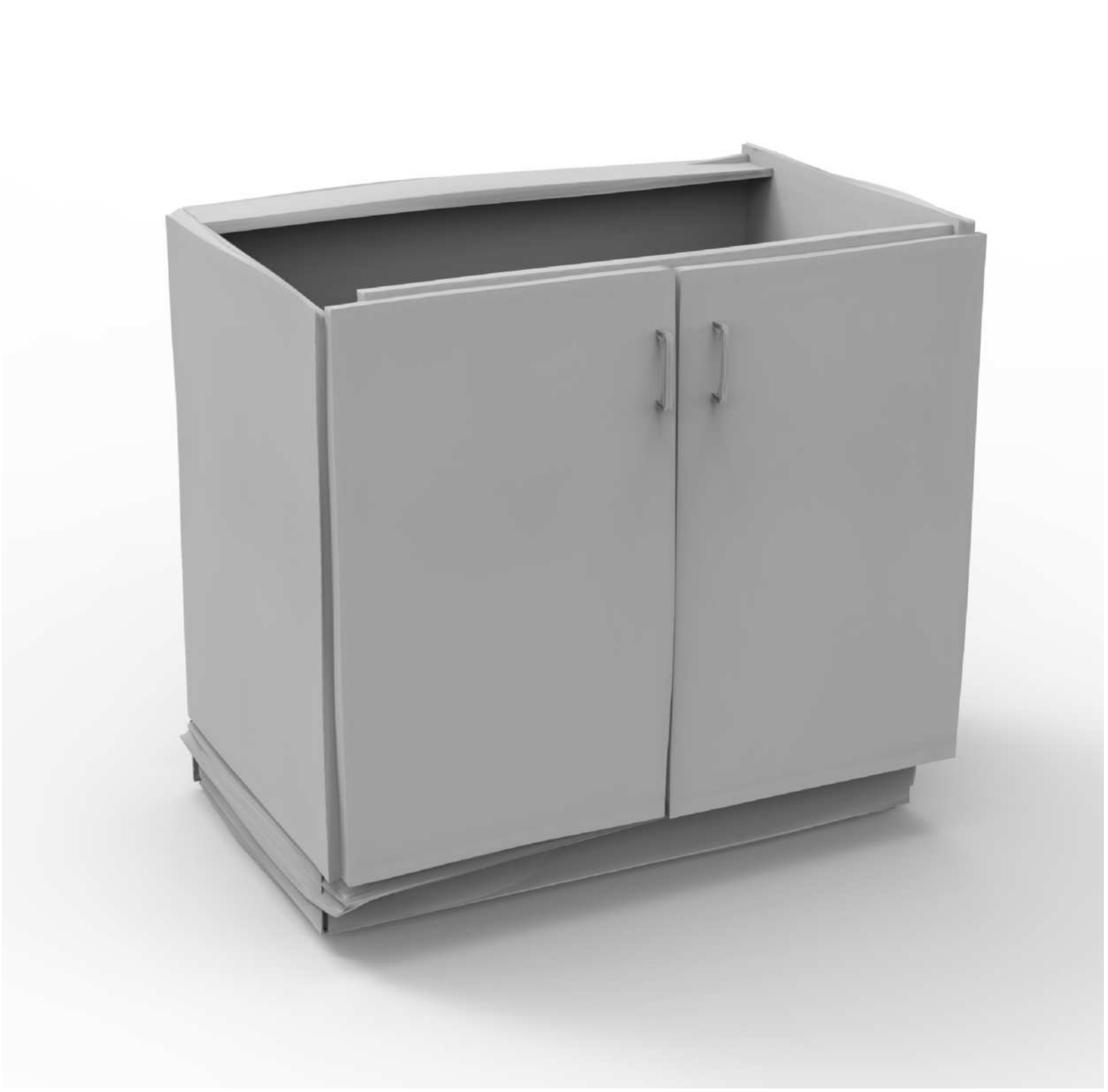}
    }
    \end{minipage}}
    \subfigure[Lamp]{
    \begin{minipage}[b]{0.24\linewidth}
    {
    \includegraphics[width=0.49\linewidth]{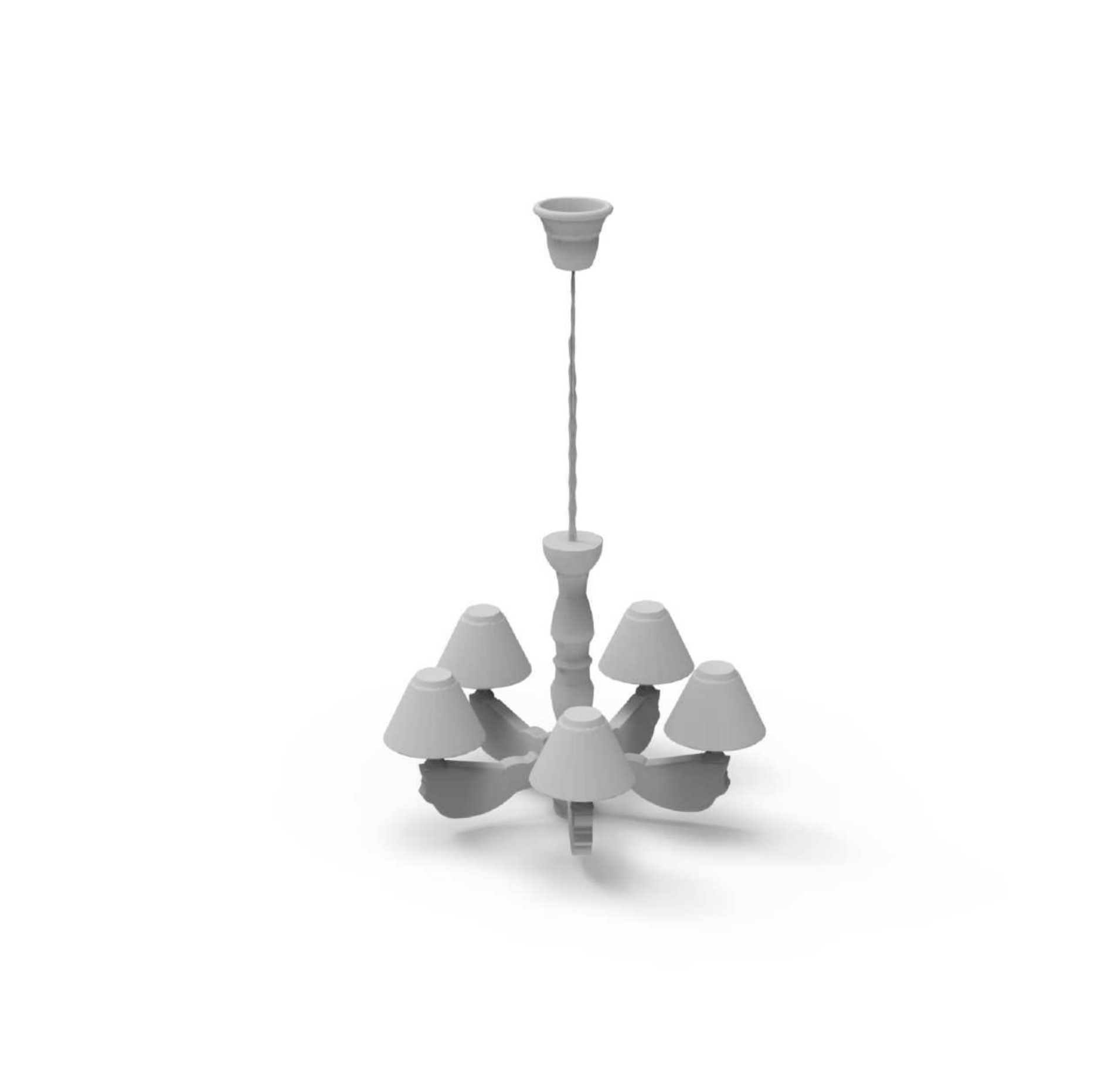}
    \includegraphics[width=0.49\linewidth]{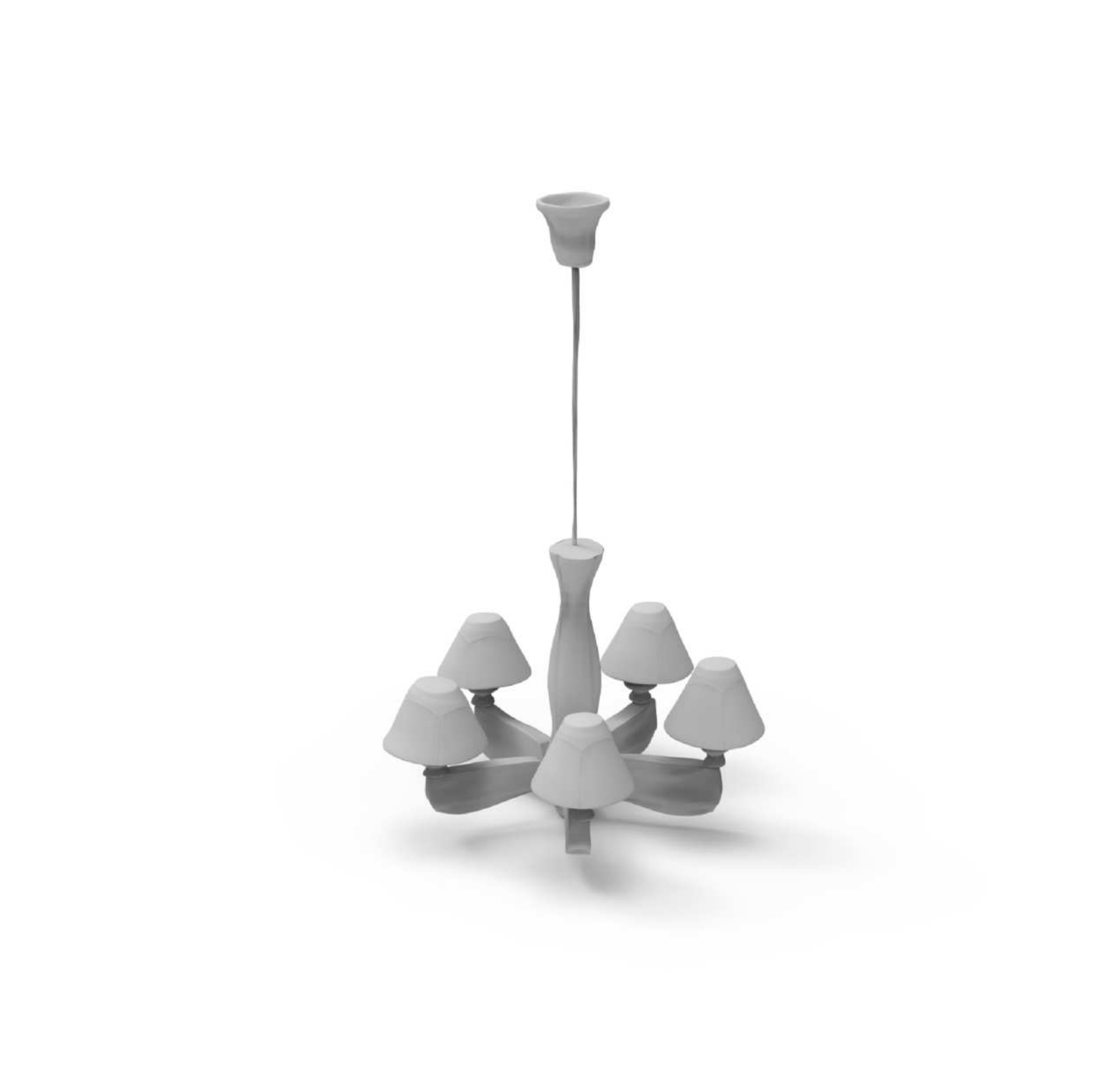}\\
    \includegraphics[width=0.49\linewidth]{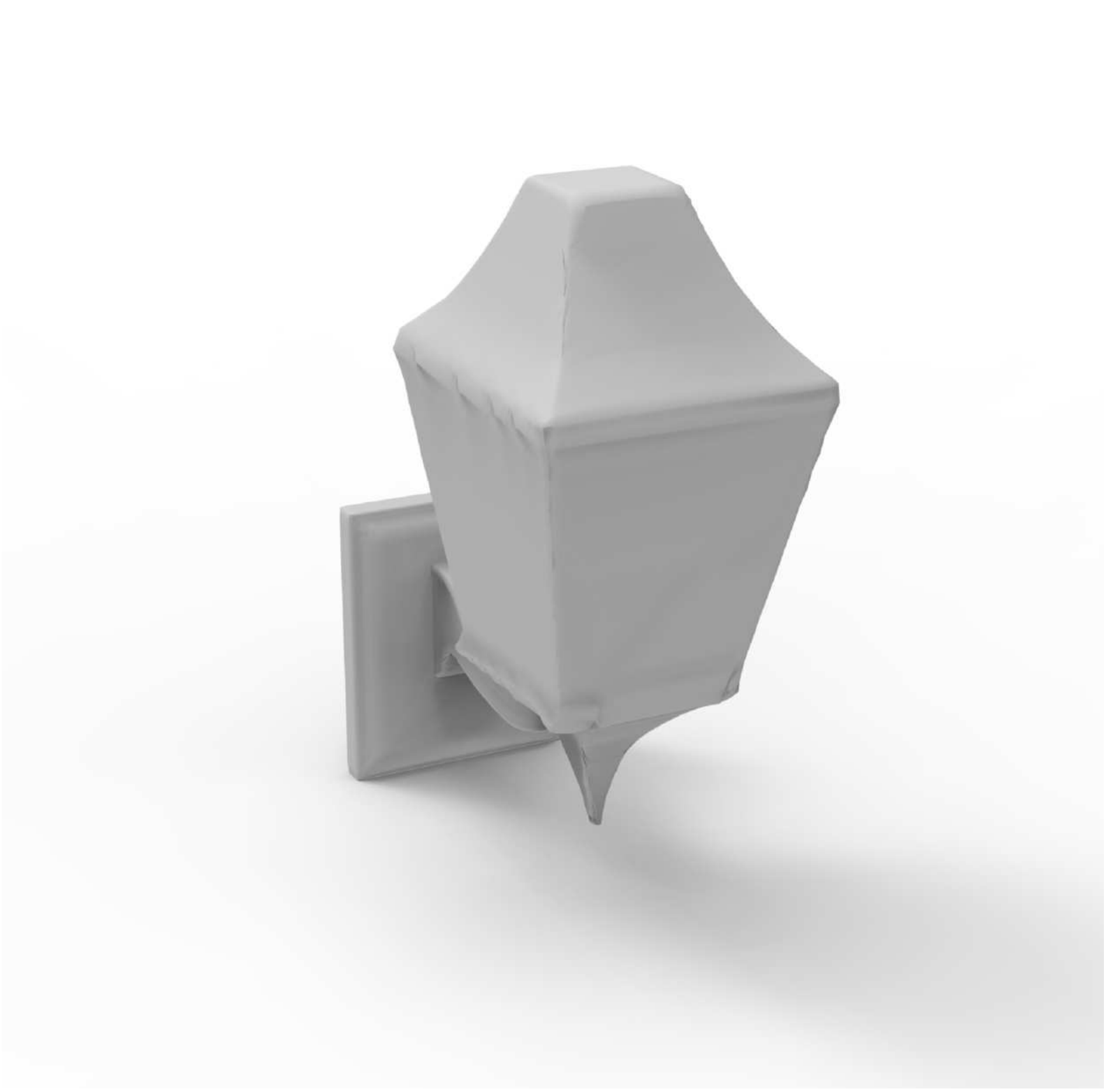}
    \includegraphics[width=0.49\linewidth]{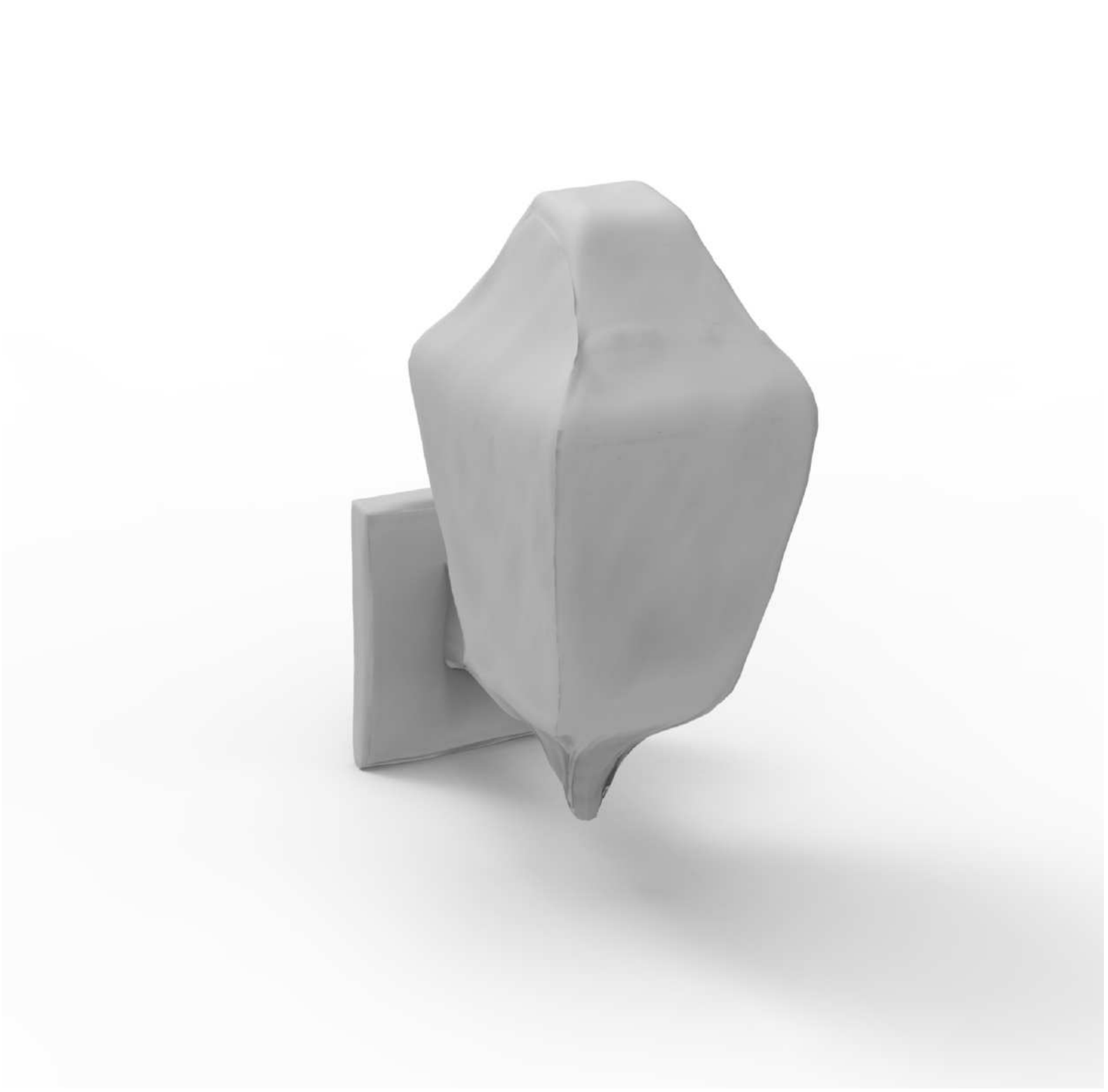}\\
    \includegraphics[width=0.48\linewidth]{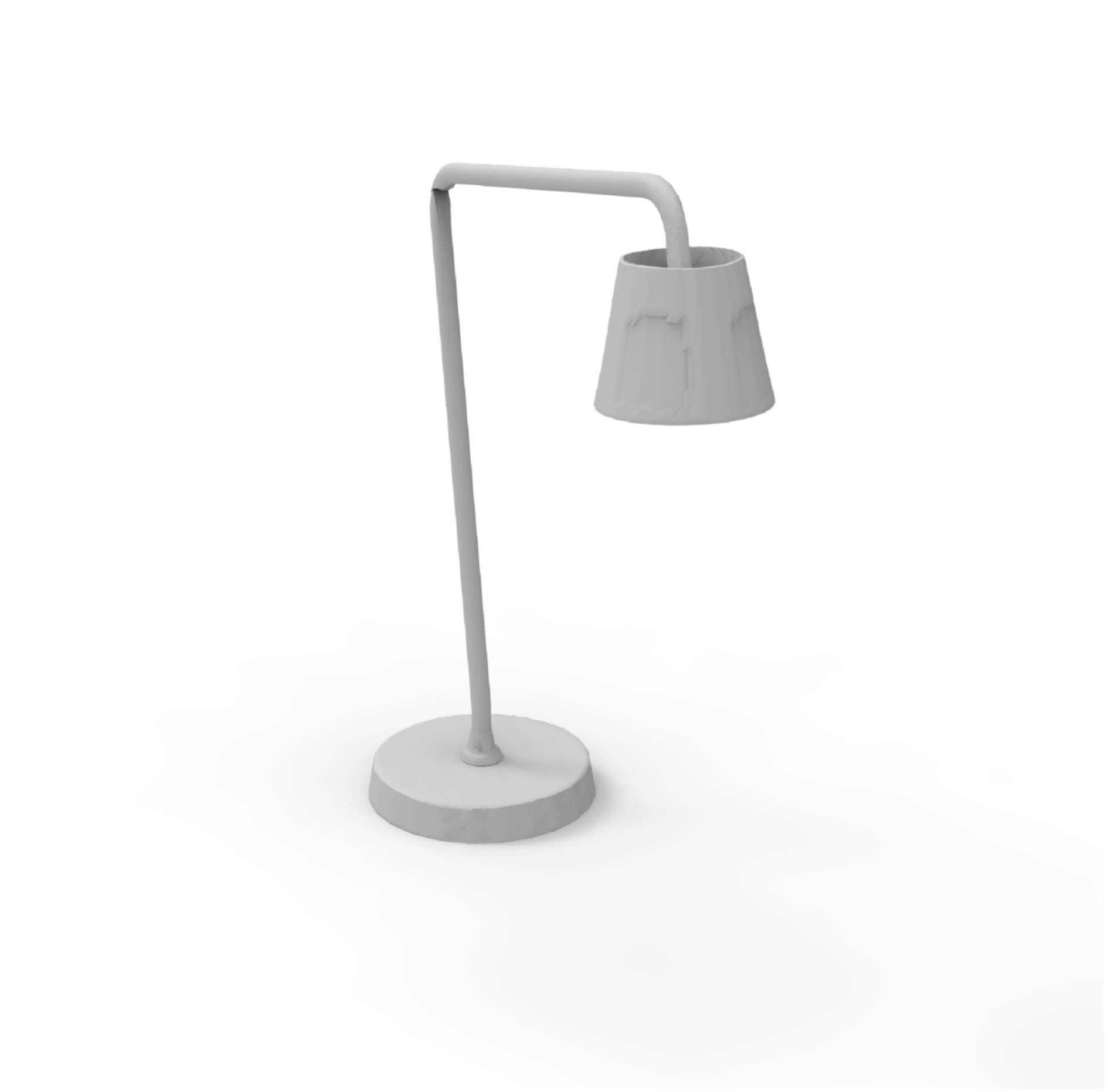}
    \includegraphics[width=0.48\linewidth]{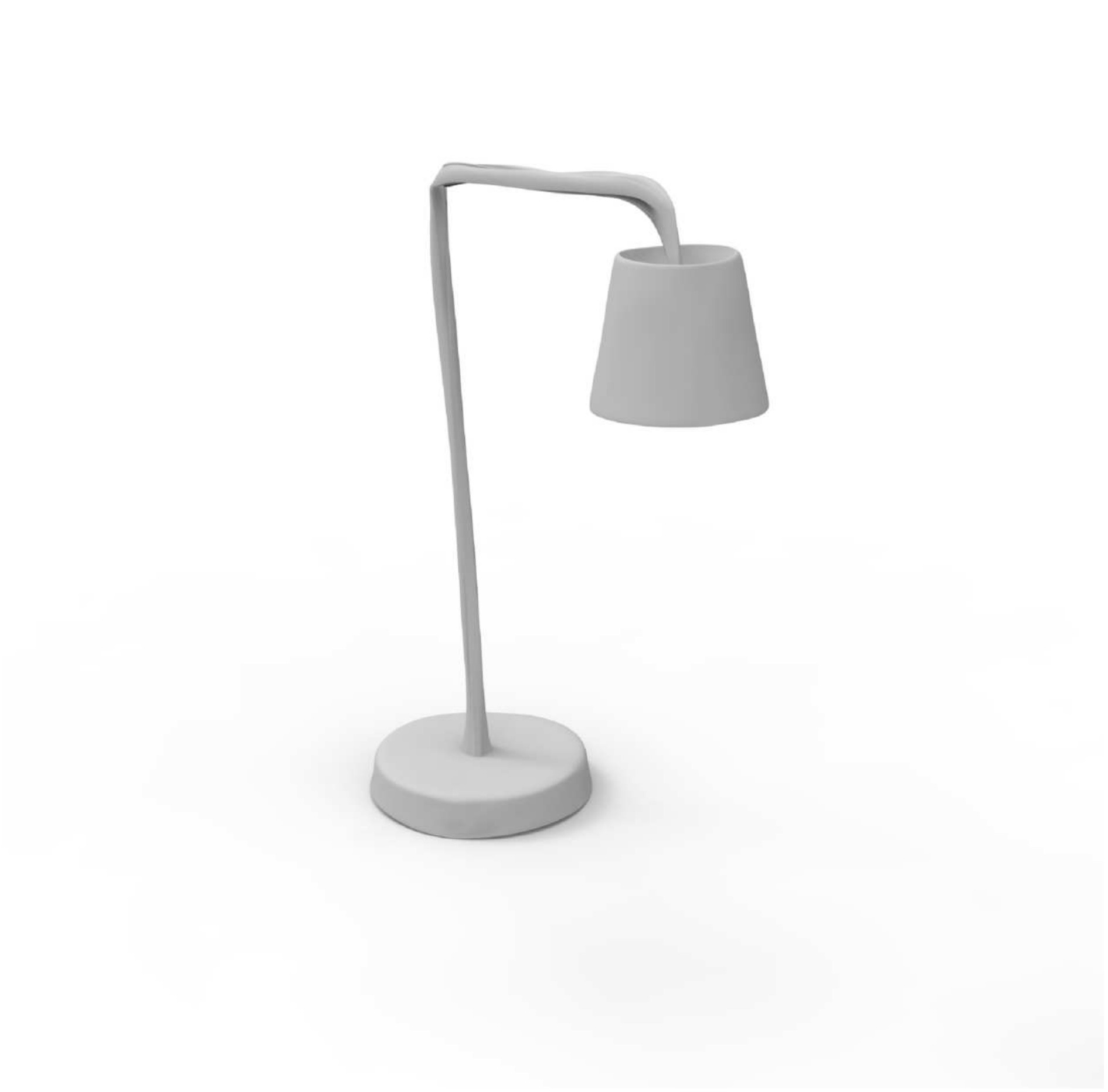}
    }
    \end{minipage}}
    \caption{\yjr{The gallery of shape reconstruction results on PartNet. For each set of results, the left column shows the ground-truth targets, and the right column presents our reconstruction results. We observe that our method can capture complex shape structures and detailed part geometry at the same time.}}
    \label{fig:reconstruction}
\end{figure*}

\begin{figure*}[t]
    \centering
    \subfigure[Shape Generation]{
    \begin{minipage}[b]{0.23\linewidth}
    \includegraphics[width=0.48\linewidth]{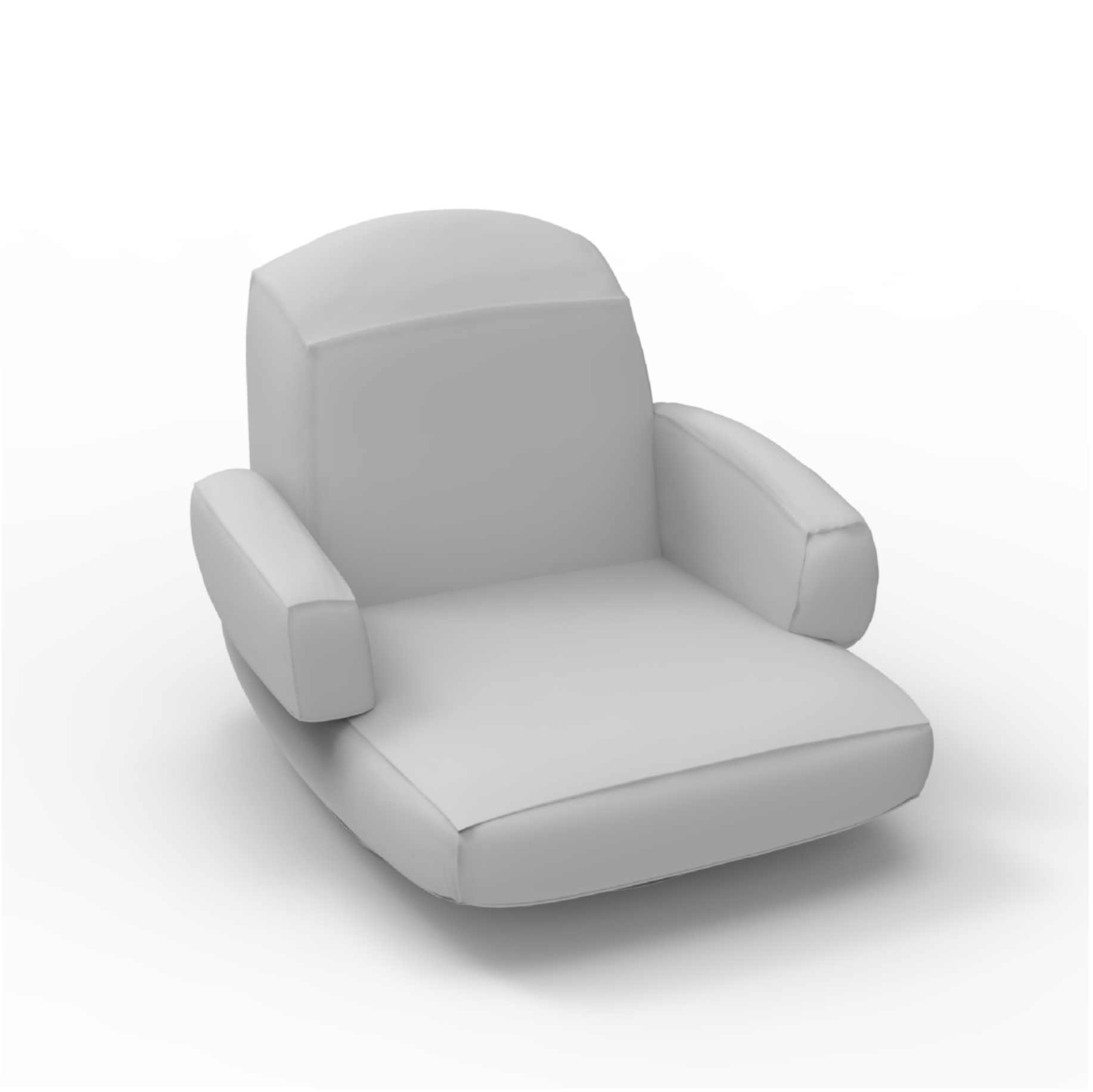}
    \includegraphics[width=0.48\linewidth]{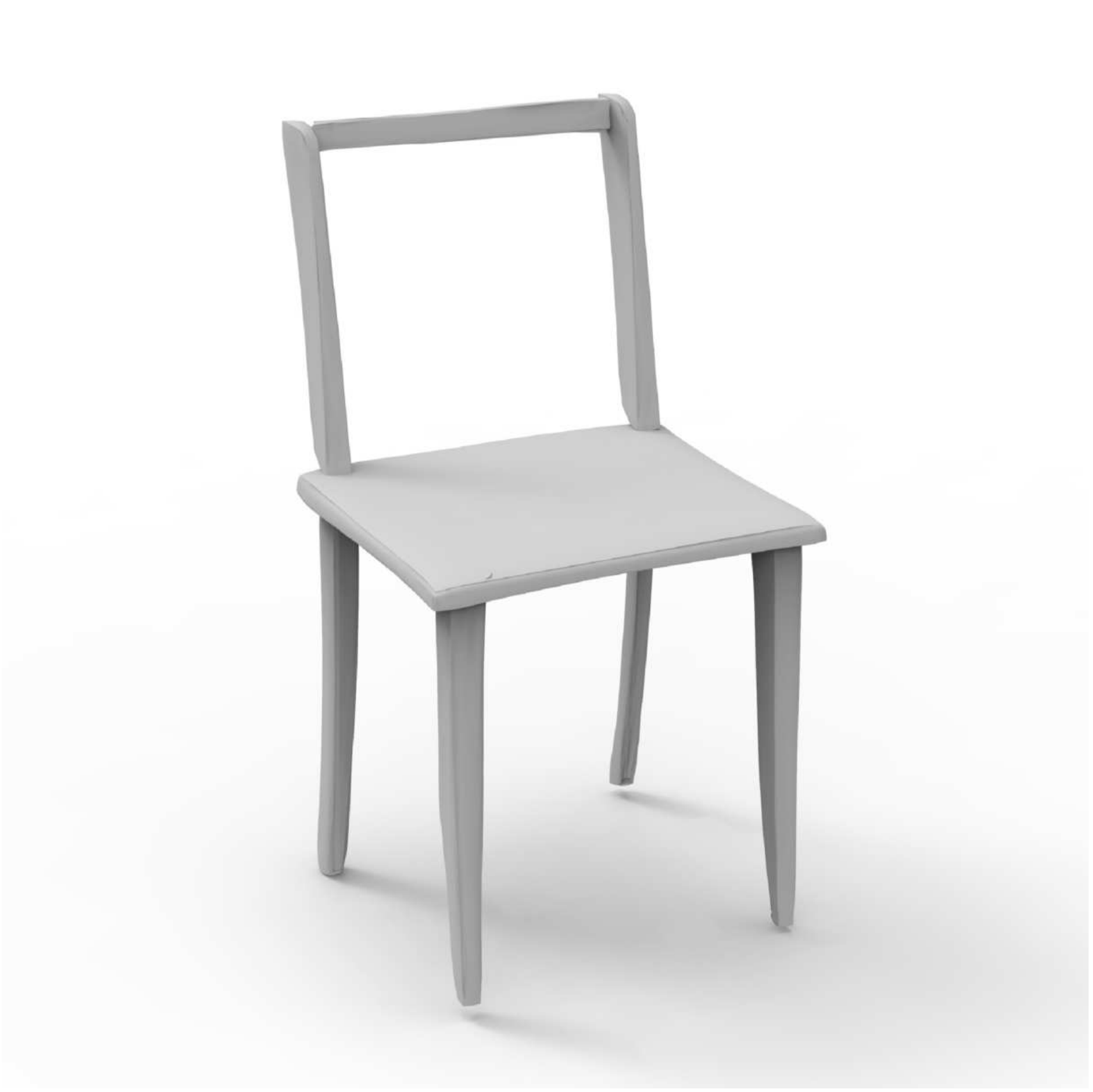}\\
    \includegraphics[width=0.48\linewidth]{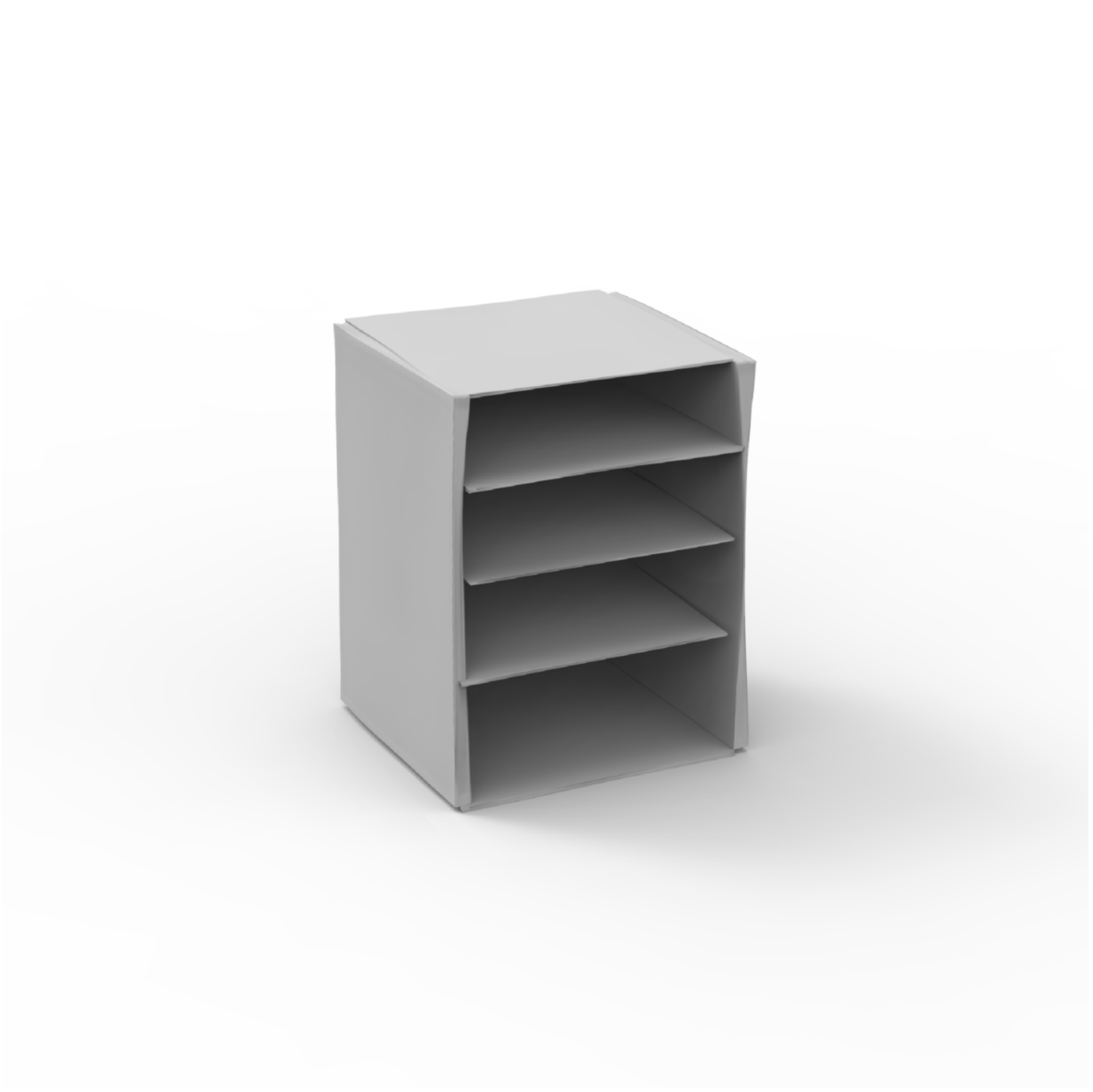}
    \includegraphics[width=0.48\linewidth]{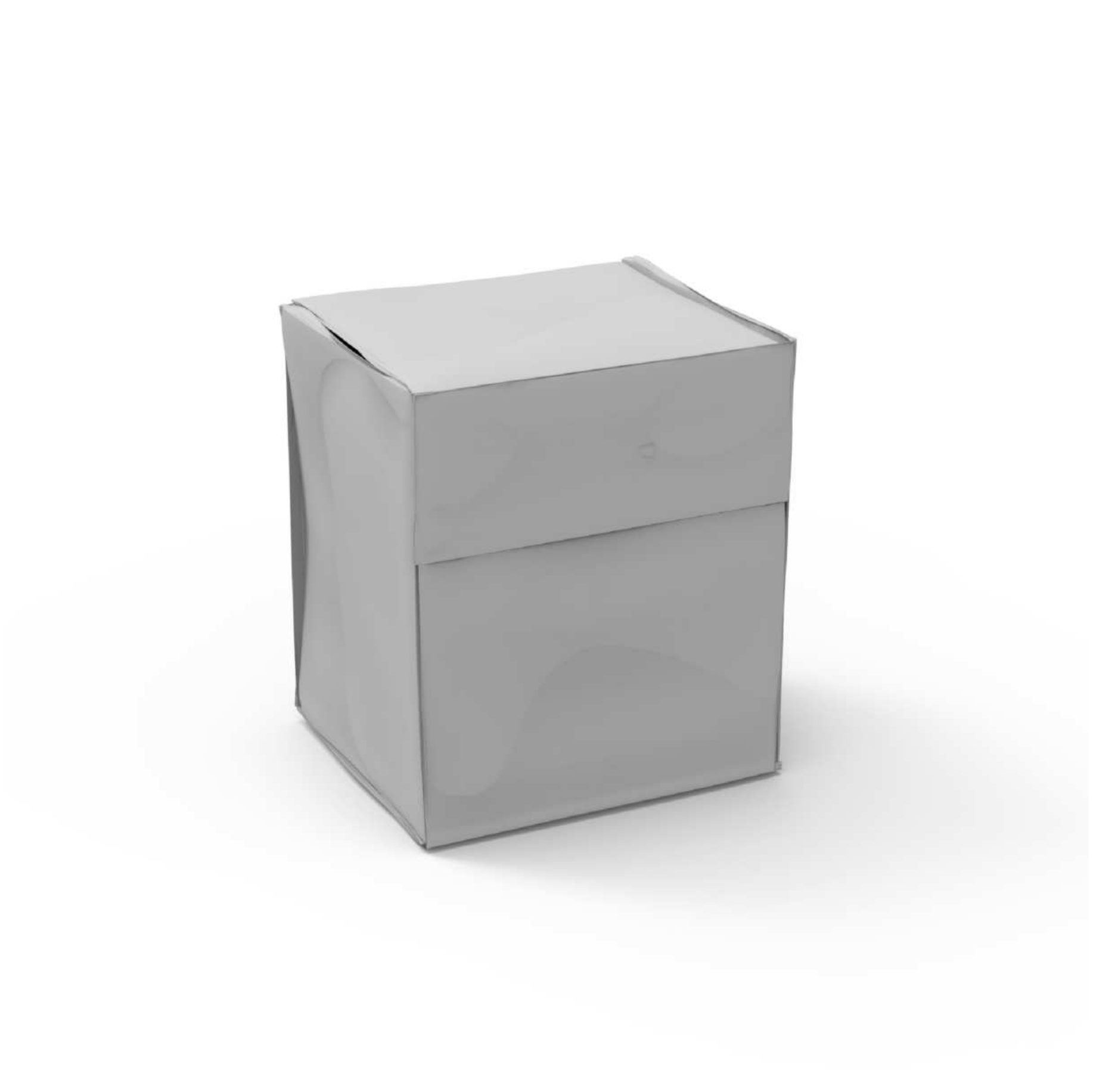}\\
    \includegraphics[width=0.48\linewidth]{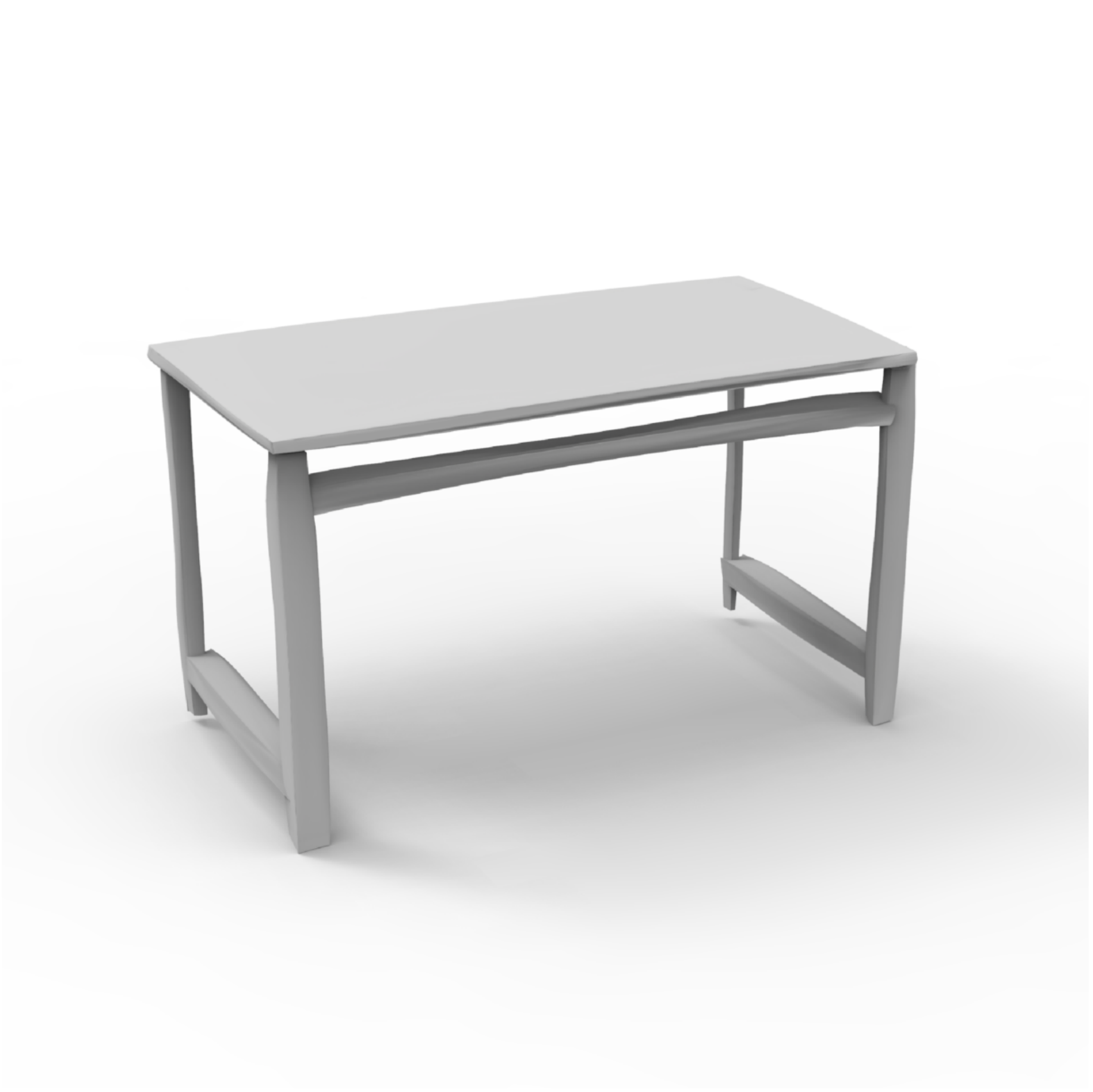}
    \includegraphics[width=0.48\linewidth]{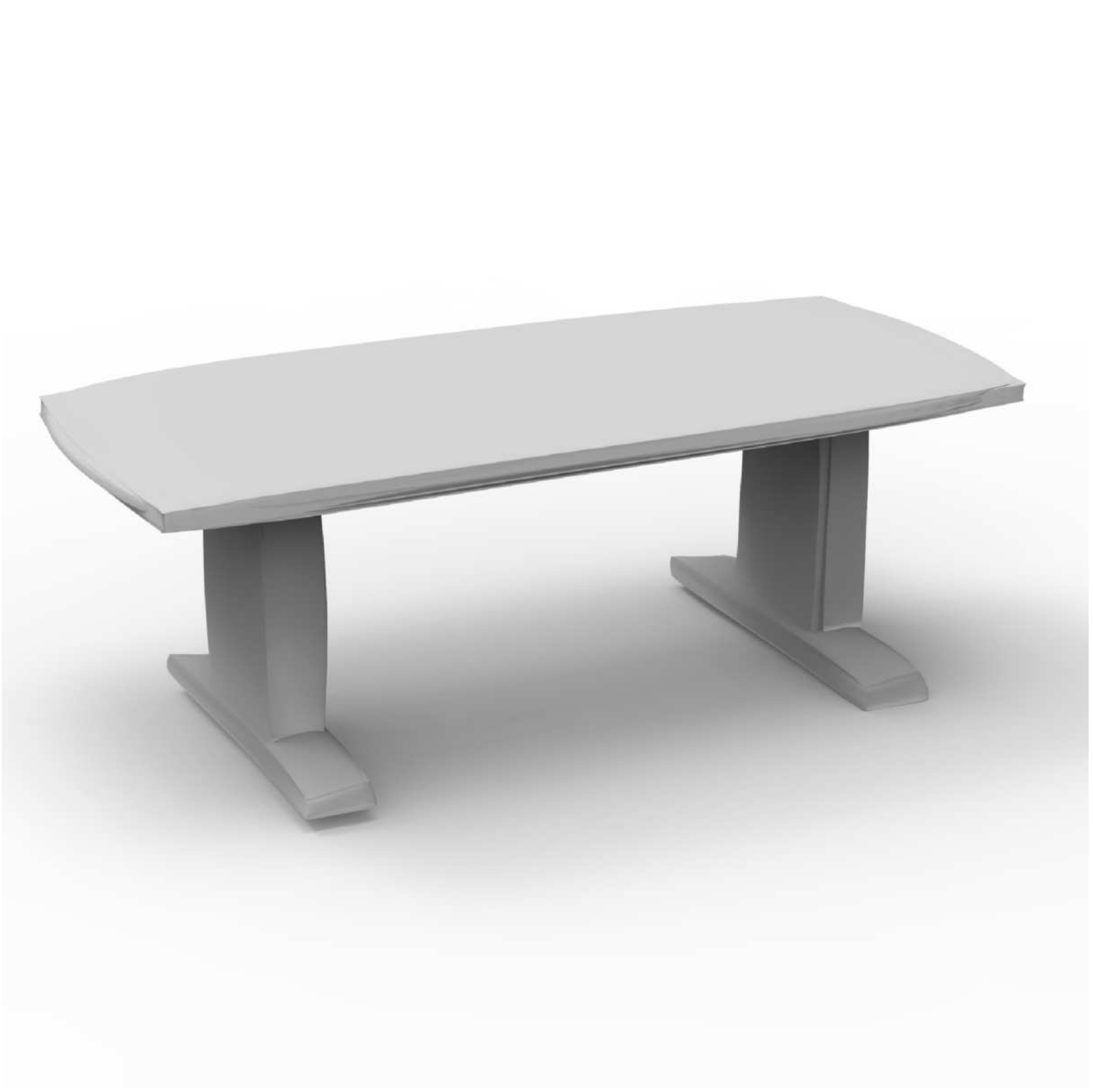}\\
    \includegraphics[width=0.48\linewidth]{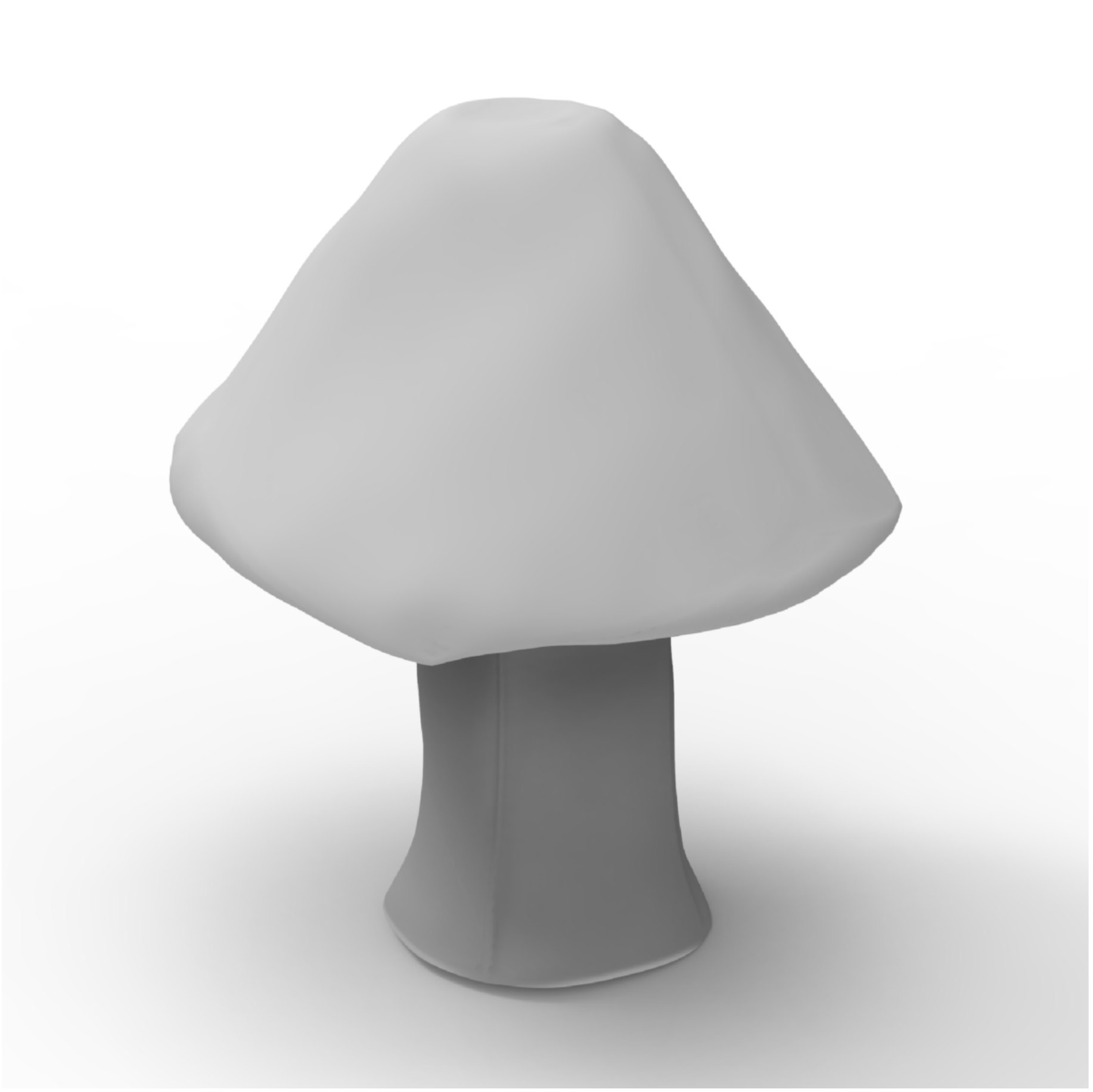}
    \includegraphics[width=0.48\linewidth]{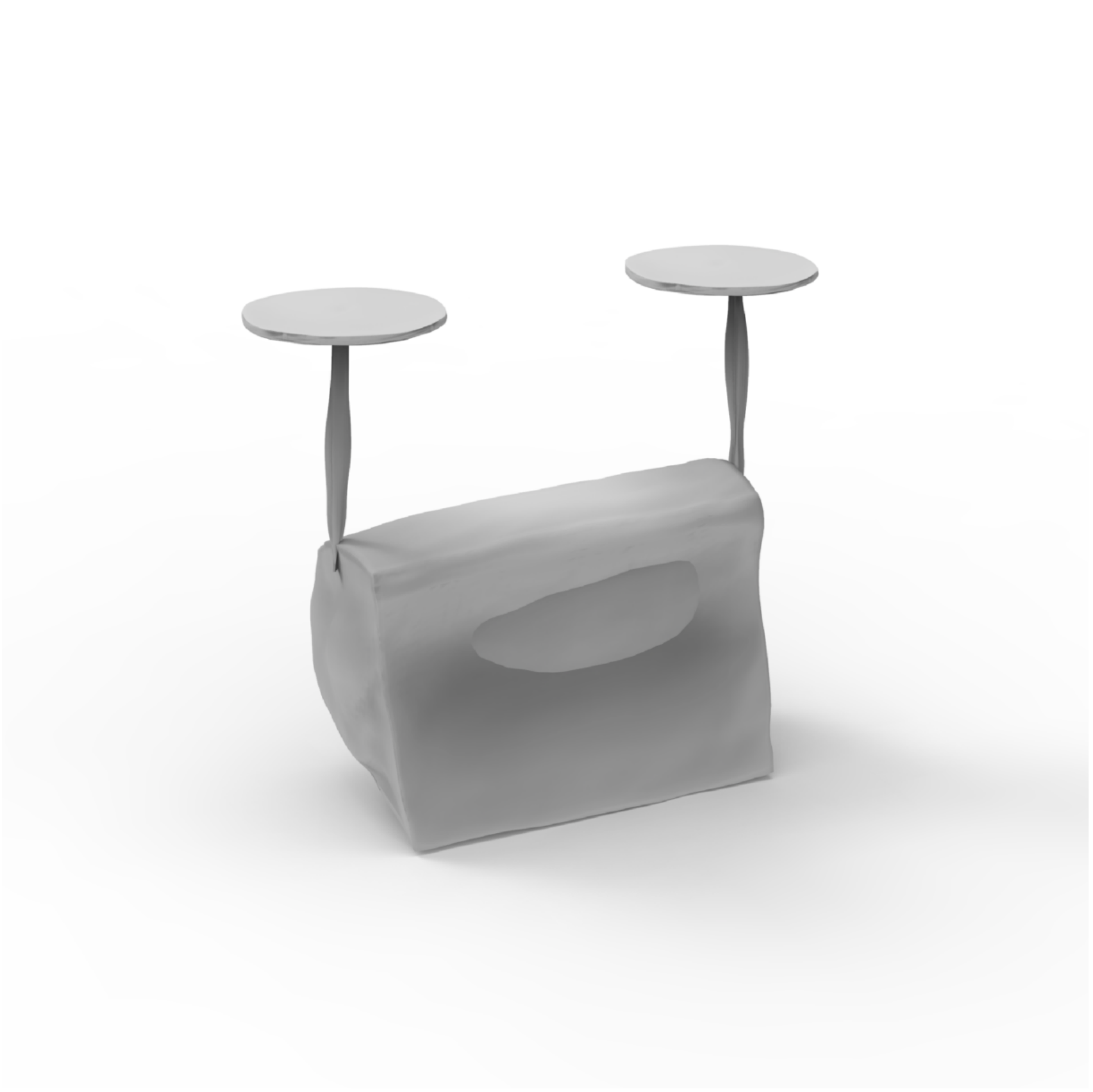}
    \end{minipage}\vline
    }
    \subfigure[Random Generation]{
    \begin{minipage}[b]{0.11\linewidth}
    \includegraphics[width=0.99\linewidth]{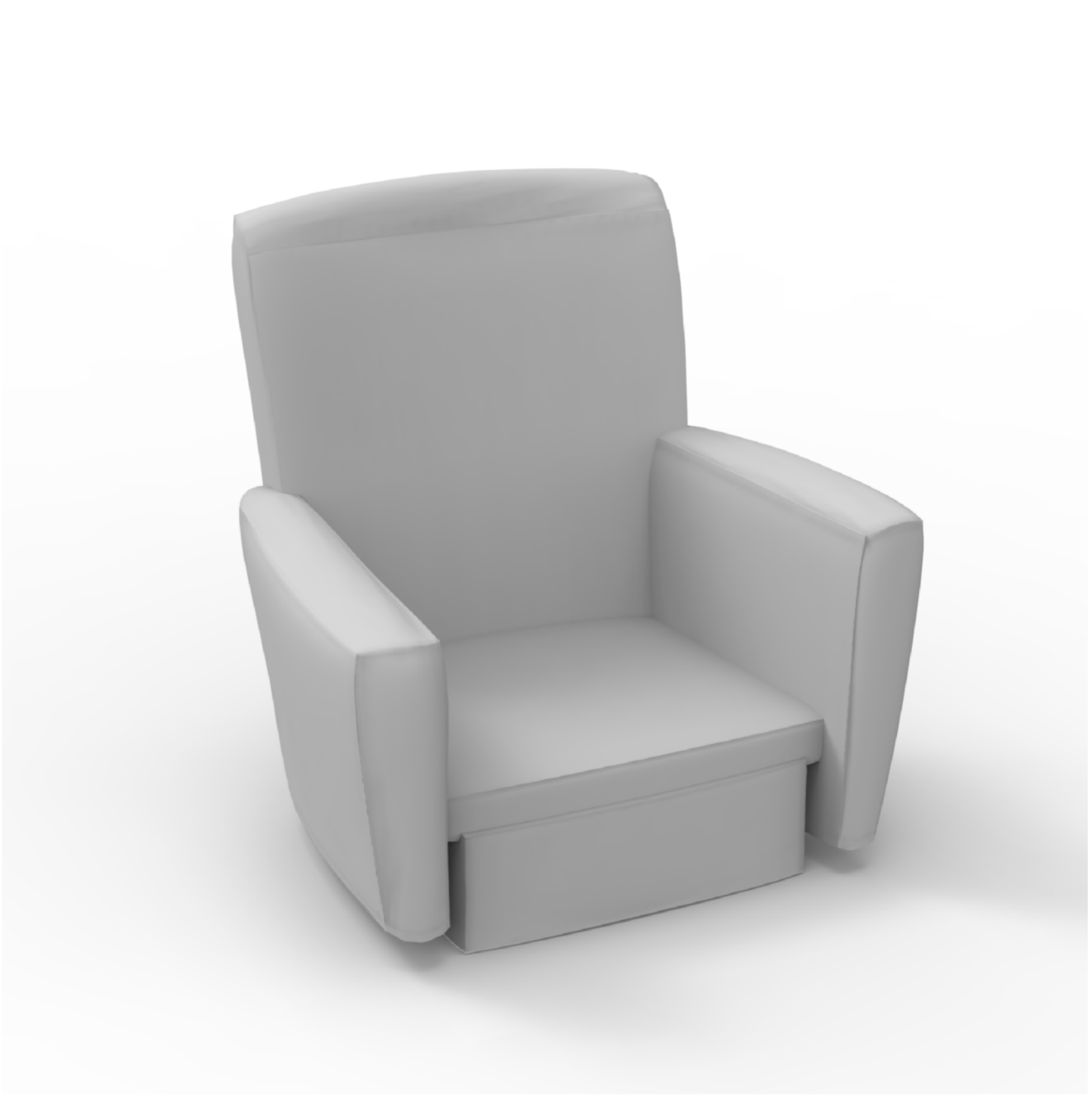}\vspace{1mm}\\
    \includegraphics[width=0.99\linewidth]{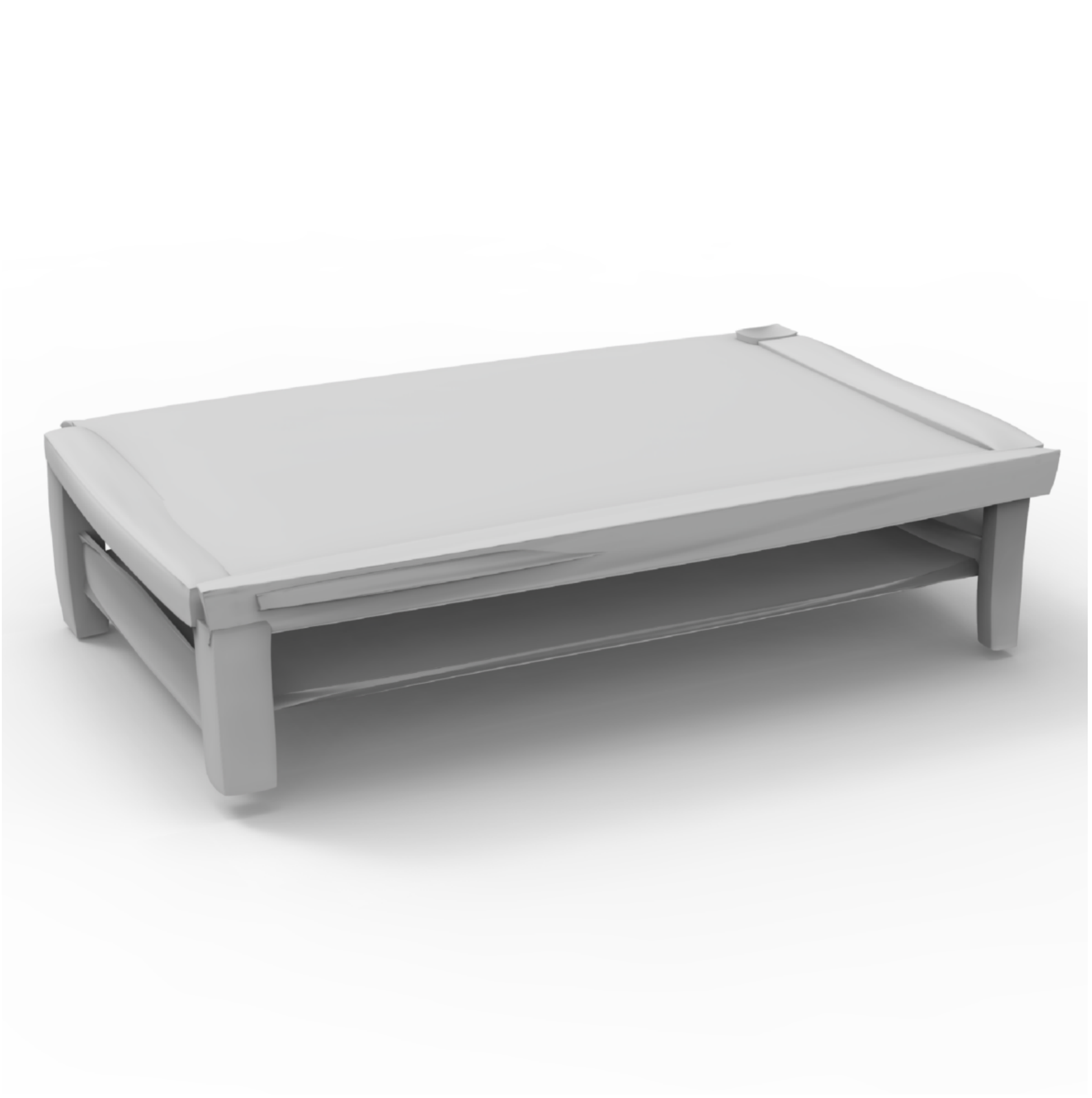}\\
    \includegraphics[width=0.99\linewidth]{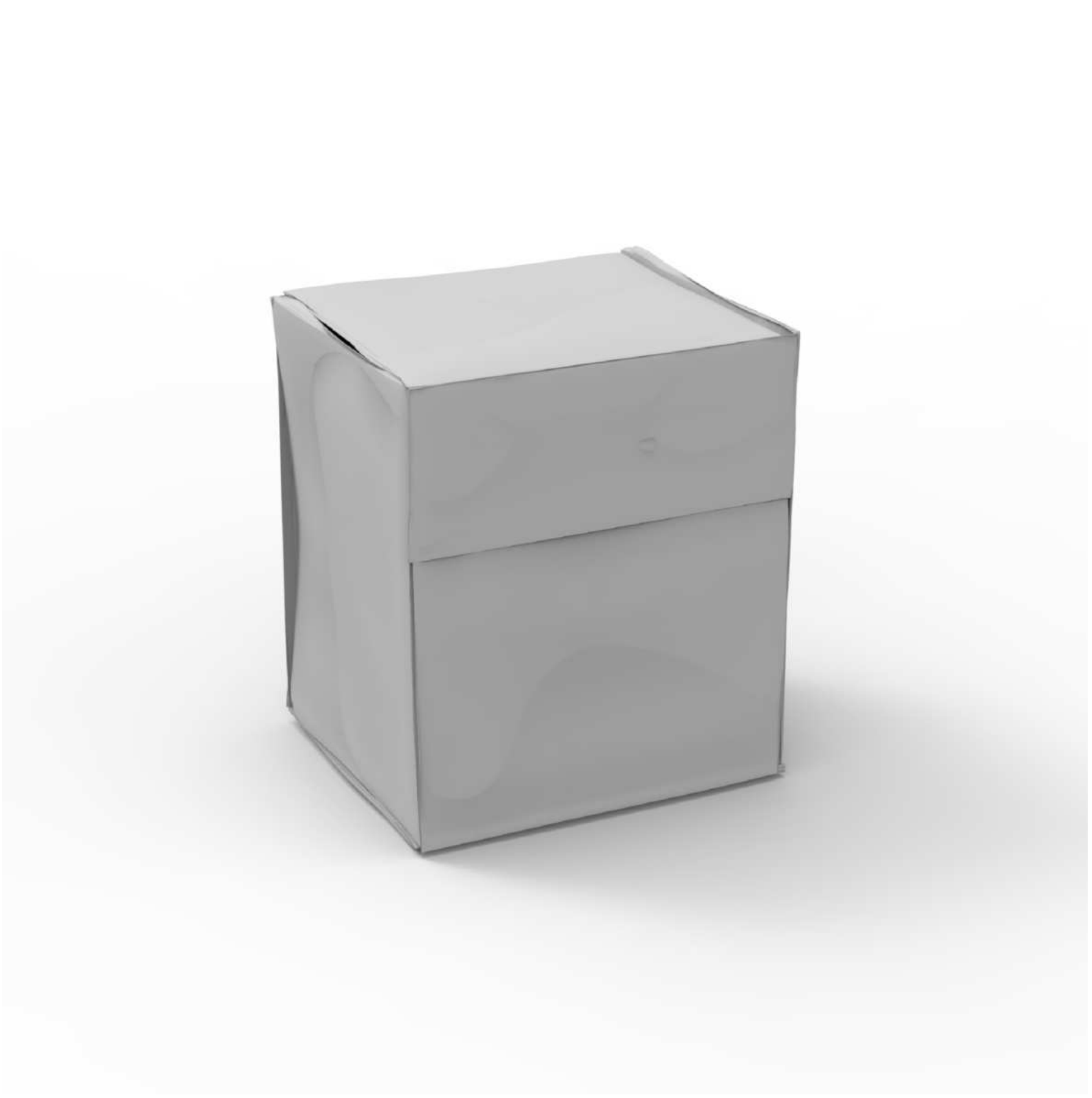}\vspace{1mm}\\
    \includegraphics[width=0.99\linewidth]{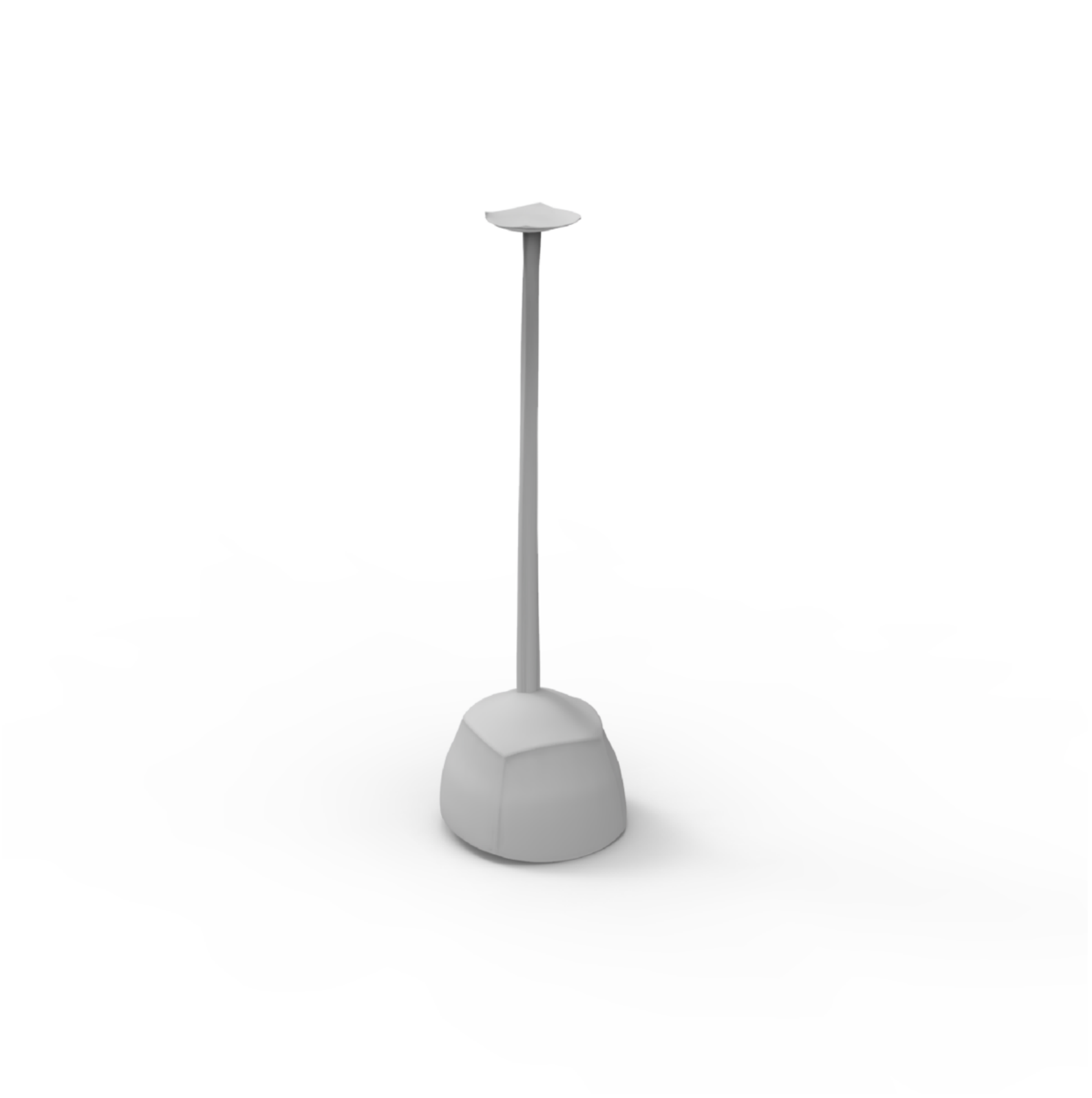}
    \end{minipage}}
    \subfigure[Top-5 retrieved shapes in training sets]{
    \begin{minipage}[b]{0.60\linewidth}
    \includegraphics[width=0.19\linewidth]{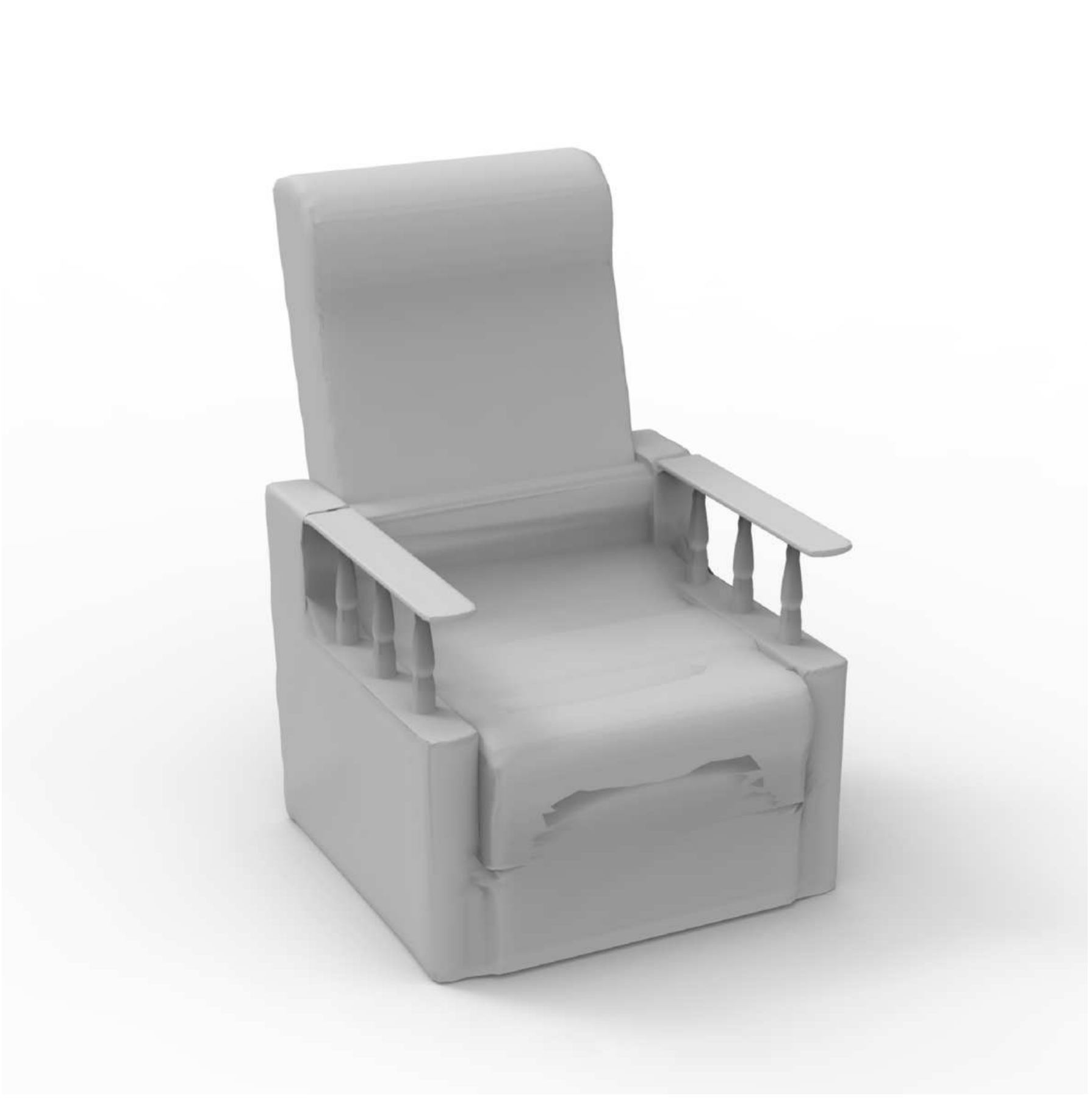}
    \includegraphics[width=0.19\linewidth]{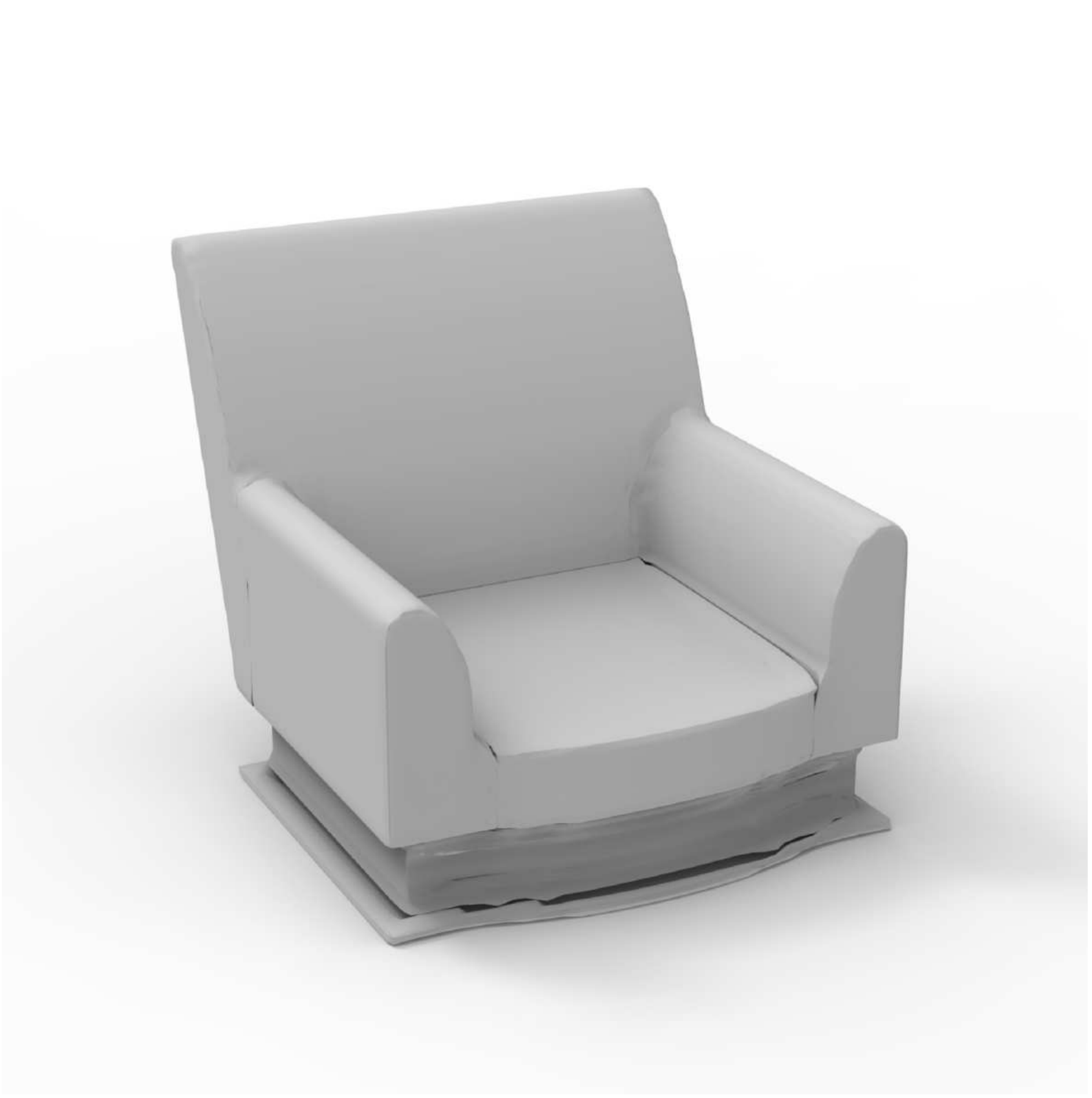}
    \includegraphics[width=0.19\linewidth]{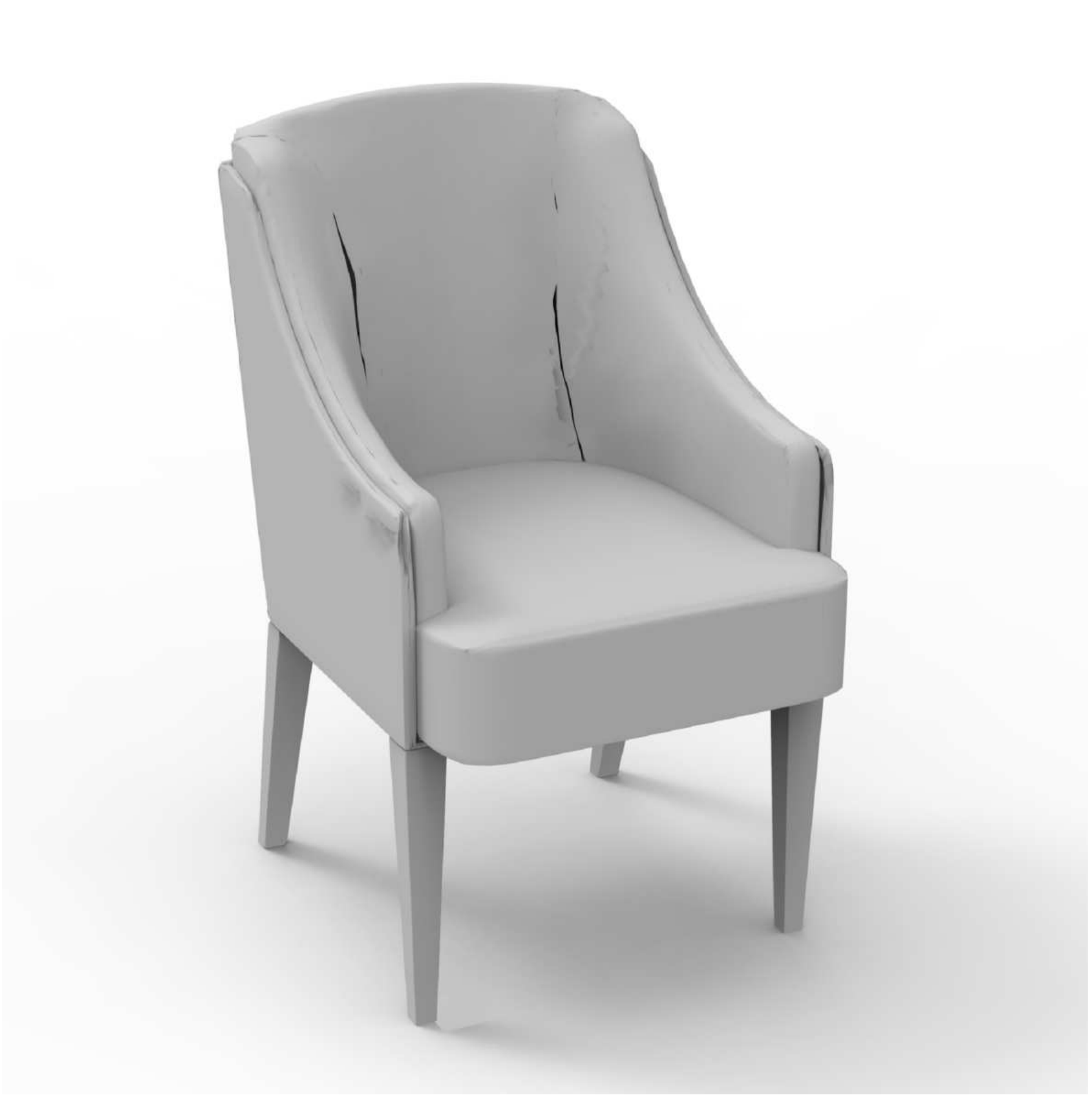}
    \includegraphics[width=0.19\linewidth]{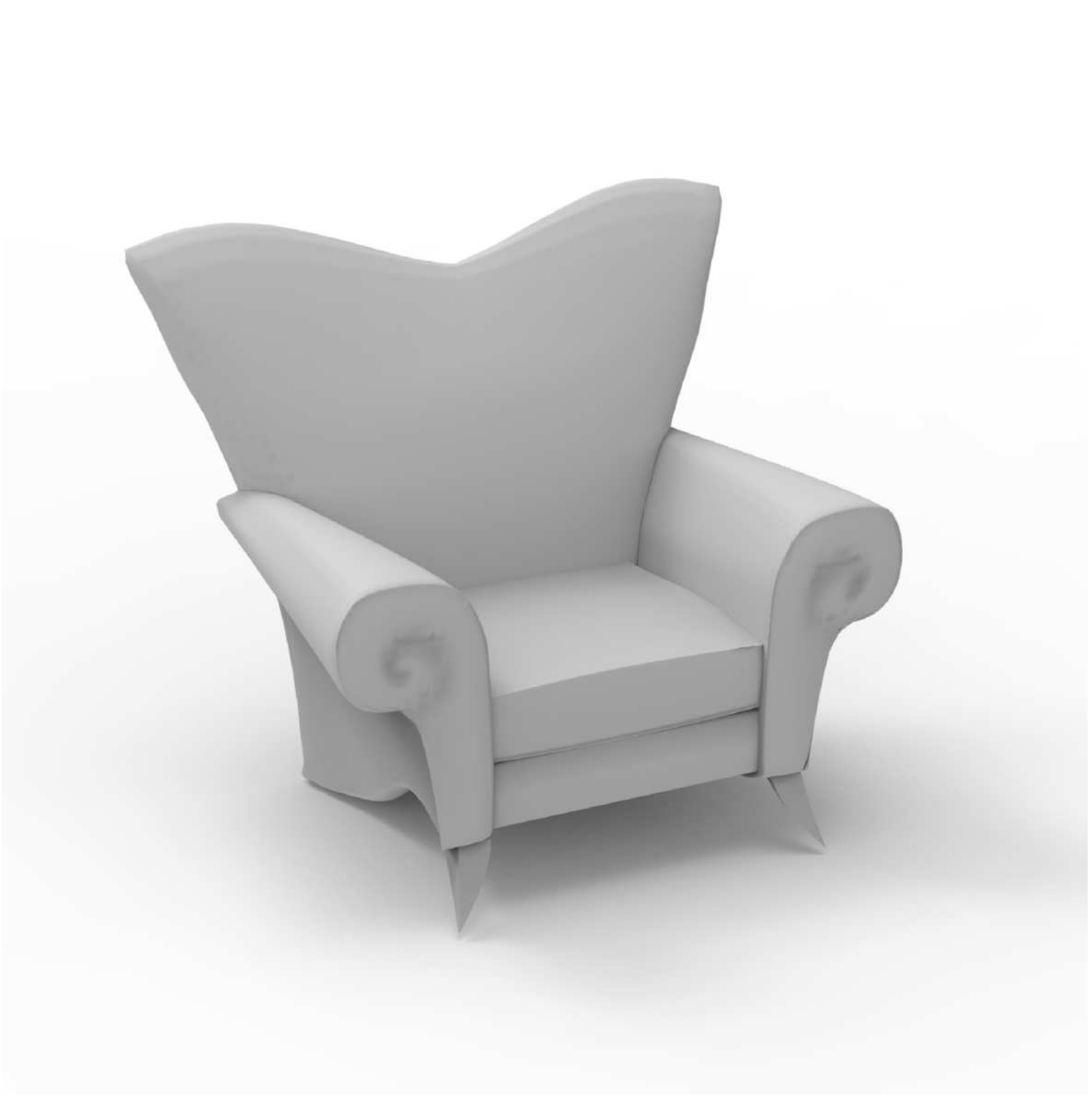}
    \includegraphics[width=0.19\linewidth]{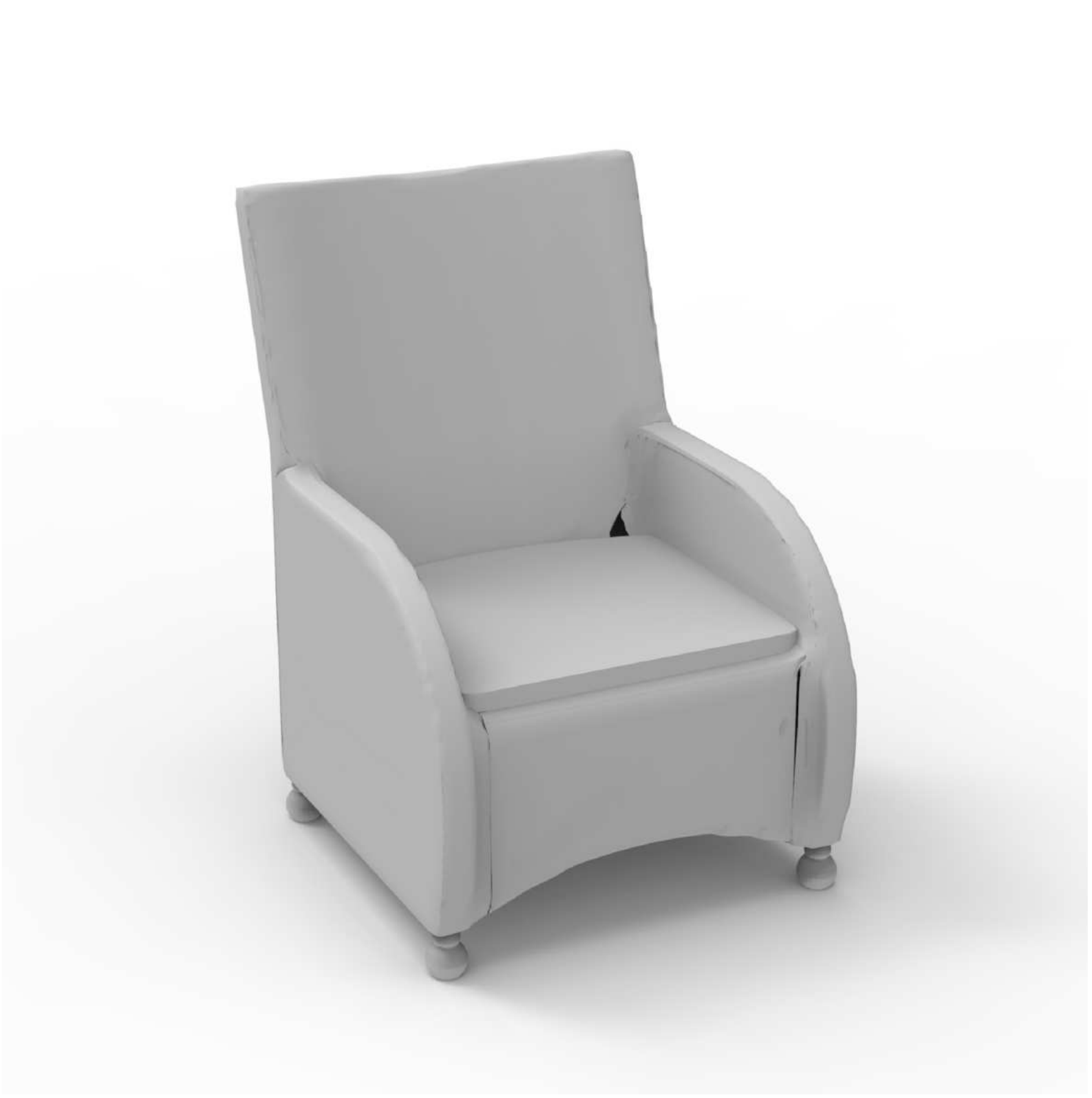}\\
    \includegraphics[width=0.19\linewidth]{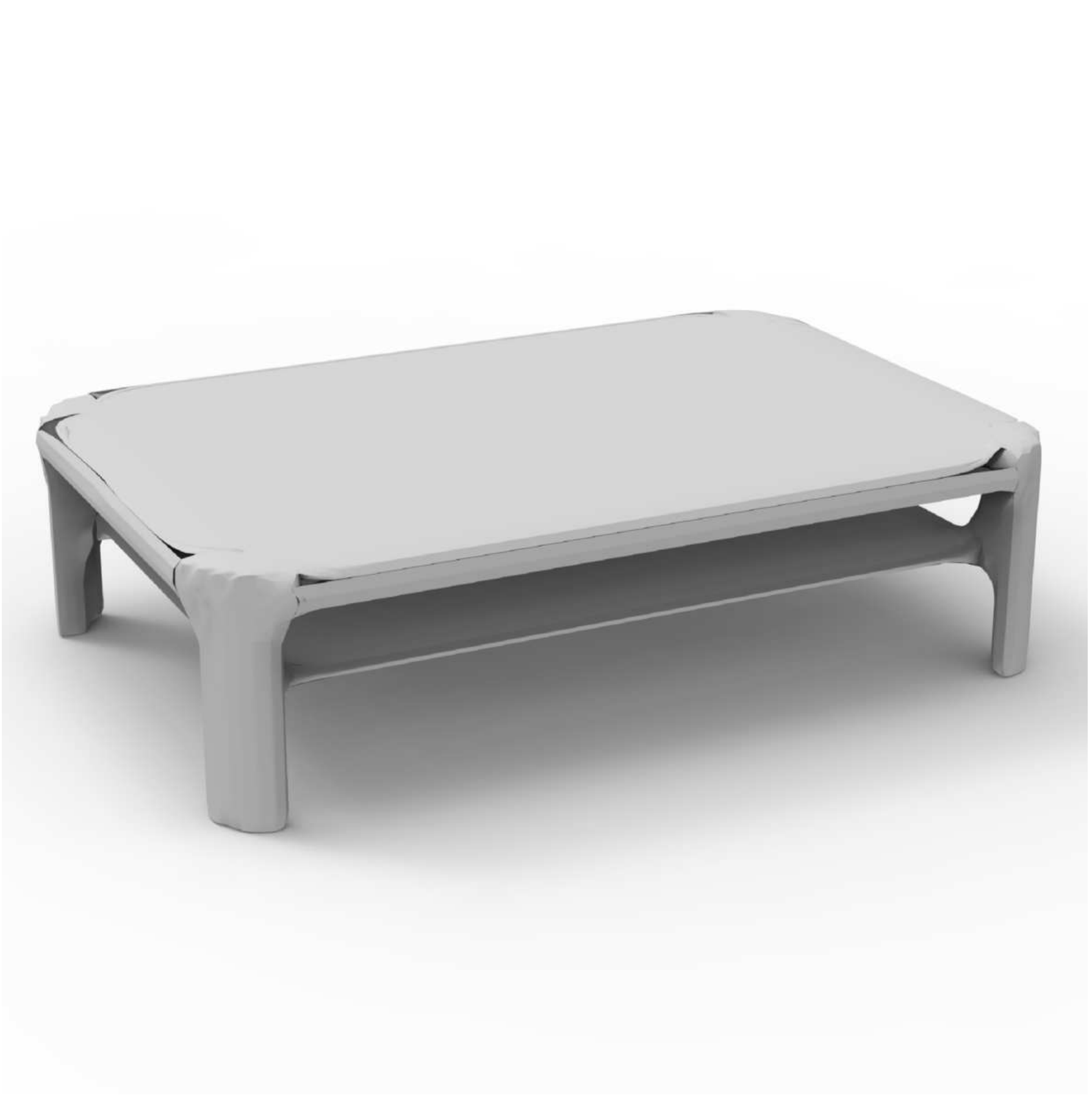}
    \includegraphics[width=0.19\linewidth]{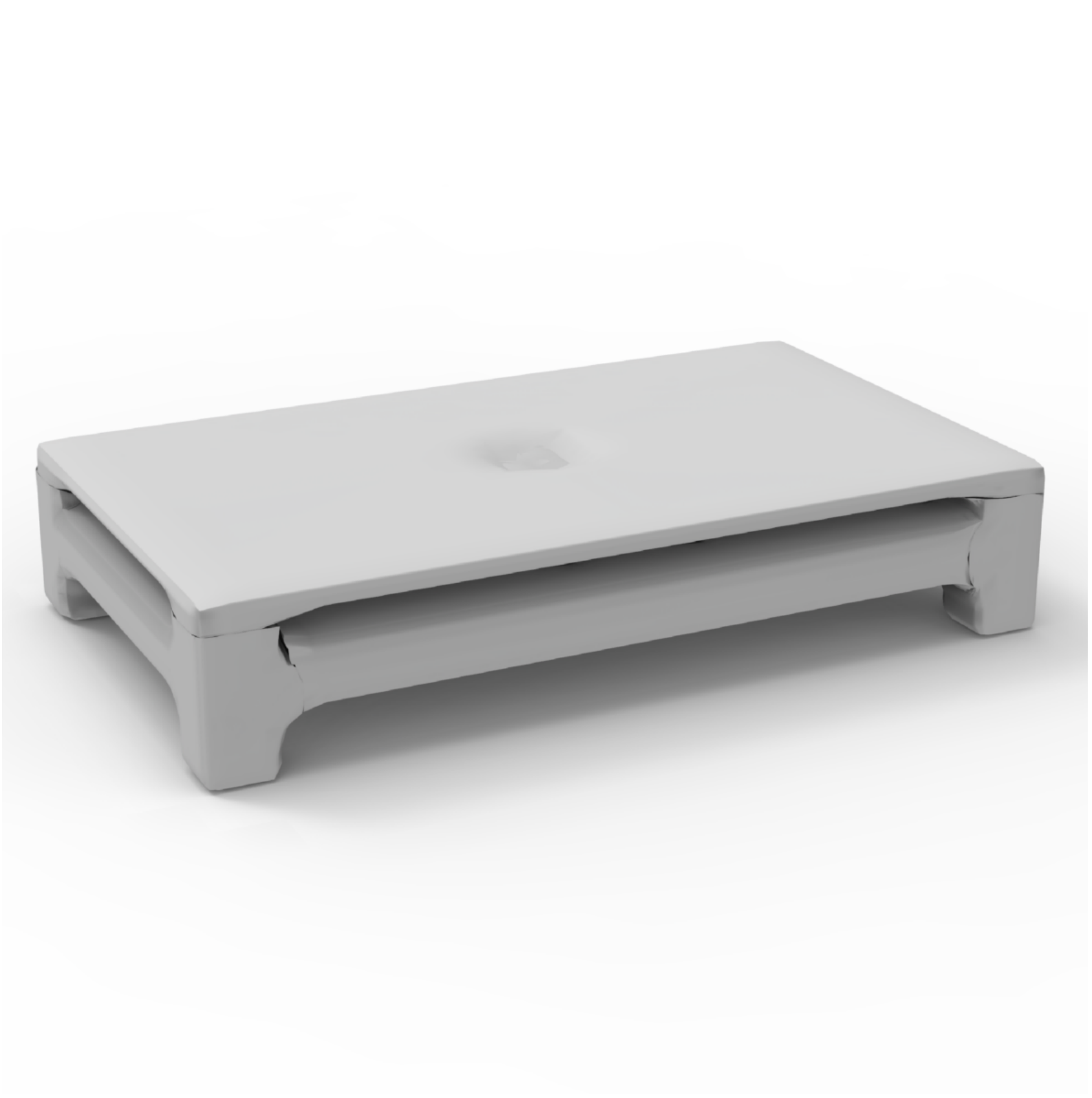}
    \includegraphics[width=0.19\linewidth]{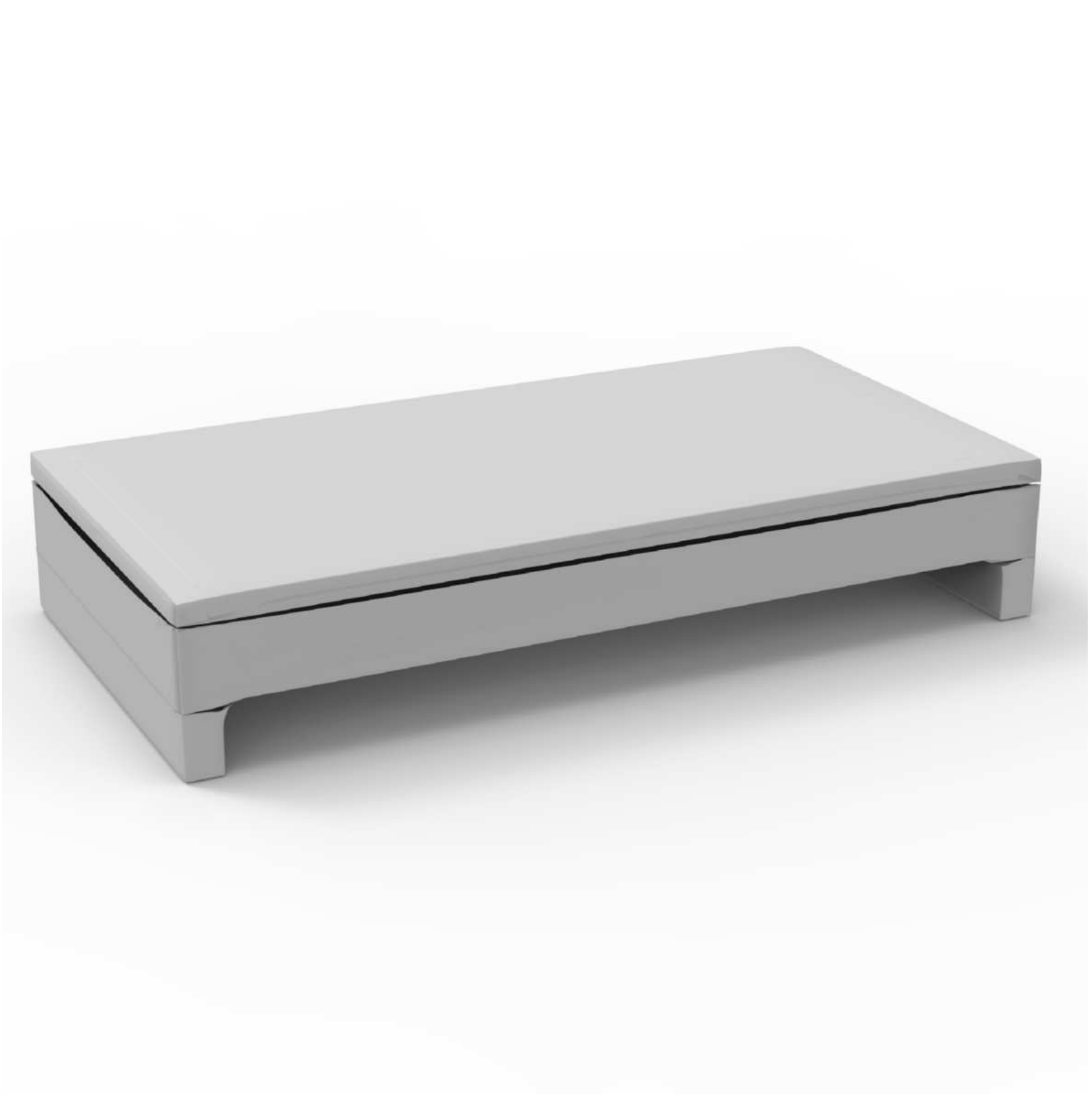}
    \includegraphics[width=0.19\linewidth]{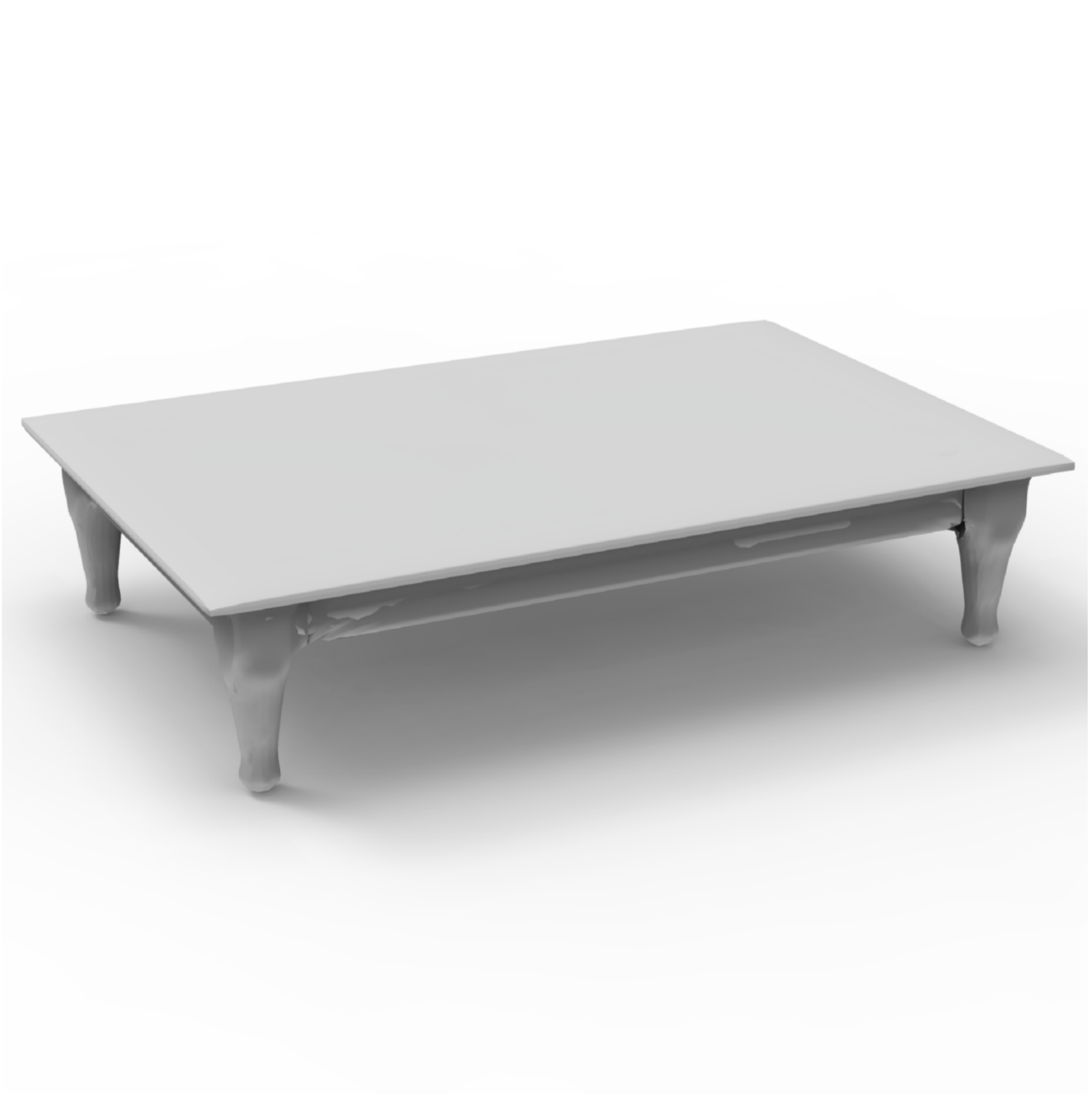}
    \includegraphics[width=0.19\linewidth]{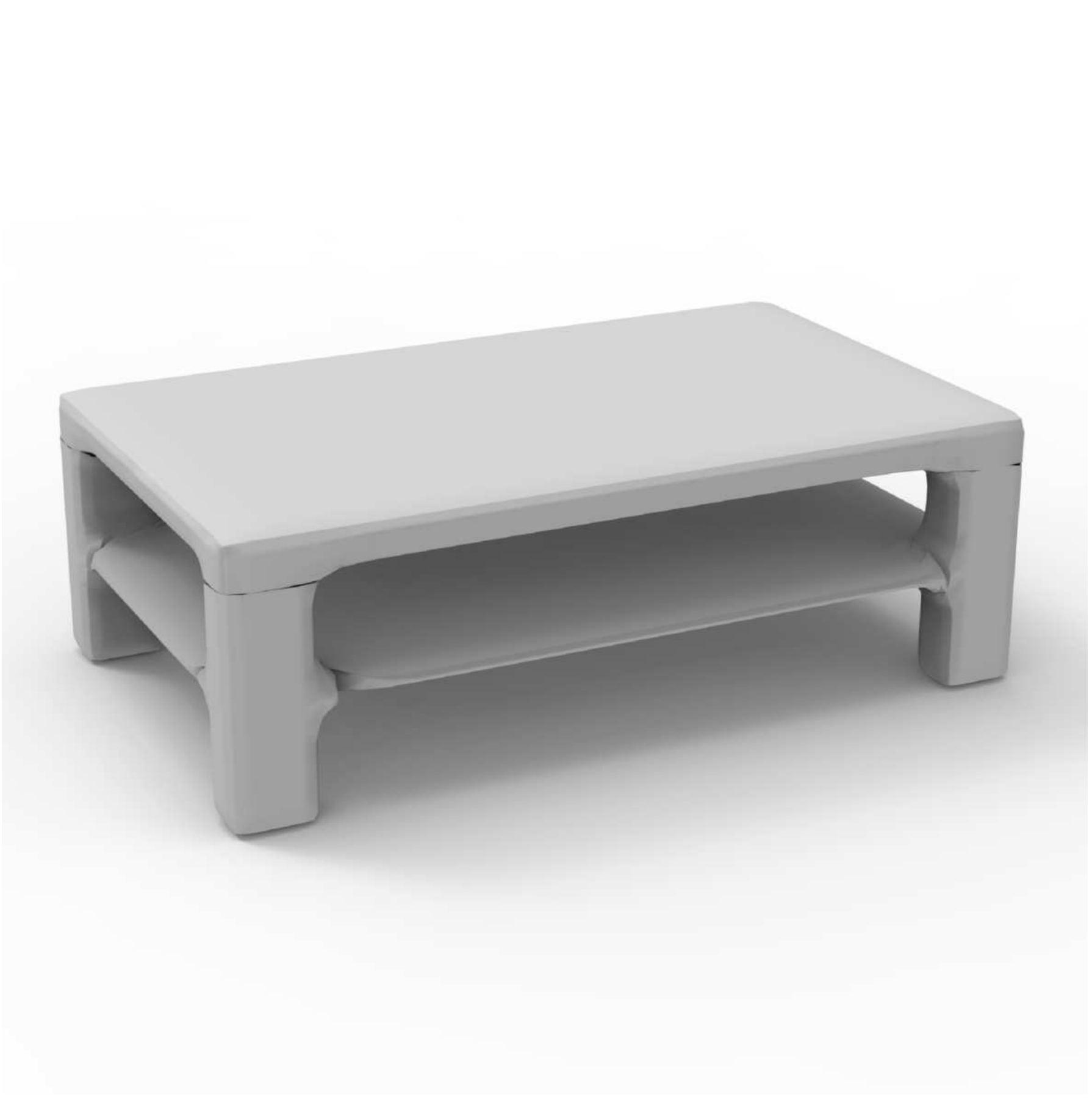}\\
    \includegraphics[width=0.19\linewidth]{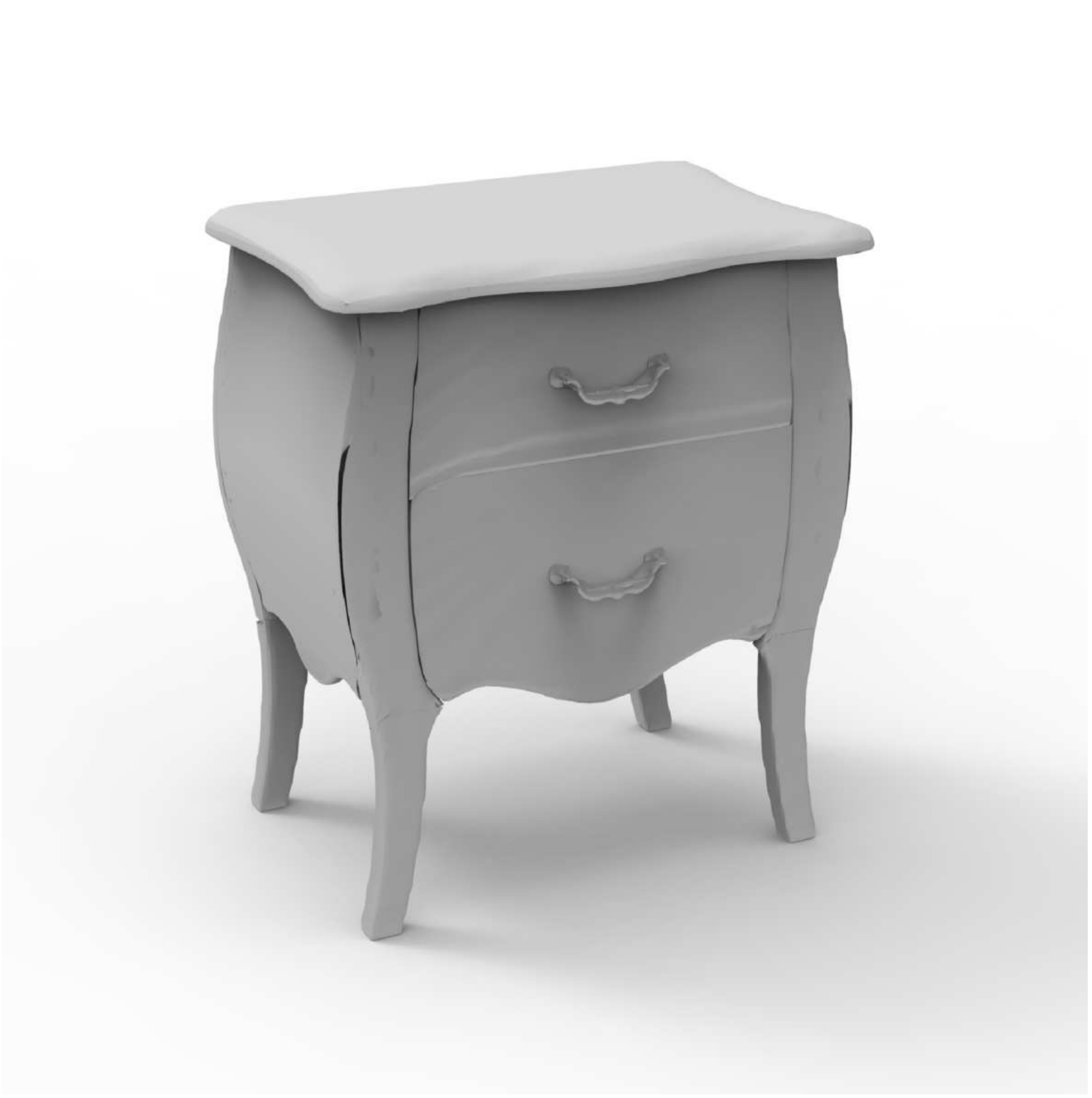}
    \includegraphics[width=0.19\linewidth]{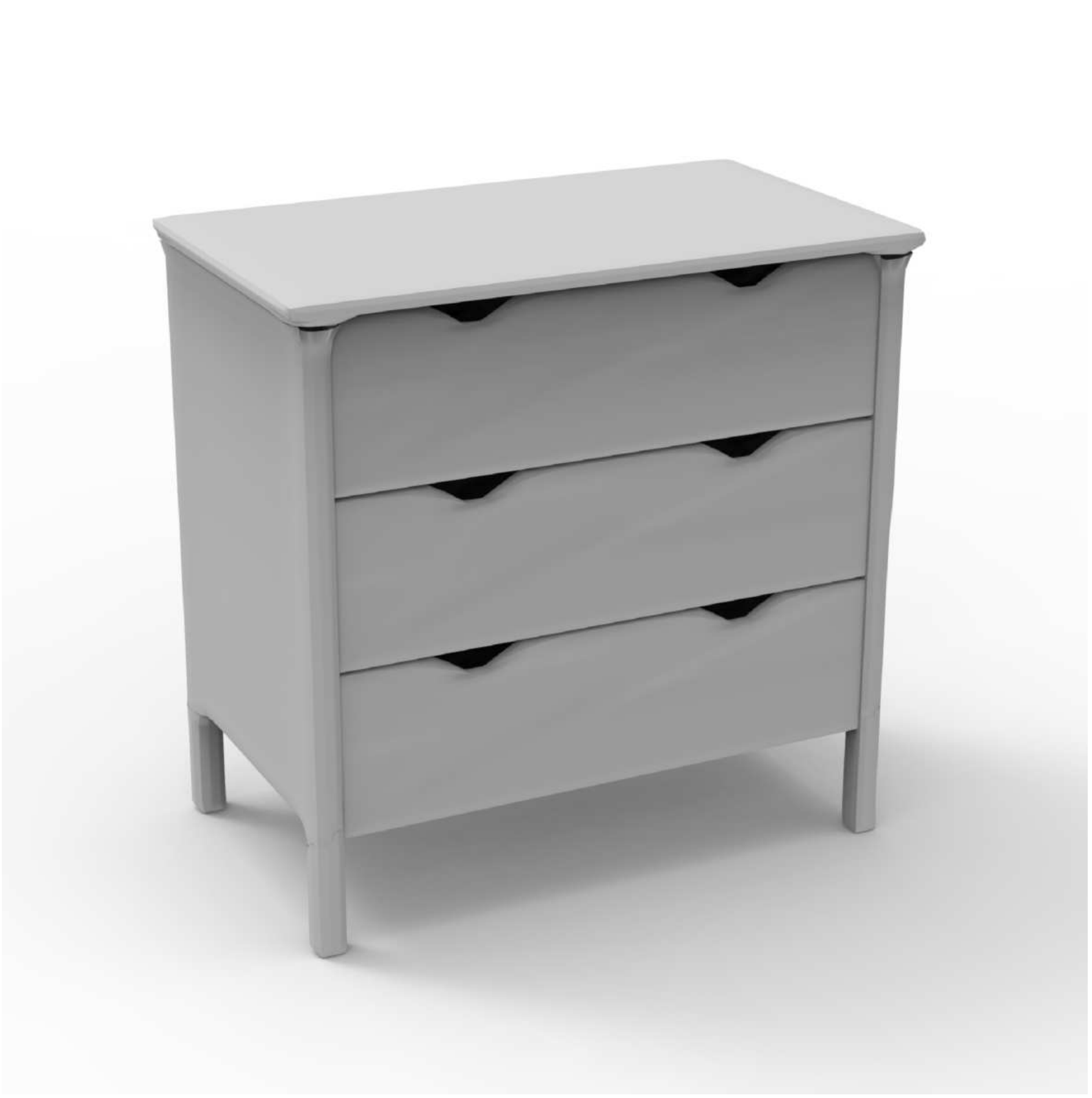}
    \includegraphics[width=0.19\linewidth]{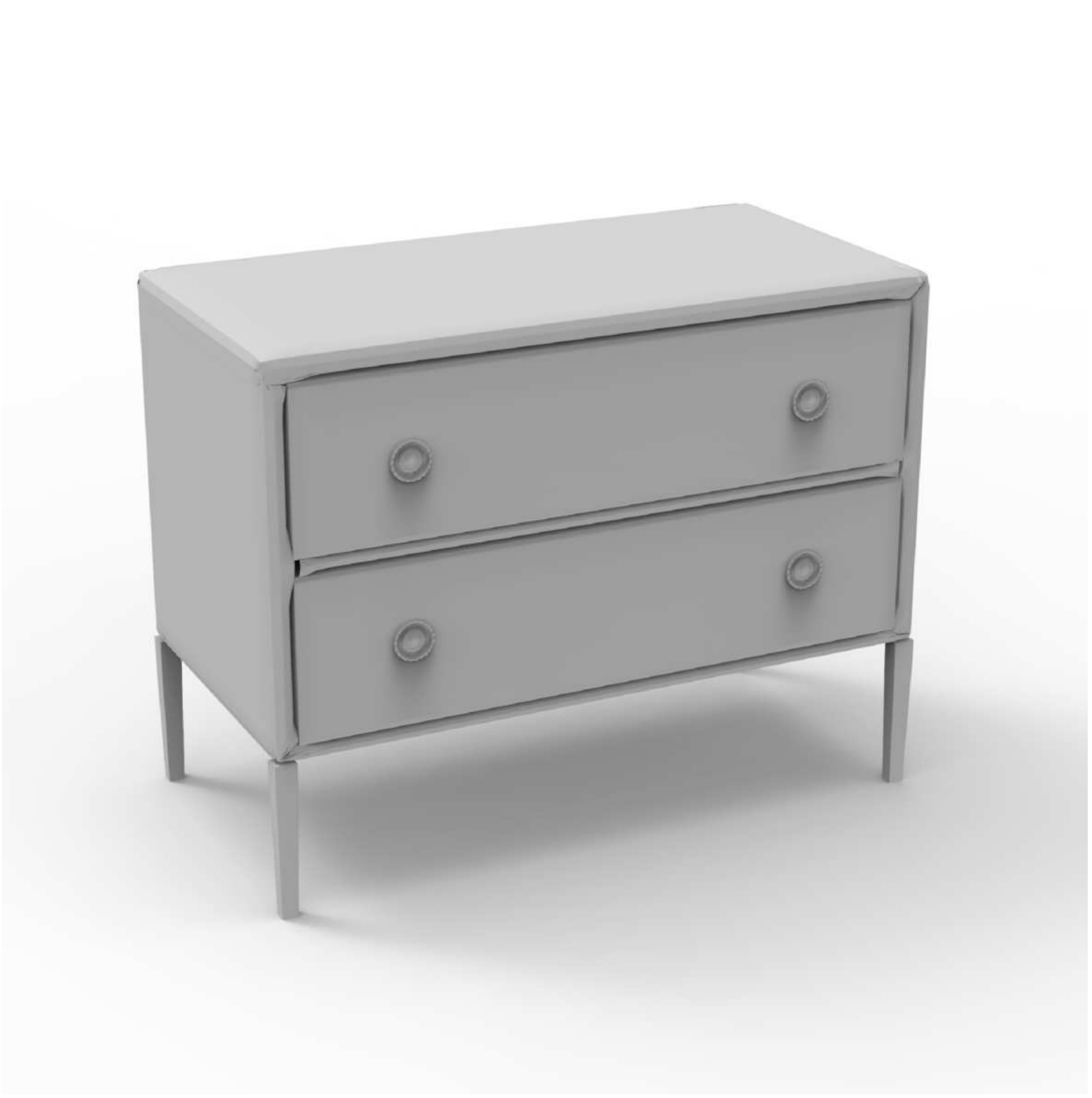}
    \includegraphics[width=0.19\linewidth]{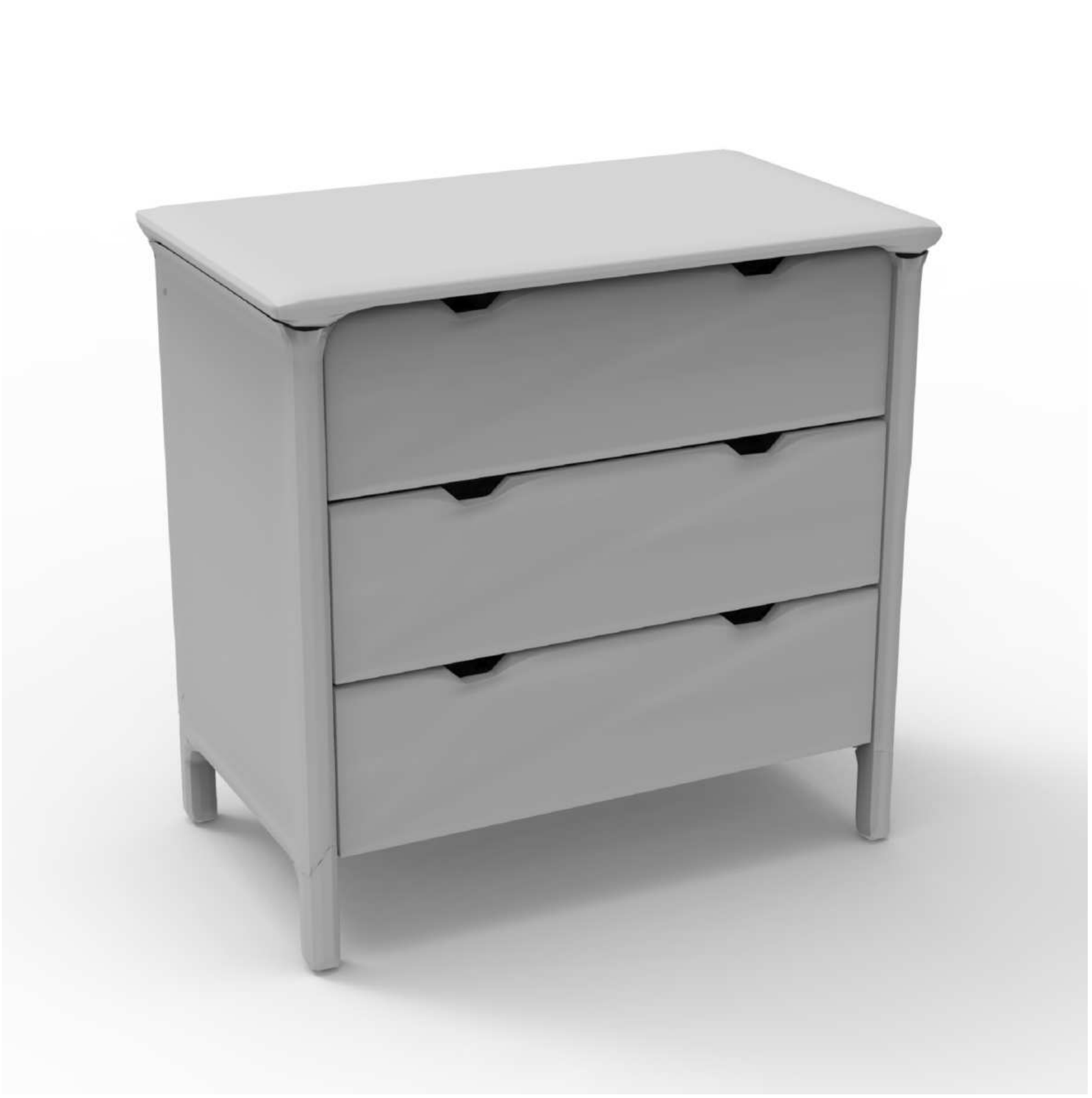}
    \includegraphics[width=0.19\linewidth]{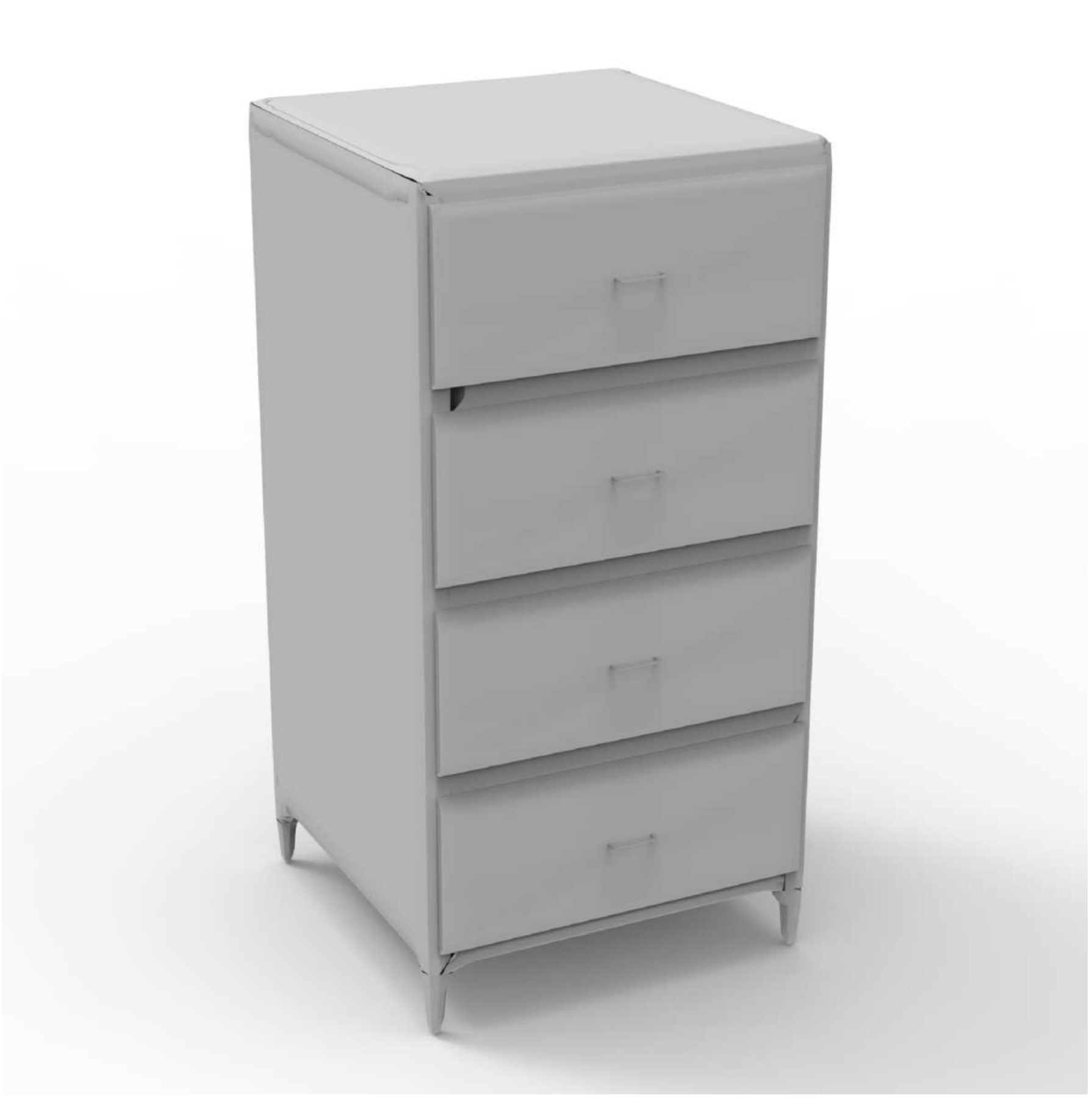}\\
    \includegraphics[width=0.19\linewidth]{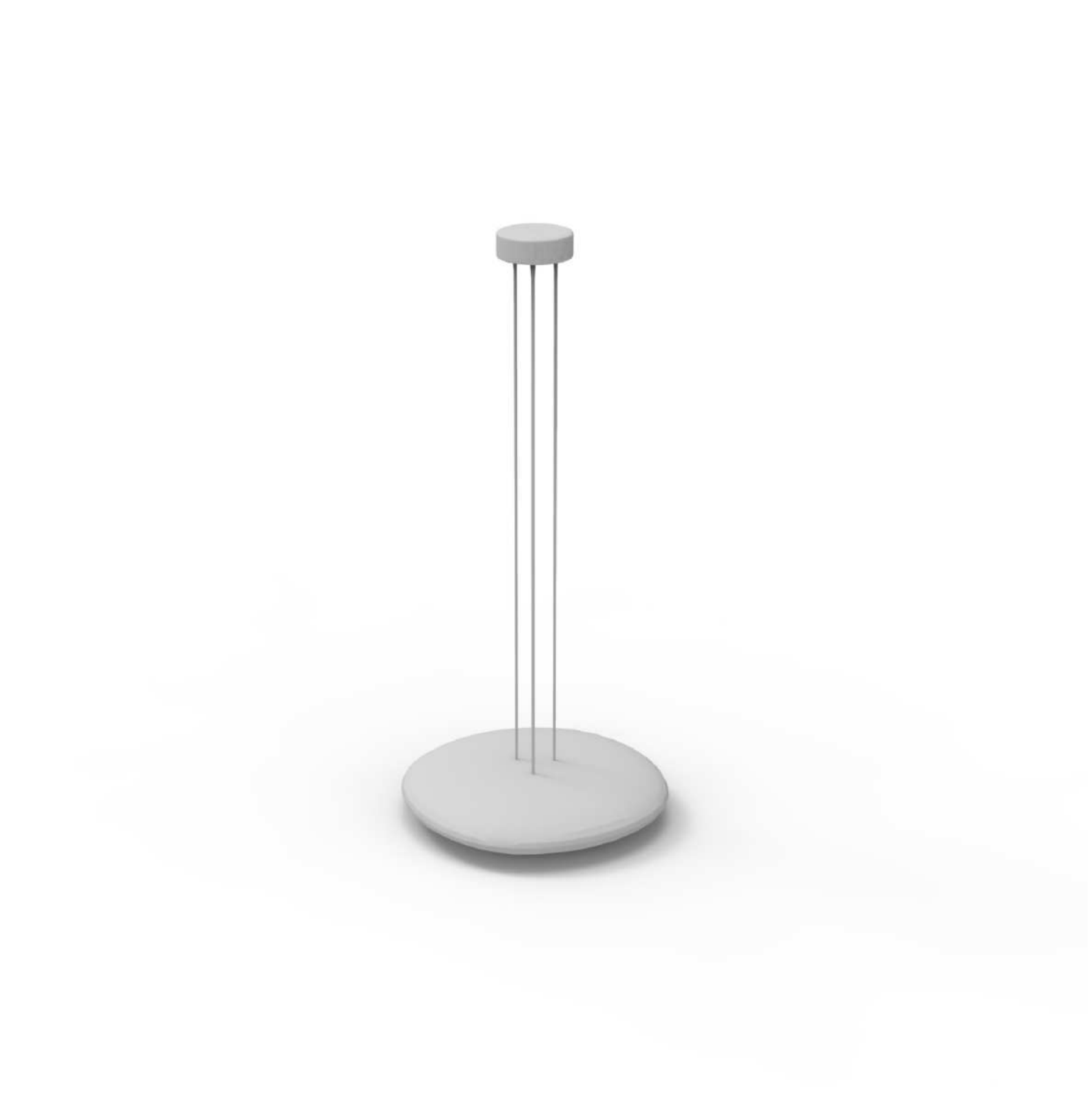}
    \includegraphics[width=0.19\linewidth]{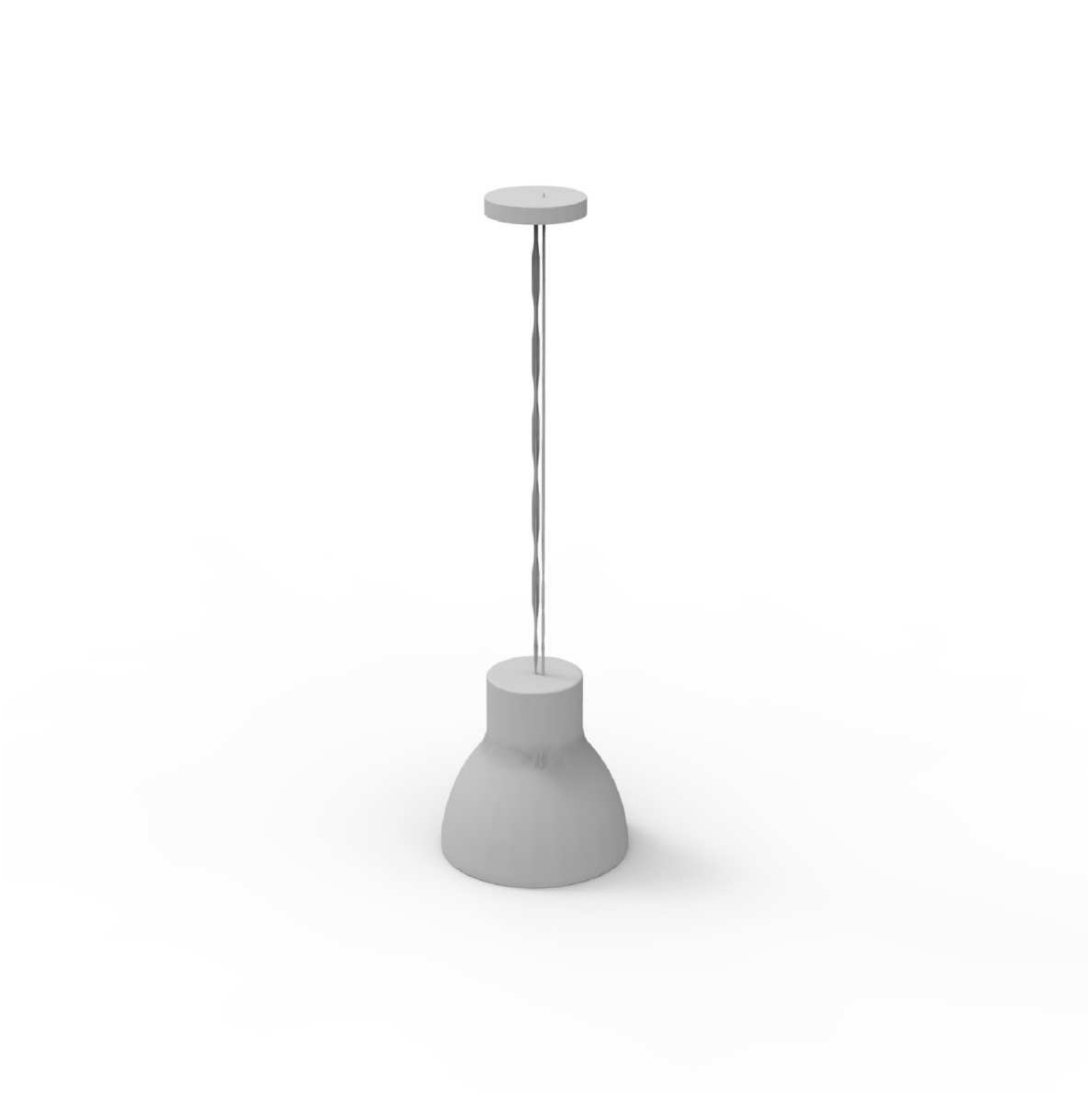}
    \includegraphics[width=0.19\linewidth]{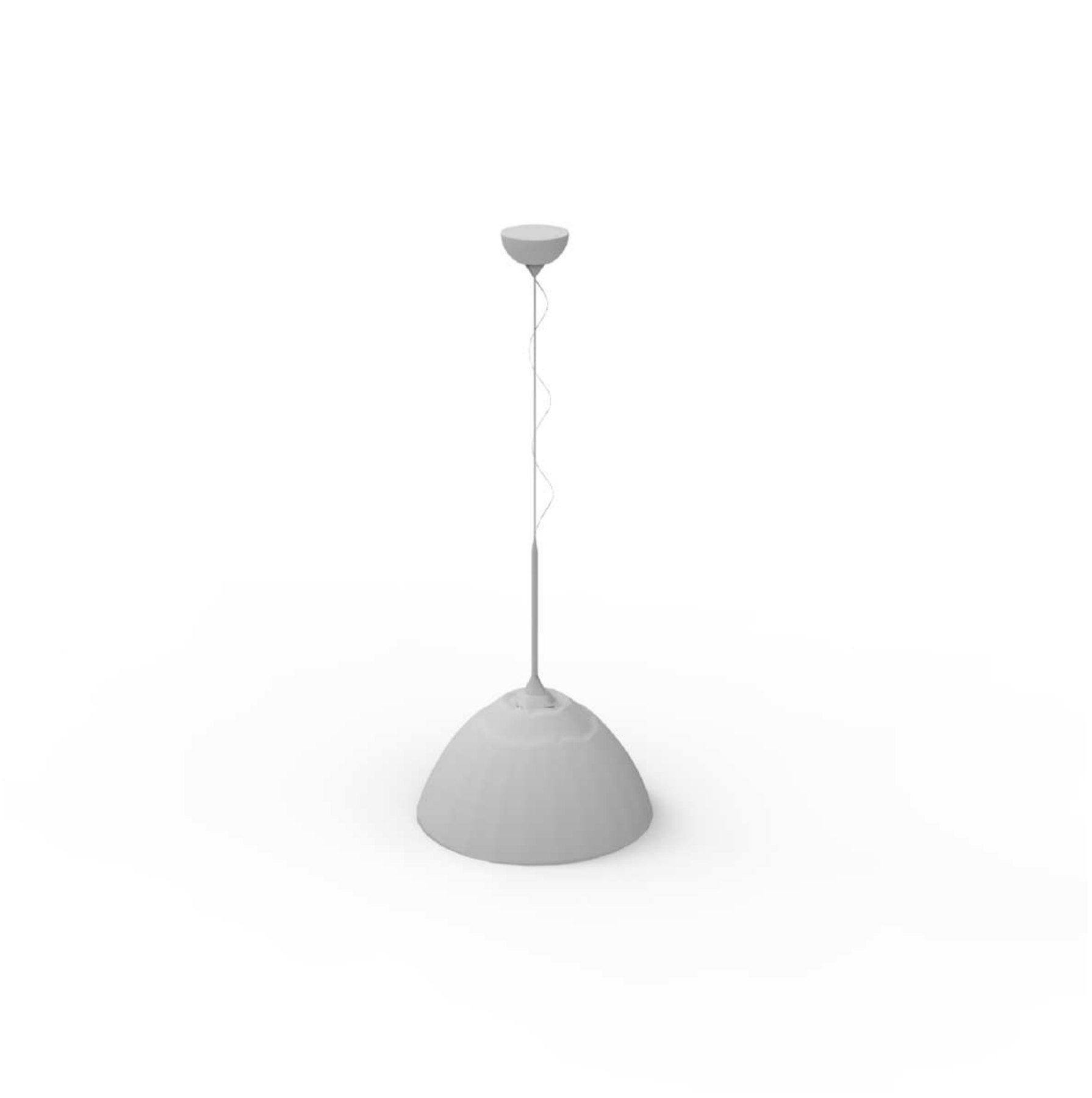}
    \includegraphics[width=0.19\linewidth]{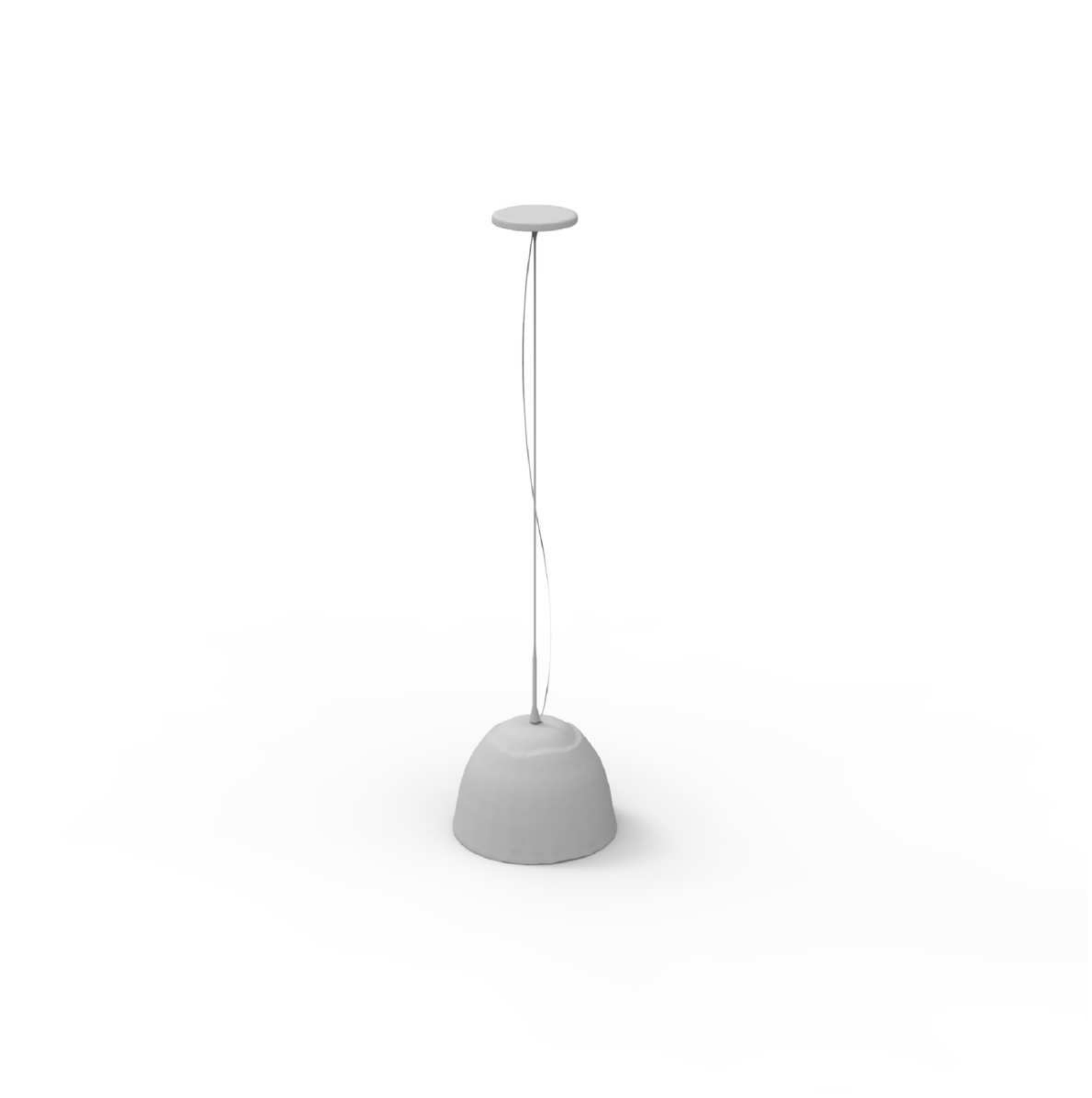}
    \includegraphics[width=0.19\linewidth]{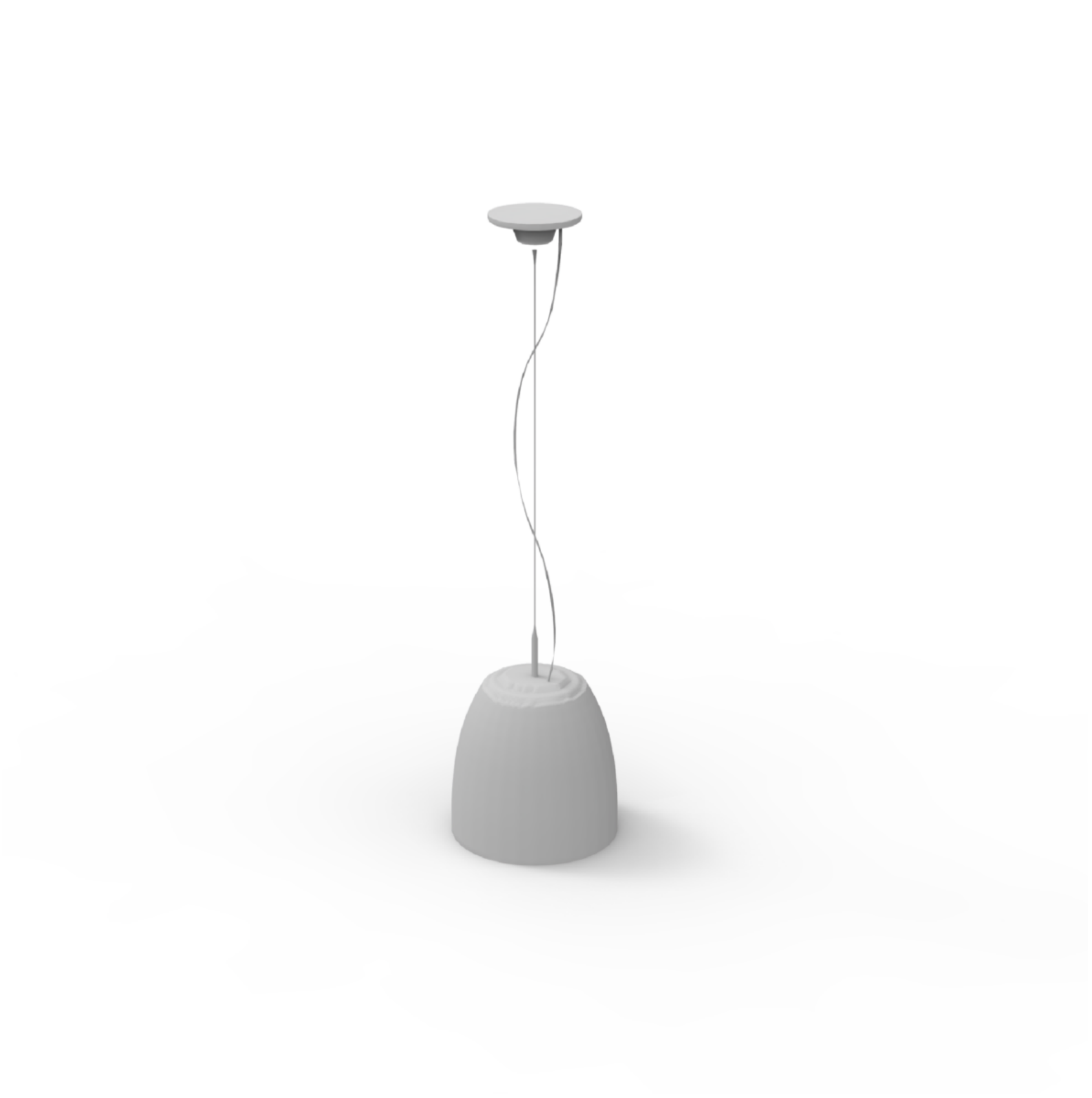}
    \end{minipage}}
    \caption{\yjr{Shape Generation. The left part is shape generation by sampling random Gaussian noises in both latent spaces of shape structure and geometry. Then we use DSG-Net to generate realistic shapes with complex structures and detailed geometry.
    The right part demonstrates the novelty of generated shapes. We present top-five retrieved shapes in the training sets, with CD as the retrieval metric, to our generated results shown in the left most column. Our generated shapes are different from the shapes in training sets, showing that DSG-Net does not simply overfit the training data.}}
    \label{fig:generation_nn}
\end{figure*}

\section{DataSet Preparation}\label{sec:data}
\yj{In this section, we summarize the dataset statistics and how to create the synthetic data for evaluation of disentangle reconstruction.
Table~\ref{tab:datastat} summarizes the data statistics.
Figure~\ref{fig:data_viz} shows example shapes in PartNet on the left, and example shapes in the synthetic dataset on the right.

For the synthetic dataset, we aim to evaluate the task of disentangled shape reconstruction quantitatively, 
The synthetic dataset contains 10,800 shapes with 54 kinds of shape structures and 200 geometric variations (different scales of boxes), in Figure~\ref{fig:data_viz} (right).
Each shape is generated by picking one shape structure and one geometric variation, granting us the access to the ground-truth shape synthesis outcome for every configuration pair.
The 54 structures are generated by enumerating structural combinations of different back types, leg styles and whether the chair has arms or not.
The 200 geometric variations are created by varying the global parameters for the part geometry (\eg the width of legs, the height of the back).}
We will release the code and data for facilitating future research.

\begin{table}[t]
\fontsize{8}{12}\selectfont
  \centering
  \caption{\yj{We summarize the data statistics of the two datasets in our experiments. We use four categories from PartNet (chairs, tables, cabinets and lamps) for the majority of our experiments and one synthetic dataset (synchairs) for evaluating disentangled shape reconstruction.}}
    \begin{tabular}{cccccc}
    \toprule[1pt]
    DataSet & Chair & Table & Cabinet & Lamp & SynChair \\
    \midrule
    \#objects & 4172  & 3967  & 667   & 653  & 10800 \\
    \#training objects & 3292  & 3285  & 481   & 478 & 8100\\
    \#test objects & 880   & 682   & 186   & 175 & 2700\\
    \bottomrule[1pt]
    \end{tabular}%
  \label{tab:datastat}%
\end{table}%

\begin{figure}[h]
\begin{minipage}[c]{\linewidth}
    \centering
    \includegraphics[width=0.19\linewidth]{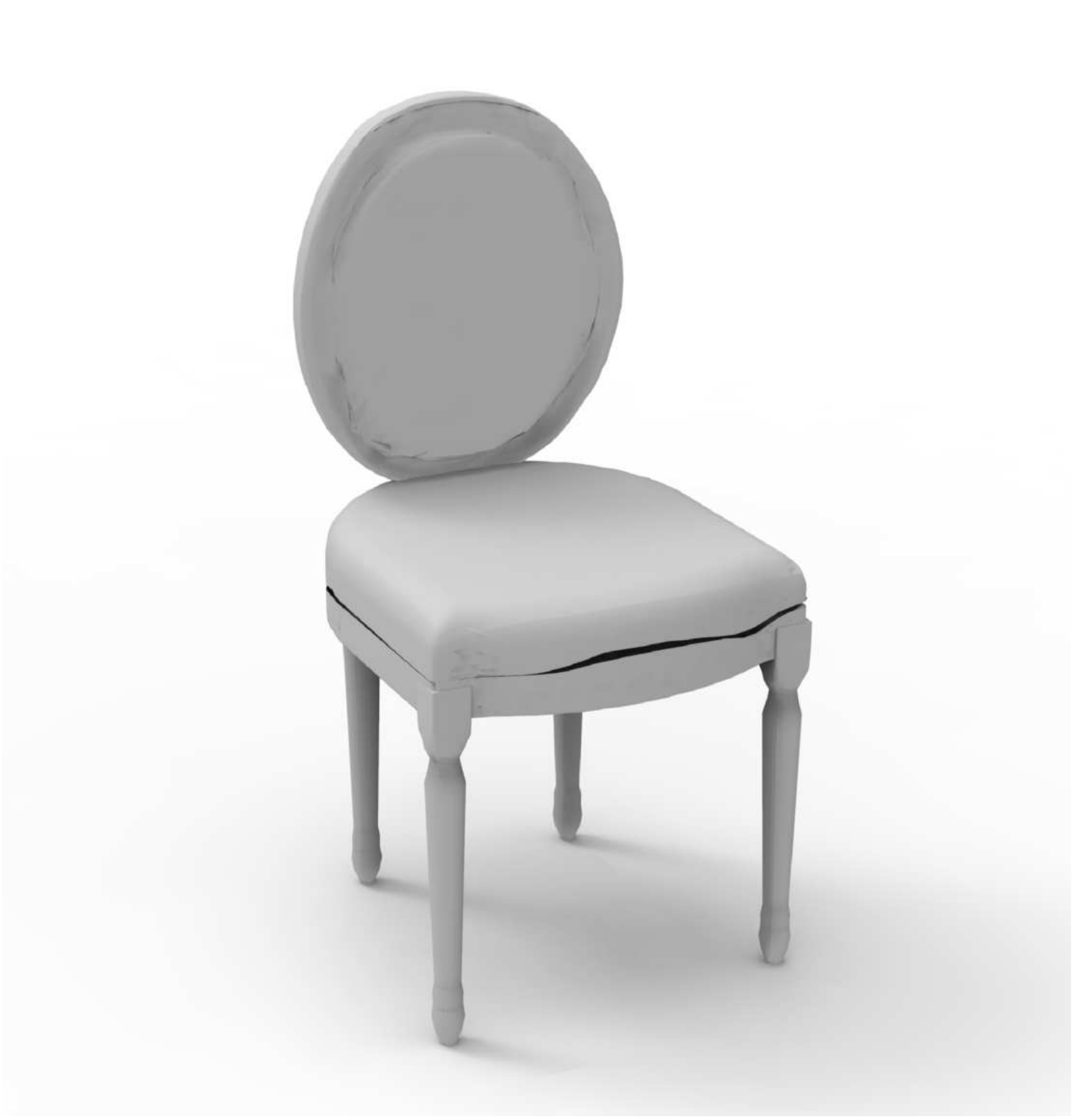}
    \includegraphics[width=0.19\linewidth]{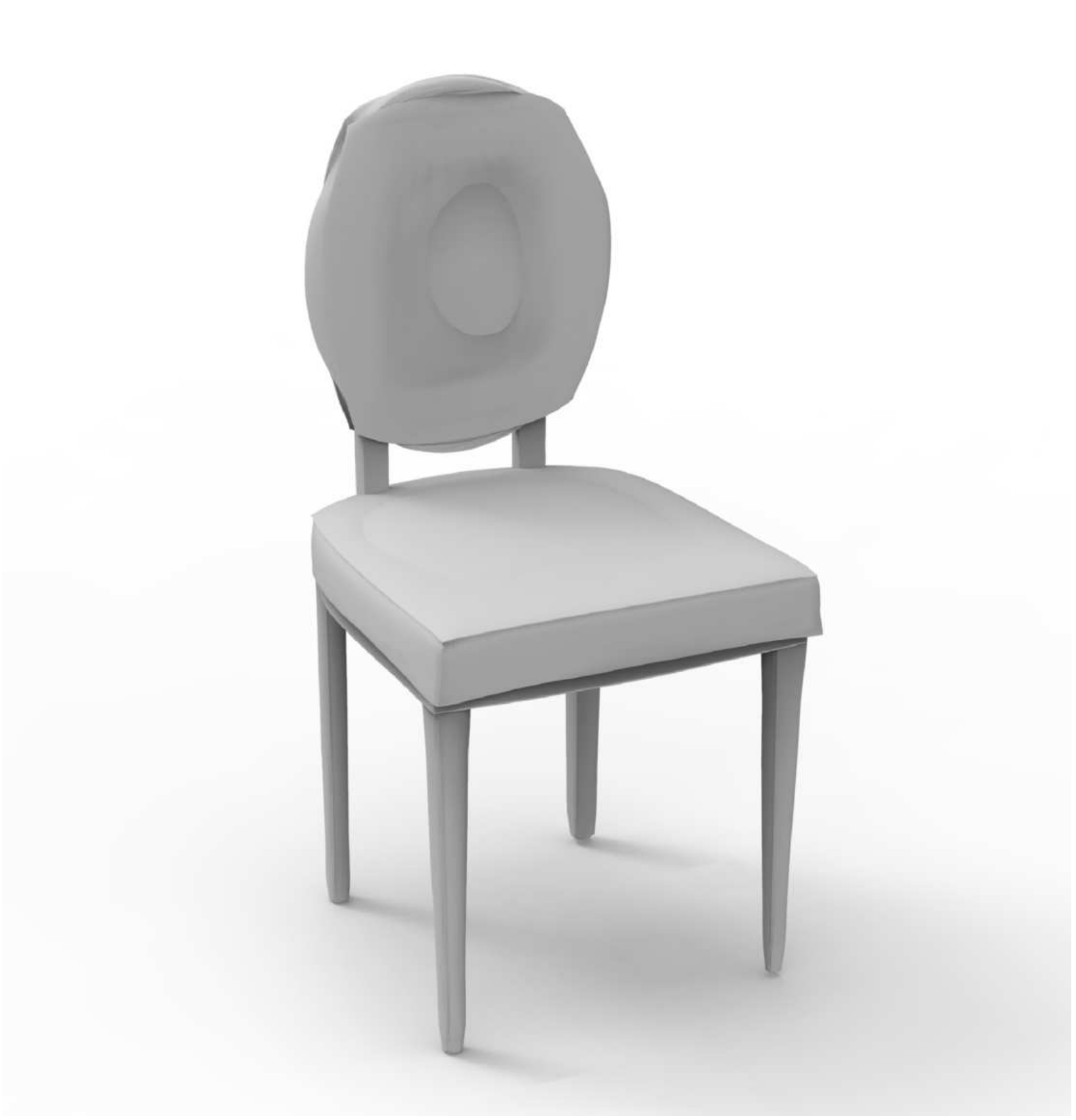}
    \includegraphics[width=0.19\linewidth]{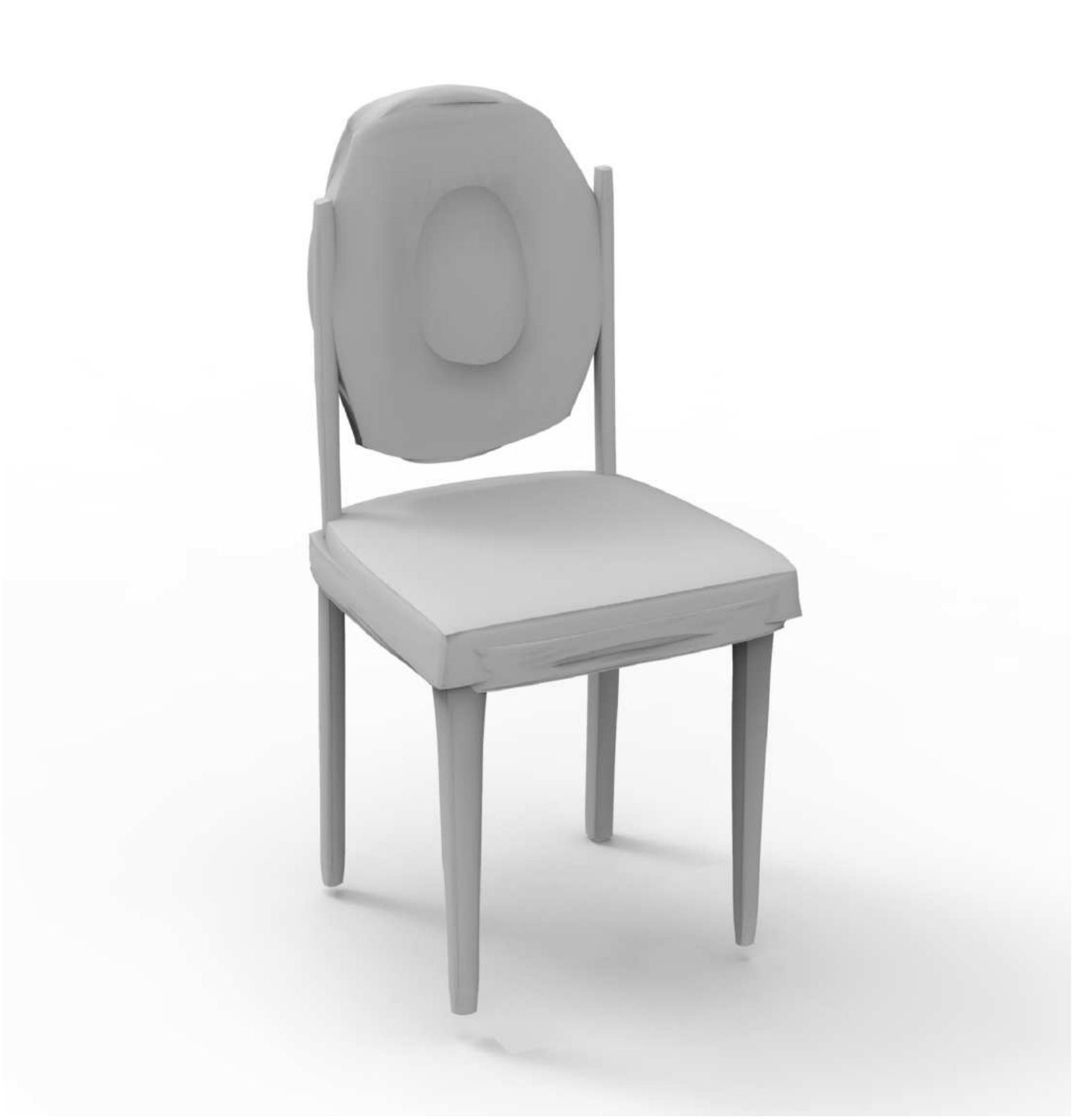}
    \includegraphics[width=0.19\linewidth]{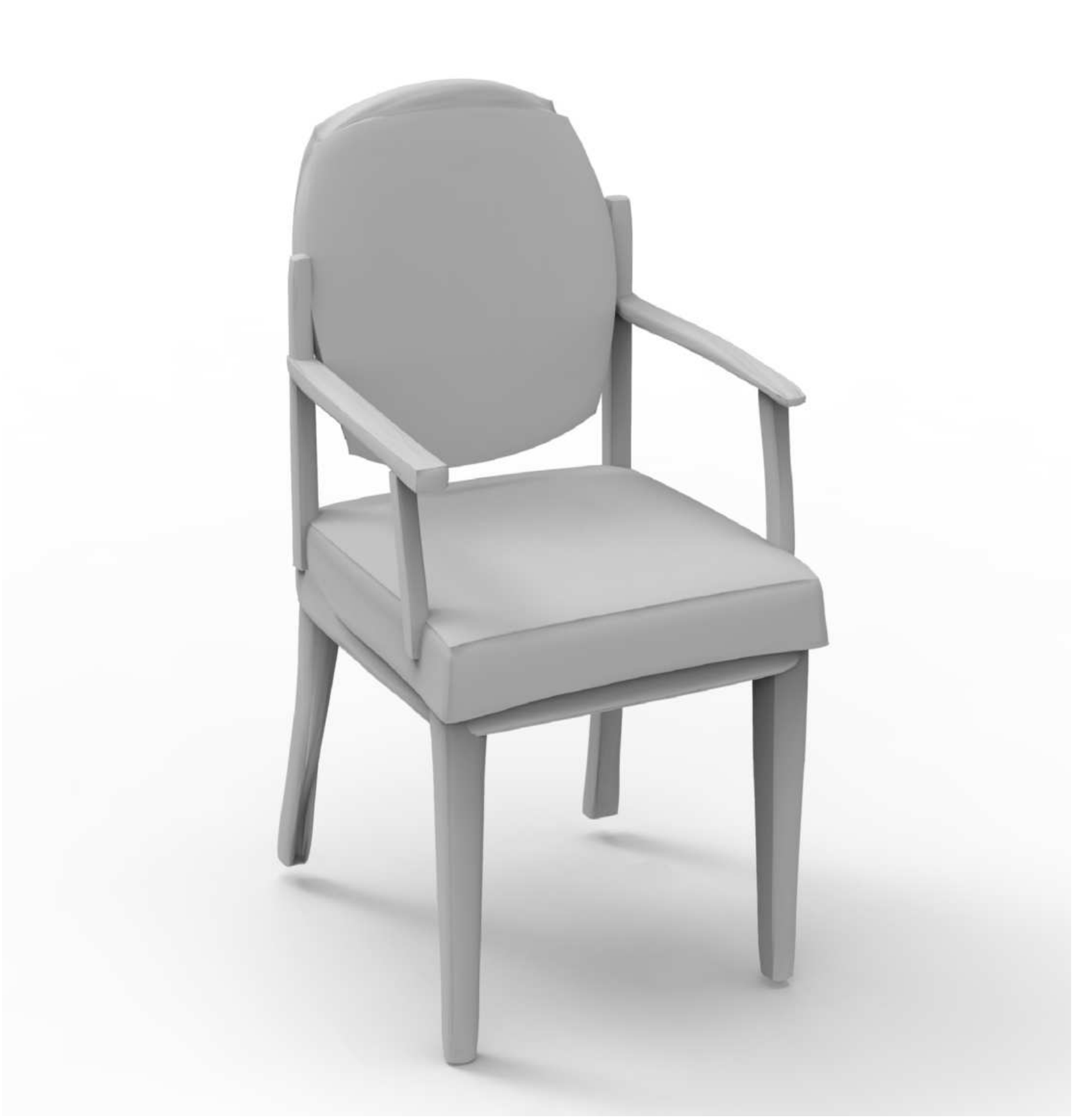}
    \includegraphics[width=0.19\linewidth]{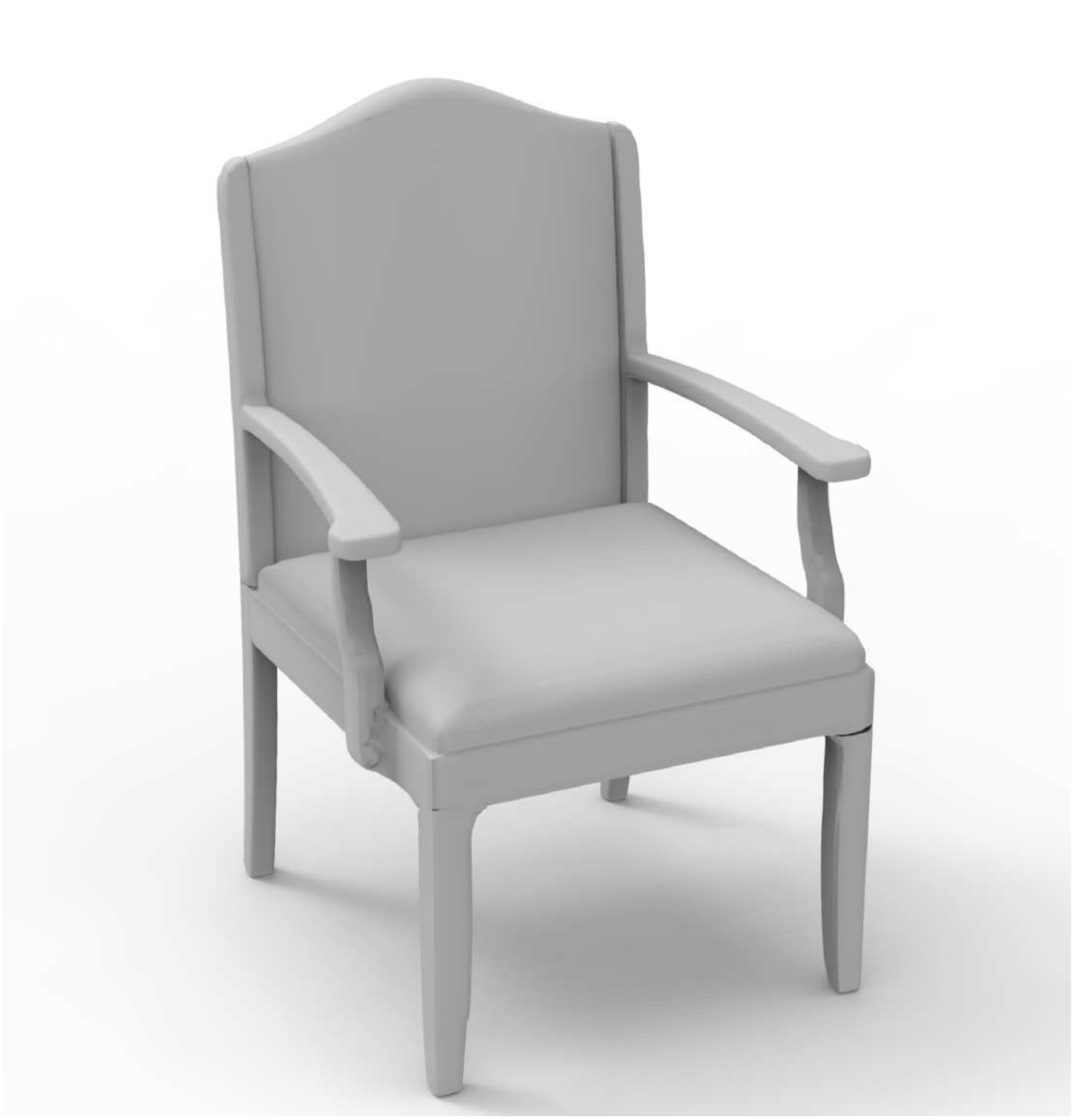}\\
    \includegraphics[width=0.19\linewidth]{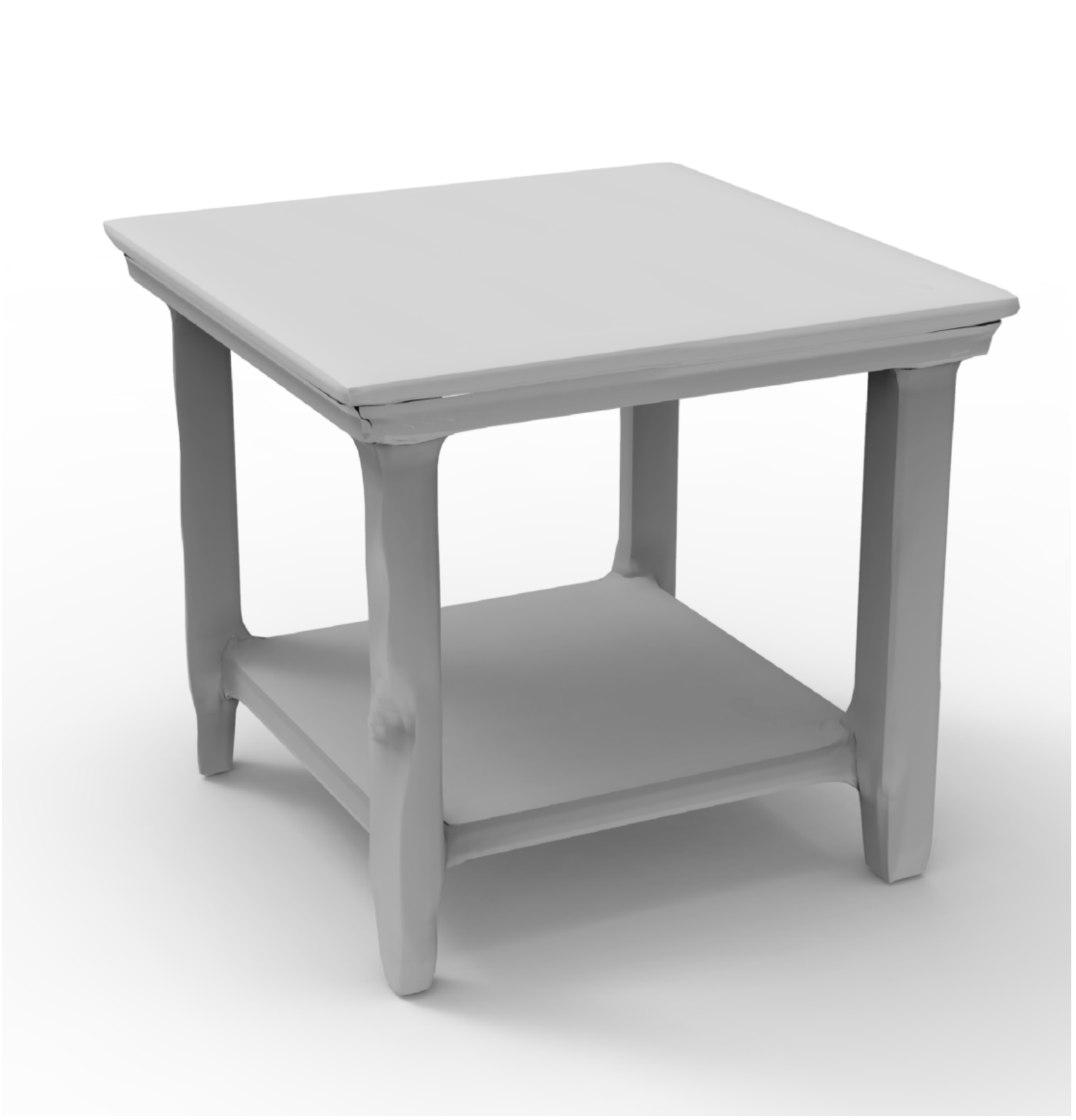}
    \includegraphics[width=0.19\linewidth]{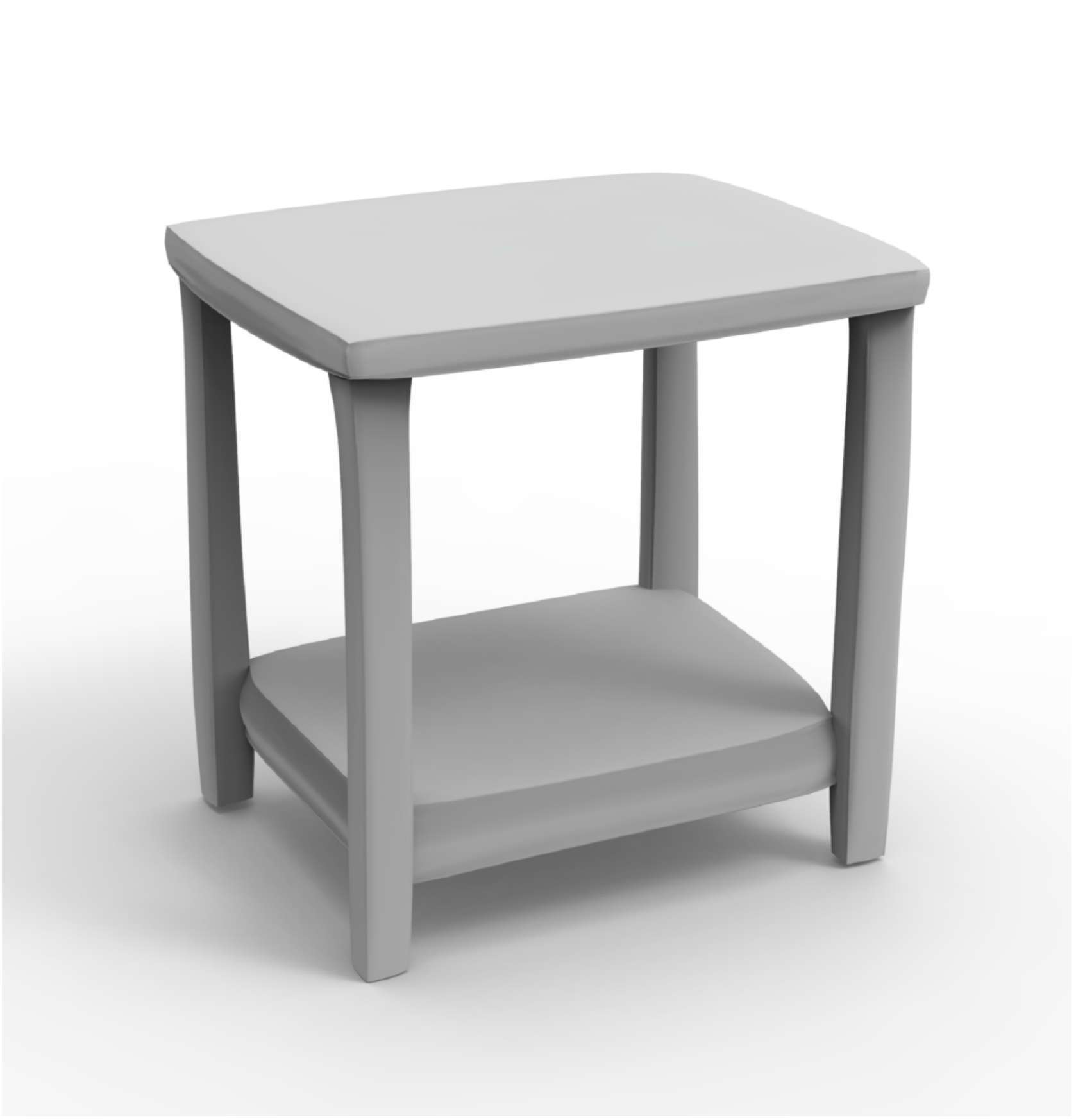}
    \includegraphics[width=0.19\linewidth]{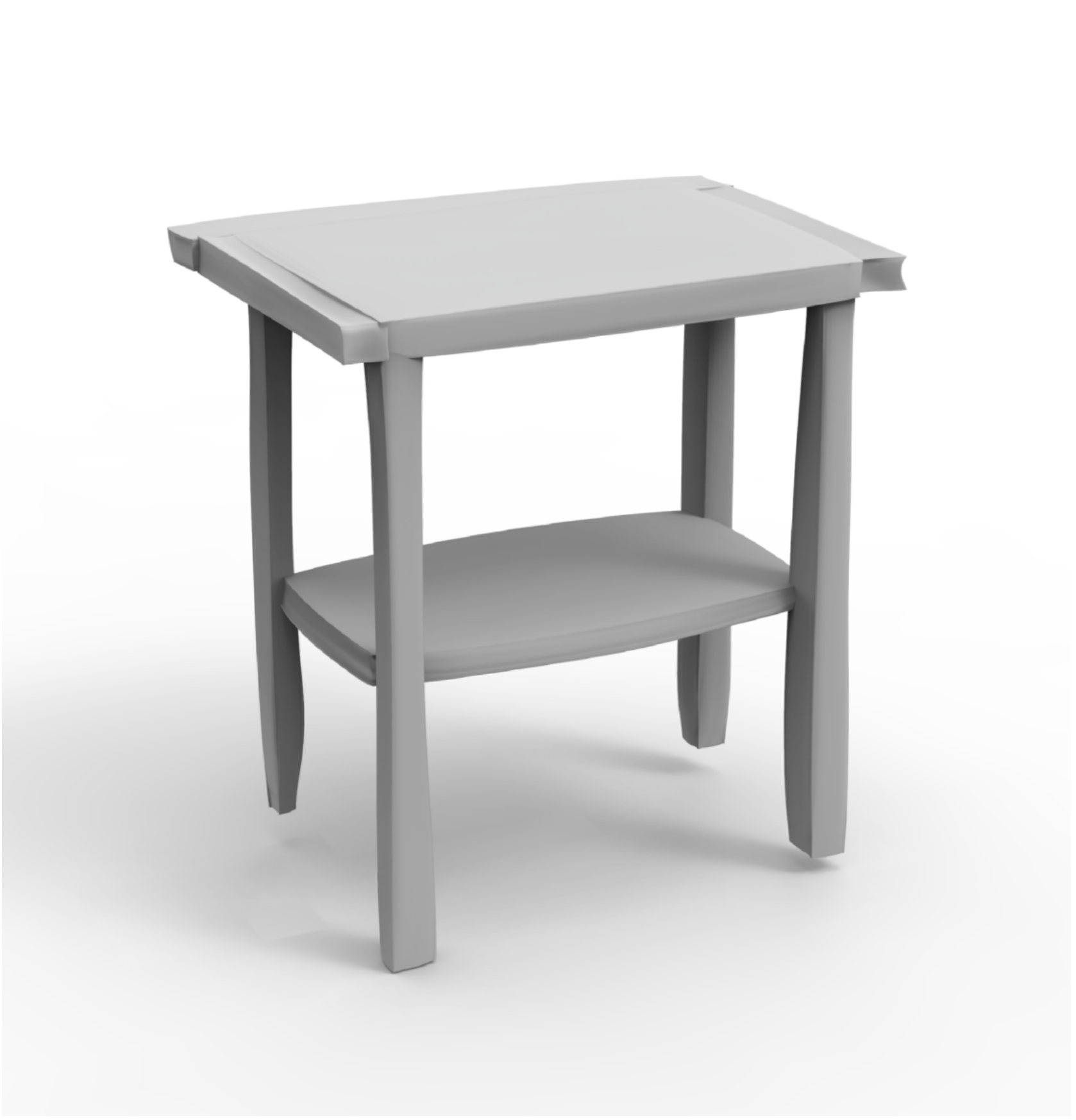}
    \includegraphics[width=0.19\linewidth]{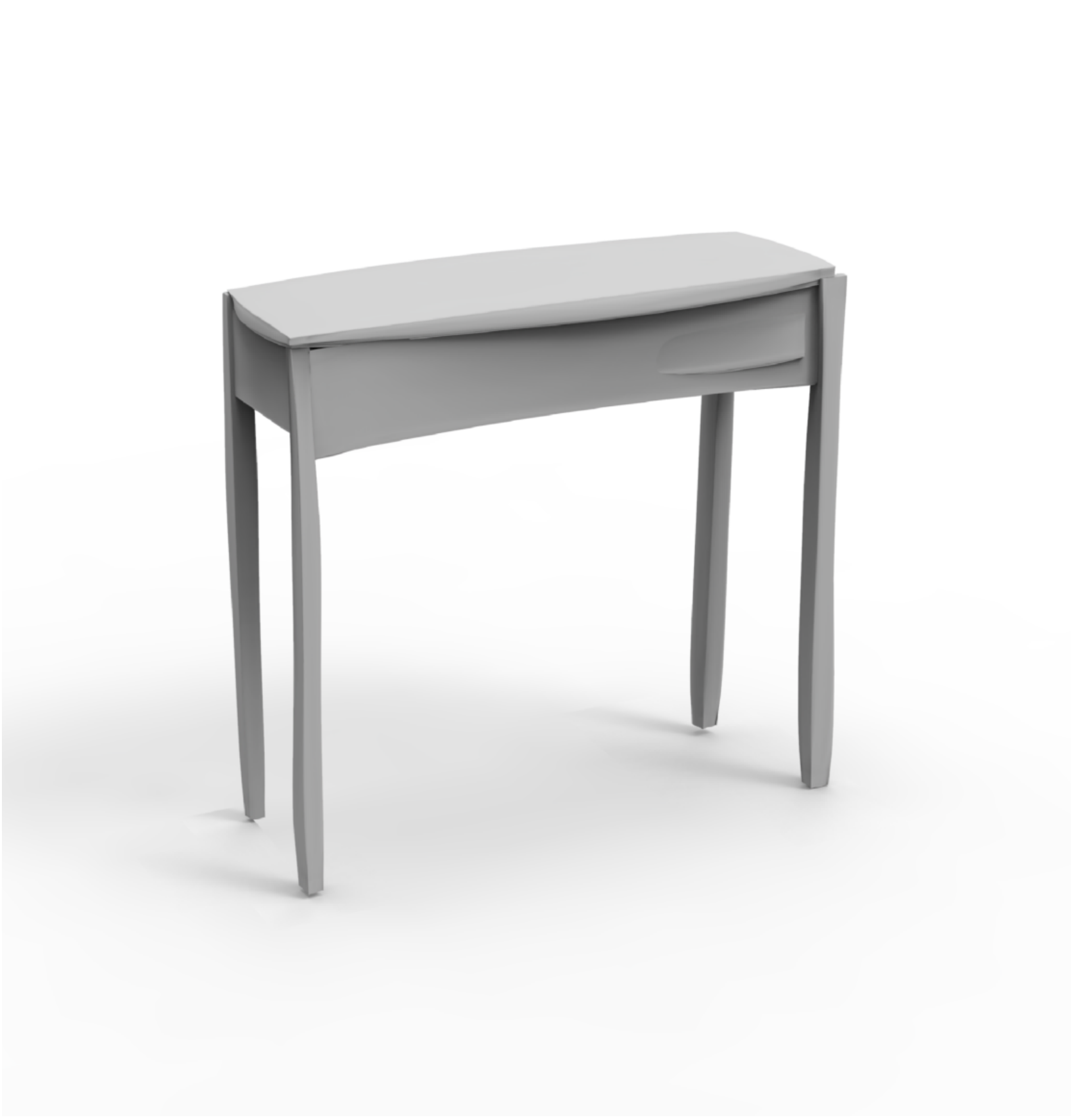}
    \includegraphics[width=0.19\linewidth]{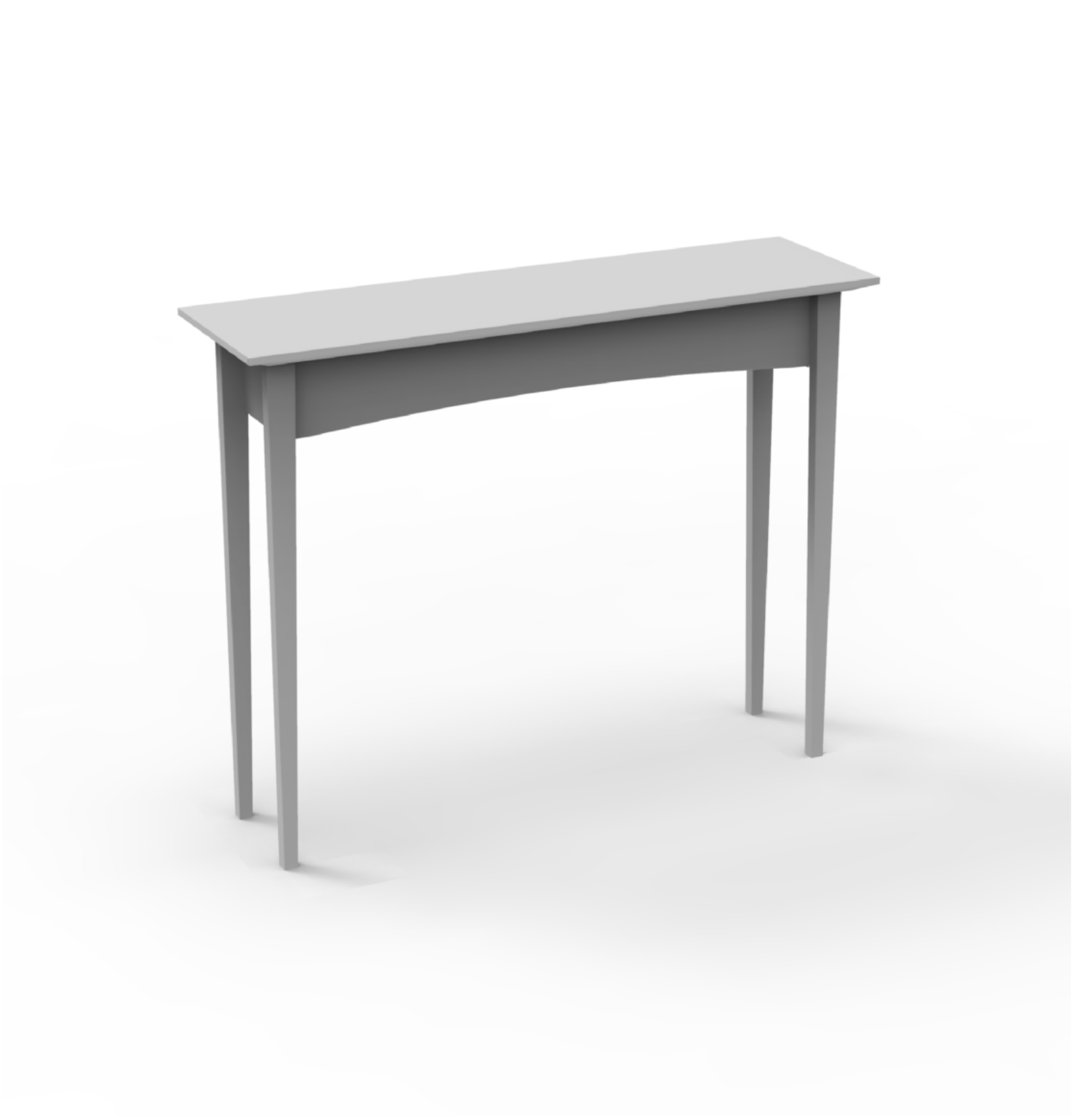}\\
    \includegraphics[width=0.19\linewidth]{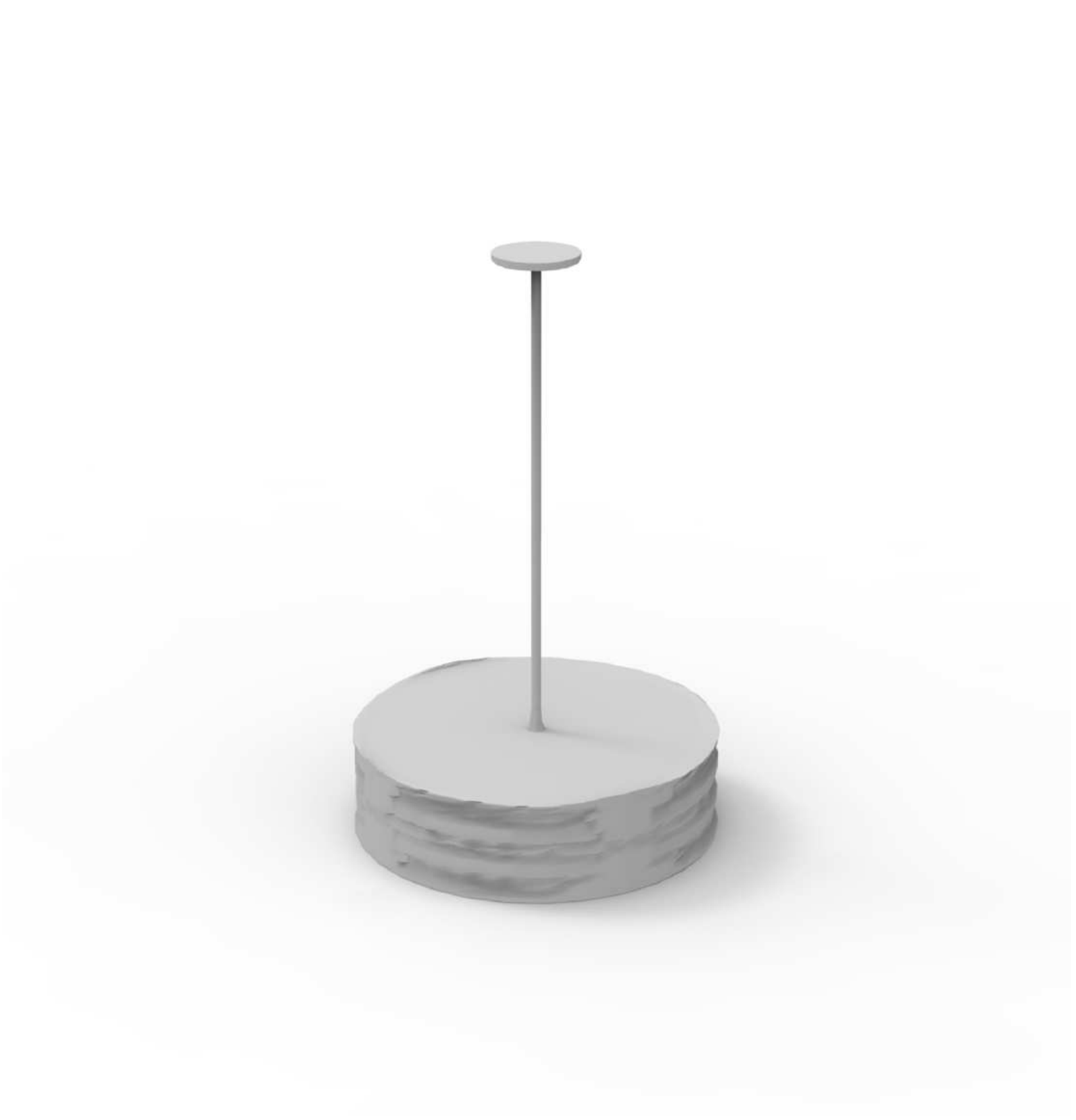}
    \includegraphics[width=0.19\linewidth]{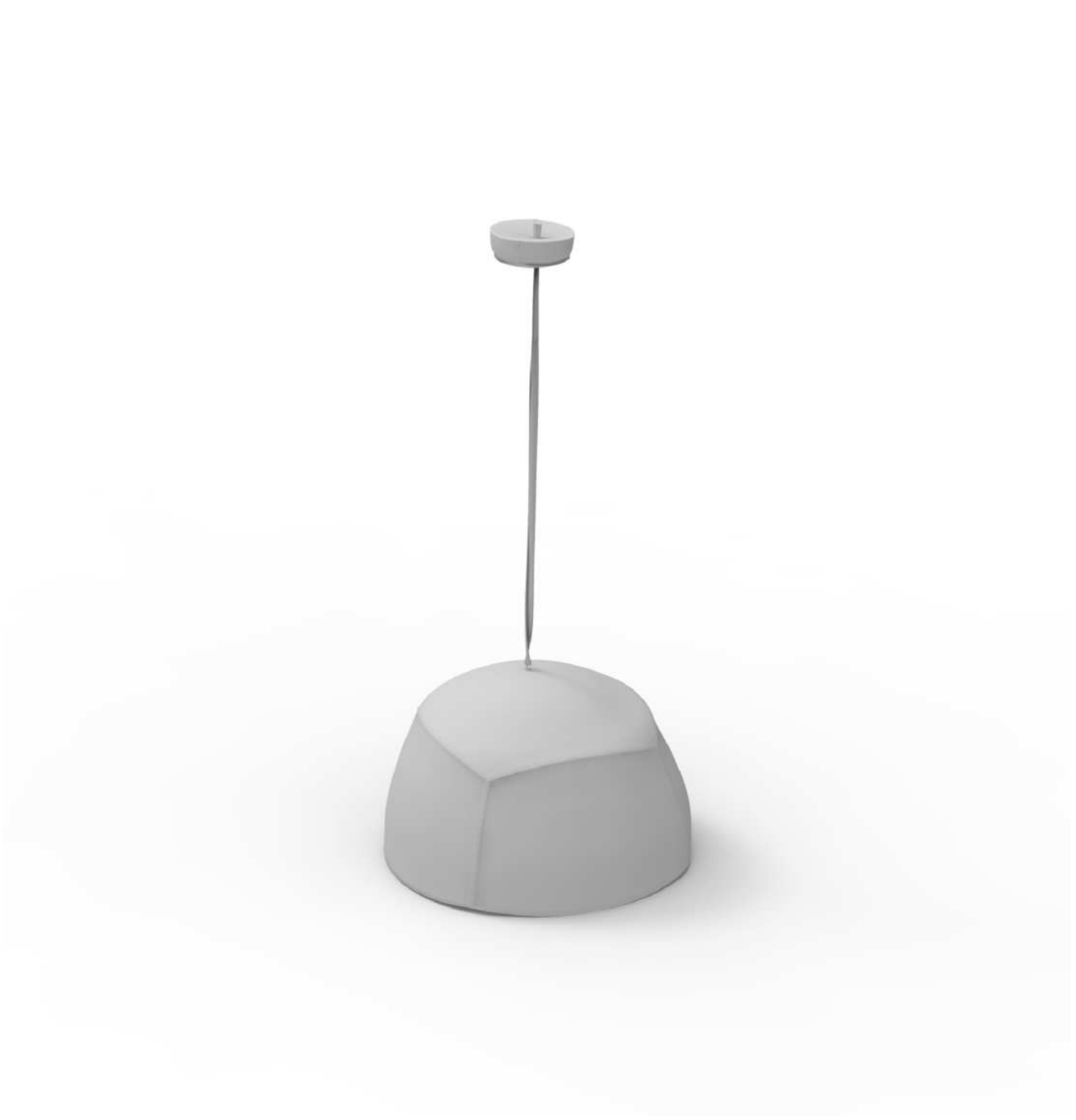}
    \includegraphics[width=0.19\linewidth]{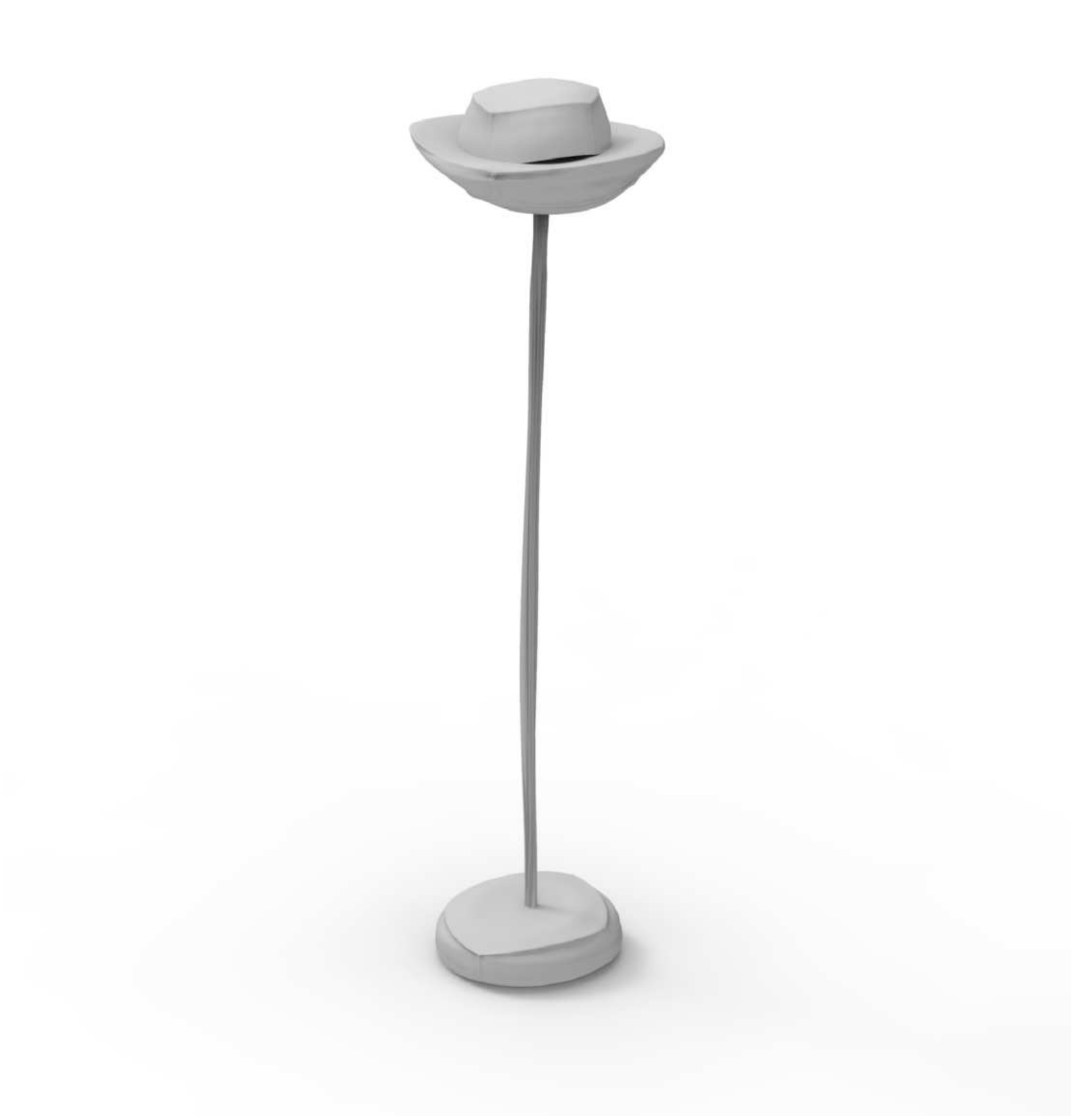}
    \includegraphics[width=0.19\linewidth]{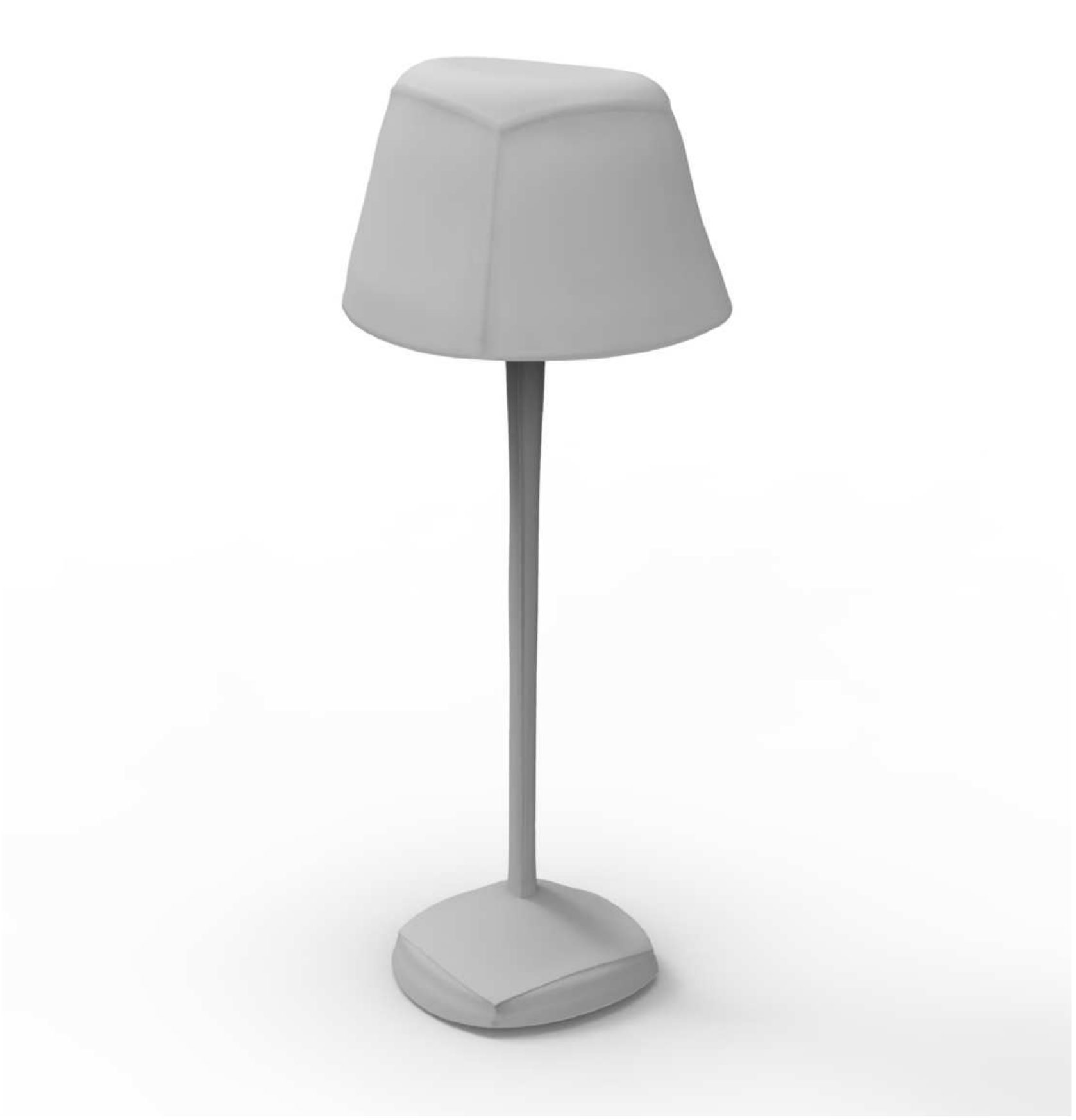}
    \includegraphics[width=0.19\linewidth]{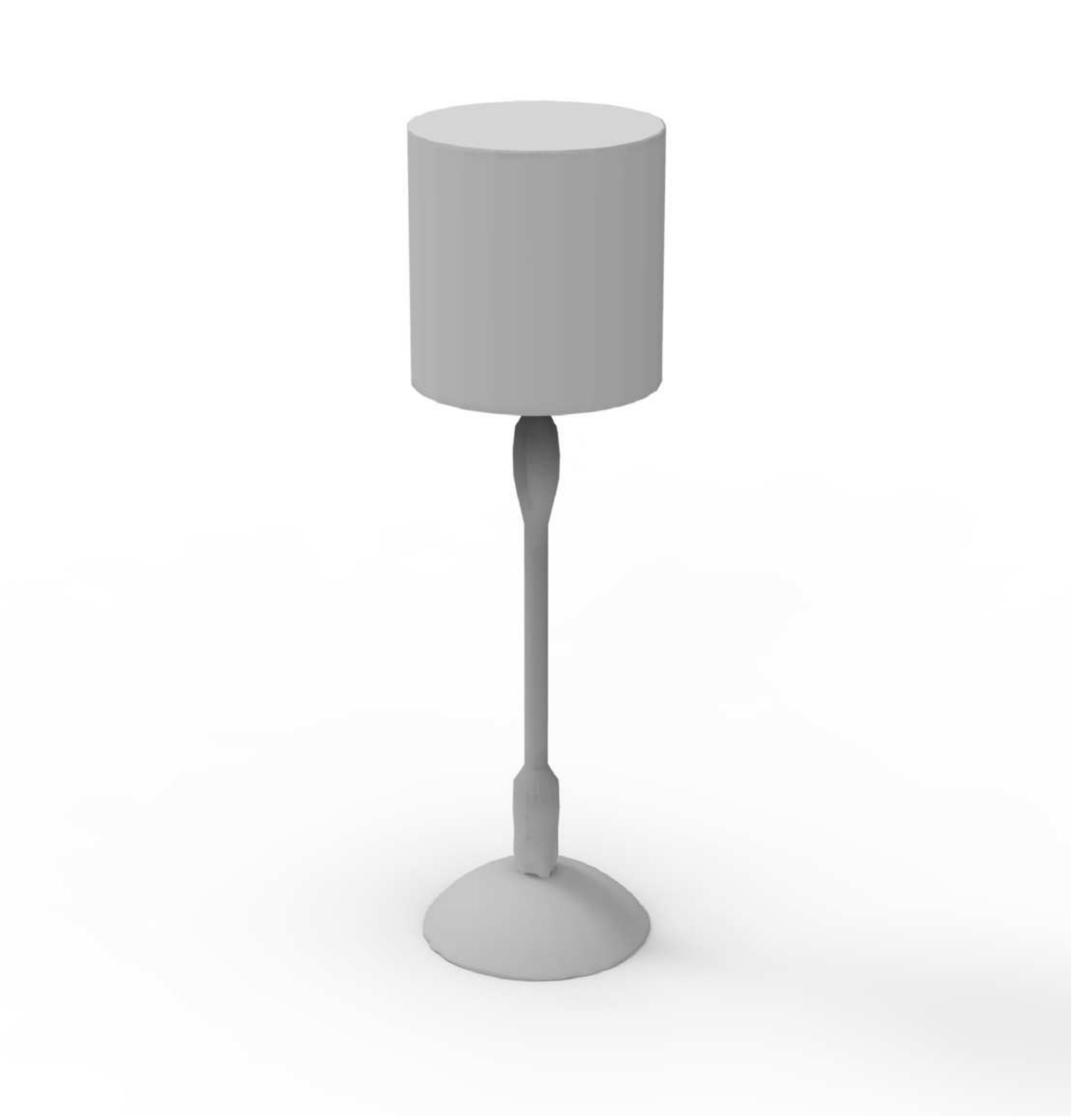}\\
    \includegraphics[width=0.19\linewidth]{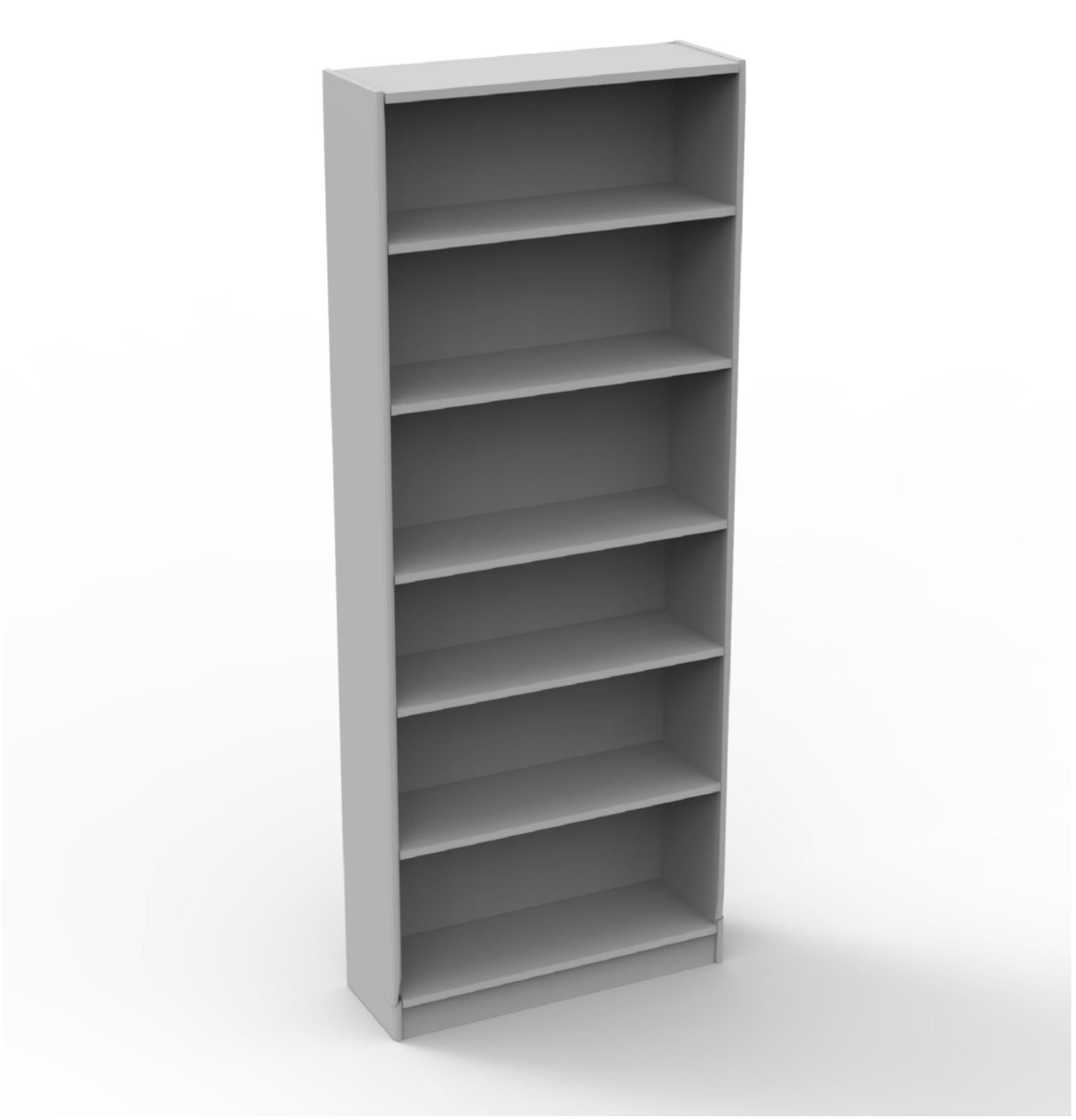}
    \includegraphics[width=0.19\linewidth]{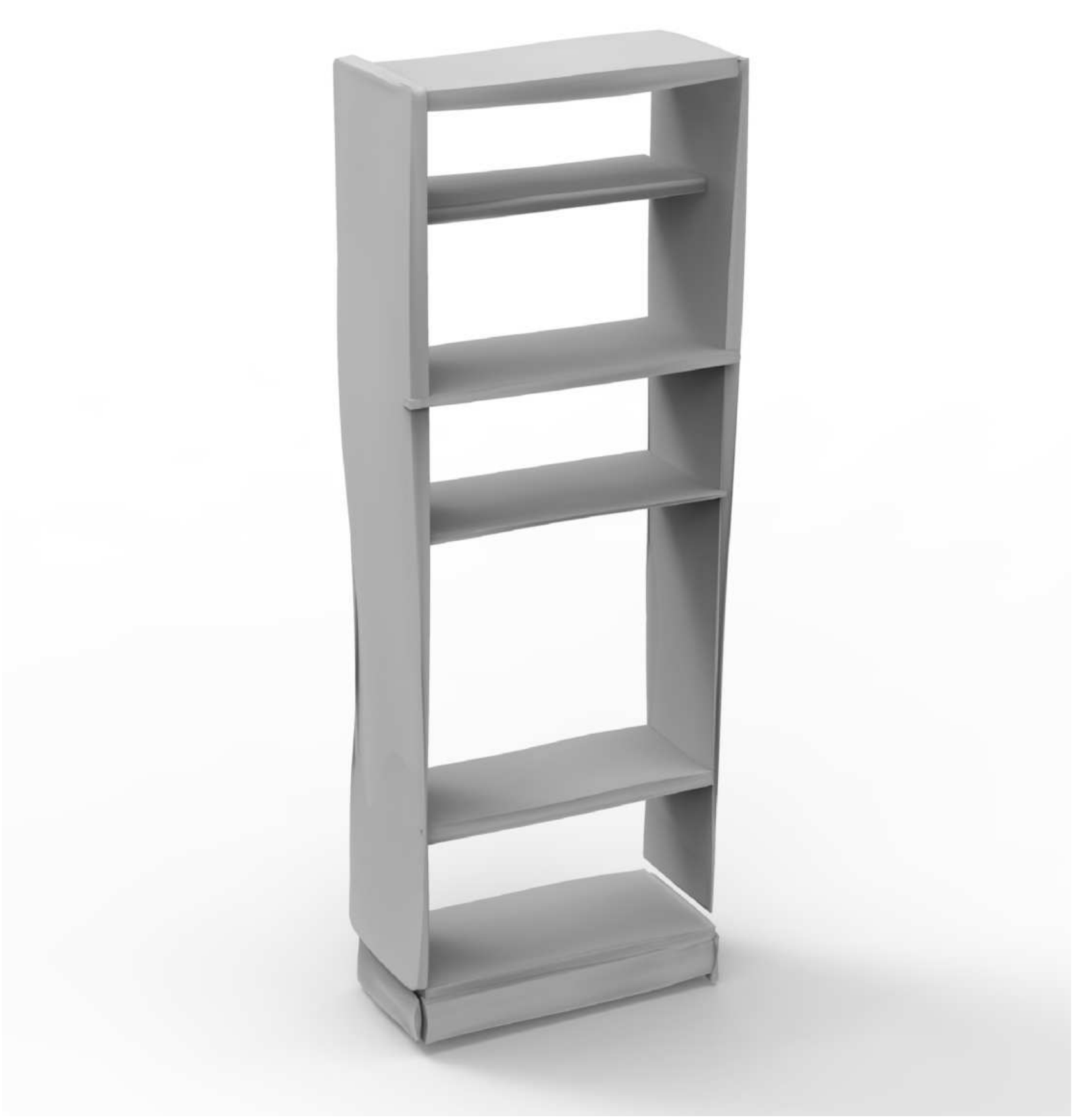}
    \includegraphics[width=0.19\linewidth]{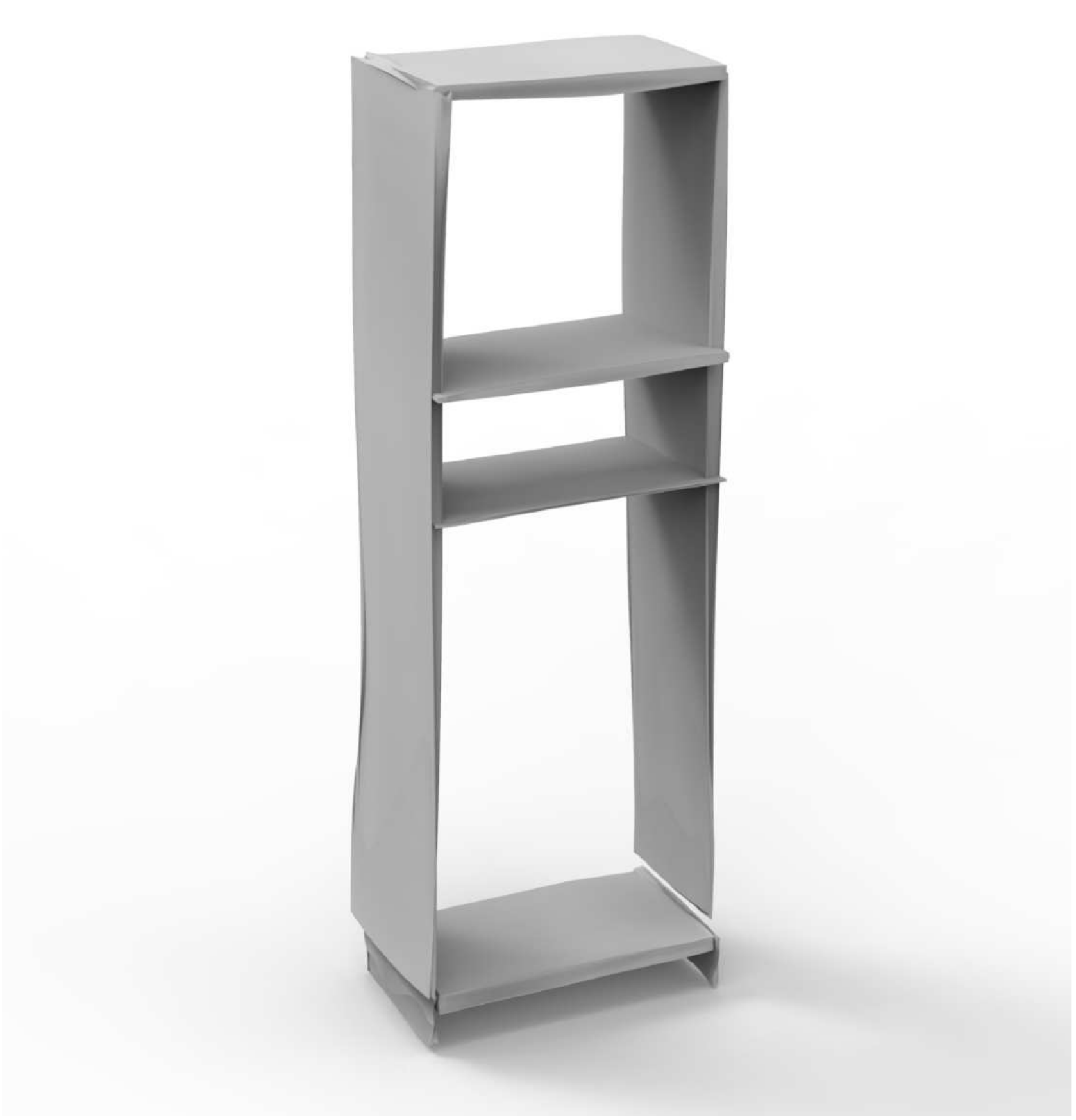}
    \includegraphics[width=0.19\linewidth]{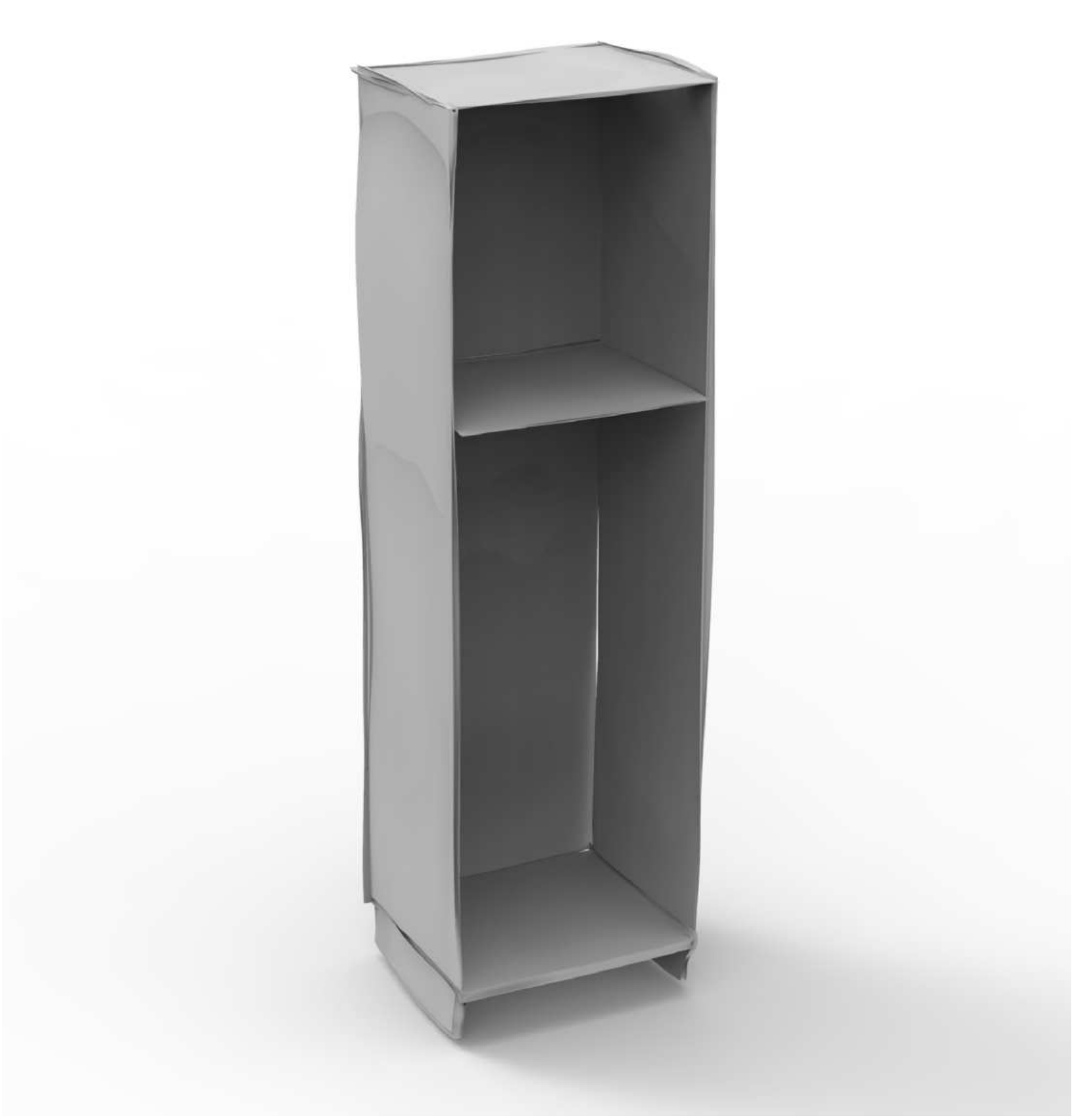}
    \includegraphics[width=0.19\linewidth]{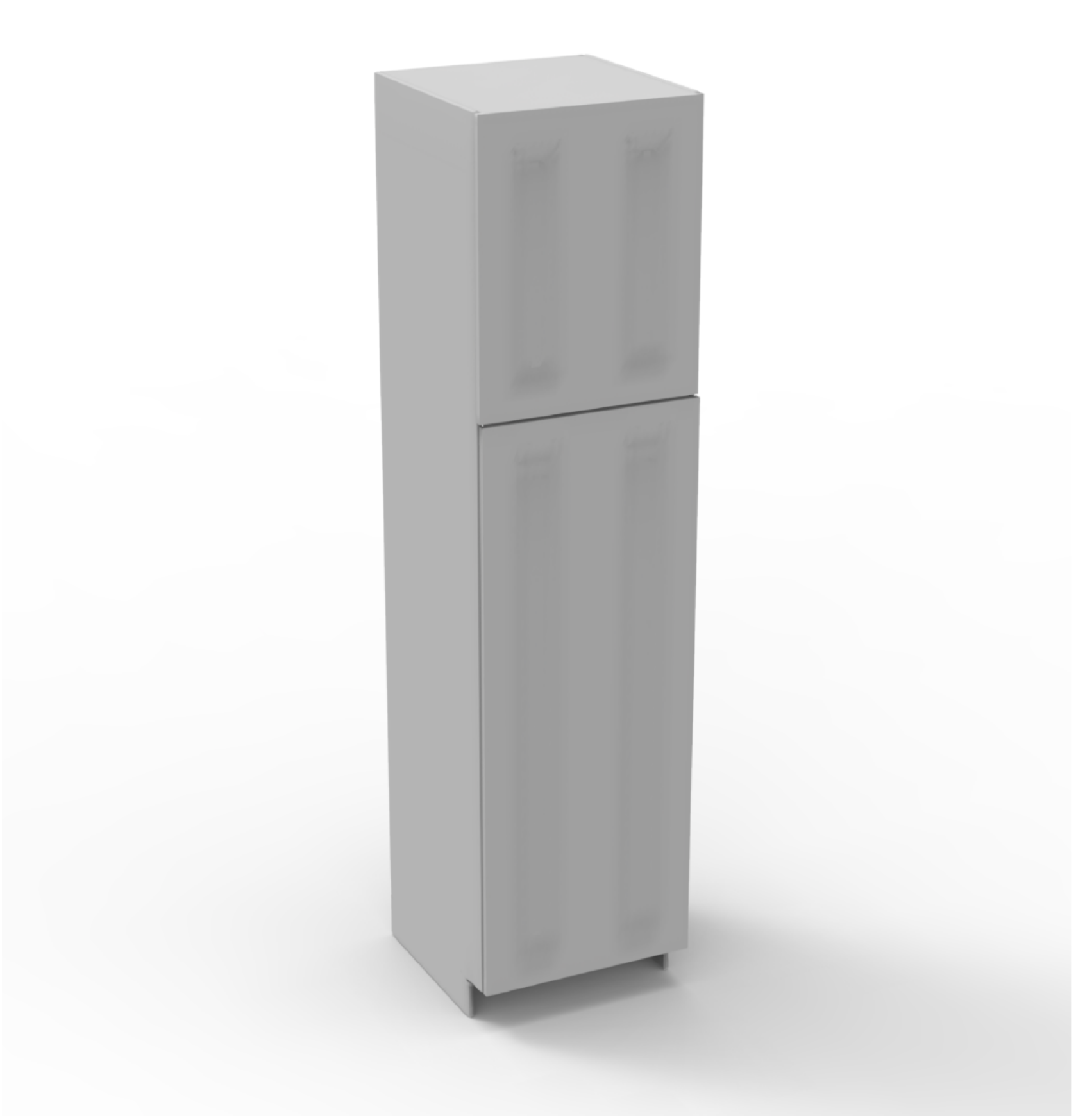}
    \vspace{-3mm}
    \caption{\yjr{Shape interpolation results. \yj{We linearly interpolate between input shape pairs (the left most and the right most shapes) jointly in the structure and geometry latent spaces. We see both continuous geometry variations and discrete structure changes. For the chair examples, in the first row, we see that the armrests become smaller and then disappear while the backrest changes from a square to round fashion in a more natural manner, while in the second row, the backrest gradually becomes square, while the supporter disappears form the first chair to the second chair.
    We observe similar behaviors for the table, lamp, and cabinet results.}}}
    \label{fig:interp}
% \end{figure}
\end{minipage}
\begin{minipage}[c]{\linewidth}
% \begin{figure}[h]
  \centering
  \begin{tikzpicture}
  \matrix[nodes={anchor=south west,inner sep=0pt}]{ 
    \node (A1) {\includegraphics[width=0.2\linewidth, cfbox=orange 1pt 1pt]{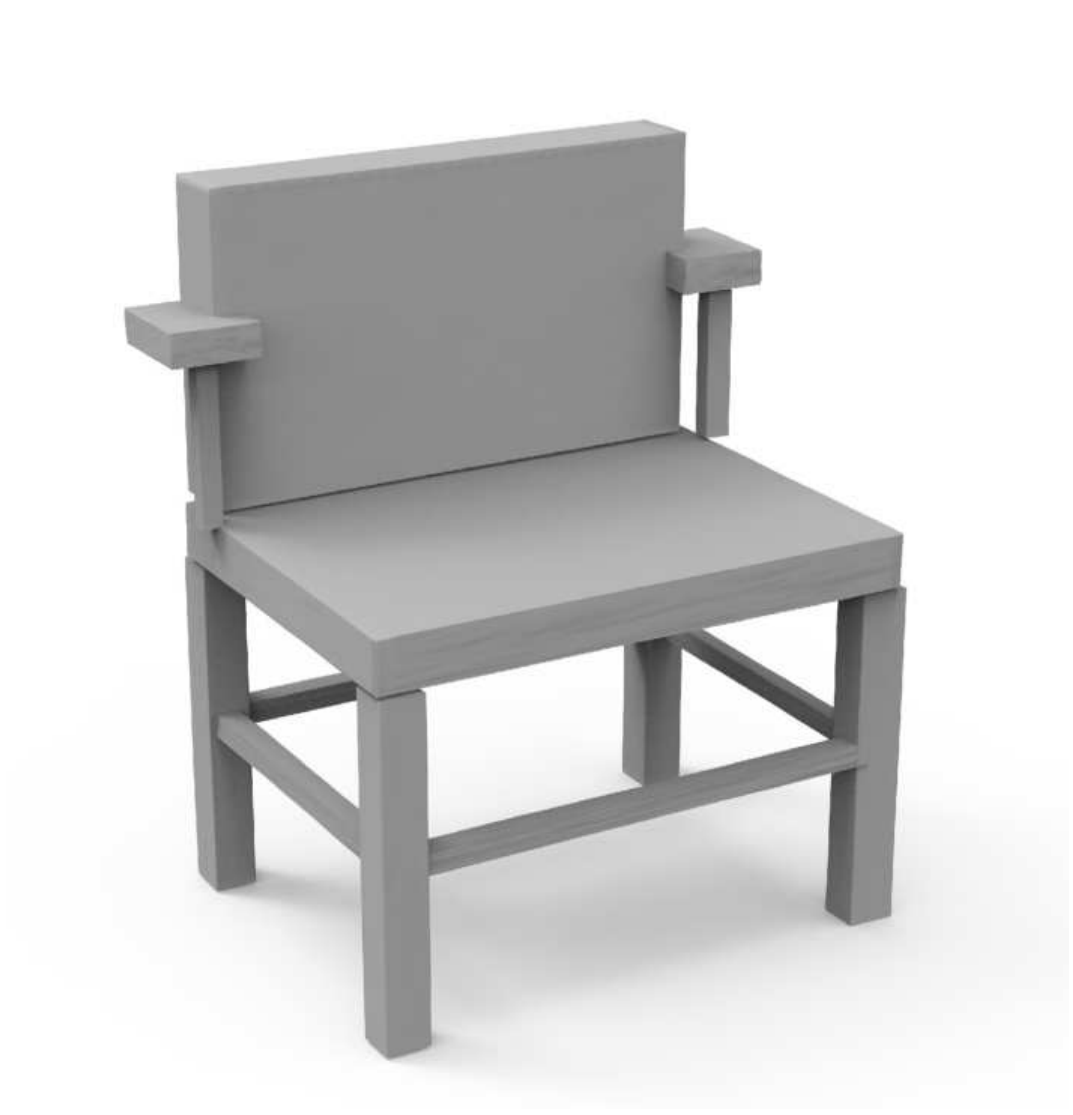}};  & 
    \node (A2) {\includegraphics[width=0.2\linewidth]{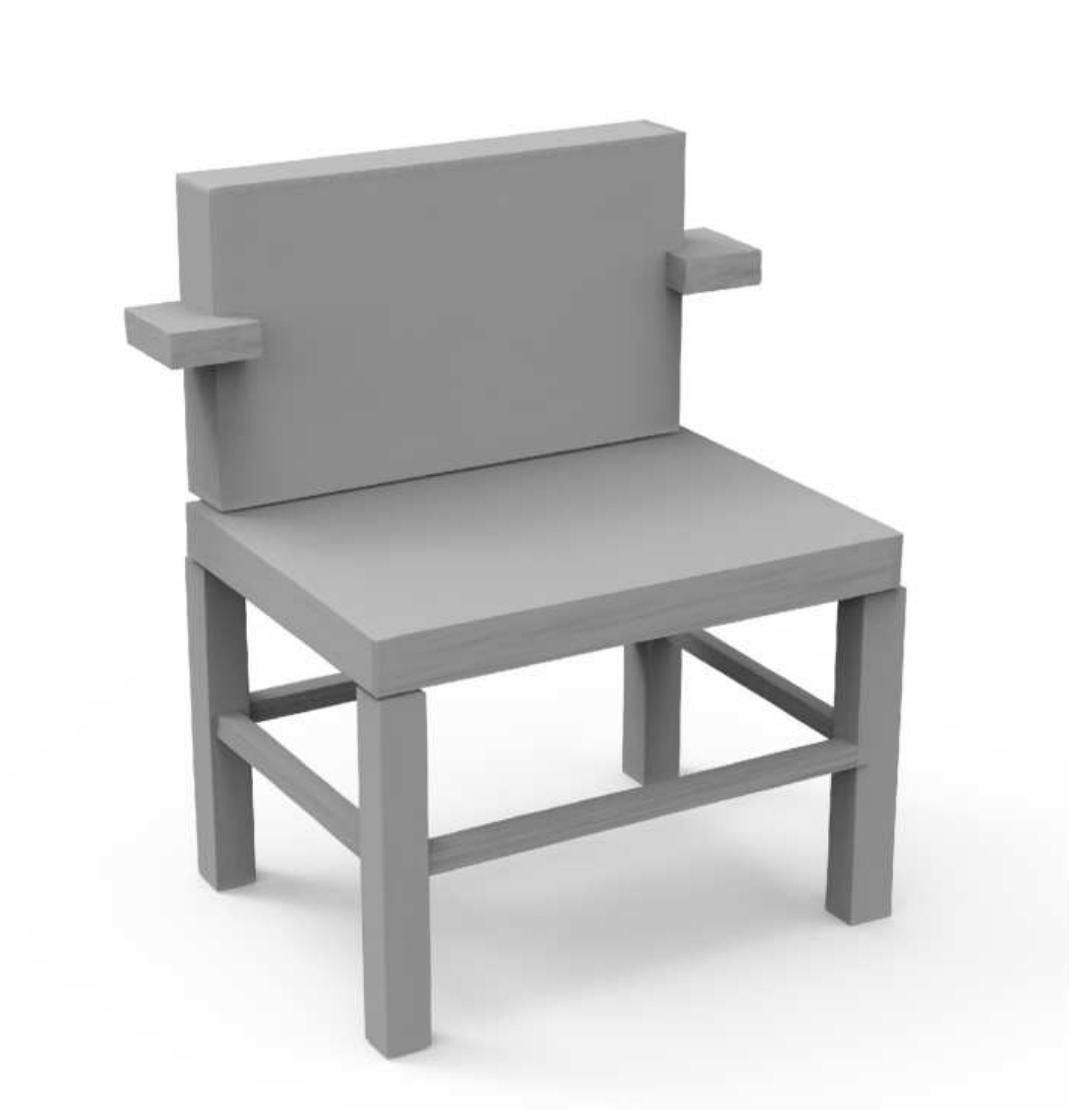}};  & 
    \node (A3) {\includegraphics[width=0.2\linewidth]{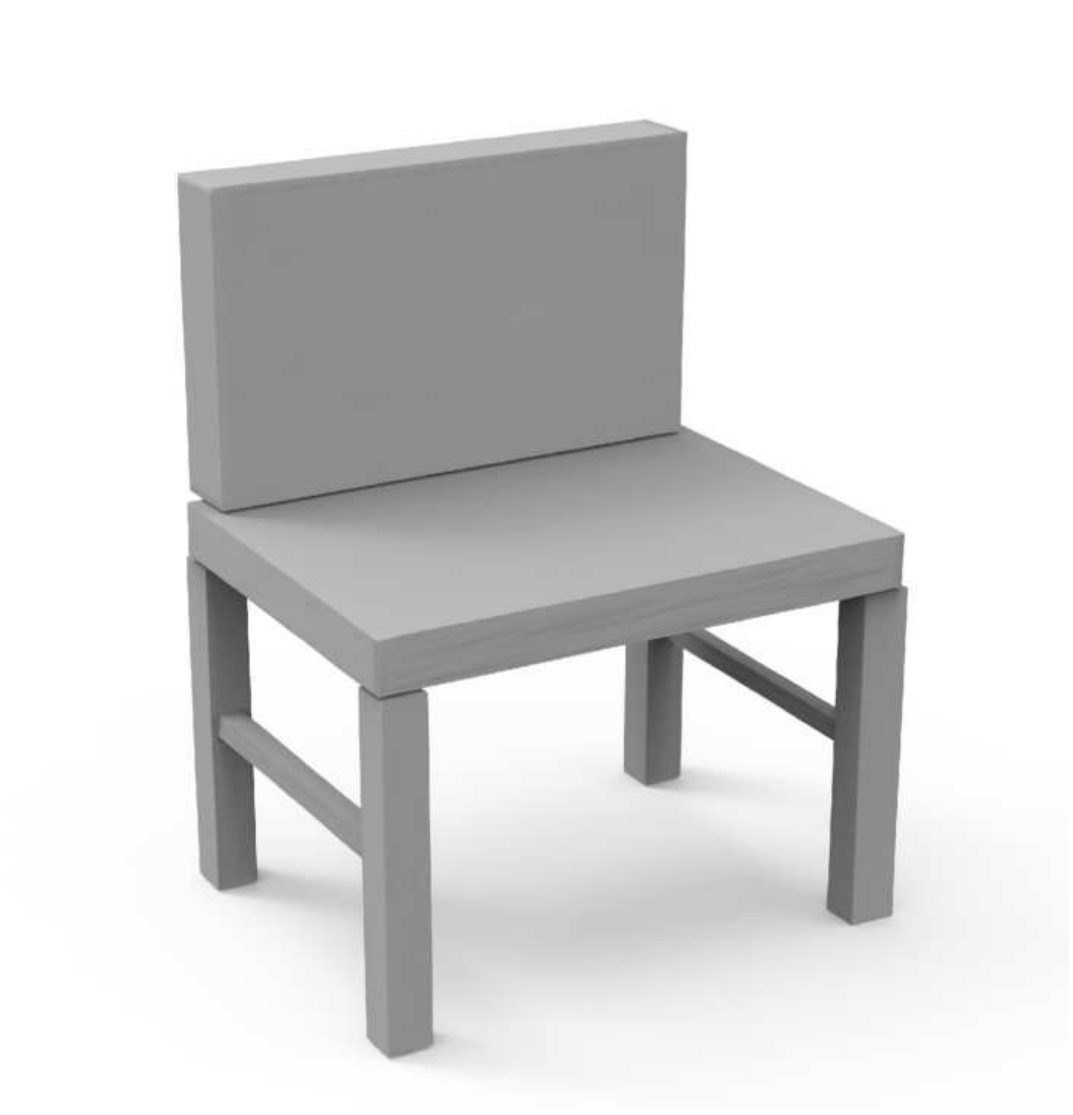}};  & 
    \node (A4) {\includegraphics[width=0.2\linewidth]{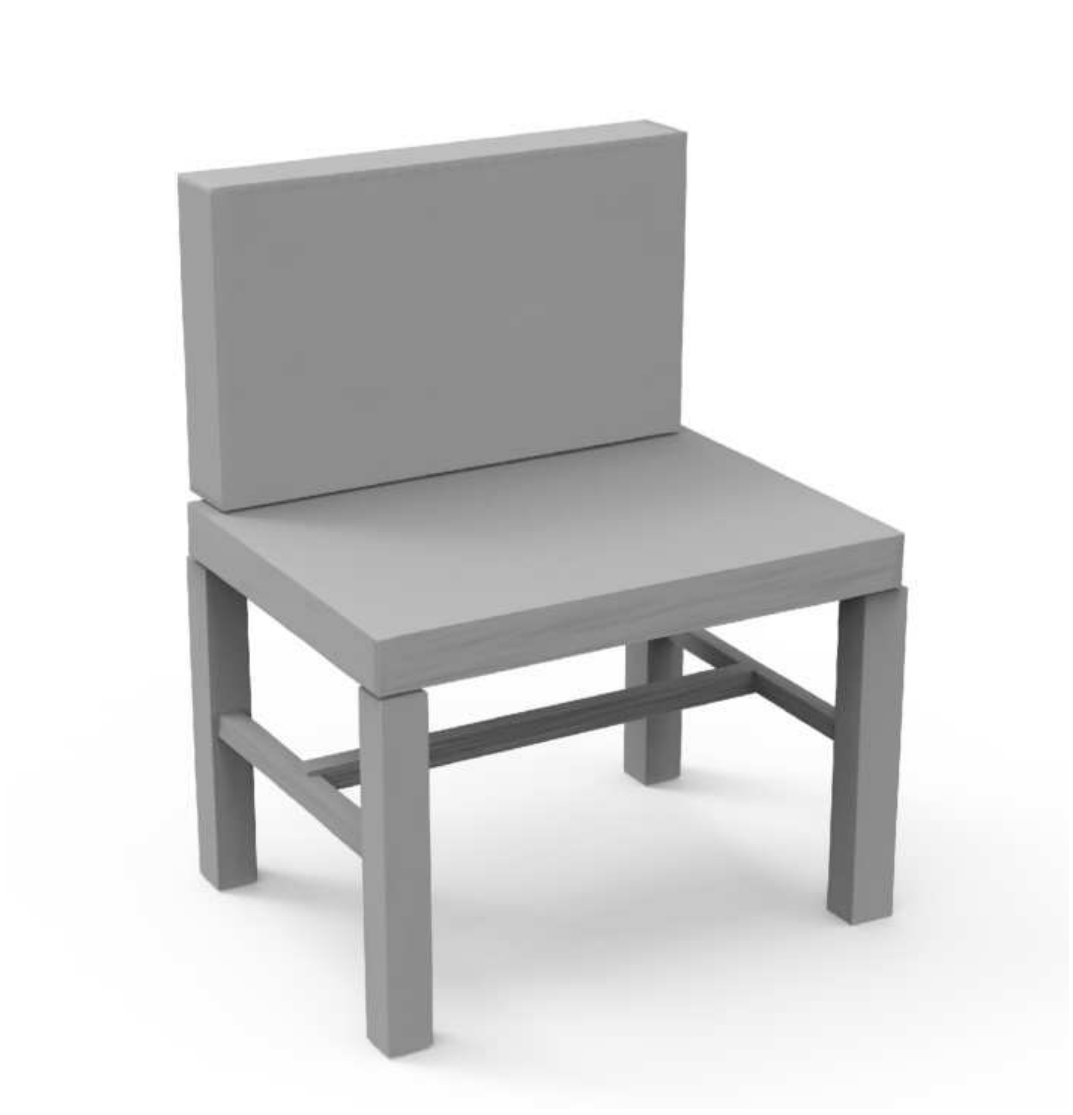}}; \\
    \node (B1) {\includegraphics[width=0.2\linewidth]{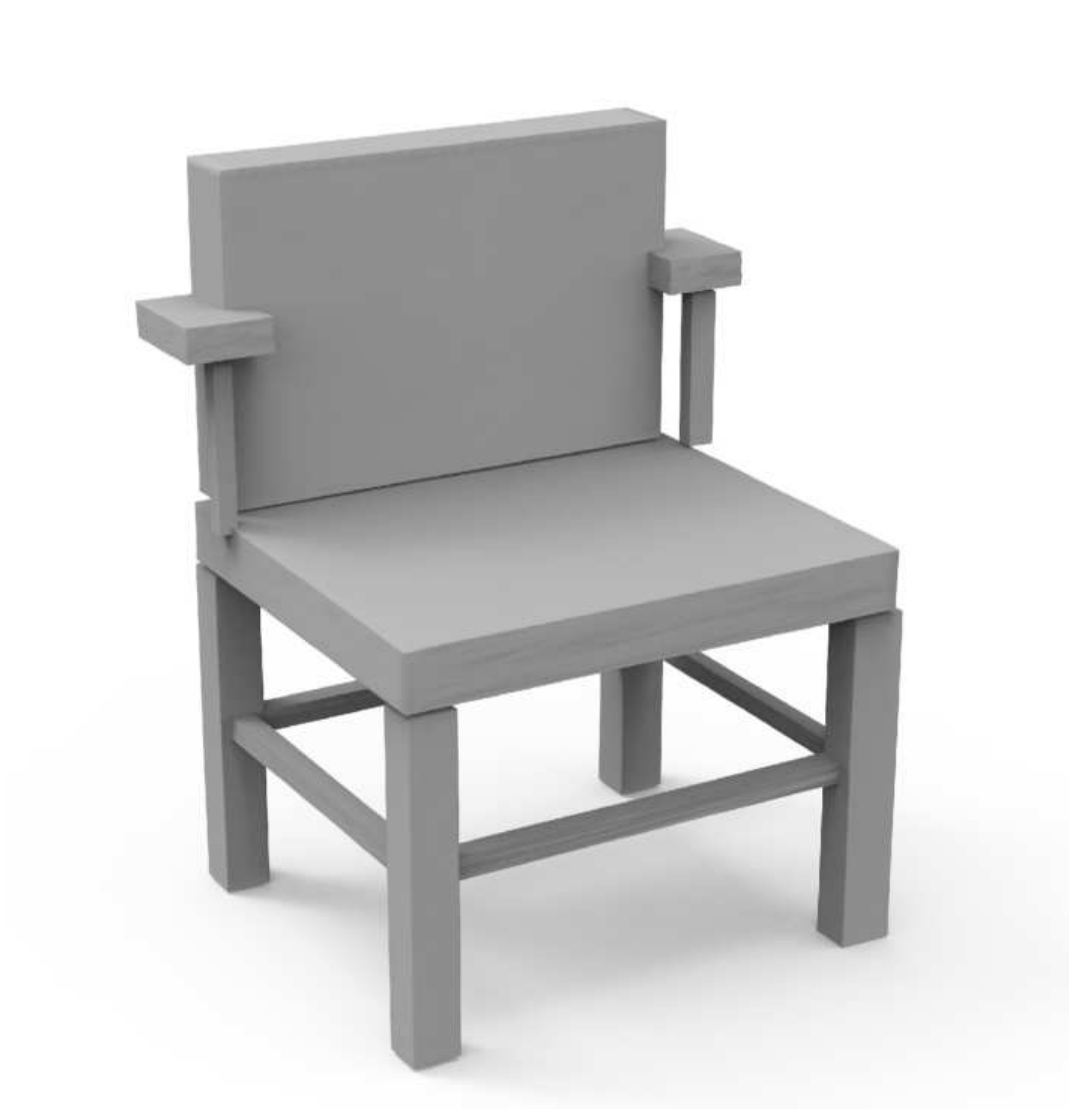}}; & 
    \node (B2) {\includegraphics[width=0.2\linewidth]{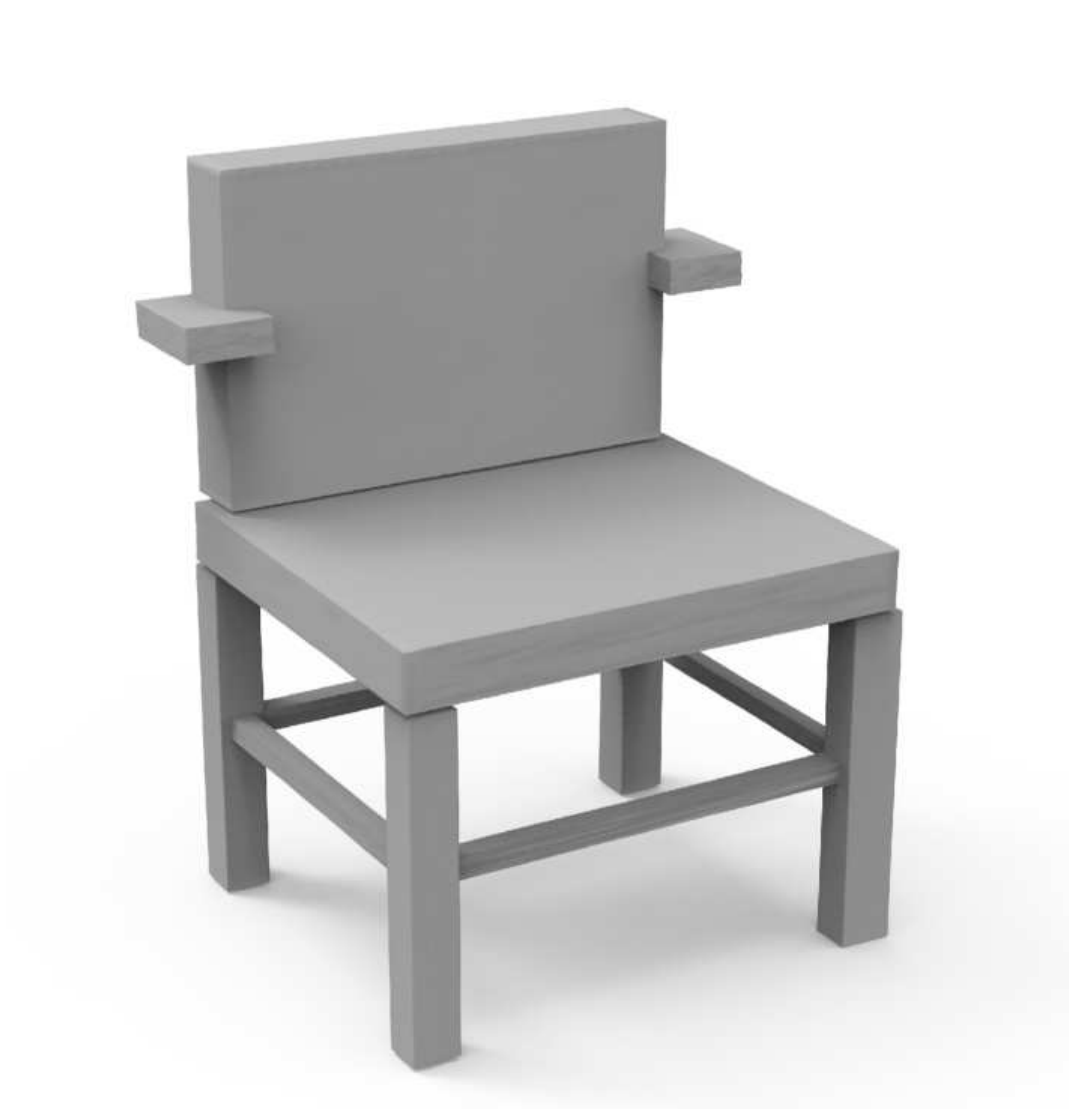}}; &
    \node (B3) {\includegraphics[width=0.2\linewidth]{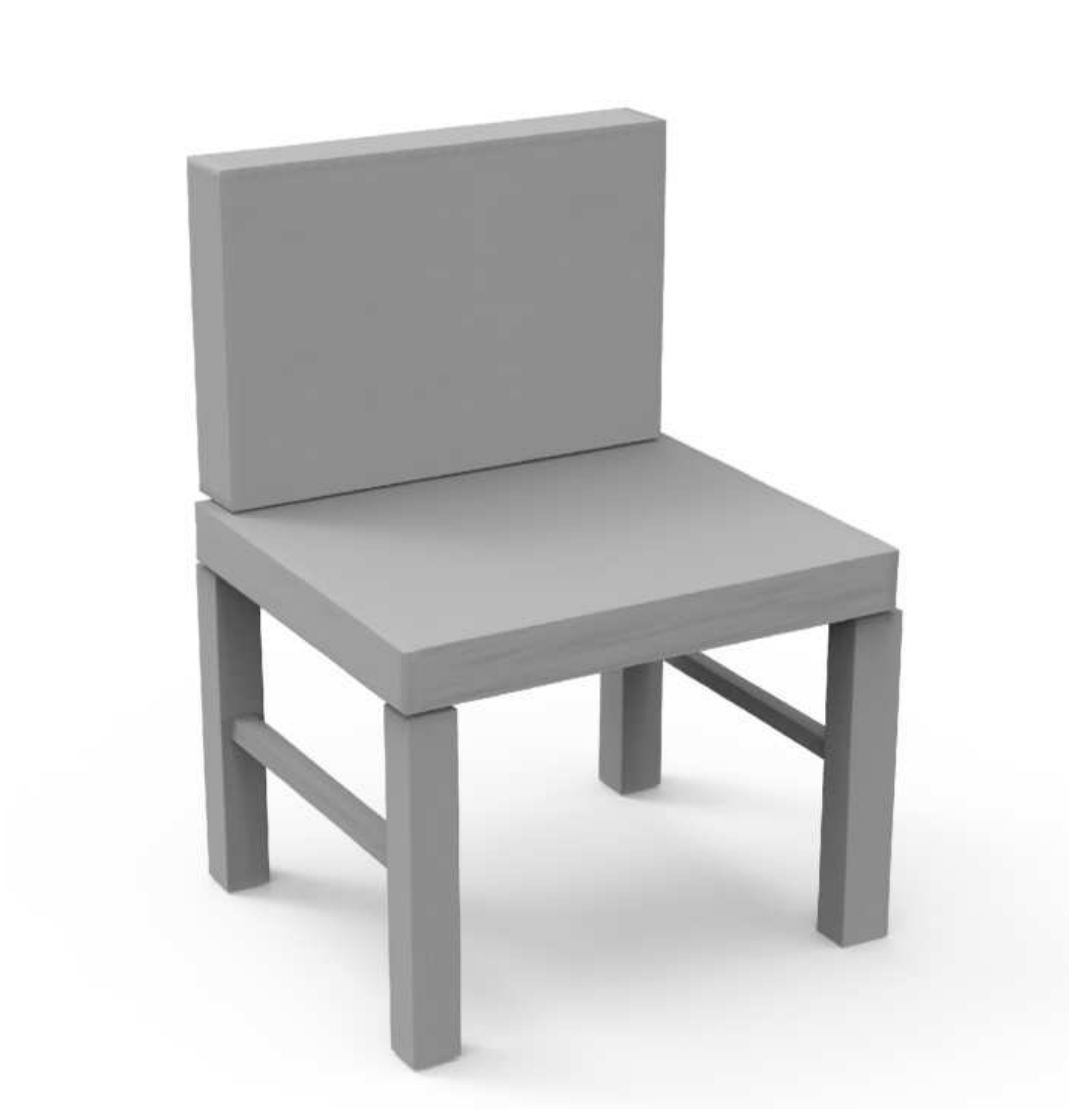}}; &
    \node (B4) {\includegraphics[width=0.2\linewidth]{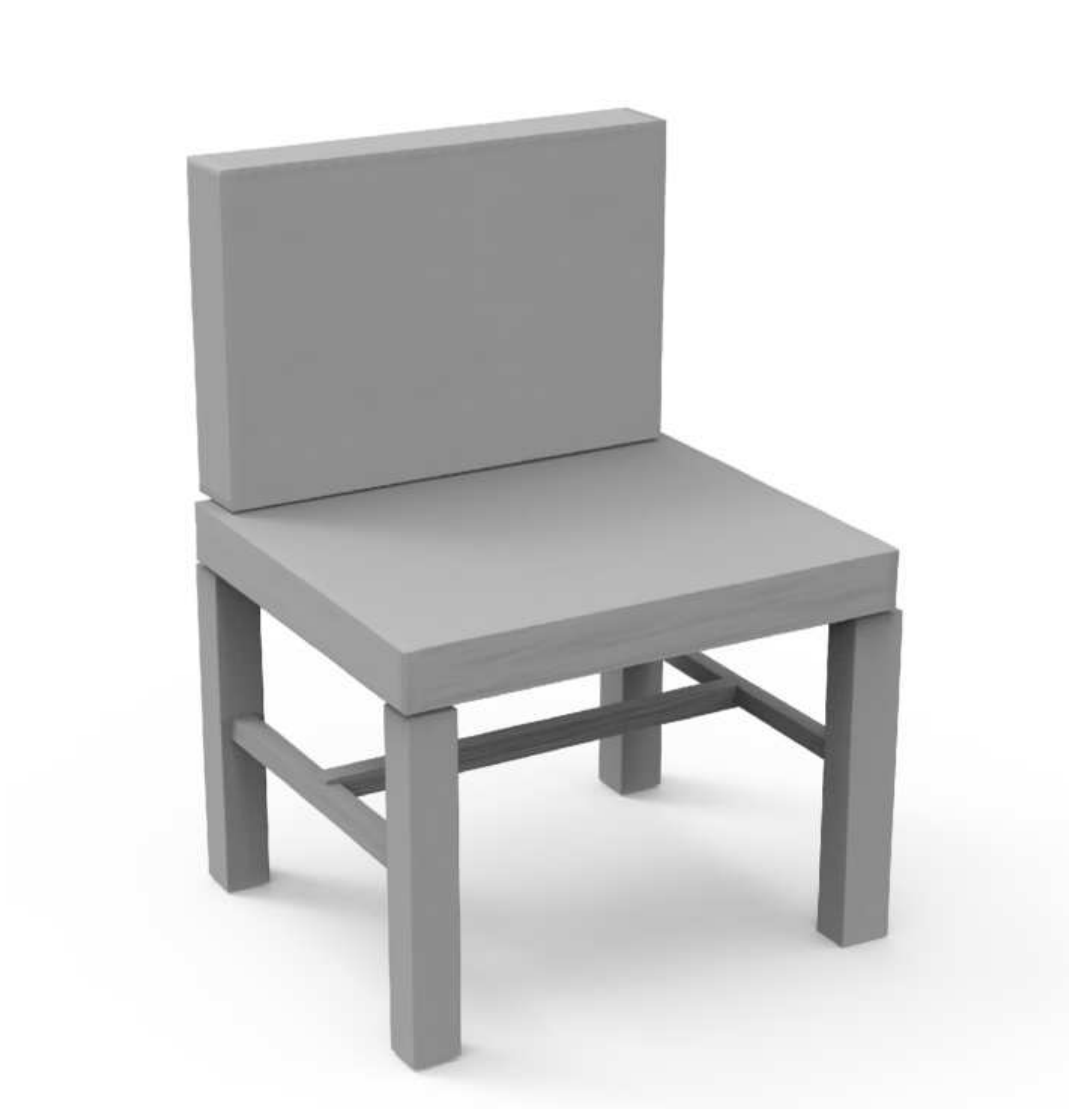}}; \\
    \node (C1) {\includegraphics[width=0.2\linewidth]{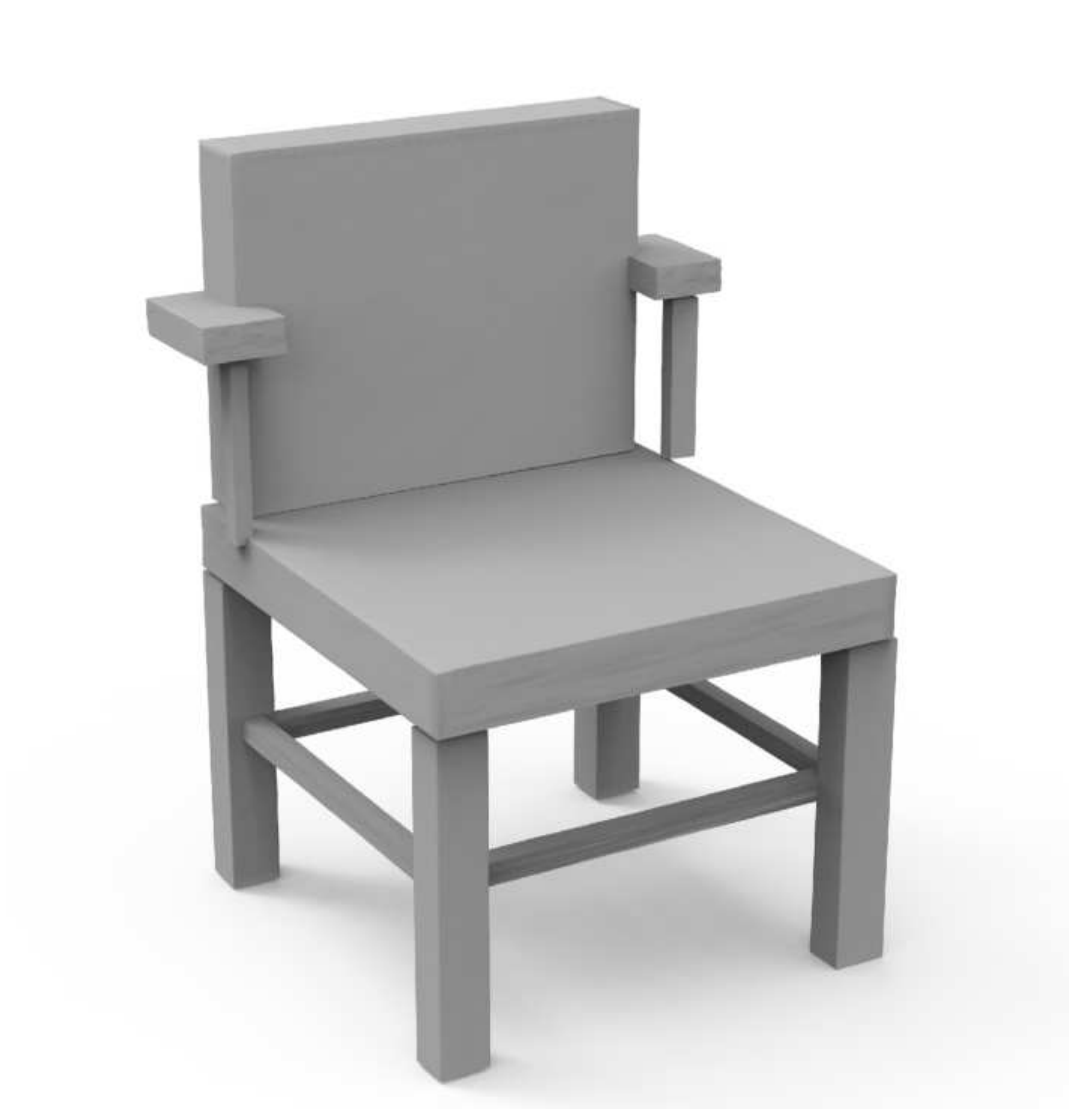}}; & 
    \node (C2) {\includegraphics[width=0.2\linewidth]{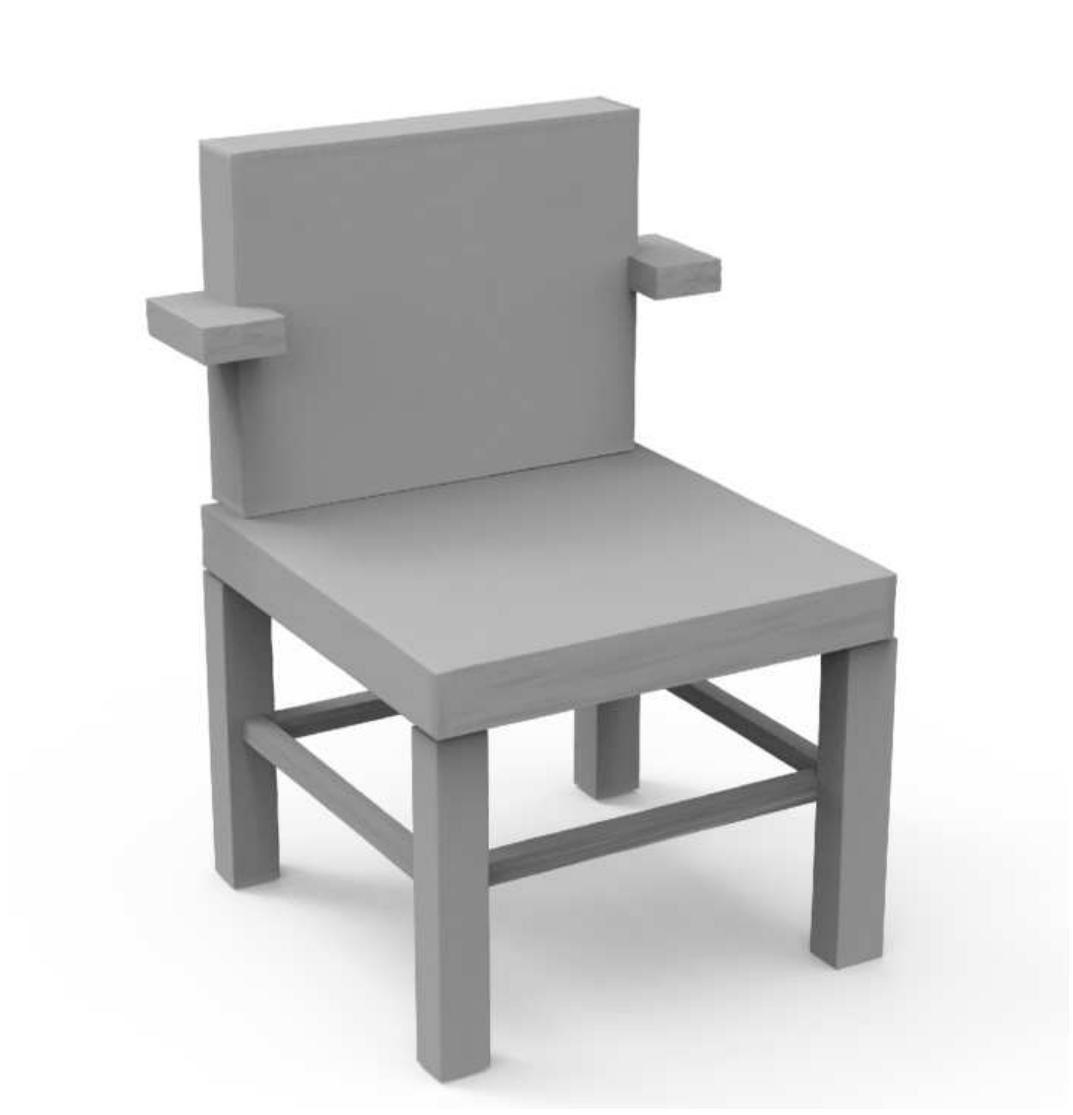}}; &
    \node (C3) {\includegraphics[width=0.2\linewidth]{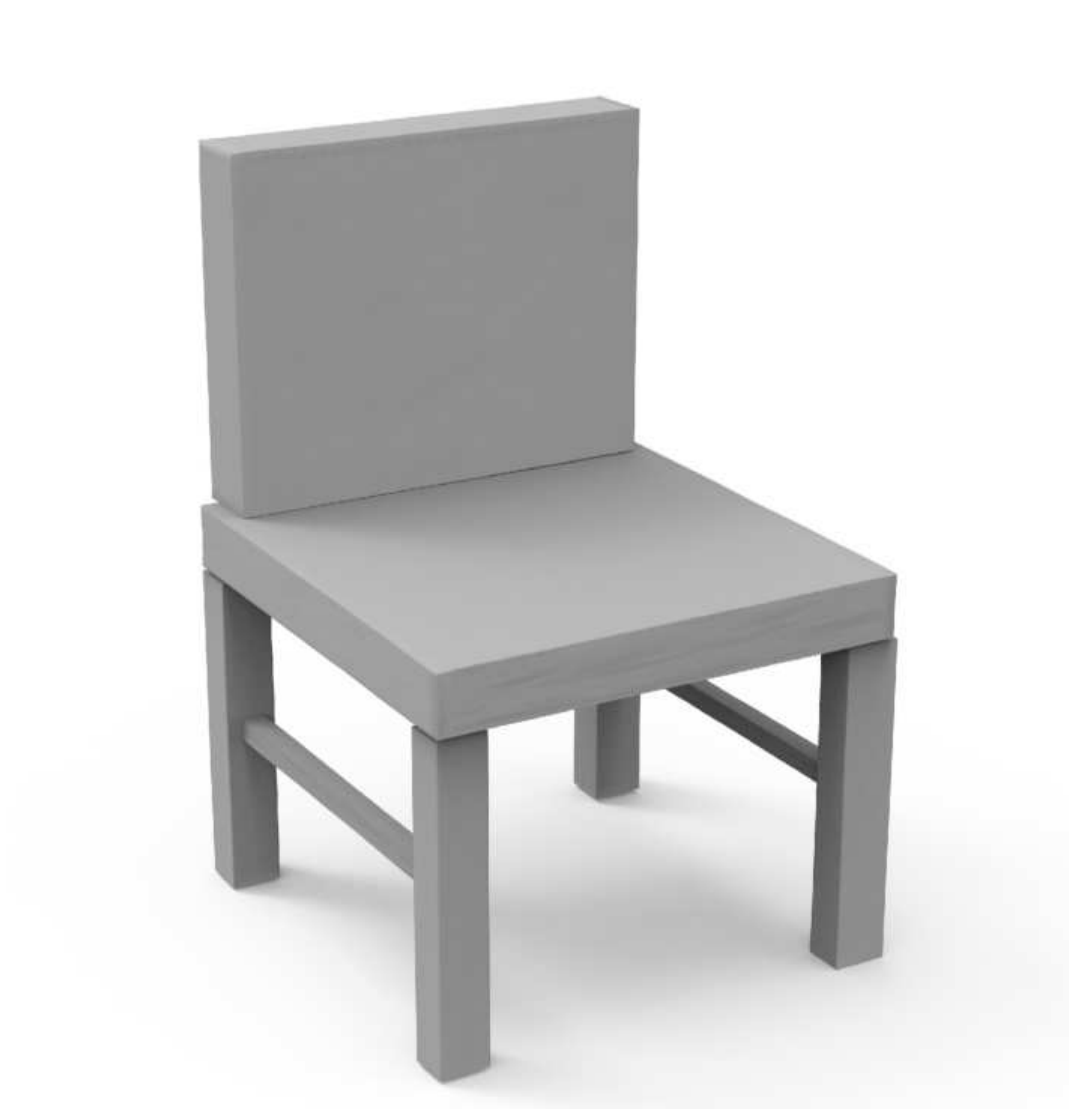}}; &
    \node (C4) {\includegraphics[width=0.2\linewidth]{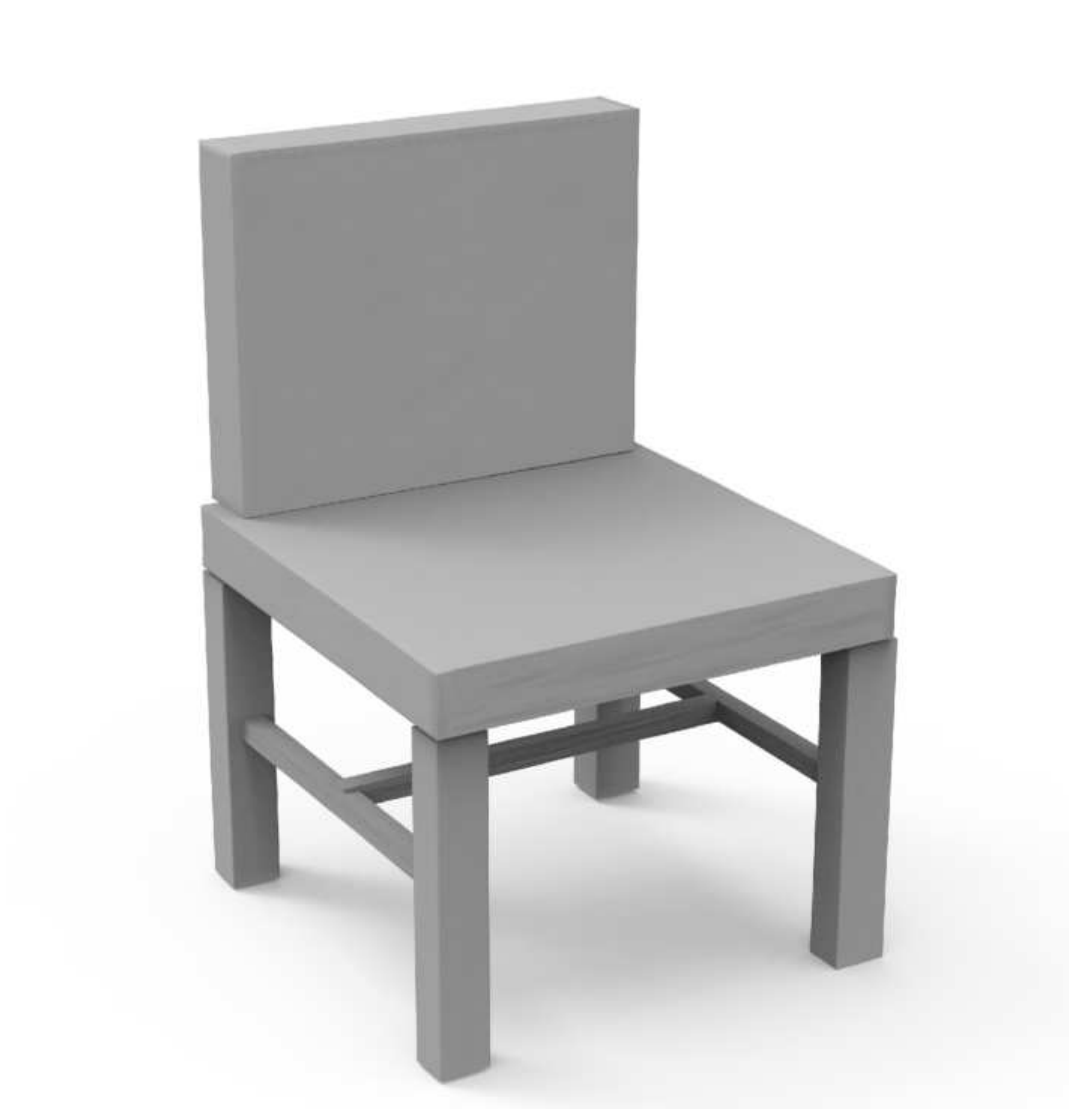}}; \\
    \node (D1) {\includegraphics[width=0.2\linewidth]{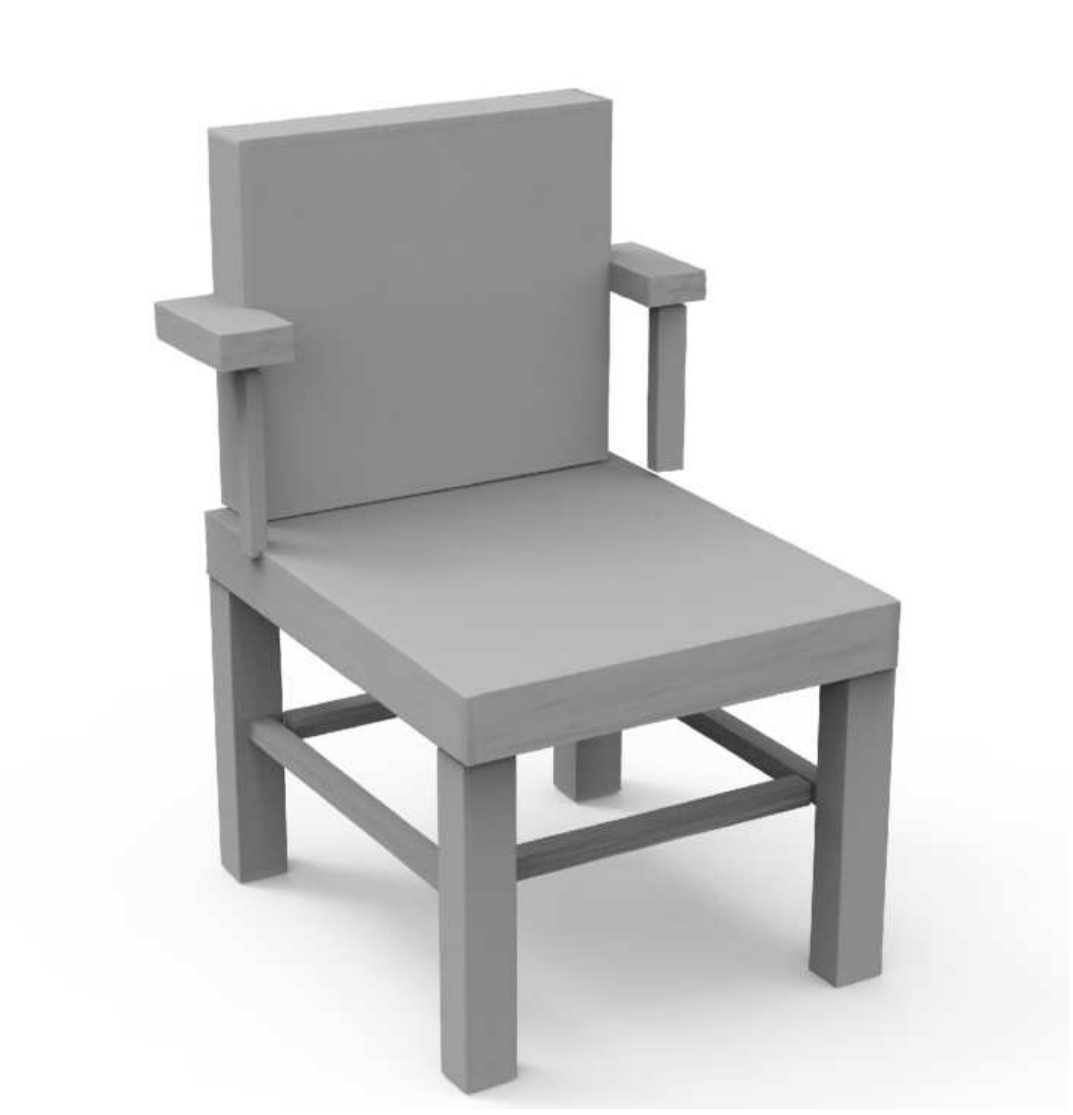}}; & 
    \node (D2) {\includegraphics[width=0.2\linewidth]{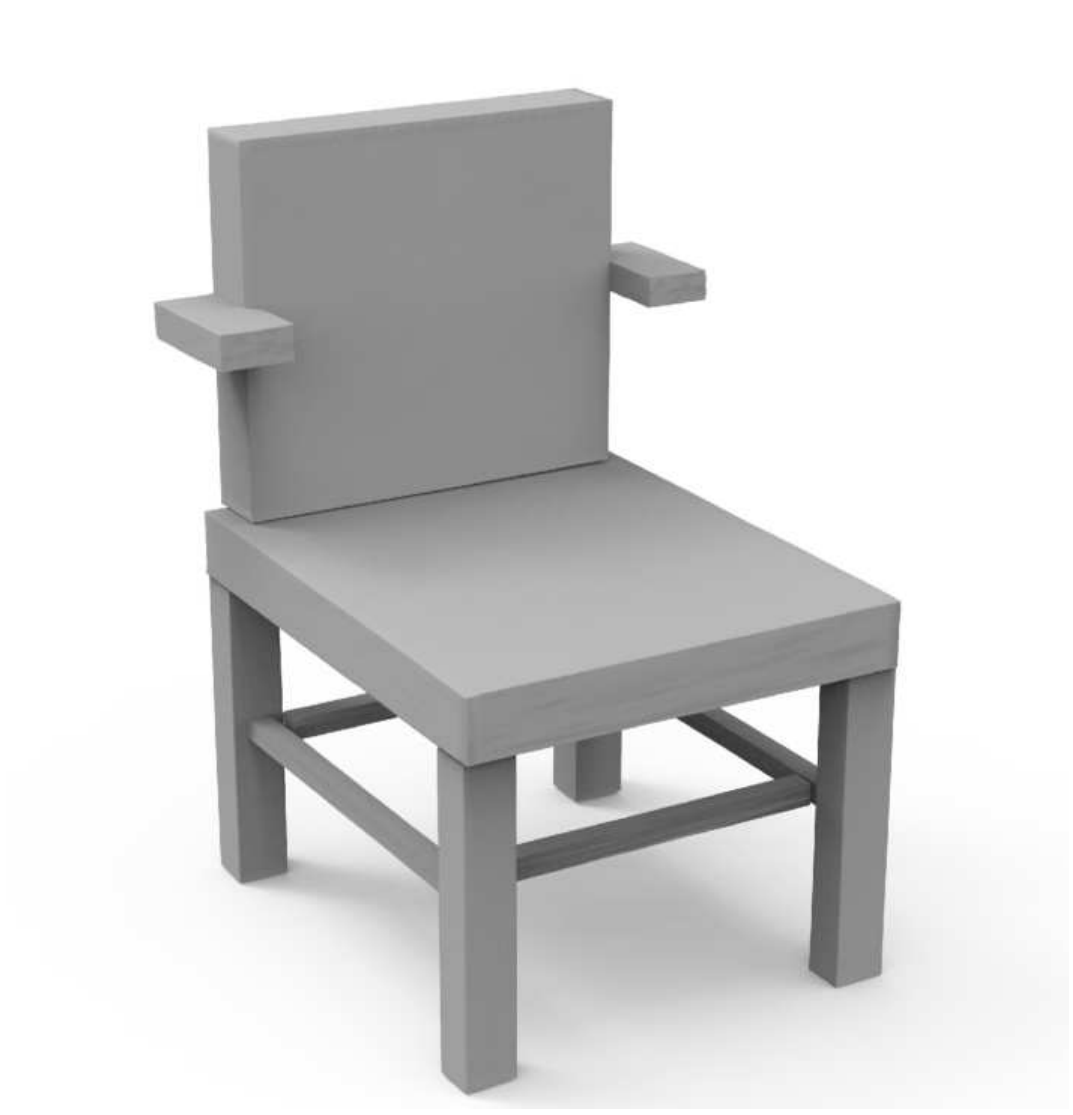}}; &
    \node (D3) {\includegraphics[width=0.2\linewidth]{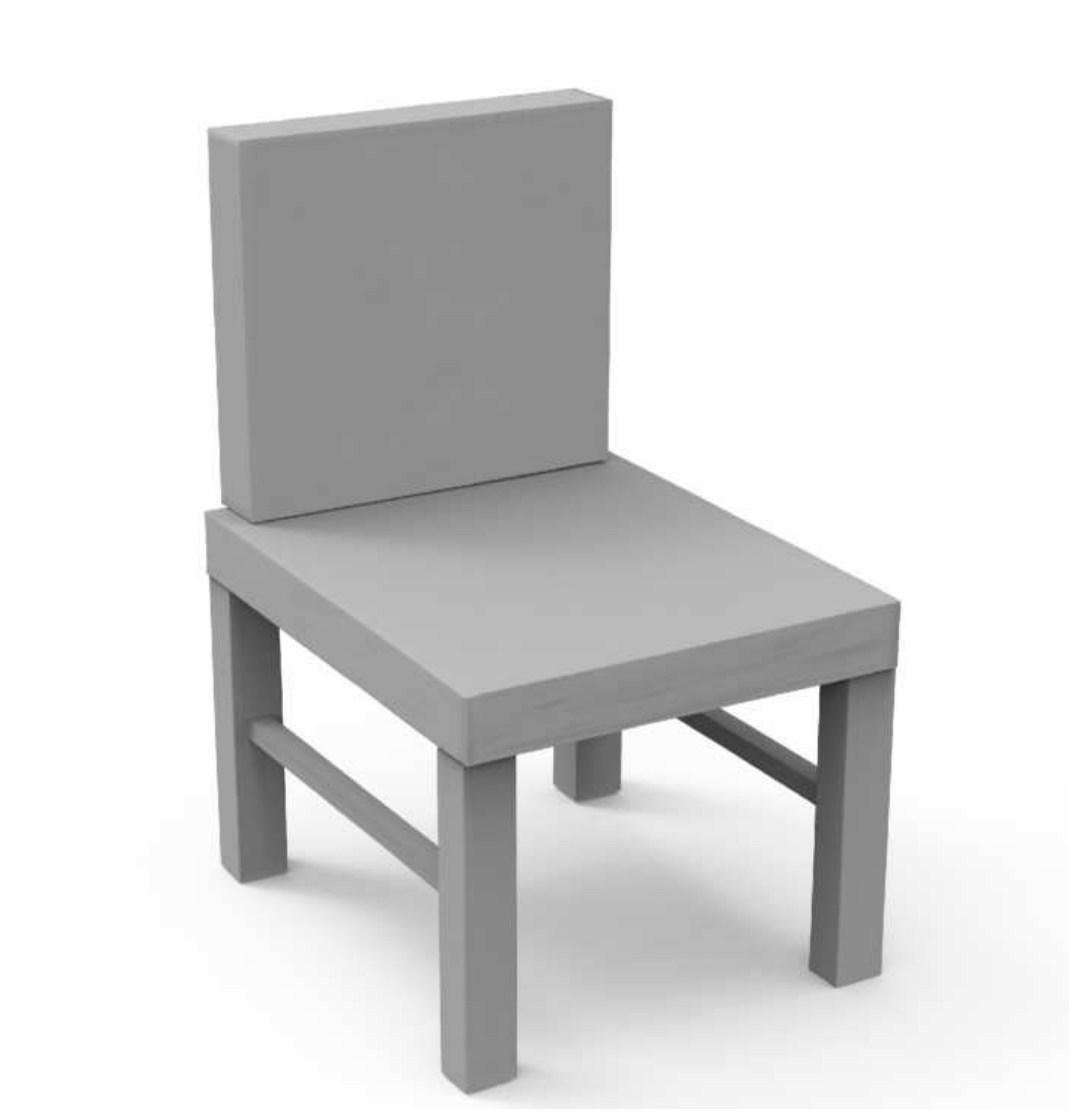}}; &
    \node (D4) {\includegraphics[width=0.2\linewidth, cfbox=orange 1pt 1pt]{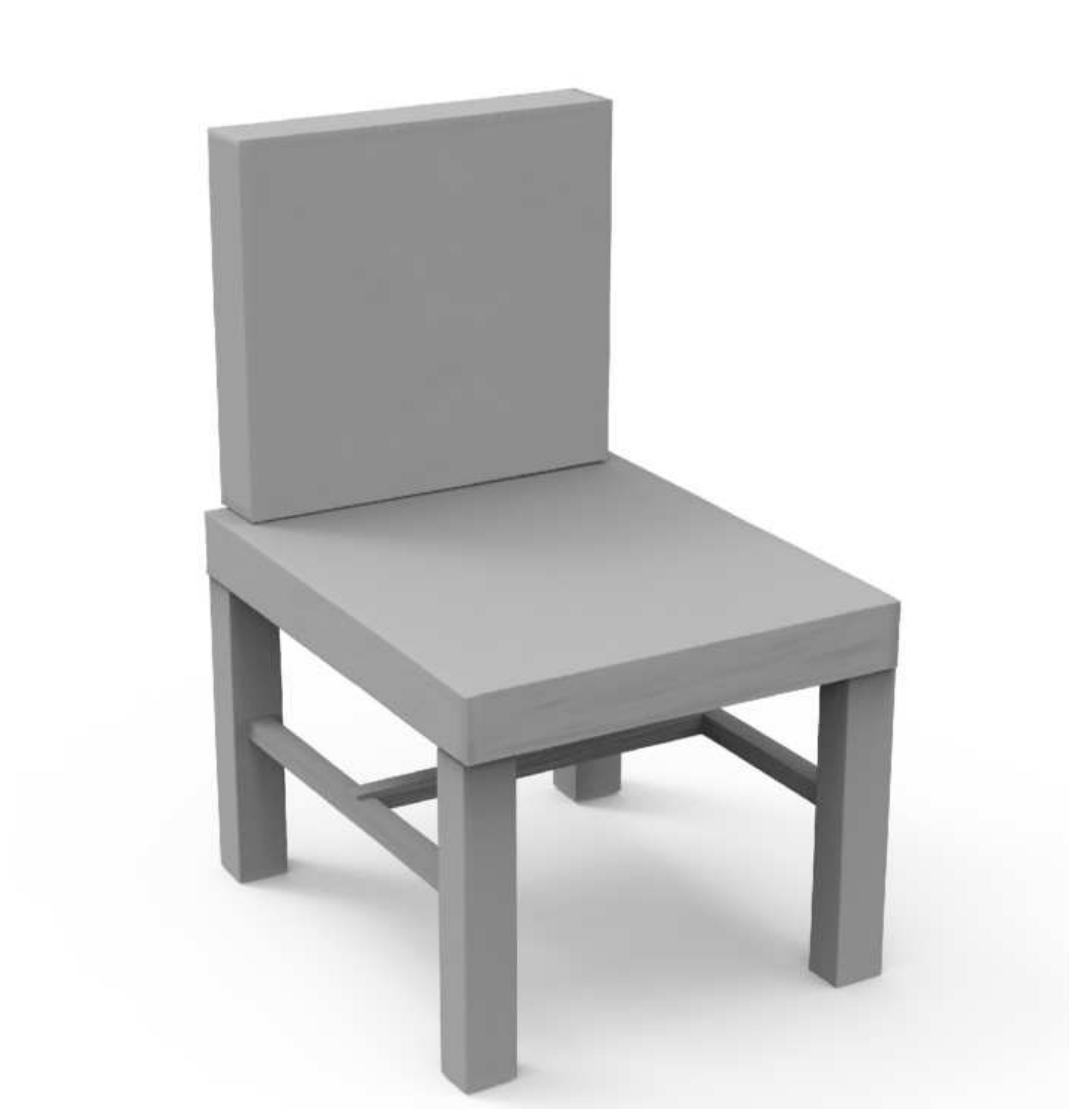}}; \\
  };
  \node[fit=(A1) (A2) (A3) (A4)
            (B1) (B2) (B3) (B4)
            (C1) (C2) (C3) (C4)
            (D1) (D2) (D3) (D4),
            inner sep=0pt,
            ] (PIC) {};

  \draw[line width=1pt,arrows={-Stealth[length=4mm]}] ([xshift=-1em,yshift=-0.5em]PIC.south west) -- ([xshift=-1em,yshift=-0.5em]PIC.south east);
  \draw[line width=1pt,arrows={-Stealth[length=4mm]}] ([yshift=-1em,xshift=-0.5em]PIC.south west) -- ([yshift=-1em,xshift=-0.5em]PIC.north west);

  \node[anchor=south] (label) [font=\fontsize{10}{10}\selectfont]at ([xshift=-3em,yshift=-2.em]A3|-PIC.south) {Structure};
  \node[anchor=center,rotate=90] (label) [font=\fontsize{10}{10}\selectfont] at ([xshift=-1.5em,yshift=3em]C1-|PIC.west) {Geometry};

\end{tikzpicture}
\vspace{-1mm}
    \caption{\yjr{Disentangled shape reconstruction and interpolation results on the synthetic data. Here, the top left and bottom right shapes (highlighted with orange boxes) are the input shapes. %
    The remaining shapes are generated automatically with our DSG-Net, where in each row, the \emph{structure} of the shapes is interpolated while keeping the geometry unchanged, whereas in each column, the \emph{geometry} is interpolated while retaining the structure. \yj{The vertical axis and horizontal axis represent the variation of structure and geometry respectively.}}}
    \label{fig:decouple_synthetic}
    \end{minipage}
\end{figure}

\begin{figure}[h]
  \centering
  \begin{tikzpicture}
  \matrix[nodes={anchor=south west,inner sep=0pt}]{ 
    \node (A1) {\includegraphics[width=0.175\linewidth, cfbox=orange 1pt 1pt]{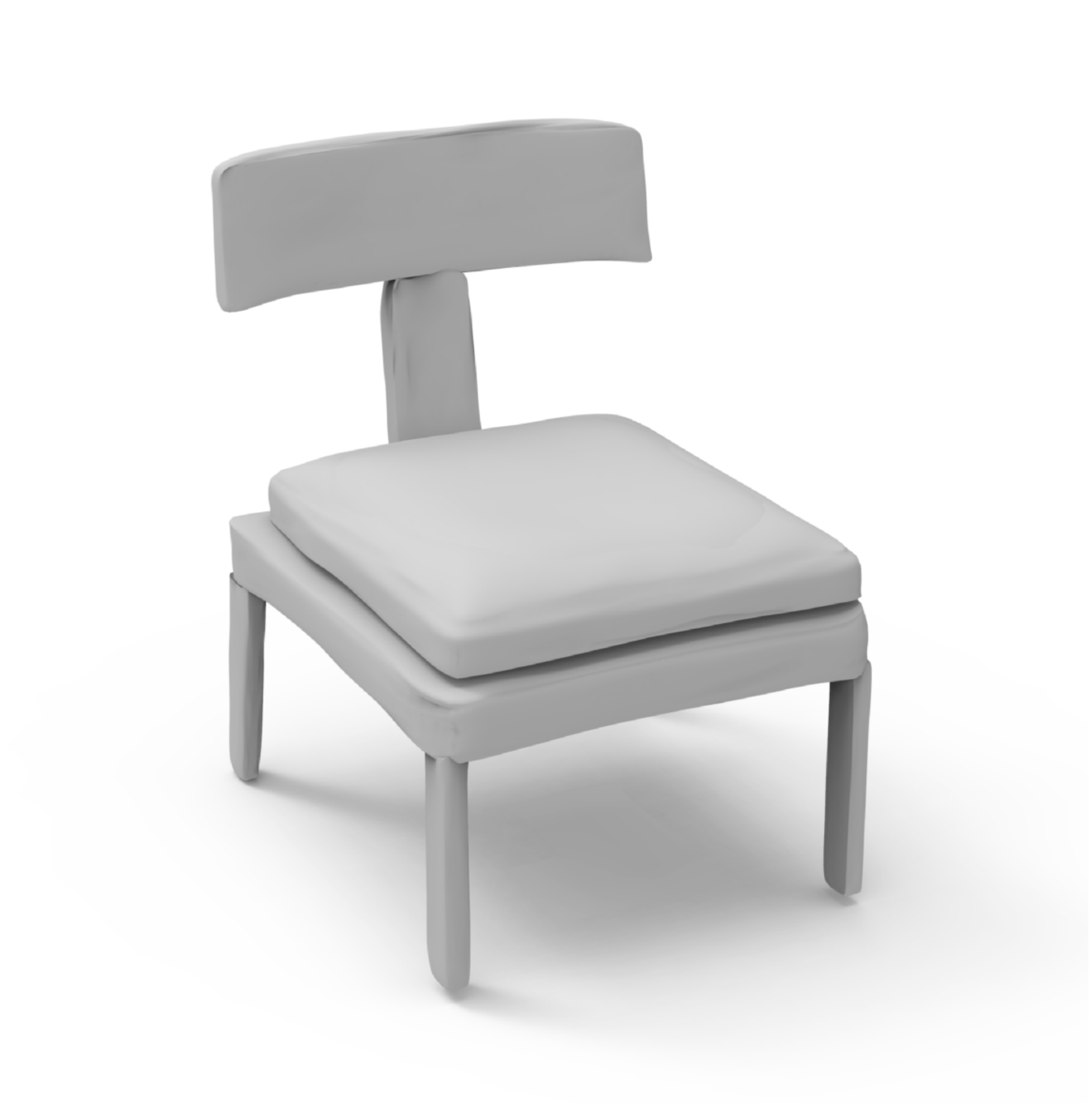}};  & 
    \node (A2) {\includegraphics[width=0.175\linewidth]{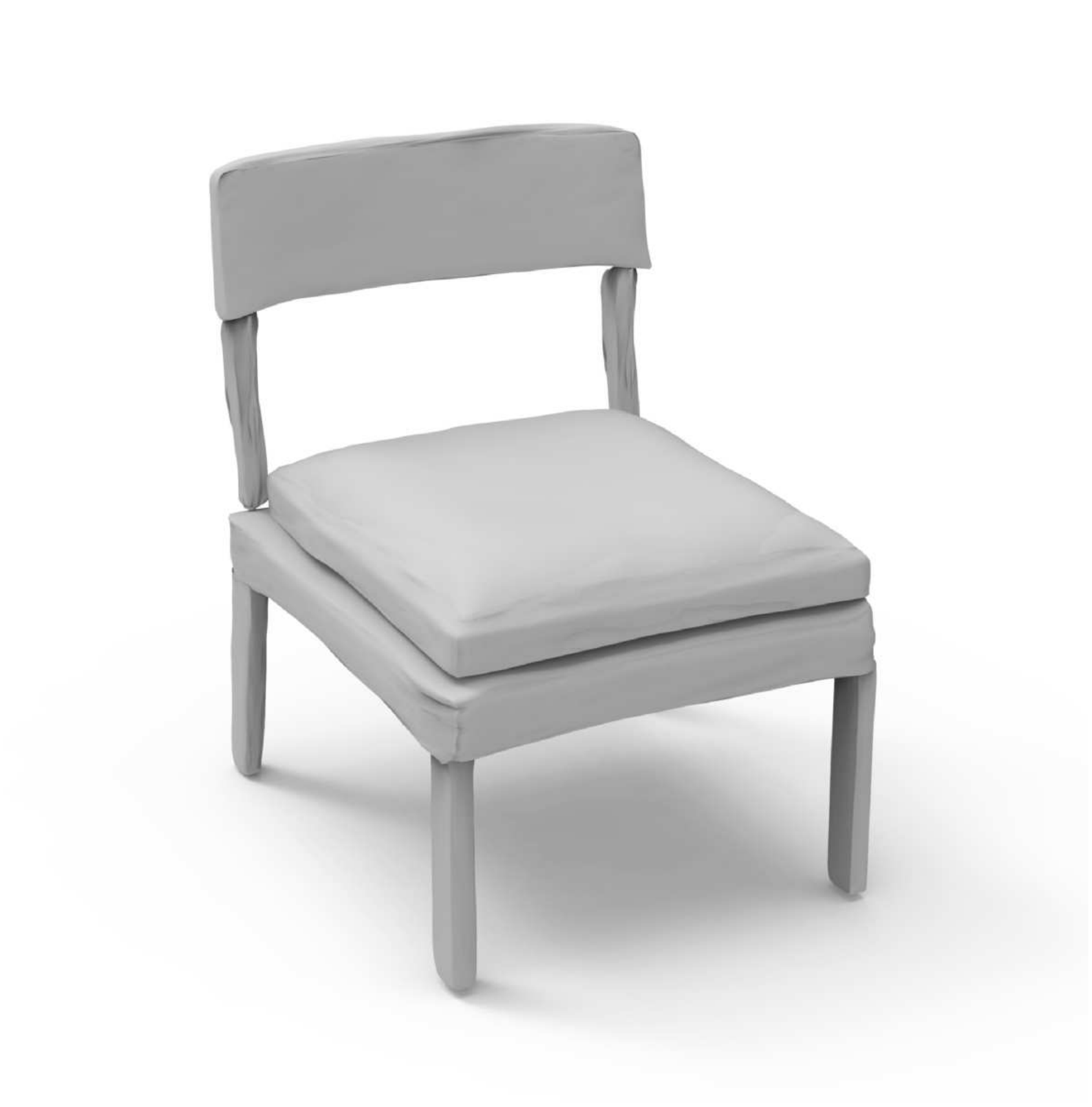}};  & 
    \node (A3) {\includegraphics[width=0.175\linewidth]{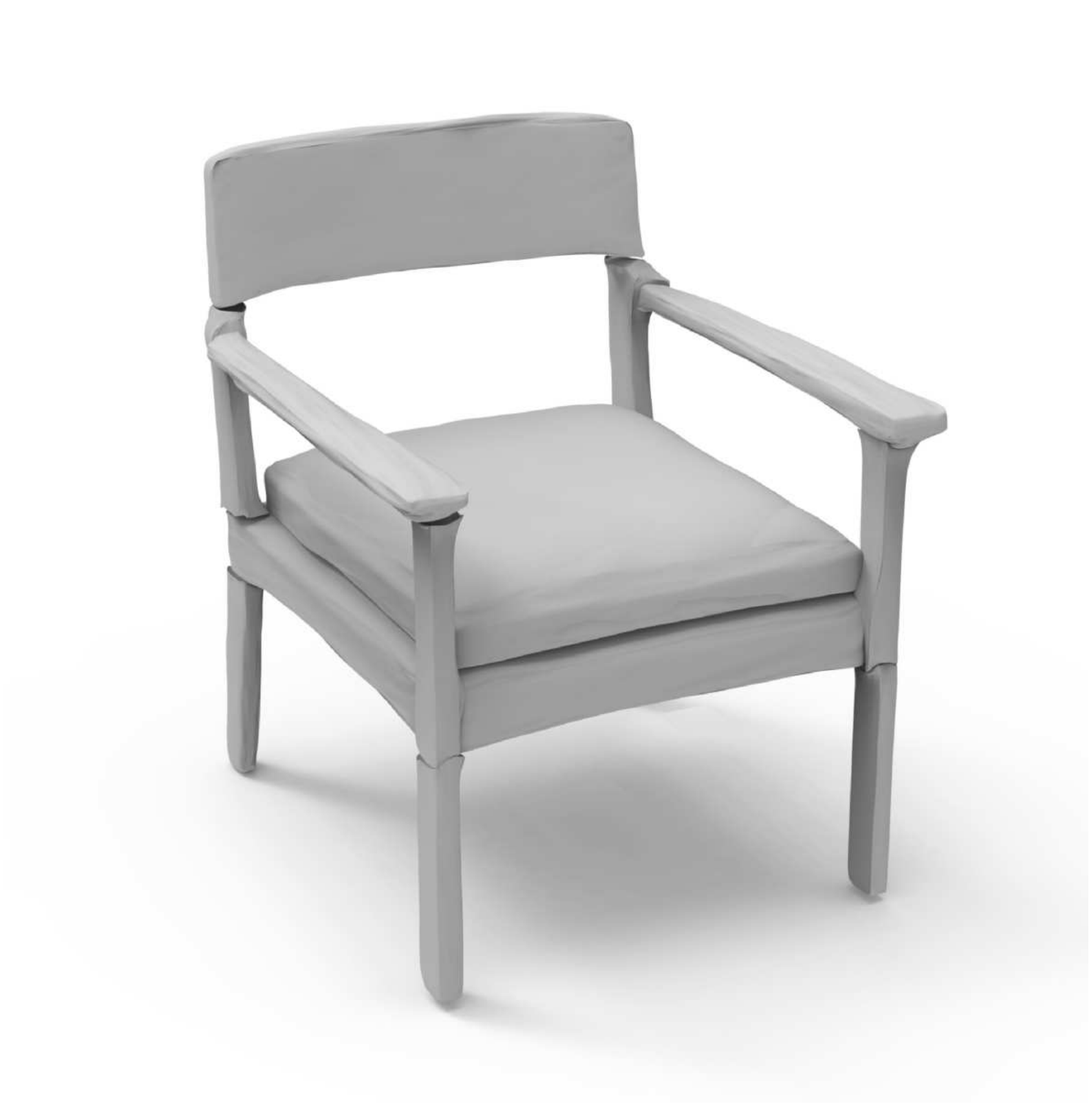}};  & 
    \node (A4) {\includegraphics[width=0.175\linewidth]{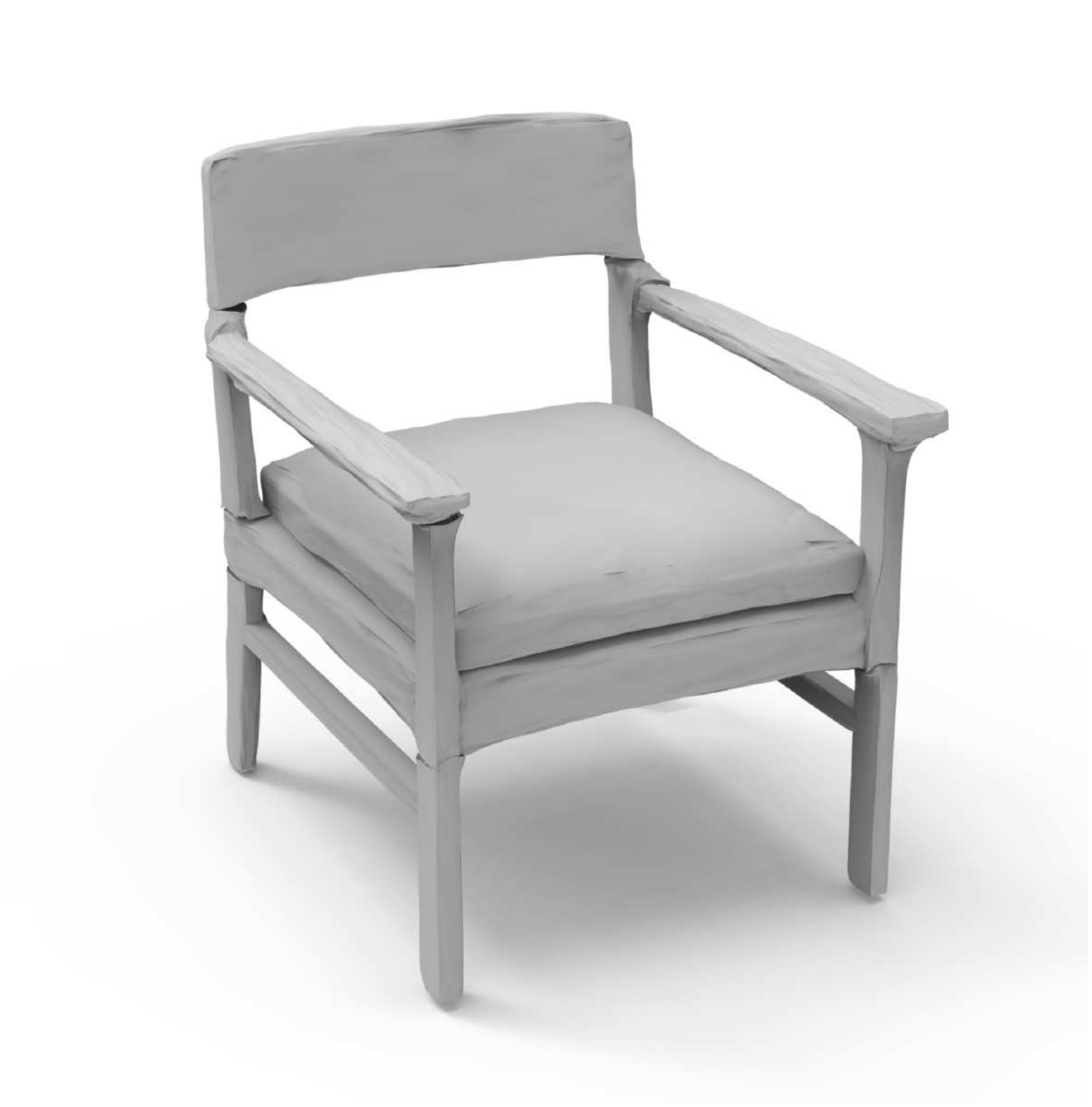}};  & 
    \node (A5) {\includegraphics[width=0.175\linewidth]{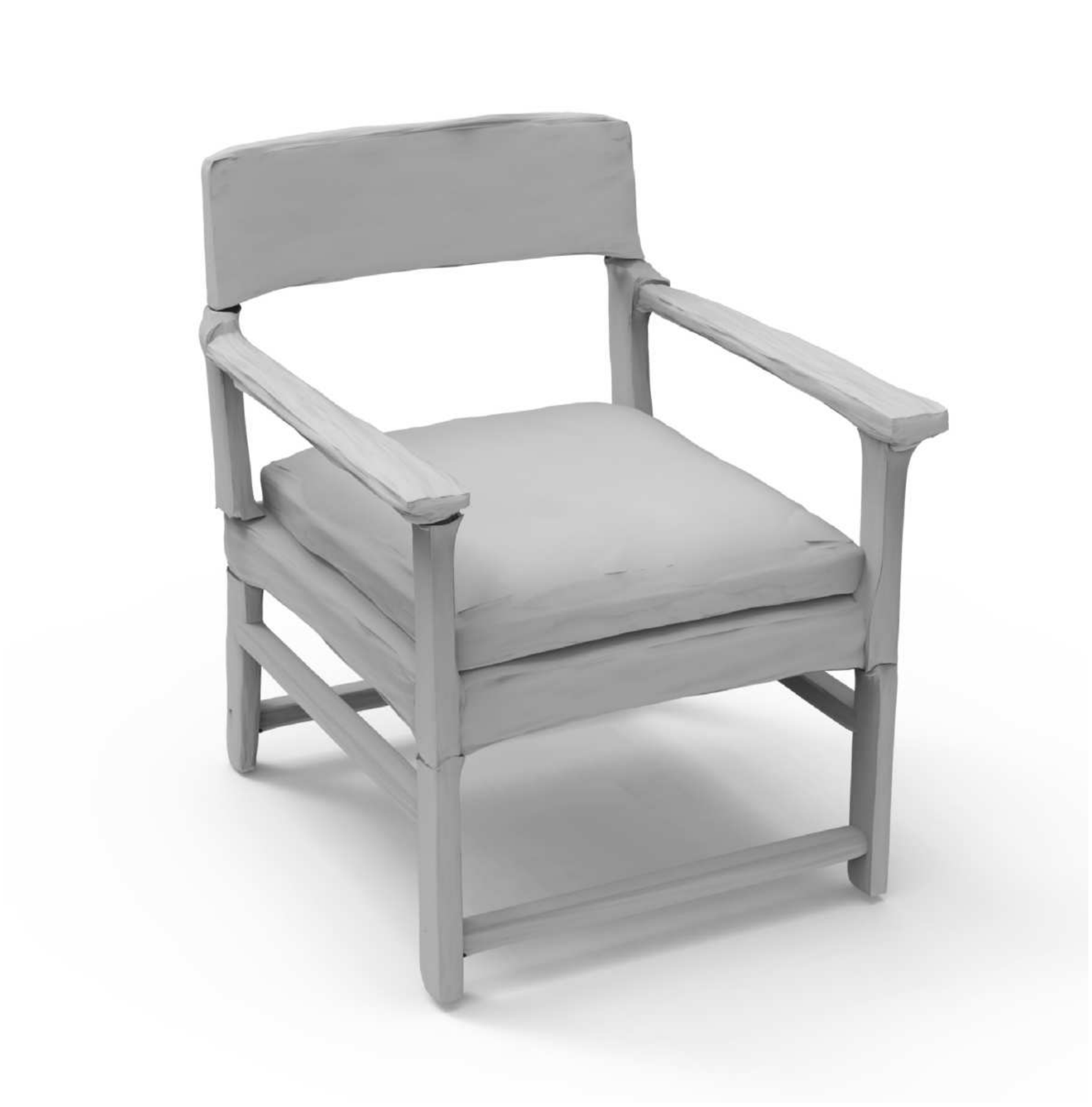}}; \\
    \node (B1) {\includegraphics[width=0.175\linewidth]{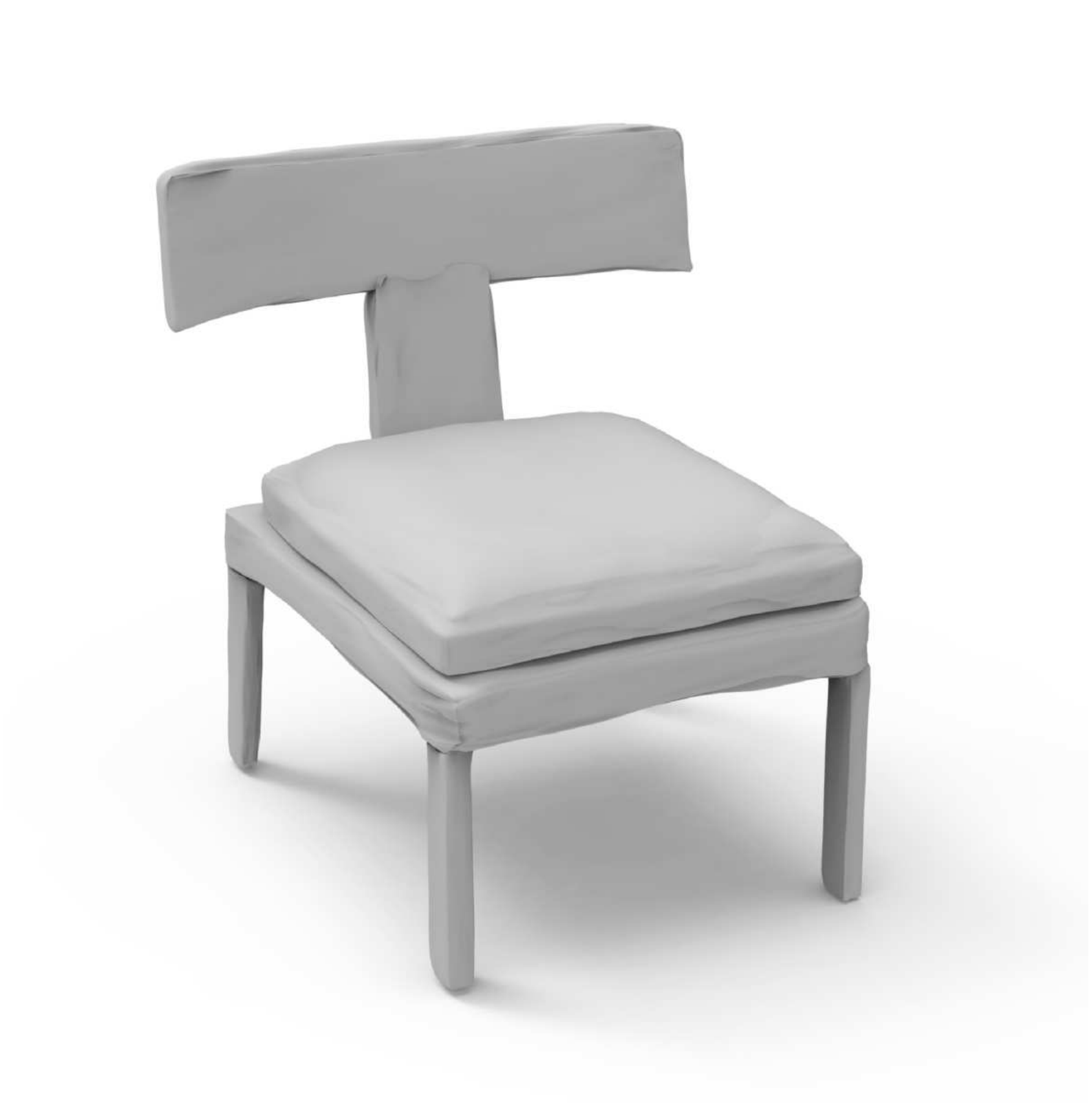}}; & 
    \node (B2) {\includegraphics[width=0.175\linewidth]{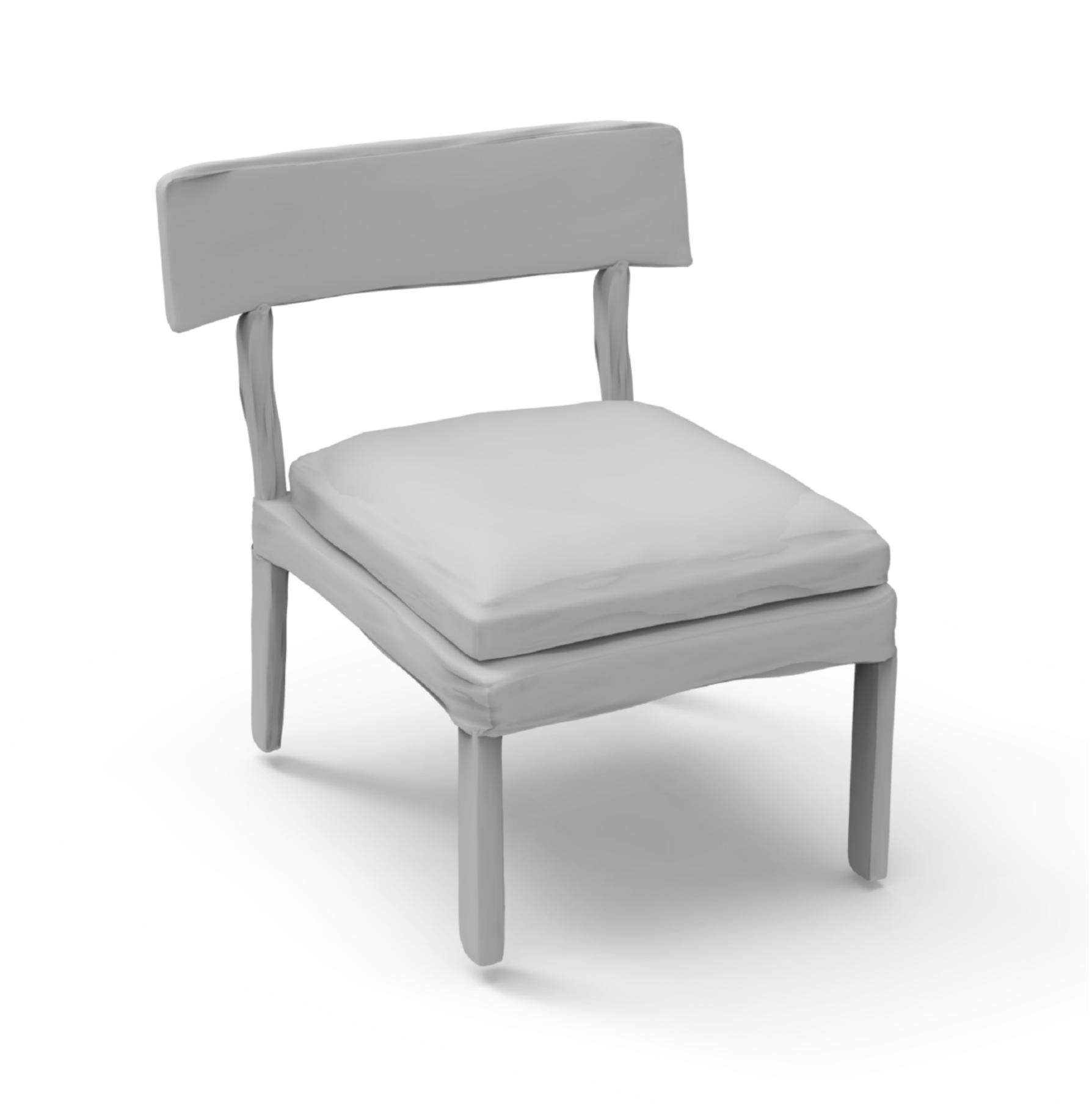}}; &
    \node (B3) {\includegraphics[width=0.175\linewidth]{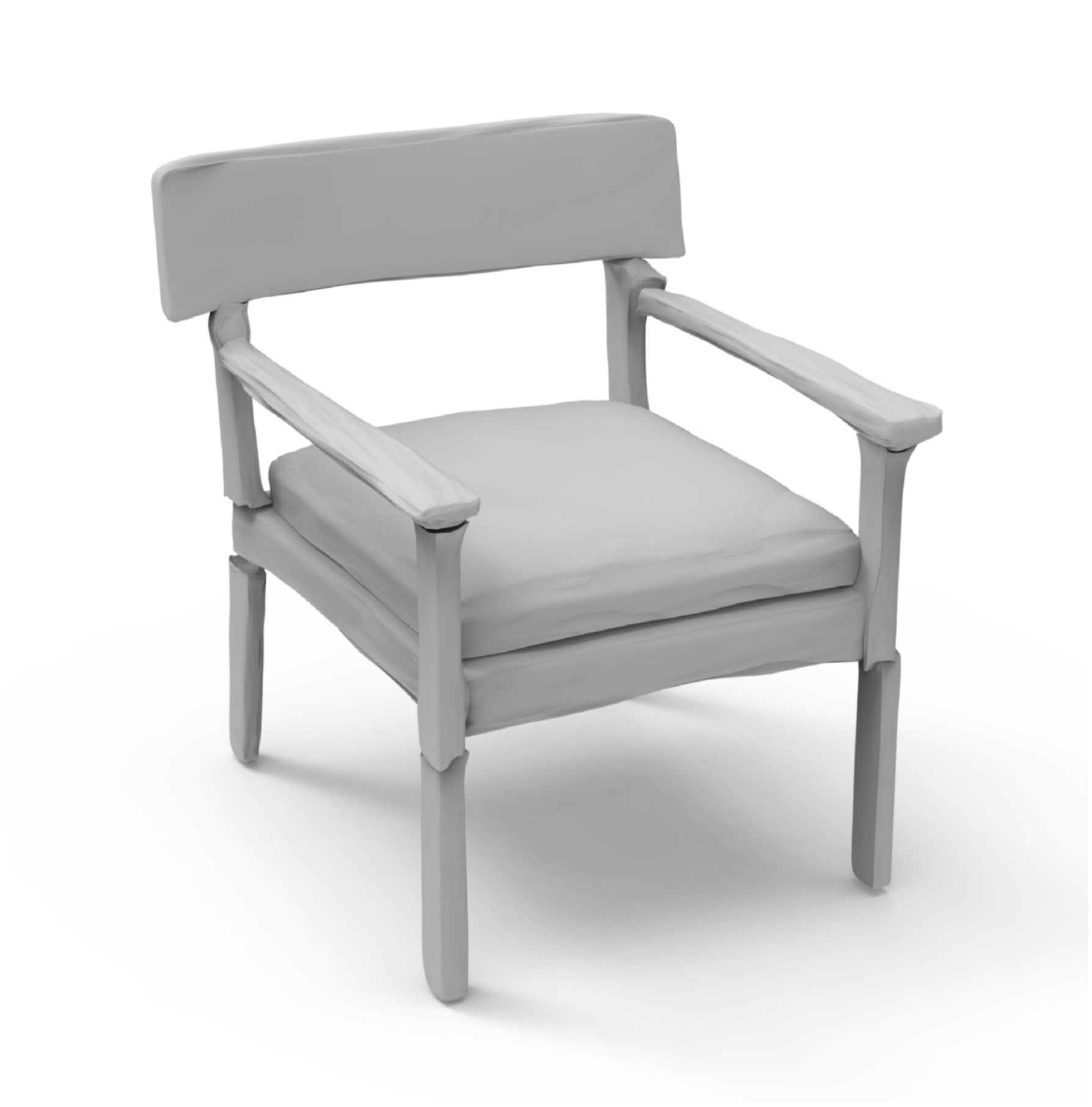}}; &
    \node (B4) {\includegraphics[width=0.175\linewidth]{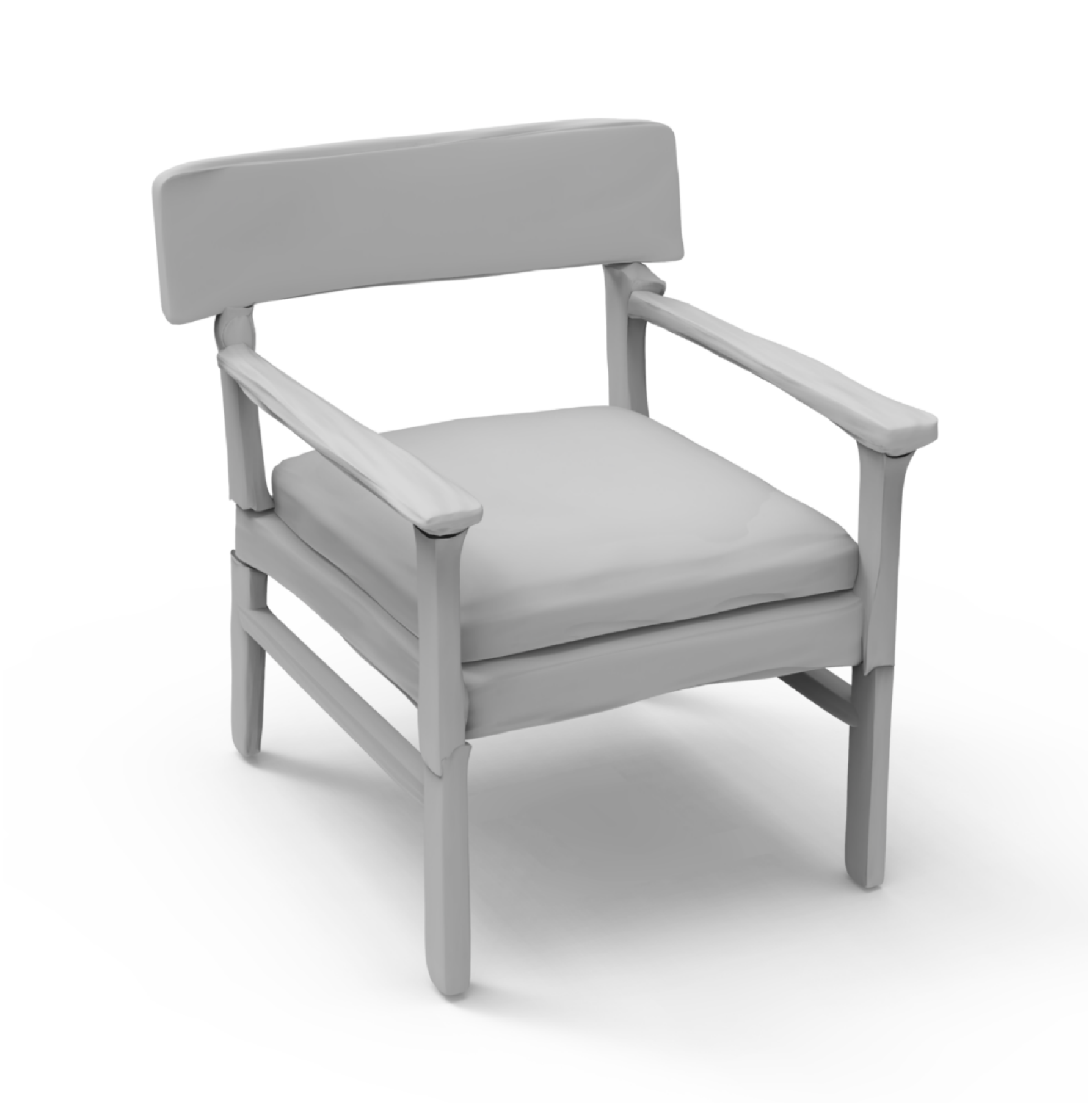}}; &
    \node (B5) {\includegraphics[width=0.175\linewidth]{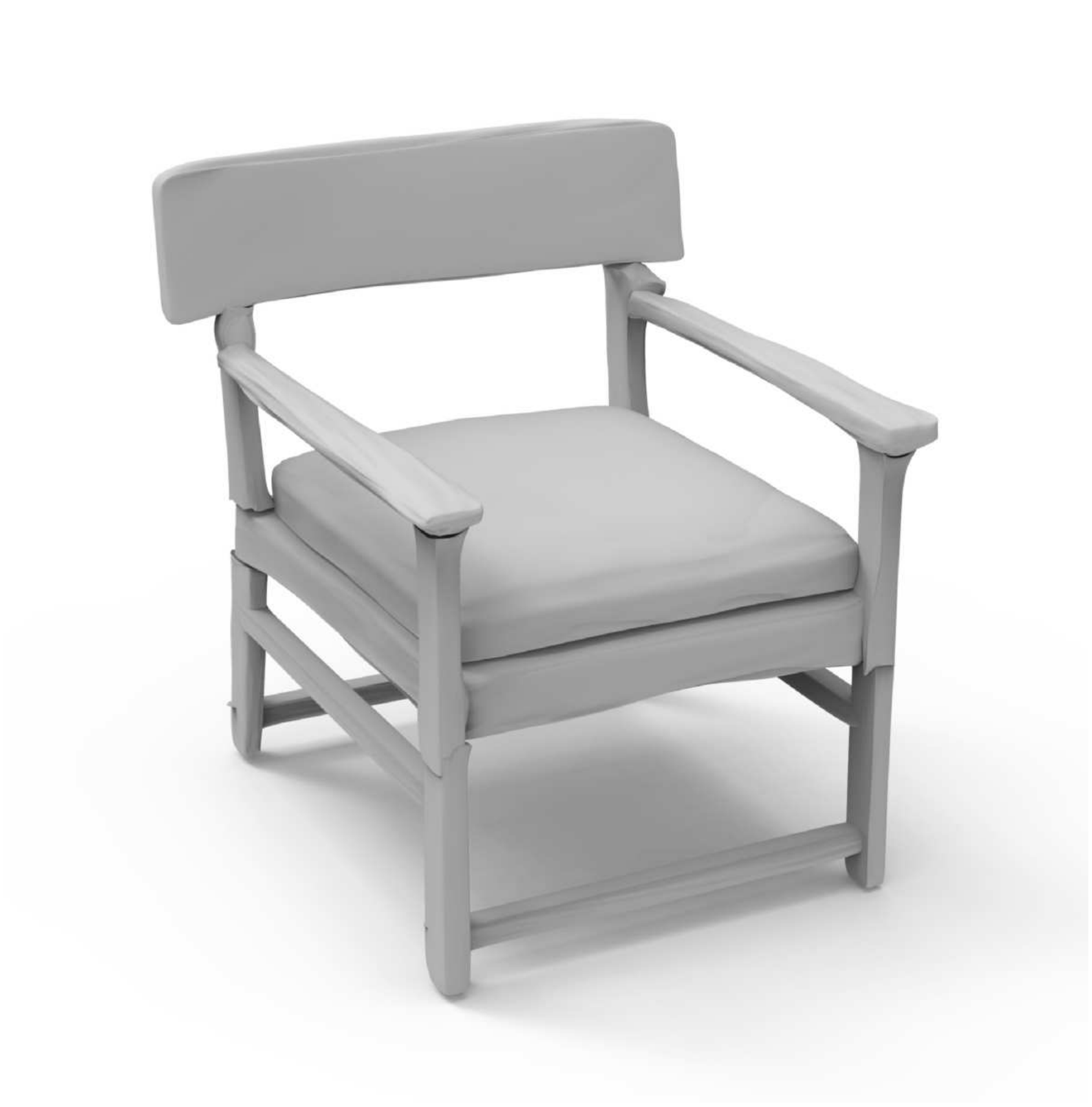}}; \\
    \node (C1) {\includegraphics[width=0.175\linewidth]{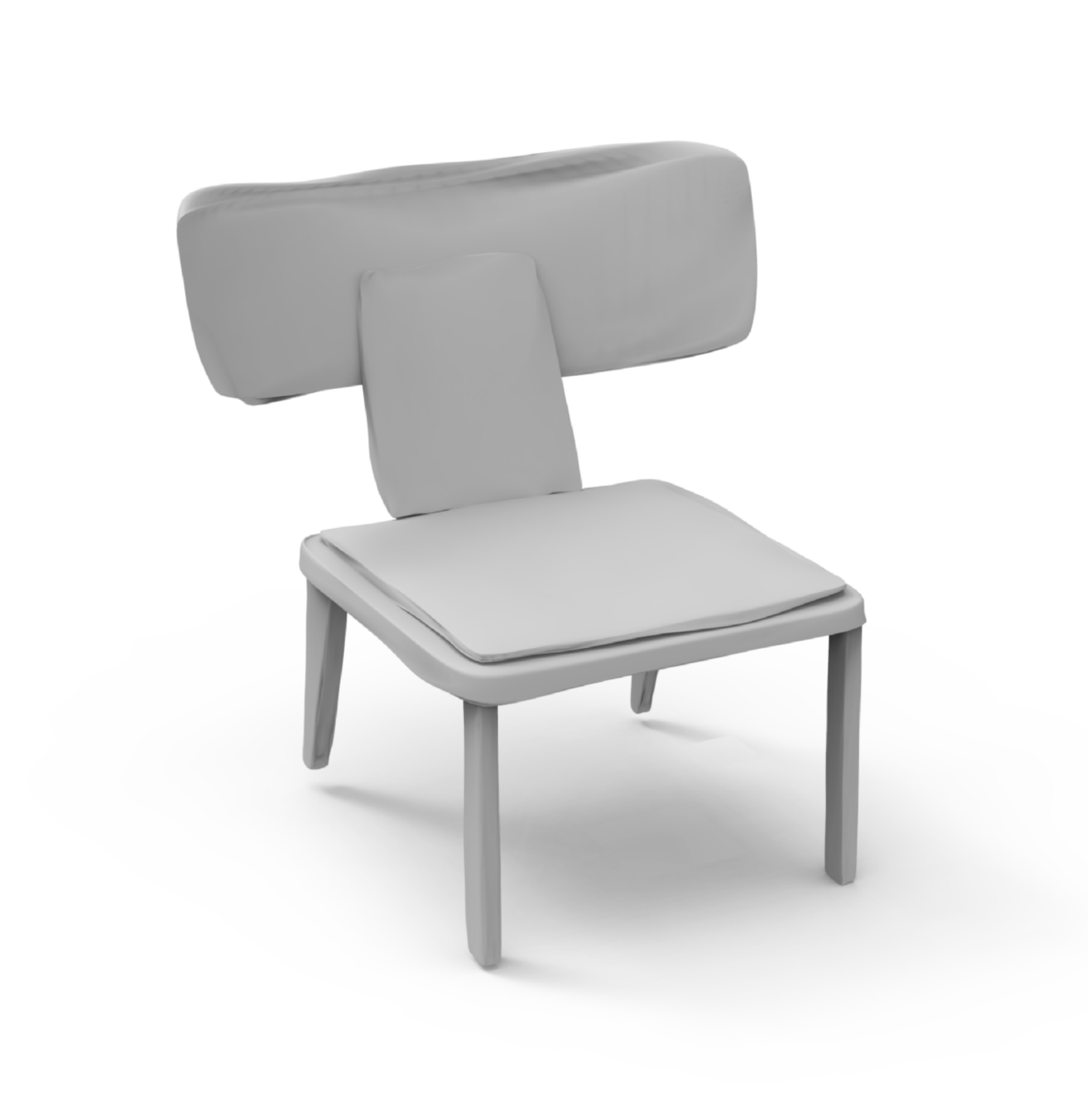}}; & 
    \node (C2) {\includegraphics[width=0.175\linewidth]{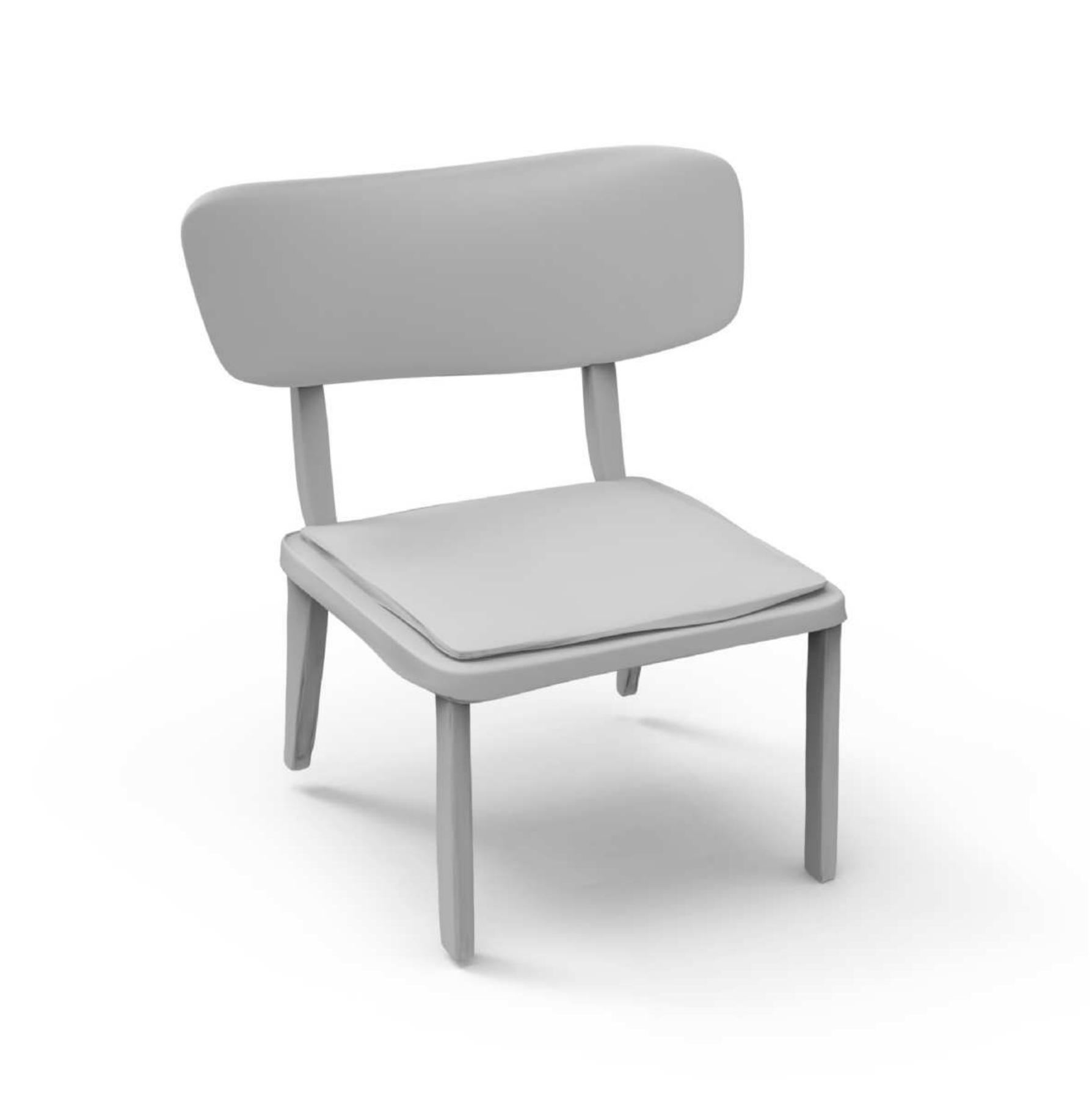}}; &
    \node (C3) {\includegraphics[width=0.175\linewidth]{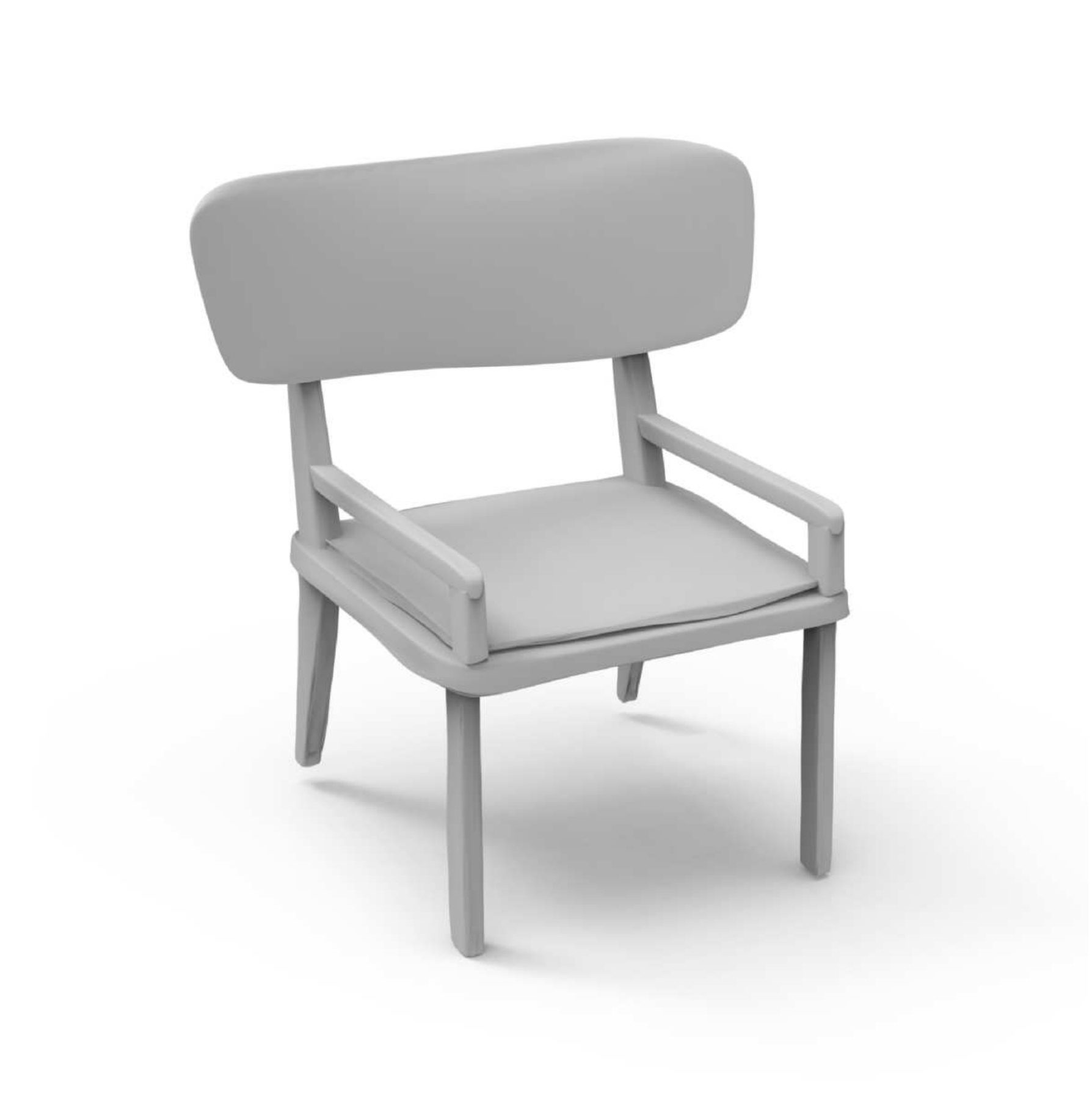}}; &
    \node (C4) {\includegraphics[width=0.175\linewidth]{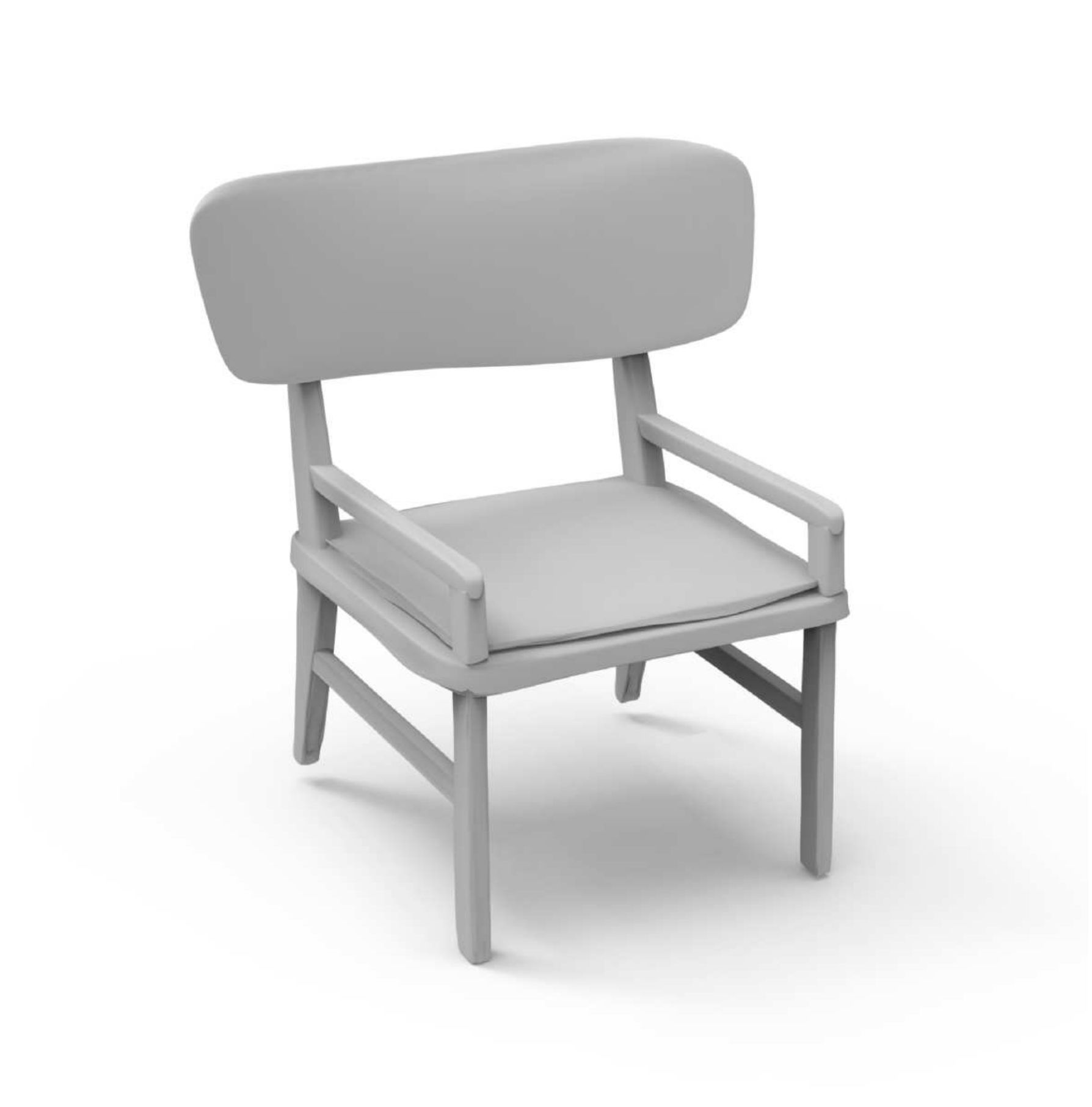}}; &
    \node (C5) {\includegraphics[width=0.175\linewidth]{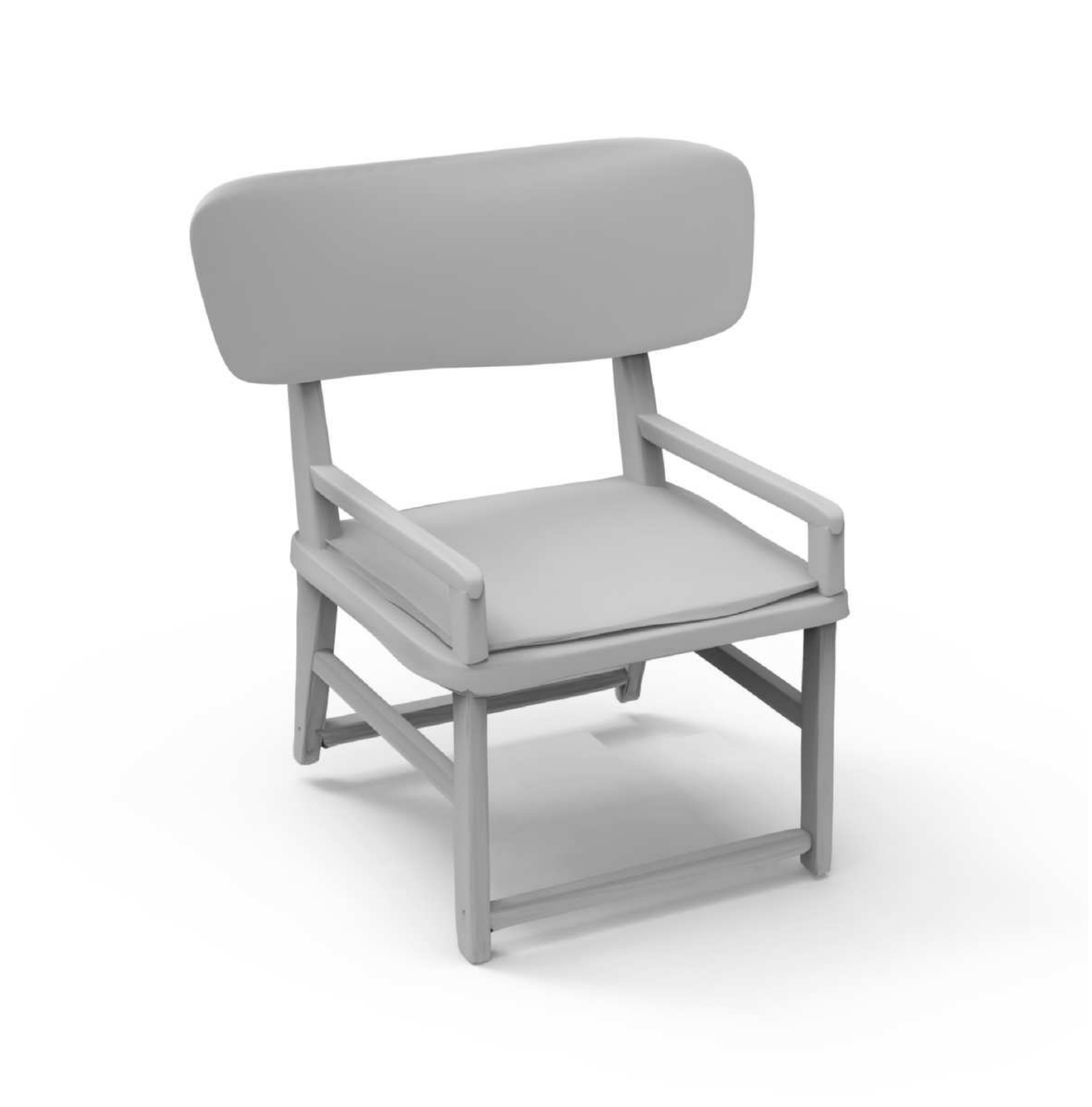}}; \\
    \node (D1) {\includegraphics[width=0.175\linewidth]{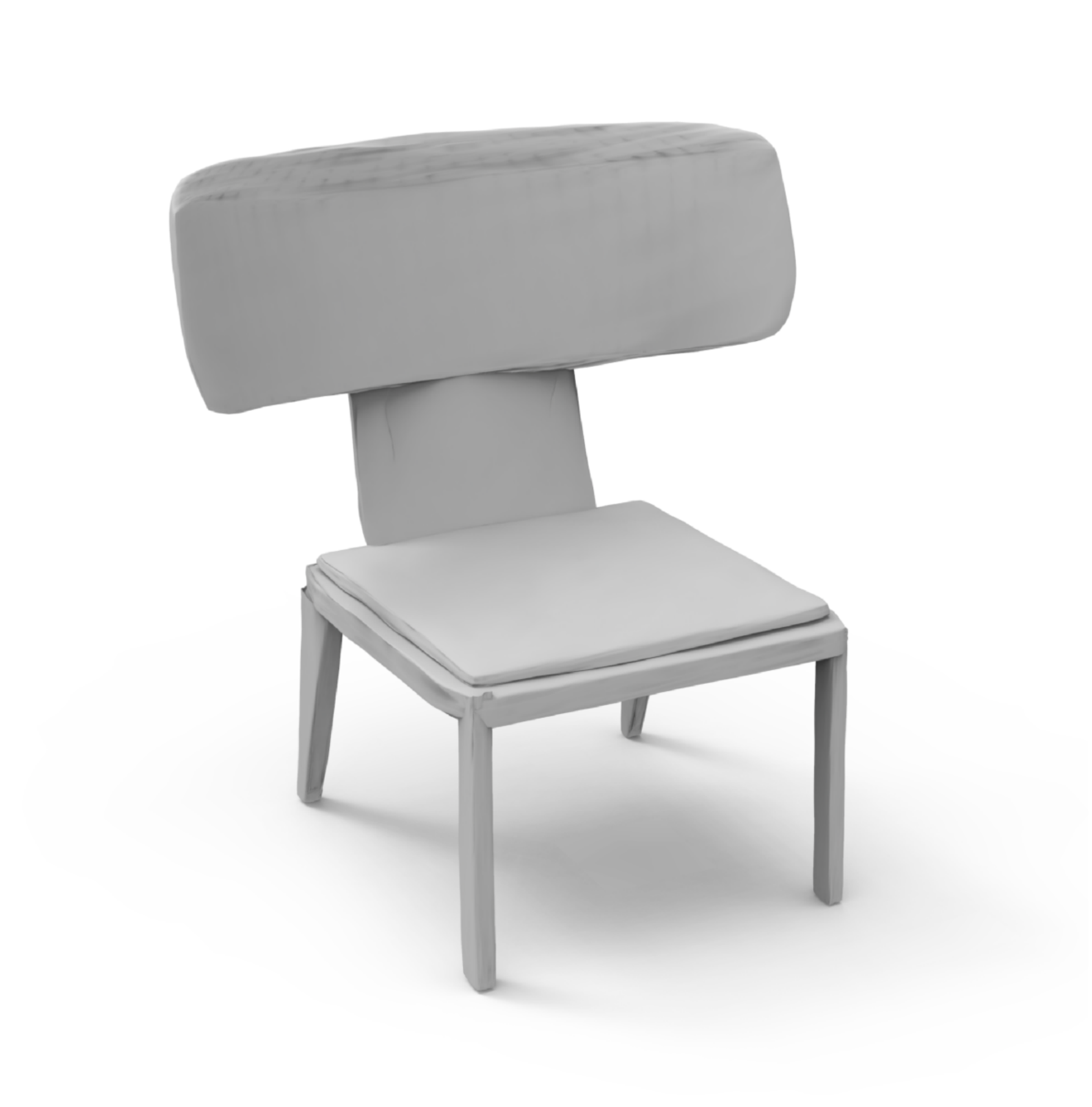}}; & 
    \node (D2) {\includegraphics[width=0.175\linewidth]{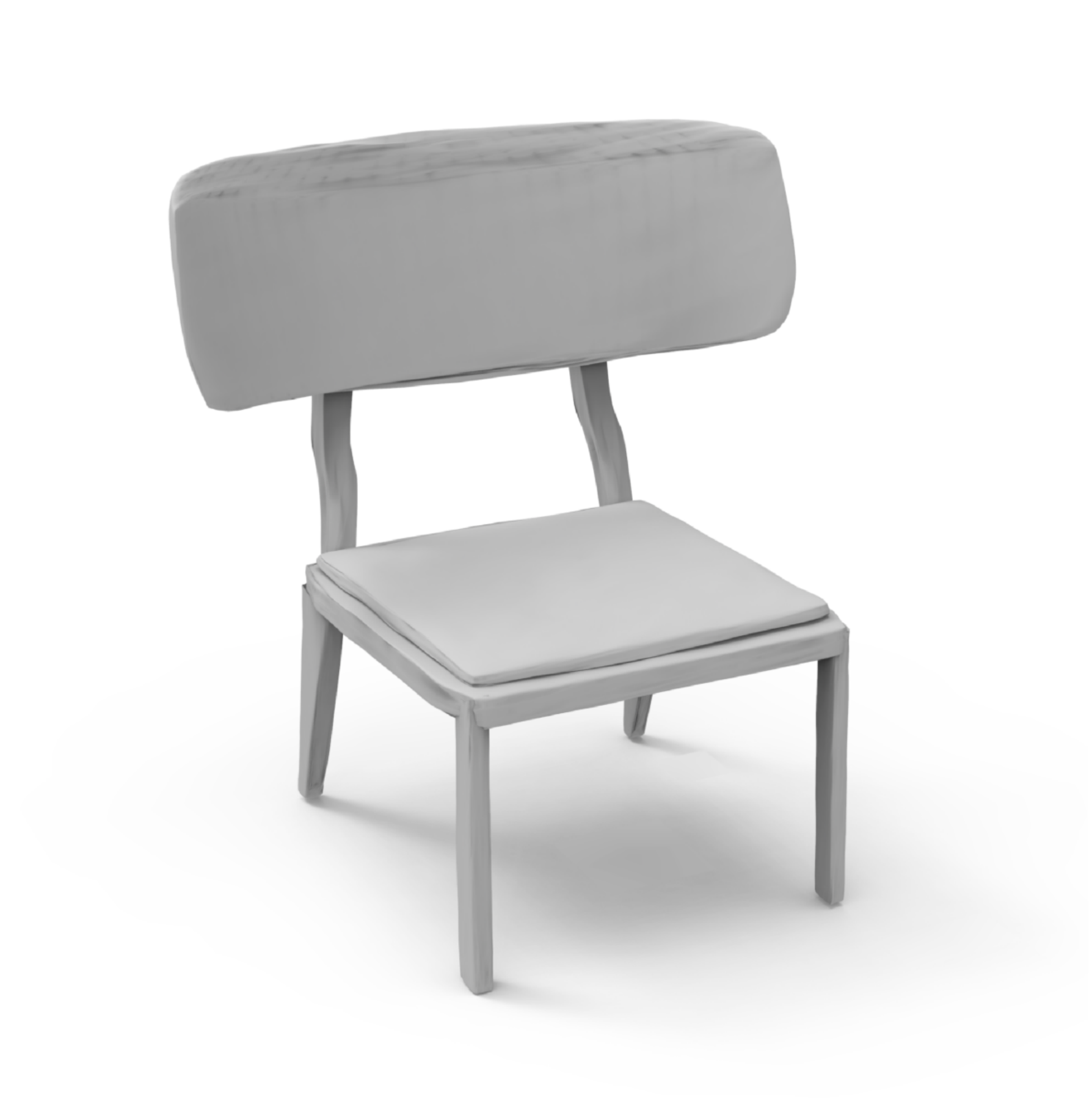}}; &
    \node (D3) {\includegraphics[width=0.175\linewidth]{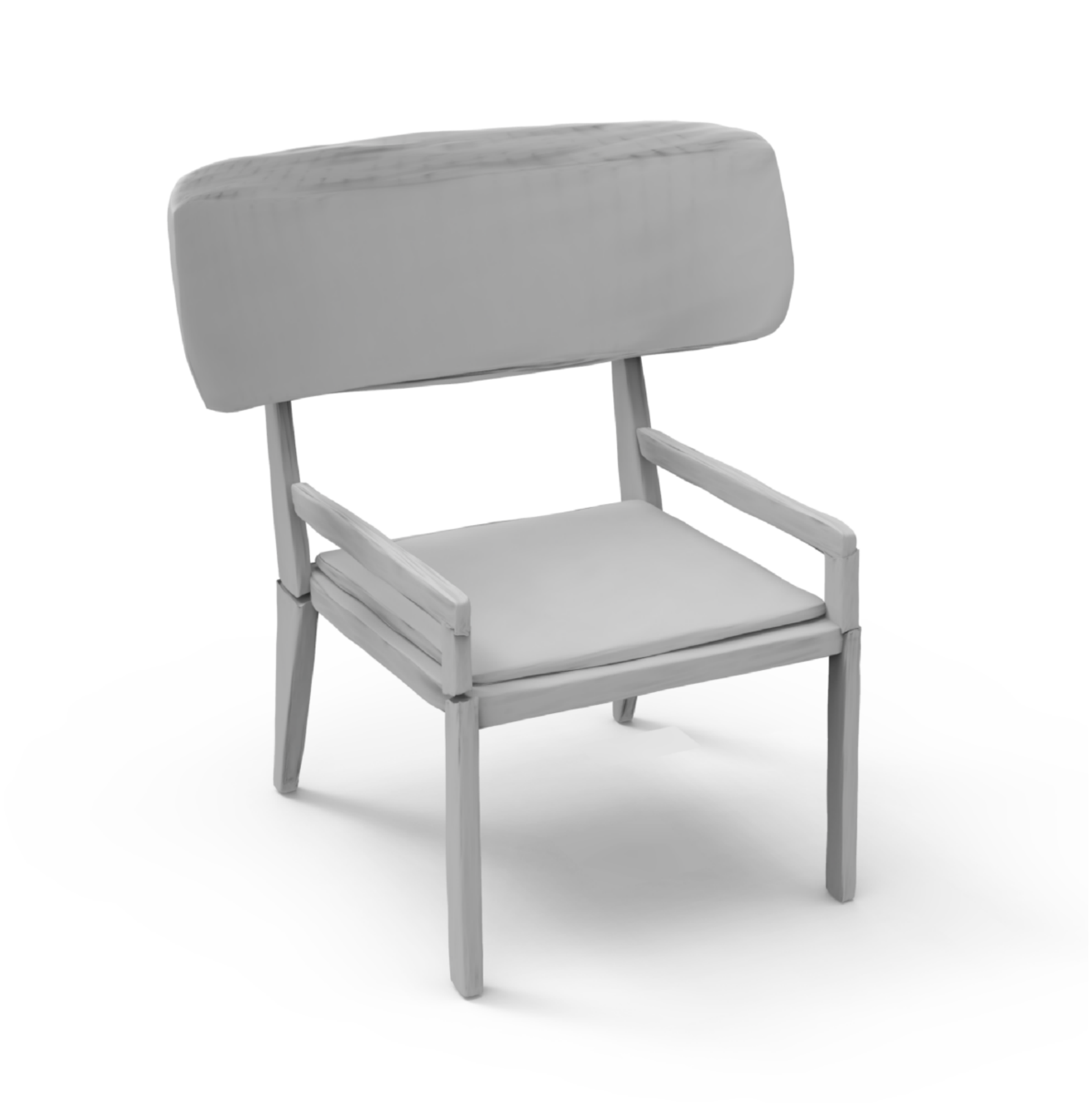}}; &
    \node (D4) {\includegraphics[width=0.175\linewidth]{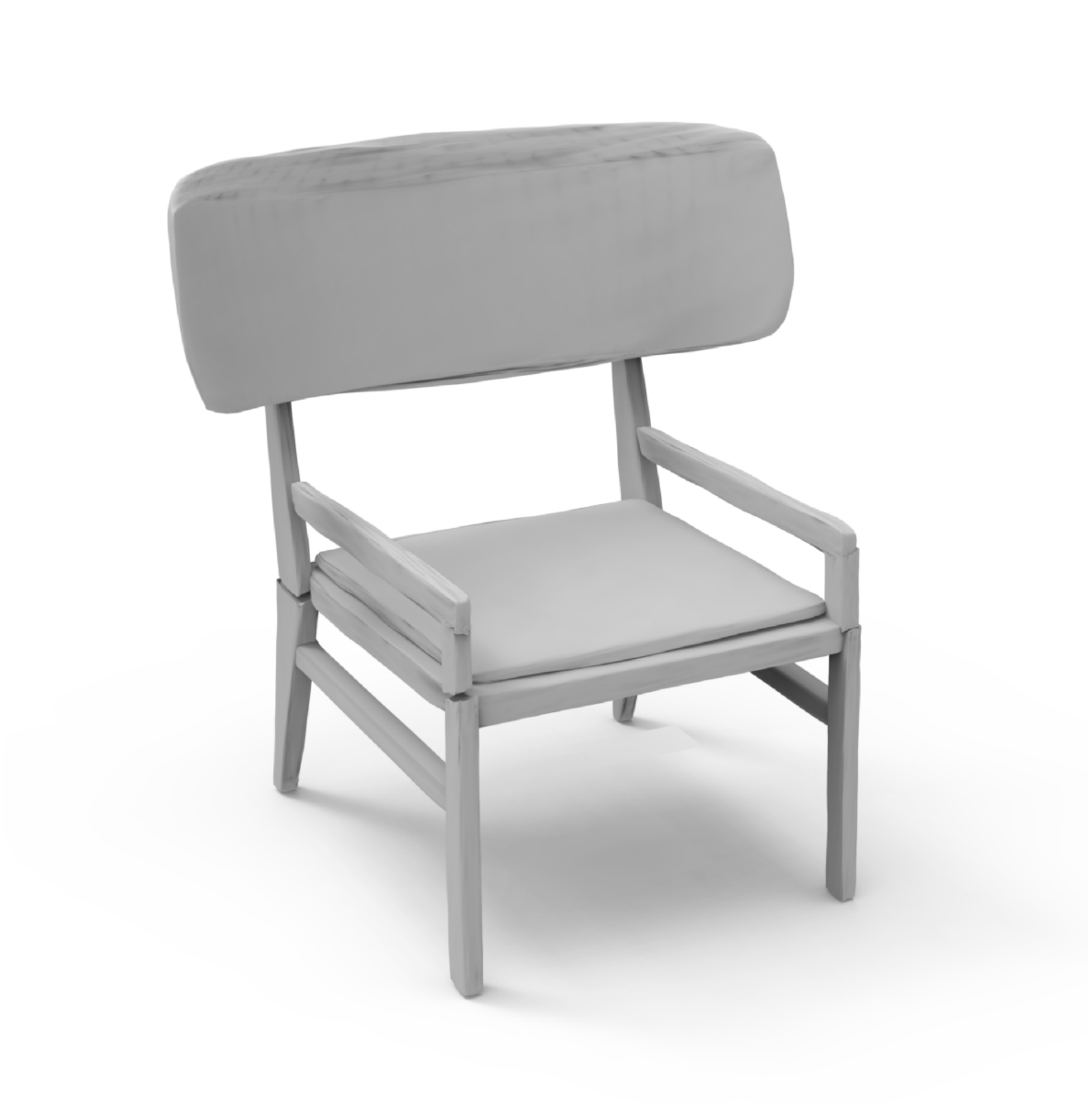}}; &
    \node (D5) {\includegraphics[width=0.175\linewidth]{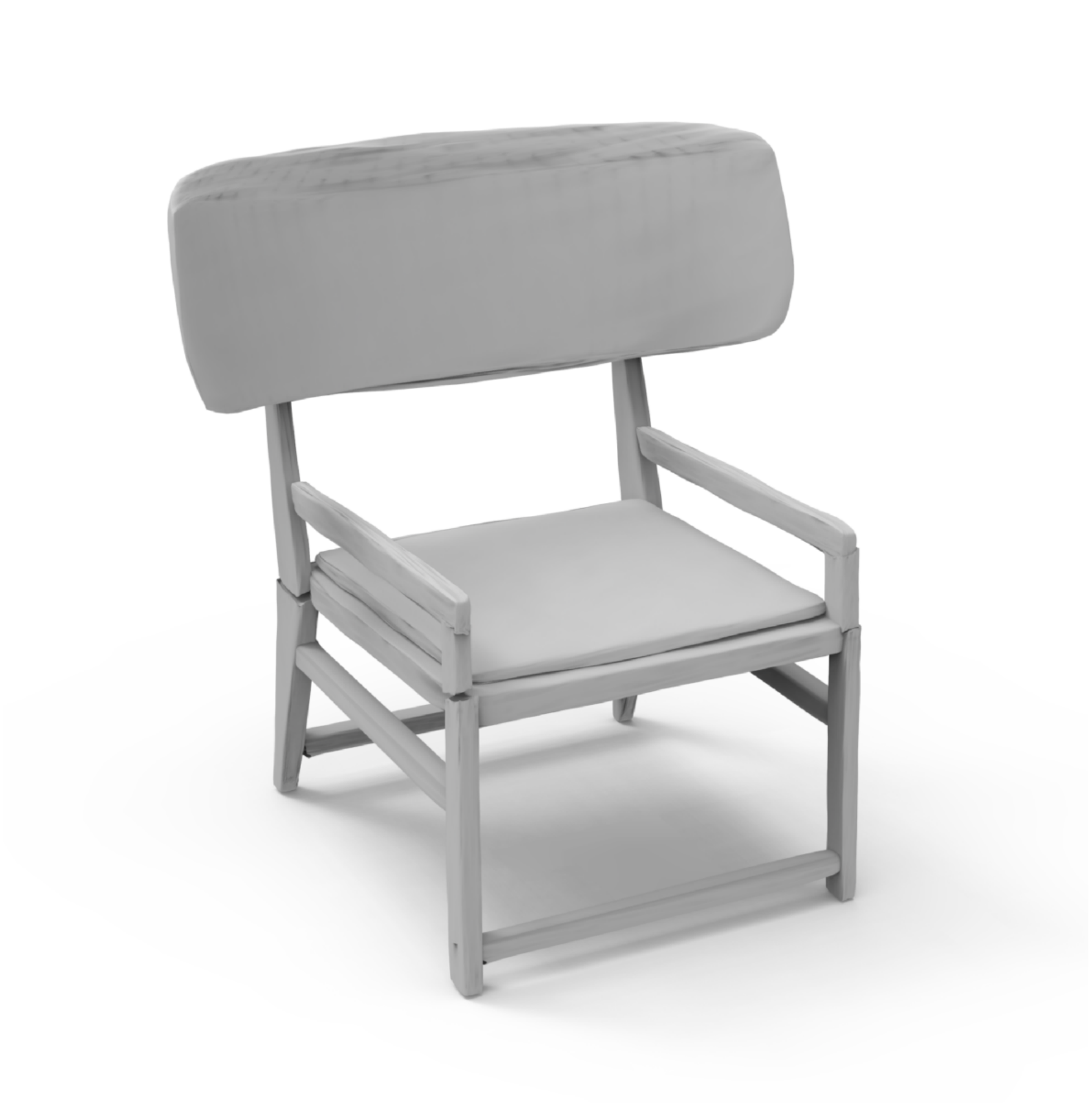}}; \\
    \node (E1) {\includegraphics[width=0.175\linewidth]{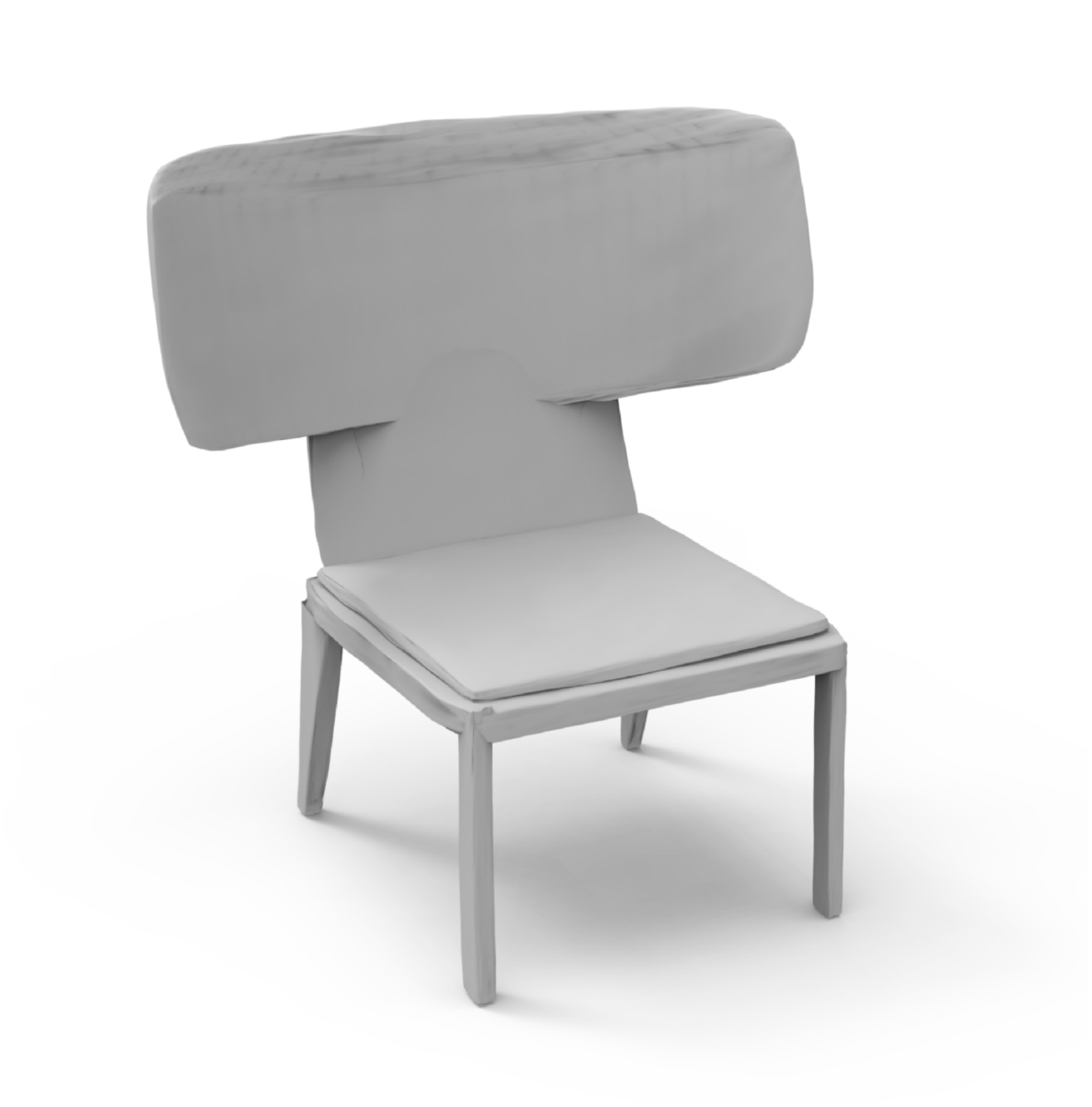}}; & 
    \node (E2) {\includegraphics[width=0.175\linewidth]{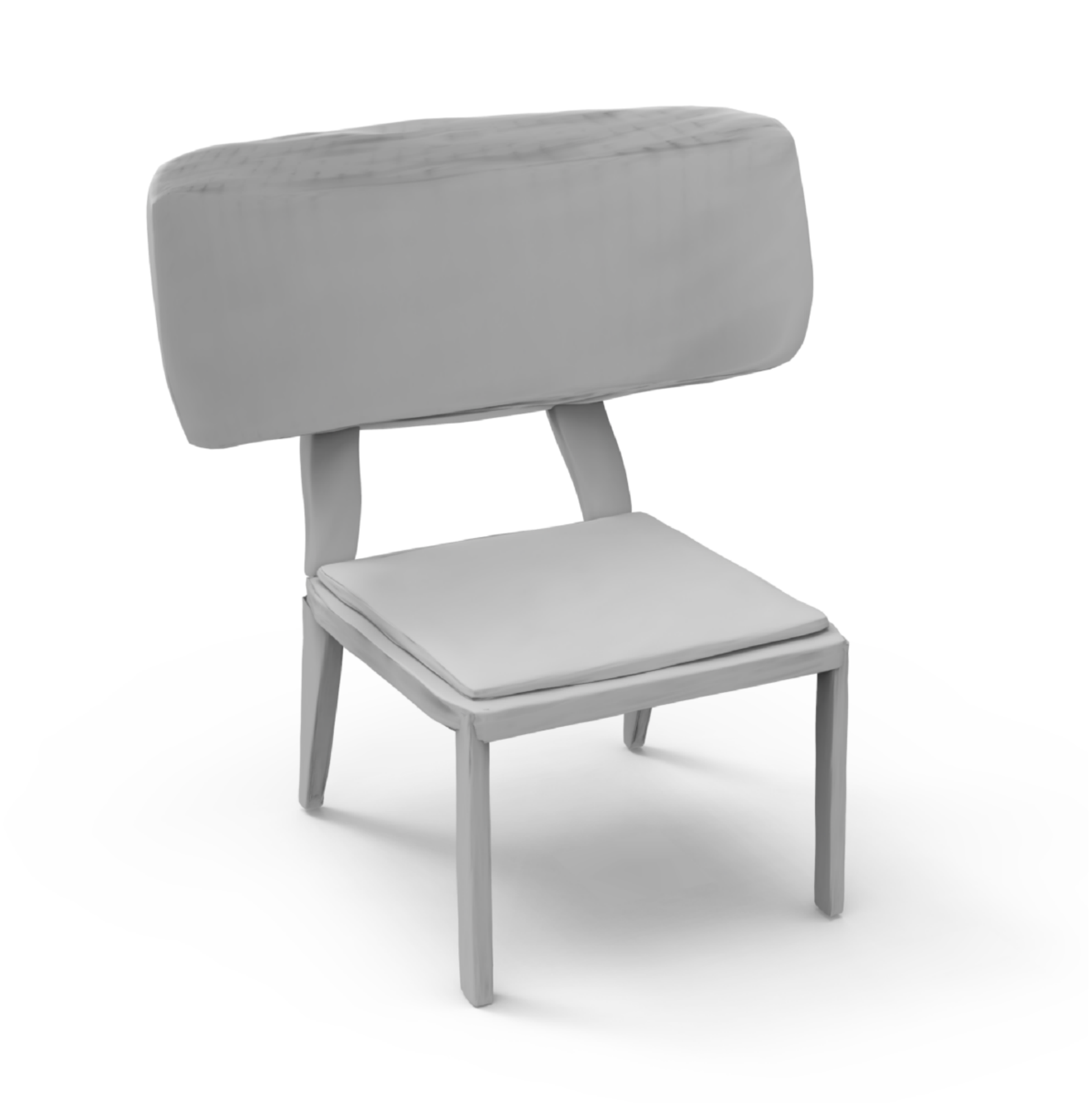}}; &
    \node (E3) {\includegraphics[width=0.175\linewidth]{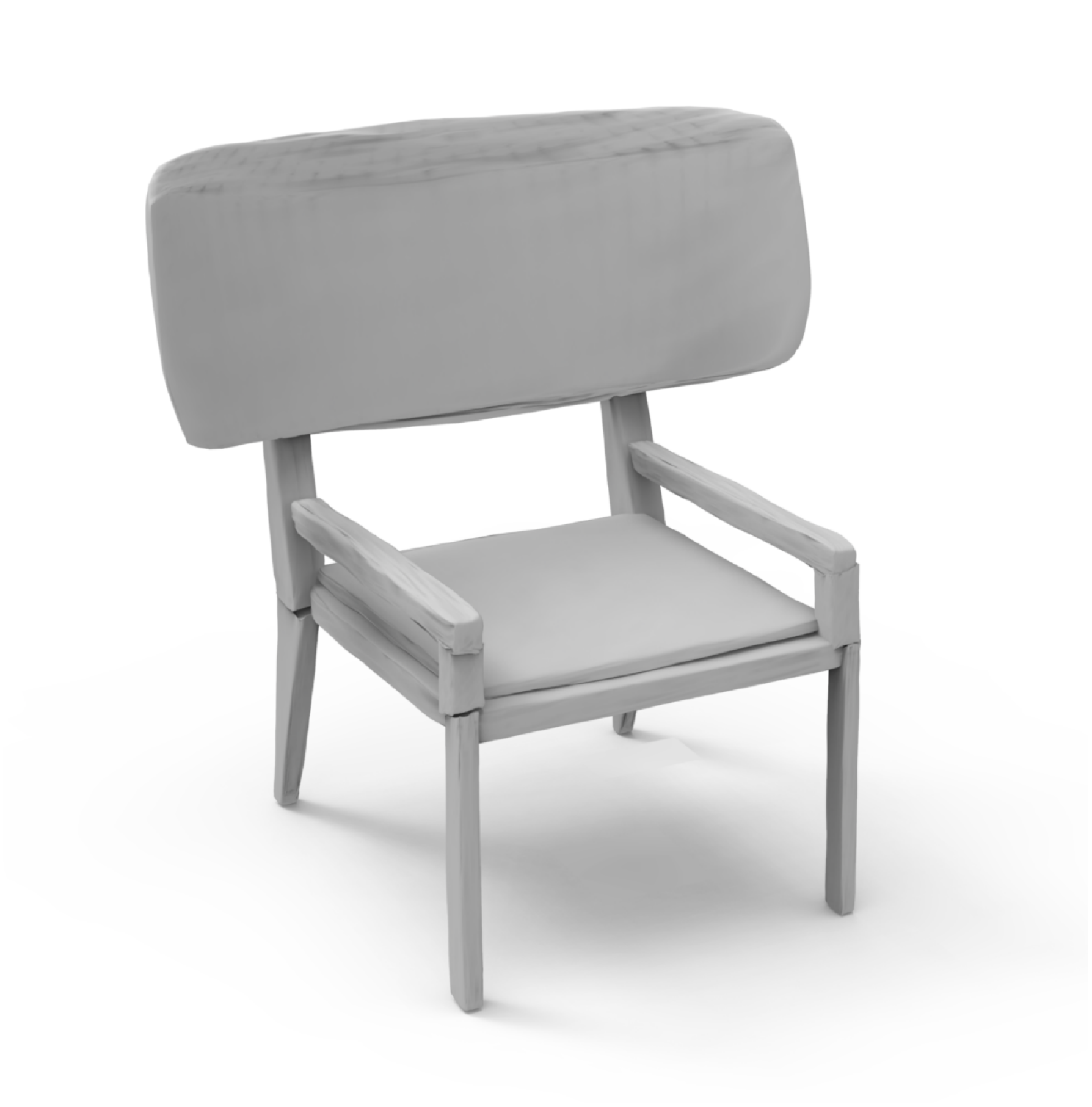}}; &
    \node (E4) {\includegraphics[width=0.175\linewidth]{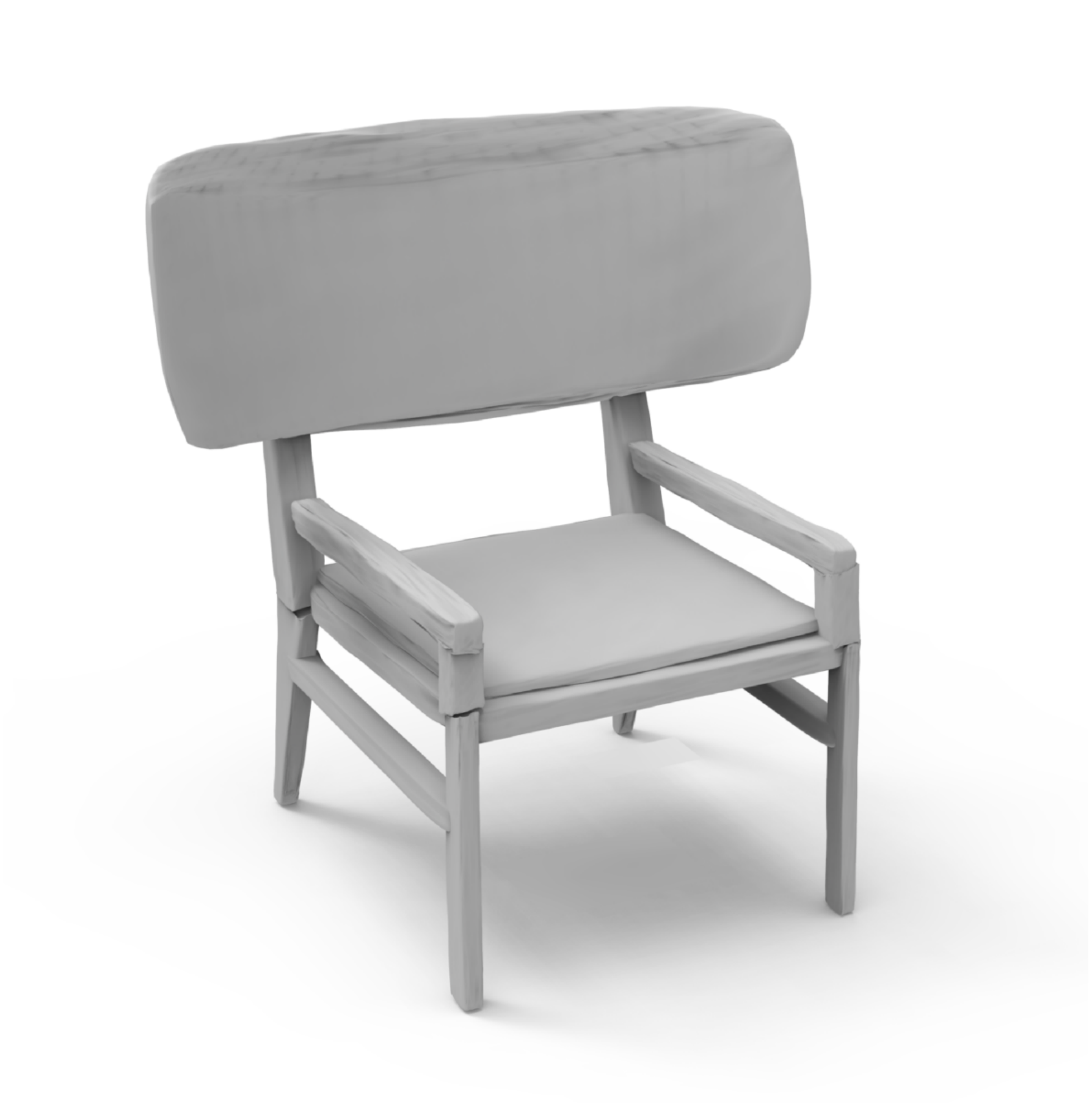}}; &
    \node (E5) {\includegraphics[width=0.175\linewidth, cfbox=orange 1pt 1pt]{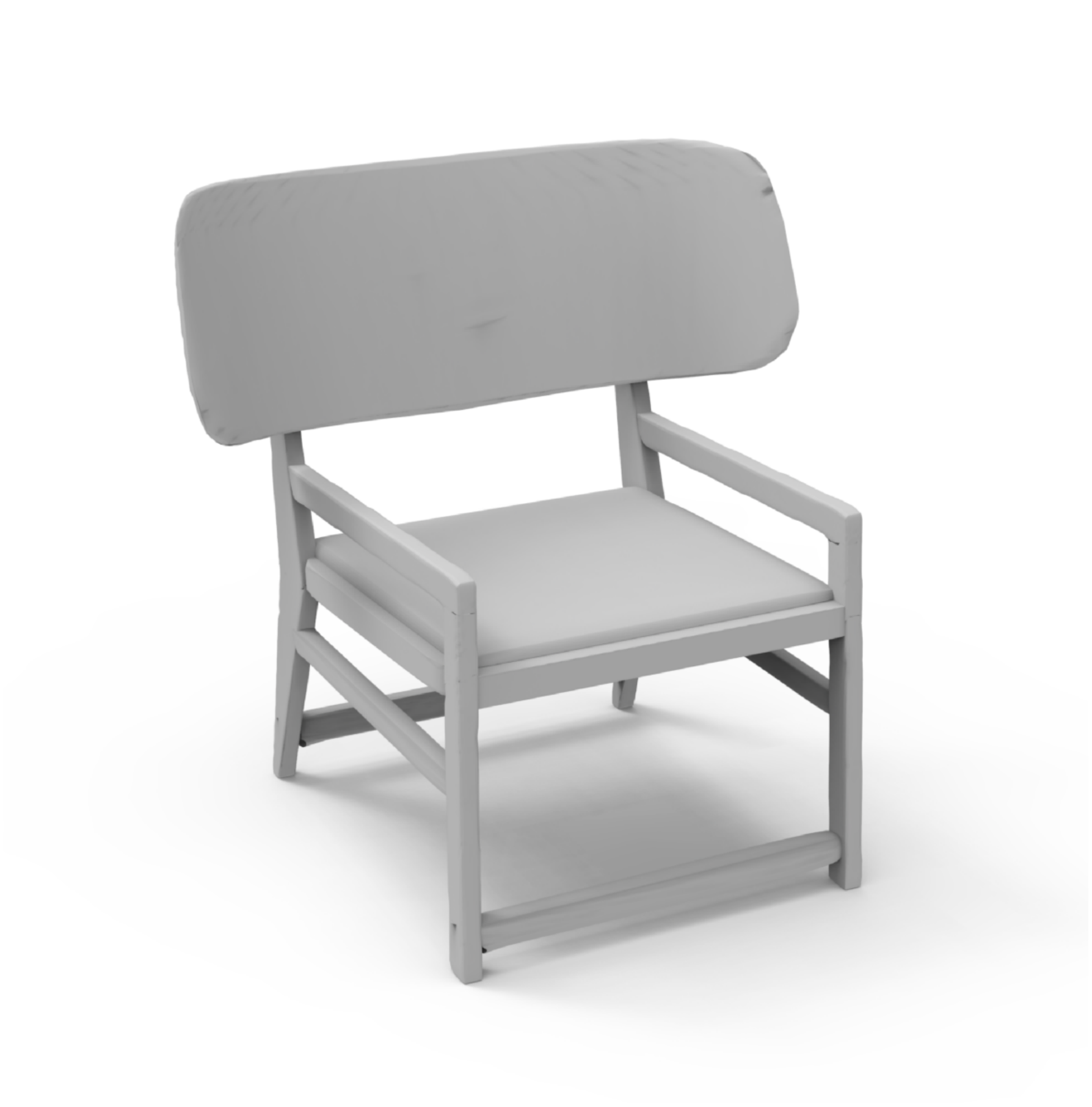}}; \\
  };
  \node[fit=(A1) (A2) (A3) (A4) (A5)
            (B1) (B2) (B3) (B4) (B5)
            (C1) (C2) (C3) (C4) (C5)
            (D1) (D2) (D3) (D4) (D5)
            (E1) (E2) (E3) (E4) (E5),
            inner sep=0pt,
            ] (PIC) {};

  \draw[line width=1pt,arrows={-Stealth[length=4mm]}] ([xshift=-1em,yshift=-0.5em]PIC.south west) -- ([xshift=-1em,yshift=-0.5em]PIC.south east);
  \draw[line width=1pt,arrows={-Stealth[length=4mm]}] ([yshift=-1em,xshift=-0.5em]PIC.south west) -- ([yshift=-1em,xshift=-0.5em]PIC.north west);

  \node[anchor=south] (label) [font=\fontsize{10}{10}\selectfont]at ([yshift=-2.em]A3|-PIC.south) {Structure};
  \node[anchor=center,rotate=90] (label) [font=\fontsize{10}{10}\selectfont] at ([xshift=-1.5em]C1-|PIC.west) {Geometry};

\end{tikzpicture}
    \caption{\yjr{Disentangled shape reconstruction and interpolation results on PartNet chairs. Here, the top left and bottom right shapes (highlighted with orange boxes) are the input shapes. %
    The remaining shapes are generated automatically with our DSG-Net, where in each row, the \emph{structure} of the shapes is interpolated while keeping the geometry unchanged, whereas in each column, the \emph{geometry} is interpolated while retaining the structure. \yj{The vertical axis and horizontal axis represent the variation of structure and geometry respectively.}}}
    \label{fig:decouple_chair}
\end{figure}

\begin{figure}[h]
  \centering
  \begin{tikzpicture}

  \matrix[nodes={anchor=south west,inner sep=0pt}]{ 
    \node (A1) {\includegraphics[width=0.17\linewidth, cfbox=orange 1pt 1pt]{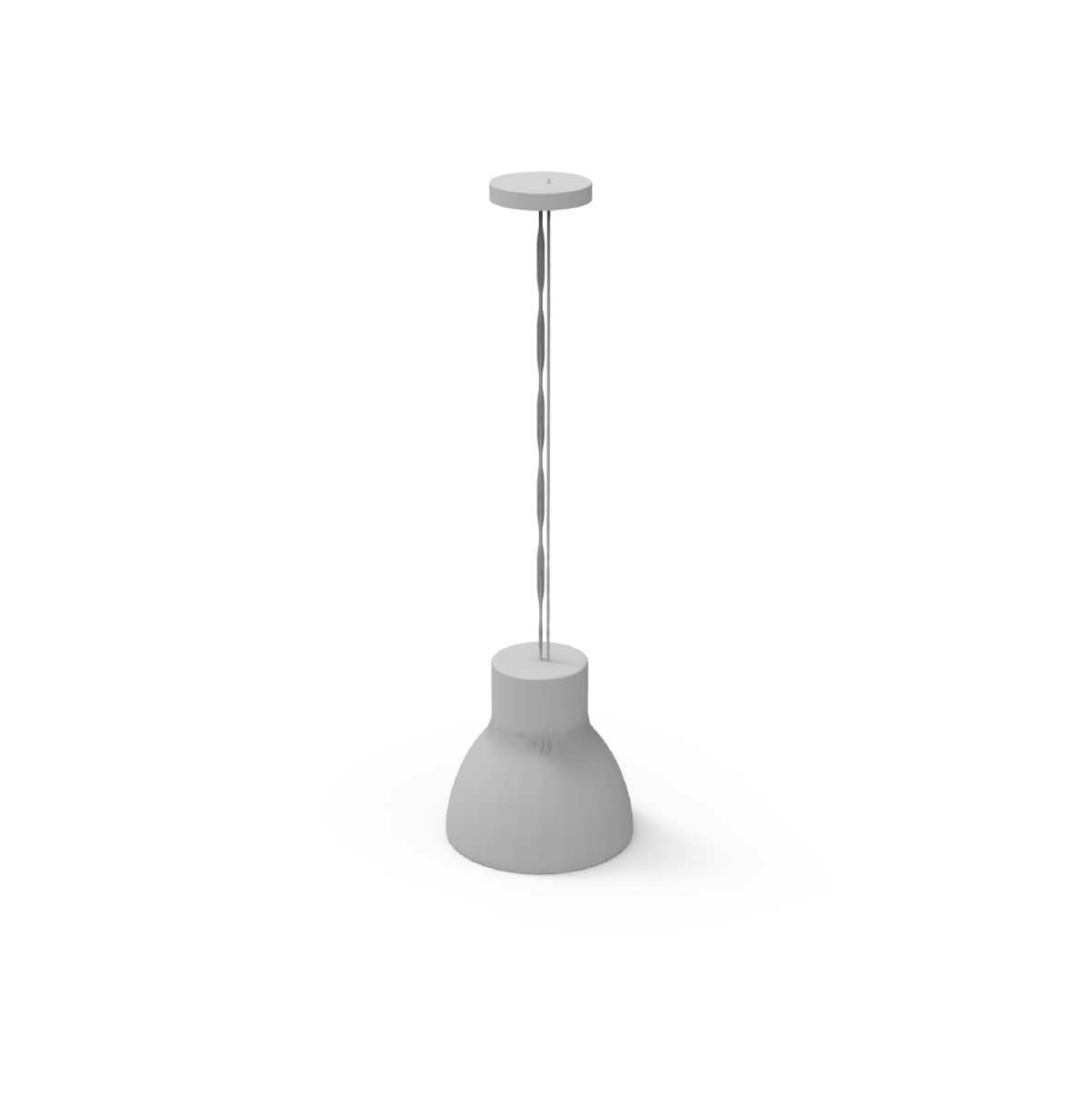}};  & 
    \node (A2) {\includegraphics[width=0.17\linewidth]{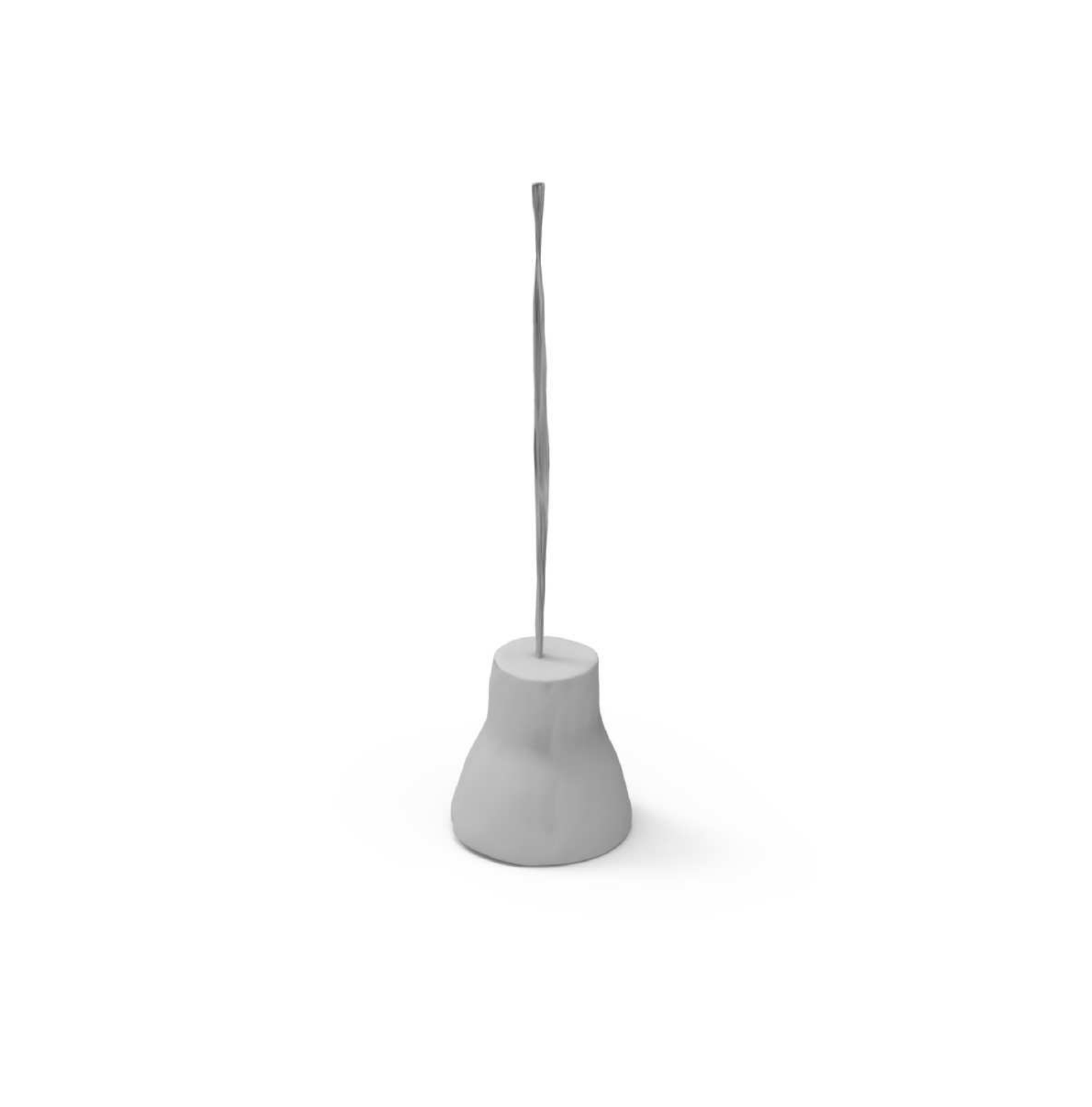}};  & 
    \node (A3) {\includegraphics[width=0.17\linewidth]{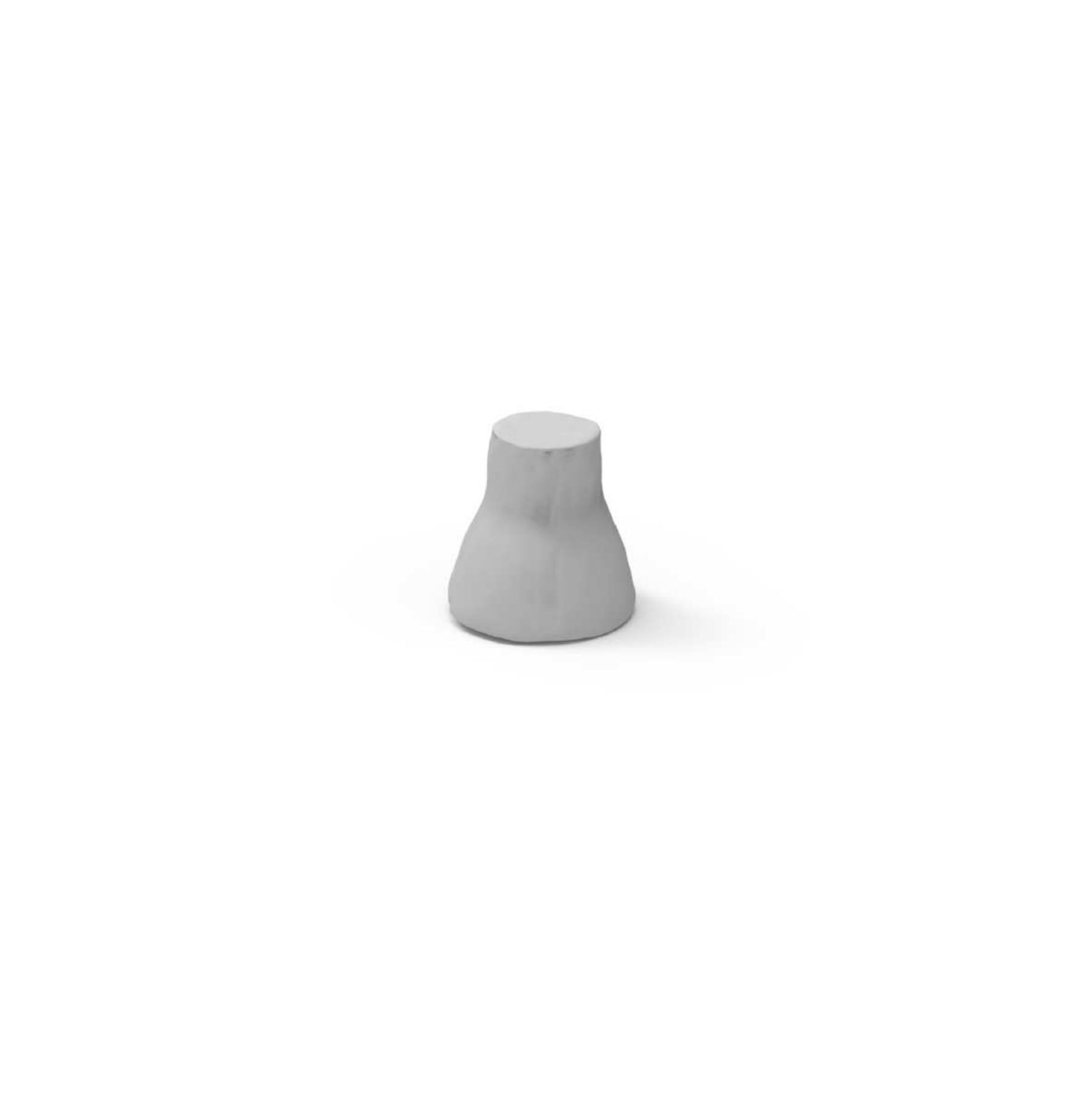}};  & 
    \node (A4) {\includegraphics[width=0.17\linewidth]{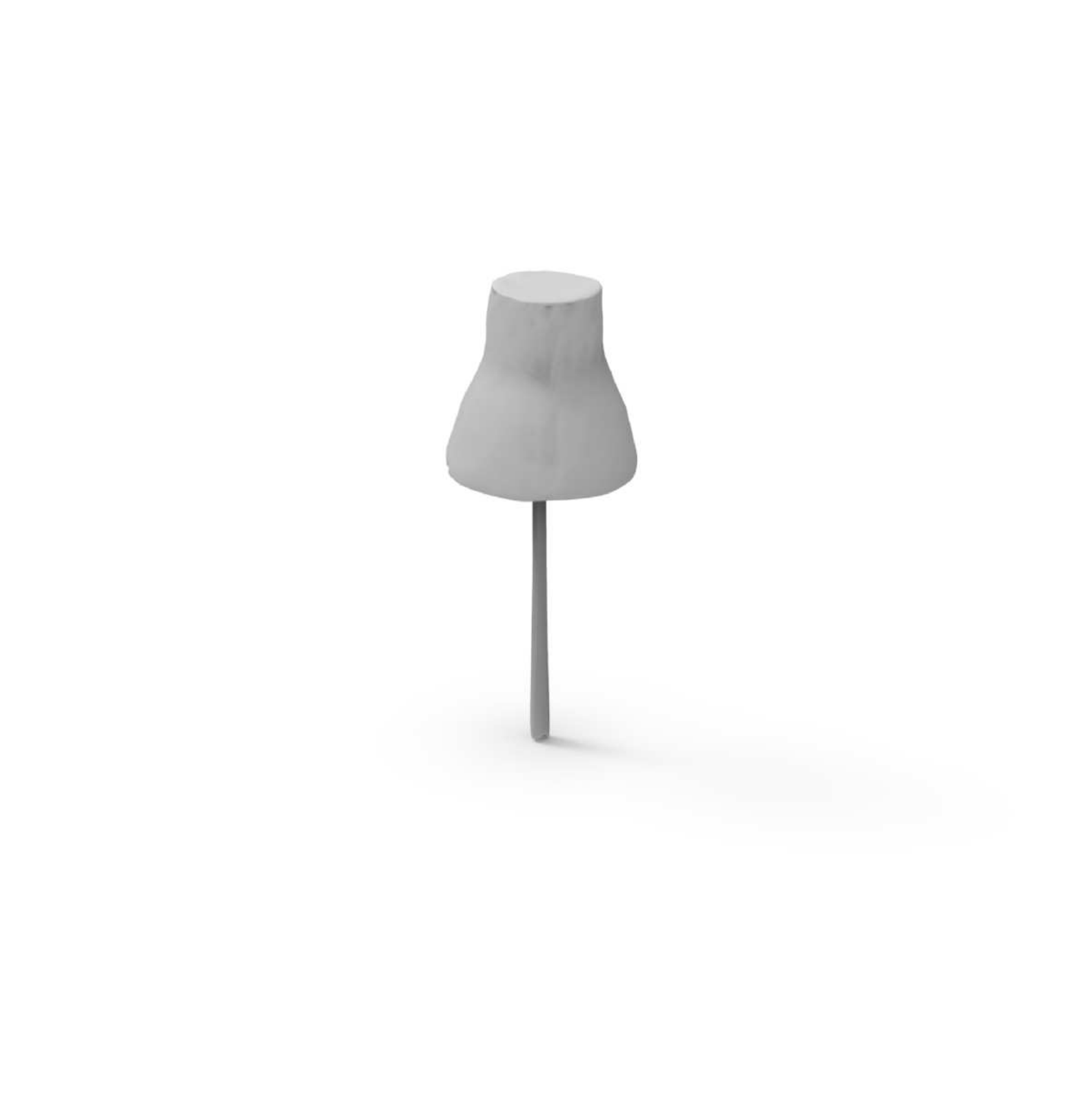}};  & 
    \node (A5) {\includegraphics[width=0.17\linewidth]{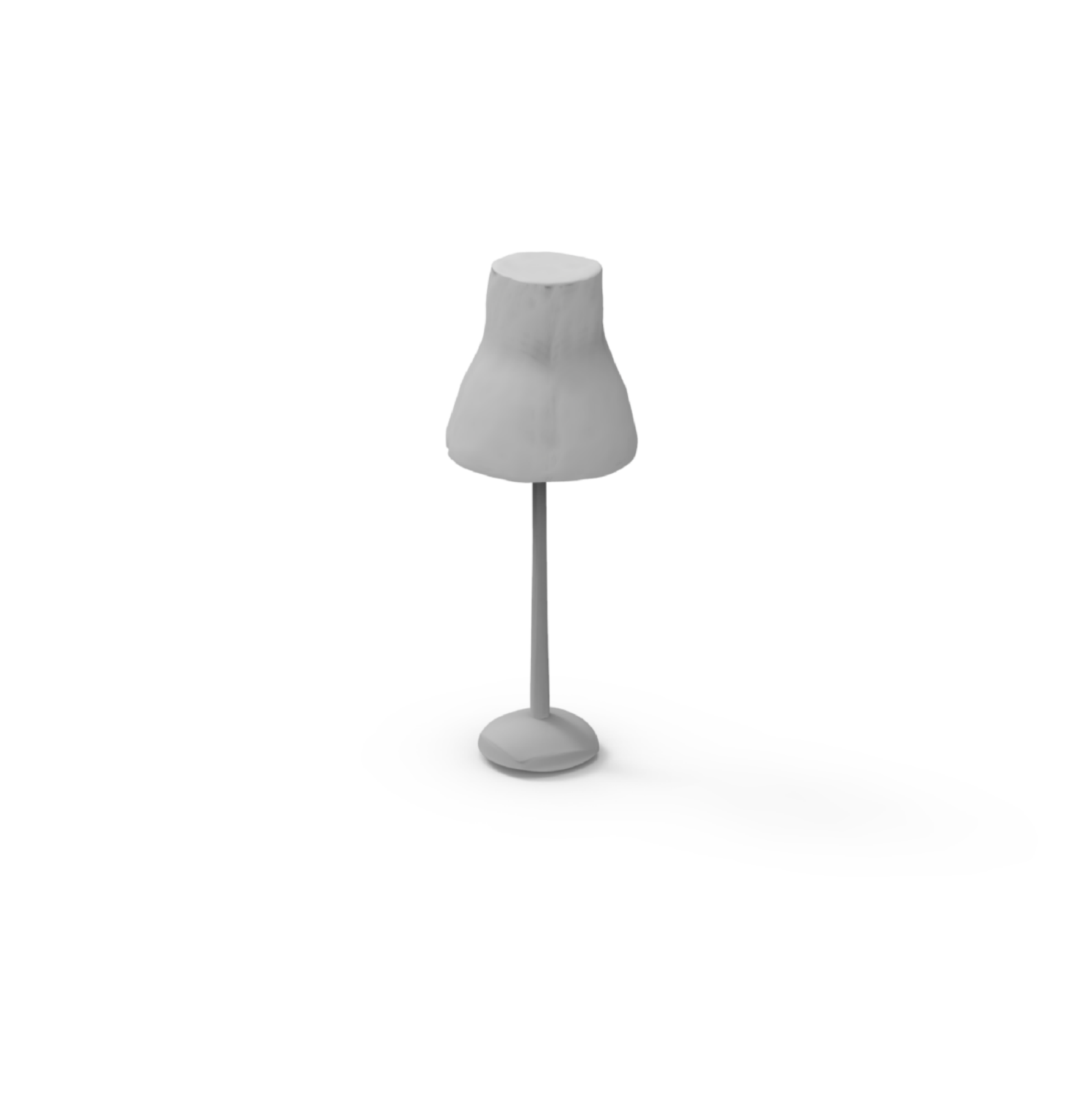}}; \\
    \node (B1) {\includegraphics[width=0.17\linewidth]{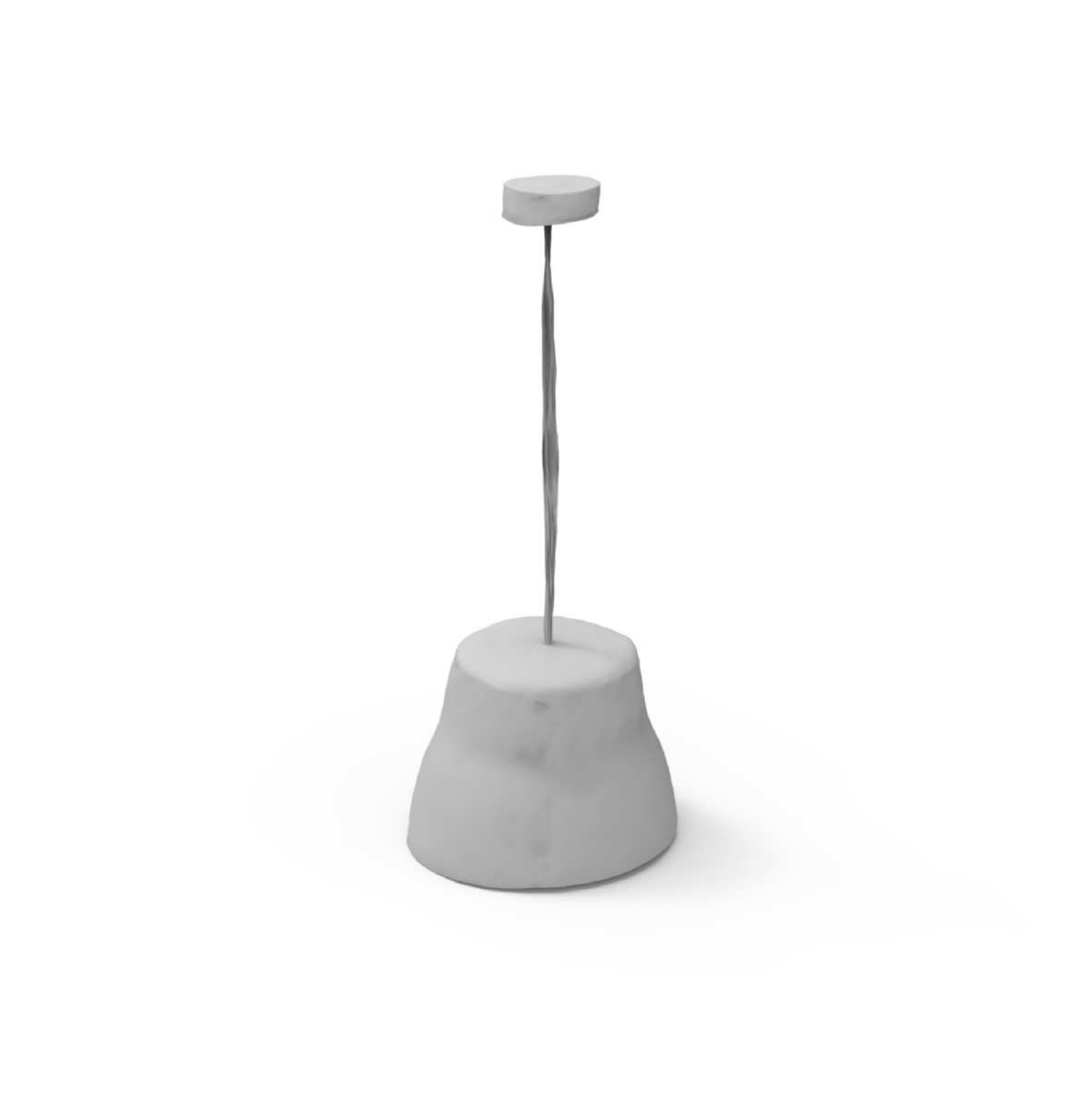}}; & 
    \node (B2) {\includegraphics[width=0.17\linewidth]{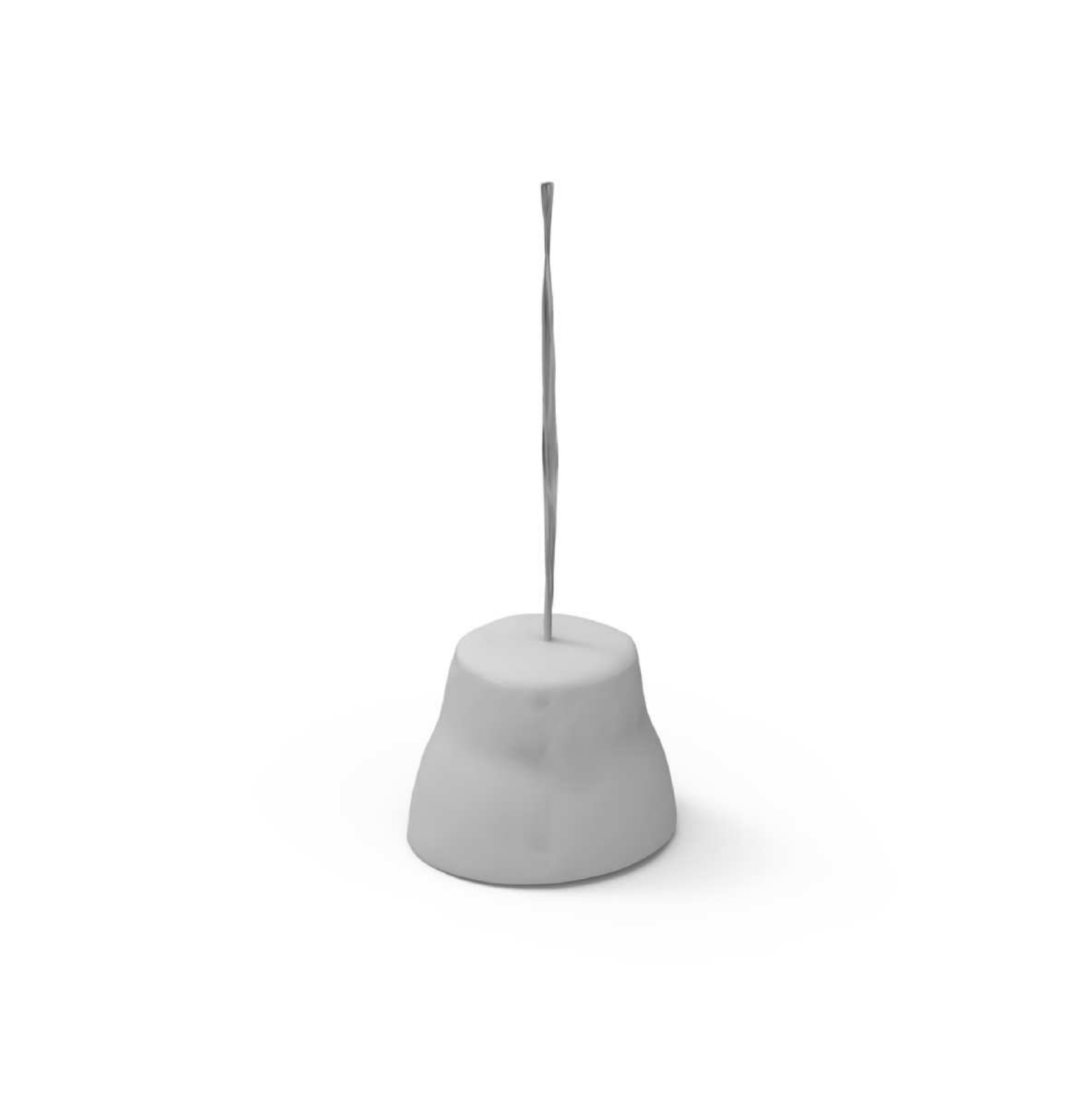}}; &
    \node (B3) {\includegraphics[width=0.17\linewidth]{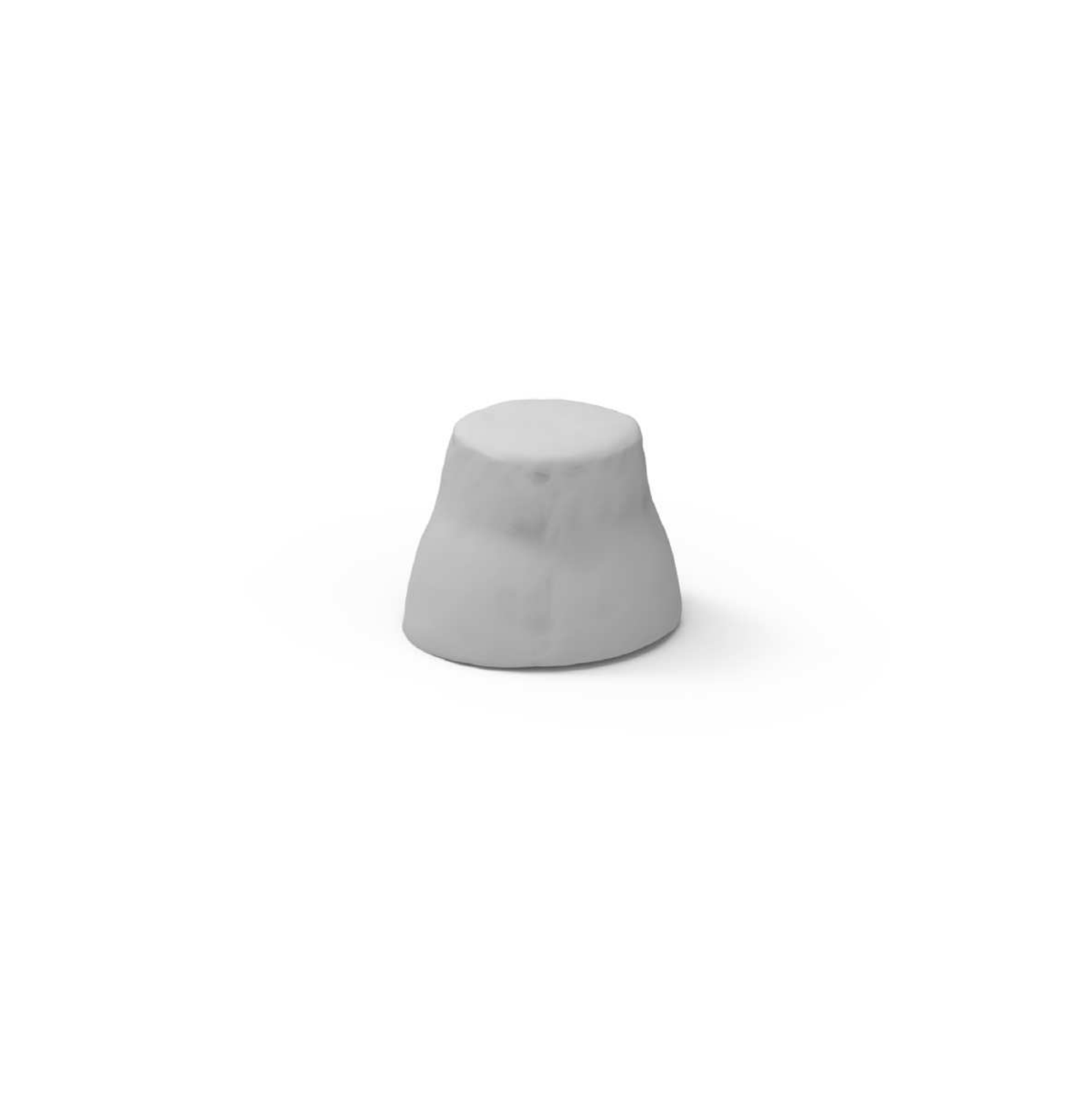}}; &
    \node (B4) {\includegraphics[width=0.17\linewidth]{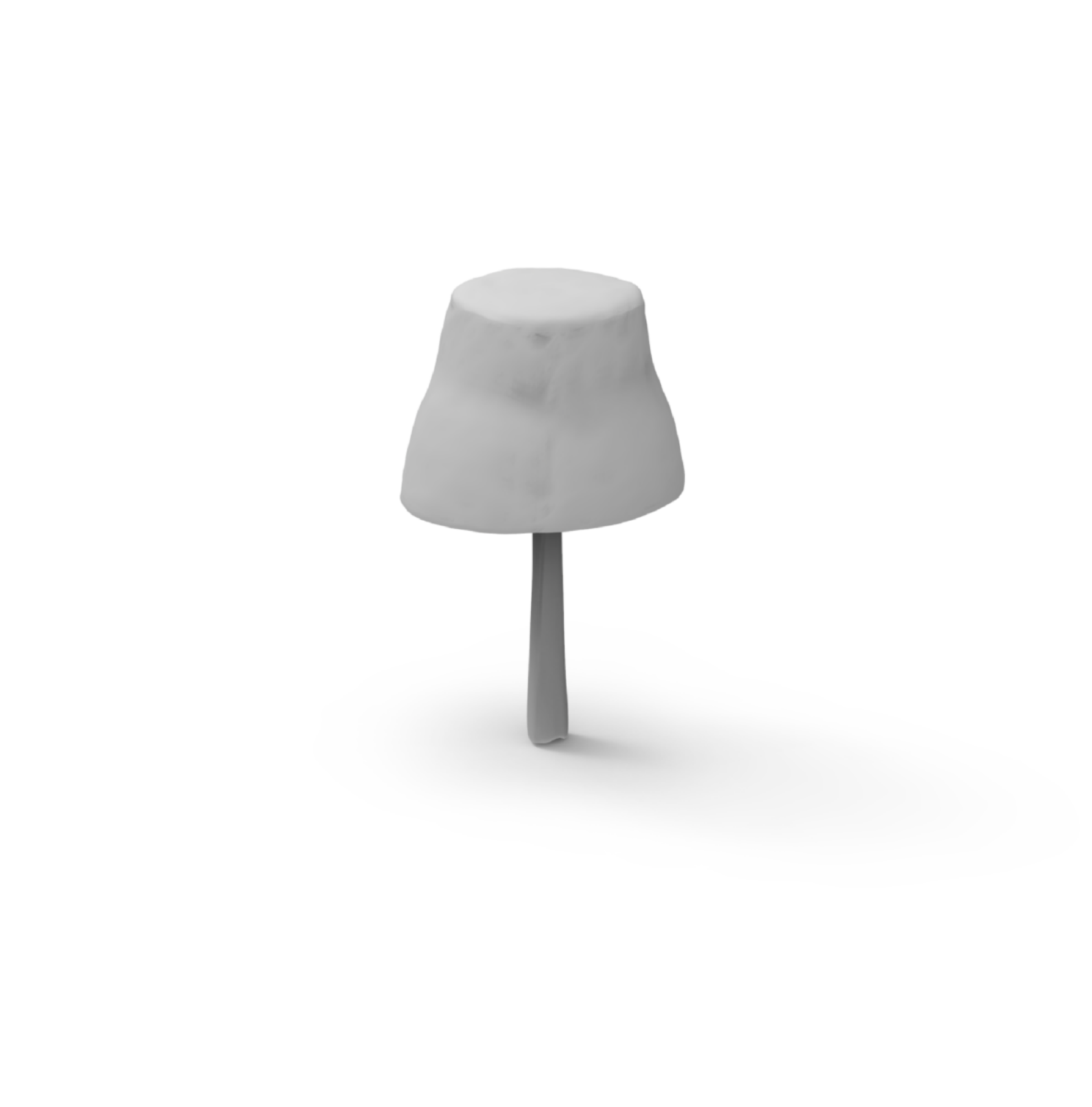}}; &
    \node (B5) {\includegraphics[width=0.17\linewidth]{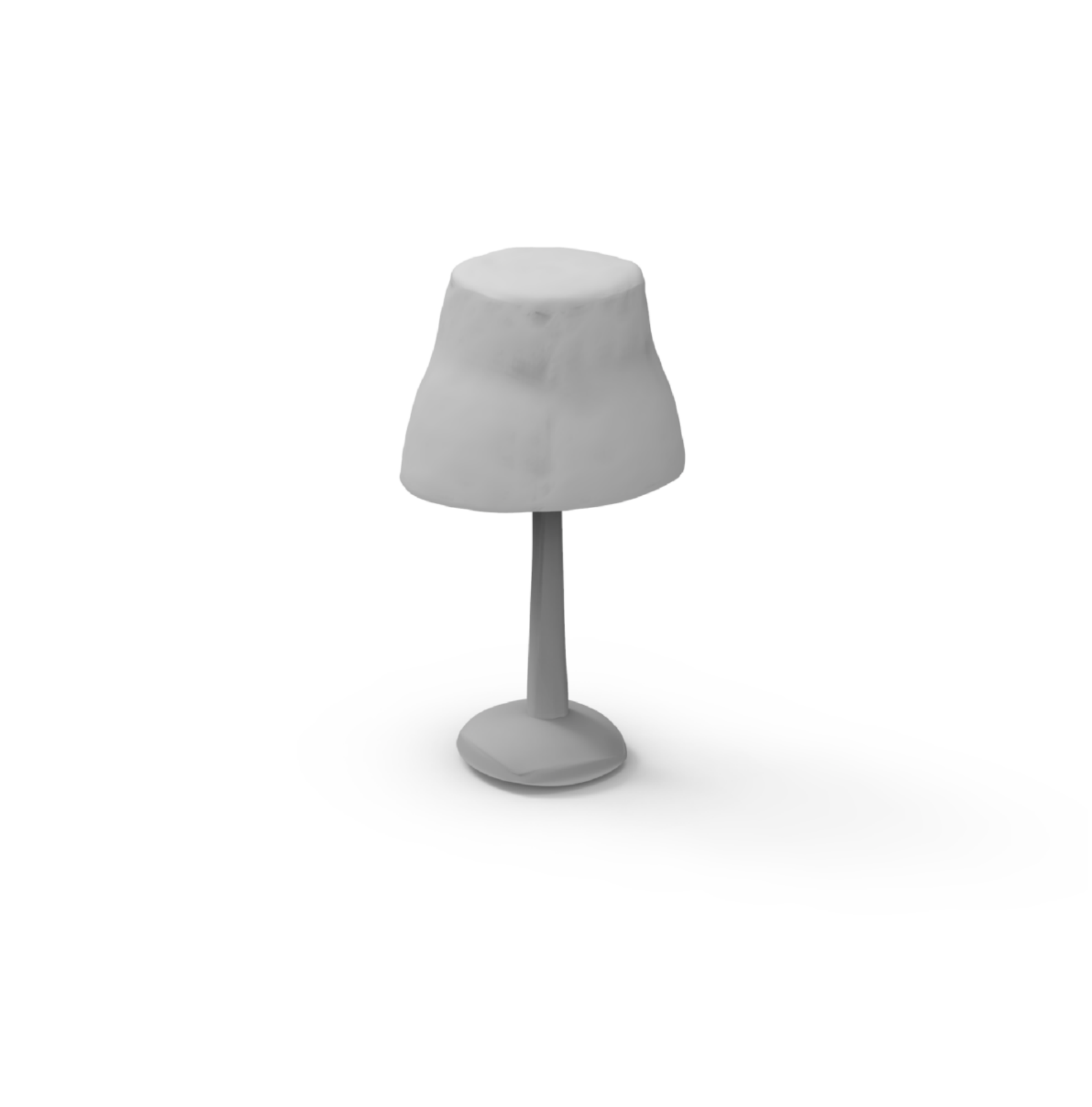}}; \\
    \node (C1) {\includegraphics[width=0.17\linewidth]{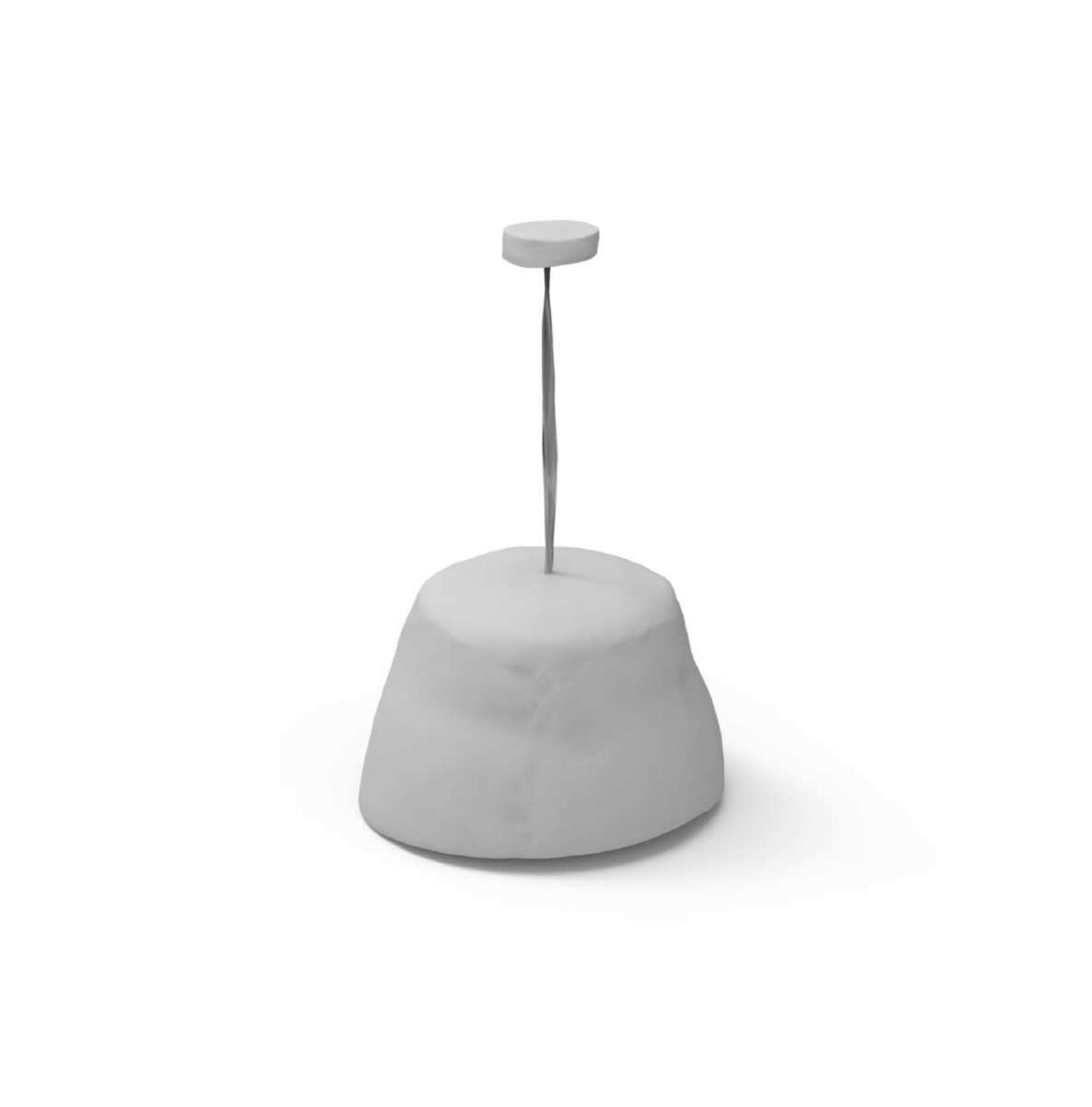}}; & 
    \node (C2) {\includegraphics[width=0.17\linewidth]{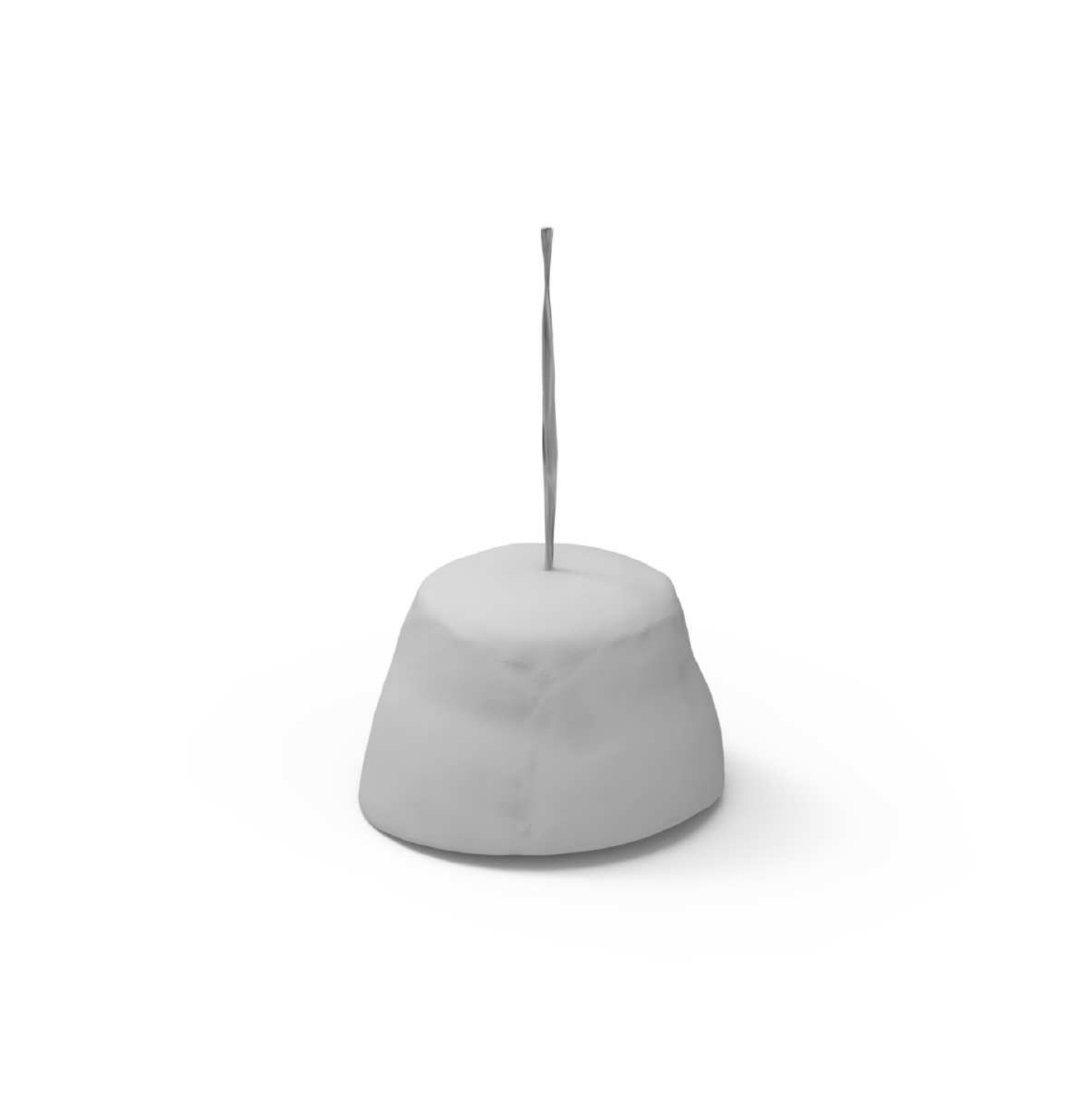}}; &
    \node (C3) {\includegraphics[width=0.17\linewidth]{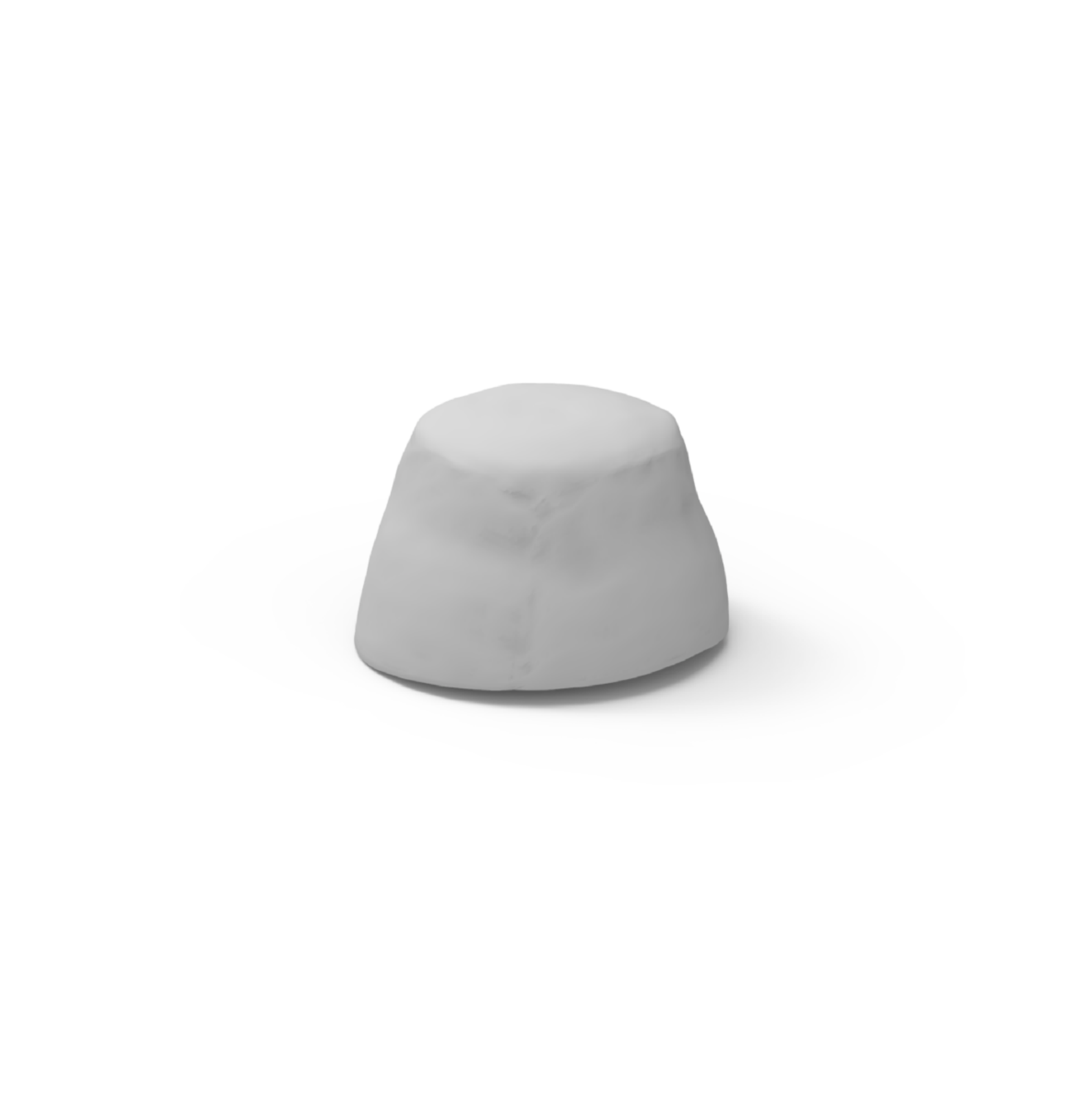}}; &
    \node (C4) {\includegraphics[width=0.17\linewidth]{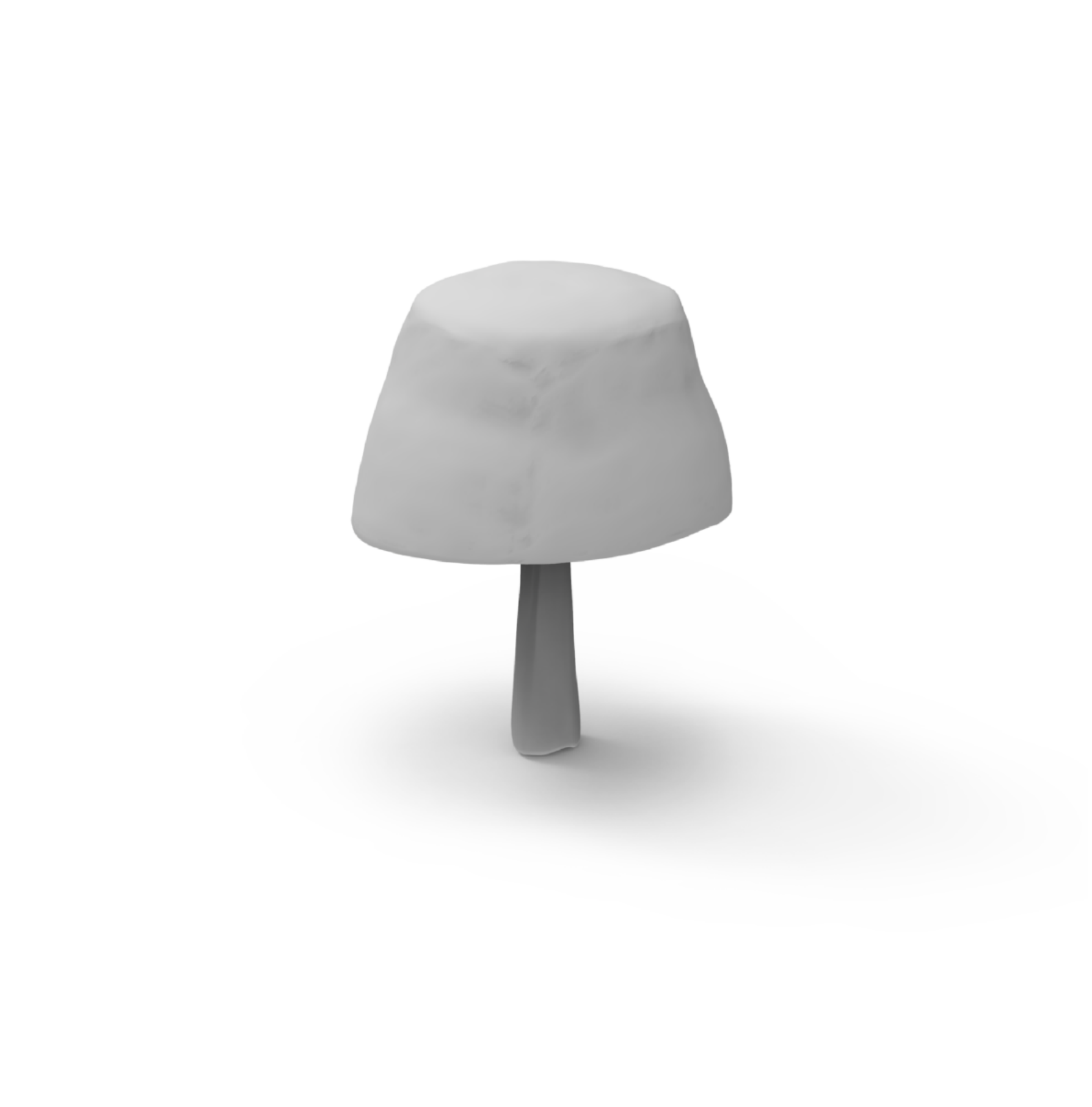}}; &
    \node (C5) {\includegraphics[width=0.17\linewidth]{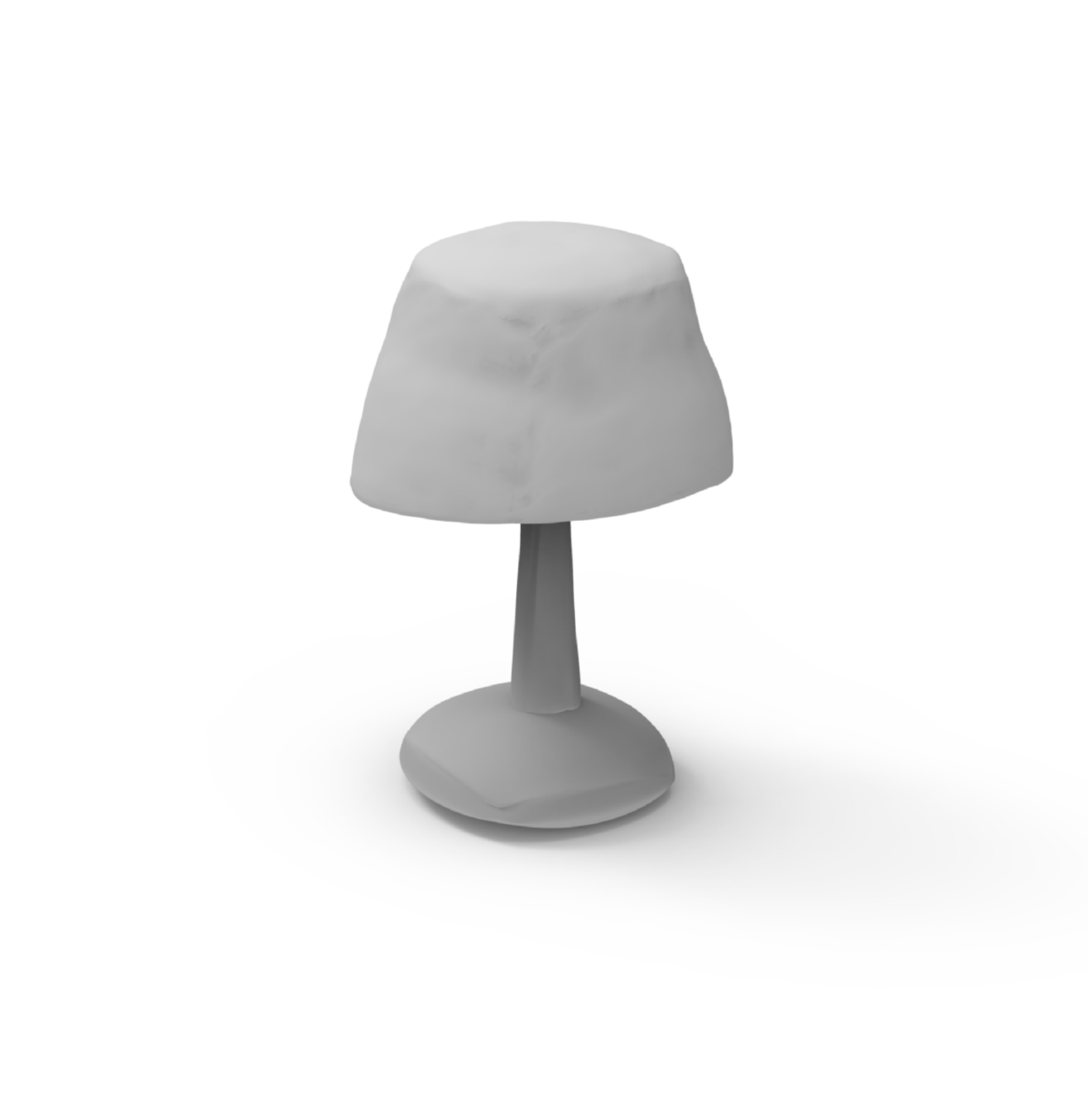}}; \\
    \node (D1) {\includegraphics[width=0.17\linewidth]{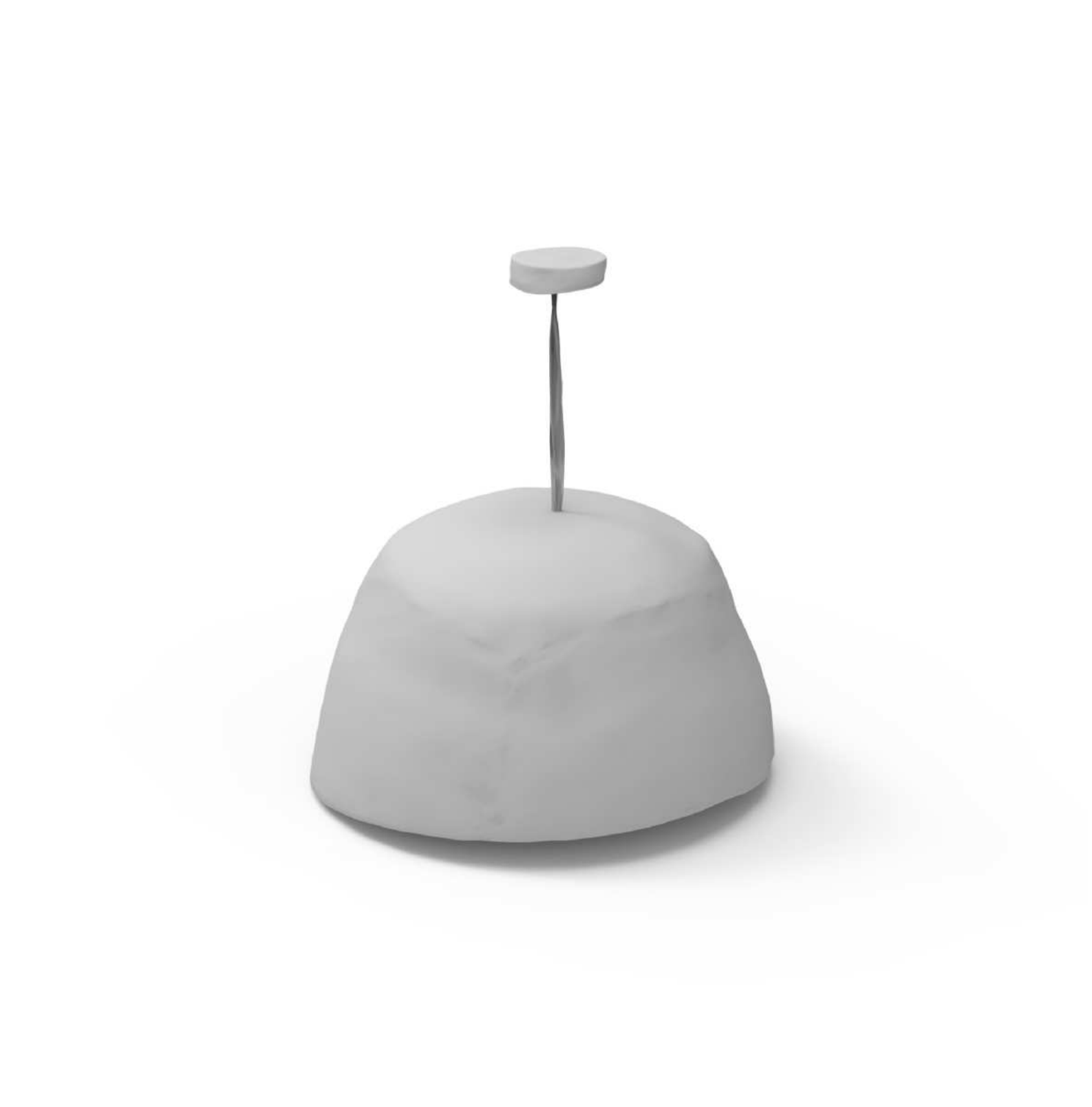}}; & 
    \node (D2) {\includegraphics[width=0.17\linewidth]{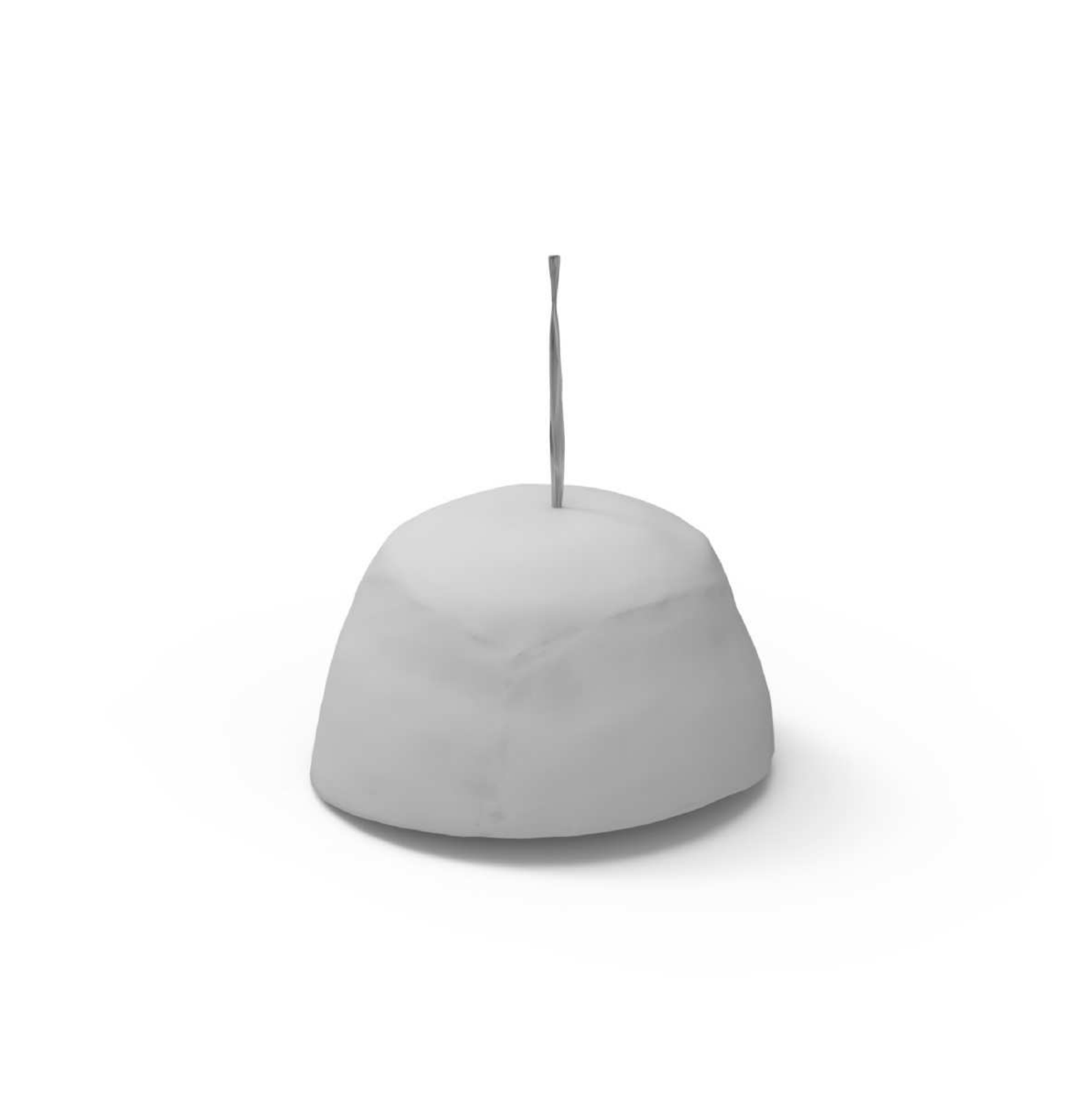}}; &
    \node (D3) {\includegraphics[width=0.17\linewidth]{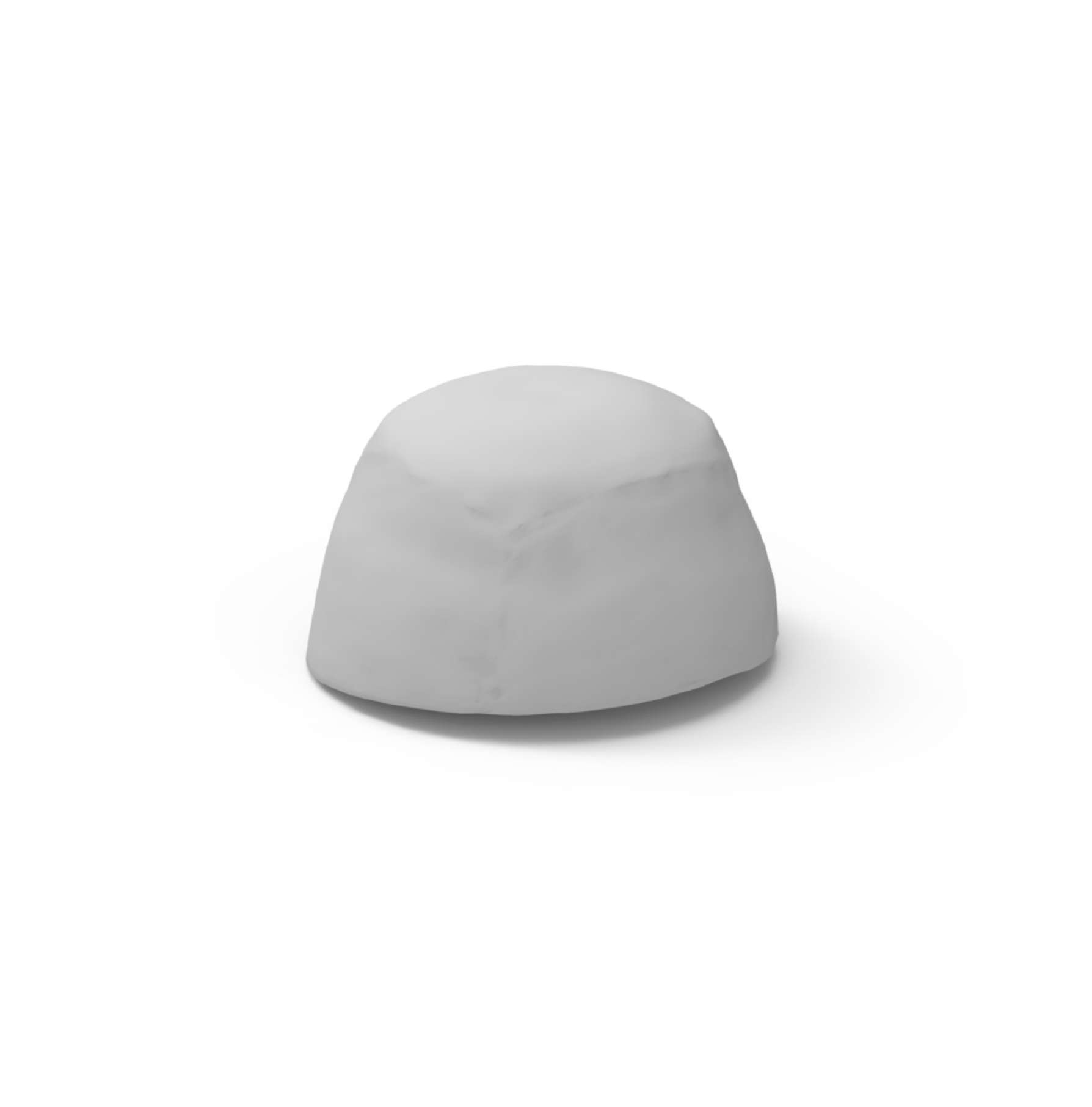}}; &
    \node (D4) {\includegraphics[width=0.17\linewidth]{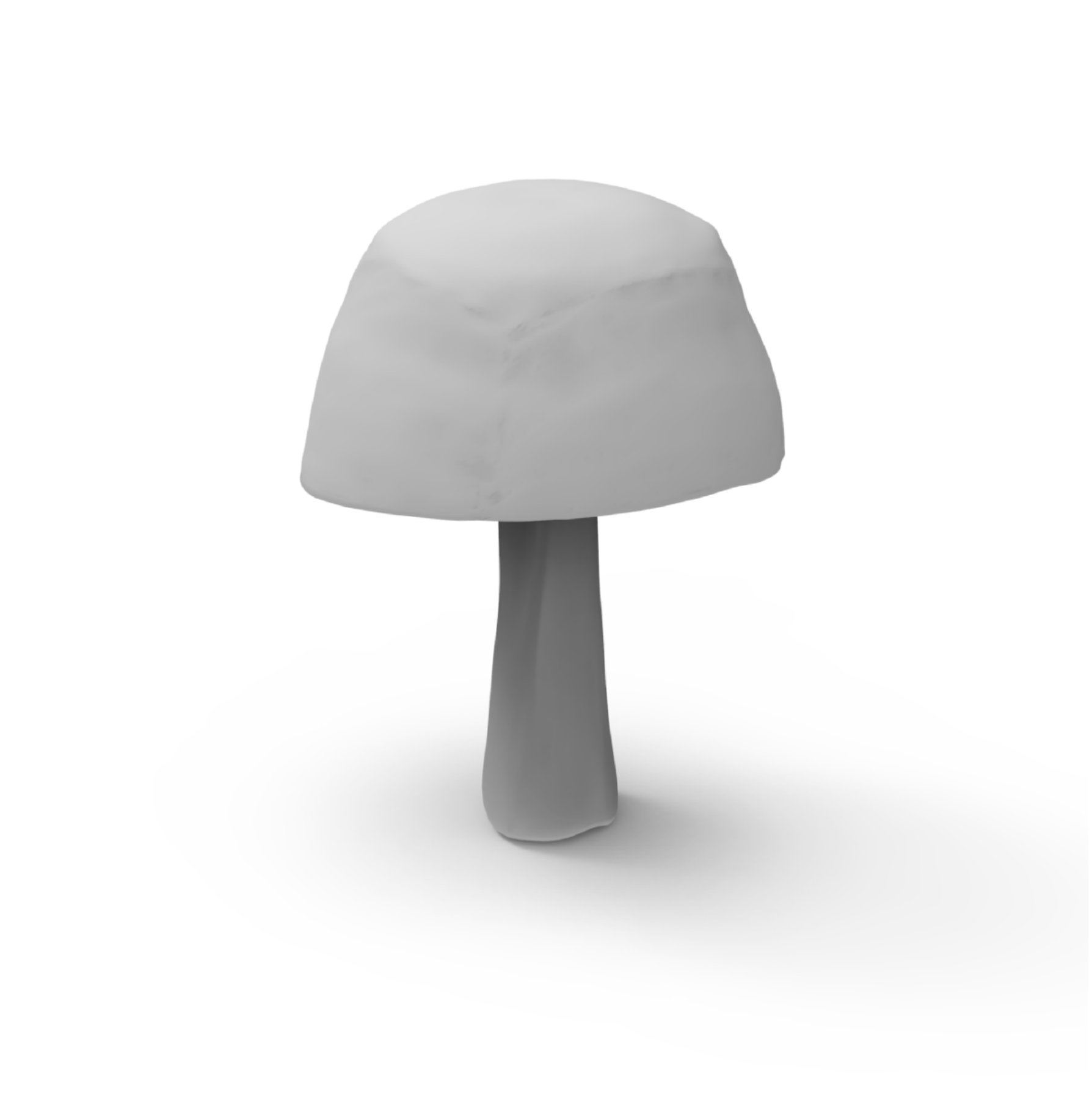}}; &
    \node (D5) {\includegraphics[width=0.17\linewidth]{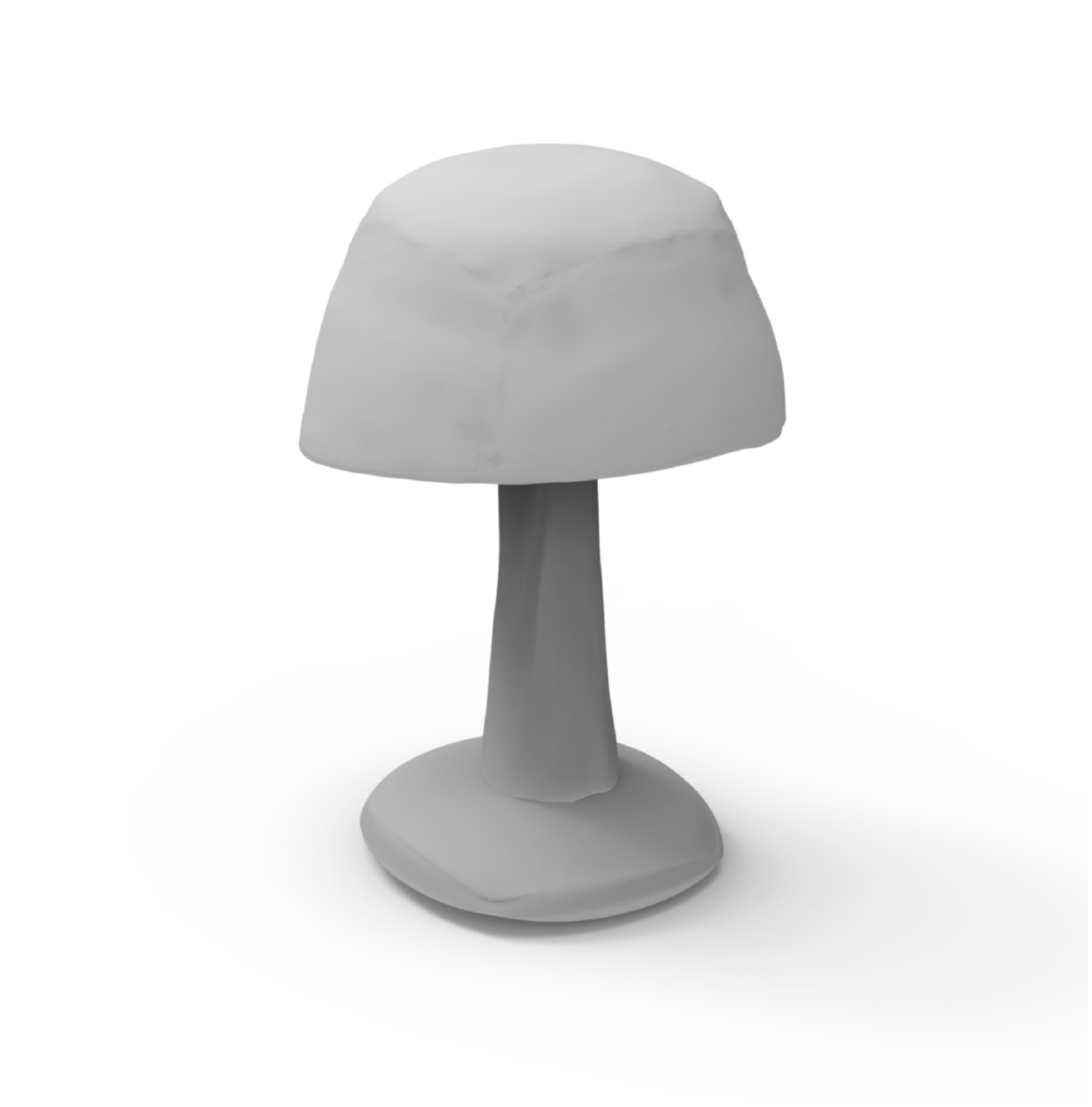}}; \\
    \node (F1) {\includegraphics[width=0.17\linewidth]{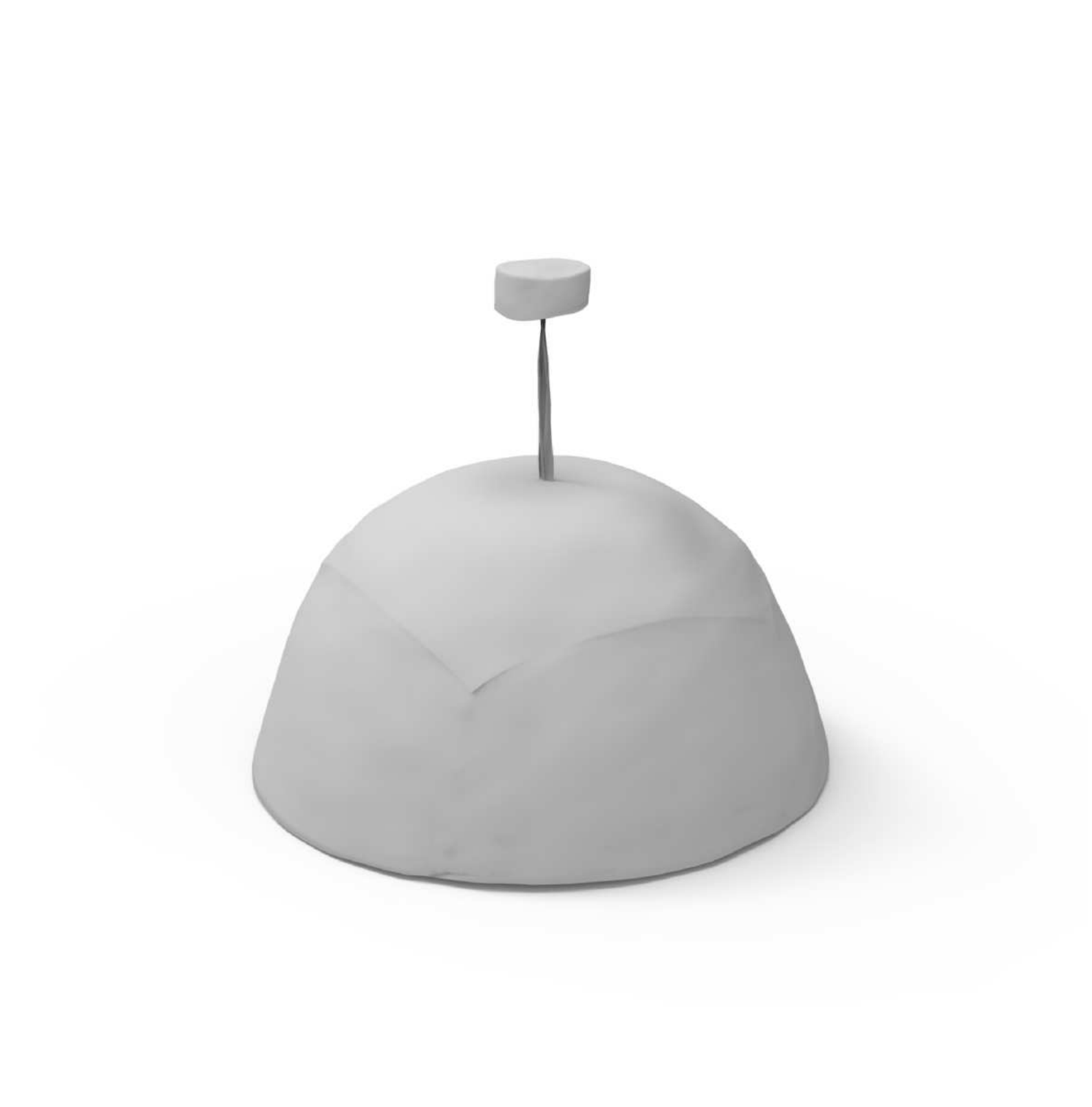}}; & 
    \node (F2) {\includegraphics[width=0.17\linewidth]{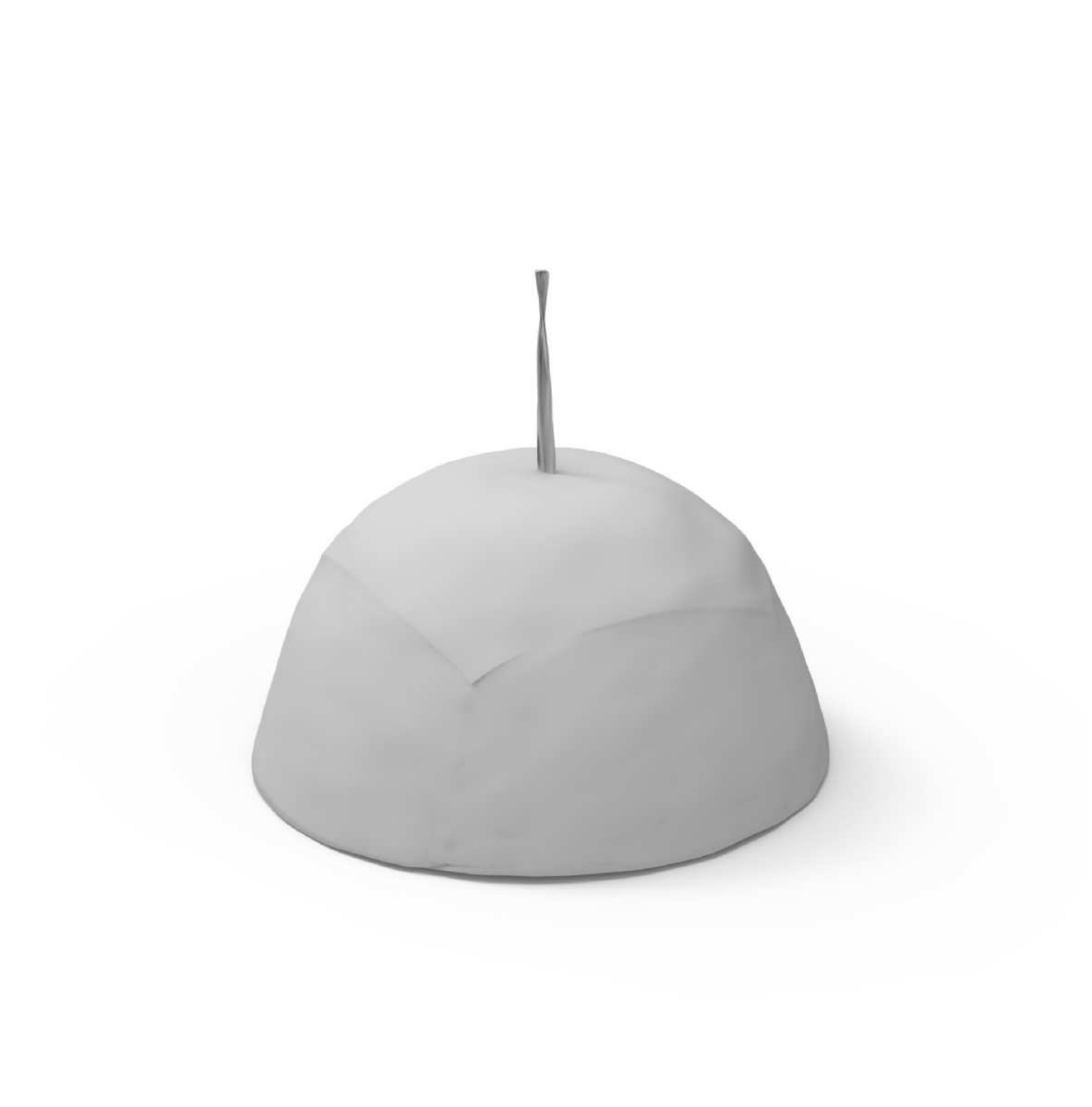}}; &
    \node (F3) {\includegraphics[width=0.17\linewidth]{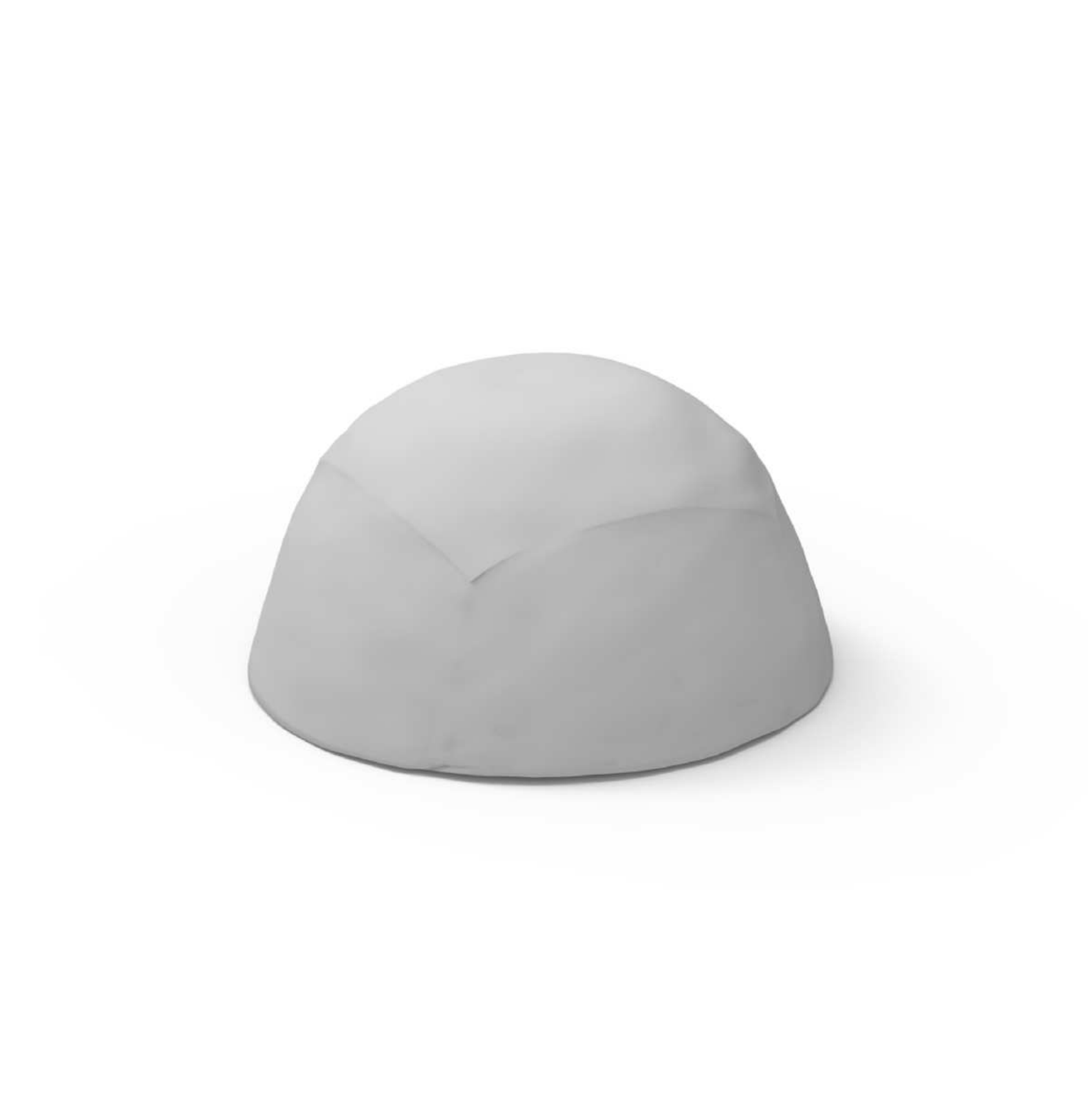}}; &
    \node (F4) {\includegraphics[width=0.17\linewidth]{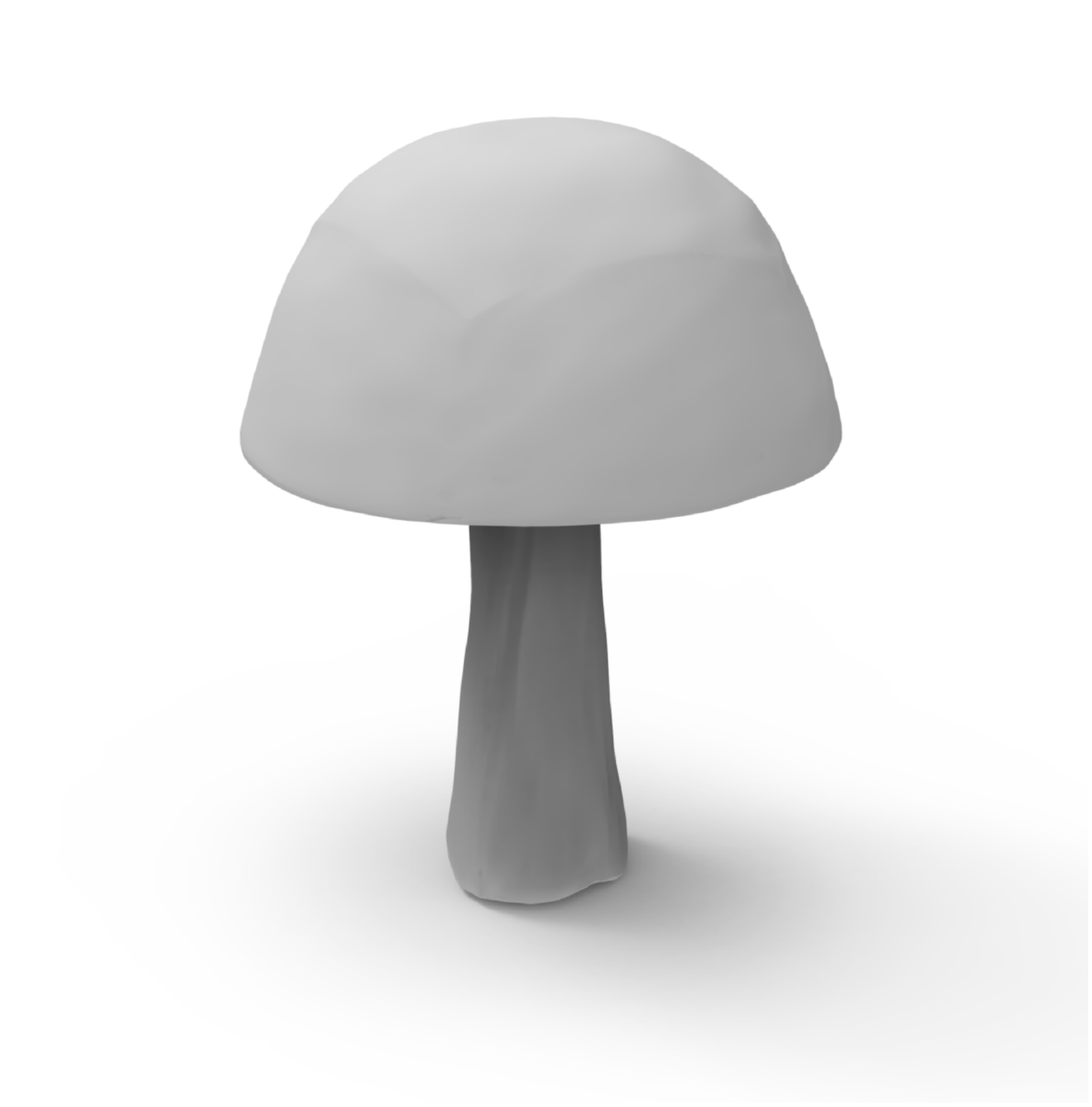}}; &
    \node (F5) {\includegraphics[width=0.17\linewidth, cfbox=orange 1pt 1pt]{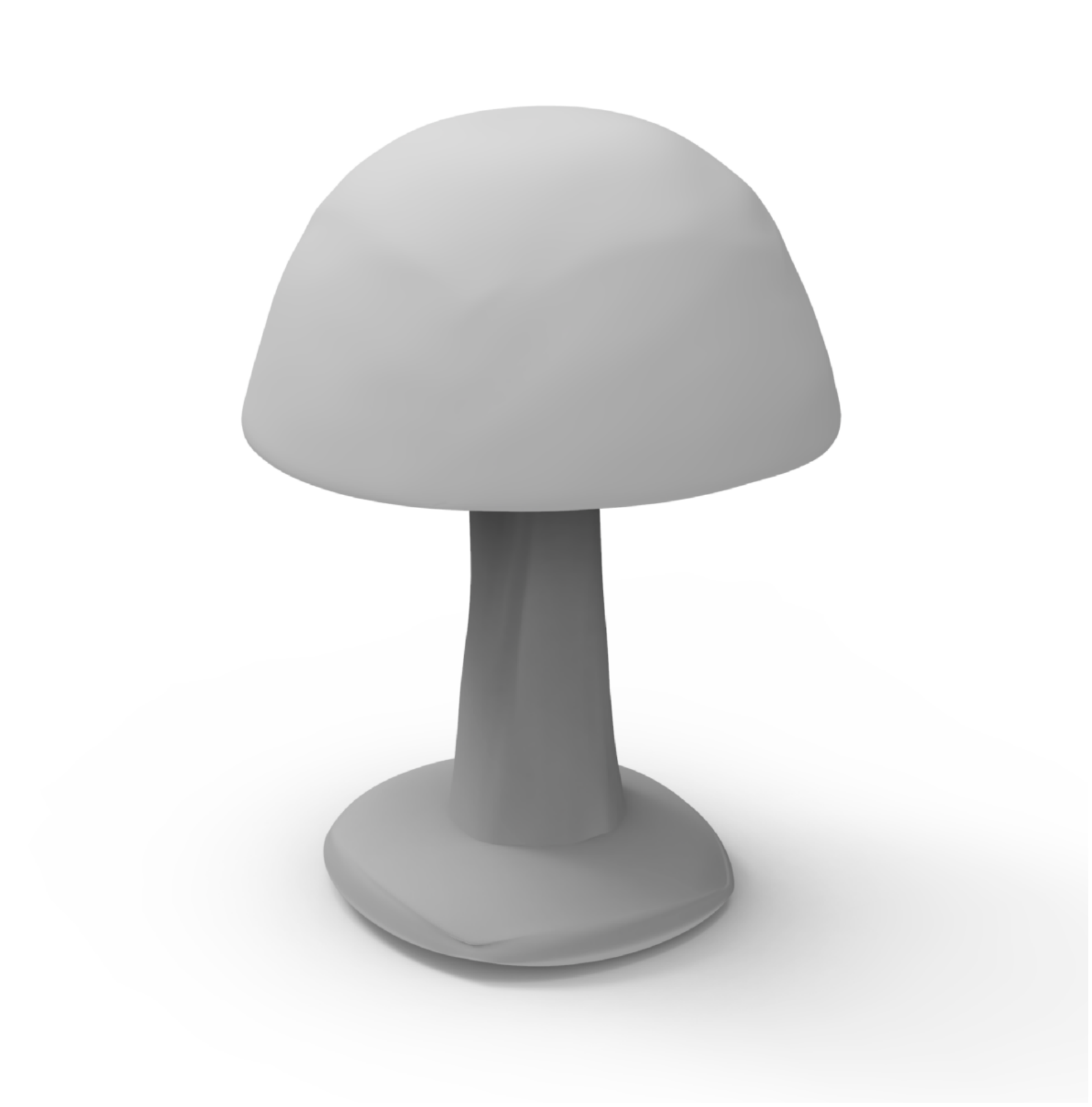}}; \\
  };
  \node[fit=(A1) (A2) (A3) (A4) (A5)
            (B1) (B2) (B3) (B4) (B5)
            (C1) (C2) (C3) (C4) (C5)
            (D1) (D2) (D3) (D4) (D5)
            (F1) (F2) (F3) (F4) (F5),
            inner sep=0pt,
            ] (PIC) {};

  \draw[line width=1pt,arrows={-Stealth[length=4mm]}] ([xshift=-1em,yshift=-0.5em]PIC.south west) -- ([xshift=-1em,yshift=-0.5em]PIC.south east);
  \draw[line width=1pt,arrows={-Stealth[length=4mm]}] ([yshift=-1em,xshift=-0.5em]PIC.south west) -- ([yshift=-1em,xshift=-0.5em]PIC.north west);

  \node[anchor=south] (label) [font=\fontsize{10}{10}\selectfont]at ([yshift=-2.em]A3|-PIC.south) {Structure};
  \node[anchor=center,rotate=90] (label) [font=\fontsize{10}{10}\selectfont] at ([xshift=-1.5em]C1-|PIC.west) {Geometry};

\end{tikzpicture}
\vspace{-3mm}
    \caption{\yjr{Disentangled shape reconstruction and interpolation results on PartNet Lamps. Here, the top left and bottom right shapes (highlighted with orange box) are the input shapes. %
    The remaining shapes are generated automatically with our DSG-Net, where in each row, the \emph{structure} of the shapes is interpolated while keeping the geometry unchanged, whereas in each column, the \emph{geometry} is interpolated while retaining the structure. \yj{The vertical axis and horizontal axis represent the variation of structure and geometry respectively.}}}
    \label{fig:decouple_lamp}
\end{figure}

\section{Training and Implementation Details}\label{sec:impl}
\yj{For our network, we train the part geometry conditional VAE and the coupled hierarchical VAEs simultaneously. The part geometry conditional VAE is used to recover the part geometric details according to the structure context. Two coupled hierarchical VAEs aim to learn two disentangled latent spaces for shape geometry and structure in a disentangled but tightly coupled manner. Training of the whole network is optimized by the Adam solver~\cite{kingma2014adam}. All learnable parameters are initialized randomly with Gaussian distribution. For the training, we set the batch size as 16 and use a learning rate that starts from 0.001 and decays every 100 steps with the decay rate set to 0.9, until the loss converges with about 1000 iterations. 
We use the same energy weights as in StructureNet~\cite{mo2019structurenet} for the structure part of the network. 
For the energy weights for the geometry part, we set 100.0 in our all experiments. 
}

Our whole network is implemented in PyTorch~\cite{paszke2017automatic}. 
The backbone network framework is borrowed from StructureNet~\cite{mo2019structurenet}. 
Most linear layers in our disentangled hierarchical graph networks are implemented as Multilayer Perceptrons (MLPs) with LeakyReLU activations.  
Our network does not use batch normalization~\cite{ioffe2015batch}.
Besides, our conditional part geometry VAE can be pre-trained or trained together with the disentangled hierarchical graph VAEs. 
Due to the difficulty of batch training of recursive neural networks, we use the same training strategy as StructureNet that the network sequentially forwards through each shape, summarizes the loss of each shape and backwards gradients for a batch of shapes. 
Empirically, our network converges after 4 -- 5 days of training with a single GeForce RTX 2080Ti and an Intel i9-9900K CPU. Memory consumption is approximately 4 GB.

% \clearpage
\section{More Results on Shape Reconstruction}\label{sec:more_recon}
In Figure~\ref{fig:reconstruction}, we show more results on shape reconstruction for PartNet Chairs, PartNet Lamps, PartNet Tables, and PartNet Cabinets. In all figures, the left parts are the input shape, and the right parts are the reconstructed shapes from our network.

\section{More Results on Shape Generation}\label{sec:more_gen}
\yjr{
In Figure~\ref{fig:generation_nn}, we show more results on shape generation for PartNet Chairs, PartNet Lamps, PartNet Tables, and PartNet Cabinets. Furthermore, we also present the top-5 retrieved shapes in the training set to demonstrate the novelty of shape generation. In all figures, the left parts are the generated shapes, and the right parts are the generated shape and its top-five retrieved shapes in the training sets, with CD as the retrieval metric.
}

\section{More Results on Shape Interpolation}\label{sec:more_inter}
\yjr{
In Figure~\ref{fig:interp}, we show more results on shape interpolation for PartNet Chairs, PartNet Lamps, PartNet Tables, and PartNet Cabinets. In all figures, the first and last shape for each row are the source shape and target shape for interpolation.
}

\section{More Results on Disentangled Shape Interpolation}\label{sec:more_disinter}

In this section, we show more results on disentangled shape interpolation for PartNet Chairs (Figure~\ref{fig:decouple_chair}), PartNet Lamps (Figure~\ref{fig:decouple_lamp}) and the synthetic chair dataset (Figure~\ref{fig:decouple_synthetic}). In all the figures, the top left and bottom right shapes (highlighted with orange boxes) show the input shapes. The vertical axis illustrates the geometry variation, while the horizontal axis presents the structure variation. The remaining shapes are automatically generated with our DSG-Net, where in each row, the \emph{structure} of shapes is interpolated while keeping the geometry unchanged, whereas in each column, the \emph{geometry} is interpolated while retaining the structure.

\section{More Results on Disentangled Shape Generation}\label{sec:more_disgen}

DSG-Net learns two disentangled latent spaces for modeling shape structure and geometry respectively, which enables a novel application of disentangled shape generation.
For any given shape input, we can extract its geometric or structural code by our encoder, and then we can explore random generation in the other latent space.
We show more disentangled shape generation on the four PartNet categories in Figure~\ref{fig:generation_pt2pc2}.

\begin{figure*}[t]
\centering

\begin{minipage}{0.14\linewidth}
\centering
\includegraphics[width=0.99\linewidth]{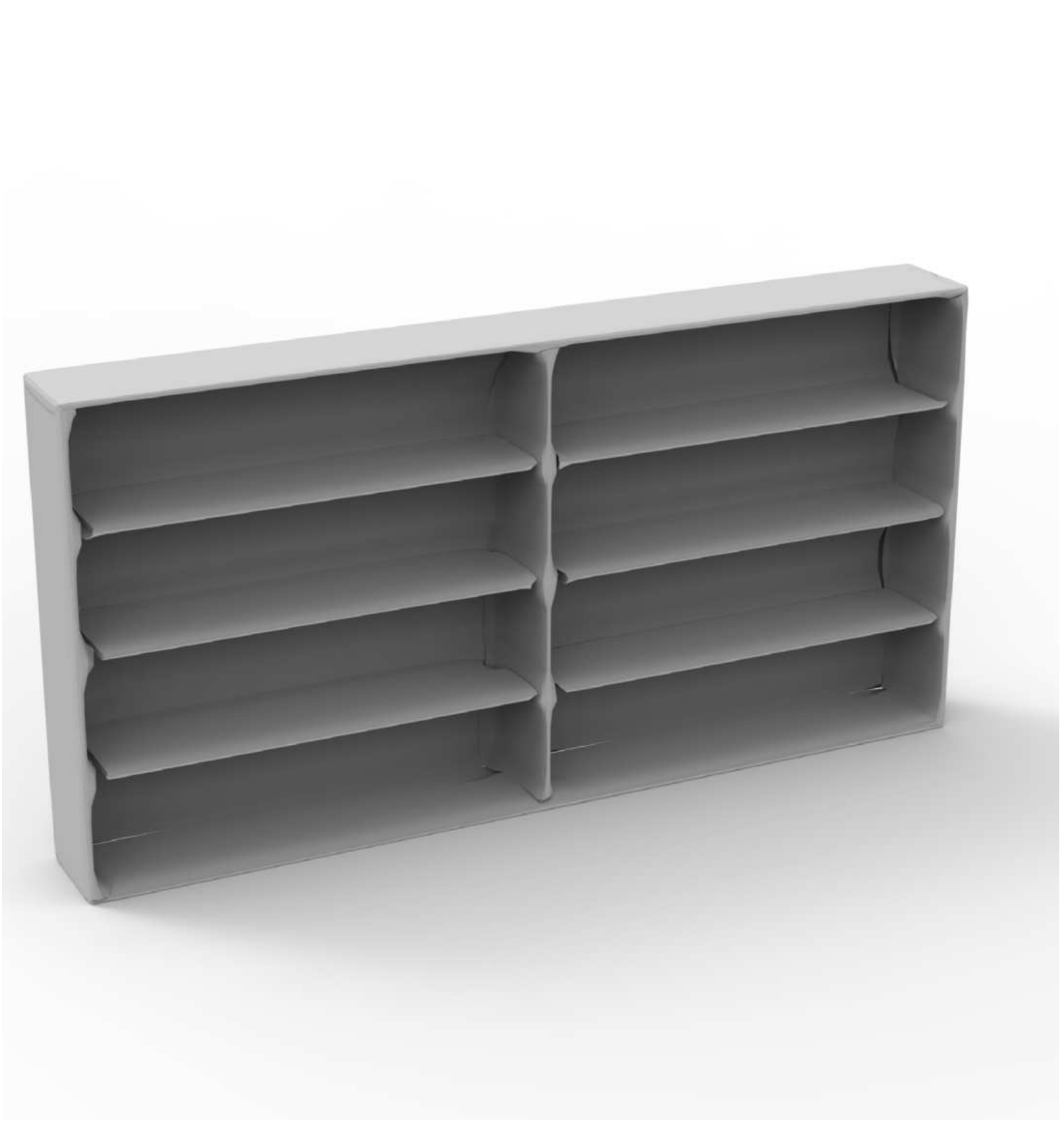}
\end{minipage}
\begin{minipage}{0.44\linewidth}
\centering
    \includegraphics[width=0.23\linewidth]{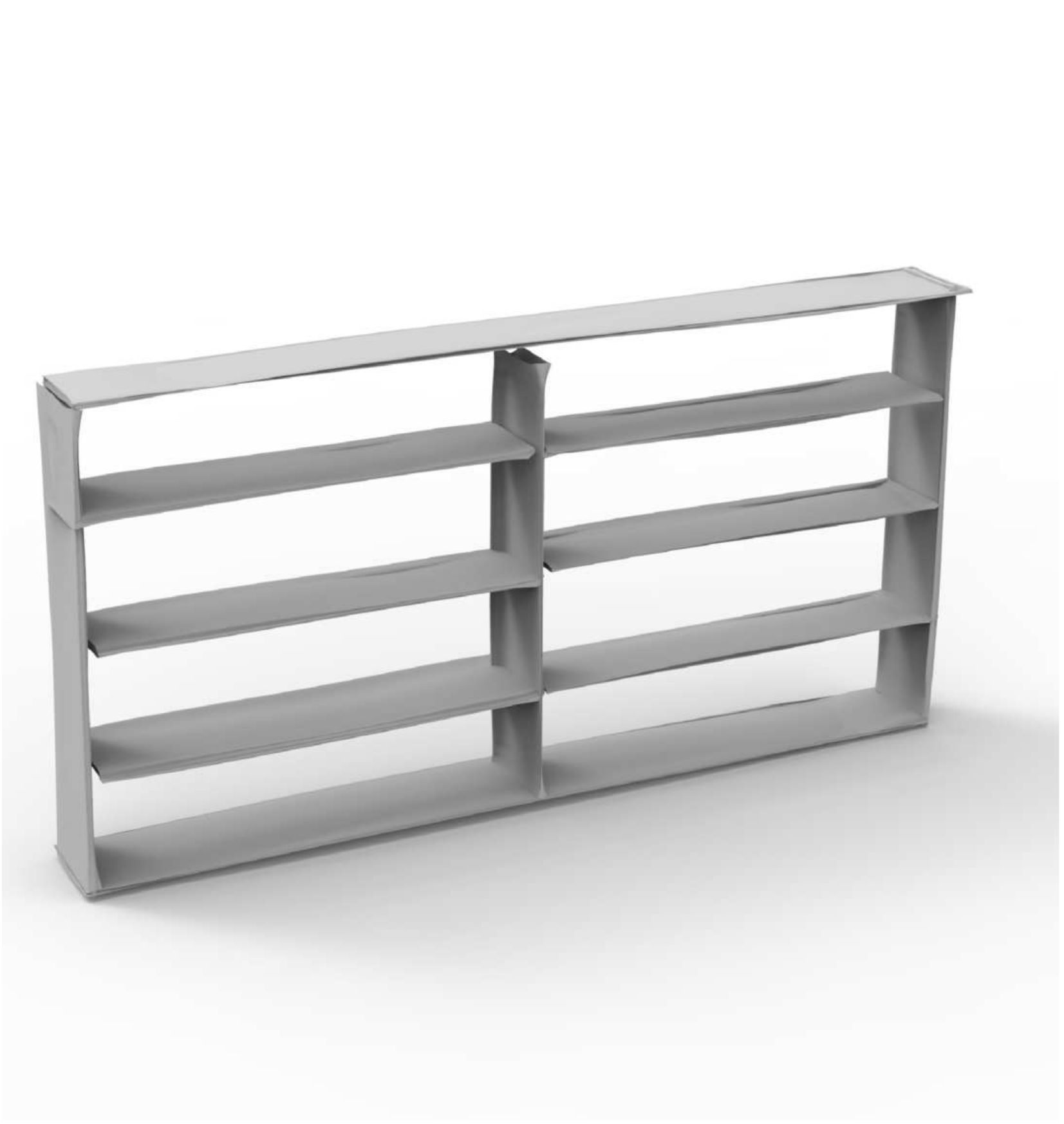}
    \includegraphics[width=0.23\linewidth]{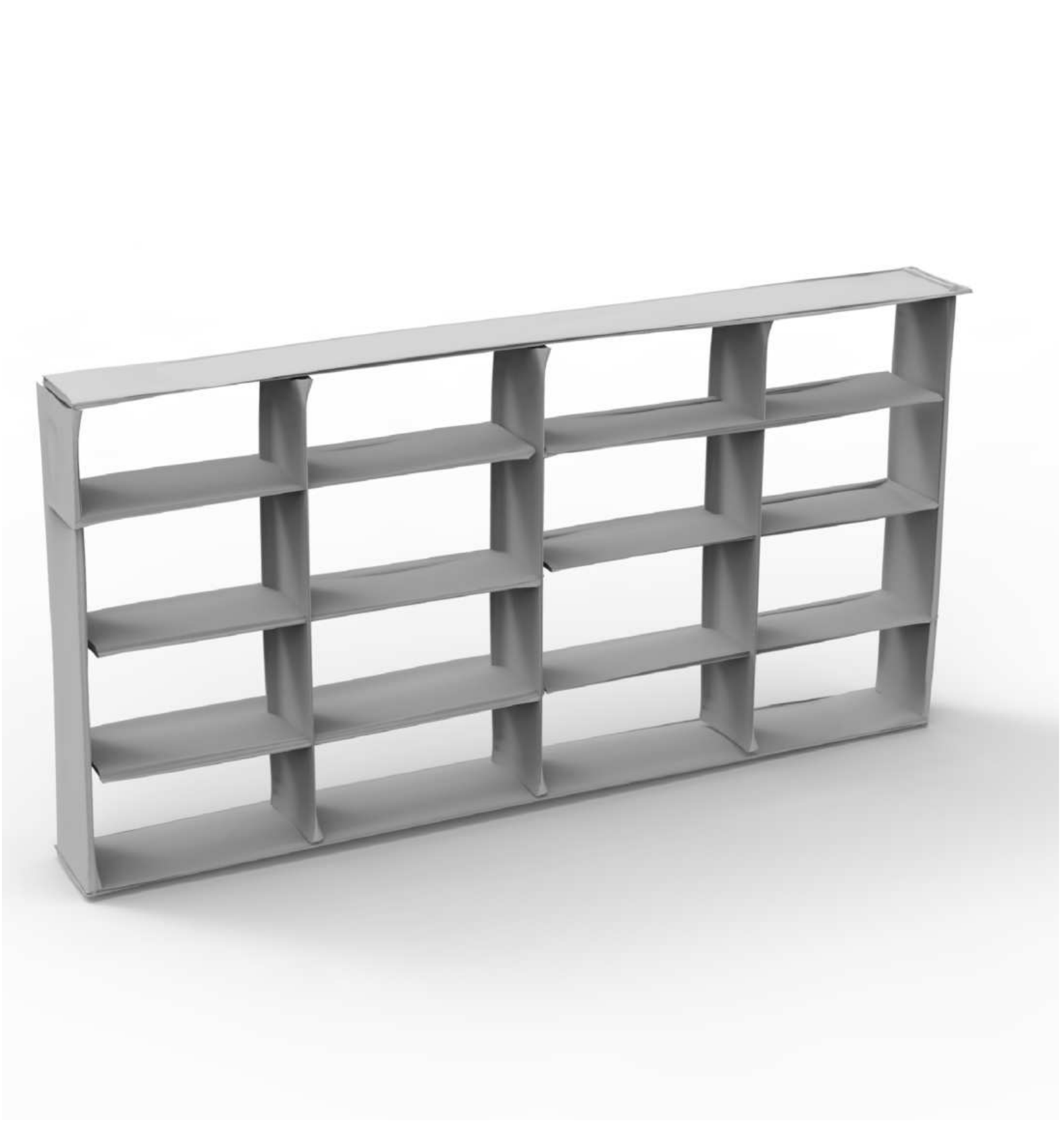}
    \includegraphics[width=0.23\linewidth]{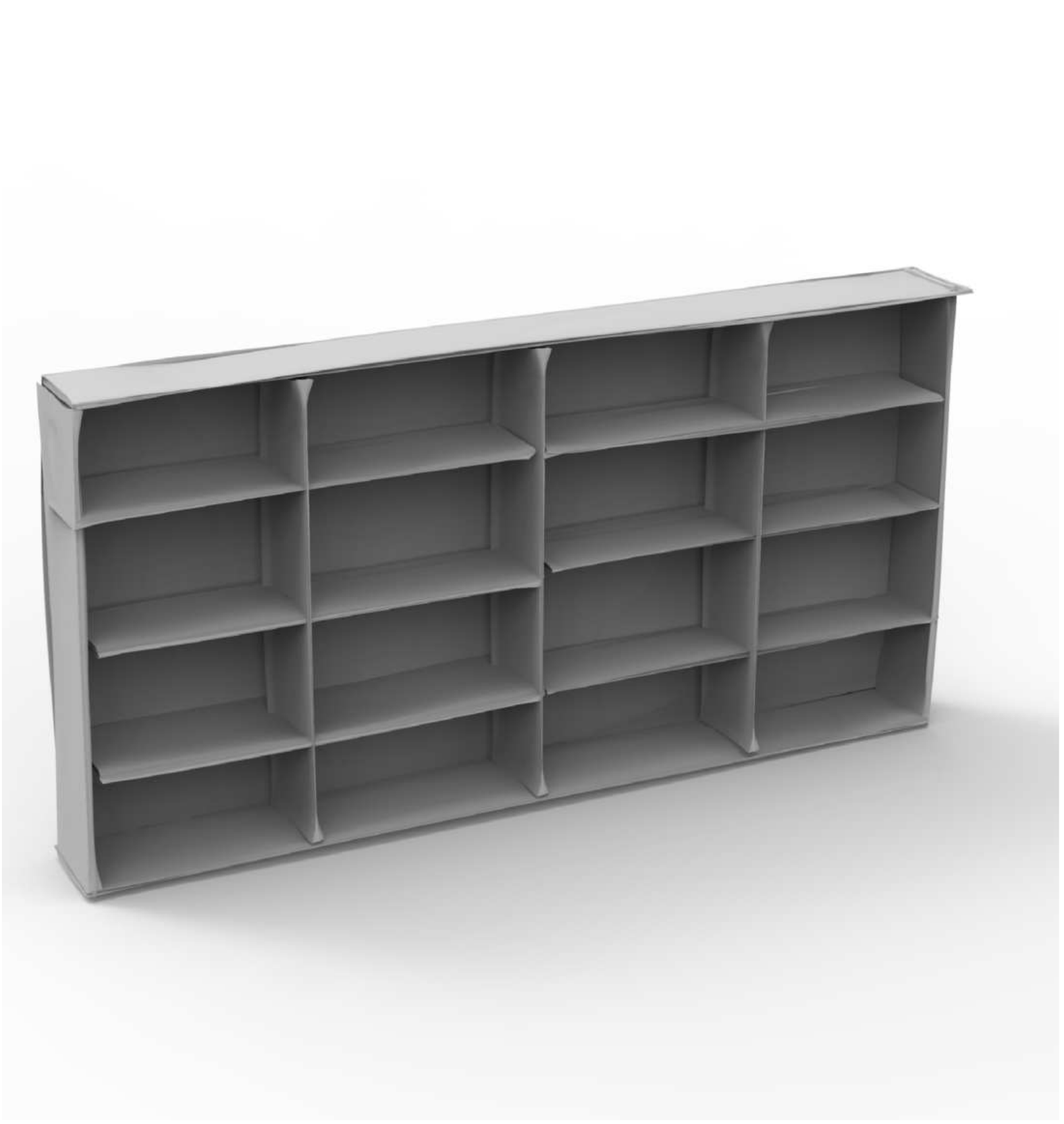}
    \includegraphics[width=0.23\linewidth]{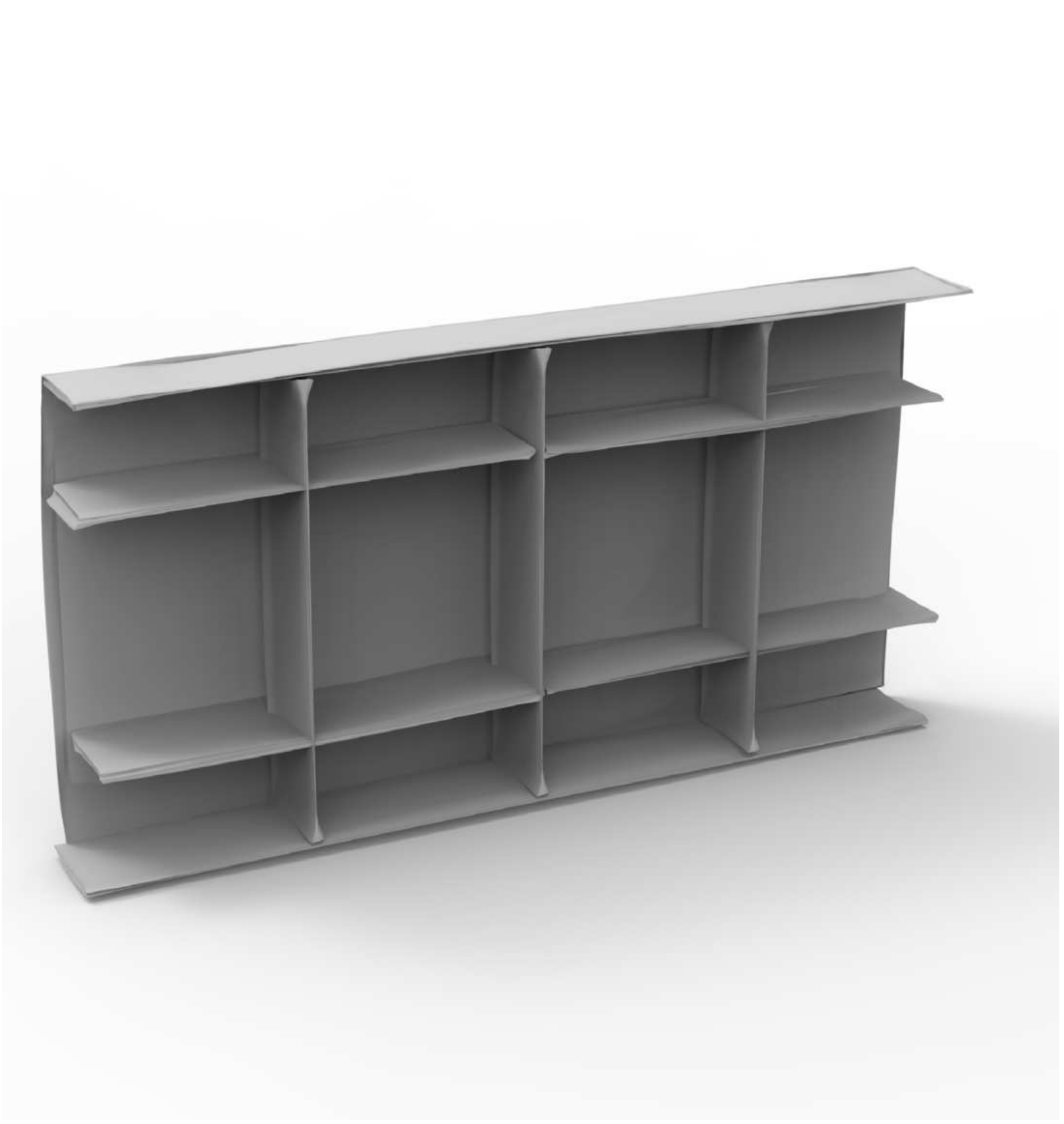}
    \\
    \includegraphics[width=0.23\linewidth]{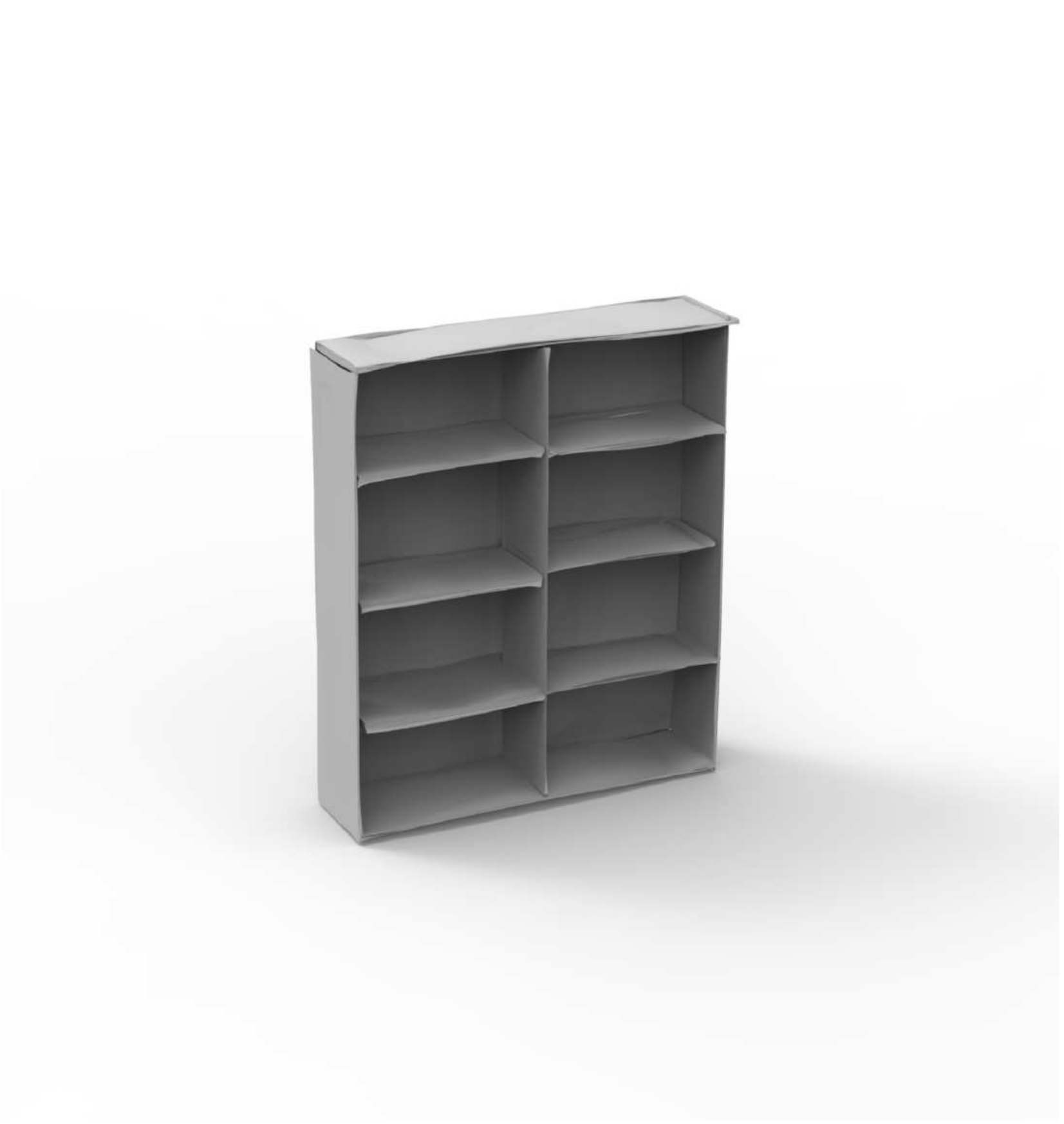}
    \includegraphics[width=0.23\linewidth]{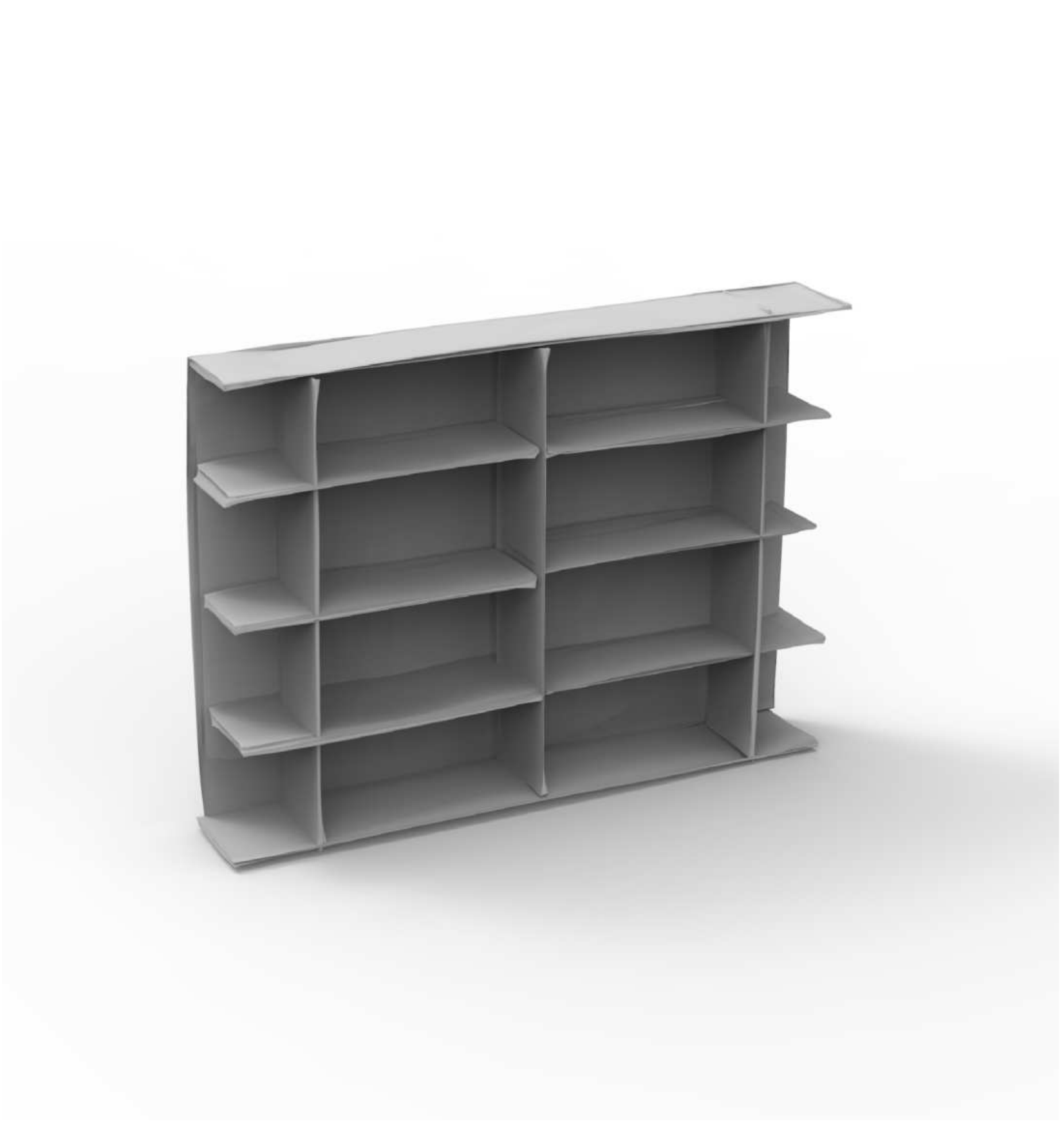}
    \includegraphics[width=0.23\linewidth]{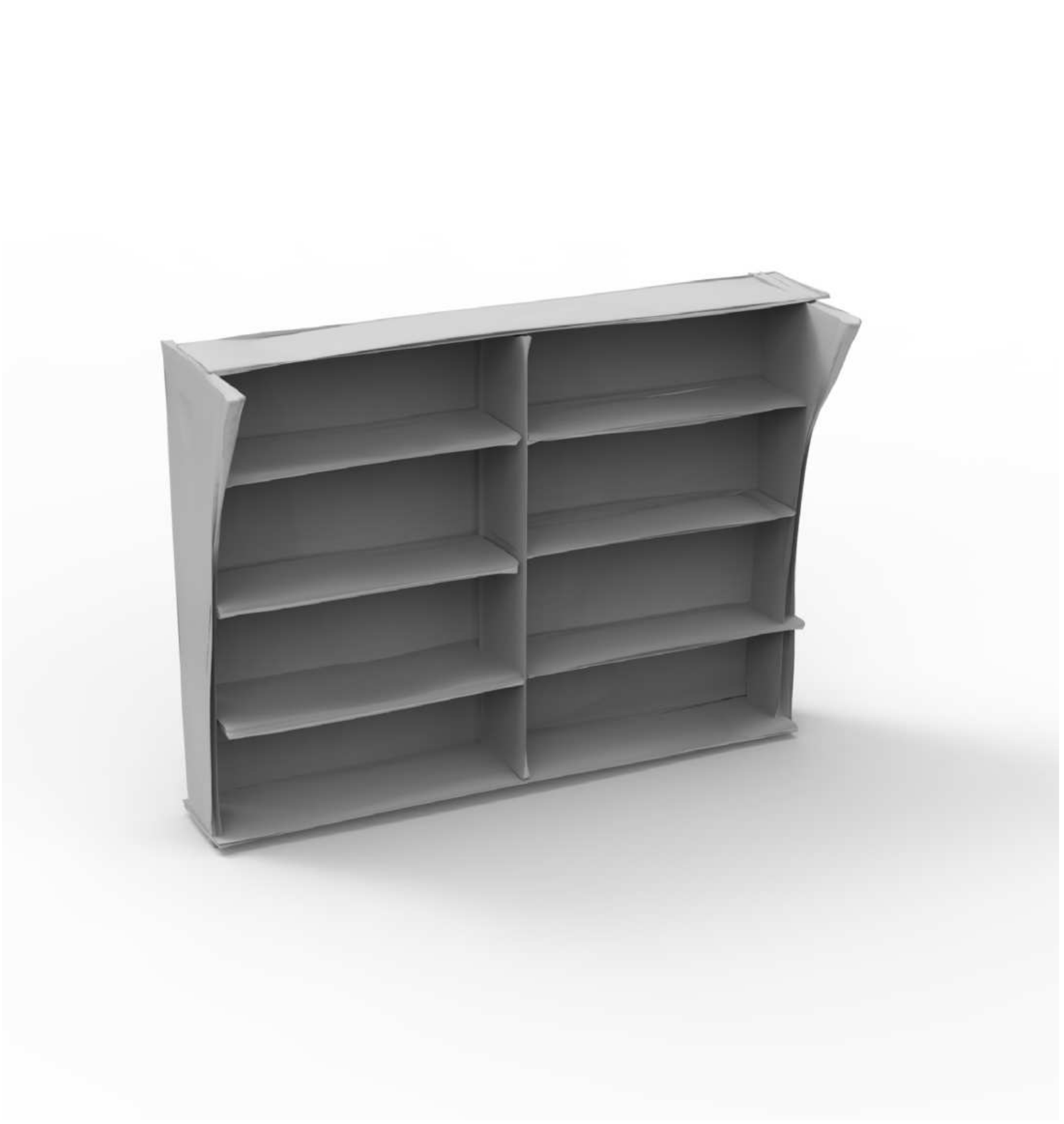}
    \includegraphics[width=0.23\linewidth]{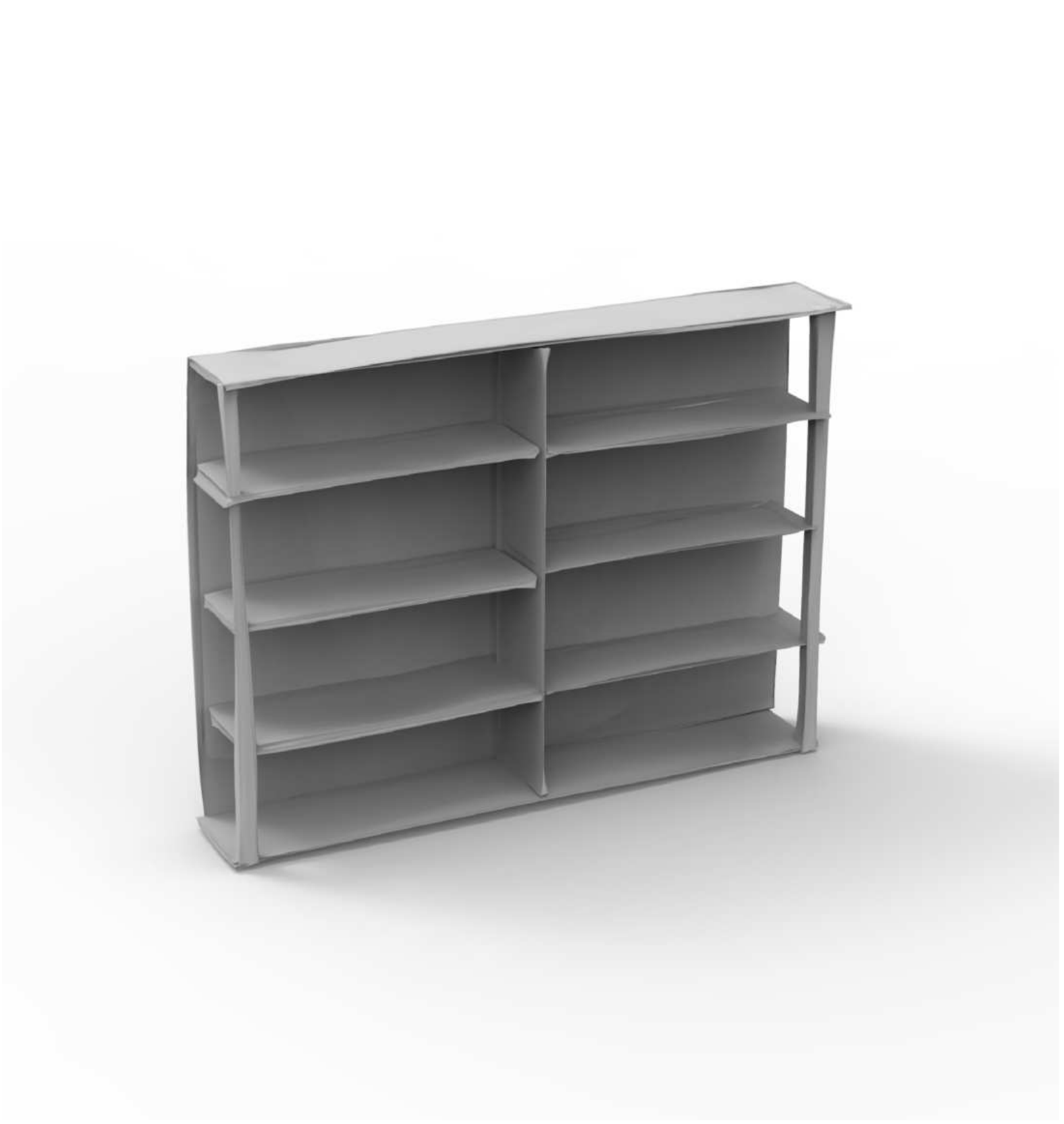}
\end{minipage}

\begin{minipage}{0.15\linewidth}
\centering
\includegraphics[width=0.99\linewidth]{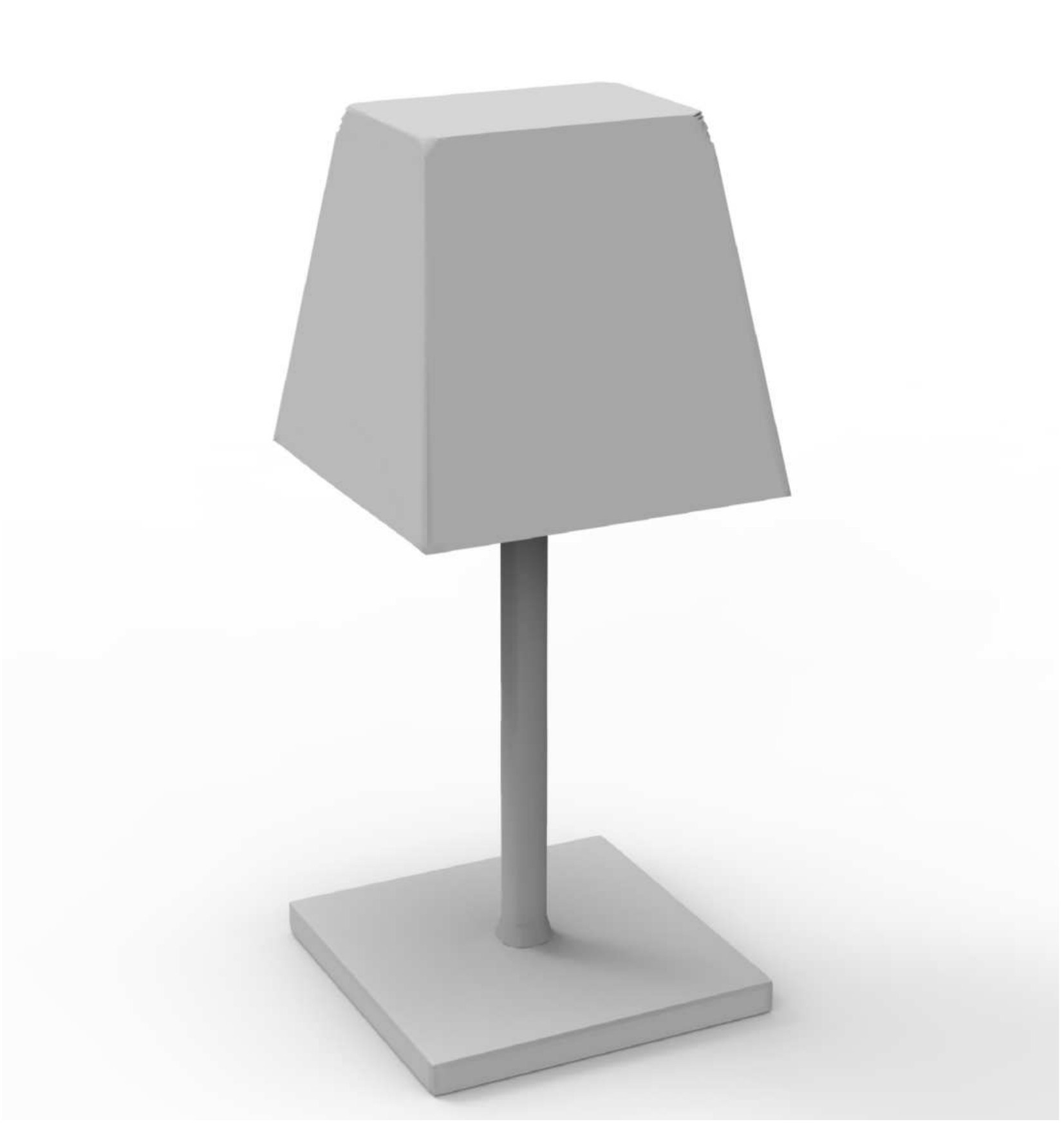}
\end{minipage}
\begin{minipage}{0.44\linewidth}
\centering
    \includegraphics[width=0.23\linewidth]{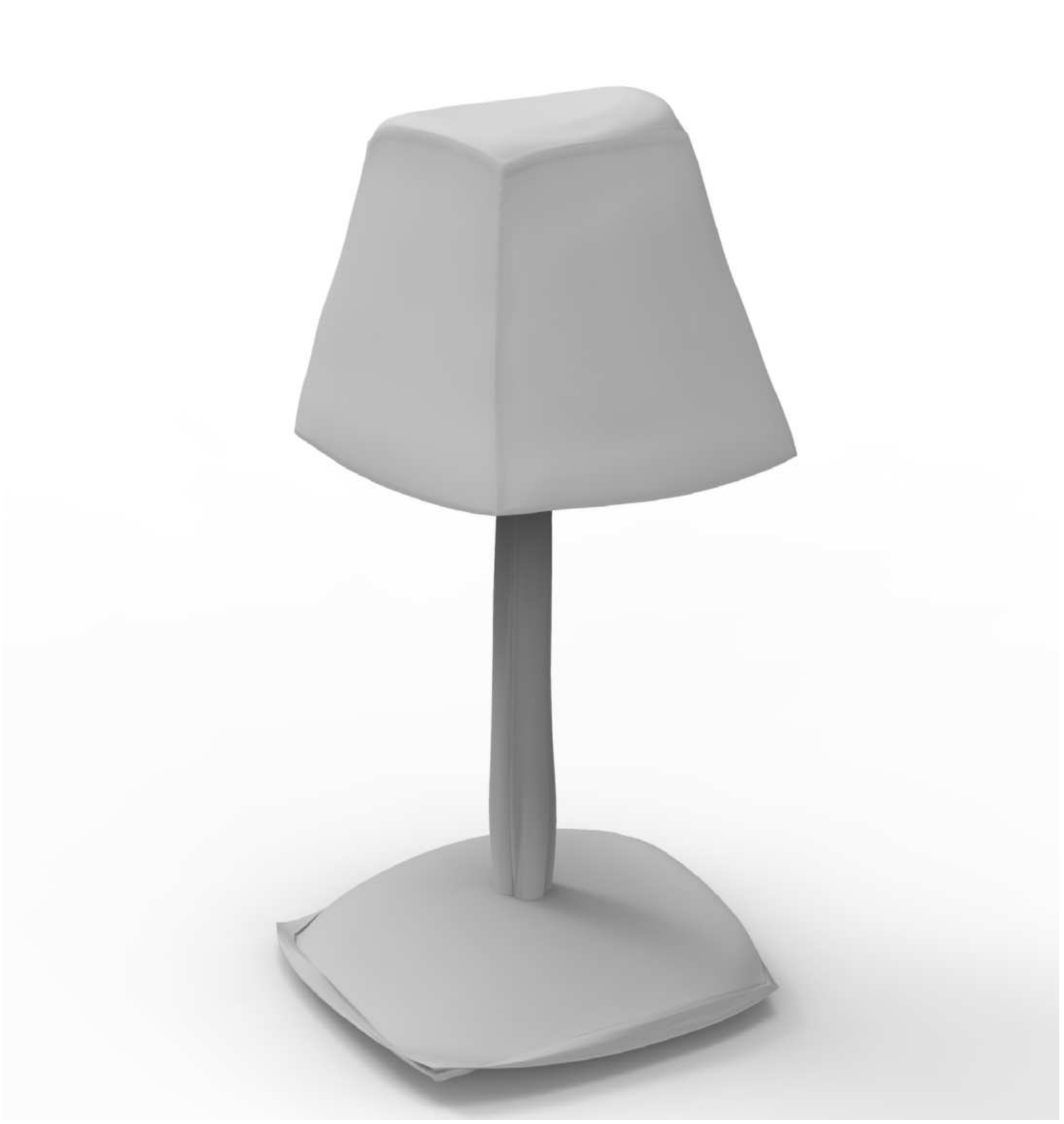}
    \includegraphics[width=0.23\linewidth]{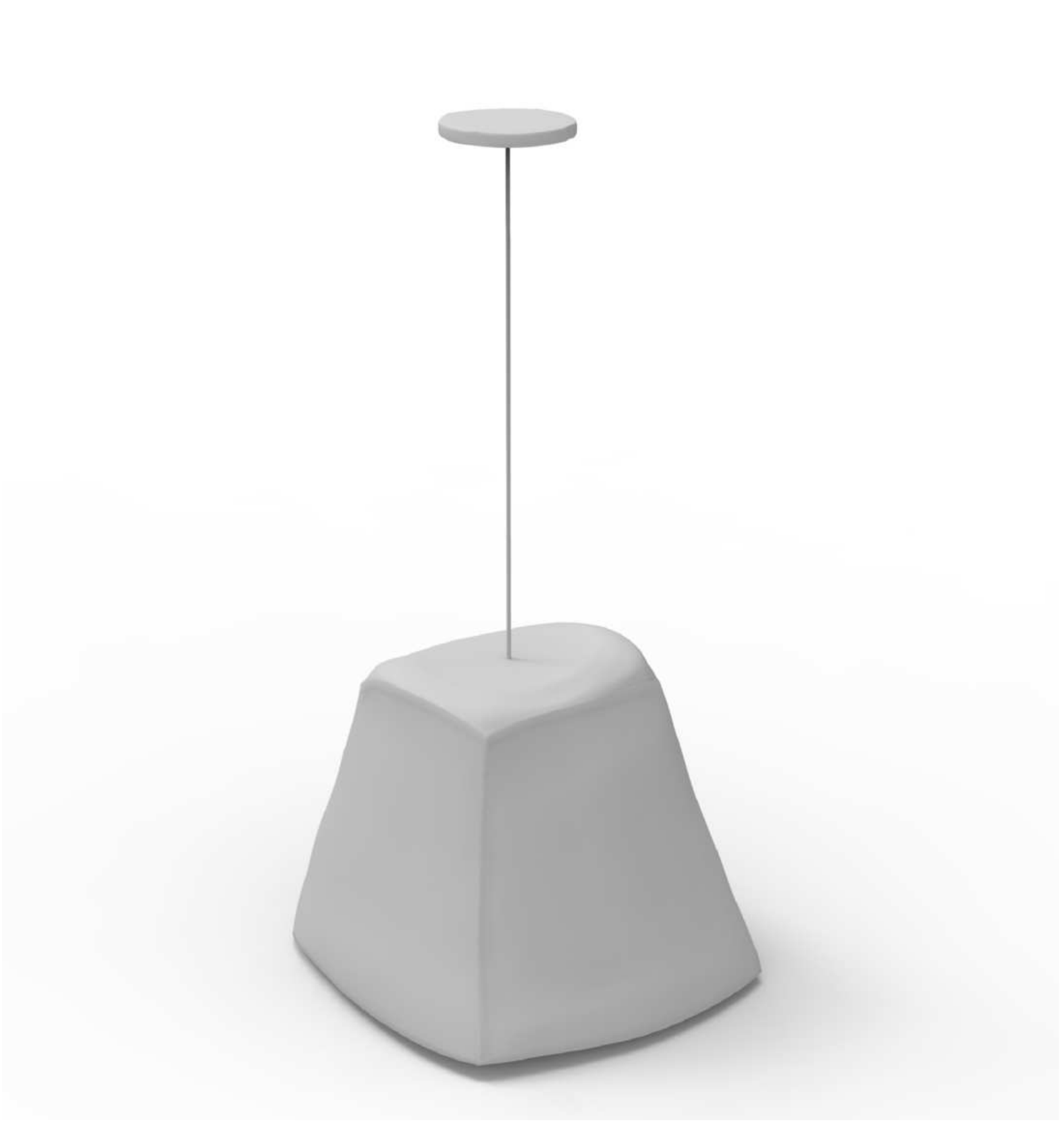}
    \includegraphics[width=0.23\linewidth]{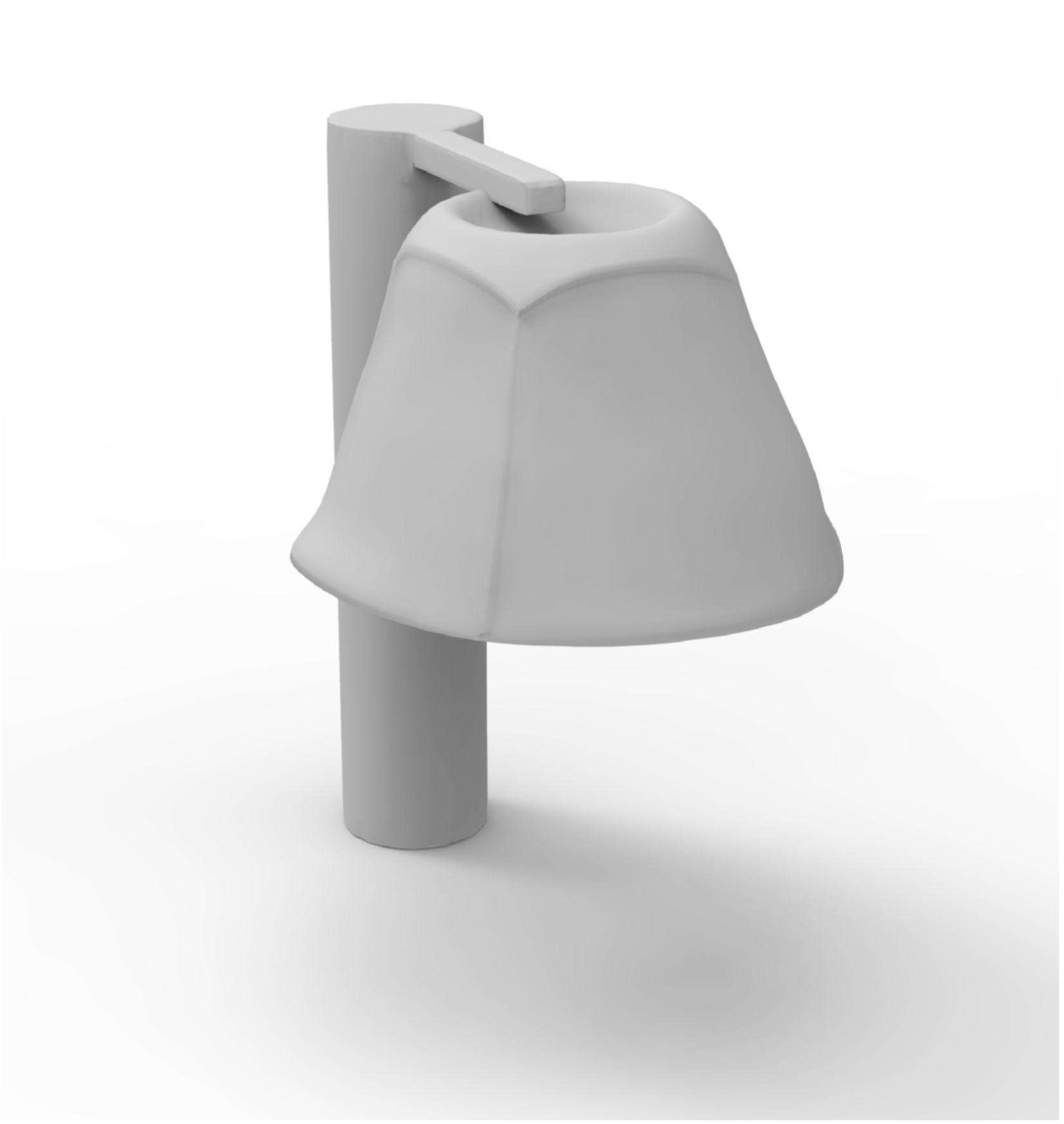}
    \includegraphics[width=0.23\linewidth]{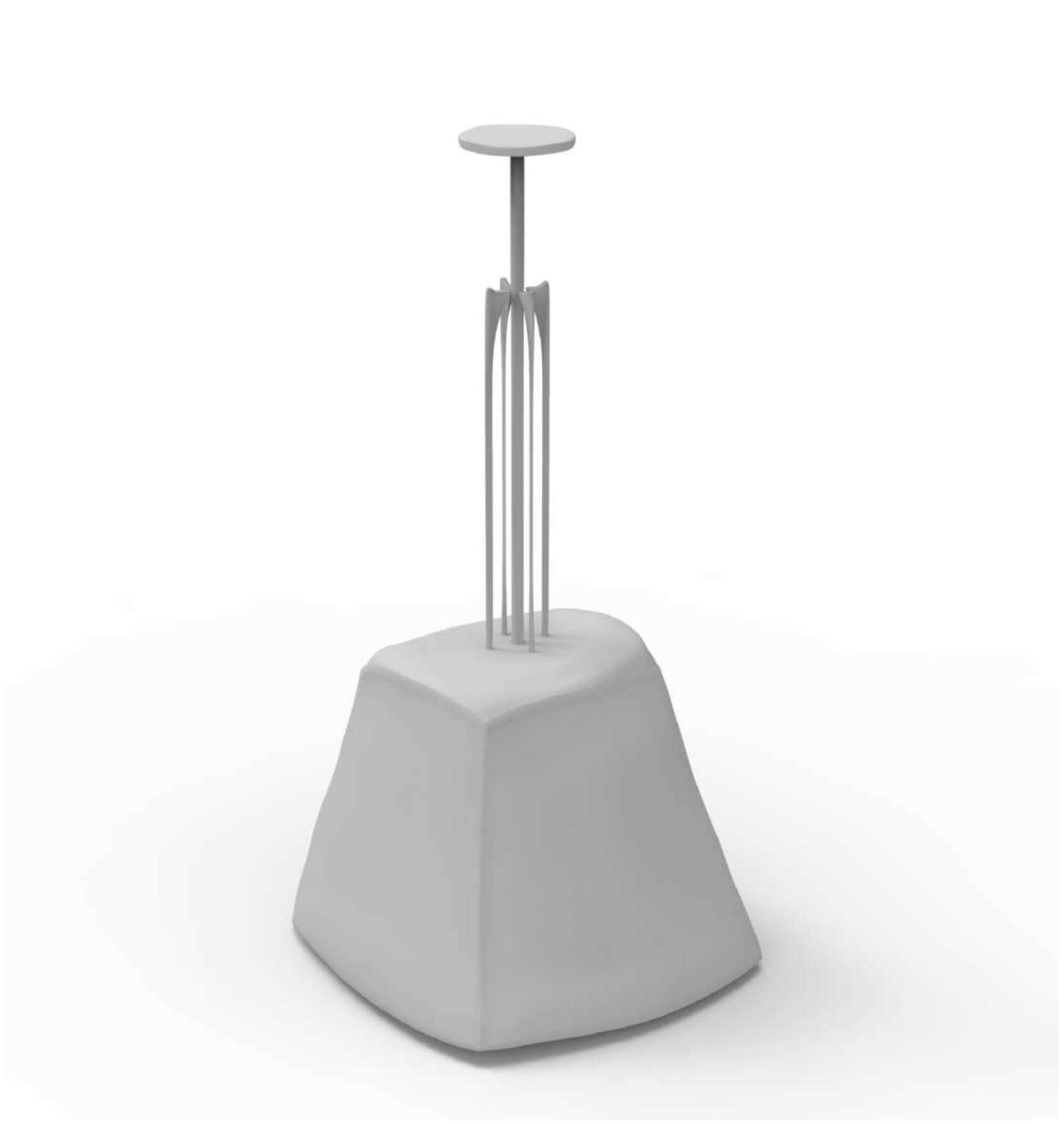}
    \\
    \includegraphics[width=0.23\linewidth]{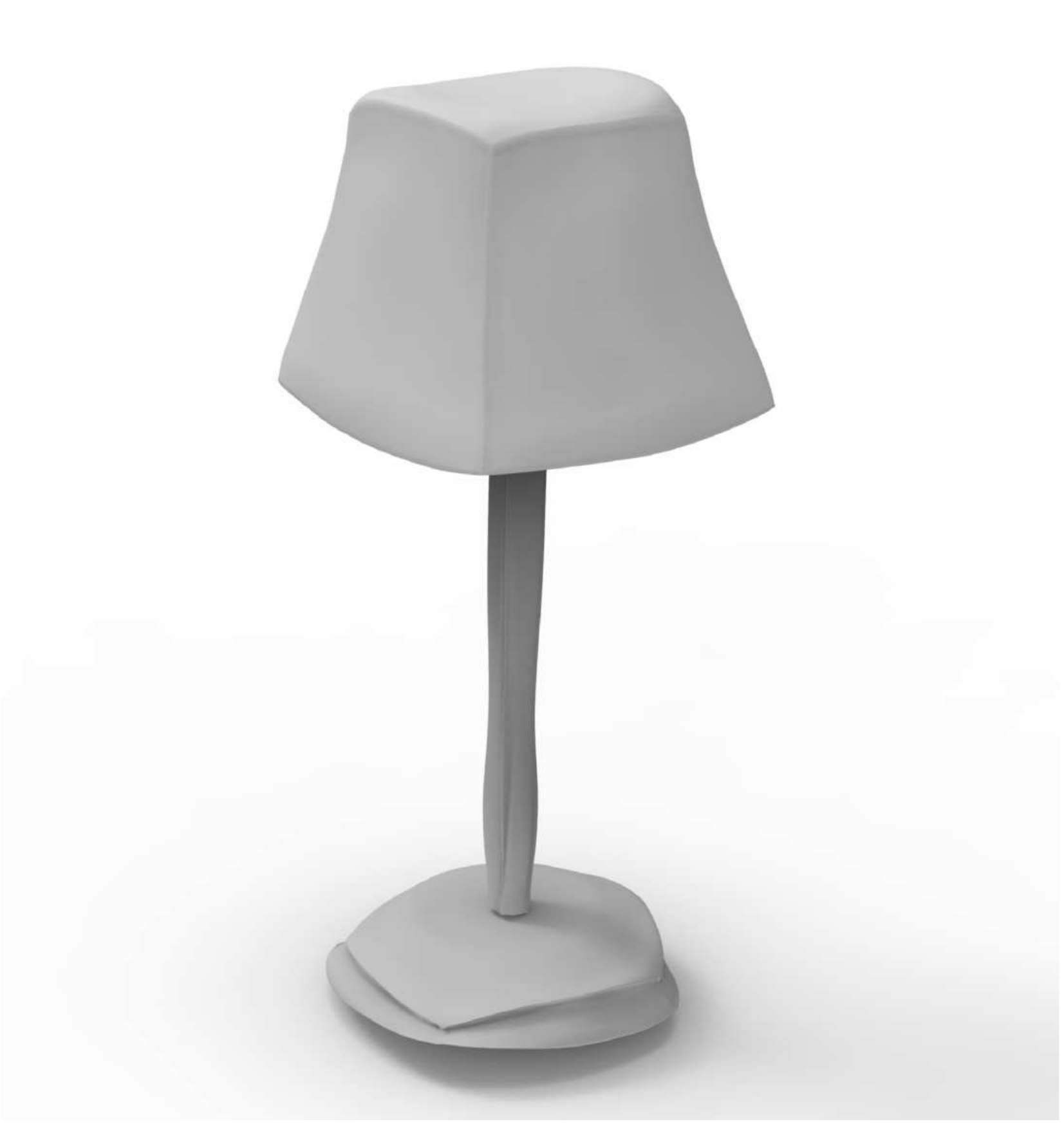}
    \includegraphics[width=0.23\linewidth]{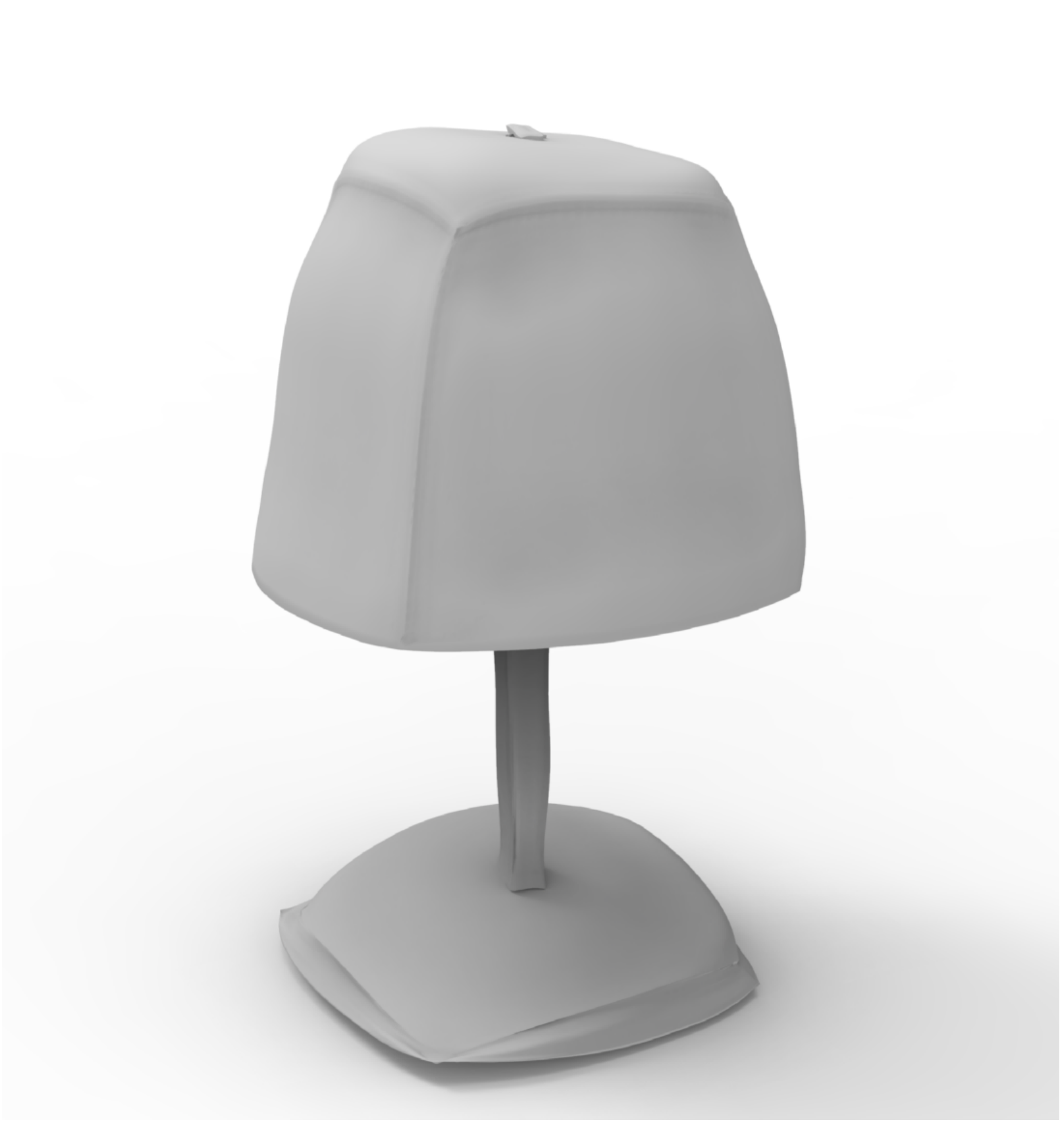}
    \includegraphics[width=0.23\linewidth]{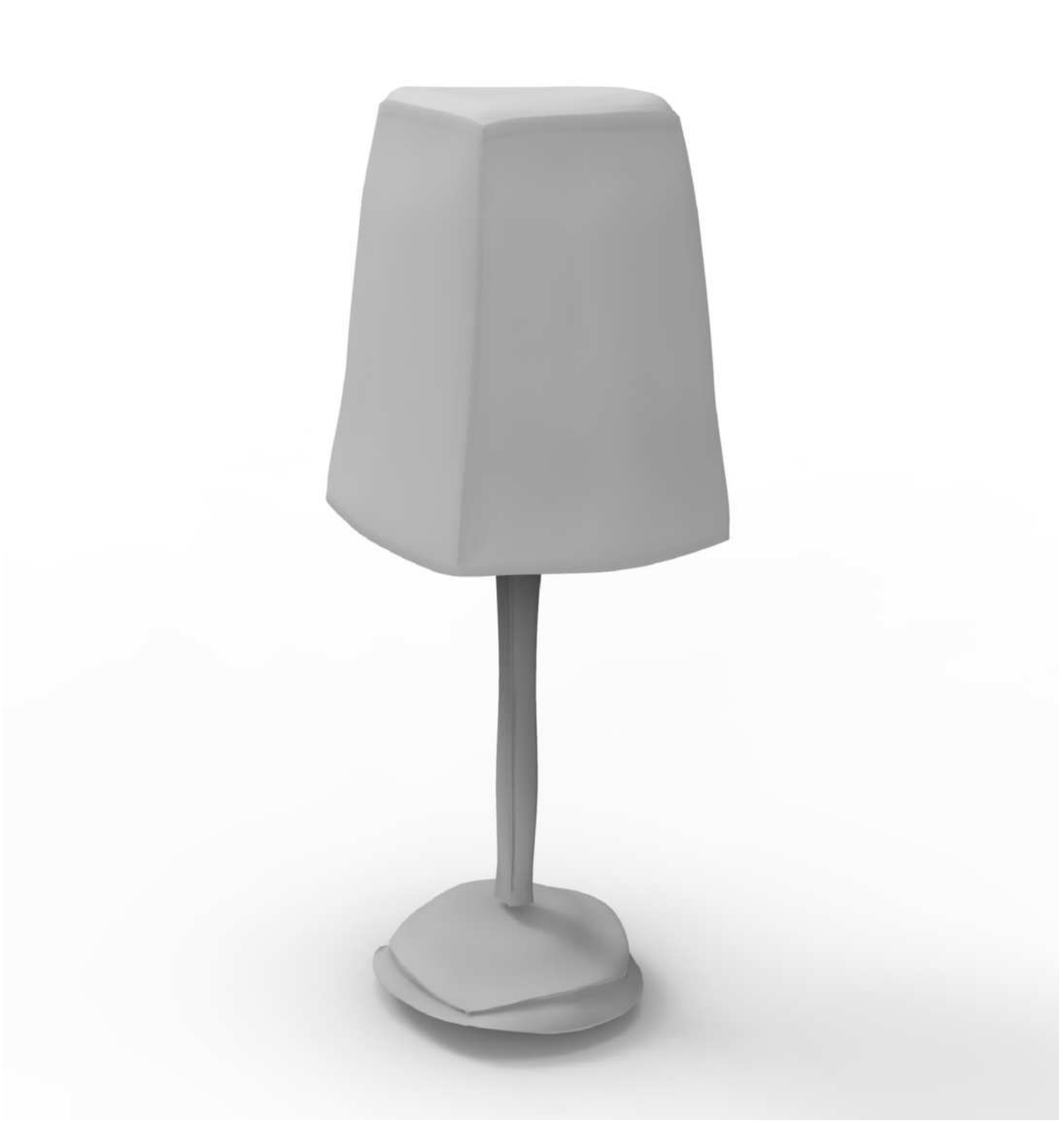}
    \includegraphics[width=0.23\linewidth]{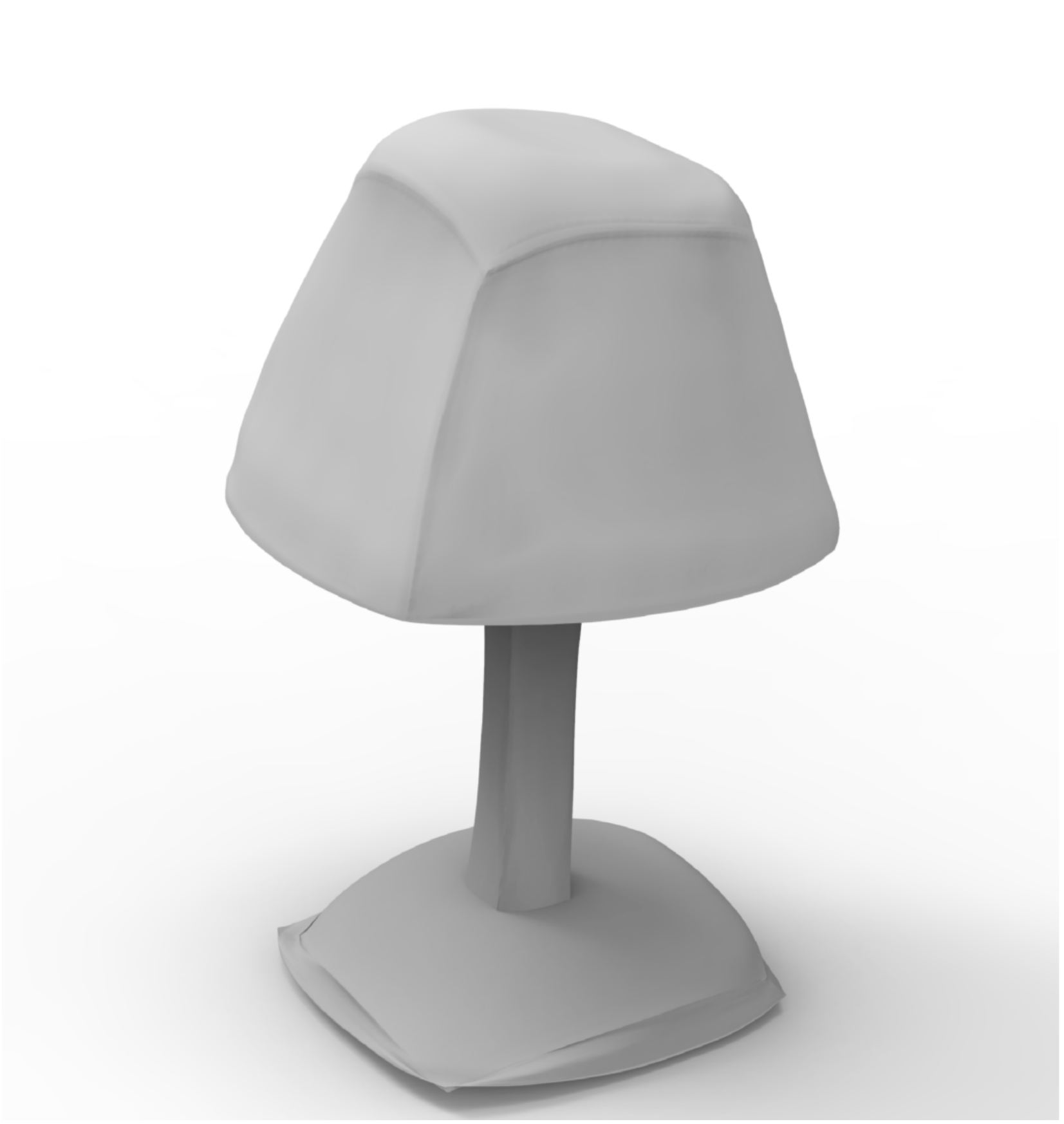}
\end{minipage}

\begin{minipage}{0.15\linewidth}
\centering
\includegraphics[width=0.99\linewidth]{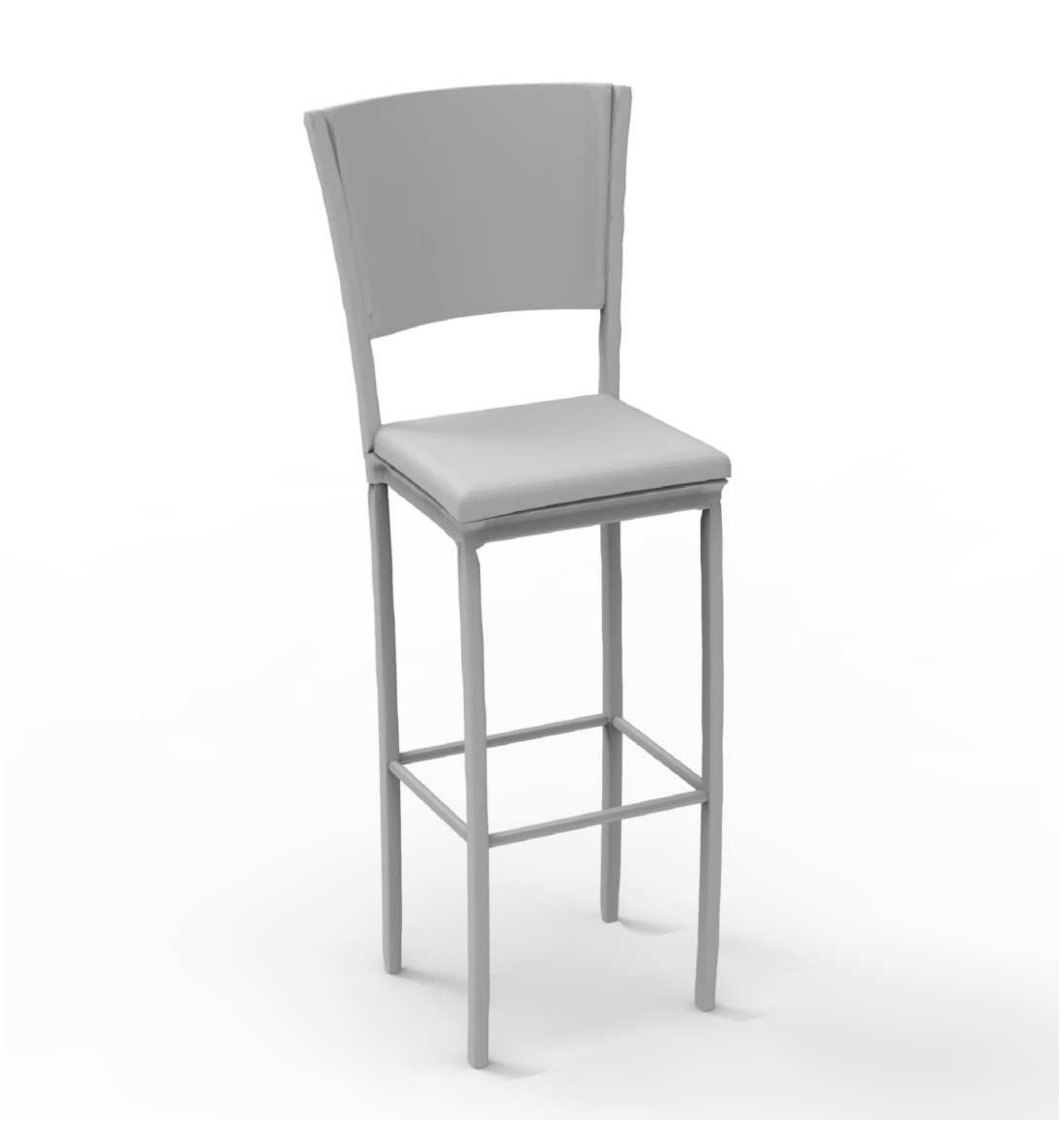}
\end{minipage}
\begin{minipage}{0.44\linewidth}
\centering
    \includegraphics[width=0.23\linewidth]{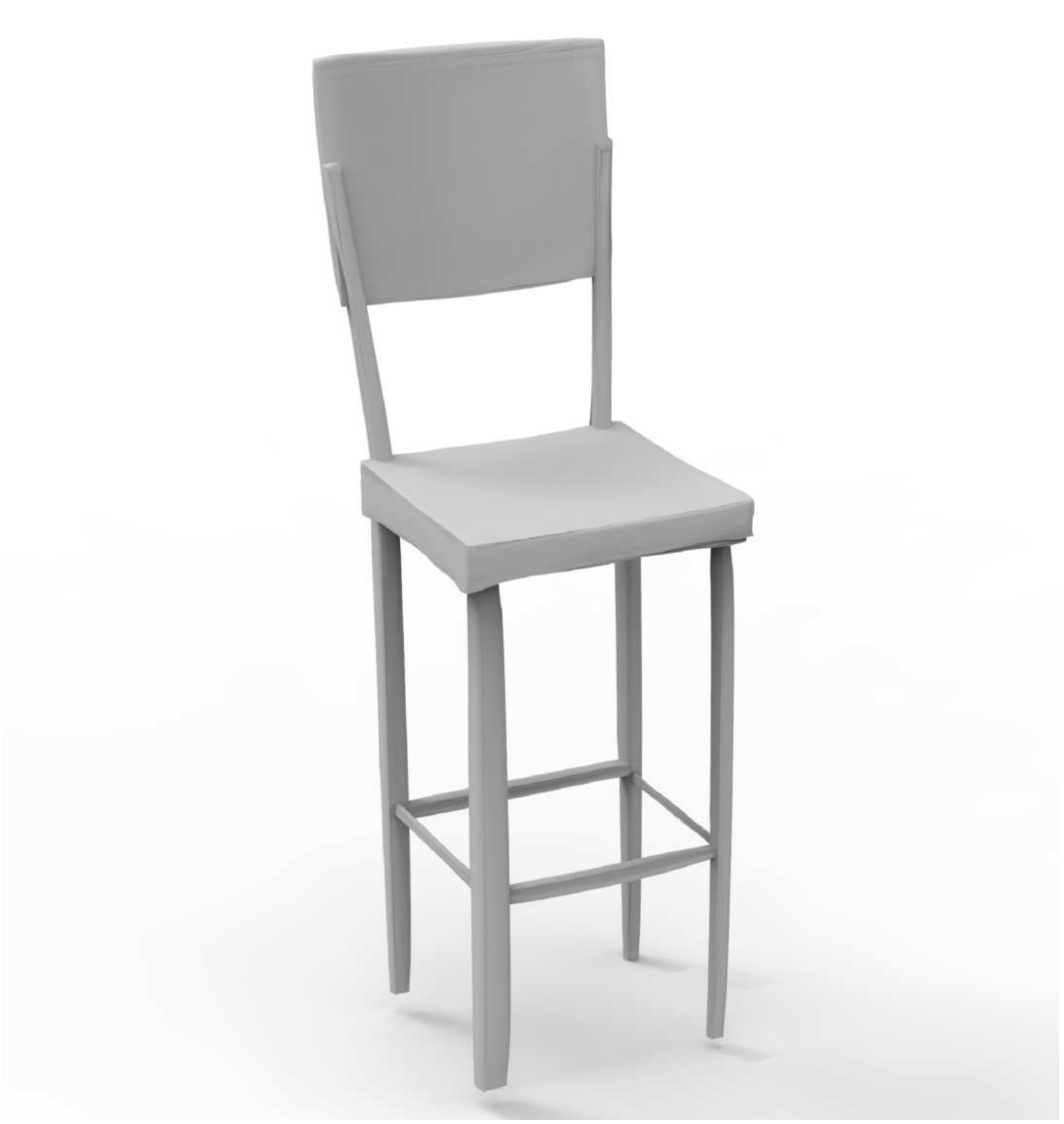}
    \includegraphics[width=0.23\linewidth]{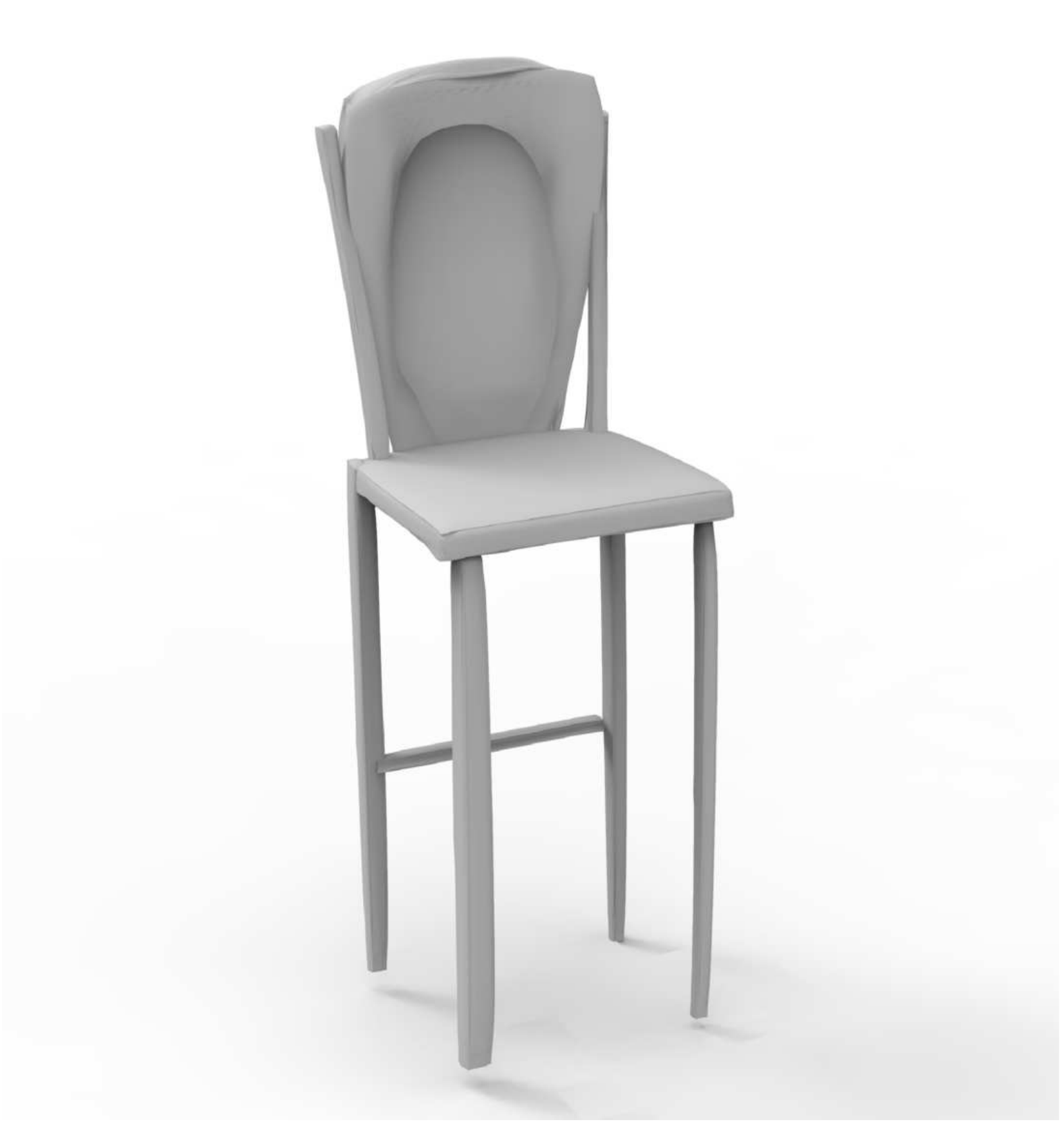}
    \includegraphics[width=0.23\linewidth]{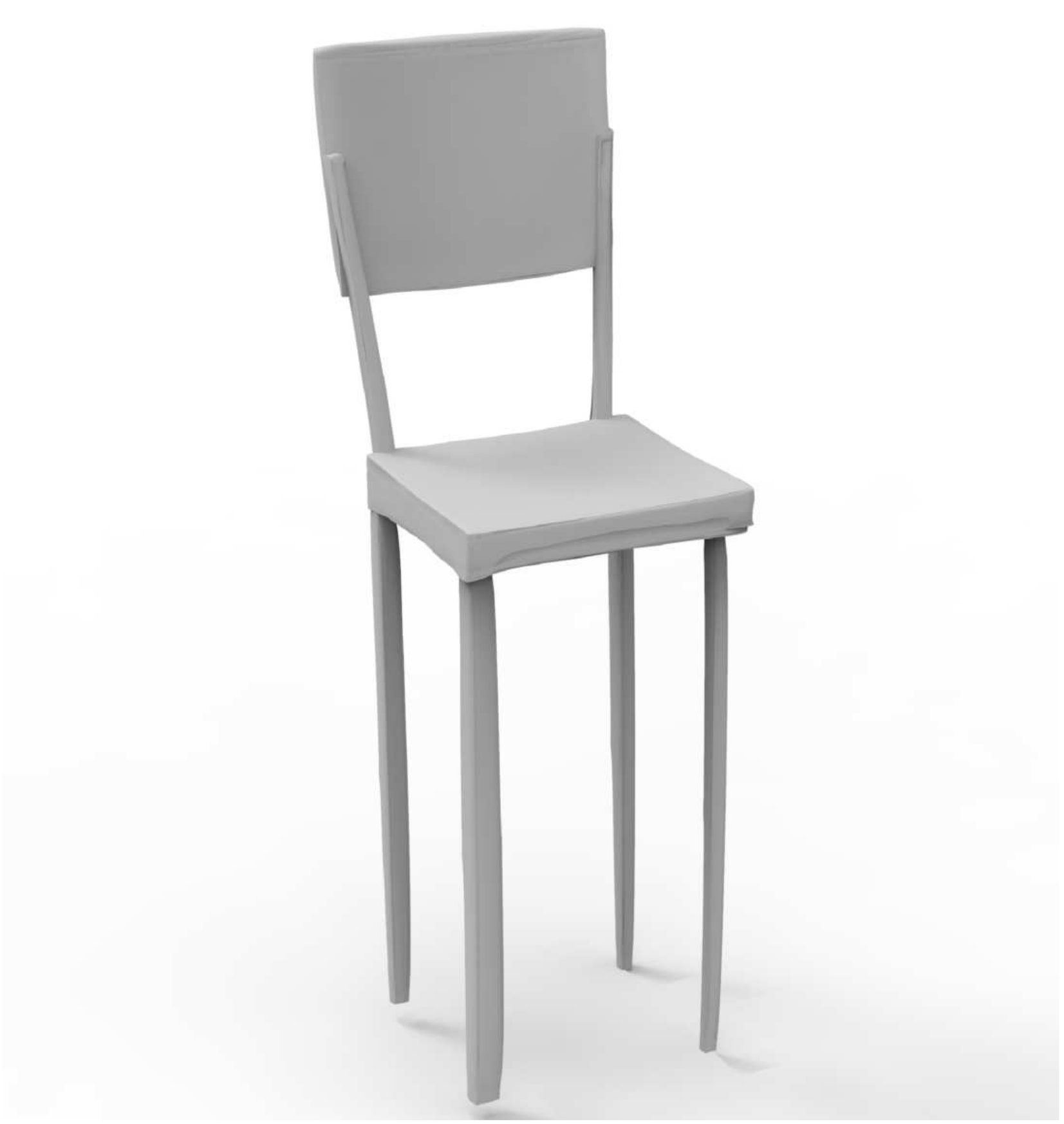}
    \includegraphics[width=0.23\linewidth]{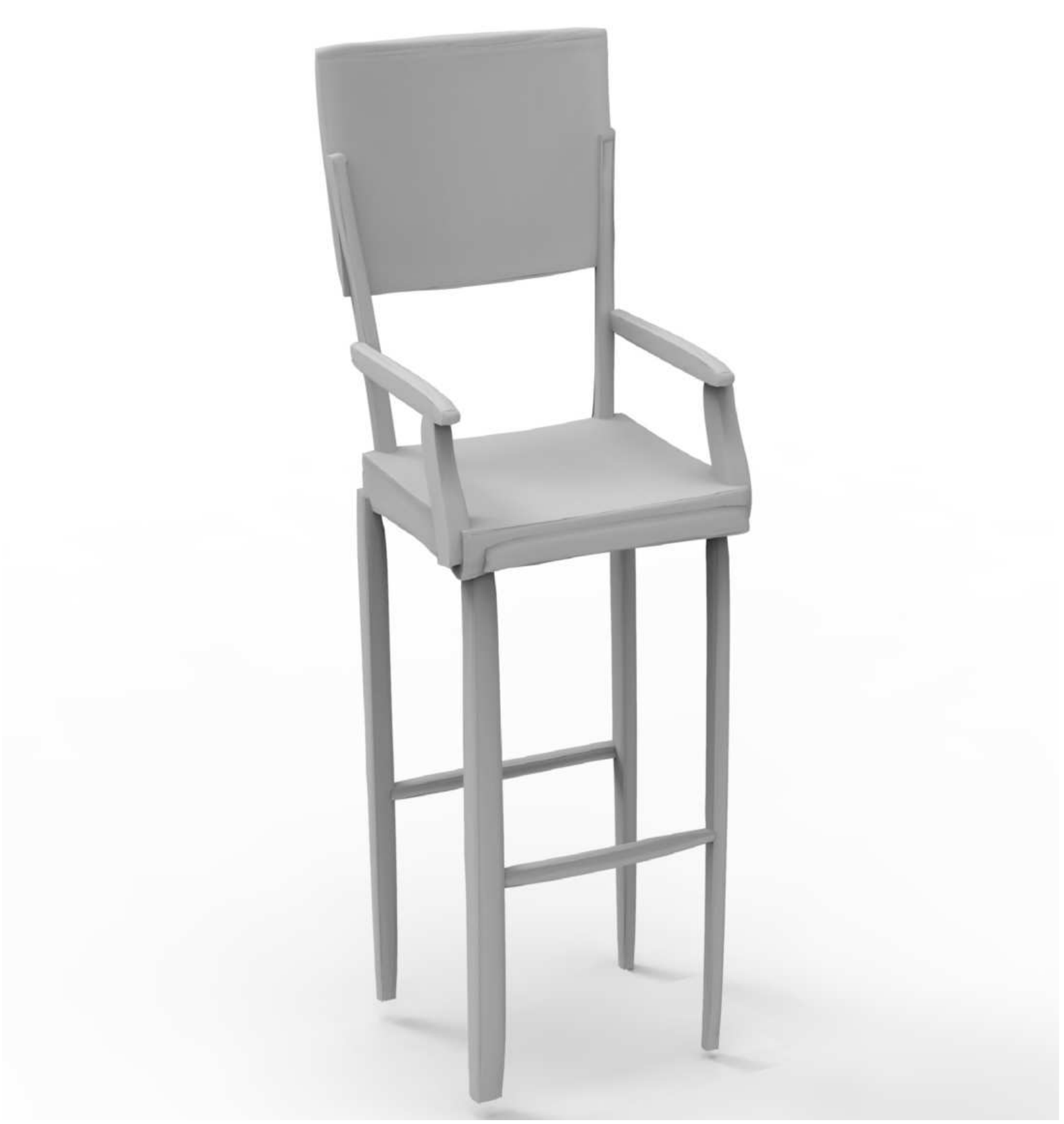}
    \\
    \includegraphics[width=0.23\linewidth]{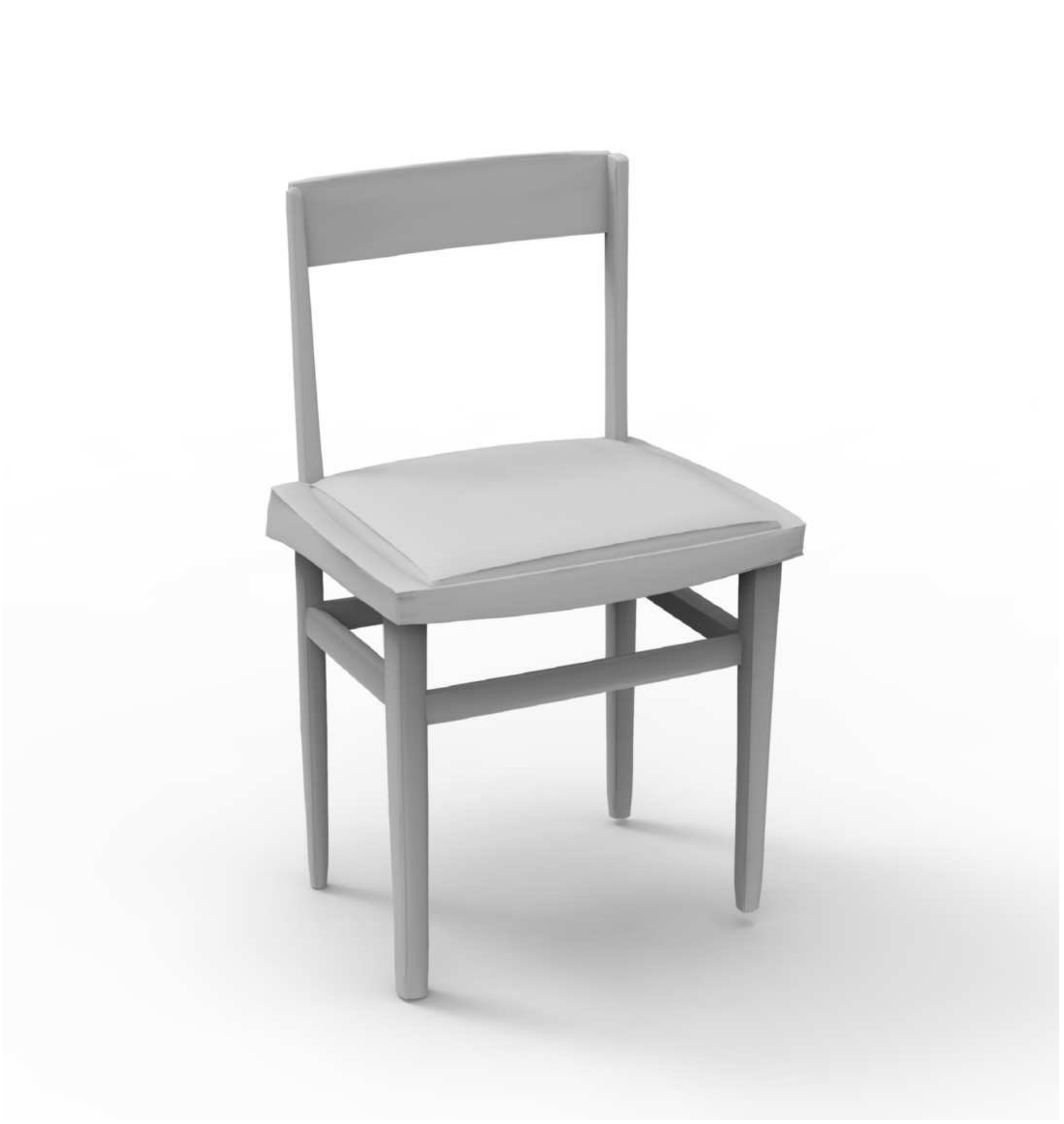}
    \includegraphics[width=0.23\linewidth]{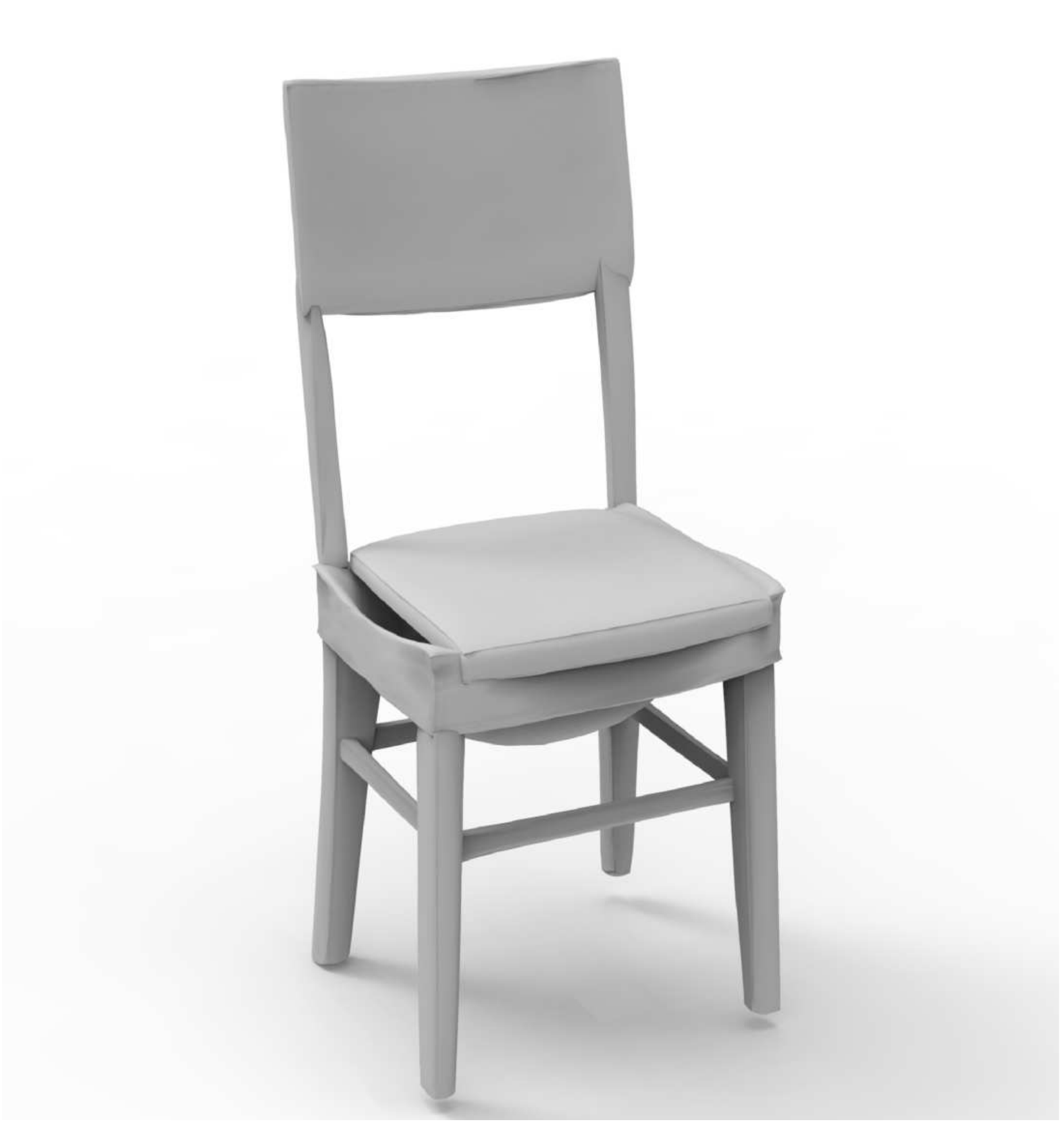}
    \includegraphics[width=0.23\linewidth]{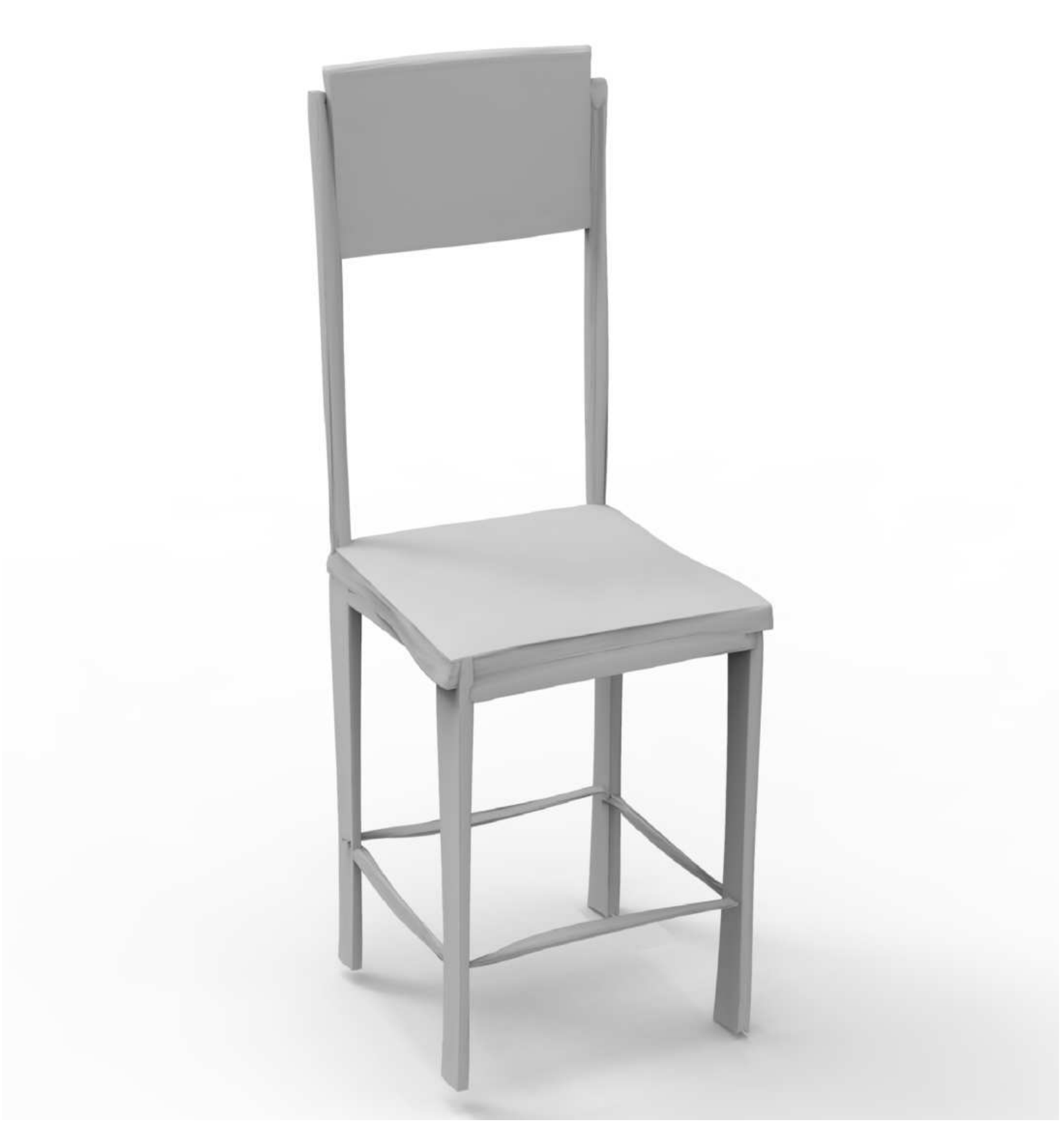}
    \includegraphics[width=0.23\linewidth]{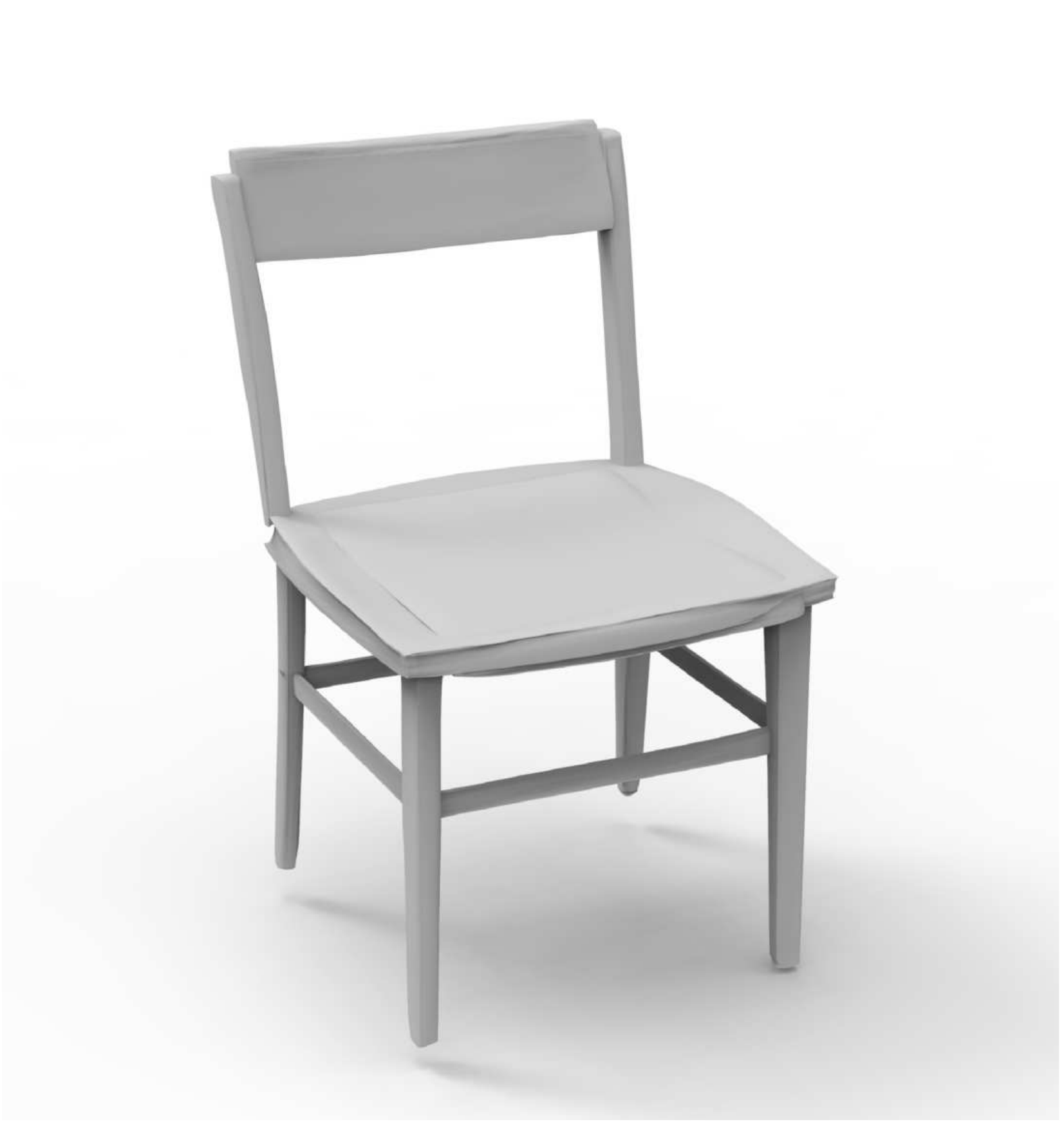}
\end{minipage}

\begin{minipage}{0.15\linewidth}
\centering
\includegraphics[width=0.99\linewidth]{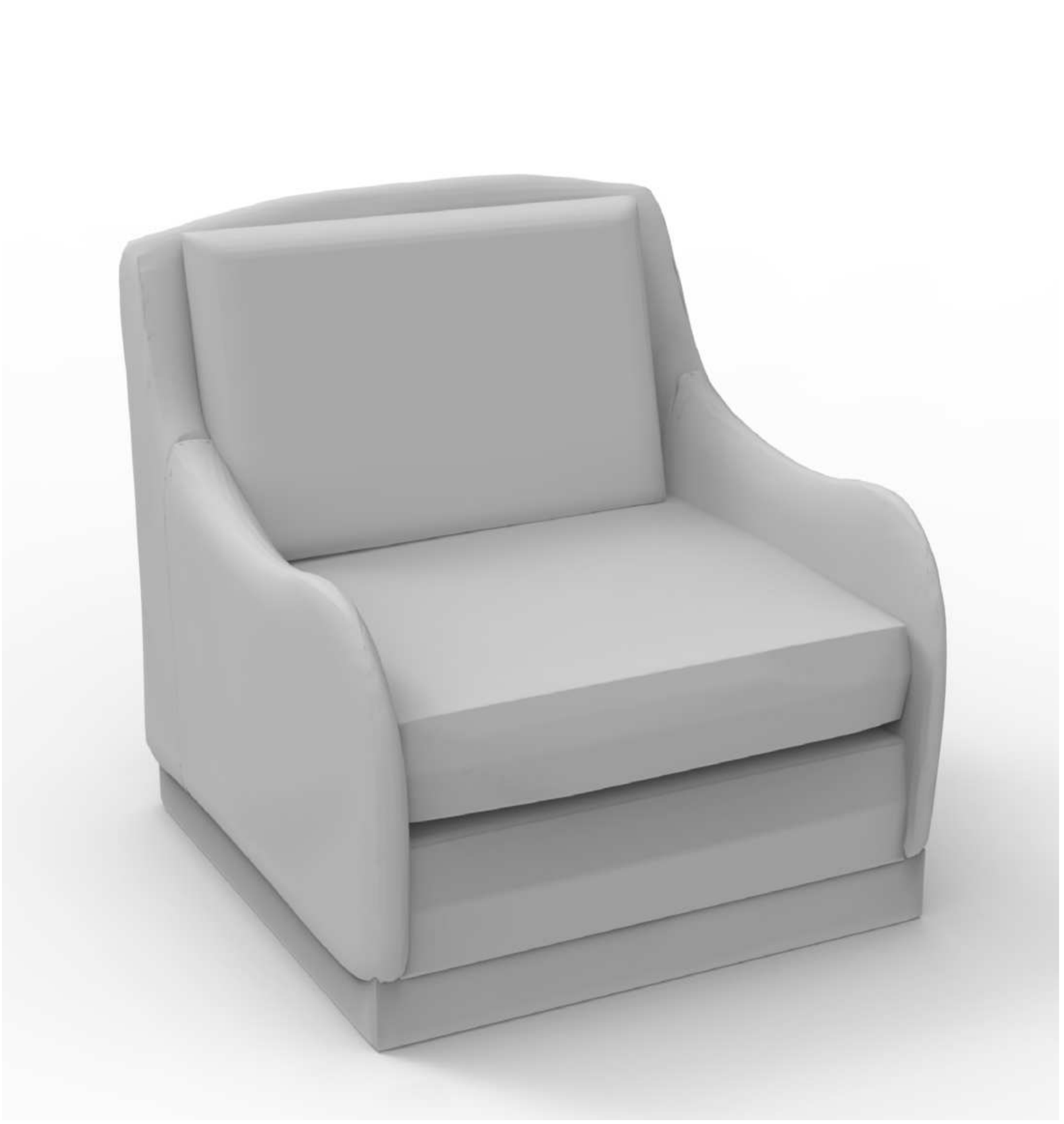}
\end{minipage}
\begin{minipage}{0.44\linewidth}
\centering
    \includegraphics[width=0.23\linewidth]{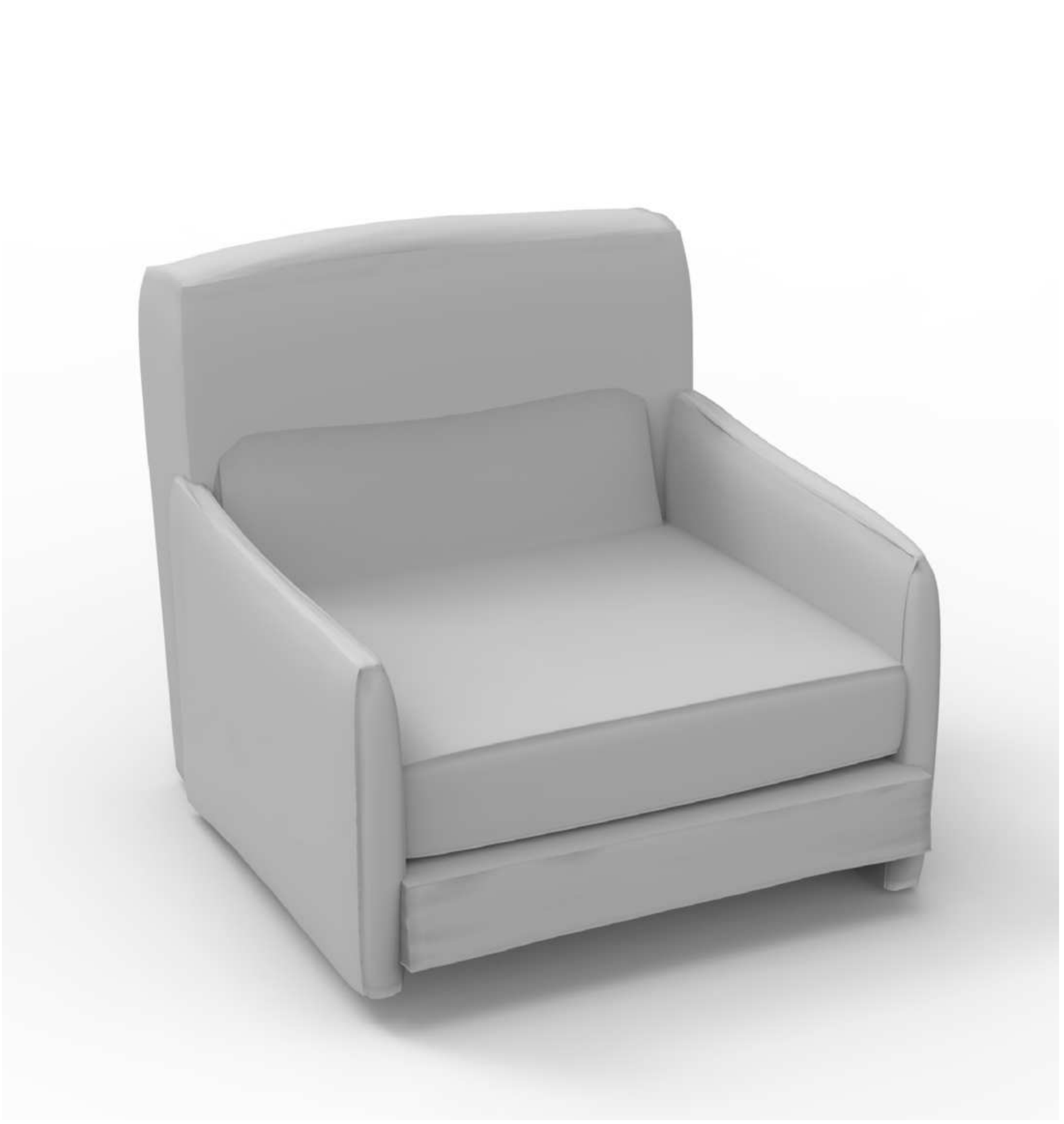}
    \includegraphics[width=0.23\linewidth]{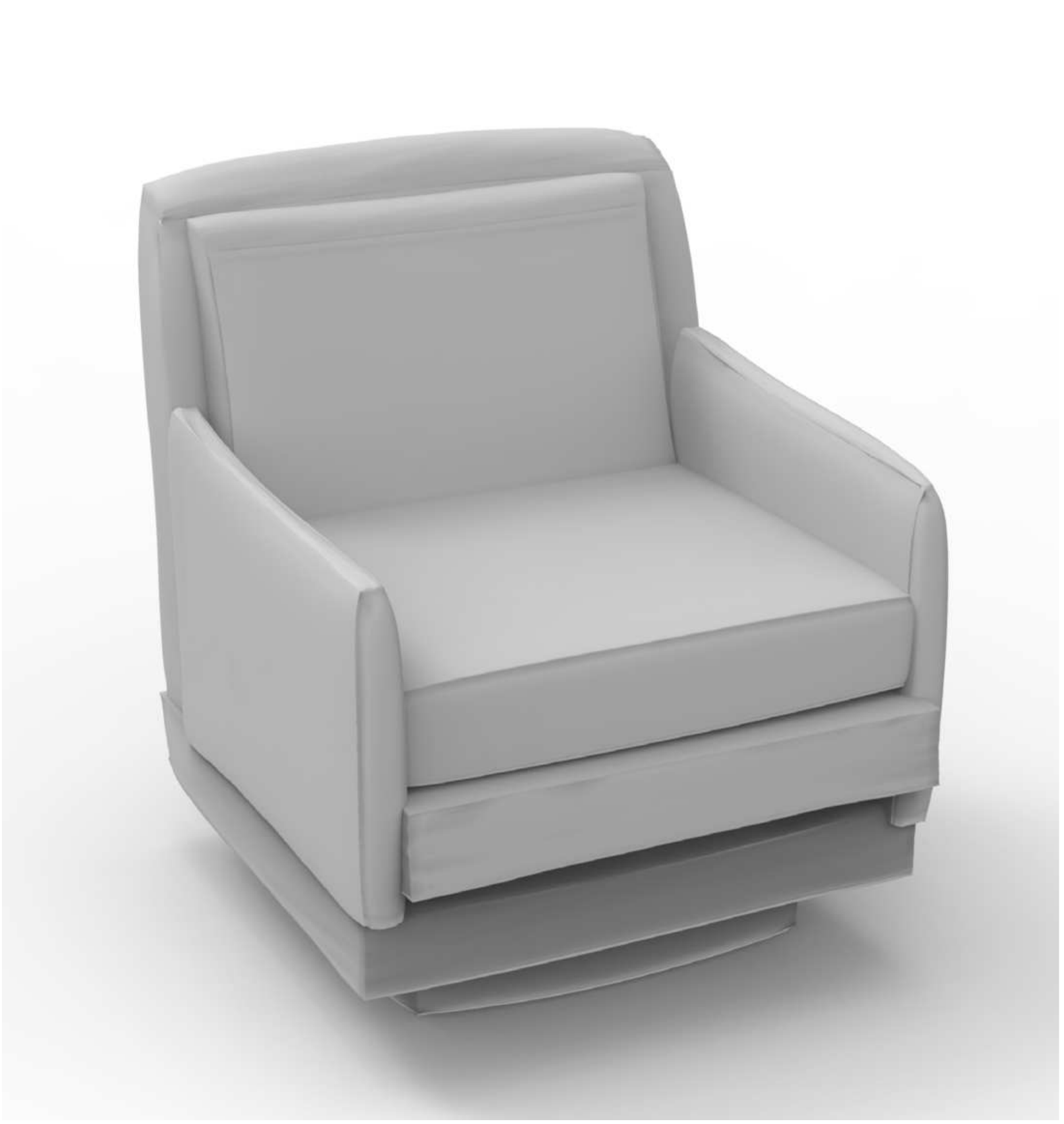}
    \includegraphics[width=0.23\linewidth]{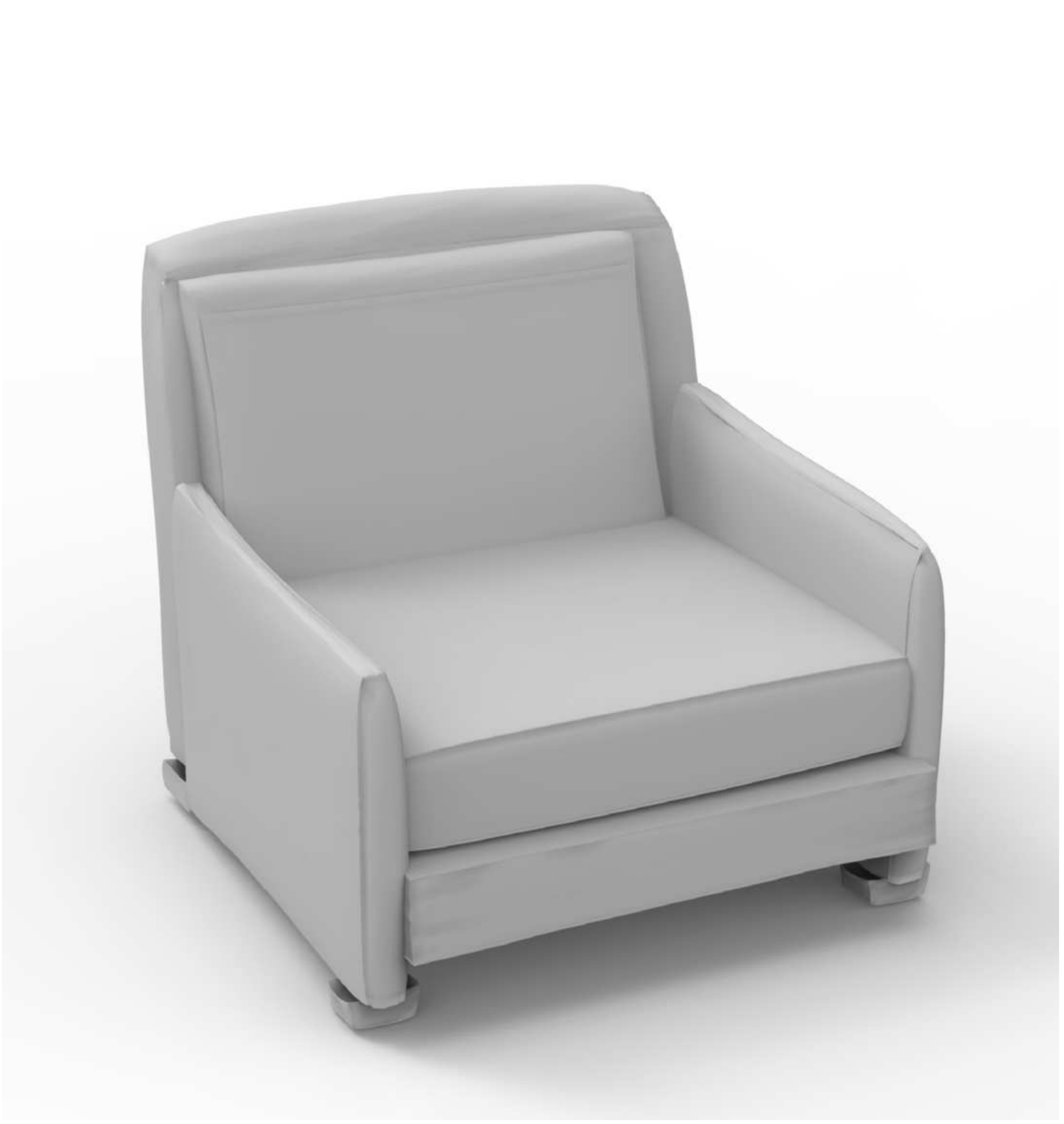}
    \includegraphics[width=0.23\linewidth]{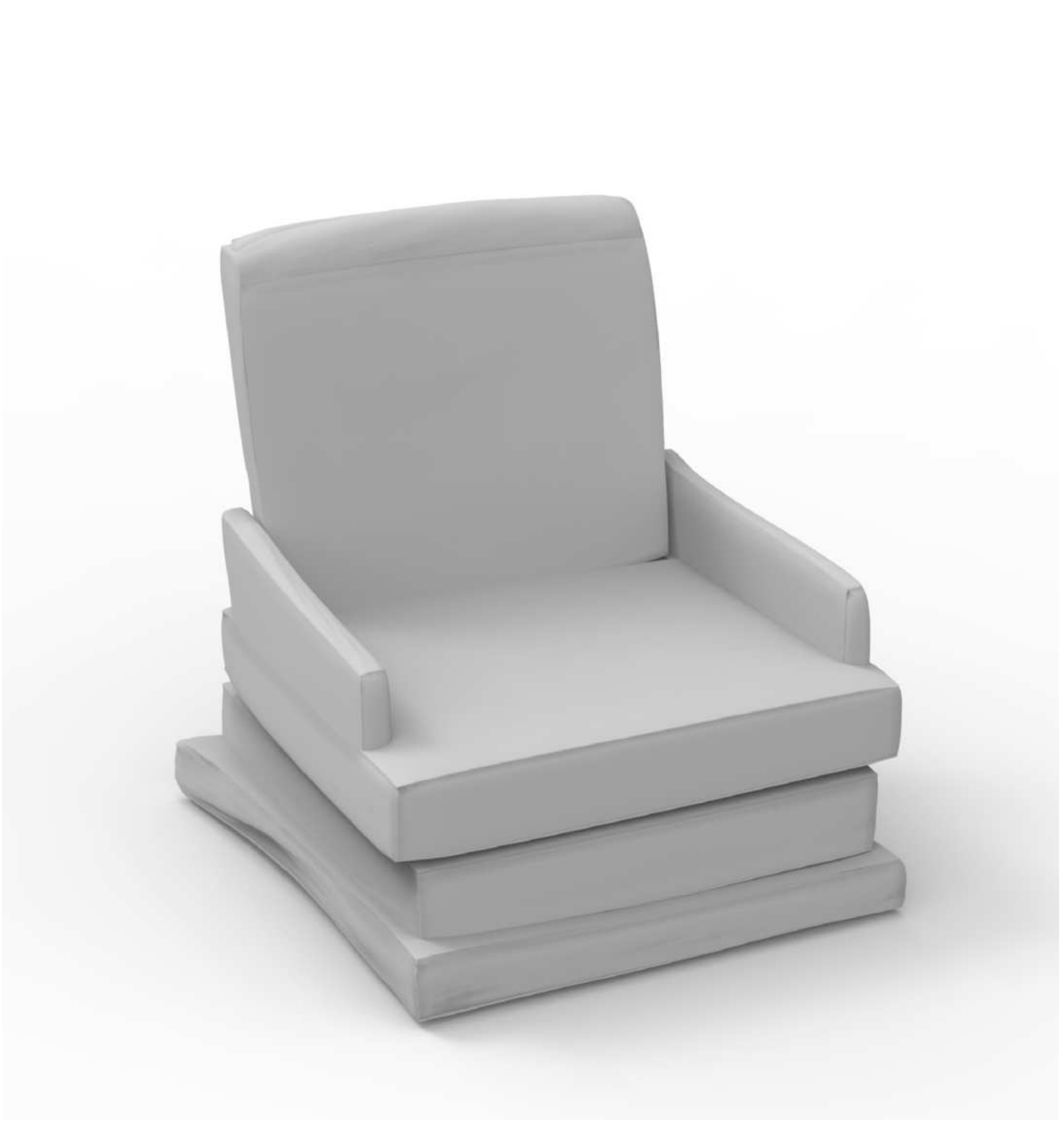}
    \\
    \includegraphics[width=0.23\linewidth]{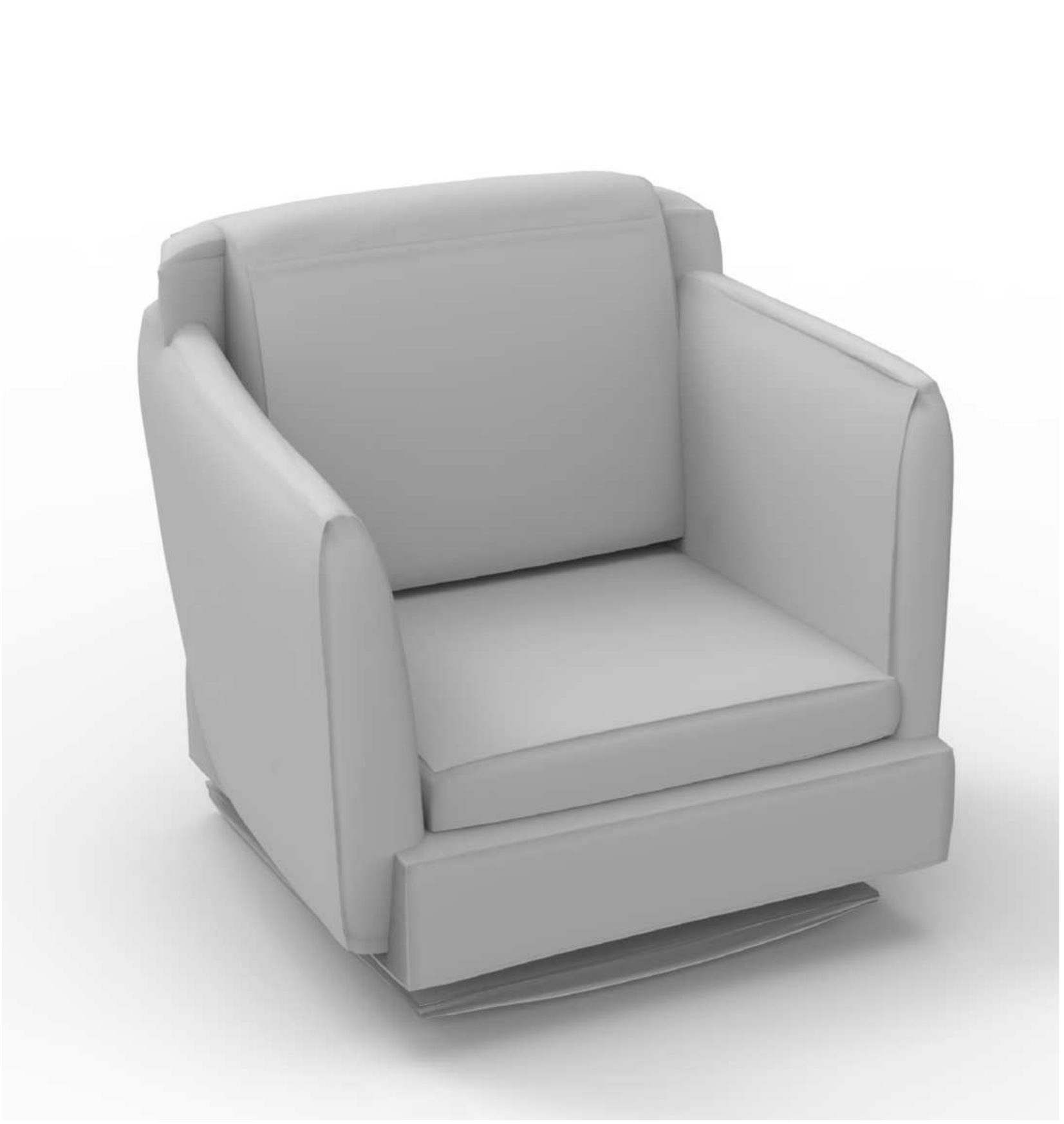}
    \includegraphics[width=0.23\linewidth]{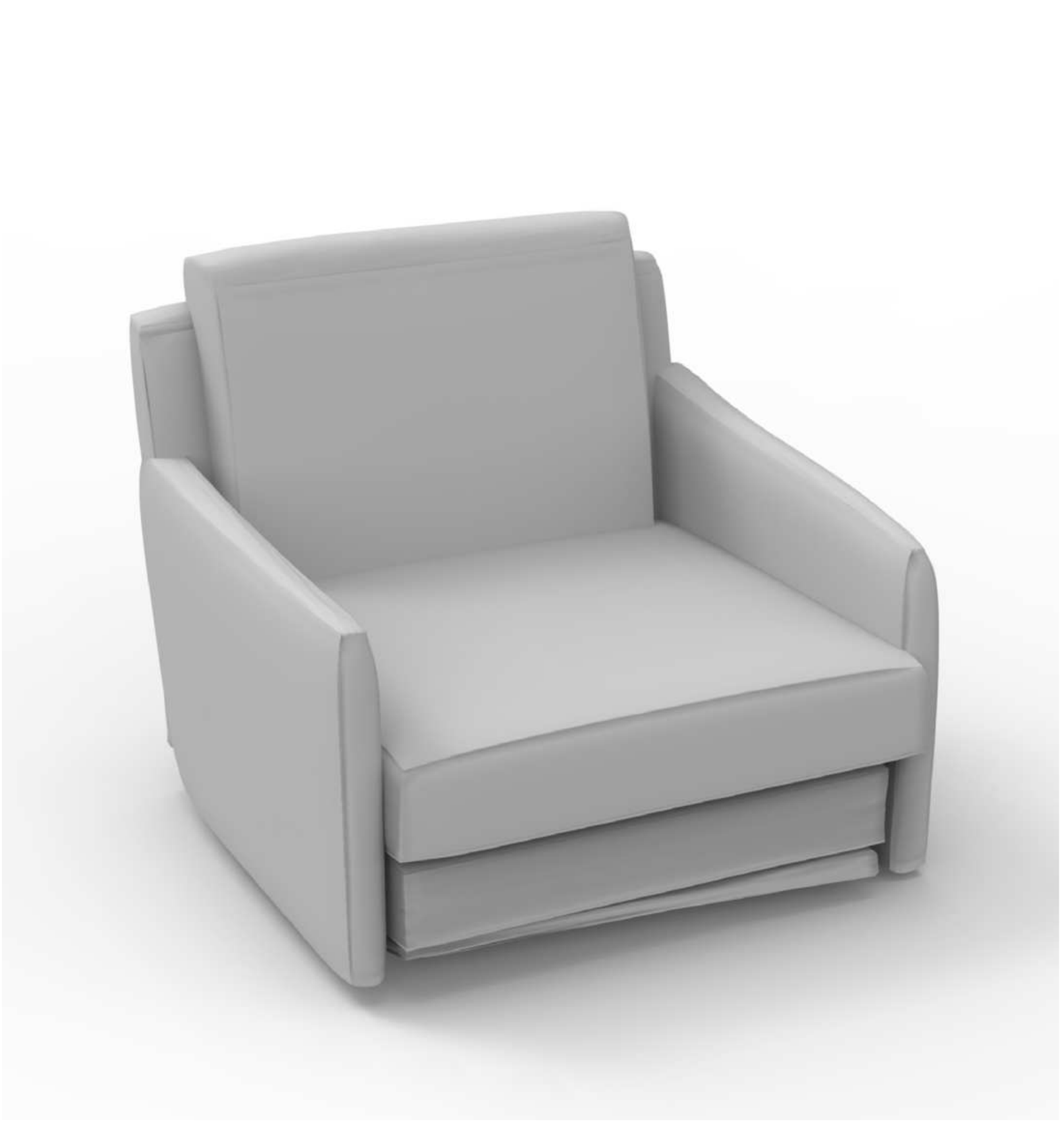}
    \includegraphics[width=0.23\linewidth]{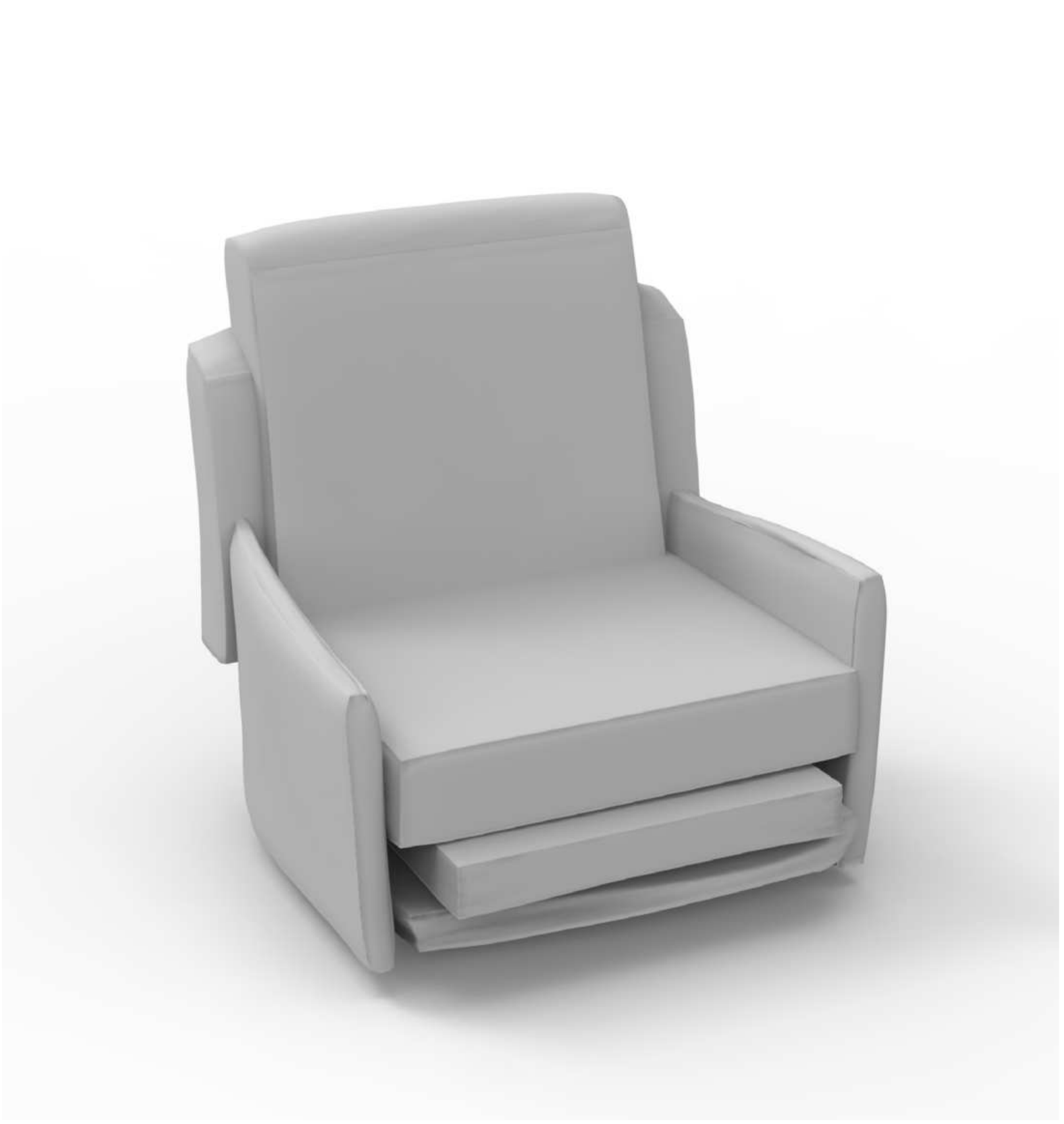}
    \includegraphics[width=0.23\linewidth]{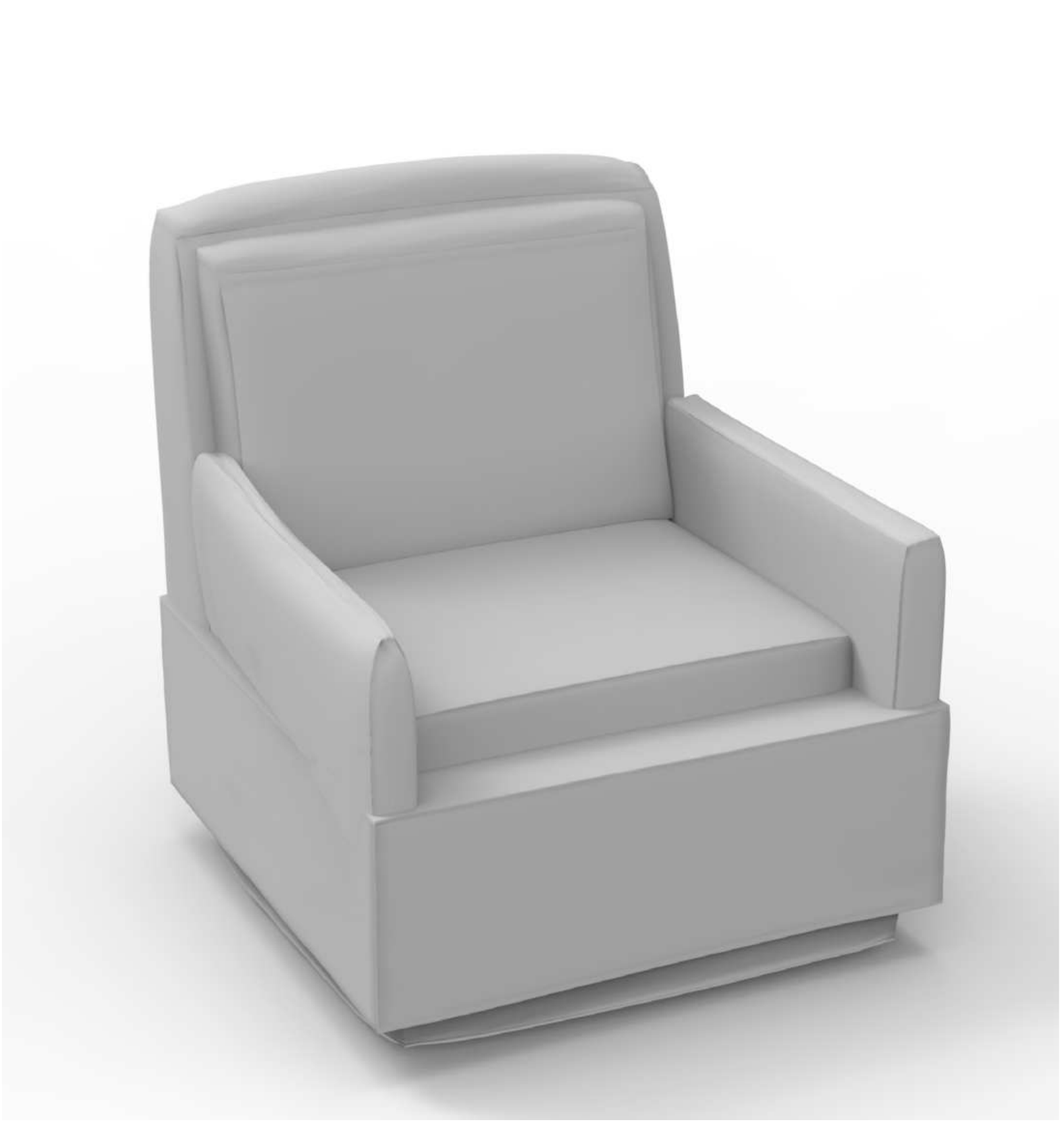}
\end{minipage}

\subfigure[Given Shape]{
\begin{minipage}{0.15\linewidth}
\centering
\includegraphics[width=0.99\linewidth]{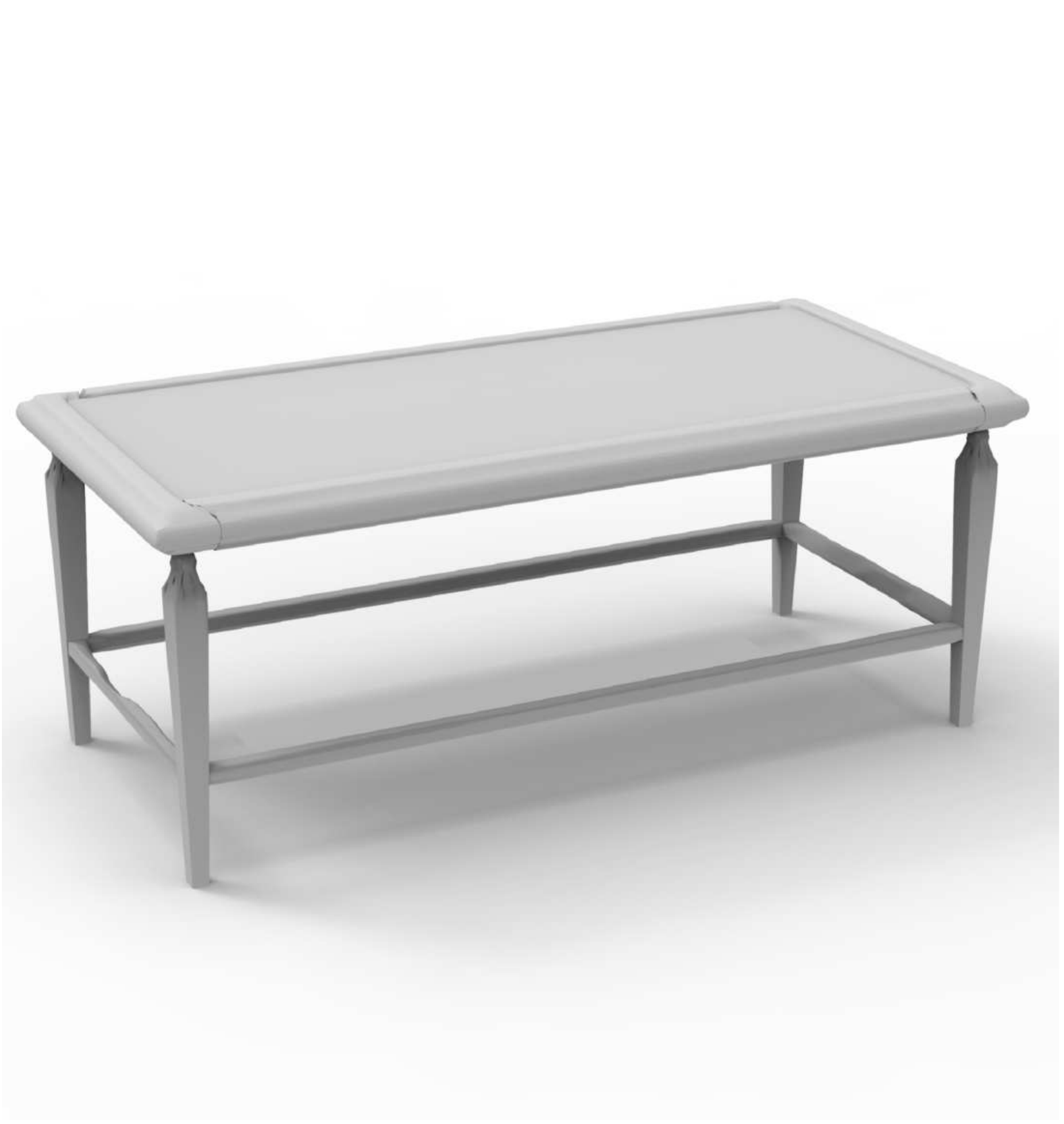}
\end{minipage}
}
\subfigure[Random generation]{
\begin{minipage}{0.44\linewidth}
\centering
    \includegraphics[width=0.23\linewidth]{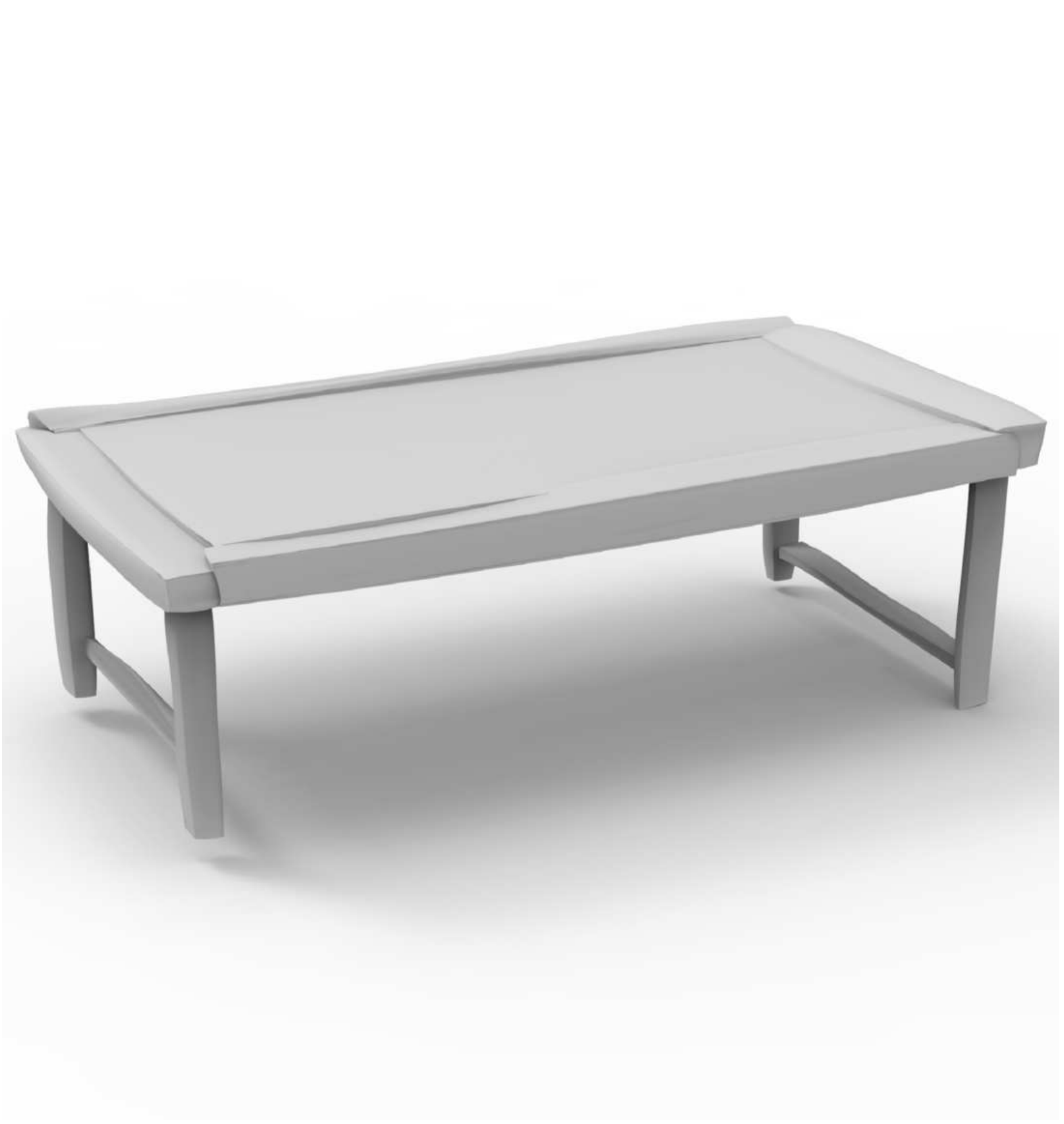}
    \includegraphics[width=0.23\linewidth]{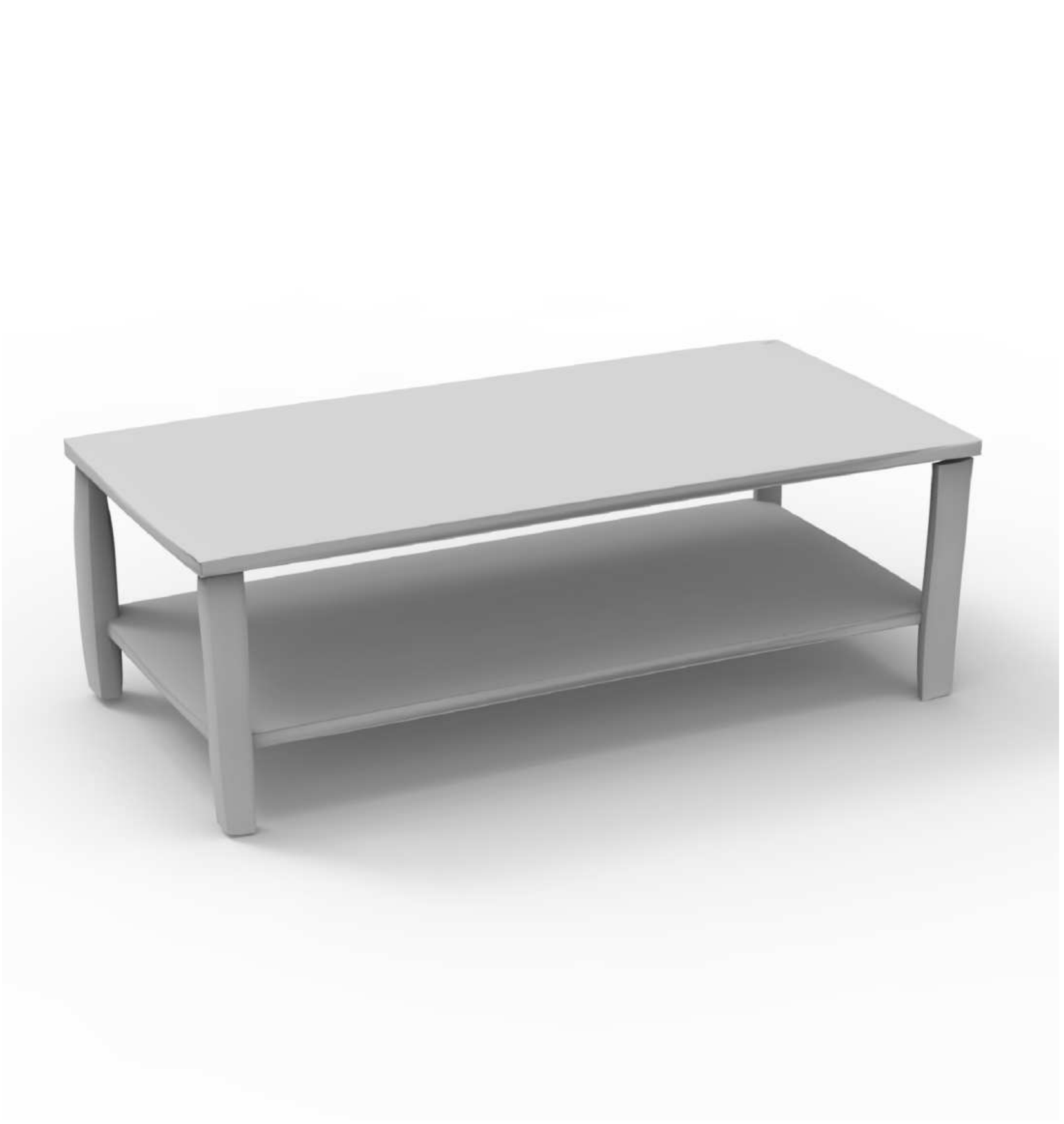}
    \includegraphics[width=0.23\linewidth]{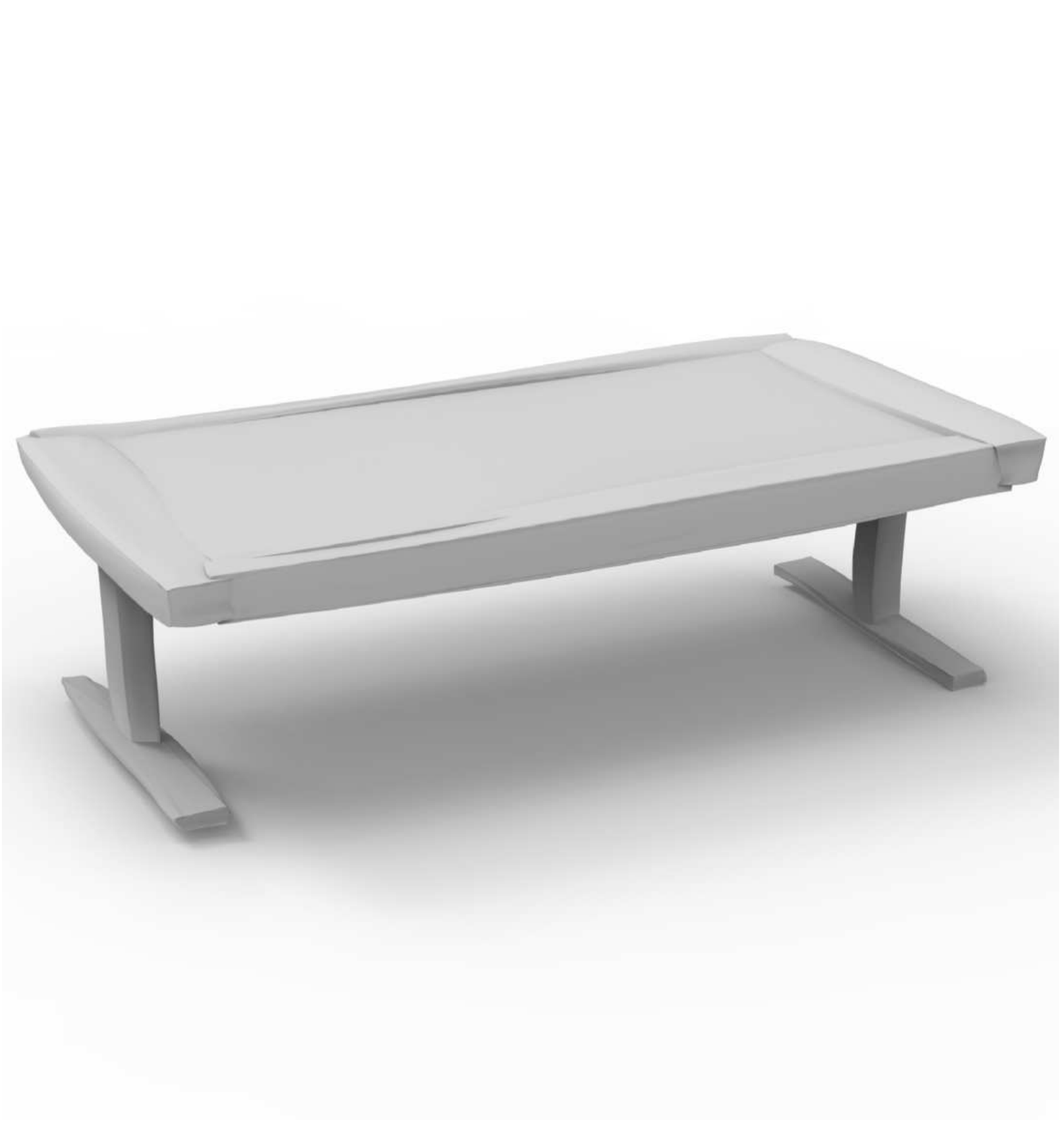}
    \includegraphics[width=0.23\linewidth]{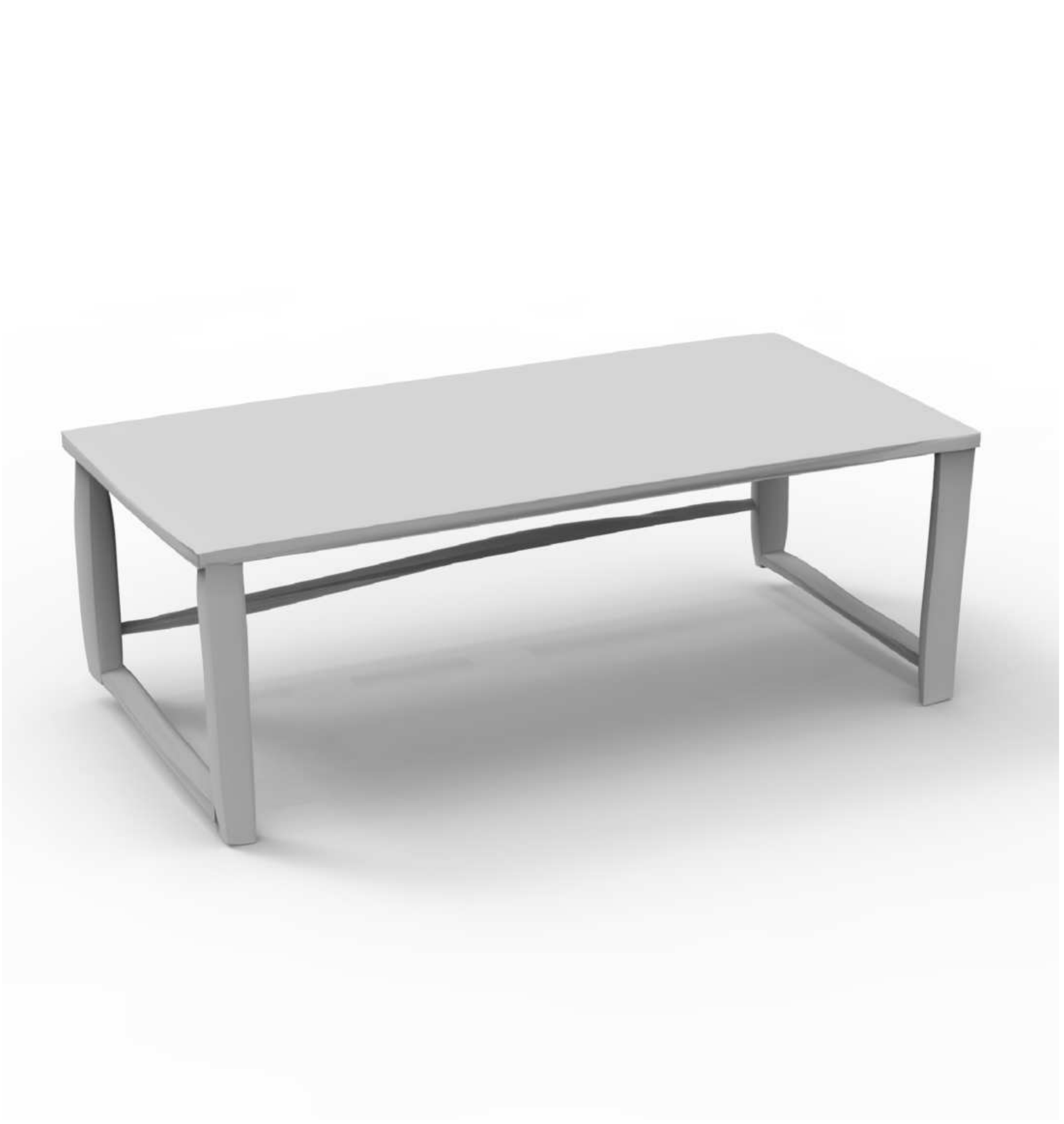}
    \\
    \includegraphics[width=0.23\linewidth]{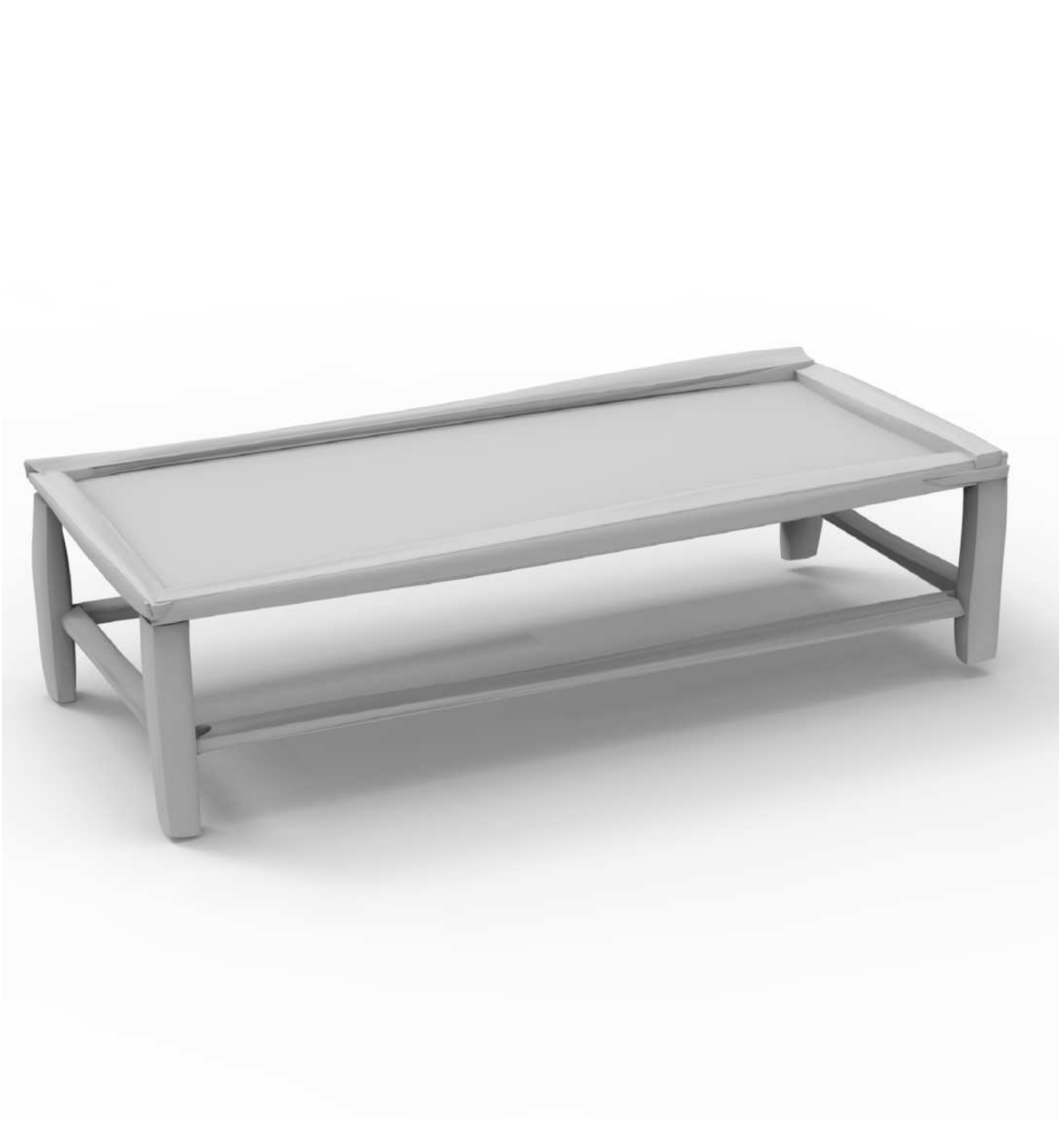}
    \includegraphics[width=0.23\linewidth]{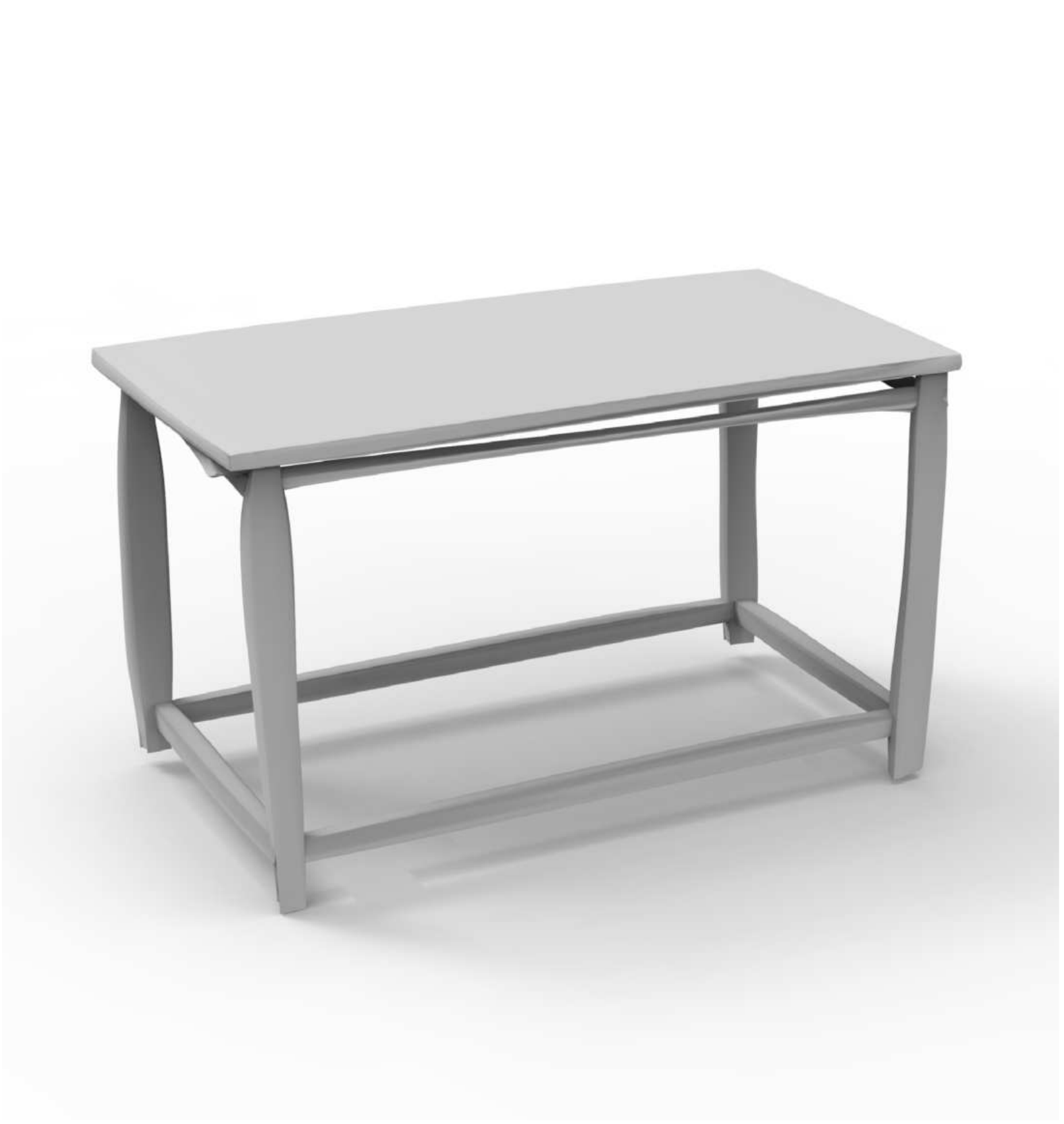}
    \includegraphics[width=0.23\linewidth]{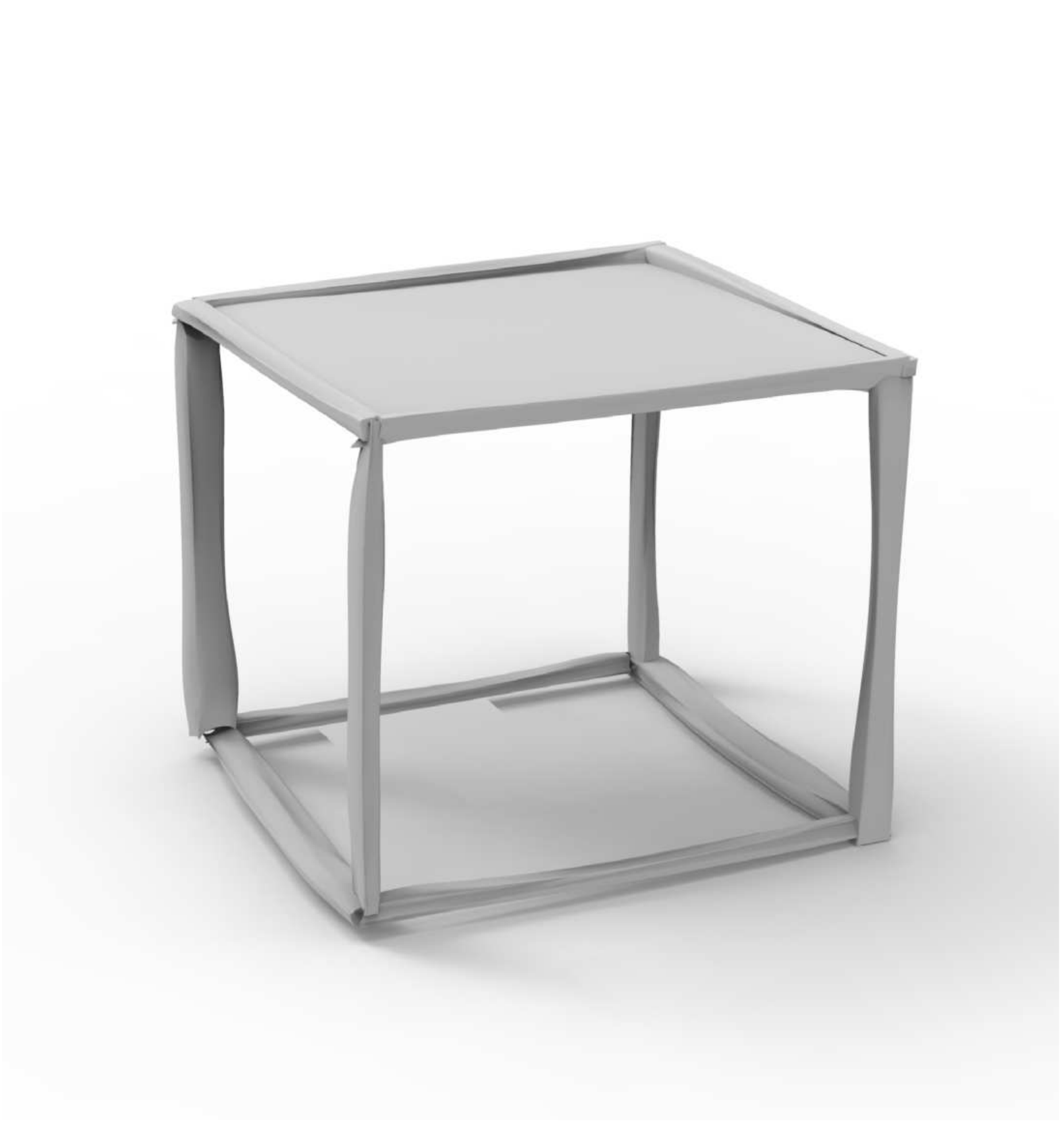}
    \includegraphics[width=0.23\linewidth]{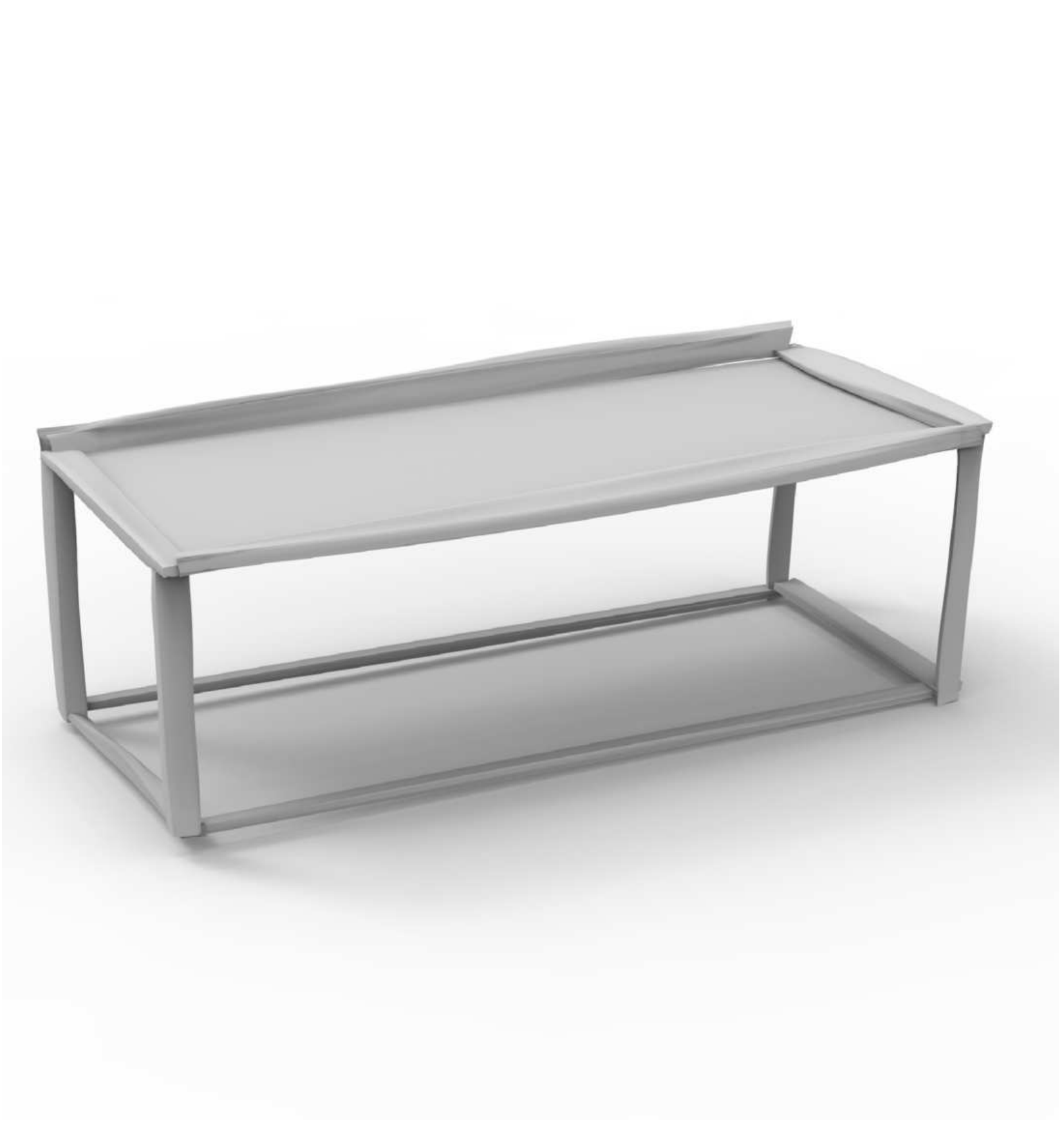}
\end{minipage}
}
    \caption{\yjr{More qualitative results for disentangled shape generation. Given an input shape (a), we extract the geometric or structural code using the DSG-Net encoder. Then, we fix one of them and randomly sample Gaussian noises in the other latent space to generate the new shapes (b). For the first row of (b), we keep the geometric code unchanged and randomly explore the structural latent space. And, for the second row, we keep the structural code unchanged and randomly sample over the geometric latent space.}}
    \label{fig:generation_pt2pc2}
\end{figure*}

% \balance

\end{document}